\documentclass[10pt]{report}
\usepackage{color}
\usepackage{epsfig}

\begin{document}

\newcommand{\s}{\mbox{$\mkern 7mu$}}

\newcommand{\ub}{\underline{u}}
\newcommand{\Cb}{\underline{C}}
\newcommand{\Lb}{\underline{L}}
\newcommand{\Lh}{\hat{L}}
\newcommand{\Lbh}{\hat{\Lb}}
\newcommand{\phib}{\underline{\phi}}
\newcommand{\Phib}{\underline{\Phi}}
\newcommand{\Db}{\underline{D}}
\newcommand{\Dh}{\hat{D}}
\newcommand{\Dbh}{\hat{\Db}}
\newcommand{\omb}{\underline{\omega}}
\newcommand{\omh}{\hat{\omega}}
\newcommand{\ombh}{\hat{\omb}}
\newcommand{\Pb}{\underline{P}}
\newcommand{\chib}{\underline{\chi}}
\newcommand{\chih}{\hat{\chi}}
\newcommand{\chibh}{\hat{\chib}}
\newcommand{\alb}{\underline{\alpha}}
\newcommand{\zeb}{\underline{\zeta}}
\newcommand{\beb}{\underline{\beta}}
\newcommand{\etb}{\underline{\eta}}
\newcommand{\Mb}{\underline{M}}
\newcommand{\oth}{\hat{\otimes}}

\newcommand{\xb}{\underline{x}}
\newcommand{\Xb}{\underline{X}}
\newcommand{\Fb}{\underline{F}}
\newcommand{\yb}{\underline{y}}

\newcommand{\xib}{\underline{\xi}}
\newcommand{\Vb}{\underline{V}}
\newcommand{\bb}{\underline{b}}
\newcommand{\Nb}{\underline{N}}

\newcommand{\fb}{\underline{f}}
\newcommand{\gb}{\underline{g}}
\newcommand{\rb}{\underline{r}}
\newcommand{\hb}{\underline{h}}
\newcommand{\ib}{\underline{i}}
\newcommand{\jb}{\underline{j}}
\renewcommand{\sb}{\underline{s}}
\newcommand{\psib}{\underline{\psi}}
\newcommand{\rhob}{\underline{\rho}}
\newcommand{\thetab}{\underline{\theta}}
\newcommand{\gammab}{\underline{\gamma}}
\newcommand{\ab}{\underline{a}}
\newcommand{\Bb}{\underline{B}}
\newcommand{\zb}{\underline{z}}
\newcommand{\mb}{\underline{m}}
\newcommand{\nb}{\underline{n}}
\newcommand{\nub}{\underline{\nu}}

\newcommand{\lb}{\underline{l}}

\newcommand{\eb}{\underline{e}}
\newcommand{\mub}{\underline{\mu}}

\newcommand{\Xib}{\underline{\Xi}}
\newcommand{\Thetab}{\underline{\Theta}}
\newcommand{\Lambdab}{\underline{\Lambda}}
\newcommand{\Kb}{\underline{K}}
\newcommand{\Ib}{\underline{I}}
\newcommand{\Wb}{\underline{W}}

\newcommand{\vphb}{\underline{\varphi}}

\newcommand{\scR}{{\cal R}\mkern-12mu /}
\newcommand{\scD}{{\cal D}\mkern-12mu /}
\newcommand{\sGa}{\Gamma\mkern-10mu /}

\newcommand{\sRic}{Ric\mkern-19mu /\,\,\,\,}

\newcommand{\sL}{{\cal L}\mkern-10mu /}

\newcommand{\sLh}{\hat{\sL}}

\newcommand{\sg}{g\mkern-9mu /}
\newcommand{\seps}{\epsilon\mkern-8mu /}
\newcommand{\sd}{d\mkern-10mu /}
\newcommand{\sR}{R\mkern-10mu /}
\newcommand{\snab}{\nabla\mkern-13mu /}
\newcommand{\sdiv}{\mbox{div}\mkern-19mu /\,\,\,\,}
\newcommand{\scurl}{\mbox{curl}\mkern-19mu /\,\,\,\,}
\newcommand{\slap}{\mbox{$\triangle  \mkern-13mu / \,$}}
\newcommand{\sGamma}{\Gamma\mkern-10mu /}

\newcommand{\somega}{\omega\mkern-10mu /}
\newcommand{\somb}{\omb\mkern-10mu /}

\newcommand{\spi}{\pi\mkern-10mu /}

\newcommand{\sJ}{J\mkern-10mu /}
\renewcommand{\sp}{p\mkern-9mu /}

\newcommand{\ih}{\hat{i}}

\newcommand{\cRb}{\underline{{\cal R}}}
\newcommand{\scRb}{\underline{\scR}}

\newcommand{\bfob}{\underline{{\bf O}}}

\newcommand{\og}{\overline{g}}
\newcommand{\oE}{\overline{E}}
\newcommand{\oD}{\overline{D}}
\newcommand{\oM}{\overline{M}}
\newcommand{\oB}{\overline{B}}
\newcommand{\oH}{\overline{H}}
\newcommand{\oU}{\overline{U}}
\newcommand{\oA}{\overline{A}}
\renewcommand{\oD}{\overline{D}}

\newcommand{\os}{\overline{s}}

\newcommand{\Th}{\hat{T}}
\newcommand{\Nh}{\hat{N}}

\newcommand{\sgh}{\hat{\sg}}

\newcommand{\tcL}{\tilde{{\cal L}}}

\newcommand{\up}[1]{\stackrel{\circ}{#1}}

\thispagestyle{empty}
\begin{center}
{\Huge \bf \sc The Formation}\\
\vspace{3mm}
{\Huge \bf \sc of Black Holes}\\
\vspace{3mm} {\Huge \bf \sc in General Relativity}
\end{center}
\bigskip
\begin{center}
{\Large \bf \sc  Demetrios Christodoulou}
\end{center}
\vspace{10mm}
\begin{center}
{\Large May 18, 2008}

\vspace{.10in}

\end{center}
\vfill

\section*{\huge{Contents}}

\vspace{10mm}

\noindent {\bf Prologue}

\vspace{5mm}

\noindent {\bf Chapter 1 : The Optical Structure Equations}\\
1.1 The basic geometric setup\\
1.2 The optical structure equations\\
1.3 The Bianchi equations\\
1.4 Canonical coordinate systems

\vspace{5mm}

\noindent {\bf Chapter 2 : The Characteristic Initial Data}\\
2.1 The characteristic initial data\\
2.2 Construction of the solution in an initial domain

\vspace{5mm}

\noindent {\bf Chapter 3 : $L^\infty$ Estimates for the Connection Coefficients}\\
3.1 Introduction\\
3.2 $L^\infty$ estimates for $\chi^\prime$\\
3.3 $L^\infty$ estimates for $\chib^\prime$\\
3.4 $L^\infty$ estimates for $\eta, \etb$\\
3.5 $L^\infty$ estimates for $\omega, \omb$\\
3.6 The smallness requirement on $\delta$

\vspace{5mm}

\noindent {\bf Chapter 4 : $L^4(S)$ Estimates for the 1st
Derivatives
of the Connection Coefficients}\\
4.1 Introduction\\
4.2 $L^4(S)$ estimates for $\snab\chi$\\
4.3 $L^4(S)$ estimates for $\snab\chib$\\
4.4 $L^4(S)$ estimates for $\snab\eta$, $\snab\etb$\\
4.5 $L^4(S)$ estimates for $\sd\omega$, $\sd\omb$\\
4.6 $L^4(S)$ estimates for $D\omega$, $\Db\omb$

\vspace{5mm}

\noindent {\bf Chapter 5 : The Uniformization Theorem}\\
5.1 Introduction. An $L^2(S)$ estimate for $K-\overline{K}$\\
5.2 Sobolev inequalities on $S$. The isoperimetric constant\\
5.3 The uniformization theorem\\
5.4 $L^p$ elliptic theory on $S$

\vspace{5mm}

\noindent {\bf Chapter 6 : $L^4(S)$ Estimates for the 2nd
Derivatives of the Connection Coefficients}\\
6.1 Introduction\\
6.2 $L^4(S)$ estimates for $\snab^{ \ 2}\chi$, $K-\overline{K}$\\
6.3 $L^4(S)$ estimates for $\snab^{ \ 2}\chib$\\
6.4 $L^4(S)$ estimates for $\snab^{ \ 2}\eta$, $\snab^{ \ 2}\etb$\\
6.5 $L^4(S)$ estimate for $\snab^{ \ 2}\omb$\\
6.6 $L^4(S)$ estimate for $\snab^{ \ 2}\omega$

\vspace{5mm}

\noindent {\bf Chapter 7 : $L^2$ Estimates for the 3rd Derivatives
of the Connection Coefficients}\\
7.1 Introduction\\
7.2 $L^2$ estimates for $\snab^{ \ 2}\eta$, $\snab^{ \ 2}\etb$\\
7.3 $L^2$ elliptic theory for generalized Hodge systems on $S$\\
7.4 $L^2$ estimates for $\snab^{ \ 3}\chi^\prime$, $\sd K$\\
7.5 $L^2$ estimates for $\snab^{ \ 3}\chib^\prime$\\
7.6 $L^2$ estimates for $\snab^{ \ 3}\eta$, $\snab^{ \ 3}\etb$\\
7.7 $L^2$ estimate for $\snab^{ \ 3}\omb$\\
7.8 $L^2$ estimates for $\snab^{ \ 2}\omega$ and $\snab^{ \ 3}\omega$\\
7.9 $L^2$ estimates for $\sd D\omega$, $\sd\Db\omb$, $D^2\omega$,
$\Db^2\omb$

\vspace{5mm}

\noindent {\bf Chapter 8 : The Multiplier Fields and the Commutation Fields}\\
8.1 Introduction\\
8.2 $L^\infty$ estimates for the deformation tensors of $L$, $K$ and $S$\\
8.3 Construction of the rotation vectorfields $O_i$\\
8.4 $L^\infty$ estimates for the $O_i$ and $\snab O_i$\\
8.5 $L^\infty$ estimates for the deformation tensors of the $O_i$

\vspace{5mm}

\noindent {\bf Chapter 9 : Estimates for the Derivatives of the Deformation Tensors of the Commutation Fields}\\
9.1 $L^4(S)$ estimates for the 1st derivatives of the deformation tensors of $L$, $S$\\
9.2 $L^4(S)$ estimates for the 1st derivatives of the deformation tensors of the $O_i$\\
9.3 $L^2$ estimates for the 2nd derivatives of the deformation tensors of $L$, $S$\\
9.4 $L^2$ estimates for the 2nd derivatives of the deformation
tensors of the $O_i$

\vspace{5mm}

\noindent {\bf Chapter 10 : The Sobolev Inequalities on the $C_u$ and $\Cb_{\ub}$}\\
10.1 Introduction\\
10.2 The Sobolev inequalties on the $C_u$\\
10.3 The Sobolev inequalities on the $\Cb_{\ub}$

\vspace{5mm}

\noindent {\bf Chapter 11 : The $S$-tangential Derivatives and the Rotational Lie Derivatives}\\
11.1 Introduction and preliminaries\\
11.2 The coercivity inequalities on the standard sphere\\
11.3 The coercivity inequalities on $S_{\ub,u}$

\vspace{5mm}

\noindent {\bf Chapter 12 : Weyl Fields and Currents. The Existence Theorem}\\
12.1 Weyl fields and Bianchi equations. Weyl currents\\
12.2 Null decompositions of Weyl fields and currents \\
12.3 The Bel-Robinson tensor. The energy-momentum density vectorfields\\
12.4 The divergence theorem in spacetime\\
12.5 The energies and fluxes. The quantity ${\cal P}_2$\\
12.6 The controlling quantity ${\cal Q}_2$. Bootstrap assumptions and the comparison lemma\\
12.7 Statement of the existence theorem. Outline of the continuity
argument

\vspace{5mm}

\noindent {\bf Chapter 13 : The Multiplier Error Estimates}\\
13.1 Preliminaries\\
13.2 The multiplier error estimates

\vspace{5mm}

\noindent {\bf Chapter 14 : The 1st Order Weyl Current Error Estimates}\\
14.1 Introduction\\
14.2 The error estimates arising from $J^1$\\
14.3 The error estimates arising from $J^2$\\
14.4 The error estimates arising from $J^3$

\vspace{5mm}

\noindent {\bf Chapter 15 : The 2nd Order Weyl Current Error Estimates}\\
15.1 The 2nd order estimates which are of the same form as the 1st
order estimates\\
15.2 The genuine 2nd order error estimates

\vspace{5mm}

\noindent {\bf Chapter 16 : The Energy-Flux Estimates. Completion of the Continuity Argument}\\
16.1 The energy-flux estimates\\
16.2 Higher order bounds\\
16.3 Completion of the continuity argument\\
16.4 Restatement of the existence theorem

\vspace{5mm}

\noindent {\bf Chapter 17 : Trapped Surface Formation}

\vspace{5mm}

\noindent {\bf Bibliography}

\pagebreak

\section*{\huge{Prologue}}

\vspace{10mm}

The story of the black hole begins with Schwarzschild's discovery [Sc] of the Schwarzschild solution in 1916, soon after Einstein's foundation
of the general theory of relativity [Ei1] and his final formulation of the field equations of gravitation [Ei2], the {\em Einstein equations}, in 1915. 
The Schwarzschild solution is a solution of the vacuum Einstein equations which is spherically symmetric and depends on a positive parameter $M$, the {\em mass}. With $r$ 
such that the area of the spheres, which are the orbits of the rotation group, is $4\pi r^2$, the solution in the coordinate system in which it was originally discovered had a singularity 
at $r=2M$. For this reason only the part which corresponds to $r>2M$ was originally thought to make sense. This part is static and represents the gravitational field 
outside a static spherically symmetric body with surface area corresponding to some $r_0>2M$. 

However, the understanding of Schwarzschild's solution gradually changed. First, in 1923 Birkoff [Bir] proved a theorem which shows that the Schwarzschild solution 
is the only spherically symmetric solution of the vacuum Einstein equations. One does not therefore need to assume that the solution is static. Thus, Schwarzschild's 
solution represents the gravitational field outside any spherically symmetric body, evolving in any manner whatever, for example undergoing 
gravitational collapse. 

Eddington [Ed], in 1924, made a coordinate change which transformed the Schwarzschild metric into a form which is not singular at $r=2M$, however he failed to take proper notice of this. Only in 1933, with Lema\^{i}tre's work [L], it was realized that the singularity at $r=2M$ is not a true singularity but rather a failure of the 
original coordinate system. Eddington's transformation was rediscovered by Finkelstein [Fi] in 1958, who realized that the hypersurface $r=2M$ is an {\em event horizon}, 
the boundary of the region of spacetime which is causally connected to infinity, and recognized the dynamic nature of the region $r<2M$. Now, Schwarzschild's 
solution is symmetric under time reversal, and one part of it, the one containing the {\em future event horizon}, the boundary of the region of spacetime which 
can send signals to infinity, is covered by one type of Eddington-Finkelstein coordinates, while the other part, the one containing the {\em past event horizon}, the boundary of the region of spacetime which can receive signals from infinity, is covered by the other type of Eddington-Finkelstein coordinates. 
Actually, only the first part is physically relevant, because only future event horizons can form dynamically, in gravitational 
collapse. Systems of coordinates that cover the complete analytic extension of the Schwarzschild solution had been provided earlier (in 1950) by Synge [Sy], and a single most 
covenient system that covers the complete analytic extension was discovered independently by Kruskal [Kr] and Szekeres [Sz] in 1960.

Meanwhile in 1939, Oppenheimer and Snyder had studied the gravitational collapse of a pressure-free fluid ball of uniform density, a uniform density ``ball of dust". 
Even though this is a highly idealized model problem, their work was very significant, being the first work on relativistic gravitational collapse. As mentioned above, the spacetime geometry in the vacuum region outside the ball is given by the 
Schwarzscild metric. Oppenheimer and Snyder analyzed the causal structure of the solution. They considered in particular an observer on the surface of the dust ball 
sending signals to a faraway stationary observer at regularly spaced intervals as judged by his own clock. They discovered that the spacing between the 
arrival times of these signals to the farway observer becomes progressively longer, tending to infinity as the radius $r_0$ corresponding to the surface of the 
ball approaches $2M$. This effect has since been called the {\em infinite redshift} effect. 
The observer on the sufrace of the dust ball may keep sending signals after $r_0$ has become less than $2M$, but these signals proceed  
to ever smaller values of $r$ until, within a finite afine parameter interval, they reach a true singularity at $r=0$. The observer on the surface of the ball reaches 
himself this singular state within a finite time interval as judged by his own clock. The concept of a future event horizon and hence of a region of spacetime 
bounded by this horizon from which no signals can be sent which reach arbitrarily large distances, was thus already implicit in Oppenheimer-Snyder work. 

The 1964 work of Penrose [P1] introduced the concept of {\em null infinity}, which made possible the precise general definition of a {\em future event horizon} 
as {\em the boundary of the causal past of future null infinity}. A turning point was reached in 1965 by the introduction by Penrose of the concept of 
a closed {\em trapped surface} and his proof of the first {\em singularity theorem}, or, more precisely, {\em incompleteness theorem} [P2]. Penrose 
defined a trapped surface as being a spacelike surface in spacetime, such that an infinitesimal virtual displacement of the surface along either family of future-directed 
null geodesic normals to the surface leads to a pointwise decrease of the area element. On the basis of this concept, Penrose proved the following theorem:

\vspace{2.5mm}

{\em A spacetime $(M,g)$ cannot be future null geodesically complete if:

1. \ \ \ $Ric(N,N)\geq 0$ for all null vectors $N$.

2. \ \ \ There is a non-compact Cauchy hypersurface $H$ in $M$.

and:

3. \ \ \ There is a closed trapped surface $S$ in $M$.}

\vspace{2.5mm}

Here $Ric$ is the Ricci curvature of $g$ and condition 1 is always satisfied by virtue of the Einstein equations and the physical positivity condition on the 
energy-momentum-stress tensor of matter. 

Once the notions of null infinity and of a closed trapped surface where introduced, it did not take long to show that a spacetime with a complete future null infinity 
which contains a closed trapped surface must contain a future event horizon, the interior of which contains the trapped surface (see [H-E], Proposition 9.2.1). 
For the ideas and methods which go into Penrose's theorem the reader may consult, besides the monograph by Hawking and Ellis just mentioned, the article 
of Penrose in [P3] as well as his monograph [P4]. Further singularity theorems, which also cover cosmological situations, where subsequently established by 
Hawking and Penrose (see [H-E]), but it is the original sigularity theorem quoted above which is of interest in the present context, as it corcerns gravitational collapse. We should also mention that the term {\em black hole} for the interior of the future event horizon was introduced by Wheeler in 1967 (see [Wh]). 

Now, the 1952 work of Choquet-Bruhat [Cho1] (see also [Cho2] and [Cho3]) had shown that any initial data set $(H, \og, k)$, where $H$ is a 3-dimensional manifold $\og$ is a Riemannian 
metric on $H$ and $k$ is a symmetric 2-covariant tensorfield on $H$ such that the pair $(\og,k)$ satisfies the so called ``constraint equations", has  
a {\em future development} $(M,g)$, namely a 4-dimensional manifold $M$ endowed with a Lorentzian metric $g$ satisfying the vacuum Einstein equations, such that 
$H$ is the past boundary of $M$, $\og$ and $k$ are the first and second fudamental forms of $H$ relative to $(M,g)$, and for each $p\in M$ each past directed causal curve initiating at $p$ terminates at a point of $H$. The constraint equations are the contracted Codazzi and twice contracted Gauss equations of the embedding of $H$ in $M$.
The subsequent 1969 work of Choquet-Bruhat and Geroch [C-G] then showed that each such an initial data set has 
a unique {\em maximal future development} $M^*$, namely a future development, in the above sense, which extends every other future development of the same initial data set. 
Geroch [Ge] subsequently showed that for any future development $(M,g)$, $M$ is diffeomorphic to $[0,\infty)\times H$. Moreover, the above theorems extend to the case where instead of 
vacuum we have suitable matter, such as a perfect fluid, or an electromagnetic field. 
In the light of the theorem of Choquet-Bruhat and Geroch condition 2 in Penrose's theorem may be replaced by the statement that $(M,g)$ is the maximal future 
development of initial data on a complete non-compact spacelike hypersurface. 

In 1990 Rendall [R] solved in a very satisfactory manner the local {\em characteristic initial value problem} for the vacuum Einstein equations 
(earlier work had been done by Choquet-Bruhat [Cho4] and by M\"{u}ller zum Hagen and Seifert [M-S]). In this 
case we have, in the role of $H$, either two null hypersurfaces $C$ and $\Cb$ intersecting in a spacelike surface $S$, $S$ being the past boundary of both  $C$ and $\Cb$, or a future null geodesic cone $C_o$ of a point $o$. The initial data on $C$ and $\Cb$ are the conformal intrinsic geometry of these null hypersurfaces, together with 
the full intrinsic geometry of $S$, the initial rate of change of the area element of $S$ under displacement along $C$ and $\Cb$, and a certain 1-form on $S$ (the  torsion). The initial data on $C_o$ are the conformal intrinsic geometry of $C_o$ 
and certain regularity conditions at $o$. In contrast to the case where the initial data are given on a spacelike hypersurface, there are no constraints, and the 
initial data can be freely specified. The theorem of Rendall then shows that any such characteristic initial data has a future development $(M,g)$, bounded in the past by a neighborhood of $S$ in $C\bigcup\Cb$ and of $o$ in $C_o$ respectively. The theorem of Choquet-Bruhat and Geroch, which applies to future developments, 
then shows that there is a unique maximal future development $(M^*,g)$ corresponding to the given characteristic initial data. 

Now the proof of the theorem of Penrose is by showing that if $M$ where complete, the boundary $\partial J^+(S)$ of the causal future $J^+(S)$ of the closed trapped surface $S$ would be compact. 
The integral curves of any timelike vectorfield on $M$ would define a continuous mapping of $\partial J^+(S)$ into $H$, $M$ being a development of $H$, and 
this mapping would have to be a homeomorphism onto its image, $\partial J^+(S)$ being compact. This leads to a contradiction with the assumption that $H$ is 
non-compact. We see that the proof makes no use of the strictly spacelike nature of $H$ other than through the assumption that $M$ is a future development of $H$. 
We may therefore replace $H$ by a complete future null geodesic cone and restate the theorem as follows. 
Here vacuum or suitable matter is assumed. We do not state the 
first condition of Penrose's because as we already mentioned, it is automatically satisfied by virtue of the physical positivity condition that the 
energy-momentum-stress tensor  
of matter satisfies.  

\vspace{2.5mm}

{\em Let us be given regular charactesistic initial data on a complete null geodesic cone $C_o$ of a point $o$. 
Let $(M^*,g)$ be the maximal future development of the data on $C_o$. Suppose that $M^*$ contains a closed trapped surface. Then $(M^*,g)$ is future null geodesically 
incomplete.}

\vspace{2.5mm}

An important remark at this point is that it is not {\em a priori} obvious that closed trapped surfaces are {\em evolutionary}. That is, it is not obvious whether closed trapped surfaces can form in evolution starting from 
initial conditions in which no such surfaces are present. What is more important, the physically interesting problem is the problem where the initial conditions are of arbitrarily {\em low compactness}, that is, arbitrarily far from already containing closed trapped surfaces, and we are asked to follow the long time evolution and show that, under suitable circumstances, closed trapped 
surfaces eventually form. Only an analysis of the dynamics of graviational collapse can achieve this aim.  

Returning to our review of the historical development of the black hole concept, a very significant development took place in 1963, shortly before the work of 
Penrose. This was the discovery by Kerr [Ke] of a two parameter family of axially symmetric solutions of the vacuum Einstein equations, with an event horizon, 
the exterior of which is a regular asymptotically flat region. The two parameters are the {\em mass} $M$, which is positive, and the {\em angular momentum} $L$ about the axis of symmetry, which is subject to the restriction $|L|\leq M^2$. Kerr's solution reduces in the spatial case of vanishing angular momentum to Schwarzschild's 
solution. The Kerr solution possesses an additional Killing field, besides the generator of rotations about the axis, however this additional Killing field, 
in contrast to the case of the Schwarzschild solution, is timelike not on the entire exterior of the horizon, but only in the exterior of a non-spacelike 
hypersurface containing the horizon. So only in this exterior region is the solution stationary in a strict sense. At every point of the region between the two hypersurfaces, called {\em ergosphere}, the additional Killing field is spacelike, 
but the plane which is the linear span of the two vectors at this point is timelike. On the horizon itself the plane becomes null and tangent to the horizon, and the null line generating this null plane defines the {\em angular velocity} of the horizon, a constant associated to the horizon. Kerr's solution is symmetric under time reversal if also the 
sign of the angular momentum is reversed, hence it possesses, besides the future even horizon, also an unphysical past event horizon, just like the the Schwarzschild 
solution.

The fascinating properties of the Kerr solution where revealed in the decade following its discovery. In particular Boyer and Lindquist [B-L] introduced a more 
convenient coordinate system and obtained the maximal analytic extension. One of these fascinating properties which concerns us here, is that the hypersurfaces 
of constant Boyer-Lindquist coordinate $t$ are complete asymptotically flat maximal spacelike hypersurfaces, their maximal future development contains closed trapped 
surfaces and, in accordance with Penrose's theorem, is incomplete. Nevertheless the future boundary of the maximal development is {\em nowhere singular}, the solution 
extending as an analytic solution across this boundary. This future boundary is a regular null hypersurface, a {\em Cauchy horizon},  illustrating the fact that incompleteness of the maximal future development does not imply a singular future boundary.

Returning again to the question of whether closed trapped surfaces are evolutionary, one may at first hand say that the question was already settled in the affirmative by the Oppenheimer-Snyder analysis. This is because hyperbolic systems of partial differential equations, such as the Einstein-Euler equations 
describing a perfect fluid in general relativity, possess the property of continuous dependence of the solution on the initial conditions. This holds at a given 
non-singular solution, for a given finite time interval. Thus, since the initial condition of a homogeneous dust ball leads to a trapped sphere within 
a finite time interval, initial conditions which are sufficiently close to this special initial condition will also lead to the formation of closed trapped surfaces within the same time interval, the condition for a closed spacelike surface to be trapped being an open condition. However, as we remarked above, the case that one is really interested 
in is that for which the initial homogeneous dust ball is of low compactness, far from already containing trapped spheres, and it is only by contracting for sufficiently long time that a trapped sphere eventually forms. In this case the closeness condition of the continuous dependence theorem may require the initial
conditions to be so unreasonably close to those of a homogeneous dust ball that the result is devoid of physical significance. 

With the above remarks in mind the author turned to the study of the gravitational collapse of an inhomogeneous dust ball [Chr1]. In this case, the initial state is still spherically 
symmetric, but the density is a function of the distance from the center of the ball. The corresponding spherically symmetric solution had already been obtained 
in closed form by Tolman in 1934 [T], in comoving coordinates, but its causal structure had not been investigated. This required integrating the equations for the 
radial null geodesics. A very different picture from the one  found by Oppenheimer and Snyder emerged from this study. The initial density being assumed a decreasing 
function of the distance from the center, so that the central density is higher than the mean density, it was found that as long as the collapse proceeds from 
an initial state of low compactness, the central density becomes infinite before a black hole has a chance to form, thus invalidating the neglect of pressure 
and casting doubt on the predictions of the model from this point on, in particular on the prediction that a black hole eventually forms. 

At this point the author turned to the spherically symmetric scalar field model [Chr2]. This is the next simplest material model after the dust model. The energy-momentum-stress tensor of matter is in this case that corresponding to a scalar field $\phi$:
\begin{equation}
T_{\mu\nu}=\partial_\mu\phi\partial_\nu\phi+\frac{1}{2}\sigma g_{\mu\nu}, \ \ \ \sigma=-(g^{-1})^{\mu\nu}\partial_\mu\phi\partial_\nu\phi
\label{p.1}
\end{equation}
The integrability condition for Einstein's equations, namely 
that $T_{\mu\nu}$ is divergence-free, is then equivalent to the wave equation for $\phi$ relative to the metric $g$. The problem had been 
given to the author by his teacher, John Archibald Wheeler, in 1968 (see [Chr3]), as a model problem through which insight into the dynamics of gravitational 
collapse would be gained. In the case of the dust model, there is no force opposing the gravitational attraction , so there is no alternative to collapse. 
This is not the case for the scalar field model and indeed in [Chr2] it was shown that if the initial data are suitably small we obtain a complete regular solution 
dispersing to infinity in the infinite future. So, for the scalar field model there is a {\em threshold} for gravitational collapse. In this paper and the papers on the scalar field that followed, the initial data where given on a complete future null geodesic cone $C_o$ extending to infinity. The initial data on $C_o$ consist of the function $\alpha_0=\left.\partial(r\phi)/\partial s\right|_{C_o}$,  $s$ being the affine parameter along the generators of $C_o$.

The next paper [Chr4] on the scalar field problem addressed the general case, when the initial data where no longer restricted by a smallness condition. The aim of this 
work was to prove the existence of a solution with a complete {\em domain of outer communications}, that is, a development possessing a complete future null 
infinity, the domain of outer communications being defined as the causal past of future null infinity. This was tantamount to proving the {\em weak cosmic censorship 
conjecture} of Penrose [P5] (called ``asymptotic future predictability" in [H-E]) in the context of the spherically symmetric scalar field model. The aim was not 
reached in this paper. What was established instead was the existence, for all regular asymptotically flat initial data, of a {\em generalized solution} 
corresponding to a complete domain of outer communications. A generalized solution had enough regularity to permit the study of the  asymptotic behavior in 
the domain of outer communications in the next paper, however no uniqueness could be claimed for these generalized solutions, so the conjecture of Penrose was left 
open. 

In [Chr5] it was shown that when the final Bondi mass, that is, the infimum of the Bondi mass at future null infinity, is different from zero, a black hole forms of mass equal to the final Bondi mass, surrounded by vacuum. The rate of growth of the redshift of light seen by faraway observers was determined and the asymptotic wave behavior at future null infinity and along the event horizon, was analyzed. However, the question of whether there exist initial conditions which lead to a 
non-zero final Bondi mass was not addressed in this paper. 

The next paper [Chr6] was a turning point in the study of the spherically symmetric scalar field problem. Because of the fact that it has provided a stepping stone 
for the present monograph, I quote its main theorem. Here $C_o$ denotes the initial future null geodesic cone. 

\vspace{2.5mm}

{\em Consider on $C_o$ an annular region bounded by two spheres $S_{1,0}$ and $S_{2,0}$ with $S_{2,0}$ in the exterior of $S_{1,0}$. Let $\delta_0$ and $\eta_0$ be the 
dimensionless size and the dimensionless mass content of the region, defined by
$$\delta_0=\frac{r_{2,0}}{r_{1,0}}-1 \ \ \ \ \ \eta_0=\frac{2(m_{2,0}-m_{1,0})}{r_{2,0}}$$
$r_{1,0},r_{2,0}$ and $m_{1,0},m_{2,0}$ being the area radii and mass contents of $S_{1,0},S_{2,0}$ respectively. Let $\Cb_1$ and $\Cb_2$ be incoming 
null hypersurfaces through $S_{1,0}$ and $S_{2,0}$ and consider the spheres $S_1$ and $S_2$ at which $\Cb_1$ and $\Cb_2$ intersect future null geodesic cones 
$C$ with vertices on the central timelike geodesic $\Gamma_0$. There are positive constants $c_0$ and $c_1$ such that if $\delta_0\leq c_0$ and 
$$\eta_0>c_1\delta_0\log\left(\frac{1}{\delta_0}\right)$$
then $S_2$ becomes trapped before $S_1$ reduces to a point on $\Gamma_0$. There is a future null geodesic cone $C^*$ with vertex on $\Gamma_0$ such that 
$S^*_2$ is a maximal sphere in $C^*$ while $r^*_1>0$. }

It was further shown that the region of trapped spheres, the {\em trapped region}, terminates at a strictly spacelike singular boundary, and contains spheres 
whose mass content is bounded from below by a positive constant depending only on $r_{1,0},r_{2,0}$, a fact which implies that the final Bondi mass is positive, 
thus connecting with the previous work. 

An important remark concerning the proof of the above theorem needs to be made here. The proof does not consider at all the region interior to the incoming null 
hypersurface $\Cb_1$. However, the implicit assumption is made that no singularities form on $\Gamma_0$ up to the vertex of $C^*$.  If a smallness condition is imposed on the restriction of the initial data to the interior of $S_{1,0}$, the argument of [Chr2] shows that this assumption indeed holds. 
Also, by virtue of the way in which the theorem was latter applied in [Chr9], the assumption in question was a priori known to hold. 

Since the spherical dust model had been disqualified by [Chr1] as establishing the dynamical formation of trapped spheres, the work [Chr6] was the first to establish 
the dynamical formation of closed trapped spheres in gravitational collapse, although off course severely limited by the restriction to spherical symmetry and 
by the fact that it concerned an idealized matter model. 

Solutions with initial data of bounded variation were considered in [Chr7] and a sharp sufficent condition on the initial data was found for the avoidance of 
singularities, namely that the total variation be sufficiently small, greatly improving the result of [Ch2]. Moreover, a sharp extension criterion for solutions 
was established, namely that if the ratio of the mass content to the radius of spheres tends to zero as we approach a point on $\Gamma_0$ from its causal past, then 
the solution extends as a regular solution to include a full neighborhood of the point. The structure of solutions of bounded variation was studied and it was shown that at each point of $\Gamma_0$ the solutions are locally scale invariant. Finally, the behavior of the solutions at the singular boundary was analyzed. 

In [Chr8] the author constructed examples of solutions corresponding to regular asymptotically flat initial data which develop singularities which are not preceeded 
by a trapped region but have future null geodesic cones expanding to infinity. It was thus established for the first time that {\em naked singularities} do, in fact,  occur in the graviational collapse of a scalar field. Also, other examples where constructed which contain {\em singular future null geodesic cones} which have 
collapsed to lines and again are not preceeded by trapped regions. 

The work on the spherically symmetric scalar field model culminated in [Chr9].  Taking the space of initial data to be the space ${\cal A}$ of absolutely continous 
functions on the non-negative real line, the theorem proved in [Chr9] was the following. 

\vspace{2.5mm}

{\em Let us denote by ${\cal R}$ the subset of ${\cal A}$ consisting of those initial data which lead to a complete maximal future development, and by ${\cal S}$ its complement in ${\cal A}$. Let also ${\cal G}\subset{\cal S}$ be the subset 
consisting of those initial data which lead to a maximal development possessing a complete future null infinity and a strictly spacelike singular future boundary. 
Then ${\cal E}={\cal S}\setminus{\cal G}$ has the following property. For each initial data $\alpha_0\in {\cal E}$ there is a function $f\in {\cal A}$, depending on 
$\alpha_0$, such that the line ${\cal L}_{\alpha_0}=\{\alpha_0+cf \ : \ c\in\Re\}$ in ${\cal A}$ is contained in ${\cal G}$, except for $\alpha_0$ itself. Moreover, the lines ${\cal L}_{\alpha_{0,1}}$, ${\cal L}_{\alpha_{0,2}}$ corresponding to distinct $\alpha_{0,1},\alpha_{0,2}\in {\cal E}$ do not intersect.}

\vspace{2.5mm}

The exceptional set ${\cal E}$ being, according this theorem, of codimension at least 1, 
the theorem established, within the spherically symmetric scalar field model, the validity not only of the weak cosmic censorship conjecture of Penrose, but also 
of his {\em strong cosmic censorship conjecture}, formulated in [P6]. This states, roughly speaking, that generic asymptotically flat initial data have a maximal development which is either complete or terminates in a totally singular future boundary. The general notion of causal boundary of a spacetime manifold  was 
defined in [G-K-P]. The relationship between the two cosmic censorship conjectures is discussed in [Chr10]. In the case of the spherically symmetric scalar field model there is no system of local coordinates in which the metric extends as a Lorentzian 
metric through any point of the singular future boundary. 

The proof of the above theorem is along the following lines. It is first shown that if $\Gamma_0$ is 
complete, the maximal future development is also complete. Thus one can assume that $\Gamma_0$ has a singular end point $e$. We then consider $\Cb_e$, the boundary 
of the causal past of $e$. This intersects the initial future null geodesic cone $C_o$ in a sphere $S_{0,e}$. Given then any sphere $S_{0,1}$ exterior to $S_{0,e}$ on $C_o$, but as close as we wish to $S_{0,e}$, we consider the incoming null hypersurface $\Cb_1$ through $S_{0,1}$. Allowing a suitable modification of the initial data as in the statement of the theorem, with $f$ a function vanishing in the interior of $S_{0,e}$ on $C_o$, it is then shown that there exists a point $p_0$ on $\Gamma_0$, 
earlier than $e$, such that the annular region on the future null geodesic cone $C_{p_0}$, with vertex at $p_0$, bounded by the intersections with $\Cb_e$ and $\Cb_1$ 
satisfies the hypotheses of the theorem of [Chr6]. It is in this part of the proof that the singular nature of the point $e$ is used. Application of the theorem 
of [Chr6] then shows that if we consider future null geodesic cones $C_p$ with vertices $p$ on the segment 
of $\Gamma_0$ between $p_0$ and $e$, and the corresponding intersections with $\Cb_e$ and $\Cb_1$, then for some $p^*$ in this segment earlier than $e$, 
$C_{p^*}\bigcap\Cb_1$ is a maximal sphere in $C_{p^*}$, and the part of $\Cb_1$ to the future of this sphere lies in a trapped region. 
We see therefore the essential role played by the formation of trapped spheres theorem of [Chr6] in the proof of the cosmic censorship conjectures in [Chr9] 
in the framework of the spherically symmetric scalar field model. 

A model, closely related to the scalar field model but with surprizing new features, was studied by Dafermos in [D1], [D2]. In this model we have in addition 
to the scalar field an electromagnetic field. The two fields are only indirectly coupled, through their interaction with the gravitational field, the 
energy-momentum stress tensor of matter being the sum of \ref{p.1} with the Maxwell energy-momentum-stress tensor for the electromagnetic field. 
By the imposition of spherical symmetry, the electromagnetic field is simply the Coulomb field corresponding to a constant charge $Q$. This is non-vanishing 
by virtue of the fact that the topology of the manifold is $\Re^2\times S^2$, like the manifold of the Schwarzschild solution, so there are spheres which are not 
homologous to zero. Dafermos showed that in this case part of the boundary of the maximal development is a Cauchy horizon, through which the metric 
can be continued in a $C^0$ manner, but at which, generically, the mass function blows up. As a consequence, generically, there is no local coordinate system in any neighborhood 
of any point on the Cauchy horizon in which the connection coefficients (Christoffel symbols) are square integrable. This means that the solution ceases to make 
sense even as a weak solution of the Einstein-Maxwell-scalar field equations if we attempt to include the boundary. The work of Dafermos illustrates how much 
care is needed in formulating the strong cosmic censorship conjecture. In particular the formulation given in [Chr10] according to which $C^0$ extensions 
through the boundary of the maximal development are generically excluded, turned out to be incorrect. Only if the condition is added that there be no extension 
{\em as a solution, even in a weak sense}, to include any part of the boundary, is the counterexample avoided. 

Before the work on the scalar field model was completed, the author introduced and studied a model which was designed to capture some of the features of 
actual stellar gravitational collapse while capitalizing to a maximum extent on the knowledge gained in the study of the dust and scalar field models. 
This was the two-phase model, introduced in [Chr11] and studied further in [Chr12] and [Chr13]. Let us recall here that 
a perfect fluid model is in general defined by specifying a function $e(n,s)$, the {\em energy per particle} as a function of $n$, the {\em number of particles per unit volume}, and $s$, the {\em entropy per particle}. This is called the {\em equation of state}. Then the {\em mass-energy density} $\rho$, the {\em pressure} $p$ and the {\em temperature} $\theta$ are given by:
\begin{equation}
\rho=ne, \ \ \ p=n^2\frac{\partial e}{\partial n}, \ \ \theta=\frac{\partial e}{\partial s}
\label{p.2}
\end{equation} 
The mechanics of 
a perfect fluid are governed by the differential conservation laws
\begin{equation}
\nabla_\nu T^{\mu\nu}=0, \ \ \ \ \ \nabla_\mu I^\mu=0
\label{p.3}
\end{equation}
where $T^{\mu\nu}$ is the energy-momentum-stress tensor
\begin{equation}
T^{\mu\nu}=\rho u^\mu u^\nu+p((g^{-1})^{\mu\nu}+u^\mu u^\nu),
\label{p.4}
\end{equation}
$u^\mu$ being the {\em fluid velocity}, and
\begin{equation}
I^\mu=nu^\mu
\label{p.5}
\end{equation}
is the {\em particle current}. In the case of the two-phase model, $p$ is a function 
of $\rho$ alone. For such fluids, called {\em barotropic}, $p$ and $\rho$ are functions of the single variable
\begin{equation}
\mu=nm(s)
\label{p.6}
\end{equation}
where $m(s)$ a positive increasing function of $s$. In the two-phase model, 
if $\rho$ is less than a critical value, which by proper choice of units we may set equal to 1, the matter is 
as soft as possible, the sound speed being equal to 0, while if $\rho$ is greater than 1, the matter is as hard as possible, the sound speed being equal to 1, 
that is, to the speed of light in vacuum. Let us recall here that the {\em sound speed} $\eta$ is in general given by:
\begin{equation}
\eta^2=\left(\frac{dp}{d\rho}\right)_s
\label{p.7}
\end{equation}
The pressure in the two-phase model is then given by:
\begin{equation}
p=\left\{\begin{array}{lll}
0&:&\mbox{if $\rho\leq 1$}\\
\rho-1&:&\mbox{if $\rho>1$}
\end{array}\right.
\label{p.8}
\end{equation}
The condition of spherical symmetry being imposed, the flow is irrotational. 
The soft phase of the two-phase model coincides with the dust model while the hard phase coincides with the scalar field 
model with the restriction that $-(g^{-1})^{\mu\nu}\partial_\nu\phi$ be a future-directed timelike vectorfield. With 
\begin{equation}
\sigma=-(g^{-1})^{\mu\nu}\partial_\mu\phi\partial_\nu\phi,
\label{p.9}
\end{equation}
the density of mass-energy $\rho$ and the fluid velocity $u^\mu$ are given by:
\begin{equation}
\rho=\frac{1}{2}(\sigma+1), \ \ \ \ \ u^\mu=-\frac{(g^{-1})^{\mu\nu}\partial_\nu\phi}{\sqrt{\sigma}}
\label{p.10}
\end{equation}
The energy-momentum-stress tensor in the hard phase is:
\begin{equation}
T_{\mu\nu}=\partial_\mu\phi\partial_\nu\phi+\frac{1}{2}(\sigma-1)g_{\mu\nu}
\label{p.11}
\end{equation}
so it differs from the standard one for a scalar field \ref{p.1}  by the term $-(1/2)g_{\mu\nu}$, the divergence-free condition on $T_{\mu\nu}$ being again 
equivalent to the wave equation for $\phi$ in the metric $g$. 

Each of the two phases is by itself incomplete, the soft phase being limited by the condition $\rho\leq 1$ and the hard phase being limited by the condition 
$\sigma\geq 1$. The soft phase turns in contraction into the hard phase, while the hard phase turns upon expansion into the soft phase. Only the two phases taken 
together constitute a complete model. The hypersurface which forms the interface between the two phases has both spacelike and timelike components. Across a 
spacelike component, the thermodynamic variables $n,s$ or $\rho,p$ and the fluid velocity $u^\mu$ are continuous, the final values of one phase providing the 
initial values for the next phase. However, across a timelike component the thermodynamic variables and the fluid velocity suffer discontinuities, determined by 
the integral form of the conservation laws. These are of an irreversible character, each point of a timelike component which is crossed by a flow line 
being a point of increase of the entropy. A timelike component of the phase boundary is therefore a {\em shock}, and the development of these shocks is a free boundary problem, which was studied in [Chr12] and [Chr13]. With initial condition an inhomogeneous dust ball at zero entropy these papers showed that the core of the ball 
turns continuously into the hard phase, however at a certain sphere a shock forms which propagates outwards absorbing the exterior part of the original dust ball. 
Behind this shock we have the hard phase at positive entropy, but the analysis was not carried further to investigate under what initial conditions a black hole 
will eventually form. We should also mention here that the two-phase model admits a one parameter family of static solutions, balls of the hard phase, surrounded 
by vacuum.

The goal of the effort in the field of relativistic gravitational collapse is the study of the formation of black holes and singularities for general 
asymptotically flat initial conditions, that is, when no symmetry conditions are imposed. 

In this connection, an interesting theorem was established by Schoen and Yau [S-Y1], as an outgrowth of their proof of the general case of the positive mass theorem 
[S-Y2] (their earlier work [S-Y1] covered the case of a maximal spacelike hypersurface of vanishing linear momentum; a different proof of the general theorem was subsequently given by Witten [Wi]). In [S-Y1] it is shown that if the energy density minus the magnitude of the momentum density of matter on a spacelike hypersurface is everywhere bounded from below by a positive constant $b$ in a region which is large enough in a suitable sense which roughly corresponds to linear dimensions of at least $b^{-1/2}$, then the spacelike hypersurface must 
contain a closed trapped surface diffeomorphic to $S^2$. Although this work does not address the problem of evolution, the constraint equations alone entering the proof, it is nevertheless 
relevant for the problem of evolution, in so far as it reduces the problem of the dymamical formation of a closed trapped surface to the problem of showing that,  under suitable circumstances, the required material energy concentration eventually occurs. 

As far as the problem of evolution itself, let us first discuss the case where the material model is a perfect 
fluid. Then, as we have seen in the spherically symmetric case, before closed trapped surfaces form, shock waves already form. Now, the general problem of shock formation in a relativistic fluid, in the physical case of 3 spatial dimensions has recently been studied by the author in the monograph [Chr14]. This work is in the framework of special relativity. We should remark here that the only 
previous result in relation to shock formation in 3 spatial dimensions was the result of Sideris [Si] which considers the non-relativistic problem of a classical ideal gas with adiabatic index $\gamma>1$. Moreover, in that work it is only shown that the solutions cannot remain $C^1$ for all time, no information being given as to the nature 
of the breakdown.  
The theorems proved in the monograph [Chr14] give, on the other hand, a detailed picture of shock formation. In particular a detailed description is 
given of the geometry of the boundary of the maximal development of the initial data and of the behavior of the solution at this boundary. The notion of maximal development 
in this context is not that relative to the background Minkowski metric $g_{\mu\nu}$, but rather the one relative to the {\em acoustical metric}
\begin{equation}
h_{\mu\nu}=g_{\mu\nu}+(1-\eta^2)u_\mu u_\nu, \ \ \ u_\mu=g_{\mu\nu}u^\nu
\label{p.12}
\end{equation}
a Lorenzian metric, the null cones of which are the sound cones. It is not appropriate to give here a complete summary of the results of [Chr14]. Instead, the following short 
discussion should suffice to give the reader a feeling for the present status of shock wave theory in the physical case of 3 spatial dimensions. In [Chr14] it 
is shown that 
the boundary of the maximal development in the above ``acoustical" sense consists of a regular part and a singular part. Each component of the regular part $\Cb$ is an incoming characteristic (relative to $h$) hypersurface which has a singular past boundary. The singular part of the boundary is the locus of points 
where the density of foliations by outgoing characteristic (relative to $h$) hypersurfaces blows up. It is the union $\partial_{-}B\bigcup B$, where each component of 
$\partial_{-}B$ is a smooth embedded surface in Minkowski spacetime, the tangent plane to which at each point is contained in the exterior of the sound cone at that point. On the other hand, each component of $B$ is a smooth embedded hypersurface in Minkowski spacetime, the tangent hyperplane to which at each point is contained in the exterior of the sound cone at that point, with the exception of a single generator of the sound cone, which lies on on the hyperplane itself. The past boundary of a component of $B$ is the corresponding component of $\partial_{-}B$. The latter is at the same time the past boundary of a component of $\Cb$. This is the surface 
where a shock begins to form. Now the maximal development in the acoustical sense, or ``maximal classical solution", is the physical solution of the problem up to $\Cb\bigcup\partial_{-}B$, but not 
up to $B$. In the last part of [Chr14] the problem of the physical continuation of the solution is set up as the {\em shock development problem}. This a free 
boundary problem 
associated to each component of $\partial_{-}B$. In this problem one is required to construct a hypersurface of discontinuity $K$, the shock, lying in the past of the corresponding componentof $B$ but having the same past boundary as the latter, namely the given component of $\partial_{-}B$, the tangent hyperplanes to $K$ and $B$ coinciding along $\partial_{-}B$. Moreover, one is required to construct a solution of the differential conservation laws in the domain in Minkowski spacetime bounded in the past by $\Cb\bigcup K$, agreeing with the maximal classical solution on $\Cb\bigcup\partial_{-}B$, while having jumps across $K$ relative to the 
data induced on $K$ by the maximal classical solution, jumps satisfying the jump conditions which follow from the integral form of the conservation laws. 
Finally, $K$ is required to be spacelike relative to the acoustical metic induced by the maximal classical solution, which holds in the past of $K$, 
and timelike relative to the new solution, which holds in the future of $K$. The maximal classical solution thus provides the boundary conditions on $\Cb\bigcup\partial_{-}B$, as well as a barrier at $B$. 

The shock development problem is only set up, not solved, in [Chr14]. The author plans to address this problem in the near future. One final result of [Chr14] 
needs to be mentioned here however. In the context of [Chr14], the solution is irrotational up to $K$. At the end of that monograph a formula is derived 
for the jump in vorticity across $K$ of a solution of the shock development problem. This formula shows that while the flow is irrotational ahead of the shock, 
it acquires vorticity immediately behind. 

This brings us to another problem. This is the problem of the long time behavior of the vorticity along the fluid flow lines. By reason of the result just quoted, this problem must be also be solved to achieve an understanding of the dynamics, even when the initial conditions are restricted to be irrotational. 
Now, even in the non-relativistic case, and even in the case that the compressibility of the fluid is neglected, this is a very difficult problem. Indeed, the 
problem, in the context of the incompressible Euler equations, of whether or not the vorticity blows up in finite time along some flow lines, is one of the 
great unsolved problems of mathematics (see [Co]). 

In conclusion, it is clear that the above basic fluid mechanical problems must be solved first, before any attempt is made to address the problem of the 
general non-spherical gravitational collapse of a perfect fluid in general relativity. 

However, once the restriction to spherical symmetry is removed, the dynamical degrees of freedom of the gravitational field itself come into play, and the thought strikes one that we may not need matter at all to form black holes. Even in vacuum closed trapped surfaces could perhaps be formed by the focusing of sufficiently strong incoming gravitational waves. It is in fact this problem which John Wheeler related to the author back in 1968: {\em the formation of black holes in pure general relativity, by the focusing of incoming gravitational waves}. And it is this problem the complete solution of which is found in the present monograph. Because of the absence of spherically symmetric solutions of 
the vacuum Einstein equations other than the Schwarzchild solution, the problem in question was far out of reach at that time, and for this reason John 
Wheeler advised the author to consider instead the spherically symmetric scalar field problem as a model problem, by solving which insights would be gained which 
would prepare us to attack the original problem. Indeed there is some analogy between scalar waves and gravitational waves, but whereas a scalar field is a fiction 
introduced only for pedagogical reasons, gravitational waves are a fundamental aspect of physical reality. We should remember here the remarks of Einstein in 
regard to the two sides of his equations. The right hand side, which involves the energy-momentum-stress tensor of matter, he called ``wood", while the left hand 
side, the Ricci curvature, he called ``marble", recalling, perhaps, the simplicity of an ancient Greek temple.  
 
We shall now state the simplest version of the theorem on the formation of closed trapped surfaces in pure general relativity which this monograph establishes.  
This is the limiting version, where we have an asymptotic characteristic initial value problem with initial data at past null infinity.  Denoting by $\ub$ the 
``advanced time", it is assumed that the initial data are trivial for $\ub\leq 0$. Our methods allow us to replace this assumption by a suitable falloff condition 
in $|\ub|$ for $\ub\leq 0$, thereby extending the theorem. This would introduce no new difficulties of principle, but would require more technical work, which would have considerably lengthened the monograph, obscuring the main new ideas. 

\vspace{2.5mm}

{\em Let $k, l$ be positive constants, $k>1$, $l<1$. Let us be given smooth asymptotic initial data at past null infinity which is trivial for advanced time 
$\ub\leq 0$. Suppose that the incoming energy per unit solid angle in each direction in the advanced time interval $[0,\delta]$ is not less than $k/8\pi$. 
Then if $\delta$ is suitably small, the maximal development of the data contains a closed trapped surface $S$ which is diffeomorphic to $S^2$ and has area}
$$\mbox{Area}(S)\geq 4\pi l^2$$

\vspace{2.5mm}

The form of the smallness assumption on $\delta$ is specified in the precise form of the theorem, stated in Chapter 17. We remark that by virtue of the 
scale invariance of the vacuum Einstein equations, the theorem holds with $k$, $l$, and $\delta$, replaced by $ak$, $al$, and $a\delta$, respectively, for any 
positive constant $a$. 

The above theorem is obtained through a theorem in which the initial data is given on a complete future null geodesic cone $C_o$. The generators 
of the cone are parametrized by an affine parameter $s$ measured from the vertex $o$ and defined so that the corresponding null geodesic vectorfield has 
projection $T$ at $o$ along a fixed unit future-directed timelike vector $T$ at $o$. It is assumed that the initial data are trivial for $s\leq r_0$, for 
some $r_0>1$. The boundary of this trivial region is then a round sphere of radius $r_0$. The advanced time $\ub$ is then defined along $C_o$ by 
\begin{equation}
\ub=r-r_0
\label{p.13}
\end{equation}
The formation of closed trapped surfaces theorem is similar in this case, the only difference being that the ``incoming energy per unit solid angle in each direction in the advanced time interval $[0,\delta]$", a notion defined only at past null infinity, is replaced by the integral 
\begin{equation}
\frac{r_0^2}{8\pi} \int_0^\delta ed\ub
\label{p.14}
\end{equation}
on the affine parameter segment $[r_0,r_0+\delta]$ of each generator of $C_o$. The function $e$ is an invariant of the conformal intrinsic geometry of $C_o$, 
given by:
\begin{equation}
e=\frac{1}{2}|\chih|_{\sg}^2
\label{p.15}
\end{equation}
where $\sg$ is the induced metric on the sections of $C_o$ corresponding to constant values of the affine parameter, and $\chih$ is the {\em shear} of these 
sections, the trace-free part of their 2nd fundamental form relative to $C_o$. The theorem for a cone $C_o$ is established for any $r_0>1$ and the smallness 
condition on $\delta$ is independent of $r_0$. The domain of dependence, in the maximal development, of the trivial region in $C_o$ is a domain in Minkowski 
spacetime bounded in the past by the trivial part of $C_o$ and in the future by $\Cb_e$, the past null geodesic cone of a point $e$ at arc length  
$2r_0$ along the timelike geodesic $\Gamma_0$ from $o$ with tangent vector $T$ at $o$. Considering then the corresponding complete timelike geodesic in Minkowski spacetime, 
fixing the origin on this geodesic to be the midpoint of the segment between $o$ and $e$, a segment of arc length $2r_0$, the limiting form of the theorem is obtained by letting $r_0\rightarrow\infty$, keeping the origin fixed, so that $o$ tends to the infinite past along the timelike geodesic.

The theorem on the formation of closed trapped surfaces in this monograph may be compared to the corresponding theorem in [Chr6] for the spherically symmetric 
scalar field problem quoted above. In sharp contrast to that theorem however, here almost all the work goes into establishing an {\em existence theorem} for 
a development of the initial data which extends far enough into the future so that trapped spheres have a evenlually chance to form within this development.  
This theorem is first stated as Theorem 12.1 in the way in which it is actually proved, and then restated as Theorem 16.1, in the way in which it can most readily be 
applied, after the proof is completed. So all chapters of this monograph, with the exception of the last (and shortest) chapter, are devoted to the proof of the existence theorem. On the other hand, there is a wealth of information in Theorem 16.1, which gives us full knowledge of the geometry of spacetime 
when closed trapped surfaces begin to form. The theorems established in this monograph thus constitute the first foray into the long time dynamics of general relativity in the large, that is, when the initial data are no longer confined to a suitably small neighborhood of Minkowskian data. However, the existence theorem 
which we establish does not cover the whole of the maximal development, and for this reason the question regarding the nature of the future boundary of the 
maximal development is left unanswered. 

We shall now give a brief discussion of the mathematical methods employed in this monograph, for, as is generally acknowledged, the methods in a mathematical 
work are often more important than the results. This monograph relies on three methods, two of which stem from the author's work with Klainerman [C-K] on 
the stability of the Minkowski spacetime, and the third method is new. We shall first summarize the first two methods. 

The work [C-K] which established the global nonlinear stability of the Minkowski spacetime of special relativity within the framework of general relativity, was 
a work within pure general relativity, concerned, like the present one, with the ``marble side" of Einstein's equations, the ``wood" side having been set equal to zero. Two where the main mathematical methods employed, and they were both new at the time when the work was composed. The first method was peculiar to 
Einstein's equations, while the second had wider application, and could, in principle, be extended to all Euler-Lagrange systems of partial differential equations of hyperbolic type. 

The first method was a way of looking at Einstein's equations which allowed estimates for the spacetime curvature to be obtained. A full exposition of this method is 
given in Chapter 12, which is also self-contained, except for Propositions 12.1, 12.5 and 12.6, which are quoted directly from [C-K]. Only the barest outline 
of the chief features will be given here. 
The method applies also in the 
presence of matter, to obtain the required estimates for the spacetime curvature. Its present form is dependent on the 4-dimensional nature of the spacetime manifold, although a generalization to higher dimensions can be found. 

Instead of considering the Einstein equations themselves, we considered the Bianchi identities in the form which they assume by virtue of the Einstein 
equations. We then introduced the general concept of a {\em Weyl field} $W$ on a 4-dimensional Lorentzian manifold $(M,g)$ to be a 4-covariant tensorfield with 
the algebraic properties of the Weyl or {\em conformal} curvature tensor. Given a Weyl field $W$ one can define a left dual $\s^* W$ as well as a right dual 
$W^*$, but as a consequence of the algebraic properties of a Weyl field the two duals coincide. Moreover, $\s^* W=W^*$ is also a Weyl field. A Weyl field is subject to equations which are analogues of Maxwell's equations for the 
electromagetic field. These are linear equations, in general inhomogeneous, which we call {\em Bianchi equations}. They are of the form:
\begin{equation}
\nabla^\alpha W_{\alpha\beta\gamma\delta}=J_{\beta\gamma\delta}
\label{p.16}
\end{equation}
the right hand side $J$, or more generally any 3-covariant tensorfield with the algebraic properties of the right hand side, we call a {\em Weyl current}. 
These equations seem at first sight to be the analogues of only half of Maxwell's equations, but it turns out that they are equivalent to the equations
\begin{equation}
\nabla_{[\alpha}W_{\beta\gamma]\delta\epsilon}=\epsilon_{\mu\alpha\beta\gamma}J^{*\mu}_{\s\s\delta\epsilon}, \ \ \ 
J^*_{\beta\gamma\delta}=\frac{1}{2}J_{\beta}^{\s\mu\nu}\epsilon_{\mu\nu\gamma\delta}
\label{p.17}
\end{equation}
which are analogues of the other half of Maxwell's equations. Here $\epsilon$ is the volume 4-form of $(M,g)$. The fundamental Weyl field is the Riemann curvature 
tensor of $(M,g)$, $(M,g)$ being a solution of the vacuum Einstein equations, and in this case the corresponding Weyl current vanishes, the Bianchi equations 
reducing to the Bianchi identities. 

Given a vectorfield $Y$ and a Weyl field $W$ or Weyl current $J$ there is a ``variation" of $W$ and $J$ with respect to $Y$, a modified Lie derivative 
$\tcL_Y W$, $\tcL_Y J$, which is also a Weyl field or Weyl current respectively. The modified Lie derivative commutes with duality. The Bianchi equations have certain conformal covariance properies which imply the following. If $J$ is the Weyl current associated to the Weyl field $W$ according to the Bianchi equations, then the Weyl current associated to $\tcL_Y W$ is 
the sum of $\tcL_Y J$ and a bilinear expression which is on one hand linear in $\s^{(Y)}\tilde{\pi}$ and its first covariant derivative and other the other hand in $W$ 
and its first covariant derivative (see Proposition 12.1). Here we denote by $\s^{(Y)}\tilde{\pi}$ the 
{\em deformation tensor} of $Y$, namely the trace-free part of the Lie derivative of the metric $g$ with respect to $Y$. This measures the rate of change of 
the conformal geometry of $(M,g)$ under the flow generated by $Y$. From the fundamental Weyl field, the Riemann curvature tensor of $(M,g)$, and a set of vector 
fields $Y_1,..., Y_n$ which we call {\em commutation fields}, derived Weyl fields of up to any given order $m$ are generated by the repeated application of 
the operators $\tcL_{Y_i}:i=1,...,n$. A basic requirement on the set of commutation fields is that it spans the tangent space to $M$ at each point. 
The Weyl currents associated to these derived Weyl fields are then determined by the deformation tensors of the commutation fields. 

Given a Weyl field $W$ there is a  4-covariant tensorfield $Q(W)$ associated to $W$, which is symmetric and trace-free in any pair of indices. It is a quadratic expression in $W$, analogous to the Maxwell energy-monentum-stress tensor for the electromagnetic field. We call $Q(W)$ 
the {\em Bel-Robinson tensor} associated to $W$, because it had been discovered by Bel and Robinson [Be] in the case of the fundamental Weyl field, the Riemann curvature 
tensor of a solution of the vacuum Einstein equations. The Bel-Robinson tensor has a remarkable positivity property: $Q(W)(X_1,X_2,X_3,X_4)$ is non-negative for any 
tetrad $X_1,X_2,X_3,X_4$ of future directed non-spacelike vectors at a point. Moreover, the divergence of $Q(W)$ is a bilinear expression which is linear in $W$ and 
in the associated Weyl current $J$ (see Proposition 12.6). Given a Weyl field $W$ and a triplet of future directed non-spacelike vectorfields $X_1,X_2,X_3$, which we call {\em multiplier fields} we 
define the {\em energy-momentum density} vectorfield $P(W;X_1,X_2,X_3)$ associated to $W$ and to the triplet $X_1,X_2,X_3$ by:
\begin{equation}
P(W;X_1,X_2,X_3)^\alpha=-Q(W)^\alpha_{\beta\gamma\delta}X_1^\beta X_2^\gamma X_3^\delta
\label{p.18}
\end{equation}
Then the divergence of $P(W;X_1,X_2,X_3)$ is the sum of $-(\mbox{div}Q(W))(X_1,X_2,X_3)$ and a bilinear expression which is linear in $Q(W)$ and in the deformation 
tensors of $X_1,X_2,X_3$. The divergence theorem in spacetime, applied to a domain which is a development of part of the initial hypersurface, then expresses the 
integral of the 3-form dual to $P(W;X_1,X_2,X_3)$ on the future boundary of this domain, in terms of the integral of the same 3-form on the past boundary of the domain, namely on the part of the initial hypersurface, and the spacetime integral of the divergence. The boundaries being {\em achronal} (that is, no pair of points on each boundary can be joined by a timelike curve) the integrals are integrals of non-negative functions, by virtue of the positivity property of $Q(W)$.  For the set of Weyl fields of order up to $m$ which are derived from the 
fundamental Weyl field, the Riemann curvature tensor of $(M,g)$, the divergences are determined by the deformation tensors of the commutation fields and their 
derivatives up to order $m$, and from the deformation tensors of the multiplier fields. And the integrals on the future boundary give control of all the derivatives of the curvature up to order $m$. This is how estimates for the spacetime curvature are obtained, once a suitable set of multiplier fields and a suitable set 
of commutation fields have been provided. 

This is precisely where the second method comes in. This method constructs the required sets of vectorfields by using the geometry of the two parameter foliation 
of spacetime by the level sets of two functions. These two functions, in the first realization of this method in [C-K], where the {\em time function $t$}, 
the level sets of which are maximal spacelike hypersurfaces $H_t$ of vanishing total momentum, and the {\em optical function $u$}, which we may think of 
as ``retarded time", the level sets of which are outgoing 
null hypersurfaces $C_u$. These are chosen so that density of the foliation of each $H_t$ by the traces of the $C_u$, that is, by the surfaces of intersection $S_{t,u}=H_t\bigcap C_u$, which are diffeomorphic to $S^2$, tends to 1 as $t\rightarrow\infty$. In other words, the $S_{t,u}$ on each $H_t$ become evenly spaced in the limit $t\rightarrow\infty$. 
It was already clear at the time of the composition of the work [C-K] that the two functions did not enter the problem on equal footing. The optical function 
$u$ placed a much more important role. This is due to the fact that the problem involved outgoing waves reaching future null infinity, and it is the outgoing 
family of null hypersurfaces $C_u$ which follows these waves. The role of the family of maximal spacelike hypersurfaces $H_t$ was to obtain a suitable family of 
sections of each $C_u$, the family $S_{t,u}$ corresponding to a given $u$, to provide the future boundary, or part of the future boundary, of domains where 
the divergence theorem is applied, and also to serve as a means by which, in the proof of the existence theorem, the method of continuity can be applied. 
The geometric entities describing the two parameter foliation of spacetime by the $S_{t,u}$ are estimated in terms of the spacetime curvature. This 
yields estimates for the deformation tensors of the multiplier fields and the commutation fields in terms of the spacetime curvature, thus connecting with the 
first method. 

Another realization of this method is found in [Chr14]. There in the role of the time function  we have the Minkowskian time coordinate $t$ which vanishes on the 
initial hyperplane. The level sets of this function are then a family $H_t$ of parallel spacelike hyperplanes in Minkowski spacetime. In the role of the optical function we have the {\em acoustical function} $u$, the level sets $C_u$ of which are outgoing characteristic hypersurfaces relative to the acoustical 
metric $h$. In this case however, these are defined by their traces $S_{0,u}$ on the initial hyperplane $H_0$, which are diffeomorphic to $S^2$. The density of the foliation of each $H_t$ 
by the traces of the $C_u$, that is, by the surfaces of intersection $S_{t,u}=H_t\bigcap C_u$, in fact blows up in finite time $t^*(u,\vartheta)$ for $(u,\vartheta)$ in an  open subset of $\Re\times S^2$, $\vartheta\in S^2$ labeling the generators of each $C_u$, and this defines the singular boundary $B$, whose past boundary $\partial_{-}B$ 
is the surface, not necessarily connected, from which shocks begin to form. The relative roles of the two functions is even clearer in this work, because  
the blow up of the density of  foliations by outgoing characteristic hypersurfaces is what characterizes shock formation. 

Returning to general relativity, a variant of the method is obtained if we place in the role of the time function $t$ another {\em optical function $\ub$}, which we 
may think of as ``advanced time", the level sets of which are incoming null hypersurfaces. This approach had its origin in the author's effort to understand the so-called ``memory effect" of gravitational waves [Chr15]. This effect is a manifestation of the nonlinear nature of the asymptotic gravitational 
laws at future null infinity. Now future null infinity is an ideal incoming null hypersurface at infinity, so the analysis required the consideration of a family 
of incoming null hypersurfaces the interiors of the traces of which on the initial spacelike hypersurface $H_0$ give an exhaustion of $H_0$. A two parameter 
family of surfaces diffeomorphic to $S^2$, the ``wave fronts", was then obtained, namely the intersections of this incoming family with the outgoing 
family of null hypersurfaces. A set of notes [Chr16] was then written up where the basic structure equations of such a ``double null" foliation where derived, 
and a rough outline was given on how one can proceed to estimate the geometric quantities associated with such a foliation in terms of the spacetime curvature. 

A double null foliation was subsequently employed by Klainerman and Nicol\`{o} in [K-N] (where the aforementioned notes are gracefully acknowledged) to provide a simpler variant of the exterior part of the proof of the stability of 
Minkowski spacetime, namely that part which considers the domain of dependence of the exterior of a compact set in the initial asymtoptically flat 
spacelike hypersurface. The developments stemming from the original work [C-K] include the work of Zipser [Z], which extended the original theorem to the 
Einstein-Maxwell equations, and most recently the work of Bieri [Bie], which extended the theorem in vacuum by requiring a smallness condition only on up to 
the 1st derivatives of the Ricci curvature of $\og_0$, the induced metric on $H_0$, and up to the 2nd derivatives of $k_0$, the 2nd fundamental form of $H_0$, 
instead of up to the 2nd and up to the 3rd derivatives respectively, as in the original theorem, and moreover with weights depending on the distance from 
an origin on $(H_0,\og_0)$, which are reduced by one power of this distance, relative to the weights assumed in [C-K]. 

In the present work, the roles of the two optical functions are reversed, because we are considering incoming rather than outgoing waves, and it is the incoming 
null hypersurfaces $\Cb_{\ub}$, the level sets of $\ub$, which follow these waves. However, in the present work, taking the other function to be the conjugate 
optical function $u$ is not merely a matter of convenience, but it is essential for what we wish to achieve. This is because the $C_u$, the level sets of $u$, are here,  
like the initial hypersurface $C_o$ itself, future null geodesic cones with vertices on the timelike geodesic $\Gamma_0$, and the trapped spheres which eventually form are sections 
$S_{\ub,u}=\Cb_{\ub}\bigcap C_u$ of ``late" $C_u$, everywhere along which those $C_u$ have negative expansion. 

We now come to the new method. This method is a method of treating the focusing of incoming waves, and like the second method it is of wider application. 
A suitable name for this method is {\em short pulse method}. Its point of departure resembles that of the short wavelength or geometric optics
approximation, in so far as it depends on the presence of a small length, but thereafter the two approaches diverge.  The short pulse method is a method which, 
in the context of Euler-Lagrange systems of partial differential equations of hyperbolic type, allows us to establish an existence theorem for a development of the initial data which is large enough so that interesting things have chance to occur within this development,  
if a nonlinear system is involved. One may ask at this point: what does it mean for a length to be small in the context of the vacuum Einstein equations? 
For, the equations are scale invariant. Here {\em small} means {\em by comparison to the area radius of the trapped sphere to be formed}. 

With initial data on a complete future null geodesic cone $C_o$, as explained above, which are trivial for $s\leq r_0$, we consider the restriction of the initial data to $s\leq r_0+\delta$. In terms of the advanced time $\ub$, we restrict attention to the interval $[0,\delta]$, the data being trivial for $\ub\leq 0$. 
The retarded time $u$ is set equal to $u_0=-r_0$ at $o$ and therefore on $C_o$, which is then also denoted $C_{u_0}$. Also, $u-u_0$ is defined along $\Gamma_0$ 
to be one half the arc length from $o$. This determines $u$ everywhere. The development whose existence we want to establish is that bounded in the future 
by the spacelike hypersurface $H_{-1}$ where $\ub+u=-1$ and by the incoming null hypersurface $\Cb_\delta$. We denote this development by $M_{-1}$. We define $L$ and $\Lb$ to be the future directed null vectorfields the integral curves of which are the generators of the $C_u$ and $\Cb_{\ub}$, parametrized by $\ub$ and $u$ respectively, so that
\begin{equation}
Lu=\Lb\ub=0, \ \ L\ub=\Lb u=1
\label{p.19}
\end{equation}
The flow $\Phi_\tau$ generated by $L$ defines a diffeomorphism of $S_{\ub,u}$ onto $S_{\ub+\tau,u}$, while the flow $\Phib_\tau$ generated by $\Lb$ defines 
a diffeomorphism of $S_{\ub,u}$ onto $S_{\ub,u+\tau}$. The positive function $\Omega$ defined by
\begin{equation}
g(L,\Lb)=-2\Omega^2
\label{p.20}
\end{equation}
may be thought of as the inverse density of the double null foliation. We denote by $\Lh$ and $\Lbh$ the corresponding normalized future directed null vectorfields
\begin{equation}
\Lh=\Omega^{-1}L, \ \  \Lbh=\Omega^{-1}\Lb, \ \ \mbox{so that $g(\Lh,\Lbh)=-2$}
\label{p.21}
\end{equation}

The first step is the analysis of the equations along the initial hypersurface $C_{u_0}$. This analysis is performed in Chapter 2, and it is particularly 
clear and simple because of the fact that $C_{u_0}$ is a null hypersurface, so we are dealing with the characteristic initial value problem and there is a way 
of formulating the problem in terms of free data which are not subject to any constraints. The full set of data which includes all the curvature components and their 
transversal derivatives, up to any given order, along $C_{u_0}$, is then determined by integrating ordinary differential equations along the generators of $C_{u_0}$. 
As we shall see in Chapter 2, the free data may be described as a 2-covariant symmetric positive definite tensor density $m$, of weight -1 and unit determinant, on $S^2$, 
depending on $\ub$. This is of the form:
\begin{equation}
m=\exp\psi
\label{p.22}
\end{equation}
where $\psi$ is a 2-dimensional symmetric trace-free matrix valued ``function" on $S^2$, depending on $\ub\in [0,\delta]$, and transforming under change of charts on $S^2$ in such a way so as to make $m$ a 2-covariant tensor density of weight -1. The transformation rule is particularly simple if stereographic charts on $S^2$ are used. Then there is a function $O$ 
defined on the intersection of the domains of the north and south polar stereographic charts on $S^2$, with values in the 2-dimensional symmetric orthogonal matrices of determinant 
-1 such that in going from the north polar chart to the south polar chart or vise-versa, $\psi\mapsto\tilde{O}\psi O$ and $m\mapsto\tilde{O}m O$. 
The crucial ansatz of the short pulse method is the following. We consider an arbitrary smooth 2-dimensional symmetric trace-free matrix valued ``function" $\psi_0$ 
on $S^2$, 
depending on $s\in[0,1]$, which extends smoothly by 0 to $s\leq 0$, and we set:
\begin{equation}
\psi(\ub,\vartheta)=\frac{\delta^{1/2}}{|u_0|}\psi_0\left(\frac{\ub}{\delta},\vartheta\right), \ \ (\ub,\vartheta)\in[0,\delta]\times S^2
\label{p.23}
\end{equation}
The analysis of the equations along $C_{u_0}$ then gives, for the components of the spacetime curvature along $C_{u_0}$:
\begin{eqnarray}
\sup_{C_{u_0}}|\alpha|&\leq& O_2(\delta^{-3/2}|u_0|^{-1})\nonumber\\
\sup_{C_{u_0}}|\beta|&\leq& O_2(\delta^{-1/2}|u_0|^{-2})\nonumber\\
\sup_{C_{u_0}}|\rho|,\sup_{C_{u_0}}|\sigma|&\leq& O_3(|u_0|^{-3})\nonumber\\
\sup_{C_{u_0}}|\beb|&\leq& O_4(\delta|u_0|^{-4})\nonumber\\
\sup_{C_{u_0}}|\alb|&\leq& O_5(\delta^{3/2}|u_0|^{-5})
\label{p.24}
\end{eqnarray}
Here $\alpha,\alb$ are the trace-free symmetric 2-covariant tensorfields on each $S_{\ub,u}$ defined by:
\begin{equation}
\alpha(X,Y)=R(X,\Lh,Y,\Lh), \ \ \ \alb(X,\Lbh,Y,\Lbh)
\label{p.25}
\end{equation}
for any pair of vectors $X,Y$ tangent to $S_{\ub,u}$ at a point, 
$\beta,\beb$ are the 1-forms on each $S_{\ub,u}$ defined by:
\begin{equation}
\beta(X)=\frac{1}{2}R(X,\Lh,\Lbh,\Lh), \ \ \ \beb(X)=\frac{1}{2}R(X,\Lbh,\Lbh,\Lh)
\label{p.26}
\end{equation}
and $\rho,\sigma$ are the functions on each $S_{\ub,u}$ defined by:
\begin{equation}
\rho=\frac{1}{4}R(\Lbh,\Lh,\Lbh,\Lh), \ \ \frac{1}{2}R(X,Y,\Lbh,\Lh)=\sigma\seps(X,Y)
\label{p.27}
\end{equation}
for any pair of vectors $X,Y$ tangent to $S_{\ub,u}$ at a point, $\seps$ being the area form of $S_{\ub,u}$. The symbol $O_k(\delta^p|u_0|^r)$ means 
the product of $\delta^p|u_0|^r$ with a non-negative non-decreasing continuous function of the $C^k$ norm of $\psi_0$ on $[0,1]\times S^2$. 
The pointwise magnitudes of tensors on $S_{\ub,u}$ are with respect to the induced metric $\sg$, which is positive definite, the surfaces being spacelike. 
The precise estimates are given in Chapter 2. We should emphasize here that the role of the ansatz \ref{p.23} is to obtain estimates of the form \ref{p.24}, 
that is with the same dependence on $\delta$ and $|u_0|$, and 
analogous estimates for the $L^4$ norms on the $S_{\ub,u_0}$, $\ub\in[0,\delta]$, of the 1st derivatives of the curvature components, and for the $L^2$ norms 
on $C_{u_0}$ of the 2nd derivatives of the curvature components (with the exception of the 2nd transversal derivative of $\alb$). If the quantities 
\begin{eqnarray}
&\sup_{C_{u_0}}\left(\delta^{3/2}|u_0||\alpha|\right)\nonumber\\
&\sup_{C_{u_0}}\left(\delta^{1/2}|u_0|^2|\beta|\right)\nonumber\\
&\sup_{C_{u_0}}\left(|u_0|^3|\rho|\right),\sup_{C_{u_0}}\left(|u_0|^3|\sigma|\right)\nonumber\\
&\sup_{C_{u_0}}\left(\delta^{-1}|u_0|^4|\beb|\right)\nonumber\\
&\sup_{C_{u_0}}\left(\delta^{-3/2}|u_0|^{9/2}|\alb|\right),
\label{p.28}
\end{eqnarray}
and analogous quantities for the 1st and 2nd derivatives, are assumed to have bounds which are independent of $|u_0|$ or $\delta$, the ansatz \ref{p.23} 
can be dispensed with, and indeed the chapters following Chapter 2 make no reference to it, until, at the end of Chapter 16, the existence theorem is restated, after it has been proven, as Theorem 16.1. However the ansatz \ref{p.23} is the simplest way to ensure that the required bounds hold, and there is no loss of generality 
involved, $\psi_0$ being an arbitrary ``function" on $[0,1]\times S^2$ with values in the 2-dimensional symmetric trace-free matrices. Note here that, since 
$|u_0|>1$, what is required of the last of \ref{p.28} is weaker than what is provided by the last of \ref{p.24}. A last remark before we proceed to the main point 
is that the last three of the estimates \ref{p.24} require more than two derivatives of $\psi_0$, so there is an apparent loss of derivatives from what would 
be expected of curvature components. This loss of derivatives is intrinsic to the characteristic initial value problem and occurs even for the wave equation 
in Minkowski spacetime (see [M]). It is due to the fact that one expresses the full data, which includes transversal derivatives of any order, in terms of the free data.  
No such loss of derivatives is present in our spacetime estimates, which are sharp, and depend only on the $L^2$ norm of up to the 2nd derivatives of 
the curvature components on the initial hypersurface (with the exception of the 2nd transversal derivative of $\alb$), precisely as in [C-K]. Nevertheless the 
initial data are assumed to be $C^\infty$ in this work, and the solutions which we construct are also $C^\infty$.  

To come to the main point, the reader should focus on the dependence on $\delta$ of the right hand sides of \ref{p.24}. This displays what we may call 
the {\em short pulse  hierarchy}. And this hierarchy is {\em nonlinear}. For, if only the linearized form of the equations was considered, a different hierarchy 
would be obtained: the exponents of $\delta$ in the first two of \ref{p.24} would be the same, but the exponents of $\delta$ in the last three of \ref{p.24} 
would instead be $1/2,3/2,5/2$, respectively. 

A question that immediately comes up when one ponders the ansatz \ref{p.24}, is why is the ``amplitude" of the pulse proportional to the square root of 
the ``length" of the pulse? (the factor $|u_0|^{-1}$ is the standard decay factor in 3 spatial dimensions, the square root of the area of the wave fronts). 
Where does this relationship come from? Obviously, there is no such relationship in a linear theory. The answer is that it comes from our desire to form trapped 
surfaces in the development $M_{-1}$. If a problem involving the focusing of incoming waves in a different context was the problem under study, for example 
the formation of electromagnetic shocks by the focusing of incoming electromagnetic waves in a nonlinear medium, the relationship between length and amplitude 
would be different and it would be dictated by the desire to form such shocks within our development.

Another remark concerning different applications of the short pulse method, in particular applications to problems of shock formation, is that it is more natural 
in these problems to use in the role of the retarded time $u$ the time function $t$ whose level sets are parallel  
spacelike hyperplanes of the background Minkowski metric, as in [Chr14]. However the analysis of the equations along an outgoing characteristic hypersurface is indispensable as a crucial step of the short pulse method, because, once the correct relationship between length and amplitude has been guessed, it is this 
analysis which yields the short pulse hierarchy. 

The short pulse hierarchy is the key to the existence theorem as well as to the trapped surface formation theorem. We must still outline however in what way do 
we establish that the short pulse hierarchy is preserved in evolution. This is off course the main step of the short pulse method. 
What we do is to reconsider the first two methods previously outlined in the light of the short pulse hierarchy. 

Let us revisit the first method. 
We take as multiplier fields the vectorfields $L$ and $K$, where
\begin{equation}
K=u^2\Lb
\label{p.29}
\end{equation}
In this monograph, as already metioned above, we take the initial data to be trivial for $\ub\leq 0$ and as a consequence the spacetime region corresponding to 
$\ub\leq 0$ is a domain in Minkowski spacetime. We may thus confine attention to the nontrivial region $\ub\geq 0$. We denote by $M^\prime_{-1}$ this non-trivial region in $M_{-1}$. To extend the theorem to the case where the data 
is non-trivial for $\ub\leq 0$ but satisfy a suitable falloff condition in $|\ub|$, in the region $\ub\leq 0$ we replace $L$ as a multiplier field by
\begin{equation}
\Lb+L=2T
\label{p.30}
\end{equation}
and redefine $K$ to be:
\begin{equation}
K=u^2\Lb+\ub^2 L
\label{p.31}
\end{equation}
Since, in any case, a smallness condition can be imposed on the part of the data corresponding to $\ub\leq 0$, we already know from the work on the stability of Minkowski spacetime 
that in the associated domain of dependence, that is, in the spacetime region $\ub\leq 0$, the solution will satisfy a corresponding smallness condition. 
In particular the said smallness condition will be satisfied along $\Cb_0$, and this suffices for us to proceed with our estimates in the region $\ub\geq 0$ with the 
multiplier fields $L$ and $K$, with $K$ as in \ref{p.29}. So all the difficulty lies in the region $M^\prime_{-1}$ where the pulse travels. 

For each of the Weyl 
fields to be specified below, we define the energy-momentum density vectorfields 
\begin{equation}
\stackrel{(n)}{P}(W) \ : \ n=0,1,2,3
\label{p.32}
\end{equation}
where:
\begin{eqnarray}
&&\stackrel{(0)}{P}(W)=P(W;L,L,L)\nonumber\\
&&\stackrel{(1)}{P}(W)=P(W;K,L,L)\nonumber\\
&&\stackrel{(2)}{P}(W)=P(W;K,K,L)\nonumber\\
&&\stackrel{(3)}{P}(W)=P(W;K,K,K)
\label{p.33}
\end{eqnarray}
As commutation fields we take $L$, $S$, defined by:
\begin{equation}
S=u\Lb+\ub L,
\label{p.34}
\end{equation}
and the three rotation fields $O_i \ : i=1,2,3$. The latter are defined according to the second method as follows. In the Minkowskian region we introduce rectangular coordinates $x^\mu \ : \mu=0,1,2,3$, taking the $x^0$ axis to be the timelike geodesic $\Gamma_0$. In the Minkowskian region, in particular on the sphere $S_{0,u_0}$, the $O_i$ are the generators of 
rotations about the $x^i \ :i=1,2,3$ spatial coordidate axes. The $O_i$ are then first defined on $C_{u_0}$ by conjugation with the flow of $L$ and then in spacetime 
by conjugation with the flow of $\Lb$. The  Weyl fields which we consider are, besides the fundamental Weyl field $R$, the Riemann curvature tensor, the following derived Weyl fields
\begin{eqnarray}
&&\mbox{1st order:} \ \ \tcL_L R,  \ \tcL_{O_i}R:i=1,2,3, \ \tcL_S R\nonumber\\
&&\mbox{2nd order:} \ \ \tcL_L\tcL_L R, \ \tcL_{O_i}\tcL_L R:i=1,2,3, \ \tcL_{O_j}\tcL_{O_i}R:i,j=1,2,3, \nonumber\\
&&\hspace{20mm} \tcL_{O_i}\tcL_S R:i=1,2,3, \ \tcL_S\tcL_S R
\label{p.35}
\end{eqnarray}
We assign to each Weyl field the index $l$ according to the number of $\tcL_L$ operators in the definition of $W$ in terms of $R$. We then define total 
2nd order energy-momentum densities
\begin{equation}
\stackrel{(n)}{P}_2 \ : \ n=0,1,2,3
\label{p.36}
\end{equation}
as the sum of $\delta^{2l}\stackrel{(n)}{P}(W)$ over all the above Weyl fields in the case $n=3$, all the above Weyl fields except those whose definition involves 
the operator $\tcL_S$ in the cases $n=0,1,2$.  We then define the total 2nd order energies $\stackrel{(n)}{E}_2(u)$ as the integrals on the $C_u$ and the 
total 2nd order fluxes $\stackrel{(n)}{F}_2(\ub)$ as the integrals on the $\Cb_{\ub}$, of the 3-forms dual to the $\stackrel{(n)}{P}_2$. Of the fluxes only 
$\stackrel{(3)}{F}_2(\ub)$ plays a role in the problem. Finally, with the exponents $q_n \ : \ n=0,1,2,3$ defined by:
\begin{equation}
q_0=1, \ \ q_1=0, \ \ q_2=-\frac{1}{2}, \ \ q_3=-\frac{3}{2},
\label{p.37}
\end{equation}
according to the short pulse hierarchy, we define the quantities
\begin{equation}
\stackrel{(n)}{{\cal E}}_2=\sup_u\left(\delta^{2q_n}\stackrel{(n)}{E}_2(u)\right) \ : \ n=0,1,2,3, \ \ \ \ \ 
\stackrel{(3)}{{\cal F}}_2=\sup_{\ub}\left(\delta^{2q_3}\stackrel{(3)}{F}_2(\ub)\right)
\label{p.38}
\end{equation}
The objective then is to obtain bounds for these quantities in terms of the initial data. 

This requires properly estimating the deformation tensor of $K$, as well as the deformations tensors of $L,S$ and the $O_i:i=1,2,3$ and their derivatives of up to 2nd 
order. In doing this, the short pulse method meshes with the second method previously described. This is the content of Chapters 3 - 9 and shall be very briefly described in the outline of the contents of each chapter which follows. 

The estimates of the error integrals, namely the integrals of the absolute values of the divergences of the $\stackrel{(n)}{P}_2$, which is the content of 
Chapters 13 - 15, then yield inequalities for the quantities \ref{p.38}. These inequalities contain, besides the initial data terms 
\begin{equation}
\stackrel{(n)}{D}=\delta^{2q_n}\stackrel{(n)}{E}_2(u_0) \ : \ n=0,1,2,3,
\label{p.39}
\end{equation}
terms of $O(\delta^p)$ for some $p>0$, which are 
innocuous, as they can be made less than or equal to 1 by subjecting $\delta$ to a suitable smallness condition, {\em but they also contain terms which 
are nonlinear in the quantities \ref{p.38}}. From such a nonlinear system of inequalities, no bounds can in general be deduced, because here, in constrast with [C-K], the initial data quantities are allowed to be arbitrarily large. However a miracle occurs: our system of inequalities is {\em reductive}. That is, the inequalities, 
taken in proper sequence, reduce to a sequence of sublinear inequalities, thus allowing us to obtain the sought for bounds. 

We remark that although the first two methods on which the present work is based stem from the work [C-K], it is only in the present work, in conjunction with 
the new method, that the full power of the original methods is revealed. 

In applying the short pulse method to problems in other areas of the field of partial differential equations of hyperbolic type, an analogue of the first method is 
needed. This is supplied in the context of Euler-Lagrange systems, that is, systems of partial differential equations derivable from an action principle, 
by the structures studied in [Chr17]. The analogue of the concept of a Weyl field is the general concept of {\em variation}, or variation through solutions. 
The analogue of the Bel-Robinson tensor is the {\em canonical stress} associated to such variations. 
In the area of continuum mechanics or the electrodynamics of continuous media, the fundamental variation is that with respect to a subgroup of the Poincare 
group of the underlying Minkowski spacetime, while the higher order variations are generated by the commutation fields, as in general relativity (see [Chr14]). 
A particularly interesting problem that may be approached on the basis of the methods which we have discussed, in conjunction with ideas from [Chr14], is the formation of 
electromagnetic shocks by the focusing of incoming electromagnetic waves in isotropic nonlinear media, that is, media with a nonlinear relationship between the electromagnetic field and the electomagnetic displacement. In this problem, unlike the problem 
of shock formation by outgoing compression waves in fluid mechanics, there is a {\em threshold} for shock formation, 
as there is a threshold for closed trapped sphere formation in the present monograph.

We shall now give a brief outline of the contents of the different chapters of this monograph and of their logical connections. The basic geometric construction,  the structure equations of the 
double null foliation called the ``optical structure equations", and the Bianchi identities, are presented in the introductory Chapter 1. 
The Einstein equations are contained in the optical structure equations. The basic geometric entities associated to the double null foliation are the inverse 
density function $\Omega$, the  metric $\sg$ induced on the surfaces $S_{\ub,u}$ and its Gauss curvature $K$, the second fundamental forms $\chi$ and $\chib$ of 
$S_{\ub,u}$ relative to $C_u$ and $\Cb_{\ub}$ respectively, the torsion forms $\eta$ and $\etb$ of $S_{\ub,u}$ relative to $C_u$ and $\Cb_{\ub}$ respectively, 
and the functions $\omega$ and $\omb$, the derivatives of $\log \Omega$ with respect to $L$ and $\Lb$ respectively. The torsion forms are given by:
\begin{equation}
\eta=\zeta+\sd\log\Omega, \ \ \ \ \ \etb=-\zeta+\sd\log\Omega
\label{p.40}
\end{equation}
where $\zeta$ may be caled {\em the} torsion. It is the obstruction to integrability of the distribution of planes orthogonal to the tangent planes to the $S_{\ub,u}$. In \ref{p.40} $\sd$ denotes the differential of the restriction of a function to any given $S_{\ub,u}$. The optical entities $\chi,\chib,\eta,\etb,\omega,\omb$ 
are called {\em connection coefficients} in the succeeding chapters, to emphasize their differential order, intermediate between the metric entities $\Omega$ and $\sg$, and the curvature entities $K$ and the spacetime curvature components.
``Canonical coordinates" are defined in the last section of Chapter 1 and play a basic role in this monograph. 

The subject of Chapter 2 is the characteristic initial data and the derivation of the estimates for the full data in terms of the free data. This is where the ansatz \ref{p.23} is introduced and the short pulse hierarchy first appears. Thus Chapter 2 is fundamental 
to the whole work. 

Chapters 3 - 7 form a unity. The subject of these chapters is the derivation of estimates for the connection coefficients in terms of certain quantities defined 
by the spacetime curvature. These chapters are in logical sequence, which extends to Chapters 8 and 9, however the place of the whole sequence of Chapters 3 - 9 
in the logic of the proof of the existence theorem, Theorem 12.1, is {\em after} Chapters 10 and 11. This is because the assumptions on which Chapters 3 - 9 rely, namely 
the boundedness of the quantities defined by the spacetime curvature, is established, in the course of the proof of Theorem 12.1, through the comparison lemmas, Lemmas 12.5 and 12.6, which make use of the results of Chapters 10 and 11. Thus, Chapter 10 represents a {\em new beginning}. The chapters following Chapter 12 are again 
in logical sequence. 

Chapters 3 - 7 are divided by Chapter 5 into the two pairs of chapters, on one hand Chapters 3 and 4, and on the other hand  Chapters 6 and 7, each of these two pairs 
forming a tighter unity. The first pair considers only the {\em propagation equations} among the optical structure equations. These are ordinary 
differential equations for the connection coefficients along the generators of the $C_u$ and the $\Cb_{\ub}$. The second pair considers {\em coupled systems}, 
{\em ordinary differential equations} along the generators of the $C_u$ or the $\Cb_{\ub}$ {\em coupled to  elliptic systems} on their $S_{\ub,u}$ sections. This allows us to obtain estimates 
for the connection coefficients which are of one order higher than those obtained through the propagation equations, and are optimal from the point of view of 
differentiability. There is however a loss of a factor of $\delta^{1/2}$ in behavior with respect to $\delta$, in comparison to the estimates obtained through the 
propagation equations, in the case of the entities $\eta$, $\etb$ and $\omega$. What is crucial, is that there is no such loss in the case of the entities $\chi$, $\chib$, and 
$\omb$, but the proof of this fact again uses the former estimates. 

In Chapter 3 the basic $L^\infty$ estimates for the connection coefficients are obtained. The last section of Chapter 3 explains the nature of smallness conditions 
on $\delta$ throughout the monograph. Chapter 4 derives $L^4$ on the surfaces $S_{\ub,u}$ for the 1st derivatives of the connection coefficients. 

Chapter 5 is concerned with the isoperimetric and Sobolev inequalities on the surfaces $S_{\ub,u}$, and with $L^p$ elliptic theory on these surfaces for $2<p<\infty$. 
The main part of the chapter is concerned with the proof of the uniformization theorem for a 2-dimensional Riemannian manifold $(S,\sg)$ with $S$ diffeomorphic to 
$S^2$, when only an $L^2$ bound on the Gauss curvature $K$ is assumed. The reason why this is required is that although the Gauss equation gives us $L^\infty$ 
control on $K$, the estimate is not suitable for our purposes because it involves the loss of a factor of $\delta^{1/2}$ in behavior with respect to $\delta$. 
Thus one can only rely on the estimate obtained by integrating a propagation equation, which although optimal from the point of view of behavior with respect to 
$\delta$, only gives us $L^2$ control on $K$. 

The $L^p$ elliptic theory on the $S_{\ub,u}$ is applied in Chapter 6, in the case $p=4$, to the elliptic systems mentioned above, to obtain $L^4$ estimates for the 2nd derivatives of the connection coefficients on the surfaces $S_{\ub,u}$. What makes possible the gain of 
one degree of differentiability by considering systems of ordinary differential equations along the gerenators of the $C_u$ or the $\Cb_{\ub}$ 
coupled to elliptic systems on the $S_{\ub,u}$ sections, is the fact that the principal terms in the propagation equations for certain optical entities vanish, 
by virtue of the Einstein equations. 
In the case of the coupled system pertaining to $\chi$ and $\chib$, these entities are simply the traces $\mbox{tr}\chi$ and $\mbox{tr}\chib$, and 
the Codazzi equations  constitute the elliptic systems for the trace-free parts $\chih$ and $\chibh$ respectively. In the case of the coupled systems pertaining 
to $\eta$ and $\etb$, the entities are found at one order of differentiation higher. They are the {\em mass aspect functions} $\mu$ and $\mub$, called by this name 
because of the fact, shown in [Chr15], that with $r$ being the area radius of the $S_{\ub,u}$, the limits of the functions $r^3\mu/8\pi$ and 
$r^3\mub/8\pi$ at past and future null infinity respectively, represent mass-energy per unit solid angle in a given direction and at a given advanced or retarded time 
respectively. The elliptic systems are Hodge systems constituted by one of the structure equations and by the definition of $\mu$ and $\mub$ in terms of $\eta$ and $\etb$ respectively. 
Moreover, the two sets of coupled systems, that for $\eta$ on the $C_u$, and that for $\etb$ on the $\Cb_{\ub}$, are themselves coupled. 
(The propagation equations for 
$\eta$ and $\etb$ studied in Chapters 3 and 4 are similarly coupled). In the case of the coupled systems pertaining to $\omega$ and $\omb$, the entities which satify 
propagation equations in which the principal terms vanish are found at one order of differentiation still higher. They are the functions $\somega$ and $\somb$ 
and the elliptic equations are simply the definitions of these functions in terms of $\omega$ and $\omb$ respectively. 

In the case of the $\chi$ system we have, besides what has already been described, also a coupling with the propagation equation for the Gauss curvature $K$, 
through the elliptic theory of Chapter 5 applied to the Codazzi elliptic system for $\chih$. 

Chapter 7 applies $L^2$ elliptic theory on the $S_{\ub,u}$ to the same coupled systems to obtain $L^2$ estimates on the $C_u$ for the third derivatives of the connection coefficients, the top order needed to obtain a closed system of inequalities in the proof of the existence theorem. 

One general remark concerning the contents of Chapters 3 -7 is that although some of the general structure was already encountered in the work on the stability of Minkowski spacetime, the estimates and their derivation are here quite different, and for two reasons. One is the obvious reason that some of the geometric properties are here very different, in view of the fact that we are no longer confined to a suitably small neighborhood of Minkowski spacetime and closed trapped surfaces eventually form. 
The second is the fact, in conncetion with the short pulse method, that behavior with respect to $\delta$ is here all-important. 
 
In Chapter 8 the multiplier fields and the commutation fields are defined and $L^\infty$ estimates for their deformation tensors are obtained. In Chapter 9 
$L^4$ estimates on the $S_{\ub,u}$ for the 1st derivatives of these deformation tensors and $L^2$ estimates on the $C_u$ for their 2nd derivatives are obtained. 

In Chapters 3 - 9 the symbol $O(\delta^p |u|^r)$ denotes the product of $\delta^p|u|^r$ with a non-negative, non-decreasing continuous function of certain 
initial data and spacetime curvature quantities $q_1,...,q_n$. The set of quantities $\{q_1,...,q_n\}$ is gradually enlarged as we proceed through the sequence 
of chapters. The set of quantities is replaced in the seventh section of Chapter 12 by a set which includes only initial data quantities, and it is in this new sense 
that the symbol $O(\delta^p |u|^r)$ is meant throughout the proof of Theorem 12.1, which occupies the four succeeding chapters. 

As we mention above, Chapter 10 represents a new beginning. The point is the following. Chapters 3 - 7 derive estimates for the connection coefficients in 
terms of quantities involving the $L^\infty$ norms on the $S_{\ub,u}$ of the curvature components, the $L^4$ norms on the $S_{\ub,u}$ of the 1st derivatives of 
the curvature components, and the $L^2$ norms on the $C_u$ of the 2nd derivatives of the curvature components (with the exception of those involving $\alb$ and 
the 2nd transversal derivatives of $\beb$). The first two are to be estimated in terms of the last through Sobolev inequalities on the $C_u$ (except for the quantities involving $\alb$, which are estimated in terms of the $L^2$ norm on the $\Cb_{\ub}$ of up to the 2nd derivatives of that component through a Sobolev inequality on the $\Cb_{\ub}$), but in establishing these Sobolev inequalities one cannot rely on the results of the preceeding chapters, otherwise the reasoning would be circular. So, the 
Sobolev inequalities on the $C_u$ and the $\Cb_{\ub}$ are instead established on the basis of certain {\em bootstrap} assumptions. The sharp form of the Sobolev inequality 
on the $C_u$ given by Proposition 10.1 fits perfectly with the short pulse method and is essential to its success. 

The subject of Chapter 11 is the {\em coercivity} properties of the operators $\sL_{O_i}:i=1,2,3$, the Lie derivatives of covariant tensorfields on the $S_{\ub,u}$ with respect the rotation fields $O_i:1=1,2,3$. These inequalities show that, for $m=1,2$ we can bound the sum of the squares of the $L^2$ norms on $S_{\ub,u}$ of up to the $m$th intrinsic to $S_{\ub,u}$ 
covariant derivatives of these tensorfields in terms of the sum of the squares of the $L^2$ norms on the $S_{\ub,u}$ of their Lie derivatives of up to $m$th order 
with respect to the set of rotation fields. This is important because only these rotational Lie derivatives of the curvature components (and the Lie derivatives of the 
curvature components with respect to $L$ and $\Lb$), not their covariant derivatives intrinsic to the $S_{\ub,u}$, are directly controlled by the energies and fluxes.
To establish the coercivity inequalities additional bootstrap assumptions are introduced. 

Chapter 12 is the central chapter of the monograph. This chapter lays out the first method and defines the energies and fluxes according to the short pulse method as 
discussed above. These definitions are followed by the comparison lemmas, Lemmas 12.5 and 12.6 which show that the quantity ${\cal Q}^\prime_2$, which bounds 
all the curvature quantities which enter the estimates for the connection coefficients and the deformation tensors of Chapters 3 - 9, is itself bounded in terms 
of the quantity:
\begin{equation}
{\cal P}_2=\max\{\stackrel{(0)}{{\cal E}}_2,\stackrel{(1)}{{\cal E}}_2,\stackrel{(2)}{{\cal E}}_2,\stackrel{(3)}{{\cal E}}_2;\stackrel{(3)}{{\cal F}}_2\}
\label{p.41}
\end{equation}
To establish the comparison lemmas additional bootstrap assumptions are introduced. The last section of Chapter 12 gives the statement of Theorem 12.1 in the way it 
is actually proved, and then gives an outline of the first and most important part of the continuity argument, that which concludes with the derivation of 
the reductive system of inequalities for the quantities $\stackrel{(0)}{{\cal E}}_2,\stackrel{(1)}{{\cal E}}_2,\stackrel{(2)}{{\cal E}}_2,\stackrel{(3)}{{\cal E}}_2$ and $\stackrel{(3)}{{\cal F}}_2$ in the first section of Chapter 16. 

Chapters 13 - 15 deal with the error estimates, namely the estimates for the {\em error integrals}, the spacetime integrals of the absolute values of the divergences of the energy-momenum density vectorfields $\stackrel{(n)}{P}_2$. There are two kinds of error integrals: the error integrals arising form the 
deformation tensors of the multiplier fields and those arising from the Weyl currents generated by the commutation fields. The first are treated in Chapter 13 
and the second in Chapers 14 and 15. Because of the delicacy of the final estimates, all the error terms are treated in a systematic fashion. All error integrals 
are estimated using Lemma 13.1. The concepts of {\em integrability index} and {\em excess index} are introduced. The integrability index $s$ being negative 
allows us to apply Lemma 13.1. The excess index $e$ then gives the exponent of $\delta$ contributed by the error term under consideration to the final 
system of inequalities for the quantities $\stackrel{(0)}{{\cal E}}_2,\stackrel{(1)}{{\cal E}}_2,\stackrel{(2)}{{\cal E}}_2,\stackrel{(3)}{{\cal E}}_2$ and $\stackrel{(3)}{{\cal F}}_2$. All error terms turn out to have a negative integrability index and a non-negative excess index. The terms with a positive
excess index contribute the innocuous terms $O(\delta^e)$ mentioned above. To the terms with zero excess index are associated {\em borderline error integrals}. 
These contribute the {\em nonlinear terms} to the final system of inequalities mentioned above. 

Chapter 16 completes the proof of the existence theorem. The reductive system of inequalities for the quantities $\stackrel{(0)}{{\cal E}}_2,\stackrel{(1)}{{\cal E}}_2,\stackrel{(2)}{{\cal E}}_2,\stackrel{(3)}{{\cal E}}_2$ and $\stackrel{(3)}{{\cal F}}_2$ is obtained in the first section of Chapter 16, and the required bounds 
for these quantities are deduced. The second section of Chapter 16 deduces the higher order bounds for the spacetime curvature components and the connection coefficients. 
The higher order estimates are of linear nature and are needed to show that the solution extends as a smooth solution. Only the roughest bounds are needed. 
The continuity argument is completed in the third section of Chapter 16. In this section the work of Choquet-Bruhat ([Cho1], [Cho2], [Cho3]) and that of Rendall [R] 
are used to obtain a smooth local extension of the solution in ``harmonic" (also called ``wave") coordinates. This is followed by an argument showing that 
in a suitably small extension, contained in the extension just mentioned, canonical null coordinates can be set up and the coordinate transformation from harmonic coordinates 
to canonical null coordinates is a smooth transformation with a smooth inverse, hence the metric extends smoothly also in canonical null coordinates. The 
proof of the existence theorem is then concluded. In the last section, the existence theorem is restated in the way in which it can most readily be applied. 

The last chapter, Chapter 17, establishes the theorem on the formation of closed trapped surfaces, achieving the aim of this monograph. 

The present monograph is off course a work in pure mathematics. However, by virtue of the fact that Einstein's theory is a physical theory, describing a fundamental 
aspect of nature, this work is also of physical significance. For those mathematicians who, by reading the present monograph, become interested in the physical basis 
of general relativity, we recommend the excellent book [M-T-W] where not only is the physical basis of the theory expained, but also a wealth of information is given 
which illustrates how the theory is applied to describe natural phenomena. 
 
As this work was been completed the author learned that his old teacher, John Wheeler, passed away. This monograph testifies to John Wheeler's enduring legacy in the scientific community.

\pagebreak

\chapter{The Optical Structure Equations}

\section{The Basic Geometric Setup}

We consider a spacetime manifold $(M,g)$, with boundary, smooth
solution of the vacuum Einstein equations, such that the past
boundary of $M$ is the future null geodesic cone $C_o$ of a point
$o$. $(M,g)$ is to be a development of initial data on $C_o$. That
is, for each point $p\in M$, each past directed causal curve
issuing from $p$ is to terminate in the past at a point on $C_o$.
The generators of $C_o$ are the future-directed null geodesics
issuing from $o$. These are affinely parametrized as follows.
There is a future-directed unit timelike vector $T$ at $o$ such
that the tangent vector at $o$ to each generator has projection
$T$ along $T$. We denote by $L^\prime$ the null geodesic
vectorfield along $C_o$, which is the tangent field to each
generator. The parameter of this vectorfield, measured from $o$,
is the affine parameter $s$. Each generator is to extend up to the
value $r_0+\delta$ of the parameter, where $r_0$ and $\delta$ are
constants, $r_0>1$, $1\geq\delta>0$. The null geodesic cone $C_o$
is to contain no conjugate or cut points. For
$r\in(0,r_0+\delta]$, let us denote by $C_o^r$ the truncated cone
obtained by restricting each generator to the interval $[0,r]$ of
the parameter. Let us also denote by $S_o^r$ the boundary of
$C_o^r$, the level surface $s=r$ of the function $s$ on $C_o$.

The restrictions of the initial data to $C_o^{r_0}$ are to be
trivial, that is, they are to coincide with the data corresponding
to a truncated cone in Minkowski spacetime. By the uniqueness
theorem, the domain of dependence of $C_o^{r_0}$ in $(M,g)$
coincides with a domain in Minkowski spacetime. The boundary of
this domain is a null hypersuface generated by the incoming null
geodesic normals from $S_o^{r_0}$. The solution extends as a
domain in Minkowski spacetime up to the past null geodesic cone
$\Cb_e$ of a point $e$ at arc length $2r_0$ along the timelike
geodesic $\Gamma_0$ from $o$ with tangent vector $T$ at $o$, the
aforementioned incoming null hypersurface from $S_o^{r_0}$
coinciding with $\Cb_e$. Let us denote this Minkowskian domain
$M_0$. Thus the past boundary of $M_0$ is $C_o^{r_0}$ and the
future boundary of $M_0$ is $\Cb_e$, $o$ and $e$ being
respectively the past and future end points of $\Gamma_0$, and the
common boundary of $C_o^{r_0}$ and $\Cb_e$ being the surface
$S_o^{r_0}$.

Let $q\in\Gamma_0$. We denote by $C_q$ the boundary of the causal
future of $q$ in $M$. Then $C_q$ is an outgoing null hypersurface
generated by the future directed null geodesics issuing from $q$.
Each of these geodesics has a future end point which is either a
point belonging to the future boundary of $M$ or else it is a
point conjugate to $q$ along the geodesic in question, or a cut
point, a point at which the given generator intersects another
generator. We then denote $C_q$ by $C_c$, $c\in[-r_0,0]$, if $q$
is at arc length $2(r_0+c)$ along $\Gamma_0$ from $o$. In
particular, $C_{-r_0}=C_o$. Now for $r\in(0,r_0]$ the boundary of
the domain of dependence of $C_o^r$ in $M_0$ is $\Cb_q$, the past
null geodesic cone of a point $q\in \Gamma_0$ at arc length $2r$
from $o$, so that $\Cb_q$ intersects $C_o$ in the surface $S_o^r$.
We denote $\Cb_q$ by $\Cb_c$, $c=r-r_0\in(-r_0,0]$. In particular,
$\Cb_0=\Cb_e$. For $r\in(r_0,r_0+\delta]$, we still denote by
$\Cb_c$, $c=r-r_0$, the boundary of the domain of dependence of
$C_o^r$ in $M$. In this case $c\in(0,\delta]$ and $\Cb_c$ is
generated by the incoming null geodesic normals to $S_o^r$, each
generator having a future end point which is either a point
belonging to the future boundary of $M$ or else it is a focal
point of the incoming null normal geodesic congruence to $S_o^r$
along the given generator, or a cut point, a point at which the
given generator intersects another generator.

We now define on $M$ the {\em optical functions} $u$ and $\ub$ by
the requirements that for each $c$ the $c$-level set of $u$ is the
outgoing null hypersurface $C_c$ and for each $c$ the $c$-level
set of $\ub$ is the incoming null hypersurface $\Cb_c$. In the
following we denote by $C_u$ and $\Cb_{\ub}$ an arbitrary level
set of $u$ and $\ub$ respectively. In this sense $u$ and $\ub$
denote not the optical functions themselves but values in the
range of the functions $u$ and $\ub$ respectively. We denote by
$S_{\ub,u}$ the spacelike surfaces of intersection:
\begin{equation}
S_{\ub,u}=\Cb_{\ub}\bigcap C_u \label{1.1}
\end{equation}
These are round spheres for $\ub\in(u,0]$, which corresponds to
the Minkowskian region $M_0$, degenerating to points for $\ub=u$,
which corresponds to $\Gamma_0$. Also, $S_{-r_0+r,-r_0}=S_o^r$.
The optical functions are increasing toward the future.

Let  $c^*\in(u_0,-1]$, $u_0=-r_0$. For $c\in(u_0+\delta,c^*]$ we
denote by $H_c$ the spacelike hypersurface:
\begin{equation}
H_c=\{S_{\ub,u} \ : \ \ub+u=c, \ \ub\in[u,\delta)\} \label{1.01}
\end{equation}
Also, for $c\in(2u_0,u_0+\delta]$ we denote by $H_c$ the spacelike
hypersurface
\begin{equation}
H_c=\{S_{\ub,u} \ : \ \ub+u=c, \ u\in[u_0,\ub]\} \label{1.02}
\end{equation}
Note that for $c\leq u_0$ $H_c$ is a spacelike hyperplane in the
Minkowskian region $M_0$.

The future boundary of $M$ is to be the union
$H_{c^*}\bigcup\Cb_\delta$. Thus, for each $c\in[u_0,c^*-\delta]$
an end point of a generator of $C_c$ which belongs to the future
boundary of $M$, belongs to $\Cb_\delta$, while for each $c\in
(c^*-\delta, c^*)$ an end point of a generator of $C_c$ which
belongs to the future boundary of $M$, belongs to $H_{c^*}$. Note
that for $c\geq c^*$ the generators of $C_c$ are contained in the
Minkowskian region $M\bigcap M_0$ hence end on $H_{c^*}$. For each
$c\in(0,\delta)$ a future end point of a generator of $\Cb_c$
which belongs to the future boundary of $M$, belongs to $H_{c^*}$.
For $c\leq 0$ the generators of $\Cb_c$ are contained in the
Minkowskian region $M\bigcap M_0$ hence end on $H_{c^*}$ if
$c>c^*/2$, on $\Gamma_0$ if $c\leq c^*/2$. We consider $M$ as
including its past boundary but not its future boundary. Thus $M$
corresponds to the parameter domain:
\begin{eqnarray}
D&=&\{(\ub,u) \ : \ \ub\in[u,\delta), \ u\in[u_0,c^*-\delta]\}
\bigcup\nonumber\\
&\s&\s\s\s\{(\ub,u) \ : \ \ub\in[u,c^*-u), \ u\in(c^*-\delta,c^*/2)\}\nonumber\\
&=&\{(\ub,u) \ : \ u\in[u_0,\ub], \ \ub\in[u_0,c^*/2)\}
\bigcup\nonumber\\
&\s&\s\s\s\{(\ub,u) \ : \ u\in[u_0,c^*-\ub), \
\ub\in[c^*/2,\delta)\} \label{1.03}
\end{eqnarray}
for $c^*\geq u_0+\delta$ and:
\begin{eqnarray}
D&=&\{(\ub,u) \ : \ \ub\in[u,c^*-u), \ u\in[u_0,c^*/2)\}\nonumber\\
&=&\{(\ub,u) \ : \ u\in[u_0,\ub], \ \ub\in[u_0,c^*/2)\}
\bigcup\nonumber\\
&\s&\s\s\s\{(\ub,u) \ : \ u\in[u_0,c^*-\ub), \ \ub\in[c^*/2,c^*-u_0)\}\nonumber\\
\label{1.03a}
\end{eqnarray}
for $c^*<u_0+\delta$. 
The timelike geodesic $\Gamma_0$ corresponds to:
\begin{equation}
A_0=\{(\ub,u) \ : \ \ub=u, \ u\in[u_0,c^*/2)\} \label{1.04}
\end{equation}
(see Figures 1.1, 1.2). 

\begin{figure}[htbp]
\begin{center}
\input{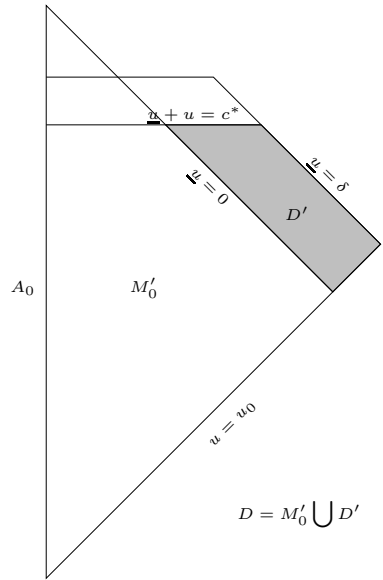}
\caption{Case $c^*\geq u_0+\delta$}
\label{fig1.1}
\end{center}
\end{figure}

\begin{figure}[htbp]
\begin{center}
\input{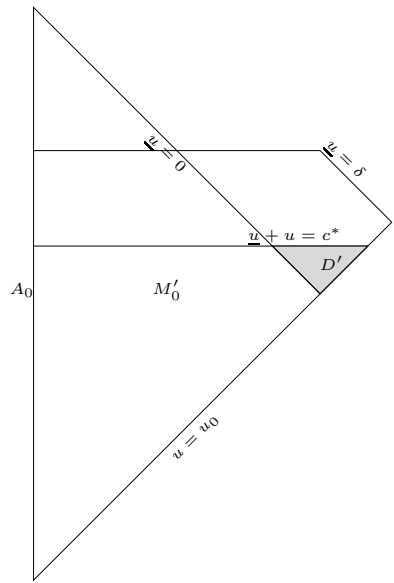}
\caption{Case $c^*< u_0+\delta$}
\label{fig1.2}
\end{center}
\end{figure}

Our approach is a continuity argument in which the basic
requirement to be imposed on $(M,g)$ is that {\em the generators
of the $C_u$ and the $\Cb_{\ub}$ have no end points in
$M\setminus\Gamma_0$}. Thus, the $C_u$ are to contain no conjugate
or cut points in $M$ and the $\Cb_{\ub}$ are to contain no focal
or cut points in $M\setminus\Gamma_0$. Then for
$u\in[u_0,c^*-\delta]$ the generators of $C_u$ have their end
points on $\Cb_\delta$ and for $u\in(c^*-\delta,c^*)$ the
generators of $C_u$ have their end points on $H_{c^*}$. Also, for
$\ub\in[0,\delta)$ the generators of $\Cb_{\ub}$ have their end
points on $H_{c^*}$. It follows that for $(\ub,u)\in D\setminus
A_0$ $S_{\ub,u}$ is a  spacelike surface embedded in $M$.

The optical functions satisfy the eikonal equation:
\begin{equation}
(g^{-1})^{\mu\nu}\partial_\mu u\partial_\nu u=0, \ \ \
(g^{-1})^{\mu\nu}\partial_\mu\ub\partial_\nu\ub=0 \label{1.2}
\end{equation}
It follows that the vectorfields $L^\prime$ and $\Lb^\prime$
defined by:
\begin{equation}
L^{\prime\mu}=-2(g^{-1})^{\mu\nu}\partial_\nu u, \ \ \
\Lb^{\prime\mu}=-2(g^{-1})^{\mu\nu}\partial_\nu\ub \label{1.3}
\end{equation}
are future directed null geodesic vectorfields:
\begin{equation}
\nabla_{L^\prime}L^\prime=0, \ \ \ \nabla_{\Lb^\prime}\Lb^\prime=0
\label{1.4}
\end{equation}
the integral curves of $L^\prime$ being the generators of each
$C_u$ and the integral curves of $\Lb^\prime$ being the generators
of each $\Cb_{\ub}$, both affinely parametrized. For, we have, in
arbitrary local coordinates,
\begin{eqnarray*}
&g_{\lambda\mu}L^{\prime\nu}\nabla_\nu
L^{\prime\mu}=-2L^{\prime\nu}\nabla_\nu\partial_\lambda u
=-2L^{\prime\nu}\nabla_\lambda\partial_\nu u\\
&=4(g^{-1})^{\nu\kappa}\partial_\kappa u\nabla_\lambda\partial_\nu
u =2\partial_\lambda((g^{-1})^{\nu\kappa}\partial_\nu
u\partial_\kappa u)=0
\end{eqnarray*}
and similarly with $u$, $L^\prime$ replaced by $\ub$, $\Lb^\prime$
respectively. Moreover, by virtue of the condition that $2(r_0+u)$
along $\Gamma_0$ coincides with  arc length along $\Gamma_0$ from
$o$, $L^\prime$ along $\Gamma_0$ has projection $T$ along $T$, the
unit future directed tangent field to $\Gamma_0$. Also, $L^\prime$
coincides along $C_{u_0}$ with the vectorfield along $C_o$
previously defined. The inner product $g(L^\prime,\Lb^\prime)$ is
then a negative function on $M$, hence there is a positive
function $\Omega$ on $M$ defined by:
\begin{equation}
-g(L^\prime,\Lb^\prime)=2\Omega^{-2} \label{1.5}
\end{equation}
We then define the normalized null vectorfields $\Lh$ and $\Lbh$
by:
\begin{equation}
\Lh=\Omega L^\prime, \ \ \ \Lbh=\Omega \Lb^\prime \label{1.6}
\end{equation}
These shall be used at each point in $M\setminus\Gamma_0$ as a
basis for the orthogonal complement of the tangent plane to the
surface $S_{\ub,u}$ through that point. We have:
\begin{equation}
g(\Lh,\Lbh)=-2 \label{1.7}
\end{equation}
We proceed to define the null vectorfields $L$ and $\Lb$ by:
\begin{equation}
L=\Omega\Lh=\Omega^2 L^\prime, \ \ \
\Lb=\Omega\Lbh=\Omega^2\Lb^\prime \label{1.8}
\end{equation}
Then from \ref{1.3} and \ref{1.5} we have:
\begin{eqnarray}
&L u=0, \ \ \ L\ub=1\nonumber\\
&\Lb u=1, \ \ \ \Lb\ub=0 \label{1.9}
\end{eqnarray}
Thus the integral curves of $L$ are the generators of each $C_u$
parametrized by $\ub$ and the integral curves of $\Lb$ are the
generators of each $\Cb_{\ub}$ parametrized by $u$. The flow
$\Phi_t$ generated by $L$ is defined on any given $C_u$ as
follows. $\Phi_0$ is the identity mapping. The mapping $\Phi_t$ is
defined at $p\in S_{\ub,u}$ by considering the unique generator of
$C_u$ through $p$: $\Phi_t(p)\in S_{\ub+t,u}$ is the point along
this generator at parameter value $\ub+t$. $\Phi_t$ is a
diffeomorphism of $S_{\ub,u}$ onto $S_{\ub+t,u}$. Similarly, the
flow $\Phib_t$ generated by $\Lb$ is defined on any given
$\Cb_{\ub}$ as follows. $\Phib_0$ is the identity mapping. The
mapping $\Phib_t$ is defined at $p\in S_{\ub,u}$ by considering
the unique generator of $\Cb_{\ub}$ through $p$: $\Phib_t(p)\in
S_{\ub,u+t}$ is the point along this generator at parameter value
$u+t$. $\Phib_t$ is a diffeomorphism of $S_{\ub,u}$ onto
$S_{\ub,u+t}$. Note that:
\begin{equation}
g(L,\Lb)=-2\Omega^2 \label{1.10}
\end{equation}

\section{The Optical Structure Equations}

Let $\xi$ be a 1-form on $M\setminus\Gamma_0$ such that
\begin{equation}
\xi(L)=\xi(\Lb)=0 \label{1.11}
\end{equation}
We then say that $\xi$ is a $S$ 1-form. Conversely, given for each
$(\ub,u)\in D$, $\ub>u$, a 1-form $\xi$ intrinsic to the surface
$S_{\ub,u}$, then along each $S_{\ub,u}$, $\xi$ extends to the
tangent bundle of $M$ over $S_{\ub,u}$ by the conditions
\ref{1.11}. We thus obtain an $S$ 1-form on $M\setminus\Gamma_0$.
Thus a $S$ 1-form may be thought of as the specification of a
1-form intrinsic to $S_{\ub,u}$ for each $(\ub,u)$. Similar
considerations apply to $p$-covariant tensorfields. Thus, a
$p$-covariant $S$ tensorfield is a $p$-covariant tensorfield on
$M\setminus\Gamma_0$ which vanishes if either $L$ or $\Lb$ is
inserted in one of its $p$ entries, and can be thought of as the
specification of a $p$-covariant tensorfield intrinsic to
$S_{\ub,u}$ for each $(\ub,u)$. Next, a $S$ vectorfield is a
vectorfield $X$ defined on $M\setminus\Gamma_0$ such that $X$ is
at each point $x\in M\setminus\Gamma_0$ tangential to the surface
$S_{\ub,u}$ through $x$. Thus a $S$ vectorfield may be thought of
as the specification of a vectorfield intrinsic to $S_{\ub,u}$ for
each $(\ub,u)$. Similarly, a $q$-contravariant $S$ tensorfield is
a $q$-contravariant tensorfield $W$ defined on
$M\setminus\Gamma_0$ such that at each point $x\in
M\setminus\Gamma_0$ $W_x$ belongs to $\otimes^q T_x S_{\ub,u}$,
and can be thought of as a $q$ contravariant vectorfield intrinsic
to $S_{\ub,u}$ for each $(\ub,u)$. Finally, a type $T^q_p$ $S$
tensorfield $\theta$ is a type $T^q_p$ tensorfield defined on
$M\setminus\Gamma_0$ such that at each $x\in M\setminus\Gamma_0$
and each $X_1, ... ,X_p\in T_x M$ we have
$\theta(X_1,...,X_p)\in\otimes^q T_x S_{\ub,u}$ and
$\theta(X_1,...,X_p)=0$ if one of $X_1,...,X_p$ is either $L$ or
$\Lb$. Thus a general type $T^q_p$ $S$ tensorfield may be thought
of as the specification of a type $T^q_p$ tensorfield intrinsic to
$S_{\ub,u}$ for each $(\ub,u)$.

Let then $\xi$ be a $S$ 1-form. We define $\sL_L\xi$ to be, for
each $(\ub,u)\in D$, $\ub>u$, the restriction to $TS_{\ub,u}$ of
${\cal L}_L\xi$, the Lie derivative of $\xi$ with respect to $L$,
a notion intrinsic to each $C_u$. Then $\sL_L\xi$ is an $S$ 1-form
as well. Note that in any case we have $({\cal L}_L\xi)(L)=0$. We
define $\sL_{\Lb}\xi$ to be, for each $(\ub,u)\in D$, $\ub>u$, the
restriction to $TS_{\ub,u}$ of ${\cal L}_{\Lb}\xi$, the Lie
derivative of $\xi$ with respect to $\Lb$, a notion intrinsic to
each $\Cb_{\ub}$. Then $\sL_{\Lb}\xi$ is an $S$ 1-form as well.
Note that in any case we have $({\cal L}_{\Lb}\xi)(\Lb)=0$. For
any $p$-covariant $S$ tensorfield $\xi$, the derivatives
$\sL_L\xi$ and $\sL_{\Lb}\xi$ are similarly defined.

Consider next the case of an $S$ vectorfield $Y$.

\noindent {\bf Lemma 1.1} \ \ \  Let $X$ be a vectorfield defined
along a given $C_u$ and tangential to its $S_{\ub,u}$ sections.
Then the vectorfield $[L,X]$, defined along $C_u$ is also
tangential to the $S_{\ub,u}$ sections. Also, let $X$ be a
vectorfield defined along a given $\Cb_{\ub}$ and tangential to
its $S_{\ub,u}$ sections. Then the vectorfield $[\Lb,X]$, defined
along $\Cb_{\ub}$ is also tangential to the $S_{\ub,u}$ sections.

\noindent {\em Proof:} \  To establish the first part of the
lemma, let $\Psi_s$ be the 1-parameter group generated by $X$.
Then $\Psi_s$ maps each $S_{\ub,u}$ into itself. Since $\Phi_t$
maps $S_{\ub,u}$ into $S_{\ub+t,u}$, it follows that
$\Psi_{-s}\circ\Phi_{-t}\circ\Psi_s\circ\Phi_t$ maps each
$S_{\ub,u}$ into itself. This must coincide to order $ts$ as $t,
s\rightarrow 0$ with $\Omega_{ts}$, where $\Omega_r$ is the
1-parameter group generated by $[L,X]$. It follows that $[L,X]$ is
tangential to the $S_{\ub,u}$. The second part of the lemma is
established in a similar manner.

If then $Y$ is a $S$ vectorfield, according to the above lemma
${\cal L}_L Y=[L,Y]$ and ${\cal L}_{\Lb} Y=[\Lb,Y]$ are also $S$
vectorfields. We thus define the derivatives $\sL_L Y$ and
$\sL_{\Lb} Y$ to be simply ${\cal L}_L Y$ and ${\cal L}_{\Lb} Y$
respectively. The case of any $q$-contravariant $S$ tensorfield
$W$ in the role of $Y$ is formally identical to the case of a $S$
vectorfield, such a tensorfield being expressible as the sum of
tensor products of $S$ vectorfields. Thus the derivatives $\sL_L
W$ and $\sL_{\Lb} W$ are defined to be simply ${\cal L}_L W$ and
${\cal L}_{\Lb} W$ respectively, which are themselves $q$
contravariant $S$ tensorfields.

Consider finally the general case of an arbitrary type $T^q_p$ $S$
tensorfield $\theta$. We may consider, in accordance with the
above, $\theta$ as being, at each surface $S_{\ub,u}$ and at each
point $x\in S_{\ub,u}$, a $p$-linear form in $T_x S_{\ub,u}$ with
values in $\otimes^q T_x S_{\ub,u}$. Then $\sL_L\theta$ is defined
by considering $\theta$ on each $C_u$ extended to $TC_u$ according
to the condition that it vanishes if one of the entries is $L$ and
setting $\sL_L\theta$ equal to the restriction to the $TS_{\ub,u}$
of the usual Lie derivative with respect to $L$ of this extension.
Similarly, $\sL_{\Lb}\theta$ is defined by considering $\theta$ on
each $\Cb_{\ub}$ extended to $T\Cb_{\ub}$ according to the
condition that it vanishes if one of the entries is $\Lb$ and
setting $\sL_{\Lb}\theta$ equal to the restriction to the
$TS_{\ub,u}$ of the usual Lie derivative with respect to $\Lb$ of
this extension. Then $\sL_L\theta$ and $\sL_{\Lb}\theta$ are
themselves type $T^q_p$ $S$ tensorfields.

In the following we use the simplified notation:
\begin{equation}
\sL_L\theta=D\theta, \ \ \ \sL_{\Lb}\theta=\Db\theta \label{1.12}
\end{equation}
In particular, for a function $f$:
\begin{equation}
Df=Lf, \ \ \ \Db f=\Lb f \label{1.13}
\end{equation}

For any function $f$ defined on $M\setminus\Gamma_0$ we denote by
$\sd f$ the $S$ 1-form which is the restriction to each surface
$S_{\ub,u}$ of $df$, the differential of $f$.

\noindent {\bf Lemma 1.2} \ \ \ For any function $f$ defined on
$M\setminus\Gamma_0$ we have:
$$D\sd f=\sd Df, \ \ \ \Db\sd f=\sd\Db f$$

\noindent {\em Proof:} \ Consider the first equality. If we
evaluate each side on $L$ or $\Lb$ then both sides vanish by
definition. On the other hand, if we evaluate on a $S$ vectorfield
$X$ the left hand side is:
$$({\cal L}_L\sd f)(X)=L((\sd f)(X))-(\sd f)([L,X])=L(Xf)-[L,X]f=X(Lf)$$
and the right hand side is:
$$(\sd Lf)(X)=X(Lf)$$
as well. This establishes the first equality. The second equality
is established in a similar manner.

Let $X$ be a $S$ vectorfield and $\xi$ a $S$ 1-form. Then
$\sL_X\xi$ is defined on each $S_{\ub,u}$ as the usual Lie
derivative of $\xi$ as a 1-form on $S_{\ub,u}$ with respect to $X$
as a vectorfield on $S_{\ub,u}$, a notion intrinsic to
$S_{\ub,u}$, which makes no reference to the ambient spacetime. If
$\xi$ is any $p$ covariant $S$ tensorfield, $\sL_X\xi$ is
similarly defined. Consider next the case of another $S$
vectorfield $Y$. Then $\sL_X Y$ is defined to be simply ${\cal
L}_X Y=[X,Y]$, which is itself a $S$ vectorfield. The case of any
$q$ contravariant $S$ tensorfield $W$ in the role of $Y$ is
formally identical to the case of a $S$ vectorfield, such a
tensorfield being expressible as the sum of tensor products of $S$
vectorfields. Thus $\sL_X W$ is defined to be simply ${\cal L}_X
W$, which is itself a $q$ contravariant $S$ tensorfield. Consider
finally the general case of an arbitrary type $T^q_p$ $S$
tensorfield $\theta$. Then $\sL_X\theta$ is again defined on each
$S_{\ub,u}$ as the usual Lie derivative of $\theta$ as a type
$T^q_p$ tensorfield on $S_{\ub,u}$ with respect to $X$ as a
vectorfield on $S_{\ub,u}$, a notion intrinsic to $S_{\ub,u}$,
which makes no reference to the ambient spacetime.

If $\eta$ and $\theta$ are any two $S$ tensorfields of any, not
necessarily the same, type, and we denote by $\eta\cdot\theta$ an
arbitrary contraction of the tensor product $\eta\otimes\theta$, then
the Leibniz rule:
\begin{equation}
\sL_X(\eta\cdot\theta)=(\sL_X\theta)\cdot\theta+\eta\cdot(\sL_X\theta)
\label{1.c1}
\end{equation}
with $X$ an arbitrary $S$ vectorfield, holds, and so do the
Leibniz rules:
\begin{eqnarray}
&&D(\eta\cdot\theta)=(D\eta)\cdot\theta+\eta\cdot(D\theta)\nonumber\\
&&\Db(\eta\cdot\theta)=(\Db\eta)\cdot\theta+\eta\cdot(\Db\theta)
\label{1.c2}
\end{eqnarray}

Let us recall that if $X$ and $Y$ are arbitrary vectorfields and
$\theta$ an arbitrary tensorfield on any manifold, we have:
$${\cal L}_X{\cal L}_Y\theta-{\cal L}_Y{\cal L}_X\theta={\cal L}_{[X,Y]}\theta$$
It then readily follows that if $X$ and $Y$ are $S$ vectorfields
and $\theta$ a $S$ tensorfield of any type, we have:
\begin{equation}
\sL_X\sL_Y\theta-\sL_Y\sL_X\theta=\sL_{[X,Y]}\theta \label{1.c3}
\end{equation}
Moreover, we have the following lemma.

\vspace{5mm}

\noindent{\bf Lemma 1.3} \ \ \ Let $X$ be a $S$ vectorfield and
$\theta$ a $S$ tensorfield of any type. We then have, noting that
by Lemma 1.1 $[L,X]$ and $[\Lb,X]$ are $S$ vectorfields,
$$D\sL_X\theta-\sL_X D\theta=\sL_{[L,X]}\theta$$
and
$$\Db\sL_X\theta-\sL_X\Db\theta=\sL_{[\Lb,X]}\theta$$

\noindent{\em Proof:} \ To establish the first part of the lemma
we may confine ourselves to a given $C_u$, considering the given
$C_u$ to be our manifold. Consider first the case that $\theta$ is
a $S$ 1-form. Then if $ Y$ is any $S$ vectorfield we have
$(\sL_X\theta)(Y)=({\cal L}_X\theta)(Y)$ however
$(\sL_X\theta)(L)=0$ while
$$({\cal L}_X\theta)(L)=-\theta([X,L])$$
Since by \ref{1.9} we have $(d\ub)(L)=1$, while $(d\ub)(Y)=0$, the
$S$ 1-form $\sL_X\theta$ is given on the manifold $C_u$ by:
\begin{equation}
\sL_X\theta={\cal L}_X\theta-fd\ub \ \ \mbox{where} \ \
f=\theta([L,X]) \label{1.c4}
\end{equation}
On the other hand, since $(D\theta)(Y)=({\cal L}_L\theta)(Y)$
while $(D\theta)(L)=0=({\cal L}_L\theta)(L)$, the $S$ 1-form
$D\theta$ is given on the manifold $C_u$ simply by:
\begin{equation}
D\theta={\cal L}_L\theta \label{1.c5}
\end{equation}
Let then again $Y$ be an arbitrary $S$ vectorfield. From
\ref{1.c4} we have:
\begin{equation}
(D\sL_X\theta)(Y)=({\cal L}_L\sL_X\theta)(Y)=({\cal L}_L({\cal
L}_X\theta-fd\ub))(Y)=({\cal L}_L{\cal L}_X\theta)(Y) \label{1.c6}
\end{equation}
the last step by virtue of the facts that $(d\ub)(Y)=0$ and ${\cal
L}_Ld\ub=dL\ub=0$, by \ref{1.9}. Also, from \ref{1.c5},
\begin{equation}
(\sL_X D\theta)(Y)=({\cal L}_X D\theta)(Y)=({\cal L}_X{\cal
L}_L\theta)(Y) \label{1.c7}
\end{equation}
Subtracting \ref{1.c7} from \ref{1.c6} we then obtain:
\begin{equation}
(D\sL_X\theta-\sL_X D\theta)(Y)=({\cal L}_L{\cal L}_X\theta-{\cal
L}_X{\cal L}_L\theta)(Y)=({\cal
L}_{[L,X]}\theta)(Y)=(\sL_{[L,X]}\theta)(Y) \label{1.c8}
\end{equation}
This is the first part of the lemma in the case of a $S$ 1-form.
The case of a $p$ covariant tensorfield is similar. The case of a
$q$ contravariant $S$ tensorfield is immediate, in view of the
fact that in this case the operators $D$ and $\sL_X$ reduce to
${\cal L}_L$ and ${\cal L}_X$ respectively. Finally, the general
case of a type $T^q_p$ $S$ tensorfield $\theta$ follows by taking
arbitrary $S$ vectorfields $Y_1,...,Y_p$, applying the lemma to
the $q$ contravariant tensorfield $\theta(Y_1,...,Y_p)$ and to
each of the vectorfields $Y_1,...,Y_p$, and using the Leibniz
rules \ref{1.c1} and the first of \ref{1.c2}. (In fact this gives
an alternative proof for the case of a $p$ covariant $S$
tensorfield.)

The second part of the lemma is established in an analogous
manner.

\vspace{5mm}

From \ref{1.4}, \ref{1.6}, \ref{1.8} we obtain:
\begin{equation}
\nabla_{\Lh}\Lh=\omh\Lh, \ \ \ \nabla_{\Lbh}\Lbh=\ombh\Lbh
\label{1.14}
\end{equation}
and:
\begin{equation}
\nabla_L L=2\omega L, \ \ \ \nabla_{\Lb}\Lb=2\omb\Lb \label{1.15}
\end{equation}
where:
\begin{equation}
\omh=\Omega^{-1}\omega, \ \ \ \ombh=\Omega^{-1}\omb \label{1.16}
\end{equation}
and:
\begin{equation}
\omega=D\log\Omega, \ \ \ \omb=\Db\log\Omega \label{1.17}
\end{equation}

The tangent hyperplane $T_p C_u$ to a given null hypersurface
$C_u$ at a point $p\in C_u$ consists of all the vectors $X$ at $p$
which are orthogonal to $\Lh_p$:
\begin{equation}
T_p C_u=\{X\in T_p M \ : \ g(X,\Lh_p)=0\} \label{1.18}
\end{equation}
We have $\Lh_p\in T_p C_u$, $\Lh_p$ being orthogonal to itself.
Similarly, the tangent hyperplane $T_p\Cb_{\ub}$ to a given null
hypersurface $\Cb_{\ub}$ at a point $p\in\Cb_{\ub}$ consists of
all the vectors $X$ at $p$ which are orthogonal to $\Lbh_p$:
\begin{equation}
T_p\Cb_{\ub}=\{X\in T_p M \ : \ g(X,\Lbh_p)=0\} \label{1.19}
\end{equation}
and we have $\Lbh_p\in T_p\Cb_{\ub}$, $\Lbh_p$ being orthogonal to
itself.

Thus the induced metrics on $C_u$ and $\Cb_{\ub}$ are degenerate.
This is in fact the definition of a {\em null hypersurface} in a
spacetime manifold $(M,g)$. On the other hand, the {\em induced
metric} $\sg$ on each surface $S_{\ub,u}$, a symmetric 2-covariant
$S$ tensorfield, is positive definite. Any vector $X\in T_p C_u$
can be uniquely decomposed into a vector collinear to $\Lh_p$ and
a vector tangent to the $S_{\ub,u}$ section through $p$ which we
denote by $P X$:
\begin{equation}
X=c\Lh_p+P X, \ \ P X\in T_p S_{\ub,u} \label{1.20}
\end{equation}
Thus we have at each $p\in C_u$ a projection $P$ of $T_p C_u$ onto
$T_p S_{\ub,u}$. If $X, Y$ is a pair of vectors tangent to $C_u$
at $p$, then:
\begin{equation}
g(X,Y)=\sg(P X,P Y) \label{1.21}
\end{equation}
Similarly, any vector $X\in T_p\Cb_{\ub}$ can be uniquely
decomposed into a vector collinear to $\Lbh_p$ and a vector
tangent to the $S_{\ub,u}$ section through $p$ which we denote by
$\Pb X$:
\begin{equation}
X=c\Lbh_p +\Pb X, \ \ \Pb X\in T_p S_{\ub,u} \label{1.22}
\end{equation}
Thus we have at each $p\in\Cb_{\ub}$ a projection $\Pb$ of
$T_p\Cb_{\ub}$ onto $T_p S_{\ub,u}$. If $X, Y$ is a pair of
vectors tangent to $\Cb_{\ub}$ at $p$, then:
\begin{equation}
g(X,Y)=\sg(\Pb X,\Pb Y) \label{1.23}
\end{equation}

The {\em 2nd fundamental form} $\chi$ of a given null hypersurface
$C_u$ is a bilinear form in $T_p C_u$ at each $p\in C_u$ defined
as follows. Let $X, Y\in T_p C_u$. Then:
\begin{equation}
\chi(X,Y)=g(\nabla_X \Lh,Y) \label{1.24}
\end{equation}
The bilinear form $\chi$ is symmetric. For, if we extend $X, Y$ to
vectorfields along $C_u$ which are tangential to $C_u$, we have,
in view of the fact that $g(X,\Lh)=g(Y,\Lh)=0$,
$$\chi(X,Y)-\chi(Y,X)=-g(\Lh,\nabla_X Y)+g(\Lh,\nabla_Y X)=-g(\Lh,[X,Y])=0$$
as $[X,Y]$ is also tangential to $C_u$ hence orthogonal to $\Lh$.
We note that the 2nd fundamental form of $C_u$ is intrinsic to
$C_u$, the vectorfield $\Lh$ being tangential to $C_u$. We have:
\begin{equation}
\chi(X,Y)=\chi(P X, P Y) \label{1.25}
\end{equation}
Thus $\chi$ is a symmetric 2-covariant $S$ tensorfield. We may
consider $\chi$ to be the 2nd fundamental form of the sections
$S_{\ub,u}$ of $C_u$ relative to $C_u$.

The {\em 2nd fundamental form} $\chib$ of a given null
hypersurface $\Cb_{\ub}$ is a bilinear form in $T_p\Cb_{\ub}$ at
each $p\in\Cb_{\ub}$ defined as follows. Let $X, Y\in
T_p\Cb_{\ub}$. Then:
\begin{equation}
\chib(X,Y)=g(\nabla_X \Lbh,Y) \label{1.26}
\end{equation}
The bilinear form $\chib$ is symmetric. For, if we extend $X, Y$
to vectorfields along $\Cb_{\ub}$ which are tangential to
$\Cb_{\ub}$, we have, in view of the fact that
$g(X,\Lbh)=g(Y,\Lbh)=0$,
$$\chib(X,Y)-\chib(Y,X)=-g(\Lbh,\nabla_X Y)+g(\Lbh,\nabla_Y X)=-g(\Lbh,[X,Y])=0$$
as $[X,Y]$ is also tangential to $\Cb_{\ub}$ hence orthogonal to
$\Lbh$. We note that the 2nd fundamental form of $\Cb_{\ub}$ is
intrinsic to $\Cb_{\ub}$, the vectorfield $\Lbh$ being tangential
to $\Cb_{\ub}$. We have:
\begin{equation}
\chib(X,Y)=\chib(\Pb X,\Pb Y) \label{1.27}
\end{equation}
Thus $\chib$ is a symmetric 2-covariant $S$ tensorfield. We may
consider $\chib$ to be the 2nd fundamental form of the sections
$S_{\ub,u}$ of $\Cb_{\ub}$ relative to $\Cb_{\ub}$.

We have:
\begin{equation}
D\sg=2\Omega\chi, \ \ \ \Db\sg=2\Omega\chib \label{1.28}
\end{equation}
To establish the first of \ref{1.28}, let $X, Y$ be vectorfields
along $C_u$ which are tangential to the $S_{\ub,u}$ sections. Then
by Lemma 1.1:
\begin{eqnarray*}
&(D\sg)(X,Y)=({\cal L}_L\sg)(X,Y)=L(\sg(X,Y))-\sg([L,X],Y)-\sg(X,[L,Y])\\
&=L(g(X,Y))-g([L,X],Y)-g(X,[L,Y])=g(\nabla_X L,Y)+g(X,\nabla_Y L)
\end{eqnarray*}
The second of \ref{1.28} is established in a similar manner. We
shall refer to \ref{1.28} as the {\em first variational formulas}.

Let now $X, Y$ be vectors tangent to $S_{\ub,u}$ at a point $p$.
We extend $X, Y$ to Jacobi fields along the generator of $C_u$
through $p$ by the conditions:
\begin{equation}
\nabla_L X=\nabla_X L, \ \ \ \nabla_L Y=\nabla_Y L; \ \ \mbox{that
is:} \ \ [L,X]=[L,Y]=0 \label{1.29}
\end{equation}
Then $X, Y$ are at each point tangential to the $S_{\ub,u}$
section through that point and we have:
\begin{eqnarray}
&(D\chi)(X,Y)=({\cal L}_L\chi)(X,Y)=L(\chi(X,Y))=L(g(\nabla_X \Lh,Y))\nonumber\\
&=\Omega g(\nabla_X\Lh, \nabla_Y\Lh)+g(\nabla_L\nabla_X\Lh,Y)
\label{1.30}
\end{eqnarray}
Now, by the definition of the spacetime curvature and \ref{1.29}:
$$\nabla_L\nabla_X\Lh=\nabla_X\nabla_L\Lh+R(L,X)\Lh$$
hence by \ref{1.14}:
\begin{equation}
g(\nabla_L\nabla_X\Lh,Y)=\omega g(\nabla_X\Lh,Y)+g(R(L,X)\Lh,Y)
=\omega\chi(X,Y)-\Omega R(X,\Lh,Y,\Lh) \label{1.31}
\end{equation}
In regard to the term $g(\nabla_X\Lh,\nabla_Y\Lh)$ we note that
the vectors $\nabla_X \Lh, \nabla_Y \Lh$ are tangential to $C_u$,
for,
$$g(\nabla_X \Lh, \Lh)=\frac{1}{2}X(g(\Lh,\Lh))=0, \ \ \ g(\nabla_Y \Lh, \Lh)=\frac{1}{2}Y(g(\Lh,\Lh))=0$$
Thus \ref{1.21} applies and we obtain:
\begin{equation}
g(\nabla_X\Lh,\nabla_Y\Lh)=\sg(P\nabla_X\Lh,P\nabla_Y\Lh)
\label{1.32}
\end{equation}
Now by the definition \ref{1.24} we have:
\begin{equation}
P\nabla_X\Lh=\chi^\sharp\cdot X, \ \ \
P\nabla_Y\Lh=\chi^\sharp\cdot Y \label{1.33}
\end{equation}
Here, if $\theta$ is any 2-covariant $S$ tensorfield we denote by
$\theta^\sharp$ the $T^1_1$-type $S$ tensorfield, which at each
point $p\in S_{\ub,u}$ is the linear transformation of $T_p
S_{\ub,u}$ corresponding, through the inner product $\sg$ at $p$,
to the bilinear form $\theta$ at $p$. That is, for any pair $X,
Y\in T_p S_{\ub,u}$ we have:
\begin{equation}
\sg(\theta^\sharp\cdot X,Y)=\theta(X,Y) \label{1.34}
\end{equation}
In terms of an arbitrary basis $(e_A \ : \ A=1,2)$ for $T_p
S_{\ub,u}$ we have:
\begin{equation}
\theta^\sharp\cdot e_A=\theta_A^{\sharp\s C} e_C, \ \ \
\theta_A^{\sharp\s C}=\theta_{AB}(\sg^{-1})^{BC} \label{1.37}
\end{equation}
where $\theta_{AB}=\theta(e_A,e_B)$, $\sg_{AB}=\sg(e_A,e_B)$ are
the components of $\theta$, $\sg$ at $p$ in this basis. Thus:
\begin{equation}
\sg(P\nabla_X\Lh,P\nabla_Y\Lh)=\sg(\chi^\sharp\cdot X,
\chi^\sharp\cdot Y)=(\chi\times\chi)(X,Y) \label{1.35}
\end{equation}
Here, if $\theta$ and $\theta^\prime$ are symmetric 2-covariant
$S$ tensorfields we denote by $\theta\times\theta^\prime$ the
2-covariant $S$ tensorfield defined by:
\begin{equation}
(\theta\times\theta^\prime)(X,Y)=\sg(\theta^\sharp\cdot
X,\theta^{\prime\sharp}\cdot Y) \ \ \mbox{: for any pair $X,Y\in
T_p S_{\ub,u}$} \label{1.133}
\end{equation}
In terms of components in an arbitrary local frame field for
$S_{\ub,u}$,
\begin{equation}
(\theta\times\theta^\prime)_{AB}=\sg_{CD}\theta^{\sharp
C}_A\theta^{\prime\sharp D}_B
=(\sg^{-1})^{CD}\theta_{AC}\theta^\prime_{BD} \label{1.134}
\end{equation}
Note that for any symmetric 2-covariant $S$ tensorfield $\theta$
we have:
\begin{equation}
\sg\times\theta=\theta\times\sg=\theta \label{1.a4}
\end{equation}

Let $\alpha$ be the spacetime curvature component which is the
symmetric 2-covariant $S$ tensorfield given by:
\begin{equation}
\alpha(X,Y)=R(X,\Lh,Y,\Lh) \ \ \mbox{: for any pair $X,Y\in T_p
S_{\ub,u}$} \label{1.38}
\end{equation}
In view of \ref{1.31}, \ref{1.32}, \ref{1.35} and \ref{1.38},
equation \ref{1.30} holding for an arbitrary pair $X,Y\in T_p
S_{\ub,u}$ implies:
\begin{equation}
D\chi=\omega\chi+\Omega(\chi\times\chi-\alpha) \label{1.39}
\end{equation}
Let us define $\chi^\prime$, the 2nd fundamental form of $C_u$
with respect to the null geodesic field $L^\prime$, by:
\begin{equation}
\chi^\prime(X,Y)=g(\nabla_X L^\prime,Y) \ \ \mbox{: for any pair
$X,Y\in T_p C_u$} \label{1.40}
\end{equation}
We then have:
\begin{equation}
\chi^\prime=\Omega^{-1}\chi \label{1.41}
\end{equation}
In terms of $\chi^\prime$ equation \ref{1.39} takes the form:
\begin{equation}
D\chi^\prime=\Omega^2 \chi^\prime\times\chi^\prime-\alpha
\label{1.42}
\end{equation}

Let $\alb$ be the spacetime curvature component which is the
symmetric 2-covariant $S$ tensorfield given by:
\begin{equation}
\alb(X,Y)=R(X,\Lbh,Y,\Lbh) \ \ \mbox{: for any pair $X,Y\in T_p
S_{\ub,u}$} \label{1.43}
\end{equation}
Proceeding in an analogous manner, with $\Cb_{\ub}, \Lb, \chib$ in
the roles of $C_u, L, \chi$ respectively we deduce the equation:
\begin{equation}
\Db\chib=\omb\chib+\Omega(\chib\times\chib-\alb) \label{1.44}
\end{equation}
Defining $\chib^\prime$, the 2nd fundamental form of $\Cb_{\ub}$
with respect to the null geodesic field $\Lb^\prime$, by:
\begin{equation}
\chib^\prime(X,Y)=g(\nabla_X \Lb^\prime,Y) \ \ \mbox{: for any
pair $X,Y\in T_p \Cb_{\ub}$} \label{1.45}
\end{equation}
we  have:
\begin{equation}
\chib^\prime=\Omega^{-1}\chib \label{1.46}
\end{equation}
In terms of $\chib^\prime$ equation \ref{1.44} takes the form:
\begin{equation}
\Db\chib^\prime=\Omega^2\chib^\prime\times\chib^\prime-\alb
\label{1.47}
\end{equation}
We shall refer to equations \ref{1.39}, \ref{1.44}, and their
equivalents, equations \ref{1.42}, \ref{1.47}, as the {\em second
variational formulas}. The first of equations \ref{1.28} and
equation \ref{1.39} or \ref{1.42} are {\em propagation equations}
along the generators of each $C_u$, while the second of equations
\ref{1.28} and equation \ref{1.44} or \ref{1.47} are {\em
propagation equations} along the generators of each $\Cb_{\ub}$.

In the following we call {\em conjugation} the formal operation of
exchanging $C_u$ and $L$ with $\Cb_{\ub}$ and $\Lb$ respectively.
We call two entities {\em conjugate} if the definition of one is
obtained from the definition of the other by conjugation. Thus
$\omb$, $\chib$ and $\alb$ are the conjugates of $\omega$, $\chi$
and $\alpha$ respectively. Also, the operator $\Db$ is the
conjugate of the operator $D$. Similarly, we call two formulas
{\em conjugate} if one is obtained from the other by conjugation.
Thus formulas \ref{1.44} and \ref{1.47} are the conjugates of
formulas \ref{1.39} and \ref{1.44} respectively. By virtue of the
symmetry of the geometric structure under investigation, if a
formula holds its conjugate must also hold.

The {\em torsion} of a given surface $S_{\ub,u}$ with respect to
the null hypersurface $C_u$ is the 1-form $\zeta$ on $S_{\ub,u}$
defined at each $p\in S_{\ub,u}$ by:
\begin{equation}
\zeta(X)=\frac{1}{2}g(\nabla_X\Lh,\Lbh) \ \ \mbox{: for any $X\in
T_p S_{\ub,u}$} \label{1.48}
\end{equation}
As this is defined for every $(\ub,u)$, $\zeta$ is defined as a
$S$ 1-form on $M\setminus\Gamma_0$. The entity conjugate to
$\zeta$ is the {\em torsion} of $S_{\ub,u}$ with respect to
$\Cb_{\ub}$, the 1-form $\zeb$ on $S_{\ub,u}$ defined at each
$p\in S_{\ub,u}$ by:
\begin{equation}
\zeb(X)=\frac{1}{2}g(\nabla_X\Lbh,\Lh) \ \ \mbox{: for any $X\in
T_p S_{\ub,u}$} \label{1.49}
\end{equation}
However, by \ref{1.7} we have, simply:
\begin{equation}
\zeb=-\zeta \label{1.50}
\end{equation}

Let $X$ be a vector tangent to $S_{\ub,u}$ at a point $p$. We
extend $X$ to a Jacobi field along the generator of $C_u$ through
$p$ by the condition (see \ref{1.29}):
\begin{equation}
\nabla_L X=\nabla_X L, \ \ \mbox{that is:} \ \ [L,X]=0
\label{1.51}
\end{equation}
Then $X$ is at each point tangential to the $S_{\ub,u}$ section
through that point and we have:
\begin{eqnarray}
&(D\zeta)(X)=({\cal L}_L\zeta)(X)=L(\zeta(X))=\frac{1}{2}L(g(\nabla_X\Lh,\Lbh))\nonumber\\
&=\frac{1}{2}\Omega
g(\nabla_X\Lh,\nabla_{\Lh}\Lbh)+\frac{1}{2}g(\nabla_L\nabla_X\Lh,\Lbh)
\label{1.52}
\end{eqnarray}
Now, by the definition of the spacetime curvature and \ref{1.51}:
$$\nabla_L\nabla_X\Lh=\nabla_X\nabla_L\Lh+R(L,X)\Lh$$
hence by \ref{1.14}:
\begin{eqnarray}
&g(\nabla_L\nabla_X\Lh,\Lbh)=g(\nabla_X\nabla_L\Lh,\Lbh)+g(R(L,X)\Lh,\Lbh)\nonumber\\
&=-2X\omega+2\omega\zeta(X)-\Omega R(X,\Lh,\Lbh,\Lh) \label{1.53}
\end{eqnarray}
In regard to the term $g(\nabla_X\Lh,\nabla_{\Lh}\Lbh)$ we recall
that the vector $\nabla_X\Lh$ is tangential to $C_u$ hence
\ref{1.21} applies. By virtue of \ref{1.33} and the definition
\ref{1.48} we then have:
\begin{equation}
\nabla_X\Lh=-\zeta(X)\Lh+\chi^\sharp\cdot X \label{1.54}
\end{equation}
By \ref{1.14} we have:
\begin{equation}
g(\Lh,\nabla_{\Lh}\Lbh)=-g(\nabla_{\Lh}\Lh,\Lbh)=2\hat{\omega}
\label{1.55}
\end{equation}
To calculate $g(Y,\nabla_{\Lh}\Lbh)$, where $Y$ is a vector at a
point $q\in C_u$ which is tangential to the $S_{\ub,u}$ section
through that point, we first extend $Y$ to a Jacobi field along
the generator of $C_u$ through $q$ by the condition:
$$\nabla_L Y=\nabla_Y L, \ \ \mbox{that is:} \ \ [L,Y]=0$$
Then $Y$ is at each point tangential to the $S_{\ub,u}$ section
through that point and we have:
\begin{equation}
g(Y,\nabla_L\Lbh)=-g(\nabla_Y L,\Lbh)=2Y\Omega-2\Omega\zeta(Y)
\label{1.56}
\end{equation}
by virtue of the definition \ref{1.48}. The equality of the first
and last members of \ref{1.56} holds in particular at $q$. Thus
this equality holds for any vector at a point of $C_u$ which is
tangential to the $S_{\ub,u}$ section through that point, in
particular the vector $\chi^\sharp\cdot X$. Taking also into
account \ref{1.55} we then conclude through \ref{1.54} that:
\begin{equation}
\Omega
g(\nabla_X\Lh,\nabla_{\Lh}\Lbh)=-2\omega\zeta(X)-2\Omega\zeta(\chi^\sharp\cdot
X)+2(\chi^\sharp\cdot X)\Omega \label{1.57}
\end{equation}
For any $T^1_1$ type $S$ tensor $\theta$ and any $S$ 1-form $\xi$
we define the $S$ 1-form $\theta\cdot\xi$ by:
\begin{equation}
(\theta\cdot\xi)(X)=\xi(\theta\cdot X) \ \ \ :\forall X\in
TS_{\ub,u} \label{1.58}
\end{equation}
or, in terms of an arbitrary local frame field $(e_A \ : \ A=1,2)$
for $S_{\ub,u}$,
\begin{equation}
(\theta\cdot\xi)(e_A)=\xi_B\theta^B_A, \ \ \
\theta(e_A)=\theta^B_A e_B, \ \ \ \xi(e_A)=\xi_A \label{1.59}
\end{equation}
The 2nd term on the right in \ref{1.57} is then
$-2\Omega(\chi^\sharp\cdot\zeta)(X)$ while the 3rd term is
$2(\chi^\sharp\cdot\sd\Omega)(X)$.

Let $\beta$ be the spacetime curvature component which is $S$
1-form given by:
\begin{equation}
\beta(X)=\frac{1}{2}R(X,\Lh,\Lbh,\Lh) \ \ \mbox{: for any $X\in
T_p S_{\ub,u}$} \label{1.60}
\end{equation}
In view of \ref{1.53}, \ref{1.57} and the above remarks, equation
\ref{1.52} holding for an arbitrary vector $X\in T_p S_{\ub,u}$
implies:
\begin{equation}
D\zeta=-\sd\omega-\chi^\sharp\cdot(\Omega\zeta-\sd\Omega)-\Omega\beta
\label{1.61}
\end{equation}

Let $\beb$ be the spacetime curvature component which is the $S$
1-form given by:
\begin{equation}
\beb(X)=\frac{1}{2}R(X,\Lbh,\Lbh,\Lh) \ \ \mbox{: for any $X\in
T_p S_{\ub,u}$} \label{1.62}
\end{equation}
The conjugate of $\beta$ is then $-\beb$. In view of \ref{1.50}
the conjugate of equation \ref{1.61} reads:
\begin{equation}
-\Db\zeta=-\sd\omb+\chib^\sharp\cdot(\Omega\zeta+\sd\Omega)+\Omega\beb
\label{1.63}
\end{equation}
Also, the conjugate of equation \ref{1.54} reads:
\begin{equation}
\nabla_X\Lbh=\zeta(X)\Lbh+\chib^\sharp\cdot X \label{1.a1}
\end{equation}

By virtue of Lemma 1.2 and the definitions \ref{1.17} we have:
\begin{equation}
\sd\omega=D\sd\log\Omega, \ \ \ \sd\omb=\Db\sd\log\Omega
\label{1.64}
\end{equation}
Thus defining the $S$ 1-form $\eta$ and its conjugate $\etb$ by:
\begin{equation}
\eta=\zeta+\sd\log\Omega, \ \ \ \etb=-\zeta+\sd\log\Omega
\label{1.65}
\end{equation}
equations \ref{1.61} and \ref{1.63} take the form:
\begin{equation}
D\eta=\Omega(\chi^\sharp\cdot\etb-\beta) \label{1.66}
\end{equation}
\begin{equation}
\Db\etb=\Omega(\chib^\sharp\cdot\eta+\beb) \label{1.67}
\end{equation}
Equation \ref{1.66} is a {\em propagation equation} along the
generators of each $C_u$, while equation \ref{1.67} is a {\em
propagation equation} along the generators of each $\Cb_{\ub}$.
The $S$ 1-form $\eta$ may be considered to be the torsion of
$S_{\ub,u}$ with respect to the null geodesic field $L^\prime$,
for,
\begin{equation}
\eta(X)=\frac{\Omega^2}{2}g(\nabla_X L^\prime,\Lb^\prime) \ \ :
\forall X\in TS_{\ub,u} \label{1.68}
\end{equation}
Similarly, the $S$ 1-form $\etb$ may be considered to be the
torsion of $S_{\ub,u}$ with respect to the null geodesic field
$\Lb^\prime$, for,
\begin{equation}
\etb(X)=\frac{\Omega^2}{2}g(\nabla_X\Lb^\prime,L^\prime) \ \ :
\forall X\in TS_{\ub,u} \label{1.69}
\end{equation}

From \ref{1.55}, \ref{1.56} and the definitions \ref{1.65} we
obtain:
\begin{equation}
\nabla_{\Lh}\Lbh=-\hat{\omega}\Lbh+2\etb^\sharp \label{1.70}
\end{equation}
the conjugate of which reads:
\begin{equation}
\nabla_{\Lbh}\Lh=-\hat{\omb}\Lh+2\eta^\sharp \label{1.71}
\end{equation}
Here $\eta^\sharp$ and $\etb^\sharp$ are the $S$ vectorfields
corresponding, through the metric $\sg$ to the $S$ 1-forms $\eta$
and $\etb$ respectively. In general if $\xi$ is an $S$ 1-form we
denote by $\xi^\sharp$ the $S$ vectorfield defined by:
\begin{equation}
\sg(\xi^\sharp, X)=\xi(X) \ \ : \forall X\in TS_{\ub,u}
\label{1.72}
\end{equation}
Thus we can write:
\begin{equation}
\xi^\sharp=\sg^{-1}\cdot\xi \label{1.73}
\end{equation}
Equations \ref{1.70} and \ref{1.71} imply:
\begin{equation}
\nabla_L\Lb=2\Omega^2\etb^\sharp, \ \ \
\nabla_{\Lb}L=2\Omega^2\eta^\sharp \label{1.74}
\end{equation}
hence:
\begin{equation}
[\Lb,L]=2\Omega^2(\eta^\sharp-\etb^\sharp)=4\Omega^2\zeta^\sharp
\label{1.75}
\end{equation}
This equation clarifies the geometrical significance of the
torsion $\zeta$. It is the {\em obstruction to integrability of
the distribution of timelike planes} which is orthogonal to the
surfaces $S_{\ub,u}$. This distribution is spanned by the
vectorfields $L$, $\Lb$. As a consequence of the commutation
formula \ref{1.75} we have the following lemma.

\vspace{5mm}

\noindent{\bf Lemma 1.4} \ \ \ Let $\theta$ be a $S$ tensorfield
of any type. We have:
$$\Db D\theta-D\Db\theta=\sL_{4\Omega^2\zeta^\sharp}\theta$$

\noindent{\em Proof:} \ Consider first the case of a $S$ 1-form.
For an arbitrary $S$ vectorfield $Y$ we have $(D\theta)(Y)=({\cal
L}_L\theta)(Y)$. Also, $(D\theta)(L)=0=({\cal L}_L\theta)(L)$.
However, $(D\theta)(\Lb)=0$, while
$$({\cal L}_L\theta)(\Lb)=-\theta([L,\Lb])$$
Since by \ref{1.9} we have $(du)(L)=0$, $(du)(\Lb)=1$, while
$(du)(Y)=0$, the $S$ 1-form $D\theta$ is given by:
\begin{equation}
D\theta={\cal L}_L\theta-hdu \ \ \mbox{where} \ \
h=\theta([\Lb,L]) \label{1.c9}
\end{equation}
Similarly, the $S$ 1-form $\Db\theta$ is given by:
\begin{equation}
\Db\theta={\cal L}_{\Lb}\theta+hd\ub \label{1.c10}
\end{equation}
Let again $Y$ be an arbitrary $S$ vectorfield. From \ref{1.c9} we
have:
\begin{equation}
(\Db D\theta)(Y)=({\cal L}_{\Lb}D\theta)(Y)=({\cal L}_{\Lb}({\cal
L}_L\theta-hdu))(Y)=({\cal L}_{\Lb}{\cal L}_L\theta)(Y)
\label{1.c11}
\end{equation}
the last step by virtue of the facts that $(du)(Y)=0$ and ${\cal
L}_{\Lb}du=d\Lb u=0$, by \ref{1.9}. Also, from \ref{1.c10} we
have:
\begin{equation}
(D\Db\theta)(Y)=({\cal L}_L\Db\theta)(Y)=({\cal L}_L({\cal
L}_{\Lb}\theta+hd\ub))(Y)=({\cal L}_L{\cal L}_{\Lb}\theta)(Y)
\label{1.c12}
\end{equation}
the last step by virtue of the facts that $(d\ub)(Y)=0$ and ${\cal
L}_L d\ub=dL\ub=0$, by \ref{1.9}. Subtracting \ref{1.c12} from
\ref{1.c11} we the obtain:
\begin{eqnarray}
&&(\Db D\theta-D\Db\theta)(Y)=({\cal L}_{\Lb}{\cal L}_L\theta-{\cal L}_L{\cal L}_{\Lb}\theta)(Y)=({\cal L}_{[\Lb,L]}\theta)(Y)\nonumber\\
&&\hspace{29mm}=({\cal
L}_{4\Omega^2\zeta^\sharp}\theta)(Y)=(\sL_{4\Omega^2\zeta^\sharp}\theta)(Y)
\label{1.c13}
\end{eqnarray}
by the commutation formula \ref{1.75}. This is the lemma in the
case of a $S$ 1-form. The case of a $p$ covariant tensorfield is
similar. The case of a $q$ contravariant $S$ tensorfield is
immediate, in view of the fact that in this case the operators $D$
and $\Db$ reduce to ${\cal L}_L$ and ${\cal L}_{\Lb}$
respectively. Finally, the general case of a type $T^q_p$ $S$
tensorfield $\theta$ follows by taking arbitrary $S$ vectorfields
$Y_1,...,Y_p$, applying the lemma to the $q$ contravariant
tensorfield $\theta(Y_1,...,Y_p)$ and to each of the vectorfields
$Y_1,...,Y_p$, and using the Leibniz rules \ref{1.c2}. (In fact
this gives an alternative proof for the case of a $p$ covariant
$S$ tensorfield.)

\vspace{5mm}

By \ref{1.15}:
\begin{equation}
\omb=-\frac{1}{4}\Omega^{-2}g(\nabla_{\Lb}\Lb,L) \label{1.76}
\end{equation}
We have:
\begin{eqnarray}
&D\omb=\frac{1}{2}\Omega^{-2}\omega g(\nabla_{\Lb}\Lb,L)-\frac{1}{4}\Omega^{-2}L(g(\nabla_{\Lb}\Lb,L))\label{1.77}\\
&=-2\omega\omb-\frac{1}{4}\Omega^{-2}g(\nabla_L\nabla_{\Lb}\Lb,L)-\frac{1}{4}\Omega^{-2}g(\nabla_{\Lb}\Lb,\nabla_L
L) \nonumber
\end{eqnarray}
By the definition of the spacetime curvature and \ref{1.75} we
have:
\begin{eqnarray}
&\nabla_L\nabla_{\Lb}\Lb-\nabla_{\Lb}\nabla_L\Lb=R(L,\Lb)\Lb+\nabla_{[L,\Lb]}\Lb\nonumber\\
&=R(L,\Lb)\Lb-4\Omega^2\nabla_{\zeta^\sharp}\Lb\label{1.78}
\end{eqnarray}
In view of the fact that by \ref{1.74} $g(\nabla_L\Lb,L)=0$, we
have:
\begin{equation}
g(\nabla_{\Lb}\nabla_L\Lb,L)=-g(\nabla_L\Lb,\nabla_{\Lb}L)=-4\Omega^4(\eta,\etb)
\label{1.79}
\end{equation}
where we have again appealed to \ref{1.74}. Here we denote by
$(\xi,\xi^\prime)$ the inner product of the $S$ 1-forms $\xi$ and
$\xi^\prime$ with respect to the induced metric $\sg$:
\begin{equation}
(\xi,\xi^\prime)=\sg^{-1}(\xi,\xi^\prime)=\sg(\xi^\sharp,\xi^{\prime\sharp})
\label{1.80}
\end{equation}
Also, in the following, we denote by $|\xi|$ the magnitude of the
$S$ 1-form $\xi$ with respect to $\sg$:
\begin{equation}
|\xi|=(\xi,\xi)^{1/2} \label{1.81}
\end{equation}
By \ref{1.49}, \ref{1.50}:
\begin{eqnarray}
&g(\nabla_{\zeta^\sharp}\Lb,L)=-2\Omega^2\zeta(\zeta^\sharp)-2\Omega\zeta^\sharp\Omega\nonumber\\
&=-\Omega^2(\eta-\etb,\eta)\label{1.82}
\end{eqnarray}
Defining the function $\rho$ to be the spacetime curvature
component:
\begin{equation}
\rho=\frac{1}{4}R(\Lbh,\Lh,\Lbh,\Lh) \label{1.83}
\end{equation}
we have:
\begin{equation}
g(R(L,\Lb)\Lb,L)=4\Omega^4\rho \label{1.84}
\end{equation}
In view of \ref{1.78}, \ref{1.79}, \ref{1.82}, \ref{1.84}, we
obtain:
\begin{equation}
-\frac{1}{4}\Omega^{-2}g(\nabla_L\nabla_{\Lb}\Lb,L)=\Omega^2(2(\eta,\etb)-|\eta|^2-\rho)
\label{1.85}
\end{equation}
Substituting in \ref{1.77} and taking also into account the fact
that by \ref{1.15}:
$$-\frac{1}{4}\Omega^{-2}g(\nabla_{\Lb}\Lb,\nabla_L L)=2\omega\omb$$
we then obtain:
\begin{equation}
D\omb=\Omega^2(2(\eta,\etb)-|\eta|^2-\rho) \label{1.86}
\end{equation}
This is a {\em propagation equation} along the generators of each
$C_u$. The conjugate equation
\begin{equation}
\Db\omega=\Omega^2(2(\eta,\etb)-|\etb|^2-\rho) \label{1.87}
\end{equation}
is a {\em propagation equation} along the generators of each
$\Cb_{\ub}$. Note that $\rho$ is conjugate to itself.

Let $\snab$ be the covariant derivative intrinsic to $S_{\ub,u}$
associated to the induced metric $\sg$. The operator $\snab$
satisfies the following. Let $X,Y$ be $S$ vectorfields. Then:
\begin{equation}
\snab_X Y=\Pi\nabla_X Y \label{1.88}
\end{equation}
Here $\Pi$ is the projection operator to the surfaces $S_{\ub,u}$,
given by:
\begin{equation}
\forall V\in T_p(M\setminus\Gamma_0) \ : \ \ \Pi
V=V+\frac{1}{2}g(V,\Lbh)\Lh+\frac{1}{2}g(V,\Lh)\Lbh\in T_p
S_{\ub,u} \label{1.89}
\end{equation}
Let $\sR$ be the intrinsic curvature of $S_{\ub,u}$ and let
$X,Y,Z,W$ be $S$ vectorfields. We have:
\begin{equation}
\sR(W,Z,X,Y)=\sg(W,\sR(X,Y)Z)=g(W,\snab_X\snab_Y Z-\snab_Y\snab_X
Z-\snab_{[X,Y]}Z) \label{1.90}
\end{equation}
Now by \ref{1.88}, \ref{1.89},
\begin{equation}
\snab_Y Z=\nabla_Y Z+\frac{1}{2}g(\nabla_Y
Z,\Lbh)\Lh+\frac{1}{2}g(\nabla_Y Z,\Lh)\Lbh \label{1.91}
\end{equation}
and:
\begin{equation}
\snab_X\snab_Y Z=\Pi\nabla_X\nabla_Y Z+\frac{1}{2}g(\nabla_Y
Z,\Lbh)\Pi\nabla_X\Lh +\frac{1}{2}g(\nabla_Y Z,\Lh)\Pi\nabla_X\Lbh
\label{1.92}
\end{equation}
Substituting \ref{1.92} and an analogous formula with $X$ and $Y$
interchanged, as well as \ref{1.91} with $Y$ replaced by $[X,Y]$,
in \ref{1.90}, and recalling that
$$g(W, \nabla_X\nabla_Y Z-\nabla_Y\nabla_X Z-\nabla_{[X,Y]}Z)=g(W,R(X,Y)Z)=R(W,Z,X,Y)$$
then yields:
\begin{eqnarray*}
&\sR(W,Z,X,Y)=R(W,Z,X,Y)+\frac{1}{2}g(\nabla_Y Z,\Lbh)g(W,\nabla_X\Lh)+\frac{1}{2}g(\nabla_Y Z,\Lh)g(W,\nabla_X\Lbh)\\
&-\frac{1}{2}g(\nabla_X
Z,\Lbh)g(W,\nabla_Y\Lh)-\frac{1}{2}g(\nabla_X
Z,\Lh)g(W,\nabla_Y\Lbh)
\end{eqnarray*}
Noting that
$$g(\nabla_Y Z,\Lbh)=-g(Z,\nabla_Y\Lbh)=-\chib(Y,Z), \ \ \ g(\nabla_Y Z,\Lh)=-g(Z,\nabla_Y\Lh)=-\chi(Y,Z)$$
and similarly with $Y$ replaced by $X$, we then obtain:
\begin{eqnarray}
&\sR(W,Z,X,Y)+\frac{1}{2}\chi(W,X)\chib(Z,Y)+\frac{1}{2}\chib(W,X)\chi(Z,Y)\nonumber\\
&-\frac{1}{2}\chi(W,Y)\chib(Z,X)-\frac{1}{2}\chib(W,Y)\chi(Z,X)=R(W,Z,X,Y)
\label{1.93}
\end{eqnarray}
Let $(e_A \ : \ A=1,2)$ be a local frame field for $S_{\ub,u}$.
Setting $W=e_A$, $Z=e_B$, $X=e_C$, $Y=e_D$, and denoting
$\sR(e_A,e_B,e_C,e_D)=\sR_{ABCD}$, $R(e_A,e_B,e_C,e_D)=R_{ABCD}$,
\ref{1.93} takes the form:
\begin{eqnarray}
&\sR_{ABCD}+\frac{1}{2}\chi_{AC}\chib_{BD}+\frac{1}{2}\chib_{AC}\chi_{BD}\nonumber\\
&-\frac{1}{2}\chi_{AD}\chib_{BC}-\frac{1}{2}\chib_{AD}\chi_{BC}=R_{ABCD}
\label{1.94}
\end{eqnarray}
Now, since $S_{\ub,u}$ is 2-dimensional we have:
\begin{equation}
\sR_{ABCD}=K\seps_{AB}\seps_{CD}=K(\sg_{AC}\sg_{BD}-\sg_{AD}\sg_{BC})
\label{1.95}
\end{equation}
where $K$ is the {\em Gauss curvature} of $S_{\ub,u}$ and
$\seps_{AB}=\seps(e_A,e_B)$ are the components of the area 2-form
$\seps$ of $S_{\ub,u}$. Also, there is a function $f$ such that:
\begin{equation}
R_{ABCD}=f\seps_{AB}\seps_{CD}=f(\sg_{AC}\sg_{BD}-\sg_{AD}\sg_{BC})
\label{1.96}
\end{equation}
Then the whole content of equations \ref{1.94} is contained in the
equation obtained by contacting \ref{1.94} with
$(1/2)(\sg^{-1})^{AC}(\sg^{-1})^{BD}$, namely the equation:
\begin{equation}
K+\frac{1}{2}\mbox{tr}\chi\mbox{tr}\chib-\frac{1}{2}(\chi,\chib)=f
\label{1.97}
\end{equation}
Here if $\theta$ is a 2-covariant $S$ tensorfield we denote by
$\mbox{tr}\theta$ its trace with repsect to $\sg$:
\begin{equation}
\mbox{tr}\theta=(\sg^{-1})^{AB}\theta_{AB} \label{1.98}
\end{equation}
Also, if $\theta$ and $\theta^\prime$ are 2-covariant $S$
tensorfields we denote by $(\theta,\theta^\prime)$ their inner
product with respect to $\sg$:
\begin{equation}
(\theta,\theta^\prime)=(\sg^{-1})^{AC}(\sg^{-1})^{BD}\theta_{AB}\theta^\prime_{CD}
\label{1.99}
\end{equation}
and we denote by $|\theta|$ the magnitude of the 2-covariant $S$
tensorfield $\theta$ with respect to $\sg$:
\begin{equation}
|\theta|=(\theta,\theta)^{1/2} \label{1.a6}
\end{equation}
Note that:
\begin{equation}
(\sg,\theta)=\mbox{tr}\theta \label{1.a7}
\end{equation}
Also, note that if $\theta$ and $\theta^\prime$ are symmetric
2-covariant $S$ tensorfields then:
\begin{equation}
\mbox{tr}(\theta\times\theta^\prime)=\mbox{tr}(\theta^\prime\times\theta)=(\theta,\theta^\prime)
\label{1.a2}
\end{equation}

Now the vacuum Einstein equations read:
\begin{equation}
Ric=0 \label{1.100}
\end{equation}
where $Ric$ is the {\em Ricci curvature} of $(M,g)$, expressed in
an arbitrary local frame field for $(M,g)$ by:
\begin{equation}
Ric_{\mu\nu}=(g^{-1})^{\kappa\lambda}R_{\kappa\mu\lambda\nu}
\label{1.101}
\end{equation}
The reciprocal metric $g^{-1}$ is expressed in terms of the
vectorfields $\Lbh$, $\Lh$ and the reciprocal induced metric
$\sg^{-1}$ by:
\begin{equation}
g^{-1}=-\frac{1}{2}\Lbh\otimes\Lh-\frac{1}{2}\Lh\otimes\Lbh+\sg^{-1}
\label{1.102}
\end{equation}
Complementing the frame field $(e_A \ : \ A=1,2)$ with the
vectorfields $e_3=\Lbh$, $e_4=\Lh$, to obtain a frame field
$(e_\mu \ : \ \mu=1,2,3,4)$ for $M$, we can express the definition
\ref{1.101} in the form:
\begin{equation}
Ric_{\mu\nu}=-\frac{1}{2}R_{3\mu 4\nu}-\frac{1}{2}R_{4\mu
3\nu}+(\sg^{-1})^{CD}R_{C\mu D\nu} \label{1.103}
\end{equation}
In particular, equation $Ric_{AB}=0$ reads:
\begin{equation}
\frac{1}{2}(R_{3A4B}+R_{4A3B})=(\sg^{-1})^{CD}R_{CADB}=f\sg_{AB}
\label{1.104}
\end{equation}
by \ref{1.96}. On the other hand, equation $Ric_{34}=0$ reads:
\begin{equation}
(\sg^{-1})^{AB}R_{A3B4}=\frac{1}{2}R_{4334}=-2\rho \label{1.105}
\end{equation}
by the definition \ref{1.83}. Therefore:
\begin{eqnarray}
&\rho=-\frac{1}{2}(\sg^{-1})^{AB}R_{A3B4}=-\frac{1}{4}(\sg^{-1})^{AB}(R_{A3B4}+R_{B3A4})\nonumber\\
&=-\frac{1}{4}(\sg^{-1})(R_{3A4B}+R_{4A3B})=-f \label{1.106}
\end{eqnarray}
Thus \ref{1.96} becomes:
\begin{equation}
R_{ABCD}=-\rho\seps_{AB}\seps_{CD}=-\rho(\sg_{AC}\sg_{BD}-\sg_{AD}\sg_{BC})
\label{1.107}
\end{equation}
and \ref{1.97} becomes:
\begin{equation}
K+\frac{1}{2}\mbox{tr}\chi\mbox{tr}\chib-\frac{1}{2}(\chi,\chib)=-\rho
\label{1.108}
\end{equation}
This is the {\em Gauss equation} of the embedding of the surfaces
$S_{\ub,u}$ in the spacetime manifold $(M,g)$.

Let now $\xi$ be a $S$ 1-form and $X,Y$ be $S$ vectorfields. Then
$$(\snab_X \xi)(Y)=X(\xi(Y))-\xi(\snab_X Y)=X(\xi(Y))-\xi(\nabla_X Y)=(\nabla_X \xi)(Y)$$
Thus $\snab_X \xi$ is the restriction to $TS_{\ub,u}$ of
$\nabla_X\xi$, and the same holds with $\xi$ being any
$p$-covariant $S$ tensorfield. Let then $X,Y,Z$ be $S$
vectorfields. We have:
\begin{eqnarray}
&(\snab_X\chi)(Y,Z)-(\snab_Y\chi)(X,Z)=(\nabla_X\chi)(Y,Z)-(\nabla_Y\chi)(X,Z)\label{1.109}\\
&=X(\chi(Y,Z))-Y(\chi(X,Z))-\chi([X,Y],Z)-\chi(Y,\nabla_X
Z)+\chi(X,\nabla_Y Z)\nonumber
\end{eqnarray}
Now, by the definition \ref{1.24}:
\begin{eqnarray}
&X(\chi(Y,Z))-Y(\chi(X,Z))=X(g(\nabla_Y\Lh,Z))-Y(g(\nabla_X\Lh,Z))\label{1.110}\\
&=g(\nabla_X\nabla_Y\Lh-\nabla_Y\nabla_X\Lh,Z)+g(\nabla_Y\Lh,\nabla_X Z)-g(\nabla_X\Lh,\nabla_Y Z)\nonumber\\
&=g(Z,R(X,Y)\Lh)+g(\nabla_{[X,Y]}\Lh,Z)+g(\nabla_Y\Lh,\nabla_X
Z)-g(\nabla_X\Lh,\nabla_Y Z)\nonumber
\end{eqnarray}
Substituting in \ref{1.109} and recalling the definition
\ref{1.24} we then obtain:
\begin{equation}
(\snab_X\chi)(Y,Z)-(\snab_Y\chi)(X,Z)=R(Z,\Lh,X,Y)+g(\nabla_Y\Lh,(\nabla_X
Z)^\bot)-g(\nabla_X\Lh,(\nabla_Y Z)^\bot) \label{1.111}
\end{equation}
where for any vector $V\in T_p(M\setminus\Gamma_0)$ we denote by
$V^\bot$ the part of $V$ which lies in the timelike plane which is
the orthogonal complement of $T_p S_{\ub,u}$. From \ref{1.89}:
\begin{equation}
V^\bot=V-\Pi V=-\frac{1}{2}g(V,\Lbh)\Lh-\frac{1}{2}g(V,\Lh)\Lbh
\label{1.112}
\end{equation}
From the definitions \ref{1.24}, \ref{1.26} we have:
$$g(\nabla_X Z,\Lbh)=-g(Z,\nabla_X\Lbh)=-\chib(X,Z), \ \ \ g(\nabla_X Z,\Lh)=-g(Z,\nabla_X\Lh)=-\chi(X,Z)$$
and similarly with $X$ replaced by $Y$. Hence from the definition
\ref{1.48}:
$$g(\nabla_Y\Lh,(\nabla_X Z)^\bot)=\chi(X,Z)\zeta(Y)$$
and similarly with $X$ and $Y$ exchanged. Substituting in
\ref{1.111} we then obtain:
\begin{equation}
(\snab_X\chi)(Y,Z)-(\snab_Y\chi)(X,Z)=R(Z,\Lh,X,Y)+\chi(X,Z)\zeta(Y)-\chi(Y,Z)\zeta(X)
\label{1.113}
\end{equation}
In terms of a local frame field $(e_A \ : A=1,2)$ for $S_{\ub,u}$
complemented by $e_3=\Lbh$, $e_4=\Lh$, setting $X=e_A$, $Y=e_B$,
$Z=e_C$, \ref{1.113} takes the form:
\begin{equation}
\snab_A\chi_{BC}-\snab_B\chi_{AC}=R_{C4AB}+\chi_{AC}\zeta_B-\chi_{BC}\zeta_A
\label{1.114}
\end{equation}
Since $S_{\ub,u}$ is 2-dimensional we have:
\begin{equation}
R_{C4AB}=\xi_C\seps_{AB} \label{1.115}
\end{equation}
where $\xi_C$ are the components of a $S$ 1-form $\xi$. The whole
content of equations \ref{1.114} is then contained in the equation
obtained by contracting with $(\sg^{-1})^{AC}$, namely the
equation:
\begin{equation}
\snab_A\chi^{\sharp
A}_B-e_B(\mbox{tr}\chi)=-\s^*\xi_B+\mbox{tr}\chi\zeta_B-\chi^{\sharp
A}_B\zeta_A \label{1.116}
\end{equation}
Here $\s^*\xi_B$ are the components of the $S$ 1-form $\s^*\xi$,
the left dual, relative to $S_{\ub,u}$, of the $S$ 1-form $\xi$:
\begin{equation}
\s^*\xi=\seps^\sharp\cdot\xi \label{1.117}
\end{equation}
where $\seps^\sharp$ is the $T^1_1$ type $S$ tensorfield
corresponding through $\sg$ to the area 2-form $\seps$:
\begin{equation}
\seps^{\sharp\s C}_A=\seps_{AB}(\sg^{-1})^{BC} \label{1.118}
\end{equation}
Defining the intrinsic divergence of a symmetric 2-covariant $S$
tensorfield $\theta$ to be the $S$ 1-form $\sdiv\theta$ with
components:
\begin{equation}
(\sdiv\theta)_B=\snab_A\theta^{\sharp A}_B, \ \ \ \theta^{\sharp
A}_B=\theta_{BC}(\sg^{-1})^{CA} \label{1.119}
\end{equation}
equation \ref{1.116} reads:
\begin{equation}
\sdiv\chi-\sd\mbox{tr}\chi=-\s^*\xi+\mbox{tr}\chi\zeta-\chi^\sharp\cdot\zeta
\label{1.120}
\end{equation}
Now from \ref{1.103} we have:
\begin{equation}
-\frac{1}{2}R_{344B}+(\sg^{-1})^{CA}R_{C4AB}=Ric_{4B}=0
\label{1.121}
\end{equation}
Thus, in view of the definition \ref{1.60} we have:
\begin{equation}
-\s^*\xi_B=(\sg^{-1})^{CA}R_{C4AB}=-\beta_B \label{1.122}
\end{equation}
hence:
\begin{equation}
\xi=-\s^*\beta, \ \ \
R_{C4AB}=-\s^*\beta_C\seps_{AB}=-\sg_{CA}\beta_B+\sg_{CB}\beta_A
\label{1.123}
\end{equation}
and \ref{1.120} becomes:
\begin{equation}
\sdiv\chi-\sd\mbox{tr}\chi+\chi^\sharp\cdot\zeta-\mbox{tr}\chi\zeta=-\beta
\label{1.124}
\end{equation}
The conjugate equation is:
\begin{equation}
\sdiv\chib-\sd\mbox{tr}\chib-\chib^\sharp\cdot\zeta+\mbox{tr}\chib\zeta=\beb
\label{1.125}
\end{equation}
Also, the conjugate of formula \ref{1.123} reads:
\begin{equation}
R_{C3AB}=\s^*\beb_C\seps_{AB}=\sg_{CA}\beb_B-\sg_{CB}\beb_A
\label{1.126}
\end{equation}
Equation \ref{1.124} and \ref{1.125} are the Codazzi equations of
the embedding of the surfaces $S_{\ub,u}$ in the spacetime
manifold $(M,g)$. In terms of the forms $\chi^\prime$,
$\chib^\prime$, $\eta$, $\etb$, the Codazzi equations take the
form:
\begin{equation}
\sdiv\chi^\prime-\sd\mbox{tr}\chi^\prime+\chi^{\prime\sharp}\cdot\eta-\mbox{tr}\chi^\prime\eta=-\Omega^{-1}\beta
\label{1.127}
\end{equation}
\begin{equation}
\sdiv\chib^\prime-\sd\mbox{tr}\chib^\prime+\chib^{\prime\sharp}\cdot\etb-\mbox{tr}\chib^\prime\etb=\Omega^{-1}\beb
\label{1.128}
\end{equation}

Let again $X,Y$ be vectors tangent to $S_{\ub,u}$ at a point $p$.
We extend $X,Y$ to Jacobi fields along the generator of $C_u$
through $p$ by the conditions \ref{1.29}. We then have:
\begin{eqnarray}
&(D(\Omega\chib))(X,Y)=({\cal L}_L(\Omega\chib))(X,Y)=L(\Omega\chib(X,Y))=L(g(\nabla_X\Lb,Y))\nonumber\\
&=g(\nabla_L\nabla_X\Lb,Y)+g(\nabla_X\Lb,\nabla_Y L) \label{1.129}
\end{eqnarray}
Now, by the definition of the spacetime curvature and \ref{1.29}:
$$\nabla_L\nabla_X\Lb=\nabla_X\nabla_L\Lb+R(L,X)\Lb$$
hence by \ref{1.74}:
\begin{eqnarray}
&g(\nabla_L\nabla_X\Lb,Y)=2g(\nabla_X(\Omega^2\etb^\sharp),Y)+g(R(L,X)\Lb,Y)\label{1.130}\\
&=2\Omega^2((\nabla_X\etb)(Y)+2(X\log\Omega)\etb(Y))-\Omega^2
R(Y,\Lbh,X,\Lh)\nonumber
\end{eqnarray}
Also, by formulas \ref{1.54}, \ref{1.a1}:
\begin{equation}
\nabla_X\Lb=\eta(X)\Lb+\Omega\chib^\sharp\cdot X, \ \ \ \nabla_Y
L=\etb(Y)L+\Omega\chi^\sharp\cdot Y \label{1.131}
\end{equation}
hence:
\begin{equation}
g(\nabla_X\Lb,\nabla_Y
L)=-2\Omega^2\eta(X)\etb(Y)+\Omega^2(\chib\times\chi)(X,Y) \label{1.132}
\end{equation}
Substituting \ref{1.130} and \ref{1.132} in \ref{1.129} and taking
into account the fact that $(\nabla_X\etb)(Y)=(\snab_X\etb)(Y)$,
and also that by the definitions \ref{1.65} we have:
\begin{equation}
2\sd\log\Omega=\eta+\etb \label{1.135}
\end{equation}
we obtain:
\begin{equation}
(D(\Omega\chib))(X,Y)=\Omega^2\{2(\snab_X\etb)(Y)+2\etb(X)\etb(Y)+(\chib\times\chi)(X,Y)-R(Y,\Lbh,X,\Lh)\}
\label{1.136}
\end{equation}
Now the restriction to $TS_{\ub,u}$ of the 2-form
$(1/2)R(\s,\s,\Lbh,\Lh)$ is a $S$ 2-form, hence proportional to
the area 2-form $\seps$. Therefore there is a function $\sigma$
such that :
\begin{equation}
\frac{1}{2}R(X,Y,\Lbh,\Lh)=\sigma\seps(X,Y) \ \ \mbox{: for any
pair $X,Y\in T_p S_{\ub,u}$} \label{1.137}
\end{equation}
The left hand side of \ref{1.137} is equal to
$$\frac{1}{2}(R(X,\Lbh,Y,\Lh)-R(Y,\Lbh,X,\Lh))$$
by the cyclic property of the curvature tensor. On the other hand,
from \ref{1.105}, \ref{1.107} we have:
\begin{equation}
\frac{1}{2}(R(X,\Lbh,Y,\Lh)+R(Y,\Lbh,X,\Lh)=-\rho\sg(X,Y)
\label{1.138}
\end{equation}
We conclude that:
\begin{equation}
R(X,\Lbh,Y,\Lh)=-\rho\sg(X,Y)+\sigma\seps(X,Y) \ \ \mbox{: for any
pair $X,Y\in T_p S_{\ub,u}$} \label{1.139}
\end{equation}
Setting $X=e_A$, $Y=e_B$, the antisymmetric part of \ref{1.136}
reads, since the left hand side is symmetric:
\begin{equation}
\snab_A\etb_B-\snab_B\etb_A=\frac{1}{2}(\chi\times\chib-\chib\times\chi)_{AB}-\sigma\seps_{AB}
\label{1.140}
\end{equation}
The whole content of \ref{1.140} is contained in the equation
obtained from \ref{1.140} by contracting with
$(1/2)\seps^{\sharp\sharp
AB}=(1/2)\seps_{CD}(\sg^{-1})^{AC}(\sg^{-1})^{BD}$, namely the
equation:
\begin{equation}
\scurl\etb=\frac{1}{2}\chi\wedge\chib-\sigma \label{1.141}
\end{equation}
Here we denote by $\scurl\xi$ the intrinsic curl of a $S$ 1-form
$\xi$:
\begin{equation}
\scurl\xi=\frac{1}{2}\seps^{\sharp\sharp
AB}(\snab_A\xi_B-\snab_B\xi_A)=\seps^{\sharp\sharp AB}\snab_A\xi_B
\label{1.142}
\end{equation}
If $\xi$ and $\xi^\prime$ are $S$ 1-forms we denote:
\begin{equation}
\xi\wedge\xi^\prime=\seps^{\sharp\sharp AB}\xi_A\xi^\prime_B
\label{1.a11}
\end{equation}
Also, if $\theta$ and $\theta^\prime$ are symmetric 2-covariant
$S$ tensorfields we denote:
\begin{equation}
\theta\wedge\theta^\prime=\seps^{\sharp\sharp
AB}(\sg^{-1})^{CD}\theta_{AC}\theta^\prime_{BD} \label{1.143}
\end{equation}
Note that:
\begin{equation}
\theta\times\theta^\prime-\theta^\prime\times\theta=\theta\wedge\theta^\prime
\seps \label{1.a3}
\end{equation}
The conjugate of $\sigma$ being $-\sigma$, the conjugate of
equation \ref{1.141} is:
\begin{equation}
\scurl\eta=-\frac{1}{2}\chi\wedge\chib+\sigma \label{1.144}
\end{equation}
Note that by the definitions \ref{1.65}:
\begin{equation}
\scurl\eta=\scurl\zeta=-\scurl\etb \label{1.145}
\end{equation}
In view of \ref{1.138} the symmetric part of \ref{1.136} is the
equation:
\begin{equation}
D(\Omega\chib)=\Omega^2\{\snab\etb+\tilde{\snab\etb}+2\etb\otimes\etb+(1/2)(\chi\times\chib+\chib\times\chi)+\rho\sg\}
\label{1.146}
\end{equation}
The conjugate equation is:
\begin{equation}
\Db(\Omega\chi)=\Omega^2\{\snab\eta+\tilde{\snab\eta}+2\eta\otimes\eta+(1/2)(\chi\times\chib+\chib\times\chi)+\rho\sg\}
\label{1.147}
\end{equation}

Now, by the definitions \ref{1.65}:
$$\eta=-\etb+2\sd\log\Omega, \ \ \ \s\etb=-\eta+2\sd\log\Omega$$
In view of Lemma 1.2 we then have:
\begin{equation}
\Db\eta=-\Db\etb+2\sd\omb, \ \ \ D\etb=-D\eta+2\sd\omega
\label{1.148}
\end{equation}
or, by \ref{1.66}, \ref{1.67},
\begin{equation}
\Db\eta=-\Omega(\chib^\sharp\cdot\eta+\beb)+2\sd\omb \label{1.149}
\end{equation}
\begin{equation}
D\etb=-\Omega(\chi^\sharp\cdot\etb-\beta)+2\sd\omega \label{1.150}
\end{equation}
The first variational formulas \ref{1.28}, the second variational
formulas \ref{1.42}, \ref{1.47}, the propagation equations
\ref{1.66}, \ref{1.67}, \ref{1.86}, \ref{1.87}, the Gauss equation
\ref{1.109}, the Codazzi equations \ref{1.127}, \ref{1.128},
equations \ref{1.141}, \ref{1.145}, and equations \ref{1.146},
\ref{1.147}, together with equations \ref{1.149}, \ref{1.150},
constitute the {\em optical structure equations}.

Let again $(e_A \ : \ A=1,2)$ be an arbitrary local frame field
for the $S_{\ub,u}$. We complement $(e_A \ : \ A=1,2)$, as in the
preceding, with the vectorfields $e_3=\Lbh$, $e_4=\Lh$, to obtain
a frame field $(e_\mu \ : \ \mu=1,2,3,4)$ for $M$. In view of
equations \ref{1.14}, \ref{1.54}, \ref{1.64}, \ref{1.70},
\ref{1.71}, and the relation \ref{1.88}, the connection
coefficients of the frame field $(e_\mu \ : \ \mu=1,2,3,4)$ are
given by the following table:
\begin{eqnarray}
&\nabla_A e_B=\snab_A e_B+\frac{1}{2}\chi_{AB}e_3+\frac{1}{2}\chib_{AB}e_4\nonumber\\
&\nabla_3 e_A=\snab_3 e_A+\eta_A e_3, \ \ \ \nabla_4 e_A=\snab_4 e_A+\etb_A e_4\nonumber\\
&\nabla_A e_3=\chib_A^{\sharp B}e_B+\zeta_A e_3, \ \ \ \nabla_A e_4=\chi_A^{\sharp B}e_B-\zeta_A e_4\nonumber\\
&\nabla_3 e_4=2\eta^{\sharp A}e_A-\ombh e_4, \ \ \ \nabla_4 e_3=2\etb^{\sharp A} e_A-\omh e_3\nonumber\\
&\nabla_3 e_3=\ombh e_3, \ \ \ \nabla_4 e_4=\omh e_4 \label{1.151}
\end{eqnarray}
Here:
$$\eta^{\sharp A}=(\sg^{-1})^{AB}\eta_B, \ \ \ \etb^{\sharp A}=\sg^{-1})^{AB}\etb_B, \ \ \
\chi_A^{\sharp B}=(\sg^{-1})^{BC}\chi_{AC}, \ \ \ \chib_A^{\sharp
B}=(\sg^{-1})^{BC}\chib_{AC}$$ Also, if $X$ is a $S$ vectorfield
we denote:
\begin{equation}
\snab_3 X=\Pi\nabla_{\Lbh}X, \ \ \ \snab_4 X=\Pi\nabla_{\Lh}Y
\label{1.152}
\end{equation}
where $\Pi$ is the projection operator to $S_{\ub,u}$, defined by
\ref{1.89}. If $W$ is any $q$-contravariant $S$ tensorfield,
$\snab_3 W$ and $\snab_4 W$ are defined in a similar manner,
extending the projection operator $\Pi$ at each $x\in
M\setminus\Gamma_0$ to the tensor product $\otimes^q T_x
S_{\ub,u}$. On the other hand if $\xi$ is any $p$-covariant $S$
tensorfield, we define $\snab_3\xi$ and $\snab_4\xi$ to be simply
the restrictions to $TS_{\ub,u}$ of $\nabla_{\Lbh}\xi$ and
$\nabla_{\Lh}\xi$ respectively. Finally, if $\theta$ is an
arbitrary type $T^q_p$ $S$ tensorfield, we define $\snab_3\theta$
and $\snab_4\theta$ to be the projections to $S_{\ub,u}$ of the
restrictions to $TS_{\ub,u}$ of $\nabla_{\Lbh}\theta$ and
$\nabla_{\Lh}\theta$ respectively.

The components of the spacetime volume form $\epsilon$ are given
by:
\begin{equation}
\epsilon_{AB34}=2\seps_{AB} \label{1.153}
\end{equation}

From \ref{1.38}, \ref{1.43}, \ref{1.60}, \ref{1.62}, \ref{1.83},
\ref{1.107}, \ref{1.123}, \ref{1.126}, \ref{1.137}, \ref{1.139},
the components of the curvature tensor of a metric which is a
solution of the vacuum Einstein equations \ref{1.100} are given
by:
\begin{eqnarray}
&R_{A3B3}=\alb_{AB} \ \ \ \ R_{A4B4}=\alpha_{AB}\nonumber\\
&R_{A334}=2\beb_A \ \ \ \ R_{A434}=2\beta_A\nonumber\\
&R_{3434}=4\rho \ \ \ \ R_{AB34}=2\sigma\seps_{AB}\nonumber\\
&R_{A3BC}=\s^*\beb_A\seps_{BC}=\sg_{AB}\beb_C-\sg_{AC}\beb_B \nonumber\\
&R_{A4BC}=-\s^*\beta_A\seps_{BC}=-\sg_{AB}\beta_C+\sg_{AC}\beta_B\nonumber\\
&R_{A4B4}=-\rho\sg_{AB}+\sigma\seps_{AB}\nonumber\\
&R_{ABCD}=-\rho\seps_{AB}\seps_{CD}=-\rho(\sg_{AC}\sg_{BD}-\sg_{AD}\sg_{BC})
\label{1.154}
\end{eqnarray}
Moreover we have:
\begin{equation}
\mbox{tr}\alb=Ric_{33}=0 \ \ \ \ \mbox{tr}\alpha=Ric_{44}=0
\label{1.155}
\end{equation}
Conversely, if the curvature tensor of a metric $g$ is of the form
\ref{1.154} with $\alb$, $\alpha$ verifying \ref{1.155}, then $g$
is a solution of the vacuum Einstein equations.

\section{The Bianchi Equations}

For a solution of the vacuum Einstein equations, the {\em Bianchi
identities}
\begin{equation}
\nabla_{[\alpha} R_{\beta\gamma]\delta\epsilon}:=\nabla_\alpha
R_{\beta\gamma\delta\epsilon} +\nabla_\beta
R_{\gamma\alpha\delta\epsilon}+\nabla_\gamma
R_{\alpha\beta\delta\epsilon}=0 \label{1.156}
\end{equation}
are equivalent to the {\em contracted Bianchi identities}:
\begin{equation}
\nabla^\alpha R_{\alpha\beta\gamma\delta}=0 \ \ \ \
\nabla^\alpha:=(g^{-1})^{\alpha\beta}\nabla_\beta \label{1.157}
\end{equation}
This is a special case of a more general fact to be discussed in
the first section of Chapter 12. We shall presently write down the
components of these equations.

\vspace{5mm}

\noindent {\bf Proposition 1.1} \ \ \ For a solution of the vacuum
Einstein equations the components of $\nabla R$, the covariant
derivative of the curvature tensor, are given by:
\begin{eqnarray*}
\nabla_4 R_{A4B4}&=&\snab_4\alpha_{AB}-2\omh\alpha_{AB}\\
\nabla_3 R_{A3B3}&=&\snab_3\alb_{AB}-2\ombh\alb_{AB}\\
\nabla_3 R_{A4B4}&=&\snab_3\alpha_{AB}+2\ombh\alpha_{AB}-4(\eta\hat{\otimes}\beta)_{AB}\\
\nabla_4 R_{A3B3}&=&\snab_4\alb_{AB}+2\omh\alb_{AB}+4(\etb\hat{\otimes}\beb)_{AB}\\
\nabla_A R_{B4C4}&=&\snab_A\alpha_{BC}-(\chi\hat{\otimes}\beta)_{ABC}+2\zeta_A\alpha_{BC}\\
\nabla_A R_{B3C3}&=&\snab_A\alb_{BC}+(\chib\hat{\otimes}\beb)_{ABC}-2\zeta_A\alb_{BC}\\
\nabla_4 R_{A434}&=&2\snab_4\beta_A-2\omh\beta_A-2\etb^{\sharp B}\alpha_{AB}\\
\nabla_3 R_{A334}&=&2\snab_3\beb_A-2\ombh\beb_A+2\eta^{\sharp B}\alb_{AB}\\
\nabla_3 R_{A434}&=&2\snab_3\beta_A+2\ombh\beta_A-6(\eta_A\rho+\s^*\eta_A\sigma)\\
\nabla_4 R_{A334}&=&2\snab_4\beb_A+2\omh\beb_A+6(\etb_A\rho-\s^*\etb_A\sigma)\\
\nabla_A
R_{B434}&=&2\snab_A\beta_B+2\zeta_A\beta_B-\chib_A^{\sharp
C}\alpha_{BC}
-3(\chi_{AB}\rho+\seps_A^{\sharp\s C}\chi_{CB}\sigma)\\
\nabla_A R_{B334}&=&2\snab_A\beb_B-2\zeta_A\beb_B+\chi_A^{\sharp
C}\alb_{BC}
+3(\chib_{AB}\rho-\seps_A^{\sharp\s C}\chib_{CB}\sigma)\\
\nabla_4 R_{3434}&=&4e_4\rho-8\etb^{\sharp A}\beta_A\\
\nabla_3 R_{3434}&=&4e_3\rho+8\eta^{\sharp A}\beb_A\\
\nabla_A R_{3434}&=&4\sd_A\rho+4(\chi_A^{\sharp B}\beb_B-\chib_A^{\sharp B}\beta_B)\\
\nabla_4 R_{AB34}&=&2\seps_{AB}e_4\sigma+4(\etb_A\beta_B-\etb_B\beta_A)\\
\nabla_3 R_{AB34}&=&2\seps_{AB}e_3\sigma+4(\eta_A\beb_B-\eta_B\beb_A)\\
\nabla_A R_{BC34}&=&2\seps_{BC}\sd_A\sigma+2(\chi_{AB}\beb_C-\chi_{AC}\beb_B)+2(\chib_{AB}\beta_C-\chib_{AC}\beta_B)\\
\nabla_4 R_{A4BC}&=&\sg_{AC}\snab_4\beta_B-\sg_{AB}\snab_4\beta_C
-\omh(\sg_{AC}\beta_B-\sg_{AB}\beta_C)+\eta_B\alpha_{AC}-\eta_C\alpha_{AB}\\
\nabla_3 R_{A3BC}&=&-\sg_{AC}\snab_3\beb_B+\sg_{AB}\snab_3\beb_C
+\ombh(\sg_{AC}\beb_B-\sg_{AB}\beb_C)+\etb_B\alb_{AC}-\etb_C\alb_{AB}\\
\nabla_3 R_{A4BC}&=&\sg_{AC}\snab_3\beta_B-\sg_{AB}\snab_3\beta_C+\ombh(\sg_{AC}\beta_B-\sg_{AB}\beta_C)\\
&\s&\s\s\s+3(-\sg_{AC}\eta_B+\sg_{AB}\eta_C)\rho-(\seps_{AC}\eta_B-\seps_{AB}\eta_C+\seps_{BC}\eta_A)\sigma\\
\nabla_4 R_{A3BC}&=&-\sg_{AC}\snab_4\beb_B+\sg_{AB}\snab_4\beb_C+\omh(-\sg_{AC}\beb_B+\sg_{AB}\beb_C)\\
&\s&\s\s\s+3(-\sg_{AC}\etb_B+\sg_{AB}\etb_C)\rho+(\seps_{AC}\etb_B-\seps_{AB}\etb_C+\seps_{BC}\etb_A)\sigma\\
\nabla_A R_{B4CD}&=&\sg_{BD}\snab_A\beta_C-\sg_{BC}\snab_A\beta_D+\zeta_A(\sg_{BD}\beta_C-\sg_{BC}\beta_D)\\
&\s&+\frac{1}{2}(\chib_{AC}\alpha_{BD}-\chib_{AD}\alpha_{AC})\\
&\s&+\frac{3}{2}(\chi_{AD}\sg_{BC}-\chi_{AC}\sg_{BD})\rho
-(\chi_{AC}\seps_{BD}-\chi_{AD}\seps_{BC}+2\chi_{AB}\seps_{CD})\sigma\\
\nabla_A R_{B3CD}&=&-\sg_{BD}\snab_A\beb_C+\sg_{BC}\snab_A\beb_D+\zeta_A(\sg_{BD}\beb_C-\sg_{BC}\beb_D)\\
&\s&+\frac{1}{2}(\chi_{AC}\alb_{BD}-\chi_{AD}\alb_{AC})\\
&\s&+\frac{3}{2}(\chib_{AD}\sg_{BC}-\chib_{AC}\sg_{BD})\rho
+(\chib_{AC}\seps_{BD}-\chib_{AD}\seps_{BC}+2\chib_{AB}\seps_{CD})\sigma\\
\nabla_4 R_{A3B4}&=&-\sg_{AB}e_4\rho+\seps_{AB}e_4\sigma
+2(\etb_A\beta_B-\etb_B\beta_A+\etb^{\sharp C}\beta_C\sg_{AB})\\
\nabla_3 R_{A3B4}&=&-\sg_{AB}e_3\rho+\seps_{AB}e_3\sigma
+2(\eta_A\beb_B-\eta_B\beb_A-\eta^{\sharp C}\beta_C\sg_{AB})\\
\nabla_A R_{B3C4}&=&-\sg_{BC}\sd_A\rho+\seps_{BC}\sd_A\sigma\\
&\s&+\chib_{AB}\beta_C-\chib_{AC}\beta_B+\sg_{BC}\chib_A^{\sharp
D}\beta_D
-\chi_{AB}\beb_C-\chi_{AC}\beb_B-\sg_{BC}\chi_A^{\sharp D}\beb_D\\
\nabla_4 R_{ABCD}&=&-(\sg_{AC}\sg_{BD}-\sg_{AD}\sg_{BC})e_4\rho\\
&\s&+\etb_A(\beta_C\sg_{DB}-\beta_D\sg_{CB})-\etb_B(\beta_C\sg_{DA}-\beta_D\sg_{CA})\\
&\s&+\etb_C(\beta_A\sg_{BD}-\beta_B\sg_{AD})-\etb_D(\beta_A\sg_{BC}-\beta_B\sg_{AC})\\
\nabla_3 R_{ABCD}&=&-(\sg_{AC}\sg_{BD}-\sg_{AD}\sg_{BC})e_3\rho\\
&\s&-\eta_A(\beb_C\sg_{DB}-\beb_D\sg_{CB})+\eta_B(\beb_C\sg_{DA}-\beb_D\sg_{CA})\\
&\s&-\eta_C(\beb_A\sg_{BD}-\beb_B\sg_{AD})+\eta_D(\beb_A\sg_{BC}-\beb_B\sg_{AC})\\
\nabla_A R_{BCDE}&=&(\sg_{BD}\sg_{CE}-\sg_{BE}\sg_{CD})\sd_A\rho\\
&\s&+\frac{1}{2}\chi_{AB}(-\beb_D\sg_{EC}+\beb_E\sg_{DC})-\frac{1}{2}\chi_{AC}(-\beb_D\sg_{EB}+\beb_E\sg_{DB})\\
&\s&+\frac{1}{2}\chi_{AD}(-\beb_B\sg_{CE}+\beb_C\sg_{BE})-\frac{1}{2}\chi_{AE}(-\beb_B\sg_{CD}+\beb_C\sg_{BD})\\
&\s&+\frac{1}{2}\chib_{AB}(\beta_D\sg_{EC}-\beta_E\sg_{DC})-\frac{1}{2}\chib_{AC}(\beta_D\sg_{EB}-\beta_E\sg_{DB})\\
&\s&+\frac{1}{2}\chib_{AD}(\beta_B\sg_{CE}-\beta_C\sg_{BE})-\frac{1}{2}\chib_{AE}(\beta_B\sg_{CD}-\beta_C\sg_{BD})
\end{eqnarray*}
Here, for $S$ 1-forms $\xi$, $\xi^\prime$ we denote by
$\xi\hat{\otimes}\xi^\prime$ the symmetric trace-free 2-covariant
$S$ tensorfield:
\begin{equation}
\xi\hat{\otimes}\xi^\prime=\xi\otimes\xi^\prime+\xi^\prime\otimes\xi-(\xi,\xi^\prime)\sg
\label{1.158}
\end{equation}
In components,
$$(\xi\hat{\otimes}\xi^\prime)_{AB}=\xi_A\xi^\prime_B+\xi_B\xi^\prime_A-(\xi,\xi^\prime)\sg_{AB}$$
Also, for a symmetric 2-covariant $S$ tensorfield $\theta$ and a
$S$ 1-form $\xi$ we denote by $\theta\hat{\otimes}\xi$ the
3-covariant $S$ tensorfield, symmetric and trace-free in the last
two entries:
\begin{equation}
\theta\hat{\otimes}\xi=\theta\otimes\xi+\tilde{\theta\otimes\xi}-\theta^\sharp\cdot\xi\sg
\label{1.159}
\end{equation}
In components,
$$(\theta\hat{\otimes}\xi)_{ABC}=\theta_{AB}\xi_C+\theta_{AC}\xi_B-\theta_A^{\sharp D}\xi_D\sg_{BC}$$

\noindent {\em Proof:} The proof is straightforward using tables
\ref{1.154}, \ref{1.151} and the definitions of the operators
$\snab$, $\snab_3$, $\snab_4$ as applied to $p$-covariant $S$
tensorfields, which imply in particular that:
$$\snab\sg=0, \ \ \ \snab\seps=0$$
and that:
\begin{equation}
\snab_3\sg=0, \ \ \ \snab_3\seps=0; \ \ \ \ \ \snab_4\sg=0, \ \ \
\snab_4\seps=0 \label{1.160}
\end{equation}
The fact that $\snab_3\seps=0$ is seen most easily by choosing the
frame field $(e_A \ : \ A=1,2)$ to be othonormal. Then
$\sg_{AB}=\delta_{AB}$, $\seps_{AB}$ is the antisymmetric
2-dimensional symbol, and:
$$0=e_3(g(e_A,e_B))=g(\nabla_3 e_A,e_B)+g(e_A,\nabla_3 e_B)=g(\snab_3 e_A,e_B)+g(e_A,\snab_3 e_B)$$
hence, setting:
$$\snab_3 e_A=\Mb_A^B e_B$$
we have:
$$\Mb_A^B+\Mb_B^A=0$$
that is, $\Mb_A^B$ is an antisymmetric matrix. Thus:
$$\Mb_A^B=\Mb\seps_{AB}$$
and we obtain:
\begin{eqnarray*}
&(\snab_3\seps)_{AB}=(\nabla_3\seps)(e_A,e_B)=e_3(\seps_{AB})-\seps(\snab_3 e_A, e_B)-\seps(e_A,\snab_3 e_B)\\
&=-\seps(\snab_3 e_A, e_B)-\seps(e_A,\snab_3
e_B)=-\Mb(\seps_{AC}\seps_{CB}+\seps_{BC}\seps_{AC})=0
\end{eqnarray*}
The fact that $\snab_4\seps=0$ is shown in a similar manner, while
the fact that $\snab_3\sg=\snab_4\sg=0$ follows trivially.

\vspace{5mm}

Using the expression \ref{1.102} for $g^{-1}$ we can write down
the contracted Bianchi identities \ref{1.157} in the form:
\begin{equation}
-\frac{1}{2}\nabla_3 R_{4\beta\gamma\delta}-\frac{1}{2}\nabla_4
R_{3\beta\gamma\delta} +(\sg^{-1})^{AB}\nabla_A
R_{B\beta\gamma\delta}=0 \label{1.161}
\end{equation}
In applying Proposition 1.1, we decompose $\chi$ and $\chib$ into
their trace-free parts, respectively $\chih$ and $\chibh$, and
their traces:
\begin{equation}
\chi=\chih+\frac{1}{2}\sg\mbox{tr}\chi, \ \ \
\chib=\chibh+\frac{1}{2}\sg\mbox{tr}\chib \label{1.162}
\end{equation}
We then make use of the following {\em basic fact}: for any pair of
trace-free symmetric 2-covariant $S$ tensorfields $\theta$,
$\theta^\prime$ we have:
\begin{equation}
\theta\times\theta^\prime+\theta^\prime\times\theta=(\theta,\theta^\prime)\sg
\label{1.163}
\end{equation}
In particular, for any trace-free symmetric 2-covariant $S$
tensorfield $\theta$ we have:
\begin{equation}
\theta\times\theta=\frac{1}{2}|\theta|^2 \sg\label{1.a5}
\end{equation}
We also make use of the fact that if $\theta$ is a trace-free
2-covariant $S$ tensorfield then $\s^*\theta$ defined by:
\begin{equation}
\s^*\theta_{AB}=\seps_A^{\sharp\s C}\theta_{CB} \label{1.164}
\end{equation}
is also a trace-free symmetric 2-covariant $S$ tensorfield. We
then conclude that the components of the contracted Bianchi
identities are the following system of equations:
\begin{eqnarray}
&\snab_3\alpha+\frac{1}{2}\mbox{tr}\chib\alpha+2\ombh\alpha-\snab\oth\beta-(4\eta+\zeta)\oth\beta
+3\chih\rho+3\s^*\chih\sigma=0\nonumber\\
&\snab_4\alb+\frac{1}{2}\mbox{tr}\chi\alb+2\omh\alb+\snab\oth\beb+(4\etb-\zeta)\oth\beb
+3\chibh\rho-3\s^*\chibh\sigma=0\nonumber\\
&\snab_4\beta+2\mbox{tr}\chi\beta-\omh\beta-\sdiv\alpha-(\etb^\sharp+2\zeta^\sharp)\cdot\alpha=0\nonumber\\
&\snab_3\beb+2\mbox{tr}\chib\beb-\ombh\beb+\sdiv\alb+(\eta^\sharp-2\zeta^\sharp)\cdot\alb=0\nonumber\\
&\snab_3\beta+\mbox{tr}\chib\beta+\ombh\beta-\sd\rho-\s^*\sd\sigma-3\eta\rho-3\s^*\eta\sigma-2\chih^\sharp\cdot\beb=0\nonumber\\
&\snab_4\beb+\mbox{tr}\chi\beb+\omh\beb+\sd\rho-\s^*\sd\sigma+3\etb\rho-3\s^*\etb\sigma-2\chibh^\sharp\cdot\beta=0\nonumber\\
&e_4\rho+\frac{3}{2}\mbox{tr}\chi\rho-\sdiv\beta-(2\etb+\zeta,\beta)+\frac{1}{2}(\chibh,\alpha)=0\nonumber\\
&e_3\rho+\frac{3}{2}\mbox{tr}\chib\rho+\sdiv\beb+(2\eta-\zeta,\beb)+\frac{1}{2}(\chih,\alb)=0\nonumber\\
&e_4\sigma+\frac{3}{2}\mbox{tr}\chi\sigma+\scurl\beta+(2\etb+\zeta)\wedge\beta-\frac{1}{2}\chibh\wedge\alpha=0\nonumber\\
&e_3\sigma+\frac{3}{2}\mbox{tr}\chib\sigma+\scurl\beb+(2\eta-\zeta)\wedge\beb+\frac{1}{2}\chih\wedge\alb=0
\label{1.b1}
\end{eqnarray}
Here if $\xi$ is an $S$ 1-form we denote by $\snab\oth\xi$ the
symmetric trace-free 2-covariant $S$ tensorfield:
$$\snab\oth\xi=\snab\xi+\tilde{\snab\xi}-\sg\sdiv\xi$$

We note that if $\xi$ is a $S$ 1-form we have:
\begin{eqnarray*}
&(D\xi)(e_A)=({\cal L}_L\xi)(e_A)=L(\xi(e_A))-\xi([L,e_A])=L(\xi(e_A))-\xi(\nabla_L e_A)+\xi(\nabla_A L)\\
&=(\nabla_L\xi)(e_A)+\xi(\nabla_A
L)=\Omega((\snab_4\xi)(e_A)+\chi_A^{\sharp B}\xi(e_B))
\end{eqnarray*}
or:
\begin{equation}
D\xi=\Omega(\snab_4\xi+\chi^\sharp\cdot\xi) \label{1.165}
\end{equation}
Similarly:
\begin{equation}
\Db\xi=\Omega(\snab_3\xi+\chib^\sharp\cdot\xi) \label{1.166}
\end{equation}
Also, if $\theta$ is a symmetric 2-covariant $S$ tensorfield we
have:
\begin{eqnarray*}
&(D\theta)(e_A,e_B)=({\cal L}_L\theta)(e_A,e_B)=L(\theta(e_A,e_B))-\theta([L,e_A],e_B)-\theta(e_A,[L,e_B])\\
&=L(\theta(e_A,e_B))-\theta(\nabla_L e_A,e_B)-\theta(e_A,\nabla_L e_B)+\theta(\nabla_A L,e_B)+\theta(e_A,\nabla_B L)\\
&=(\nabla_L\theta)(e_A,e_B)+\theta(\nabla_A L,e_B)+\theta(e_A,\nabla_B L)\\
&=\Omega((\snab_4\theta)(e_A,e_B)+\chi_A^{\sharp
C}\theta(e_C,e_B)+\chi_B^{\sharp C}\theta(e_A,e_C))
\end{eqnarray*}
or:
\begin{equation}
D\theta=\Omega(\snab_4\theta+\chi\times\theta+\theta\times\chi)
\label{1.167}
\end{equation}
Similarly:
\begin{equation}
\Db\theta=\Omega(\snab_3\theta+\chib\times\theta+\theta\times\chib)
\label{1.168}
\end{equation}
In the case that $\theta$ is trace-free, decomposing $\chi$ and
$\chib$ as in \ref{1.162} and using \ref{1.163}, we obtain,
denoting by $\Dh\theta$ and $\Dbh\theta$ the trace-free parts of
$D\theta$ and $\Db\theta$ respectively, simply:
\begin{equation}
\Dh\theta=\Omega(\snab_4\theta+\mbox{tr}\chi\theta), \ \ \
\Dbh\theta=\Omega(\snab_3\theta+\mbox{tr}\chib\theta)
\label{1.169}
\end{equation}

Taking into account \ref{1.165}, \ref{1.166} and \ref{1.169}, we
arrive at the form of the Bianchi identities given by the
following proposition.

\vspace{5mm}

\noindent{\bf Proposition 1.2} \ \ \ For a solution of the vaccuum
Einstein equations the Bianchi identities take the form of the
following system of equations:
\begin{eqnarray*}
&\Dbh\alpha-\frac{1}{2}\Omega\mbox{tr}\chib\alpha+2\omb\alpha+\Omega\{-\snab\oth\beta-(4\eta+\zeta)\oth\beta+3\chih\rho+3\s^*\chih\sigma\}=0\\
&\Dh\alb-\frac{1}{2}\Omega\mbox{tr}\chi\alb+2\omega\alb+\Omega\{\snab\oth\beb+(4\etb-\zeta)\oth\beb+3\chibh\rho-3\s^*\chibh\sigma\}=0\\
&D\beta+\frac{3}{2}\Omega\mbox{tr}\chi\beta-\Omega\chih^\sharp\cdot\beta-\omega\beta-\Omega\{\sdiv\alpha+(\etb^\sharp+2\zeta^\sharp)\cdot\alpha\}=0\\
&\Db\beb+\frac{3}{2}\Omega\mbox{tr}\chib\beb-\Omega\chibh^\sharp\cdot\beb-\omb\beb+\Omega\{\sdiv\alb+(\eta^\sharp-2\zeta^\sharp)\cdot\alb\}=0\\
&\Db\beta+\frac{1}{2}\Omega\mbox{tr}\chib\beta-\Omega\chibh\cdot\beta+\omb\beta-\Omega\{\sd\rho+\s^*\sd\sigma+3\eta\rho+3\s^*\eta\sigma
+2\chih^\sharp\cdot\beb\}=0\\
&D\beb+\frac{1}{2}\Omega\mbox{tr}\chi\beb-\Omega\chih^\sharp\cdot\beb+\omega\beb+\Omega\{\sd\rho-\s^*\sd\sigma+3\etb\rho-3\s^*\etb\sigma
-2\chibh^\sharp\cdot\beta\}=0\\
&D\rho+\frac{3}{2}\Omega\mbox{tr}\chi\rho-\Omega\left\{\sdiv\beta+(2\etb+\zeta,\beta)-\frac{1}{2}(\chibh,\alpha)\right\}=0\\
&\Db\rho+\frac{3}{2}\Omega\mbox{tr}\chib\rho+\Omega\left\{\sdiv\beb+(2\eta-\zeta,\beb)+\frac{1}{2}(\chih,\alb)\right\}=0\\
&D\sigma+\frac{3}{2}\Omega\mbox{tr}\chi\sigma+\Omega\left\{\scurl\beta+(2\etb+\zeta,\s^*\beta)-\frac{1}{2}\chibh\wedge\alpha\right\}=0\\
&\Db\sigma+\frac{3}{2}\Omega\mbox{tr}\chib\sigma+\Omega\left\{\scurl\beb+(2\eta-\zeta,\s^*\beb)+\frac{1}{2}\chih\wedge\alb\right\}=0
\end{eqnarray*}

\section{Canonical coordinate systems}

Given local coordinates $(\vartheta^A:A=1,2)$ on a domain
$U\subset S_{0,u_0}$, we can define coordinates
$(\ub,u;\vartheta^A:A=1,2)$ in the domain
\begin{equation}
M_U=\bigcup_{(\ub,u)\in D\setminus
A_0}\Phib_{u-u_0}(\Phi_{\ub}(U))\subset M\setminus\Gamma_0
\label{1.170}
\end{equation}
as follows. Given a point $p\in M_U$, then $p\in S_{\ub,u}$ for a
unique $(\ub,u)\in D\setminus A_0$. The coordinates $(\ub,u)$ are
then assigned to $p$ accordingly. Moreover,
$p=\Phib_{u-u_0}(\Phi_{\ub}(q))$ for a unique $q\in U\subset
S_{0,u_0}$. Let $(\vartheta^1,\vartheta^2)$ be the coordinates of
$q$. The coordinates $(\vartheta^1,\vartheta^2)$ are then assigned
to $p$ accordingly. The coordinates $(\ub,u;\vartheta^A:A=1,2)$ on
$M_U\subset M\setminus\Gamma_0$ are thus canonically associated to
the coordinates $(\vartheta^A:A=1,2)$ on $U\subset S_{0,u_0}$.
Also, if $\{U_1,U_2\}$ is a covering of $S_{0,u_0}$, a sphere in
Euclidean 3-dimensional space, then $\{M_{U_1}, M_{U_2}\}$ is  a
covering of $M\setminus \Gamma_0$. In these ``canonical"
coordinates the vectorfields $\Lb$ and $L$ are given by:
\begin{equation}
\Lb=\frac{\partial}{\partial u}, \ \ \
L=\frac{\partial}{\partial\ub}+b^A\frac{\partial}{\partial\vartheta^A}
\label{1.171}
\end{equation}
where:
\begin{equation}
b^A(\ub,u_0,\vartheta)=0 \ \ \
(\vartheta:=(\vartheta^1,\vartheta^2)) \label{1.172}
\end{equation}
According to the commutation relation \ref{1.75} we have:
\begin{equation}
\frac{\partial b^A}{\partial u}=4\Omega^2\zeta^{\sharp A}
\label{1.173}
\end{equation}
where:
\begin{equation}
\zeta^{\sharp A}=(\sg^{-1})^{AB}\zeta_B \label{1.174}
\end{equation}
and:
\begin{equation}
\sg_{AB}=\sg\left(\frac{\partial}{\partial\vartheta^A},\frac{\partial}{\partial\vartheta^B}\right),
\ \ \
\zeta_A=\zeta\left(\frac{\partial}{\partial\vartheta^A}\right)
\label{1.175}
\end{equation}
are the components of $\sg$, $\zeta$ in the coordinate frame field
$(\partial/\partial\vartheta^A:A=1,2)$ for the $S_{\ub,u}$. Taking
also into account \ref{1.10} and the fact that:
\begin{equation}
g\left(\Lb,\frac{\partial}{\partial\vartheta^A}\right)=g\left(L,\frac{\partial}{\partial\vartheta^A}\right)=0
\label{1.176}
\end{equation}
the vectorfield $\partial/\partial\vartheta^A$ being tangential to
the surfaces $S_{\ub,u}$, we conclude that the spacetime metric
$g$ is given on $M_U$ by:
\begin{equation}
g=-2\Omega^2(du\otimes d\ub+d\ub\otimes
du)+\sg_{AB}(d\vartheta^A-b^A d\ub)\otimes(d\vartheta^B-b^B d\ub)
\label{1.177}
\end{equation}
The volume form $\epsilon=d\mu_g$ of $g$ is then given by:
\begin{equation}
\epsilon=2\Omega^2\sqrt{\mbox{det}\sg}du\wedge d\ub\wedge
d\vartheta^1\wedge d\vartheta^2 \label{1.179}
\end{equation}
or:
\begin{equation}
d\mu_g=2\Omega^2 d\mu_{\sg}\wedge du\wedge d\ub \label{1.180}
\end{equation}
where:
\begin{equation}
d\mu_{\sg}=\seps=\sqrt{\mbox{det}\sg}d\vartheta^1\wedge\vartheta^2
\label{1.181}
\end{equation}
is the area form of $S_{\ub,u}$.

In the Minkowskian region $M_0$ we may introduce Cartesian
coordinates $(x^0,x^1,x^2,x^3)$ such that the $x^0$ axis coincides
with $\Gamma_0$ and $x^0=0$ at the point $e$. The Cartesian
coordinates are then unique up to an $SO(3)$ rotation in
$(x^1,x^2,x^3)$. The surface $S_{0,u}$ on the boundary $\Cb_0$ of
$M_0$, is the sphere of radius $|u|$ in the hyperplane $x^0=u$:
\begin{equation}
|x|=|u|, \ \ \mbox{where} \ \ |x|=\sqrt{(x^1)^2+(x^2)^2+(x^3)^2}
\label{1.182}
\end{equation}
The coordinates $(x^1,x^2,x^3)$ define stereographic coordinates
$(\vartheta^1,\vartheta^2)$ on $S_{0,u_0}$. We have two
stereographic charts, the north polar chart, induced by the
projection from the south pole $q_2=(0,0,-|u_0|)$ to the plane
$x^3=|u_0|$, and the south polar chart, induced by the projection
from the north pole $q_1=(0,0,|u_0|)$ to the plane $x^3=-|u_0|$.
The domain of the north polar chart is $U_1=S_{0,u_0}\setminus
q_2$ and the chart is the mapping of $U_1$ onto $\Re^2$ by
$(x^1,x^2,x^3)\in U_1\mapsto (\vartheta^1,\vartheta^2)\in \Re^2$
by:
\begin{equation}
\vartheta^1=\frac{2x^1}{|x|+x^3}, \ \ \
\vartheta^2=\frac{2x^2}{|x|+x^3} \label{1.183}
\end{equation}
the north pole $p_1$ being mapped to the origin in $\Re^2$. The
domain of the south polar chart is $U_2=S_{0,u_0}\setminus q_1$
and the chart is the mapping of $U_2$ onto $\Re^2$ by
$(x^1,x^2,x^3)\in U_2\mapsto (\vartheta^1,\vartheta^2)\in \Re^2$
by:
\begin{equation}
\vartheta^1=\frac{2x^1}{|x|-x^3}, \ \ \
\vartheta^2=\frac{2x^2}{|x|-x^3} \label{1.184}
\end{equation}
the south pole $p_2$ being mapped to the origin in $\Re^2$. The
image by either chart of the intersection $U_1\bigcap
U_2=S_{0,u_0}\setminus\{p_1,p_2\}$ is $\Re^2\setminus (0,0)$. The
transformation from north polar coordinates to south polar
coordinates is the analytic mapping $f$ of $\Re^2$ onto itself
given by:
\begin{equation}
\vartheta\mapsto f(\vartheta)=\frac{4\vartheta}{|\vartheta|^2}, \
\ \mbox{where} \ \ \vartheta=(\vartheta^1,\vartheta^2), \
|\vartheta|=\sqrt{(\vartheta^1)^2+(\vartheta^2)^2} \label{1.185}
\end{equation}
The equator $x^3=0$ of $S_{0,u_0}$ is mapped by either chart onto
the circle of radius 2 in $\Re^2$. Note that:
\begin{equation}
f=f^{-1} \label{1.186}
\end{equation}
so the inverse transformation, from south polar coordinates to
north polar coordinates is identical in form. Let
$\sigma\in(0,1)$. Then, restricting the domains $U_1$ and $U_2$
to:
\begin{eqnarray}
&&U_{1,\sigma}=\{(x^1,x^2,x^3)\in S_{0,u_0} \ : \ x^3>-\sigma|u_0|\}, \nonumber\\
&&U_{2,\sigma}=\{(x^1,x^2,x^3)\in S_{0,u_0} \ : \
x^3<\sigma|u_0|\} \label{1.187}
\end{eqnarray}
$\{U_{1,\sigma},U_{2,\sigma}\}$ is a covering of $S_{0,u_0}$. The
image by the north polar chart of $U_{1,\sigma}$ and the image by
the south polar chart of $U_{2,\sigma}$ is $D_{2\rho}$, the open
disk of radius $2\rho$ in $\Re^2$, where:
\begin{equation}
\rho=\sqrt{\frac{1+\sigma}{1-\sigma}}>1 \label{1.188}
\end{equation}
The image by either chart of $U_{1,\sigma}\bigcap U_{2,\sigma}$ is
the open annulus:
\begin{equation}
A_\rho=\{\vartheta\in \Re^2 \ : \
\frac{2}{\rho}<|\vartheta|<2\rho\} \label{1.189}
\end{equation}
The mapping $f$ maps $A_\rho$ onto itself and has bounded
derivatives of all orders on $A_\rho$.

The stereographic coordinates on $U_1$, $U_2$ induce canonical
coordinates on $M_{U_1},M_{U_2}$ which are unique up to the
original $SO(3)$ rotation. The induced metric
$\left.\sg\right|_{S_{0,u_0}}$ on $S_{0,u_0}$ is given by:
\begin{equation}
\left.\sg\right|_{S_{0,u_0}}=|u_0|^2\up{\sg} \label{1.190}
\end{equation}
where $\up{\sg}$ is the standard metric on $S^2$. In the canonical
coordinates induced by the stereographic coordinates \ref{1.190}
reads:
\begin{equation}
\sg_{AB}(0,u_0,\vartheta)=|u_0|^2\up{\sg}_{AB}(\vartheta)
\label{1.191}
\end{equation}
where $\up{\sg}_{AB}(\vartheta)$ is given in both charts by:
\begin{equation}
\up{\sg}_{AB}(\vartheta)=\frac{\delta_{AB}}{\left(1+\frac{1}{4}|\vartheta|^2\right)^2}
\label{1.192}
\end{equation}

\chapter{The Characteristic Initial Data}

\section{The characteristic initial data}

In the following we shall show how characteristic initial data is
set up on $C_o$. As mentioned in Chapter 1, the restrictions of
the initial data to $C_o^{r_0}$ are trivial, that is, they
coincide with the data corresponding to a truncated cone in
Minkowski spacetime. We thus confine ourselves to
\begin{equation}
C_o\setminus(C_o^{r_0}\setminus S_o^{r_0}) \label{2.1}
\end{equation}
where the data is non-trivial. From now on we denote by $C_{u_0}$,
the non-trivial part \ref{2.1} of $C_o$, rather than the
whole of $C_o$. Recall that $\Cb_0$ is the outer boundary of the
Minkowskian region $M_0$. On $\Cb_0$ we have the trivial initial
data induced by the Minkowski metric on $M_0$. The two null
hypersurfaces $C_{u_0}$ and $\Cb_0$ intersect at
$S_o^{r_0}=S_{0,u_0}$:
\begin{equation}
\Cb_0\bigcap C_{u_0}=S_{0,u_0} \label{2.2}
\end{equation}
a round sphere of radius $r_0=|u_0|$ in Euclidean 3-dimensional
space, the boundary of $H_{u_0}\bigcap M_0$, a ball of radius
$r_0$ in Euclidean 3-dimensional space.

The characteristic initial data on $C_{u_0}$ is to be the
specification of the conformal geometry of $C_{u_0}$, in the
manner to be presently described. Let us denote again by
$\up{\sg}$ the standard metric on $S^2$, given in stereographic
coordinates by \ref{1.192}. The induced metric
$\left.\sg\right|_{S_{\ub,u_0}} \ : \ub\in[0,\delta]$ may be
expressed the form:
\begin{equation}
\left.\sg\right|_{S_{\ub,u_0}}=(\left.\phi\right|_{S_{\ub,u_0}})^2\left.\sgh\right|_{S_{\ub,u_0}}
\label{2.3}
\end{equation}
where the metric $\left.\sgh\right|_{S_{\ub,u_0}}$ is subject to
the requirement that
$\left.\Phi^*_{\ub}\sgh\right|_{S_{\ub,u_0}}$, a metric on
$S_{0,u_0}$, has the same area form as
$\left.\sg\right|_{S_{0,u_0}}$:
\begin{equation}
d\mu_{\left.\Phi^*_{\ub}\sgh\right|_{S_{\ub,u_0}}}=d\mu_{\left.\sg\right|_{S_{0,u_0}}}
\label{2.6}
\end{equation}
The positive function $\left.\phi\right|_{S_{\ub,u_0}}$ is then
defined by:
\begin{equation}
d\mu_{\left.\sg\right|_{S_{\ub,u_0}}}=(\left.\phi\right|_{S_{\ub,u_0}})^2
d\mu_{\left.\sgh\right|_{S_{\ub,u_0}}} \label{2.4}
\end{equation}
In terms of canonical coordinates on $C_{u_0}$ induced by
stereographic coordinates on $S_{0,u_0}$, the requirement
\ref{2.6} reads, in view of \ref{1.190} - \ref{1.192},
\begin{equation}
\sqrt{\mbox{det}\sgh(\ub,u_0,\vartheta)}=|u_0|^2w^2(\vartheta)
\label{2.7}
\end{equation}
where
\begin{equation}
w(\vartheta)=\frac{1}{1+\frac{1}{4}|\vartheta|^2} \label{2.8}
\end{equation}
in both stereographic charts. Thus, $\sgh$ is given by:
\begin{equation}
\sgh_{AB}(\ub,u_0,\vartheta)=|u_0|^2w^2(\vartheta)m_{AB}(\ub,u_0,\vartheta)
\label{2.9}
\end{equation}
where $m$ satisfies:
\begin{equation}
\mbox{det}m=1 \label{2.10}
\end{equation}
and \ref{2.4} becomes:
\begin{equation}
\sqrt{\mbox{det}\sg(\ub,u_0,\vartheta)}=\phi^2(\ub,u_0,\vartheta)\sqrt{\mbox{det}\sgh(\ub,u_0,\vartheta)}
\label{2.5}
\end{equation}

The characteristic initial data on $C_{u_0}$ consists of the
specification of $\sgh$ or equivalently of $m$ along $C_{u_0}$.
Now, in changing in $U_1\bigcap U_2$ from the north polar
stereogeraphic chart to the south polar stereographic chart, or
vice-versa, the components of $\sgh$ must transform as the
components of a 2-covariant $S$ tensofield. Thus, with
\begin{equation}
\vartheta^\prime=f(\vartheta), \ \ \ \vartheta=f(\vartheta^\prime)
\label{2.11}
\end{equation}
(see \ref{1.185}, \ref{1.186}), we must have:
\begin{equation}
\sgh_{AB}(\ub,u_0,\vartheta)=T^C_A(\vartheta)T^D_B(\vartheta)\sgh^\prime_{CD}(\ub,u_0,\vartheta^\prime)
\label{2.12}
\end{equation}
where:
\begin{equation}
T^C_A(\vartheta)=\frac{\partial
f^C(\vartheta)}{\partial\vartheta^A}=\frac{4}{|\vartheta|^2}\left(\delta_{CA}-2\frac{\vartheta^C\vartheta^A}{|\vartheta|^2}\right)
\label{2.13}
\end{equation}
In terms of matrix notation, with $\tilde{T}$ the transpose of
$T$, the transformation rule \ref{2.12} takes the form:
\begin{equation}
\sgh(\ub,u_0,\vartheta)=\tilde{T}(\vartheta)\sgh^\prime(\ub,u_0,\vartheta^\prime)T(\vartheta)
\label{2.14}
\end{equation}
The components of $\up{\sg}$, given by \ref{1.192}, similarly
satisfy:
\begin{equation}
\up{\sg}(\vartheta)=\tilde{T}(\vartheta)\up{\sg^\prime}(\vartheta^\prime)T(\vartheta)
\label{2.15}
\end{equation}
In view of the definition \ref{2.8}, we conclude that:
\begin{equation}
\tilde{T}(\vartheta)T(\vartheta)=\frac{w^2(\vartheta)}{w^{\prime
2}(\vartheta)}I \label{2.16}
\end{equation}
where:
\begin{equation}
w^\prime(\vartheta)=w(\vartheta^\prime) \ \ : \
\vartheta^\prime=f(\vartheta) \label{2.17}
\end{equation}
and $I$ is the identity matrix. Note also from \ref{2.13} that:
\begin{equation}
\tilde{T}(\vartheta)=T(\vartheta) \label{2.18}
\end{equation}
and:
\begin{equation}
\mbox{det}T<0, \ \ \ |\mbox{det}T|=\frac{w^2}{w^{\prime 2}} \label{2.19}
\end{equation}
(The last follows also from \ref{2.16}.) 
Substituting \ref{2.9} in \ref{2.14} and noting \ref{2.17} we conclude that the
components of $m$ must transform according to the rule:
\begin{equation}
w^2(\vartheta)m(\ub,u_0,\vartheta)=w^{\prime
2}(\vartheta)\tilde{T}(\vartheta)m^\prime(\ub,u_0,\vartheta^\prime)T(\vartheta)
\label{2.20}
\end{equation}
which in view of \ref{2.19} we may write simply as:
\begin{equation}
m=|\mbox{det}T|^{-1}\tilde{T}m^\prime T \label{2.21}
\end{equation}
Taking determinants we obtain
\begin{equation}
\mbox{det}m^\prime=\mbox{det}m \label{2.22}
\end{equation}
Thus the condition \ref{2.10} on $m$ implies the same condition on
$m^\prime$ and vice-versa. The transformation rule \ref{2.21} shows $m$ to be a tensor
density of weight -1. Setting:
\begin{equation}
O=\frac{T}{|\mbox{det}T|^{1/2}}
\label{2.b1}
\end{equation}
The matrix $O$ is according to \ref{2.13} given by:
\begin{equation}
O^C_A(\vartheta)=\delta_{CA}-2\frac{\vartheta^C\vartheta^A}{|\vartheta|^2}
\label{2.b2}
\end{equation}
By \ref{2.16}, \ref{2.19} $O$ is a symmetric orthogonal matrix of determinant $-1$:
\begin{equation}
\tilde{O}O=I, \ \ \ \tilde{O}=O, \ \ \ \mbox{det}O=-1
\label{2.b3}
\end{equation}
In terms of $O$ the transformation rule \ref{2.20} reads:
\begin{equation}
m(\ub,u_0,\vartheta)=\tilde{O}(\vartheta)m^\prime(\ub,u_0,\vartheta^\prime)O(\vartheta)
\label{2.b4}
\end{equation}
which we may write simply as:
\begin{equation}
m=\tilde{O}m^\prime O
\label{2.b5}
\end{equation}

Now the matrix $m$ at a given point on $C_{u_0}$ is a
2-dimensional positive definite symmetric unimodular matrix. Such
a matrix has the form
\begin{equation}
m=\left(\begin{array}{cc}
Z+X&Y\\
Y&Z-X
\end{array}\right)
\ \ \ \mbox{where} \ \ \ Z^2-X^2-Y^2=1 \ \ \mbox{and} \ \ Z>0
\label{2.23}
\end{equation}
Thus these matrices constitute the upper unit hyperboloid $H_1^+$
in $\Re^3$. The exponential mapping $\exp$ is an analytic mapping
defined on the 4-dimensional linear space of all 2-dimensional
matrices $A$:
\begin{equation}
\exp A=I+\sum_{n=1}^\infty\frac{1}{n!}A^n \label{2.24}
\end{equation}
The exponential map restricted to the 2-dimensional subspace
$\hat{S}$ of all symmetric trace-free 2-dimensional matrices is an
analytic diffeomorphism of $\hat{S}$ onto $H^+_1$. A matrix
$A\in\hat{S}$ is of the form:
\begin{equation}
A=\left(\begin{array}{cc}
a&b\\
b&-a
\end{array}\right)
\label{2.25}
\end{equation}
If $O$ is an orthogonal transformation and $A\mapsto \tilde{O}AO$,
then:
$$\exp (\tilde{O}AO)=\tilde{O}(\exp A)O$$
thus if $A$ is brought by a suitable orthogonal transformation to
diagonal form:
$$A=\left(\begin{array}{cc}
\lambda&0\\
0&-\lambda
\end{array}\right)$$
then
$$\exp A=\left(\begin{array}{cc}
e^\lambda&0\\
0&e^{-\lambda}
\end{array}\right)$$

In view of the above, we may express:
\begin{equation}
m=\exp \psi \label{2.26}
\end{equation}
where $\psi$ belongs at each point to $\hat{S}$.

\vspace{5mm}

\noindent{\bf Lemma 2.1} \ \ \ The transformation rule:
$$\psi=\tilde{O}\psi^\prime O$$
for $\psi$, is equivalent to the transformation rule \ref{2.b5}
for $m$.

\noindent{\em Proof:} \ Let $\psi$ transform according to the
above rule. Then by \ref{2.b3} we have:
$$\psi^n=\tilde{O}\psi^\prime(O\tilde{O}\psi^\prime)^{n-1}O
=\tilde{O}\psi^{\prime n}O$$ 
Therefore:
$$\exp \psi=I+\sum_{n=1}^\infty\frac{1}{n!}\psi^n=
\tilde{O}\left(I+\sum_{n=1}^\infty\frac{1}{n!}\psi^{\prime n}\right)O
=\tilde{O}(\exp \psi^\prime)O$$ that is:
$$m=\tilde{O}m^\prime O$$
as required. The converse then follows from the fact that $\exp$
is an analytic diffeomorphism of $\hat{S}$ onto $H_1^+$, so in particular it is one to one.

\vspace{5mm}

Let us denote the coordinates of the north polar chart by
$(\vartheta^1,\vartheta^2)$ and those of the south polar chart by
$(\vartheta^{\prime 1},\vartheta^{\prime 2})$. The transformation rule of
Lemma 2.1 takes the form:
\begin{equation}
\psi(\ub,u_0,\vartheta)=\tilde{O}(\vartheta)\psi^\prime(\ub,u_0,\vartheta^\prime)O(\vartheta)
 \ \ \mbox{: where $\vartheta^\prime=f(\vartheta)$}
\label{2.27}
\end{equation}
Also, noting from \ref{2.13} and \ref{1.185} that:
\begin{equation}
O(\vartheta^\prime)=O(\vartheta) \ \ \mbox{: where $\vartheta^\prime=f(\vartheta)$}
\label{2.28}
\end{equation}
the inverse transformation may be written in similar form (recall
\ref{1.186} and the symmetry of $O$):
\begin{equation}
\psi^\prime(\ub,u_0,\vartheta^\prime)=\tilde{O}(\vartheta^\prime)\psi(\ub,u_0,\vartheta)O(\vartheta^\prime)
 \ \ \mbox{: where $\vartheta=f(\vartheta^\prime)$}
\label{2.29}
\end{equation}

We shall restrict the domain of the north polar chart to
$\oU_{1,\sigma}$ and that of the south polar chart to
$\oU_{2,\sigma}$ (see \ref{1.187}) where $\sigma\in(0,1)$. The
image by the north polar chart of $\oU_{1,\sigma}$ and the image
by the south polar chart of $\oU_{2,\sigma}$ is $\oD_{2\rho}$, the
closed disk of radius $2\rho$ in $\Re^2$, where $\rho$ is given by
\ref{1.188}:
\begin{equation}
\rho=\sqrt{\frac{1+\sigma}{1-\sigma}}>1 \label{2.30}
\end{equation}
This closed disk is the range of the coordinates
$(\vartheta^1,\vartheta^2)$ as well as the range of the
coordinates $(\vartheta^{\prime 1},\vartheta^{\prime 2})$. The
image by either chart of $\oU_{1,\sigma}\bigcap\oU_{2,\sigma}$ is
the closed annulus:
\begin{equation}
\oA_{\rho}=\{\vartheta\in\Re^2 \ :
\frac{2}{\rho}\leq|\vartheta|\leq 2\rho\}=\{\vartheta^\prime\in
\Re^2 \ : \frac{2}{\rho}\leq|\vartheta^\prime|\leq 2\rho\}
\label{2.31}
\end{equation}

By \ref{2.9}, \ref{2.26} and the transformation rule \ref{2.27},
or \ref{2.29}, the specification of a smooth conformal metric $\sgh$ on
$C_{u_0}$ is equivalent to the specification of a pair of smooth
mappings
\begin{equation}
\psi_{u_0} \ : \ [0,\delta]\times\oD_{2\rho}\rightarrow\hat{S}, \
\ \ \psi^\prime_{u_0} \ : \
[0,\delta]\times\oD_{2\rho}\rightarrow\hat{S} \label{2.32}
\end{equation}
such that with
\begin{equation}
\psi(\ub,u_0,\vartheta)=\psi_{u_0}(\ub,\vartheta), \ \ \
\psi^\prime(\ub,u_0,\vartheta^\prime)=\psi^\prime_{u_0}(\ub,\vartheta^\prime)
\label{2.33}
\end{equation}
the transformation rule \ref{2.27}, or \ref{2.29}, holds on
$[0,\delta]\times\oA_{\rho}$. Then with:
\begin{eqnarray}
&&m(\ub,u_0,\vartheta)=\exp(\psi(\ub,u_0,\vartheta)) \ : \ (\ub,\vartheta)\in [0,\delta]\times\oD_{2\rho}\nonumber\\
&&m^\prime(\ub,u_0,\vartheta^\prime)=\exp(\psi^\prime(\ub,u_0,\vartheta^\prime))
\ : \ (\ub,\vartheta^\prime)\in [0,\delta]\times\oD_{2\rho}
\label{2.34}
\end{eqnarray}
the conformal metric $\sgh$ is given, in canonical coordinates, in the
domain in $C_{u_0}$ corresponding to $\oU_{1,\sigma}$ by:
\begin{equation}
\sgh_{AB}(\ub,u_0,\vartheta)=|u_0|^2 w^2(\vartheta)m_{AB}(\ub,u_0,\vartheta)
\label{2.35}
\end{equation}
and in the domain in $C_{u_0}$ corresponding to $\oU_{2,\sigma}$
by:
\begin{equation}
\sgh^\prime_{AB}(\ub,u_0,\vartheta^\prime)=|u_0|^2
w^2(\vartheta^\prime)m^\prime_{AB}(\ub,u_0,\vartheta^\prime)
\label{2.36}
\end{equation}

{\em A  crucial aspect of our setup is the following definition of
the mappings \ref{2.32}.} We consider a pair of smooth mappings
\begin{equation}
\psi_0 \ : \ [0,1]\times\oD_{2\rho}\rightarrow\hat{S}, \ \ \
\psi^\prime_0 \ : \ [0,1]\times\oD_{2\rho}\rightarrow\hat{S}
\label{2.37}
\end{equation}
such that the transformation rule \ref{2.27}, \ref{2.29}, that is:
\begin{equation}
\psi_0(s,\vartheta)=\tilde{O}(\vartheta)\psi^\prime_0(s,\vartheta^\prime)O(\vartheta)
 \ \ \mbox{: where $\vartheta^\prime=f(\vartheta)$}
\label{2.38}
\end{equation}
or:
\begin{equation}
\psi^\prime_0(s,\vartheta^\prime)=\tilde{O}(\vartheta^\prime)\psi_0(s,\vartheta)O(\vartheta^\prime)
 \ \ \mbox{: where $\vartheta=f(\vartheta^\prime)$}
\label{2.39}
\end{equation}
holds for all $(s,\vartheta)\in [0,1]\times\oA_{\rho}$,
$(s,\vartheta^\prime)\in [0,1]\times\oA_{\rho}$. We require the
mappings $\psi_0$ and $\psi^\prime_0$ to extend smoothly by zero
to $s<0$. In other words we require:
\begin{equation}
\left(\frac{\partial^n\psi_0}{\partial s^n}\right)(0,\vartheta)=0
\ : \ \forall \vartheta\in \oD_{2\rho}, \ \ \ \ \
\left(\frac{\partial^n\psi^\prime_0}{\partial
s^n}\right)(0,\vartheta^\prime)=0 \ : \ \forall\vartheta^\prime\in
\oD_{2\rho} \label{2.40}
\end{equation}
for all non-negative integers $n$. This is required so that the
initial data on $C_{u_0}$ are smooth extensions of the trivial
data on $C_o^{r_0}$. We then set:
\begin{equation}
\psi_{u_0}(\ub,\vartheta)=\frac{\delta^{1/2}}{|u_0|}\psi_0\left(\frac{\ub}{\delta},\vartheta\right),
\ \ \
\psi^\prime_{u_0}(\ub,\vartheta^\prime)=\frac{\delta^{1/2}}{|u_0|}\psi^\prime_0\left(\frac{\ub}{\delta},\vartheta^\prime\right)
\label{2.41}
\end{equation}
The mappings $\psi_{u_0}$, $\psi^\prime_{u_0}$ so defined
automatically satisfy the transformation rule \ref{2.27},
\ref{2.29}.

In the following, for any multiplet of real numbers of the form
$\Xi^{B_1...B_q}_{A_1...A_p} \ : \ A_1,...,A_p=1,2;
B_1,...,B_q=1,2$ we denote by $|\Xi|$ the ``magnitude":
\begin{equation}
|\Xi|=\sqrt{\sum_{A_1,...,A_p;B_1,...,B_q}(\Xi^{B_1...B_q}_{A_1...A_p})^2}
\label{2.42}
\end{equation}
Let $\|\psi_0\|_{C^k([0,1]\times\oD_{2\rho})}$ and
$\|\psi^\prime_0\|_{C^k([0,1]\times\oD_{2\rho})}$ be the $C^k$
norms on $[0,1]\times\oD_{2\rho}$ of $\psi_0$ and $\psi^\prime_0$
respectively:
\begin{eqnarray}
&&\|\psi_0\|_{C^k([0,1]\times\oD_{2\rho})}=\max_{m+n\leq k}\sup_{(s,\vartheta)\in[0,1]\times\oD_{2\rho}}\left|\frac{\partial^m}{\partial\vartheta^m}\frac{\partial^n}{\partial s^n}\psi_0(s,\vartheta)\right|\nonumber\\
&&\|\psi^\prime_0\|_{C^k([0,1]\times\oD_{2\rho})}=\max_{m+n\leq
k}\sup_{(s,\vartheta)\in[0,1]\times\oD_{2\rho}}\left|\frac{\partial^m}{\partial\vartheta^m}\frac{\partial^n}{\partial
s^n}\psi^\prime_0(s,\vartheta)\right| \label{2.43}
\end{eqnarray}
Here, we denote by $\partial^m/\partial\vartheta^m$ the multiplet
of differential operators:
$$\frac{\partial^m}{\partial\vartheta^{A_1}...\partial\vartheta^{A_m}}$$
We then set:
\begin{equation}
M_k=\max\{\|\psi_0\|_{C^k([0,1]\times\oD_{2\rho})},\|\psi^\prime_0\|_{C^k([0,1]\times\oD_{2\rho})}\}
\label{2.44}
\end{equation}

Let $\|m-I\|_{C^k_\delta([0,\delta]\times\oD_{2\rho})}$ and
$\|m^\prime-I\|_{C^k_\delta([0,\delta]\times\oD_{2\rho})}$ be the
following weighted $C^k$ norms on $[0,\delta]\times\oD_{2\rho}$ of
$m-I$ and $m^\prime-I$ respectively:
\begin{eqnarray}
&&\|m-I\|_{C^k_\delta([0,\delta]\times\oD_{2\rho})}=\max_{m+n\leq
k}\sup_{(\ub,\vartheta)\in[0,\delta]\times\oD_{2\rho}}\delta^n\left|\frac{\partial^m}{\partial\vartheta^m}\frac{\partial^n}{\partial\ub^n}(m-I)(\ub,u_0,\vartheta)
\right|\nonumber\\
&&\|m^\prime-I\|_{C^k_\delta([0,\delta]\times\oD_{2\rho})}=\max_{m+n\leq k}\sup_{(\ub,\vartheta)\in[0,\delta]\times\oD_{2\rho}}\delta^n\left|\frac{\partial^m}{\partial\vartheta^m}\frac{\partial^n}{\partial\ub^n}(m-I)(\ub,u_0,\vartheta)\right|\nonumber\\
&&\label{2.45}
\end{eqnarray}
The definitions \ref{2.40}, \ref{2.32} and \ref{2.33} imply that
$m$ and $m^\prime$ are smooth functions on
$[0,\delta]\times\oD_{2\rho}$ and for each non-negative integer
$k$ we have:
\begin{equation}
\max\{\|m-I\|_{C^k_\delta([0,\delta]\times\oD_{2\rho})},\|m^\prime-I\|_{C^k_\delta([0,\delta]\times\oD_{2\rho})}\}
\leq \delta^{1/2}|u_0|^{-1}F_k(M_k) \label{2.46}
\end{equation}
where $F_k$ is a non-negative non-decreasing continuous function
on the non-negative real line.

Let $\xi$ be a $C^k$ $T^q_p$ type $S$ tensorfield on $C_{u_0}$,
represented in the north polar stereographic chart by the $C^k$
component functions
$\xi^{B_1...B_q}_{A_1...A_p}(\ub,u_0,\vartheta)$ and in the south
polar stereographic chart by the $C^k$ component functions
$\xi^{\prime B_1...B_q}_{A_1...A_p}(\ub,u_0,\vartheta)$. We define
the weighted $C^k$ norms of the component functions by:
\begin{eqnarray}
&&\|\xi\|_{C^k_\delta([0,\delta]\times\oD_{2\rho})}=\max_{m+n\leq k}\sup_{(\ub,\vartheta)\in[0,\delta]\times\oD_{2\rho}}\left(\delta^n\left|\frac{\partial^m}{\partial\vartheta^m}\frac{\partial^n}{\partial\ub^n}\xi(\ub,u_0,\vartheta)\right|\right)\nonumber\\
&&\|\xi^\prime\|_{C^k_\delta([0,\delta]\times\oD_{2\rho})}=\max_{m+n\leq k}\sup_{(\ub,\vartheta)\in[0,\delta]\times\oD_{2\rho}}\left(\delta^n\left|\frac{\partial^m}{\partial\vartheta^m}\frac{\partial^n}{\partial\ub^n}\xi^\prime(\ub,u_0,\vartheta)\right|\right)\nonumber\\
&&\label{2.47}
\end{eqnarray}
We then define the weighted $C^k$ norm of $\xi$ on $C_{u_0}$ by:
\begin{equation}
\|\xi\|_{C^k_\delta(C_{u_0})}=\max\{\|\xi\|_{C^k_\delta([0,\delta]\times\oD_{2\rho})},
\|\xi^\prime\|_{C^k_\delta([0,\delta]\times\oD_{2\rho})}\}
\label{2.48}
\end{equation}
Moreover, we say that a smooth $T^q_p$ type $S$ tensorfield $\xi$
on $C_{u_0}$ is ${\cal M}_l(\delta^r|u_0|^s)$, for real numbers
$r,s$ and non-negative integer $l$,
\begin{equation}
\xi={\cal M}_l(\delta^r|u_0|^s) \label{2.49}
\end{equation}
if for every non-negative integer $k$ we have:
\begin{equation}
\|\xi\|_{C^k_\delta(C_{u_0})}\leq \delta^r|u_0|^s F_k(M_{k+l})
\label{2.50}
\end{equation}
where $F_k$ is a non-negative non-decreasing continuous function
on the non-negative real line.

It follows from \ref{2.46} and \ref{2.9}, in view of \ref{2.8} and
\ref{1.192}, that $\sgh$ is a smooth metric on $C_{u_0}$ and for
each non-negative integer $k$ we have:
\begin{equation}
\||u_0|^{-2}\sgh-\up{\sg}\|_{C^k_\delta(C_{u_0})}\leq
\delta^{1/2}|u_0|^{-1}F_k(M_k) \label{2.51}
\end{equation}
where $F_k$ is a non-negative non-decreasing continuous function
on the non-negative real line. Thus, in accordance with the above
definition we have:
\begin{equation}
|u_0|^{-2}\sgh-\up{\sg}={\cal M}_0(\delta^{1/2}|u_0|^{-1})
\label{2.52}
\end{equation}
Let $\hat{K}$ be the Gauss curvature of $\sgh$. Then $\hat{K}$ is
a smooth function on $C_{u_0}$ and from \ref{2.51} we deduce that
for every non-negative integer $k$ we have:
\begin{equation}
\|\hat{K}-|u_0|^{-2}\|_{C^k_\delta(C_{u_0})}\leq\delta^{1/2}|u_0|^{-3}F_k(M_{k+2})
\label{2.53}
\end{equation}
where $F_k$ is a (different) non-negative non-decreasing
continuous function on the non-negative real line. Thus, in
accordance with the above definition we have:
\begin{equation}
\hat{K}-|u_0|^{-2}={\cal M}_2(\delta^{1/2}|u_0|^{-3}) \label{2.54}
\end{equation}

Consider now equation \ref{2.3}. In canonical coordinates we have:
\begin{equation}
\sg_{AB}(\ub,u_0,\vartheta)=\phi^2(\ub,u_0,\vartheta)\sgh_{AB}(\ub,u_0,\vartheta)
\label{2.55}
\end{equation}
Differentiating with respect to $\ub$ we obtain:
\begin{equation}
\frac{\partial\sg_{AB}}{\partial\ub}=\phi^2\frac{\partial\sgh_{AB}}{\partial\ub}+2\phi\frac{\partial
f}{\partial\ub}\sgh_{AB} \label{2.56}
\end{equation}
Now, along $C_{u_0}$ we have:
\begin{equation}
\Omega=1 \ \ \mbox{: along $C_{u_0}$} \label{2.57}
\end{equation}
$\ub$ coinciding up to the additive constant $r_0$ with the affine
parameter $s$ along the generators of $C_{u_0}$, hence
$$1=L\ub=\Omega^2 L^\prime\ub=\Omega^2 L^\prime s=\Omega^2 \ \ \mbox{: along $C_{u_0}$}$$
Thus, the first of equations \ref{1.28} becomes along $C_{u_0}$:
\begin{equation}
D\sg=2\chi \label{2.58}
\end{equation}
or, in canonical coordinates,
\begin{equation}
\frac{\partial}{\partial\ub}\sg_{AB}=2\chi_{AB}=2\chih_{AB}+\mbox{tr}\chi\sg_{AB}
\label{2.59}
\end{equation}
in terms of the decomposition \ref{1.162}. Now, from \ref{2.9} we
have:
\begin{equation}
\frac{\partial\sgh}{\partial\ub}(\ub,u_0,\vartheta)=|u_0|^2w^2(\vartheta)\frac{\partial
m_{AB}}{\partial\ub}(\ub,u_0,\vartheta) \label{2.60}
\end{equation}
while by \ref{2.10}:
\begin{equation}
\mbox{det}\sgh(\ub,u_0,\vartheta)=|u_0|^4 w^4(\vartheta)
\label{2.61}
\end{equation}
hence:
\begin{equation}
(\sgh^{-1})^{AB}\frac{\partial\sgh_{AB}}{\partial\ub}=\frac{\partial}{\partial\ub}\mbox{log
det}\sgh=0 \label{2.62}
\end{equation}
Comparing then \ref{2.59} with \ref{2.56} we conclude that:
\begin{equation}
\chih_{AB}=\frac{1}{2}\phi^2\frac{\partial\sgh}{\partial\ub} \ \ \
\ \ \mbox{tr}\chi=\frac{2}{\phi}\frac{\partial\phi}{\partial\ub}
\label{2.63}
\end{equation}
Denoting pointwise magnitudes of $S$ tensorfields relative to
$\sg$ by $|\s|_{\sg}$ to avoid confusion with the notation
\ref{2.42}, let us define on $C_{u_0}$ the function:
\begin{equation}
e=\frac{1}{2}|\chih|^2_{\sg}=\frac{1}{2}(\sg^{-1})^{AC}(\sg^{-1})^{BD}\chih_{AB}\chih_{CD}
\label{2.64}
\end{equation}
By \ref{2.55} and the first of \ref{2.63} we obtain:
\begin{equation}
e=\frac{1}{8}(\sgh^{-1})^{AC}(\sgh^{-1})^{BD}\frac{\partial\sgh_{AB}}{\partial\ub}\frac{\partial\sgh_{CD}}{\partial\ub}
\label{2.65}
\end{equation}
This is a smooth non-negative function on $C_{u_0}$ which depends
only on $\sgh$. From \ref{2.51} we deduce that for every
non-negative integer $k$ we have:
\begin{equation}
\|e\|_{C^k_\delta(C_{u_0})}\leq \delta^{-1}|u_0|^{-2}F_k(M_{k+1})
\label{2.66}
\end{equation}
where $F_k$ is a (different) non-negative non-decreasing
continuous function on the non-negative real line. Thus, in
accordance with the notation \ref{2.49} we have:
\begin{equation}
e={\cal M}_1(\delta^{-1}|u_0|^{-2}) \label{2.67}
\end{equation}

Equation \ref{2.58} implies:
\begin{equation}
D\sg^{-1}=-2\chi^{\sharp\sharp} \ \ \mbox{: on $C_{u_0}$}
\label{2.68}
\end{equation}
where $\chi^{\sharp\sharp}$ is the symmetric 2-contravariant $S$
tensorfield with components:
\begin{equation}
\chi^{\sharp\sharp AB}=\chi_{CD}(\sg^{-1})^{AC}(\sg^{-1})^{BD}
\label{2.69}
\end{equation}
It follows that for any 2-covariant $S$ tensorfield $\theta$ on
$C_{u_0}$ we have:
\begin{equation}
D\mbox{tr}\theta=\mbox{tr}D\theta-2(\chi,\theta) \ \ \mbox{: on
$C_{u_0}$} \label{2.70}
\end{equation}
Thus in view of \ref{1.a2}, the second of \ref{1.155}, and
\ref{2.57}, the trace of the second variation equation \ref{1.42}
reads, along $C_{u_0}$:
\begin{equation}
D\mbox{tr}\chi=-|\chi|_{\sg}^2=-\frac{1}{2}(\mbox{tr}\chi)^2-|\chih|_{\sg}^2
\ \ \mbox{: along $C_{u_0}$} \label{2.71}
\end{equation}
Substituting from \ref{2.64} and the second of \ref{2.63} this
equation becomes the following linear second order ordinary
differential equation for the function $\phi$ along the generators
of $C_{u_0}$:
\begin{equation}
\frac{\partial^2\phi}{\partial\ub^2}+e\phi=0 \label{2.72}
\end{equation}
The initial conditions on $S_{0,u_0}$ for this equation are the
following. First, by \ref{2.4} and \ref{2.6} at $\ub=0$ we have:
\begin{equation}
\left.\phi\right|_{S_{0,u_0}}=1 \label{2.73}
\end{equation}
Then by the second of \ref{2.63} at $\ub=0$:
\begin{equation}
\left.\frac{\partial\phi}{\partial\ub}\right|_{S_{0,u_0}}=\frac{1}{2}\left.\mbox{tr}\chi\right|_{S_{0,u_0}}=\frac{1}{|u_0|}
\label{2.74}
\end{equation}
the last being the mean curvature of  a round sphere of radius
$r_0=|u_0|$ in Euclidean 3-dimensional space with respect to the
outer normal.

The fact that $e$ is non-negative, together with the fact that
$\phi$ must be positive, implies that along each generator of
$C_{u_0}$ $\phi$ is a concave function of $\ub$. Equation
\ref{2.72} together with the initial conditions \ref{2.73},
\ref{2.74} has a unique smooth solution on $C_{u_0}$. However,
this solution may not be everywhere positive on $C_{u_0}$.
Consider a generator of $C_{u_0}$, corresponding to some
$\vartheta\in S^2$. Then either $\phi$ is positive everywhere on
this generator, that is $\phi(\ub,u_0,\vartheta)>0$ for all
$\ub\in[0,\delta]$, or there is a $\ub^*(\vartheta)\in [0,\delta]$
such that $\phi(\ub,u_0,\vartheta)>0$ for all
$\ub\in[0,\ub^*(\vartheta))$, but
$\phi(\ub^*(\vartheta),u_0,\vartheta)=0$. (The point with
coordinates $(\ub^*(\vartheta),u_0,\vartheta)$ may be a point
conjugate to $o$ along the given generator, or a truly singular
point.) If there exists a generator, that is a $\vartheta\in S^2$,
where the second alternative holds, then the initial data are {\em
singular}.

\vspace{5mm}

{\bf Lemma 2.2} \ \ \ The initial data are regular if
$\delta|u_0|^{-2}$ is suitably small depending on $M_1$.

\noindent{\em Proof:} \ For, $\phi(\ub,u_0,\vartheta)$ is concave
on $[0,\delta]$ if the first alternative holds, on
$[0,\ub^*(\vartheta)]$ if the second alternative holds, hence, in
view of the initial conditions \ref{2.73}, \ref{2.74} we have:
\begin{equation}
\phi(\ub,u_0,\vartheta)\leq 1+\frac{\ub}{|u_0|} \label{2.75}
\end{equation}
for all $\ub\in[0,\delta]$ if the first alternative holds, for all
$\ub\in[0,\ub^*(\vartheta)]$ if the second alternative holds.
Integrating \ref{2.72} once and using the initial condition
\ref{2.74} we obtain:
\begin{equation}
\frac{\partial\phi}{\partial\ub}(\ub,u_0,\vartheta)=\frac{1}{|u_0|}-\int_0^{\ub}\phi(\ub^\prime,u_0,\vartheta)e(\ub^\prime,u_0,\vartheta)
d\ub^\prime \label{2.76}
\end{equation}
Substituting \ref{2.75} and the fact that by \ref{2.66} with
$k=0$:
\begin{equation}
\sup_{C_{u_0}}e\leq\delta^{-1}|u_0|^{-2}F_0(M_1) \label{2.77}
\end{equation}
we obtain:
\begin{equation}
\frac{\partial\phi}{\partial\ub}(\ub,u_0,\vartheta)\geq\frac{1}{|u_0|}-\left(1+\frac{\delta}{|u_0|}\right)\frac{F_0(M_1)}{|u_0|^2}
\label{2.78}
\end{equation}
for all $\ub\in[0,\delta]$ if the first alternative holds, for all
$\ub\in[0,\ub^*(\vartheta)]$ if the second alternative holds.
Integrating again and using the initial condition \ref{2.73} we
then obtain:
\begin{equation}
\phi(\ub,u_0,\vartheta)\geq
1+\frac{\ub}{|u_0|}\left\{1-\left(1+\frac{\delta}{|u_0|}\right)\frac{F_0(M_1)}{|u_0|}\right\}
\label{2.79}
\end{equation}
for all $\ub\in[0,\delta]$ if the first alternative holds, for all
$\ub\in[0,\ub^*(\vartheta)]$ if the second alternative holds. In
particular, if the second alternative holds, then setting
$\ub=\ub^*(\vartheta)$ we must have:
\begin{equation}
0\geq
1+\frac{\ub^*}{|u_0|}\left\{1-\left(1+\frac{\delta}{|u_0|}\right)\frac{F_0(M_1)}{|u_0|}\right\}
\label{2.80}
\end{equation}
However, since $|u_0|>1$, $\delta\leq 1$, $\ub^*\leq \delta$, this
is impossible if:
\begin{equation}
2\delta|u_0|^{-2}F_0(M_1)<1 \label{2.b6}
\end{equation}

Therefore, under this smallness condition on $\delta$, the second
alternative is ruled out and the initial data are regular.

\vspace{5mm}

In the following we shall in fact impose the stronger condition:
\begin{equation}
4\delta F_0(M_1)\leq 1 \label{2.81}
\end{equation}
Since then \ref{2.75} and \ref{2.79} hold for all
$\ub\in[0,\delta]$, we have:
\begin{equation}
\sup_{C_{u_0}}\left|\phi-1-\frac{\ub}{|u_0|}\right|\leq
2\delta|u_0|^{-2}F_0(M_1)\leq\frac{1}{2} \label{2.82}
\end{equation}
Substituting \ref{2.82} and \ref{2.77} in \ref{2.72} then yields:
\begin{equation}
\sup_{C_{u_0}}\left|\frac{\partial\phi}{\partial\ub}-\frac{1}{|u_0|}\right|\leq
|u_0|^{-2}F^\prime_0(M_1) \label{2.83}
\end{equation}
where $F^\prime_0$ is a non-negative non-decreasing continuous
function on the non-negative real line.

We proceed to estimate $\partial^m \phi/\partial\vartheta^m$. Here
it is convenient to use multi-index notation. If $a=(a_1,a_2)$ is
a pair of non-negative integers, we denote:
\begin{equation}
\left(\frac{\partial}{\partial\vartheta}\right)^a=\frac{\partial^{|a|}}{\partial(\vartheta^1)^{a_1}\partial(\vartheta^2)^{a_2}}
\label{2.84}
\end{equation}
a differential operator of order $|a|$, where $|a|$ is the
``length"' of the multi-index:
\begin{equation}
|a|=a_1+a_2 \label{2.85}
\end{equation}
If $a$ and $b$ are two multi-indices, we say that $b\leq a$ if
$b_1\leq a_1$ and $b_2\leq a_2$. Then $a-b$ is also a multi-index,
of length $|a|-|b|$. We also denote:
\begin{equation}
a!=a_1!a_2! \ \ \ \ \
\left(\begin{array}{l}a\\b\end{array}\right)=\frac{a!}{b!(a-b)!}
\label{2.86}
\end{equation}
If $f,g$ is a pair of smooth functions on
$[0,\delta]\times\oD_{2\rho}$, we have the Leibniz rule:
\begin{equation}
\left(\frac{\partial}{\partial\vartheta}\right)^a= \sum_{b\leq
a}\left(\begin{array}{l}a\\b\end{array}\right)
\left(\frac{\partial}{\partial\vartheta}\right)^b
f\left(\frac{\partial}{\partial\vartheta}\right)^{a-b}g
\label{2.87}
\end{equation}

We work in each of the two stereographic charts separately.
Applying $(\partial/\partial\vartheta)^a$ to \ref{2.72} we obtain:
\begin{equation}
\frac{\partial^2}{\partial\ub^2}\left(\frac{\partial}{\partial\vartheta}\right)^a\phi+e\left(\frac{\partial}{\partial\vartheta}\right)^a\phi
=f_a \label{2.88}
\end{equation}
where:
\begin{equation}
f_a=-\sum_{b\leq a, |b|\neq
0}\left(\frac{\partial}{\partial\vartheta}\right)^b e
\left(\frac{\partial}{\partial\vartheta}\right)^{a-b}\phi
\label{2.89}
\end{equation}
Applying $(\partial/\partial\vartheta)^a$ for $|a|\neq 0$ to the
initial conditions \ref{2.73}, \ref{2.74} we obtain simply:
\begin{equation}
\left(\frac{\partial}{\partial\vartheta}\right)^a\phi(0,u_0,\vartheta)=\left(\frac{\partial}{\partial\vartheta}\right)^a\frac{\partial\phi}{\partial\ub}(0,u_0,\vartheta)=0
\ \  : \ \mbox{ for  $|a|\neq 0$} \label{2.90}
\end{equation}

\vspace{5mm}

\noindent{\bf Lemma 2.3} \ \ \ If condition \ref{2.81} is
satisfied, then for each non-zero multi-index $a$ we have:
$$\sup_{(\ub,\vartheta)\in[0,\delta]\times\oD_{2\rho}}\left|\left(\frac{\partial}{\partial\vartheta}\right)^a\phi(\ub,u_0,\vartheta\right|
\leq \delta|u_0|^{-2}F_{|a|}(M_{|a|+1})$$ and:
$$\sup_{(\ub,\vartheta)\in[0,\delta]\times\oD_{2\rho}}\left|\left(\frac{\partial}{\partial\vartheta}\right)^a\frac{\partial\phi}{\partial\ub}
(\ub,u_0,\vartheta)\right|\leq
|u_0|^{-2}F^\prime_{|a|}(M_{|a|+1})$$ where $F_k$, $F^\prime_k$
are non-negative non-decreasing continuous functions on the
non-negative real line.

\noindent{\em Proof:} \ The proof is by induction on $|a|$. Let
the lemma hold for all $|a|=1,...,k-1$ (no assumption if $k=1$).
Then given any $a$ with $|a|=k$, by \ref{2.66}, \ref{2.82} and the
inductive hypothesis, there is a non-negative non-decreasing
continous function $G_{|a|}$ such that:
\begin{equation}
\sup_{(\ub,\vartheta)\in[0,\delta]\times\oD_{2\rho}}|f_a|\leq
\delta^{-1}|u_0|^{-2}G_{|a|}(M_{|a|+1}) \label{2.91}
\end{equation}
We integrate \ref{2.88} twice, using the initial conditions
\ref{2.90}, to obtain:
\begin{equation}
\left(\frac{\partial}{\partial\vartheta}\right)^a\phi(\ub,u_0,\vartheta)=
\int_0^{\ub}\int_0^{\ub^\prime}\left\{f_a-e\left(\frac{\partial}{\partial\vartheta}\right)^a\phi\right\}(\ub^{\prime\prime},u_0,\vartheta)
d\ub^{\prime\prime}d\ub^\prime \label{2.92}
\end{equation}
It follows that:
\begin{eqnarray}
&&\left|\left(\frac{\partial}{\partial\vartheta}\right)^a\phi(\ub,u_0,\vartheta)\right|
\leq\int_0^{\ub}\int_0^{\ub^\prime}|f_a(\ub^{\prime\prime},u_0,\vartheta)|d\ub^{\prime\prime}d\ub^\prime\nonumber\\
&&\hspace{20mm}+\int_0^{\ub}\int_0^{\ub^\prime}e\left|\left(\frac{\partial}{\partial\vartheta}\right)^a\phi\right|(\ub^{\prime\prime},u_0,\vartheta)d\ub^{\prime\prime}d\ub^\prime
\label{2.93}
\end{eqnarray}
By \ref{2.91} the first double integral on the right is bounded
by:
\begin{equation}
\frac{1}{2}\delta|u_0|^{-2}G_{|a|}(M_{|a|+1}) \label{2.94}
\end{equation}
Setting
\begin{equation}
x_a(\ub)=\sup_{\ub^\prime\in[0,\ub]}\left|\left(\frac{\partial}{\partial\vartheta}\right)^a\phi(\ub^\prime,u_0,\vartheta)\right|
\label{2.95}
\end{equation}
then by \ref{2.77} the second double integral on the right in
\ref{2.93} is bounded by:
\begin{equation}
\frac{1}{2}\delta|u_0|^{-2}F_0(M_1)x_a(\ub) \label{2.96}
\end{equation}
$x_a(\ub)$ being a non-decreasing function of $\ub$. Hence we
obtain:
\begin{equation}
\left|\left(\frac{\partial}{\partial\vartheta}\right)^a\phi(\ub,u_0,\vartheta)\right|\leq
\frac{1}{2}\delta|u_0|^{-2} F_0(M_1)x_a(\ub)
+\frac{1}{2}\delta|u_0|^{-2}G_{|a|}(M_{|a|+1}) \label{2.97}
\end{equation}
This inequality holds also with $\ub$ replaced by
$\ub^\prime\in[0,\ub]$ on the left, the right hand side being a
non-decreasing function of $\ub$. Therefore taking on the left the
supremum with respect to $\ub^\prime\in[0,\ub]$ yields:
\begin{equation}
x_a(\ub)\leq \frac{1}{2}\delta|u_0|^{-2} F_0(M_1)x_a(\ub)
+\frac{1}{2}\delta|u_0|^{-2}G_{|a|}(M_{|a|+1}) \label{2.98}
\end{equation}
which by \ref{2.81} implies:
\begin{equation}
x_a(\ub)\leq\frac{4}{7}\delta|u_0|^{-2}G_{|a|}(M_{|a|+1})
\label{2.99}
\end{equation}
Setting $\ub=\delta$ this is the inequality of the lemma for
$\phi$ with
\begin{equation}
F_k=\frac{4}{7}G_k \label{2.100}
\end{equation}
Integrating \ref{2.88} once using the second of the initial
conditions \ref{2.90} we obtain:
\begin{equation}
\left(\frac{\partial}{\partial\vartheta}\right)^a\frac{\partial\phi}{\partial\ub}(\ub,u_0,\vartheta)=
\int_0^{\ub}\left\{f_a-e\left(\frac{\partial}{\partial\vartheta}\right)^a\phi\right\}(\ub^\prime,u_0,\vartheta)
d\ub^\prime \label{2.101}
\end{equation}
which implies:
\begin{eqnarray}
&&\left|\left(\frac{\partial}{\partial\vartheta}\right)^a\frac{\partial\phi}{\partial\ub}(\ub,u_0,\vartheta)\right|
\leq\int_0^{\ub}|f_a(\ub^\prime,u_0,\vartheta)|d\ub^\prime\nonumber\\
&&\hspace{20mm}+\int_0^{\ub}e\left|\left(\frac{\partial}{\partial\vartheta}\right)^a\phi\right|(\ub^\prime,u_0,\vartheta)d\ub^\prime
\label{2.102}
\end{eqnarray}
By \ref{2.91} the first integral on the right is bounded by:
\begin{equation}
|u_0|^{-2}G_{|a|}(M_{|a|+1}) \label{2.103}
\end{equation}
while the second integral on the right is bounded by:
\begin{equation}
\delta^{-1}|u_0|^{-2}F_0(M_1)x_a(\ub) \label{2.104}
\end{equation}
Substituting the bound \ref{2.99} for the latter gives the
inequality of the lemma for $\partial\phi/\partial\ub$  with
\begin{equation}
F^\prime_k=\left(1+\frac{4}{7}F_0\right)G_k \label{2.105}
\end{equation}
($F_0(M_{k+1})\geq F_0(M_1)$, $F_0$ being a non-decreasing
function). This completes the inductive step and therefore the
proof of the lemma.

\vspace{5mm}

Applying repeatedly $\partial/\partial\ub$ to equation \ref{2.72}
and using the estimates \ref{2.82}, \ref{2.83} and \ref{2.66} we
deduce:
\begin{equation}
\sup_{C_{u_0}}\left|\frac{\partial^n\phi}{\partial\ub^n}\right|\leq
\delta^{1-n}|u_0|^{-2}F_{n,0}(M_{n-1}) \ \ : \ \mbox{for $n\geq
2$} \label{2.106}
\end{equation}
where the $F_{n,0}$ are non-negative non-decreasing continuous
functions on the non-negative real line.

Similarly, applying repeatedly $\partial/\partial\ub$ to equation
\ref{2.88} and using Lemma 2.3 and the bound \ref{2.77} we deduce:
\begin{eqnarray}
&&\sup_{(\ub,\vartheta)\in[0,\delta]\times\oD_{2\rho}}\left|\left(\frac{\partial}{\partial\vartheta}\right)^a\frac{\partial^n\phi}{\partial\ub^n}(\ub,u_0,\vartheta)\right|\leq \delta^{1-n}|u_0|^{-2}F^{\prime\prime}_{n,|a|}(M_{|a|+n-1}) \nonumber\\
&& \hspace{15mm} : \ \mbox{for $n\geq 2, |a|\neq 0$} \label{2.107}
\end{eqnarray}
where the $F_{n,k}$ are non-negative non-decreasing continuous
functions on the real line.

Putting together the results for the two stereographic charts we
then obtain:
\begin{eqnarray}
&&\left\|\phi-1-\frac{\ub}{|u_0|}\right\|_{C^k_\delta(C_{u_0})}\leq \delta|u_0|^{-2}F_{0,k}(M_{k+1})\nonumber\\
&&\left\|\frac{\partial\phi}{\partial\ub}-\frac{1}{|u_0|}\right\|_{C^k_\delta(C_{u_0})}\leq |u_0|^{-2}F_{1,k}(M_{k+1})\nonumber\\
&&\left\|\frac{\partial^2\phi}{\partial\ub^2}\right\|_{C^k_\delta(C_{u_0})}\leq\delta^{-1}|u_0|^{-2}F_{2,k}(M_{k+1})
\label{2.108}
\end{eqnarray}
where the $F_{0,k}$, $F_{1,k}$, $F_{2,k}$ are new non-negative
no-decreasing continuous functions on the real line. Thus in
accordance with the notation \ref{2.49} we have:
\begin{eqnarray}
&&\phi-1-\frac{\ub}{|u_0|}={\cal M}_1(\delta|u_0|^{-2})\nonumber\\
&&\frac{\partial\phi}{\partial\ub}-\frac{1}{|u_0|}={\cal M}_1(|u_0|^{-2})\nonumber\\
&&\frac{\partial^2\phi}{\partial\ub^2}={\cal
M}_1(\delta^{-1}|u_0|^{-2}) \label{2.109}
\end{eqnarray}
From \ref{2.3}, \ref{2.52} and \ref{2.109} we conclude that:
\begin{equation}
|u_0|^{-2}\sg-\up{\sg}={\cal M}_1(\delta^{1/2}|u_0|^{-1})
\label{2.110}
\end{equation}
Also, from \ref{2.63}, \ref{2.52} and \ref{2.109} we conclude
that:
\begin{equation}
\chih={\cal M}_1(\delta^{-1/2}|u_0|) \ \ \ \ \
\mbox{tr}\chi-\frac{2}{|u_0|}={\cal M}_1(|u_0|^{-2}) \label{2.111}
\end{equation}

Now, since $\up{\sg}$ is given in each stereographic chart by
\ref{1.192}, that is:
\begin{equation}
\up{\sg}_{AB}(\vartheta)=w^2(\vartheta)\delta_{AB} \label{2.112}
\end{equation}
where $w$ is the function \ref{2.7}, if we denote by
$\lambda(\ub,u_0,\vartheta)$ and $\Lambda(\ub,u_0,\vartheta)$
respectively the smallest and largest eigenvalues of the matrix
$|u_0|^{-2}\sg_{AB}(\ub,u_0,\vartheta)$, then \ref{2.110} implies
that if $\delta$ is suitably small depending on $M_1$ we have:
\begin{equation}
\sup_{(\ub,\vartheta)\in[0,\delta]\times\oD_{2\rho}}\max\{|\lambda(\ub,u_0,\vartheta)-w^2(\vartheta)|,
|\Lambda(\ub,u_0,\vartheta)-w^2(\vartheta)|\}\leq\frac{1}{2}\frac{1}{(1+\rho^2)^2}
\label{2.113}
\end{equation}
Since the function $w$ satisfies:
\begin{equation}
\frac{1}{1+\rho^2}\leq w(\vartheta)\leq 1 \ \  : \
\forall\vartheta\oD_{2\rho} \label{2.114}
\end{equation}
it follows that:
\begin{eqnarray}
&&\frac{1}{2(1+\rho^2)^2}\leq\lambda(\ub,u_0,\vartheta)\leq\Lambda(\ub,u_0,\vartheta)\leq 1+\frac{1}{2(1+\rho^2)^2}\nonumber\\
&&\hspace{20mm} : \ \mbox{for all
$(\ub,\vartheta)\in[0,\delta]\times\oD_{2\rho}$} \label{2.115}
\end{eqnarray}
Let $\xi$ be a $T^q_p$ type $S$ tensorfield on $C_{u_0}$. Then by
\ref{2.115} there is a positive constant $C_{p,q}$ depending only
on $p,q$ and the initial choice of $\rho$ such that:
\begin{eqnarray}
&&C^{-1}_{p,q}|u_0|^{q-p}|\xi|\leq\nonumber\\
&&\hspace{10mm}|\xi|_{\sg}=\left(\sg_{A_1 C_1}...\sg_{A_q C_q}(\sg^{-1})^{B_1 D_1}...(\sg^{-1})^{B_p D_p}\xi^{A_1...A_q}_{B_1...B_p}\xi^{C_1...C_q}_{D_1...D_p}\right)^{1/2}\nonumber\\
&&\hspace{40mm}\leq C_{p,q}|u_0|^{q-p}|\xi| \label{2.116}
\end{eqnarray}
Let then $\xi$ be a $C^k$ $T^q_p$ type $S$ tensorfield on
$C_{u_0}$. We define the weighted invariant $C^k$ norm of $\xi$
by:
\begin{equation}
\|\xi\|_{{\bf C}^k_\delta(C_{u_0})}=\max_{m+n\leq
k}\sup_{C_{u_0}}\left(\delta^n|u_0|^m|\snab^m D^n\xi|_{\sg}\right)
\label{2.117}
\end{equation}
where $\snab$ is the covariant derivative on the $S_{\ub,u_0}$
with respect to $\left.\sg\right|_{S_{\ub,u_0}}$. It follows from
\ref{2.110} and \ref{2.116} that there is a non-negative
non-decreasing continuous function $F_k$ such that:
\begin{equation}
\|\xi\|_{{\bf C}^k_\delta(C_{u_0})}\leq
|u_0|^{q-p}F_k(M_{k+1})\|\xi\|_{C^k_\delta(C_{u_0})} \label{2.118}
\end{equation}
and:
\begin{equation}
\|\xi\|_{C^k_\delta(C_{\ub_0})}\leq
|u_0|^{p-q}F_k(M_{k+1})\|\xi\|_{{\bf C}^k_\delta(C_{u_0})}
\label{2.119}
\end{equation}
We say that a smooth $T^q_p$ type $S$ tensorfield $\xi$ on
$C_{u_0}$ is ${\bf M}_l(\delta^r|u_0|^s)$, for real numbers $r,s$,
and non-negative integer $l$,
\begin{equation}
\xi={\bf M}_l(\delta^r|u_0|^s) \label{2.120}
\end{equation}
if for every non-negative integer $k$ we have:
\begin{equation}
\|\xi\|_{{\bf C}^k_\delta(C_{u_0})}\leq \delta^r|u_0|^s
F_k(M_{k+l}) \label{2.121}
\end{equation}
where $F_k$ is a non-negative non-decreasing continuous function.
Then \ref{2.118}, \ref{2.119} imply the following lemma.

\vspace{5mm}

\noindent{\bf Lemma 2.4} Let $\xi$ be a smooth $T^q_p$ type $S$
tensorfield on $C_{u_0}$. Then for $l\geq 1$ $\xi={\cal
M}_l(\delta^r|u_0|^s)$ implies $\xi={\bf
M}_l(\delta^r|u_0|^{s+q-p})$ and conversely.

\vspace{5mm}

In particular \ref{2.111} imply:
\begin{equation}
\chih={\bf M}_1(\delta^{-1/2}|u_0|^{-1}) \ \ \ \ \
\mbox{tr}\chi-\frac{2}{|u_0|}={\bf M_1}(|u_0|^{-2}) \label{2.122}
\end{equation}
Also, \ref{2.54} implies:
\begin{equation}
\hat{K}-|u_0|^{-2}={\bf M}_2(\delta^{1/2}|u_0|^{-3}) \label{2.123}
\end{equation}
Now $K$, the Gauss curvature of $\sg$, is given in terms of
$\hat{K}$ and $\phi$ by:
\begin{equation}
K=\phi^{-2}(\hat{K}-\slap_{\sgh}\log\phi) \label{2.124}
\end{equation}
while by the first of \ref{2.109}:
\begin{equation}
\phi-1-\frac{\ub}{|u_0|}={\bf M}_1(\delta|u_0|^{-2}) \label{2.125}
\end{equation}
It then follows that:
\begin{equation}
K-|u_0|^{-2}={\bf M}_3(\delta^{1/2}|u_0|^{-3}) \label{2.126}
\end{equation}

We turn to consider the torsion $\zeta$ on $C_{u_0}$ (see
\ref{1.48}). By \ref{2.57} and \ref{1.65} we have:
\begin{equation}
\eta=\zeta, \ \ \ \etb=-\zeta \ \ \ \mbox{: along $C_{u_0}$}
\label{2.127}
\end{equation}
Thus the propagation equation \ref{1.66} becomes, along $C_{u_0}$:
\begin{equation}
D\zeta+\chi^\sharp\cdot\zeta=-\beta \label{2.128}
\end{equation}
On the other hand, the Codazzi equation \ref{1.124} reads:
\begin{equation}
\sdiv\chi-\sd\mbox{tr}\chi+\chi^\sharp\cdot\zeta-\mbox{tr}\chi\cdot\zeta=-\beta
\label{2.129}
\end{equation}
Substituting for $-\beta$ from \ref{2.129} into \ref{2.128} then
yields the following propagation equation for $\zeta$ along the
generators of $C_{u_0}$:
\begin{equation}
D\zeta+\mbox{tr}\chi\zeta=\xi \label{2.130}
\end{equation}
where $\xi$ is the $S$ 1-form:
\begin{equation}
\xi=\sdiv\chi-\sd\mbox{tr}\chi \label{2.131}
\end{equation}
By \ref{2.122}:
\begin{equation}
\xi={\bf M}_2(\delta^{-1/2}|u_0|^{-2}) \ \ \mbox{or,
equivalently,} \ \ \xi={\cal M}_2(\delta^{-1/2}|u_0|^{-1})
\label{2.132}
\end{equation}
Equation \ref{2.130} is a linear first order ordinary differential
equation for $\zeta$, which in each stereographic chart reads:
\begin{equation}
\frac{\partial\zeta_A}{\partial\ub}+\mbox{tr}\chi\zeta_A=\xi_A
\label{2.133}
\end{equation}
The initial condition on $S_{0,\ub}$ is simply:
\begin{equation}
\left.\zeta\right|_{S_{0,u_0}}=0 \label{2.134}
\end{equation}
$S_{0,u_0}$ being a sphere lying on the spacelike hyperplane
$x^0=u_0$ with future directed unit normal $\Th=\partial/\partial
x^0$ at the boundary of the Minkowskian region  $M_0$, the
vectorfields $\Lh$ and $\Lbh$ having projection $\Th$ along $\Th$
in $M_0$ . Equation \ref{2.133} together with the initial
condition \ref{2.134} has a unique smooth solution $\zeta$ on
$C_{u_0}$. This equation and the equations to follow as well as
the derivative equations, obtained by applying
$(\partial/\partial\vartheta)^a$, for some multi-index $a$ to the
original equations, are all equations of the form:
\begin{equation}
\frac{dx}{dt} +a\cdot x=b \label{2.a1}
\end{equation}
on the interval $[0,\delta]$, where the unkown $x$ is a function
of $t$ with values in $\Re^n$ for some $n$, $a$ is a given smooth
function of $t$ with values in the $n$ dimensional matrices, and
$b$ is a given smooth function of $t$ with values in $\Re^n$.
Denoting by $|\s|$, as in \ref{2.42}, the Euclidean norm on
$\Re^n$, and by $<\s,\s>$ the Euclidean inner product on $\Re^n$,
equation \ref{2.a1} implies:
\begin{equation}
\frac{d}{dt}|x|^2=2(-<x,a\cdot x>+<x,b>)\leq 2|x|(|a||x|+|b|)
\label{2.a2}
\end{equation}
where $|a(t)|$ is the Euclidean norm of $a(t)$ in $\Re^{n^2}$, as
in \ref{2.42}. It then follows that (see Lemma 3.1 of the next
chapter):
\begin{equation}
|x(t)|\leq e^{A(t)}\left\{|x(0)|+\int_0^t
e^{-A(t^\prime)}|b(t^\prime)|dt^\prime\right\} \label{2.a3}
\end{equation}
where:
\begin{equation}
A(t)=\int_0^t|a(t^\prime)|dt^\prime \label{2.a4}
\end{equation}
In this manner, using the second of \ref{2.111} and the second of
\ref{2.132} we deduce, in the present case:
\begin{equation}
\zeta={\cal M}_2(\delta^{1/2}|u_0|^{-1}) \ \ \mbox{hence}  \ \
\zeta={\bf M}_2(\delta^{1/2}|u_0|^{-2}) \label{2.135}
\end{equation}

We turn to $\mbox{tr}\chib$. Now, the first of \ref{1.28} implies:
\begin{equation}
D\sg^{-1}=-2\Omega\chi^{\sharp\sharp} \label{2.136}
\end{equation}
where $\chi^{\sharp\sharp}$ is the symmetric 2-contravariant $S$
tensorfield with components:
\begin{equation}
\chi^{\sharp\sharp AB}=\chi_{CD}(\sg^{-1})^{AC}(\sg^{-1})^{BD}
\label{2.137}
\end{equation}
It follows that for any 2-covariant $S$ tensorfield $\theta$ we
have:
\begin{equation}
D\mbox{tr}\theta=\mbox{tr}D\theta-2\Omega(\chi,\theta)
\label{2.138}
\end{equation}
Taking into account \ref{2.138} and \ref{1.a2} the trace of
equation \ref{1.146}, is the equation:
\begin{equation}
D(\Omega\mbox{tr}\chib)=\Omega^2\left\{2\sdiv\etb+2|\etb|_{\sg}^2-(\chi,\chib)+2\rho\right\}
\label{2.139}
\end{equation}
Substituting for $\rho$ from the Gauss equation \ref{1.108} we
then obtain the equation:
\begin{equation}
D(\Omega\mbox{tr}\chib)=\Omega^2\left\{2\sdiv\etb+2|\etb|_{\sg}^2-2K-\mbox{tr}\chi\mbox{tr}\chib\right\}
\label{2.140}
\end{equation}
which along $C_{u_0}$ becomes the following propagation equation
for $\mbox{tr}\chib$ along the generators of $C_{u_0}$:
\begin{equation}
D\mbox{tr}\chib+\mbox{tr}\chi\mbox{tr}\chib=2\lambda \label{2.141}
\end{equation}
where $\lambda$ is the function:
\begin{equation}
\lambda=-K-\sdiv\zeta+|\zeta|_{\sg}^2 \label{2.142}
\end{equation}
By \ref{2.126} and \ref{2.135}:
\begin{equation}
\lambda+|u_0|^{-2}={\bf M}_3(\delta^{1/2}|u_0|^{-3}) \ \ \mbox{or,
equivalently,} \ \ \lambda+|u_0|^{-2}={\cal
M}_3(\delta^{1/2}|u_0|^{-3}) \label{2.143}
\end{equation}
Equation \ref{2.141} is a linear first order ordinary differential
equation for $\mbox{tr}\chib$, which in each stereographic chart
reads:
\begin{equation}
\frac{\partial\mbox{tr}\chib}{\partial\ub}+\mbox{tr}\chi\mbox{tr}\chib=2\lambda
\label{2.144}
\end{equation}
The initial condition on $S_{0,u_0}$ is:
\begin{equation}
\left.\mbox{tr}\chib\right|_{S_{0,u_0}}=-\frac{2}{|u_0|}
\label{2.145}
\end{equation}
the last being the mean curvature of  a round sphere of radius
$r_0=|u_0|$ in Euclidean 3-dimensional space with respect to the
inner normal. Equation \ref{2.144} together with the initial
condition \ref{2.145} has a unique smooth solution
$\mbox{tr}\chib$ on $C_{u_0}$. Setting:
\begin{equation}
\mbox{tr}\chib=-\frac{2}{|u_0|}+\frac{2\ub}{|u_0|^2}+\nub
\label{2.146}
\end{equation}
and:
\begin{equation}
\mbox{tr}\chi=\frac{2}{|u_0|}+\nu \label{2.147}
\end{equation}
equation \ref{2.144} becomes the following linear first order
ordinary differential equation for $\nub$:
\begin{equation}
\frac{\partial\nub}{\partial\ub}+\mbox{tr}\chi\nub=2\lambda^\prime
\label{2.148}
\end{equation}
where:
\begin{equation}
\lambda^\prime=\lambda+|u_0|^{-2}-\frac{2\ub}{|u_0|^3}+\left(1-\frac{\ub}{|u_0|}\right)\frac{\nu}{|u_0|}
\label{2.149}
\end{equation}
and the initial condition \ref{2.145} becomes simply:
\begin{equation}
\left.\nub\right|_{S_{0,u_0}}=0 \label{2.150}
\end{equation}
Now, according to the second of \ref{2.111}:
\begin{equation}
\nu={\cal M}_1(|u_0|^{-2}) \label{2.151}
\end{equation}
which together with \ref{2.143} yields:
\begin{equation}
\lambda^\prime={\cal M}_3(|u_0|^{-3}) \label{2.152}
\end{equation}
Using this and the second of \ref{2.111} we readily deduce from
\ref{2.149}, \ref{2.150}:
\begin{equation}
\nub={\cal M}_3(\delta|u_0|^{-3}) \ \ \mbox{hence also} \ \
\nub={\bf M}_3(\delta|u_0|^{-3}) \label{2.153}
\end{equation}

We turn to $\chibh$. Equation \ref{1.146} becomes along $C_{u_0}$:
\begin{equation}
D\chib=-\snab\zeta-\tilde{\snab\zeta}+2\zeta\otimes\zeta+\frac{1}{2}(\chi\times\chib+\chib\times\chi)+\rho\sg
\label{2.154}
\end{equation}
Decomposing $\chib$ and $\chi$ as in \ref{1.162} and using the
first of equations \ref{1.28} we obtain:
\begin{equation}
D\chib=D\chibh+\mbox{tr}\chib\chih+\frac{1}{2}(D\mbox{tr}\chib+\mbox{tr}\chi\mbox{tr}\chib)\sg
\label{2.155}
\end{equation}
and by \ref{2.138} applied to $\chibh$ we have:
\begin{equation}
\mbox{tr}D\chibh=2(\chih,\chibh) \label{2.156}
\end{equation}
Substituting then \ref{2.155} on the left in \ref{2.154} and
taking the trace free parts of both sides, we obtain, taking into
account the identity \ref{1.163}, the following propagation
equation for $\chibh$ along the generators of $C_{u_0}$:
\begin{equation}
D\chibh-(\chih,\chibh)\sg-\frac{1}{2}\mbox{tr}\chi\chibh=\theta
\label{2.157}
\end{equation}
where $\theta$ is the trace-free symmetric 2-covariant $S$
tensorfield:
\begin{equation}
\theta=-\snab\oth\zeta+\zeta\oth\zeta-\frac{1}{2}\mbox{tr}\chib\chih
\label{2.158}
\end{equation}
By \ref{2.135}, \ref{2.146}, \ref{2.153}, and the first of
\ref{2.122}:
\begin{equation}
\theta={\bf M}_3(\delta^{-1/2}|u_0|^{-2}) \ \ \mbox{or,
equivalently,} \ \ \theta={\cal M}_3(\delta^{-1/2}) \label{2.159}
\end{equation}
Equation \ref{2.159} is a linear first order ordinary differential
equation for $\chibh$ which in each stereographic chart reads:
\begin{equation}
\frac{\partial\chibh_{AB}}{\partial\ub}-(\chih,\chibh)\sg_{AB}-\frac{1}{2}\mbox{tr}\chi\chibh_{AB}=\theta_{AB}
\label{2.160}
\end{equation}
where:
$$(\chih,\chibh)=(\sg^{-1})^{CE}(\sg^{-1})^{DF}\chih_{CD}\chibh_{EF}$$
The initial condition on $S_{0,u_0}$ is simply:
\begin{equation}
\left.\chibh\right|_{S_{0,u_0}}=0 \label{2.161}
\end{equation}
$S_{0,u_0}$ being a round sphere in Euclidean 3-dimensional space,
hence umbilical. Equation \ref{2.160} together with the initial
condition \ref{2.161} has a unique smooth solution on $C_{u_0}$.
Writing:
$$(\chih,\chibh)=\chih^{\sharp\sharp}\cdot\chibh, \ \ \ (\chih^{\sharp\sharp})^{AB}=(\sg^{-1})^{AC}(\sg^{-1})^{BD}\chih_{CD}$$
and noting that by \ref{2.116}:
\begin{equation}
|\chih^{\sharp\sharp}|\leq C|u_0|^{-4}|\chih| \label{2.162}
\end{equation}
while since $|\sg|_{\sg}=\sqrt{2}$ we have:
\begin{equation}
|\sg|\leq C|u_0|^2 , \label{2.163}
\end{equation}
and using \ref{2.111} and \ref{2.159}, we deduce:
\begin{equation}
\chibh={\cal M}_3(\delta^{1/2}) \ \ \mbox{hence} \ \ \chibh={\bf
M}_3(\delta^{1/2}|u_0|^{-2}) \label{2.164}
\end{equation}

We now consider the curvature components along $C_{u_0}$. Consider
equation \ref{1.42} along $C_{u_0}$:
\begin{equation}
D\chi=\chi\times\chi-\alpha \label{2.165}
\end{equation}
Decomposing $\chi$ as in \ref{1.162} and using the first of
equations \ref{1.28} we obtain:
\begin{equation}
D\chi=D\chih+\mbox{tr}\chi\chih+\frac{1}{2}(D\mbox{tr}\chi+(\mbox{tr}\chi)^2)\sg
\label{2.166}
\end{equation}
and by \ref{2.138} applied to $\chih$ we have:
\begin{equation}
\mbox{tr}D\chih=2|\chih|^2 \label{2.167}
\end{equation}
Substituting then \ref{2.166} on the left in \ref{2.165} and
taking the trace-free parts of both sides, we obtain, taking into
account the identity \ref{1.a5},
\begin{equation}
D\chih-|\chih|^2_{\sg}\sg=-\alpha \label{2.168}
\end{equation}
The first of \ref{2.122} then yields:
\begin{equation}
\alpha={\bf M}_2(\delta^{-3/2}|u_0|^{-1}) \label{2.169}
\end{equation}

Next, $\beta$ is given by \ref{2.129}. Using \ref{2.122} and
\ref{2.135} we then obtain:
\begin{equation}
\beta={\bf M}_2(\delta^{-1/2}|u_0|^{-2}) \label{2.170}
\end{equation}

Next, $\rho$ is given by the Gauss equation \ref{1.118} which we
may write in the form:
\begin{equation}
K+\frac{1}{4}\mbox{tr}\chi\mbox{tr}\chib-\frac{1}{2}(\chih,\chibh)=-\rho
\label{2.171}
\end{equation}
Using \ref{2.126}, \ref{2.122}, \ref{2.146}, \ref{2.153}, and
\ref{2.164}, we then obtain:
\begin{equation}
\rho={\bf M}_3(|u_0|^{-3}) \label{2.172}
\end{equation}
Also, $\sigma$ is given by equation \ref{1.141} (see \ref{1.145})
which reads:
\begin{equation}
\scurl\zeta+\frac{1}{2}\chih\wedge\chibh=\sigma \label{2.173}
\end{equation}
Using \ref{2.122}, \ref{2.135}, and \ref{2.164}, we then obtain:
\begin{equation}
\sigma={\bf M}_3(|u_0|^{-3}) \label{2.174}
\end{equation}

To derive appropriate estimates for the remaining curvature
components, namely for $\beb$ and $\alb$, we appeal to the Bianchi
indentities, given by Proposition 1.2. The sixth Bianchi identity
reads, along $C_{u_0}$:
\begin{equation}
D\beb+\frac{1}{2}\mbox{tr}\chi\beb-\chih^\sharp\cdot\beb=\kappa
\label{2.175}
\end{equation}
where $\kappa$ is the $S$ 1-form:
\begin{equation}
\kappa=-\sd\rho+\s^*\sd\sigma+3\zeta\rho-3\s^*\zeta\sigma+2\chibh^\sharp\cdot\beta
\label{2.176}
\end{equation}
By \ref{2.172}, \ref{2.174}, \ref{2.135}, \ref{2.164}, and
\ref{2.170}:
\begin{equation}
\kappa={\bf M}_4(|u_0|^{-4}) \ \ \mbox{or, equivalently,} \ \
\kappa={\cal M}_4(|u_0|^{-3}) \label{2.177}
\end{equation}
Equation \ref{2.175} is a linear first order ordinary differential
equation for $\beb$ along he generators of $C_{u_0}$, which in
each stereographic chart reads:
\begin{equation}
\frac{\partial\beb_A}{\partial\ub}+\frac{1}{2}\mbox{tr}\chi\beb_A-\chih^{\sharp
B}_A\beb_B=\kappa_A \label{2.178}
\end{equation}
The initial condition on $C_{u_0}$ is simply:
\begin{equation}
\left.\beb\right|_{S_{0,u_0}}=0 \label{2.179}
\end{equation}
$S_{0,u_0}$ lying on the boundary of the Minkowskian region $M_0$.
Equation \ref{2.175} together with the initial condition
\ref{2.179} has a unique smooth solution $\beb$ on $C_{u_0}$.
Using \ref{2.111} and \ref{2.177} we deduce:
\begin{equation}
\beb={\cal M}_4(\delta|u_0|^{-3}) \ \ \mbox{hence} \ \ \beb={\bf
M}_4(\delta|u_0|^{-4}) \label{2.180}
\end{equation}

In view of the fact that by \ref{2.138} applied to $\alb$ we have,
along $C_{u_0}$:
\begin{equation}
\mbox{tr}D\alb=2(\chih,\alb)\sg \label{2.181}
\end{equation}
the second of the Bianchi identities of Proposition 1.2 reads,
along $C_{u_0}$:
\begin{equation}
D\alb-(\chih,\alb)\sg-\frac{1}{2}\mbox{tr}\chi\alb=\tau
\label{2.182}
\end{equation}
where $\tau$ is the trace-free symmetric 2-covariant $S$
tensorfield:
\begin{equation}
\tau=-\snab\oth\beb+5\zeta\oth\beb-3\chibh\rho+3\s^*\chibh\sigma
\label{2.183}
\end{equation}
By \ref{2.180}, \ref{2.135}, \ref{2.164}, \ref{2.172}, and
\ref{2.174}:
\begin{equation}
\tau={\bf M}_5(\delta^{1/2}|u_0|^{-5}) \ \ \mbox{or,
equivalently,} \ \ \tau={\cal M}_5(\delta^{1/2}|u_0|^{-3})
\label{2.184}
\end{equation}
the principal term in $\tau$ being the term $-\snab\oth\beb$, but
the leading terms in behavior with respect to $\delta$ being the
terms $-3\chibh\rho+\s^*\chibh\sigma$. Equation \ref{2.182} is a
linear first order ordinary differential equation for $\alb$ along
the generators of $C_{u_0}$, which in each stereographic chart
reads:
\begin{equation}
\frac{\partial\alb_{AB}}{\partial\ub}-(\chih,\alb)\sg_{AB}-\frac{1}{2}\mbox{tr}\chi\alb_{AB}=\tau_{AB}
\label{2.185}
\end{equation}
where:
$$(\chih,\alb)=(\sg^{-1})^{CE}(\sg^{-1})^{DF}\chih_{CD}\alb_{EF}$$
The initial condition on $S_{0,u_0}$ is simply:
\begin{equation}
\left.\alb\right|_{S_{0,u_0}}=0 \label{2.186}
\end{equation}
$S_{0,u_0}$ lying on the boundary of the Minkowskian region $M_0$.
Equation \ref{2.185} together with the initial condition
\ref{2.186} has a unique smooth solution $\alb$ on $C_{u_0}$.
Noting \ref{2.163} and the fact that by \ref{2.116}:
\begin{equation}
|(\chih,\alb)|\leq|\chih|_{\sg}|\alb|_{\sg}\leq
C|u_0|^{-4}|\chih||\alb| \label{2.187}
\end{equation}
and using \ref{2.111} and \ref{2.184}, we deduce:
\begin{equation}
\alb={\cal M}_5(\delta^{3/2}|u_0|^{-3}) \ \ \mbox{hence} \ \
\alb={\bf M}_5(\delta^{3/2}|u_0|^{-5}) \label{2.188}
\end{equation}

Of the connection coefficients only $\omb$ remains to be
appropriately estimated along $C_{u_0}$. This satisfies the
propagation equation \ref{1.86}, which along $C_{u_0}$ becomes:
\begin{equation}
D\omb=h \label{2.189}
\end{equation}
where $h$ is the function:
\begin{equation}
h=-3|\zeta|^2_{\sg}-\rho \label{2.190}
\end{equation}
By \ref{2.135} and \ref{2.172}:
\begin{equation}
h={\bf M}_3(|u_0|^{-3}) \ \ \mbox{or, equivalently,} \ \ h={\cal
M}_3(|u_0|^{-3}) \label{2.191}
\end{equation}
Equation \ref{2.189} reads in each stereographic chart, simply:
\begin{equation}
\frac{\partial\omb}{\partial\ub}=h \label{2.192}
\end{equation}
The initial condition on $S_{0,u_0}$ is simply:
\begin{equation}
\left.\omb\right|_{S_{0,u_0}}=0 \label{2.193}
\end{equation}
For, $u$ is an affine parameter along the generators of $\Cb_0$,
hence:
\begin{equation}
\omb=0 \ \mbox{: along $\Cb_0$} \label{2.194}
\end{equation}
Equation \ref{2.192} together with the initial condition
\ref{2.193} has a unique smooth solution $\omb$ on $C_{u_0}$ and
by \ref{2.191} we have:
\begin{equation}
\omb={\cal M}_3(\delta|u_0|^{-3}) \ \ \mbox{hence} \ \ \omb={\bf
M}_3(\delta|u_0|^{-3}) \label{2.195}
\end{equation}

Our construction requires that we also appropriately estimate
$\Db\omb$ and $\Db\alb$ along $C_{u_0}$. To estimate the first we
consider again equation \ref{1.86}. Applying $\Db$ to this
equation and using the commutation formula \ref{1.75} we obtain:
\begin{eqnarray}
&&D\Db\omb=-4\Omega^2\zeta^\sharp\cdot\sd\omb+\Omega^2\left\{2(\etb-\eta,\Db\eta)+2(\eta,\Db\etb)-\Db\rho\right\}\nonumber\\
&&\hspace{13mm}+2\Omega^2\omb(2(\eta,\etb)-|\eta|_{\sg}^2-\rho)+2\Omega^3(-2\eta\cdot\chib\cdot\etb^\sharp+\eta^\sharp\cdot\chib\cdot\eta^\sharp)
\nonumber\\
&&\label{2.196}
\end{eqnarray}
Substituting for $\Db\eta$ from \ref{1.149}, for $\Db\etb$ from
\ref{1.67}, and for $\Db\rho$ from the eighth Bianchi identity of
Proposition 1.2, and evaluating the result along $C_{u_0}$, yields
the following propagation equation for $\Db\omb$ along $C_{u_0}$:
\begin{equation}
D\Db\omb=\dot{h} \label{2.197}
\end{equation}
where $\dot{h}$ is the function:
\begin{eqnarray}
&&\dot{h}=-12(\zeta,\sd\omb)-2\omb(3|\zeta|_{\sg}^2+\rho)+\sdiv\beb\nonumber\\
&&\hspace{7mm}+\frac{3}{2}\mbox{tr}\chib\rho+7(\zeta,\beb)+\frac{1}{2}(\chih,\alb)+12\zeta^\sharp\cdot\chib\cdot\zeta^\sharp
\label{2.198}
\end{eqnarray}
By \ref{2.195}, \ref{2.135}, \ref{2.164}, \ref{2.172},
\ref{2.180}, and \ref{2.188}:
\begin{equation}
\dot{h}={\bf M}_5(|u_0|^{-4} \ \ \mbox{or, equivalently,} \ \
\dot{h}={\cal M}_5(|u_0|^{-5}) \label{2.199}
\end{equation}
Equation \ref{2.199} reads in each stereographic chart, simply:
\begin{equation}
\frac{\partial\Db\omb}{\partial\ub}=\dot{h} \label{2.200}
\end{equation}
In view of \ref{2.194} the initial condition on $S_{0,u_0}$ is
simply:
\begin{equation}
\left.\Db\omb\right|_{S_{0,u_0}}=0 \label{2.201}
\end{equation}
Equation \ref{2.200} together with the initial condition
\ref{2.201} has a unique smooth solution $\Db\omb$ on $C_{u_0}$
and by \ref{2.199} we have:
\begin{equation}
\Db\omb={\cal M}_5(\delta|u_0|^{-4}) \ \ \mbox{hence} \ \
\omb={\bf M}_5(\delta|u_0|^{-4}) \label{2.202}
\end{equation}

Finally, to obtain an appropriate estimate for $\Db\alb$ along
$C_{u_0}$ we consider again the second Bianchi identity of
Proposition 1.2. Applying $\Db$ to this identity and taking into account Lemma 1.4 
we obtain:
\begin{eqnarray}
&&D\Db\alb-(\Omega\chih,\Db\alb)\sg-\frac{1}{2}\Omega\mbox{tr}\chi\Db\alb+2\omega\Db\alb=-4\sL_{\Omega^2\zeta^{\sharp}}\alb\nonumber\\
&&\hspace{10mm}+(\Db(\Omega\chih),\alb)\sg-\frac{1}{2}\Db(\Omega\mbox{tr}\chi)\alb+2\Omega\chib(\Omega\chih,\alb)-4\Omega^2(\chib,\chih\times\alb)\sg\nonumber\\
&&\hspace{10mm}-2(\Db\omega)\alb+\Omega\omb\left\{-\snab\oth\beb+(\zeta-4\etb)\oth\beb-3\chibh\rho+3\s^*\chibh\sigma\right\}\nonumber\\
&&\hspace{10mm}+\Omega\left\{-\snab\oth\Db\beb+2\hat{\Db\sGamma}\cdot\beb+2\Omega\chibh\sdiv\beb\right.\nonumber\\
&&\hspace{10mm}+(\Db\zeta-4\Db\etb)\oth\beb+(\zeta-4\etb)\oth\Db\beb-2\Omega\chibh(\zeta-4\etb,\beb)\nonumber\\
&&\hspace{10mm}\left.-3(\Db\chibh)\rho+3(\Db\s^*\chibh)\sigma-3\chibh\Db\rho+3\s^*\chibh\Db\sigma\right\}
\label{2.204}
\end{eqnarray}
Here $\Db\Gamma$ is the Lie derivative with respect to $\Lb$ of
the induced connection on $S_{\ub,u}$, given by (see Lemma 4.1):
\begin{equation}
(\Db\sGamma)^C_{AB}=(\sg^{-1})^{CD}(\snab_A(\Omega\chib)_{BD}+\snab_B(\Omega\chib)_{AD}-\snab_D(\Omega\chib)_{AB})
\label{2.205}
\end{equation}
and $\hat{\Db\sGamma}$ is the trace-free part of $\Db\sGamma$ with
respect to the lower indices:
\begin{equation}
(\hat{\Db\sGamma})^C_{AB}=(\Db\sGamma)^C_{AB}-\frac{1}{2}\sg_{AB}(\sg^{-1})^{DE}(\Db\sGamma)^C_{DE}
\label{2.206}
\end{equation}
The terms $2\hat{\Db\sGamma}\cdot\beb+2\Omega\chibh\sdiv\beb$
represent the commutator:
$$-\Db\snab\oth\beb+\snab\oth\Db\beb$$
(see Lemma 4.1). We substitute for $\Db(\Omega\chi)$ from
\ref{1.147}, for $\Db\omega$ from \ref{1.87}, for $\Db\zeta$ from
\ref{1.63}, for $\Db\etb$ from \ref{1.67}, and for $\Db\chib$ from
\ref{1.44}, and evaluate the results on $C_{u_0}$. Also, we
substitute for $\Db\beb$ from the fourth Bianchi identity of
Proposition 1.2, which reads, along $C_{u_0}$:
\begin{equation}
\Db\beb+\frac{3}{2}\mbox{tr}\chib\beb-\chibh^\sharp\cdot\beb-\omb\beb=-\sdiv\alb+\zeta
\label{2.207}
\end{equation}
and for $\Db\rho$ and $\Db\sigma$ from the eighth and tenth
Bianchi identities of Proposition 1.2, respectively, evaluated
along $C_{u_0}$. Taking into account the fact that for any
trace-free symmetric 2-covariant $S$ tensorfield $\theta$ we have:
\begin{equation}
\snab\oth\sdiv\theta=\slap\theta-2K\theta \label{2.208}
\end{equation}
and the fact that by \ref{2.57} we have:
\begin{equation}
\omega=0 \ \ \mbox{: along $C_{u_0}$} \label{2.209}
\end{equation}
we deduce the following propagation equation for $\Db\alb$ along
the generators of $C_{u_0}$:
\begin{equation}
D\Db\alb-(\chih,\Db\alb)\sg-\frac{1}{2}\mbox{tr}\chi\Db\alb=\dot{\tau}
\label{2.210}
\end{equation}
where $\dot{\tau}$ is the symmetric 2-covariant $S$ tensorfield:
\begin{eqnarray}
&&\dot{\tau}=\slap\alb-2K\alb-4\sL_{\zeta^\sharp}\alb+\snab\oth((3/2)\mbox{tr}\chib\beb-\chibh^\sharp\cdot\beb-\omb\beb-\zeta^\sharp\cdot\alb)
\nonumber\\
&&\hspace{5mm}-5\zeta\oth(\sdiv\alb-\zeta^\sharp\cdot\alb+(3/2)\mbox{tr}\chib\beb-\chibh^\sharp\cdot\beb-\omb\beb)\nonumber\\
&&\hspace{5mm}+3\chibh(\sdiv\beb+(3/2)\mbox{tr}\chib\rho+(\zeta,\beb)+(1/2)(\chih,\alb))\nonumber\\
&&\hspace{5mm}-3\s^*\chibh(\scurl\beb+(3/2)\mbox{tr}\chib\sigma+(\zeta,\s^*\beb)+(1/2)\chih\wedge\alb)\nonumber\\
&&\hspace{5mm}+2\chibh\sdiv\beb+\omb(-\snab\oth\beb+5\zeta\oth\beb-3\chibh\rho+3\chibh\sigma)\nonumber\\
&&\hspace{5mm}+2\hat{\Db\sGamma}\cdot\beb+(\sd\omb-5\chib^\sharp\cdot\zeta-5\beb)\oth\beb\nonumber\\
&&\hspace{5mm}+((\snab\oth\zeta+\zeta\oth\zeta+(1/2)\mbox{tr}\chib\chih-(1/2)\mbox{tr}\chi\chibh),\alb)\sg\nonumber\\
&&\hspace{5mm}-(\sdiv\zeta+|\zeta|_{\sg}^2-(1/2)(\chi,\chib)+\rho)\alb+2\chib(\chih,\alb)-4(\chib,\chih\times\alb)\sg\nonumber\\
&&\hspace{5mm}+2(3|\zeta|^2+\rho)\alb-10\chibh(\zeta,\beb)-3(\omb\chibh+|\chibh|_{\sg}^2\sg-\alb)\rho
+3(\omb\s^*\chibh-\s^*\alb)\sigma\nonumber\\
&&\label{2.211}
\end{eqnarray}
Using \ref{2.122}, \ref{2.126}, \ref{2.135}, \ref{2.146},
\ref{2.153}, \ref{2.164}, \ref{2.172}, \ref{2.174}, \ref{2.180},
\ref{2.188}, and \ref{2.195}, we find:
\begin{equation}
\dot{\tau}={\bf M}_7(\delta^{1/2}|u_0|^{-6}) \ \ \mbox{or,
equivalently,} \ \ \dot{\tau}={\cal M}_7(\delta^{1/2}|u_0|^{-4})
\label{2.212}
\end{equation}
the principal term in $\dot{\tau}$ being the term $\slap\alb$, but
the leading terms in behavior with respect to $\delta$ being the
terms $(9/2)\mbox{tr}\chib(\chibh\rho-\s^*\chibh\sigma)$. Equation
\ref{2.210} is a linear first order ordinary differential equation
for $\Db\alb$ along the generators of $C_{u_0}$, which in each
stereographic chart reads:
\begin{equation}
\frac{\partial(\Db\alb)_{AB}}{\partial\ub}-(\chih,\Db\alb)\sg_{AB}-\frac{1}{2}\mbox{tr}\chi(\Db\alb)_{AB}=\dot{\tau}_{AB}
\label{2.213}
\end{equation}
where:
$$(\chih,\Db\alb)=(\sg^{-1})^{CE}(\sg^{-1})^{DF}\chih_{CD}(\Db\alb)_{EF}$$
The initial condition on $S_{0,u_0}$ is simply:
\begin{equation}
\left.\Db\alb\right|_{S_{0,u_0}}=0 \label{2.214}
\end{equation}
$S_{0,u_0}$ lying on the boundary of the Minkowskian region $M_0$.
Equation \ref{2.210} together with the initial condition
\ref{2.214} has a unique smooth solution $\Db\alb$ on $C_{u_0}$.
Equation \ref{2.210} is in fact identical in form to equation
\ref{2.185} but with $\dot{\tau}$ in the role of $\tau$. In view
of \ref{2.212}, we deduce:
\begin{equation}
\Db\alb={\cal M}_7(\delta^{3/2}|u_0|^{-4}) \ \ \mbox{hence} \ \
\Db\alb={\bf M}_7(\delta^{3/2}|u_0|^{-6}) \label{2.215}
\end{equation}

Finally it is clear that the procedure which we have discussed
leads inductively to the conclusion that $\Db^n\omb$ and
$\Db^n\alb$ are smooth on $C_{u_0}$ for all non-negative integers
$n$ and, together with the results already obtained, this implies,
through the optical structure equations and the Bianchi
identities, that the transversal derivatives of any order of the
all the connection coefficients and of all the curvature
components are smooth on $C_{u_0}$. However, no further estimates
are needed.

\section{Construction of the solution in an initial domain}

Our setup, in summary, is the following. We have two null
hypersurfaces, $\Cb_0$ and $C_{u_0}$ intersecting at a sphere
$S_{0,u_0}$. The freely specifiable initial data on each of these
hypersurfaces consists of the conformal metric $\sgh$. In our case
the data is trivial on $\Cb_0$. Moreover, the full induced metric
$\sg$ is given on $S_{0,u_0}$, in our case the metric of a round
sphere of radius $|u_0|$ in Euclidean 3-dimensional space. Also,
the traces $\mbox{tr}\chib$ and $\mbox{tr}\chi$ of the 2nd
fundamental forms of $S_{0,u_0}$ relative to $\Cb_0$ and $C_{u_0}$
are given. In our case these are the mean curvatures of a round
sphere of radius $|u_0|$ in Euclidean 3-dimensonal space relative
to the inner and outer normals, respectively. Finally, the torsion
$\zeta$ of $S_{0,u_0}$ is given, which in our case vanishes, being
the torsion of a surface lying on a spacelike hyperplane in
Minkowski spacetime.

With initial data on two interesecting null hypersufaces set up in
this way, the theorem of Rendall [R] gives a spacetime neighborbood
$U$ of $S_{0,u_0}$, bounded in the past by the two null
hypersurfaces, and a unique smooth solution of the vacuum
Einstein equations in $U$, $(U,g)$ being a development of the
initial data on $U\bigcap(\Cb_0\bigcup C_{u_0})$. The solution is
obtained in ``harmonic" (also called ``wave") coordinates, adapted
to the two null hypersurfaces.

The construction to be presented in the third section of Chapter
16, then produces a smooth solution in canonical coordinates, as
defined in Chapter 1, in a domain $M^\prime\subset U$, of the form
$M^\prime=M\setminus(M_0\setminus\Cb_0)$, with $M$, as in Chapter
1, corresponding to some $c^*>u_0$ but sufficiently close to
$u_0$.

\chapter{$L^\infty$ Estimates for the Connection Coefficients}

\section{Introduction}

As we discussed in Chapter 1, we are assuming that we have a
spacetime manifold $(M,g)$ which is a smooth solution of the
vacuum Einstein equations such that the domain $M_0\bigcap M$ is
isometric to a domain in Minkowski spacetime. From now on we
restrict attention to the non-trivial part of the spacetime
manifold $(M,g)$:
\begin{equation}
M^\prime=M\setminus
\left(M_0\setminus\Cb_0\right)=\bigcup_{(\ub,u)\in
D^\prime}S_{\ub,u} \label{3.1}
\end{equation}
where $D^\prime$ is the parameter domain (see \ref{1.03}):
\begin{eqnarray}
&&D^\prime=\{(\ub,u) \ : \ \ub\in[0,\delta), \
u\in[u_0,c^*-\delta]\}
\bigcup\nonumber\\
&&\hspace{30mm}\{(\ub,u) \ : \ \ub\in[0,c^*-u), \ u\in(c^*-\delta,c^*)\}\nonumber\\
&&\hspace{5mm}=\{(\ub,u) \ : \ u\in[u_0,c^*-\ub), \ \ub\in[0,\delta)\}
\label{3.01}
\end{eqnarray}
for $c^*\geq u_0+\delta$, and:
\begin{eqnarray}
&&D^\prime=\{(\ub,u) \ : \ \ub\in[0,c^*-u), \ u\in[u_0,c^*)\}\nonumber\\
&&\hspace{5mm}=\{(\ub,u) \ : \ u\in[u_0,c^*-\ub), \ \ub\in[0,c^*-u_0)\}
\label{3.01a}
\end{eqnarray}
for $c^*<u_0+\delta$ (see Figures 1.1, 1.2).

The inner boundary of $M^\prime$ and outer boundary of the
Minkowskian region $M_0$ is the incoming null hypersurface
$\Cb_0$. We denote in the following by $C_u$ the part lying in
$M^\prime$ of what was denoted in Chapter 1 by $C_u$.

We recall from Chapter 1 that according to the fundamental
requirement on $(M^\prime,g)$ of the continuity argument, {\em the
generators of the $C_u$ and the $\Cb_{\ub}$ have no end points in
$M^\prime$}. It follows that for each $(\ub,u)\in D^\prime$
$S_{\ub,u}$ is a spacelike surface embedded in $M^\prime$.

The objective of the present chapter is to derive $L^\infty$
estimates for the connection coefficients in $M^\prime$ on the
basis of $L^\infty$ bounds for the curvature components in
$M^\prime$ and the following basic bootstrap assumption:

\vspace{5mm}

\ \ \ {\bf A0:} \ $\frac{1}{2}\leq\Omega\leq 2$ \ : \ in
$M^\prime$

\vspace{5mm}

The arguments of the present chapter, as well as those of the next
chapter, rely only on the propagation equations, equations
\ref{1.28}, \ref{1.42}, \ref{1.47}, \ref{1.66}, \ref{1.67},
\ref{1.86}, \ref{1.87} of Chapter 1. What is assumed in regard to
the curvature components in the present chapter is that the
following quantities are finite:
\begin{eqnarray}
{\cal R}_0^\infty(\alpha)&=&\sup_{M^\prime}(|u|\delta^{3/2}|\alpha|)\nonumber\\
{\cal R}_0^\infty(\beta)&=&\sup_{M^\prime}(|u|^2\delta^{1/2}|\beta|)\nonumber\\
{\cal R}_0^\infty(\rho)&=&\sup_{M^\prime}(|u|^3|\rho|)\nonumber\\
{\cal R}_0^\infty(\sigma)&=&\sup_{M^\prime}(|u|^3|\sigma|)\nonumber\\
{\cal R}_0^\infty(\beb)&=&\sup_{M^\prime}(|u|^4\delta^{-1}|\beb|)\nonumber\\
{\cal
R}_0^\infty(\alb)&=&\sup_{M^\prime}(|u|^{9/2}\delta^{-3/2}|\alb|)
\label{3.2}
\end{eqnarray}
By the results of Chapter 2, the corresponding quantities on
$C_{u_0}$, obtained by replacing the supremum on $M^\prime$ by the
supremum on $C_{u_0}$, are all bounded by a non-negative
non-decreasing continous function of $M_5$, the quantity requiring
$M_k$ with the highest $k$ being the one corresponding to $\alb$.

Throughout this monograph we shall denote by $C$ various numerical
constants which are {\em independent of $u_0$, $u_1$, $c^*$ or
$\delta$}.

We shall first derive an $L^\infty$ estimate for $\chi^\prime$,
then an $L^\infty$ estimate for $\chib^\prime$. After that we
shall derive $L^\infty$ estimates for $\eta, \etb$ together, as
the propagation equations \ref{1.66}, \ref{1.67} are coupled.
Finally, we shall derive $L^\infty$ estimates for $\omb$ and
$\omega$.

Now the first variational formulas \ref{1.28} imply:
\begin{equation}
D\sg^{-1}=-2\Omega\chi^{\sharp\sharp}, \ \ \
\Db\sg^{-1}=-2\Omega\chib^{\sharp\sharp} \label{3.3}
\end{equation}
where $\chi^{\sharp\sharp}$, $\chib^{\sharp\sharp}$, are the
symmetric 2-contravariant $S$ tensorfields with components, in an
arbitrary local frame field $(e_A \ : \ A=1,2)$ for $S_{\ub,u}$,
\begin{equation}
\chi^{\sharp\sharp AB}=\chi_{CD}(\sg^{-1})^{AC}(\sg^{-1})^{BD}, \
\ \ \ \chib^{\sharp\sharp
AB}=\chib_{CD}(\sg^{-1})^{AC}(\sg^{-1})^{BD} \label{3.4}
\end{equation}
It follows that for any 2-covariant $S$ tensorfield $\theta$ we
have:
\begin{equation}
D\mbox{tr}\theta=\mbox{tr}D\theta-2\Omega(\chi,\theta), \ \ \ \
\Db\mbox{tr}\theta=\mbox{tr}\Db\theta-2\Omega(\chib,\theta)
\label{3.5}
\end{equation}
Thus in view of \ref{1.a2} and equations \ref{1.155}, the traces of
the second variation equations \ref{1.42} and \ref{1.47} read:
\begin{equation}
D\mbox{tr}\chi^\prime=-\Omega^2|\chi^\prime|^2=-\frac{1}{2}\Omega^2(\mbox{tr}\chi^\prime)^2-\Omega^2|\chih^\prime|^2
\label{3.6}
\end{equation}
\begin{equation}
\Db\mbox{tr}\chib^\prime=-\Omega^2|\chib^\prime|^2=-\frac{1}{2}\Omega^2(\mbox{tr}\chib^\prime)^2-\Omega^2|\chibh^\prime|^2
\label{3.7}
\end{equation}
Decomposing $\chi^\prime$ according to \ref{1.162}, we have, by
the first of equations \ref{1.28},
\begin{eqnarray*}
D\chih^\prime&=&D\chi^\prime-\Omega^2\mbox{tr}\chi^\prime\chi^\prime-\frac{1}{2}\sg D\mbox{tr}\chi^\prime\\
&=&D\chi^\prime-\Omega^2\mbox{tr}\chi^\prime\chih^\prime
-\frac{1}{2}\left(D\mbox{tr}\chi^\prime+\Omega^2(\mbox{tr}\chi^\prime)^2\right)\sg
\end{eqnarray*}
Substituting for $D\chi^\prime$ from equation \ref{1.42} and
noting that by \ref{1.a4} and the identity \ref{1.a5}:
\begin{eqnarray*}
\chi^\prime\times\chi^\prime&=&\chih^\prime\times\chih^\prime+\mbox{tr}\chi^\prime\chih^\prime
+\frac{1}{4}\sg(\mbox{tr}\chi^\prime)^2\\
&=&\mbox{tr}\chi^\prime\chih^\prime+\left(\frac{1}{2}|\chih^\prime|^2+\frac{1}{4}(\mbox{tr}\chi^\prime)^2\right)\sg
\end{eqnarray*}
we conclude that we have simply:
\begin{equation}
\Dh\chih^\prime=-\alpha \label{3.8}
\end{equation}
Similarly we have:
\begin{equation}
\Dbh\chibh^\prime=-\alb \label{3.9}
\end{equation}

In the following we shall make repeated use of the following
elementary lemma on ordinary differential inequalities.

\vspace{5mm}

\noindent {\bf Lemma 3.1} \ \ \ Let $v$ be a non-negative function
defined on an interval $I$ with initial point $t_0$ and let $v^2$
be weakly differentiable in $I$ and satisfy the ordinary
differential inequality
$$\frac{d}{dt}(v^2)\leq 2v(av+b)$$
where $a, b$ are integrable functions on $I$ with $b$
non-negative. We then have:
$$v(t)\leq e^{A(t)}\left\{v(t_0)+\int_{t_0}^t e^{-A(t^\prime)}b(t^\prime)dt^\prime\right\} \ \ : \ \forall t\in I$$
where
$$A(t)=\int_{t_0}^ta(t^\prime)dt^\prime$$

\noindent {\em Proof:} \ Given a positive real number
$\varepsilon$ we set:
\begin{equation}
v_\varepsilon=\sqrt{\varepsilon^2+v^2} \label{3.b1}
\end{equation}
Then $v_\varepsilon$ is weakly differentiable on $I$ and
satisfies:
\begin{equation}
\frac{dv_\varepsilon}{dt}=\frac{1}{2v_\varepsilon}\frac{d}{dt}(v^2)
\leq a\frac{v^2}{v_\varepsilon}+b\frac{v}{v_\varepsilon}
\label{3.b2}
\end{equation}
Now, we have:
$$v_\varepsilon-\frac{v^2}{v_\varepsilon}=\frac{\varepsilon^2}{\sqrt{\varepsilon^2+v^2}}$$
hence:
\begin{equation}
0\leq v_\varepsilon-\frac{v^2}{v_\varepsilon}\leq\varepsilon
\label{3.b3}
\end{equation}
Therefore, writing:
\begin{equation}
a\frac{v^2}{v_\varepsilon}=av_\varepsilon-a\left(v_\varepsilon-\frac{v^2}{v_\varepsilon}\right)
\label{3.b4}
\end{equation}
we have:
\begin{equation}
-a\left(v_\varepsilon-\frac{v^2}{v_\varepsilon}\right)\leq\varepsilon|a|
\label{3.b5}
\end{equation}
Also, since
$$\frac{v}{v_\varepsilon}\leq 1,$$
we have:
\begin{equation}
b\frac{v}{v_\varepsilon}\leq b \label{3.b6}
\end{equation}
Substituting \ref{3.b4} - \ref{3.b6} in \ref{3.b2} we obtain:
\begin{equation}
\frac{dv_\varepsilon}{dt}\leq av_\varepsilon+b+\varepsilon|a|
\label{3.b7}
\end{equation}
Integrating from the initial point $t_0$, yields, for all $t\in
I$:
\begin{eqnarray}
v_\varepsilon&\leq&e^{A(t)}\left\{\sqrt{\varepsilon^2+v^2(t_0)}+\int_{t_0}^t
e^{-A(t^\prime)}b(t^\prime)dt^\prime\right.
\nonumber\\
&\s&\s\s\s\left.+\varepsilon\int_{t_0}^te^{-A(t^\prime)}|a|(t^\prime)dt^\prime\right\}
\label{3.b8}
\end{eqnarray}
This holds for every $\varepsilon>0$. Taking then the limit
$\varepsilon\rightarrow 0$ yields the lemma.

\section{$L^\infty$ Estimates for $\chi^\prime$}

\noindent {\bf Proposition 3.1} \ \ \ Under the bootstrap
assumption {\bf A0} there are numerical constants $C$ such that if
$$\delta\leq\frac{1}{C(1+{\cal R}_0^\infty(\alpha))^2}$$
then the following inequalities hold on $M^\prime:$
$$\frac{1}{|u|}\geq\frac{1}{2}\mbox{tr}\chi^\prime\geq\frac{1}{|u|}
-\frac{C}{|u|^2}\left[\delta+({\cal R}_0^\infty(\alpha))^2\right]
\ \ \mbox{: on $M^\prime$}$$
$$|\chih^\prime|\leq\frac{C{\cal R}_0^\infty(\alpha)}{|u|\delta^{1/2}} \ \ \mbox{: on $M^\prime$}$$

\noindent {\em Proof:} \ Equations \ref{3.6} and \ref{3.8}
together with the first of equations \ref{1.28} constitute a
nonlinear system of ordinary differential equations along each
generator of each $C_u$ for $\sg$, $\mbox{tr}\chi^\prime$,
$\chih^\prime$, if we think of $\Omega$ and $\alpha$ as given. The
initial conditions for this system are given on $\Cb_0$, the outer
boundary of the Minkowskian region $M_0$. Along $\Cb_0$ we have:
\begin{equation}
\Omega=1 \ \ \mbox{: along $\Cb_0$} \label{3.10}
\end{equation}
$u$ being an affine parameter along the generators of $\Cb_0$,
and:
\begin{equation}
\mbox{tr}\chi^\prime=\frac{2}{|u|}, \ \ \ \chih^\prime=0 \ \ \
\mbox{: along $\Cb_0$} \label{3.11}
\end{equation}
$S_{0,u}$ being a Euclidean sphere of radius $|u|$.

Consider then a given generator of a given $C_u$. We set:
\begin{equation}
\frac{1}{2}\mbox{tr}\chi^\prime=\frac{1}{|u|}-x \label{3.12}
\end{equation}
Integrating equation \ref{3.6} from $\Cb_0$ along the given
generator using as an initial condition the first of \ref{3.11},
which takes the form
\begin{equation}
x=0 \ \ \mbox{: on $\Cb_0$}, \label{3.13}
\end{equation}
yields:
\begin{equation}
x=\int_0^{\ub}\Omega^2\left(\frac{1}{4}(\mbox{tr}\chi^\prime)^2+\frac{1}{2}|\chih^\prime|^2\right)d\ub^\prime
\label{3.14}
\end{equation}
Thus $x\geq 0$.

For positive numerical constants $a$ and $k$ let ${\cal P}(t)$ be
the property:
$${\cal P}(t) \ : \ \ x\leq\frac{1}{|u|^2}\left[a\delta+k({\cal R}_0^\infty(\alpha))^2\right]
\ \ \mbox{: for all $\ub\in [0,t]$}$$ The constants $a$ and $k$
shall be chosen in the sequel. ${\cal P}(0)$ is true and it
follows by continuity that ${\cal P}(t)$ is true for sufficiently
small positive $t$. Let $t^*$ be the least upper bound of the set
of values of $t\in [0,\min\{\delta,c^*-u\})$ for which ${\cal
P}(t)$ is true. If $t^*=\min\{\delta,c^*-u\}$ then ${\cal P}(t)$
is true for all $t\in(0,\min\{\delta,c^*-u\})$, that is, the
estimate holds everywhere along the generator under consideration.

Suppose then that $t^*<\min\{\delta,c^*-u\}$. Then by continuity
${\cal P}(t^*)$ is true. Consequently:
\begin{equation}
\frac{1}{2}|\mbox{tr}\chi^\prime|\leq
|u|^{-1}\max\{1,|u|^{-1}\left[ a\delta+k({\cal
R}_0^\infty(\alpha))^2\right]\} \ \ \mbox{: for all $\ub\in
[0,t^*]$} \label{3.15}
\end{equation}
along the generator in question. Since $|u|\geq |c^*|\geq 1$,
setting
\begin{equation}
y=1+k({\cal R}_0^\infty(\alpha))^2 \label{3.16}
\end{equation}
then, provided that
\begin{equation}
a\delta\leq 1 \label{3.17}
\end{equation}
we have
$$\max\{1,|u|^{-1}\left[ a\delta+k({\cal R}_0^\infty(\alpha))^2\right]\}\leq y$$
hence:
\begin{equation}
|\mbox{tr}\chi^\prime|\leq \frac{2y}{|u|} \label{3.18}
\end{equation}

Now, by \ref{3.3}:
\begin{eqnarray*}
&D|\chih^\prime|^2=D(\chih^\prime, \chih^\prime)=D((\sg^{-1})^{AC}(\sg^{-1})^{BD}\chih^\prime_{AB}\chih^\prime_{CD})\\
&=-4\Omega^2\chi^{\prime\sharp\sharp
AC}\chih^\prime_{AB}\chih_C^{\prime\sharp B}
+2\chih^{\sharp\sharp AB}(D\chih^\prime)_{AB}\\
&=-4\Omega^2(\chi^\prime,\chih^\prime\times\chih^\prime)+2(\chih^\prime,D\chih^\prime)
\end{eqnarray*}
Since by the identity \ref{1.a5}
$$\chih^\prime\times\chih^\prime=\frac{1}{2}|\chih^\prime|^2\sg$$
and $(\chi^\prime,\sg)=\mbox{tr}\chi^\prime$, while
$(\chih^\prime,D\chih^\prime)=(\chih^\prime,\Dh\chih^\prime)$, we
obtain, by equation \ref{3.8}:
\begin{equation}
D|\chih^\prime|^2+2\Omega^2\mbox{tr}\chi^\prime|\chih^\prime|^2=-2(\chih^\prime,\alpha)
\label{3.19}
\end{equation}
Since
$$|(\chih^\prime,\alpha)|\leq |\chih^\prime||\alpha|$$
\ref{3.19} implies:
\begin{equation}
D|\chih^\prime|^2+2\Omega^2\mbox{tr}\chi^\prime|\chih^\prime|^2\leq
2|\chih^\prime||\alpha| \label{3.20}
\end{equation}
This is an ordinary differential inequality along the generator
under consideration. Applying Lemma 3.1 with initial point the
point of intersection with $\Cb_0$, yields, in view of the second
of \ref{3.11}:
\begin{equation}
|\chih^\prime|\leq e^{-F}\int_0^{\ub}e^{F}|\alpha|d\ub^\prime
\label{3.21}
\end{equation}
where:
\begin{equation}
F=\int_0^{\ub}\Omega^2\mbox{tr}\chi^\prime d\ub^\prime
\label{3.22}
\end{equation}
By virtue of \ref{3.18} and the bootstrap assumption {\bf A0} we
have, recalling that $\ub\in[0,\delta]$,
\begin{equation}
|F(\ub)-F(\ub^\prime)|=\left|\int_{\ub^\prime}^{\ub}\Omega^2\mbox{tr}\chi^\prime
d\ub^{\prime\prime}\right|
\leq\int_0^{\ub}\Omega^2|\mbox{tr}\chi^\prime|
d\ub^\prime\leq\frac{8y\delta}{|u|} \label{3.23}
\end{equation}
Hence:
\begin{equation}
|F(\ub)-F(\ub^\prime)|\leq\log 2 \label{3.24}
\end{equation}
provided that:
\begin{equation}
\delta\leq\frac{\log 2}{8y} \label{3.25}
\end{equation}
Substituting \ref{3.24} and the first of the definitions \ref{3.2}
in \ref{3.21} we obtain, recalling again that $\ub\in [0,\delta]$,
\begin{equation}
|\chih^\prime|\leq \frac{2{\cal
R}_0^\infty(\alpha)}{|u|\delta^{1/2}} \label{3.26}
\end{equation}

Let us now evaluate \ref{3.14} at $\ub=t^*$. In view of the bounds
\ref{3.18} and \ref{3.26} and assumption {\bf A0} we obtain:
\begin{equation}
x(t^*)\leq\int_0^{t^*}\frac{4}{|u|^2}\left(y^2+\frac{2({\cal
R}_0^\infty(\alpha))^2}{\delta}\right)d\ub \leq
\frac{4}{|u|^2}(y^2\delta+2({\cal R}_0^\infty(\alpha))^2)
\label{3.27}
\end{equation}
since $t^*\in (0,\delta]$. Suppose then that:
\begin{equation}
(4y^2-a)\delta<8({\cal R}_0^\infty(\alpha))^2 \label{3.28}
\end{equation}
The inequality \ref{3.27} would then imply that:
\begin{equation}
x(t^*)<\frac{1}{|u|^2}\left[a\delta+16({\cal
R}_0^\infty(\alpha))^2\right] \label{3.29}
\end{equation}
Therefore, setting
\begin{equation}
k=16 \label{3.30}
\end{equation}
the inequality defining property ${\cal P}$ would not be saturated
at $\ub=t^*$. Then, by continuity, property ${\cal P}$ would be
true for some $t>t^*$ contradicting the definition of $t^*$.

Let us set
\begin{equation}
a=8 \label{3.31}
\end{equation}
Then \ref{3.25} implies \ref{3.17}. Also, in view of the choice
\ref{3.30}, definition \ref{3.16} becomes:
\begin{equation}
y=1+16({\cal R}_0^\infty(\alpha))^2 \label{3.32}
\end{equation}
and condition \ref{3.28} takes the form:
\begin{equation}
8(y^2-2)\delta<y-1 \label{3.33}
\end{equation}
This is a non-trivial condition for $y>\sqrt{2}$. Comparing with
\ref{3.25}, the minimum value of the function
$$\frac{(y-1)y}{y^2-2} \ \ \mbox{on $(\sqrt{2},\infty)$}$$
is
$$\frac{1+(3/4)\sqrt{2}}{1+\sqrt{2}}>\log 2$$
Therefore \ref{3.25} implies \ref{3.28} as well. It follows that
with the choices \ref{3.30} and \ref{3.31},
$t^*=\min\{\delta,c^*-u\}$, ${\cal P}(t)$ is true for all
$t\in(0,\min\{\delta,c^*-u\})$, hence also the estimate \ref{3.26}
holds for all $\ub\in [0,\min\{\delta,c^*-u\})$, provided that
$\delta$ satisfies the smallness condition \ref{3.25}. Finally,
the smallness condition
$$\delta\leq\frac{\log 2}{128(1+{\cal R}_0^\infty(\alpha))^2}$$
implies condition \ref{3.25}. This completes the proof of the
proposition.

\vspace{5mm}

\section{$L^\infty$ Estimates for $\chib^\prime$}

We proceed to derive an $L^\infty$ estimate for $\chib^\prime$.
Now along $C_{u_0}$ we have:
\begin{equation}
\Omega=1 \ \ \mbox{: along $C_{u_0}$} \label{3.34}
\end{equation}
$\ub$ coinciding up to the additive constant $r_0$ with the affine
parameter $s$ along the generators of $C_{u_0}$.

Consider any $(\ub_1,u_1)\in D^\prime$. Let $D_1$ be the parameter
subdomain:
\begin{equation}
D_1=[0,\ub_1]\times[u_0,u_1]\subset D^\prime \label{3.02}
\end{equation}
and let $M_1$ be the corresponding subdomain of $M^\prime$:
\begin{equation}
M_1=\bigcup_{(\ub,u)\in D_1}S_{\ub,u} \label{3.03}
\end{equation}
In the following lemmas we fix attention to $M_1$. We shall derive
uniform estimates in $M_1$ which are independent of
$(\ub_1,u_1)\in D^\prime$. Uniform estimates in $M^\prime$ will
then follow. We denote by $C_u^{\ub_1}$ the part of $C_u$ which
corresponds to $\ub\leq \ub_1$. Thus for $u\in[u_0,u_1]$
$C_u^{\ub_1}$ is the part of $C_u$ lying in $M_1$.

Let $s^*$ be the least upper bound of the set of values of $s\in
[u_0,u_1]$ such that:
\begin{equation}
|\Omega-1|\leq\frac{1}{18}|u|^{-2} \ \ \mbox{: on $C_u^{\ub_1}$,
for all $u\in [u_0,s]$} \label{3.35}
\end{equation}
Then by continuity $s^*>u_0$ and we have:
\begin{equation}
|\Omega-1|\leq\frac{1}{18}|u|^{-2} \ \ \mbox{: on $C_u^{\ub_1}$,
for all $u\in [u_0,s^*]$} \label{3.36}
\end{equation}

Let us denote:
\begin{equation}
{\cal
D}_0^\infty(\chibh)=\sup_{C_{u_0}}(|u_0|^2\delta^{-1/2}|\chibh|)
\label{3.37}
\end{equation}
Here the supremum is meant to be taken on all of $C_{u_0}$, not only on the part which lies in 
$M^\prime$. This is the same for $c^*\geq u_0+\delta$, but not for $c^*\in(u_0,u_0+\delta)$. In the latter case 
the part of $C_{u_0}$ lying in $M^\prime$ is the part $C^{c^*-u_0}_{u_0}$. 
By the results of Chapter 2, ${\cal D}_0^\infty(\chibh)$ is
bounded by a non-negative non-decreasing continuous function of
$M_3$.

Also, let us denote by $M_1^{s^*}$ the subset of $M_1$ where
$u\leq s^*$:
\begin{equation}
M_1^{s^*}=\bigcup_{u\in[u_0,s^*]}C_u^{\ub_1} \label{3.38}
\end{equation}

\vspace{5mm}

\noindent {\bf Lemma 3.2} \ \ \ Suppose that on $C_{u_0}$
$\mbox{tr}\chib<0$ and:
$$\left|\frac{2}{\mbox{tr}\chib}+|u_0|\right|<\frac{1}{9} \ \ \ \mbox{: on $C_{u_0}$}$$
Then $\mbox{tr}\chib^\prime<0$ on $M^\prime$ and there are
numerical constants $C$ such that if
$$C({\cal D}_0^\infty(\chibh)+{\cal R}_0^\infty(\alb))^2\delta\leq 1$$
the following inequalities hold on $M_1^{s^*}$:
$$\left|\frac{2}{\mbox{tr}\chib^\prime}+|u|\right|\leq\frac{1}{3} \ \ \ \mbox{: on $M_1^{s^*}$}$$
$$|\chibh^\prime|\leq C\left(\frac{\delta^{1/2}}{|u|^2}{\cal D}_0^\infty(\chibh)
+\frac{\delta^{3/2}}{|u|^{7/2}}{\cal R}_0^\infty(\alb)\right) \ \
\ \mbox{: on $M_1^{s^*}$}$$

\noindent {\em Proof :} \ Equations \ref{3.7} and \ref{3.9}
together with the second of equations \ref{1.28} constitute a
nonlinear system of ordinary differential equations along each
generator of each $\Cb_{\ub}$ for $\sg$, $\mbox{tr}\chib^\prime$,
$\chibh^\prime$, if we think of $\Omega$ and $\alb$ as given. The
initial conditions for this system are given on $C_{u_0}$. Since
by \ref{3.7} $\Db\mbox{tr}\chi^\prime\leq 0$ while by the
assumptions of the lemma $\mbox{tr}\chib^\prime=\mbox{tr}\chib<0$
on $C_{u_0}$, it follows that $\mbox{tr}\chib^\prime<0$ in
$M^\prime$.

Consider then a given generator of a given $\Cb_{\ub}$,
$\ub\in[0,\ub_1]$.  We set:
\begin{equation}
\xb=-\frac{2}{\mbox{tr}\chi^\prime} \label{3.39}
\end{equation}
Then by \ref{3.7} $\xb$ satisfies:
\begin{equation}
\Db\xb=-\Omega^2-\frac{1}{2}\Omega^2\xb^2|\chibh^\prime|^2
\label{3.40}
\end{equation}
Integrating this equation from $C_{u_0}$ along the given generator
yields:
\begin{equation}
\xb(u)=\xb(u_0)-\int_{u_0}^u\Omega^2
du^\prime-\int_{u_0}^u\frac{1}{2}\Omega^2\xb^2|\chibh^\prime|^2
du^\prime \label{3.41}
\end{equation}

Let $\underline{\cal P}(t)$ be the property:
$$\underline{\cal P}(t) \ : \ \ |\xb-|u||\leq \frac{1}{3}
\ \ \mbox{: for all $u\in [u_0,t]$}$$ Then by the assumptions of
the lemma $\underline{\cal P}(0)$ is true and it follows by
continuity that $\underline{\cal P}(t)$ is true for sufficiently
small positive $t$. Let $t^*$ be the least upper bound of the set
of values of $t\in[u_0,s^*]$ for which $\underline{\cal P}(t)$ is
true. Then by continuity $\underline{\cal P}(t^*)$ is true. It
follows that:
\begin{equation}
\frac{1}{|u|+(1/3)}\leq\frac{1}{\xb}\leq\frac{1}{|u|-(1/3)} \ \
\mbox{: for all $u\in [u_0,t^*]$} \label{3.42}
\end{equation}
or:
\begin{equation}
-\frac{(1/3)}{|u|(|u|+(1/3))}\leq \frac{1}{\xb}-\frac{1}{|u|}\leq
\frac{(1/3)}{|u|(|u|-(1/3))} \ \ \mbox{: for all $u\in [u_0,t^*]$}
\label{3.43}
\end{equation}
which, since $|u|\geq 1$, implies:
\begin{equation}
\left|\frac{1}{\xb}-\frac{1}{|u|}\right|\leq \frac{1}{2|u|^2}
 \ \ \mbox{: for all $u\in [u_0,t^*]$}
\label{3.44}
\end{equation}
that is:
\begin{equation}
\left|\frac{1}{2}\mbox{tr}\chib^\prime+\frac{1}{|u|}\right|\leq
\frac{1}{2|u|^2}
 \ \ \mbox{: for all $u\in [u_0,t^*]$}
\label{3.45}
\end{equation}
Also, we have:
\begin{equation}
\frac{2|u|}{3}\leq\xb(u)\leq\frac{4|u|}{3} \ \ \mbox{: for all
$u\in [u_0,t^*]$} \label{3.46}
\end{equation}
Moreover, \ref{3.45} and \ref{3.46} together with \ref{3.36}
imply:
\begin{equation}
\left|\Omega^2\mbox{tr}\chib^\prime+\frac{2}{|u|}\right|\leq\frac{3}{2|u|^2}
\ \ \mbox{: for all $u\in [u_0,t^*]$} \label{3.47}
\end{equation}

Now, following an argument similar to that leading from \ref{3.8}
to \ref{3.20}, we derive from equation \ref{3.9} the inequality:
\begin{equation}
\Db|\chibh^\prime|^2+2\Omega^2\mbox{tr}\chib^\prime|\chibh^\prime|^2\leq
2|\chibh^\prime||\alb| \label{3.48}
\end{equation}
This is an ordinary differential inequality along the generator
under consideration. Applying Lemma 3.1 with initial point the
point of intersection with $C_{u_0}$, yields:
\begin{equation}
|\chibh^\prime(u)|\leq e^{-\Fb(u)}\left(|\chibh^\prime(u_0)|
+\int_{u_0}^u e^{\Fb(u^\prime)}|\alb|(u^\prime) du^\prime\right)
\label{3.49}
\end{equation}
where:
\begin{equation}
\Fb(u)=\int_{u_0}^u\Omega^2\mbox{tr}\chib^\prime du^\prime
\label{3.50}
\end{equation}
By virtue of \ref{3.47} we have, recalling that $|u|\geq 1$,
\begin{eqnarray*}
&&\left|\Fb(u)-\Fb(u^\prime)+\int_{u^\prime}^u
\frac{2}{|u^{\prime\prime}|}du^{\prime\prime}\right|
=\left|\int_{u^\prime}^u\left((\Omega^2\mbox{tr}\chib^\prime)(u^{\prime\prime})
+\frac{2}{|u^{\prime\prime}|}\right)du^{\prime\prime}\right|\\
&&\leq\int_{u^\prime}^u\left|(\Omega^2\mbox{tr}\chib^\prime)(u^{\prime\prime})
+\frac{2}{|u^{\prime\prime}|}\right|du^{\prime\prime}
\leq\int_{u_0}^u\left|(\Omega^2\mbox{tr}\chib^\prime)(u^\prime)
+\frac{2}{|u^\prime|}\right|du^\prime\\
&&\hspace{30mm}\leq\int_{u_0}^u\frac{3}{2|u^\prime|^2}du^\prime\leq\frac{3}{2|u|}\leq\frac{3}{2}
\end{eqnarray*}
that is:
\begin{equation}
\left|\Fb(u)-\Fb(u^\prime)+\log\left(\frac{|u^\prime|^2}{|u|^2}\right)\right|\leq
\frac{3}{2} \label{3.51}
\end{equation}
In particular, setting $u^\prime=u_0$,
\begin{equation}
\left|\Fb(u)+\log\left(\frac{|u_0|^2}{|u|^2}\right)\right|\leq\frac{3}{2}
\label{3.52}
\end{equation}
Inequalities \ref{3.51}, \ref{3.52} imply:
\begin{equation}
e^{-\Fb(u)}\leq e^{3/2}\frac{|u_0|^2}{|u|^2}, \ \ \
e^{-\Fb(u)+\Fb(u^\prime)}\leq e^{3/2}\frac{|u^\prime|^2}{|u|^2}
\label{3.53}
\end{equation}
Substituting \ref{3.53}, the last of the definitions \ref{3.2} and
the definition \ref{3.37}, in \ref{3.49}, noting that by
\ref{3.34} $\chibh^\prime=\chibh$ on $C_{u_0}$, we obtain:
\begin{equation}
|\chibh^\prime(u)|\leq
e^{3/2}\frac{\delta^{1/2}}{|u|^2}\left({\cal D}_0^\infty(\chibh)
+\frac{2\delta}{3|u|^{3/2}}{\cal R}_0^\infty(\alb)\right) \ \
\mbox{: for all $u\in [u_0,t^*]$} \label{3.54}
\end{equation}

Let us now evaluate \ref{3.41} at $u=t^*$:
\begin{equation}
\xb(t^*)=\xb(u_0)-\int_{u_0}^{t^*}\Omega^2
du-\int_{u_0}^{t^*}\frac{1}{2}\Omega^2\xb^2|\chibh^\prime|^2 du
\label{3.55}
\end{equation}
By virtue of \ref{3.36} we have, recalling that $t^*\leq -1$,
\begin{eqnarray}
&&\int_{u_0}^{t^*}\Omega^2
du\leq\int_{u_0}^{t^*}\left(1+(1/18)|u|^{-2}\right)^2 du
\leq \int_{u_0}^{t^*}\left(1+(1/6)|u|^{-2}\right)du\nonumber\\
&&\hspace{30mm}\leq |u_0|-|t^*|+(1/6)|t^*|^{-1} \label{3.56}
\end{eqnarray}
and:
\begin{eqnarray}
&&\int_{u_0}^{t^*}\Omega^2
du\geq\int_{u_0}^{t^*}\left(1-(1/18)|u|^{-2}\right)^2 du
\geq \int_{u_0}^{t^*}\left(1-(1/9)|u|^{-2}\right)du\nonumber\\
&&\hspace{30mm}\geq |u_0|-|t^*|-(1/9)|t^*|^{-1} \label{3.57}
\end{eqnarray}
Also, by virtue of \ref{3.54}, \ref{3.46} and \ref{3.36} we have:
\begin{eqnarray}
&&0\leq\int_{u_0}^{t^*}\frac{1}{2}\Omega^2\xb^2|\chibh^\prime|^2 du
\leq e^3\delta
\int_{u_0}^{t^*}\left({\cal D}_0^\infty(\chibh)+(2/3)|u|^{-3/2}\delta{\cal R}_0^\infty(\alb)\right)^2 |u|^{-2}du\nonumber\\
&&\leq e^3\delta\int_{u_0}^{t^*}\left({\cal
D}_0^\infty(\chibh)+\delta{\cal
R}_0^\infty(\alb)\right)^2|u|^{-2}du
\leq e^3\delta\left({\cal D}_0^\infty(\chibh)+\delta{\cal R}_0^\infty(\alb)\right)^2|t^*|^{-1}\nonumber\\
&&\hspace{40mm}\leq \frac{1}{18}|t^*|^{-1} \label{3.58}
\end{eqnarray}
the last step provided that:
\begin{equation}
18 e^3\left({\cal D}_0^\infty(\chibh)+{\cal
R}_0^\infty(\alb)\right)^2\delta\leq 1 \label{3.59}
\end{equation}
In view of \ref{3.56}, \ref{3.57}, \ref{3.58}, we conclude from
\ref{3.55} that:
\begin{equation}
\xb(u_0)-|u_0|-\frac{2}{9}|t^*|^{-1}\leq
\xb(t^*)-|t^*|\leq\xb(u_0)-|u_0|+\frac{1}{9}|t^*|^{-1}
\label{3.60}
\end{equation}
According to the assumptions of the lemma
\begin{equation}
|\xb(u_0)-|u_0||<\frac{1}{9} \label{3.61}
\end{equation}
In view of the fact that $|t^*|\geq 1$, it follows that:
\begin{equation}
|\xb(t^*)-|t^*||<\frac{1}{3} \label{3.62}
\end{equation}
Thus the inequality defing property $\underline{\cal P}$ is not
saturated at $u=t^*$. Hence by continuity, property
$\underline{\cal P}$ is true for some $t>t^*$ contradicting the
definition of $t^*$, unless $t^*=s^*$. This completes the proof of
the lemma. Note from \ref{3.45} that we have shown that
\begin{equation}
\left|\frac{1}{2}\mbox{tr}\chib^\prime+\frac{1}{|u|}\right|\leq
\frac{1}{2|u|^2}
 \ \ \mbox{: for all $u\in [u_0,s^*]$}
\label{3.63}
\end{equation}

\vspace{5mm}

\section{$L^\infty$ Estimates for $\eta,\etb$}

We proceed to derive $L^\infty$ estimates for $\eta,\etb$. These
satisfy the {\em coupled} system \ref{1.68} and \ref{1.69} of
ordinary differential equations along the generators of each $C_u$
and each $\Cb_{\ub}$ respectively. This is a linear system for
$\eta, \etb$, if we think of $\Omega$, $\sg$, $\chi$, $\chib$ and
$\beta, \beb$ as given. The initial condition for \ref{1.68} is on
$\Cb_0$ where we have:
\begin{equation}
\eta=0 \ \ \mbox{: along $\Cb_0$} \label{3.64}
\end{equation}
$\Cb_0$ being the outer boundary of the Minkowskian region $M_0$.
The initial condition for \ref{1.69} is on $C_{u_0}$ where, as
we recall from Chapter 2,  we have:
\begin{equation}
\etb=-\eta \ \ \mbox{: along $C_{u_0}$} \label{3.65}
\end{equation}

\vspace{5mm}

\noindent {\bf Lemma 3.3} \ \ \ Under the assumptions of
Proposition 3.1 and Lemma 3.2 the following estimates hold on
$M_1^{s^*}$:
$$|\eta|\leq \frac{C\delta^{1/2}}{|u|^2}{\cal R}_0^\infty(\beta)+\frac{C\delta^{3/2}}{|u|^3}{\cal R}_0^\infty(\beb)$$
$$|\etb|\leq \frac{C\delta^{1/2}}{|u|^2}{\cal R}_0^\infty(\beta)+\frac{C\delta}{|u|^3}{\cal R}_0^\infty(\beb)$$

\noindent {\em Proof:} \ By \ref{3.3} we have:
\begin{eqnarray}
&D|\eta|^2=D(\eta,\eta)=D((\sg^{-1})^{AB}\eta_A\eta_B)\nonumber\\
&=-2\Omega\chi^{\sharp\sharp AB}\eta_A\eta_B+2\eta^{\sharp A}(D\eta)_A\nonumber\\
&=-2\Omega\chi(\eta^\sharp,\eta^\sharp)+2(\eta, D\eta)
\label{3.66}
\end{eqnarray}
Since
$$|\chi(\eta^\sharp,\eta^\sharp)|\leq |\chi||\eta|^2, \ \ \ |(\eta, D\eta)|\leq |\eta||D\eta|,$$
and from equation \ref{1.68}:
$$|D\eta|\leq \Omega(|\chi||\etb|+|\beta|),$$
\ref{3.66} implies:
\begin{equation}
D|\eta|^2\leq 2\Omega|\eta|\{|\chi|(|\eta|+|\etb|)+|\beta|\}
\label{3.67}
\end{equation}
Thinking of $|\etb|$ as given, this is an ordinary differential
inequality for $|\eta|$ along the generators of the $C_u$.
Consider a given generator of a given $C_u$. We apply Lemma 3.1
with initial point the point of intersection with $\Cb_0$. In view
of the initial condition \ref{3.64} we obtain:
\begin{equation}
|\eta(\ub)|\leq\int_0^{\ub}\exp\left(\int_{\ub^\prime}^{\ub}(\Omega|\chi|)(\ub^{\prime\prime})d\ub^{\prime\prime}\right)
(\Omega(|\chi||\etb|+|\beta|))(\ub^\prime)d\ub^\prime \label{3.b9}
\end{equation}
Now, in view of the smallness condition on $\delta$ of Proposition
3.1, the estimates of that proposition yield:
\begin{equation}
|\chi|\leq \frac{C(1+{\cal R}_0^\infty(\alpha))}{\delta^{1/2}|u|}
\ \ \mbox{: on $M^\prime$} \label{3.68}
\end{equation}
Hence, the integral in the exponential in \ref{3.b9} is bounded
by:
$$\int_{\ub^\prime}^{\ub}\frac{C(1+{\cal R}_0^\infty(\alpha))}{\delta^{1/2}|u|}d\ub^{\prime\prime}
=\frac{C(1+{\cal
R}_0^\infty(\alpha))}{\delta^{1/2}|u|}(\ub-\ub^\prime)\leq
C(1+{\cal R}_0^\infty(\alpha))\delta^{1/2}$$ The last is not
greater than $\log 2$ provided that $\delta$ satisfies a smallness
condition of the same form as that required in Proposition 3.1.
Substituting then the bound \ref{3.68} and the second of the
definitions \ref{3.2} in \ref{3.b9} we obtain, along the generator
under consideration, the inequality:
\begin{equation}
|\eta(\ub)|\leq \frac{C(1+{\cal
R}_0^\infty(\alpha))}{\delta^{1/2}|u|}\int_0^{\ub}|\etb(\ub^\prime)|d\ub^\prime
+\frac{C\delta^{1/2}{\cal R}_0^\infty(\beta)}{|u|^2} \label{3.70}
\end{equation}
Let us define:
\begin{equation}
y(\ub,u)=\sup_{S_{\ub,u}}|\eta|, \ \ \
\yb(\ub,u)=\sup_{S_{\ub,u}}|\etb| \label{3.71}
\end{equation}
Then \ref{3.70} implies:
\begin{equation}
|\eta(\ub)|\leq \frac{C(1+{\cal
R}_0^\infty(\alpha))}{\delta^{1/2}|u|}\int_0^{\ub}\yb(\ub^\prime,u)d\ub^\prime
+\frac{C\delta^{1/2}{\cal R}_0^\infty(\beta)}{|u|^2} \label{3.72}
\end{equation}
Considering the generator of $C_u$ through each point $p\in
S_{\ub,u}$ and taking the supremum over $p\in S_{\ub,u}$ we then
obtain:
\begin{equation}
y(\ub,u)\leq \frac{C(1+{\cal
R}_0^\infty(\alpha))}{\delta^{1/2}|u|}\int_0^{\ub}\yb(\ub^\prime,u)d\ub^\prime
+\frac{C\delta^{1/2}{\cal R}_0^\infty(\beta)}{|u|^2} \label{3.73}
\end{equation}

By \ref{3.3} we have, in analogy with \ref{3.66},
\begin{equation}
\Db|\etb|^2=-2\Omega\chib(\etb^\sharp,\etb^\sharp)+2(\etb, D\etb)
\label{3.74}
\end{equation}
We now use the fact that
$$\chib(\etb^\sharp,\etb^\sharp)=\frac{1}{2}\mbox{tr}\chib|\etb|^2+\chibh(\etb^\sharp,\etb^\sharp)
\geq \frac{1}{2}\mbox{tr}\chib|\etb|^2-|\chibh||\etb|^2$$ together
with the fact that:
$$|(\etb,\Db\etb)|\leq |\etb||\Db\etb|$$
and equation \ref{1.69} to deduce:
\begin{equation}
\Db|\etb|^2+2\Omega\left(\frac{1}{2}\mbox{tr}\chib-|\chibh|\right)|\etb|^2\leq
2\Omega(|\chib||\eta|+|\beb|)|\etb| \label{3.75}
\end{equation}
Thinking of $|\eta|$ as given, this is an ordinary differential
inequality for $|\etb|$ along the generators of the $\Cb_{\ub}$.
Consider a given generator of a given $\Cb_{\ub}$,
$\ub\in[0,\ub_1]$. We apply Lemma 3.1 with initial point the point
of intersection with $C_{u_0}$ to obtain:
\begin{equation}
|\etb(u)|\leq e^{\tilde{\Fb}(u)}\left\{|\etb(u_0)|+\int_{u_0}^u
e^{-\tilde{\Fb}(u^\prime)}
(\Omega(|\chib||\eta|+|\beb|))(u^\prime)du^\prime\right\}
\label{3.76}
\end{equation}
where:
\begin{equation}
\tilde{\Fb}(u)=\int_{u_0}^u\Omega\left(-\frac{1}{2}\mbox{tr}\chib+|\chibh|)\right)du^\prime
\label{3.77}
\end{equation}
Let $u\leq s^*$. Then by Lemma 3.2 we have:
\begin{eqnarray*}
&&\left|\tilde{\Fb}(u)-\tilde{\Fb}(u^\prime)-\int_{u^\prime}^u\frac{1}{|u^{\prime\prime}|}du^{\prime\prime}\right|
=\left|\int_{u^\prime}^u\left(-\frac{1}{2}(\Omega\mbox{tr}\chib)(u^{\prime\prime})-\frac{1}{|u^{\prime\prime}|}
+(\Omega|\chibh|)(u^{\prime\prime})\right)du^{\prime\prime}\right|\\
&&\hspace{20mm}\leq\int_{u^\prime}^u\left|\frac{1}{2}(\Omega\mbox{tr}\chib)(u^{\prime\prime})
+\frac{1}{|u^{\prime\prime}|}\right|du^{\prime\prime}
+\int_{u^\prime}^u(\Omega|\chibh|)(u^{\prime\prime})du^{\prime\prime}\\
&&\hspace{20mm}\leq\int_{u_0}^u\frac{3}{4|u^\prime|^2}du^\prime+\int_{u_0}^u\frac{C\delta^{1/2}
({\cal D}_0^\infty(\chibh)+\delta{\cal R}_0^\infty(\alb))}{|u^\prime|^2}du^\prime\\
&&\hspace{30mm}\leq\int_{u_0}^u\frac{3}{2|u^\prime|^2}du^\prime\leq\frac{3}{2|u|}\leq\frac{3}{2}
\end{eqnarray*}
the last line by virtue of a smallness condition on $\delta$ of
the same form as that required in Lemma 3.2. We have thus
obtained:
\begin{equation}
\left|\tilde{\Fb}(u)-\tilde{\Fb}(u^\prime)-\log\left(\frac{|u^\prime|}{|u|}\right)\right|\leq\frac{3}{2}
\label{3.78}
\end{equation}
In particular, setting $u^\prime=u_0$,
\begin{equation}
\left|\tilde{\Fb}(u)-\log\left(\frac{|u_0|}{|u|}\right)\right|\leq\frac{3}{2}
\label{3.79}
\end{equation}
Also, by Lemma 3.2:
\begin{equation}
|\chib|\leq\frac{C}{|u|} \ \ \mbox{: on $M_1^{s^*}$} \label{3.80}
\end{equation}
under the smallness condition on $\delta$ required in Lemma 3.2.
Let us set:
\begin{equation}
{\cal D}_0^\infty(\etb)=\sup_{C^{\ub_1}_{u_0}}(|u_0|^2\delta^{-1/2}|\etb|)
\label{3.81}
\end{equation}
Here $C^{\ub_1}_{u_0}$, is the part of $C_{u_0}$ corresponding to $\ub\in[0,\ub_1]$.  
Substituting the bounds \ref{3.78}, \ref{3.79}, \ref{3.80}, the
next to last of the definitions \ref{3.2}, the first of
definitions \ref{3.71} and definition \ref{3.81} in \ref{3.76} we
deduce for $u\in[u_0,s^*]$ the inequality:
\begin{eqnarray}
|\etb(u)|&\leq&C\frac{|u_0|}{|u|}|\etb(u_0)|+C\int_{u_0}^u\frac{|u^\prime|}{|u|}\left(\frac{|\eta(u^\prime)|}{|u^\prime|}
+\frac{\delta{\cal R}_0^\infty(\beb)}{|u^\prime|^4}\right)du^\prime\nonumber\\
&\leq&\frac{C\delta^{1/2}}{|u||u_0|}{\cal
D}_0^\infty(\etb)+\frac{C}{|u|}\int_{u_0}^u
y(\ub,u^\prime)du^\prime +\frac{C\delta}{|u|^3}{\cal
R}_0^\infty(\beb) \label{3.82}
\end{eqnarray}
Considering the generator of $\Cb_{\ub}$ through each point $p\in
S_{\ub,u}$ and taking the supremum over $p\in S_{\ub,u}$ we then
obtain, for each $(\ub,u)\in[0,\ub_1]\times[u_0,s^*]$:
\begin{equation}
\yb(\ub,u)\leq \frac{C\delta^{1/2}}{|u||u_0|}{\cal
D}_0^\infty(\etb)+\frac{C}{|u|}\int_{u_0}^u
y(\ub,u^\prime)du^\prime +\frac{C\delta}{|u|^3}{\cal
R}_0^\infty(\beb) \label{3.83}
\end{equation}
This implies, for each $(\ub,u)\in[0,\ub_1]\times[u_0,s^*]$:
\begin{eqnarray}
\int_0^{\ub}\yb(\ub^\prime,u)d\ub^\prime&\leq&\frac{C\delta^{3/2}}{|u||u_0|}{\cal
D}_0^\infty(\etb)
+\frac{C\delta^2}{|u|^3}{\cal R}_0^\infty(\beb)\nonumber\\
&\s&+\frac{C}{|u|}\int_{u_0}^u\left\{\int_0^{\ub}y(\ub^\prime,u^\prime)d\ub^\prime\right\}du^\prime
\label{3.84}
\end{eqnarray}
Hence, setting:
\begin{equation}
z(u)=\sup_{\ub\in[0,\ub_1]}y(\ub,u)=\sup_{C_u^{\ub_1}}|\eta|
\label{3.85}
\end{equation}
we obtain, for each $(\ub,u)\in[0,\ub_1]\times[u_0,s^*]$:
\begin{eqnarray}
\int_0^{\ub}\yb(\ub^\prime,u)d\ub^\prime&\leq&\frac{C\delta^{3/2}}{|u||u_0|}{\cal
D}_0^\infty(\etb)
+\frac{C\delta^2}{|u|^3}{\cal R}_0^\infty(\beb)\nonumber\\
&\s&+\frac{C\delta}{|u|}\int_{u_0}^u z(u^\prime)du^\prime
\label{3.86}
\end{eqnarray}
Substituting in \ref{3.73} we then obtain, for each
$(\ub,u)\in[0,\ub_1]\times[u_0,s^*]$:
\begin{eqnarray}
y(\ub,u)&\leq&\frac{C\delta^{1/2}}{|u|^2}{\cal R}_0^\infty(\beta)
+\frac{C\delta^{3/2}}{|u|^3}({\cal D}_0^\infty(\etb)+{\cal R}_0^\infty(\beb))\nonumber\\
&\s&+\frac{C\delta^{1/2}(1+{\cal
R}_0^\infty(\alpha))}{|u|^2}\int_{u_0}^u z(u^\prime)du^\prime
\label{3.87}
\end{eqnarray}
Taking the supremum over $\ub\in [0,\ub_1]$ then yields the linear
integral inequality:
\begin{eqnarray}
z(u)&\leq&\frac{C\delta^{1/2}}{|u|^2}{\cal R}_0^\infty(\beta)
+\frac{C\delta^{3/2}}{|u|^3}({\cal D}_0^\infty(\etb)+{\cal R}_0^\infty(\beb))\nonumber\\
&\s&+\frac{C\delta^{1/2}(1+{\cal
R}_0^\infty(\alpha))}{|u|^2}\int_{u_0}^u z(u^\prime)du^\prime\nonumber\\
&:&\forall u\in[u_0,s^*]\label{3.88}
\end{eqnarray}
Setting
\begin{equation}
Z(u)=\int_{u_0}^u z(u^\prime) du^\prime, \ \ \mbox{we have
$Z(u_0)=0$} \label{3.89}
\end{equation}
and \ref{3.88} takes the form:
\begin{equation}
\frac{dZ}{du}\leq aZ+b \label{3.90}
\end{equation}
where
\begin{eqnarray}
&&a(u)=\frac{C\delta^{1/2}(1+{\cal R}_0^\infty(\alpha))}{|u|^2}\label{3.91}\\
&&b(u)=\frac{C\delta^{1/2}}{|u|^2}{\cal R}_0^\infty(\beta)
+\frac{C\delta^{3/2}}{|u|^3}({\cal D}_0^\infty(\etb)+{\cal
R}_0^\infty(\beb))\label{3.92}
\end{eqnarray}
Integrating \ref{3.90} from $u_0$ we obtain:
\begin{equation}
Z(u)\leq\int_{u_0}^u e^{\int_{u^\prime}^u
a(u^{\prime\prime})du^{\prime\prime}}b(u^\prime)du^\prime \ \ \ :
\ \forall u\in[u_0,s^*] \label{3.93}
\end{equation}
We have:
\begin{equation}
\int_{u^\prime}^u a(u^{\prime\prime})du^{\prime\prime}\leq
\frac{C\delta^{1/2}(1+{\cal R}_0^\infty(\alpha))}{|u|} \leq\log 2
\label{3.94}
\end{equation}
the last step by virtue of a smallness condition on $\delta$ of
the same form as that required in Proposition 3.1. Therefore:
\begin{eqnarray}
Z(u)&\leq& 2\int_{u_0}^ub(u^\prime)du^\prime\nonumber\\
&\leq&\frac{C\delta^{1/2}}{|u|}{\cal R}_0^\infty(\beta)
+\frac{C\delta^{3/2}}{|u|^2}({\cal D}_0^\infty(\etb)+{\cal
R}_0^\infty(\beb)) \nonumber\\
&:&\forall u\in[u_0,s^*]\label{3.95}
\end{eqnarray}
and substituting in \ref{3.87} yields:
\begin{equation}
y(\ub,u)\leq \frac{C\delta^{1/2}}{|u|^2}{\cal R}_0^\infty(\beta)
+\frac{C\delta^{3/2}}{|u|^3}({\cal D}_0^\infty(\etb)+{\cal
R}_0^\infty(\beb)) \ \ \ \ : \ \forall
(\ub,u)\in[0,\ub_1]\times[u_0,s^*] \label{3.96}
\end{equation}
taking into account a smallness condition on $\delta$ of the same
form as that required in the proof of Proposition 3.1. Also, since
$$\int_{u_0}^u y(\ub,u^\prime)du^\prime\leq Z(u),$$
substituting \ref{3.95} in \ref{3.83} yields:
\begin{equation}
\yb(\ub,u)\leq\frac{C\delta^{1/2}}{|u|^2}\left(\frac{|u|}{|u_0|}{\cal
D}_0^\infty(\etb)+{\cal R}_0^\infty(\beta)\right)
+\frac{C\delta}{|u|^3}({\cal D}_0^\infty(\etb)+{\cal
R}_0^\infty(\beb)) \label{3.97}
\end{equation}
To complete the proof of the lemma we must derive an appropriate
estimate for ${\cal D}_0^\infty(\etb)$. To do this we consider
\ref{3.67} on $C_{u_0}$. Since along $C_{u_0}$ condition
\ref{3.65} holds, \ref{3.67} simplifies on $C_{u_0}$ to:
\begin{equation}
D|\eta|^2\leq 2\Omega|\eta|\{2|\chi||\eta|+|\beta|\} \label{3.98}
\end{equation}
This is an ordinary differential inequality along the generators
of $C_{u_0}$, to which Lemma 3.1 applies. Taking into account
\ref{3.68} and a smallness condition on $\delta$ of the same form
as that required in the proof of Proposition 3.1, we deduce:
\begin{equation}
{\cal D}_0^\infty(\etb)\leq C{\cal R}_0^\infty(\beta) \label{3.99}
\end{equation}
Substituting then \ref{3.99} in \ref{3.96} and \ref{3.97}, in view
of the definitions \ref{3.71}, yields the lemma.

\vspace{5mm}

\section{$L^\infty$ Estimates for $\omega, \omb$}

We proceed to derive an $L^\infty$ estimate for $\omb$ on
$M_1^{s^*}$. Using this estimate we shall derive an $L^\infty$
estimate for $\log\Omega$ in $M_1^{s^*}$ which improves the bound
\ref{3.36}. This shall enable us to show that $s^*=u_1$ so that
the previous lemmas actually hold on $M_1$ and, since
$(\ub_1,u_1)\in D^\prime$ is arbitrary, the estimates hold on all
of $M^\prime$. We shall then derive an $L^\infty$ estimate for
$\omega$ on $M^\prime$. Finally, using the improved estimate for
$\log\Omega$ we shall derive an estimate for
$\mbox{tr}\chib^\prime$ in $M^\prime$ which improves the estimate
given by Lemma 3.2.

\vspace{5mm}

\noindent {\bf Lemma 3.4} \ \ \ Under the assumptions of
Proposition 3.1 and Lemma 3.2 the following estimate holds on
$M_1^{s^*}$:
$$|\omb|\leq \frac{C\delta}{|u|^3}{\cal R}_0^\infty(\rho)
+\frac{C\delta^2}{|u|^4}({\cal R}_0^\infty(\beta)+{\cal
R}_0^\infty(\beb))^2$$

\noindent {\em Proof:} \ Equation \ref{1.86} is a linear ordinary
differential equation for $\omb$ along the generators of the
$C_u$. Substituting the bounds for $\eta, \etb$ of Lemma 3.3 as
well as the third of the definitions \ref{3.2} we obtain:
\begin{eqnarray}
&&D|\omb|\leq |D\omb|\leq \Omega(2|\eta|^2+|\etb|^2+|\rho|)\nonumber\\
&&\leq \frac{C}{|u|^3}{\cal
R}_0^\infty(\rho)+\frac{C\delta}{|u|^4}({\cal
R}_0^\infty(\beta)+{\cal R}_0^\infty(\beb))^2 \label{3.100}
\end{eqnarray}
Integrating this inequality from $\Cb_0$ along each generator of
each $C_u^{\ub_1}$, $u\in[u_0,s^*]$, using the fact that by
\ref{3.10}:
\begin{equation}
\omb=0 \ \ \mbox{: along $\Cb_0$} \label{3.a1}
\end{equation}
yields the lemma.

\vspace{5mm}

\noindent {\bf Lemma 3.5} \ \ \ Under the assumptions of
Proposition 3.1 and Lemma 3.2 there is a numerical constant $C$
such that if $\delta$ satisfies the smallness condition:
$$C\left[{\cal R}_0^\infty(\rho)+({\cal R}_0^\infty(\beta)+{\cal R}_0^\infty(\beb))^2\right]\delta<1$$
then we have $s^*=u_1$ and the estimate
$$\log\Omega\leq\frac{C\delta}{|u|^2}{\cal R}_0^\infty(\rho)
+\frac{C\delta^2}{|u|^3}({\cal R}_0^\infty(\beta)+{\cal
R}_0^\infty(\beb))^2$$ holds on $M_1$.

\noindent {\em Proof:} \ According to the second of the
definitions \ref{1.17} we have:
\begin{equation}
\Db\log\Omega=\omb \label{3.101}
\end{equation}
This is a linear ordinary differential equation for $\log\Omega$
along the generators of the $\Cb_{\ub}$. Integrating this equation
from $C_{u_0}$ along a given generator of a given $\Cb_{\ub}$,
$\ub\in[0,\ub_1]$, using the initial condition \ref{3.34} we
obtain:
\begin{equation}
|\log\Omega(u)|\leq\int_{u_0}^u|\omb(u^\prime)| du^\prime
\label{3.102}
\end{equation}
Substituting the estimate for $\omb$ of Lemma 3.4 we conclude that
on $M_1^{s^*}$ we have:
\begin{equation}
|\log\Omega|\leq \frac{C\delta}{|u|^2}{\cal R}_0^\infty(\rho)
+\frac{C\delta^2}{|u|^3}({\cal R}_0^\infty(\beta)+{\cal
R}_0^\infty(\beb))^2 \label{3.103}
\end{equation}
Since $|\Omega-1|\leq\max\{\Omega,1\}|\log\Omega|$, this implies
that on $M_1^{s^*}$:
\begin{equation}
|\Omega-1|\leq \frac{C\delta}{|u|^2}\left[{\cal R}_0^\infty(\rho)
+\delta({\cal R}_0^\infty(\beta)+{\cal R}_0^\infty(\beb))^2\right]
\label{3.104}
\end{equation}
Thus, if we require $\delta$ to satisfy the smallness condition:
\begin{equation}
C\delta\left[{\cal R}_0^\infty(\rho) +({\cal
R}_0^\infty(\beta)+{\cal R}_0^\infty(\beb))^2\right]<\frac{1}{18}
\label{3.105}
\end{equation}
we have:
\begin{equation}
|\Omega-1|<\frac{1}{18}|u|^{-2} \ \ \mbox{: on $C_u^{\ub_1}$, for
all $u\in [u_0,s^*]$} \label{3.106}
\end{equation}
hence by continuity \ref{3.36} holds for some $s>s^*$
contradicting the definition of $s^*$, unless $s^*=u_1$. This
completes the proof of the lemma.

\vspace{5mm}

Since by the above lemma $M_1^{s^*}=M_1$, the results of Lemmas
3.2, 3.3 and 3.4 hold on $M_1$. Moreover, since $(\ub_1,u_1)\in
D^\prime$ is arbitrary we conclude:

\vspace{5mm}

\noindent {\bf Proposition 3.2} \ \ \  Under the assumptions of
Proposition 3.1 and Lemmas 3.2 and 3.5, the estimates of Lemmas
3.2 - 3.5 hold on all of $M^\prime$.

\vspace{5mm}

\noindent {\bf Proposition 3.3} \ \ \ Under the assumptions of
Proposition 3.1 and Lemmas 3.1 and 3.4 the following estimate
holds on $M^\prime$:
$$|\omega|\leq \frac{C}{|u|^2}{\cal R}_0^\infty(\rho)
+\frac{C\delta}{|u|^3}({\cal R}_0^\infty(\beta)+{\cal
R}_0^\infty(\beb))^2$$

\noindent {\em Proof:} \ Equation \ref{1.87} is a linear ordinary
differential equation for $\omega$ along the generators of the
$\Cb_{\ub}$. Substituting the estimates for $\eta, \etb$ of Lemma
3.3 as well as the third of the definitions \ref{3.2} we obtain:
\begin{eqnarray}
&&\Db|\omega|\leq |\Db\omega|\leq \Omega(2|\etb|^2+|\eta|^2+|\rho|)\nonumber\\
&&\leq \frac{C}{|u|^3}{\cal
R}_0^\infty(\rho)+\frac{C\delta}{|u|^4}({\cal
R}_0^\infty(\beta)+{\cal R}_0^\infty(\beb))^2 \label{3.107}
\end{eqnarray}
Integrating this inequality from $C_{u_0}$ along each generator of
each $\Cb_{\ub}$ using the fact that by \ref{3.34}:
\begin{equation}
\omega=0 \ \ \mbox{: along $C_{u_0}$} \label{3.a2}
\end{equation}
yields the proposition.

\vspace{5mm}

Now, if the initial data along $C_0$ where trivial we would have:
$$\mbox{tr}\chib=\frac{2}{u-\ub}=-\frac{2}{|u_0|}\left(1-\frac{\ub}{|u_0|}+O\left(\frac{\delta^2}{|u_0|^2}\right)\right)
\ \ \ \mbox{: on $C_{u_0}$}$$ Motivated by the comparison with the
trivial case we define:
\begin{equation}
{\cal
D}_0^\infty(\mbox{tr}\chib)=\sup_{C_{u_0}}\left(|u_0|^3\delta^{-1}\left|\frac{1}{2}\mbox{tr}\chib
+\frac{1}{|u_0|}\left(1-\frac{\ub}{|u_0|}\right)\right|\right)
\label{3.108}
\end{equation}
Here, as in \ref{3.37}, the supremum is meant to be taken on all of $C_{u_0}$, not only on the part which lies in 
$M^\prime$. This is the same for $c^*\geq u_0+\delta$, but not for $c^*\in(u_0,u_0+\delta)$. In the latter case 
the part of $C_{u_0}$ lying in $M^\prime$ is the part $C^{c^*-u_0}_{u_0}$. 
By the results of Chapter 2 (see \ref{2.146}, \ref{2.153}) ${\cal
D}_0^\infty(\mbox{tr}\chib)$ is bounded by a non-negative
non-decreasing continuous function of $M_3$.

\vspace{5mm}

\noindent {\bf Proposition 3.4} \ \ \ Under the assumptions of
Proposition 3.1 and Lemmas 3.2 and 3.5 the following estimate
holds on $M^\prime$:
\begin{eqnarray*}
\left|\frac{1}{2}\mbox{tr}\chib^\prime+\frac{1}{|u|}\left(1-\frac{\ub}{|u|}\right)\right|
&\leq& C\delta|u|^{-3}\left[{\cal
D}_0^\infty(\mbox{tr}\chib)+({\cal D}_0^\infty(\chibh))^2
+{\cal R}_0^\infty(\rho)\right]\\
&\s&+\delta^2|u|^{-3}\left[(1+{\cal
D}_0^\infty(\mbox{tr}\chib))^2+{\cal D}_0^\infty(\chibh){\cal
R}_0^\infty(\alb)
\right.\\
&\s&\s\s\s\s\s\s\s\s\s\left.+({\cal R}_0^\infty(\beta)+{\cal
R}_0^\infty(\beb)+{\cal R}_0^\infty(\alb))^2\right]
\end{eqnarray*}
This improves the estimate for $\mbox{tr}\chib^\prime$ given by
Lemma 3.2.

\noindent {\em Proof:} \ We revisit equation \ref{3.41} which we
write in the form, subtracting $|u|+\ub$ from both sides,
\begin{eqnarray}
\xb(u)-|u|-\ub&=&\xb(u_0)-|u_0|-\ub-\int_{u_0}^u\Omega^2 du^\prime+|u_0|-|u|\nonumber\\
&\s&-\int_{u_0}^u\frac{1}{2}\Omega^2\xb^2|\chibh^\prime|^2
du^\prime \label{3.109}
\end{eqnarray}
Using the result of Lemma 3.5 we now estimate:
\begin{eqnarray}
&&\left|\int_{u_0}^u\Omega^2
du^\prime-|u_0|+|u|\right|=\left|\int_{u_0}^u(\Omega^2-1)du^\prime\right|
\leq\int_{u_0}^u|\Omega^2-1|du^\prime\nonumber\\
&&\leq C\int_{u_0}^u|\log\Omega|du^\prime\leq
\frac{C\delta}{|u|}{\cal R}_0^\infty(\rho)
+\frac{C\delta^2}{|u|^2}({\cal R}_0^\infty(\beta)+{\cal
R}_0^\infty(\beb))^2 \label{3.110}
\end{eqnarray}
Also, using the results of Lemma 3.2 we estimate, as in \ref{3.58}
\begin{eqnarray}
&&0\leq\int_{u_0}^u\frac{1}{2}\Omega^2\xb^2|\chibh^\prime|^2
du^\prime \leq C\delta\int_{u_0}^u\left({\cal D}_0^\infty(\chibh)
+|u^\prime|^{-3/2}\delta{\cal R}_0^\infty(\alb)\right)^2|u^\prime|^{-2}du^\prime\nonumber\\
&&\hspace{30mm}\leq \frac{C\delta}{|u|}\left({\cal D}_0^\infty(\chibh)
+|u|^{-3/2}\delta{\cal R}_0^\infty(\alb)\right)^2\label{3.111}
\end{eqnarray}
Moreover, setting
\begin{equation}
\mbox{tr}\chib_{|C_{u_0}}=-\frac{2}{|u_0|}\left(1-\frac{\ub}{|u_0|}+\varepsilon\right)
\label{3.112}
\end{equation}
we have, by definition \ref{3.108},
\begin{equation}
|\varepsilon|\leq \delta|u_0|^{-2}{\cal
D}_0^\infty(\mbox{tr}\chib) \label{3.113}
\end{equation}
Since
\begin{eqnarray*}
\xb(u_0)-|u_0|-\ub&=&\left\{\frac{1}{\left(1-\frac{\ub}{|u_0|}+\varepsilon\right)}
-\left(1+\frac{\ub}{|u_0|}-\varepsilon\right)-\varepsilon\right\}|u_0|\\
&=&\frac{\left(\frac{\ub}{|u_0|}-\varepsilon\right)^2|u_0|}{\left(1-\frac{\ub}{|u_0|}+\varepsilon\right)}
-\varepsilon|u_0|\\
&=&\left(\frac{\ub}{|u_0|}-\varepsilon\right)^2\xb(u_0)-\varepsilon|u_0|
\end{eqnarray*}
we obtain, in view of \ref{3.61},
\begin{eqnarray}
|\xb(u_0)-|u_0|-\ub|&\leq&|\varepsilon|+C\left(\frac{\ub}{|u_0|}-\varepsilon\right)^2|u_0|\nonumber\\
&\leq&\delta|u_0|^{-1}{\cal
D}_0^\infty(\mbox{tr}\chib)+C\delta^2|u_0|^{-1}(1+{\cal
D}_0^\infty(\mbox{tr}\chib))^2
\nonumber\\
&\s&\label{3.114}
\end{eqnarray}
Substituting \ref{3.110}, \ref{3.111}, \ref{3.114} in \ref{3.109}
yields:
\begin{eqnarray}
|\xb-|u|-\ub|&\leq& C\delta|u|^{-1}\left[{\cal
D}_0^\infty(\mbox{tr}\chib)+({\cal D}_0^\infty(\chibh))^2
+{\cal R}_0^\infty(\rho)\right]\nonumber\\
&\s&+\delta^2|u|^{-1}\left[(1+{\cal
D}_0^\infty(\mbox{tr}\chib))^2+{\cal D}_0^\infty(\chibh){\cal
R}_0^\infty(\alb)
\right.\nonumber\\
&\s&\s\s\s\s\s\s\s\s\s\left.+({\cal R}_0^\infty(\beta)+{\cal
R}_0^\infty(\beb)+{\cal R}_0^\infty(\alb))^2\right] \label{3.115}
\end{eqnarray}
Since
$$\frac{1}{2}\mbox{tr}\chib^\prime+\frac{1}{|u|+\ub}=\frac{\xb-|u|-\ub}{(|u|+\ub)\xb}$$
and by Lemma 3.2
$$\xb\geq |u|-\frac{1}{3}\geq\frac{2|u|}{3},$$
the proposition follows.

\section{The smallness requirement on $\delta$}

Let us now review the assumptions of Lemma 3.2, in the light of
the definition \ref{3.108}. Consider first the assumption that
$\mbox{tr}\chib<0$ on $C_{u_0}$. Now, the condition
\begin{equation}
(1+{\cal D}_0^\infty(\mbox{tr}\chib))\delta<1 \label{3.116}
\end{equation}
implies:
\begin{equation}
{\cal
D}_0^\infty(\mbox{tr}\chib)<|u_0|^2\left(\frac{1}{\delta}-\frac{1}{|u_0|}\right)
\label{3.117}
\end{equation}
The last implies:
\begin{equation}
\left|\frac{1}{2}\mbox{tr}\chib+\frac{1}{|u_0|}\left(1-\frac{\ub}{|u_0|}\right)\right|<\frac{1}{|u_0|}\left(1-\frac{\delta}{|u_0|}\right)
\label{3.118}
\end{equation}
which in turn implies that $\mbox{tr}\chib<0$ on $C_{u_0}$.
Therefore the assumption that $\mbox{tr}\chib<0$ on $C_{u_0}$ may
be replaced by the smallness condition \ref{3.116} on $\delta$.

Consider next the assumption that
\begin{equation}
\left|\frac{2}{\mbox{tr}\chib}+|u_0|\right|<\frac{1}{9} \ \
\mbox{: on $C_{u_0}$} \label{3.119}
\end{equation}
Suppose that for some constant $\varepsilon_0\in(0,1)$ we have:
\begin{equation}
\left|\frac{1}{2}\mbox{tr}\chib+\frac{1}{|u_0|}\right|<\frac{\varepsilon_0}{|u_0|^2}
\ \ \mbox{: on $C_{u_0}$} \label{3.120}
\end{equation}
It follows in a straightforward manner that we then also have:
\begin{equation}
\left|\frac{2}{\mbox{tr}\chib}+|u_0|\right|<\frac{\varepsilon_0}{1-\varepsilon_0}
\ \ \mbox{: on $C_{u_0}$} \label{3.121}
\end{equation}
Thus, setting $\varepsilon_0=1/10$, \ref{3.119} follows if we
assume that:
\begin{equation}
\left|\frac{1}{2}\mbox{tr}\chib+\frac{1}{|u_0|}\right|<\frac{1}{10}\frac{1}{|u_0|^2}
\ \ \mbox{: on $C_{u_0}$} \label{3.122}
\end{equation}
On the other hand from the definition \ref{3.108} we obtain:
\begin{equation}
\left|\frac{1}{2}\mbox{tr}\chib+\frac{1}{|u_0|}\right|\leq\frac{\delta{\cal
D}_0^\infty(\mbox{tr}\chib)}{|u_0|^3}+\frac{\delta}{|u_0|^2}
 \ \ \mbox{: on $C_{u_0}$}
\label{3.123}
\end{equation}
and the right hand side of \ref{3.123} is less than
$(1/10)|u_0|^{-2}$ if
\begin{equation}
(1+{\cal D}_0^\infty(\mbox{tr}\chib))\delta<\frac{1}{10}
\label{3.124}
\end{equation}
This being a stronger condition of the form \ref{3.116} it impies
both assumptions of Lemma 3.2.

To summarize, the assumptions of the present chapter are, besides
the basic bootstrap assumption {\bf A0} and the finiteness of the
quantities \ref{3.2}, the following smallness conditions on
$\delta$:
\begin{eqnarray}
&&C(1+{\cal R}_0^\infty(\alpha))^2\delta\leq 1 \ \ \ \mbox{: from Proposition 3.1}\nonumber\\
&&C(1+{\cal D}_0^\infty(\mbox{tr}\chib)\delta\leq 1 \ \ \ \mbox{: from Lemma 3.2, through \ref{3.124}}\nonumber\\
&&C({\cal D}_0^\infty(\chibh)+{\cal R}_0^\infty(\alb))^2\delta\leq 1 \ \ \ \mbox{: from Lemma 3.2}\nonumber\\
&&C[{\cal R}_0^\infty(\rho)+({\cal R}_0^\infty(\beta)+{\cal
R}_0^\infty(\beb))^2]\delta\leq 1 \ \ \mbox{: from Lemma 3.5}
\label{3.125}
\end{eqnarray}
Definining
\begin{equation}
{\cal D}_0^\infty=\max\{{\cal D}_0^\infty(\mbox{tr}\chib),{\cal
D}_0^\infty(\chibh)\} \label{3.126}
\end{equation}
\begin{equation}
{\cal R}_0^\infty=\max\{{\cal R}_0^\infty(\alpha),{\cal
R}_0^\infty(\beta), {\cal R}_0^\infty(\rho),{\cal
R}_0^\infty(\sigma),{\cal R}_0^\infty(\beb),{\cal
R}_0^\infty(\alb)\} \label{3.127}
\end{equation}
the above smallness conditions on delta may be simplified to:
\begin{eqnarray}
&&C(1+{\cal R}_0^\infty)^2\delta\leq 1\nonumber\\
&&C(1+{\cal D}_0^\infty)\delta\leq 1\nonumber\\
&&C({\cal D}_0^\infty+{\cal R}_0^\infty)^2\delta\leq 1\nonumber\\
&&C{\cal R}_0^\infty(1+{\cal R}_0^\infty)]\delta\leq 1
\label{3.128}
\end{eqnarray}
A number $n$ of smallness conditions on $\delta$ of the form:
\begin{equation}
F_1({\cal D}_0^\infty,{\cal R}_0^\infty)\delta\leq 1, \ . \ . \ .
\ , F_n({\cal D}_0^\infty,{\cal R}_0^\infty)\delta\leq 1
\label{3.129}
\end{equation}
where $F_1,..., F_n$ are non-negative non-decreasing continuous
functions defined on the non-negative quadrant $\{(x,y)\in \Re^2 \
: \ x\geq 0 \ \mbox{and} \ y\geq 0\}$, may be replaced by a single
smallness condition of the same form:
\begin{equation}
F({\cal D}_0^\infty,{\cal R}_0^\infty)\delta\leq 1 \label{3.130}
\end{equation}
where $F$ is the upper envelope of the functions $F_1,...,F_n$:
\begin{equation}
F(x,y)=\max\{F_1(x,y),...,F_n(x,y)\} \ \ : \ \forall \{(x,y)\in
\Re^2 \ : \ x\geq 0 \ \mbox{and} \ y\geq 0\} \label{3.131}
\end{equation}
also a non-negative non-decreasing continuous function on the
non-negative quadarant. In conclusion, we may say that the results
of the present chapter hold {\em if $\delta$ is suitably small
depending on ${\cal D}_0^\infty$ and ${\cal R}_0^\infty$}, and the
last phrase means that there is a non-negative non-decreasing
continous function  $F$ on the non-negative quadarant such that
\ref{3.130} holds.

More generally the phrase {\em if $\delta$ is suitably small
depending on the non-negative quantities $q_1,...,q_n$} shall mean
throughout this monograph the following: {\em if there is a
non-negative non-decreasing continous function $F$ on
$[0,\infty)^n$ such that:}
\begin{equation}
\delta F(q_1,...,q_n)\leq 1 \label{3.132}
\end{equation}
Note that we call a function $F$ defined on $[0,\infty)^n$ {\em
non-decreasing} if $y_i\geq x_i$ for all $i=1,...,n$ implies
$F(y_1,...,y_n)\geq F(x_1,...,x_n)$. Let then $G_1,...,G_n$ be
non-negative non-decreasing continous functions on $[0,\delta)^m$.
These define a continuous mapping:
\begin{equation}
G \ : \ [0,\infty)^m\rightarrow [0,\infty)^n \ \ \ \mbox{by:} \ \
\ (r_1,...r_m)\mapsto(G_1(r_1,...,r_m),...G_n(r_1,...,r_m))
\label{3.133}
\end{equation}
The composition $F\circ G$ is then a non-negative non-decreasing
continuous function on $[0,\infty)^m$. In particular, taking
\begin{eqnarray}
&&F(x_1,...,x_n)=x_1+...+x_n, \ \ \mbox{or} \nonumber\\
&&F_(x_1,...,x_n)=x_1...x_n, \ \ \mbox{or} \nonumber\\
&&F_(x_1,...,x_n)=\max\{x_1,...,x_n\} \label{3.134}
\end{eqnarray}
we conclude that the sum, product and upper envelope of
non-negative non-decreasing continous functions on $[0,\infty)^m$
is itself a non-negative non-decreasing continous function on
$[0,\infty)^m$. Moreover, if the non-negative quantities
$q_1,...,q_n$ satisfy bounds in terms of the non-negative
quantities $r_1,...,r_m$ of the form:
\begin{equation}
q_1\leq G_1(r_1,...,r_m), \ . \ . \ . \ , q_n\leq G_n(r_1,...,r_m)
\label{3.135}
\end{equation}
and $\delta$ satisfies the smallness condition:
\begin{equation}
\delta (F\circ G)(r_1,...,r_m)\leq 1 \label{3.136}
\end{equation}
then $\delta$ satisfies the smallness condition \ref{3.132}. We
may therefore say that $\delta$ is suitably small depending on
$q_1,...,q_n$, provided that $\delta$ is suitably small depending
on $r_1,...,r_m$.

\chapter{$L^4(S)$ Estimates for the 1st Derivatives of the Connection Coefficients}

\section{Introduction}

The objective of the present chapter is to derive $L^4(S)$
estimates for the 1st derivatives of the connection coefficients in
$M^\prime$ on the basis of $L^4(S)$ bounds for the 1st derivatives of
the curvature components in $M^\prime$. In this and the next three
chapters the fundamental bootstrap assumption {\bf A0} of Chapter
3 is understood to hold.

The arguments of the present chapter, like those of Chapter 3,
rely only on the propagation equations, equations \ref{1.28},
\ref{1.42}, \ref{1.47}, \ref{1.66}, \ref{1.67}, \ref{1.86},
\ref{1.87} of Chapter 1. What is assumed in regard to the
curvature components in the present chapter is, besides the
assumptions of Chapter 3 (see \ref{3.2}), that the following
quantities are finite:
\begin{eqnarray}
&&\scR_1^4(\alpha)=\sup_{(\ub,u)\in D^\prime}
\left(|u|^{3/2}\delta^{3/2}\|\snab\alpha\|_{L^4(S_{\ub,u})}\right)\nonumber\\
&&\scR_1^4(\beta)=\sup_{(\ub,u)\in D^\prime}
\left(|u|^{5/2}\delta^{1/2}\|\snab\beta\|_{L^4(S_{\ub,u})}\right)\nonumber\\
&&\scR_1^4(\rho)=\sup_{(\ub,u)\in D^\prime}
\left(|u|^{7/2}\|\sd\rho\|_{L^4(S_{\ub,u})}\right)\nonumber\\
&&\scR_1^4(\sigma)=\sup_{(\ub,u)\in D^\prime}
\left(|u|^{7/2}\|\sd\sigma\|_{L^4(S_{\ub,u})}\right)\nonumber\\
&&\scR_1^4(\beb)=\sup_{(\ub,u)\in D^\prime}
\left(|u|^{9/2}\delta^{-1}\|\snab\beta\|_{L^4(S_{\ub,u})}\right)\nonumber\\
&&\scR_1^4(\alb)=\sup_{(\ub,u)\in D^\prime}
\left(|u|^{5}\delta^{-3/2}\|\snab\alb\|_{L^4(S_{\ub,u})}\right)
\label{4.1}
\end{eqnarray}
and:
\begin{eqnarray}
&&{\cal R}_0^4(D\rho)=\sup_{(\ub,u)\in D^\prime}
\left(|u|^{5/2}\delta\|D\rho\|_{L^4(S_{\ub,u})}\right)\nonumber\\
&&{\cal R}_0^4(D\sigma)=\sup_{(\ub,u)\in D^\prime}
\left(|u|^{5/2}\delta\|D\sigma\|_{L^4(S_{\ub,u})}\right)
\label{4.2}
\end{eqnarray}
By the results of Chapter 2, the corresponding quantities on
$C_{u_0}$, obtained by replacing the supremum on $D^\prime$ by the
supremum on $([0,\delta]\times\{u_0\})\bigcap D^\prime$, are all
bounded by a non-negative non-decreasing continuous function of
$M_6$, the quantity requiring $M_k$ with the highest $k$ being the
one corresponding to $\snab\alb$.

We shall first derive an $L^4(S)$ estimate for $\snab\chi^\prime$,
then an $L^4(S)$ estimate for $\snab\chib^\prime$. After that we
shall derive $L^4(S)$ estimates for $\snab\eta$, $\snab\etb$
together, as the propagation equations \ref{1.66}, \ref{1.67} are
coupled. After that we shall derive $L^4(S)$ estimates for
$\sd\omb$ and $\sd\omega$. Finally, we shall derive $L^4(S)$
estimates for $D\omega$ and $\Db\omb$. The $L^4(S)$ estimates for
$\snab\eta$, $\snab\etb$ imply through equations \ref{1.146},
\ref{1.147} $L^4(S)$ estimates for $\Db\chi$, $D\chib$, while the
$L^4(S)$ estimates for $\sd\omega$, $\sd\omb$ imply through
equations \ref{1.149}, \ref{1.150} $L^4(S)$ estimates for
$\Db\eta$, $D\etb$. In fact, taking into account \ref{3.5} and
\ref{1.a2} the traces of equations \ref{1.146}, \ref{1.147} are
the equations:
\begin{equation}
D(\Omega\mbox{tr}\chib)=\Omega^2\left\{2\sdiv\etb+2|\etb|^2-(\chi,\chib)+2\rho\right\}
\label{4.c1}
\end{equation}
\begin{equation}
\Db(\Omega\mbox{tr}\chi)=\Omega^2\left\{2\sdiv\eta+2|\eta|^2-(\chi,\chib)+2\rho\right\}
\label{4.c2}
\end{equation}
while, in view of the fact that
$$\Dh(\Omega\chibh)=\hat{D(\Omega\chib)}-\Omega^2\mbox{tr}\chib\chih, \ \ \
\Dbh(\Omega\chih)=\hat{\Db(\Omega\chi)}-\Omega^2\mbox{tr}\chi\chibh,$$
taking into account the identity \ref{1.163} (see also
\ref{1.a4}), the trace-free parts of equations \ref{1.146},
\ref{1.147} are seen to reduce to the equations:
\begin{equation}
\Dh(\Omega\chibh)=\Omega^2\left\{\snab\oth\etb+\etb\oth\etb+\frac{1}{2}\mbox{tr}\chi\chibh-\frac{1}{2}\mbox{tr}\chib\chih\right\}
\label{4.c3}
\end{equation}
\begin{equation}
\Dbh(\Omega\chih)=\Omega^2\left\{\snab\oth\eta+\eta\oth\eta+\frac{1}{2}\mbox{tr}\chib\chih-\frac{1}{2}\mbox{tr}\chi\chibh
\right\} \label{4.c4}
\end{equation}

We begin with seven lemmas.

\vspace{5mm}

\noindent {\bf Lemma 4.1} \ \ \ Let $\theta$ be an arbitrary
$p$-covariant $S$ tensorfield on $M^\prime$. We have, with respect
to an arbitrary local frame field for the $S_{\ub,u}$,
$$(D\snab\theta-\snab D\theta)_{AB_1 ... B_p}=-\sum_{i=1}^p(D\sGa)^C_{AB_i}\theta_{B_1...\stackrel{C}{>B_i<}...B_p}$$
where $D\sGa$ is the Lie derivative with respect to $L$ of the
induced connection on the surfaces $S_{\ub,u}$, a $T^1_2$-type $S$
tensorfield, symmetric in the lower indices and given by:
$$\sg_{CD}(D\sGa)^D_{AB}=\snab_A(\Omega\chi)_{BC}+\snab_B(\Omega\chi)_{AC}-\snab_{C}(\Omega\chi)_{AB}$$
Similarly,
$$(\Db\snab\theta-\snab\Db\theta)_{AB_1 ... B_p}=-\sum_{i=1}^p(\Db\sGa)^C_{AB_i}
\theta_{B_1...\stackrel{C}{>B_i<}...B_p}$$ where $\Db\sGa$ is the
Lie derivative with respect to $\Lb$ of the induced connection on
the surfaces $S_{\ub,u}$, a $T^1_2$-type $S$ tensorfield,
symmetric in the lower indices and given by:
$$\sg_{CD}(\Db\sGa)^D_{AB}=\snab_A(\Omega\chib)_{BC}+\snab_B(\Omega\chib)_{AC}-\snab_C(\Omega\chib)_{AB}$$
Here
$$\stackrel{C}{>B_i<}$$
signifies that the index $B_i$ is missing and in its place we have
the index $C$.

\vspace{2.5mm}

\noindent {\em Proof:} \ To derive the formula for
$D\snab\theta-\snab D\theta$ we consider a given outgoing null
hypersurface $C_u$ and work in a Jacobi field frame defined in a
neighborhood ${\cal V}$ of a generator $\gamma_p$ of $C_u$,
corresponding to a point $p\in S_{0,u}$, of the form:
\begin{equation}
{\cal V}=\bigcup_{q\in{\cal U}}\gamma_q \label{4.3}
\end{equation}
where ${\cal U}$ is a neighborhood of the point $p$ in $S_{0,u}$
and $\gamma_q$ denotes the generator of $C_u$ corresponding to the
point $q\in S_{0,u}$. A Jacobi field frame $(e_A : A=1,2)$ is
characterized by the conditions:
\begin{equation}
[L,e_A]=0 \ \ : \ A=1,2 \label{4.4}
\end{equation}
It then follows by the Jacobi identity
$$[L,[e_A,e_B]]+[e_A,[e_B,L]]+[e_B,[L,e_A]]=0$$
that also:
\begin{equation}
[L,[e_A,e_B]]=0 \label{4.5}
\end{equation}
We have:
\begin{equation}
(D\snab\theta)(e_A,e_{B_1},...,e_{B_p})=L((\snab\theta)(e_A,e_{B_1},...,e_{B_p}))
\label{4.6}
\end{equation}
Also,
\begin{eqnarray*}
(\snab\theta)(e_A,e_{B_1},...,e_{B_p})&=&(\snab_{e_A}\theta)(e_{B_1},...,e_{B_p})\\
&=&e_A(\theta(e_{B_1},...e_{B_p}))-\sum_i\theta(e_{B_1},...\stackrel{\snab_{e_A}e_{B_i}}{>e_{B_i}<}...,e_{B_p})
\end{eqnarray*}
where
$$\stackrel{\snab_{e_A} e_{B_i}}{>e_{B_i}<}$$
signifies that the $i$th entry $e_{B_i}$ is replaced by
$\snab_{e_A}e_{B_i}$. Since
\begin{equation}
\snab_{e_A}e_B=\sGa^C_{AB}e_C \label{4.7}
\end{equation}
where $\sGa^C_{AB}$ are the coefficients of the induced connection
on the surfaces $S_{\ub,u}$ in the frame field $(e_A : A=1,2)$, we
obtain:
\begin{equation}
(\snab\theta)(e_A,e_{B_1},...,e_{B_p})=e_A(\theta_{B_1 ... B_p})
-\sum_{i=1}^p\sGa^C_{AB_i}\theta_{B_1...\stackrel{C}{>B_i<}...B_p}
\label{4.8}
\end{equation}
Now, by \ref{4.4},
\begin{equation}
L(e_A(\theta_{B_1 ... B_p}))=e_A(L(\theta_{B_1 ...
B_p}))=e_A((D\theta)_{B_1 ... B_p}) \label{4.9}
\end{equation}
and, denoting
\begin{equation}
L(\sGa^C_{AB})=(D\sGa)^C_{AB} \label{4.10}
\end{equation}
we have:
\begin{eqnarray}
L(\sGa^C_{AB_i}\theta_{B_1...\stackrel{C}{>B_i<}...B_p})&=&(L(\sGa^C_{AB_i}))\theta_{B_1...\stackrel{C}{>B_i<}...B_p}
+\sGa^C_{AB_i}L(\theta_{B_1...\stackrel{C}{>B_i<}...B_p})\nonumber\\
&=&(D\sGa)^C_{AB_i}\theta_{B_1...\stackrel{C}{>B_i<}...B_p}+\sGa^C_{AB_i}(D\theta)_{B_1...\stackrel{C}{>B_i<}...B_p}
\nonumber\\
&\s&\label{4.11}
\end{eqnarray}
Substituting \ref{4.8} on the right hand side in \ref{4.6} and
taking into account \ref{4.9} and \ref{4.11}, the right hand side
of \ref{4.6} becomes:
$$(\snab D\theta)(e_A,e_{B_1},...,e_{B_p})-\sum_{i=1}^p(D\sGa)^C_{AB_i}\theta_{B_1...\stackrel{C}{>B_i<}...B_p}$$
We conclude that:
\begin{equation}
(D\snab\theta-\snab D\theta)_{AB_1 ...
B_p}=-\sum_{i=1}^p(D\sGa)^C_{AB_i}\theta_{B_1...\stackrel{C}{>B_i<}...B_p}
\label{4.12}
\end{equation}
To obtain the formula for $D\sGa$ we consider that:
\begin{eqnarray*}
&e_A(\sg_{BC})=e_A(\sg(e_B,e_C))=\sg(\snab_{e_A}e_B,e_C)+\sg(e_B,\snab_{e_A}e_C)\\
&=\sg(\sGa^D_{AB}e_D,e_C)+\sg(e_B,\sGa^D_{AC}e_D)=\sg_{DC}\sGa^D_{AB}+\sg_{BD}\sGa^D_{AC}
\end{eqnarray*}
hence:
\begin{equation}
L(e_A(\sg_{BC}))=(L(\sg_{DC}))\sGa^D_{AB}+(L(\sg_{BD}))\sGa^D_{AC}
+\sg_{DC}(D\sGa)^D_{AB}+\sg_{BD}(D\sGa)^D_{AC} \label{4.13}
\end{equation}
Now, according to \ref{1.28}:
\begin{equation}
L(\sg_{AB})=(D\sg)_{AB}=2\Omega\chi_{AB} \label{4.14}
\end{equation}
hence by \ref{4.4}:
\begin{equation}
L(e_A(\sg_{BC}))=e_A(L(\sg_{BC}))=e_A(2\Omega\chi_{BC})
\label{4.15}
\end{equation}
Thus \ref{4.13} reads:
$$e_A(2\Omega\chi_{BC})=2\Omega\chi_{DC}\sGa^D_{AB}+2\Omega\chi_{BD}\sGa^D_{AC}
+\sg_{DC}(D\sGa)^D_{AB}+\sg_{BD}(D\sGa)^D_{AC}$$ which, since
$$e_A(\Omega\chi_{AB})-\sGa^D_{AB}\Omega\chi_{DC}-\sGa^D_{AC}\Omega\chi_{BD}=\snab_A(\Omega\chi)_{BC},$$
is equivalent to:
\begin{equation}
\snab_A(\Omega\chi)_{BC}=\frac{1}{2}\sg_{DC}(D\sGa)^D_{AB}+\frac{1}{2}\sg_{BD}(D\sGa)^D_{AC}
\label{4.16}
\end{equation}
Denoting this formula by $\{ABC\}$, we form the sum
$\{ABC\}+\{BAC\}-\{CAB\}$ to obtain:
\begin{eqnarray}
&&\snab_A(\Omega\chi)_{BC}+\snab_B(\Omega\chi)_{AC}-\snab_C(\Omega\chi)_{AB}=
\frac{1}{2}\sg_{CD}((D\sGa)^D_{AB}+(D\sGa)^D_{BA})\nonumber\\
&&\s\s\s\s\s\s\s\s\s\s
+\frac{1}{2}\sg_{AD}((D\sGa)^D_{BC}-(D\sGa)^D_{CB})
+\frac{1}{2}\sg_{BD}((D\sGa)^D_{AC}-(D\sGa)^D_{CA})\nonumber\\
&&\label{4.17}
\end{eqnarray}
Now, we have:
\begin{equation}
[e_A,e_B]=\snab_{e_A}e_B-\snab_{e_B}e_A=(\sGa^C_{AB}-\sGa^C_{BA})e_C
\label{4.18}
\end{equation}
Also, for arbitrary functions $f^A : A=1,2$ we have, by \ref{4.4},
\begin{equation}
[L,f^A e_A]=(Lf^A)e_A \label{4.19}
\end{equation}
It follows that:
$$((D\sGa)^C_{AB}-(D\sGa)^C_{BA})e_C=[L,[e_A,e_B]]=0$$
by virtue of \ref{4.5}. Thus:
\begin{equation}
(D\sGa)^C_{AB}-(D\sGa)^C_{BA}=0 \label{4.20}
\end{equation}
Taking this into account, \ref{4.17} reduces to:
\begin{equation}
\snab_A(\Omega\chi)_{BC}+\snab_B(\Omega\chi)_{AC}-\snab_C(\Omega\chi)_{AB}=
\sg_{CD}(D\sGa)^D_{AB} \label{4.21}
\end{equation}
We have thus established the first part of the lemma. The second
part then follows by conjugation.

\vspace{5mm}

In the following we shall use a simplified notation when writing
down components of $S$ tensorfields relative to an arbitrary frame
field for the $S_{\ub,u}$. That is we shall use the components of
$\sg^{-1}$ to raise indices. Thus, if $\theta$ is a 2-covariant
$S$ tensorfield we denote the components of the corresponding
$T^1_1$-type $S$ tensorfield $\theta^\sharp$ defined by \ref{1.34}
simply by  $\theta_A^{\s B}$:
\begin{equation}
\theta_A^{\s B}=\theta_{AC}(\sg^{-1})^{CB} \label{4.22}
\end{equation}
and the components of the corresponding 2-contravariant $S$
tensorfield $\theta^{\sharp\sharp}$ simply by $\theta^{AB}$:
\begin{equation}
\theta^{AB}=(\sg^{-1})^{AC}(\sg^{-1})^{BD}\theta_{CD} \label{4.23}
\end{equation}
More generally, if $\theta$ is a $p$-covariant $S$ tensorfield and
$\{i_1,...,i_k\}$ is an ordered subset of $\{1,...,p\}$, we denote
\begin{equation}
\theta_{A_1...>A_{i_1}<...>A_{i_k}<...A_p}^{\s\s\s\s\s\s
B_{i_1}.\s.\s.\s B_{i_k}\s\s\s\s}
=(\sg^{-1})^{A_{i_1}B_{i_1}}...(\sg^{-1})^{A_{i_k}B_{i_k}}\theta_{A_1...A_{i_1}...A_{i_k}...A_p}
\label{4.24}
\end{equation}
Note that according to this notation we have
$\sg^{AB}=(\sg^{-1})^{AB}$. However to avoid confusion we shall
keep the notation $(\sg^{-1})^{AB}$ for the components of the
reciprocal metric.

Let $\theta, \theta^\prime$ be a pair of arbitrary type $T^q_p$
$S$ tensorfields. Their inner product is defined by:
\begin{equation}
(\theta,\theta^\prime)=\sg_{A_1 B_1}...\sg_{A_q
B_q}(\sg^{-1})^{C_1 D_1}...(\sg^{-1})^{C_p D_p}
\theta^{A_1...A_q}_{C_1...C_p}\theta^{\prime
B_1...B_q}_{D_1...D_p} \label{4.25}
\end{equation}
The magnitude of a $T^q_p$ $S$ tensorfield $\theta$ is accordingly
defined by:
\begin{equation}
|\theta|=(\theta,\theta)^{1/2} \label{4.26}
\end{equation}

\vspace{5mm}

\noindent {\bf Lemma 4.2} \ \ \ Let $\theta$ be an arbitrary $p$
covariant $S$ tensorfield on $M^\prime$. We have:
$$D(|\theta|^2)+p\Omega\mbox{tr}\chi|\theta|^2=2(\theta,D\theta)-
2\sum_{i=1}^p\Omega\chih^{A_i}_{B_i}\theta^{A_1...\stackrel{B_i}{>A_i<}...A_p}\theta_{A_1...A_p}$$
and:
$$\Db(|\theta|^2)+p\Omega\mbox{tr}\chib|\theta|^2=2(\theta,\Db\theta)-
2\sum_{i=1}^p\Omega\chibh^{A_i}_{B_i}\theta^{A_1...\stackrel{B_i}{>A_i<}...A_p}\theta_{A_1...A_p}$$

\vspace{2.5mm}

\noindent {\em Proof:} We have:
$$|\theta|^2=(\sg^{-1})^{A_1 B_1}...(\sg^{-1})^{A_p B_p}\theta_{A_1...A_p}\theta_{B_1...B_p}$$
Thus, in view of the first of \ref{2.3}, which reads
$$(D\sg^{-1})^{AB}=-2\Omega\chi^{AB},$$
and the fact that the operator $D$ satisfies the Leibniz rule, we
obtain:
\begin{eqnarray}
D(|\theta|^2)&=&(\sg^{-1})^{A_1 B_1} ... (\sg^{-1})^{A_p B_p}
\{(D\theta_{A_1...A_p})\theta_{B_1...B_p}+\theta_{A_1...A_p}(D\theta_{B_1...B_p})\}\nonumber\\
&\s&+\sum_{i=1}^p(\sg^{-1})^{A_1
B_1}...\stackrel{(-2\Omega\chi^{A_i B_i})}{>(\sg^{-1})^{A_i
B_i}<}...(\sg^{-1})^{A_p B_p}
\theta_{A_1...A_p}\theta_{B_1...B_p}\nonumber\\
&=&2\theta^{A_1...A_p}(D\theta)_{A_1...A_p}
-2\sum_{i=1}^p\Omega\chi^{A_i
B_i}\theta^{A_1...>A_i<...A_p}_{\s\s\s\s\s\s
B_i\s\s\s\s\s\s}\theta_{A_1...A_p} \label{4.27}
\end{eqnarray}
Substituting the decomposition
$$\chi^{A_i B_i}=\chih^{A_i B_i}+\frac{1}{2}\mbox{tr}\chi(\sg^{-1})^{A_i B_i}$$
yields the first part of the lemma. The second part then follows
by conjugation.

\vspace{5mm}

Let $\theta$ be an arbitrary $p$ covariant $S$ tensorfield on
$M^\prime$. Consider a given $C_u$. The flow $\Phi_t$ generated by
$L$ on $C_u$ (see Chapter 1) defines a diffeomorphism $\Phi_{\ub}$
of $S_{0,u}$ onto $S_{\ub,u}$. Then
\begin{equation}
\theta(\ub)=\Phi_{\ub}^*\left.\theta\right|_{S_{\ub,u}}
\label{4.28}
\end{equation}
is a 1-parameter family of $p$ covariant $S$ tensorfields on $S_{0,u}$ and we
have:
\begin{equation}
\frac{\partial}{\partial\ub}\theta(\ub)=\dot{\theta}(\ub)
\label{4.29}
\end{equation}
where
\begin{equation}
\dot{\theta}(\ub)=\Phi_{\ub}^*\left.D\theta\right|_{S_{\ub,u}}
\label{4.30}
\end{equation}
Similarly, consider a given $\Cb_{\ub}$. The flow $\Phib_t$
generated by $\Lb$ on $\Cb_{\ub}$ (see Chapter 1) defines a
diffeomorphism $\Phib_{u-u_0}$ of $S_{\ub,u_0}$ onto $S_{\ub,u}$.
Then
\begin{equation}
\theta(u)=\Phib_{u-u_0}^*\left.\theta\right|_{S_{\ub,u}}
\label{4.31}
\end{equation}
is a 1-parameter family of $p$ covariant $S$ tensorfields on $S_{\ub,u_0}$ and
we have:
\begin{equation}
\frac{\partial}{\partial u}\theta(u)=\dot{\theta}(u) \label{4.32}
\end{equation}
where
\begin{equation}
\dot{\theta}(u)=\Phib_{u-u_0}^*\left.\Db\theta\right|_{S_{\ub,u}}
\label{4.33}
\end{equation}

\vspace{5mm}

\noindent {\bf Lemma 4.3} \ \ \ Consider a given $C_u$ and let
$\rho$ be a non-negative function on $C_u$ and
$\rho(\ub)=\Phi_{\ub}^*\left.\rho\right|_{S_{\ub,u}}=\left.\rho\right|_{S_{\ub,u}}\circ\Phi_{\ub}$
be the corresponding 1-parameter family of functions on $S_{0,u}$.
Then if $\delta$ is suitably small depending on ${\cal R}_0^\infty$ (see \ref{3.127}) we have, for each
$p\geq 1$:
$$2^{-1/p}\|\rho(\ub)\|_{L^p(S_{0,u})}\leq\|\rho\|_{L^p(S_{\ub,u})}\leq 2^{1/p}\|\rho(\ub)\|_{L^p(S_{0,u})}$$
In particular, taking $\rho=1$ we have, since
$\mbox{Area}(S_{0,u})=4\pi |u|^2$,
$$2\pi |u|^2\leq\mbox{Area}(S_{\ub,u})\leq 8\pi |u|^2$$
Also, consider a given $\Cb_{\ub}$ and let $\rho$ be a
non-negative function on $\Cb_{\ub}$ and
$\rho(u)=\Phib_{u-u_0}^*\left.\rho\right|_{S_{\ub,u}}=\left.\rho\right|_{S_{\ub,u}}\circ\Phib_{u-u_0}$
be the corresponding 1-parameter family of functions on
$S_{\ub,u_0}$. Then if $\delta$ is suitably small depending on ${\cal D}_0^\infty$, ${\cal R}_0^\infty$ 
(see \ref{3.126}, \ref{3.127}) we have, for each $p\geq 1$:
$$e^{-3/2p}\left(\frac{|u|^2}{|u_0|^2}\right)^{1/p}\|\rho(u)\|_{L^p(S_{\ub,u_0})}\leq
\|\rho\|_{L^p(S_{\ub,u})}\leq
e^{3/2p}\left(\frac{|u|^2}{|u_0|^2}\right)^{1/p}\|\rho(u)\|_{L^p(S_{\ub,u_0})}$$

\noindent {\em Proof:} \ Consider first a non-negative function
$\rho$ on $C_u$. We have:
\begin{equation}
\|\rho\|_{L^p(S_{\ub,u})}=\left(\int_{S_{\ub,u}}\rho^p
d\mu_{\sg}\right)^{1/p} \label{4.34}
\end{equation}
and:
\begin{equation}
\|\rho(\ub)\|_{L^p(S_{0,u})}=\left(\int_{S_{0,u}}(\rho(\ub))^p
d\mu_{\sg}\right)^{1/p} \label{4.35}
\end{equation}
Now, $\Phi_{\ub}$ being a diffeomorphism of $S_{0,u}$ onto
$S_{\ub,u}$, we have:
\begin{equation}
\int_{S_{\ub,u}}\rho^p
d\mu_{\sg}=\int_{S_{0,u}}(\Phi_{\ub}^*\rho)^p\Phi_{\ub}^*
d\mu_{\sg} \label{4.36}
\end{equation}
Also,
$$\Phi_{\ub}^* \left.d\mu_{\sg}\right|_{S_{\ub,u}}=d\mu_{\Phi_{\ub}^* \left.\sg\right|_{S_{\ub,u}}}$$
and:
\begin{eqnarray}
&&\frac{\partial}{\partial\ub}\Phi_{\ub}^*
\left.d\mu_{\sg}\right|_{S_{\ub,u}}
=\Phi_{\ub}^*\left.(Dd\mu_{\sg})\right|_{S_{\ub,u}}=\Phi_{\ub}^*\left.(\Omega\mbox{tr}\chi
d\mu_{\sg})\right|_{S_{\ub,u}}
\nonumber\\
&&\hspace{20mm}=(\left.(\Omega\mbox{tr}\chi)\right|_{S_{\ub,u}}\circ\Phi_{\ub})\Phi_{\ub}^*
\left.d\mu_{\sg}\right|_{S_{\ub,u}} \label{4.37}
\end{eqnarray}
Evaluating this at any given point $q\in S_{0,u}$ and integrating
with respect to $\ub$, noting that
$$\Phi_0^* \left.d\mu_{\sg}\right|_{S_{0,u}}=\left.d\mu_{\sg}\right|_{S_{0,u}}$$
$\Phi_0$ being the identity mapping on $S_{0,u}$, we obtain:
\begin{equation}
(\Phi_{\ub}^*\left. d\mu_{\sg}\right|_{S_{\ub,u}})(q)=
\exp\left(\int_0^{\ub}(\Omega\mbox{tr}\chi)(\Phi_{\ub^\prime}(q))d\ub^\prime\right)
\left.d\mu_{\sg}\right|_{S_{0,u}}(q) \ \ \ : \ \forall q\in
S_{0,u} \label{4.38}
\end{equation}
Here
$$\ub^\prime\mapsto\Phi_{\ub^\prime}(q)$$
is the generator of $C_u$ through $q$. Thus, the integral in the
exponential is the function $F(\ub)$ defined by \ref{3.22}. If $\delta$ is suitably small depending on ${\cal R}_0^\infty$  we have (see \ref{3.24}):
\begin{equation}
|F(\ub)|\leq\log 2 \label{4.39}
\end{equation}
It follows that:
\begin{equation}
2^{-1} \left.d\mu_{\sg}\right|_{S_{0,u}}\leq \Phi_{\ub}^*\left.
d\mu_{\sg}\right|_{S_{\ub,u}} \leq 2
\left.d\mu_{\sg}\right|_{S_{0,u}} \label{4.40}
\end{equation}
In view of \ref{4.34} - \ref{4.36} the first part of the lemma
then follows.

Consider next a non-negative function $\rho$ on $\Cb_{\ub}$. We
have:
\begin{equation}
\|\rho\|_{L^p(S_{\ub,u})}=\left(\int_{S_{\ub,u}}\rho^p
d\mu_{\sg}\right)^{1/p} \label{4.41}
\end{equation}
and:
\begin{equation}
\|\rho(u)\|_{L^p(S_{\ub,u_0})}=\left(\int_{S_{\ub,u_0}}(\rho(u))^p
d\mu_{\sg}\right)^{1/p} \label{4.42}
\end{equation}
Now, $\Phib_{u-u_0}$ being a diffeomorphism of $S_{\ub,u_0}$ onto
$S_{\ub,u}$, we have:
\begin{equation}
\int_{S_{\ub,u}}\rho^p
d\mu_{\sg}=\int_{S_{\ub,u_0}}(\Phib_{u-u_0}^*\rho)^p\Phib_{u-u_0}^*
d\mu_{\sg} \label{4.43}
\end{equation}
Also,
$$\Phib_{u-u_0}^* \left.d\mu_{\sg}\right|_{S_{\ub,u}}=d\mu_{\Phib_{u-u_0}^* \left.\sg\right|_{S_{\ub,u}}}$$
and:
\begin{eqnarray}
&&\frac{\partial}{\partial u}\Phib_{u-u_0}^*
\left.d\mu_{\sg}\right|_{S_{\ub,u}} =\Phib_{u-u_0}^*\left.(\Db
d\mu_{\sg})\right|_{S_{\ub,u}}
=\Phib_{u-u_0}^*\left.(\Omega\mbox{tr}\chib
d\mu_{\sg})\right|_{S_{\ub,u}}
\nonumber\\
&&\hspace{20mm}=(\left.(\Omega\mbox{tr}\chib)\right|_{S_{\ub,u}}\circ\Phib_{u-u_0})\Phib_{u-u_0}^*
\left.d\mu_{\sg}\right|_{S_{\ub,u}} \label{4.44}
\end{eqnarray}
Evaluating this at any given point $q\in S_{\ub,u_0}$ and
integrating with respect to $u$, noting that
$$\Phib_0^* \left.d\mu_{\sg}\right|_{S_{\ub,u_0}}=\left.d\mu_{\sg}\right|_{S_{\ub,u_0}}$$
$\Phib_0$ being the identity mapping on $S_{\ub,u_0}$, we obtain:
\begin{equation}
(\Phib_{u-u_0}^*\left. d\mu_{\sg}\right|_{S_{\ub,u}})(q)=
\exp\left(\int_{u_0}^{u}(\Omega\mbox{tr}\chib)(\Phib_{u^\prime-u_0}(q))du^\prime\right)
\left.d\mu_{\sg}\right|_{S_{\ub,u_0}}(q) \ \ \ : \ \forall q\in
S_{\ub,u_0} \label{4.45}
\end{equation}
Here
$$u^\prime\mapsto\Phib_{u^\prime-u_0}(q)$$
is the generator of $\Cb_{\ub}$ through $q$. Thus, the integral in
the exponential is the function $\Fb(u)$ defined by \ref{3.50}.
If $\delta$ is suitably small depending on ${\cal D}_0^\infty$, ${\cal R}_0^\infty$ we have (see \ref{3.52}):
\begin{equation}
\left|\Fb(u)+\log\left(\frac{|u_0|^2}{|u|^2}\right)\right|\leq\frac{3}{2}
\label{4.46}
\end{equation}
It follows that:
\begin{equation}
e^{-3/2}\frac{|u|^2}{|u_0|^2}\left.d\mu_{\sg}\right|_{S_{\ub,u_0}}\leq
\Phib_{u-u_0}^*\left. d\mu_{\sg}\right|_{S_{\ub,u}} \leq
e^{3/2}\frac{|u|^2}{|u_0|^2}\left.d\mu_{\sg}\right|_{S_{\ub,u_0}}
\label{4.47}
\end{equation}
In view of \ref{4.41} - \ref{4.43} the second part of the lemma
then follows.

\vspace{5mm}

\noindent{\bf Lemma 4.4} \ \ \ Let $\psi$ be a non-negative
function vanishing on $\Cb_0$ such that $\psi^2$ is weakly
differentiable along the generators of $C_u$ and satisfies the
inequality:
$$D\psi^2+\lambda\Omega\mbox{tr}\chi\psi^2\leq 2l\Omega|\chih|\psi^2+2\psi\rho$$
where $\lambda$ is a real number, $\rho$ is a non-negative
function and $l$ a non-negative constant. Then, if $\delta$ is suitably small depending on ${\cal R}_0^\infty$, for each $p\geq 1$ there is a
positive constant $C$ depending only on $p$, $\lambda$, $l$ such
that:
$$\|\psi\|_{L^p(S_{\ub,u})}\leq C\int_0^{\ub}\|\rho\|_{L^p(S_{\ub^\prime,u})}d\ub^\prime$$

\noindent{\em Proof:} \ Consider the generator of $C_u$ through an
arbitrary point $q\in S_{0,u}$:
$$\ub^\prime\mapsto\Phi_{\ub^\prime}(q)$$
We apply Lemma 3.1 to obtain:
\begin{equation}
\psi(\Phi_{\ub}(q))\leq\int_0^{\ub}\exp\left(\int_{\ub^\prime}^{\ub}
f_{\lambda,l}(\Phi_{\ub^{\prime\prime}}(q))d\ub^{\prime\prime}\right)
\rho(\Phi_{\ub^\prime}(q))d\ub^\prime \ \ \ : \ \forall q\in
S_{0,u} \label{4.b1}
\end{equation}
where:
\begin{equation}
f_{\lambda,l}=-\frac{\lambda}{2}\Omega\mbox{tr}\chi+l\Omega|\chih|
\label{4.b2}
\end{equation}
With the notations of Lemma 4.3, \ref{4.b2} reads:
\begin{equation}
\psi(\ub)\leq\int_0^{\ub}\exp\left(\int_{\ub^\prime}^{\ub}f_{\lambda,l}(\ub^{\prime\prime})d\ub^{\prime\prime}\right)
\rho(\ub^\prime)d\ub^\prime \label{4.b3}
\end{equation}
an inequality for a 1-parameter family of non-negative functions
on $S_{0,u}$.

Now, by Proposition 3.1:
\begin{equation}
f_{\lambda,l}\leq C|u|^{-1}\{l\delta^{-1/2}{\cal
R}_0^\infty(\alpha)+|\lambda|[1+({\cal R}_0^\infty(\alpha))^2]\}
\label{4.b4}
\end{equation}
hence:
\begin{equation}
\int_{\ub^\prime}^{\ub}f_{\lambda,l}(\ub^{\prime\prime})d\ub^{\prime\prime}
\leq C\delta^{1/2}|u|^{-1}\{l{\cal
R}_0^\infty(\alpha)+|\lambda|\delta^{1/2}[1+({\cal
R}_0^\infty(\alpha))^2]\} \leq l+|\lambda| \label{4.b5}
\end{equation}
the last step provided that $\delta$ is suitably small depending
on ${\cal R}_0^\infty(\alpha)$. We conclude that:
\begin{equation}
\psi(\ub)\leq C\int_0^{\ub}\rho(\ub^\prime)d\ub^\prime
\label{4.b6}
\end{equation}
It follows that:
\begin{equation}
\|\psi(\ub)\|_{L^p(S_{0,u})}\leq
C\int_0^{\ub}\|\rho(\ub^\prime)\|_{L^p(S_{0,u})}d\ub^\prime
\label{4.b7}
\end{equation}
Now by the first part of Lemma 4.3:
$$\|\psi\|_{L^p(S_{\ub,u})}\leq 2^{1/p}\|\psi(\ub)\|_{L^p(S_{0,u})}$$
and:
$$\|\rho(\ub^\prime)\|_{L^p(S_{0,u})}\leq 2^{1/p}\|\rho\|_{L^p(S_{\ub^\prime,u})}$$
Therefore \ref{4.b7} implies:
\begin{equation}
\|\psi\|_{L^p(S_{\ub,u})}\leq
C\int_0^{\ub}\|\rho\|_{L^p(S_{\ub^\prime,u})}d\ub^\prime
\label{4.b8}
\end{equation}
for a new constant $C$, depending on $p$, and the lemma is proved.

\vspace{5mm}

\noindent{\bf Lemma 4.5} \ \ \ Let $\psib$ be a non-negative
function such that $\psib^2$ is weakly differentiable along the
generators of $\Cb_{\ub}$ and satisfies the inequality:
$$\Db\psib^2+\lambda\Omega\mbox{tr}\chib\psib^2\leq 2l\Omega|\chibh|\psib^2+2\psib\rhob$$
where $\lambda$ is a real number, $\rhob$ is a non-negative
function and $l$ a non-negative constant. Then, if $\delta$ is suitably small depending on 
${\cal D}_0^\infty$, ${\cal R}_0^\infty$,
for each $p\geq 1$ there is a positive constant $C$ depending only
on $p$, $\lambda$, $l$ such that:
\begin{eqnarray*}
|u|^{\lambda-(2/p)}\|\psib\|_{L^p(S_{\ub,u})}&\leq& C|u_0|^{\lambda-(2/p)}\|\psib\|_{L^p(S_{\ub,u_0})}\\
&\s&+C\int_{u_0}^u|u^\prime|^{\lambda-(2/p)}\|\rhob\|_{L^p(S_{\ub,u^\prime})}du^\prime
\end{eqnarray*}

\noindent{\em Proof:} \ Consider the generator of $\Cb_{\ub}$
through an arbitrary point $q\in S_{\ub,u_0}$:
$$u^\prime\mapsto\Phib_{u^\prime-u_0}(q)$$
We apply Lemma 3.1 to obtain:
\begin{eqnarray}
\psib(\Phib_{u-u_0}(q))&\leq&\exp\left(\int_{u_0}^u\fb_{\lambda,l}(\Phib_{u^\prime-u_0}(q))du^\prime\right)
\psib(q)\label{4.b9}\\
&\s&+\int_{u_0}^u\exp\left(\int_{u^\prime}^u
\fb_{\lambda,l}(\Phib_{u^{\prime\prime}-u_0}(q))du^{\prime\prime}\right)
\rhob(\Phib_{u^\prime-u_0}(q))du^\prime\nonumber\\
&:& \ \forall q\in S_{\ub,u_0}\nonumber
\end{eqnarray}
where:
\begin{equation}
\fb_{\lambda,l}=-\frac{\lambda}{2}\Omega\mbox{tr}\chib+l\Omega|\chibh|
\label{4.b10}
\end{equation}
With the notations of Lemma 4.3, \ref{4.b9} reads:
\begin{eqnarray}
\psib(u)&\leq&\exp\left(\int_{u_0}^u\fb_{\lambda,l}(u^\prime)du^\prime\right)\psib(u_0)\label{4.b11}\\
&\s&+\int_{u_0}^u\exp\left(\int_{u^\prime}^u\fb_{\lambda,l}(u^{\prime\prime})du^{\prime\prime}\right)
\rhob(u^\prime)du^\prime\nonumber
\end{eqnarray}
an inequality for a 1-parameter family of non-negative functions
on $S_{\ub,u_0}$.

Now by Lemmas 3.2 and 3.5 and Proposition 3.4:
\begin{equation}
\fb_{\lambda,l}-\frac{\lambda}{|u|}\left(1-\frac{\ub}{|u|}\right)\leq
|\lambda|\delta|u|^{-3}F_1 +l\delta^{1/2}|u|^{-2}F_2 \label{4.b12}
\end{equation}
Here $F_1$ and $F_2$ are non-negative non-decreasing continuous
functions of the quantities ${\cal D}_0^\infty$, ${\cal
R}_0^\infty$. 
It follows that if $\delta$ is suitably small depending on ${\cal
D}_0^\infty$ and ${\cal R}_0^\infty$ it holds:
\begin{equation}
\fb_{\lambda,l}-\frac{\lambda}{|u|}\leq\frac{|\lambda|+l}{|u|^2}
\label{4.b13}
\end{equation}
Hence:
\begin{equation}
\int_{u^\prime}^u\fb_{\lambda,l}(u^{\prime\prime})du^{\prime\prime}\leq
C+\log \left(\frac{|u^\prime|}{|u|}\right)^\lambda \label{4.b14}
\end{equation}
In particular,
\begin{equation}
\int_{u_0}^u\fb_{\lambda,l}(u^\prime)du^\prime\leq
C+\log\left(\frac{|u_0|}{|u|}\right)^\lambda \label{4.b15}
\end{equation}
Substituting \ref{4.b14}, \ref{4.b15} in \ref{4.b11} we conclude
that:
\begin{equation}
|u|^\lambda\psib(u)\leq
C|u_0|^\lambda\psib(u_0)+C\int_{u_0}^u|u^\prime|^\lambda\rhob(u^\prime)du^\prime
\label{4.b16}
\end{equation}
It follows that:
\begin{equation}
|u|^\lambda\|\psib(u)\|_{L^p(S_{\ub,u_0})}\leq
C|u_0|^\lambda\|\psib(u_0)\|_{L^p(S_{\ub,u_0})}
+C\int_{u_0}^u|u^\prime|^\lambda\|\rhob(u^\prime)\|_{L^p(S_{\ub,u_0})}du^\prime
\label{4.b17}
\end{equation}
Now by the second part of Lemma 4.3:
$$\|\psib\|_{L^p(S_{\ub,u})}\leq e^{3/2p}\left(\frac{|u|}{|u_0|}\right)^{2/p}\|\psib(u)\|_{L^p(S_{\ub,u_0})}$$
and:
$$\|\rhob(u^\prime)\|_{L^p(S_{\ub,u_0})}\leq
e^{3/2p}\left(\frac{|u_0|}{|u^\prime|}\right)^{2/p}\|\rhob\|_{L^p(S_{\ub,u^\prime})}$$
Therefore \ref{4.b17} implies:
\begin{eqnarray}
|u|^{\lambda-(2/p)}\|\psi\|_{L^p(S_{\ub,u})}&\leq& C|u_0|^{\lambda-(2/p)}\|\psib\|_{L^p(S_{\ub,u_0})}\nonumber\\
&\s&
+C\int_0^{\ub}|u^\prime|^{\lambda-(2/p)}\|\rhob\|_{L^p(S_{\ub,u^\prime})}du^\prime
\label{4.b18}
\end{eqnarray}
for a new constant $C$, depending on $p$, and the lemma is proved.

\vspace{5mm}

\noindent {\bf Lemma 4.6} \ \ \ Let $\theta$ be a $r$ covariant
$S$ tensorfield on $M^\prime$ vanishing on $\Cb_0$ and satisfying
along the generators of $C_u$ the propagation equation:
$$D\theta=\frac{\nu}{2}\Omega\mbox{tr}\chi\theta+\gamma\cdot\theta+\xi$$
where $\nu$ is a real number, $\xi$ is a $r$ covariant $S$
tensorfield on $M^\prime$, and $\gamma$ is a type $T^r_r$ $S$
tensorfield on $M^\prime$ satisfying, pointwise,
$$|\gamma|\leq m\Omega|\chih|$$
where $m$ is a positive constant. Then, if $\delta$ is suitably small depending on ${\cal R}_0^\infty$, for each $p\geq 1$ there is a positive constant
$C$ depending only on $p, r, \nu, m$ such that:
$$\|\theta\|_{L^p(S_{\ub,u})}\leq C\int_0^{\ub}\|\xi\|_{L^p(S_{\ub^\prime,u})}d\ub^\prime$$

\noindent{\em Proof:} \ By virtue of the first part of Lemma 4.2
we have:
\begin{equation}
D|\theta|^2+r\Omega\mbox{tr}\chi|\theta|^2\leq
2(\theta,D\theta)+2r\Omega|\chih||\theta|^2 \label{4.b19}
\end{equation}
By the propagation equation,
\begin{eqnarray}
(\theta,D\theta)&=&\frac{\nu}{2}\Omega\mbox{tr}\chi|\theta|^2+(\theta,\gamma\cdot\theta)+(\theta,\xi)\nonumber\\
&\leq&\frac{\nu}{2}\Omega\mbox{tr}\chi|\theta|^2+m\Omega|\chih||\theta|^2+|\theta||\xi|
\label{4.b20}
\end{eqnarray}
Substituting \ref{4.b20} in \ref{4.b19} we obtain the inequality:
\begin{equation}
D|\theta|^2+(r-\nu)\Omega\mbox{tr}\chi|\theta|^2\leq
2(m+r)\Omega|\chih||\theta|^2+2|\theta||\xi| \label{4.b21}
\end{equation}
Setting then:
$$\psi=|\theta|, \ \ \ \rho=|\xi|, \ \ \ \lambda=r-\nu, \ \ \ l=m+r$$
and applying Lemma 4.4, yields the lemma.

\vspace{5mm}

\noindent {\bf Lemma 4.7} \ \ \ Let $\thetab$ be a $r$ covariant
$S$ tensorfield on $M^\prime$ satisfying along the generators of
$\Cb_{\ub}$ the propagation equation:
$$\Db\thetab=\frac{\nu}{2}\Omega\mbox{tr}\chib\thetab+\gammab\cdot\thetab+\xib$$
where $\nu$ is a real number, $\xib$ is a $r$ covariant $S$
tensorfield on $M^\prime$, and $\gammab$ is a type $T^r_r$ $S$
tensorfield on $M^\prime$ satisfying, pointwise,
$$|\gammab|\leq m\Omega|\chibh|$$
where $m$ is a positive constant. Then, if $\delta$ is suitably small depending on ${\cal D}_0^\infty$, ${\cal R}_0^\infty$, for each $p\geq
1$ there is a positive constant $C$ depending only on $p, r, \nu,
m$ such that:
\begin{eqnarray*}
|u|^{r-\nu-(2/p)}\|\thetab\|_{L^p(S_{\ub,u})}&\leq&
C|u_0|^{r-\nu-(2/p)}\|\thetab\|_{L^p(S_{\ub,u_0})}\\
&\s& +
C\int_{u_0}^u|u^\prime|^{r-\nu-(2/p)}\|\xib\|_{L^p(S_{\ub,u^\prime})}du^\prime
\end{eqnarray*}

\noindent{\em Proof:} \ By virtue of the second part of Lemma 4.2
we have:
\begin{equation}
\Db|\thetab|^2+r\Omega\mbox{tr}\chib|\thetab|^2\leq
2(\thetab,\Db\thetab)+2r\Omega|\chibh||\thetab|^2 \label{4.b22}
\end{equation}
By the propagation equation,
\begin{eqnarray}
(\thetab,\Db\thetab)&=&\frac{\nu}{2}\Omega\mbox{tr}\chib|\thetab|^2+(\thetab,\gammab\cdot\thetab)+(\thetab,\xib)\nonumber\\
&\leq&\frac{\nu}{2}\Omega\mbox{tr}\chib|\thetab|^2+m\Omega|\chibh||\thetab|^2+|\thetab||\xib|
\label{4.b23}
\end{eqnarray}
Substituting \ref{4.b23} in \ref{4.b22} we obtain the inequality:
\begin{equation}
\Db|\thetab|^2+(r-\nu)\Omega\mbox{tr}\chib|\thetab|^2\leq
2(m+r)\Omega|\chibh||\thetab|^2+2|\thetab||\xib| \label{4.b24}
\end{equation}
Setting then:
$$\psib=|\thetab|, \ \ \ \rhob=|\xib|, \ \ \ \lambda=r-\nu, \ \ \ l=m+r$$
and applying Lemma 4.5, yields the lemma.

\vspace{5mm}

According to the results of Propositions 3.1 - 3.4 we have, on
$M^\prime$:
\begin{eqnarray}
&&\delta^{-1}|u|^2\log\Omega\leq C{\cal R}_0^\infty(\rho)+O(\delta|u|^{-1})\nonumber\\
&&|u|^2\left|\frac{1}{2}\mbox{tr}\chi-\frac{1}{|u|}\right|\leq C({\cal R}_0^\infty(\alpha))^2+O(\delta)\nonumber\\
&&\delta^{1/2}|u||\chih|\leq C{\cal R}_0^\infty(\alpha)\nonumber\\
&&\delta^{-1}|u|^3\left|\frac{1}{2}\mbox{tr}\chib
+\frac{1}{|u|}\left(1-\frac{\ub}{|u|}\right)\right|\leq C\{{\cal
D}_0^\infty(\mbox{tr}\chib)+({\cal D}_0^\infty(\chibh))^2+{\cal
R}_0^\infty(\rho)\}
+O(\delta)\nonumber\\
&&\delta^{-1/2}|u|^2|\chibh|\leq C{\cal D}_0^\infty(\chibh)+O(\delta|u|^{-3/2})\nonumber\\
&&\delta^{-1/2}|u|^2|\eta|\leq C{\cal R}_0^\infty(\beta)+O(\delta|u|^{-1})\nonumber\\
&&\delta^{-1/2}|u|^2|\etb|\leq C{\cal
R}_0^\infty(\beta)+O(\delta^{1/2}|u|^{-1}) \label{4.49}
\end{eqnarray}
Here we denote by $O(\delta^p|u|^r)$, for real numbers $p,r$, the
product of $\delta^p|u|^r$ with a non-negative non-decreasing
continuous function of the quantities ${\cal D}_0^\infty$, ${\cal
R}_0^\infty$.

\section{$L^4(S)$ estimates for $\snab\chi$}

\noindent {\bf Proposition 4.1} \ \ \ We have:
\begin{eqnarray*}
\|\sd\mbox{tr}\chi\|_{L^4(S_{\ub,u})}&\leq& C|u|^{-5/2}{\cal
R}_0^\infty(\alpha)\scR_1^4(\alpha)
+O(\delta^{1/2}|u|^{-5/2})\\
\|\snab\chih\|_{L^4(S_{\ub,u})}&\leq& C\delta^{-1/2}|u|^{-3/2}\scR_1^4(\alpha)+O(|u|^{-5/2})\\
&:&\forall (\ub,u)\in D^\prime
\end{eqnarray*}
provided that $\delta$ is suitably small depending on ${\cal
D}_0^\infty$, ${\cal R}_0^\infty$.

\noindent {\em Proof:} \ We consider equations \ref{3.6} and
\ref{3.8}. In view of the first of \ref{3.3}, equation \ref{3.8}
is equivalent to:
\begin{equation}
D\chih^\prime=-\alpha+\Omega^2|\chih^\prime|^2\sg \label{4.52}
\end{equation}
We apply Lemma 1.2 to equation \ref{3.6} and Lemma 4.1 to equation
\ref{4.52} to obtain the following linear system of propagation
equations for $\sd\mbox{tr}\chi^\prime$, $\snab\chih^\prime$ along
each generator of each $C_u$:
\begin{eqnarray}
D\sd\mbox{tr}\chi^\prime&=&f\cdot\sd\mbox{tr}\chi^\prime+g\cdot\snab\chih^\prime+r\nonumber\\
D\snab\chih^\prime&=&h\cdot\sd\mbox{tr}\chi^\prime+i\cdot\snab\chih^\prime+s
\label{4.53}
\end{eqnarray}
Here $f, g, r, h, i, s$ are respectively $T^1_1, T^3_1, T^0_1,
T^1_3, T^3_3, T^0_3$ type $S$ tensorfields, the components of
which with respect to an arbitrary local frame field for the
$S_{\ub,u}$ are given by:
\begin{eqnarray}
&&f_A^B=-\Omega^2\mbox{tr}\chi^\prime\delta_A^B\nonumber\\
&&g_A^{BCD}=-2\Omega^2\chih^{\prime CD}\delta_A^B\nonumber\\
&&r_A=-2\Omega\sd_A\Omega\left[\frac{1}{2}(\mbox{tr}\chi^\prime)^2+|\chih^\prime|^2\right]\nonumber\\
&&h^D_{ABC}=-\frac{1}{2}\Omega^2(2\chih^\prime_{BC}\delta_A^D+\chih^\prime_{AC}\delta_B^D+\chih^\prime_{AB}\delta_C^D
-\sg_{AB}\chih^{\prime D}_C-\sg_{AC}\chih^{\prime D}_B)\nonumber\\
&&i_{ABC}^{DEF}=2\Omega^2\delta_A^D\sg_{BC}\chih^{\prime EF}\nonumber\\
&&\hspace{13mm}-\frac{1}{2}\Omega^2\left[\delta_A^D(\delta_B^E\chih^{\prime
F}_C+\delta_B^F\chih^{\prime E}_C
+\delta_C^E\chih^{\prime F}_B+\delta_C^F\chih^{\prime E}_B)\right.\nonumber\\
&&\hspace{23mm}+\delta_A^E(\delta_B^D\chih^{\prime
F}_C+\delta_C^D\chih^{\prime F}_B)
+\delta_A^F(\delta_B^D\chih^{\prime E}_C+\delta_C^D\chih^{\prime E}_B)\nonumber\\
&&\hspace{23mm}\left.-\chih^{\prime
D}_B(\delta_A^E\delta_C^F+\delta_A^F\delta_C^E)
-\chih^{\prime D}_C(\delta_A^E\delta_B^F+\delta_A^F\delta_B^E)\right]\nonumber\\
&&s_{ABC}=-\snab_A\alpha_{BC}+2\Omega|\chih^\prime|^2\sd_A\Omega\sg_{BC}\nonumber\\
&&\hspace{13mm}-2\Omega(\chih^{\prime
D}_B\sd_A\Omega+\chih^{\prime
D}_A\sd_B\Omega-\chih^\prime_{AB}\sd^D\Omega)\chih^\prime_{DC}
\nonumber\\
&&\hspace{13mm}-2\Omega(\chih^{\prime
D}_C\sd_A\Omega+\chih^{\prime
D}_A\sd_C\Omega-\chih^\prime_{AC}\sd^D\Omega)\chih^\prime_{DB}
\nonumber\\
&&\hspace{13mm}-\Omega\mbox{tr}\chi^\prime\left[2\chih^\prime_{BC}\sd_A\Omega+\chih^\prime_{AB}\sd_C\Omega
+\chih^\prime_{AC}\sd_B\Omega\right.\nonumber\\
&&\hspace{33mm}\left.-(\sg_{AB}\chih^\prime_{DC}+\sg_{AC}\chih^\prime_{DB})\sd^D\Omega\right]
\label{4.54}
\end{eqnarray}
Thus \ref{4.53} reads, in an arbitrary local frame field for the
$S_{\ub,u}$,
\begin{eqnarray*}
&&(D\sd\mbox{tr}\chi^\prime)_A=f_A^B(\sd\mbox{tr}\chi^\prime)_B+g_A^{BCD}(\snab\chih^\prime)_{BCD}+r_A\\
&&(D\snab\chih^\prime)_{ABC}=h^D_{ABC}(\sd\mbox{tr}\chi^\prime)_D+i_{ABC}^{DEF}(\snab\chih^\prime)_{DEF}+s_{ABC}
\end{eqnarray*}

We have, pointwise:
\begin{equation}
|g|,|h|,|i|\leq C\Omega|\chih| \label{4.56}
\end{equation}
Also, recalling that, from \ref{1.65},
\begin{equation}
\sd\log\Omega=\frac{1}{2}(\eta+\etb) \label{4.57}
\end{equation}
we have, by \ref{4.49}, the pointwise bounds:
\begin{eqnarray}
|r|&\leq&C\delta^{-1/2}|u|^{-4}\{{\cal R}_0^\infty(\beta)({\cal R}_0^\infty(\alpha))^2+O(\delta^{1/2})\}\nonumber\\
|s|&\leq&|\snab\alpha|+C\delta^{-1/2}|u|^{-4}\{{\cal
R}_0^\infty(\beta)({\cal R}_0^\infty(\alpha))^2+O(\delta^{1/2})\}
\label{4.58}
\end{eqnarray}

By Lemma 4.2 with the $S$ 1-form $\sd\mbox{tr}\chi^\prime$ in the
role of $\theta$,
$$D(|\sd\mbox{tr}\chi^\prime|^2)+\Omega\mbox{tr}\chi|\sd\mbox{tr}\chi^\prime|^2=
2(\sd\mbox{tr}\chi^\prime,D\sd\mbox{tr}\chi^\prime)-2\Omega\chih^{AB}(\sd\mbox{tr}\chi^\prime)_A(\sd\mbox{tr}\chi^\prime)_B$$
From the first of \ref{4.53} and the first of \ref{4.54}:
\begin{eqnarray*}
(\sd\mbox{tr}\chi^\prime,D\sd\mbox{tr}\chi^\prime)&=&-\Omega\mbox{tr}\chi|\sd\mbox{tr}\chi^\prime|^2
+(\sd\mbox{tr}\chi^\prime, r+g\cdot\snab\chih^\prime)\\
&\leq&-\Omega\mbox{tr}\chi|\sd\mbox{tr}\chi^\prime|^2+|\sd\mbox{tr}\chi^\prime|(|r|+|g||\snab\chih^\prime|)
\end{eqnarray*}
Since also
$|\chih^{AB}(\sd\mbox{tr}\chi^\prime)_A(\sd\mbox{tr}\chi^\prime)_B|\leq
|\chih||\sd\mbox{tr}\chi^\prime|^2$, we obtain:
\begin{equation}
D(|\sd\mbox{tr}\chi^\prime|^2)+3\Omega\mbox{tr}\chi|\sd\mbox{tr}\chi^\prime|^2
\leq
2|\sd\mbox{tr}\chi^\prime|(|r|+|g||\snab\chih^\prime|)+2\Omega|\chih||\sd\mbox{tr}\chi^\prime|^2
\label{4.59}
\end{equation}

By Lemma 4.2 with the 3-covariant $S$ tensorfield
$\snab\chih^\prime$ in the role of $\theta$,
\begin{eqnarray*}
D(|\snab\chih^\prime|^2)+3\Omega\mbox{tr}\chi|\snab\chih^\prime|^2&=&2(\snab\chih^\prime,D\snab\chih^\prime)
-2\Omega\chih^A_D(\snab\chih^\prime)^{DBC}(\snab\chih^\prime)_{ABC}\\
&\s&-2\Omega\chih^B_D(\snab\chih^\prime)^{ADC}(\snab\chih^\prime)_{ABC}
-2\Omega\chih^C_D(\snab\chih^\prime)^{ABD}(\snab\chih^\prime)_{ABC}
\end{eqnarray*}
From the second of \ref{4.53}:
\begin{eqnarray*}
(\snab\chih^\prime,D\snab\chih^\prime)&=&(\snab\chih^\prime, s+h\cdot\sd\mbox{tr}\chi^\prime+i\cdot\snab\chih^\prime)\\
&\leq&|\snab\chih^\prime|(|s|+|h||\sd\mbox{tr}\chi^\prime|+|i||\snab\chih^\prime|)
\end{eqnarray*}
Since also each of
$\chih^A_D(\snab\chih^\prime)^{DBC}(\snab\chih^\prime)_{ABC}$,
$\chih^B_D(\snab\chih^\prime)^{ADC}(\snab\chih^\prime)_{ABC}$,
$\chih^C_D(\snab\chih^\prime)^{ABD}(\snab\chih^\prime)_{ABC}$ is
bounded in absolute value by $|\chih||\snab\chih^\prime|^2$ (the
last two are equal in view of the symmetry of $\snab\chih^\prime$
in the last two indices) we obtain:
\begin{equation}
D(|\snab\chih^\prime|^2)+3\Omega\mbox{tr}\chi|\snab\chih^\prime|^2\leq
2|\snab\chih^\prime|(|s|+|h||\sd\mbox{tr}\chi^\prime|+|i||\snab\chih^\prime|)
+6\Omega|\chih||\snab\chih^\prime|^2 \label{4.60}
\end{equation}

Let us define the non-negative function:
\begin{equation}
\psi=\sqrt{|\sd\mbox{tr}\chi^\prime|^2+|\snab\chih^\prime|^2}
\label{4.61}
\end{equation}
By \ref{4.59}, \ref{4.60}, and the pointwise bounds \ref{4.56},
$\psi^2$ satisfies the inequality:
\begin{equation}
D(\psi^2)+3\Omega\mbox{tr}\chi\psi^2\leq
2C\Omega|\chih|\psi^2+2\psi(|r|+|s|) \label{4.62}
\end{equation}
We then apply Lemma 4.4 with $p=4$, $\lambda=3$, $l=C$, and
$\rho=|r|+|s|$, to obtain:
\begin{equation}
\|\psi\|_{L^4(S_{\ub,u})}\leq
C\int_0^{\ub}\left\{\|r\|_{L^4(S_{\ub^\prime,u})}+\|s\|_{L^4(S_{\ub^\prime,u})}
\right\}d\ub^\prime \label{4.70}
\end{equation}
Now for any function $\phi\in L^\infty(S_{\ub,u})$ and any $p\geq
1$ we have, by Lemma 4.3,
\begin{equation}
\|\phi\|_{L^p(S_{\ub,u})}\leq
(\mbox{Area}(S_{\ub,u}))^{1/p}\|\phi\|_{L^\infty(S_{\ub,u})} \leq
(8\pi|u|^2)^{1/p}\|\phi\|_{L^\infty(S_{\ub,u})} \label{4.71}
\end{equation}
Consequently, the inequalities \ref{4.58} imply, in view of the
first of the definitions \ref{4.1},
\begin{eqnarray}
\|r\|_{L^4(S_{\ub^\prime,u})}&\leq& C\delta^{-1/2}|u|^{-7/2}\{{\cal R}_0^\infty(\beta)({\cal R}_0^\infty(\alpha))^2+O(\delta^{1/2})\}\nonumber\\
\|s\|_{L^4(S_{\ub^\prime,u})}&\leq& \delta^{-3/2}|u|^{-3/2}\scR_1^4(\alpha)\nonumber\\
&\s&+C\delta^{-1/2}|u|^{-7/2}\{{\cal R}_0^\infty(\beta)({\cal
R}_0^\infty(\alpha))^2+O(\delta^{1/2})\} \label{4.72}
\end{eqnarray}
Substituting in \ref{4.70} then yields:
\begin{equation}
\|\psi\|_{L^4(S_{\ub,u})}\leq
C\delta^{-1/2}|u|^{-3/2}\scR_1^4(\alpha)
+C\delta^{1/2}|u|^{-7/2}\{{\cal R}_0^\infty(\beta)({\cal
R}_0^\infty(\alpha))^2+O(\delta^{1/2})\} \label{4.73}
\end{equation}

Going back to  \ref{4.59}, that inequality implies:
\begin{equation}
D(|\sd\mbox{tr}\chi^\prime|^2)+3\Omega\mbox{tr}\chi|\sd\mbox{tr}\chi^\prime|^2
\leq 2|\sd\mbox{tr}\chi^\prime|(|r|+C\Omega|\chih|\psi)
\label{4.74}
\end{equation}
Applying Lemma 4.4 with $|\sd\mbox{tr}\chi^\prime|$ in the role of
$\psi$ and $|r|+C\Omega|\chih|\psi$ in the role of $\rho$ yields,
taking $p=4$:
\begin{eqnarray}
\|\sd\mbox{tr}\chi^\prime\|_{L^4(S_{\ub,u})}&\leq&C\int_0^{\ub}\|r\|_{L^4(S_{\ub^\prime,u})}d\ub^\prime\label{4.75}\\
&\s&+C\delta^{-1/2}|u|^{-1}{\cal
R}_0^\infty(\alpha)\int_0^{\ub}\|\psi\|_{L^4(S_{\ub^\prime,u})}d\ub^\prime\nonumber
\end{eqnarray}
Substituting the estimate \ref{4.73} as well as the first of the
estimates \ref{4.72} then yields:
\begin{equation}
\|\sd\mbox{tr}\chi^\prime\|_{L^4(S_{\ub,u})}\leq C|u|^{-5/2}{\cal
R}_0^\infty(\alpha)\scR_1^4(\alpha)
+C\delta^{1/2}|u|^{-7/2}\{{\cal R}_0^\infty(\beta)({\cal
R}_0^\infty(\alpha))^2+O(\delta^{1/2})\} \label{4.76}
\end{equation}
On the other hand, in view of the definition \ref{4.61}, the
estimate \ref{4.73} directly implies:
\begin{equation}
\|\snab\chih^\prime\|_{L^4(S_{\ub,u})}\leq
C\delta^{-1/2}|u|^{-3/2}\scR_1^4(\alpha)
+C\delta^{1/2}|u|^{-7/2}\{{\cal R}_0^\infty(\beta)({\cal
R}_0^\infty(\alpha))^2+O(\delta^{1/2})\} \label{4.77}
\end{equation}
Since
$$\sd\mbox{tr}\chi=\sd(\Omega\mbox{tr}\chi^\prime)=\Omega(\sd\mbox{tr}\chi^\prime+(\sd\log\Omega)\mbox{tr}\chi^\prime)$$
$$\snab\chih=\snab(\Omega\chih^\prime)=\Omega(\snab\chih^\prime+(\sd\log\Omega)\otimes\chih^\prime)$$
by \ref{4.57} and the bounds \ref{4.49} the estimates \ref{4.76},
\ref{4.77} imply the estimates of the proposition.

\section{$L^4(S)$ estimates for $\snab\chib$}

With the notations of Section 3 of Chapter 3, we consider any
$(\ub_1,u_1)\in D^\prime$ and fix attention in the following
lemmas to the parameter subdomain $D_1$ and the corresponding
subdomain $M_1$ of $M^\prime$ (see \ref{3.02}, \ref{3.03}).

With a positive constant $k$ to be appopriately chosen in the
sequel, let $s^*$ be the least upper bound of the set of values of
$s\in[u_0,u_1]$ such that:
\begin{equation}
\|\sd\log\Omega\|_{L^4(S_{\ub,u})}\leq k\delta|u|^{-5/2} \ \ \ : \
\mbox{for all $(\ub,u)\in [0,\ub_1]\times[u_0,s]$} \label{4.78}
\end{equation}
Then by continuity $s^*>u_0$ (recall that by \ref{3.34}
$\sd\log\Omega$ vanishes along $C_{u_0}$) and we have:
\begin{equation}
\|\sd\log\Omega\|_{L^4(S_{\ub,u})}\leq k\delta|u|^{-5/2} \ \ \ : \
\mbox{for all $(\ub,u)\in [0,\ub_1]\times[u_0,s^*]$} \label{4.79}
\end{equation}

Let us denote:
\begin{equation}
\scD_1^4(\mbox{tr}\chib)=|u_0|^{7/2}\delta^{-1}\sup_{\ub\in[0,\delta]}\|\sd\mbox{tr}\chib\|_{L^4(S_{\ub,u_0})}
\label{4.80}
\end{equation}
\begin{equation}
\scD_1^4(\chibh)=|u_0|^{5/2}\delta^{-1/2}\sup_{\ub\in[0,\delta]}\|\snab\chibh\|_{L^4(S_{\ub,u_0})}
\label{4.81}
\end{equation}
Here, as in \ref{3.37}, \ref{3.108}, we are considering all of $C_{u_0}$, not only the part which lies in 
$M^\prime$. This is the same for $c^*\geq u_0+\delta$, but not for $c^*\in(u_0,u_0+\delta)$. In the latter case 
the part of $C_{u_0}$ lying in $M^\prime$ is the part $C^{c^*-u_0}_{u_0}$, so $\ub$ would be restricted to the interval 
$[0,c^*-u_0]$. 
By the results of Chapter 2, $\scD_1^4(\mbox{tr}\chib)$ and
$\scD_1^4(\chibh)$ are bounded by a non-negative non-decreasing
continuous function of $M_4$.

Also, let us denote by $D_1^{s^*}$ the subset of $D_1$ where
$u\leq s^*$:
\begin{equation}
D_1^{s^*}=[0,\ub_1]\times[u_0,s^*] \label{4.82}
\end{equation}
and by $M_1^{s^*}$ the corresponding subset of $M_1$.

\vspace{5mm}

\noindent {\bf Lemma 4.8} \ \ \ We have:
\begin{eqnarray*}
\|\sd\mbox{tr}\chib\|_{L^4(S_{\ub,u})}
&\leq& C\delta |u|^{-7/2}\left\{k+(1+{\cal D}_0^\infty(\chibh)+O(\delta))(\scD_1^4(\mbox{tr}\chib)+\scD_1^4(\chibh))\right\}\\
&\s&+C\delta^{3/2}|u|^{-9/2}({\cal D}_0^\infty(\chibh)+O(\delta))^2({\cal R}_0^\infty(\beta)+O(\delta^{1/2}))\\
&\s&+C\delta^2|u|^{-5}({\cal
D}_0^\infty(\chibh)+O(\delta))\scR_1^4(\alb)
\end{eqnarray*}
\begin{eqnarray*}
\|\snab\chibh\|_{L^4(S_{\ub,u})}&\leq& C\delta^{1/2}|u|^{-5/2}(\scD_1^4(\mbox{tr}\chib)+\scD_1^4(\chibh))\\
&\s&+C\delta|u|^{-7/2}\{k+({\cal D}_0^\infty(\chibh)+O(\delta))({\cal R}_0^\infty(\beta)+O(\delta^{1/2}))\}\nonumber\\
&\s&+C\delta^{3/2}|u|^{-4}\scR_1^4(\alb)
\end{eqnarray*}
for all $(\ub,u)\in D_1^{s^*}$, provided that $\delta$ is suitably
small depending on ${\cal D}_0^\infty$, ${\cal R}_0^\infty$.

\noindent {\em Proof:} \ We consider equations \ref{3.7} and
\ref{3.9}. In view of the second of \ref{3.3}, equation \ref{3.9}
is equivalent to:
\begin{equation}
\Db\chibh^\prime=-\alb+\Omega^2|\chibh^\prime|^2\sg \label{4.83}
\end{equation}
We apply Lemma 1.2 to equation \ref{3.7} and Lemma 4.1 to equation
\ref{4.83} to obtain the following linear system of propagation
equations for $\sd\mbox{tr}\chib^\prime$, $\snab\chibh^\prime$
along each generator of each $\Cb_{\ub}$:
\begin{eqnarray}
\Db\sd\mbox{tr}\chib^\prime&=&\fb\cdot\sd\mbox{tr}\chib^\prime+\gb\cdot\snab\chibh^\prime+\rb\nonumber\\
\Db\snab\chibh^\prime&=&\hb\cdot\sd\mbox{tr}\chib^\prime+\ib\cdot\snab\chibh^\prime+\sb
\label{4.84}
\end{eqnarray}
Here $\fb,\gb,\rb,\hb,\ib,\sb$ are respectively $T^1_1, T^3_1,
T^0_1, T^1_3, T^3_3, T^0_3$ type $S$ tensorfields, the components
of which with respect to an arbitrary local frame field for the
$S_{\ub,u}$ are given by:
\begin{eqnarray}
&&\fb_A^B=-\Omega^2\mbox{tr}\chib^\prime\delta_A^B\nonumber\\
&&\gb_A^{BCD}=-2\Omega^2\chibh^{\prime CD}\delta_A^B\nonumber\\
&&\rb_A=-2\Omega\sd_A\Omega\left[\frac{1}{2}(\mbox{tr}\chib^\prime)^2+|\chibh^\prime|^2\right]\nonumber\\
&&\hb^D_{ABC}=-\frac{1}{2}\Omega^2(2\chibh^\prime_{BC}\delta_A^D+\chibh^\prime_{AC}\delta_B^D+\chibh^\prime_{AB}\delta_C^D
-\sg_{AB}\chibh^{\prime D}_C-\sg_{AC}\chibh^{\prime D}_B)\nonumber\\
&&\ib_{ABC}^{DEF}=2\Omega^2\delta_A^D\sg_{BC}\chibh^{\prime EF}\nonumber\\
&&\hspace{13mm}-\frac{1}{2}\Omega^2\left[\delta_A^D(\delta_B^E\chibh^{\prime
F}_C+\delta_B^F\chibh^{\prime E}_C
+\delta_C^E\chibh^{\prime F}_B+\delta_C^F\chibh^{\prime E}_B)\right.\nonumber\\
&&\hspace{13mm}+\delta_A^E(\delta_B^D\chibh^{\prime
F}_C+\delta_C^D\chibh^{\prime F}_B)
+\delta_A^F(\delta_B^D\chibh^{\prime E}_C+\delta_C^D\chibh^{\prime E}_B)\nonumber\\
&&\hspace{13mm}\left.-\chibh^{\prime
D}_B(\delta_A^E\delta_C^F+\delta_A^F\delta_C^E)
-\chibh^{\prime D}_C(\delta_A^E\delta_B^F+\delta_A^F\delta_B^E)\right]\nonumber\\
&&\sb_{ABC}=-\snab_A\alb_{BC}+2\Omega|\chibh^\prime|^2\sd_A\Omega\sg_{BC}\nonumber\\
&&\hspace{13mm}-2\Omega(\chibh^{\prime
D}_B\sd_A\Omega+\chibh^{\prime
D}_A\sd_B\Omega-\chibh^\prime_{AB}\sd^D\Omega)\chibh^\prime_{DC}
\nonumber\\
&&\hspace{13mm}-2\Omega(\chibh^{\prime
D}_C\sd_A\Omega+\chibh^{\prime
D}_A\sd_C\Omega-\chibh^\prime_{AC}\sd^D\Omega)\chibh^\prime_{DB}
\nonumber\\
&&\hspace{13mm}-\Omega\mbox{tr}\chib^\prime\left[2\chibh^\prime_{BC}\sd_A\Omega+\chibh^\prime_{AB}\sd_C\Omega
+\chibh^\prime_{AC}\sd_B\Omega\right.\nonumber\\
&&\hspace{33mm}\left.-(\sg_{AB}\chibh^\prime_{DC}+\sg_{AC}\chibh^\prime_{DB})\sd^D\Omega\right]
\label{4.85}
\end{eqnarray}
Thus \ref{4.84} reads, in an arbitrary local frame field for the
$S_{\ub,u}$,
\begin{eqnarray*}
(\Db\sd\mbox{tr}\chib^\prime)_A&=&\fb_A^B(\sd\mbox{tr}\chib^\prime)_B+\gb_A^{BCD}(\snab\chibh^\prime)_{BCD}+\rb_A\\
(\Db\snab\chibh^\prime)_{ABC}&=&\hb^D_{ABC}(\sd\mbox{tr}\chib^\prime)_D+\ib^{DEF}_{ABC}(\snab\chibh^\prime)_{DEF}
+\sb_{ABC}
\end{eqnarray*}

We have the pointwise bounds:
\begin{equation}
|\gb|,|\hb|,|\ib|\leq C\Omega|\chibh| \label{4.87}
\end{equation}
Moreover, by virtue of \ref{4.79} and the bounds \ref{4.49}:
\begin{equation}
\|\rb\|_{L^4(S_{\ub,u})}\leq Ck\delta|u|^{-9/2} \ \ \ : \ \forall
(\ub,u)\in D_1^{s^*} \label{4.88}
\end{equation}
provided that $\delta$ is suitably small depending on ${\cal
D}_0^\infty$, ${\cal R}_0^\infty$. Also, from the sixth of the
definitions \ref{4.1} and the $L^\infty$ bound for $\sd\log\Omega$
which results from the expression \ref{4.57}, we obtain:
\begin{equation}
\|\sb\|_{L^4(S_{\ub,u})}\leq\delta^{3/2}|u|^{-5}\scR_1^4(\alb)
+C\delta|u|^{-9/2}({\cal D}_0^\infty(\chibh)+O(\delta))({\cal
R}_0^\infty(\beta)+O(\delta^{1/2})) \label{4.89}
\end{equation}
provided that $\delta$ is suitably small depending on ${\cal
D}_0^\infty$, ${\cal R}_0^\infty$.

Now, following an argument similar to that leading from equations
\ref{4.53} to the inequalities \ref{4.59} and \ref{4.60}, we
deduce from equations \ref{4.84}, using Lemma 4.2, the
inequalities:
\begin{equation}
\Db(|\sd\mbox{tr}\chib^\prime|^2)+3\Omega\mbox{tr}\chib|\sd\mbox{tr}\chib^\prime|^2
\leq
2|\sd\mbox{tr}\chib^\prime|(|\rb|+|\gb||\snab\chibh^\prime|)+2\Omega|\chibh||\sd\mbox{tr}\chib^\prime|^2
\label{4.90}
\end{equation}
and:
\begin{equation}
\Db(|\snab\chibh^\prime|^2)+3\Omega\mbox{tr}\chib|\snab\chibh^\prime|^2
\leq
2|\snab\chibh^\prime|(|\sb|+|\hb||\sd\mbox{tr}\chib^\prime|+|\ib||\snab\chibh^\prime|)
+6\Omega|\chibh||\snab\chibh^\prime|^2 \label{4.91}
\end{equation}

Let us define the non-negative function:
\begin{equation}
\psib=\sqrt{|\sd\mbox{tr}\chib^\prime|^2+|\snab\chibh^\prime|^2}
\label{4.92}
\end{equation}
By \ref{4.90}, \ref{4.91}, and the pointwise bounds \ref{4.87},
$\psib^2$ satisfies the inequality:
\begin{equation}
\Db(\psib^2)+3\Omega\mbox{tr}\chib\psib^2\leq
2C\Omega|\chibh|\psib^2+2\psib(|\rb|+|\sb|) \label{4.93}
\end{equation}
We then apply Lemma 4.5 with $p=4$, $\lambda=3$, $l=C$, and
$\rhob=|\rb|+|\sb|$, to obtain:
\begin{eqnarray}
|u|^{5/2}\|\psib\|_{L^4(S_{\ub,u})}&\leq& C|u_0|^{5/2}\|\psib\|_{L^4(S_{\ub,u_0})}\label{4.103}\\
&\s&+C\int_{u_0}^u|u^\prime|^{5/2}\left\{\|\rb\|_{L^4(S_{\ub,u^\prime})}+\|\sb\|_{L^4(S_{\ub,u^\prime})}\right\}du^\prime
\nonumber
\end{eqnarray}
Substituting the bounds \ref{4.88}, \ref{4.89} and noting that by
the definitions \ref{4.80}, \ref{4.81} and \ref{4.92}:
\begin{equation}
\|\psib\|_{L^4(S{\ub,u_0})}\leq
\delta^{1/2}|u_0|^{-5/2}(\scD_1^4(\mbox{tr}\chib)+\scD_1^4(\chibh))
\label{4.104}
\end{equation}
we conclude that, for all $(\ub,u)\in D_1^{s^*}$:
\begin{eqnarray}
\|\psib\|_{L^4(S_{\ub,u})}&\leq& C\delta^{1/2}|u|^{-5/2}(\scD_1^4(\mbox{tr}\chib)+\scD_1^4(\chibh))\nonumber\\
&\s&+C\delta|u|^{-7/2}\{k+({\cal D}_0^\infty(\chibh)+O(\delta))({\cal R}_0^\infty(\beta)+O(\delta^{1/2}))\}\nonumber\\
&\s&+C\delta^{3/2}|u|^{-4}\scR_1^4(\alb)\label{4.105}
\end{eqnarray}

Going back to \ref{4.90}, that inequality implies:
\begin{equation}
\Db(|\sd\mbox{tr}\chib^\prime|^2)+3\Omega\mbox{tr}\chib|\sd\mbox{tr}\chib^\prime|^2
\leq 2|\sd\mbox{tr}\chib^\prime|(|\rb|+C\Omega|\chibh|\psib)
\label{4.106}
\end{equation}
Applying Lemma 4.5 with $|\sd\mbox{tr}\chib^\prime|$ in the role
of $\psib$ and $|\rb|+C\Omega|\chibh|\psib$ in the role of $\rhob$
yields, taking $p=4$:
\begin{eqnarray}
|u|^{5/2}\|\sd\mbox{tr}\chib^\prime\|_{L^4(S_{\ub,u})}
&\leq& C|u_0|^{5/2}\|\sd\mbox{tr}\chib^\prime\|_{L^4(S_{\ub,u_0})}\label{4.107}\\
&\s&+C\int_{u_0}^u|u^\prime|^{5/2}\|\rb\|_{L^4(S_{\ub,u^\prime})}du^\prime\nonumber\\
&\s&+C\delta^{1/2}({\cal
D}_0^\infty(\chibh)+O(\delta))\int_{u_0}^u|u^\prime|^{1/2}\|\psib\|_{L^4(S_{\ub,u^\prime})}du^\prime
\nonumber
\end{eqnarray}
Substituting the bounds \ref{4.88} and \ref{4.105} and recalling
the definition \ref{4.80}, we then obtain:
\begin{eqnarray}
\|\sd\mbox{tr}\chib^\prime\|_{L^4(S_{\ub,u})} &\leq& C\delta
|u|^{-7/2}\left\{k+(1+{\cal
D}_0^\infty(\chibh)+O(\delta))(\scD_1^4(\mbox{tr}\chib)+\scD_1^4(\chibh))\right\}
\nonumber\\
&\s&+C\delta^{3/2}|u|^{-9/2}({\cal D}_0^\infty(\chibh)+O(\delta))^2({\cal R}_0^\infty(\beta)+O(\delta^{1/2}))\nonumber\\
&\s&+C\delta^2|u|^{-5}({\cal
D}_0^\infty(\chibh)+O(\delta))\scR_1^4(\alb) \label{4.108}
\end{eqnarray}
for all $(\ub,u)\in D_1^{s^*}$, provided that $\delta$ is suitably
small depending on ${\cal D}_0^\infty$, ${\cal R}_0^\infty$. On
the other hand, in view of the definition \ref{4.92}, the estimate
\ref{4.105} directly implies:
\begin{eqnarray}
\|\snab\chibh^\prime\|_{L^4(S_{\ub,u})}&\leq& C\delta^{1/2}|u|^{-5/2}(\scD_1^4(\mbox{tr}\chib)+\scD_1^4(\chibh))\nonumber\\
&\s&+C\delta|u|^{-7/2}\{k+({\cal D}_0^\infty(\chibh)+O(\delta))({\cal R}_0^\infty(\beta)+O(\delta^{1/2}))\}\nonumber\\
&\s&+C\delta^{3/2}|u|^{-4}\scR_1^4(\alb)\label{4.109}
\end{eqnarray}
for all $(\ub,u)\in D_1^{s^*}$. Since
$$\sd\mbox{tr}\chib=\sd(\Omega\mbox{tr}\chib^\prime)=
\Omega(\sd\mbox{tr}\chib^\prime+(\sd\log\Omega)\mbox{tr}\chib^\prime)$$
$$\snab\chibh=\snab(\Omega\chibh^\prime)=\Omega(\snab\chibh^\prime+(\sd\log\Omega)\otimes\chibh^\prime)$$
by \ref{4.79} the estimates \ref{4.108} and \ref{4.109} imply the
estimates of the lemma.

\section{$L^4(S)$ estimates for $\snab\eta$, $\snab\etb$}

We proceed to derive $L^4(S)$ estimates for $\snab\eta$,
$\snab\etb$ in $M_1^{s^*}$. We consider equations \ref{1.68} and
\ref{1.69}. Applying Lemma 4.1 to these equations we obtain the
following linear system of propagation equations for $\snab\eta$,
$\snab\etb$ along the generators of each $C_u$ and each
$\Cb_{\ub}$ respectively:
\begin{eqnarray}
D\snab\eta&=&a\cdot\snab\etb+b\nonumber\\
\Db\snab\etb&=&\ab\cdot\snab\eta+\bb \label{4.110}
\end{eqnarray}
Here $a,\ab$ are $T^2_2$ type $S$ tensorfields while $b,\bb$ are
$T^0_2$ type $S$ tensorfields, the components of which with
respect to an arbitrary local frame field for the $S_{\ub,u}$ are
given by:
\begin{eqnarray}
a_{AB}^{CD}&=&\Omega\delta_A^C\chi_B^D\nonumber\\
\ab_{AB}^{CD}&=&\Omega\delta_A^C\chib_B^D\nonumber\\
b_{AB}&=&\Omega\{(\snab_A\chi_B^C)\etb_C+(\sd_A\log\Omega)\chi_B^C\etb_C\}\nonumber\\
&\s&-(D\sGamma)^C_{AB}\eta_C-\Omega\{\snab_A\beta_B+(\sd_A\log\Omega)\beta_B\}\nonumber\\
\bb_{AB}&=&\Omega\{(\snab_A\chib_B^C)\eta_C+(\sd_A\log\Omega)\chib_B^C\eta_C\}\nonumber\\
&\s&-(\Db\sGamma)^C_{AB}\etb_C+\Omega\{\snab_A\beb_B+(\sd_A\log\Omega)\beb_B\}
\label{4.111}
\end{eqnarray}
Thus \ref{4.110} read, in an arbitrary local frame field for the
$S_{\ub,u}$,
\begin{eqnarray*}
(D\snab\eta)_{AB}&=&a_{AB}^{CD}(\snab\etb)_{CD}+b_{AB}\\
(\Db\snab\etb)_{AB}&=&\ab_{AB}^{CD}(\snab\eta)_{CD}+\bb_{AB}
\end{eqnarray*}
We decompose:
\begin{eqnarray}
a_{AB}^{CD}&=&\frac{1}{2}\Omega\delta_A^C\delta_B^D\mbox{tr}\chi+\Omega\delta_A^C\chih_B^D\nonumber\\
\ab_{AB}^{CD}&=&\frac{1}{2}\Omega\delta_A^C\delta_B^D\mbox{tr}\chib+\Omega\delta_A^C\chibh_B^D
\label{4.112}
\end{eqnarray}
Then the above equations take the form:
\begin{eqnarray}
(D\snab\eta)_{AB}&=&\frac{1}{2}\Omega\mbox{tr}\chi(\snab\etb)_{AB}+\Omega\chih_B^C(\snab\etb)_{AC}+b_{AB}\nonumber\\
(\Db\snab\etb)_{AB}&=&\frac{1}{2}\Omega\mbox{tr}\chib(\snab\eta)_{AB}+\Omega\chibh_B^C(\snab\eta)_{AC}+\bb_{AB}
\label{4.113}
\end{eqnarray}

To the first of \ref{4.113} we apply Lemma 4.6 with $\snab\eta$ in
the role of $\theta$ and
$$\frac{1}{2}\Omega\mbox{tr}\chi\snab\etb+\Omega\chih\cdot\snab\etb+b$$
in the role of $\xi$. Then $r=2$, $\nu=0$, $\gamma=0$. Noting that
\begin{equation}
\Omega\left(\frac{1}{2}|\mbox{tr}\chi|+|\chih|\right)\leq
\tilde{a} \label{4.118}
\end{equation}
where
\begin{equation}
\tilde{a}=C\delta^{-1/2}|u|^{-1}({\cal
R}_0^\infty(\alpha)+O(\delta^{1/2})) \label{4.119}
\end{equation}
we obtain, taking $p=4$:
\begin{eqnarray}
\|\snab\eta\|_{L^4(S_{\ub,u})}&\leq& C\tilde{a}(u)\int_0^{\ub}\|\snab\etb\|_{L^4(S{\ub^\prime,u})}d\ub^\prime\nonumber\\
&\s&+C\int_0^{\ub}\|b\|_{L^4(S_{\ub,u})}d\ub^\prime \label{4.130}
\end{eqnarray}

To the second of \ref{4.113} we apply Lemma 4.7 with $\snab\etb$
in the role of $\thetab$ and
$$\frac{1}{2}\Omega\mbox{tr}\chib\snab\eta+\Omega\chibh\cdot\snab\eta+\bb$$
in the role of $\xib$. Then $r=2$, $\nu=0$, $\gammab=0$. Noting
that
\begin{equation}
\Omega\left(\frac{1}{2}|\mbox{tr}\chib|+|\chibh|\right)\leq
\tilde{\ab} \label{4.123}
\end{equation}
where
\begin{equation}
\tilde{\ab}=C|u|^{-1}(1+O(\delta^{1/2})) \label{4.124}
\end{equation}
we obtain, taking $p=4$:
\begin{eqnarray}
|u|^{3/2}\|\snab\etb\|_{L^4(S_{\ub,u})}&\leq& C|u_0|^{3/2}\|\snab\etb\|_{L^4(S_{\ub,u_0})}\nonumber\\
&\s&+C\int_{u_0}^u|u^\prime|^{3/2}\tilde{\ab}(u^\prime)\|\snab\eta\|_{L^4(S_{\ub,u^\prime})}du^\prime\nonumber\\
&\s&+C\int_{u_0}^u|u^\prime|^{3/2}\|\bb\|_{L^4(S_{\ub,u^\prime})}du^\prime
\label{4.136}
\end{eqnarray}
From \ref{4.124} we obtain:
\begin{equation}
\tilde{\ab}\leq C|u|^{-1} \label{4.137}
\end{equation}
(for a new numerical constant $C$) provided that $\delta$ is
suitably small depending on ${\cal D}_0^\infty$, ${\cal
R}_0^\infty$. Also, we define:
\begin{equation}
\scD_1^4(\etb)=|u_0|^{5/2}\delta^{-1/2}\sup_{\ub\in[0,\ub_1]}\|\snab\etb\|_{L^4(S_{\ub,u_0})}
\label{4.138}
\end{equation} 
Substituting the above in \ref{4.136} yields:
\begin{eqnarray}
|u|^{3/2}\|\snab\etb\|_{L^4(S_{\ub,u})}&\leq&
C|u_0|^{-1}\delta^{1/2}\scD_1^4(\etb)
+C\int_{u_0}^u|u^\prime|^{1/2}\|\snab\eta\|_{L^4(S_{\ub,u^\prime})}du^\prime\nonumber\\
&\s&+C\int_{u_0}^u|u^\prime|^{3/2}\|\bb\|_{L^4(S_{\ub,u^\prime})}du^\prime
\label{4.139}
\end{eqnarray}
for all $(\ub,u)\in D_1^{s^*}$.

The inequalities \ref{4.130}, \ref{4.139} constitute a linear
system of integral inequalities for the quantities
$\|\snab\eta\|_{L^4(S_{\ub,u})}$, $\|\snab\etb\|_{L^4(S_{\ub,u})}$
in the domain $D_1^{s^*}$. Setting:
\begin{equation}
z(u)=\sup_{\ub\in[0,\ub_1]}\|\snab\eta\|_{L^4(S_{\ub,u})}
\label{4.140}
\end{equation}
and:
\begin{equation}
\Bb(u)=\int_{u_0}^u|u^\prime|^{3/2}\sup_{\ub\in[0,\ub_1]}\|\bb\|_{L^4(S_{\ub,u^\prime})}du^\prime
\label{4.141}
\end{equation}
\ref{4.139} implies, for all $(\ub,u)\in D_1^{s^*}$:
\begin{eqnarray}
|u|^{3/2}\int_0^{\ub}\|\snab\etb\|_{L^4(S_{\ub^\prime,u})}d\ub^\prime&\leq&
C\delta^{3/2}|u_0|^{-1}\scD_1^4(\etb)+C\delta\Bb(u)\nonumber\\
&\s&+C\delta\int_{u_0}^u|u^\prime|^{1/2}z(u^\prime)du^\prime
\label{4.142}
\end{eqnarray}
Substituting in \ref{4.130} we then obtain, for all $(\ub,u)\in
D_1^{s^*}$:
\begin{eqnarray*}
\|\snab\eta\|_{L^4(S_{\ub,u})}&\leq&C|u|^{-3/2}\tilde{a}(u)(\delta^{3/2}|u_0|^{-1}\scD_1^4(\etb)+\delta\Bb(u))\\
&\s&+C\int_0^{\ub}\|b\|_{L^4(S_{\ub^\prime,u})}d\ub^\prime
+C|u|^{-3/2}\tilde{a}(u)\delta\int_{u_0}^u|u^\prime|^{1/2}z(u^\prime)du^\prime
\end{eqnarray*}
Taking the supremum over $\ub\in[0,\ub_1]$ then yields the linear
integral inequality:
\begin{equation}
z(u)\leq
\lambda(u)+\nu(u)\int_{u_0}^u|u^\prime|^{1/2}z(u^\prime)du^\prime
\label{4.143}
\end{equation}
where:
\begin{equation}
\lambda(u)=C\delta\sup_{\ub\in[0,\ub_1]}\|b\|_{L^4(S_{\ub,u})}
+C\delta|u|^{-3/2}\tilde{a}(u)(\delta^{1/2}|u_0|^{-1}\scD_1^4(\etb)+\Bb(u))
\label{4.144}
\end{equation}
and:
\begin{equation}
\nu(u)=C\delta|u|^{-3/2}\tilde{a}(u) \label{4.145}
\end{equation}
Setting
\begin{equation}
Z(u)=\int_{u_0}^u|u^\prime|^{1/2}z(u^\prime)du^\prime, \ \ \
\mbox{we have $Z(u_0)=0$} \label{4.146}
\end{equation}
and \ref{4.143} takes the form:
\begin{equation}
\frac{dZ}{du}\leq|u|^{1/2}(\lambda+\nu Z) \label{4.147}
\end{equation}
Integrating from $u_0$ we obtain:
\begin{equation}
Z(u)\leq
\int_{u_0}^u\exp\left(\int_{u^\prime}^u|u^{\prime\prime}|^{1/2}\nu(u^{\prime\prime})du^{\prime\prime}\right)
|u^\prime|^{1/2}\lambda(u^\prime)du^\prime \ \ \ : \ \forall
u\in[u_0,s^*] \label{4.148}
\end{equation}
From \ref{4.145} and \ref{4.119} we have:
\begin{eqnarray}
\int_{u^\prime}^u|u^{\prime\prime}|^{1/2}\nu(u^{\prime\prime})du^{\prime\prime}&=&
C\delta\int_{u^\prime}^u|u^{\prime\prime}|^{-1}\tilde{a}(u^{\prime\prime})du^{\prime\prime}\nonumber\\
&\leq&C|u|^{-1}\delta^{1/2}({\cal
R}_0^\infty(\alpha)+O(\delta^{1/2}))
\nonumber\\
&\leq& 1 \label{4.149}
\end{eqnarray}
provided that $\delta$ is suitably small depending on ${\cal
D}_0^\infty$, ${\cal R}_0^\infty$. Therefore:
\begin{equation}
Z(u)\leq C\int_{u_0}^u|u^\prime|^{1/2}\lambda(u^\prime)du^\prime \
\ \ : \ \forall u\in[u_0,s^*] \label{4.150}
\end{equation}
and substituting in \ref{4.143} yields:
\begin{equation}
z(u)\leq\lambda(u)+C\nu(u)\int_{u_0}^u|u^\prime|^{1/2}\lambda(u^\prime)du^\prime
\ \ \ : \ \forall u\in[u_0,s^*] \label{4.151}
\end{equation}
Moreover, since
$$\int_{u_0}^u|u^\prime|^{1/2}\|\snab\eta\|_{L^4(S_{\ub,u^\prime})}du^\prime$$
setting, in analogy with \ref{4.140},
\begin{equation}
\zb(u)=\sup_{\ub\in[0,\ub_1]}\|\snab\etb\|_{L^4(S_{\ub,u})}
\label{4.152}
\end{equation}
substituting \ref{4.150} in \ref{4.139} and taking the supremum
over $\ub\in[0,\ub_1]$ yields:
\begin{equation}
|u|^{3/2}\zb(u)\leq C|u_0|^{-1}\delta^{1/2}\scD_1^4(\etb)+C\Bb(u)
+C\int_{u_0}^u|u^\prime|^{1/2}\lambda(u^\prime)du^\prime \ \ \ : \
\forall u\in[u_0,s^*] \label{4.153}
\end{equation}

We shall now derive an appropriate estimate for $\scD_1^4(\etb)$.
To do this we recall that along $C_{u_0}$ condition \ref{3.65}
holds:
\begin{equation}
\etb=-\eta \ \ \ : \ \mbox{along $C_{u_0}$} \label{4.154}
\end{equation}
Thus, the first of \ref{4.110} simplifies along $C_{u_0}$ to:
\begin{equation}
D\snab\eta=-a\cdot\snab\eta+b \ \ \ : \ \mbox{along $C_{u_0}$}
\label{4.155}
\end{equation}
Equivalently, the first of \ref{4.113} simplifies along $C_{u_0}$
to:
\begin{equation}
(D\snab\eta)_{AB}+\frac{1}{2}\Omega\mbox{tr}\chi(\snab\eta)_{AB}=-\Omega\chih_B^C(\snab\eta)_{AC}+b_{AB}
\label{4.156}
\end{equation}
We apply Lemma 4.6 with $\snab\eta$ in the role of $\theta$ and
$b$ in the role of $\xi$. Here $r=2$, $\nu=-1$, $\gamma=-\Omega
I\otimes\chih^\sharp$. Taking $p=4$ we obtain:
\begin{equation}
\|\snab\eta\|_{L^4(S_{\ub,u_0})}\leq
C\int_0^{\ub}\|b\|_{L^4(S_{\ub^\prime,u_0})}d\ub^\prime \label{4.158}
\end{equation}
which, in view of \ref{4.154}, implies:
\begin{equation}
\scD_1^4(\etb)\leq
C|u_0|^{5/2}\delta^{1/2}\sup_{\ub\in[0,\ub_1]}\|b\|_{L^4(S_{\ub,u_0})}
\label{4.159}
\end{equation}

We shall now derive bounds for $b$ and $\bb$ in $L^4(S_{\ub,u})$.
From the expressions for $D\sGamma$, $\Db\sGamma$ of Lemma 4.1 we
have, pointwise on $S_{\ub,u}$,
$$|D\sGamma|\leq 3|\snab(\Omega\chi)|, \ \ \ |\Db\sGamma|\leq 3|\snab(\Omega\chib)|$$
hence:
\begin{equation}
\|D\sGamma\|_{L^4(S_{\ub,u})}\leq
3\|\snab(\Omega\chi)\|_{L^4(S_{\ub,u})}, \ \ \
\|\Db\sGamma\|_{L^4(S_{\ub,u})}\leq
3\|\snab(\Omega\chib)\|_{L^4(S_{\ub,u})} \label{4.160}
\end{equation}
Writing
$$\snab(\Omega\chi)=\Omega(\snab\chi+(\sd\log\Omega)\otimes\chi)$$
from Proposition 4.1, \ref{4.57} and the bounds \ref{4.49} we
obtain:
\begin{eqnarray}
\|\snab(\Omega\chi)\|_{L^4(S_{\ub,u})}&\leq&C\delta^{-1/2}|u|^{-3/2}\scR_1^4(\alpha)+O(|u|^{-5/2})\nonumber\\
&:&\forall (\ub,u)\in D^\prime \label{4.161}
\end{eqnarray}
provided that $\delta$ is suitably small depending on ${\cal
D}_0^\infty$, ${\cal R}_0^\infty$. Also, writing
$$\snab(\Omega\chib)=\Omega(\snab\chib+(\sd\log\Omega)\otimes\chib)$$
from Lemma 4.8 and \ref{4.79} we obtain:
\begin{eqnarray}
\|\snab(\Omega\chib)\|_{L^4(S_{\ub,u})}&\leq&C\delta^{1/2}|u|^{-5/2}(\scD_1^4(\mbox{tr}\chib)+\scD_1^4(\chibh))\nonumber\\
&\s&+C\delta|u|^{-7/2}k+O(\delta|u|^{-7/2})+C\delta^{3/2}|u|^{-4}\scR_1^4(\alb)\nonumber\\
&:&\forall (\ub,u)\in D_1^{s^*} \label{4.162}
\end{eqnarray}
Let us denote from now on by $O(\delta^p|u|^r)$, for real numbers
$p$, $r$, the product of $\delta^p|u|^r$ with a non-negative
non-decreasing continuous function of the quantities ${\cal
D}_0^\infty, {\cal R}_0^\infty$, {\em and} $\scD_1^4, \scR_1^4$, where:
\begin{equation}
\scD_1^4=\max\{\scD_1^4(\mbox{tr}\chib),\scD_1^4(\chibh)\}
\label{4.163}
\end{equation}
\begin{equation}
\scR_1^4=\max\{\scR_1^4(\alpha),\scR_1^4(\beta),\scR_1^4(\rho),
\scR_1^4(\sigma),\scR_1^4(\beb),\scR_1^4(\alb)\} \label{4.164}
\end{equation}
With this notation \ref{4.162} simplifies to:
\begin{eqnarray}
\|\snab(\Omega\chib)\|_{L^4(S_{\ub,u})}&\leq&C\delta^{1/2}|u|^{-5/2}(\scD_1^4(\mbox{tr}\chib)+\scD_1^4(\chibh))\nonumber\\
&\s&+C\delta|u|^{-7/2}k+O(\delta|u|^{-7/2})\nonumber\\
&:&\forall (\ub,u)\in D_1^{s^*} \label{4.165}
\end{eqnarray}
Note that the dependence on the constant $k$ is, and must be, kept
explicit.

Using \ref{4.161}, the first of \ref{4.160}, the second of the
definitions \ref{4.1}, and \ref{4.57} together with the bounds
\ref{4.49}, we obtain from the third of \ref{4.111}:
\begin{eqnarray}
\|b\|_{L^4(S_{\ub,u})}&\leq&C\delta^{-1/2}|u|^{-5/2}\scR_1^4(\beta)+O(|u|^{-7/2})\nonumber\\
&:&\forall (\ub,u)\in D^\prime \label{4.166}
\end{eqnarray}
Also, using \ref{4.165}, the second of \ref{4.160}, the fifth of
the definitions \ref{4.1}, and \ref{4.57} together with the bounds
\ref{4.49}, we obtain from the fourth of \ref{4.111}:
\begin{eqnarray}
\|\bb\|_{L^4(S_{\ub,u})}&\leq&C\delta|u|^{-9/2}\left\{\scR_1^4(\beb)
+({\cal R}_0^\infty(\beta)+O(\delta^{1/2}))\scD_1^4(\mbox{tr}\chib)+\scD_1^4(\chibh))\right\}\nonumber\\
&\s&+C\delta^{3/2}|u|^{-11/2}({\cal R}_0^\infty(\beta)+O(\delta^{1/2}))k+O(\delta^{3/2}|u|^{-11/2})\nonumber\\
&:&\forall (\ub,u)\in D_1^{s^*}\label{4.167}
\end{eqnarray}
Hence (see \ref{4.141}):
\begin{eqnarray}
\Bb(u)&\leq&C\delta|u|^{-2}\left\{\scR_1^4(\beb)
+({\cal R}_0^\infty(\beta)+O(\delta^{1/2}))(\scD_1^4(\mbox{tr}\chib)+\scD_1^4(\chibh))\right\}\nonumber\\
&\s&+C\delta^{3/2}|u|^{-3}({\cal R}_0^\infty(\beta)+O(\delta^{1/2}))k+O(\delta^{3/2}|u|^{-3})\nonumber\\
&:&\forall u\in[u_0,s^*]\label{4.168}
\end{eqnarray}
Substituting \ref{4.159}, \ref{4.166}, and \ref{4.168}, in
\ref{4.144} and recalling \ref{4.119} we obtain:
\begin{eqnarray}
\lambda(u)&\leq&C\delta^{1/2}|u|^{-5/2}\scR_1^4(\beta)+O(\delta|u|^{-7/2})\nonumber\\
&\s&+C\delta^2|u|^{-11/2}({\cal
R}_0^\infty(\alpha)+O(\delta^{1/2}))({\cal
R}_0^\infty(\beta)+O(\delta^{1/2}))k
\nonumber\\
&:&\forall u\in[u_0,s^*] \label{4.170}
\end{eqnarray}
Substituting \ref{4.170} in \ref{4.151} and noting that from
\ref{4.145} and \ref{4.119}:
\begin{equation}
|u|^{3/2}\nu(u)=C\delta\tilde{a}(u)\leq C\delta^{1/2}|u|^{-1}
({\cal R}_0^\infty(\alpha)+O(\delta^{1/2}))\leq 1 \label{4.171}
\end{equation}
provided that $\delta$ is suitably small depending on ${\cal
D}_0^\infty$, ${\cal R}_0^\infty$, yields:
\begin{eqnarray}
z(u)&\leq&C\delta^{1/2}|u|^{-5/2}\scR_1^4(\beta)+O(\delta|u|^{-7/2})\nonumber\\
&\s&+C\delta^2|u|^{-11/2}({\cal
R}_0^\infty(\alpha)+O(\delta^{1/2}))({\cal
R}(\beta)+O(\delta^{1/2}))k
\nonumber\\
&:&\forall u\in[u_0,s^*] \label{4.172}
\end{eqnarray}
Also, substituting \ref{4.159}, \ref{4.166}, and \ref{4.168},
\ref{4.170}, in \ref{4.153}, yields:
\begin{eqnarray}
\zb(u)&\leq&C\delta^{1/2}|u|^{-5/2}\scR_1^4(\beta)+O(\delta|u|^{-7/2})\nonumber\\
&\s&+C\delta^{3/2}|u|^{-9/2}({\cal R}_0^\infty(\beta)+O(\delta^{1/2}))k\nonumber\\
&:&\forall u\in[u_0,s^*] \label{4.173}
\end{eqnarray}
In conclusion, recalling the definitions \ref{4.140}, \ref{4.152}
of $z(u)$, $\zb(u)$, we have established the following lemma.

\vspace{5mm}

\noindent {\bf Lemma 4.9} \ \ \ We have:
\begin{eqnarray*}
\|\snab\eta\|_{L^4(S_{\ub,u})}&\leq&C\delta^{1/2}|u|^{-5/2}\scR_1^4(\beta)+O(\delta|u|^{-7/2})\\
&\s&+C\delta^2|u|^{-11/2}({\cal
R}_0^\infty(\alpha)+O(\delta^{1/2}))({\cal
R}_0^\infty(\beta)+O(\delta^{1/2}))k
\end{eqnarray*}
\begin{eqnarray*}
\|\snab\etb\|_{L^4(S_{\ub,u})}&\leq&C\delta^{1/2}|u|^{-5/2}\scR_1^4(\beta)+O(\delta|u|^{-7/2})\nonumber\\
&\s&+C\delta^{3/2}|u|^{-9/2}({\cal
R}_0^\infty(\beta)+O(\delta^{1/2}))k
\end{eqnarray*}
for all $(\ub,u)\in D_1^{s^*}$, provided that $\delta$ is suitably
small depending on ${\cal D}_0^\infty$, ${\cal R}_0^\infty$.

\section{$L^4(S)$ estimates for $\sd\omega$, $\sd\omb$}

We proceed to derive an $L^4(S)$ estimate for $\sd\omb$ on
$M_1^{s^*}$. Using this estimate we shall derive an $L^4(S)$
estimate for $\sd\log\Omega$ in $M_1^{s^*}$ which, with a suitable
choice of the constant $k$, improves the bound \ref{4.79}. This
shall enable us to show that $s^*=u_1$ so that the previous lemmas
actually hold on the entire parameter domain $D_1$, and, since
$(\ub_1,u_1)\in D^\prime$ is arbitrary, the estimates hold on all
of $D^\prime$, that is, on all of $M^\prime$. We shall then derive
an $L^4(S)$ estimate for $\sd\omega$ on $M^\prime$.

\vspace{5mm}

\noindent {\bf Lemma 4.10} \ \ \ We have:
\begin{eqnarray*}
\|\sd\omb\|_{L^4(S_{\ub,u})}&\leq&C\delta|u|^{-7/2}\scR_1^4(\rho)+O(\delta^{3/2}|u|^{-9/2})\nonumber\\
&\s&+C\delta^3|u|^{-13/2}({\cal
R}_0^\infty(\beta)+O(\delta^{1/2}))^2 k
\end{eqnarray*}
for all $(\ub,u)\in D_1^{s^*}$, provided that $\delta$ is suitably
small depending on ${\cal D}_0^\infty$, ${\cal R}_0^\infty$.

\noindent{\em Proof:} \ We apply $\sd$ to equation \ref{1.86} to
obtain, by virtue of Lemma 1.2, the following propagation equation
for $\sd\omb$ along the generators of each $C_u$:
\begin{equation}
D\sd\omb=\Omega^2 l \label{4.174}
\end{equation}
where:
\begin{eqnarray}
l&=&-2(\eta-\etb)\cdot\snab\eta+2\eta\cdot\snab\etb-\sd\rho\nonumber\\
&\s&+(\eta+\etb)(-|\eta|^2+2(\eta,\etb)-\rho) \label{4.175}
\end{eqnarray}
Here we have also used \ref{4.57}. To \ref{4.174} we apply Lemma
4.6 with $\sd\omb$ in the role of $\theta$ and $\Omega^2 l$ in the
role of $\xi$. Here $r=1$, $\nu=0$, $\gamma=0$. Taking $p=4$ we
obtain:
\begin{equation}
\|\sd\omb\|_{L^4(S_{\ub,u})}\leq
C\int_0^{\ub}\|l\|_{L^4(S_{\ub^\prime,u})}d\ub^\prime \ \ \ : \
\forall (\ub,u)\in D^\prime \label{4.177}
\end{equation}
Using the $L^4(S)$ estimates for $\snab\eta$, $\snab\etb$ of Lemma
4.9 as well as the third of the definitions \ref{4.1} we obtain,
for all $(\ub,u)\in D_1^{s^*}$,
\begin{eqnarray}
\|l\|_{L^4(S_{\ub,u})}&\leq&|u|^{-7/2}\scR_1^4(\rho)+O(\delta^{1/2}|u|^{-9/2})\nonumber\\
&\s&+C\delta^2|u|^{-13/2}({\cal
R}_0^\infty(\beta)+O(\delta^{1/2}))^2 k \label{4.178}
\end{eqnarray}
provided that $\delta$ is suitably small depending on ${\cal
D}_0^\infty$, ${\cal R}_0^\infty$. Substituting this in
\ref{4.177} yields the lemma.

\vspace{5mm}

\noindent {\bf Lemma 4.11} \ \ \ There is a numerical constant $C$
such that with
$$k=C\scR_1^4(\rho)+O(\delta^{1/2})$$
we have $s^*=u_1$ and the estimate:
$$\|\sd\log\Omega\|_{L^4(S_{\ub,u})}\leq C\delta|u|^{-5/2}\scR_1^4(\rho)+O(\delta^{3/2}|u|^{-7/2})$$
holds for all $(\ub,u)\in D_1$, provided that $\delta$ is suitably
small depending on ${\cal D}_0^\infty$, ${\cal R}_0^\infty$.

\noindent {\em Proof:} \ By Lemma 1.2 and the second of the
definitions \ref{1.17}:
\begin{equation}
\Db\sd\log\Omega=\sd\omb \label{4.179}
\end{equation}
To this we apply Lemma 4.7 with $\sd\log\Omega$ in the role of
$\thetab$ and $\sd\omb$ in the role of $\xib$. Here $r=1$,
$\nu=0$, $\gammab=0$. Taking $p=4$ and  taking into account the
fact that $\sd\log\Omega$ vanishes on $C_{u_0}$ (see \ref{3.34})
we obtain:
\begin{equation}
|u|^{1/2}\|\sd\log\Omega\|_{L^4(S_{\ub,u})}\leq
C\int_{u_0}^u|u^\prime|^{1/2}\|\sd\omb\|_{L^4(S_{\ub,u})}du^\prime
\label{4.181}
\end{equation}
Substituting the $L^4(S)$ estimate for $\sd\omb$ of Lemma 4.10 we
obtain:
\begin{eqnarray}
\|\sd\log\Omega\|_{L^4(S_{\ub,u})}&\leq&C\delta|u|^{-5/2}\scR_1^4(\rho)+O(\delta^{3/2}|u|^{-7/2})\nonumber\\
&\s&+C\delta^3|u|^{-11/2}({\cal R}_0^\infty(\beta)+O(\delta^{1/2}))^2 k\nonumber\\
&:&\forall (\ub,u)\in D_1^{s^*} \label{4.182}
\end{eqnarray}
In reference to this estimate, let:
\begin{eqnarray}
a&=&C\scR_1^4(\rho)+O(\delta^{1/2})\nonumber\\
b&=&C({\cal R}_0^\infty(\beta)+O(\delta^{1/2}))^2 \label{4.183}
\end{eqnarray}
The estimate \ref{4.182} implies:
\begin{equation}
\|\sd\log\Omega\|_{L^4(S_{\ub,u})}\leq \delta(a+\delta^2
bk)|u|^{-5/2} \ \ \ : \ \forall(\ub,u)\in D_1^{s^*} \label{4.184}
\end{equation}
Choosing:
\begin{equation}
k=2a \label{4.185}
\end{equation}
we have:
\begin{equation}
a+\delta^2 bk<2a \ \ \ \mbox{provided that $2b\delta^2<1$}
\label{4.186}
\end{equation}
The last is a smallness condition on $\delta$ depending on ${\cal
D}_0^\infty$, ${\cal R}_0^\infty$. The estimate \ref{4.184} then
implies:
\begin{equation}
\|\sd\log\Omega\|_{L^4(S_{\ub,u})}< k\delta|u|^{-5/2} \ \ \ : \
\forall (\ub,u)\in D_1^{s^*}=[0,\ub_1]\times[0,s^*] \label{4.187}
\end{equation}
hence by continuity \ref{4.78} holds for some $s>s^*$
contradicting the definition of $s^*$, unless $s^*=u_1$. This
completes the proof of the lemma.

\vspace{5mm}

Since by the above lemma $D_1^{s^*}=D_1$, with $k$ as in the
statement of Lemma 4.11, the results of Lemmas 4.8, 4.9 and 4.10
hold for all $(\ub,u)\in D_1$. Since $(\ub_1,u_1)$ is arbitrary,
these results hold for all $(\ub,u)\in D^\prime$. Moreover, by
virtue of the $L^4(S)$ estimate for $\sd\log\Omega$ of Lemma 4.11,
we can now estimate (see third of \ref{4.85}):
\begin{equation}
\|\rb\|_{L^4(S_{\ub,u})}\leq
C\delta|u|^{-9/2}\scR_1^4(\rho)+O(\delta^{3/2}|u|^{-11/2})
\label{4.188}
\end{equation}
Using this bound in place of the bound \ref{4.88}, we deduce,
following the argument leading to the estimate \ref{4.108},
\begin{eqnarray}
\|\sd\mbox{tr}\chib^\prime\|_{L^4(S_{\ub,u})}&\leq&C\delta|u|^{-7/2}\left\{\scR_1^4(\rho)
+(1+{\cal D}_0^\infty(\chibh)+O(\delta))(\scD_1^4(\mbox{tr}\chib)+\scD_1^4(\chibh))\right\}\nonumber\\
&\s&+O(\delta^{3/2}|u|^{-9/2}) \ \ \ : \ \forall (\ub,u)\in
D^\prime \label{4.189}
\end{eqnarray}
Also, applying Lemma 4.7, taking $p=4$, with $\snab\chibh^\prime$ in the role of $\thetab$, $\ib$ in the role of $\gammab$  
and $\hb\cdot\sd\chib^\prime+\sb$ in the role of $\xib$, to the second of \ref{4.84}, in which case $r=3$, $\nu=0$, 
and using the estimates \ref{4.89} and \ref{4.189}, we deduce:
\begin{equation}
\|\snab\chibh^\prime\|_{L^4(S_{\ub,u})}\leq
C\delta^{1/2}|u|^{-5/2}\scD_1^4(\chibh)+O(\delta|u|^{-7/2})
\label{4.86}
\end{equation}
We thus arrive at the following proposition.

\vspace{5mm}

\noindent {\bf Proposition 4.2} \ \ \ The following estimates hold
for all $(\ub,u)\in D^\prime$:
\begin{eqnarray*}
\|\sd\mbox{tr}\chib\|_{L^4(S_{\ub,u})}&\leq&C\delta|u|^{-7/2}\left\{\scR_1^4(\rho)
+(1+{\cal D}_0^\infty(\chibh)+O(\delta))(\scD_1^4(\mbox{tr}\chib)+\scD_1^4(\chibh))\right\}\\
&\s&+O(\delta^{3/2}|u|^{-9/2})\\
\|\snab\chibh\|_{L^4(S_{\ub,u})}&\leq&C\delta^{1/2}|u|^{-5/2}\scD_1^4(\chibh)
+O(\delta|u|^{-7/2})\\
\|\snab\eta\|_{L^4(S_{\ub,u})}&\leq&C\delta^{1/2}|u|^{-5/2}\scR_1^4(\beta)+O(\delta|u|^{-7/2})\\
\|\snab\etb\|_{L^4(S_{\ub,u})}&\leq&C\delta^{1/2}|u|^{-5/2}\scR_1^4(\beta)+O(\delta|u|^{-7/2})\\
\|\sd\omb\|_{L^4(S_{\ub,u})}&\leq&C\delta|u|^{-7/2}\scR_1^4(\rho)+O(\delta^{3/2}|u|^{-9/2})
\end{eqnarray*}
provided that $\delta$ is suitably small depending on ${\cal
D}_0^\infty$, ${\cal R}_0^\infty$.

\vspace{5mm}

\noindent {\bf Proposition 4.3} \ \ \ We have:
$$\|\sd\omega\|_{L^4(S_{\ub,u})}\leq C|u|^{-5/2}\scR_1^4(\rho)+O(\delta^{1/2}|u|^{-7/2})
\ \ \ : \ \forall (\ub,u)\in D^\prime$$ provided that $\delta$ is
suitably small depending on ${\cal D}_0^\infty$, ${\cal
R}_0^\infty$.

\noindent {\em Proof:} \ We apply $\sd$ to equation \ref{1.87} to
obtain, by virtue of Lemma 1.2, the following propagation equation
for $\sd\omega$ along the generators of each $\Cb_{\ub}$:
\begin{equation}
\Db\sd\omega=\Omega^2\lb \label{4.190}
\end{equation}
where:
\begin{eqnarray}
\lb&=&2(\eta-\etb)\cdot\snab\etb+2\etb\cdot\snab\eta-\sd\rho\nonumber\\
&\s&+(\eta+\etb)(-|\etb|^2+2(\eta,\etb)-\rho) \label{4.191}
\end{eqnarray}
(Equation \ref{4.190} is the conjugate of equation \ref{4.174}.)
To \ref{4.190} we apply Lemma 4.7 with $\sd\omega$ in the role of
$\thetab$ and $\Omega^2\lb$ in the role of $\xib$. Here $r=1$,
$\nu=0$, $\gammab=0$. Recalling that $\omega$ vanishes on
$C_{u_0}$, we obtain, taking $p=4$,
\begin{equation}
|u|^{1/2}\|\sd\omega\|_{L^4(S_{\ub,u})}\leq
C\int_{u_0}^u|u^\prime|^{1/2}\|\lb\|_{L^4(S_{\ub,u^\prime})}du^\prime
\label{4.192}
\end{equation}
Substituting the $L^4(S)$ estimates for $\snab\eta, \snab\etb$ of
Proposition 4.2 as well as the third of the definitions \ref{4.1}
we obtain, for all $(\ub,u)\in D^\prime$:
\begin{equation}
\|\lb\|_{L^4(S_{\ub,u})}\leq
|u|^{-7/2}\scR_1^4(\rho)+O(\delta^{1/2}|u|^{-9/2}) \label{4.193}
\end{equation}
Substituting this in \ref{4.192} yields the proposition.

\section{$L^4(S)$ estimates for $D\omega, \Db\omb$}

\noindent {\bf Proposition 4.4} \ \ \ The following estimates hold
for all $(\ub,u)\in D^\prime$:
\begin{eqnarray*}
\|D\omega\|_{L^4(S_{\ub,u})}&\leq&C\delta^{-1}|u|^{-3/2}{\cal R}_0^4(D\rho)+O(|u|^{-5/2})\\
\|\Db\omb\|_{L^4(S_{\ub,u})}&\leq&C\delta|u|^{-7/2}{\cal
R}_0^\infty(\rho)+O(\delta^2|u|^{-9/2})
\end{eqnarray*}
provided that $\delta$ is suitably small depending on ${\cal
D}_0^\infty$, ${\cal R}_0^\infty$.

\noindent {\em Proof:} \ By the commutation formula \ref{1.75} we
have:
\begin{equation}
\Db D\omega=D\Db\omega+2\Omega^2(\eta-\etb)^\sharp\cdot\sd\omega
\label{4.194}
\end{equation}
and $\Db\omega$ is given by equation \ref{1.86}. Applying $D$ to
equation \ref{1.86} and using the first of equations \ref{1.28} we
obtain:
\begin{eqnarray}
D\Db\omega&=&\Omega^2\{2((\eta-\etb),D\etb)+2(\etb,D\eta)\nonumber\\
&\s&-4\Omega(\eta,\chi^\sharp\cdot\etb)+2\Omega(\etb,\chi^\sharp\cdot\etb)\nonumber\\
&\s&+2\omega(2(\eta,\etb)-|\etb|^2-\rho)-D\rho\} \label{4.195}
\end{eqnarray}
Substituting for $D\eta$, $D\etb$ from \ref{1.66}, \ref{1.150}, we
obtain the following propagation equation for $D\omega$:
\begin{equation}
\Db(D\omega)=\Omega^2\nb \label{4.196}
\end{equation}
where:
\begin{eqnarray}
\nb&=&-D\rho+6(\eta-\etb)^\sharp\cdot\sd\omega+2\Omega(\eta-2\etb,\beta)\label{4.197}\\
&\s&-6\Omega(\eta-\etb,\chi^\sharp\cdot\etb)+2\omega(2(\eta,\etb)-|\etb|^2-\rho)\nonumber
\end{eqnarray}
To \ref{4.196} we apply Lemma 4.7 with $D\omega$ in the role of
$\thetab$ and $\Omega^2\nb$ in the role of $\xib$. Here $r=0$,
$\nu=0$, $\gammab=0$. Taking into account the fact that $D\omega$
vanishes on $C_{u_0}$ and setting $p=4$ we obtain:
\begin{equation}
|u|^{-1/2}\|D\omega\|_{L^4(S_{\ub,u})}\leq
C\int_{u_0}^u|u^\prime|^{-1/2}\|\nb\|_{L^4(S_{\ub,u^\prime})}du^\prime
\label{4.198}
\end{equation}
Using the $L^4(S)$ estimate for $\sd\omega$ of Proposition 4.3 as
well as the first of the definitions \ref{4.2} we obtain, for all
$(\ub,u)\in D^\prime$,
\begin{equation}
\|\nb\|_{L^4(S_{\ub,u})}\leq C\delta^{-1}|u|^{-3/2}{\cal
R}_0^4(D\rho)+O(|u|^{-5/2}) \label{4.199}
\end{equation}
Substituting this in \ref{4.198} yields the $L^4(S)$ estimate for
$D\omega$ of the proposition.

Next, we deduce the conjugate of equation \ref{4.196}, which is
the following propagation equation for $\Db\omb$:
\begin{equation}
D(\Db\omb)=\Omega^2 n \label{4.200}
\end{equation}
where:
\begin{eqnarray}
n&=&-\Db\rho-6(\eta-\etb)^\sharp\cdot\sd\omb-2\Omega(\etb-2\eta,\beb)\label{4.201}\\
&\s&+6\Omega(\eta-\etb,\chib^\sharp\cdot\eta)+2\omb(2(\eta,\etb)-|\eta|^2-\rho)\nonumber
\end{eqnarray}
To \ref{4.200} we apply Lemma 4.6 with $\Db\omb$ in the role of
$\theta$ and $\Omega^2 n$ in the role of $\xi$. Here $r=0$,
$\nu=0$, $\gamma=0$. Setting $p=4$ we obtain:
\begin{equation}
\|\Db\omb\|_{L^4(S_{\ub,u})}\leq
C\int_0^{\ub}\|n\|_{L^4(S_{\ub^\prime,u})}d\ub^\prime
\label{4.202}
\end{equation}
Using the $L^4(S)$ estimate for $\sd\omb$ of Proposition 4.2 as
well as the third of the definitions \ref{4.2} we obtain, for all
$(\ub,u)\in D^\prime$,
\begin{equation}
\|n\|_{L^4(S_{\ub,u})}\leq
\|\Db\rho\|_{L^4(S_{\ub,u})}+O(\delta|u|^{-9/2}) \label{4.203}
\end{equation}
Now, according to the eighth of the Bianchi identities of
Proposition 1.2:
\begin{equation}
\Db\rho+\frac{3}{2}\Omega\mbox{tr}\chib\rho=-\Omega\left\{\sdiv\beb+(2\eta-\zeta,\beb)+\frac{1}{2}(\chih,\alb)\right\}
\label{4.204}
\end{equation}
Hence, by the results of Chapter 3 and the fifth of the
definitions \ref{4.1} we have:
\begin{equation}
\|\Db\rho\|_{L^4(S_{\ub,u})}\leq C|u|^{-7/2}{\cal
R}_0^\infty(\rho)+O(\delta|u|^{-9/2}) \label{4.205}
\end{equation}
Substituting this in \ref{4.203} and the result in \ref{4.202}
yields the $L^4(S)$ estimate for $\Db\omb$ of the proposition.

\chapter{The Uniformization Theorem}

\section{Introduction. An $L^2(S)$ estimate for $K-\overline{K}$}

The arguments of the next two chapters use elliptic theory on the
surfaces $S_{\ub,u}$. In particular, the estimates of the next
chapter use $L^4$ elliptic theory on these surfaces. To derive the
necessary estimates we rely on the conformal properties of the
elliptic systems under consideration and make essential use of the
uniformization theorem to reduce the problem to one on the
standard sphere. The uniformization theorem states that the
2-dimensional Riemannian manifold $(S_{\ub,u},\sg)$ is conformal
to the standard sphere. Now, the uniformization theorem has been
established in [C-K] under the assumption that the Gauss curvature
$K$ of $\sg$ belongs to $L^\infty(S_{\ub,u})$ and the bounds for
the conformal factor derived there depend on
$\|K-\overline{K}\|_{L^\infty(S_{\ub,u})}$. Here, for an arbitrary
function $f$ defined on $M^\prime$ we denote by $\overline{f}$ the
function which on each $S_{\ub,u}$ is the mean value of $f$ on
$S_{\ub,u}$. By the Gauss Bonnet theorem:
\begin{equation}
\int_{S_{\ub,u}}K d\mu_{\sg}=4\pi \label{5.1}
\end{equation}
hence:
\begin{equation}
\overline{K}=\frac{4\pi}{\mbox{Area}(S_{\ub,u})} \label{5.2}
\end{equation}
We may derive an $L^\infty$ bound for $K$ on $S_{\ub,u}$ through
the Gauss equation \ref{1.108}, which we may write in the form:
\begin{equation}
K+\frac{1}{4}\mbox{tr}\chi\mbox{tr}\chib-\frac{1}{2}(\chih,\chibh)=-\rho
\label{5.3}
\end{equation}
In fact, from the 2nd, 3rd, 4th and 5th of the bounds \ref{4.49}
and the 3rd of the definitions \ref{3.2} we have:
\begin{equation}
\left|K-\frac{1}{|u|^2|}\right|\leq C|u|^{-3} \{{\cal
R}_0^\infty(\alpha)({\cal R}_0^\infty(\alpha)+{\cal
D}_0^\infty(\chih))+{\cal R}_0^\infty(\rho)\}+O(\delta|u|^{-3})
\label{5.5}
\end{equation}
This estimate is however not suitable for our present purposes,
because in the present situation the quantities ${\cal
D}_0^\infty$ and ${\cal R}_0^\infty$ are allowed to be as large as
we wish. The appropriate bound for $K$ is the $L^2$ bound to be
derived in the present section. In the next section an appropriate
version of the uniformization theorem will be deduced, which
relies only on such an $L^2$ bound for $K$.

We begin with the following proposition. We define:
\begin{equation}
\scR_2(\alpha)=\sup_{u\in [u_0,c^*)}\left(|u|^2\delta\|\snab^{ \
2}\alpha\|_{L^2(C_u)}\right) \label{5.6}
\end{equation}
where the $L^2$ norm on $C_u$ is with respect to the measure:
$$d\mu_{\sg} d\ub$$
Thus, for any $S$ tensorfield $\theta$ we have:
\begin{equation}
\|\theta\|^2_{L^2(C_u)}=\int_{C_u}|\theta|^2=\left\{\begin{array}{lll}
\int_0^\delta\left(\int_{S_{\ub,u}}|\theta|^2d\mu_{\sg}\right)d\ub&:&u\in[u_0,c^*-\delta]\\
\int_0^{c^*-u}\left(\int_{S_{\ub,u}}|\theta|^2d\mu_{\sg}\right)d\ub&:&u\in(c^*-\delta,c^*)
\end{array}\right.
\label{5.7}
\end{equation}
for $c^*\geq u_0+\delta$, and:
\begin{equation}
\|\theta\|^2_{L^2(C_u)}=\int_{C_u}|\theta|^2=\int_0^{c^*-u}\left(\int_{S_{\ub,u}}|\theta^2|d\mu_{\sg}\right) \ : \ 
u\in[u_0,c^*)
\label{5.7a}
\end{equation}
for $c^*<u_0+\delta$ 
(see \ref{3.01}, \ref{3.01a} and Figures 1.1, 1.2). By the results of Chapter 2, the quantity corresponding to $\scR_2(\alpha)$ on  
$C_{u_0}$, obtained by replacing the supremum on $[u_0,c^*)$ by
the value at $u_0$,
is bounded by a non-negative non-decreasing 
continuous function of $M_4$.

\vspace{5mm}

\noindent{\bf Proposition 5.1} \ \ \ We have:
$$\sqrt{|\snab^{ \ 2}\mbox{tr}\chi^\prime\|^2_{L^2(S_{\ub,u})}+\|\snab^{ \ 2}\chih^\prime\|^2_{L^2(S_{\ub,u})}}
\leq \delta^{-1/2}|u|^{-2}\scR_2(\alpha)+O(|u|^{-3})$$ for all
$(\ub,u)\in D^\prime$, provided that $\delta$ is suitably small
depending on ${\cal D}_0^\infty$, ${\cal R}_0^\infty$.

\noindent {\em Proof:} \ We apply $\snab$ to the system
\ref{4.53}. In view of Lemma 4.1 we obtain the following linear
system of propagation equations for $\snab^{ \
2}\mbox{tr}\chi^\prime$, $\snab^{ \ 2}\chih^\prime$ along the
generators of the $C_u$:
\begin{eqnarray}
D\snab^{ \ 2}\mbox{tr}\chi^\prime&=&f\cdot\snab^{ \ 2}\mbox{tr}\chi^\prime+g\cdot\snab^{ \ 2}\chih^\prime+r^\prime\nonumber\\
D\snab^{ \ 2}\chih^\prime&=&h\cdot\snab^{ \
2}\mbox{tr}\chi^\prime+i\cdot\snab^{ \ 2}\chih^\prime+s^\prime
\label{5.8}
\end{eqnarray}
where:
\begin{eqnarray}
r^\prime&=&\snab r +
(\snab(\Omega\chi),\sd\mbox{tr}\chi^\prime)+\snab
f\cdot\sd\mbox{tr}\chi^\prime
+\snab g\cdot\snab\chih^\prime\nonumber\\
s^\prime&=&\snab s +(\snab(\Omega\chi),\snab\chih^\prime)+\snab
h\cdot\sd\mbox{tr}\chi^\prime +\snab i\cdot\snab\chih^\prime
\label{5.9}
\end{eqnarray}
Here we denote by $(\s, \s)$ a bilinear expression with
coefficients which depend only on $\sg$.

By Lemma 4.2 with the (symmetric) 2-covariant $S$ tensorfield
$\snab^{ \ 2}\mbox{tr}\chi^\prime$ in the role of $\theta$,
$$D(|\snab^{ \ 2}\mbox{tr}\chi^\prime|^2)+2\Omega\mbox{tr}\chi|\snab^{ \ 2}\mbox{tr}\chi^\prime|^2
=2(\snab^{ \ 2}\mbox{tr}\chi^\prime,D\snab^{ \
2}\mbox{tr}\chi^\prime) +2(\Omega\chih,\snab^{ \
2}\mbox{tr}\chi^\prime,\snab^{ \ 2}\mbox{tr}\chi^\prime)$$ Here we
denote by $(\s,\s,\s)$ a trilinear expression with coefficients
which depend only on $\sg$. From the first of \ref{5.8} and the
first of \ref{4.54}:
\begin{eqnarray*}
(\snab^{ \ 2}\mbox{tr}\chi^\prime,D\snab^{ \
2}\mbox{tr}\chi^\prime)&=&-\Omega\mbox{tr}\chi|\snab^{ \
2}\mbox{tr}\chi^\prime|^2
+(\snab^{ \ 2}\mbox{tr}\chi^\prime,r^\prime+g\cdot\snab^{ \ 2}\chih^\prime)\\
&\leq&-\Omega\mbox{tr}\chi|\snab^{ \
2}\mbox{tr}\chi^\prime|^2+|\snab^{ \ 2}\mbox{tr}\chi^\prime|
(|r^\prime|+|g||\snab^{ \ 2}\chih^\prime|)
\end{eqnarray*}
Since also $|(\Omega\chih,\snab^{ \ 2}\mbox{tr}\chi^\prime,\snab^{
\ 2}\mbox{tr}\chi^\prime)| \leq C\Omega|\chih||\snab^{ \
2}\mbox{tr}\chi^\prime|^2$, we obtain:
\begin{equation}
D(|\snab^{ \
2}\mbox{tr}\chi^\prime|^2)+4\Omega\mbox{tr}\chi|\snab^{ \
2}\mbox{tr}\chi^\prime|^2 \leq 2|\snab^{ \
2}\mbox{tr}\chi^\prime|(|r^\prime|+|g||\snab^{ \
2}\chih^\prime|)+2C\Omega|\chih||\snab^{ \
2}\mbox{tr}\chi^\prime|^2 \label{5.10}
\end{equation}

By Lemma 4.2 with the 4-covariant $S$ tensorfield $\snab^{ \
2}\chih^\prime$ in the role of $\theta$,
$$D(|\snab^{ \ 2}\chih^\prime|^2)+4\Omega\mbox{tr}\chi|\snab^{ \ 2}\chih^\prime|^2=2(\snab^{ \ 2}\chih^\prime,D\snab^{ \ 2}\chih^\prime)
+2(\Omega\chih,\snab^{ \ 2}\chih^\prime,\snab^{ \
2}\chih^\prime)$$ From the second of \ref{5.8}:
\begin{eqnarray*}
(\snab^{ \ 2}\chih^\prime,D\snab^{ \ 2}\chih^\prime)&=&(\snab^{ \
2}\chih^\prime, s^\prime+h\cdot\snab^{ \ 2}\mbox{tr}\chi^\prime
+i\cdot\snab^{ \ 2}\chih^\prime)\\
&\leq&|\snab^{ \ 2}\chih^\prime|(|s^\prime|+|h||\snab^{ \
2}\mbox{tr}\chi^\prime|+|i||\snab^{ \ 2}\chih^\prime|)
\end{eqnarray*}
Since also $|(\Omega\chih,\snab^{ \ 2}\chih^\prime,\snab^{ \
2}\chih^\prime)|\leq C\Omega|\chih||\snab^{ \ 2}\chih^\prime|^2$,
we obtain:
\begin{eqnarray}
D(|\snab^{ \ 2}\chih^\prime|^2)+4\Omega\mbox{tr}\chi|\snab^{ \
2}\chih^\prime|^2
&\leq& 2|\snab^{ \ 2}\chih^\prime|(|s^\prime|+|h||\snab^{ \ 2}\mbox{tr}\chi^\prime|+|i||\snab^{ \ 2}\chih^\prime|)\nonumber\\
&\s&+2C\Omega|\chih||\snab^{ \ 2}\chih^\prime|^2 \label{5.11}
\end{eqnarray}

Let us define the non-negative function:
\begin{equation}
\psi^\prime=\sqrt{|\snab^{ \ 2}\mbox{tr}\chi^\prime|^2+|\snab^{ \
2}\chih^\prime|^2} \label{5.12}
\end{equation}
By \ref{5.10}, \ref{5.11}, and the pointwise bounds \ref{4.56},
$\psi^{\prime 2}$ satisfies the inequality:
\begin{equation}
D(\psi^{\prime 2})+4\Omega\mbox{tr}\chi\psi^{\prime 2}\leq
2C\Omega|\chih|\psi^{\prime 2}
+2\psi^\prime(|r^\prime|+|s^\prime|) \label{5.13}
\end{equation}
We then apply Lemma 4.4 with $p=2$, $\lambda=4$, $l=C$, and
$\rho=|r^\prime|+|s^\prime|$, to obtain:
\begin{equation}
\|\psi^\prime\|_{L^2(S_{\ub,u})}\leq
C\int_0^{\ub}\left\{\|r^\prime\|_{L^2(S_{\ub^\prime,u})}
+\|s^\prime\|_{L^2(S_{\ub^\prime,u})}\right\}d\ub^\prime
\label{5.14}
\end{equation}

In reference to the first of \ref{5.9} we have, for all
$(\ub,u)\in D^\prime$:
\begin{equation}
\|(\snab(\Omega\chi),\sd\mbox{tr}\chi^\prime)\|_{L^2(S_{\ub,u})}\leq
C\|\snab(\Omega\chi)\|_{L^4(S_{\ub,u})}\|\sd\mbox{tr}\chi^\prime\|_{L^4(S_{\ub,u})}\leq
O(\delta^{-1/2}|u|^{-4}) \label{5.15}
\end{equation}
by Proposition 4.1, and:
\begin{eqnarray}
&\|\snab f\cdot\sd\mbox{tr}\chi^\prime\|_{L^2(S_{\ub,u})}\leq
C\|\snab(\Omega\mbox{tr}\chi)\|_{L^4(S_{\ub,u})}\|\sd\mbox{tr}\chi^\prime\|_{L^4(S_{\ub,u})}\leq O(|u|^{-5})\nonumber\\
&\|\snab g\cdot\snab\chih^\prime\|_{L^2(S_{\ub,u})}\leq
C\|\snab(\Omega\chih)\|_{L^4(S_{\ub,u})}\|\snab\chih^\prime\|_{L^4(S_{\ub,u})}\leq
O(\delta^{-1}|u|^{-3}) \label{5.16}
\end{eqnarray}
by the first two of \ref{4.54} and Proposition 4.1. Also, in view
of the fact that from \ref{4.57} we can express:
\begin{equation}
\snab^{ \ 2}\log\Omega=\frac{1}{2}(\snab\eta+\snab\etb)
\label{5.17}
\end{equation}
we obtain, by Proposition 4.1 and the 3rd and 4th of the estimates
of Proposition 4.2:
\begin{equation}
\|\snab r\|_{L^2(S_{\ub,u})}\leq O(\delta^{-1/2}|u|^{-4})
\label{5.18}
\end{equation}
Consequently:
\begin{equation}
\|r^\prime\|_{L^2(S_{\ub,u})}\leq O(\delta^{-1}|u|^{-3})
\label{5.19}
\end{equation}

In reference to the second of \ref{5.9} we have, for all
$(\ub,u)\in D^\prime$:
\begin{equation}
\|(\snab(\Omega\chi),\snab\chih^\prime)\|_{L^2(S_{\ub,u})}\leq
C\|\snab(\Omega\chi)\|_{L^4(S_{\ub,u})}\|\snab\chih^\prime\|_{L^4(S_{\ub,u})}\leq
O(\delta^{-1}|u|^{-3}) \label{5.20}
\end{equation}
by Proposition 4.1, and:
\begin{eqnarray}
&\|\snab h\cdot\sd\mbox{tr}\chi^\prime\|_{L^2(S_{\ub,u})}\leq
C\|\snab(\Omega\chih)\|_{L^4(S_{\ub,u})}\|\sd\mbox{tr}\chi^\prime\|_{L^4(S_{\ub,u})}\leq
O(\delta^{-1/2}|u|^{-4})
\nonumber\\
&\|\snab i\cdot\snab\chih^\prime\|_{L^2(S_{\ub,u})}\leq
C\|\snab(\Omega\chih)\|_{L^4(S_{\ub,u})}\|\snab\chih^\prime\|_{L^4(S_{\ub,u})}\leq
O(\delta^{-1}|u|^{-3}) \label{5.21}
\end{eqnarray}
by the third and fourth of \ref{4.54} and Proposition 4.1. Also,
in view of \ref{5.17}, we obtain, by Proposition 4.1 and the 3rd
and 4th of the estimates of Proposition 4.2:
\begin{equation}
\|\snab s\|_{L^2(S_{\ub,u})}\leq \|\snab^{ \
2}\alpha\|_{L^2(S_{\ub,u})}+O(\delta^{-1/2}|u|^{-4}) \label{5.22}
\end{equation}
Consequently:
\begin{equation}
\|s^\prime\|_{L^2(S_{\ub,u})}\leq \|\snab^{ \
2}\alpha\|_{L^2(S_{\ub,u})}+O(\delta^{-1}|u|^{-3}) \label{5.23}
\end{equation}

Substituting the estimates \ref{5.19} and \ref{5.23} in \ref{5.14}
and noting that:
$$\int_0^{\ub}\|\snab^{ \ 2}\alpha\|_{L^2(S_{\ub^\prime,u})}d\ub^\prime
\leq \ub^{1/2}\left(\int_0^{\ub}\|\snab^{ \
2}\alpha|^2_{L^2(S_{\ub^\prime,u})}d\ub^\prime\right)^{1/2} \leq
\delta^{1/2}\|\snab^{ \ 2}\alpha\|_{L^2(C_u)}$$ we obtain, in view
of the definition \ref{5.6},
\begin{equation}
\|\psi^\prime\|_{L^2(S_{\ub,u})}\leq
\delta^{-1/2}|u|^{-2}\scR_2(\alpha)+O(|u|^{-3}) \label{5.24}
\end{equation}
for all $(\ub,u)\in D^\prime$. Since
$$\|\psi^\prime\|^2_{L^2(S_{\ub,u})}=\|\snab^{ \ 2}\mbox{tr}\chi^\prime\|^2_{L^2(S_{\ub,u})}
+\|\snab^{ \ 2}\chih^\prime\|^2_{L^2(S_{\ub,u})}$$ \ref{5.24} is
the estimate of the proposition.

\vspace{5mm}

Let now $\sR^C_{ADB}$ be the components of the intrinsic curvature
of $S_{\ub,u}$ with respect to a Jacobi field frame defined in a
neighborhood of a generator of $C_u$, as in the proof of the first
part of Lemma 4.1. Then we have:
\begin{equation}
(D\sR)^C_{ADB}=D(\sR^C_{ADB})=\snab_D D\sGamma^C_{BA}-\snab_B
D\sGamma^C_{DA} \label{5.25}
\end{equation}
hence, if $\sRic_{AB}=\sR^C_{ACB}$ are the components of the Ricci
curvature of $S_{\ub,u}$ in the same frame,
\begin{equation}
(D\sRic)_{AB}=D(\sRic_{AB})=\snab_C D\sGamma^C_{BA}-\snab_B
D\sGamma^C_{CA} \label{5.26}
\end{equation}
Since $S_{\ub,u}$ is 2-dimensional,
\begin{equation}
\sR^C_{ADB}=K(\delta^C_D\sg_{AB}-\delta^C_B\sg_{AD}), \ \ \
\sRic_{AB}=K\sg_{AB}, \ \ \ K=(\sg^{-1})^{AB}\sRic_{AB}
\label{5.27}
\end{equation}
where $K$ is the Gauss curvature of $S_{\ub,u}$. Thus, we have (by
the first of \ref{3.3}):
\begin{eqnarray*}
DK&=&\frac{1}{2}D((\sg^{-1})^{AB}\sRic_{AB})\\
&=&-\Omega\chi^{AB}\sRic_{AB}+\frac{1}{2}(\sg^{-1})^{AB}(D\sRic)_{AB}\\
&=&-\Omega\mbox{tr}\chi K+\frac{1}{2}(\sg^{-1})^{AB}(D\sRic)_{AB}
\end{eqnarray*}
Hence, from \ref{5.26},
$$DK+\Omega\mbox{tr}\chi K=\frac{1}{2}(\sg^{-1})^{AB}\snab_C D\sGamma^C_{AB}
-\frac{1}{2}\snab^A D\sGamma^B_{BA}$$ Substituting the expression
for $D\Gamma$ of the first part of Lemma 4.1 we then obtain:
$$DK+\Omega\mbox{tr}\chi K=\snab^B\snab^A(\Omega\chi_{AB})-\slap(\Omega\mbox{tr}\chi)$$
or:
\begin{eqnarray}
DK+\Omega\mbox{tr}\chi K&=&\sdiv\sdiv(\Omega\chi)-\slap(\Omega\mbox{tr}\chi)\nonumber\\
&=&\sdiv\sdiv(\Omega\chih)-\frac{1}{2}\slap(\Omega\mbox{tr}\chi)
\label{5.28}
\end{eqnarray}
This is the propagation equation for the Gauss curvature along the
generators of the $C_u$.

By \ref{5.2},
$$D\overline{K}=-\frac{4\pi}{(\mbox{Area}(S_{\ub,u}))^2}D\mbox{Area}(S_{\ub,u})$$
and we have
\begin{equation}
D\mbox{Area}(S_{\ub,u})=\int_{S_{\ub,u}}\Omega\mbox{tr}\chi
d\mu_{\sg}=\overline{\Omega\mbox{tr}\chi}\mbox{Area}(S_{\ub,u})
\label{5.29}
\end{equation}
hence:
\begin{equation}
D\overline{K}=-\overline{\Omega\mbox{tr}\chi}\s\overline{K}
\label{5.30}
\end{equation}

From \ref{5.28} and \ref{5.30} we obtain the following propagation
equation for $K-\overline{K}$ along the generators of the $C_u$:
\begin{eqnarray}
D(K-\overline{K})+\Omega\mbox{tr}\chi(K-\overline{K})&=&-(\Omega\mbox{tr}\chi-\overline{\Omega\mbox{tr}\chi})\overline{K}
\nonumber\\
&\s&+\sdiv\sdiv(\Omega\chih)-\frac{1}{2}\slap(\Omega\mbox{tr}\chi)
\label{5.31}
\end{eqnarray}
To this equation we apply Lemma 4.6, taking $p=2$. Here $r=0$,
$\nu=-2$, $\gamma=0$. We obtain:
\begin{eqnarray}
\|K-\overline{K}\|_{L^2(S_{\ub,u})}&\leq&
C\int_0^{\ub}\|(\Omega\mbox{tr}\chi-\overline{\Omega\mbox{tr}\chi})\overline{K}\|_{L^2(S_{\ub^\prime,u})}d\ub^\prime
\label{5.32}\\
&\s&+C\int_0^{\ub}\|\sdiv\sdiv(\Omega\chih)-\frac{1}{2}\slap(\Omega\mbox{tr}\chi)\|_{L^2(S_{\ub^\prime,u})}d\ub^\prime
\nonumber
\end{eqnarray}

From the first two of the bounds \ref{4.49}:
\begin{equation}
\left|\Omega\mbox{tr}\chi-\frac{2}{|u|}\right|\leq C|u|^{-2}({\cal
R}_0^\infty(\alpha))^2+O(\delta|u|^{-2}) \label{5.33}
\end{equation}
hence also:
$$\left|\overline{\Omega\mbox{tr}\chi}-\frac{2}{|u|}\right|
\leq C|u|^{-2}({\cal R}_0^\infty(\alpha))^2+O(\delta|u|^{-2})$$ It
follows that:
\begin{equation}
|\Omega\mbox{tr}\chi-\overline{\Omega\mbox{tr}\chi}|\leq
C|u|^{-2}({\cal R}_0^\infty(\alpha))^2+O(\delta|u|^{-2})
\label{5.34}
\end{equation}
Together with \ref{5.2} and the lower and upper bounds for
$\mbox{Area}(S_{\ub,u})$ of Lemma 4.3 this implies that:
$$\|(\Omega\mbox{tr}\chi-\overline{\Omega\mbox{tr}\chi})\overline{K}\|_{L^2(S_{\ub,u})}\leq O(|u|^{-3})$$
therefore the first integral on the right in \ref{5.32} is bounded
by $O(\delta|u|^{-3})$. Also, by Proposition 5.1 together with
expression \ref{5.17} and Proposition 4.2:
$$\|\sdiv\sdiv(\Omega\chih)-\frac{1}{2}\slap(\Omega\mbox{tr}\chi)\|_{L^2(S_{\ub,u})}
\leq C\delta^{-1/2}|u|^{-2}\scR_2(\alpha)+O(|u|^{-3})$$ therefore
the second integral on the right in \ref{5.32} is bounded by
$C\delta^{1/2}|u|^{-2}\scR_2(\alpha)+O(\delta|u|^{-3})$. We thus
arrive at the following proposition.

\vspace{5mm }

\noindent {\bf Proposition 5.2} \ \ \ We have:
$$\|K-\overline{K}\|_{L^2(S_{\ub,u})}\leq C\delta^{1/2}|u|^{-2}\scR_2(\alpha)+O(\delta|u|^{-3})$$
for all $(\ub,u)\in D^\prime$, provided that $\delta$ is suitably
small depending on ${\cal D}_0^\infty$, ${\cal R}_0^\infty$.

\vspace{5mm}

\section{Sobolev inequalities on $S$. The isoperimetric constant}

In the following we shall make use of the {\em Sobolev
inequalities} of $S_{\ub,u}$. These inequalities are all derived
from the {\em isoperimetric Sobolev inequality} (see [O]) of $S_{\ub,u}$: If
$f$ is an arbitrary function which is integrable and with
integrable derivative on $S_{\ub,u}$, then $f$ is
square-integrable on $S_{\ub,u}$ and we have:
\begin{equation}
\int_{S_{\ub,u}}(f-\overline{f})^2
d\mu_{\sg}\leq\mbox{I}(S_{\ub,u})\left(\int_{S_{\ub,u}}|\sd
f|d\mu_{\sg}\right)^2 \label{5.35}
\end{equation}
Here $\mbox{I}(S_{\ub,u})$ is the {\em isoperimetric constant} of
$S_{\ub,u}$:
\begin{equation}
\mbox{I}(S_{\ub,u})=\sup_U\frac{\min\{\mbox{Area}(U),\mbox{Area}(U^c)\}}{(\mbox{Perimeter}(\partial
U))^2} \label{5.36}
\end{equation}
where the supremum is over all domains $U$ with $C^1$ boundary
$\partial U$ in $S_{\ub,u}$, and $U^c=S_{\ub,u}\setminus U$
denotes the complement of $U$ in $S_{\ub,u}$.

\vspace{5mm}

\noindent{\bf Lemma 5.1} \ \ \ Let $(S,\sg)$ be a compact
2-dimensional Riemannian manifold and $\xi$ a tensorfield on $S$,
of arbitrary type, which is square integrable and with square
integrable first covariant derivative. Then for each $2<p<\infty$
$\xi\in L^p(S)$ and we have:
$$(\mbox{Area}(S))^{-1/p}\|\xi\|_{L^p(S)}\leq C_p\sqrt{\mbox{I}^\prime(S)}\|\xi\|_{W_1^2(S)}$$
Here $C_p$ is a numerical constant depending only on $p$,
$$\mbox{I}^\prime(S)=\max\{\mbox{I}(S),1\}$$
where $\mbox{I}(S)$ is the isoperimetric constant of $S$, and we
denote:
$$\|\xi\|_{W_1^2(S)}=\|\snab\xi\|_{L^2(S)}+(\mbox{Area}(S))^{-1/2}\|\xi\|_{L^2(S)}$$
where $\snab$ is the covariant derivative operator associated to
the metric $\sg$.

\noindent{\em Proof:} \ We first establish the lemma in the case
of functions $\phi$ on $S$. In reference to \ref{5.35}, which
holds for an arbitrary compact 2-dimensional manifold $(S,\sg)$,
since
\begin{equation}
\|\overline{f}\|_{L^2(S)}=|\overline{f}|(\mbox{Area}(S))^{1/2}, \
\ \ |\overline{f}|\leq (\mbox{Area}(S))^{-1}\|f\|_{L^1(S)}
\label{5.a1}
\end{equation}
\ref{5.35} implies:
\begin{equation}
\|f\|_{L^2(S)}\leq \sqrt{\mbox{I}^\prime(S)}\|f\|_{W_1^1(S)}
\label{5.a2}
\end{equation}
where we denote:
\begin{equation}
\|f\|_{W_1^1(S)}=\|\sd
f\|_{L^1(S)}+(\mbox{Area}(S))^{-1/2}\|f\|_{L^1(S)} \label{5.a3}
\end{equation}
$\sd f$ being the differential of $f$ on $S$.

We now rescale the metric $\sg$ on $S$, setting:
\begin{equation}
\tilde{\sg}=(\mbox{Area}(S))^{-1}\sg \ \ \mbox{so that} \ \
d\mu_{\tilde{\sg}}=(\mbox{Area}(S))^{-1}d\mu_{\sg} \label{5.a4}
\end{equation}
to a metric $\tilde{\sg}$ of unit area. Taking into account the
fact that relative to the new metric, we have, in arbitrary local
coordinates on $S$,
\begin{equation}
|\sd\phi|_{\tilde{\sg}}^2=(\tilde{\sg}^{-1})^{AB}\partial_A\phi\partial_B\phi
=\mbox{Area}(S)(\sg^{-1})^{AB}\partial_A\phi\partial_B\phi=\mbox{Area}(S)|\sd\phi|_{\sg}^2
\label{5.a5}
\end{equation}
we obtain:
\begin{equation}
\|\phi\|_{W_1^2(S,\sg)}=\|\phi\|_{W_1^2(S,\tilde{\sg})}
\label{5.a6}
\end{equation}
and we have:
\begin{equation}
\|\phi\|_{W_1^2(S,\tilde{\sg})}=\|\sd\phi\|_{L^2(S,\tilde{\sg})}+\|\phi\|_{L^2(S,\tilde{\sg})}
\label{5.a7}
\end{equation}
Moreover, \ref{5.a2} and \ref{5.a3} read relative to
$\tilde{\sg}$:
\begin{equation}
\|f\|_{L^2(S,\tilde{\sg})}\leq\sqrt{\mbox{I}^\prime(S)}\|f\|_{W_1^1(S,\tilde{\sg})}
\label{5.a8}
\end{equation}
(the isoperimetric constant is invariant under rescalings) and:
\begin{equation}
\|f\|_{W_1^1(S,\tilde{\sg})} \label{5.a9}
\end{equation}

We now set:
\begin{equation}
\tilde{\phi}=\frac{1}{\sqrt{\mbox{I}^\prime(S)}}\frac{|\phi|}{\|\phi\|_{W_1^2(S,\tilde{\sg})}}
\label{5.a10}
\end{equation}
Then $\tilde{\phi}\geq 0$ and:
\begin{equation}
\|\tilde{\phi}\|_{W_1^2(S,\tilde{\sg})}=\frac{1}{\sqrt{\mbox{I}^\prime(S)}}
\label{5.a11}
\end{equation}
Taking $f=\tilde{\phi}^k$, $k>1$, in \ref{5.a8} we obtain:
\begin{equation}
\|\tilde{\phi}^k\|_{L^2(S,\tilde{\sg})}\leq\sqrt{\mbox{I}^\prime(S)}\|\tilde{\phi}^k\|_{W_1^1(S,\tilde{\sg})}
\label{5.a12}
\end{equation}
Since $\sd(\tilde{\phi}^k)=k\tilde{\phi}^{k-1}\sd\tilde\phi$  we
have:
$$\|\sd(\tilde{\phi}^k)\|_{L^1(S,\tilde{\sg})}\leq k\|\tilde{\phi}^{k-1}\|_{L^2(S,\tilde{\sg})}
\|\sd\tilde{\phi}\|_{L^2(S,\tilde{\sg})}$$ and:
$$\|\tilde{\phi}\|_{L^1(S,\tilde{\sg})}=\|\tilde{\phi}^{k-1}\tilde{\phi}\|_{L^1(S,\tilde{\sg})}
\leq\|\tilde{\phi}^{k-1}\|_{L^2(S,\tilde{\sg})}\|\tilde{\phi}\|_{L^2(S,\tilde{\sg})}$$
hence, adding,
\begin{equation}
\|\tilde{\phi}^k\|_{W_1^1(S,\tilde{\sg})}\leq
k\|\tilde{\phi}^{k-1}\|_{L^2(S,\tilde{\sg})}
\|\tilde\phi\|_{W_1^2(S,\tilde{\sg})} \label{5.a13}
\end{equation}
Substituting \ref{5.a11} in \ref{5.a13} and the result in
\ref{5.a12} yields:
$$\|\tilde{\phi}^k\|_{L^2(S,\tilde{\sg})}\leq k\|\tilde{\phi}^{k-1}\|_{L^2(S,\tilde{\sg})}$$
which is equivalent to:
\begin{equation}
\|\tilde{\phi}\|_{L^{2k}(S,\tilde{\sg})}\leq
k^{1/k}\|\tilde{\phi}\|_{L^{2(k-1)}(S,\tilde{\sg})}^{1-(1/k)}
\label{5.a14}
\end{equation}

We now set:
$$k=2,3,4,...$$
For $k=2$ we have $2(k-1)=2$ and taking $f=\tilde{\phi}$ in
\ref{5.a8} we obtain, by \ref{5.a10} and \ref{5.a11},
\begin{equation}
\|\tilde{\phi}\|_{L^2(S,\tilde{\sg})}=\frac{1}{\sqrt{I}^\prime(S)}
\frac{\|\phi\|_{L^2(S,\tilde{\sg})}}{\|\phi\|_{W_1^2(S,\tilde{\sg})}}
\leq\frac{\|\phi\|_{W_1^1(S,\tilde{\sg})}}{\|\phi\|_{W_1^2(S,\tilde{\sg})}}\leq
1 \label{5.a15}
\end{equation}
From \ref{5.a14} we then obtain:
$$\|\tilde{\phi}\|_{L^4(S,\tilde{\sg})}\leq 2^{1/2}$$
and in general, for any $k=2,3,4,...$ the recursive inequality
\ref{5.a14} implies:
\begin{equation}
\|\tilde{\phi}\|_{L^{2k}(S,\tilde{\sg})}\leq C_k \label{5.a16}
\end{equation}
where $C_k$ is defined by the recursion:
$$C_k=k^{1/k}C_{k-1}, \ \ \ C_1=1$$
hence:
\begin{equation}
C_k=\prod_{j=2}^k j^{1/j} \label{5.a17}
\end{equation}
Moreover, since $(S,\tilde{\sg})$ has unit area the norm
$\|f\|_{L^p(S,\tilde{\sg})}$ for a given function $f$ is a
non-decreasing function of the exponent $p$. Therefore \ref{5.a16}
implies that for any $2<p<\infty$:
\begin{equation}
\|\tilde{\phi}\|_{L^p(S,\tilde{\sg})}\leq C^\prime_p \label{5.a18}
\end{equation}
where $C^\prime_p=C_k$ with $k$ the smallest integer such that
$p\leq 2k$.

In view of the definition \ref{5.a10}, the inequality \ref{5.a18}
is equivalent to:
\begin{equation}
\|\phi\|_{L^p(S,\tilde{\sg})}\leq
C^\prime_p\sqrt{\mbox{I}^\prime(S)}\|\phi\|_{W_1^2(S,\tilde{\sg})}
\label{5.a19}
\end{equation}
Morever, by \ref{5.a6} and the fact that:
\begin{equation}
\|\phi\|_{L^p(S,\tilde{\sg})}=(\mbox{Area}(S))^{-1/p}\|\phi\|_{L^p(S,\sg)}
\label{5.a20}
\end{equation}
the inequality \ref{5.a19} is equivalent to:
\begin{equation}
(\mbox{Area}(S))^{-1/p}\|\phi\|_{L^p(S,\sg)}\leq
C^\prime_p\sqrt{\mbox{I}^\prime(S)}\|\phi\|_{W_1^2(S,\sg)}
\label{5.a21}
\end{equation}
This establishes the lemma in the case of functions $\phi$. To
deduce the lemma for tensorfields $\xi$ of arbitrary type, we set:
\begin{equation}
\phi=\sqrt{\varepsilon^2+|\xi|^2_{\sg}} \label{5.a22}
\end{equation}
where $\varepsilon$ is a positive constant. We then have:
$$\sd\phi=\frac{(\xi,\snab\xi)_{\sg}}{\sqrt{\varepsilon^2+|\xi|^2_{\sg}}}$$
hence:
\begin{equation}
|\sd\phi|_{\sg}\leq |\snab\xi|_{\sg} \label{5.a23}
\end{equation}
The general form of the lemma then follows taking the limit
$\varepsilon\rightarrow 0$.

\vspace{5mm}

\noindent{\bf Lemma 5.2} \ \ \ Let $(S,\sg)$ be a compact
2-dimensional Riemannian manifold and $\xi$ a tensorfield on $S$,
of arbitrary type, which belongs to $L^p(S)$ and with first
covariant derivative which also belongs to $L^p(S)$, for some
$p>2$. Then $\xi\in L^\infty(S)$ and we have:
$$\sup_S|\xi|\leq C_p\sqrt{I^\prime(S)}(\mbox{Area}(S))^{(1/2)-(1/p)}\|\xi\|_{W_1^p(S)}$$
Here $C_p$ is a numerical constant depending only on $p$, and we
denote:
$$\|\xi\|_{W_1^p(S)}=\|\snab\xi\|_{L^p(S)}+(\mbox{Area}(S))^{-1/2}\|\xi\|_{L^p(S)}$$

\noindent{\em Proof:} \ The proof is an adapation of an argument found in [G-T]. We first establish the lemma in the case
of functions $\phi$ on $S$. Following the proof of Lemma 5.1 we
now set:
\begin{equation}
\tilde{\phi}=\frac{1}{\sqrt{\mbox{I}^\prime(S)}}\frac{|\phi|}{\|\phi\|_{W_1^p(S,\tilde{\sg})}}
\label{5.a24}
\end{equation}
Then $\tilde{\phi}\geq 0$ and:
\begin{equation}
\|\tilde{\phi}\|_{W_1^p(S,\tilde{\sg})}=\frac{1}{\sqrt{\mbox{I}^\prime(S)}}
\label{5.a25}
\end{equation}
Taking $f=\tilde{\phi}^k$, $k>1$ in \ref{5.a8} we obtain:
\begin{equation}
\|\tilde{\phi}^k\|_{L^2(S,\tilde{\sg})}\leq
\sqrt{\mbox{I}^\prime(S)}\|\tilde{\phi}^k\|_{W_1^1(S,\tilde{\sg})}
\label{5.a26}
\end{equation}
Since $\sd(\tilde{\phi}^k)=k\tilde{\phi}^{k-1}\sd\tilde{\phi}$, we
now have, by H\"{o}lder's inequality:
\begin{equation}
\|\sd(\tilde{\phi}^k)\|_{L^1(S,\tilde{\sg})}\leq
k\|\tilde{\phi}^{k-1}\|_{L^{p^\prime}(S,\tilde{\sg})}
\|\sd\tilde{\phi}\|_{L^p(S,\tilde{\sg})} \label{5.a27}
\end{equation}
and:
\begin{equation}
\|\tilde{\phi}^k\|_{L^1(S,\tilde{\sg})}=\|\tilde{\phi}^{k-1}\tilde{\phi}\|_{L^1(S,\tilde{\sg})}
\leq\|\tilde{\phi}^{k-1}\|_{L^{p^\prime}(S,\tilde{\sg})}\|\tilde{\phi}\|_{L^p(S,\tilde{\sg})}
\label{5.a28}
\end{equation}
where $p^\prime$ is the conjugate exponent:
\begin{equation}
\frac{1}{p}+\frac{1}{p^\prime}=1 \label{5.a29}
\end{equation}
Hence, adding,
\begin{equation}
\|\tilde{\phi}^k\|_{W_1^1(S,\tilde{\sg})}\leq
k\|\tilde{\phi}^{k-1}\|_{L^{p^\prime}(S,\tilde{\sg})}
\|\tilde{\phi}\|_{W_1^p(S,\tilde{\sg})} \label{5.a30}
\end{equation}
Substituting \ref{5.a25} in \ref{5.a30} and the result in
\ref{5.a26} then yields:
$$\|\tilde{\phi}^k\|_{L^2(S,\tilde{\sg})}\leq k\|\tilde{\phi}^{k-1}\|_{L^{p^\prime}(S,\tilde{\sg})}$$
which is equivalent to:
\begin{equation}
\|\tilde{\phi}\|_{L^{2k}(S,\tilde{\sg})}\leq
k^{1/k}\|\tilde{\phi}\|_{L^{p^\prime(k-1)}(S,\tilde{\sg})}^{1-(1/k)}
\label{5.a31}
\end{equation}
This implies:
\begin{equation}
\|\tilde{\phi}\|_{L^{2k}(S,\tilde{\sg})}\leq
k^{1/k}\|\tilde{\phi}\|_{L^{p^\prime k}(S,\tilde{\sg})}^{1-(1/k)}
\label{5.a32}
\end{equation}
for, by virtue of the fact that $S$ has unit area with respect to
$\tilde{\sg}$, the norm $\|f\|_{L^p(S,\tilde{\sg})}$ for a given
function $f$ is a non-decreasing function of the exponent $p$. The
ratio of the exponent on the left in \ref{5.a32} to the exponent
on the right is $2/p^\prime>1$.

We now set:
\begin{equation}
k=\left(\frac{2}{p^\prime}\right)^n \ \ : \ n=1,2,3,...
\label{5.a33}
\end{equation}
For $n=1$ the exponent on the right in \ref{5.a32} is 2, and
taking $f=\phi$ in \ref{5.a8} we obtain, by \ref{5.a25}:
\begin{equation}
\|\tilde{\phi}\|_{L^2(S,\tilde{\sg})}=\frac{1}{\sqrt{\mbox{I}^\prime(S)}}
\frac{\|\phi\|_{L^2(S,\tilde{\sg})}}{\|\phi\|_{W_1^p(S,\tilde{\sg})}}
\leq\frac{\|\phi\|_{W_1^1(S,\tilde{\sg})}}{\|\phi\|_{W_1^p(S,\tilde{\sg})}}\leq
1 \label{5.a34}
\end{equation}
It follows from \ref{5.a32} - \ref{5.a34}, by induction on $n$,
that for every $n=1,2,3,...$:
\begin{equation}
\|\tilde{\phi}\|_{L^{2(2/p^\prime)^n}(S,\tilde{\sg})}\leq
\left(\frac{2}{p^\prime}\right)^{\sum_{m=1}^n m(2/p^\prime)^{-m}}
\label{5.a35}
\end{equation}
Taking the limit $n\rightarrow\infty$ we then obtain:
\begin{equation}
\sup_S(\tilde{\phi})\leq C_p \label{5.a36}
\end{equation}
where $C_p$ is the constant:
\begin{equation}
C_p=\left(\frac{2}{p^\prime}\right)^{\sum_{m=1}^\infty
m(2/p^\prime)^{-m}}=
\left(\frac{2}{p^\prime}\right)^{\frac{(2/p^\prime)}{((2/p^\prime)-1)^2}}
\label{5.a37}
\end{equation}

In view of the definition \ref{5.a24}, \ref{5.a37} is equivalent
to:
\begin{equation}
\sup_S|\phi|\leq
C_p\sqrt{\mbox{I}^\prime(S)}\|\phi\|_{W_1^p(S,\tilde{\sg})}
\label{5.a38}
\end{equation}
Now, by \ref{5.a4}, \ref{5.a5}, we have:
\begin{eqnarray}
&&\|\phi\|_{W_1^p(S,\tilde{\sg})}=\|\sd\phi\|_{L^p(S,\tilde{\sg})}+\|\phi\|_{L^p(S,\tilde{\sg})}\nonumber\\
&&=(\mbox{Area}(S))^{(1/2)-(1/p)}\{\|\sd\phi\|_{L^p(S,\sg)}+(\mbox{Area}(S))^{-1/2}\|\phi\|_{L^p(S,\sg)}\}\nonumber\\
&&=(\mbox{Area}(S))^{(1/2)-(1/p)}\|\phi\|_{W_1^p(S,\sg)}
\label{5.a39}
\end{eqnarray}
We thus conclude that:
\begin{equation}
\sup_S|\phi|\leq
C_p\sqrt{\mbox{I}^\prime(S)}(\mbox{Area}(S))^{(1/2)-(1/p)}\|\phi\|_{W_1^p(S,\sg)}
\label{5.a40}
\end{equation}
This establishes the lemma in the case of functions $\phi$. To
deduce the lemma for tensorfields $\xi$ of arbitrary type, we
define $\phi$ according to \ref{5.a22}. Then in view of
\ref{5.a23} the general form of the lemma follows taking the limit
$\varepsilon\rightarrow 0$.

\vspace{5mm}

We see from Lemmas 5.1 and 5.2 that the Sobolev inequalities of
$S_{\ub,u}$ thus involve the isoperimetric constant
$\mbox{I}(S_{\ub,u})$ of $S_{\ub,u}$. Consequently, to make use of
these inequalities  we must first obtain an upper bound for
$\mbox{I}(S_{\ub,u})$. Considering a given $C_u$, the derivation
of this upper bound shall be based on upper and lower bounds for
the eigenvalues of the metric
$\sg(\ub)=\left.\Phi^*_{\ub}\sg\right|_{S_{\ub,u}}$ with respect
to the metric $\sg(0)=\left.\sg\right|_{S_{0,u}}$.

From \ref{4.28} - \ref{4.30} and the first of \ref{1.28} we have:
\begin{equation}
\frac{\partial}{\partial\ub}\sg(\ub)=2(\Omega\chi)(\ub)
\label{5.37}
\end{equation}
We consider the eigenvalues of $\sg(\ub)$ with respect to
$\sg(0)$. Let $\lambda(\ub)$ be the smallest eigenvalue and
$\Lambda(\ub)$ be the largest eigenvalue.

\vspace{5mm}

\noindent {\bf Lemma 5.3} \ \ \ Under the assumptions of
Proposition 3.1 we have:
$$\lambda(\ub)\geq 1-O(\delta^{1/2}|u|^{-1}), \ \ \ \Lambda(\ub)\leq 1+O(\delta^{1/2}|u|^{-1})$$
for all $(\ub,u)\in D^\prime$.

\noindent{\em Proof:} \ Let us define:
\begin{equation}
\mu(\ub)=\frac{d\mu_{\sg(\ub)}}{d\mu_{\sg(0)}}=\sqrt{\lambda(\ub)\Lambda(\ub)}
\label{5.38}
\end{equation}
and:
\begin{equation}
\nu(\ub)=\frac{1}{\mu(\ub)}\sup_{|X|_{\sg(0)}=1}\sg(\ub)(X,X)=\frac{\Lambda(\ub)}{\mu(\ub)}
=\sqrt{\frac{\Lambda(\ub)}{\lambda(\ub)}} \label{5.39}
\end{equation}
We have:
\begin{equation}
\frac{\partial}{\partial\ub}\mu(\ub)=\frac{1}{2}
\mbox{tr}_{\sg(\ub)}\left(\frac{\partial}{\partial\ub}\sg(\ub)\right)\mu(\ub)=(\Omega\mbox{tr}\chi)(\ub)\mu(\ub)
\label{5.40}
\end{equation}
Integrating with respect to $\ub$ and noting that $\mu(0)=1$
yields:
\begin{equation}
\mu(\ub)=\exp\left(\int_0^{\ub}(\Omega\mbox{tr}\chi)(\ub^\prime)d\ub^\prime\right)
\label{5.41}
\end{equation}
In view of the second of the bounds \ref{4.49} we then obtain:
\begin{equation}
|\mu(\ub)-1|\leq O(\delta|u|^{-1}) \label{5.42}
\end{equation}
To estimate $\nu(\ub)$ we set:
\begin{equation}
\hat{\sg}(\ub)=\frac{\sg(\ub)}{\mu(\ub)} \label{5.43}
\end{equation}
Then according to the definition \ref{5.39}:
\begin{equation}
\sup_{|X|_{\sg(0)}=1}\hat{\sg}(\ub)(X,X)=\nu(\ub) \label{5.44}
\end{equation}
From \ref{5.37} and \ref{5.40} we have:
\begin{eqnarray*}
\frac{\partial}{\partial\ub}\hat{\sg}(\ub)&=&\frac{1}{\mu(\ub)}\frac{\partial}{\partial\ub}\sg(\ub)
-\frac{\sg(\ub)}{(\mu(\ub))^2}\frac{\partial}{\partial\ub}\mu(\ub)\\
&=&\frac{2}{\mu(\ub)}(\Omega\chi)(\ub)-\frac{\sg(\ub)}{\mu(\ub)}(\Omega\mbox{tr}\chi)(\ub)
\end{eqnarray*}
or, simply:
\begin{equation}
\frac{\partial}{\partial\ub}\hat{\sg}(\ub)=\frac{2}{\mu(\ub)}(\Omega\chih)(\ub)
\label{5.45}
\end{equation}
Consider any tangent vector $X$ to $S_{0,u}$ such that
$|X|_{\sg(0)}=1$. Then:
\begin{equation}
\frac{\partial}{\partial\ub}\hat{\sg}(\ub)(X,X)=\frac{2}{\mu(\ub)}(\Omega\chih)(\ub)(X,X)
\label{5.46}
\end{equation}
Integrating with respect to $\ub$ and noting that
$\hat{\sg}(0)(X,X)=1$ yields:
\begin{equation}
\hat{\sg}(\ub)(X,X)=1+2\int_0^{\ub}\frac{(\Omega\chih)(\ub^\prime)(X,X)}{\mu(\ub^\prime)}d\ub^\prime
\label{5.47}
\end{equation}
hence:
\begin{equation}
\hat{\sg}(\ub)(X,X)\leq
1+2\int_0^{\ub}\frac{|(\Omega\chih)(\ub^\prime)(X,X)|}{\mu(\ub^\prime)}d\ub^\prime
\label{5.48}
\end{equation}
Now if $\xi$ is an arbitrary type $T^q_p$ tensorfield on $S_{0,u}$
then, in components with respect to an arbitrary local frame field
on $S_{0,u}$,
\begin{equation}
|\xi|_{\sg(\ub)}^2=(\sg(\ub))_{A_1 B_1} . . . (\sg(\ub))_{A_q
B_q}(\sg^{-1}(\ub))^{C_1 D_1} . . . (\sg^{-1}(\ub))^{C_p D_p}
\xi^{A_1 ... A_q}_{C_1 ... C_p} \xi^{B_1 ... B_q}_{D_1 ... D_p}
\label{5.49}
\end{equation}
It follows that:
\begin{equation}
(\lambda(\ub))^q(\Lambda(\ub))^{-p}|\xi|_{\sg(0)}^2\leq
|\xi|_{\sg(\ub)}^2 \leq
(\Lambda(\ub))^q(\lambda(\ub))^{-p}|\xi|_{\sg(0)}^2 \label{5.50}
\end{equation}
In particular, taking $\xi=(\Omega\chih)(\ub)$  ($q=0$, $p=2$), we
have:
\begin{equation}
|(\Omega\chih)(\ub)|_{\sg(\ub)}\geq
(\Lambda(\ub))^{-1}|(\Omega\chih)(\ub)|_{\sg(0)} \label{5.51}
\end{equation}
hence:
\begin{equation}
|(\Omega\chih)(\ub)(X,X)|\leq
|(\Omega\chih)(\ub)|_{\sg(0)}|X|_{\sg(0)}^2=|(\Omega\chih)(\ub)|_{\sg(0)}
\leq \Lambda(\ub)|(\Omega\chih)(\ub)|_{\sg(\ub)} \label{5.52}
\end{equation}
Substituting in \ref{5.48}, taking the supremum over $X\in T_q
S_{0,u}$ such that $|X|_{\sg(0)}=1$ at each $q\in S_{0,u}$, and
recalling \ref{5.39} and \ref{5.44} we obtain the following linear
integral inequality for $\nu(\ub)$:
\begin{equation}
\nu(\ub)\leq
1+2\int_0^{\ub}|(\Omega\chih)(\ub^\prime)|_{\sg(\ub^\prime)}\nu(\ub^\prime)d\ub^\prime
\label{5.53}
\end{equation}
which implies:
\begin{equation}
\nu(\ub)\leq\exp\left(2\int_0^{\ub}|(\Omega\chih)(\ub^\prime)|_{\sg(\ub^\prime)}d\ub^\prime\right)
\label{5.54}
\end{equation}
In view of the third of the bounds \ref{4.49} we then obtain
(recall that by the definition \ref{5.39} $\nu(\ub)\geq 1$):
\begin{equation}
1\leq\nu(\ub)\leq 1+O(\delta^{1/2}|u|^{-1}) \label{5.55}
\end{equation}
Since from the definitions \ref{5.38}, \ref{5.39} we can express:
\begin{equation}
\lambda(\ub)=\frac{\mu(\ub)}{\nu(\ub)}, \ \ \
\Lambda(\ub)=\mu(\ub)\nu(\ub) \label{5.56}
\end{equation}
the lemma follows from the bounds \ref{5.42} and \ref{5.55}.

\vspace{5mm}

\noindent{\bf Lemma 5.4} \ \ \ Under the assumptions of
Proposition 3.1 we have:
$$\mbox{I}(S_{\ub,u})\leq\frac{1}{\pi}$$
for all $(\ub,u)\in D^\prime$, provided that $\delta$ is suitably
small depending on ${\cal R}_0^\infty(\alpha)$. In particular, the
constant $\mbox{I}^\prime(S_{\ub,u})$ appearing in the Sobolev
inequalities, Lemmas 5.1 and 5.2 is equal to 1.

\noindent{\em Proof:} \ A domain $U_{\ub}\subset S_{\ub,u}$ with
$C^1$ boundary $\partial U_{\ub}$ is the image by $\Phi_{\ub}$ of
a domain $U_0\subset S_{0,u}$ with $C^1$ boundary $\partial U_0$.

We have:
\begin{equation}
\mbox{Perimeter}(\partial U_{\ub})=\int_{\partial
U_{\ub}}ds=\int_{\partial U_0}ds(\ub), \ \ \
\mbox{Perimeter}(\partial U_0)=\int_{\partial U_0}ds(0)
\label{5.57}
\end{equation}
Here $ds$ is the element of arc length of $\partial U_{\ub}$ with
respect to the metric $\left.\sg\right|_{S_{\ub,u}}$ on
$S_{\ub,u}$, while $ds(\ub)$ is the element of arc length of
$\partial U_0$ with respect to the metric $\sg(\ub)$ on $S_{0,u}$
and $ds(0)$ the element of arc length of $\partial U_0$ with
respect to the metric $\sg(0)=\left.\sg\right|_{S_{0,u}}$ on
$S_{0,u}$. Now:
\begin{equation}
ds(\ub)\geq \sqrt{\lambda(\ub)}ds(0) \label{5.58}
\end{equation}
hence:
\begin{equation}
\mbox{Perimeter}(\partial U_{\ub})\geq
\inf_{S_{0,u}}\sqrt{\lambda(\ub)}\mbox{Perimeter}(\partial U_0)
\label{5.59}
\end{equation}
Lemma 5.3 then implies:
\begin{equation}
\frac{\mbox{Perimeter}(\partial
U_{\ub})}{\mbox{Perimeter}(\partial U_0)}\geq
1-O(\delta^{1/2}|u|^{-1}) \label{5.60}
\end{equation}
provided that $\delta$ is suitably small depending on ${\cal
R}_0^\infty(\alpha)$ (recall the second and third of the bounds
\ref{4.49} which come from Proposition 3.1).

On the other hand we have:
\begin{equation}
\mbox{Area}(U_{\ub})=\int_{U_{\ub}}d\mu_{\sg}=\int_{U_0}d\mu_{\sg(\ub)},
\ \ \ \mbox{Area}(U_0)=\int_{U_0}d\mu_{\sg(0)} \label{5.61}
\end{equation}
where $d\mu_{\sg}$ is the area element of $S_{\ub,u}$ with respect
to the induced metric $\left.\sg\right|_{S_{\ub,u}}$, while
$d\mu_{\sg(\ub)}$ is the area element on $S_{0,u}$ induced by the
metric $\sg(\ub)$ and $d\mu_{\sg(0)}$ the area element on
$S_{0,u}$ induced by the metric
$\sg(0)=\left.\sg\right|_{S_{0,u}}$. By \ref{5.38}:
\begin{equation}
d\mu_{\sg(\ub)}=\mu(\ub)d\mu_{\sg(0)} \label{5.62}
\end{equation}
hence:
\begin{equation}
\mbox{Area}(U_{\ub})\leq\sup_{S_{0,u}}\mu(\ub)\mbox{Area}(U_0)
\label{5.63}
\end{equation}
The bound \ref{5.42} then implies:
\begin{equation}
\frac{\mbox{Area}(U_{\ub})}{\mbox{Area}(U_0)}\leq
1+O(\delta|u|^{-1}) \label{5.64}
\end{equation}

The inequalities \ref{5.60} and \ref{5.64} together imply:
\begin{equation}
\frac{\mbox{Area}(U_{\ub})}{(\mbox{Perimeter}(\partial
U_{\ub}))^2} \leq
(1+O(\delta^{1/2}|u|^{-1})\frac{\mbox{Area}(U_0)}{(\mbox{Perimeter}(\partial
U_0))^2} \label{5.65}
\end{equation}
provided that $\delta$ is suitably small depending on ${\cal
R}_0^\infty(\alpha)$ (recall the second and third of the bounds
\ref{4.49} which come from Proposition 3.1). Similarly we obtain:
\begin{equation}
\frac{\mbox{Area}(U^c_{\ub})}{(\mbox{Perimeter}(\partial
U_{\ub}))^2} \leq
(1+O(\delta^{1/2}|u|^{-1})\frac{\mbox{Area}(U^c_0)}{(\mbox{Perimeter}(\partial
U_0))^2} \label{5.66}
\end{equation}
Therefore:
\begin{equation}
\frac{\min\{\mbox{Area}(U_{\ub}),\mbox{Area}(U_{\ub}^c)\}}{(\mbox{Perimeter}(\partial
U_{\ub}))^2} \leq (1+O(\delta^{1/2}|u|^{-1}))
\frac{\min\{\mbox{Area}(U_{\ub}),\mbox{Area}(U_{\ub}^c)\}}{(\mbox{Perimeter}(\partial
U_{\ub}))^2} \label{5.67}
\end{equation}
Taking the supremum over all domains $U_{\ub}$ with $C^1$ boundary
$\partial U_{\ub}$ in $S_{\ub,u}$ we conclude that:
\begin{equation}
\mbox{I}(S_{\ub,u})\leq
(1+O(\delta^{1/2}|u|^{-1}))\mbox{I}(S_{0,u}) \label{5.68}
\end{equation}
Now:
\begin{equation}
\mbox{I}(S_{0,u})=\frac{1}{2\pi} \label{5.69}
\end{equation}
$S_{0,u}$ being a round sphere in Euclidean space, the supremum
being achieved for a hemisphere. The lemma then follows.

\section{The uniformization theorem}

We now turn to the uniformization problem. Given a 2-dimensional
compact Riemannian manifold $(S,\sg)$ with $S$ diffeomorphic to
$S^2$ the problem is to find a suitable function $\phi$ such that
the conformally related metric
\begin{equation}
\up{g}=e^{-2\phi}\sg \label{5.70}
\end{equation}
has Gauss curvature $\up{K}=1$, so that $(S,\up{\sg})$ is
isometric to the standard sphere (the unit sphere in Euclidean
space). In [C-K] a solution to this problem was obtained in the
form of the following proposition (Lemma 2.3.2 of [C-K]).

\vspace{5mm}

\noindent{\bf Proposition 5.3} \ \ \ Let $(S,\sg)$ be a
2-dimensional compact Riemannian manifold with $S$ diffeomorphic
to $S^2$. Then there exists a conformal transformation of the
metric
$$\up{\sg}=\Omega^2\sg$$
such that $\up{K}=1$. Moreover the conformal factor $\Omega$ can
be chosen such that the quantities $\Omega_m^{-1}$, $\Omega_M$,
$\Omega_1$ have upper bounds which depend only on upper bounds for
$$\sup_S r^2|K|, \ \ \ \frac{\mbox{diam}(S)}{r}$$
Here, we denote:
$$r=\sqrt{\frac{\mbox{Area}(S)}{4\pi}}$$
$$\Omega_m=\inf_S r\Omega, \ \ \ \Omega_M=\sup_S r\Omega$$
$$\Omega_1=\sup_S\Omega^{-2}|\sd\Omega|$$
Also, for every $2\leq p<\infty$ the quantity
$$\Omega_{2,p}=\left(\int_S\Omega^{-3p+2}|\snab^{ \ 2}\Omega|^p\right)^{1/p}$$
has a bound which depends only on upper bounds for 
$$\sup_S r^2|K|, \ \ \ \frac{\mbox{diam}(S)}{r}$$
and on $p$. In addition, if with
$$k_m=\inf_S r^2 K, \ \ \ k_M=\sup_S r^2 K$$
we have $k_m>0$ then the bounds on $\Omega_m^{-1}$, $\Omega_M$,
$\Omega_1$ depend only on upper bounds for $k_m^{-1}$, $k_M$, and
the bound on $\Omega_{2,p}$ depends only on upper bounds for $k_m^{-1}$, $k_M$ and on 
$p$.

\vspace{5mm}

As we have remarked in the introduction to the present chapter, we
cannot directly apply here the above proposition, because we do
not possess the appropriate $L^\infty$ bounds on $K$. We proceed
instead as follows. We first define the function $\omega$ to be
the solution of
\begin{equation}
\slap_{\sg}\omega=K-\overline{K} \ \ \ \mbox{such that
$\overline{\omega}=0$} \label{5.71}
\end{equation}
Then the conformally related metric
\begin{equation}
\sg^\prime=e^{2\omega}\sg \label{5.72}
\end{equation}
has Gauss curvature $K^\prime$ given by:
\begin{equation}
K^\prime=e^{-2\omega}(K-\slap_{\sg}\omega)=e^{-2\omega}\overline{K}
\label{5.73}
\end{equation}
Since, according to \ref{5.2},
\begin{equation}
\overline{K}=\frac{1}{r^2} \label{5.74}
\end{equation}
we shall obtain the required $L^\infty$ bounds for $K^\prime$ once
we derive an appropriate $L^\infty$ bound for $\omega$.

\vspace{5mm}

\noindent{\bf Lemma 5.5} \ \ \ We have, for all $(\ub,u)\in
D^\prime$:
$$\sup_{S_{\ub,u}}|\omega|\leq \frac{1}{2}\log 2$$
provided that $\delta$ is suitably small depending on ${\cal
D}_0^\infty$, ${\cal R}_0^\infty$, $\scD_1^4$, $\scR_1^4$, and
$\scR_2(\alpha)$.

\noindent{\em Proof:} We start from the following basic integral
identity for the Laplacian $\slap_{\sg}$ acting on functions
$\phi$ on the compact 2-dimensional manifold $(S,\sg)$ (see
Chapter 2 of [C-K]):
\begin{equation}
\int_S\{|\snab^{ \
2}\phi|^2+K|\sd\phi|^2\}d\mu_{\sg}=\int_S|\slap\phi|^2 d\mu_{\sg}
\label{5.75}
\end{equation}
We apply this to the function $\omega$ obtaining, by virtue of
equation \ref{5.71},
\begin{equation}
\int_S\{|\snab^{ \ 2}\omega|^2+K|\sd\omega|^2\}d\mu_{\sg}=\int_S|K-\overline{K}|^2
d\mu_{\sg} \label{5.76}
\end{equation}
Writing $K=\overline{K}+K-\overline{K}$ we estimate:
\begin{equation}
\int_S|K-\overline{K}||\sd\omega|^2\leq\|K-\overline{K}\|_{L^2(S)}\|\sd\omega\|_{L^4(S)}^2
\label{5.77}
\end{equation}
Now, by Lemma 5.1 with $p=4$ applied to $\sd\omega$ and Lemma 5.4:
\begin{equation}
\|\sd\omega\|_{L^4(S)}^2\leq Cr\{\|\snab^{ \
2}\omega\|_{L^2(S)}^2+r^{-2}\|\sd\omega\|_{L^2(S)}^2\}
\label{5.78}
\end{equation}
In view of \ref{5.74}, \ref{5.76} then implies:
\begin{equation}
\left(1-Cr\|K-\overline{K}\|_{L^2(S)}\right)\left(\|\snab^{ \
2}\omega\|_{L^2(S)}^2+r^{-2}\|\sd\omega\|_{L^2(S)}^2\right)\leq
\|K-\overline{K}\|_{L^2(S)}^2
\label{5.79}
\end{equation}
Now, by Proposition 5.2:
\begin{equation}
r^2\|K-\overline{K}\|_{L^2(S)}\leq
C\delta^{1/2}\scR_2(\alpha)+O(\delta) \label{5.80}
\end{equation}
It follows that if $\delta$ is suitably small depending on ${\cal
D}_0^\infty$, ${\cal R}_0^\infty$, $\scD_1^4$, $\scR_1^4$, and
$\scR_2(\alpha)$, then, in reference to the factror on the
left in \ref{5.79} we have:
$$Cr\|K-\overline{K}\|_{L^2(S)}\leq\frac{1}{2}$$
and \ref{5.79} implies:
\begin{equation}
\|\snab^{ \
2}\omega\|_{L^2(S)}^2+r^{-2}\|\sd\omega\|_{L^2(S)}^2\leq
2\|K-\overline{K}\|_{L^2(S)}^2 \label{5.81}
\end{equation}
We now apply Lemma 5.1 with $p=q$ to $\sd\omega$. In view of Lemma
5.4 we obtain:
\begin{equation}
\|\sd\omega\|_{L^q(S)}^2\leq C_q r^{4/q}\{\|\snab^{ \
2}\omega\|_{L^2(S)}^2+r^{-2}\|\sd\omega\|_{L^2(S)}\} \label{5.82}
\end{equation}
Here $q$ is any real number greater than 2. Substituting
\ref{5.81} then yields:
\begin{eqnarray}
\|\sd\omega\|_{L^q(S)}&\leq& C_q r^{2/q}\|K-\overline{K}\|_{L^2(S)}\label{5.83}\\
&\leq& C_q
r^{(2/q)-2}\{C\delta^{1/2}\scR_2(\alpha)+O(\delta)\}\nonumber
\end{eqnarray}
by \ref{5.80}. Next, we apply Lemma 5.1 with $p=q$ to $\omega$ to
obtain:
\begin{equation}
\|\omega\|_{L^q(S)}^2\leq C_q
r^{4/q}\{\|\sd\omega\|_{L^2(S)}^2+r^{-2}\|\omega\|_{L^2(S)}^2\}
\label{5.84}
\end{equation}
Now, since $\overline{\omega}=0$ the isoperimetric inequality
\ref{5.35} reduces in the case of $\omega$ to:
\begin{equation}
\|\omega\|_{L^2(S)}^2\leq \mbox{I}(S)\|\sd\omega\|_{L^1(S)}^2
\label{5.85}
\end{equation}
Since
$$\|\sd\omega\|_{L^1(S)}^2\leq\mbox{Area}(S)\|\sd\omega\|_{L^2(S)}^2$$
\ref{5.85} implies:
\begin{equation}
r^{-2}\|\omega\|_{L^2(S)}^2\leq
4\pi\mbox{I}(S)\|\sd\omega\|_{L^2(S)}^2 \label{5.86}
\end{equation}
Substituting this in \ref{5.84} then yields, in view of Lemma 5.4,
\begin{equation}
\|\omega\|_{L^q(S)}^2\leq C_q r^{4/q}\|\sd\omega\|_{L^2(S)}^2
\label{5.87}
\end{equation}
Hence, from \ref{5.81}:
\begin{eqnarray}
\|\omega\|_{L^q(S)}&\leq& C_q r^{(2/q)+1}\|K-\overline{K}\|_{L^2(S)}\label{5.88}\\
&\leq& C_q r^{(2/q)-1}\{C\delta^{1/2}\scR_2(\alpha)+O(\delta)\}
\label{5.89}
\end{eqnarray}
Combining \ref{5.83} with \ref{5.89} we obtain:
\begin{equation}
\|\omega\|_{W_1^q(S)}\leq C_q
r^{(2/q)-2}\{C\delta^{1/2}\scR_2(\alpha)+O(\delta)\} \label{5.90}
\end{equation}
This holds for any $2<q<\infty$. Taking then $q=4$ and applying
Lemma 5.2 with $p=q$ we obtain, in view of Lemma 5.4,
\begin{equation}
\sup_{S}|\omega|\leq
Cr^{-1}\{C\delta^{1/2}\scR_2(\alpha)+O(\delta)\} \label{5.91}
\end{equation}
The lemma then follows.

\vspace{5mm}

According to the above lemma we have:
\begin{equation}
\frac{1}{\sqrt{2}}\leq e^\omega\leq\sqrt{2} \label{5.92}
\end{equation}
It follows that with
$$r=\sqrt{\frac{\mbox{Area}(S,\sg)}{4\pi}}, \ \ \ r^\prime=\sqrt{\frac{\mbox{Area}(S,\sg^\prime)}{4\pi}}$$
we have:
\begin{equation}
\frac{1}{\sqrt{2}}\leq\frac{r^\prime}{r}\leq\sqrt{2} \label{5.93}
\end{equation}
Since according to \ref{5.73}, \ref{5.74}
\begin{equation}
K^\prime=\frac{e^{-2\omega}}{r^2} \label{5.94}
\end{equation}
the inequalities \ref{5.92} and \ref{5.93} imply:
\begin{equation}
\frac{1}{4}\leq r^{\prime 2} K^\prime\leq 4 \label{5.95}
\end{equation}
Therefore the quantities
\begin{equation}
k^\prime_m=\inf_S r^{\prime 2}K^\prime, \ \ \ k^\prime_M=\sup_S
r^{\prime 2} K^\prime \label{5.96}
\end{equation}
satisfy the bounds:
\begin{equation}
k^\prime_m\geq\frac{1}{4}, \ \ \ k^\prime_M\leq 4 \label{5.97}
\end{equation}
We can then apply Proposition 5.3 to the Riemannian manifold
$(S,\sg^\prime)$ to conclude that there exists a conformal factor
$\Omega^\prime$ such that the metric
\begin{equation}
\up{\sg}=\Omega^{\prime 2}\sg^\prime \label{5.98}
\end{equation}
has Gauss curvature $\up{K}=1$ and the quantities $\Omega^{\prime
-1}_m$, $\Omega^\prime_M$, $\Omega^\prime_1$, where
\begin{equation}
\Omega^\prime_m=\inf_S r^\prime\Omega^\prime, \ \ \
\Omega^\prime_M=\sup_S r^\prime\Omega^\prime, \label{5.99}
\end{equation}
\begin{equation}
\Omega^\prime_1=\sup_S\Omega^{\prime
-2}|\sd\Omega^\prime|_{\sg^\prime}, \label{5.100}
\end{equation}
are bounded by numerical constants, while for every $2\leq
p<\infty$ the quantity
\begin{equation}
\Omega^\prime_{2,p}=\left(\int_S\Omega^{\prime -3p+2} |\snab^{\
\prime 2}\Omega^\prime|_{\sg^\prime}^p
d\mu_{\sg^\prime}\right)^{1/p} \label{5.101}
\end{equation}
is bounded by a numerical constant depending only on $p$.

Setting
\begin{equation}
\Omega^\prime=e^{-\phi^\prime}=r^{\prime -1}e^{-\psi^\prime}
\label{5.102}
\end{equation}
we have:
\begin{equation}
\Omega^\prime_m=\inf_S e^{-\psi^\prime}, \ \ \
\Omega^\prime_M=\sup_S e^{-\psi^\prime} \label{5.103}
\end{equation}
therefore the bounds on $\Omega^\prime_m$, $\Omega^\prime_M$ are
equivalent to a bound for
\begin{equation}
\sup_S|\psi^\prime| \label{5.104}
\end{equation}
Moreover, we have:
\begin{equation}
\Omega^{\prime -2}|\sd\Omega^\prime|_{\sg^\prime}=r^\prime
e^{\psi^\prime}|\sd\psi^\prime|_{\sg^\prime} \label{5.105}
\end{equation}
therefore, modulo the bound for \ref{5.104}, the bound for
$\Omega^\prime_1$ is equivalent to a bound for
\begin{equation}
\sup_S\left(r^\prime|\sd\psi^\prime|_{\sg^\prime}\right)
\label{5.106}
\end{equation}
Finally, by \ref{5.102} we have:
\begin{equation}
\Omega^\prime_{2,p}=r^{\prime 2-(2/p)}\left(\int_S
e^{(2p-2)\psi^\prime}|\snab^{\ \prime 2}\psi^\prime-
\sd\psi^\prime\otimes\sd\psi^\prime|_{\sg^\prime}^p
d\mu_{\sg^\prime}\right)^{1/p} \label{5.107}
\end{equation}
It follows that, modulo the bounds for \ref{5.106} and
\ref{5.104}, the bound for $\Omega^\prime_{2,p}$ is equivalent to
a bound for
\begin{equation}
r^{\prime 2-(2/p)}\|\snab^{\ \prime
2}\psi^\prime\|_{L^p(S,\sg^\prime)} \label{5.108}
\end{equation}

Combining the conformal transformations \ref{5.72} and \ref{5.98}
we conclude that:
\begin{equation}
\sg=e^{2\phi}\up{\sg}=r^2 e^{2\psi}\up{\sg} \label{5.109}
\end{equation}
where:
\begin{equation}
\phi=\phi^\prime-\omega \ \ \mbox{or} \ \
\psi=\log(r^\prime/r)+\psi^\prime-\omega \label{5.110}
\end{equation}
Lemma 5.5 together with the inequalities \ref{5.93} and the bound
for \ref{5.104} then yield a bound for
\begin{equation}
\sup_S|\psi| \label{5.111}
\end{equation}
by a numerical constant.

Let us define:
\begin{equation}
k_{1,q}=r^{1-(2/q)}\|\sd\psi\|_{L^q(S,\sg)} \label{5.112}
\end{equation}
Since by \ref{5.110}
\begin{equation}
\sd\psi=\sd\psi^\prime-\sd\omega, \label{5.113}
\end{equation}
and by the bound for \ref{5.106} and the inequalities \ref{5.92}
and \ref{5.93}
$$r^{1-(2/q)}\|\sd\psi^\prime\|_{L^q(S,\sg)}\leq C,$$
while by the estimate \ref{5.83}
\begin{eqnarray}
r^{1-(2/q)}\|\sd\omega\|_{L^q(S,\sg)}&\leq& C_q r^{-1}\{C\delta^{1/2}\scR_2(\alpha)+O(\delta)\}\nonumber\\
&\leq& C^\prime_q, \label{5.114}
\end{eqnarray}
the last step provided that $\delta$ satisfies the smallness
condition of Lemma 5.5, we conclude that, under the assumptions of
Lemma 5.5, for every $2\leq q<\infty$ $k_{1,q}$ is bounded by a
numerical constant depending only on $q$. We have thus arrived at
the following proposition.

\vspace{5mm} \noindent{\bf Proposition 5.4} \ \ \ Let the
assumptions of Lemma 5.5 hold. Then for each $(\ub,u)\in D^\prime$
the induced metric $\sg$ on $S_{\ub,u}$ can be expressed in the
form:
$$\sg=r^2 e^{2\psi}\up{\sg}$$
where the metric $\up{\sg}$ has Gauss curvature $\up{K}=1$, thus
$(S_{\ub,u},\up{\sg})$ is isometric to the standard sphere.
Moreover there is numerical constant $C$, and, for each $2\leq
q<\infty$ a numerical constant $C_q$ depending only on $q$, such
that for all $(\ub,u)\in D^\prime$:
$$\sup_{S_{\ub,u}}|\psi|\leq C$$
and:
$$k_{1,q}:=r^{1-(2/q)}\|\sd\psi\|_{L^q(S,\sg)}\leq C_q$$
where:
$$r=\sqrt{\frac{\mbox{Area}(S_{\ub,u},\sg)}{4\pi}}$$

\vspace{5mm}

We shall now proceed to derive a bound for the quantity
\begin{equation}
k_{1,\infty}=r\sup_S|\sd\psi|_{\sg} \label{5.115}
\end{equation}
as well as for the quantity $k_{2,4}$, where, for any $1\leq
p\leq\infty$:
\begin{equation}
k_{2,p}=r^{2-(2/p)}\|\snab^{ \ 2}\psi\|_{L^p(S,\sg)},
\label{5.116}
\end{equation}
under the assumption that $K\in L^4(S)$.

Now we already have a bound for \ref{5.106} by a numerical
constant, and, in view of the inequalities \ref{5.92} and
\ref{5.93}, this is equivalent to a bound for
\begin{equation}
r\sup_S|\sd\psi^\prime|_{\sg} \label{5.117}
\end{equation}
by a numerical constant. Therefore, in view of \ref{5.113}, we
shall obtain a bound for $k_{1,\infty}$ once we derive a bound for
\begin{equation}
r\sup_S|\sd\omega|_{\sg} \label{5.118}
\end{equation}
Also, for any $2\leq p <\infty$ we already have a bound for
\ref{5.108} by a numerical constant depending only on $p$, and, in
view of the inequalities \ref{5.92} and \ref{5.93}, this is
equivalent to a bound for
\begin{equation}
r^{2-(2/p)}\|\snab^{ \ \prime 2}\psi^\prime\|_{L^p(S,\sg)}
\label{5.119}
\end{equation}
for any $2\leq p<\infty$ by a numerical constant depending only
on $p$. Now, we have:
\begin{equation}
\snab^{ \ \prime 2}\psi^\prime=\snab^{ \
2}\psi^\prime-\triangle^\prime\cdot\sd\psi^\prime \label{5.120}
\end{equation}
where $\triangle^\prime=\sGamma^\prime-\sGamma$ is the difference
of the connections associated to $\sg^\prime$ and $\sg$, a type
$T^1_2$ tensorfield on $S$, symmetric in the lower entries, and
expressed in local coordinates on $S$ by:
\begin{equation}
\triangle^{\prime
C}_{AB}=\delta_A^C\partial_B\omega+\delta_B^C\partial_B\omega-(\sg^{-1})^{CD}\sg_{AB}\partial_D\omega
\label{5.121}
\end{equation}
We have, pointwise:
\begin{equation}
|\triangle^\prime|_{\sg}\leq C|\sd\omega|_{\sg} \label{5.122}
\end{equation}
hence:
\begin{equation}
|\snab^{ \ 2}\psi^\prime|_{\sg}\leq |\snab^{ \ \prime
2}\psi^\prime|_{\sg}+C|\sd\omega|_{\sg}|\sd\psi^\prime|_{\sg}
\label{5.123}
\end{equation}
In view of the bound for \ref{5.117} it follows that:
\begin{equation}
\|\snab^{ \ 2}\psi^\prime|_{L^p(S,\sg)}\leq \|\snab^{ \ \prime
2}\psi^\prime\|_{L^p(S,\sg)} +Cr^{-1}\|\sd\omega\|_{L^p(S,\sg)}
\label{5.124}
\end{equation}
Thus, in view of \ref{5.114} with $q=p$, the bound for \ref{5.119}
implies a bound for
\begin{equation}
r^{2-(2/p)}\|\snab^{ \ 2}\psi^\prime\|_{L^p(S,\sg)} \label{5.125}
\end{equation}
for every $2\leq p<\infty$ by a numerical constant depending only
on $p$. Therefore, in view of \ref{5.113} we shall obtain a bound
for $k_{2,4}$ once we derive a bound for:
\begin{equation}
r^{3/2}\|\snab^{ \ 2}\omega\|_{L^4(S,\sg)} \label{5.126}
\end{equation}

To derive bounds for \ref{5.118} and \ref{5.126} in terms of
$\|K-\overline{K}\|_{L^4(S,\sg)}$, we consider again equation
\ref{5.71}. Now, the equation
\begin{equation}
\slap_{\sg}\omega=\rho \label{5.127}
\end{equation}
for a function $\omega$ on $S$ is conformally covariant; that is,
if $\sg=e^{2\phi}\up{\sg}$ then $\omega$ satisfies:
\begin{equation}
\slap_{\up{\sg}}\omega=\up{\rho} \ \ \mbox{where} \
\up{\rho}=e^{2\phi}\rho \label{5.128}
\end{equation}
Now, with $\phi=\log r+\psi$ and $\psi$ being as in Proposition
5.4, $(S,\up{\sg})$ is isometric to the standard sphere, therefore
the standard $L^p$ Calderon-Zygmund inequalities hold on
$(S,\up{\sg})$:
\begin{equation}
\|\up{\snab^{ \
2}}\omega\|_{L^p(S,\up{\sg})}+\|\sd\omega\|_{L^p(S,\up{\sg})}\leq
C_p\|\up{\rho}\|_{L^p(S,\up{\sg})} \label{5.129}
\end{equation}
for any $1<p<\infty$, the constant $C_p$ depending only on $p$.

We have:
\begin{equation}
|\up{\rho}|^p d\mu_{\up{\sg}}=e^{(2p-2)\phi}|\rho|^p d\mu_{\sg}
=r^{2p-2}e^{(2p-2)\psi}|\rho|^p d\mu_{\sg} \label{5.130}
\end{equation}
hence:
\begin{equation}
\|\up{\rho}\|_{L^p(S,\up{\sg})}\leq
r^{2-(2/p)}e^{(2-(2/p))\psi_M}\|\rho\|_{L^p(S,\up{\sg})}
\label{5.131}
\end{equation}
Here and in the following we denote:
\begin{equation}
\psi_M=\sup_S\psi, \ \ \ \psi_m=\inf_S\psi, \ \ \
\mbox{osc}\psi=\psi_M-\psi_m \label{5.132}
\end{equation}

If $\xi$ is a $q$-covariant tensorfield on $S$, then
\begin{equation}
|\xi|^p_{\sg}d\mu_{\sg}=e^{-(pq-2)\phi}|\xi|^p_{\up{\sg}}d\mu_{\up{\sg}}=
r^{-(pq-2)}e^{-(pq-2)\psi}|\xi|^p_{\up{\sg}}d\mu_{\up{\sg}}
\label{5.133}
\end{equation}
It follows that if $pq\geq 2$ we have:
\begin{eqnarray}
&&r^{-(q-(2/p))}e^{-(q-(2/p))\psi_M}\|\xi\|_{L^p(S,\up{\sg})}\nonumber\\
&&\hspace{3cm}\leq\|\xi\|_{L^p(S,\sg)}\leq\label{5.134}\\
&&\hspace{4.5cm}r^{-(q-(2/p))}e^{-(q-(2/p))\psi_m}\|\xi\|_{L^p(S,\up{\sg})}\nonumber
\end{eqnarray}
Applying this to $\xi=\sd\omega$ $(q=1)$ and to $\xi=\up{\snab^{ \
2}}\omega$ $(q=2)$ we obtain, for $2\leq p<\infty$,
\begin{equation}
\|\sd\omega\|_{L^p(S,\sg)}\leq
r^{-(1-(2/p))}e^{-(1-(2/p))\psi_m}\|\sd\omega\|_{L^p(S,\up{\sg})}
\label{5.135}
\end{equation}
\begin{equation}
\|\up{\snab^{ \ 2}}\omega\|_{L^p(S,\sg)}\leq
r^{-(2-(2/p))}e^{-(2-(2/p))\psi_m}\|\up{\snab}^2\omega\|_{L^p(S,\up{\sg})}
\label{5.136}
\end{equation}
Now, we have:
\begin{equation}
\snab^{ \ 2}\omega=\up{\snab^{ \
2}}\omega+\up{\triangle}\cdot\sd\omega \label{5.137}
\end{equation}
where $\up{\triangle}=\up{\sGamma}-\sGamma$ is the difference of
the connections associated to $\up{\sg}$ and $\sg$, a type $T^1_2$
tensorfield on $S$, symmetric in the lower entries, and expressed
in local coordinates on $S$ by:
\begin{eqnarray}
\up{\triangle^C}_{AB}&=&-\delta_A^C\partial_B\phi-\delta_B^C\partial_B\phi+(\sg^{-1})^{CD}\sg_{AB}\partial_D\phi\nonumber\\
&=&-\delta_A^C\partial_B\psi-\delta_B^C\partial_B\psi+(\sg^{-1})^{CD}\sg_{AB}\partial_D\psi
\label{5.138}
\end{eqnarray}
We have, pointwise:
\begin{equation}
|\up{\triangle}|_{\sg}\leq C|\sd\psi|_{\sg} \label{5.139}
\end{equation}
It follows that:
\begin{eqnarray}
\|\up{\triangle}\cdot\sd\omega\|_{L^p(S,\sg)}&\leq& C\|\sd\psi\|_{L^{2p}(S,\sg)}\|\sd\omega\|_{L^{2p}(S,\sg)}\nonumber\\
&=&Cr^{-1+(1/p)}k_{1,2p}\|\sd\omega\|_{L^{2p}(S,\sg)}
\label{5.140}
\end{eqnarray}
From \ref{5.137}, \ref{5.136}, \ref{5.129} and \ref{5.131} we then
obtain:
\begin{eqnarray}
\|\snab^{ \ 2}\omega\|_{L^p(S,\sg)}&\leq&C_p e^{(2-(2/p))\mbox{osc}\psi}\|\rho\|_{L^p(S,\sg)}\nonumber\\
&\s&+Cr^{-1+(1/p)}k_{1,2p}\|\sd\omega\|_{L^{2p}(S,\sg)}
\label{5.141}
\end{eqnarray}
We apply this to equation \ref{5.71} taking $p=4$. Here
$\rho=K-\overline{K}$ and from Proposition 5.4 we have:
\begin{equation}
\sup_{S}|\psi|\leq C, \ \ \ k_{1,8}\leq C \label{5.142}
\end{equation}
while from \ref{5.83} with $q=8$:
\begin{eqnarray}
\|\sd\omega\|_{L^8(S,\sg)}&\leq& Cr^{-7/4}\{C\delta^{1/2}\scR_2(\alpha)+O(\delta)\}\nonumber\\
&\leq& C^\prime \label{5.143}
\end{eqnarray}
provided that $\delta$ is suitably small depending on ${\cal
D}_0^\infty$, ${\cal R}_0^\infty$, $\scD_1^4$, $\scR_1^4$, and
$\scR_2(\alpha)$. We then obtain:
\begin{equation}
r^{3/2}\|\snab^{ \ 2}\omega\|_{L^4(S,\sg)}\leq
Cr^{3/2}\|K-\overline{K}\|_{L^4(S,\sg)}+C r^{-1} \label{5.144}
\end{equation}

Combining \ref{5.144} with the bound for \ref{5.125} for $p=4$
yields:
\begin{equation}
k_{2,4}\leq C\{1+r^{3/2}\|K-\overline{K}\|_{L^4(S,\sg)}\}
\label{5.145}
\end{equation}
Also, by Lemma 5.2 with $p=4$ applied to $\sd\omega$, Lemma 5.4,
and the estimates \ref{5.83} with $q=4$ and \ref{5.144} we obtain:
\begin{equation}
r\sup_S|\sd\omega|\leq Cr^{3/2}\|K-\overline{K}\|_{L^4(S,\sg)} +C
r^{-1} \label{5.146}
\end{equation}
hence, combining with the bound for \ref{5.117} yields:
\begin{equation}
k_{1,\infty}\leq C\{1+r^{3/2}\|K-\overline{K}\|_{L^4(S,\sg)}\}
\label{5.147}
\end{equation}

\section{$L^p$ elliptic theory on $S$}

Consider now the equation
\begin{equation}
\sdiv_{\sg}\theta=f \label{5.148}
\end{equation}
for a trace-free symmetric 2-covariant tensorfield $\theta$ on
$S$. Here $f$ is a given 1-form on $S$. This is an elliptic
equation for $\theta$. In fact, the following integral identity
holds (see Chapter 2 of [C-K]):
\begin{equation}
\int_S\{|\snab\theta|^2_{\sg}+2K|\theta|^2_{\sg}\}d\mu_{\sg}=2\int_S|f|^2_{\sg}d\mu_{\sg}
\label{5.149}
\end{equation}
Equation \ref{5.148} is conformally covariant; that is, if
$\sg=e^{2\phi}\up{\sg}$ then $\theta$ satisfies:
\begin{equation}
\sdiv_{\up{\sg}}\theta=\up{f} \ \ \mbox{where} \ \
\up{f}=e^{2\phi}f \label{5.150}
\end{equation}
Recall that with $\phi=\log r+\psi$ and $\psi$ being as in
Proposition 5.4, $(S,\up{\sg})$ is isometric to the standard
sphere. The operator $\sdiv_{\up{\sg}}$ from trace-free symmetric
2-covariant tensorfields to 1-forms on the standard sphere being a
first order elliptic operator with vanishing kernel, the following
$L^p$ Calderon-Zygmund inequalities hold on $(S,\up{\sg})$:
\begin{equation}
\|\up{\snab}\theta\|_{L^p(S,\up{\sg})}+\|\theta\|_{L^p(S,\up{\sg})}\leq
C_p\|\up{f}\|_{L^p(S,\up{\sg})} \label{5.151}
\end{equation}
and:
\begin{equation}
\|\up{\snab^{ \ 2}}\theta\|_{L^p(S,\up{\sg})}\leq
C_p\{\|\up{\snab}\up{f}\|_{L^p(S,\up{\sg})}+\|\up{f}\|_{L^p(S,\up{\sg})}\}
\label{5.152}
\end{equation}
for any $1<p<\infty$, the constants $C_p$ depending only on $p$.

Applying \ref{5.134} to $\xi=\theta$ $(q=2)$, to
$\xi=\up{\snab}\theta$ $(q=3)$, and to $\xi=\up{\snab^{ \
2}}\theta$ $(q=4)$, we obtain:
\begin{eqnarray}
\|\theta\|_{L^p(S,\sg)}&\leq& r^{-(2-(2/p))}e^{-(2-(2/p))\psi_m}\|\theta\|_{L^p(S,\up{\sg})}\nonumber\\
\|\up{\snab}\theta\|_{L^p(S,\sg)}&\leq& r^{-(3-(2/p))}e^{-(3-(2/p))\psi_m}\|\up{\snab}\theta\|_{L^p(S,\up{\sg})}\nonumber\\
\|\up{\snab^{ \ 2}}\theta\|_{L^p(S,\sg)}&\leq&
r^{-(4-(2/p))}e^{-(4-(2/p))\psi_m} \|\up{\snab^{ \
2}}\theta\|_{L^p(S,\up{\sg})} \label{5.153}
\end{eqnarray}
Also, applying \ref{5.134} to $\xi=\up{f}$ $(q=1)$ and to
$\xi=\up{\snab}\up{f}$ $(q=2)$ we obtain, for $p\geq 2$,
\begin{eqnarray}
\|\up{f}\|_{L^p(S,\up{\sg})}&\leq& r^{1-(2/p)}e^{(1-(2/p))\psi_M}\|\up{f}\|_{L^p(S,\sg)}\nonumber\\
\|\up{\snab}\up{f}\|_{L^p(S,\up{\sg})}&\leq&
r^{2-(2/p)}e^{(1-(2/p))\psi_M}\|\up{\snab}\up{f}\|_{L^p(S,\sg)}
\label{5.154}
\end{eqnarray}
Moreover, since $\up{f}=r^2 e^{2\psi}f$, we have:
$$\up{\snab}\up{f}=r^2 e^{2\psi}(\up{\snab}f+2\sd\psi\otimes f)$$
hence, recalling the definition \ref{5.115}:
\begin{eqnarray}
\|\up{f}\|_{L^p(S,\sg)}&\leq& r^2 e^{2\psi_M}\|f\|_{L^p(S,\sg)}\nonumber\\
\|\up{\snab}\up{f}\|_{L^p(S,\sg)}&\leq& r^2
e^{2\psi_M}\{\|\up{\snab}f\|_{L^p(S,\sg)}
+2r^{-1}k_{1,\infty}\|f\|_{L^p(S,\sg)}\}\label{5.155}
\end{eqnarray}
In view of \ref{5.153} - \ref{5.155} the inequalities \ref{5.151}
and \ref{5.152} imply, for any $2\leq p<\infty$:
\begin{equation}
\|\theta\|_{L^p(S,\sg)}\leq C_p r
e^{(2-(2/p))\mbox{osc}\psi}e^{\psi_M}\|f\|_{L^p(S,\sg)}
\label{5.156}
\end{equation}
\begin{equation}
\|\up{\snab}\theta\|_{L^p(S,\sg)}\leq C_p
e^{(3-(2/p))\mbox{osc}\psi}\|f\|_{L^p(S,\sg)} \label{5.157}
\end{equation}
and:
\begin{eqnarray}
&&\|\up{\snab^{ \ 2}}\theta\|_{L^p(S,\sg)}\leq\nonumber\\
&&\hspace{1cm}C_p e^{(4-(2/p))\mbox{osc}\psi}
\left\{\|\up{\snab}f\|_{L^p(S,\sg)}+r^{-1}(e^{-\psi_M}+k_{1,\infty})\|f\|_{L^p(S,\sg)}\right\}\nonumber\\
&&\label{5.158}
\end{eqnarray}

We have, in arbitrary local coordinates on $S$:
\begin{equation}
\snab_A\theta_{BC}-\up{\snab}_A\theta_{BC}=\up{\triangle^D}_{AB}\theta_{DC}+\up{\triangle^D}_{AC}\theta_{BD}
\label{5.159}
\end{equation}
where $\up{\triangle}$, the difference of connections associated
to $\up{\sg}$ and $\sg$ is given by \ref{5.138}. It follows that:
\begin{equation}
\|\snab\theta\|_{L^p(S,\sg)}\leq \|\up{\snab}\theta\|_{L^p(S,\sg)}
+C\|\up{\triangle}\|_{L^\infty(S,\sg)}\|\theta\|_{L^p(S,\sg)}
\label{5.160}
\end{equation}
and from \ref{5.139} and the definition \ref{5.115} we obtain:
\begin{equation}
\|\up{\triangle}\|_{L^\infty(S,\sg)}\leq Cr^{-1} k_{1,\infty}
\label{5.161}
\end{equation}
From \ref{5.160}, \ref{5.156} and \ref{5.157} we then obtain:
\begin{equation}
\|\snab\theta\|_{L^p(S)}\leq C_p
e^{(3-(2/p))\mbox{osc}\psi}(1+e^{\psi_m}k_{1,\infty})\|f\|_{L^p(S,\sg)}
\label{5.162}
\end{equation}

Next, we have:
\begin{eqnarray*}
&&\snab_A\snab_B\theta_{CD}-\up{\snab}_A\up{\snab}_B\theta_{CD}\\
&&\hspace{1cm}=
\snab_A(\up{\snab}_B\theta_{CD}+\up{\triangle^E}_{BC}\theta_{ED}+\up{\triangle^E}_{BD}\theta_{CE})
-\up{\snab}_A\up{\snab}_B\theta_{CD}\\
&&\hspace{1cm}=\up{\triangle^E}_{AB}\up{\snab}_E\theta_{CD}+\up{\triangle^E}_{AC}\up{\snab}_B\theta_{ED}
+\up{\triangle^E}_{AD}\up{\snab}_B\theta_{CE}\\
&&\hspace{1cm}+\up{\triangle^E}_{BC}\snab_A\theta_{ED}+\up{\triangle^E}_{BD}\snab_A\theta_{CE}
+\theta_{ED}\snab_A\up{\triangle^E}_{BC}+\theta_{CE}\snab_A\up{\triangle^E}_{BD}
\end{eqnarray*}
Substituting from \ref{5.159} we then obtain:
\begin{equation}
\snab^{ \ 2}\theta-\up{\snab^{ \
2}}\theta=(\up{\triangle},\snab\theta)+(\snab\up{\triangle},\theta)
+(\up{\triangle},\up{\triangle},\theta) \label{5.163}
\end{equation}
where:
\begin{eqnarray}
(\up{\triangle},\snab\theta)_{ABCD}&=&\up{\triangle^E}_{AB}\snab_E\theta_{CD}+\up{\triangle^E}_{AC}\snab_B\theta_{ED}
+\up{\triangle^E}_{AD}\snab_B\theta_{CE}\nonumber\\
&\s&+\up{\triangle^E}_{BC}\snab_A\theta_{ED}+\up{\triangle^E}_{BD}\snab_A\theta_{CE}\nonumber\\
(\snab\up{\triangle},\theta)_{ABCD}&=&\theta_{ED}\snab_A\up{\triangle^E}_{BC}+\theta_{CE}\snab_A\up{\triangle^E}_{BD}
\nonumber\\
(\up{\triangle},\up{\triangle},\theta)_{ABCD}&=&-(\up{\triangle^E}_{AB}\up{\triangle^F}_{EC}
+\up{\triangle^E}_{AC}\up{\triangle^F}_{BE})\theta_{FD}\nonumber\\
&\s&-(\up{\triangle^E}_{AB}\up{\triangle^F}_{ED}+\up{\triangle^E}_{AD}\up{\triangle^F}_{BE})\theta_{CF}\nonumber\\
&\s&-(\up{\triangle^E}_{AC}\up{\triangle^F}_{BD}+\up{\triangle^E}_{AD}\up{\triangle^F}_{BC})\theta_{EF}
\label{5.164}
\end{eqnarray}
Now, from \ref{5.161},
\begin{eqnarray}
\|(\up{\triangle},\snab\theta)\|_{L^p(S,\sg)}&\leq&
C\|\up{\triangle}\|_{L^\infty(S,\sg)}\|\snab\theta\|_{L^p(S,\sg)}
\nonumber\\
&\leq& C^\prime r^{-1}k_{1,\infty}\|\snab\theta\|_{L^p(S,\sg)}
\label{5.165}
\end{eqnarray}
Moreover,
$$\|(\snab\up{\triangle},\theta)\|_{L^p(S,\sg)}\leq C\|\snab\up{\triangle}\|_{L^p(S,\sg)}\|\theta\|_{L^\infty(S,\sg)}$$
and from \ref{5.138} and the definition \ref{5.116},
\begin{equation}
\|\snab\up{\triangle}\|_{L^p(S,\sg)}\leq Cr^{-(2-(2/p))}k_{2,p}
\label{5.166}
\end{equation}
hence:
\begin{equation}
\|(\snab\up{\triangle},\theta)\|_{L^p(S,\sg)}\leq
Cr^{-(2-(2/p))}k_{2,p}\|\theta\|_{L^\infty(S,\sg)} \label{5.167}
\end{equation}
Also,
$$\|(\up{\triangle},\up{\triangle},\theta)\|_{L^p(S,\sg)}\leq C\|\up{\triangle}\|^2_{L^{2p}(S,\sg)}
\|\theta\|_{L^\infty(S,\sg)}$$ and from \ref{5.138} and the
definition \ref{5.112},
\begin{equation}
\|\up{\triangle}\|_{L^{2p}(S,\sg)}\leq Cr^{-(1-(1/p))}k_{1,2p}
\label{5.168}
\end{equation}
hence:
\begin{equation}
\|(\up{\triangle},\up{\triangle},\theta)\|_{L^p(S,\sg)}\leq
Cr^{-(2-(2/p))}(k_{1,2p})^2\|\theta\|_{L^\infty(S,\sg)}
\label{5.169}
\end{equation}
In view of \ref{5.165}, \ref{5.167}, \ref{5.169}, we conclude from
\ref{5.163} that:
\begin{eqnarray}
&&\|\snab^{ \ 2}\theta-\up{\snab^{ \ 2}}\theta\|_{L^p(S,\sg)}\leq Cr^{-1}k_{1,\infty}\|\snab\theta\|_{L^p(S,\sg)}\nonumber\\
&&\hspace{30mm}+Cr^{-(2-(2/p))}\left[k_{2,p}+(k_{1,2p})^2\right]\|\theta\|_{L^\infty(S,\sg)}
\label{5.170}
\end{eqnarray}

We also have:
\begin{equation}
\up{\snab}_A f_B=\snab_A f_B-\up{\triangle^C}_{AB}f_C
\label{5.171}
\end{equation}
hence:
\begin{eqnarray}
\|\up{\snab}f\|_{L^p(S,\sg)}&\leq&\|\snab f\|_{L^p(S,\sg)}+\|\up{\triangle}\|_{L^\infty(S,\sg)}\|f\|_{L^p(S,\sg)}\nonumber\\
&\leq&\|\snab
f\|_{L^p(S,\sg)}+Cr^{-1}k_{1,\infty}\|f\|_{L^p(S,\sg)}
\label{5.172}
\end{eqnarray}

In view of \ref{5.170} and \ref{5.172} we conclude from
\ref{5.158} that:
\begin{eqnarray}
&&\|\snab^{ \ 2}\theta\|_{L^p(S,\sg)}\leq\nonumber\\
&&\hspace{1cm}C_p e^{(4-(2/p))\mbox{osc}\psi}
\left\{\|\snab f\|_{L^p(S,\sg)}+r^{-1}(e^{-\psi_M}+k_{1,\infty})\|f\|_{L^p(S,\sg)}\right\}\nonumber\\
&&\hspace{1cm}+C\left\{r^{-1}k_{1,\infty}\|\snab\theta\|_{L^p(S,\sg)}
+r^{-(2-(2/p))}\left[k_{2,p}+(k_{1,2p})^2\right]\|\theta\|_{L^\infty(S,\sg)}\right\}\nonumber\\
&&\label{5.173}
\end{eqnarray}
We have thus established the following lemma.

\vspace{5mm}

\noindent{\bf Lemma 5.6} \ \ \ Let $\theta$ be a trace-free
symmetric 2-covariant tensorfield on $S$ satisfying the equation
$$\sdiv_{\sg}\theta=f$$
where $f$ is a given 1-form on $S$. Then with $\psi$ being the
function which appears in Proposition 5.4 we have, for every
$2\leq p<\infty$ the estimates:
$$\|\theta\|_{L^p(S,\sg)}\leq C_p r e^{(2-(2/p))\mbox{osc}\psi}e^{\psi_M}\|f\|_{L^p(S,\sg)}$$
$$\|\snab\theta\|_{L^p(S)}\leq C_p e^{(3-(2/p))\mbox{osc}\psi}(1+e^{\psi_m}k_{1,\infty})\|f\|_{L^p(S,\sg)}$$
\begin{eqnarray*}
&&\|\snab^{ \ 2}\theta\|_{L^p(S,\sg)}\leq\\
&&\hspace{1cm}C_p e^{(4-(2/p))\mbox{osc}\psi}
\left\{\|\snab f\|_{L^p(S,\sg)}+r^{-1}(e^{-\psi_M}+k_{1,\infty})\|f\|_{L^p(S,\sg)}\right\}\\
&&\hspace{1cm}+C\left\{r^{-1}k_{1,\infty}\|\snab\theta\|_{L^p(S,\sg)}
+r^{-(2-(2/p))}\left[k_{2,p}+(k_{1,2p})^2\right]\|\theta\|_{L^\infty(S,\sg)}\right\}
\end{eqnarray*}
$C$ and $C_p$ being numerical constants, the last depending on
$p$.

\vspace{5mm}

Setting now $p=4$, we have the bounds \ref{5.142}, \ref{5.145} and
\ref{5.147}. Substituting these bounds in Lemma 5.6 and taking
into account the fact that from Lemma 4.3
\begin{equation}
\frac{1}{\sqrt{2}}\leq\frac{r}{|u|}\leq\sqrt{2} \label{5.174}
\end{equation}
we deduce the following lemma.

\vspace{5mm}

\noindent{\bf Lemma 5.7} \ \ \ Let $\theta$ be a trace-free
2-covariant $S$ tensorfield in $M^\prime$ satisfying the equation
$$\sdiv\theta=f$$
where $f$ is a given $S$ 1-form. Then, provided that $\delta$ is
suitably small depending on ${\cal D}_0^\infty$, ${\cal
R}_0^\infty$, $\scD_1^4$, $\scR_1^4$, and $\scR_2(\alpha)$,
$\theta$ satisfies the estimate:
\begin{eqnarray*}
&&\|\snab^{ \ 2}\theta\|_{L^4(S_{\ub,u})}\leq C\|\snab f\|_{L^4(S_{\ub,u})}\\
&&\hspace{1cm}+C\{1+|u|^{3/2}\|K-\overline{K}\|_{L^4(S_{\ub,u})}\}\cdot\\
&&\hspace{2cm}\cdot\left\{|u|^{-1}\left(\|f\|_{L^4(S_{\ub,u})}
+\|\snab\theta\|_{L^4(S_{\ub,u})}\right)+|u|^{-3/2}\|\theta\|_{L^\infty(S_{\ub,u})}\right\}
\end{eqnarray*}
for every $(\ub,u)\in D^\prime$. Here $C$ is a numerical constant.

\vspace{5mm}

Lemma 5.7 will be applied to the system consisting of the propagation equations for $K-\overline{K}$, $\mbox{tr}\chi^\prime$ 
and the Codazzi elliptic system for $\chih^\prime$. For this application it is crucial that the quantity 
$\|K-\overline{K}\|_{L^4(S_{\ub,u})}$ enters the estimate linearly. After the appropriate estimate for this quantity has been established, 
a different version of the lemma, the version given below as Lemma 5.9, will be applied. 

From Lemma 5.6 we have, for every $2\leq p<\infty$:
\begin{equation}
\|\theta\|_{W_1^p(S,\sg)}\leq C_p
e^{(3-(2/p))\mbox{osc}\psi}\left[1+e^{\psi_m}(1+k_{1,\infty})\right]\|f\|_{L^p(S,\sg)}
\label{5.175}
\end{equation}
and if $p>2$ we have, by Lemma 5.2:
\begin{equation}
\|\theta\|_{L^\infty(S,\sg)}\leq
C_p\sqrt{I^\prime(S,\sg)}r^{1-(2/p)}\|\theta\|_{W_1^p(S,\sg)}
\label{5.176}
\end{equation}
Hence, if $p>2$ we have:
\begin{equation}
\|\theta\|_{L^\infty(S,\sg)}\leq C_p
\sqrt{I^\prime(S,\sg)}r^{1-(2/p)}
e^{(3-(2/p))\mbox{osc}\psi}\left[1+e^{\psi_m}(1+k_{1,\infty})\right]\|f\|_{L^p(S,\sg)}
\label{5.177}
\end{equation}
Substituting \ref{5.175}, \ref{5.177} in the last estimate of Lemma 5.6 we obtain the
following version of that lemma.

\vspace{5mm}

\noindent{\bf Lemma 5.8} \ \ \ Let $\theta$ be a trace-free
symmetric 2-covariant tensorfield on $S$ satisfying the equation
$$\sdiv_{\sg}\theta=f$$
where $f$ is a given 1-form on $S$. Then with $\psi$ being the
function which appears in Proposition 5.4 we have, for every
$2<p<\infty$ the estimate:
\begin{eqnarray*}
&&r\|\snab^{ \ 2}\theta\|_{L^p(S,\sg)}+\|\snab\theta\|_{L^p(S,\sg)}+r^{-1}\|\theta\|_{L^p(S,\sg)}\\
&&\hspace{2.5cm}\leq C_p
e^{(3-(2/p))\mbox{osc}\psi}\left\{e^{\mbox{osc}\psi}r\|\snab
f\|_{L^p(S,\sg)} +l_{2,p}\|f\|_{L^p(S,\sg)}\right\}
\end{eqnarray*}
where
\begin{eqnarray*}
&&l_{2,p}=e^{-\psi_m}+e^{\mbox{osc}\psi}k_{1,\infty}+k_{1,\infty}(1+e^{\psi_m}k_{1,\infty})\\
&&\hspace{1cm}+\left(1+e^{\psi_m}(1+k_{1,\infty})\right)\left\{1+\sqrt{I^\prime(S,\sg)}\left[k_{2,p}+(k_{1,2p})^2\right]
\right\}
\end{eqnarray*}
and $C_p$ is a numerical constant depending only on $p$.

\vspace{5mm}

Setting $p=4$, substituting the bounds \ref{5.142}, \ref{5.145},
\ref{5.147}, and taking into account \ref{5.174} as well as Lemma
5.4 we deduce the following.

\vspace{5mm}

\noindent{\bf Lemma 5.9} \ \ \ Let $\theta$ be a trace-free
2-covariant $S$ tensorfield in $M^\prime$ satisfying the equation
$$\sdiv\theta=f$$
where $f$ is a given $S$ 1-form. Then, provided that $\delta$ is
suitably small depending on ${\cal D}_0^\infty$, ${\cal
R}_0^\infty$, $\scD_1^4$, $\scR_1^4$, and $\scR_2(\alpha)$,
$\theta$ satisfies the estimate:
\begin{eqnarray*}
&&|u|\|\snab^{ \ 2}\theta\|_{L^4(S_{\ub,u})}+\|\snab\theta\|_{L^4(S_{\ub,u})}+|u|^{-1}\|\theta\|_{L^4(S_{\ub,u})}\\
&&\hspace{1cm}\leq C\left\{|u|\|\snab f\|_{L^4(S_{\ub,u})}
+\left(1+|u|^{3/2}\|K-\overline{K}\|_{L^4(S_{\ub,u})}\right)^2\|f\|_{L^4(S_{\ub,u})}\right\}
\end{eqnarray*}
for all $(\ub,u)\in D^\prime$.

\vspace{5mm}

Consider now the system of equations:
\begin{eqnarray}
\sdiv_{\sg}\theta&=&f\nonumber\\
\scurl_{\sg}\theta&=&g \label{5.178}
\end{eqnarray}
for a 1-form $\theta$ on $S$. Here $(f,g)$ is a given pair of
functions on $S$. This is an elliptic system for $\theta$. In
fact, the following integral identity holds (see Chapter 2 of
[C-K]):
\begin{equation}
\int_S\{|\snab\theta|_{\sg}^2+K|\theta\|_{\sg}^2\}d\mu_{\sg}=\int_S\{f^2+g^2\}d\mu_{\sg}
\label{5.179}
\end{equation}
The system \ref{5.178} is conformally covariant; that is, if
$\sg=e^{2\phi}\up{\sg}$ then $\theta$ satisfies:
\begin{eqnarray}
\sdiv_{\up{\sg}}\theta&=&\up{f}\nonumber\\
\scurl_{\up{\sg}}\theta&=&\up{g} \ \ \ \mbox{where} \ \
\up{f}=e^{2\phi}f, \ \up{g}=e^{2\phi}g \label{5.180}
\end{eqnarray}
Recall again that with $\phi=\log r+\psi$ and $\psi$ being as in
Proposition 5.4, $(S,\up{\sg})$ is isometric to the standard
sphere. The operator $(\sdiv_{\up{\sg}}, \scurl_{\up{\sg}})$ from
1-forms to pairs of functions on the standard sphere being a
first order elliptic operator with vanishing kernel, the following
$L^p$ Calderon-Zygmund inequalities hold on $(S,\up{\sg})$:
\begin{equation}
\|\up{\snab}\theta\|_{L^p(S,\up{\sg})}+\|\theta\|_{L^p(S,\up{\sg})}\leq
C_p\{\|\up{f}\|_{L^p(S,\up{\sg})}+\|\up{g}\|_{L^p(S,\up{\sg})}\}
\label{5.181}
\end{equation}
and:
\begin{equation}
\|\up{\snab^{ \ 2}}\theta\|_{L^p(S,\up{\sg})}\leq
C_p\{\|\sd\up{f}\|_{L^p(S,\up{\sg})}+\|\sd\up{g}\|_{L^p(S,\up{\sg})}
+\|\up{f}\|_{L^p(S,\up{\sg})}+\|\up{g}\|_{L^p(S,\up{\sg})}\}
\label{5.182}
\end{equation}

Applying \ref{5.134} to $\xi=\theta$ $(q=1)$, to
$\xi=\up{\snab}\theta$ $(q=2)$, and to $\xi=\up{\snab^{ \
2}}\theta$ $(q=3)$ we obtain, for $p\geq 2$,
\begin{eqnarray}
\|\theta\|_{L^p(S,\sg)}&\leq&r^{-(1-(2/p))}e^{-(1-(2/p))\psi_m}\|\theta\|_{L^p(S,\up{\sg})}\nonumber\\
\|\up{\snab}\theta\|_{L^p(S,\sg)}&\leq&r^{-(2-(2/p))}e^{-(2-(2/p))\psi_m}\|\up{\snab}\theta\|_{L^p(S,\up{\sg})}\nonumber\\
\|\up{\snab^{ \
2}}\theta\|_{L^p(S,\sg)}&\leq&r^{-(3-(2/p))}e^{-(3-(2/p))\psi_m}\|\up{\snab^{
\ 2}}\theta\|_{L^p(S,\up{\sg})} \label{5.183}
\end{eqnarray}
Also, since $\up{f}=r^2 e^{2\psi}f$, $\up{g}=r^2 e^{2\psi}g$ we
have:
\begin{eqnarray}
\|\up{f}\|_{L^p(S,\up{\sg})}&\leq& r^{2-(2/p)}e^{(2-(2/p))\psi_M}\|\up{f}\|_{L^p(S,\sg)}\nonumber\\
\|\up{g}\|_{L^p(S,\up{\sg})}&\leq&
r^{2-(2/p)}e^{(2-(2/p))\psi_M}\|\up{g}\|_{L^p(S,\sg)}
\label{5.184}
\end{eqnarray}
Also, applying \ref{5.134} to $\xi=\sd\up{f}, \sd\up{g}$ $(q=1)$
we obtain, for $q\geq 2$,
\begin{eqnarray*}
\|\sd\up{f}\|_{L^p(S,\up{\sg})}&\leq& r^{1-(2/p)}e^{(1-(2/p))\psi_M}\|\sd\up{f}\|_{L^p(S,\sg)}\\
\|\sd\up{g}\|_{L^p(S,\up{\sg})}&\leq&
r^{1-(2/p)}e^{(1-(2/p))\psi_M}\|\sd\up{g}\|_{L^p(S,\sg)}
\end{eqnarray*}
which, since
$$\sd\up{f}=r^2 e^{2\psi}(\sd f +2f\sd\psi), \ \ \ \sd\up{g}=r^2 e^{2\psi}(\sd g +2g\sd\psi),$$
yields, recalling the definition \ref{5.115}:
\begin{eqnarray}
\|\sd\up{f}\|_{L^p(S,\up{\sg})}&\leq&
r^{3-(2/p)}e^{(3-(2/p))\psi_M}\{\|\sd f\|_{L^p(S,\sg)}
+2r^{-1}k_{1,\infty}\|f\|_{L^p(S,\sg)}\}\nonumber\\
\|\sd\up{g}\|_{L^p(S,\up{\sg})}&\leq&
r^{3-(2/p)}e^{(3-(2/p))\psi_M}\{\|\sd g\|_{L^p(S,\sg)}
+2r^{-1}k_{1,\infty}\|g\|_{L^p(S,\sg)}\}\nonumber\\
&\s&\label{5.185}
\end{eqnarray}
In view of \ref{5.183} - \ref{5.185} the inequalities \ref{5.181}
and \ref{5.182} imply, for any $2\leq p<\infty$:
\begin{equation}
\|\theta\|_{L^p(S,\sg)}\leq C_p
re^{(1-(2/p))\mbox{osc}\psi}e^{\psi_M}\{\|f\|_{L^p(S,\sg)}+\|g\|_{L^p(S,\sg)}\}
\label{5.186}
\end{equation}
\begin{equation}
\|\up{\snab}\theta\|_{L^p(S,\sg)}\leq C_p
e^{(2-(2/p))\mbox{osc}\psi}\{\|f\|_{L^p(S,\sg)}+\|g\|_{L^p(S,\sg)}\}
\label{5.187}
\end{equation}
and:
\begin{eqnarray}
&&\|\up{\snab^{ \ 2}}\theta\|_{L^p(S,\sg)}\leq
C_p e^{(3-(2/p))\mbox{osc}\psi}\left\{\|\sd f\|_{L^p(S,\sg)}+\|\sd g\|_{L^p(S,\sg)}\right.\nonumber\\
&&\hspace{4cm} \left.
+r^{-1}(e^{-\psi_M}+k_{1,\infty})(\|f\|_{L^p(S,\sg)}+\|g\|_{L^p(S,\sg)})\right\}
\nonumber\\
&&\label{5.188}
\end{eqnarray}

We have, in arbitrary local coordinates on $S$:
\begin{equation}
\snab_A\theta_B-\up{\snab}_A\theta_B=\up{\triangle^C}_{AB}\theta_C
\label{5.189}
\end{equation}
Hence:
\begin{equation}
\|\snab\theta\|_{L^p(S,\sg)}\leq\|\up{\snab}\theta\|_{L^p(S,\sg)}+\|\up{\triangle}\|_{L^\infty(S,\sg)}\|\theta\|_{L^p(S,\sg)}
\label{5.190}
\end{equation}
From \ref{5.186}, \ref{5.187} and \ref{5.161} we then obtain:
\begin{equation}
\|\snab\theta\|_{L^p(S,\sg)}\leq C_p
e^{(2-(2/p))\mbox{osc}\psi}(1+e^{\psi_m}k_{1,\infty})\{\|f\|_{L^p(S,\sg)}
+\|g\|_{L^p(S,\sg)}\} \label{5.191}
\end{equation}

Also, proceeding as in the case of a trace-free 2-covariant
tensorfield on $S$, we deduce in the present case of a 1-form on
$S$, in analogy with \ref{5.170},
\begin{eqnarray}
\|\snab^{ \ 2}\theta-\up{\snab^{ \ 2}}\theta\|_{L^p(S,\sg)}&\leq&Cr^{-1}k_{1,\infty}\|\theta\|_{L^p(S,\sg)}\label{5.192}\\
&\s&+Cr^{-(2-(2/p))}\left[k_{2,p}+(k_{1,2p})^2\right]\|\theta\|_{L^\infty(S,\sg)}
\nonumber
\end{eqnarray}
which, together with \ref{5.188} yields:
\begin{eqnarray}
&&\|\snab^{ \ 2}\theta\|_{L^p(S,\sg)}\leq\nonumber\\
&&\hspace{1cm} C_p e^{(3-(2/p))\mbox{osc}\psi}
\left\{\|\sd f\|_{L^p(S,\sg)}+\|\sd g\|_{L^p(S,\sg)}\right.\nonumber\\
&&\hspace{4cm}\left.+r^{-1}(e^{-\psi_M}+k_{1,\infty})
\left(\|f\|_{L^p(S,\sg)}+\|g\|_{L^p(S,\sg)}\right)\right\}\nonumber\\
&&\hspace{1cm}+C\left\{r^{-1}k_{1,\infty}\|\snab\theta\|_{L^p(S,\sg)}
+r^{-(2-(2/p))}\left[k_{2,p}+(k_{1,2p})^2\right]\|\theta\|_{L^\infty(S,\sg)}\right\}\nonumber\\
&&\label{5.193}
\end{eqnarray}

Now, from \ref{5.186} and \ref{5.191} we have, for every $2\leq
p<\infty$:
\begin{eqnarray}
\|\theta\|_{W_1^p(S,\sg)}&\leq&C_p e^{(2-(2/p))\mbox{osc}\psi}\left[1+e^{\psi_m}(1+k_{1,\infty})\right]\cdot\nonumber\\
&\s&\hspace{3cm}\cdot\{\|f\|_{L^p(S,\sg)}+\|g\|_{L^p(S,\sg)}\}
\label{5.194}
\end{eqnarray}
and if $p>2$ we have, by Lemma 5.2:
\begin{equation}
\|\theta\|_{L^\infty(S,\sg)}\leq
C_p\sqrt{I^\prime(S,\sg)}r^{1-(2/p)}\|\theta\|_{W_1^p(S,\sg)}
\label{5.195}
\end{equation}
Hence, if $p>2$ we have:
\begin{eqnarray}
\|\theta\|_{L^\infty(S,\sg)}&\leq&C_p
\sqrt{I^\prime(S,\sg)}r^{1-(2/p)}
e^{(2-(2/p))\mbox{osc}\psi}\left[1+e^{\psi_m}(1+k_{1,\infty})\right]\cdot\nonumber\\
&\s&\hspace{3cm}\cdot\{\|f\|_{L^p(S,\sg)}+\|g\|_{L^p(S,\sg)}\}
\label{5.196}
\end{eqnarray}
Substituting this as well as \ref{5.191} in \ref{5.193} we obtain
the following lemma.

\vspace{5mm}

\noindent{\bf Lemma 5.10} \ \ \ Let $\theta$ be a 1-form on $S$
satisfying the system
\begin{eqnarray*}
\sdiv_{\sg}\theta&=&f\\
\scurl_{\sg}\theta&=&g
\end{eqnarray*}
where $f$ and $g$ are given functions on $S$. Then with $\psi$
being the function which appears in Proposition 5.4 we have, for
every $2<p<\infty$ the estimate:
\begin{eqnarray*}
&&r\|\snab^{ \ 2}\theta\|_{L^p(S,\sg)}+\|\snab\theta\|_{L^p(S,\sg)}+r^{-1}\|\theta\|_{L^p(S,\sg)}\\
&&\hspace{2.5cm}\leq C_p
e^{(2-(2/p))\mbox{osc}\psi}\left\{e^{\mbox{osc}\psi}
r\left(\|\sd f\|_{L^p(S,\sg)}+\|\sd g\|_{L^p(S,\sg)}\right)\right.\\
&&\hspace{6cm}\left.+l_{2,p}\left(\|f\|_{L^p(S,\sg)}+\|g\|_{L^p(S,\sg)}\right)\right\}
\end{eqnarray*}
where
\begin{eqnarray*}
&&l_{2,p}=e^{-\psi_m}+e^{\mbox{osc}\psi}k_{1,\infty}+k_{1,\infty}(1+e^{\psi_m}k_{1,\infty})\\
&&\hspace{1cm}+\left(1+e^{\psi_m}(1+k_{1,\infty})\right)\left\{1+\sqrt{I^\prime(S,\sg)}\left[k_{2,p}+(k_{1,2p})^2\right]
\right\}
\end{eqnarray*}
and $C_p$ is a numerical constant depending only on $p$.

\vspace{5mm}

Setting $p=4$, substituting the bounds \ref{5.142}, \ref{5.145},
\ref{5.147}, and taking into account \ref{5.174} as well as Lemma
5.4 we deduce the following.

\vspace{5mm}

\noindent{\bf Lemma 5.11} \ \ \ Let $\theta$ be a $S$ 1-form in
$M^\prime$ satisfying the system
\begin{eqnarray*}
\sdiv\theta&=&f\\
\scurl\theta&=&g
\end{eqnarray*}
where $f$ and $g$ are given functions. Then, provided that
$\delta$ is suitably small depending on ${\cal D}_0^\infty$,
${\cal R}_0^\infty$, $\scD_1^4$, $\scR_1^4$, and $\scR_2(\alpha)$,
$\theta$ satisfies the estimate:
\begin{eqnarray*}
&&|u|\|\snab^{ \ 2}\theta\|_{L^4(S_{\ub,u})}+\|\snab\theta\|_{L^4(S_{\ub,u})}+|u|^{-1}\|\theta\|_{L^4(S_{\ub,u})}\\
&&\hspace{1cm}\leq C\left\{|u|\left(\|\sd f\|_{L^4(S_{\ub,u})}+\|\sd g\|_{L^4(S_{\ub,u})}\right)\right.\\
&&\hspace{1.8cm}\left.+\left(1+|u|^{3/2}\|K-\overline{K}\|_{L^4(S_{\ub,u})}\right)^2
\left(\|f\|_{L^4(S_{\ub,u})}+\|g\|_{L^4(S_{\ub,u})}\right)\right\}
\end{eqnarray*}
for all $(\ub,u)\in D^\prime$.

\chapter{$L^4(S)$ Estimates for the 2nd
Derivatives of the Connection Coefficients}

\section{Introduction}

The optical estimates of Chapters 3 and 4 relied only on the
propagation equations among the optical structure equations. These
are ordinary differential equations along the generators of the
null hypersurfaces $C_u$ or $\Cb_{\ub}$. The estimates loose one
degree of differentiability, because of the presense of principal
terms on the right hand sides of the propagation equations. The
present chapter and the next, on the other hand, consider systems
of optical structure equations which consist of elliptic equations
on the $S_{\ub,u}$ sections of the null hypersurfaces $C_u$ or
$\Cb_{\ub}$ coupled to ordinary differential equations along the
generators of these hypersurfaces. The ordinary differential
equations considered here, in contrast to the remaining
propagation equations, do not contain principal terms on the right
hand side, {\em by virtue of the Einstein equations}. This
approach, which allows us to obtain optical estimates which are
optimal from the point of view of differentiability, was
introduced in [C-K] and plays a basic role in the present work as
well. There is however in this approach, as we shall see, a loss
of a factor of $\delta^{1/2}$ in behavior with respect to
$\delta$, in comparison to the estimates which rely only on the
propagation equations, in the case of $\eta$, $\etb$, and
$\omega$. What is crucial, is that there is no such loss in the
case of $\chi$, $\chib$, and $\omb$, but the proof of this fact
uses again the former estimates.

The estimates on the present chapter use $L^4$ elliptic theory on
the $S_{\ub,u}$ sections, the results derived in the previous
chapter through the uniformization theorem. The first step is the
estimate for $\snab^{ \ 2}\chi$ and, at the same time, the
estimate for $K-\overline{K}$, both in $L^4(S)$. This uses Lemma
5.7 applied to the Codazzi equation \ref{1.127}. The statement of
Lemma 5.7 shows that the two estimates must be coupled.

\section{$L^4(S)$ estimates for $\snab^{ \ 2}\chi$, $K-\overline{K}$}

We consider the propagation equation for
$\sd\mbox{tr}\chi^\prime$, the first of \ref{4.53}, which reads
(see \ref{4.54}):
\begin{equation}
D\sd\mbox{tr}\chi^\prime+\Omega\mbox{tr}\chi\sd\mbox{tr}\chi^\prime=-2\Omega(\chih,\snab\chih^\prime)+r
\label{6.1}
\end{equation}
where:
\begin{equation}
r=-2(\sd\log\Omega)\left[\frac{1}{2}(\mbox{tr}\chi)^2+|\chih|^2\right]
\label{6.2}
\end{equation}
and we denote:
\begin{equation}
(\chih,\snab\chih^\prime)_A=\chih^{BC}\snab_A\chih^\prime_{BC},
\label{6.3}
\end{equation}
in conjunction with the Codazzi equation \ref{1.127}, in the form:
\begin{equation}
\sdiv\chih^\prime=\frac{1}{2}\sd\mbox{tr}\chi^\prime+i \label{6.4}
\end{equation}
where:
\begin{equation}
i=\Omega^{-1}\left(-\beta+\frac{1}{2}\mbox{tr}\chi\eta-\chih^\sharp\cdot\eta\right),
\label{6.5}
\end{equation}
an elliptic equation for $\chih^\prime$ on each $S_{\ub,u}$
section.

\vspace{5mm}

\noindent{\bf Proposition 6.1} \ \ \ We have:
\begin{eqnarray*}
&&\|K-\overline{K}\|_{L^4(S_{\ub,u})}\leq
C\delta^{1/2}|u|^{-5/2}(\scR_1^4(\alpha)+\scR_1^4(\beta)+{\cal R}_0^\infty(\alpha)+{\cal R}_0^\infty(\beta))\\
&&\hspace{30mm}+O(\delta|u|^{-7/2})\\
&&\|\snab^{ \ 2}\mbox{tr}\chi\|_{L^4(S_{\ub,u})}\leq\\
&&\hspace{15mm}C|u|^{-7/2}(\scR_1^4(\alpha)+{\cal
R}_0^\infty(\alpha))
(\scR_1^4(\alpha)+\scR_1^4(\beta)+{\cal R}_0^\infty(\alpha)+{\cal R}_0^\infty(\beta))\\
&&\hspace{15mm}+O(\delta^{1/2}|u|^{-7/2})\\
&&|u|\|\snab^{ \
2}\chih\|_{L^4(S_{\ub,u})}+\|\snab\chih\|_{L^4(S_{\ub,u})}
+|u|^{-1}\|\chih\|_{L^4(S_{\ub,u})}\\
&&\hspace{30mm}\leq C\delta^{-1/2}|u|^{-3/2}(\scR_1^4(\beta)+{\cal
R}_0^\infty(\beta))+O(|u|^{-5/2})
\end{eqnarray*}
for all $(\ub,u)\in D^\prime$, provided that $\delta$ is suitably
small depending on ${\cal D}_0^\infty$, ${\cal R}_0^\infty$,
$\scD_1^4$, $\scR_1^4$, and $\scR_2(\alpha)$. In particular, under
such a smallness condition on $\delta$ we have:
$$|u|^{3/2}\|K-\overline{K}\|_{L^4(S_{\ub,u})}\leq 1$$

\noindent{\em Proof:} \ We apply Lemma 5.7 to equation \ref{6.4}
to obtain:
\begin{eqnarray}
&&\|\snab^{ \ 2}\chih^\prime\|_{L^4(S_{\ub,u})}\leq C(\|\snab^{ \
2}\mbox{tr}\chi^\prime\|_{L^4(S_{\ub,u})}
+\|\snab i\|_{L^4(S_{\ub,u})})\nonumber\\
&&\hspace{1cm}+C\{1+|u|^{3/2}\|K-\overline{K}\|_{L^4(S_{\ub,u})}\}\cdot\nonumber\\
&&\hspace{2cm}\left\{|u|^{-1}\left(\|\sd\mbox{tr}\chi^\prime\|_{L^4(S_{\ub,u})}+\|\snab\chih^\prime\|_{L^4(S_{\ub,u})}
+\|i\|_{L^4(S_{\ub,u})}\right)\right.\nonumber\\
&&\hspace{5cm}\left.
+|u|^{-3/2}\|\chih^\prime\|_{L^\infty(S_{\ub,u})}\right\}
\label{6.9}
\end{eqnarray}
Consider the second factor of the second term on the right in
\ref{6.9}. From Proposition 4.1, or, directly from the estimates
\ref{4.76} and \ref{4.77} we have:
\begin{eqnarray}
\|\sd\mbox{tr}\chi^\prime\|_{L^4(S_{\ub,u})}
&\leq& C|u|^{-5/2}{\cal R}_0^\infty(\alpha)\scR_1^4(\alpha)+O(\delta^{1/2}|u|^{-7/2})\label{6.10}\\
\|\snab\chih^\prime\|_{L^4(S_{\ub,u})} &\leq&
C\delta^{-1/2}|u|^{-3/2}\scR_1^4(\alpha)+O(\delta^{1/2}|u|^{-7/2})\label{6.11}
\end{eqnarray}
and by the third of \ref{4.49}:
\begin{equation}
\|\chih\|_{L^\infty(S_{\ub,u})}\leq C\delta^{-1/2}|u|^{-1}{\cal
R}_0^\infty(\alpha) \label{6.12}
\end{equation}
Also, using the estimates of Chapter 3 we obtain:
\begin{eqnarray}
\|i\|_{L^4(S_{\ub,u})}&\leq&C|u|^{1/2}\|i\|_{L^\infty(S_{\ub,u})}\nonumber\\
&\leq& C^\prime\delta^{-1/2}|u|^{-3/2}{\cal
R}_0^\infty(\beta)+O(|u|^{-5/2}) \label{6.13}
\end{eqnarray}
It follows from the above that:
\begin{eqnarray}
&&|u|^{-1}\left(\|\sd\mbox{tr}\chi^\prime\|_{L^4(S_{\ub,u})}+\|\snab\chih^\prime\|_{L^4(S_{\ub,u})}
+\|i\|_{L^4(S_{\ub,u})}\right)\nonumber\\
&&\hspace{3cm}+|u|^{-3/2}\|\chih^\prime\|_{L^\infty(S_{\ub,u})}\leq
\delta^{-1/2}|u|^{-5/2}l(u) \label{6.14}
\end{eqnarray}
where:
\begin{equation}
l(u)=C(\scR_1^4(\alpha)+{\cal R}_0^\infty(\alpha)+{\cal
R}_0^\infty(\beta))+O(\delta^{1/2}|u|^{-1}) \label{6.15}
\end{equation}
Using the results of Chapters 3 and 4 we deduce:
\begin{equation}
\|\snab i\|_{L^4(S_{\ub,u})}\leq
C\delta^{-1/2}|u|^{-5/2}\scR_1^4(\beta)+O(|u|^{-7/2}) \label{6.16}
\end{equation}
Substituting \ref{6.14} and \ref{6.16} in \ref{6.9} we obtain:
\begin{eqnarray}
&&\|\snab^{ \ 2}\chih^\prime\|_{L^4(S_{\ub,u})}\leq C\|\snab^{ \
2}\mbox{tr}\chi^\prime\|_{L^4(S_{\ub,u})}
+C\delta^{-1/2}|u|^{-1}l(u)\|K-\overline{K}\|_{L^4(S_{\ub,u})}\nonumber\\
&&\hspace{1cm}+C\delta^{-1/2}|u|^{-5/2}(\scR_1^4(\alpha)+\scR_1^4(\beta)+{\cal
R}_0^\infty(\alpha)+{\cal R}_0^\infty(\beta))
+O(|u|^{-7/2})\nonumber\\
&&\label{6.17}
\end{eqnarray}

We now revisit the propagation equation for $K-\overline{K}$,
equation \ref{5.31}. We apply Lemma 4.6 to this equation, taking
$p=4$. Here $r=0$, $\nu=-2$, $\gamma=0$. We obtain:
\begin{eqnarray}
\|K-\overline{K}\|_{L^4(S_{\ub,u})}&\leq&
C\int_0^{\ub}\|(\Omega\mbox{tr}\chi-\overline{\Omega\mbox{tr}\chi})\overline{K}\|_{L^4(S_{\ub^\prime,u})}d\ub^\prime
\label{6.18}\\
&\s&+C\int_0^{\ub}\|\sdiv\sdiv(\Omega\chih)-\frac{1}{2}\slap(\Omega\mbox{tr}\chi)\|_{L^4(S_{\ub^\prime,u})}d\ub^\prime
\nonumber
\end{eqnarray}
By \ref{5.34} the integrant of the first integral on the right in
\ref{6.18} is bounded by $O(|u|^{-7/2})$. The integrant of the
second integral on the right in \ref{6.18} is bounded by:
\begin{eqnarray*}
&&C\left(\|\snab^{ \ 2}\mbox{tr}\chi^\prime\|_{L^4(S)}+\|\snab^{ \ 2}\chih^\prime\|_{L^4(S)}\right)\\
&&\hspace{1cm}+C\left\{\|\sd\log\Omega\|_{L^\infty(S)}\left(\|\sd\mbox{tr}\chi^\prime\|_{L^4(S)}
+\|\snab\chih^\prime\|_{L^4(S)}\right)\right.\\
&&\hspace{2cm}\left.+\|\snab^{ \
2}\log\Omega\|_{L^4(S)}(\|\mbox{tr}\chi^\prime\|_{L^\infty(S)}+\|\chih^\prime\|_{L^\infty(S)})\right\}
\end{eqnarray*}
Using the estimates of Proposition 4.1, the expression \ref{5.17}
for $\snab^{ \ 2}\log\Omega$ and the estimates of Proposition 4.2
for $\snab\eta$, $\snab\etb$ in $L^4(S)$, as well as the estimates
of Chapter 3, the lower order terms
 are seen to be bounded by $O(|u|^{-7/2})$. Thus \ref{6.18} implies:
\begin{eqnarray}
\|K-\overline{K}\|_{L^4(S_{\ub,u})}&\leq&C\int_0^{\ub}\left\{\|\snab^{
\ 2}\mbox{tr}\chi^\prime\|_{L^4(S_{\ub^\prime,u})}
+\|\snab^{ \ 2}\chih^\prime\|_{L^4(S_{\ub^\prime,u})}\right\}d\ub^\prime\nonumber\\
&\s&+O(\delta|u|^{-7/2}) \label{6.19}
\end{eqnarray}

Substituting \ref{6.19} in \ref{6.17} yields a linear integral
inequality, for fixed $u$, for the quantity:
\begin{equation}
x(\ub)=\|\snab^{ \ 2}\chih^\prime\|_{L^4(S_{\ub,u})} \label{6.20}
\end{equation}
of the form:
\begin{equation}
x(\ub)\leq a\int_0^{\ub}x(\ub^\prime)d\ub^\prime+b(\ub)
\label{6.21}
\end{equation}
Here $a$ is the non-negative constant:
\begin{equation}
a=C\delta^{-1/2}|u|^{-1}l(u) \label{6.22}
\end{equation}
and $b(\ub)$ is the non-negative function:
\begin{eqnarray}
b(\ub)&=&C\|\snab^{ \ 2}\mbox{tr}\chi^\prime\|_{L^4(S_{\ub,u})}
+C\delta^{-1/2}|u|^{-1}l(u)\int_0^{\ub}\|\snab^{ \ 2}\mbox{tr}\chi^\prime\|_{L^4(S_{\ub^\prime,u})}d\ub^\prime\nonumber\\
&\s&+C\delta^{-1/2}|u|^{-5/2}(\scR_1^4(\alpha)+\scR_1^4(\beta)+{\cal
R}_0^\infty(\alpha)+{\cal R}_0^\infty(\beta))
+O(|u|^{-7/2})\nonumber\\
&\s&\label{6.23}
\end{eqnarray}
The inequality \ref{6.21} implies:
\begin{equation}
x(\ub)\leq a\int_0^{\ub}
e^{a(\ub-\ub^\prime)}b(\ub^\prime)d\ub^\prime+b(\ub) \label{6.24}
\end{equation}
Here
\begin{equation}
a(\ub-\ub^\prime)\leq a\delta=C\delta^{1/2}|u|^{-1}l(u)\leq \log 2
\label{6.25}
\end{equation}
the last step provided that $\delta$ is suitably small depending
on ${\cal D}_0^\infty$, ${\cal R}_0^\infty$, $\scD_1^4$,
$\scR_1^4$. Hence \ref{6.24} implies:
\begin{equation}
x(\ub)\leq 2a\int_0^{\ub}b(\ub^\prime)d\ub^\prime+b(\ub)
\label{6.26}
\end{equation}
From \ref{6.23} and \ref{6.25},
\begin{eqnarray}
\int_0^{\ub}b(\ub^\prime)d\ub^\prime&\leq&
C\int_0^{\ub}\|\snab^{ \ 2}\mbox{tr}\chi^\prime\|_{L^4(S_{\ub^\prime,u})}d\ub^\prime\label{6.27}\\
&\s&+C\delta^{1/2}|u|^{-5/2}(\scR_1^4(\alpha)+\scR_1^4(\beta)+{\cal
R}_0^\infty(\alpha)+{\cal R}_0^\infty(\beta))
+O(\delta|u|^{-7/2})\nonumber
\end{eqnarray}
Therefore, substituting in \ref{6.26} we obtain:
\begin{eqnarray}
\|\snab^{ \ 2}\chih^\prime\|_{L^4(S_{\ub,u})}&\leq&C\|\snab^{ \ 2}\mbox{tr}\chi^\prime\|_{L^4(S_{\ub,u})}\nonumber\\
&\s&+C\delta^{-1/2}|u|^{-1}l(u)\int_0^{\ub}\|\snab^{ \ 2}\mbox{tr}\chi^\prime\|_{L^4(S_{\ub^\prime,u})}d\ub^\prime\nonumber\\
&\s&+C\delta^{-1/2}|u|^{-5/2}(\scR_1^4(\alpha)+\scR_1^4(\beta)+{\cal
R}_0^\infty(\alpha)+{\cal R}_0^\infty(\beta))
\nonumber\\
&\s&+O(|u|^{-7/2})\label{6.28}
\end{eqnarray}

We now turn to the propagation equation \ref{6.1}. We apply Lemma
4.1 to this equation to deduce the following propagation equation
for $\snab^{ \ 2}\mbox{tr}\chi^\prime$:
\begin{equation}
D\snab^{ \ 2}\mbox{tr}\chi^\prime+\Omega\mbox{tr}\chi\snab^{ \
2}\mbox{tr}\chi^\prime =-2\Omega(\chih,\snab^{ \ 2}\chih^\prime)+e
\label{6.6}
\end{equation}
where:
\begin{equation}
e_{AB}=-(D\sGamma)^C_{AB}\sd_C\mbox{tr}\chi^\prime-\sd_A(\Omega\mbox{tr}\chi)\sd_B\mbox{tr}\chi^\prime
-2\snab_A(\Omega\chih^{CD})\snab_B\chih^\prime_{CD}+\snab_A r_B
\label{6.7}
\end{equation}
and we denote:
\begin{equation}
(\chih,\snab^{ \
2}\chih^\prime)_{AB}=\chih^{CD}\snab_A\snab_B\chih^\prime_{CD}
\label{6.8}
\end{equation}
To estimate in $L^4(S)$ the first three terms on the right in
\ref{6.7} we place the first factors in $L^4(S)$ and the second
factors in $L^\infty(S)$. We then use Lemma 5.2 with $p=4$ to
bound the $L^\infty(S)$ norms of the second factors in terms of
their $W_1^4(S)$ norms. We have, in view of Lemma 5.4, in regard
to the first term on the right in \ref{6.7}:
\begin{eqnarray}
&&\|D\sGamma\cdot\sd\mbox{tr}\chi^\prime\|_{L^4(S_{\ub,u})}\leq
\|D\sGamma\|_{L^4(S_{\ub,u})}\|\sd\mbox{tr}\chi^\prime\|_{L^\infty(S_{\ub,u})}\nonumber\\
&&\hspace{1cm}\leq
C\|D\sGamma\|_{L^4(S_{\ub,u})}\{|u|^{1/2}\|\snab^{ \
2}\mbox{tr}\chi^\prime\|_{L^4(S_{\ub,u})}
+|u|^{-1/2}\|\sd\mbox{tr}\chi^\prime\|_{L^4(S_{\ub,u})}\}\nonumber\\
&&\label{6.29}
\end{eqnarray}
Similarly, the $L^4(S)$ norm of the second term on the right in
\ref{6.7} is bounded by:
\begin{equation}
C\|\sd(\Omega\mbox{tr}\chi)\|_{L^4(S_{\ub,u})}\{|u|^{1/2}\|\snab^{
\ 2}\mbox{tr}\chi^\prime\|_{L^4(S_{\ub,u})}
+|u|^{-1/2}\|\sd\mbox{tr}\chi^\prime\|_{L^4(S_{\ub,u})}\}
\label{6.30}
\end{equation}
and the $L^4(S)$ norm of the third term on the right in \ref{6.7}
is bounded by:
\begin{equation}
C\|\snab(\Omega\chih)\|_{L^4(S_{\ub,u})}\{|u|^{1/2}\|\snab^{ \
2}\chih^\prime\|_{L^4(S_{\ub,u})}
+|u|^{-1/2}\|\snab\chih^\prime\|_{L^4(S_{\ub,u})}\} \label{6.31}
\end{equation}
Now, from Lemma 4.1 we have, pointwise:
\begin{equation}
|D\sGamma|\leq C(|\sd(\Omega\mbox{tr}\chi)|+|\snab(\Omega\chih)|)
\label{6.32}
\end{equation}
hence:
\begin{eqnarray}
\|D\sGamma\|_{L^4(S_{\ub,u})}&\leq&
C\{\|\sd(\Omega\mbox{tr}\chi)\|_{L^4(S_{\ub,u})}+\|\snab(\Omega\chih)\|_{L^4(S_{\ub,u})}\}\nonumber\\
&\leq& C\delta^{-1/2}|u|^{-3/2}\scR_1^4(\alpha)+O(|u|^{-5/2})
\label{6.33}
\end{eqnarray}
by Proposition 4.1. Substituting also the estimates \ref{6.10},
\ref{6.11}, we conclude that the first three terms on the right in
\ref{6.7} are bounded in $L^4(S)$ norm by:
\begin{eqnarray}
&&\left(C\delta^{-1/2}|u|^{-1}\scR_1^4(\alpha)+O(|u|^{-2})\right)
\left\{\|\snab^{ \ 2}\mbox{tr}\chi^\prime\|_{L^4(S_{\ub,u})}
+\|\snab^{ \ 2}\chih^\prime\|_{L^4(S_{\ub,u})}\right.\nonumber\\
&&\hspace{5cm}\left. +C\delta^{-1/2}|u|^{-5/2}\scR_1^4(\alpha)+O(|u|^{-7/2})\right\}\nonumber\\
&&\label{6.34}
\end{eqnarray}
Using the estimates of Proposition 4.1, the expression \ref{5.17}
for $\snab^{ \ 2}\log\Omega$ and the estimates of Proposition 4.2
for $\snab\eta$, $\snab\etb$ in $L^4(S)$, as well as the estimates
of Chapter 3, we obtain, in regard to the last term on the right
in \ref{6.7}:
\begin{equation}
\|\snab r\|_{L^4(S_{\ub,u})}\leq O(\delta^{-1/2}|u|^{-9/2})
\label{6.35}
\end{equation}
We then conclude that:
\begin{eqnarray}
&&\|e\|_{L^4(S_{\ub,u})}\leq\nonumber\\
&&\hspace{5mm}\left(C\delta^{-1/2}|u|^{-1}\scR_1^4(\alpha)+O(|u|^{-2})\right)
\left\{\|\snab^{ \ 2}\mbox{tr}\chi^\prime\|_{L^4(S_{\ub,u})}
+\|\snab^{ \ 2}\chih^\prime\|_{L^4(S_{\ub,u})}\right\}\nonumber\\
&&\hspace{15mm}+C\delta^{-1}|u|^{-7/2}(\scR_1^4(\alpha))^2+O(\delta^{-1/2}|u|^{-9/2})
\label{6.36}
\end{eqnarray}
We also estimate in $L^4(S)$ the first term on the right in
\ref{6.6}:
\begin{equation}
\|\Omega(\chih,\snab^{ \ 2}\chih^\prime)\|_{L^4(S_{\ub,u})}\leq
C\delta^{-1/2}|u|^{-1}{\cal R}_0^\infty(\alpha) \|\snab^{ \
2}\chih^\prime\|_{L^4(S_{\ub,u})} \label{6.37}
\end{equation}
by the third of \ref{4.49}. Substituting finally in \ref{6.36} and
\ref{6.37} the estimate \ref{6.28} we conclude that:
\begin{eqnarray}
&&\|-2\Omega(\chih,\snab^{ \ 2}\chih^\prime)+e\|_{L^4(S_{\ub,u})}\leq\nonumber\\
&&\hspace{10mm}C\delta^{-1/2}|u|^{-1}m(u)\left\{\|\snab^{ \ 2}\mbox{tr}\chi^\prime\|_{L^4(S_{\ub,u})}\right.\nonumber\\
&&\hspace{25mm}\left.+C\delta^{-1/2}|u|^{-1}l(u)
\int_0^{\ub}\|\snab^{ \ 2}\mbox{tr}\chi^\prime\|_{L^4(S_{\ub^\prime,u})}d\ub^\prime\right\}\nonumber\\
&&\hspace{15mm}+C\delta^{-1}|u|^{-7/2}(\scR_1^4(\alpha)+{\cal
R}_0^\infty(\alpha))(\scR_1^4(\alpha)+\scR_1^4(\beta)+{\cal
R}_0^\infty(\alpha)
+{\cal R}_0^\infty(\beta))\nonumber\\
&&\hspace{15mm}+O(\delta^{-1/2}|u|^{-9/2}) \label{6.38}
\end{eqnarray}
where:
\begin{equation}
m(u)=C(\scR_1^4(\alpha)+{\cal
R}_0^\infty(\alpha))+O(\delta^{1/2}|u|^{-1}) \label{6.39}
\end{equation}

We now apply Lemma 4.6 with $p=4$ to \ref{6.6}. Here $r=2$,
$\nu=-2$, $\gamma=0$. We obtain:
\begin{equation}
\|\snab^{ \ 2}\mbox{tr}\chi^\prime\|_{L^4(S_{\ub,u})}\leq
C\int_0^{\ub}\|-2\Omega(\chih,\snab^{ \
2}\chih^\prime)+e\|_{L^4(S_{\ub^\prime,u})}d\ub^\prime
\label{6.40}
\end{equation}
Substituting the bound \ref{6.38} then yields a linear integral
inequality for $\|\snab^{ \ 2}\mbox{tr}\chi^\prime\|_{L^4(S)}$.
Since for a non-negative function $f(\ub)$ on $[0,\delta]$ we
have:
\begin{eqnarray}
&&\int_0^{\ub}\left\{\int_0^{\ub^\prime}f(\ub^{\prime\prime})d\ub^{\prime\prime}\right\}d\ub^\prime
\leq\int_0^{\ub}\left\{\int_0^{\ub}f(\ub^{\prime\prime})d\ub^{\prime\prime}\right\}d\ub^\prime\nonumber\\
&&\hspace{2cm}=\ub\int_0^{\ub}f(\ub^{\prime\prime})d\ub^{\prime\prime}
\leq\delta\int_0^{\ub}f(\ub^\prime)d\ub^\prime\label{6.a1}
\end{eqnarray}
the contribution of the term in parenthesis in \ref{6.38}
involving the integral is bounded by
\begin{equation}
C\delta^{1/2}|u|^{-1}l(u) \label{6.41}
\end{equation}
times the contribution of the first term in parenthesis in
\ref{6.38}, and the factor \ref{6.41} is less than or equal to 1
provided that $\delta$ is suitably small depending on ${\cal
D}_0^\infty$, ${\cal R}_0^\infty$, $\scD_1^4$, $\scR_1^4$.
Therefore the resulting integral inequality simplifies to:
\begin{eqnarray}
&&\|\snab^{ \ 2}\mbox{tr}\chi^\prime\|_{L^4(S_{\ub,u})}\leq
C\delta^{-1/2}|u|^{-1}m(u)\int_0^{\ub}\|\snab^{ \ 2}\mbox{tr}\chi^\prime\|_{L^4(S_{\ub^\prime,u})}d\ub^\prime\nonumber\\
&&\hspace{15mm}+C|u|^{-7/2}(\scR_1^4(\alpha)+{\cal
R}_0^\infty(\alpha))
(\scR_1^4(\alpha)+\scR_1^4(\beta)+{\cal R}_0^\infty(\alpha)+{\cal R}_0^\infty(\beta))\nonumber\\
&&\hspace{15mm}+O(\delta^{1/2}|u|^{-9/2}) \label{6.42}
\end{eqnarray}
This is a linear integral inequality of the form \ref{6.21}, but
with $b$ a positive constant. Noting that
\begin{equation}
C\delta^{1/2}|u|^{-1}m(u)\leq 1 \label{6.43}
\end{equation}
provided that $\delta$ is suitably small depending on ${\cal
D}_0^\infty$, ${\cal R}_0^\infty$, $\scD_1^4$, $\scR_1^4$, the
integral inequality \ref{6.42} implies:
\begin{eqnarray}
&&\|\snab^{ \ 2}\mbox{tr}\chi^\prime\|_{L^4(S_{\ub,u})}\leq\nonumber\\
&&\hspace{15mm}C|u|^{-7/2}(\scR_1^4(\alpha)+{\cal
R}_0^\infty(\alpha))
(\scR_1^4(\alpha)+\scR_1^4(\beta)+{\cal R}_0^\infty(\alpha)+{\cal R}_0^\infty(\beta))\nonumber\\
&&\hspace{15mm}+O(\delta^{1/2}|u|^{-9/2}) \label{6.44}
\end{eqnarray}

Substituting the estimate \ref{6.44} in \ref{6.28} we obtain:
\begin{eqnarray}
\|\snab^{ \ 2}\chih^\prime\|_{L^4(S_{\ub,u})}&\leq&
C\delta^{-1/2}|u|^{-5/2}(\scR_1^4(\alpha)+\scR_1^4(\beta)+{\cal
R}_0^\infty(\alpha)+{\cal R}_0^\infty(\beta))
\nonumber\\
&\s&+O(|u|^{-7/2})\label{6.45}
\end{eqnarray}
Substituting then the estimates \ref{6.44} and \ref{6.45} in
\ref{6.19} yields:
\begin{eqnarray}
\|K-\overline{K}\|_{L^4(S_{\ub,u})}&\leq&
C\delta^{1/2}|u|^{-5/2}(\scR_1^4(\alpha)+\scR_1^4(\beta)+{\cal
R}_0^\infty(\alpha)+{\cal R}_0^\infty(\beta))
\nonumber\\
&\s&+O(\delta|u|^{-7/2})\label{6.46}
\end{eqnarray}
It follows that:
\begin{equation}
|u|^{3/2}\|K-\overline{K}\|_{L^4(S_{\ub,u})}\leq 1 \label{6.47}
\end{equation}
provided that $\delta$ is suitably small depending on ${\cal
D}_0^\infty$, ${\cal R}_0^\infty$, $\scD_1^4$, $\scR_1^4$. We now
apply Lemma 5.9 to equation \ref{6.4}. In view of \ref{6.47} the
conclusion of that lemma simplifies and we obtain:
\begin{eqnarray}
&&|u|\|\snab^{ \
2}\chih^\prime\|_{L^4(S_{\ub,u})}+\|\snab\chih^\prime\|_{L^4(S_{\ub,u})}
+|u|^{-1}\|\chih^\prime\|_{L^4(S_{\ub,u})}\nonumber\\
&&\hspace{1cm}\leq C\left\{|u|(\|\snab^{ \
2}\mbox{tr}\chi^\prime\|_{L^4(S_{\ub,u})}
+\|\snab i\|_{L^4(S_{\ub,u})})\right.\nonumber\\
&&\hspace{3cm}\left.+\|\sd\mbox{tr}\chi^\prime\|_{L^4(S_{\ub,u})}+\|i\|_{L^4(S_{\ub,u})}\right\}
\label{6.48}
\end{eqnarray}
Substituting on the right the estimates \ref{6.10}, \ref{6.44},
\ref{6.13}, \ref{6.16} then yields:
\begin{eqnarray}
&&|u|\|\snab^{ \
2}\chih^\prime\|_{L^4(S_{\ub,u})}+\|\snab\chih^\prime\|_{L^4(S_{\ub,u})}
+|u|^{-1}\|\chih^\prime\|_{L^4(S_{\ub,u})}\nonumber\\
&&\hspace{2cm}\leq C\delta^{-1/2}|u|^{-3/2}(\scR_1^4(\beta)+{\cal
R}_0^\infty(\beta))+O(|u|^{-5/2}) \label{6.49}
\end{eqnarray}
In view of the estimates \ref{6.44}, \ref{6.46}, \ref{6.49}, the
proposition follows if we also take into account the expression
\ref{5.17} for $\snab^{ \ 2}\log\Omega$ and the estimates of
Proposition 4.2 for $\snab\eta$, $\snab\etb$ in $L^4(S)$.

\vspace{5mm}

\section{$L^4(S)$ estimates for $\snab^{ \ 2}\chib$}

We now consider any $(\ub_1,u_1)\in D^\prime$ and fix attention in
the following lemmas to the parameter subdomain $D_1$ and the
corresponding subdomain $M_1$ of $M^\prime$ (see \ref{3.02},
\ref{3.03}).

With a positive constant $k$ to be appropriately chosen in the
sequel, let $s^*$ be the least upper bound of the set of values of
$s\in[u_0,u_1]$ such that:
\begin{equation}
\|\snab^{ \ 2}\log\Omega\|_{L^4(S_{\ub,u})}\leq k\delta|u|^{-7/2}\
\ \ : \ \mbox{for all $(\ub,u)\in [0,\ub_1]\times[u_0,s]$}
\label{6.50}
\end{equation}
Then by continuity $s^*>u_0$ (recall that by \ref{3.34} $\snab^{ \
2}\log\Omega$ vanishes along $C_{u_0}$) and we have:
\begin{equation}
\|\snab^{ \ 2}\log\Omega\|_{L^4(S_{\ub,u})}\leq k\delta|u|^{-7/2}\
\ \ : \ \mbox{for all $(\ub,u)\in [0,\ub_1]\times[u_0,s^*]$}
\label{6.51}
\end{equation}
In the estimates to follow the dependence on the constant $k$ is
made explicit.

Let us denote:
\begin{equation}
\scD_2^4(\mbox{tr}\chib)=|u_0|^{9/2}\delta^{-1}\sup_{\ub\in[0,\delta]}\|\snab^{
\ 2}\mbox{tr}\chib\|_{L^4(S_{\ub,u_0})} \label{6.79}
\end{equation}
Here, as in \ref{3.37}, \ref{3.108}, \ref{4.80}, \ref{4.81}, we are considering all of $C_{u_0}$, not only the part which lies in 
$M^\prime$. 
By the results of Chapter 2 this quantity is bounded by a non-negative non-decreasing continuous function of $M_5$.  
Let us denote by $D_1^{s^*}$ the subset of $D_1$ where
$u\leq s^*$:
\begin{equation}
D_1^{s^*}=[0,\ub_1]\times[u_0,s^*] \label{6.53}
\end{equation}
and by $M_1^{s^*}$ the corresponding subset of $M_1$.

We consider the propagation equation for
$\sd\mbox{tr}\chib^\prime$, the first of \ref{4.84}, which reads
(see \ref{4.85}):
\begin{equation}
\Db\sd\mbox{tr}\chib^\prime+\Omega\mbox{tr}\chib\sd\mbox{tr}\chib^\prime=-2\Omega(\chibh,\snab\chibh^\prime)+\rb
\label{6.54}
\end{equation}
where:
\begin{equation}
\rb=-2(\sd\log\Omega)\left[\frac{1}{2}(\mbox{tr}\chib)^2+|\chibh|^2\right]
\label{6.55}
\end{equation}
and we denote:
\begin{equation}
(\chibh,\snab\chibh^\prime)_A=\chibh^{BC}\snab_A\chibh^\prime_{BC},
\label{6.56}
\end{equation}
in conjunction with the Codazzi equation \ref{1.128}, in the form:
\begin{equation}
\sdiv\chibh^\prime=\frac{1}{2}\sd\mbox{tr}\chib^\prime+\ib
\label{6.57}
\end{equation}
where:
\begin{equation}
\ib=\Omega^{-1}\left(\beb+\frac{1}{2}\mbox{tr}\chib\etb-\chibh^\sharp\cdot\etb\right),
\label{6.58}
\end{equation}
an elliptic equation for $\chibh^\prime$ on each $S_{\ub,u}$
section. Equations \ref{6.54} and \ref{6.57} are the conjugates of
equations \ref{6.1} and \ref{6.4} respectively.

\vspace{5mm}

\noindent{\bf Lemma 6.1} \ \ \ We have:
\begin{eqnarray*}
&&\|\snab^{ \ 2}\mbox{tr}\chib\|_{L^4(S_{\ub,u})}\leq C\delta|u|^{-9/2}\left\{\scD_2^4(\mbox{tr}\chib)+k+(\scD_1^4(\mbox{tr}\chib)+\scD_1^4(\chibh))^2\right.\\
&&\hspace{3cm}\left.+(\scD_1^4(\mbox{tr}\chib)+\scD_1^4(\chibh)+{\cal D}_0^\infty(\chibh))(\scR_1^4(\beta)+{\cal R}_0^\infty(\beta))\right\}\\
&&\hspace{3cm}+O(\delta^{3/2}|u|^{-11/2})
\end{eqnarray*}
\begin{eqnarray*}
&&|u|\|\snab^{ \
2}\chibh\|_{L^4(S_{\ub,u})}+\|\snab\chibh\|_{L^4(S_{\ub,u})}
+|u|^{-1}\|\chibh\|_{L^4(S_{\ub,u})}\\
&&\hspace{1cm}\leq C\delta^{1/2}|u|^{-5/2}(\scR_1^4(\beta)+{\cal
R}_0^\infty(\beta))+C\delta|u|^{-7/2}k+O(\delta|u|^{-7/2})
\end{eqnarray*}
for all $(\ub,u)\in D_1^{s^*}$, provided that $\delta$ is suitably
small depending on ${\cal D}_0^\infty$, ${\cal R}_0^\infty$,
$\scD_1^4$, $\scR_1^4$, and $\scR_2(\alpha)$.

\noindent{\em Proof:} \ We apply Lemma 5.9 to equation \ref{6.57}.
In view of the last conclusion of Proposition 6.1 the conclusion
of Lemma 5.9 simplifies and we obtain:
\begin{eqnarray}
&&|u|\|\snab^{ \
2}\chibh^\prime\|_{L^4(S_{\ub,u})}+\|\snab\chibh^\prime\|_{L^4(S_{\ub,u})}
+|u|^{-1}\|\chibh^\prime\|_{L^4(S_{\ub,u})}\nonumber\\
&&\hspace{1cm}\leq C\left\{|u|(\|\snab^{ \
2}\mbox{tr}\chib^\prime\|_{L^4(S_{\ub,u})}
+\|\snab\ib\|_{L^4(S_{\ub,u})})\right.\nonumber\\
&&\hspace{3cm}\left.+\|\sd\mbox{tr}\chib^\prime\|_{L^4(S_{\ub,u})}+\|\ib\|_{L^4(S_{\ub,u})}\right\}
\label{6.59}
\end{eqnarray}
Using Proposition 4.2 and the results of Chapter 3 we deduce:
\begin{equation}
\|\snab\ib\|_{L^4(S_{\ub,u})}\leq
C\delta^{1/2}|u|^{-7/2}\scR_1^4(\beta)+O(\delta|u|^{-9/2})
\label{6.60}
\end{equation}
the leading contribution to $\snab\ib$ in behavior with respect to
$\delta$ coming from the term
$(1/2)\Omega^{-1}\mbox{tr}\chib\snab\etb$. Also, using the
estimates of Chapter 3 we obtain:
\begin{eqnarray}
\|\ib\|_{L^4(S_{\ub,u})}&\leq& C|u|^{1/2}\|\ib\|_{L^\infty(S_{\ub,u})}\nonumber\\
&\leq& C^\prime\delta^{1/2}|u|^{-5/2}{\cal R}_0^\infty(\beta)
\label{6.61}
\end{eqnarray}
the leading contribution to $\ib$ in behavior with respect to
$\delta$ coming from the term
$(1/2)\Omega^{-1}\mbox{tr}\chib\etb$. Substituting \ref{6.60},
\ref{6.61} and the estimate \ref{4.189} in \ref{6.59} we obtain:
\begin{eqnarray}
&&|u|\|\snab^{ \
2}\chibh^\prime\|_{L^4(S_{\ub,u})}+\|\snab\chibh^\prime\|_{L^4(S_{\ub,u})}
+|u|^{-1}\|\chibh^\prime\|_{L^4(S_{\ub,u})}\nonumber\\
&&\hspace{1cm}\leq C|u|\|\snab^{ \ 2}\mbox{tr}\chib^\prime\|_{L^4(S_{\ub,u})}\nonumber\\
&&\hspace{1.5cm} +C\delta^{1/2}|u|^{-5/2}(\scR_1^4(\beta)+{\cal
R}_0^\infty(\beta))+O(\delta|u|^{-7/2}) \label{6.62}
\end{eqnarray}

We now turn to the propagation equation \ref{6.54}. We apply Lemma
4.1 to this equation to deduce the following propagation equation
for $\snab^{ \ 2}\mbox{tr}\chib^\prime$:
\begin{equation}
\Db\snab^{ \
2}\mbox{tr}\chib^\prime+\Omega\mbox{tr}\chib^\prime\snab^{ \
2}\mbox{tr}\chib^\prime= -2\Omega(\chibh,\snab^{ \
2}\chibh^\prime)+\eb \label{6.63}
\end{equation}
where:
\begin{equation}
\eb_{AB}=-(\Db\sGamma)^C_{AB}\sd_C\mbox{tr}\chib^\prime-\sd_A(\Omega\mbox{tr}\chib)\sd_B\mbox{tr}\chib^\prime
-2\snab_A(\Omega\chibh^{CD})\snab_B\chibh^\prime_{CD}+\snab_A\rb_B
\label{6.64}
\end{equation}
and we denote:
\begin{equation}
(\chibh,\snab^{ \
2}\chibh^\prime)_{AB}=\chibh^{CD}\snab_A\snab_B\chibh^\prime_{CD}
\label{6.65}
\end{equation}
To estimate in $L^4(S)$ the first three terms on the right in
\ref{6.64} we place the first factors in $L^4(S)$ and the second
factors in $L^\infty(S)$. We then use Lemma 5.2 with $p=4$ to
bound the $L^\infty(S)$ norms of the second factors in terms of
their $W_1^4(S)$ norms. We have, in view of Lemma 5.4, in regard
to the first term on the right in \ref{6.64}:
\begin{eqnarray}
&&\|\Db\sGamma\cdot\sd\mbox{tr}\chib^\prime\|_{L^4(S_{\ub,u})}\leq
\|\Db\sGamma\|_{L^4(S_{\ub,u})}\|\sd\mbox{tr}\chib^\prime\|_{L^\infty(S_{\ub,u})}\nonumber\\
&&\hspace{1cm}\leq
C\|\Db\sGamma\|_{L^4(S_{\ub,u})}\{|u|^{1/2}\|\snab^{ \
2}\mbox{tr}\chib^\prime\|_{L^4(S_{\ub,u})}
+|u|^{-1/2}\|\sd\mbox{tr}\chib^\prime\|_{L^4(S_{\ub,u})}\}\nonumber\\
&&\label{6.66}
\end{eqnarray}
Similarly, the $L^4(S)$ norm of the second term on the right in
\ref{6.64} is bounded by:
\begin{equation}
C\|\sd(\Omega\mbox{tr}\chib)\|_{L^4(S_{\ub,u})}\{|u|^{1/2}\|\snab^{
\ 2}\mbox{tr}\chib^\prime\|_{L^4(S_{\ub,u})}
+|u|^{-1/2}\|\sd\mbox{tr}\chib^\prime\|_{L^4(S_{\ub,u})}\}
\label{6.67}
\end{equation}
and the $L^4(S)$ norm of the third term on the right in \ref{6.64}
is bounded by:
\begin{equation}
C\|\snab(\Omega\chibh)\|_{L^4(S_{\ub,u})}\{|u|^{1/2}\|\snab^{ \
2}\chibh^\prime\|_{L^4(S_{\ub,u})}
+|u|^{-1/2}\|\snab\chibh^\prime\|_{L^4(S_{\ub,u})}\} \label{6.68}
\end{equation}
Now, from Lemma 4.1 we have, pointwise:
\begin{equation}
|\Db\sGamma|\leq
C(|\sd(\Omega\mbox{tr}\chib)|+|\snab(\Omega\chibh)|) \label{6.69}
\end{equation}
hence:
\begin{eqnarray}
\|\Db\sGamma\|_{L^4(S_{\ub,u})}&\leq&
C\{\|\sd(\Omega\mbox{tr}\chib)\|_{L^4(S_{\ub,u})}+\|\snab(\Omega\chibh)\|_{L^4(S_{\ub,u})}\}\nonumber\\
&\leq&
C\delta^{1/2}|u|^{-5/2}(\scD_1^4(\mbox{tr}\chib)+\scD_1^4(\chibh))+O(\delta|u|^{-7/2})
\label{6.70}
\end{eqnarray}
by Proposition 4.2 and Lemma 4.11. Using Proposition 4.2 and Lemma
4.11 we then conclude that the first three terms on the right in
\ref{6.64} are bounded in $L^4(S)$ norm by:
\begin{eqnarray}
&&C\left(\delta^{1/2}|u|^{-2}(\scD_1^4(\mbox{tr}\chib)+\scD_1^4(\chibh))+O(\delta|u|^{-3})\right)\cdot\nonumber\\
&&\hspace{2cm}\cdot\left\{\|\snab^{ \
2}\mbox{tr}\chib^\prime\|_{L^4(S_{\ub,u})}
+\|\snab^{ \ 2}\chibh^\prime\|_{L^4(S_{\ub,u})}\right.\nonumber\\
&&\hspace{2cm}\s\left.+C\delta^{1/2}|u|^{-7/2}(\scD_1^4(\mbox{tr}\chib)+\scD_1^4(\chibh))+O(\delta|u|^{-9/2})\right\}
\label{6.71}
\end{eqnarray}

The last term on the right in \ref{6.64}, $\snab\rb$, contains the
term:
\begin{equation}
-2\snab^{ \
2}\log\Omega\left[\frac{1}{2}(\mbox{tr}\chib)^2+|\chib|^2\right]
\label{6.72}
\end{equation}
We estimate this in $L^4(S)$ using \ref{6.51}. We obtain a bound
by:
$$Ck\delta|u|^{-11/2}$$
for all $(\ub,u)\in D_1^{s^*}$. The remaining terms in $\snab\rb$
are bounded in $L^4(S)$ using Proposition 4.2 and Lemma 4.11 by
$O(\delta|u|^{-15/2})$. Thus:
\begin{equation}
\|\snab\rb\|_{L^4(S_{\ub,u})}\leq
Ck\delta|u|^{-11/2}+O(\delta^2|u|^{-15/2}) \ \ \ : \ \forall
(\ub,u)\in D_1^{s^*} \label{6.73}
\end{equation}
We then conclude that:
\begin{eqnarray}
&&\|\eb\|_{L^4(S_{\ub,u})}\leq\nonumber\\
&&\hspace{1cm}C\left(\delta^{1/2}|u|^{-2}(\scD_1^4(\mbox{tr}\chib)+\scD_1^4(\chibh))+O(\delta|u|^{-3})\right)\cdot\nonumber\\
&&\hspace{3cm}\cdot\left\{\|\snab^{ \
2}\mbox{tr}\chib^\prime\|_{L^4(S_{\ub,u})}
+\|\snab^{ \ 2}\chibh^\prime\|_{L^4(S_{\ub,u})}\right\}\nonumber\\
&&\hspace{2cm}\s+C\delta|u|^{-11/2}\left\{k+(\scD_1^4(\mbox{tr}\chib)+\scD_1^4(\chibh))^2\right\}
+O(\delta^{3/2}|u|^{-13/2})\nonumber\\
&&\label{6.74}
\end{eqnarray}
for all $(\ub,u)\in D_1^{s^*}$. We also estimate in $L^4(S)$ the
first term on the right in \ref{6.63}:
\begin{equation}
\|\Omega(\chibh,\snab^{ \ 2}\chibh^\prime)\|_{L^4(S_{\ub,u})}\leq
C\left(\delta^{1/2}|u|^{-2}({\cal
D}_0^\infty(\chibh)+O(\delta^{3/2}|u|^{-7/2})\right) \|\snab^{ \
2}\chibh^\prime\|_{L^4(S_{\ub,u})} \label{6.75}
\end{equation}
by Proposition 3.2 (Lemma 3.2). Substituting finally in \ref{6.74}
and \ref{6.75} the estimate \ref{6.62} we conclude that:
\begin{eqnarray}
&&\|-2\Omega(\chibh,\snab^{ \ 2}\chibh^\prime)+\eb\|_{L^4(S_{\ub,u})}\leq\nonumber\\
&&\hspace{2cm} C\delta^{1/2}|u|^{-2}\mb(u)\|\snab^{ \ 2}\mbox{tr}\chib^\prime\|_{L^4(S_{\ub,u})}\nonumber\\
&&\hspace{1cm}+C\delta|u|^{-11/2}\left\{k+(\scD_1^4(\mbox{tr}\chib)+\scD_1^4(\chibh))^2\right.\nonumber\\
&&\hspace{2cm}\left.+(\scD_1^4(\mbox{tr}\chib)+\scD_1^4(\chibh)+{\cal
D}_0^\infty(\chibh))(\scR_1^4(\beta)+{\cal
R}_0^\infty(\beta))\right\}
\nonumber\\
&&\hspace{2cm}+O(\delta^{3/2}|u|^{-13/2}) \label{6.76}
\end{eqnarray}
for all $(\ub,u)\in D_1^{s^*}$. Here:
\begin{equation}
\mb(u)=C(\scD_1^4(\mbox{tr}\chib)+\scD_1^4(\chibh)+{\cal
D}_0^\infty(\chibh))+O(\delta^{1/2}|u|^{-1}) \label{6.77}
\end{equation}

We now apply Lemma 4.7 with $p=4$ to \ref{6.63}. Here $r=2$,
$\nu=-2$, $\gammab=0$. We obtain:
\begin{eqnarray}
|u|^{7/2}\|\snab^{ \ 2}\mbox{tr}\chib^\prime\|_{L^4(S_{\ub,u})}
&\leq&C|u_0|^{7/2}\|\snab^{ \ 2}\mbox{tr}\chib^\prime\|_{L^4(S_{\ub,u_0})}\label{6.78}\\
&\s&+C\int_{u_0}^u|u^\prime|^{7/2}\|-2\Omega(\chibh,\snab^{ \
2}\chibh^\prime)+\eb\|_{L^4(S_{\ub,u^\prime})}du^\prime \nonumber
\end{eqnarray}
for all $(\ub,u)\in D_1^{s^*}$. Substituting \ref{6.76} and
recalling the definition \ref{6.79} yields a linear integral
inequality, for fixed $\ub$, for the quantity:
\begin{equation}
\xb(u)=|u|^{7/2}\|\snab^{ \
2}\mbox{tr}\chib^\prime\|_{L^4(S_{\ub,u})} \label{6.80}
\end{equation}
of the form:
\begin{equation}
\xb(u)\leq\int_{u_0}^u\ab(u^\prime)\xb(u^\prime)du^\prime+\bb(u)
\label{6.81}
\end{equation}
Here $\ab$ is the non-negative function:
\begin{equation}
\ab(u)=C\delta^{1/2}|u|^{-2}\mb(u) \label{6.82}
\end{equation}
and $\bb$ is the non-negative non-decreasing function:
\begin{eqnarray}
&&\bb(u)=C\delta|u_0|^{-1}\scD_2^4(\mbox{tr}\chib)\nonumber\\
&&\hspace{1cm}+\int_{u_0}^u\left\{C\delta|u|^{-2}\left[k+(\scD_1^4(\mbox{tr}\chib)+\scD_1^4(\chibh))^2\right.\right.\nonumber\\
&&\hspace{2cm}\left.+(\scD_1^4(\mbox{tr}\chib)+\scD_1^4(\chibh)+{\cal
D}_0^\infty(\chibh))(\scR_1^4(\beta)+{\cal
R}_0^\infty(\beta))\right]
\nonumber\\
&&\hspace{3cm}\left.+O(\delta^{3/2}|u|^{-3})\right\}du^\prime
\label{6.83}
\end{eqnarray}
Setting
\begin{equation}
\Xb(u)=\int_{u_0}^u a(u^\prime)\xb(u^\prime)du^\prime, \ \
\mbox{we have $\Xb(u_0)=0$} \label{6.84}
\end{equation}
and \ref{6.81} takes the form:
\begin{equation}
\frac{d\Xb}{du}\leq \ab(\Xb+\bb) \label{6.85}
\end{equation}
It follows that:
\begin{eqnarray}
\Xb(u)&\leq&\int_{u_0}^u e^{\int_{u^\prime}^u
\ab(u^{\prime\prime})du^{\prime\prime}}\ab(u^\prime)\bb(u^\prime)du^\prime
\label{6.86}\\
&=&e^{\int_{u_0}^u
\ab(u^\prime)du^\prime}\left\{\bb(u_0)-\bb(u)e^{-\int_{u_0}^u
\ab(u^\prime)du^\prime} +\int_{u_0}^u
e^{-\int_{u_0}^{u^\prime}\ab(u^{\prime\prime})du^{\prime\prime}}\frac{d\bb}{du^\prime}(u^\prime)du^\prime\right\}\nonumber
\end{eqnarray}
Hence:
\begin{eqnarray}
\xb(u)&\leq&\Xb(u)+\bb(u)\nonumber\\
&\leq&e^{\int_{u_0}^u \ab(u^\prime)du^\prime}\bb(u_0)+\int_{u_0}^u
e^{\int_{u^\prime}^u\ab(u^{\prime\prime})du^{\prime\prime}}
\frac{d\bb}{du^\prime}(u^\prime)du^\prime\nonumber\\
&\s&\label{6.87}
\end{eqnarray}
Now, if $\delta$ is suitably small depending on ${\cal
D}_0^\infty$, ${\cal R}_0^\infty$, $\scD_1^4$, $\scR_1^4$, we have
(see \ref{6.82} and \ref{6.77}):
\begin{equation}
\int_{u_0}^u\ab(u^\prime)du^\prime\leq\log 2 \label{6.88}
\end{equation}
therefore \ref{6.87} simplifies to:
\begin{equation}
\xb(u)\leq
2\bb(u_0)+2\int_{u_0}^u\frac{d\bb}{du^\prime}du^\prime=2\bb(u)
\label{6.89}
\end{equation}
In view of the definitions \ref{6.80}, \ref{6.83} we conclude
that:
\begin{eqnarray}
&&\|\snab^{ \ 2}\mbox{tr}\chib^\prime\|_{L^4(S_{\ub,u})}\leq
C\delta|u|^{-9/2}\left\{\scD_2^4(\mbox{tr}\chib)
+k+(\scD_1^4(\mbox{tr}\chib)+\scD_1^4(\chibh))^2\right.\nonumber\\
&&\hspace{3cm}\left.+(\scD_1^4(\mbox{tr}\chib)+\scD_1^4(\chibh)+{\cal
D}_0^\infty(\chibh))(\scR_1^4(\beta)+{\cal
R}_0^\infty(\beta))\right\}
\nonumber\\
&&\hspace{3cm}+O(\delta^{3/2}|u|^{-11/2})\label{6.90}
\end{eqnarray}
for all $(\ub,u)\in D_1^{s^*}$.

Substituting the estimate \ref{6.90} in \ref{6.62} we obtain:
\begin{eqnarray}
&&|u|\|\snab^{ \
2}\chibh^\prime\|_{L^4(S_{\ub,u})}+\|\snab\chibh^\prime\|_{L^4(S_{\ub,u})}
+|u|^{-1}\|\chibh^\prime\|_{L^4(S_{\ub,u})}\nonumber\\
&&\hspace{1cm}\leq
C\delta|u|^{-7/2}k+C\delta^{1/2}|u|^{-5/2}(\scR_1^4(\beta)+{\cal
R}_0^\infty(\beta))+O(\delta|u|^{-7/2})
\nonumber\\
&&\label{6.91}
\end{eqnarray}
for all $(\ub,u)\in D_1^{s^*}$. In view of the estimates
\ref{6.90}, \ref{6.91}, the lemma follows taking also into account
\ref{6.51}, the estimates of Proposition 4.2, as well as the
estimates of Chapter 3.

\vspace{5mm}

\section{$L^4(S)$ estimates for $\snab^{ \ 2}\eta$, $\snab^{ \ 2}\etb$}

The {\em mass aspect function} $\mu$ of the surface $S_{\ub,u}$
considered as a section of the null hypersurface $C_u$ is defined
by:
\begin{equation}
\mu=K+\frac{1}{4}\mbox{tr}\chi\mbox{tr}\chib-\sdiv\eta
\label{6.92}
\end{equation}
Similarly, the {\em mass aspect function} $\mub$ of the surface
$S_{\ub,u}$ considered as a section of the null hypersurface
$\Cb_{\ub}$ is defined by:
\begin{equation}
\mub=K+\frac{1}{4}\mbox{tr}\chi\mbox{tr}\chib-\sdiv\etb
\label{6.93}
\end{equation}
Substituting for the Gauss curvature $K$ from the Gauss equation
\ref{1.108}, the above definitions take the form:
\begin{eqnarray}
\mu&=&-\rho+\frac{1}{2}(\chih,\chibh)-\sdiv\eta\label{6.94}\\
\mub&=&-\rho+\frac{1}{2}(\chih,\chibh)-\sdiv\etb\label{6.95}
\end{eqnarray}
If we consider $\mu$ as given, equation \ref{6.94} together with
equation \ref{1.144} constitute an elliptic system for $\eta$ on
each $S_{\ub,u}$ section of $C_u$:
\begin{eqnarray}
\sdiv\eta&=&-\rho+\frac{1}{2}(\chih,\chibh)-\mu\nonumber\\
\scurl\eta&=&\sigma-\frac{1}{2}\chih\wedge\chibh \label{6.96}
\end{eqnarray}
Similarly, if we consider $\mub$ as given, equation \ref{6.95}
together with equation \ref{1.141} constitute an elliptic system
for $\etb$ on each $S_{\ub,u}$ section   of $\Cb_{\ub}$:
\begin{eqnarray}
\sdiv\etb&=&-\rho+\frac{1}{2}(\chih,\chibh)-\mub\nonumber\\
\scurl\etb&=&-\sigma+\frac{1}{2}\chih\wedge\chibh \label{6.97}
\end{eqnarray}
(Note from \ref{1.143} that if $\theta$ is any symmetric
2-covariant $S$ tensorfield we have:
$$\sg\wedge\theta=\theta\wedge\sg=0$$
Hence, if $\theta$, $\theta^\prime$ is any pair of symmetric
2-covariant $S$ tensorfields, then:
$$\theta\wedge\theta^\prime=\hat{\theta}\wedge\hat{\theta}^\prime$$
where $\hat{\theta}$ and $\hat{\theta}^\prime$ are the trace-free
parts of $\theta$ and $\theta^\prime$ respectively.) The elliptic
system \ref{6.96} is to be considered in conjunction with a
propagation equation for $\mu$ along the generators of $C_u$.
Similarly, the elliptic system \ref{6.97} is to be considered in
conjunction with a propagation equation for $\mub$ along the
generators of $\Cb_{\ub}$. We shall presently derive these
propagation equations.

We consider the expression \ref{6.92} for $\mu$. The Gauss
curvature $K$ satisfies along the generators of $C_u$ the
propagation equation \ref{5.28}:
\begin{equation}
DK=-\Omega\mbox{tr}\chi
K+\sdiv\sdiv(\Omega\chih)-\frac{1}{2}\slap(\Omega\mbox{tr}\chi)
\label{6.98}
\end{equation}
Taking the trace of equation \ref{1.146} and noting \ref{1.a2},
the definition \ref{6.95}, and the fact that by the first of
\ref{3.3} if $\theta$ is any 2-covariant $S$ tensorfield we have:
\begin{equation}
D\mbox{tr}\theta=\mbox{tr}(D\theta)-2\Omega(\chi,\theta)
\label{6.99}
\end{equation}
we obtain the equation:
\begin{equation}
D(\Omega\mbox{tr}\chib)=\Omega^2\left\{-\frac{1}{2}\mbox{tr}\chi\mbox{tr}\chib-2\mub+2|\etb|^2\right\}
\label{6.100}
\end{equation}
Writing
$\mbox{tr}\chi\mbox{tr}\chib=\mbox{tr}\chi^\prime(\Omega\mbox{tr}\chib)$
and using also the propagation equation \ref{3.6} we then deduce:
\begin{equation}
\frac{1}{4}D(\mbox{tr}\chi\mbox{tr}\chib)=\Omega\left\{-\frac{1}{4}(\mbox{tr}\chi)^2\mbox{tr}\chib
-\frac{1}{4}\mbox{tr}\chib|\chih|^2-\frac{1}{2}\mbox{tr}\chi\mub+\frac{1}{2}\mbox{tr}\chi|\etb|^2\right\}
\label{6.101}
\end{equation}
Next, we revisit the propagation equation \ref{1.66} for $\eta$:
\begin{equation}
D\eta=\Omega\left(\chih^\sharp\cdot\etb+\frac{1}{2}\mbox{tr}\chi\etb-\beta\right)
\label{6.102}
\end{equation}
Now, from \ref{6.4}, \ref{6.5} we can express $\beta$ in the form:
$$-\beta=\Omega\left(\sdiv\chih^\prime-\frac{1}{2}\sd\mbox{tr}\chi^\prime\right)
-\frac{1}{2}\mbox{tr}\chi\eta+\chih^\sharp\cdot\eta$$ or:
\begin{eqnarray}
-\Omega\beta&=&\sdiv(\Omega\chih)-\frac{1}{2}\sd(\Omega\mbox{tr}\chi)\label{6.103}\\
&\s&+\Omega\left\{-2\chih^\sharp\cdot\sd\log\Omega+\mbox{tr}\chi\sd\log\Omega
+\chih^\sharp\cdot\eta-\frac{1}{2}\mbox{tr}\chi\eta\right\}\nonumber
\end{eqnarray}
Substituting in \ref{6.102} and recalling that, from \ref{1.65},
$$\eta+\etb=2\sd\log\Omega$$
yields:
\begin{equation}
D\eta=\sdiv(\Omega\chih)-\frac{1}{2}\sd(\Omega\mbox{tr}\chi)+\Omega\mbox{tr}\chi\etb
\label{6.104}
\end{equation}
From Lemma 4.1 for any $S$ 1-form $\xi$ we have:
\begin{equation}
D\snab\xi-\snab D\xi=-D\sGamma\cdot\xi \label{6.105}
\end{equation}
and:
\begin{equation}
\mbox{tr}(D\sGamma\cdot\xi)=2(\sdiv(\Omega\chih),\xi)
\label{6.106}
\end{equation}
Also, since $\sdiv\xi=\mbox{tr}\snab\xi$, we have, from
\ref{6.99},
$$D\sdiv\xi-\mbox{tr}(D\snab\xi)=-2\Omega(\chi,\snab\xi)$$
Thus, taking the trace of \ref{6.105} we obtain:
\begin{equation}
D\sdiv\xi-\sdiv
D\xi=-2\sdiv(\Omega\chih^\sharp\cdot\xi)-\Omega\mbox{tr}\chi\sdiv\xi
\label{6.107}
\end{equation}
Applying $\sdiv$ to \ref{6.102} and substituting in \ref{6.107}
with $\eta$ in the role of $\xi$ we then obtain:
\begin{eqnarray}
-D\sdiv\eta&=&-\sdiv\sdiv(\Omega\chih)+\frac{1}{2}\slap(\Omega\mbox{tr}\chi)\nonumber\\
&\s&+\Omega\mbox{tr}\chi\sdiv\eta+\sdiv j \label{6.108}
\end{eqnarray}
where $j$ is the $S$ 1-form:
\begin{equation}
j=\Omega(2\chih^\sharp\cdot\eta-\mbox{tr}\chi\etb) \label{6.109}
\end{equation}
Adding finally \ref{6.98}, \ref{6.101} and \ref{6.108} we obtain
the desired propagation equation for $\mu$:
\begin{eqnarray}
D\mu&=&-\Omega\mbox{tr}\chi\mu-\frac{1}{2}\Omega\mbox{tr}\chi\mub\nonumber\\
&\s&+\Omega\left(-\frac{1}{4}\mbox{tr}\chib|\chih|^2+\frac{1}{2}\mbox{tr}\chi|\etb|^2\right)\nonumber\\
&\s&+\sdiv j \label{6.110}
\end{eqnarray}
In a simmilar manner we derive the following propagation equation
for $\mub$:
\begin{eqnarray}
\Db\mub&=&-\Omega\mbox{tr}\chib\mub-\frac{1}{2}\Omega\mbox{tr}\chib\mu\nonumber\\
&\s&+\Omega\left(-\frac{1}{4}\mbox{tr}\chi|\chibh|^2+\frac{1}{2}\mbox{tr}\chib|\eta|^2\right)\nonumber\\
&\s&+\sdiv\jb \label{6.111}
\end{eqnarray}
where $\jb$ is the $S$ 1-form:
\begin{equation}
\jb=\Omega(2\chibh\cdot\etb-\mbox{tr}\chib\eta) \label{6.112}
\end{equation}
Equation \ref{6.111} is simply the conjugate of equation
\ref{6.110}. We note the crucial fact that the right hand sides of
these equations do not contain principal terms (such terms would
involve the first derivatives of the curvature or the second
derivatives of the connection coefficients).

In the following we denote by $O(\delta^p|u|^r)$, for real numbers
$p$, $r$, the product of $\delta^p |u|^r$ with a non-negative
non-decreasing continuous function of the quantities ${\cal
D}_0^\infty$, ${\cal R}_0^\infty$, $\scD_1^4$, $\scR_1^4$, {\em
and} $\scD_2^4(\mbox{tr}\chib)$.

\vspace{5mm}

\noindent{\bf Lemma 6.2} \ \  We have:
\begin{eqnarray*}
&&|u|\|\snab^{ \
2}\eta\|_{L^4(S_{\ub,u})}+\|\snab\eta\|_{L^4(S_{\ub,u})}
+|u|^{-1}\|\eta\|_{L^4(S_{\ub,u})}\\
&&\hspace{1cm}\leq C|u|^{-5/2}\tilde{A}+O(\delta^{1/2}|u|^{-5/2})\\
&&\hspace{1cm}\s+O(\delta^{5/2}|u|^{-11/2})k
\end{eqnarray*}
\begin{eqnarray*}
&&|u|\|\snab^{ \
2}\etb\|_{L^4(S_{\ub,u})}+\|\snab\etb\|_{L^4(S_{\ub,u})}
+|u|^{-1}\|\etb\|_{L^4(S_{\ub,u})}\\
&&\hspace{1cm}\leq C|u|^{-5/2}\tilde{A}+O(\delta^{1/2}|u|^{-5/2})\\
&&\hspace{1cm}\s+O(\delta^{3/2}|u|^{-9/2})k
\end{eqnarray*}
for all $(\ub,u)\in D_1^{s^*}$, provided that $\delta$ is suitably
small depending on ${\cal D}_0^\infty$, ${\cal R}_0^\infty$,
$\scD_1^4$, $\scR_1^4$, and $\scR_2(\alpha)$. Moreover, the
coefficients of $k$ depend only on ${\cal D}_0^\infty$, ${\cal
R}_0^\infty$. Here:
\begin{eqnarray*}
\tilde{A}&=&\scR_1^4(\rho)+\scR_1^4(\sigma)+{\cal
R}_0^\infty(\rho)+{\cal R}_0^\infty(\sigma)
+\scR_1^4(\alpha){\cal R}_0^\infty(\alpha)\\
&\s&+(\scR_1^4(\alpha)+{\cal R}_0^\infty(\alpha)){\cal
D}_0^\infty(\chibh)+
(\scD_1^4(\mbox{tr}\chib)+\scD_1^4(\chibh)){\cal
R}_0^\infty(\alpha)
\end{eqnarray*}

\noindent{\em Proof:} \ We apply Lemma 5.11 to the systems
\ref{6.96} and \ref{6.97}. In view of the last conclusion of
Proposition 6.1 the conclusion of Lemma 5.11 simplifies and we
obtain:
\begin{eqnarray}
&&|u|\|\snab^{ \ 2}\eta\|_{L^4(S_{\ub,u})}+\|\snab\eta\|_{L^4(S_{\ub,u})}+|u|^{-1}\|\eta\|_{L^4(S_{\ub,u})}\nonumber\\
&&\hspace{1cm}\leq
C\left\{|u|\left(\|\sd\mu\|_{L^4(S_{\ub,u})}+\|\sd\rho\|_{L^4(S_{\ub,u})}
+\|\sd\sigma\|_{L^4(S_{\ub,u})}\right.\right.\nonumber\\
&&\hspace{2cm}\left.+\|\sd(\chih,\chibh)\|_{L^4(S_{\ub,u})}+\|\sd(\chih\wedge\chibh)\|_{L^4(S_{\ub,u})}\right)
\nonumber\\
&&\hspace{2cm}+\|\mu\|_{L^4(S_{\ub,u})}+\|\rho\|_{L^4(S_{\ub,u})}+\|\sigma\|_{L^4(S_{\ub,u})}\nonumber\\
&&\hspace{2cm}\left.+\|(\chih,\chibh)\|_{L^4(S_{\ub,u})}+\|\chih\wedge\chibh\|_{L^4(S_{\ub,u})}\right\}
\label{6.113}
\end{eqnarray}
and:
\begin{eqnarray}
&&|u|\|\snab^{ \ 2}\etb\|_{L^4(S_{\ub,u})}+\|\snab\etb\|_{L^4(S_{\ub,u})}+|u|^{-1}\|\etb\|_{L^4(S_{\ub,u})}\nonumber\\
&&\hspace{1cm}\leq
C\left\{|u|\left(\|\sd\mub\|_{L^4(S_{\ub,u})}+\|\sd\rho\|_{L^4(S_{\ub,u})}
+\|\sd\sigma\|_{L^4(S_{\ub,u})}\right.\right.\nonumber\\
&&\hspace{2cm}\left.+\|\sd(\chih,\chibh)\|_{L^4(S_{\ub,u})}+\|\sd(\chih\wedge\chibh)\|_{L^4(S_{\ub,u})}\right)
\nonumber\\
&&\hspace{2cm}+\|\mub\|_{L^4(S_{\ub,u})}+\|\rho\|_{L^4(S_{\ub,u})}+\|\sigma\|_{L^4(S_{\ub,u})}\nonumber\\
&&\hspace{2cm}\left.+\|(\chih,\chibh)\|_{L^4(S_{\ub,u})}+\|\chih\wedge\chibh\|_{L^4(S_{\ub,u})}\right\}
\label{6.114}
\end{eqnarray}
We have:
\begin{eqnarray}
&&\|\sd(\chih,\chibh)\|_{L^4(S_{\ub,u})}+\|\sd(\chih\wedge\chibh)\|_{L^4(S_{\ub,u})}\nonumber\\
&&\hspace{1cm}\leq
C\left\{\|\snab\chih\|_{L^4(S_{\ub,u})}\|\chibh\|_{L^\infty(S_{\ub,u})}
+\|\snab\chibh\|_{L^4(S_{\ub,u})}\|\chih\|_{L^\infty(S_{\ub,u})}\right\}\nonumber\\
&&\hspace{1cm}\leq C|u|^{-7/2}[\scR_1^4(\alpha){\cal
D}_0^\infty(\chibh)+
(\scD_1^4(\mbox{tr}\chib)+\scD_1^4(\chibh)){\cal R}_0^\infty(\alpha)]\nonumber\\
&&\hspace{1cm}\s+O(\delta^{1/2}|u|^{-9/2})\label{6.115}
\end{eqnarray}
by Propositions 4.1 and 4.2 and the results of Chapter 3.  Also,
\begin{eqnarray}
&&\|(\chih,\chibh)\|_{L^4(S_{\ub,u})}+\|\chih\wedge\chibh\|_{L^4(S_{\ub,u})}\nonumber\\
&&\hspace{1cm}\leq C|u|^{1/2}\|\chih\|_{L^\infty(S_{\ub,u})}\|\chibh\|_{L^\infty(S_{\ub,u})}\nonumber\\
&&\hspace{1cm}\leq C|u|^{-5/2}{\cal R}_0^\infty(\alpha){\cal
D}_0^\infty(\chibh)+O(\delta|u|^{-9/2}) \label{6.116}
\end{eqnarray}
by the results of Chapter 3. Moreover, by Proposition 4.2 and the
results of Chapter 3 we can estimate:
\begin{equation}
\|\mu\|_{L^4(S_{\ub,u})}, \|\mub\|_{L^4(S_{\ub,u})} \leq
C|u|^{-5/2}[{\cal R}_0^\infty(\rho)+{\cal R}_0^\infty(\alpha){\cal
D}_0^\infty(\chibh)] +O(\delta^{1/2}|u|^{-5/2}) \label{6.117}
\end{equation}
Substituting the above in \ref{6.113}, \ref{6.114} and taking also
into account the third and fourth of the definitions \ref{4.1} we
obtain:
\begin{eqnarray}
&&|u|\|\snab^{ \ 2}\eta\|_{L^4(S_{\ub,u})}+\|\snab\eta\|_{L^4(S_{\ub,u})}+|u|^{-1}\|\eta\|_{L^4(S_{\ub,u})}\nonumber\\
&&\hspace{1cm}\leq
C|u|\|\sd\mu\|_{L^4(S_{\ub,u})}+C|u|^{-5/2}A+O(\delta^{1/2}|u|^{-5/2})
\label{6.118}
\end{eqnarray}
and:
\begin{eqnarray}
&&|u|\|\snab^{ \ 2}\etb\|_{L^4(S_{\ub,u})}+\|\snab\etb\|_{L^4(S_{\ub,u})}+|u|^{-1}\|\etb\|_{L^4(S_{\ub,u})}\nonumber\\
&&\hspace{1cm}\leq
C|u|\|\sd\mub\|_{L^4(S_{\ub,u})}+C|u|^{-5/2}A+O(\delta^{1/2}|u|^{-5/2})
\label{6.119}
\end{eqnarray}
where:
\begin{eqnarray}
A&=&\scR_1^4(\rho)+\scR_1^4(\sigma)+{\cal R}_0^\infty(\rho)+{\cal R}_0^\infty(\sigma)\nonumber\\
&\s&+(\scR_1^4(\alpha)+{\cal R}_0^\infty(\alpha)){\cal
D}_0^\infty(\chibh)+
(\scD_1^4(\mbox{tr}\chib)+\scD_1^4(\chibh)){\cal
R}_0^\infty(\alpha) \label{6.120}
\end{eqnarray}

We now turn to the propagation equations \ref{6.110}, \ref{6.111}.
In view of Lemma 1.2 we deduce the following propagation equations
for $\sd\mu$, $\sd\mub$:
\begin{equation}
D\sd\mu=-\Omega\mbox{tr}\chi\sd\mu-\frac{1}{2}\Omega\mbox{tr}\chi\sd\mub+h
\label{6.121}
\end{equation}
\begin{equation}
\Db\sd\mub=-\Omega\mbox{tr}\chib\sd\mub-\frac{1}{2}\Omega\mbox{tr}\chib\sd\mu+\hb
\label{6.122}
\end{equation}
where $h$, $\hb$ are the $S$ 1-forms:
\begin{eqnarray}
h&=&\sd\sdiv j-\left(\mu+\frac{1}{2}\mub\right)\sd(\Omega\mbox{tr}\chi)\nonumber\\
&\s&+\sd\left[\Omega\left(-\frac{1}{4}\mbox{tr}\chib|\chih|^2+\frac{1}{2}\mbox{tr}\chi|\etb|^2\right)\right]
\label{6.123}
\end{eqnarray}
\begin{eqnarray}
\hb&=&\sd\sdiv\jb-\left(\mub+\frac{1}{2}\mu\right)\sd(\Omega\mbox{tr}\chib)\nonumber\\
&\s&+\sd\left[\Omega\left(-\frac{1}{4}\mbox{tr}\chi|\chibh|^2+\frac{1}{2}\mbox{tr}\chib|\eta|^2\right)\right]
\label{6.124}
\end{eqnarray}

The $S$ 1-form $\sd\sdiv j$, the first term on the right in
\ref{6.123} is of the form:
\begin{eqnarray}
\sd\sdiv j&=&(\Omega\chih,\snab^{ \ 2}\eta)+(\Omega\mbox{tr}\chi,\snab^{ \ 2}\etb)\nonumber\\
&\s&+(\snab(\Omega\chih),\snab\eta)+(\sd(\Omega\mbox{tr}\chi),\snab\etb)\nonumber\\
&\s&+(\snab^{ \ 2}(\Omega\chih),\eta)+(\snab^{ \
2}(\Omega\mbox{tr}\chi),\etb) \label{6.125}
\end{eqnarray}
The first two terms, placing the first factors in $L^\infty(S)$
and the second factors in $L^4(S)$, are bounded in $L^4(S)$ by:
\begin{equation}
C\delta^{-1/2}|u|^{-1}{\cal R}_0^\infty(\alpha)\|\snab^{ \
2}\eta\|_{L^4(S_{\ub,u})} +C(|u|^{-1}+O(|u|^{-2}))\|\snab^{ \
2}\etb\|_{L^4(S_{\ub,u})} \label{6.126}
\end{equation}
To estimate the second two terms in $L^4(S)$, we place the first
factors in $L^4(S)$ and the second factors in $L^\infty(S)$. We
then use Lemma 5.2 with $p=4$ to bound the $L^\infty(S)$ norms of
the second factors in terms of their $W_1^4(S)$ norms. Using also
Propositions 4.1 and 4.2 we then obtain a bound by:
\begin{eqnarray}
&&(C\delta^{-1/2}|u|^{-1}\scR_1^4(\alpha)+O(|u|^{-2}))\|\snab^{ \
2}\eta\|_{L^4(S_{\ub,u})}
+O(|u|^{-2})\|\snab^{ \ 2}\etb\|_{L^4(S_{\ub,u})}\nonumber\\
&&+O(|u|^{-9/2}) \label{6.127}
\end{eqnarray}
Finally, to estimate the last two terms in $L^4(S)$, we place the
first factors in $L^4(S)$ using Proposition 6.1 and the second
factors in $L^\infty(S)$ using the results of Chapter 3. We then
obtain a bound by $O(|u|^{-9/2})$. Combining the above results
yields:
\begin{eqnarray}
&&\|\sd\sdiv j\|_{L^4(S_{\ub,u})}\leq\nonumber\\
&&\hspace{1cm} C\delta^{-1/2}|u|^{-1}(\scR_1^4(\alpha)+{\cal
R}_0^\infty(\alpha)
+O(\delta^{1/2}|u|^{-1}))\|\snab^{ \ 2}\eta\|_{L^4(S_{\ub,u})}\nonumber\\
&&\hspace{1cm} +C|u|^{-1}(1+O(|u|^{-1})\|\snab^{ \
2}\etb\|_{L^4(S_{\ub,u})}+O(|u|^{-9/2}) \label{6.128}
\end{eqnarray}

The $S$ 1-form $\sd\sdiv\jb$, the first term on the right in
\ref{6.124} is of the form:
\begin{eqnarray}
\sd\sdiv\jb&=&(\Omega\chibh,\snab^{ \ 2}\etb)+(\Omega\mbox{tr}\chib,\snab^{ \ 2}\eta)\nonumber\\
&\s&+(\snab(\Omega\chibh),\snab\etb)+(\sd(\Omega\mbox{tr}\chib),\snab\eta)\nonumber\\
&\s&+(\snab^{ \ 2}(\Omega\chibh),\etb)+(\snab^{ \
2}(\Omega\mbox{tr}\chib),\eta) \label{6.129}
\end{eqnarray}
The first two terms, placing the first factors in $L^\infty(S)$
and the second factors in $L^4(S)$, are bounded in $L^4(S)$ by:
\begin{eqnarray}
&&C\delta^{1/2}|u|^{-2}({\cal D}_0^\infty(\chibh)+O(\delta|u|^{-3/2}))\|\snab^{ \ 2}\etb\|_{L^4(S_{\ub,u})}\nonumber\\
&&+C|u|^{-1}(1+O(\delta|u|^{-1}))\|\snab^{ \
2}\eta\|_{L^4(S_{\ub,u})} \label{6.130}
\end{eqnarray}
To estimate the second two terms in $L^4(S)$, we place the first
factors in $L^4(S)$ and the second factors in $L^\infty(S)$. We
then use Lemma 5.2 with $p=4$ to bound the $L^\infty(S)$ norms of
the second factors in terms of their $W_1^4(S)$ norms. Using also
Proposition 4.2 we then obtain a bound by:
\begin{eqnarray}
&&C\delta^{1/2}|u|^{-2}(\scD_1^4(\mbox{tr}\chib)+\scD_1^4(\chibh)+O(\delta^{1/2}|u|^{-1}))
\|\snab^{ \ 2}\etb\|_{L^4(S_{\ub,u})}\nonumber\\
&&+O(\delta|u|^{-3})\|\snab^{ \
2}\eta\|_{L^4(S_{\ub,u})}+O(\delta|u|^{-11/2}) \label{6.131}
\end{eqnarray}
Finally, to estimate the last two terms in $L^4(S)$, we place the
first factors in $L^4(S)$ and the second factors in $L^\infty(S)$
using the results of Chapter 3 to obtain a bound by:
\begin{eqnarray}
&&C\delta^{1/2}|u|^{-2}({\cal
R}_0^\infty(\beta)+O(\delta^{1/2}|u|^{-1})) \|\snab^{ \
2}(\Omega\chibh)\|_{L^4(S_{\ub,u})}
\nonumber\\
&&+C\delta^{1/2}|u|^{-2}({\cal R}_0^\infty(\beta)
+O(\delta|u|^{-1}))\|\snab^{ \ 2}(\Omega\mbox{tr}\chib)\|_{L^4(S_{\ub,u})}\nonumber\\
&&\leq O(\delta^{3/2}|u|^{-13/2})k+O(\delta|u|^{-11/2})
\label{6.132}
\end{eqnarray}
for all $(\ub,u)\in D_1^{s^*}$, by Lemma 6.1, taking also into
account \ref{6.51}, the estimates of Proposition 4.2 for
$\snab\chibh$, as well as the estimates of Chapter 3. Note that
the coefficient of $k$ depends only on ${\cal D}_0^\infty$, ${\cal
R}_0^\infty$. Combining the above results yields:
\begin{eqnarray}
&&\|\sd\sdiv\jb\|_{L^4(S_{\ub,u})}\leq\nonumber\\
&&\hspace{1cm}C\delta^{1/2}|u|^{-2}(\scD_1^4(\mbox{tr}\chib)+\scD_1^4(\chibh)+{\cal
D}_0^\infty(\chibh)+O(\delta^{1/2}|u|^{-1}))
\|\snab^{ \ 2}\etb\|_{L^4(S{\ub,u})}\nonumber\\
&&\hspace{2cm}+C|u|^{-1}(1+O(\delta|u|^{-2}))\|\snab^{ \ 2}\eta\|_{L^4(S_{\ub,u})}\nonumber\\
&&\hspace{2cm}+O(\delta^{3/2}|u|^{-13/2})k+O(\delta|u|^{-11/2})
\label{6.133}
\end{eqnarray}

Consider next the second terms in each of \ref{6.123},
\ref{6.124}. Consider the expressions \ref{6.94}, \ref{6.95}. By
Lemma 5.2 with $p=4$ and Proposition 4.2 we have:
\begin{eqnarray}
\|\snab\eta\|_{L^\infty(S_{\ub,u})}&\leq& C|u|^{1/2}\|\snab\eta\|_{W_1^4(S_{\ub,u})}\nonumber\\
&\leq& C|u|^{1/2}\|\snab^{ \
2}\eta\|_{L^4(S_{\ub,u})}+O(\delta^{1/2}|u|^{-3}) \label{6.134}
\end{eqnarray}
\begin{eqnarray}
\|\snab\etb\|_{L^\infty(S_{\ub,u})}&\leq& C|u|^{1/2}\|\snab\etb\|_{W_1^4(S_{\ub,u})}\nonumber\\
&\leq& C|u|^{1/2}\|\snab^{ \
2}\etb\|_{L^4(S_{\ub,u})}+O(\delta^{1/2}|u|^{-3}) \label{6.135}
\end{eqnarray}
Thus, taking also into account the results of Chapter 3 we obtain:
\begin{eqnarray}
\|\mu\|_{L^\infty(S_{\ub,u})}
&\leq& C|u|^{-3}({\cal R}_0^\infty(\rho)+{\cal R}_0^\infty(\alpha){\cal D}_0^\infty(\chibh))\nonumber\\
&\s&+C|u|^{1/2}\|\snab^{ \
2}\eta\|_{L^4(S_{\ub,u})}+O(\delta^{1/2}|u|^{-3}) \label{6.136}
\end{eqnarray}
\begin{eqnarray}
\|\mub\|_{L^\infty(S_{\ub,u})}
&\leq& C|u|^{-3}({\cal R}_0^\infty(\rho)+{\cal R}_0^\infty(\alpha){\cal D}_0^\infty(\chibh))\nonumber\\
&\s&+C|u|^{1/2}\|\snab^{ \
2}\etb\|_{L^4(S_{\ub,u})}+O(\delta^{1/2}|u|^{-3}) \label{6.137}
\end{eqnarray}
Placing then the first factors in the second terms in each of
\ref{6.123}, \ref{6.124} in $L^\infty(S)$ and the second factors
in $L^4(S)$ using the results of Chapter 4 we obtain that the
second term in \ref{6.123} is bounded in $L^4(S)$ norm by:
\begin{equation}
C|u|^{-2}[{\cal
R}_0^\infty(\alpha)\scR_1^4(\alpha)+O(\delta^{1/2})] (\|\snab^{ \
2}\eta\|_{L^4(S_{\ub,u})}+\|\snab^{ \
2}\etb\|_{L^4(S_{\ub,u})})+O(|u|^{-11/2}) \label{6.138}
\end{equation}
while the second term in \ref{6.124} is bounded in $L^4(S)$ norm
by:
\begin{eqnarray}
&&C\delta|u|^{-3}[\scR_1^4(\rho)+(1+{\cal
D}_0^\infty(\chibh))(\scD_1^4(\mbox{tr}\chib)+\scD_1^4(\chibh))
+O(\delta^{1/2})]\cdot\nonumber\\
&&\hspace{18mm}\cdot(\|\snab^{ \
2}\eta\|_{L^4(S_{\ub,u})}+\|\snab^{ \ 2}\etb\|_{L^4(S_{\ub,u})})
+O(\delta|u|^{-13/2}) \label{6.139}
\end{eqnarray}

By the results of Chapters 3 and 4 the last term in \ref{6.123} is
bounded in $L^4(S)$ by:
\begin{equation}
C\delta^{-1}|u|^{-7/2}{\cal
R}_0^\infty(\alpha)\scR_1^4(\alpha)+O(\delta^{-1/2}|u|^{-9/2})
\label{6.140}
\end{equation}
while the last term in \ref{6.124} is bounded in $L^4(S)$ by:
\begin{equation}
O(\delta|u|^{-11/2}) \label{6.141}
\end{equation}

Combining the results \ref{6.128}, \ref{6.138} and \ref{6.140}
yields:
\begin{eqnarray}
\|h\|_{L^4(S_{\ub,u})}&\leq&C\delta^{-1/2}|u|^{-1}(\scR_1^4(\alpha)+{\cal
R}_0^\infty(\alpha)
+O(\delta^{1/2}|u|^{-1}))\|\snab^{ \ 2}\eta\|_{L^4(S_{\ub,u})}\nonumber\\
&\s&+C|u|^{-1}(1+O(|u|^{-1}))\|\snab^{ \ 2}\etb\|_{L^4(S_{\ub,u})}\nonumber\\
&\s&+C\delta^{-1}|u|^{-7/2}{\cal
R}_0^\infty(\alpha)\scR_1^4(\alpha)+O(\delta^{-1/2}|u|^{-9/2})
\label{6.142}
\end{eqnarray}
Combining the results \ref{6.133}, \ref{6.139} and \ref{6.141}
yields:
\begin{eqnarray}
\|\hb\|_{L^4(S_{\ub,u})}&\leq&C\delta^{1/2}|u|^{-2}(\scD_1^4(\mbox{tr}\chib)+\scD_1^4(\chibh)
+{\cal D}_0^\infty(\chibh)\nonumber\\
&\s&\hspace{3cm}+O(\delta^{1/2}|u|^{-1}))\|\snab^{ \ 2}\etb\|_{L^4(S_{\ub,u})}\nonumber\\
&\s&+C|u|^{-1}(1+O(\delta|u|^{-1})\|\snab^{ \ 2}\eta\|_{L^4(S_{\ub,u})}\nonumber\\
&\s&+O(\delta^{3/2}|u|^{-13/2})k+O(\delta|u|^{-11/2})
\label{6.143}
\end{eqnarray}
for all $(\ub,u)\in D_1^{s^*}$. Substituting for $\|\snab^{ \
2}\eta\|_{L^4(S_{\ub,u})}$, $\|\snab^{ \
2}\etb\|_{L^4(S_{\ub,u})}$ the estimates \ref{6.118}, \ref{6.119}
we then obtain:
\begin{eqnarray}
\|h\|_{L^4(S_{\ub,u})}&\leq&C\delta^{-1/2}|u|^{-1}(\scR_1^4(\alpha)+{\cal
R}_0^\infty(\alpha)
+O(\delta^{1/2}|u|^{-1}))\|\sd\mu\|_{L^4(S_{\ub,u})}\nonumber\\
&\s&+C|u|^{-1}(1+O(|u|^{-1}))\|\sd\mub\|_{L^4(S_{\ub,u})}\nonumber\\
&\s&+C\delta^{-1}|u|^{-7/2}{\cal
R}_0^\infty(\alpha)\scR_1^4(\alpha)+O(\delta^{-1/2}|u|^{-9/2})
\label{6.144}
\end{eqnarray}
\begin{eqnarray}
\|\hb\|_{L^4(S_{\ub,u})}&\leq&C\delta^{1/2}|u|^{-2}(\scD_1^4(\mbox{tr}\chib)+\scD_1^4(\chibh)
+{\cal D}_0^\infty(\chibh)\nonumber\\
&\s&\hspace{3cm}+O(\delta^{1/2}|u|^{-1}))\|\sd\mub\|_{L^4(S_{\ub,u})}\nonumber\\
&\s&+C|u|^{-1}(1+O(\delta|u|^{-1}))\|\sd\mu\|_{L^4(S_{\ub,u})}+C|u|^{-9/2}A\nonumber\\
&\s&+O(\delta^{3/2}|u|^{-13/2})k+O(\delta^{1/2}|u|^{-9/2})\nonumber\\
&\s&\label{6.145}
\end{eqnarray}
where the coefficient of $k$ depends only on ${\cal D}_0^\infty$,
${\cal R}_0^\infty$.

We now turn to the propagation equations \ref{6.121}, \ref{6.122}.
To \ref{6.121} we apply Lemma 4.6 with $p=4$. Here $r=1$,
$\nu=-2$,$\gamma=0$, and we obtain:
\begin{equation}
\|\sd\mu\|_{L^4(S_{\ub,u})}\leq
C\int_0^{\ub}\left\|-\frac{1}{2}\Omega\mbox{tr}\chi\sd\mub+h\right\|_{L^4(S_{\ub^\prime,u})}
d\ub^\prime \label{6.146}
\end{equation}
To \ref{6.122} we apply Lemma 4.7 with $p=4$. Here again $r=1$,
$\nu=-2$, $\gammab=0$, and we obtain:
\begin{eqnarray}
&&|u|^{5/2}\|\sd\mub\|_{L^4(S_{\ub,u})}\leq C|u_0|^{5/2}\|\sd\mub\|_{L^4(S_{\ub,u_0})}\nonumber\\
&&\hspace{25mm}+C\int_{u_0}^u|u^\prime|^{5/2}\left\|-\frac{1}{2}\Omega\mbox{tr}\chib\sd\mu+\hb\right
\|_{L^4(S_{\ub,u^\prime})}du^\prime \label{6.147}
\end{eqnarray}
Substituting in \ref{6.146} and \ref{6.147} the estimates
\ref{6.144} and \ref{6.145} yields the following system of linear
integral inequalities for the quantities
$\|\sd\mu\|_{L^4(S_{\ub,u})}$, $\|\sd\mub\|_{L^4(S_{\ub,u})}$ on
the domain $D_1^{s^*}$:
\begin{eqnarray}
\|\sd\mu\|_{L^4(S_{\ub,u})}&\leq& a(u)\int_0^{\ub}\|\sd\mu\|_{L^4(S_{\ub^\prime,u})}d\ub^\prime\label{6.148}\\
&\s&+b(u)\int_0^{\ub}\|\sd\mub\|_{L^4(S_{\ub^\prime,u})}d\ub^\prime+f(u)\nonumber
\end{eqnarray}
\begin{eqnarray}
|u|^{5/2}\|\sd\mub\|_{L^4(S_{\ub,u})}&\leq&
\int_{u_0}^u|u^\prime|^{5/2}\ab(u^\prime)\|\sd\mub\|_{L^4(S_{\ub,u^\prime})}du^\prime\label{6.149}\\
&\s&+\int_{u_0}^u|u^\prime|^{5/2}\bb(u^\prime)\|\sd\mu\|_{L^4(S_{\ub,u^\prime})}du^\prime+|u|^{5/2}\fb(u)\nonumber
\end{eqnarray}
Here:
\begin{eqnarray}
a(u)&=&C\delta^{-1/2}|u|^{-1}(\scR_1^4(\alpha)+{\cal
R}_0^\infty(\alpha)+O(\delta^{1/2}|u|^{-1}))
\label{6.150}\\
b(u)&=&C|u|^{-1}(1+O(|u|^{-1}))
\label{6.151}\\
f(u)&=&C|u|^{-7/2}{\cal
R}_0^\infty(\alpha)\scR_1^4(\alpha)+O(\delta^{1/2}|u|^{-9/2})
\label{6.152}\\
\ab(u)&=&C\delta^{1/2}|u|^{-2}(\scD_1^4(\mbox{tr}\chib)+\scD_1^4(\chibh)+{\cal
D}_0^\infty(\chibh)+O(\delta^{1/2}|u|^{-1}))
\label{6.153}\\
\bb(u)&=&C|u|^{-1}(1+O(\delta|u|^{-1})) \label{6.154}
\end{eqnarray}
and:
\begin{eqnarray}
\fb(u)&=&C|u|^{-5/2}|u_0|^{-1}\scD_1^4(\mub)+C|u|^{-7/2}A\nonumber\\
&\s&+O(\delta^{3/2}|u|^{-11/2})k+O(\delta^{1/2}|u|^{-7/2})
\label{6.155}
\end{eqnarray}
where we have defined:
\begin{equation}
\scD_1^4(\mub)=|u_0|^{7/2}\sup_{\ub\in[0,\ub_1]}\|\sd\mub\|_{L^4(S_{\ub,u_0})}
\label{6.156}
\end{equation}
and the coefficient of $k$ depends only on ${\cal D}_0^\infty$,
${\cal R}_0^\infty$.

Consider first the integral inequality \ref{6.148}. At fixed $u$,
considering $\|\sd\mub\|_{L^4(S_{\ub,u})}$ as given, this
inequality is of the form \ref{6.21}, with $a(u)$ in the role of
the constant $a$ and
$$b(u)\int_0^{\ub}\|\sd\mub\|_{L^4(S_{\ub^\prime,u})}d\ub^\prime+f(u)$$
in the role of the non-negative function $b(\ub)$. If, as is the
case here, the function $b(\ub)$ is non-decreasing, the result
\ref{6.24} implies:
\begin{equation}
x(\ub)\leq e^{a\ub}b(\ub) \label{6.157}
\end{equation}
Moreover, if
\begin{equation}
a\delta\leq\log 2 \label{6.158}
\end{equation}
\ref{6.157} in turn implies:
\begin{equation}
x(\ub)\leq 2b(\ub) \label{6.159}
\end{equation}
Condition \ref{6.158} in the present case reads:
\begin{equation}
a(u)\delta=C\delta^{1/2}|u|^{-1}(\scR_1^4(\alpha)+{\cal
R}_0^\infty(\alpha)+O(\delta^{1/2}|u|^{-1}))\leq\log 2
\label{6.160}
\end{equation}
and is indeed satisfied provided that $\delta$ is suitably small
depending on ${\cal D}_0^\infty$, ${\cal R}_0^\infty$, $\scD_1^4$,
$\scR_1^4$. The result \ref{6.159} then takes the form:
\begin{equation}
\|\sd\mu\|_{L^4(S_{\ub,u})}\leq
2b(u)\int_0^{\ub}\|\sd\mub\|_{L^4(S_{\ub^\prime,u})}d\ub^\prime+2f(u)
\label{6.161}
\end{equation}

Consider next the integral inequality \ref{6.149}. At fixed $\ub$,
considering $\|\sd\mu\|_{L^4(S_{\ub,u})}$ as given, this
inequality is of the form \ref{6.81} with $\ab(u)$ in the role of
the non-negative function $\ab$ and
$$\int_{u_0}^u|u^\prime|^{5/2}\bb(u^\prime)\|\sd\mu\|_{L^4(S_{\ub,u^\prime})}du^\prime+|u|^{5/2}\fb(u)$$
in the role of the non-negative non-decreasing function $\bb$.
Note in particular from \ref{6.155} that the function
$|u|^{5/2}\fb(u)$ is indeed non-decreasing. Recall that the result
\ref{6.87} implies the result \ref{6.89} provided that condition
\ref{6.88} holds. In view of \ref{6.153} this condition holds if
\begin{equation}
C\delta^{1/2}(\scD_1^4(\mbox{tr}\chib)+\scD_1^4(\chibh)+{\cal
D}_0^\infty(\chibh)+O(\delta^{1/2}))\leq\log 2 \label{6.162}
\end{equation}
which is indeed satisfied if $\delta$ is suitably small depending
on ${\cal D}_0^\infty$, ${\cal R}_0^\infty$, $\scD_1^4$,
$\scR_1^4$. The result \ref{6.89} then takes the form:
\begin{equation}
|u|^{5/2}\|\sd\mub\|_{L^4(S_{\ub,u})}\leq
2\int_{u_0}^u|u^\prime|^{5/2}\bb(u^\prime)\|\sd\mu\|_{L^4(S_{\ub,u^\prime})}du^\prime+2|u|^{5/2}\fb(u)
\label{6.163}
\end{equation}
We have thus reduced the system of integral inequalities
\ref{6.148}, \ref{6.149} on $D_1^{s^*}$ to the system \ref{6.161},
\ref{6.163}.

Let us set:
\begin{eqnarray}
y(u)&=&\sup_{\ub\in[0,\ub_1]}\|\sd\mu\|_{L^4(S_{\ub,u})}\label{6.164}\\
\yb(u)&=&\sup_{\ub\in[0,\ub_1]}\|\sd\mub\|_{L^4(S_{\ub,u})}\label{6.165}
\end{eqnarray}
Since
\begin{eqnarray*}
&&\int_0^{\ub}\left\{\int_{u_0}^u|u^\prime|^{5/2}\bb(u^\prime)\|\sd\mu\|_{L^4(S_{\ub^\prime,u^\prime})}du^\prime
\right\}d\ub^\prime\\
&&=\int_{u_0}^u\left\{\int_0^{\ub}\|\sd\mu\|_{L^4(S_{\ub^\prime,u^\prime})}d\ub^\prime\right\}
|u^\prime|^{5/2}\bb(u^\prime)du^\prime\\
&&\leq\delta\int_{u_0}^u
|u^\prime|^{5/2}\bb(u^\prime)y(u^\prime)du^\prime
\end{eqnarray*}
replacing $\ub$ by $\ub^\prime$ in \ref{6.163} and integrating
with respect to $\ub^\prime$ on $[0,\ub]$ yields:
\begin{equation}
\int_0^{\ub}\|\sd\mub\|_{L^4(S_{\ub^\prime,u})}d\ub^\prime\leq
2\delta|u|^{-5/2}\int_{u_0}^u
|u^\prime|^{5/2}\bb(u^\prime)y(u^\prime)du^\prime+2\delta\fb(u)
\label{6.166}
\end{equation}
Substituting in \ref{6.161} we then obtain, for all $(\ub,u)\in
D_1^{s^*}$:
\begin{eqnarray*}
\|\sd\mu\|_{L^4(S_{\ub,u})}&\leq&
4\delta |u|^{-5/2}b(u)\int_{u_0}^u |u^\prime|^{5/2}\bb(u^\prime)y(u^\prime)du^\prime\\
&\s&+4\delta b(u)\fb(u)+2f(u)
\end{eqnarray*}
Taking the supremum over $\ub\in[0,\ub_1]$ then yields the linear
integral inequality:
\begin{equation}
y(u)\leq m(u)+n(u)\int_{u_0}^u
|u^\prime|^{5/2}\bb(u^\prime)y(u^\prime)du^\prime \label{6.167}
\end{equation}
where:
\begin{equation}
m(u)=4\delta b(u)\fb(u)+2f(u) \label{6.168}
\end{equation}
and:
\begin{equation}
n(u)=4\delta |u|^{-5/2}b(u) \label{6.169}
\end{equation}
From \ref{6.151}, \ref{6.152} and \ref{6.155} we have:
\begin{eqnarray}
m(u)&\leq&C|u|^{-7/2}{\cal R}_0^\infty(\alpha)\scR_1^4(\alpha)+O(\delta^{1/2}|u|^{-9/2})\nonumber\\
&\s&+C\delta|u|^{-9/2}(1+O(|u|^{-1}))\scD_1^4(\mub)\nonumber\\
&\s&+O(\delta^{5/2}|u|^{-13/2})k\label{6.170}\\
n(u)&\leq&C\delta|u|^{-7/2}(1+O(|u|^{-1}))\label{6.171}
\end{eqnarray}
for new numerical constants $C$. Again, the coefficient of $k$
depends only on ${\cal D}_0^\infty$, ${\cal R}_0^\infty$. Setting
\begin{equation}
Y(u)=\int_{u_0}^u
|u^\prime|^{5/2}\bb(u^\prime)y(u^\prime)du^\prime, \ \ \ \mbox{we
have $Y(u_0)=0$} \label{6.172}
\end{equation}
and \ref{6.167} takes the form:
\begin{equation}
\frac{dY}{du}\leq |u|^{5/2}\bb(u)(m+nY) \label{6.173}
\end{equation}
Integrating from $u_0$ we obtain:
\begin{equation}
Y(u)\leq\int_{u_0}^u\exp\left(\int_{u^\prime}^u|u^{\prime\prime}|^{5/2}\bb(u^{\prime\prime})
n(u^{\prime\prime})du^{\prime\prime}\right)|u^\prime|^{5/2}\bb(u^\prime)m(u^\prime)du^\prime
\label{6.174}
\end{equation}
for all $u\in [u_0,s^*]$. From \ref{6.154} and \ref{6.171} we
have:
\begin{eqnarray}
&&\int_{u^\prime}^u|u^{\prime\prime}|^{5/2}\bb(u^{\prime\prime})n(u^{\prime\prime})du^{\prime\prime}\label{6.175}\\
&&\hspace{1cm}\leq
C\delta\int_{u^\prime}^u|u^{\prime\prime}|^{-2}(1+O(|u^{\prime\prime}|^{-1}))du^{\prime\prime}
\leq C\delta|u|^{-1}(1+O(|u|^{-1}))\leq\log 2\nonumber
\end{eqnarray}
the last step provided that $\delta$ is suitably small depending
on ${\cal D}_0^\infty$, ${\cal R}_0^\infty$, $\scD_1^4$,
$\scR_1^4$. Therefore:
\begin{equation}
Y(u)\leq
2\int_{u_0}^u|u^\prime|^{5/2}\bb(u^\prime)m(u^\prime)du^\prime
\label{6.176}
\end{equation}
and substituting in \ref{6.167} yields:
\begin{equation}
y(u)\leq
m(u)+2n(u)\int_{u_0}^u|u^\prime|^{5/2}\bb(u^\prime)m(u^\prime)du^\prime
\label{6.177}
\end{equation}
Moreover, since
$$\int_{u_0}^u|u^\prime|^{5/2}\bb(u^\prime)\|\sd\mu\|_{L^4(S_{\ub,u^\prime})}du^\prime\leq Y(u)$$
substituting \ref{6.176} in \ref{6.163} and taking the supremum
over $\ub\in[0,\ub_1]$ yields:
\begin{equation}
|u|^{5/2}\yb(u)\leq
2|u|^{5/2}\fb(u)+4\int_{u_0}^u|u^\prime|^{5/2}\bb(u^\prime)m(u^\prime)du^\prime
\label{6.178}
\end{equation}

We shall now derive an appropriate estimate for $\scD_1^4(\mub)$.
Recall that on $C_{u_0}$ condition \ref{3.65} holds:
$$\eta+\etb=0 \ \ \mbox{: on $C_{u_0}$}$$
Thus from the definitions \ref{6.94}, \ref{6.95} we have:
\begin{equation}
\mu+\mub=-2\rho+(\chih,\chibh) \ \ \mbox{: on $C_{u_0}$}
\label{6.179}
\end{equation}
Hence (see \ref{6.115}):
\begin{eqnarray}
&&\|\sd\mub\|_{L^4(S_{\ub,u_0})}\leq\|\sd\mu\|_{L^4(S_{\ub,u_0})}\nonumber\\
&&\hspace{1cm}+C|u_0|^{-7/2}[\scR_1^4(\rho)+\scR_1^4(\alpha){\cal
D}_0^\infty(\chibh)
+(\scD_1^4(\mbox{tr}\chib)+\scD_1^4(\chibh)){\cal R}_0^\infty(\alpha)]\nonumber\\
&&\hspace{2cm}+O(\delta^{1/2}|u_0|^{-9/2}) \label{6.180}
\end{eqnarray}
Substituting in \ref{6.148} at $u=u_0$ then yields the following
linear integral inequality for $\|\sd\mu\|_{L^4(S_{\ub,u_0})}$:
\begin{equation}
\|\sd\mu\|_{L^4(S_{\ub,u_0})}\leq
\tilde{a}(u_0)\int_0^{\ub}\|\sd\mu\|_{L^4(S_{\ub^\prime,u_0})}d\ub^\prime+\tilde{f}(u_0)
\label{6.181}
\end{equation}
where:
\begin{equation}
\tilde{a}(u_0)=a(u_0)+b(u_0) \label{6.182}
\end{equation}
\begin{eqnarray}
\tilde{f}(u_0)&=&f(u_0)+\delta b(u_0)\left\{C|u_0|^{-7/2}[\scR_1^4(\rho)+\scR_1^4(\alpha){\cal D}_0^\infty(\chibh)\right.\nonumber\\
&\s&\hspace{2cm}\left.+(\scD_1^4(\mbox{tr}\chib)+\scD_1^4(\chibh)){\cal
R}_0^\infty(\alpha)]
+O(\delta^{1/2}|u_0|^{-9/2})\right\}\nonumber\\
&\leq&C|u_0|^{-7/2}{\cal
R}_0^\infty(\alpha)\scR_1^4(\alpha)+O(\delta^{1/2}|u_0|^{-9/2})
\label{6.183}
\end{eqnarray}
The integral inequality \ref{6.181} implies:
\begin{equation}
\|\sd\mu\|_{L^4(S_{\ub,u_0})}\leq
e^{\delta\tilde{a}(u_0)}\tilde{f}(u_0) \label{6.184}
\end{equation}
Since (see \ref{6.150}, \ref{6.151})
\begin{equation}
\delta\tilde{a}(u_0)\leq\delta^{1/2}|u_0|^{-1}(\scR_1^4(\alpha)+{\cal
R}_0^\infty(\alpha)+O(\delta^{1/2}))\leq\log 2 \label{6.185}
\end{equation}
the last step provided that $\delta$ is suitably small depending
on ${\cal D}_0^\infty$, ${\cal R}_0^\infty$, $\scD_1^4$,
$\scR_1^4$, \ref{6.184} in turn implies:
\begin{eqnarray}
\|\sd\mu\|_{L^4(S_{\ub,u_0})}&\leq&2\tilde{f}(u_0)\nonumber\\
&\leq&C|u_0|^{-7/2}{\cal
R}_0^\infty(\alpha)\scR_1^4(\alpha)+O(\delta^{1/2}|u_0|^{-9/2})
\label{6.186}
\end{eqnarray}
It then follows through \ref{6.180} that:
\begin{eqnarray}
\scD_1^4(\mub)&\leq&C\left[\scR_1^4(\rho)+\right.\nonumber\\
&\s&\s\s\s\left.\scR_1^4(\alpha)({\cal D}_0^\infty(\chibh)+{\cal
R}_0^\infty(\alpha))
+(\scD_1^4(\mbox{tr}\chib)+\scD_1^4(\chibh)){\cal R}_0^\infty(\alpha)\right]\nonumber\\
&\s&\s\s\s+O(\delta^{1/2}|u_0|^{-1}) \label{6.187}
\end{eqnarray}
Substituting \ref{6.187} in \ref{6.170}, the latter simplifies to:
\begin{eqnarray}
m(u)&\leq&C|u|^{-7/2}{\cal R}_0^\infty(\alpha)\scR_1^4(\alpha)+O(\delta^{1/2}|u|^{-9/2})\nonumber\\
&\s&+O(\delta^{5/2}|u|^{-13/2})k\label{6.188}
\end{eqnarray}
Also, substituting \ref{6.187} in \ref{6.155} we obtain:
\begin{eqnarray}
\fb(u)&\leq&C|u|^{-7/2}\tilde{A}\nonumber\\
&\s&+O(\delta^{3/2}|u|^{-11/2})k+O(\delta^{1/2}|u|^{-7/2})
\label{6.189}
\end{eqnarray}
where:
\begin{eqnarray}
\tilde{A}&=&\scR_1^4(\rho)+\scR_1^4(\sigma)+{\cal
R}_0^\infty(\rho)+{\cal R}_0^\infty(\sigma)
+\scR_1^4(\alpha){\cal R}_0^\infty(\alpha)\nonumber\\
&\s&+(\scR_1^4(\alpha)+{\cal R}_0^\infty(\alpha)){\cal
D}_0^\infty(\chibh)+
(\scD_1^4(\mbox{tr}\chib)+\scD_1^4(\chibh)){\cal
R}_0^\infty(\alpha) \label{6.190}
\end{eqnarray}
Substituting the definition \ref{6.154} and the bounds \ref{6.171}
and \ref{6.188} in \ref{6.177} then yields:
\begin{equation}
y(u)\leq C|u|^{-7/2}{\cal
R}_0^\infty(\alpha)\scR_1^4(\alpha)+O(\delta^{1/2}|u|^{-9/2})
+O(\delta^{5/2}|u|^{-13/2})k\label{6.191}
\end{equation}
provided that $\delta$ is suitably small depending on ${\cal
D}_0^\infty$, ${\cal R}_0^\infty$, $\scD_1^4$, $\scR_1^4$. Also,
substituting \ref{6.189}, the definition \ref{6.154} and
\ref{6.188} in \ref{6.178} yields:
\begin{equation}
\yb(u)\leq C|u|^{-7/2}\tilde{A}
+O(\delta^{3/2}|u|^{-11/2})k+O(\delta^{1/2}|u|^{-7/2})
\label{6.192}
\end{equation}
provided again that $\delta$ is suitably small depending on ${\cal
D}_0^\infty$, ${\cal R}_0^\infty$, $\scD_1^4$, $\scR_1^4$.
Moreover, the coefficients of $k$ in \ref{6.191}, \ref{6.192}
depend only on ${\cal D}_0^\infty$, ${\cal R}_0^\infty$.

In view of the definitions \ref{6.164}, \ref{6.165}, the estimates
\ref{6.191}, \ref{6.192} imply through \ref{6.118}, \ref{6.119}
the estimates of the lemma.

\vspace{5mm}

\section{$L^4(S)$ estimate for $\snab^{ \ 2}\omb$}

We proceed to derive an $L^4(S)$ estimate for $\snab^{ \ 2}\omb$
on $M_1^{s^*}$. Using this estimate we shall derive an $L^4(S)$
estimate for $\snab^{ \ 2}\log\Omega$ in $M_1^{s^*}$ which, with a
suitable choice of the constant $k$, improves the bound
\ref{6.51}. This shall enable us to show that $s^*=u_1$ so that
the previous lemmas actually hold on the entire parameter domain
$D_1$, and, since $(\ub_1,u_1)\in D^\prime$ is arbitrary, the
estimates hold on all of $D^\prime$, that is, on all of
$M^\prime$.

We introduce the function:
\begin{equation}
\somb=\slap\omb-\sdiv(\Omega\beb) \label{6.193}
\end{equation}
We shall derive a propagation equation for $\somb$ along the
generators of the $C_u$. Since
$$\slap\omb=\mbox{tr}\snab^{ \ 2}\omb$$
\ref{6.99} applies and we obtain:
\begin{equation}
D\slap\omb=\mbox{tr}(D\snab^{ \ 2}\omb)-2\Omega(\chi,\snab^{ \
2}\omb) \label{6.194}
\end{equation}
On the other hand, since $\snab^{ \ 2}\omb=\snab(\sd\omb)$, by
Lemma 4.1 and Lemma 1.2 we have:
\begin{equation}
(D\snab^{ \ 2}\omb)_{AB}=(\snab^{ \
2}D\omb)_{AB}-(D\sGamma)^C_{AB}(\sd\omb)_C \label{6.195}
\end{equation}
and:
\begin{equation}
(\sg^{-1})^{AB}(D\sGamma)^C_{AB}=2(\sdiv(\Omega\chih))^C
\label{6.196}
\end{equation}
We thus obtain:
\begin{equation}
D\slap\omb=\slap
D\omb-\Omega\mbox{tr}\chi\slap\omb-2\Omega(\chih,\snab^{ \
2}\omb)-2(\sdiv(\Omega\chih),\sd\omb) \label{6.197}
\end{equation}
Moreover, from \ref{1.86} we have:
\begin{eqnarray}
\slap D\omb&=&\slap[\Omega^2(2(\eta,\etb)-|\eta|^2)]\nonumber\\
&\s&-\Omega^2\slap\rho-2(\sd(\Omega^2),\sd\rho)-\rho\slap(\Omega^2)
\label{6.198}
\end{eqnarray}
The principal term on the right hand side is the term
$-\Omega^2\slap\rho$.

Noting that
$$D(\Omega\beb)=\Omega(D\beb+\omega\beb)$$
the sixth Bianchi identity of Proposition 1.2 reads:
\begin{equation}
D(\Omega\beb)+\frac{1}{2}\mbox{tr}\chi(\Omega\beb)=\Omega^2\{-\sd\rho+\s^*\sd\sigma-3\etb\rho+3\s^*\etb\sigma
+\chih^\sharp\cdot\beb+2\chibh^\sharp\cdot\beta\} \label{6.201}
\end{equation}
By \ref{6.107} applied to the $S$ 1-form $\Omega\beb$:
\begin{equation}
D\sdiv(\Omega\beb)=\sdiv
D(\Omega\beb)-\Omega\mbox{tr}\chi\sdiv(\Omega\beb)-2\sdiv(\Omega^2\chih^\sharp\cdot\beb)
\label{6.202}
\end{equation}
Substituting from \ref{6.201} then yields:
\begin{eqnarray}
D\sdiv(\Omega\beb)&=&-\frac{3}{2}\Omega\mbox{tr}\chi\sdiv(\Omega\beb)-\frac{1}{2}(\sd(\Omega\mbox{tr}\chi),\Omega\beb)
\nonumber\\
&\s&-\Omega^2\slap\rho-(\sd(\Omega^2),\sd\rho)+(\sd(\Omega^2),\s^*\sd\sigma)\nonumber\\
&\s&+\sdiv[\Omega^2(-\chih^\sharp\cdot\beb+2\chibh^\sharp\cdot\beta-3\etb\rho+3\s^*\etb\sigma)]
\label{6.203}
\end{eqnarray}
Note that $\sdiv(\s^*\sd\sigma)=0$ identically, the Hessian of a
function being symmetric. The principal term on the right hand
side is the term $-\Omega^2\slap\rho$ and cancels the principal
term on the right hand side of \ref{6.198} in considering the
propagation equation for $\somb$.

Combining \ref{6.197}, \ref{6.198} and \ref{6.203}, we obtain the
following propagation equation for the function $\somb$ along the
generators of the $C_u$:
\begin{equation}
D\somb+\Omega\mbox{tr}\chi\somb=-2\Omega(\chih,\snab^{ \
2}\omb)+\mb \label{6.204}
\end{equation}
where $\mb$ is the function:
\begin{eqnarray}
\mb&=&-2(\sdiv(\Omega\chih),\sd\omb)+\frac{1}{2}\sdiv(\Omega^2\mbox{tr}\chi\beb)\nonumber\\
&\s&-(\sd(\Omega^2),\sd\rho)-(\sd(\Omega^2),\s^*\sd\sigma)-\rho\slap(\Omega^2)\nonumber\\
&\s&+\slap[\Omega^2(2(\eta,\etb)-|\eta|^2)]\nonumber\\
&\s&+\sdiv[\Omega^2(\chih^\sharp\cdot\beb-2\chibh^\sharp\cdot\beta+3\etb\rho-3\s^*\etb\sigma)]
\label{6.205}
\end{eqnarray}
We note the crucial fact that the right hand side of equation
\ref{6.204} does not contain principal terms (such terms would
involve the second derivatives of the curvature or the third
derivatives of the connection coefficients). The propagation
equation \ref{6.204} is to be considered in conjunction with the
elliptic equation
\begin{equation}
\slap\omb=\somb+\sdiv(\Omega\beb) \label{6.206}
\end{equation}
which is simply a re-writing of the definition \ref{6.193}.

\vspace{5mm}

\noindent{\bf Lemma 6.3} \ \ \ We have:
\begin{eqnarray*}
\|\snab^{ \ 2}\omb\|_{L^4(S_{\ub,u})}&\leq&C\delta|u|^{-9/2}[\scR_1^4(\beb)+{\cal D}_0^\infty(\chibh)\scR_1^4(\beta)+(\scD_1^4(\mbox{tr}\chib)+\scD_1^4(\chibh)){\cal R}_0^\infty(\beta)]\nonumber\\
&\s&+O(\delta^{3/2}|u|^{-11/2})+O(\delta^3|u|^{-15/2})k
\end{eqnarray*}
for all $(\ub,u)\in D_1^{s^*}$, provided that $\delta$ is suitably
small depending on ${\cal D}_0^\infty$, ${\cal R}_0^\infty$,
$\scD_1^4$, $\scR_1^4$, and $\scR_2(\alpha)$. Moreover, the
coefficient of $k$ depends only on ${\cal D}_0^\infty$, ${\cal
R}_0^\infty$.

\noindent{\em Proof:} \ We shall first obtain a bound for
$\|\mb\|_{L^4(S_{\ub,u})}$ for all $(\ub,u)\in D_1^{s^*}$. The
first term on the right in \ref{6.205} is bounded in $L^4(S)$ by:
$$C\|\snab(\Omega\chih)\|_{L^\infty(S_{\ub,u})}\|\sd\omb\|_{L^4(S_{\ub,u})}$$
and by Proposition 6.1 and Lemma 5.2 with $p=4$:
\begin{equation}
\|\snab(\Omega\chih)\|_{L^\infty(S_{\ub,u})}\leq
C\delta^{-1/2}|u|^{-2}(\scR_1^4(\beta)+{\cal R}_0^\infty(\beta))
+O(|u|^{-3}) \label{6.207}
\end{equation}
Taking also into account the $L^4(S)$ bound for $\sd\omb$ of
Proposition 4.2 we then conclude that the first term on the right
in \ref{6.205} is bounded in $L^4(S)$ by:
\begin{equation}
O(\delta^{1/2}|u|^{-11/2}) \label{6.208}
\end{equation}
The second term on the right in \ref{6.205} is bounded in $L^4(S)$
by:
\begin{equation}
O(\delta|u|^{-11/2}) \label{6.209}
\end{equation}
Writing (see \ref{1.65})
$$\sd\log\Omega=\frac{1}{2}(\eta+\etb)$$
and using the results of Chapter 3, the third and fourth terms on
the right in \ref{6.205} are bounded in $L^4(S)$ by:
\begin{equation}
O(\delta^{1/2}|u|^{-11/2}) \label{6.210}
\end{equation}
Also, writing
$$\slap\log\Omega=\frac{1}{2}(\sdiv\eta+\sdiv\etb)$$
and using the results of Chapter 4, the fifth term on the right in
\ref{6.205} is in $L^4(S)$ similarly bounded. The principal part
of the sixth term on the right in \ref{6.205} is:
$$2\Omega^2((\etb-\eta,\slap\eta)+(\eta,\slap\etb))$$
This part is bounded in $L^4(S)$ using Lemma 6.2 and the results
of Chapter 3 by:
\begin{equation}
O(\delta^2|u|^{-15/2})k+O(\delta^{1/2}|u|^{-11/2}) \label{6.211}
\end{equation}
where the coefficient of $k$ depends only on ${\cal D}_0^\infty$,
${\cal R}_0^\infty$. There is also a part of the form:
$$2\Omega^2(2(\snab\eta,\snab\etb)-|\snab\eta|^2)$$
This is estimated in $L^4(S)$ by placing one factor of $\snab\eta$
in $L^\infty(S)$ using Lemma 6.2 in conjunction with Lemma 5.2 with $p=4$, which gives:
\begin{equation}
\|\snab\eta\|_{L^\infty(S_{\ub,u})}\leq
C|u|^{-3}\tilde{A}+O(\delta^{1/2}|u|^{-3})
+O(\delta^{5/2}|u|^{-6})k \label{6.212}
\end{equation}
and the other factor of $\snab\eta$ or $\snab\etb$ in $L^4(S)$
using the results of Chapter 4. We then obtain an $L^4(S)$ bound
for this part by:
\begin{equation}
O(\delta^3|u|^{-17/2})k+O(\delta^{1/2}|u|^{-11/2}) \label{6.213}
\end{equation}
where the coefficient of $k$ is independent of
$\scD_2^4(\mbox{tr}\chib)$. The remainder of the sixth term on the
right in \ref{6.205} is of lower order. Finally, we have the
seventh term on the right in \ref{6.205}. The part:
$$-2\sdiv(\Omega^2\chibh^\sharp\cdot\beta)$$
gives the leading contribution in behavior with respect to
$\delta$. Using the results of Chapters 3 and 4 this part is seen
to be bounded in $L^4(S)$:
\begin{equation}
C|u|^{-9/2}[{\cal
D}_0^\infty(\chibh)\scR_1^4(\beta)+(\scD_1^4(\mbox{tr}\chib)+\scD_1^4(\chibh)){\cal
R}_0^\infty(\beta)] +O(\delta^{1/2}|u|^{-11/2}) \label{6.214}
\end{equation}
while the remainder of the seventh term is bounded in $L^4(S)$ by:
\begin{equation}
O(\delta^{1/2}|u|^{-11/2}) \label{6.215}
\end{equation}
Collecting the above results (\ref{6.208} - \ref{6.211},
\ref{6.213} - \ref{6.215}) we conclude that:
\begin{eqnarray}
\|\mb\|_{L^4(S_{\ub,u})}&\leq&C|u|^{-9/2}[{\cal D}_0^\infty(\chibh)\scR_1^4(\beta)+(\scD_1^4(\mbox{tr}\chib)+\scD_1^4(\chibh)){\cal R}_0^\infty(\beta)]\nonumber\\
&\s&+O(\delta^2|u|^{-15/2})k+O(\delta^{1/2}|u|^{-11/2})
\label{6.216}
\end{eqnarray}
provided that $\delta$ is suitably small depending on ${\cal
D}_0^\infty$, ${\cal R}_0^\infty$, $\scD_1^4$, $\scR_1^4$, the
coefficient of $k$ being independent of
$\scD_2^4(\mbox{tr}\chib)$.

Consider now equation \ref{6.206}.  Setting $\theta=\sd\omb$, the
$S$ 1-form $\theta$ satisfies a system of the form \ref{5.178}
with $f=\somb+\sdiv(\Omega\beb)$, $g=0$. The estimate \ref{5.191}
holds. Substituting in this estimate the bound \ref{5.147} for
$k_{1,\infty}$, we obtain, by virtue of the last conclusion of
Proposition 6.1, the estimate:
\begin{equation}
\|\snab\theta\|_{L^p(S_{\ub,u})}\leq
C_p\{\|f\|_{L^p(S_{\ub,u})}+\|g\|_{L^p(S_{\ub,u})}\} \label{6.217}
\end{equation}
In the present case taking $p=4$ this yields:
\begin{equation}
\|\snab^{ \ 2}\omb\|_{L^4(S_{\ub,u})}\leq
C\|\somb\|_{L^4(S_{\ub,u})}+C\delta|u|^{-9/2}\scR_1^4(\beb)
+O(\delta^2|u|^{-13/2}) \label{6.218}
\end{equation}
It follows that in reference to the first term on the right in
\ref{6.204} we have:
\begin{equation}
\|\Omega(\chih,\snab^{ \ 2}\omb)\|_{L^4(S_{\ub,u})}\leq
C\delta^{-1/2}|u|^{-1}{\cal
R}_0^\infty(\alpha)\|\somb\|_{L^4(S_{\ub,u})}
+O(\delta^{1/2}|u|^{-11/2}) \label{6.219}
\end{equation}
We now apply Lemma 4.6 with $p=4$ to the propagation equation
\ref{6.204}. Here $r=0$, $\nu=-2$, $\gamma=0$ and we obtain:
\begin{equation}
\|\somb\|_{L^4(S_{\ub,u})}\leq
C\int_0^{\ub}\left\|-2\Omega(\chih,\snab^{ \
2}\omb)+\mb\right\|_{L^4(S_{\ub^\prime,u})} d\ub^\prime
\label{6.220}
\end{equation}
Substituting the estimates \ref{6.220} and \ref{6.216} we obtain
the following linear integral inequality for the quantity
$\|\somb\|_{L^4(S_{\ub,u})}$ on the domain $D_1^{s^*}$:
\begin{equation}
\|\somb\|_{L^4(S_{\ub,u})}\leq
a(u)\int_0^{\ub}\|\somb\|_{L^4(S_{\ub^\prime,u})}d\ub^\prime+b(u)
\label{6.221}
\end{equation}
Here:
\begin{eqnarray}
a(u)&=&C\delta^{-1/2}|u|^{-1}{\cal R}_0^\infty(\alpha)\label{6.222}\\
b(u)&=&C\delta|u|^{-9/2}[{\cal D}_0^\infty(\chibh)\scR_1^4(\beta)+(\scD_1^4(\mbox{tr}\chib)+\scD_1^4(\chibh)){\cal R}_0^\infty(\beta)]\nonumber\\
&\s&+O(\delta^{3/2}|u|^{-11/2})+O(\delta^3|u|^{-15/2})k\label{6.223}
\end{eqnarray}
where the coefficient of $k$ is independent of
$\scD_2^4(\mbox{tr}\chib)$. At fixed $u$ the inequality
\ref{6.221} is of the form \ref{6.21}, with $a(u)$ in the role of
the constant $a$ and the constant $b(u)$ in the role of the
non-negative function $b(\ub)$. Then \ref{6.157} holds, which,
since
$$a(u)\delta=C\delta^{1/2}|u|^{-1}{\cal R}_0^\infty(\alpha)\leq\log 2$$
provided that $\delta$ is suitably small depending on ${\cal
R}_0^\infty(\alpha)$, implies \ref{6.159}, that is:
\begin{eqnarray}
\|\somb\|_{L^4(S_{\ub,u})}&\leq&C\delta|u|^{-9/2}[{\cal
D}_0^\infty(\chibh)\scR_1^4(\beta)
+(\scD_1^4(\mbox{tr}\chib)+\scD_1^4(\chibh)){\cal R}_0^\infty(\beta)]\nonumber\\
&\s&+O(\delta^{3/2}|u|^{-11/2})+O(\delta^3|u|^{-15/2})k\label{6.224}
\end{eqnarray}
where the coefficent of $k$ is independent of
$\scD_2^4(\mbox{tr}\chib)$. Substituting in \ref{6.218} then
yields the lemma.

\vspace{5mm}

\noindent{\bf Lemma 6.4} \ \ \ There is a numerical constant $C$
such that with
$$k=C[{\cal D}_0^\infty(\chibh)\scR_1^4(\beta)+(\scD_1^4(\mbox{tr}\chib)+\scD_1^4(\chibh)){\cal R}_0^\infty(\beta)]
+O(\delta^{1/2})$$ we have $s^*=u_1$ and the estimate:
\begin{eqnarray*}
\|\snab^{ \
2}\log\Omega\|_{L^4(S_{\ub,u})}&\leq&C\delta|u|^{-7/2}[\scR_1^4(\beb)+{\cal
D}_0^\infty(\chibh)\scR_1^4(\beta)\\&\s&+(\scD_1^4(\mbox{tr}\chib)+\scD_1^4(\chibh)){\cal
R}_0^\infty(\beta)]+O(\delta^{3/2}|u|^{-9/2})
\end{eqnarray*}
holds for all $(\ub,u)\in D_1$, provided that $\delta$ is suitably
small depending on ${\cal D}_0^\infty$, ${\cal R}_0^\infty$,
$\scD_1^4$, $\scR_1^4$, and $\scR_2(\alpha)$.

\noindent{\em Proof:} \ By Lemma 4.1 and the second of the
definitions \ref{1.17}:
\begin{equation}
\Db\snab^{ \ 2}\log\Omega=\snab^{ \
2}\omb-\Db\sGamma\cdot\sd\log\Omega \label{6.225}
\end{equation}
and we have:
$$(\Db\sGamma)^C_{AB}=\snab_A(\Omega\chib)_B^{\s C}+\snab_B(\Omega\chib)_A^{\s C}-\snab^C(\Omega\chib)_{AB}$$
By Lemma 6.1 and Lemma 5.2 with $p=4$ (and the estimate of
Proposition 4.2 for $\mbox{tr}\chib$):
\begin{equation}
\|\snab(\Omega\chib)\|_{L^\infty(S_{\ub,u})}\leq
C\delta^{1/2}|u|^{-3}(\scR_1^4(\beta)+{\cal R}_0^\infty(\beta))
+C\delta|u|^{-4}k+O(\delta|u|^{-4}) \label{6.226}
\end{equation}
hence also:
\begin{equation}
\|\Db\sGamma\|_{L^\infty(S_{\ub,u})}\leq
C\delta^{1/2}|u|^{-3}(\scR_1^4(\beta)+{\cal R}_0^\infty(\beta))
+C\delta|u|^{-4}k+O(\delta|u|^{-4}) \label{6.227}
\end{equation}
for all $(\ub,u)\in D_1^{s^*}$. Using then also Lemma 4.11 we can
estimate:
\begin{eqnarray}
\|\Db\sGamma\cdot\sd\log\Omega\|_{L^4(S_{\ub,u})}&\leq&\|\Db\sGamma\|_{L^\infty(S_{\ub,u})}\|\sd\log\Omega\|_{L^4(S_{\ub,u})}
\nonumber\\
&\leq&O(\delta^{3/2}|u|^{-11/2})+O(\delta^2|u|^{-13/2})k
\label{6.228}
\end{eqnarray}
where the coefficient of $k$ is independent of
$\scD_2^4(\mbox{tr}\chib)$. We now apply Lemma 4.7 with $p=4$ to
equation \ref{6.225}. Here $r=2$, $\nu=0$, $\gammab=0$. Taking
into account the fact that $\snab^{ \ 2}\log\Omega$ vanishes on
$C_{u_0}$ we obtain:
\begin{equation}
|u|^{3/2}\|\snab^{ \ 2}\log\Omega\|_{L^4(S_{\ub,u})}\leq
\int_{u_0}^u|u^\prime|^{3/2} \left\|\snab^{ \
2}\omb-\Db\sGamma\cdot\sd\log\Omega\right\|_{L^4(S_{\ub,u^\prime})}du^\prime
\label{6.229}
\end{equation}
Substituting the estimate \ref{6.228} and the estimate for
$\snab^{ \ 2}\omb$ in $L^4(S)$ of Lemma 6.3 we then obtain:
\begin{eqnarray}
\|\snab^{ \ 2}\log\Omega\|_{L^4(S_{\ub,u})}&\leq&C\delta|u|^{-7/2}[\scR_1^4(\beb)+{\cal D}_0^\infty(\chibh)\scR_1^4(\beta)\nonumber\\
&\s&+(\scD_1^4(\mbox{tr}\chib)+\scD_1^4(\chibh)){\cal R}_0^\infty(\beta)]+O(\delta^{3/2}|u|^{-9/2})\nonumber\\
&\s&+O(\delta^2|u|^{-11/2})k \label{6.230}
\end{eqnarray}
where the coefficient of $k$ is independent of
$\scD_2^4(\mbox{tr}\chib)$. In reference to this estimate, let:
\begin{eqnarray}
a&=&C[\scR_1^4(\beb)+{\cal D}_0^\infty(\chibh)\scR_1^4(\beta)+(\scD_1^4(\mbox{tr}\chib)+\scD_1^4(\chibh)){\cal R}_0^\infty(\beta)]+O(\delta^{1/2})\nonumber\\
b&=&O(1) \label{6.231}
\end{eqnarray}
so that the coefficient of $k$ in \ref{6.230} is
$\delta^2|u|^{-11/2}b$, $b$ being independent of
$\scD_2^4(\mbox{tr}\chib)$. The estimate \ref{6.230} then implies:
\begin{equation}
\|\snab^{ \ 2}\log\Omega\|_{L^4(S_{\ub,u})}\leq\delta(a+\delta
bk)|u|^{-7/2} \ \ : \ \forall (\ub,u)\in D_1^{s^*} \label{6.232}
\end{equation}
Choosing:
\begin{equation}
k=2a \label{6.233}
\end{equation}
we have:
\begin{equation}
a+\delta bk<2a \ \ \mbox{provided that $2b\delta<1$} \label{6.234}
\end{equation}
The last is a smallness condition on $\delta$ depending on ${\cal
D}_0^\infty$, ${\cal R}_0^\infty$, $\scD_1^4$, $\scR_1^4$. The
estimate \ref{6.232} then implies:
\begin{equation}
\|\snab^{ \ 2}\log\Omega\|_{L^4(S_{\ub,u})}< k\delta|u|^{-7/2} \ \
\ : \ \forall (\ub,u)\in D_1^{s^*}=[0,\ub_1]\times[0,s^*]
\label{6.235}
\end{equation}
hence by continuity \ref{6.50} holds for some $s>s^*$
contradicting the definition of $s^*$, unless $s^*=u_1$. This
completes the proof of the lemma.

\vspace{5mm}

Since by the above lemma $D_1^{s^*}=D_1$, with $k$ as in the
statement of Lemma 6.4, the results of Lemmas 6.1, 6.2 and 6.3
hold for all $(\ub,u)\in D_1$. Since $(\ub_1,u_1)$ is arbitrary,
these results hold for all $(\ub,u)\in D^\prime$. Moreover, by
virtue of the $L^4(S)$ estimate for $\snab^{ \ 2}\log\Omega$ of
Lemma 6.4, we can now estimate (see \ref{6.72}):
\begin{eqnarray}
&&\|\snab\rb\|_{L^4(S_{\ub,u})}\leq\nonumber\\
&&\hspace{1cm}C\delta|u|^{-11/2}C[{\cal
D}_0^\infty(\chibh)\scR_1^4(\beta)
+(\scD_1^4(\mbox{tr}\chib)+\scD_1^4(\chibh)){\cal R}_0^\infty(\beta)]\nonumber\\
&&\hspace{1cm}+O(\delta^{3/2}|u|^{-13/2}) \label{6.236}
\end{eqnarray}
Using this bound in place of the bound \ref{6.73}, we deduce,
following the argument leading to the estimate \ref{6.90},
\begin{equation}
\|\snab^{ \ 2}\mbox{tr}\chib^\prime\|_{L^4(S_{\ub,u})}\leq
C\delta|u|^{-9/2}B+O(\delta^{3/2}|u|^{-11/2}) \ \ : \ \forall
(\ub,u)\in D_1 \label{6.237}
\end{equation}
where:
\begin{eqnarray}
B&=&\scD_2^4(\mbox{tr}\chib)
+(\scD_1^4(\mbox{tr}\chib)+\scD_1^4(\chibh))^2\label{6.238}\\
&\s&+(\scD_1^4(\mbox{tr}\chib)+\scD_1^4(\chibh)+{\cal
D}_0^\infty(\chibh))(\scR_1^4(\beta)+{\cal
R}_0^\infty(\beta))+\scR_1^4(\beb)\nonumber
\end{eqnarray}
We thus arrive at the following proposition.

\vspace{5mm}

\noindent{\bf Proposition 6.2} \ \ \ The following estimates hold
for all $(\ub,u)\in D^\prime$:
\begin{eqnarray*}
&&\|\snab^{ \ 2}\mbox{tr}\chib\|_{L^4(S_{\ub,u})}\leq C\delta|u|^{-9/2}B+O(\delta^{3/2}|u|^{-11/2})\\
&&|u|\|\snab^{ \
2}\chibh\|_{L^4(S_{\ub,u})}+\|\snab\chibh\|_{L^4(S_{\ub,u})}
+|u|^{-1}\|\chibh\|_{L^4(S_{\ub,u})}\\
&&\hspace{1cm}\leq C\delta^{1/2}|u|^{-5/2}(\scR_1^4(\beta)+{\cal R}_0^\infty(\beta))+O(\delta|u|^{-7/2})\\
&&|u|\|\snab^{ \
2}\eta\|_{L^4(S_{\ub,u})}+\|\snab\eta\|_{L^4(S_{\ub,u})}
+|u|^{-1}\|\eta\|_{L^4(S_{\ub,u})}\\
&&\hspace{1cm}\leq C|u|^{-5/2}\tilde{A}+O(\delta^{1/2}|u|^{-5/2})\\
&&|u|\|\snab^{ \
2}\etb\|_{L^4(S_{\ub,u})}+\|\snab\etb\|_{L^4(S_{\ub,u})}
+|u|^{-1}\|\etb\|_{L^4(S_{\ub,u})}\\
&&\hspace{1cm}\leq C|u|^{-5/2}\tilde{A}+O(\delta^{1/2}|u|^{-5/2})\\
&&\|\snab^{ \ 2}\omb\|_{L^4(S_{\ub,u})}\leq C\delta|u|^{-9/2}[\scR_1^4(\beb)+{\cal D}_0^\infty(\chibh)\scR_1^4(\beta)+(\scD_1^4(\mbox{tr}\chib)+\scD_1^4(\chibh)){\cal R}_0^\infty(\beta)]\nonumber\\
&&\hspace{1cm}+O(\delta^{3/2}|u|^{-11/2})
\end{eqnarray*}
provided that $\delta$ is suitably small depending on ${\cal
D}_0^\infty$, ${\cal R}_0^\infty$, $\scD_1^4$, $\scR_1^4$, and
$\scR_2(\alpha)$. Here $B$ and $\tilde{A}$ are defined by
\ref{6.238} and \ref{6.190} respectively.

\section{$L^4(S)$ estimate for $\snab^{ \ 2}\omega$}

\noindent{\bf Proposition 6.3} \ \ \ The following estimate holds
for all $(\ub,u)\in D^\prime$:
$$\|\snab^{ \ 2}\omega\|_{L^4(S_{\ub,u})}\leq C\delta^{-1/2}|u|^{-5/2}\scR_1^4(\beta)+O(|u|^{-7/2})$$
provided that $\delta$ is suitably small depending on ${\cal
D}_0^\infty$, ${\cal R}_0^\infty$, $\scD_1^4$, $\scR_1^4$, and
$\scR_2(\alpha)$.

\noindent{\em Proof:} \ We introduce the conjugate of the function
$\somb$ (see \ref{6.193}):
\begin{equation}
\somega=\slap\omega+\sdiv(\Omega\beta) \label{6.239}
\end{equation}
Proceeding as in the derivation of the propagation equation
\ref{6.204} we derive the following propagation equation for
$\somega$, the conjugate of \ref{6.204}:
\begin{equation}
\Db\somega+\Omega\mbox{tr}\chib\somega=-2\Omega(\chibh,\snab^{ \
2}\omega)+m \label{6.240}
\end{equation}
where $m$ is the function:
\begin{eqnarray}
m&=&-2(\sdiv(\Omega\chibh),\sd\omega)-\frac{1}{2}\sdiv(\Omega\mbox{tr}\chib\beta)\nonumber\\
&\s&-(\sd(\Omega^2),\sd\rho)+(\sd(\Omega^2),\s^*\sd\sigma)-\rho\slap(\Omega^2)\nonumber\\
&\s&+\slap[\Omega^2(2(\eta,\etb)-|\etb|^2)]\nonumber\\
&\s&+\sdiv[\Omega^2(-\chibh^\sharp\cdot\beta+2\chih^\sharp\cdot\beb+3\eta\rho+3\s^*\eta\sigma)]
\label{6.241}
\end{eqnarray}
The propagation equation \ref{6.240} is to be considered in
conjunction with the elliptic equation
\begin{equation}
\slap\omega=\somega-\sdiv(\Omega\beta) \label{6.242}
\end{equation}

We shall first obtain a bound for $\|m\|_{L^4(S_{\ub,u})}$. The
first term on the right in \ref{6.241} is bounded in $L^4(S)$ by:
$$C\|\snab(\Omega\chibh)\|_{L^\infty(S_{\ub,u})}\|\sd\omega\|_{L^4(S_{\ub,u})}$$
and by Proposition 6.2 and Lemma 5.2 with $p=4$:
\begin{equation}
\|\snab(\Omega\chibh)\|_{L^\infty(S_{\ub,u})}\leq
C\delta^{1/2}|u|^{-3}(\scR_1^4(\beta)+{\cal R}_0^\infty(\beta))
+O(\delta|u|^{-4}) \label{6.243}
\end{equation}
Taking also into account the $L^4(S)$ bound for $\sd\omega$ of
Proposition 4.3 we then conclude that the first term on the right
in \ref{6.241} is bounded in $L^4(S)$ by:
\begin{equation}
O(\delta^{1/2}|u|^{-11/2}) \label{6.244}
\end{equation}
The second term on the right in \ref{6.241} is bounded in $L^4(S)$
by:
\begin{equation}
C\delta^{-1/2}|u|^{-7/2}\scR_1^4(\beta)+O(\delta^{1/2}|u|^{-11/2})
\label{6.245}
\end{equation}
The third, fourth and fifth terms on the right in \ref{6.241} are
bounded in $L^4(S)$ by:
\begin{equation}
O(\delta|u|^{-13/2}) \label{6.246}
\end{equation}
by Lemmas 6.4 and 4.11, which in conjunction with Lemma 5.2 with
$p=4$ imply:
\begin{equation}
\|\sd\log\Omega\|_{L^\infty(S_{\ub,u})}\leq O(\delta|u|^{-3})
\label{6.247}
\end{equation}
The sixth term on the right in \ref{6.241} is bounded in $L^4(S)$
by:
\begin{equation}
O(\delta^{1/2}|u|^{-11/2}) \label{6.248}
\end{equation}
using Lemma 6.4, Proposition 6.2, the estimate \ref{6.247} and the
results of Chapters 3 and 4. Finally, using the results of
Chapters 3 and 4 we deduce that the seventh term on the right in
\ref{6.241} is bounded in $L^4(S)$ by:
\begin{equation}
O(|u|^{-9/2}) \label{6.249}
\end{equation}
Collecting the above results (\ref{6.244} - \ref{6.246},
\ref{6.248}) we conclude that:
\begin{equation}
\|m\|_{L^4(S_{\ub,u})}\leq
C\delta^{-1/2}|u|^{-7/2}\scR_1^4(\beta)+O(|u|^{-9/2})
\label{6.250}
\end{equation}

Consider now equation \ref{6.242}. Setting $\theta=\sd\omega$, the
$S$ 1-form $\theta$ satisfies a system of the form \ref{5.178}
with $f=\somega-\sdiv(\Omega\beta)$, $g=0$. The estimate
\ref{6.217} holds. In the present case taking $p=4$ yields:
\begin{equation}
\|\snab^{ \ 2}\omega\|_{L^4(S_{\ub,u})}\leq
C\|\somega\|_{L^4(S_{\ub,u})}+C\delta^{-1/2}|u|^{-5/2}\scR_1^4(\beta)
+O(\delta^{1/2}|u|^{-9/2}) \label{6.251}
\end{equation}
It follows that in reference to the first term on the right in
\ref{6.240} we have:
\begin{equation}
\|\Omega(\chibh,\snab^{ \ 2}\omega)\|_{L^4(S_{\ub,u})}\leq
C\delta^{1/2}|u|^{-2} ({\cal
D}_0^\infty(\chibh)+O(\delta|u|^{-3/2}))\|\somega\|_{L^4(S_{\ub,u})}+O(|u|^{-9/2})
\label{6.252}
\end{equation}
We now apply Lemma 4.7 with $p=4$ to the propagation equation
\ref{6.240}. Here $r=0$, $\nu=-2$, $\gammab=0$ and we obtain:
\begin{eqnarray}
|u|^{3/2}\|\somega\|_{L^4(S_{\ub,u})}&\leq&C|u_0|^{3/2}\|\somega\|_{L^4(S_{\ub,u_0})}\label{6.253}\\
&\s&+C\int_{u_0}^u|u^\prime|^{3/2}\left\|-2\Omega(\chibh,\snab{ \
2}\omega)+m\right\|_{L^4(S_{\ub,u^\prime})}du^\prime \nonumber
\end{eqnarray}
Substituting the estimates \ref{6.252} and \ref{6.250} and noting
that in view of the fact that $\omega$ vanishes on $C_{u_0}$ we
have, from \ref{6.239},
\begin{equation}
\|\somega\|_{L^4(S_{\ub,u_0})}\leq
C\delta^{-1/2}|u_0|^{-5/2}\scR_1^4(\beta)
+O(\delta^{1/2}|u_0|^{-9/2}), \label{6.254}
\end{equation}
we obtain the following linear integral inequality for the
quantity $|u|^{3/2}\|\somega\|_{L^4(S_{\ub,u})}$:
\begin{equation}
|u|^{3/2}\|\somega\|_{L^4(S_{\ub,u})}\leq\int_{u_0}^u\ab(u^\prime)|u^\prime|^{3/2}\|\somega\|_{L^4(S_{\ub,u^\prime})}
du^\prime+\bb(u) \label{6.255}
\end{equation}
where:
\begin{eqnarray}
\ab(u)&=&C\delta^{1/2}|u|^{-2}
({\cal D}_0^\infty(\chibh)+O(\delta|u|^{-3/2}))\label{6.256}\\
\bb(u)&=&C\delta^{-1/2}|u|^{-1}\scR_1^4(\beta)+O(|u|^{-2})\label{6.257}
\end{eqnarray}
At fixed $\ub$ the inequality \ref{6.255} is of the form
\ref{6.81} with the quantity
$|u|^{3/2}\|\somega\|_{L^4(S_{\ub,u})}$ in the role of $\xb(u)$.
Note that here the function $\bb(u)$ is non-decreasing. Moreover,
we have
$$\int_{u_0}^u\ab(u^\prime)du^\prime\leq\log 2$$
provided that $\delta$ is suitably small depending on ${\cal
D}_0^\infty$, ${\cal R}_0^\infty$. It follows that \ref{6.89}
holds, that is:
\begin{equation}
\|\somega\|_{L^4(S_{\ub,u})}\leq
C\delta^{-1/2}|u|^{-5/2}\scR_1^4(\beta)+O(|u|^{-7/2})
\label{6.258}
\end{equation}
Substituting finally this in \ref{6.251} yields the lemma.

\chapter{$L^2$ Estimates for the 3rd Derivatives of the Connection
Coefficients}

\section{Introduction}

In the present chapter we shall assume that besides the quantity
$\scR_2(\alpha)$ defined by \ref{5.6} also the following
quantities are finite:
\begin{eqnarray}
\scR_2(\beta)&=&\sup_{u\in[u_0,c^*)}(|u|^3\|\snab^{ \ 2}\beta\|_{L^2(C_u)})\nonumber\\
\scR_2(\rho)&=&\sup_{u\in[u_0,c^*)}(|u|^4\delta^{-1/2}\|\snab^{ \ 2}\rho\|_{L^2(C_u)})\nonumber\\
\scR_2(\sigma)&=&\sup_{u\in[u_0,c^*)}(|u|^4\delta^{-1/2}\|\snab^{ \ 2}\sigma\|_{L^2(C_u)})\nonumber\\
\scR_2(\beb)&=&\sup_{u\in[u_0,c^*)}(|u|^5\delta^{-3/2}\|\snab^{ \
2}\beb\|_{L^2(C_u)}) \label{7.01}
\end{eqnarray}
and:
\begin{eqnarray}
\scR_1(D\rho)&=&\sup_{u\in[u_0,c^*)}(|u|^3\delta^{1/2}\|\sd D\rho\|_{L^2(C_u)})\nonumber\\
\scR_1(D\sigma)&=&\sup_{u\in[u_0,c^*)}(|u|^3\delta^{1/2}\|\sd D\sigma\|_{L^2(C_u)})\nonumber\\
{\cal R}_0(D^2\rho)&=&\sup_{u\in[u_0,c^*)}(|u|^2\delta^{3/2}\|D^2\rho\|_{L^2(C_u)})\nonumber\\
{\cal
R}_0(D^2\sigma)&=&\sup_{u\in[u_0,c^*)}(|u|^2\delta^{3/2}\|D^2\sigma\|_{L^2(C_u)})
\label{7.02}
\end{eqnarray}
and also:
\begin{equation}
\scR_1(\Db\beb)=\sup_{u\in[u_0,c^*)}(|u|^5\delta^{-3/2}\|\snab\Db\beb\|_{L^2(C_u)})
\label{7.03}
\end{equation}
By the results of Chapter 2, the corresponding quantities on
$C_{u_0}$, obtained by replacing the supremum on $[u_0,c^*)$ by
the value at $u_0$, are all bounded by a non-negative
non-decreasing continuous function of $M_7$, the quantity
requiring $M_k$ with the highest $k$ being the one corresponding
to $\snab\Db\beb$, which is expressed in terms of $\snab\sdiv\alb$
by the fourth Bianchi identity of Proposition 1.2.

\section{$L^2$ estimates for $\snab^{ \ 2}\eta$, $\snab^{ \ 2}\etb$}

As we have already remarked,  the estimates of Chapter 6 for
$\snab^{ \ 2}\eta$, $\snab^{ \ 2}\etb$ in $L^4(S)$, loose a factor
of $\delta^{1/2}$ in behavior with respect to $\delta$ in
comparison with the $L^4(S)$ estimates for $\snab\eta$,
$\snab\etb$ of Chapter 4. As a preliminary step we shall presently
derive $L^2$ estimates for $\snab^{ \ 2}\eta$, $\snab^{ \ 2}\etb$,
using only the propagation equations, which are similar to the
estimates of Chapter 4 in behavior with respect to $\delta$. We
consider any $(\ub_1,u_1)\in D^\prime$ and fix attention to the
parameter subdomain $D_1$ and the corresponding subdomain $M_1$ of
$M^\prime$ (see \ref{3.02}, \ref{3.03}).

We apply Lemma 4.1 to the propagation equations \ref{4.113} to
obtain:
\begin{eqnarray}
(D\snab^{ \ 2}\eta)_{ABC}&=&\frac{1}{2}\Omega\mbox{tr}\chi(\snab^{
\ 2}\etb)_{ABC}+\Omega\chih_C^D(\snab^{ \ 2}\etb)_{ABD}
+b^\prime_{ABC}\nonumber\\
(\Db\snab^{ \
2}\etb)_{ABC}&=&\frac{1}{2}\Omega\mbox{tr}\chib(\snab^{ \
2}\eta)_{ABC}+\Omega\chibh_C^D(\snab^{ \ 2}\eta)_{ABD}
+\bb^\prime_{ABC} \label{7.1}
\end{eqnarray}
where:
\begin{eqnarray}
b^\prime_{ABC}&=&-(D\sGamma)^D_{AB}\snab_D\eta_C-(D\sGamma)^D_{AC}\snab_B\eta_D\nonumber\\
&\s&+\frac{1}{2}\sd_A(\Omega\mbox{tr}\chi)\snab_B\etb_C+\snab_A(\Omega\chih_C^D)\snab_B\etb_D\nonumber\\
&\s&+\snab_Ab_{BC}\nonumber\\
\bb^\prime_{ABC}&=&-(\Db\sGamma)^D_{AB}\snab_D\etb_C-(\Db\sGamma)^D_{AC}\snab_B\etb_D\nonumber\\
&\s&+\frac{1}{2}\sd_A(\Omega\mbox{tr}\chib)\snab_B\eta_C+\snab_A(\Omega\chibh_C^D)\snab_B\eta_D\nonumber\\
&\s&+\snab_A\bb_{BC} \label{7.2}
\end{eqnarray}

To the first of \ref{7.1} we apply Lemma 4.6 with $\snab^{ \
2}\eta$ in the role of $\theta$ and
$$\frac{1}{2}\Omega\mbox{tr}\chi\snab^{ \ 2}\etb +\Omega\chih\cdot\snab^{ \ 2}\etb+b^\prime$$
in the role of $\xi$. Then $r=3$, $\nu=0$, $\gamma=0$. In view of
\ref{4.118}, \ref{4.119} we obtain, taking $p=2$:
\begin{eqnarray}
\|\snab^{ \
2}\eta\|_{L^2(S_{\ub,u})}&\leq&C\tilde{a}(u)\int_0^{\ub}\|\snab^{
\ 2}\etb\|_{L^2(S_{\ub^\prime,u})}d\ub^\prime
\nonumber\\
&\s&+C\int_0^{\ub}\|b^\prime\|_{L^2(S_{\ub^\prime,u})}d\ub^\prime
\label{7.3}
\end{eqnarray}

To the second of \ref{7.1} we apply Lemma 4.7 with $\snab^{ \
2}\etb$ in the role of $\thetab$ and
$$\frac{1}{2}\Omega\mbox{tr}\chib\snab^{ \ 2}\eta+\Omega\chibh\cdot\snab^{ \ 2}\eta+\bb$$
in the role of $\xib$. Then $r=3$, $\nu=0$, $\gammab=0$. In view
of \ref{4.123}, \ref{4.124} we obtain, taking $p=2$:
\begin{eqnarray}
|u|^2\|\snab^{ \ 2}\etb\|_{L^2(S_{\ub,u})}&\leq&C|u_0|^2\|\snab^{
\ 2}\etb\|_{L^2(S_{\ub,u_0})}
+C\int_{u_0}^u|u^\prime|^2\tilde{\ab}\|\snab^{ \ 2}\eta\|_{L^2(S_{\ub,u^\prime})}du^\prime\nonumber\\
&\s&+C\int_{u_0}^u|u^\prime|^2\|\bb^\prime\|_{L^2(S_{\ub,u^\prime})}du^\prime
\label{7.4}
\end{eqnarray}
Defining:
\begin{equation}
\scD_2(\etb)=|u_0|^3\delta^{-1/2}\sup_{\ub\in[0,\ub_1]}\|\snab^{
\ 2}\etb\|_{L^2(S_{\ub,u_0})} \label{7.5}
\end{equation}
and taking into account \ref{4.137}, the inequality \ref{7.4}
reduces to:
\begin{eqnarray}
|u|^2\|\snab^{ \
2}\etb\|_{L^2(S_{\ub,u})}&\leq&C|u_0|^{-1}\delta^{1/2}\scD_2(\etb)
+C\int_{u_0}^u|u^\prime|\|\snab^{ \ 2}\eta\|_{L^2(S_{\ub,u^\prime})}du^\prime\nonumber\\
&\s&+C\int_{u_0}^u|u^\prime|^2\|\bb^\prime\|_{L^2(S_{\ub,u^\prime})}du^\prime
\label{7.6}
\end{eqnarray}

The inequalities \ref{7.3}, \ref{7.6} constitute a system of
linear integral inequalities for the quantities $\|\snab^{ \
2}\eta\|_{L^2(S_{\ub,u})}$, $\|\snab^{ \
2}\etb\|_{L^2(S_{\ub,u})}$ on the domain $D_1$. Setting:
\begin{equation}
z^\prime(u)=\sup_{\ub\in[0,\ub_1]}\|\snab^{ \
2}\eta\|_{L^2(S_{\ub,u})} \label{7.7}
\end{equation}
replacing $\ub$ by $\ub^\prime\in[0,\ub]$ in \ref{7.6} and
integrating with respect to $\ub^\prime$ on $[0,\ub]$ yields, for
all $(\ub,u)\in D_1$:
\begin{eqnarray}
|u|^2\int_0^{\ub}\|\snab^{ \
2}\etb\|_{L^2(S_{\ub^\prime,u})}d\ub^\prime&\leq&
C\delta^{3/2}|u_0|^{-1}\scD_2(\etb)
+C\delta\int_{u_0}^u|u^\prime|z^\prime(u^\prime)du^\prime\nonumber\\
&\s&+C\int_{u_0}^u|u^\prime|^2\int_0^{\ub_1}\|\bb^\prime\|_{L^2(S_{\ub^\prime,u^\prime})}d\ub^\prime
du^\prime \label{7.8}
\end{eqnarray}
Substituting \ref{7.8} in \ref{7.3} and taking the supremum over
$\ub\in[0,\ub_1]$ then yields the linear integral inequality:
\begin{equation}
z^\prime(u)\leq\lambda^\prime(u)+\nu^\prime(u)\int_{u_0}^u|u^\prime|z^\prime(u^\prime)du^\prime
\label{7.9}
\end{equation}
where:
\begin{eqnarray}
\lambda^\prime(u)&=&C|u|^{-2}\tilde{a}(u)\left(\delta^{3/2}|u_0|^{-1}\scD_2(\etb)
+\int_{u_0}^u|u^\prime|^2\int_0^{\ub_1}\|\bb^\prime\|_{L^2(S_{\ub^\prime,u^\prime})}d\ub^\prime du^\prime\right)\nonumber\\
&\s&+C\int_0^{\ub_1}\|b^\prime\|_{L^2(S_{\ub^\prime,u})}d\ub^\prime
\label{7.10}
\end{eqnarray}
and:
\begin{equation}
\nu^\prime(u)=C\delta|u|^{-2}\tilde{a}(u) \label{7.11}
\end{equation}
Setting
\begin{equation}
Z^\prime(u)=\int_{u_0}^u|u^\prime|z^\prime(u^\prime)du^\prime, \ \
\mbox{we have $Z^\prime(u_0)=0$} \label{7.12}
\end{equation}
and \ref{7.9} takes the form:
\begin{equation}
\frac{dZ^\prime}{du}\leq |u|(\lambda^\prime+\nu^\prime Z^\prime)
\label{7.13}
\end{equation}
Integrating from $u_0$ we obtain:
\begin{equation}
Z^\prime(u)\leq\int_{u_0}^u\exp\left(\int_{u^\prime}^u|u^{\prime\prime}|\nu^\prime(u^{\prime\prime})du^{\prime\prime}\right)
|u^\prime|\lambda^\prime(u^\prime)du^\prime \ \ : \ \forall
u\in[u_0,u_1] \label{7.14}
\end{equation}
From \ref{7.11} and \ref{4.149} we have:
\begin{equation}
\int_{u^\prime}^u|u^{\prime\prime}|\nu^\prime(u^{\prime\prime})du^{\prime\prime}
=C\delta\int_{u^\prime}^u|u^{\prime\prime}|^{-1}\tilde{a}(u^{\prime\prime})du^{\prime\prime}\leq
1 \label{7.15}
\end{equation}
provided that $\delta$ is suitably small depending on ${\cal
D}_0^\infty$, ${\cal R}_0^\infty$. Therefore:
\begin{equation}
Z^\prime(u)\leq
C\int_{u_0}^u|u^\prime|\lambda^\prime(u^\prime)du^\prime \ \ : \
\forall u\in[u_0,u_1] \label{7.16}
\end{equation}
and \ref{7.9} reads:
\begin{equation}
z^\prime(u)\leq\lambda^\prime(u)+\nu^\prime(u)Z^\prime(u)
\label{7.17}
\end{equation}
Moreover, taking the $L^2$ norm of \ref{7.6} with respect to $\ub$
on $[0,\ub_1]$, we obtain, in view of the definitions \ref{7.7}
and \ref{7.12},
\begin{eqnarray}
|u|^2\|\snab^{ \ 2}\etb\|_{L^2(C^{\ub_1}_u)}&\leq&C\delta|u_0|^{-1}\scD_2(\etb)+C\delta^{1/2}Z^\prime(u)\nonumber\\
&\s&+C\int_{u_0}^u|u^\prime|^2\|\bb^\prime\|_{L^2(C^{\ub_1}_{u^\prime})}du^\prime
\label{7.18}
\end{eqnarray}
Here we denote by $C^{\ub_1}_u$ the part of $C_u$ which
corresponds to $\ub\leq\ub_1$. This follows since for any $S$
tensorfield $\theta$ we have:
\begin{equation}
\|\theta\|^2_{L^2(C^{\ub_1}_u)}=\int_{C^{\ub_1}_u}|\theta|^2=
\int_0^{\ub_1}\left(\int_{S_{\ub,u}}|\theta|^2d\mu_{\sg}\right)d\ub
=\int_0^{\ub_1}\|\theta\|^2_{L^2(S_{\ub,u})}d\ub \label{7.19}
\end{equation}
(see \ref{5.7}, \ref{5.7a}). In deriving \ref{7.18} we have also used the fact
that
\begin{equation}
\|\theta\|_{L^2(C^{\ub_1}_u)}\leq
\ub_1^{1/2}\sup_{\ub\in[0,\ub_1]}\|\theta\|_{L^2(S_{\ub,u})} \leq
\delta^{1/2}\sup_{\ub\in[0,\ub_1]}\|\theta\|_{L^2(S_{\ub,u})}
\label{7.20}
\end{equation}
to bound
$$\int_{u_0}^u|u^\prime|\|\snab^{ \ 2}\eta\|_{L^2(C^{\ub_1}_{u^\prime})}du^\prime\leq \delta^{1/2}Z^\prime(u)$$

We shall now derive an appropriate estimate for $\scD_2(\etb)$.
To do this we recall that along $C_{u_0}$ we have:
\begin{equation}
\etb=-\eta \ \ : \ \mbox{along $C_{u_0}$} \label{7.a1}
\end{equation}
Thus, the first of \ref{7.1} simplifies along $C_{u_0}$ to:
\begin{equation}
(D\snab^{ \ 2}\eta)_{ABC}+\frac{1}{2}\Omega\mbox{tr}\chi(\snab^{ \
2}\eta)_{ABC}=-\Omega\chih_C^D(\snab^{ \ 2}\eta)_{ABD}
+b^\prime_{ABC} \label{7.a2}
\end{equation}
We apply to Lemma 4.6 with $\snab^{ \ 2}\eta$ in the role of
$\theta$ and $b^\prime$ in the role of $\xi$. Here $r=3$,
$\nu=-1$, $\gamma=-\Omega I\otimes\chih^\sharp$. Taking $p=2$ we
obtain:
\begin{equation}
\|\snab^{ \ 2}\eta\|_{L^2(S_{\ub,u_0})}\leq
C\int_{0}^{\ub}\|b^\prime\|_{L^2(S_{\ub^\prime,u_0})}d\ub^\prime
\label{7.a3}
\end{equation}
which, in view of \ref{7.a1} implies:
\begin{equation}
\scD_2(\etb)\leq
C|u_0|^3\delta^{-1/2}\int_{0}^{\ub_1}\|b^\prime\|_{L^2(S_{\ub^\prime,u_0})}d\ub^\prime
\label{7.a4}
\end{equation}

Consider the integrals involving $b^\prime$, $\bb^\prime$ in
\ref{7.10}, \ref{7.18} and \ref{7.a4}. The principal terms in $b$
and $\bb$ from the point of view of differentiability are the
terms $-\Omega\snab\beta$ and $\Omega\snab\beb$ respectively (see
\ref{4.111}). Thus the principal parts of $b^\prime$ and
$\bb^\prime$ are:
\begin{eqnarray}
\stackrel{(P)}{b^\prime}&=&-\Omega\snab^{ \ 2}\beta\nonumber\\
\stackrel{(P)}{\bb^\prime}&=&\Omega\snab^{ \ 2}\beb \label{7.21}
\end{eqnarray}
and we have:
\begin{eqnarray}
\int_0^{\ub_1}\|\stackrel{(P)}{b^\prime}\|_{L^2(S_{\ub^\prime,u})}d\ub^\prime&\leq&
C\int_0^{\ub_1}\|\snab^{ \ 2}\beta\|_{L^2(S_{\ub^\prime,u})}d\ub^\prime\nonumber\\
&\leq&C\ub_1^{1/2}\left(\int_0^{\ub_1}\|\snab^{ \ 2}\beta\|^2_{L^2(S_{\ub^\prime,u})}d\ub^\prime\right)^{1/2}\nonumber\\
&\leq&C\delta^{1/2}\|\snab^{ \ 2}\beta\|_{L^2(C_u)} \ \leq \
C\delta^{1/2}|u|^{-3}\scR_2(\beta) \label{7.22}
\end{eqnarray}
and:
\begin{eqnarray}
\int_0^{\ub_1}\|\stackrel{(P)}{\bb^\prime}\|_{L^2(S_{\ub^\prime,u})}d\ub^\prime&\leq&
C\int_0^{\ub_1}\|\snab^{ \ 2}\beb\|_{L^2(S_{\ub^\prime,u})}d\ub^\prime\nonumber\\
&\leq&C\ub^{1/2}\left(\int_0^{\ub_1}\|\snab^{ \ 2}\beb\|_{L^2(S_{\ub^\prime,u})}d\ub^\prime\right)^{1/2}\nonumber\\
&\leq&C\delta^{1/2}\|\snab^{ \ 2}\beb\|_{L^2(C_u)} \ \leq
C\delta^2|u|^{-5}\scR_2(\beb) \label{7.23}
\end{eqnarray}
hence:
\begin{equation}
\int_{u_0}^u|u^\prime|^2\int_0^{\ub_1}\|\stackrel{(P)}{\bb^\prime}\|_{L^2(S_{\ub^\prime,u^\prime})}d\ub^\prime
du^\prime \leq C\delta^2|u|^{-2}\scR_2(\beb) \label{7.24}
\end{equation}
Also:
\begin{equation}
\|\stackrel{(P)}{\bb^\prime}\|_{L^2(C^{\ub_1}_u)}\leq C\|\snab^{ \
2}\beb\|_{L^2(C_u)} \leq C\delta^{3/2}|u|^{-5}\scR_2(\beb)
\label{7.25}
\end{equation}
Writing
\begin{eqnarray}
b^\prime&=&\stackrel{(P)}{b^\prime}+\stackrel{(N)}{b^\prime}\nonumber\\
\bb^\prime&=&\stackrel{(P)}{\bb^\prime}+\stackrel{(N)}{\bb^\prime}
\label{7.26}
\end{eqnarray}
the non-principal parts $\stackrel{(N)}{b^\prime}$ and
$\stackrel{(N)}{\bb^\prime}$ can be estimated in $L^2(S)$ using
the results of Chapters 3, 4 and 6. We obtain:
\begin{eqnarray}
\|\stackrel{(N)}{b^\prime}\|_{L^2(S_{\ub,u})}&\leq& O(|u|^{-4})\nonumber\\
\|\stackrel{(N)}{\bb^\prime}\|_{L^2(S_{\ub,u})}&\leq&
O(\delta|u|^{-5}) \label{7.27}
\end{eqnarray}
Combining with \ref{7.22} - \ref{7.25} we then obtain:
\begin{equation}
\int_0^{\ub_1}\|b^\prime\|_{L^2(S_{\ub^\prime,u})}d\ub^\prime\leq
C\delta^{1/2}|u|^{-3}\scR_2(\beta) +O(\delta|u|^{-4}) \label{7.28}
\end{equation}
\begin{equation}
\int_0^{\ub_1}\|\bb^\prime\|_{L^2(S_{\ub^\prime,u})}d\ub^\prime\leq
C\delta^2|u|^{-5}\scR_2(\beb) +O(\delta^2|u|^{-5}) \label{7.29}
\end{equation}
\begin{equation}
\int_{u_0}^u|u^\prime|^2\int_0^{\ub_1}\|\bb^\prime\|_{L^2(S_{\ub^\prime,u^\prime})}d\ub^\prime
du^\prime \leq C\delta^2|u|^{-2}\scR_2(\beb)+O(\delta^2|u|^{-2})
\label{7.30}
\end{equation}
\begin{equation}
\|\bb^\prime\|_{L^2(C^{\ub_1}_u)} \leq
C\delta^{3/2}|u|^{-5}\scR_2(\beb)+O(\delta^{3/2}|u|^{-5})
\label{7.31}
\end{equation}
In particular:
\begin{equation}
\int_0^{\ub_1}\|b^\prime\|_{L^2(S_{\ub^\prime,u_0})}d\ub^\prime\leq
C\delta^{1/2}|u_0|^{-3}\scR_2(\beta) +O(\delta|u_0|^{-4})
\label{7.32}
\end{equation}
hence, from \ref{7.a4}:
\begin{equation}
\scD_2(\etb)\leq C\scR_2(\beta)+O(\delta^{1/2}|u_0|^{-1})
\label{7.33}
\end{equation}

Substituting the estimates \ref{7.33}, \ref{7.30} and \ref{7.28}
in \ref{7.10} (and recalling \ref{4.119}) we obtain:
\begin{equation}
\lambda^\prime(u)\leq
C\delta^{1/2}|u|^{-3}\scR_2(\beta)+O(\delta|u|^{-4}) \label{7.34}
\end{equation}
Here and in the following we denote by $O(\delta^p|u|^r)$, for
real numbers $p$, $r$, the product of $\delta^p|u|^r$ with a
non-negative non-decreasing continuous function of the quantities
${\cal D}_0^\infty$, ${\cal R}_0^\infty$, $\scD_1^4$, $\scR_1^4$,
$\scD_2^4(\mbox{tr}\chib)$,  {\em and} $\scR_2$, where:
\begin{equation}
\scR_2=\max\{\scR_2(\alpha),\scR_2(\beta),\scR_2(\rho),\scR_2(\sigma),\scR_2(\beb)\}
\label{7.35}
\end{equation}
The bound \ref{7.34} implies (see \ref{7.16}):
\begin{equation}
Z^\prime(u)\leq
C\delta^{1/2}|u|^{-1}\scR_2(\beta)+O(\delta|u|^{-2}) \label{7.36}
\end{equation}
hence from \ref{7.17} (recalling \ref{7.11} and \ref{4.119}) we
obtain:
\begin{equation}
z^\prime(u)\leq
C\delta^{1/2}|u|^{-3}\scR_2(\beta)+O(\delta|u|^{-4}) \ \ : \
\forall u\in[u_0,u_1] \label{7.37}
\end{equation}
Also, substituting the estimates \ref{7.33}, \ref{7.36} and
\ref{7.31} in \ref{7.18} we obtain:
\begin{equation}
\|\snab^{ \ 2}\etb\|_{L^2(C^{\ub_1}_u)}\leq
C\delta|u|^{-3}\scR_2(\beta)+O(\delta^{3/2}|u|^{-4}) \ \ : \
\forall u\in[u_0,u_1] \label{7.38}
\end{equation}
In view of the definition \ref{7.7} and the fact that
$(\ub_1,u_1)\in D^\prime$ is arbitrary the estimates \ref{7.37}
and \ref{7.38} yield the following Proposition:

\vspace{5mm}

\noindent{\bf Proposition 7.1} \ \ \ We have:
$$\|\snab^{ \ 2}\eta\|_{L^2(S_{\ub,u})}\leq C\delta^{1/2}|u|^{-3}\scR_2(\beta)+O(\delta|u|^{-4})
\ \ : \ \forall (\ub,u)\in D^\prime$$
$$\|\snab^{ \ 2}\etb\|_{L^2(C_u)}\leq C\delta|u|^{-3}\scR_2(\beta)+O(\delta^{3/2}|u|^{-4}) \ \ : \ \forall u\in [u_0,c^*)$$
provided that $\delta$ is suitably small depending on ${\cal
D}_0^\infty$, ${\cal R}_0^\infty$, $\scD_1^4$, $\scR_1^4$, and
$\scR_2(\alpha)$.

\vspace{5mm}

\section{$L^2$ elliptic theory for generalized Hodge systems on $S$}

To estimate in $L^2$ the third derivatives of the connection
cefficients we shall make use of the following two lemmas from
Chapter 2 of [C-K] (Lemmas 2.2.2 and 2.2.3) for geneneralized Hodge
systems on a 2-dimensional compact Riemannian manifold $(S,\sg)$.
For the sake of a self-contained presentation, we shall also give
the proofs of these lemmas. Let $\xi$ be a totally symmetric $r+1$
covariant tensorfield on $S$. We define $\sdiv\xi$ and $\scurl\xi$
to be the totally symmetric $r$ covariant tensorfields on $S$,
given in a arbitrary local frame field by:
\begin{eqnarray}
(\sdiv\xi)_{A_1...A_r}&=&\snab^B\xi_{BA_1...A_r}\\
(\scurl\xi)_{A_1...A_r}&=&\seps^{BC}\snab_B\xi_{CA_1...A_r}
\label{7.39}
\end{eqnarray}
We also define $\mbox{tr}\xi$ to be the totally symmetric $r-1$
covariant tensorfield on $S$, given in an arbitrary local frame
field by:
\begin{equation}
(\mbox{tr}\xi)_{A_1...A_{r-1}}=(\sg^{-1})^{BC}\xi_{BCA_1...A_{r-1}}
\label{7.40}
\end{equation}
If $r=0$, then by convention $\mbox{tr}\xi=0$.

\vspace{5mm}

\noindent{\bf Lemma 7.1} \ \ \ Let $\xi$ be a totally symmetric
$r+1$ covariant tensorfield on a 2-dimensional compact Riemannian
manifold $(S,\sg)$, satisfying the generalized Hodge system:
\begin{eqnarray*}
\sdiv\xi&=&f\\
\scurl\xi&=&g\\
\mbox{tr}\xi&=&h
\end{eqnarray*}
where $f, g$ are given totally symmetric $r$ covariant
tensorfields on $S$ and $h$ is a given totally symmetric $r-1$
covariant tensorfield on $S$. Then
$$\int_S\{|\snab\xi|^2+(r+1)K|\xi|^2\}d\mu_{\sg}=\int_S\{|f|^2+|g|^2+rK|h|^2\}d\mu_{\sg}$$

\noindent{\em Proof}: According to the definition of $\scurl\xi$:
$$\snab_C\xi_{BA_1...A_r}=\snab_B\xi_{CA_1...A_r}-\seps_{BC}g_{A_1...A_r}$$
Differentiating and commuting covariant derivatives we obtain:
\begin{eqnarray}
\snab_D\snab_C\xi_{BA_1...A_r}&=&\snab_D\snab_B\xi_{CA_1...A_r}-\seps_{BC}\snab_D g_{A_1...A_r}\nonumber\\
&=&\snab_B\snab_D\xi_{CA_1...A_r}-\sum_{i=1}^r\sR_{A_i MBD}\xi^M_{CA_1...>A_i<...A_r}\nonumber\\
&\s&-\sR_{CMBD}\xi^M_{A_1...A_r}-\seps_{BC}\snab_D g_{A_1...A_r}
\label{7.41}
\end{eqnarray}
where
$$\sR_{ABCD}=(\sg_{AC}\sg_{BD}-\sg_{AD}\sg_{BC})K$$
is the curvature tensor of $(S,\sg)$. Taking the trace of
\ref{7.41} by contracting with $(\sg^{-1})^{CD}$ we deduce, using
the first and third of the equations of the Hodge system,
\begin{eqnarray}
\slap\xi_{BA_1...A_r}&=&\snab_B f_{A_1...A_r}-\seps_{BC}\snab^C g_{A_1...A_r}\nonumber\\
&\s&+(r+1)K\xi_{BA_1...A_r}-\sum_{i=1}^r\sg_{A_i
B}Kh_{A_1...>A_i<...A_r} \label{7.42}
\end{eqnarray}
Contracting then with $\xi^{BA_1...A_r}$ to form on the left hand
side the inner product $(\xi,\slap\xi)$, integrating on $S$, and
integrating by parts on the left, yields the lemma.

\vspace{5mm}

Let $\xi$ be a totally symmetric $r+1$ covariant tensorfield on
$S$. The symmetrized covariant derivative of $\xi$, denoted by
$(\snab\xi)^s$, is the totally symmetric $r+2$ covariant
tensorfield on $S$ given by:
\begin{equation}
(\snab\xi)^s_{BA_1...A_r}=\frac{1}{r+2}\left(\snab_B\xi_{A_1...A_{r+1}}
+\sum_{i=1}^{r+1}\snab_{A_i}\xi_{A_1...\stackrel{B}{>A_i<}...A_{r+1}}\right)
\label{7.43}
\end{equation}
Also, we denote by $(\s^*\xi)^s$ the symmetrized dual of $\xi$, a
totally symmetric $r+1$ covariant tensorfield on $S$ given by:
\begin{equation}
(\s^*\xi)^s_{A_1...A_{r+1}}=\frac{1}{r+1}\sum_{i=1}^{r+1}\seps_{A_i}^{\s
B}\xi_{A_1...\stackrel{B}{>A_i<}...A_{r+1}} \label{7.44}
\end{equation}

\vspace{5mm}

\noindent{\bf Lemma 7.2} \ \ \ Let $\xi$ be a totally symmetric
$r+1$ covariant tensorfield on a 2-dimensional Riemannian manifold
$(S,\sg)$ satisfying the generalized Hodge system of Lemma 7.1 for
given $f,g,h$. Then $\xi^\prime=(\snab\xi)^s$ satisfies a similar
system:
\begin{eqnarray*}
\sdiv\xi^\prime&=&f^\prime\\
\scurl\xi^\prime&=&g^\prime\\
\mbox{tr}\xi^\prime&=&h^\prime
\end{eqnarray*}
where
\begin{eqnarray*}
f^\prime&=&(\snab f)^s-\frac{1}{r+2}(\s^*\snab g)^s\\
&\s&+(r+1)K\xi-\frac{2K}{r+1}(\sg\otimes^s h)\\
g^\prime&=&\frac{r+1}{r+2}(\snab g)^s+(r+1)K(\s^*\xi)^s\\
h^\prime&=&\frac{2}{r+2}f+\frac{r}{r+2}(\snab h)^s
\end{eqnarray*}
Here, we denote by $(\s^*\snab g)^s$ is the totally symmetric
$r+1$  covariant tensorfield given by:
$$(\s^*\snab g)^s_{A_1...A_{r+1}}=\frac{1}{r+1}\sum_{i=1}^{r+1}\seps_{A_i}^{\s B}\snab_B g_{A_1...>A_i<...A_{r+1}}$$
Also, we denote by $\sg\otimes^s h$ the symmetrized tensor product
of $\sg$ and $h$, a totally symmetric $r+1$ covariant tensorfield
given by:
$$(\sg\otimes^s h)_{A_1...A_{r+1}}=\sum_{i<j=1,...,r+1}\sg_{A_i A_j}h_{A_1...>A_i<...>A_j<...A_{r+1}}$$

\noindent{\em Proof:} \ From \ref{7.43} we have:
\begin{equation}
(\sdiv\xi^\prime)_{A_1...A_{r+1}}=\frac{1}{r+2}\left\{\slap\xi_{A_1...A_{r+1}}+
\sum_{i=1}^{r+1}\snab^B\snab_{A_i}\xi_{A_1...\stackrel{B}{>A_i<}...A_{r+1}}\right\}
\label{7.45}
\end{equation}
Consider first the sum on the right hand side. Commuting covariant
derivatives we deduce:
\begin{eqnarray}
\snab^B\snab_{A_i}\xi_{A_1...\stackrel{B}{>A_i<}...A_{r+1}}
&=&\snab_{A_i}f_{A_1...A_{r+1}}+\sR^B_{\s MBA_i}\xi^M_{A_1...>A_i<...A_{r+1}}\nonumber\\
&\s&+\sum_{\stackrel{j=1,...,r+1}{j\neq i}}\sR_{A_jMBA_i}\xi^{MB}_{A_1...>A_j<...>A_i<...A_{r+1}}\nonumber\\
&=&\snab_{A_i}f_{A_1...>A_i<...A_{r+1}}
+(r+1)K\xi_{A_1...A_{r+1}}\label{7.46}\\
&\s&-\sum_{\stackrel{j=1,...,r+1}{j\neq i}}K\sg_{A_j
A_i}h_{A_1...>A_j<...>A_i<...A_{k+1}} \nonumber
\end{eqnarray}
It follows that:
\begin{eqnarray}
\sum_{i=1}^{r+1}\snab^B\snab_{A_i}\xi_{A_1...\stackrel{B}{>A_i<}...A_{r+1}}
&=&(r+1)(\snab f)^s_{A_1...A_{r+1}}\label{7.47}\\
&\s&+(r+1)^2 K\xi_{A_1...A_{r+1}}-2K(\sg\otimes^s
h)_{A_1...A_{r+1}}\nonumber
\end{eqnarray}
We turn to the first term in parenthesis on the right in
\ref{7.45}. This is given by \ref{7.42} with $B$ replaced by
$A_{r+1}$. Now the left hand side of \ref{7.42} is totally
symmetric. Therefore we may totally symmetrize the right hand side
to obtain a totally symmetric formula for $\slap\xi$. The formula
which we obtain in this way is:
\begin{equation}
\slap\xi=(\snab f)^s-(\s^*\snab
g)^s+(r+1)K\xi-\frac{2K}{r+1}(\sg\otimes^s h) \label{7.48}
\end{equation}
Combining \ref{7.48} with \ref{7.47} yields the formula for
$f^\prime$ of the lemma.

From \ref{7.43} we have:
\begin{equation}
(\scurl\xi^\prime)_{A_1...A_{r+1}}=\frac{1}{r+2}\left\{\seps^{BC}\snab_B\snab_C\xi_{A_1...A_{r+1}}
+\sum_{i=1}^{r+1}\seps^{BC}\snab_B\snab_{A_i}\xi_{A_1...\stackrel{C}{>A_i<}...A_{r+1}}\right\}
\label{7.49}
\end{equation}
In regard to the first term in parenthesis on the right we have:
\begin{eqnarray}
\seps^{BC}\snab_B\snab_C\xi_{A_1...A_{r+1}}&=&\frac{1}{2}\seps^{BC}(\snab_B\snab_C\xi_{A_1...A_{r+1}}
-\snab_C\snab_B\xi_{A_1...A_{r+1}})\nonumber\\
&=&\frac{1}{2}\seps^{BC}\sum_{i=1}^{r+1}\sR_{A_i MBC}\xi^M_{A_1...>A_i<...A_{r+1}}\nonumber\\
&=&(r+1)K(\s^*\xi)^s_{A_1...A_{r+1}}\label{7.50}
\end{eqnarray}
Consider next the sum on the right in \ref{7.49}. Commuting
covariant derivatives we deduce:
\begin{eqnarray}
&&\seps^{BC}\snab_B\snab_{A_i}\xi_{A_1...\stackrel{C}{>A_i<}...A_{r+1}}=\nonumber\\
&&\hspace{1cm}\seps^{BC}\left\{\snab_{A_i}\snab_B\xi_{A_1...\stackrel{C}{>A_i<}...A_{r+1}}
+\sR_{CMBA_i}\xi^M_{A_1...>A_i<...A_{r+1}}\right.\nonumber\\
&&\hspace{2cm}\left.+\sum_{\stackrel{j=1,...,r+1}{j\neq
i}}\sR_{A_j MBA_i}\xi^{MC}_{A_1...>A_j<...>A_i<...A_{r+1}}
\right\}\nonumber\\
&&=\snab_{A_i}g_{A_1...>A_i<...A_{r+1}}\label{7.51}\\
&&\s-K\seps^{BC}\left\{\sg_{CA_i}\xi_{A_1...\stackrel{B}{>A_i<}...A_{r+1}}
-\sum_{\stackrel{j=1,...,r+1}{j\neq i}}\sg_{A_j
B}\xi_{A_1...\stackrel{A_i}{>A_j<}...\stackrel{C}{>A_i<}...A_{r+1}}
\right\}\nonumber
\end{eqnarray}
It follows that:
\begin{equation}
\sum_{i=1}^{r+1}\seps^{BC}\snab_B\snab_C\xi_{A_1...A_{r+1}}=(r+1)(\snab
g)^s +(r+1)^2 K(\s^*\xi)^s \label{7.52}
\end{equation}
Combining \ref{7.52} with \ref{7.50} we obtain the formula for
$g^\prime$ of the lemma. Finally, the formula for $h^\prime$ is
obtained in a straightforward manner.

\vspace{5mm}

Consider now, as in Lemma 5.6, a trace-free symmetric 2-covariant
tensorfield on $S$ satisfying the equation
\begin{equation}
\sdiv_{\sg}\theta=f \label{7.53}
\end{equation}
where $f$ is a given 1-form on $S$. Then, noting the identity:
\begin{equation}
\seps_{AB}\seps_{CD}=\sg_{AC}\sg_{CD}-\sg_{AD}\sg_{BC}
\label{7.54}
\end{equation}
we have:
\begin{eqnarray*}
&&\s^*(\scurl\theta)_A=\seps_A^{\s B}(\scurl\theta)_B=\seps_A^{\s B}\seps^{CD}\snab_C\theta_{DB}\\
&&\s=(\delta_A^C(\sg^{-1})^{BD}-\delta_A^D(\sg^{-1})^{BC})\snab_C\theta_{DB}=\snab_A\mbox{tr}\theta-\snab^B\theta_{AB}
=-(\sdiv\theta)_A
\end{eqnarray*}
that is:
\begin{equation}
\s^*\scurl\theta=-\sdiv\theta, \ \ \ \scurl\theta=\s^*\sdiv\theta
\label{7.55}
\end{equation}
Therefore $\theta$ satisfies the generalized Hodge system:
\begin{eqnarray}
\sdiv\theta&=&f\nonumber\\
\scurl\theta&=&\s^*f\nonumber\\
\mbox{tr}\theta&=&0 \label{7.56}
\end{eqnarray}
Defining as in Lemma 7.2 $\theta^\prime$ to be the totally
symmetric 3-covariant tensorfield on $S$:
\begin{equation}
\theta^\prime=(\snab\theta)^s
\end{equation}
that is:
\begin{equation}
\theta^\prime_{ABC}=\frac{1}{3}(\snab_A\theta_{BC}+\snab_B\theta_{AC}+\snab_C\theta_{AB})
\label{7.58}
\end{equation}
we have:
\begin{equation}
\snab_A\theta_{BC}=\theta^\prime_{ABC}+\frac{1}{3}(2\snab_A\theta_{BC}-\snab_B\theta_{AC}-\snab_C\theta_{AB})
\label{7.59}
\end{equation}
Since
$$\snab_A\theta_{BC}-\snab_B\theta_{AC}=\seps_{AB}(\scurl\theta)_C=\seps_{AB}\s^*f_C$$
\ref{7.59} becomes the following expression for $\snab\theta$ in
terms of $\theta^\prime$ and $f$:
\begin{equation}
\snab_A\theta_{BC}=\theta^\prime_{ABC}+\frac{1}{3}(\seps_{AB}\s^*f_C+\seps_{AC}\s^*f_B)
\label{7.60}
\end{equation}
According to Lemma 7.2 $\theta^\prime$ satisfies the generalized
Hodge system:
\begin{eqnarray}
\sdiv\theta^\prime&=&f^\prime\nonumber\\
\scurl\theta^\prime&=&g^\prime\nonumber\\
\mbox{tr}\theta^\prime&=&h^\prime \label{7.61}
\end{eqnarray}
where:
\begin{eqnarray}
f^\prime&=&(\snab f)^s-\frac{1}{3}(\s^*\snab\s^*f)^s+2K\theta\nonumber\\
&=&\frac{2}{3}(\snab f)^s-\frac{1}{3}\sg\sdiv f+2K\theta\nonumber\\
g^\prime&=&\frac{2}{3}(\snab\s^*f)^s+2K(\s^*\theta)^s\nonumber\\
h^\prime&=&\frac{2}{3}f\label{7.62}
\end{eqnarray}

Defining next as in Lemma 7.2 $\theta^{\prime\prime}$ to be the
totally symmetric 4-covariant tensorfield on $S$:
\begin{equation}
\theta^{\prime\prime}=(\snab\theta^\prime)^s \label{7.63}
\end{equation}
that is:
\begin{equation}
\theta^{\prime\prime}_{ABCD}=\frac{1}{4}(\snab_A\theta^\prime_{BCD}+\snab_B\theta^\prime_{ACD}+\snab_C\theta^\prime_{ABD}
+\snab_D\theta^\prime_{ABC}) \label{7.64}
\end{equation}
we have:
\begin{equation}
\snab_A\theta^\prime_{BCD}=\theta^{\prime\prime}_{ABCD}+\frac{1}{4}(3\snab_A\theta^\prime_{BCD}-\snab_B\theta^\prime_{ACD}
-\snab_C\theta^\prime_{ABD}-\snab_D\theta^\prime_{ABC})
\label{7.65}
\end{equation}
Since
$$\snab_A\theta^\prime_{BCD}-\snab_B\theta^\prime_{ACD}=\seps_{AB}(\scurl\theta^\prime)_{CD}=\seps_{AB}g^\prime_{CD}$$
\ref{7.65} becomes the following expression for
$\snab\theta^\prime$ in terms of $\theta^{\prime\prime}$ and
$g^\prime$:
\begin{equation}
\snab_A\theta^\prime_{BCD}=\theta^{\prime\prime}_{ABCD}+\frac{1}{4}(\seps_{AB}g^\prime_{CD}+\seps_{AC}g^\prime_{BD}
+\seps_{AD}g^\prime_{BC}) \label{7.66}
\end{equation}
According to Lemma 7.2 $\theta^{\prime\prime}$ satisfies the
generalized Hodge system:
\begin{eqnarray}
\sdiv\theta^{\prime\prime}&=&f^{\prime\prime}\nonumber\\
\scurl\theta^{\prime\prime}&=&g^{\prime\prime}\nonumber\\
\mbox{tr}\theta^{\prime\prime}&=&h^{\prime\prime} \label{7.67}
\end{eqnarray}
where:
\begin{eqnarray}
f^{\prime\prime}&=&(\snab f^\prime)^s-\frac{1}{4}(\s^*\snab g^\prime)^s\nonumber\\
&\s&+3K\theta^\prime-\frac{2K}{3}(\sg\otimes^s h^\prime)\nonumber\\
g^{\prime\prime}&=&\frac{3}{4}(\snab g^\prime)^s+3K(\s^*\theta^\prime)^s\nonumber\\
h^{\prime\prime}&=&\frac{1}{2}f^\prime+\frac{1}{2}(\snab
h^\prime)^s \label{7.68}
\end{eqnarray}

Substituting \ref{7.62} in \ref{7.68} we deduce the following
pointwise bounds:
\begin{eqnarray}
|f^{\prime\prime}|&\leq&C\{|\snab^{ \ 2} f|+|K||\snab\theta|+|\sd K||\theta|\}\nonumber\\
|g^{\prime\prime}|&\leq&C\{|\snab^{ \ 2} f|+|K||\snab\theta|+|\sd K||\theta|\}\nonumber\\
|h^{\prime\prime}|&\leq&C\{|\snab f|+|K||\theta|\} \label{7.69}
\end{eqnarray}

According to Lemma 7.1 $\theta^{\prime\prime}$ satisfies the
following integral identity on $S$:
\begin{equation}
\int_S\{|\snab\theta^{\prime\prime}|^2+4K|\theta^{\prime\prime}|^2\}d\mu_{\sg}
=\int_S\{|f^{\prime\prime}|^2+|g^{\prime\prime}|^2+3K|h^{\prime\prime}|^2\}d\mu_{\sg}
\label{7.70}
\end{equation}
In regard to the second term in the integrant on the left hand
side we write:
\begin{equation}
\int_S
K|\theta^{\prime\prime}|^2d\mu_{\sg}=\int_S\overline{K}|\theta^{\prime\prime}|^2
d\mu_{\sg} +\int_S(K-\overline{K})|\theta^{\prime\prime}|^2
d\mu_{\sg} \label{7.71}
\end{equation}
From this point on we take $S$ to be diffeomorphic to $S^2$. Then
according to the Gauss-Bonnet theorem:
\begin{equation}
\int_S Kd\mu_{\sg}=4\pi \label{7.a5}
\end{equation}
Denoting as in Chapter 5
$$r=\sqrt{\frac{\mbox{Area}(S)}{4\pi}} \ \ \ \mbox{so that} \ \ \overline{K}=\frac{1}{r^2}$$
we estimate the second integral on the right in \ref{7.71} by:
$$\|K-\overline{K}\|_{L^2(S)}\|\theta^{\prime\prime}\|^2_{L^4(S)}\leq C\mbox{I}^\prime(S)r\|K-\overline{K}\|_{L^2(S)}
\{\|\snab\theta^{\prime\prime}\|^2_{L^2(S)}+r^{-2}\|\theta^{\prime\prime}\|^2_{L^2(S)}\}$$
by Lemma 5.1 with $p=4$. Thus, the left hand side of \ref{7.70} is
greater than or equal to:
\begin{equation}
\left[1-4C\mbox{I}^\prime(S)r\|K-\overline{K}\|_{L^2(S)}\right]\left[\|\snab\theta^{\prime\prime}\|^2_{L^2(S)}
+r^{-2}\|\theta^{\prime\prime}\|^2_{L^2(S)}\right] \label{7.72}
\end{equation}
Hence, if
\begin{equation}
4C\mbox{I}^\prime(S)r\|K-\overline{K}\|_{L^2(S)}\leq\frac{1}{2}
\label{7.73}
\end{equation}
the left hand side of \ref{7.70} is greater than or equal to:
\begin{equation}
\frac{1}{2}\left[\|\snab\theta^{\prime\prime}\|^2_{L^2(S)}
+r^{-2}\|\theta^{\prime\prime}\|^2_{L^2(S)}\right] \label{7.74}
\end{equation}
Substituting the pointwise bounds \ref{7.69} on the right hand
side of \ref{7.70} we then conclude that:
\begin{eqnarray}
\|\snab\theta^{\prime\prime}\|_{L^2(S)}&\leq&C\left\{\|\snab^{ \
2}f\|_{L^2(S)}+\||K||\snab\theta|\|_{L^2(S)}
+\||\sd K||\theta|\|_{L^2(S)}\right\}\nonumber\\
&\s&+C\left(\int_S|K|\{|\snab
f|^2+|K|^2|\theta|^2\}d\mu_{\sg}\right)^{1/2} \label{7.75}
\end{eqnarray}
We estimate:
\begin{eqnarray}
\||K||\snab\theta|\|_{L^2(S)}&\leq&\|K\|_{L^4(S)}\|\snab\theta\|_{L^4(S)}\nonumber\\
&\leq&Cr^{-3/2}\left[1+r^{3/2}\|K-\overline{K}\|_{L^4(S)}\right]\|\snab\theta\|_{L^4(S)}
\label{7.76}
\end{eqnarray}
noting that $\|\overline{K}\|^4_{L^4(S)}=4\pi r^{-6}$. Also:
\begin{eqnarray}
&&\||\sd K||\theta|\|_{L^2(S)}\leq\|\sd K\|_{L^2(S)}\|\theta\|_{L^\infty(S)}\nonumber\\
&&\hspace{15mm}\leq C\sqrt{\mbox{I}^\prime(S)}r^{1/2}\|\sd
K\|_{L^2(S)}(\|\snab\theta\|_{L^4(S)}+r^{-1}\|\theta\|_{L^4(S)})
\label{7.77}
\end{eqnarray}
by Lemma 5.2 with $p=4$. Moreover, the last term on the right in
\ref{7.75} is bounded by:
\begin{eqnarray}
&&C\left(\|K\|_{L^2(S)}\|\snab f\|^2_{L^4(S)}+\|K\|^3_{L^3(S)}\|\theta\|^2_{L^\infty(S)}\right)^{1/2}\label{7.78}\\
&&\leq C\left(\|K\|^{1/2}_{L^2(S)}\|\snab f\|_{L^4(S)}+\|K\|^{3/2}_{L^3(S)}\|\theta\|_{L^\infty(S)}\right)\nonumber\\
&&\leq C^\prime r^{-1/2}\left[1+r\|K-\overline{K}\|_{L^2(S)}\right]^{1/2}\|\snab f\|_{L^4(S)}\nonumber\\
&&\s+C^\prime\sqrt{\mbox{I}^\prime(S)}r^{-3/2}\left[1+r^{3/2}\|K-\overline{K}\|_{L^4(S)}\right]^{3/2}
(\|\snab\theta\|_{L^4(S)}+r^{-1}\|\theta\|_{L^4(S)})\nonumber
\end{eqnarray}
noting that $\|\overline{K}\|^2_{L^2(S)}=4\pi r^{-2}$, while
$\|K\|_{L^3(S)}\leq (4\pi r^2)^{1/12}\|K\|_{L^4(S)}$. We
substitute \ref{7.76} - \ref{7.78} in \ref{7.75}. Assuming that
\begin{equation}
r^{3/2}\|K-\overline{K}\|_{L^4(S)}\leq 1 \label{7.79}
\end{equation}
noting that $r\|K-\overline{K}\|_{L^2(S)}\leq
(4\pi)^{1/4}r^{3/2}\|K-\overline{K}\|_{L^4(S)}$ and recalling from
the statement of Lemma 5.1 that by definition
$\mbox{I}^\prime(S)\geq 1$, we then obtain:
\begin{eqnarray}
\|\snab\theta^{\prime\prime}\|_{L^2(S)}&\leq&C\{\|\snab^{ \ 2} f\|_{L^2(S)}+r^{-1/2}\|\snab f\|_{L^4(S)}\label{7.80}\\
&\s&+\sqrt{\mbox{I}^\prime(S)}r^{-3/2}(1+r^2\|\sd
K\|_{L^2(S)})(\|\snab\theta\|_{L^4(S)}+r^{-1}\|\theta\|_{L^4(S)})\}
\nonumber
\end{eqnarray}

Now, from \ref{7.60}, \ref{7.66}, we have, pointwise:
\begin{eqnarray}
|\snab^{ \ 3}\theta|&\leq&|\snab^{ \ 2}\theta^\prime|+C|\snab^{ \ 2}f|\nonumber\\
|\snab^{ \
2}\theta^\prime|&\leq&|\snab\theta^{\prime\prime}|+C|\snab
g^\prime| \label{7.81}
\end{eqnarray}
while from \ref{7.62} we have, pointwise:
\begin{equation}
|\snab g^\prime|\leq C\{|\snab^{ \ 2}f|+|K||\snab\theta|+|\sd
K||\theta|\} \label{7.82}
\end{equation}
Therefore:
\begin{eqnarray}
\|\snab^{ \ 3}\theta\|_{L^2(S)}&\leq&\|\snab\theta^{\prime\prime}\|_{L^2(S)}\label{7.83}\\
&\s&+C\left\{\|\snab^{ \
2}f\|_{L^2(S)}+\||K||\snab\theta\|_{L^2(S)} +\||\sd
K||\theta|\|_{L^2(S)}\right\}\nonumber
\end{eqnarray}
Substituting the estimates \ref{7.76}, \ref{7.77} and noting the
assumption \ref{7.79} we then obtain:
\begin{eqnarray}
\|\snab^{ \ 3}\theta\|_{L^2(S)}&\leq&
C\{\|\snab\theta^{\prime\prime}\|_{L^2(S)}+\|\snab^{ \ 2} f\|_{L^2(S)}\label{7.84}\\
&\s&+\sqrt{\mbox{I}^\prime(S)}r^{-3/2}(1+r^2\|\sd
K\|_{L^2(S)})(\|\snab\theta\|_{L^4(S)}+r^{-1}\|\theta\|_{L^4(S)})\}
\nonumber
\end{eqnarray}
Combining \ref{7.84} with \ref{7.80} we establish the following
lemma.

\vspace{5mm}

\noindent{\bf Lemma 7.3} \ \ \ Let $\theta$ be a trace-free
symmetric 2-covariant tensorfield on a 2-dimensional Riemannian
manifold $(S,\sg)$, with $S$ diffeomorphic to $S^2$, satisfying
the equation
$$\sdiv\theta=f$$
where $f$ is a given 1-form on $S$. Then there is a numerical
constant $C$ such that if
$$4C\mbox{I}^\prime(S)r\|K-\overline{K}\|_{L^2(S)}\leq\frac{1}{2}$$
while
$$r^{3/2}\|K-\overline{K}\|_{L^4(S)}\leq 1$$
the following estimate holds:
\begin{eqnarray*}
\|\snab^{ \ 3}\theta\|_{L^2(S)}&\leq&C\{\|\snab^{ \ 2} f\|_{L^2(S)}+r^{-1/2}\|\snab f\|_{L^4(S)}\\
&\s&+\sqrt{\mbox{I}^\prime(S)}r^{-3/2}(1+r^2\|\sd
K\|_{L^2(S)})(\|\snab\theta\|_{L^4(S)}+r^{-1}\|\theta\|_{L^4(S)})\}
\end{eqnarray*}

\vspace{5mm}

Consider now , as in Lemma 5.9, a 1-form on $S$ satisfying the
Hodge system:
\begin{eqnarray}
\sdiv\theta&=&f\nonumber\\
\scurl\theta&=&g\label{7.85}
\end{eqnarray}
where $(f,g)$ is a given pair of functions on $S$. Defining as in
Lemma 7.2 $\theta^\prime$ to be the symmetric 2-covariant
tensorfield on $S$:
\begin{equation}
\theta^\prime=(\snab\theta)^s \label{7.86}
\end{equation}
that is:
\begin{equation}
\theta^\prime_{AB}=\frac{1}{2}(\snab_A\theta_B+\snab_B\theta_A)
\label{7.87}
\end{equation}
we have:
\begin{equation}
\snab_A\theta_B=\theta^\prime_{AB}+\frac{1}{2}(\snab_A\theta_B-\snab_B\theta_A)
\label{7.88}
\end{equation}
Since
$$\snab_A\theta_B-\snab_B\theta_A=\seps_{AB}\scurl\theta=\seps_{AB}g$$
\ref{7.88} becomes the following expression for $\snab\theta$ in
terms of $\theta^\prime$ and $g$:
\begin{equation}
\snab_A\theta_B=\theta^\prime_{AB}+\frac{1}{2}\seps_{AB}g
\label{7.89}
\end{equation}
According to Lemma 7.2 $\theta^\prime$ satisfies the generalized
Hodge system:
\begin{eqnarray}
\sdiv\theta^\prime&=&f^\prime\nonumber\\
\scurl\theta^\prime&=&g^\prime\nonumber\\
\mbox{tr}\theta^\prime&=&h^\prime\label{7.90}
\end{eqnarray}
where:
\begin{eqnarray}
f^\prime&=&\sd f-\frac{1}{2}\s^*\sd g+K\theta\nonumber\\
g^\prime&=&\frac{1}{2}\sd g+K\s^*\theta\nonumber\\
h^\prime&=&f\label{7.91}
\end{eqnarray}
We have the pointwise bounds:
\begin{eqnarray}
|f^\prime|&\leq& C\{|\sd f|+|\sd g|+|K||\theta|\}\nonumber\\
|g^\prime|&\leq& C\{|\sd g|+|K||\theta|\}\nonumber\\
|h^\prime|&=& |f|\label{7.92}
\end{eqnarray}

According to Lemma 7.1 $\theta^\prime$ satisfies the following
integral identity on $S$:
\begin{equation}
\int_S\{|\snab\theta^\prime|^2+2K|\theta^\prime|^2\}d\mu_{\sg}
=\int_S\{|f^\prime|^2+|g^\prime|^2+K|h^\prime|^2\}d\mu_{\sg}
\label{7.93}
\end{equation}
From this point on we take again $S$ to be diffeomorphic to $S^2$.
Following the argument leading to \ref{7.72} we deduce that the
left hand side of \ref{7.93} is greater than or equal to:
\begin{equation}
[1-2C\mbox{I}^\prime(S)r\|K-\overline{K}\|_{L^2(S)}]\left[\|\snab\theta^\prime\|^2_{L^2(S)}
+r^{-2}\|\theta^\prime\|^2_{L^2(S)}\right] \label{7.94}
\end{equation}
Hence if
\begin{equation}
2C\mbox{I}^\prime(S)r\|K-\overline{K}\|_{L^2(S)}\leq\frac{1}{2}
\label{7.95}
\end{equation}
the left hand side of \ref{7.93} is greater than or equal to:
\begin{equation}
\frac{1}{2}\left[\|\snab\theta^\prime\|^2_{L^2(S)}+r^{-2}\|\theta^\prime\|^2_{L^2(S)}\right]
\label{7.96}
\end{equation}
Substituting the pointwise bounds \ref{7.92} on the right hand
side of \ref{7.93} we then conclude that:
\begin{eqnarray}
\|\snab\theta^\prime\|_{L^2(S)}&\leq& C\{\|\sd f\|_{L^2(S)}+\|\sd g\|_{L^2(S)}+\||K||\theta|\|_{L^2(S)}\}\nonumber\\
&\s&+C\left(\int_S|K||f|^2 d\mu_{\sg}\right)^{1/2} \label{7.97}
\end{eqnarray}
We estimate (see \ref{7.76}):
\begin{eqnarray}
\||K||\theta|\|_{L^2(S)}&\leq&\|K\|_{L^4(S)}\|\theta\|_{L^4(S)}\nonumber\\
&\leq&Cr^{-3/2}\left[1+r^{3/2}\|K-\overline{K}\|_{L^4(S)}\right]\|\theta\|_{L^4(S)}
\label{7.98}
\end{eqnarray}
Moreover, the last term on the right in \ref{7.97} is bounded by:
\begin{equation}
C\|K\|_{L^2(S)}^{1/2}\|f\|_{L^4(S)}\leq C^\prime
r^{-1/2}\left[1+r\|K-\overline{K}\|_{L^2(S)}\right]^{1/2}\|f\|_{L^4(S)}
\label{7.99}
\end{equation}
(see \ref{7.78}). It follows that under the assumption \ref{7.79}
the inequality \ref{7.97} implies:
\begin{equation}
\|\snab\theta^\prime\|_{L^2(S)}\leq C\{\|\sd f\|_{L^2(S)}+\|\sd
g\|_{L^2(S)}+r^{-3/2}|\theta\|_{L^4(S)} +r^{-1/2}\|f\|_{L^4(S)}\}
\label{7.100}
\end{equation}
Now from \ref{7.97} we have, pointwise:
\begin{equation}
|\snab^{ \ 2}\theta|\leq |\snab\theta^\prime|+C|\sd g|
\label{7.101}
\end{equation}
therefore:
\begin{equation}
\|\snab^{ \
2}\theta\|_{L^2(S)}\leq\|\snab\theta^\prime\|_{L^2(S)}+C\|\sd
g\|_{L^2(S)} \label{7.102}
\end{equation}
In view of \ref{7.100}, \ref{7.102} we conclude that:
\begin{equation}
\|\snab^{ \ 2}\theta\|_{L^2(S)}\leq C\{\|\sd f\|_{L^2(S)}+\|\sd
g\|_{L^2(S)} +r^{-1/2}\|f\|_{L^4(S)}+r^{-3/2}\|\theta\|_{L^4(S)}\}
\label{7.103}
\end{equation}

Defining next as in Lemma 7.2 $\theta^{\prime\prime}$ to be the
totally symmetric 3-covariant tensorfield on $S$:
\begin{equation}
\theta^{\prime\prime}=(\snab\theta^\prime)^s \label{7.104}
\end{equation}
that is:
\begin{equation}
\theta^{\prime\prime}_{ABC}=\frac{1}{3}(\snab_A\theta^\prime_{BC}+\snab_B\theta^\prime_{AC}+\snab_C\theta^\prime_{AB})
\label{7.105}
\end{equation}
we have:
\begin{equation}
\snab_A\theta^\prime_{BC}=\theta^{\prime\prime}_{ABC}+\frac{1}{3}(2\snab_A\theta^\prime_{BC}-\snab_B\theta^\prime_{AC}
-\snab_C\theta^\prime_{AB}) \label{7.106}
\end{equation}
Since
$$\snab_A\theta^\prime_{BC}-\snab_B\theta^\prime_{AC}=\seps_{AB}(\scurl\theta^\prime)_C=\seps_{AB}g^\prime_C$$
\ref{7.106} becomes the following expression for
$\snab\theta^\prime$ in terms of $\theta^{\prime\prime}$ and
$g^\prime$:
\begin{equation}
\snab_A\theta^\prime_{BC}=\theta^{\prime\prime}_{ABC}+\frac{1}{3}(\seps_{AB}g^\prime_C+\seps_{AC}g^\prime_B)
\label{7.107}
\end{equation}
According to Lemma 7.2 $\theta^{\prime\prime}$ satisfies the Hodge
system:
\begin{eqnarray}
\sdiv\theta^{\prime\prime}&=&f^{\prime\prime}\nonumber\\
\scurl\theta^{\prime\prime}&=&g^{\prime\prime}\nonumber\\
\mbox{tr}\theta^{\prime\prime}&=&h^{\prime\prime}\label{7.108}
\end{eqnarray}
where:
\begin{eqnarray}
f^{\prime\prime}&=&(\snab f^\prime)^s-\frac{1}{3}(\s^*\snab g^\prime)^s\nonumber\\
&\s&+2K\theta^\prime-K(\sg\otimes^s h^\prime)\nonumber\\
g^{\prime\prime}&=&\frac{2}{3}(\snab g^\prime)^s+2K(\s^*\theta^\prime)^s\nonumber\\
h^{\prime\prime}&=&\frac{2}{3}f^\prime+\frac{1}{3}(\snab
h^\prime)^s \label{7.109}
\end{eqnarray}
Substituting \ref{7.91} in \ref{7.109} we deduce the following
pointwise bounds:
\begin{eqnarray}
|f^{\prime\prime}|&\leq&C\{|\snab^{ \ 2}f|+|\snab^{ \ 2}g|+|K||\snab\theta|+|\sd K||\theta|\}\nonumber\\
|g^{\prime\prime}|&\leq&C\{|\snab^{ \ 2}g|+|K||\snab\theta|+|\sd K||\theta|\}\nonumber\\
|h^{\prime\prime}|&\leq&C\{|\sd f|+|\sd g|+|K||\theta|\}
\label{7.110}
\end{eqnarray}

According to Lemma 7.1 $\theta^{\prime\prime}$ satisfies the
following integral identity on $S$:
\begin{equation}
\int_S\{|\snab\theta^{\prime\prime}|^2+3K|\theta^{\prime\prime}|^2\}d\mu_{\sg}
=\int_S\{|f^{\prime\prime}|^2+|g^{\prime\prime}|^2+3K|h^{\prime\prime}|^2\}d\mu_{\sg}
\label{7.111}
\end{equation}
Following the argument leading to \ref{7.72} we deduce that the
left hand side of \ref{7.111} is greater than or equal to:
\begin{equation}
\left[1-3C\mbox{I}^\prime(S)r\|K-\overline{K}\|_{L^2(S)}\right]\left[\|\snab\theta^{\prime\prime}\|^2_{L^2(S)}
+r^{-2}\|\theta^{\prime\prime}\|^2_{L^2(S)}\right] \label{7.112}
\end{equation}
Hence, if
\begin{equation}
3C\mbox{I}^\prime(S)r\|K-\overline{K}\|_{L^2(S)}\leq\frac{1}{2}
\label{7.113}
\end{equation}
the left hand side of \ref{7.111} is greater than or equal to:
\begin{equation}
\frac{1}{2}\left[\|\snab\theta^{\prime\prime}\|^2_{L^2(S)}+r^{-2}\|\theta^{\prime\prime}\|^2_{L^2(S)}\right]
\label{7.114}
\end{equation}Substituting the pointwise bounds \ref{7.110} on the right hand side of \ref{7.111} we then conclude that:
\begin{eqnarray}
\|\snab\theta^{\prime\prime}\|_{L^2(S)}&\leq&C\{\|\snab^{ \ 2}f\|_{L^2(S)}+\|\snab^{ \ 2}g\|_{L^2(S)}\}\nonumber\\
&\s&+C\{\||K||\snab\theta|\|_{L^2(S)}+\||\sd K||\theta|\|_{L^2(S)}\}\label{7.115}\\
&\s&+C\left(\int_S|K|\{|\snab f|^2+|\snab
g|^2+|K|^2|\theta|^2\}d\mu_{\sg}\right)^{1/2}\nonumber
\end{eqnarray}
Proceeding as in the argument leading from \ref{7.75} to
\ref{7.80} we deduce:
\begin{eqnarray}
\|\snab\theta^{\prime\prime}\|_{L^2(S)}&\leq&C\{\|\snab^{ \ 2}f\|_{L^2(S)}+\|\snab^{ \ 2}g\|_{L^2(S)}\label{7.116}\\
&\s&+r^{-1/2}(\|\snab f\|_{L^4(S)}+\|\snab g\|_{L^4(S)})\nonumber\\
&\s&+\sqrt{\mbox{I}^\prime(S)}r^{-3/2}(1+r^2\|\sd
K\|_{L^2(S)})(\|\snab\theta\|_{L^4(S)}+r^{-1}\|\theta\|_{L^4(S)})\}
\nonumber
\end{eqnarray}
under the assumption \ref{7.79}.

Now, from \ref{7.89}, \ref{7.107} we have, pointwise:
\begin{eqnarray}
|\snab^{ \ 3}\theta|&\leq&|\snab^{ \ 2}\theta^\prime|+C|\snab^{ \ 2}g|\nonumber\\
|\snab^{ \
2}\theta^\prime|&\leq&|\snab\theta^{\prime\prime}|+C|\snab
g^\prime| \label{7.117}
\end{eqnarray}
while from \ref{7.91} we have, pointwise:
\begin{equation}
|\snab g^\prime|\leq C\{|\snab^{ \ 2}g|+|K||\snab\theta|+|\sd
K||\theta|\} \label{7.118}
\end{equation}
Therefore:
\begin{eqnarray}
\|\snab^{ \ 3}\theta\|_{L^2(S)}&\leq&\|\snab\theta^{\prime\prime}\|\label{7.119}\\
&\s&+C\{\|\snab^{ \
2}g\|_{L^2(S)}+\||K||\snab\theta|\|_{L^2(S)}+\||\sd
K||\theta|\|_{L^2(S)}\}\nonumber
\end{eqnarray}
Substituting the analogues of the estimates \ref{7.76}, \ref{7.77}
and noting the assumption \ref{7.79} we then obtain:
\begin{eqnarray}
\|\snab^{ \ 3}\theta\|_{L^2(S)}&\leq&C\{\|\snab\theta^{\prime\prime}\|_{L^2(S)}+\|\snab^{ \ 2}g\|_{L^2(S)}\label{7.120}\\
&\s&+\sqrt{\mbox{I}^\prime(S)}r^{-3/2}(1+r^2\|\sd
K\|_{L^2(S)})(\|\snab\theta\|_{L^4(S)}+r^{-1}\|\theta\|_{L^4(S)})\}
\nonumber
\end{eqnarray}
Combining \ref{7.120} with \ref{7.116} and recalling \ref{7.103}
we conclude that the following lemma has been established.

\vspace{5mm}

\noindent{\bf Lemma 7.4} \ \ \ Let $\theta$ be a 1-form on a
2-dimensional Riemannian manifold $(S,\sg)$, with $S$
diffeomorphic to $S^2$, satisfying the Hodge system
\begin{eqnarray*}
\sdiv\theta&=&f\\
\scurl\theta&=&g
\end{eqnarray*}
where $(f,g)$ is a given pair of functions on $S$. Then there is a
numerical constant $C$ such that if
$$3C\mbox{I}^\prime(S)r\|K-\overline{K}\|_{L^2(S)}\leq\frac{1}{2}$$
while
$$r^{3/2}\|K-\overline{K}\|_{L^4(S)}\leq 1$$
the following estimates hold:
\begin{eqnarray*}
\|\snab^{ \ 2}\theta\|_{L^2(S)}&\leq&C\{\|\sd f\|_{L^2(S)}+\|\sd
g\|_{L^2(S)}
+r^{-1/2}\|f\|_{L^4(S)}+r^{-3/2}\|\theta\|_{L^4(S)}\}\\
\|\snab^{ \ 3}\theta\|_{L^2(S)}&\leq&C\{\|\snab^{ \
2}f\|_{L^2(S)}+\|\snab^{ \ 2}g\|_{L^2(S)}
+r^{-1/2}(\|\snab f\|_{L^4(S)}+\|\snab g\|_{L^4(S)})\\
&\s&+\sqrt{\mbox{I}^\prime(S)}r^{-3/2}(1+r^2\|\sd
K\|_{L^2(S)})(\|\snab\theta\|_{L^4(S)}+r^{-1}\|\theta\|_{L^4(S)})\}
\end{eqnarray*}

\vspace{5mm}

\section{$L^2(S)$ estimates for $\snab^{ \ 3}\chi^\prime$, $\sd K$}

\noindent{\bf Proposition 7.2} \ \ \ We have:
\begin{eqnarray*}
\|\sd K\|_{L^2(S_{\ub,u})}&\leq&C\delta^{1/2}|u|^{-3}(\scR_2(\beta)+\scR_1^4(\beta)+{\cal R}_0^\infty(\beta))\\
&\s&+O(\delta|u|^{-4})
\end{eqnarray*}
\begin{eqnarray*}
&&\|\snab^{ \ 3}\mbox{tr}\chi^\prime\|_{L^2(S_{\ub,u})}\leq\\
&&\hspace{1cm}+C|u|^{-4}[{\cal R}_0^\infty(\alpha)\scR_2(\beta)+
(\scR_1^4(\alpha)+{\cal R}_0^\infty(\alpha))(\scR_1^4(\beta)+{\cal R}_0^\infty(\beta))]\\
&&\hspace{1cm}+O(\delta^{1/2}|u|^{-5})
\end{eqnarray*}
for all $(\ub,u)\in D^\prime$, and:
\begin{eqnarray*}
\|\snab^{ \ 3}\chih^\prime\|_{L^2(C_u)}&\leq& C|u|^{-3}(\scR_2(\beta)+\scR_1^4(\beta)+{\cal R}_0^\infty(\beta))\\
&\s&+O(\delta^{1/2}|u|^{-4})
\end{eqnarray*}
for all $u\in[u_0,c^*)$, provided that $\delta$ is suitably small
depending on ${\cal D}_0^\infty$, ${\cal R}_0^\infty$, $\scD_1^4$,
$\scR_1^4$, and $\scR_2$. In particular, under such a smallness
condition on $\delta$ we have:
$$|u|^2\|\sd K\|_{L^2(S_{\ub,u})}\leq 1$$

\noindent{\em Proof:} \ We apply Lemma 7.3 with $\chih^\prime$ in
the role of $\theta$ to the Codazzi equation \ref{6.4}. The
assumptions of Lemma 7.3 are satisfied by virtue of the first
estimate of Proposition 6.1 provided that $\delta$ is suitably
small depending on ${\cal D}_0^\infty$, ${\cal R}_0^\infty$,
$\scD_1^4$, $\scR_1^4$ and $\scR_2(\alpha)$. Taking also into
account Lemma 5.4 we obtain:
\begin{eqnarray}
&&\|\snab^{ \ 3}\chih^\prime\|_{L^2(S_{\ub,u})}\leq
C\left\{\|\snab^{ \ 3}\mbox{tr}\chi^\prime\|_{L^2(S_{\ub,u})}
+\|\snab^{ \ 2}i\|_{L^2(S_{\ub,u})}\right.\label{7.121}\\
&&\hspace{2cm}\left.+|u|^{-1/2}(\|\snab^{ \
2}\mbox{tr}\chi^\prime\|_{L^4(S_{\ub,u})}
+\|\snab i\|_{L^4(S_{\ub,u})})\right\}\nonumber\\
&&\hspace{1cm}+C|u|^{-3/2}\left[1+|u|^2\|\sd
K\|_{L^2(S_{\ub,u})}\right]\left\{\|\snab\chih^\prime\|_{L^4(S_{\ub,u})}
+|u|^{-1}\|\chih^\prime\|_{L^4(S_{\ub,u})}\right\}\nonumber
\end{eqnarray}
Now, by the estimate \ref{6.49},
\begin{equation}
\|\snab\chih^\prime\|_{L^4(S_{\ub,u})}+|u|^{-1}\|\chih^\prime\|_{L^4(S_{\ub,u})}
\leq C\delta^{-1/2}|u|^{-3/2}(\scR_1^4(\beta)+{\cal
R}_0^\infty(\beta))+O(|u|^{-5/2}) \label{7.122}
\end{equation}
Also, by the estimates \ref{6.16} and \ref{6.44},
\begin{equation}
|u|^{-1/2}(\|\snab^{ \
2}\mbox{tr}\chi^\prime\|_{L^4(S_{\ub,u})}+\|\snab
i\|_{L^4(S_{\ub,u})}) \leq
C\delta^{-1/2}|u|^{-3}\scR_1^4(\beta)+O(|u|^{-4}) \label{7.123}
\end{equation}
Moreover, using Propositions 5.1 and 7.1 as well as the results of
Chapters 3 and 4 we deduce:
\begin{equation}
\|\snab^{ \ 2}i\|_{L^2(S_{\ub,u})}\leq C\|\snab^{ \
2}\beta\|_{L^2(S_{\ub,u})}+O(|u|^{-4}) \label{7.124}
\end{equation}
Substituting \ref{7.122} - \ref{7.124} in \ref{7.121} we obtain:
\begin{eqnarray}
\|\snab^{ \ 3}\chih^\prime\|_{L^2(S_{\ub,u})}&\leq&
C\left\{\|\snab^{ \ 3}\mbox{tr}\chi^\prime\|_{L^2(S_{\ub,u})}
+\|\snab^{ \ 2}\beta\|_{L^2(S_{\ub,u})}\right\}\nonumber\\
&\s&+C\delta^{-1/2}|u|^{-1}l(u)\|\sd K\|_{L^2(S_{\ub,u})}\label{7.125}\\
&\s&+C\delta^{-1/2}|u|^{-3}(\scR_1^4(\beta)+{\cal
R}_0^\infty(\beta))+O(|u|^{-4})\nonumber
\end{eqnarray}
where:
\begin{equation}
l(u)=C(\scR_1^4(\beta)+{\cal
R}_0^\infty(\beta))+O(\delta^{1/2}|u|^{-1}) \label{7.126}
\end{equation}

We now revisit the propagation equation for the Gauss curvature,
equation \ref{5.28}. Applying $\sd$ to this equation we obtain, by
virtue of Lemma 1.2, the following propagation equation for $\sd
K$:
\begin{equation}
D\sd K+\Omega\mbox{tr}\chi\sd
K=-K\sd(\Omega\mbox{tr}\chi)+\sd\sdiv\sdiv(\Omega\chih)
-\frac{1}{2}\sd\slap(\Omega\mbox{tr}\chi) \label{7.127}
\end{equation}
We apply Lemma 4.6 to this equation, taking $p=2$. Here $r=1$,
$\nu=-2$, $\gamma=0$. We obtain:
\begin{eqnarray}
\|\sd
K\|_{L^2(S_{\ub,u})}&\leq&C\int_0^{\ub}\|K\sd(\Omega\mbox{tr}\chi)\|_{L^2(S_{\ub^\prime,u})}d\ub^\prime
\label{7.128}\\
&\s&+C\int_0^{\ub}\|\sd\sdiv\sdiv(\Omega\chih)-\frac{1}{2}\sd\slap(\Omega\mbox{tr}\chi)\|_{L^2(S_{\ub^\prime,u})}d\ub^\prime\nonumber
\end{eqnarray}
The integrant the first integral on the right in \ref{7.128} is
bounded by:
\begin{eqnarray}
\|K\sd(\Omega\mbox{tr}\chi)\|_{L^2(S)}&\leq&\overline{K}\|\sd(\Omega\mbox{tr}\chi)\|_{L^2(S)}
+\|K-\overline{K}\|_{L^4(S)}\|\sd(\Omega\mbox{tr}\chi)\|_{L^4(S)}\nonumber\\
&\leq&O(|u|^{-4}) \label{7.129}
\end{eqnarray}
by the first estimate of Proposition 6.1 and the results of
Chapters 3 and 4. The integrant of the second integral on the
right in \ref{7.128} is bounded by:
\begin{eqnarray*}
&&C(\|\snab^{ \ 3}\mbox{tr}\chi^\prime\|_{L^2(S)}+\|\snab^{ \ 3}\chih^\prime\|_{L^2(S)})\\
&&\hspace{1cm}+C\{\|\sd\log\Omega\|_{L^\infty(S)}(\|\snab^{ \
2}\mbox{tr}\chi^\prime\|_{L^2(S)}
+\|\snab^{ \ 2}\chih^\prime\|_{L^2(S)})\\
&&\hspace{2cm}+\|\snab^{ \
2}\log\Omega\|_{L^4(S)}(\|\sd\mbox{tr}\chi^\prime\|_{L^4(S)}
+\|\snab\chih^\prime\|_{L^4(S)})\\
&&\hspace{2cm}+\|\snab^{ \
3}\log\Omega\|_{L^2(S)}(\|\mbox{tr}\chi^\prime\|_{L^\infty(S)}
+\|\chih^\prime\|_{L^\infty(S)})\}
\end{eqnarray*}
Using the estimates of Proposition 6.1 and Lemma 6.4 and writing
(see \ref{5.17}):
\begin{equation}
\snab^{ \ 3}\log\Omega=\frac{1}{2}(\snab^{ \ 2}\eta+\snab^{ \
2}\etb) \label{7.a6}
\end{equation}
the lower order terms are seen to be bounded by:
$$C\delta^{-1/2}|u|^{-1}({\cal R}_0^\infty(\alpha)+O(\delta^{1/2}))(\|\snab^{ \ 2}\eta\|_{L^2(S)}
+\|\snab^{ \ 2}\etb\|_{L^2(S)})+O(\delta^{1/2}|u|^{-5})$$
Substituting the above in \ref{7.128} yields:
\begin{eqnarray}
\|\sd K\|_{L^2(S_{\ub,u})}&\leq&C\int_0^{\ub}\left\{\|\snab^{ \
3}\mbox{tr}\chi^\prime\|_{L^2(S_{\ub^\prime,u})}
+\|\snab^{ \ 3}\chih^\prime\|_{L^2(S_{\ub^\prime,u})}\right\}\nonumber\\
&\s&+C|u|^{-1}({\cal R}_0^\infty(\alpha)+O(\delta^{1/2}))(\|\snab^{ \ 2}\eta\|_{L^2(C_u)}+\|\snab^{ \ 2}\etb\|_{L^2(C_u)})\nonumber\\
&\s&+O(\delta|u|^{-4})\label{7.130}
\end{eqnarray}
Substituting the estimates of Proposition 7.1 this simplifies to:
\begin{eqnarray}
\|\sd K\|_{L^2(S_{\ub,u})}&\leq&C\int_0^{\ub}\left\{\|\snab^{ \
3}\mbox{tr}\chi^\prime\|_{L^2(S_{\ub^\prime,u})}
+\|\snab^{ \ 3}\chih^\prime\|_{L^2(S_{\ub^\prime,u})}\right\}\nonumber\\
&\s&+O(\delta|u|^{-4})\label{7.131}
\end{eqnarray}

Substituting \ref{7.131} in \ref{7.125} yields a linear integral
inequality, for fixed $u$, for the quantity:
\begin{equation}
x(\ub)=\|\snab^{ \ 3}\chih^\prime\|_{L^2(S_{\ub,u})} \label{7.132}
\end{equation}
of the form:
\begin{equation}
x(\ub)\leq a\int_0^{\ub}x(\ub^\prime)d\ub^\prime+b(\ub)
\label{7.133}
\end{equation}
Here $a$ is the non-negative constant:
\begin{equation}
a=C\delta^{-1/2}|u|^{-1}l(u) \label{7.134}
\end{equation}
and $b(\ub)$ is the non-negative function:
\begin{eqnarray}
b(\ub)&=&C\|\snab^{ \ 3}\mbox{tr}\chi^\prime\|_{L^2(S_{\ub,u})}
+C\delta^{-1/2}|u|^{-1}l(u)\int_0^{\ub}\|\snab^{ \
3}\mbox{tr}\chi^\prime\|_{L^2(S_{\ub^\prime,u})}d\ub^\prime
\nonumber\\
&\s&+C\|\snab^{ \ 2}\beta\|_{L^2(S_{\ub,u})}
+C\delta^{-1/2}|u|^{-3}(\scR_1^4(\beta)+{\cal R}_0^\infty(\beta))+O(|u|^{-4})\nonumber\\
&\s&\label{7.135}
\end{eqnarray}
Since
\begin{equation}
a\delta=C\delta^{1/2}|u|^{-1}l(u)\leq\log 2 \label{7.136}
\end{equation}
provided that $\delta$ is suitably small depending on ${\cal
D}_0^\infty$, ${\cal R}_0^\infty$, $\scD_1^4$, $\scR_1^4$, we
deduce, following the same argument as that leading from
\ref{6.21} to \ref{6.26},
\begin{equation}
x(\ub)\leq 2a\int_0^{\ub}b(\ub^\prime)d\ub^\prime+b(\ub)
\label{7.137}
\end{equation}
From \ref{7.135} and \ref{7.136} (see \ref{6.a1}),
\begin{eqnarray}
\int_0^{\ub}b(\ub^\prime)d\ub^\prime&\leq&
C\int_0^{\ub}\|\snab^{ \ 3}\mbox{tr}\chi^\prime\|_{L^2(S_{\ub^\prime,u})}d\ub^\prime\label{7.138}\\
&\s&+C\delta^{1/2}|u|^{-3}(\scR_2(\beta)+\scR_1^4(\beta)+{\cal
R}_0^\infty(\beta))+O(\delta|u|^{-4})\nonumber
\end{eqnarray}
recalling the first of the definitions \ref{7.01}. Therefore,
substituting in \ref{7.137} we obtain:
\begin{eqnarray}
\|\snab^{ \ 3}\chih^\prime\|_{L^2(S_{\ub,u})}&\leq&C\|\snab^{ \ 3}\mbox{tr}\chi^\prime\|_{L^2(S_{\ub,u})}\nonumber\\
&\s&+C\delta^{-1/2}|u|^{-1}l(u)\int_0^{\ub}\|\snab^{ \
3}\mbox{tr}\chi^\prime\|_{L^2(S_{\ub^\prime,u})}d\ub^\prime
\nonumber\\
&\s&+C\|\snab^{ \ 2}\beta\|_{L^2(S_{\ub,u})}
+C\delta^{-1/2}|u|^{-3}(\scR_1^4(\beta)+{\cal R}_0^\infty(\beta))\nonumber\\
&\s&+O(|u|^{-4})\label{7.139}
\end{eqnarray}

We now turn to the propagation equation \ref{6.6}. We apply Lemma
4.1 to this equation to deduce the following propagation equation
for $\snab^{ \ 3}\mbox{tr}\chi^\prime$:
\begin{equation}
D\snab^{ \ 3}\mbox{tr}\chi^\prime+\Omega\mbox{tr}\chi\snab^{ \
3}\mbox{tr}\chi^\prime =-2\Omega(\chih,\snab^{ \
3}\chih^\prime)+e^\prime \label{7.140}
\end{equation}
where:
\begin{eqnarray}
e^\prime_{ABC}&=&-(D\sGamma)^D_{AB}(\snab^{ \
2}\mbox{tr}\chi^\prime)_{DC}
-(D\sGamma)^D_{AC}(\snab^{ \ 2}\mbox{tr}\chi^\prime)_{BD}\nonumber\\
&\s&-\sd_A(\Omega\mbox{tr}\chi)(\snab^{ \ 2}\mbox{tr}\chi^\prime)_{BC}-2\snab_A(\Omega\chih^{DE})\snab_B\snab_C\chih^\prime_{DE}\nonumber\\
&\s&+\snab_A e_{BC}\label{7.141}
\end{eqnarray}
and we denote:
\begin{equation}
(\chih,\snab^{ \
3}\chih^\prime)_{ABC}=\chih^{DE}\snab_A\snab_B\snab_C\chih^\prime_{DE}
\label{7.142}
\end{equation}
To estimate in $L^2(S)$ the first four terms on the right in
\ref{7.141} we place each of the two factors of each term in
$L^4(S)$ using Propositions 4.1 and 6.1. This gives a bound in
$L^2(S)$ for the sum of these terms by:
\begin{equation}
C\delta^{-1}|u|^{-4}\scR_1^4(\alpha)(\scR_1^4(\beta)+{\cal
R}_0^\infty(\beta))+O(\delta^{-1/2}|u|^{-5}) \label{7.143}
\end{equation}
The contribution to $\|\snab e\|_{L^2(S)}$ of the first three
terms in $e$ as given by \ref{6.7} is similarly bounded. Finally,
the contribution of the last term in $e$ is $\|\snab^{ \ 2}
r\|_{L^2(S)}$. Now $r$ is given by \ref{6.2}. Using the estimates
of Proposition 6.1 and Lemma 6.4 as well as the expression
\ref{7.a6} and the results of Chapters 3 and 4 we deduce:
\begin{eqnarray}
&&\|\snab^{ \ 2}r\|_{L^2(S_{\ub,u})}\leq\nonumber\\
&&\hspace{15mm}C\delta^{-1}|u|^{-2}[({\cal R}_0^\infty(\alpha))^2+O(\delta)]
(\|\snab^{ \ 2}\eta\|_{L^2(S_{\ub,u})}+\|\snab^{ \ 2}\etb\|_{L^2(S_{\ub,u})})\nonumber\\
&&\hspace{25mm}+O(|u|^{-6}) \label{7.144}
\end{eqnarray}
Combining the above results we obtain:
\begin{eqnarray}
\|e^\prime\|_{L^2(S_{\ub,u})}&\leq&C\delta^{-1}|u|^{-2}[({\cal
R}_0^\infty(\alpha))^2+O(\delta)]
(\|\snab^{ \ 2}\eta\|_{L^2(S_{\ub,u})}+\|\snab^{ \ 2}\etb\|_{L^2(S_{\ub,u})})\nonumber\\
&\s&+C\delta^{-1}|u|^{-4}\scR_1^4(\alpha)(\scR_1^4(\beta)+{\cal
R}_0^\infty(\beta))+O(\delta^{-1/2}|u|^{-5}) \label{7.145}
\end{eqnarray}
We also estimate in $L^2(S)$ the first term on the right in
\ref{7.140}:
\begin{eqnarray}
&&\|\Omega(\chih,\snab^{ \ 3}\chih^\prime)\|_{L^2(S_{\ub,u})}\leq\nonumber\\
&&\hspace{1cm}C\delta^{-1/2}|u|^{-1}{\cal R}_0^\infty(\alpha)
\left\{\|\snab^{ \ 3}\mbox{tr}\chi^\prime\|_{L^2(S_{\ub,u})}\right.\nonumber\\
&&\hspace{2cm}\left.+C\delta^{-1/2}|u|^{-1}l(u)
\int_0^{\ub}\|\snab^{ \
3}\mbox{tr}\chi^\prime\|_{L^2(S_{\ub^\prime,u})}d\ub^\prime
\right\}\nonumber\\
&&\hspace{1cm}+C\delta^{-1/2}|u|^{-1}{\cal
R}_0^\infty(\alpha)\|\snab^{ \ 2}\beta\|_{L^2(S_{\ub,u})}
\nonumber\\
&&\hspace{1cm}+C\delta^{-1}|u|^{-4}{\cal R}_0^\infty(\alpha)(\scR_1^4(\beta)+{\cal R}_0^\infty(\beta))+O(\delta^{-1/2}|u|^{-5})\label{7.146}
\end{eqnarray}
by the estimate \ref{7.139}.

We now apply Lemma 4.6 with $p=2$ to \ref{7.140}. Here $r=3$,
$\nu=-2$, $\gamma=0$. We obtain:
\begin{equation}
\|\snab^{ \ 3}\mbox{tr}\chi^\prime\|_{L^2(S_{\ub,u})}\leq
C\int_0^{\ub} \|-2\Omega(\chih,\snab^{ \
3}\chih^\prime)+e^\prime\|_{L^2(S_{\ub^\prime,u})}d\ub^\prime
\label{7.147}
\end{equation}
Substituting the bounds \ref{7.146}, \ref{7.145} then yields a
linear integral inequality for $\|\snab^{ \
3}\mbox{tr}\chi^\prime\|_{L^2(S)}$. Recalling \ref{6.a1} the
contribution of the term in parenthesis in \ref{7.146} involving
the integral is bounded by
\begin{equation}
C\delta^{1/2}|u|^{-1}l(u) \label{7.148}
\end{equation}
times the contribution of the first term in parenthesis in
\ref{7.146}, and the factor \ref{7.148} is less than or equal to 1
provided that $\delta$ is suitably small depending on ${\cal
D}_0^\infty$, ${\cal R}_0^\infty$,  $\scD_1^4$, $\scR_1^4$. Taking
also into account Proposition 7.1 and the first of the definitions
\ref{7.01}, noting that
$$\int_0^{\ub}\|\snab^{ \ 2}\etb\|_{L^2(S_{\ub^\prime})}d\ub^\prime\leq\delta^{1/2}\|\snab^{ \ 2}\etb\|_{L^2(C_u)},$$
$$\int_0^{\ub}\|\snab^{ \ 2}\beta\|_{L^2(S_{\ub^\prime,u})}d\ub^\prime
\leq\delta^{1/2}\|\snab^{ \ 2}\beta\|_{L^2(C_u)},$$ the resulting
integral inequality simplifies to:
\begin{eqnarray}
&&\|\snab^{ \ 3}\mbox{tr}\chi^\prime\|_{L^2(S_{\ub,u})}\leq
C\delta^{-1/2}{\cal R}_0^\infty(\alpha)
\int_0^{\ub}\|\snab^{ \ 3}\mbox{tr}\chi^\prime\|_{L^2(S_{\ub^\prime,u})}d\ub^\prime\nonumber\\
&&\hspace{1cm}+C|u|^{-4}[{\cal R}_0^\infty(\alpha)\scR_2(\beta)+
(\scR_1^4(\alpha)+{\cal R}_0^\infty(\alpha))(\scR_1^4(\beta)+{\cal R}_0^\infty(\beta))]\nonumber\\
&&\hspace{1cm}+O(\delta^{1/2}|u|^{-5})\label{7.149}
\end{eqnarray}
This is a linear integral inequality of the form \ref{6.18}, but
with $b$ a positive constant. Noting that
$$C\delta^{1/2}|u|^{-1}{\cal R}_0^\infty(\alpha)\leq 1$$
provided that $\delta$ is suitably small depending on ${\cal
R}_0^\infty(\alpha)$, the integral inequality \ref{7.149} implies:
\begin{eqnarray}
&&\|\snab^{ \ 3}\mbox{tr}\chi^\prime\|_{L^2(S_{\ub,u})}\leq\nonumber\\
&&\hspace{1cm}+C|u|^{-4}[{\cal R}_0^\infty(\alpha)\scR_2(\beta)+
(\scR_1^4(\alpha)+{\cal R}_0^\infty(\alpha))(\scR_1^4(\beta)+{\cal R}_0^\infty(\beta))]\nonumber\\
&&\hspace{1cm}+O(\delta^{1/2}|u|^{-5})\label{7.150}
\end{eqnarray}
This is the second conclusion of the proposition. Substituting
this in \ref{7.139} and taking the $L^2$ norm with respect to
$\ub$ on $[0,\delta)$ if $u\in[u_0,c^*-\delta)$, on $[0,c^*-u)$ if
$u\in(c^*-\delta,c^*)$, in the case $c^*\geq u_0+\delta$, and on $[0,c^*-u)$, in the case $c^*<u_0+\delta$ (see \ref{5.7}, \ref{5.7a}), the third conclusion
follows. Finally, substituting the second and third conclusions in
\ref{7.131} and noting that
$$\int_0^{\ub}\|\snab^{ \ 3}\chih^\prime\|_{L^2(S_{\ub^\prime,u})}d\ub^\prime\leq
\delta^{1/2}\|\snab^{ \ 3}\chih^\prime\|_{L^2(C_u)},$$ yields the
first conclusion of the proposition.

\vspace{5mm}

\section{$L^2$ estimates for $\snab^{ \ 3}\chib^\prime$}

We now consider any $(\ub_1,u_1)\in D^\prime$ and fix attention in
the following lemmas to the parameter subdomain $D_1$ and the
corresponding subdomain $M_1$ of $M^\prime$ (see \ref{3.02},
\ref{3.03}).

With a positive constant $k$ to be appropriately chosen in the
sequel, let $s^*$ be the least upper bound of the set of values of
$s\in[u_0,u_1]$ such that:
\begin{equation}
\|\snab^{ \ 3}\log\Omega\|_{L^2(C_u^{\ub_1})}\leq
k\delta^{3/2}|u|^{-4}\ \ \ : \ \mbox{for all $u\in[u_0,s]$}
\label{7.151}
\end{equation}
Then by continuity $s^*>u_0$ (recall that by \ref{3.34} $\snab^{ \
3}\log\Omega$ vanishes along $C_{u_0}$) and we have:
\begin{equation}
\|\snab^{ \ 3}\log\Omega\|_{L^2(C_u^{\ub_1})}\leq
k\delta^{3/2}|u|^{-4}\ \ \ : \ \mbox{for all $u\in[u_0,s^*]$}
\label{7.152}
\end{equation}
In the estimates to follow the dependence on the constant $k$ is
made explicit.

Let us denote:
\begin{equation}
\scD_3(\mbox{tr}\chib)=|u_0|^5\delta^{-3/2}\|\snab^{ \
3}\mbox{tr}\chib\|_{L^2(C_{u_0})} \label{7.153}
\end{equation}
Here, as in \ref{3.37}, \ref{3.108}, \ref{4.80}, \ref{4.81}, and \ref{6.79}, we are considering all of $C_{u_0}$, not only the part which lies in 
$M^\prime$.
By the results of Chapter 2 this quantity is bounded by a non-negative non-decreasing continous function of $M_6$. 
Let us denote by $D_1^{s^*}$ the subset of $D_1$ where
$u\leq s^*$:
\begin{equation}
D_1^{s^*}=[0,\ub_1]\times[u_0,s^*] \label{7.154}
\end{equation}
and by $M_1^{s^*}$ the corresponding subset of $M_1$.

\vspace{5mm}

\noindent{\bf Lemma 7.5} \ \ \ We have:
\begin{eqnarray*}
&&\|\snab^{ \ 3}\mbox{tr}\chib^\prime\|_{L^2(C_u^{\ub_1})}\leq
C\delta^{3/2}|u|^{-5}\left\{\scD_3(\mbox{tr}\chib)\right.\\
&&\hspace{1cm}\left.+(\scD_1^4(\mbox{tr}\chib)+\scD_1^4(\chibh)+{\cal
D}_0^\infty(\chibh))
(\scR_1^4(\beta)+{\cal R}_0^\infty(\beta))+{\cal D}_0^\infty(\chibh)\scR_2(\beta)+k\right\}\\
&&\hspace{3cm}+O(\delta^2|u|^{-6})
\end{eqnarray*}
\begin{eqnarray*}
&&\|\snab^{ \ 3}\chibh^\prime\|_{L^2(C_u^{\ub_1})}\leq
C\delta|u|^{-4}(\scR_2(\beta)+\scR_1^4(\beta)+{\cal R}_0^\infty(\beta))\\
&&\hspace{3cm}+C\delta^{3/2}|u|^{-5}k+O(\delta^{3/2}|u|^{-5})
\end{eqnarray*}
for all $u\in[u_0,s^*]$, provided that $\delta$ is suitably small
depending on ${\cal D}_0^\infty$, ${\cal R}_0^\infty$, $\scD_1^4$,
$\scR_1^4$, and $\scR_2$.

\noindent{\em Proof:} \ We apply Lemma 7.3 with $\chibh^\prime$ in
the role of $\theta$ to the Codazzi equation \ref{6.57}. In view
of Lemma 5.4 and the last conclusion of Proposition 7.2 the
conclusion of Lemma 7.3 simplifies and we obtain:
\begin{eqnarray}
&&\|\snab^{ \ 3}\chibh^\prime\|_{L^2(S_{\ub,u})}\leq
C\left\{\|\snab^{ \ 3}\mbox{tr}\chib^\prime\|_{L^2(S_{\ub,u})}
+\|\snab^{ \ 2}\ib\|_{L^2(S_{\ub,u})}\right.\label{7.155}\\
&&\hspace{2cm}\left.+|u|^{-1/2}(\|\snab^{ \
2}\mbox{tr}\chib^\prime\|_{L^4(S_{\ub,u})}
+\|\snab \ib\|_{L^4(S_{\ub,u})})\right\}\nonumber\\
&&\hspace{1cm}+C|u|^{-3/2}\left\{\|\snab\chibh^\prime\|_{L^4(S_{\ub,u})}
+|u|^{-1}\|\chibh^\prime\|_{L^4(S_{\ub,u})}\right\}\nonumber
\end{eqnarray}
Now, by the estimate \ref{6.91} (and Lemma 6.4),
\begin{equation}
\|\snab\chibh^\prime\|_{L^4(S_{\ub,u})}+|u|^{-1}\|\chibh^\prime\|_{L^4(S_{\ub,u})}\leq
C\delta^{1/2}|u|^{-5/2} (\scR_1^4(\beta)+{\cal
R}_0^\infty(\beta))+O(\delta|u|^{-7/2}) \label{7.156}
\end{equation}
Also, by the estimates \ref{6.60}, \ref{6.90} (and Lemma 6.4),
\begin{equation}
|u|^{-1/2}(\|\snab^{ \
2}\mbox{tr}\chib^\prime\|_{L^4(S_{\ub,u})}+\|\snab
\ib\|_{L^4(S_{\ub,u})}) \leq
C\delta^{1/2}|u|^{-4}\scR_1^4(\beta)+O(\delta|u|^{-5})
\label{7.157}
\end{equation}
Moreover, using the estimates for $\snab^{ \ 2}\mbox{tr}\chib$,
$\snab^{ \ 2}\chibh$ of Proposition 6.2 as well as the  results of
Chapters 3 and 4 we deduce:
\begin{eqnarray}
\|\snab^{ \ 2}\ib\|_{L^2(S_{\ub,u})}&\leq&C\|\snab^{ \
2}\beb\|_{L^2(S_{\ub,u})}
+C|u|^{-1}\|\snab^{ \ 2}\etb\|_{L^2(S_{\ub,u})}\nonumber\\
&\s&+O(\delta|u|^{-5})\label{7.158}
\end{eqnarray}
Substituting \ref{7.156} - \ref{7.158} in \ref{7.155} we obtain:
\begin{eqnarray}
\|\snab^{ \
3}\chibh^\prime\|_{L^2(S_{\ub,u})}&\leq&C\left\{\|\snab^{ \
3}\mbox{tr}\chib^\prime\|_{L^2(S_{\ub,u})}\right.
\nonumber\\
&\s&\left.+\|\snab^{ \ 2}\beb\|_{L^2(S_{\ub,u})}+|u|^{-1}\|\snab^{ \ 2}\etb\|_{L^2(S_{\ub,u})}\right\}\nonumber\\
&\s&+C\delta^{1/2}|u|^{-4}(\scR_1^4(\beta)+{\cal
R}_0^\infty(\beta))+O(\delta|u|^{-5})\label{7.159}
\end{eqnarray}

We now turn to the propagation equation \ref{6.63}. We apply Lemma
4.1 to this equation to deduce the following propagation equation
for $\snab^{ \ 3}\mbox{tr}\chib^\prime$:
\begin{equation}
\Db\snab^{ \ 3}\mbox{tr}\chib^\prime+\Omega\mbox{tr}\chib\snab^{ \
3}\mbox{tr}\chib^\prime= -2\Omega(\chibh,\snab^{ \
3}\chibh^\prime)+\eb^\prime \label{7.160}
\end{equation}
where:
\begin{eqnarray}
e^\prime_{ABC}&=&-(\Db\sGamma)^D_{AB}(\snab^{ \
2}\mbox{tr}\chib^\prime)_{DC}
-(\Db\sGamma)^D_{AC}(\snab^{ \ 2}\mbox{tr}\chib^\prime)_{BD}\nonumber\\
&\s&-\sd_A(\Omega\mbox{tr}\chib)(\snab^{ \
2}\mbox{tr}\chib^\prime)_{BC}
-2\snab_A(\Omega\chibh^{DE})\snab_B\snab_C\chibh^\prime_{DE}\nonumber\\
&\s&+\snab_A\eb_{BC}\label{7.161}
\end{eqnarray}
and we denote:
\begin{equation}
(\chibh,\snab^{ \
3}\chibh^\prime)_{ABC}=\chibh^{DE}\snab_A\snab_B\snab_C\chibh^\prime_{DE}
\label{7.162}
\end{equation}
To estimate in $L^2(S)$ the first four terms on the right in
\ref{7.161} we place each of the two factors of each term in
$L^4(S)$ using the estimates of Propositions 4.2 and 6.2
pertaining to $\chib$. This gives a bound in $L^2(S)$ for the sum
of these terms by:
\begin{equation}
C\delta|u|^{-6}(\scD_1^4(\mbox{tr}\chib)+\scD_1^4(\chibh))(\scR_1^4(\beta)+{\cal
R}_0^\infty(\beta))+O(\delta^{3/2}|u|^{-7}) \label{7.163}
\end{equation}
The contribution to $\|\snab\eb\|_{L^2(S)}$ of the first three
terms in $\eb$ as given by \ref{6.64} is similarly bounded.
Finally, the contribution of the last term in $\eb$ is $\|\snab^{
\ 2}\rb\|_{L^2(S)}$. Now $\rb$ is given by \ref{6.55}. Using the
estimate of Lemma 6.4 and the estimates of Proposition 6.2
pertaining to $\chib$ as well as the results of Chapters 3 and 4
we deduce:
\begin{equation}
\|\snab^{ \ 2}\rb\|_{L^2(S_{\ub,u})}\leq C|u|^{-2}\|\snab^{ \
3}\log\Omega\|_{L^2(S_{\ub,u})}+O(\delta^2|u|^{-8}) \label{7.164}
\end{equation}
Combining the above results we obtain:
\begin{eqnarray}
\|\eb^\prime\|_{L^2(S_{\ub,u})}&\leq&C|u|^{-2}\|\snab^{ \ 3}\log\Omega\|_{L^2(S_{\ub,u})}\nonumber\\
&\s&+C\delta|u|^{-6}(\scD_1^4(\mbox{tr}\chib)+\scD_1^4(\chibh))(\scR_1^4(\beta)+{\cal R}_0^\infty(\beta))\nonumber\\
&\s&+O(\delta^{3/2}|u|^{-7})\label{7.165}
\end{eqnarray}
We also estimate in $L^2(S)$ the first term on the right in
\ref{7.160}:
\begin{eqnarray}
&&\|\Omega(\chibh,\snab^{ \ 3}\chibh^\prime)\|_{L^2(S_{\ub,u})}\leq\nonumber\\
&&\hspace{1cm}C\delta^{1/2}|u|^{-2}({\cal
D}_0^\infty(\chibh)+O(\delta|u|^{-3/2}))
\left\{\|\snab^{ \ 3}\mbox{tr}\chib^\prime\|_{L^2(S_{\ub,u})}\right.\nonumber\\
&&\hspace{2cm}\left.+\|\snab^{ \ 2}\beb\|_{L^2(S_{\ub,u})}+|u|^{-1}\|\snab^{ \ 2}\etb\|_{L^2(S_{\ub,u})}\right\}\nonumber\\
&&\hspace{15mm}+C\delta|u|^{-6}{\cal
D}_0^\infty(\chibh)(\scR_1^4(\beta)+{\cal R}_0^\infty(\beta))
+O(\delta^{3/2}|u|^{-7})\label{7.166}
\end{eqnarray}

We now apply Lemma 4.7 with $p=2$ to \ref{7.160}. Here $r=3$,
$\nu=-2$, $\gammab=0$. We obtain:
\begin{eqnarray}
|u|^4\|\snab^{ \ 3}\mbox{tr}\chib^\prime\|_{L^2(S_{\ub,u})}&\leq&
C|u_0|^4\|\snab^{ \ 3}\mbox{tr}\chib^\prime\|_{L^2(S_{\ub,u_0})}\label{7.167}\\
&\s&+C\int_{u_0}^u|u^\prime|^4\|-2\Omega(\chibh,\snab^{ \
3}\chibh^\prime)+\eb^\prime\|_{L^2(S_{\ub,u^\prime})}du^\prime
\nonumber
\end{eqnarray}
Substituting the bounds \ref{7.165}, \ref{7.166} yields a linear
integral inequality, for fixed $\ub$, for the quantity:
\begin{equation}
\xb(\ub,u)=|u|^4\|\snab^{ \
3}\mbox{tr}\chib^\prime\|_{L^2(S_{\ub,u})} \label{7.168}
\end{equation}
of the form \ref{6.81}, that is:
$$\xb(\ub,u)\leq\int_{u_0}^u\ab(u^\prime)\xb(\ub,u^\prime)du^\prime+\bb(\ub,u)$$
with $\ab$ the non-negative function:
\begin{equation}
\ab(u)=C\delta^{1/2}|u|^{-2}({\cal
D}_0^\infty(\chib)+O(\delta|u|^{-3/2}) \label{7.169}
\end{equation}
and $\bb(\ub)$ the non-negative non-decreasing function, depending
on $\ub$,
\begin{eqnarray}
&&\bb(\ub,u)=C|u_0|^4\|\snab^{ \ 3}\mbox{tr}\chib\|_{L^2(S_{\ub,u_0})}\nonumber\\
&&\hspace{1cm}+C\delta|u|^{-1}(\scD_1^4(\mbox{tr}\chib)+\scD_1^4(\chibh)+{\cal
D}_0^\infty(\chibh))
(\scR_1^4(\beta)+{\cal R}_0^\infty(\beta))\nonumber\\
&&\hspace{1cm}+C\delta^{1/2}\int_{u_0}^u|u^\prime|^2({\cal D}_0^\infty(\chibh)+O(\delta|u^\prime|^{-3/2}))\cdot\nonumber\\
&&\hspace{3cm}\cdot\left\{\|\snab^{ \
2}\beb\|_{L^2(S_{\ub,u^\prime})}+|u^\prime|^{-1}\|\snab^{ \
2}\etb\|_{L^2(S_{\ub,u^\prime})}
\right\}du^\prime\nonumber\\
&&\hspace{1cm}+C\int_{u_0}^u|u^\prime|^2\|\snab^{ \
3}\log\Omega\|_{L^2(S_{\ub,u^\prime})}du^\prime
+O(\delta^{3/2}|u|^{-2})\label{7.170}
\end{eqnarray}
Since
\begin{equation}
\int_{u_0}^u\ab(u^\prime)du^\prime\leq\log 2 \label{7.171}
\end{equation}
provided that $\delta$ is suitably small depending on ${\cal
D}_0^\infty$, ${\cal R}_0^\infty$, we deduce, following the same
argument as that leading from \ref{6.81} to \ref{6.89},
\begin{equation}
\xb(\ub,u)\leq 2\bb(\ub,u) \label{7.172}
\end{equation}
We take the $L^2$ norm with respect to $\ub$ on $[0,\ub_1]$ of
this inequality to obtain:
\begin{equation}
\|\xb(\cdot,u)\|_{L^2([0,\ub_1])}\leq
2\|\bb(\cdot,u)\|_{L^2([0,\ub_1])} \label{7.173}
\end{equation}
Since for any $S$ tensorfield $\theta$ we have:
\begin{equation}
\|\|\theta\|_{L^2(S_{\cdot,u})}\|_{L^2([0,\ub_1])}=\left(\int_0^{\ub_1}\|\theta\|^2_{L^2(S_{\ub,u})}d\ub\right)^{1/2}
=\|\theta\|_{L^2(C_u^{\ub_1})} \label{7.174}
\end{equation}
the inequality \ref{7.173}, implies, in view of the definitions
\ref{7.168}, \ref{7.170},
\begin{eqnarray}
&&|u|^4\|\snab^{ \
3}\mbox{tr}\chib^\prime\|_{L^2(C_u^{\ub_1})}\leq
C|u_0|^4\|\snab^{ \ 3}\mbox{tr}\chib\|_{L^2(C_{u_0}^{\ub_1})}\nonumber\\
&&\hspace{1cm}+C\delta^{3/2}|u|^{-1}(\scD_1^4(\mbox{tr}\chib)+\scD_1^4(\chibh)+{\cal
D}_0^\infty(\chibh))
(\scR_1^4(\beta)+{\cal R}_0^\infty(\beta))\nonumber\\
&&\hspace{1cm}+C\delta^{1/2}\int_{u_0}^u|u^\prime|^2({\cal D}_0^\infty(\chibh)+O(\delta|u^\prime|^{-3/2}))\cdot\nonumber\\
&&\hspace{3cm}\cdot\left\{\|\snab^{ \
2}\beb\|_{L^2(C_{u^\prime}^{\ub_1})}+|u^\prime|^{-1}\|\snab^{ \
2}\etb\|_{L^2(C_{u^\prime}^{\ub_1})}
\right\}du^\prime\nonumber\\
&&\hspace{1cm}+C\int_{u_0}^u|u^\prime|^2\|\snab^{ \
3}\log\Omega\|_{L^2(C_{u^\prime}^{\ub_1})}du^\prime
+O(\delta^2|u|^{-2})\label{7.175}
\end{eqnarray}
By the second conclusion of Proposition 7.1:
\begin{equation}
\|\snab^{ \ 2}\etb\|_{L^2(C_u^{\ub_1})}\leq\|\snab^{ \
2}\etb\|_{L^2(C_u)}\leq
C\delta|u|^{-3}\scR_2(\beta)+O(\delta^{3/2}|u|^{-4}) \label{7.176}
\end{equation}
By the fourth of the definitions \ref{7.01}:
\begin{equation}
\|\snab^{ \ 2}\beb\|_{L^2(C_u^{\ub_1})}\leq\|\snab^{ \
2}\beb\|_{L^2(C_u)}\leq \delta^{3/2}|u|^{-5}\scR_2(\beb)
\label{7.177}
\end{equation}
Moreover, by the definition \ref{7.153}:
\begin{equation}
\|\snab^{ \ 3}\mbox{tr}\chib\|_{L^2(C_{u_0}^{\ub_1})}\leq\|\snab^{
\ 3}\mbox{tr}\chib\|_{L^2(C_{u_0})} \leq
\delta^{3/2}|u_0|^{-5}\scD_3(\mbox{tr}\chib) \label{7.178}
\end{equation}
Substituting \ref{7.176} - \ref{7.178} in \ref{7.175} and taking
into account \ref{7.152} we conclude that:
\begin{eqnarray}
&&|u|^4\|\snab^{ \
3}\mbox{tr}\chib^\prime\|_{L^2(C_u^{\ub_1})}\leq
C\delta^{3/2}|u_0|^{-1}\scD_3(\mbox{tr}\chib)\nonumber\\
&&\hspace{1cm}+C\delta^{3/2}|u|^{-1}\left[(\scD_1^4(\mbox{tr}\chib)+\scD_1^4(\chibh)+{\cal
D}_0^\infty(\chibh))
(\scR_1^4(\beta)+{\cal R}_0^\infty(\beta))\right.\nonumber\\
&&\hspace{4cm}\left.+{\cal D}_0^\infty(\chibh)\scR_2(\beta)+k\right]\nonumber\\
&&\hspace{4cm}+O(\delta^2|u|^{-2})\label{7.179}
\end{eqnarray}
for all $u\in[u_0,s^*]$. This yields the first conclusion of the
lemma.

We take the $L^2$ norm of the inequality \ref{7.159} with respect
to $\ub$ on $[0,\ub_1]$ to obtain:
\begin{eqnarray}
\|\snab^{ \
3}\chibh^\prime\|_{L^2(C_u^{\ub_1})}&\leq&C\left\{\|\snab^{ \
3}\mbox{tr}\chib^\prime\|_{L^2(C_u^{\ub_1})}
\right.\nonumber\\
&\s&\left.+\|\snab^2\beb\|_{L^2(C_u^{\ub_1})}+|u|^{-1}\|\snab^{ \ 2}\etb\|_{L^2(C_u^{\ub_1})}\right\}\nonumber\\
&\s&+C\delta|u|^{-4}(\scR_1^4(\beta)+{\cal
R}_0^\infty(\beta))+O(\delta^{3/2}|u|^{-5}) \label{7.180}
\end{eqnarray}
Substituting \ref{7.176}, \ref{7.177} and the estimate \ref{7.179}
then yields the second conclusion of the lemma.

\vspace{5mm}

\section{$L^2$ estimates for $\snab^{ \ 3}\eta$, $\snab^{ \ 3}\etb$}

In the following we denote by $O(\delta^p|u|^r)$, for real numbers
$p$, $r$, the product of $\delta^p |u|^r$ with a non-negative
non-decreasing continuous function of the quantities ${\cal
D}_0^\infty$, ${\cal R}_0^\infty$, $\scD_1^4$, $\scR_1^4$,
$\scD_2^4(\mbox{tr}\chib)$, $\scR_2$, {\em and}
$\scD_3(\mbox{tr}\chib)$.

\vspace{5mm}

\noindent{\bf Lemma 7.6} \ \ \ We have:
\begin{eqnarray*}
&&\|\snab^{ \ 3}\eta\|_{L^2(C^{\ub_1}_u)}\leq
C\delta^{1/2}|u|^{-4}\tilde{M}+O(\delta|u|^{-4})
+O(\delta^3|u|^{-7})k\\
&&\|\snab^{ \ 3}\etb\|_{L^2(C^{\ub_1}_u)}\leq
C\delta^{1/2}|u|^{-4}\tilde{\Mb}+O(\delta|u|^{-4})
+O(\delta^2|u|^{-6})k
\end{eqnarray*}
for all $u\in[u_0,s^*]$, provided that $\delta$ is suitably small
depending on ${\cal D}_0^\infty$, ${\cal R}_0^\infty$, $\scD_1^4$,
$\scR_1^4$, and $\scR_2$. Moreover, the coefficients of $k$ depend
only on ${\cal D}_0^\infty$, ${\cal R}_0^\infty$. Here,
\begin{eqnarray*}
\tilde{M}&=&\scR_2(\rho)+\scR_2(\sigma)+\scR_1^4(\rho)+\scR_1^4(\sigma)\\
&\s&+({\cal R}_0^\infty(\alpha)+\scR_1^4(\beta)+{\cal
R}_0^\infty(\beta))
(\scR_1^4(\beta)+{\cal R}_0^\infty(\beta))\\
&\s&+(\scR_1^4(\alpha)+{\cal R}_0^\infty(\alpha))\scR_1^4(\alpha)\\
\tilde{\Mb}&=&\tilde{M}+\tilde{A}
\end{eqnarray*}
and $\tilde{A}$ is given by \ref{6.190}.

\noindent{\em Proof:} \ We apply Lemma 7.4 with $\eta$ and $\etb$
in the role of $\theta$ to the Hodge systems \ref{6.96} and
\ref{6.97} respectively. The assumptions of Lemma 7.4 are
satisfied by virtue of the first estimate of Proposition 6.1
provided that $\delta$ is suitably small depending on ${\cal
D}_0^\infty$, ${\cal R}_0^\infty$, $\scD_1^4$, $\scR_1^4$ and
$\scR_2(\alpha)$. Moreover, in view of Lemma 5.4 and the last
conclusion of Proposition 7.2 the conclusion of Lemma 7.4
simplifies and we obtain:
\begin{eqnarray}
&&\|\snab^{ \ 3}\eta\|_{L^2(S_{\ub,u})}\leq C\left\{\|\snab^{ \
2}\rho\|_{L^2(S_{\ub,u})}
+\|\snab^{ \ 2}\sigma\|_{L^2(S_{\ub,u})}+\|\snab^{ \ 2}\mu\|_{L^2(S_{\ub,u})}\right.\nonumber\\
&&\hspace{3cm}\left.+\|\snab^{ \
2}(\chih,\chibh)\|_{L^2(S_{\ub,u})}+\|\snab^{ \
2}(\chih\wedge\chibh)\|_{L^2(S_{\ub,u})}
\right\}\nonumber\\
&&\hspace{2cm}+C|u|^{-1/2}\left\{\|\sd\rho\|_{L^4(S_{\ub,u})}+\|\sd\sigma\|_{L^4(S_{\ub,u})}+\|\sd\mu\|_{L^4(S_{\ub,u})}
\right.\nonumber\\
&&\hspace{3cm}\left.+\|\sd(\chih,\chibh)\|_{L^4(S_{\ub,u})}+\|\sd(\chih\wedge\chibh)\|_{L^4(S_{\ub,u})}\right\}\nonumber\\
&&\hspace{3cm}+C|u|^{-3/2}\left\{\|\snab\eta\|_{L^4(S_{\ub,u})}+|u|^{-1}\|\eta\|_{L^4(S_{\ub,u})}\right\}
\label{7.181}
\end{eqnarray}
and:
\begin{eqnarray}
&&\|\snab^{ \ 3}\etb\|_{L^2(S_{\ub,u})}\leq C\left\{\|\snab^{ \
2}\rho\|_{L^2(S_{\ub,u})}
+\|\snab^{ \ 2}\sigma\|_{L^2(S_{\ub,u})}+\|\snab^{ \ 2}\mub\|_{L^2(S_{\ub,u})}\right.\nonumber\\
&&\hspace{3cm}\left.+\|\snab^{ \
2}(\chih,\chibh)\|_{L^2(S_{\ub,u})}+\|\snab^{ \
2}(\chih\wedge\chibh)\|_{L^2(S_{\ub,u})}
\right\}\nonumber\\
&&\hspace{2cm}+C|u|^{-1/2}\left\{\|\sd\rho\|_{L^4(S_{\ub,u})}+\|\sd\sigma\|_{L^4(S_{\ub,u})}+\|\sd\mub\|_{L^4(S_{\ub,u})}
\right.\nonumber\\
&&\hspace{3cm}\left.+\|\sd(\chih,\chibh)\|_{L^4(S_{\ub,u})}+\|\sd(\chih\wedge\chibh)\|_{L^4(S_{\ub,u})}\right\}\nonumber\\
&&\hspace{3cm}+C|u|^{-3/2}\left\{\|\snab\etb\|_{L^4(S_{\ub,u})}+|u|^{-1}\|\etb\|_{L^4(S_{\ub,u})}\right\}
\label{7.182}
\end{eqnarray}
Now, by the results of Chapters 3 and 4,
\begin{equation}
\|\snab\eta\|_{L^4(S_{\ub,u})}+|u|^{-1}\|\eta\|_{L^4(S_{\ub,u})}\leq
C\delta^{1/2}|u|^{-5/2} (\scR_1^4(\beta)+{\cal
R}_0^\infty(\beta))+O(\delta|u|^{-7/2}) \label{7.183}
\end{equation}
and:
\begin{equation}
\|\snab\etb\|_{L^4(S_{\ub,u})}+|u|^{-1}\|\etb\|_{L^4(S_{\ub,u})}\leq
C\delta^{1/2}|u|^{-5/2} (\scR_1^4(\beta)+{\cal
R}_0^\infty(\beta))+O(\delta|u|^{-7/2}) \label{7.184}
\end{equation}
Also, recalling the definitions \ref{6.164}, \ref{6.165}, by the
estimates \ref{6.191}, \ref{6.192} and Lemma 6.4 we have:
\begin{equation}
\|\sd\mu\|_{L^4(S_{\ub,u})}\leq C|u|^{-7/2}{\cal
R}_0^\infty(\alpha)\scR_1^4(\alpha)+O(\delta^{1/2}|u|^{-9/2})
\label{7.185}
\end{equation}
and:
\begin{equation}
\|\sd\mub\|_{L^4(S_{\ub,u})}\leq
C|u|^{-7/2}\tilde{A}+O(\delta^{1/2}|u|^{-7/2}) \label{7.186}
\end{equation}
where $\tilde{A}$ is given by \ref{6.190}. Moreover, using the
estimates of Propositions 6.1 and 6.2 pertaining to $\chih$ and
$\chibh$ we deduce:
\begin{eqnarray}
&&\|\snab^{ \ 2}(\chih,\chibh)\|_{L^2(S_{\ub,u})}+\|\snab^{ \ 2}(\chih\wedge\chibh)\|_{L^2(S_{\ub,u})}\nonumber\\
&&\hspace{2cm}\leq C|u|^{-4}(\scR_1^4(\beta)+{\cal
R}_0^\infty(\beta))^2+O(\delta^{1/2}|u|^{-5}) \label{7.187}
\end{eqnarray}
and (using also Lemma 5.2):
\begin{eqnarray}
&&\|\sd(\chih,\chibh)\|_{L^4(S_{\ub,u})}+\|\sd(\chih\wedge\chibh)\|_{L^4(S_{\ub,u})}\nonumber\\
&&\hspace{2cm}\leq C|u|^{-7/2}(\scR_1^4(\beta)+{\cal
R}_0^\infty(\beta))^2+O(\delta^{1/2}|u|^{-5}) \label{7.188}
\end{eqnarray}
Substituting the above in \ref{7.181}, \ref{7.182} we obtain:
\begin{eqnarray}
&&\|\snab^{ \ 3}\eta\|_{L^2(S_{\ub,u})}\leq C\left\{\|\snab^{ \
2}\mu\|_{L^2(S_{\ub,u})}
+\|\snab^{ \ 2}\rho\|_{L^2(S_{\ub,u})}+\|\snab^{ \ 2}\sigma\|_{L^2(S_{\ub,u})}\right\}\nonumber\\
&&\hspace{3cm}+C|u|^{-4}M+O(\delta^{1/2}|u|^{-4}) \label{7.189}
\end{eqnarray}
and:
\begin{eqnarray}
&&\|\snab^{ \ 3}\etb\|_{L^2(S_{\ub,u})}\leq C\left\{\|\snab^{ \
2}\mub\|_{L^2(S_{\ub,u})}
+\|\snab^{ \ 2}\rho\|_{L^2(S_{\ub,u})}+\|\snab^{ \ 2}\sigma\|_{L^2(S_{\ub,u})}\right\}\nonumber\\
&&\hspace{3cm}+C|u|^{-4}\Mb+O(\delta^{1/2}|u|^{-4}) \label{7.190}
\end{eqnarray}
Here:
\begin{equation}
M=\scR_1^4(\rho)+\scR_1^4(\sigma)+(\scR_1^4(\beta)+{\cal
R}_0^\infty(\beta))^2+{\cal R}_0^\infty(\alpha)\scR_1^4(\alpha)
\label{7.191}
\end{equation}
and:
\begin{equation}
\Mb=\scR_1^4(\rho)+\scR_1^4(\sigma)+(\scR_1^4(\beta)+{\cal
R}_0^\infty(\beta))^2+\tilde{A} \label{7.192}
\end{equation}

Next, we apply Lemma 4.1 to the propagation equations \ref{6.121},
\ref{6.122} to obtain the following propagation equations for
$\snab^{ \ 2}\mu$, $\snab^{ \ 2}\mub$:
\begin{equation}
D\snab^{ \ 2}\mu=-\Omega\mbox{tr}\chi\snab^{ \
2}\mu-\frac{1}{2}\Omega\mbox{tr}\chi\snab^{ \ 2}\mub+h^\prime
\label{7.193}
\end{equation}
\begin{equation}
\Db\snab^{ \ 2}\mub=-\Omega\mbox{tr}\chib\snab^{ \
2}\mub-\frac{1}{2}\Omega\mbox{tr}\chib\snab^{ \ 2}\mu+\hb^\prime
\label{7.194}
\end{equation}
where $h^\prime$,$\hb^\prime$ are the symmetric 2-covariant $S$
tensorfields:
\begin{eqnarray}
h^\prime&=&-D\sGamma\cdot\sd\mu-\left(\mu+\frac{1}{2}\mub\right)\snab^{ \ 2}(\Omega\mbox{tr}\chi)\nonumber\\
&\s&-\sd(\Omega\mbox{tr}\chi)\otimes\left(\sd\mu+\frac{1}{2}\sd\mub\right)
-\left(\sd\mu+\frac{1}{2}\sd\mub\right)\otimes\sd(\Omega\mbox{tr}\chi)\nonumber\\
&\s&+\snab^{ \
2}\left[\Omega\left(-\frac{1}{4}\mbox{tr}\chib|\chih|^2+\frac{1}{2}\mbox{tr}\chi|\etb|^2\right)\right]
+\snab^{ \ 2}\sdiv j \label{7.195}
\end{eqnarray}
\begin{eqnarray}
\hb^\prime&=&-\Db\sGamma\cdot\sd\mub-\left(\mub+\frac{1}{2}\mu\right)\snab^{ \ 2}(\Omega\mbox{tr}\chib)\nonumber\\
&\s&-\sd(\Omega\mbox{tr}\chib)\otimes\left(\sd\mub+\frac{1}{2}\sd\mu\right)
-\left(\sd\mub+\frac{1}{2}\sd\mu\right)\otimes\sd(\Omega\mbox{tr}\chib)\nonumber\\
&\s&+\snab^{ \
2}\left[\Omega\left(-\frac{1}{4}\mbox{tr}\chi|\chibh|^2+\frac{1}{2}\mbox{tr}\chib|\eta|^2\right)\right]
+\snab^{ \ 2}\sdiv\jb \label{7.196}
\end{eqnarray}

We first estimate in $L^2(S)$ the first four terms on the right in
\ref{7.195} and \ref{7.196}. We have:
$$(D\sGamma)^C_{AB}=\snab_A(\Omega\chi)_B^{\s C}+\snab_B(\Omega\chi)_A^{\s C}-\snab^C(\Omega\chi)_{AB}$$
By Proposition 6.1 and Lemma 5.2 with $p=4$ (and the estimate of
Proposition 4.1 for $\mbox{tr}\chi$):
\begin{equation}
\|\snab(\Omega\chi)\|_{L^\infty(S_{\ub,u})}\leq
C\delta^{-1/2}|u|^{-2}(\scR_1^4(\beta)+{\cal R}_0^\infty(\beta))
+O(|u|^{-3}) \label{7.197}
\end{equation}
hence also:
\begin{equation}
\|D\sGamma\|_{L^\infty(S_{\ub,u})}\leq
C\delta^{-1/2}|u|^{-2}(\scR_1^4(\beta)+{\cal R}_0^\infty(\beta))
+O(|u|^{-3}) \label{7.198}
\end{equation}
On the other hand, from \ref{6.227} (and Lemma 6.4):
\begin{equation}
\|\Db\sGamma\|_{L^\infty(S_{\ub,u})}\leq
C\delta^{1/2}|u|^{-3}(\scR_1^4(\beta)+{\cal R}_0^\infty(\beta))
+O(\delta|u|^{-4}) \label{7.199}
\end{equation}
We can then estimate the first term on the right in each of
\ref{7.195}, \ref{7.196} by placing the first factor in
$L^\infty(S)$ using \ref{7.198}, \ref{7.199} and the second factor
in $L^4(S)$ using \ref{7.185}, \ref{7.186}. This gives:
\begin{equation}
\|D\sGamma\cdot\sd\mu\|_{L^2(S_{\ub,u})}\leq
O(\delta^{-1/2}|u|^{-5}), \ \ \
\|\Db\sGamma\cdot\sd\mub\|_{L^2(S_{\ub,u})}\leq
O(\delta^{1/2}|u|^{-6}) \label{7.200}
\end{equation}
By \ref{6.136}, \ref{6.137} and the estimates of Proposition 6.2
pertaining to $\eta$, $\etb$:
\begin{equation}
\|\mu\|_{L^\infty(S_{\ub,u})}, \|\mub\|_{L^\infty(S_{\ub,u})} \
\leq C|u|^{-3}({\cal R}_0^\infty(\rho) +{\cal
R}_0^\infty(\alpha){\cal
D}_0^\infty(\chib)+\tilde{A})+O(\delta^{1/2}|u|^{-3})
\label{7.201}
\end{equation}
We can then estimate the second term on the right in each of
\ref{7.195}, \ref{7.196} by placing the first factor in $L^\infty$
and the second factor in $L^4(S)$ using Propositions 6.1 and 6.2.
We obtain in this way bounds in $L^2(S)$ for the second term on
the right in each of \ref{7.195}, \ref{7.196} by $O(|u|^{-6})$ and
$O(\delta|u|^{-7})$ respectively. The third and fourth terms on
the right in each of \ref{7.195}, \ref{7.196} are estimated in a
similar manner as the first terms. We obtain in this way bounds in
$L^2(S)$ for the third and fourth terms on the right in each of
\ref{7.195}, \ref{7.196} by $O(|u|^{-6})$ and $O(\delta|u|^{-7})$
respectively.

We proceed to the fifth terms on the right in each of \ref{7.195},
\ref{7.196}. Using Proposition 6.1, the estimates of Proposition
6.2 for $\mbox{tr}\chib$ and $\etb$, as well as the results of
Chapters 3 and 4, we can estimate the fifth term on the right in
\ref{7.195} in $L^2(S)$ by:
\begin{equation}
C\delta^{-1}|u|^{-4}\left[{\cal
R}_0^\infty(\alpha)(\scR_1^4(\beta)+{\cal
R}_0^\infty(\beta))+(\scR_1^4(\alpha))^2\right]
+O(\delta^{-1/2}|u|^{-5}) \label{7.202}
\end{equation}
Also, using the estimate of Proposition 6.1 for $\mbox{tr}\chi$,
the estimates of Proposition 6.2 for $\mbox{tr}\chib$, $\chibh$
and $\eta$, as well as the results of Chapters 3 and 4, we can
estimate the fifth term on the right in \ref{7.196} in $L^2(S)$
by:
\begin{equation}
O(\delta^{1/2}|u|^{-6}) \label{7.203}
\end{equation}

We turn to the last terms on the right in \ref{7.195} and
\ref{7.196} the symmetric 2-covariant $S$ tensorfields $\snab^{ \
2}\sdiv j$ and $\snab^{ \ 2}\sdiv\jb$. These are of the form (see
\ref{6.125}, \ref{6.129}):
\begin{eqnarray}
\snab^{ \ 2}\sdiv j&=&(\Omega\chih,\snab^{ \ 3}\eta)+(\Omega\mbox{tr}\chi,\snab^{ \ 3}\etb)\nonumber\\
&\s&+(\snab(\Omega\chih),\snab^{ \ 2}\eta)+(\sd(\Omega\mbox{tr}\chi),\snab^{ \ 2}\etb)\nonumber\\
&\s&+(\snab^{ \ 2}(\Omega\chih),\snab\eta)+(\snab^{ \ 2}(\Omega\mbox{tr}\chi),\snab\etb)\nonumber\\
&\s&+(\snab^{ \ 3}(\Omega\chih),\eta)+(\snab^{ \
3}(\Omega\mbox{tr}\chi),\etb) \label{7.204}
\end{eqnarray}
and:
\begin{eqnarray}
\snab^{ \ 2}\sdiv\jb&=&(\Omega\chibh,\snab^{ \ 3}\etb)+(\Omega\mbox{tr}\chib,\snab^{ \ 3}\eta)\nonumber\\
&\s&+(\snab(\Omega\chibh),\snab^{ \ 2}\etb)+(\sd(\Omega\mbox{tr}\chib),\snab^{ \ 2}\eta)\nonumber\\
&\s&+(\snab^{ \ 2}(\Omega\chibh),\snab\etb)+(\snab^{ \ 2}(\Omega\mbox{tr}\chib),\snab\eta)\nonumber\\
&\s&+(\snab^{ \ 3}(\Omega\chibh),\etb)+(\snab^{ \
3}(\Omega\mbox{tr}\chib),\eta) \label{7.205}
\end{eqnarray}

The first pair of terms on the right in \ref{7.204}, placing the
first factors in $L^\infty(S)$ and the second factors in $L^2(S)$,
are bounded in $L^2(S)$ by:
\begin{equation}
C\delta^{-1/2}|u|^{-1}{\cal R}_0^\infty(\alpha)\|\snab^{ \
3}\eta\|_{L^2(S_{\ub,u})} +C(|u|^{-1}+O(|u|^{-2}))\|\snab^{ \
3}\etb\|_{L^2(S_{\ub,u})} \label{7.206}
\end{equation}
To estimate the second pair of terms on the right in \ref{7.204}
in $L^2(S)$, we place the first factors in $L^\infty(S)$ using
\ref{7.197} and the second factors in $L^2(S)$ to obtain a bound
by:
\begin{eqnarray}
&&C\delta^{-1/2}|u|^{-2}(\scR_1^4(\beta)+{\cal R}_0^\infty(\beta)+O(\delta^{1/2}|u|^{-1}))\cdot\nonumber\\
&&\hspace{3cm}\cdot (\|\snab^{ \ 2}\eta\|_{L^2(S_{\ub,u})}+\|\snab^{ \ 2}\etb\|_{L^2(S_{\ub,u})})\nonumber\\
&&\leq C\delta^{-1/2}|u|^{-2}(\scR_1^4(\beta)+{\cal
R}_0^\infty(\beta)+O(\delta^{1/2}|u|^{-1}))
\|\snab^{ \ 2}\etb\|_{L^2(S_{\ub,u})}\nonumber\\
&&\hspace{3cm}+O(|u|^{-5}) \label{7.207}
\end{eqnarray}
taking into account the first estimate of Proposition 7.1. To
estimate the third pair of terms on the right in \ref{7.204} in
$L^2(S)$, we place the first factors in $L^4(S)$ using Proposition
6.1 and the second factors in $L^4(S)$ using the results of
Chapter 4 to obtain a bound by:
\begin{equation}
O(|u|^{-5}) \label{7.208}
\end{equation}
Finally, the last pair of terms on the right in \ref{7.204},
placing the first factors in $L^2(S)$ and the second factors in
$L^\infty(S)$ using the results of Chapter 3, are bounded in
$L^2(S)$ by:
\begin{equation}
C\delta^{1/2}|u|^{-2}({\cal
R}_0^\infty(\beta)+O(\delta^{1/2}|u|^{-1})) (\|\snab^{ \
3}(\Omega\mbox{tr}\chi)\|_{L^2(S_{\ub,u})}+\|\snab^{ \
3}(\Omega\chih)\|_{L^2(S_{\ub,u})}) \label{7.209}
\end{equation}
Moreover, expressing $\snab^{ \ 3}\log\Omega$ as in \ref{7.a6} and
using the estimates of Proposition 6.1 and Lemma 6.4 and the fact
that:
\begin{equation}
\|\sd\log\Omega\|_{L^\infty(S_{\ub,u})}\leq O(\delta|u|^{-3})
\label{7.210}
\end{equation}
which follows from Lemmas 4.11 and 6.4 through Lemma 5.2, we can
estimate:
\begin{eqnarray}
&&\|\snab^{ \ 3}(\Omega\mbox{tr}\chi)\|_{L^2(S_{\ub,u})}+\|\snab^{ \ 3}(\Omega\chih)\|_{L^2(S_{\ub,u})}\nonumber\\
&&\hspace{1cm}\leq C\left\{\|\snab^{ \
3}\mbox{tr}\chi^\prime\|_{L^2(S_{\ub,u})}
+\|\snab^{ \ 3}\chih^\prime\|_{L^2(S_{\ub,u})}\right.\nonumber\\
&&\hspace{1cm}\s\left.+\|\snab^{ \ 3}\log\Omega\|_{L^2(S_{\ub,u})}
(\|\mbox{tr}\chi^\prime\|_{L^\infty(S_{\ub,u})}+\|\chih^\prime\|_{L^\infty(S_{\ub,u})})\right\}\nonumber\\
&&\hspace{1cm}\s+O(\delta^{1/2}|u|^{-5})
\nonumber\\
&&\hspace{1cm}\leq C\left\{\|\snab^{ \
3}\mbox{tr}\chi^\prime\|_{L^2(S_{\ub,u})}
+\|\snab^{ \ 3}\chih^\prime\|_{L^2(S_{\ub,u})}\right\}\nonumber\\
&&\hspace{1cm}\s+C\delta^{-1/2}|u|^{-1}({\cal
R}_0^\infty(\alpha)+O(\delta^{1/2}))(\|\snab^{ \
2}\eta\|_{L^2(S_{\ub,u})}
+\|\snab^{ \ 2}\etb\|_{L^2(S_{\ub,u})})\nonumber\\
&&\hspace{1cm}\s+O(\delta^{1/2}|u|^{-5})\nonumber\\
&&\hspace{1cm}\leq C\|\snab^{ \ 3}\chih^\prime\|_{L^2(S_{\ub,u})}
+C\delta^{-1/2}|u|^{-1}({\cal R}_0^\infty(\alpha)+O(\delta^{1/2}))\|\snab^{ \ 2}\etb\|_{L^2(S_{\ub,u})}\nonumber\\
&&\hspace{1cm}\s+O(|u|^{-4})\label{7.211}
\end{eqnarray}
the last step by virtue of the second estimate of Proposition 7.2
and the first estimate of Proposition 7.1. Combining the above
results \ref{7.206} - \ref{7.208} and \ref{7.209}, \ref{7.211} we
conclude that:
\begin{eqnarray}
&&\|\snab^{ \ 2}\sdiv j\|_{L^2(S_{\ub,u})}\leq
C\delta^{-1/2}|u|^{-1}{\cal R}_0^\infty(\alpha)\|\snab^{ \ 3}\eta\|_{L^2(S_{\ub,u})}\nonumber\\
&&\hspace{2cm}+C|u|^{-1}(1+O(|u|^{-1}))\|\snab^{ \ 3}\etb\|_{L^2(S_{\ub,u})}\nonumber\\
&&\hspace{1cm}+O(\delta^{-1/2}|u|^{-2})\|\snab^{ \
2}\etb\|_{L^2(S_{\ub,u})} +O(\delta^{1/2}|u|^{-2})\|\snab^{ \
3}\chih^\prime\|_{L^2(S_{\ub,u})}
\nonumber\\
&&\hspace{2cm}+O(|u|^{-5}) \label{7.212}
\end{eqnarray}

The first pair of terms on the right in \ref{7.205}, placing the
first factors in $L^\infty(S)$ and the second factors in $L^2(S)$,
are bounded in $L^2(S)$ by:
\begin{eqnarray}
&&C\delta^{1/2}|u|^{-2}({\cal D}_0^\infty(\chibh)+O(\delta|u|^{-3/2}))\|\snab^{ \ 3}\etb\|_{L^2(S_{\ub,u})}\nonumber\\
&&+C|u|^{-1}(1+O(\delta|u|^{-1}))\|\snab^{ \
3}\eta\|_{L^2(S_{\ub,u})} \label{7.213}
\end{eqnarray}
To estimate the second pair of terms on the right in \ref{7.205}
in $L^2(S)$, we place the first factors in $L^\infty(S)$ using
\ref{6.226} (and Lemma 6.4) and the second factors in $L^2(S)$ to
obtain a bound by:
\begin{eqnarray}
&&C\delta^{1/2}|u|^{-3}(\scR_1^4(\beta)+{\cal R}_0^\infty(\beta)+O(\delta^{1/2}|u|^{-1}))\cdot\nonumber\\
&&\hspace{3cm}\cdot (\|\snab^{ \ 2}\eta\|_{L^2(S_{\ub,u})}+\|\snab^{ \ 2}\etb\|_{L^2(S_{\ub,u})})\nonumber\\
&&\leq C\delta^{1/2}|u|^{-3}(\scR_1^4(\beta)+{\cal
R}_0^\infty(\beta)+O(\delta^{1/2}|u|^{-1}))
\|\snab^{ \ 2}\etb\|_{L^2(S_{\ub,u})}\nonumber\\
&&\hspace{3cm}+O(\delta|u|^{-6}) \label{7.214}
\end{eqnarray}
taking into account the first estimate of Proposition 7.1. To
estimate the third pair of terms on the right in \ref{7.205} in
$L^2(S)$, we place the first factors in $L^4(S)$ using Proposition
6.2 and the second factors in $L^4(S)$ using the results of
Chapter 4 to obtain a bound by:
\begin{equation}
O(\delta|u|^{-6}) \label{7.215}
\end{equation}
Finally, the last pair of terms on the right in \ref{7.205},
placing the first factors in $L^2(S)$ and the second factors in
$L^\infty(S)$ using the results of Chapter 3, are bounded in
$L^2(S)$ by:
\begin{equation}
C\delta^{1/2}|u|^{-2}|({\cal
R}_0^\infty(\beta)+O(\delta^{1/2}|u|^{-1})) (\|\snab^{ \
3}(\Omega\mbox{tr}\chib)\|_{L^2(S_{\ub,u})}+\|\snab^{ \
3}(\Omega\chibh)\|_{L^2(S_{\ub,u})}) \label{7.216}
\end{equation}
Moreover, taking into account the estimates of Proposition 6.2,
Lemma 6.4 and \ref{7.210}, we can estimate:
\begin{eqnarray}
&&\|\snab^{ \ 3}(\Omega\mbox{tr}\chib)\|_{L^2(S_{\ub,u})}+\|\snab^{ \ 3}(\Omega\chibh)\|_{L^2(S_{\ub,u})}\nonumber\\
&&\hspace{1cm}\leq C\left\{\|\snab^{ \
3}\mbox{tr}\chib^\prime\|_{L^2(S_{\ub,u})}
+\|\snab^{ \ 3}\chibh^\prime\|_{L^2(S_{\ub,u})}\right.\nonumber\\
&&\hspace{1cm}\s\left.+\|\snab^{ \
3}\log\Omega\|_{L^2(S_{\ub,u})}(\|\mbox{tr}\chib^\prime\|_{L^\infty(S_{\ub,u})}
+\|\chibh^\prime\|_{L^\infty(S_{\ub,u})})\right\}\nonumber\\
&&\hspace{1cm}\s+O(\delta^{3/2}|u|^{-6})\nonumber\\
&&\hspace{1cm}\leq C\left\{\|\snab^{ \
3}\mbox{tr}\chib^\prime\|_{L^2(S_{\ub,u})}
+\|\snab^{ \ 3}\chibh^\prime\|_{L^2(S_{\ub,u})}\right\}\nonumber\\
&&\hspace{1cm}\s+C(|u|^{-1}+O(\delta^{1/2}|u|^{-2}))\|\snab^{ \ 3}\log\Omega\|_{L^2(S_{\ub,u})}\nonumber\\
&&\hspace{1cm}\s+O(\delta^{3/2}|u|^{-6}) \label{7.217}
\end{eqnarray}
where the coefficient of $\|\snab^{ \ 3}\log\Omega\|_{L^2(S_{\ub,u})}$
depends only on ${\cal D}_0^\infty$, ${\cal R}_0^\infty$.
Combining the above results \ref{7.213} - \ref{7.217} we conclude
that:
\begin{eqnarray}
&&\|\snab^{ \ 2}\sdiv\jb\|_{L^2(S_{\ub,u})}\leq
C\delta^{1/2}|u|^{-2}({\cal
D}_0^\infty(\chibh)+O(\delta|u|^{-3/2}))
\|\snab^{ \ 3}\etb\|_{L^2(S_{\ub,u})}\nonumber\\
&&\hspace{2cm}+C|u|^{-1}(1+O(\delta|u|^{-1}))\|\snab^{ \ 3}\eta\|_{L^2(S_{\ub,u})}\nonumber\\
&&\hspace{2cm}+O(\delta^{1/2}|u|^{-3})\|\snab^{ \ 2}\etb\|_{L^2(S_{\ub,u})}\nonumber\\
&&\hspace{1cm}+O(\delta^{1/2}|u|^{-2})(\|\snab^{ \
3}\mbox{tr}\chib^\prime\|_{L^2(S_{\ub,u})}
+\|\snab^{ \ 3}\chibh^\prime\|_{L^2(S_{\ub,u})})\nonumber\\
&&\hspace{1cm}+O(\delta^{1/2}|u|^{-3})\|\snab^{ \
3}\log\Omega\|_{L^2(S_{\ub,u})} +O(\delta|u|^{-6})\label{7.218}
\end{eqnarray}
where the coefficient of $\|\snab^{ \ 3}\log\Omega\|_{L^2(S_{\ub,u})}$
depends only on ${\cal D}_0^\infty$, ${\cal R}_0^\infty$.

Combining the bound \ref{7.121} with the previously obtained
bounds on the first five terms on the right in \ref{7.195} we
obtain:
\begin{eqnarray}
\|h^\prime\|_{L^2(S_{\ub,u})}&\leq&C\delta^{-1/2}|u|^{-1}{\cal
R}_0^\infty(\alpha)\|\snab^{ \ 3}\eta\|_{L^2(S_{\ub,u})}
\nonumber\\
&\s&+C|u|^{-1}(1+O(|u|^{-1}))\|\snab^{ \ 3}\etb\|_{L^2(S_{\ub,u})}\nonumber\\
&\s&+O(\delta^{-1/2}|u|^{-2})\|\snab^{ \ 2}\etb\|_{L^2(S_{\ub,u})}\nonumber\\
&\s&+O(\delta^{1/2}|u|^{-2})\|\snab^{ \ 3}\chih^\prime\|_{L^2(S_{\ub,u})}\nonumber\\
&\s&+C\delta^{-1}|u|^{-4}\left[{\cal
R}_0^\infty(\alpha)(\scR_1^4(\beta)+{\cal R}_0^\infty(\beta))
+(\scR_1^4(\alpha))^2\right]\nonumber\\
&\s&+O(\delta^{-1/2}|u|^{-5})\label{7.219}
\end{eqnarray}
Combining the bound \ref{7.218} with the previously obtained
bounds on the first five terms on the right in \ref{7.196} we
obtain:
\begin{eqnarray}
\|\hb^\prime\|_{L^2(S_{\ub,u})}&\leq&C\delta^{1/2}|u|^{-2}({\cal
D}_0^\infty(\chibh)+O(\delta|u|^{-3/2}))
\|\snab^{ \ 3}\etb\|_{L^2(S_{\ub,u})}\nonumber\\
&\s&+C|u|^{-1}(1+O(\delta|u|^{-1}))\|\snab^{ \ 3}\eta\|_{L^2(S_{\ub,u})}\nonumber\\
&\s&+O(\delta^{1/2}|u|^{-3})\|\snab^{ \ 2}\etb\|_{L^2(S_{\ub,u})}\nonumber\\
&\s&+O(\delta^{1/2}|u|^{-2})(\|\snab^{ \
3}\mbox{tr}\chib^\prime\|_{L^2(S_{\ub,u})}
+\|\snab^{ \ 3}\chibh^\prime\|_{L^2(S_{\ub,u})})\nonumber\\
&\s&+O(\delta^{1/2}|u|^{-3})\|\snab^{ \
3}\log\Omega\|_{L^2(S_{\ub,u})}+O(\delta^{1/2}|u|^{-6})
\label{7.220}
\end{eqnarray}
where the coefficient of $\|\snab^{ \ 3}\log\Omega\|_{L^2(S_{\ub,u})}$
depends only on ${\cal D}_0^\infty$, ${\cal R}_0^\infty$.
Substituting in \ref{7.219}, \ref{7.220} the estimates
\ref{7.199}, \ref{7.200} yields:
\begin{eqnarray}
\|h^\prime\|_{L^2(S_{\ub,u})}&\leq&C\delta^{-1/2}|u|^{-1}{\cal
R}_0^\infty(\alpha)\|\snab^{ \ 2}\mu\|_{L^2(S_{\ub,u})}
\nonumber\\&\s&+C|u|^{-1}(1+O(|u|^{-1}))\|\snab^{ \ 2}\mub\|_{L^2(S_{\ub,u})}\nonumber\\
&\s&+C\delta^{-1/2}|u|^{-1}({\cal
R}_0^\infty(\alpha)+O(\delta^{1/2}))(\|\snab^{ \
2}\rho\|_{L^2(S_{\ub,u})}
+\|\snab^{ \ 2}\sigma\|_{L^2(S_{\ub,u})})\nonumber\\
&\s&+O(\delta^{-1/2}|u|^{-2})\|\snab^{ \ 2}\etb\|_{L^2(S_{\ub,u})}
+O(\delta^{1/2}|u|^{-2})\|\snab^{ \ 3}\chih^\prime\|_{L^2(S_{\ub,u})}\nonumber\\
&\s&+C\delta^{-1}|u|^{-4}\left[{\cal
R}_0^\infty(\alpha)(\scR_1^4(\beta)+{\cal R}_0^\infty(\beta))
+(\scR_1^4(\alpha))^2\right]\nonumber\\
&\s&+O(\delta^{-1/2}|u|^{-5})\label{7.221}
\end{eqnarray}
and:
\begin{eqnarray}
\|\hb^\prime\|_{L^2(S_{\ub,u})}&\leq&C\delta^{1/2}|u|^{-2}({\cal
D}_0^\infty(\chibh)+O(\delta|u|^{-3/2}))
\|\snab^{ \ 2}\mub\|_{L^2(S_{\ub,u})}\nonumber\\
&\s&+C|u|^{-1}(1+O(\delta|u|^{-1}))\|\snab^{ \ 2}\mu\|_{L^2(S_{\ub,u})}\nonumber\\
&\s&+C|u|^{-1}(1+O(\delta^{1/2}|u|^{-1}))(\|\snab^{ \
2}\rho\|_{L^2(S_{\ub,u})}
+\|\snab^{ \ 2}\sigma\|_{L^2(S_{\ub,u})})\nonumber\\
&\s&+O(\delta^{1/2}|u|^{-3})\|\snab^{ \ 2}\etb\|_{L^2(S_{\ub,u})}\nonumber\\
&\s&+O(\delta^{1/2}|u|^{-2}(\|\snab^{ \
3}\mbox{tr}\chib^\prime\|_{L^2(S_{\ub,u})}
+\|\snab^{ \ 3}\chibh^\prime\|_{L^2(S_{\ub,u})})\nonumber\\
&\s&+O(\delta^{1/2}|u|^{-3})\|\snab^{ \ 3}\log\Omega\|_{L^2(S_{\ub,u})}\nonumber\\
&\s&+C|u|^{-5}M+O(\delta^{1/2}|u|^{-5})\label{7.222}
\end{eqnarray}
where the coefficient of $\|\snab^{ \ 3}\log\Omega\|_{L^2(S_{\ub,u})}$
depends only on ${\cal D}_0^\infty$, ${\cal R}_0^\infty$.

We now turn to the propagation equations \ref{7.193}, \ref{7.194}.
To \ref{7.193} we apply Lemma 4.6 with $p=2$. Here $r=2$,
$\nu=-2$, $\gamma=0$, and we obtain:
\begin{equation}
\|\snab^{ \ 2}\mu\|_{L^2(S_{\ub,u})}\leq
C\int_0^{\ub}\left\|-\frac{1}{2}\Omega\mbox{tr}\chi\snab^{ \
2}\mub +h^\prime\right\|_{L^2(S_{\ub^\prime,u})}d\ub^\prime
\label{7.223}
\end{equation}
To \ref{7.194} we apply Lemma 4.7 with $p=2$. Here again $r=2$,
$\nu=-2$, $\gammab=0$, and we obtain:
\begin{eqnarray}
|u|^3\|\snab^{ \ 2}\mub\|_{L^2(S_{\ub,u})}&\leq&C|u_0|^3\|\snab^{ \ 2}\mub\|_{L^2(S_{\ub,u_0})}\label{7.224}\\
&\s&+C\int_{u_0}^u|u^\prime|^3\left\|-\frac{1}{2}\Omega\mbox{tr}\chib\snab^{
\ 2}\mu
+\hb^\prime\right\|_{L^2(S_{\ub,u^\prime})}du^\prime\nonumber
\end{eqnarray}
Substituting in \ref{7.223} and \ref{7.224} the estimates
\ref{7.221} and \ref{7.222} yields the following system of linear
integral inequalities for the quantities $\|\snab^{ \
2}\mu\|_{L^2(S_{\ub,u})}$, $\|\snab^{ \ 2}\mub\|_{L^2(S_{\ub,u})}$
on the domain $D_1^{s^*}$:
\begin{eqnarray}
\|\snab^{ \
2}\mu\|_{L^2(S_{\ub,u})}&\leq&a(u)\int_0^{\ub}\|\snab^{ \
2}\mu\|_{L^2(S_{\ub^\prime,u})}d\ub^\prime
\nonumber\\
&\s&+b(u)\int_0^{\ub}\|\snab^{ \ 2}\mub\|_{L^2(S_{\ub^\prime,u})}d\ub^\prime\nonumber\\
&\s&+f(u) \label{7.225}
\end{eqnarray}
\begin{eqnarray}
|u|^3\|\snab^{ \
2}\mub\|_{L^2(S_{\ub,u})}&\leq&\int_{u_0}^u|u^\prime|^3\ab(u^\prime)
\|\snab^{ \ 2}\mub\|_{L^2(S_{\ub,u^\prime})}du^\prime\nonumber\\
&\s&+\int_{u_0}^u|u^\prime|^3\bb(u^\prime)\|\snab^{ \ 2}\mu\|_{L^2(S_{\ub,u^\prime})}du^\prime\nonumber\\
&\s&+|u|^3\fb(\ub,u) \label{7.226}
\end{eqnarray}

In \ref{7.225},
\begin{eqnarray}
a(u)&=&C\delta^{-1/2}|u|^{-1}{\cal R}_0^\infty(\alpha)\label{7.227}\\
b(u)&=&C|u|^{-1}(1+O(|u|^{-1}))\label{7.228}\\
f(u)&=&C|u|^{-4}\left[{\cal
R}_0^\infty(\alpha)(\scR_1^4(\beta)+{\cal R}_0^\infty(\beta))
+(\scR_1^4(\alpha))^2\right]\nonumber\\
&\s&+O(\delta^{1/2}|u|^{-5})\label{7.229}
\end{eqnarray}
and we have taken into account the following three facts. First,
that:
\begin{eqnarray*}
&&\int_0^{\ub}(\|\snab^{ \
2}\rho\|_{L^2(S_{\ub^\prime,u})}+\|\snab^{ \
2}\sigma\|_{L^2(S_{\ub^\prime,u})})d\ub^\prime
\\
&&\hspace{1cm}\leq\delta^{1/2}(\|\snab^{ \ 2}\rho\|_{L^2(C_u)}+\|\snab^{ \ 2}\sigma\|_{L^2(C_u)})\\
&&\hspace{1cm}\leq\delta|u|^{-4}(\scR_2(\rho)+\scR_2(\sigma))
\end{eqnarray*}
by the 2nd and 3rd of the definitions \ref{7.01}. Second, that:
\begin{eqnarray*}
&&\int_0^{\ub}\|\snab^{ \ 2}\etb\|_{L^2(S_{\ub^\prime,u})}d\ub^\prime\nonumber\\
&&\hspace{1cm}\leq\delta^{1/2}\|\snab^{ \ 2}\etb\|_{L^2(C_u)}\nonumber\\
&&\hspace{1cm}\leq
C\delta^{3/2}|u|^{-3}\scR_2(\beta)+O(\delta^2|u|^{-4})
\end{eqnarray*}
by the 2nd estimate of Proposition 7.1. Third, that:
\begin{eqnarray*}
&&\int_0^{\ub}\|\snab^{ \ 3}\chih^\prime\|_{L^2(S_{\ub^\prime,u})}d\ub^\prime\\
&&\hspace{1cm}\leq C\delta^{1/2}\|\snab^{ \ 3}\chih^\prime\|_{L^2(C_u)}\\
&&\hspace{1cm}\leq
C\delta^{1/2}|u|^{-3}(\scR_2(\beta)+\scR_1^4(\beta)+{\cal
R}_0^\infty(\beta)) +O(\delta|u|^{-4})
\end{eqnarray*}
by the 3rd estimate of Proposition 7.2.

In \ref{7.226},
\begin{eqnarray}
\ab(u)&=&C\delta^{1/2}|u|^{-2}({\cal D}_0^\infty(\chibh)+O(\delta|u|^{-3/2}))\label{7.230}\\
\bb(u)&=&C|u|^{-1}(1+O(\delta|u|^{-1}))\label{7.231}\\
|u|^3\fb(\ub,u)&=&\int_{u_0}^u\left\{C|u^\prime|^2(1+O(\delta^{1/2}|u^\prime|^{-1}))
(\|\snab^{ \ 2}\rho\|_{L^2(S_{\ub,u^\prime})}+\|\snab^{ \ 2}\sigma\|_{L^2(S_{\ub,u^\prime})})\right.\nonumber\\
&\s&\hspace{15mm}+O(\delta^{1/2})\|\snab^{ \ 2}\etb\|_{L^2(S_{\ub,u^\prime})}\nonumber\\
&\s&\hspace{15mm}+O(\delta^{1/2}|u^\prime|)(\|\snab^{ \
3}\mbox{tr}\chib^\prime\|_{L^2(S_{\ub,u^\prime})}
+\|\snab^{ \ 3}\chibh^\prime\|_{L^2(S_{\ub,u^\prime})})\nonumber\\
&\s&\hspace{15mm}\left.+O(\delta^{1/2})\|\snab^{ \ 3}\log\Omega\|_{L^2(S_{\ub,u^\prime})}\right\}du^\prime\nonumber\\
&\s&+C|u|^{-1}M+O(\delta^{1/2}|u|^{-1})+C|u_0|^3\|\snab^{ \
2}\mub\|_{L^2(S_{\ub,u_0})}\label{7.232}
\end{eqnarray}
where the coefficient of $\|\snab^{ \ 3}\log\Omega\|_{L^2(S_{\ub,u})}$
depends only on ${\cal D}_0^\infty$, ${\cal R}_0^\infty$.

Consider first the integral inequality \ref{7.225}. At fixed $u$,
considering $\|\snab^{ \ 2}\mub\|_{L^2(S_{\ub,u})}$ as given, this
inequality is of the form \ref{6.18}, with $a(u)$ in the role of
the constant $a$ and
$$b(u)\int_0^{\ub}\|\snab^{ \ 2}\mub\|_{L^2(S_{\ub^\prime,u})}d\ub^\prime+f(u)$$
in the role of the non-negative function $b(\ub)$. If, as is the
case here, the function $b(\ub)$ is non-decreasing and \ref{6.158}
holds then the result \ref{6.21} implies \ref{6.159}. Condition
\ref{6.158} in the present case reads:
$$a(u)\delta=C\delta^{1/2}|u|^{-1}{\cal R}_0^\infty(\alpha)\leq\log 2$$
and is indeed satisfied provided that $\delta$ is suitably small
depending on ${\cal R}_0^\infty$. The result \ref{6.159} then
takes the form:
\begin{eqnarray}
\|\snab^{ \ 2}\mu\|_{L^2(S_{\ub,u})}&\leq&
2b(u)\int_0^{\ub}\|\snab^{ \
2}\mub\|_{L^2(S_{\ub^\prime,u})}d\ub^\prime
\nonumber\\
&\s&+2f(u) \label{7.233}
\end{eqnarray}

Consider next the integral inequality \ref{7.226}. At fixed $\ub$,
considering $\|\snab^{ \ 2}\mu\|_{L^2(S_{\ub,u})}$ as given, this
inequality is of the form \ref{6.81} with $\ab(u)$ in the role of
the non-negative function $\ab$ and
$$\int_{u_0}^u|u^\prime|^3\bb(u^\prime)\|\snab^{ \ 2}\mu\|_{L^2(S_{\ub,u^\prime})}du^\prime
+|u|^3\fb(\ub,u)$$ in the role of the non-negative non-decreasing
function $\bb$. Note in particular from \ref{7.232} that the
function $|u|^3\fb(\ub,u)$ is indeed non-decreasing in $u$. Recall
that the result \ref{6.87} implies the result \ref{6.89} provided
that the condition \ref{6.88} holds. In view of \ref{7.230} this
condition holds if
$$C\delta^{1/2}({\cal D}_0^\infty(\chibh)+O(\delta))\leq\log 2$$
which is indeed satisfied if $\delta$ is suitably small depending
on ${\cal D}_0^\infty$, ${\cal R}_0^\infty$. The result \ref{6.89}
then takes the form:
\begin{eqnarray}
|u|^3\|\snab^{ \ 2}\mub\|_{L^2(S_{\ub,u})}&\leq& 2\int_{u_0}^u|u^\prime|^3\bb(u^\prime)\|\snab^{ \ 2}\mu\|_{L^2(S_{\ub,u^\prime})}du^\prime\nonumber\\
&\s&+2|u|^3\fb(\ub,u) \label{7.234}
\end{eqnarray}
We have thus reduced the system of integral inequalities
\ref{7.225}, \ref{7.226} on $D_1^{s^*}$ to the system \ref{7.233},
\ref{7.234}.

Let us now define:
\begin{eqnarray}
y(u)=\left(\int_0^{\ub_1}\|\snab^{ \
2}\mu\|^2_{L^2(S_{\ub,u})}d\ub\right)^{1/2}=\|\snab^{ \
2}\mu\|_{L^2(C^{\ub_1}_u)}
\label{7.235}\\
\yb(u)=\left(\int_0^{\ub_1}\|\snab^{ \
2}\mub\|^2_{L^2(S_{\ub,u})}d\ub\right)^{1/2}=\|\snab^{ \
2}\mub\|_{L^2(C^{\ub_1}_u)} \label{7.236}
\end{eqnarray}
Taking the $L^2$ norm of \ref{7.234} with respect to $\ub$ on
$[0,\ub_1]$ then yields:
\begin{equation}
|u|^3\yb(u)\leq
2\int_{u_0}^u|u^\prime|^3\bb(u^\prime)y(u^\prime)du^\prime+2|u|^3\gb(u)
\label{7.237}
\end{equation}
where:
\begin{equation}
\gb(u)=\|\fb(\cdot,u)\|_{L^2([0,\ub_1])} \label{7.238}
\end{equation}
From \ref{7.232} we have:
\begin{eqnarray}
|u|^3\gb(u)&=&\int_{u_0}^u\left\{C|u^\prime|^2(1+O(\delta^{1/2}|u^\prime|^{-1}))\cdot\right.\nonumber\\
&\s&\hspace{15mm}\cdot(\|\snab^{ \
2}\rho\|_{L^2(C^{\ub_1}_{u^\prime})}
+\|\snab^{ \ 2}\sigma\|_{L^2(C^{\ub_1}_{u^\prime})})\nonumber\\
&\s&\hspace{15mm}+O(\delta^{1/2})\|\snab^{ \ 2}\etb\|_{L^2(C^{\ub_1}_{u^\prime})}\nonumber\\
&\s&\hspace{15mm}+O(\delta^{1/2}|u^\prime|)(\|\snab^{ \
3}\mbox{tr}\chib^\prime\|_{L^2(C^{\ub_1}_{u^\prime})}
+\|\snab^{ \ 3}\chibh^\prime\|_{L^2(C^{\ub_1}_{u^\prime})})\nonumber\\
&\s&\hspace{15mm}\left.+O(\delta^{1/2})\|\snab^{ \ 3}\log\Omega\|_{L^2(C^{\ub_1}_{u^\prime})}\right\}du^\prime\nonumber\\
&\s&+C\delta^{1/2}|u|^{-1}M+O(\delta|u|^{-1})+C|u_0|^{-1}\delta^{1/2}\scD_2(\mub)\label{7.239}
\end{eqnarray}
where the coefficient of
$\|\snab^{ \ 3}\log\Omega\|_{L^2(C{\ub_1}_{u^\prime})}$ in the integral
depends only on ${\cal D}_0^\infty$, ${\cal R}_0^\infty$. By
virtue of the 2nd and 3rd of the definitions \ref{7.01}, the 2nd
estimate of Proposition 7.1, Lemma 7.5, and \ref{7.152}, we then
obtain:
\begin{eqnarray}
|u|^3\gb(u)
&\leq&C|u|^{-1}\delta^{1/2}M^\prime+O(\delta|u|^{-1})+O(\delta^2|u|^{-3})k\nonumber\\
&\s&+C|u_0|^{-1}\delta^{1/2}\scD_2(\mub) \label{7.240}
\end{eqnarray}
where the coefficient of $k$ depends only on ${\cal D}_0^\infty$,
${\cal R}_0^\infty$. In the above we have defined:
\begin{equation}
\scD_2(\mub)=\delta^{-1/2}|u_0|^4\|\snab^{ \
2}\mub\|_{L^2(C^{\ub_1}_{u_0})} \label{7.241}
\end{equation}
and:
\begin{equation}
M^\prime=M+\scR_2(\rho)+\scR_2(\sigma) \label{7.242}
\end{equation}

Let us set:
\begin{equation}
z(u)=\sup_{\ub\in[0,\ub_1]}\|\snab^{ \ 2}\mu\|_{L^2(S_{\ub,u})}
\label{7.243}
\end{equation}
Going back to \ref{7.233} and taking the supremum with respect to
$\ub$ on $[0,\ub_1]$ we obtain:
\begin{equation}
z(u)\leq 2b(u)\int_0^{\ub_1}\|\snab^{ \
2}\mub\|_{L^2(S_{\ub,u})}d\ub+2f(u) \label{7.244}
\end{equation}
Now,
\begin{equation}
\int_0^{\ub_1}\|\snab^{ \
2}\mub\|_{L^2(S_{\ub,u})}d\ub\leq\delta^{1/2}\yb(u) \label{7.245}
\end{equation}
therefore substituting the bound \ref{7.237} for $\yb(u)$ we
obtain:
\begin{equation}
\int_0^{\ub_1}\|\snab^{ \ 2}\mub\|_{L^2(S_{\ub,u})}d\ub\leq
2\delta^{1/2}|u|^{-3}
\left[\int_{u_0}^u|u^\prime|^3\bb(u^\prime)y(u^\prime)du^\prime+|u|^3\gb(u)\right]
\label{7.246}
\end{equation}
Since
\begin{equation}
y(u)\leq\delta^{1/2} z(u) \label{7.247}
\end{equation}
\ref{7.246} implies:
\begin{equation}
\int_0^{\ub_1}\|\snab^{ \ 2}\mub\|_{L^2(S_{\ub,u})}d\ub\leq
2\delta^{1/2}|u|^{-3}
\left[\delta^{1/2}\int_{u_0}^u|u^\prime|^3\bb(u^\prime)z(u^\prime)du^\prime+|u|^3\gb(u)\right]
\label{7.248}
\end{equation}
Substituting the bound \ref{7.248} in \ref{7.244} then yields the
following linear integral inequality for $z(u)$:
\begin{eqnarray}
z(u)&\leq&4\delta|u|^{-3}b(u)\int_{u_0}^u|u^\prime|^3\bb(u^\prime)z(u^\prime)du^\prime\nonumber\\
&\s&+4\delta^{1/2}b(u)\gb(u)+2f(u)\label{7.249}
\end{eqnarray}
Moreover, setting:
\begin{equation}
Z(u)=\int_{u_0}^u|u^\prime|^3\bb(u^\prime)z(u^\prime)du^\prime
\label{7.250}
\end{equation}
we have, replacing $u$ by $u^\prime$ in \ref{7.247} multiplying by
$|u^\prime|^3\bb(u^\prime)$ and integrating with respect to
$u^\prime$ on $[u_0,u]$,
\begin{equation}
\int_{u_0}^u|u^\prime|^3\bb(u^\prime)y(u^\prime)du^\prime\leq\delta^{1/2}Z(u)
\label{7.251}
\end{equation}
therefore substituting in \ref{7.237} yields:
\begin{equation}
|u|^3\yb(u)\leq 2\delta^{1/2}Z(u)+2|u|^3\gb(u) \label{7.252}
\end{equation}
Now, the inequality \ref{7.249} takes in terms of the function $Z$
the form:
\begin{equation}
\frac{dZ}{du}\leq 4\delta b\bb Z+w \label{7.253}
\end{equation}
where:
\begin{equation}
w(u)=4\delta^{1/2}b(u)\bb(u)|u|^3\gb(u)+2\bb(u)|u|^3 f(u)
\label{7.254}
\end{equation}
Integrating \ref{7.253} from $u_0$, noting that $Z(u_0)=0$,
yields:
\begin{equation}
Z(u)\leq\int_{u_0}^u\exp\left(4\delta\int_{u^\prime}^u(b\bb)(u^{\prime\prime})du^{\prime\prime}\right)w(u^\prime)du^\prime
\label{7.255}
\end{equation}
Now from the definitions \ref{7.228}, \ref{7.231},
$$(b\bb)(u)=C|u|^{-2}(1+O(|u|^{-1}))$$
hence
$$4\delta\int_{u^\prime}^u(b\bb)(u^{\prime\prime})du^{\prime\prime}\leq\log 2$$
provided that $\delta$ is suitably small depending on ${\cal
D}_0^\infty$, ${\cal R}_0^\infty$, and \ref{7.255} simplifies to:
\begin{equation}
Z(u)\leq 2\int_{u_0}^u w(u^\prime)du^\prime \label{7.256}
\end{equation}

We shall now derive an appropriate estimate for $\scD_2(\mub)$.
Recall that on $C_{u_0}$ \ref{6.179} holds. It follows that:
\begin{eqnarray}
\|\snab^{ \ 2}\mub\|_{L^2(S_{\ub,u_0})}&\leq&\|\snab^{ \
2}\mu\|_{L^2(S_{\ub,u_0})}
+2\|\snab^{ \ 2}\rho\|_{L^2(S_{\ub,u_0})}\nonumber\\
&\s&+C|u_0|^{-4}({\cal R}_1^4(\beta)+{\cal
R}_0^\infty(\beta))^2+O(\delta^{1/2}|u_0|^{-5}) \label{7.257}
\end{eqnarray}
(see \ref{7.187}). Substituting in \ref{7.225} at $u=u_0$ then
yields the following linear integral inequality for $\snab^{ \
2}\mu\|_{L^2(S_{\ub,u_0})}$:
\begin{equation}
\|\snab^{ \
2}\mu\|_{L^2(S_{\ub,u_0})}\leq\tilde{a}(u_0)\int_0^{\ub}\|\snab^{
\ 2}\mu\|_{L^2(S_{\ub^\prime,u_0})}d\ub^\prime+\tilde{f}(u_0)
\label{7.258}
\end{equation}
where:
\begin{eqnarray}
\tilde{a}(u_0)&=&a(u_0)+b(u_0)\nonumber\\
&\leq&C\delta^{-1/2}|u_0|^{-1}{\cal
R}_0^\infty(\alpha)+O(|u_0|^{-1})
\label{7.259}\\
\tilde{f}(u_0)&=&f(u_0)+b(u_0)\left\{C\delta|u_0|^{-4}[\scR_2(\rho)\right.\nonumber\\
&\s&\hspace{2cm}\left.+(\scR_1^4(\beta)+{\cal R}_0^\infty(\beta))^2]+O(\delta^{3/2}|u_0|^{-5})\right\}\nonumber\\
&\leq&C|u_0|^{-4}[{\cal R}_0^\infty(\alpha)(\scR_1^4(\beta)+{\cal
R}_0^\infty(\beta))+(\scR_1^4(\alpha))^2]
+O(\delta^{1/2}|u_0|^{-5})\nonumber\\
&\s&\label{7.260}
\end{eqnarray}
(see \ref{7.237} - \ref{7.239}). The integral inequality
\ref{7.258} implies:
\begin{equation}
\|\snab^{ \ 2}\mu\|_{L^2(S_{\ub,u_0})}\leq
e^{\delta\tilde{a}(u_0)}\tilde{f}(u_0) \label{7.261}
\end{equation}
Since
$$\delta\tilde{a}(u_0)\leq\log 2$$
provided that $\delta$ is suitably small depending on ${\cal
D}_0^\infty$, ${\cal R}_0^\infty$, \ref{7.261} in turn implies:
\begin{eqnarray}
\|\snab^{ \ 2}\mu\|_{L^2(S_{\ub,u_0})}&\leq&2\tilde{f}(u_0)\nonumber\\
&\leq&C|u_0|^{-4}[{\cal R}_0^\infty(\alpha)(\scR_1^4(\beta)+{\cal R}_0^\infty(\beta))+(\scR_1^4(\alpha))^2]\nonumber\\
&\s&+O(\delta^{1/2}|u_0|^{-5}) \label{7.262}
\end{eqnarray}
It then follows through \ref{7.257} that:
\begin{eqnarray}
\scD_2(\mub)&\leq&C\left[\scR_2(\rho)+(\scR_1^4(\beta)+{\cal R}_0^\infty(\beta))^2\right.\nonumber\\
&\s&+\left.{\cal R}_0^\infty(\alpha)(\scR_1^4(\beta)+{\cal R}_0^\infty(\beta))+(\scR_1^4(\alpha))^2\right]\nonumber\\
&\s&+O(\delta^{1/2}|u_0|^{-1}) \label{7.263}
\end{eqnarray}

Substituting \ref{7.263} in \ref{7.240} we obtain:
\begin{equation}
|u|^3\gb(u)\leq
C\delta^{1/2}|u|^{-1}\tilde{M}+O(\delta|u|^{-1})+O(\delta^2|u|^{-3})k
\label{7.264}
\end{equation}
where (see \ref{7.191}):
\begin{eqnarray}
\tilde{M}&=&\scR_2(\rho)+\scR_2(\sigma)+\scR_1^4(\rho)+\scR_1^4(\sigma)\nonumber\\
&\s&+({\cal R}_0^\infty(\alpha)+\scR_1^4(\beta)+{\cal
R}_0^\infty(\beta))
(\scR_1^4(\beta)+{\cal R}_0^\infty(\beta))\nonumber\\
&\s&+(\scR_1^4(\alpha)+{\cal
R}_0^\infty(\alpha))\scR_1^4(\alpha)\label{7.265}
\end{eqnarray}
and the coefficient of $k$ depends only on ${\cal D}_0^\infty$,
${\cal R}_0^\infty$. Substituting \ref{7.264} and \ref{7.229} in
\ref{7.254} we obtain:
\begin{eqnarray}
w(u)&\leq&C|u|^{-2}\left[{\cal R}_0^\infty(\alpha)(\scR_1^4(\beta)+{\cal R}_0^\infty(\beta))+(\scR_1^4(\alpha))^2\right]\nonumber\\
&\s&+O(\delta^{1/2}|u|^{-3})+O(\delta^{5/2}|u|^{-5})k
\label{7.266}
\end{eqnarray}
hence, from \ref{7.256}:
\begin{eqnarray}
Z(u)&\leq&C|u|^{-1}\left[{\cal R}_0^\infty(\alpha)(\scR_1^4(\beta)+{\cal R}_0^\infty(\beta))+(\scR_1^4(\alpha))^2\right]\nonumber\\
&\s&+O(\delta^{1/2}|u|^{-2})+O(\delta^{5/2}|u|^{-4})k
\label{7.267}
\end{eqnarray}
where the coefficient of $k$ depends only on ${\cal D}_0^\infty$,
${\cal R}_0^\infty$. Substituting \ref{7.267} (see \ref{7.250}) as
well as \ref{7.264} and \ref{7.229} in \ref{7.249} then yields:
\begin{eqnarray}
z(u)&\leq&C|u|^{-4}\left[{\cal R}_0^\infty(\alpha)(\scR_1^4(\beta)+{\cal R}_0^\infty(\beta)+(\scR_1^4(\alpha))^2\right]\nonumber\\
&\s&+O(\delta^{1/2}|u|^{-5})+O(\delta^{5/2}|u|^{-7})k
\label{7.268}
\end{eqnarray}
where the coefficient of $k$ depends only on ${\cal D}_0^\infty$,
${\cal R}_0^\infty$. Also, substituting \ref{7.267} and
\ref{7.264} in \ref{7.252} yields:
\begin{equation}
\yb(u)\leq
C\delta^{1/2}|u|^{-4}\tilde{M}+O(\delta|u|^{-4})+O(\delta^2|u|^{-6})k
\label{7.269}
\end{equation}
where the coefficient of $k$ depends only on ${\cal D}_0^\infty$,
${\cal R}_0^\infty$.

Finally, taking the $L^2$ norm of \ref{7.189} with respect to
$\ub$ on $[0,\ub_1]$ we obtain, in view of the definition
\ref{7.235} and the 2nd and 3rd of the definitions \ref{7.01},
\begin{eqnarray}
\|\snab^{ \ 3}\eta\|_{L^2(C^{\ub_1}_u)}&\leq&Cy(u)+C\delta^{1/2}|u|^{-4}(\scR_2(\rho)+\scR_2(\sigma))\nonumber\\
&\s&+C\delta^{1/2}|u|^{-4}M+O(\delta|u|^{-4}) \label{7.270}
\end{eqnarray}
In view of \ref{7.247}, the bound \ref{7.268} implies in
conjunction with \ref{7.270} the first estimate of the lemma.
Also, taking the $L^2$ norm of \ref{7.190} with respect to $\ub$
on $[0,\ub_1]$ we obtain, in view of the definition \ref{7.236}
and the 2nd and 3rd of the definitions \ref{7.01},
\begin{eqnarray}
\|\snab^{ \ 3}\etb\|_{L^2(C^{\ub_1}_u)}&\leq&C\yb(u)+C\delta^{1/2}|u|^{-4}(\scR_2(\rho)+\scR_2(\sigma))\nonumber\\
&\s&+C\delta^{1/2}|u|^{-4}\Mb+O(\delta|u|^{-4}) \label{7.271}
\end{eqnarray}
Substituting the bound \ref{7.269} then yields the second estimate
of the lemma.

\vspace{5mm}

\section{$L^2$ estimate for $\snab^{ \ 3}\omb$}

We proceed to derive an estimate for $\snab^{ \ 3}\omb$ in
$L^2(C^{\ub_1}_u)$ uniform in $u$ for all $u\in[0,s^*]$. Using
this estimate we shall derive an estimate for $\snab^{ \
3}\log\Omega$ in $L^2(C^{\ub_1}_u)$ uniform in $u$ for all
$u\in[0,s^*]$, which, with a suitable choice of the constant $k$,
improves the bound \ref{7.152}. This shall enable us to show that
$s^*=u_1$ so that the previous lemmas actually hold on the entire
parameter domain $D_1$, and, since $(\ub_1,u_1)\in D^\prime$ is
arbitrary, the estimates hold on all of $D^\prime$.

\vspace{5mm}

\noindent{\bf Lemma 7.7} \ \ \ We have:
$$\|\snab^{ \ 3}\omb\|_{L^2(C^{\ub_1}_u)}\leq C\delta^{3/2}|u|^{-5}\Nb+O(\delta^2|u|^{-6})+O(\delta^{7/2}|u|^{-8})k$$
for all $u\in[u_0,s^*]$, provided that $\delta$ is suitably small
depending on ${\cal D}_0^\infty$, ${\cal R}_0^\infty$, $\scD_1^4$,
$\scR_1^4$, and $\scR_2$. Moreover, the coefficient of $k$ depends
only on ${\cal D}_0^\infty$, ${\cal R}_0^\infty$. Here,
\begin{eqnarray*}
\Nb&=&\scR_2(\beb)+\scR_1^4(\beb)+\scR_1^4(\rho)+{\cal D}_0^\infty(\chibh)(\scR_2(\beta)+\scR_1^4(\beta))\\
&\s&+(\scD_1^4(\mbox{tr}\chib)+\scD_1^4(\chibh)){\cal
R}_0^\infty(\beta) +(\scR_1^4(\beta)+{\cal R}_0^\infty(\beta))^2
\end{eqnarray*}

\noindent{\em Proof:} \ Consider the propagation equation for the
function $\somb$, equation \ref{6.204}. Applying $\sd$ to this
equation we obtain, in view of Lemma 1.2, the following
propagation equation for the $S$ 1-form $\sd\somb$:
\begin{equation}
D\sd\somb+\Omega\mbox{tr}\chi\sd\somb=-2\Omega(\chih,\snab^{ \
3}\omb)+\mb^\prime \label{7.272}
\end{equation}
where:
\begin{equation}
\mb^\prime=-\sd(\Omega\mbox{tr}\chi)\somb-2(\snab(\Omega\chih),\snab^{
\ 2}\omb)+\sd\mb \label{7.273}
\end{equation}

We shall first estimate $\mb^\prime$ in $L^2(S)$. The first two
terms on the right in \ref{7.273} are estimated in $L^2(S)$ by
placing each of the two factors in $L^4(S)$ using the estimate of
Proposition 6.2 pertaining to $\omb$ and the results of Chapter 4.
We obtain a bound for these terms in $L^2(S)$ by:
\begin{equation}
O(\delta^{1/2}|u|^{-6}) \label{7.274}
\end{equation}
To estimate $\sd\mb$ in $L^2(S)$, we consider the expression
\ref{6.205} for $\mb$. In regard to the first term on the right in
\ref{6.205} we write:
$$\sd(\sdiv(\Omega\chih),\sd\omb)=(\sd\sdiv(\Omega\chih),\sd\omb)+(\sdiv(\Omega\chih),\snab^{ \ 2}\omb)$$
Placing each of the two factors of the two terms on the right in
$L^4(S)$ using Proposition 6.1, the estimate of Proposition 6.2
for $\omb$, and the results of Chapter 4 we obtain an $L^2(S)$
bound for the contribution of the first term on the right in
\ref{6.205} to $\sd\mb$ by:
\begin{equation}
O(\delta^{1/2}|u|^{-6}) \label{7.275}
\end{equation}
In regard to the second term on the right in \ref{6.205} we write:
\begin{eqnarray*}
\sd\sdiv(\Omega^2\mbox{tr}\chi\beb)&=&\Omega^2\mbox{tr}\chi\sd\sdiv\beb+(\snab^{ \ 2}(\Omega\mbox{tr}\chi),\beb)\\
&\s&+\sd(\Omega^2\mbox{tr}\chi)\sdiv\beb+(\sd(\Omega^2\mbox{tr}\chi),\snab\beb)
\end{eqnarray*}
The first term on the right is bounded in $L^2(S)$ by:
\begin{equation}
O(|u|^{-1})\|\snab^{ \ 2}\beb\|_{L^2(S_{\ub,u})} \label{7.276}
\end{equation}
while the remaining terms are bounded in $L^2(S)$ by:
\begin{equation}
O(\delta|u|^{-7}) \label{7.277}
\end{equation}
using Proposition 6.1 and the results of Chapter 4. The
contributions to $\sd\mb$ of the third and fourth terms on the
right in \ref{6.205} are bounded in $L^2(S)$ by:
\begin{equation}
O(\delta|u|^{-3})(\|\snab^{ \ 2}\rho\|_{L^2(S_{\ub,u})}+\|\snab^{
\ 2}\sigma\|_{L^2(S_{\ub,u})})+O(\delta|u|^{-7}) \label{7.278}
\end{equation}
by Lemmas 6.4 and 4.11, which in conjunction with Lemma 5.2 with
$p=4$ imply:
\begin{equation}
\|\sd\log\Omega\|_{L^\infty(S_{\ub,u})}\leq O(\delta|u|^{-3})
\label{7.279}
\end{equation}
To estimate the contribution of the fifth term on the right in
\ref{6.205} we write:
$$\slap\log\Omega=\frac{1}{2}(\sdiv\eta+\sdiv\etb)$$
and use the first estimate of Proposition 7.1. We obtain in this
way an $L^2(S)$ bound for the contribution in question by:
\begin{equation}
O(|u|^{-3})\|\snab^{ \
2}\etb\|_{L^2(S_{\ub,u})}+O(\delta^{1/2}|u|^{-6}) \label{7.280}
\end{equation}
The principal part of the contribution of the sixth term on the
right in \ref{6.205} to $\sd\mb$ is:
$$2\Omega^2((\eta-\etb,\sd\slap\eta)+(\eta,\sd\slap\etb))$$
This part is bounded in $L^2(S)$ by:
\begin{equation}
O(\delta^{1/2}|u|^{-2})(\|\snab^{ \
3}\eta\|_{L^2(S_{\ub,u})}+\|\snab^{ \ 3}\etb\|_{L^2(S_{\ub,u})})
\label{7.281}
\end{equation}
while the remainder is bounded by:
\begin{equation}
O(\delta^{1/2}|u|^{-6}) \label{7.282}
\end{equation}
using the estimates for $\eta$, $\etb$ of Proposition 6.2 and the
results of Chapters 3 and 4. Finally, the contribution to $\sd\mb$
of the last term on the right in \ref{6.205} is:
\begin{equation}
\sd\sdiv[\Omega^2(\chih^\sharp\cdot\beb-2\chibh^\sharp\cdot\beta+2\etb\rho-3\s^*\etb\sigma)]
\label{7.283}
\end{equation}
The part:
$$-2\sd\sdiv(\Omega^2\chibh^\sharp\cdot\beta)$$
gives the leading contribution in behavior with respect to
$\delta$. Using the estimate of Proposition 6.2 pertaining to
$\chibh$, which in conjunction with Lemma 5.2 with $p=4$ implies:
\begin{equation}
\|\snab\chibh\|_{L^\infty(S_{\ub,u})} \leq
C\delta^{1/2}|u|^{-3}(\scR_1^4(\beta)+{\cal
R}_0^\infty(\beta))+O(\delta|u|^{-4}) \label{7.284}
\end{equation}
together with the results of Chapter 3, the contribution in
question is bounded in $L^2(S)$ by:
\begin{eqnarray}
&&C\delta^{1/2}|u|^{-2}({\cal
D}_0^\infty(\chibh)+O(\delta|u|^{-3/2}))
\|\snab^{ \ 2}\beta\|_{L^2(S_{\ub,u})}\nonumber\\
&&+C|u|^{-5}(\scR_1^4(\beta)+{\cal
R}_0^\infty(\beta))^2+O(\delta^{1/2}|u|^{-6}) \label{7.285}
\end{eqnarray}
The remainder of \ref{7.283} is bounded in $L^2(S)$ by:
\begin{eqnarray}
&&O(\delta^{-1/2}|u|^{-1})\|\snab^{ \ 2}\beb\|_{L^2(S_{\ub,u})}\nonumber\\
&&+O(\delta^{1/2}|u|^{-2})
(\|\snab^{ \ 2}\rho\|_{L^2(S_{\ub,u})}+\|\snab^{ \ 2}\sigma\|_{L^2(S_{\ub,u})})\nonumber\\
&&+O(|u|^{-3})\|\snab^{ \
2}\etb\|_{L^2(S_{\ub,u})}+O(\delta^{1/2}|u|^{-6}) \label{7.286}
\end{eqnarray}
Collecting the above results (\ref{7.274} - \ref{7.278},
\ref{7.280} - \ref{7.282}, \ref{7.285} - \ref{7.286}) we conclude
that:
\begin{eqnarray}
\|\mb^\prime\|_{L^2(S_{\ub,u})}&\leq&O(\delta^{1/2}|u|^{-2})(\|\snab^{
\ 3}\eta\|_{L^2(S_{\ub,u})}
+\|\snab^{ \ 3}\etb\|_{L^2(S_{\ub,u})})\nonumber\\
&\s&+O(|u|^{-3})\|\snab^{ \ 2}\etb\|_{L^2(S_{\ub,u})}\nonumber\\
&\s&+C\delta^{1/2}|u|^{-2}({\cal
D}_0^\infty(\chibh)+O(\delta|u|^{-3/2}))
\|\snab^{ \ 2}\beta\|_{L^2(S_{\ub,u})}\nonumber\\
&\s&+O(\delta^{-1/2}|u|^{-1})\|\snab^{ \ 2}\beb\|_{L^2(S_{\ub,u})}\nonumber\\
&\s&+O(\delta^{1/2}|u|^{-2})
(\|\snab^{ \ 2}\rho\|_{L^2(S_{\ub,u})}+\|\snab^{ \ 2}\sigma\|_{L^2(S_{\ub,u})})\nonumber\\
&\s&+C|u|^{-5}(\scR_1^4(\beta)+{\cal
R}_0^\infty(\beta))^2+O(\delta^{1/2}|u|^{-6}) \label{7.287}
\end{eqnarray}

Consider now equation \ref{6.206}. Setting $\theta=\sd\omb$, the
$S$ 1-form $\theta$ satisfies a Hodge system of the type
considered in Lemma 7.4 with $f=\somb+\sdiv(\Omega\beb)$, $g=0$.
By the first estimate of Lemma 7.4 we then have:
\begin{eqnarray}
\|\snab^{ \ 3}\omb\|_{L^2(S_{\ub,u})}&\leq&C\left\{\|\sd\somb\|_{L^2(S_{\ub,u})}+\|\snab^{ \ 2}\beb\|_{L^2(S_{\ub,u})}\right\}\nonumber\\
&\s&+C\delta|u|^{-5}[\scR_1^4(\beb)+\scR_1^4(\rho)+{\cal D}_0^\infty(\chibh)\scR_1^4(\beta)\nonumber\\
&\s&\hspace{2cm}+(\scD_1^4(\mbox{tr}\chib)+\scD_1^4(\chibh)){\cal R}_0^\infty(\beta)]\nonumber\\
&\s&+O(\delta^{3/2}|u|^{-6}) \label{7.288}
\end{eqnarray}
using the estimate \ref{6.224} (and Lemma 6.4) as well as the
results of Chapter 4. It follows that in reference to the first
term on the right in \ref{7.272} we have:
\begin{eqnarray}
\|\Omega(\chih,\snab^{ \
3}\omb)\|_{L^2(S_{\ub,u})}&\leq&C\delta^{-1/2}|u|^{-1}{\cal
R}_0^\infty(\alpha)
\left\{\|\sd\somb\|_{L^2(S_{\ub,u})}+\|\snab^{ \ 2}\beb\|_{L^2(S_{\ub,u})}\right\}\nonumber\\
&\s&+O(\delta^{1/2}|u|^{-6}) \label{7.289}
\end{eqnarray}
We now apply Lemma 4.6 with $p=2$ to the propagation equation
\ref{7.272}. Here $r=1$, $\nu=-2$, $\gamma=0$ and we obtain:
\begin{equation}
\|\sd\somb\|_{L^2(S_{\ub,u})}\leq
C\int_0^{\ub}\left\|-2\Omega(\chih,\snab^{ \ 3}\omb)
+\mb^\prime\right\|_{L^2(S_{\ub^\prime,u})}d\ub^\prime
\label{7.290}
\end{equation}
We substitute the estimates \ref{7.289} and \ref{7.287}. For
$\ub\in[0,\ub_1]$ we have:
\begin{eqnarray}
&&\int_0^{\ub}(\|\snab^{ \ 3}\eta\|_{L^2(S_{\ub^\prime,u})}+\|\snab^{ \ 3}\etb\|_{L^2(S_{\ub^\prime,u})})d\ub^\prime\nonumber\\
&&\hspace{2cm}\leq \delta^{1/2}(\|\snab^{ \ 3}\eta\|_{L^2(C^{\ub_1}_u)}+\|\snab^{ \ 3}\etb\|_{L^2(C^{\ub_1}_u)}\nonumber\\
&&\hspace{2cm}\leq O(\delta|u|^{-4})+O(\delta^{5/2}|u|^{-6})k
\label{7.291}
\end{eqnarray}
by Lemma 7.6, the coefficient of $k$ depending only on ${\cal
D}_0^\infty$, ${\cal R}_0^\infty$,
\begin{equation}
\int_0^{\ub}\|\snab^{ \
2}\etb\|_{L^2(S_{\ub^\prime,u})}d\ub^\prime \leq
\delta^{1/2}\|\snab^{ \ 2}\etb\|_{L^2(C^{\ub_1}_u)}\leq
O(\delta^{3/2}|u|^{-3}) \label{7.292}
\end{equation}
by the 2nd estimate of Proposition 7.1, and:
\begin{eqnarray}
&&\int_0^{\ub_1}\|\snab^{ \
2}\beta\|_{L^2(S_{\ub^\prime,u})}d\ub^\prime
\leq\delta^{1/2}\|\snab^{ \ 2}\beta\|_{L^2(C^{\ub_1}_u)}\leq\delta^{1/2}|u|^{-3}\scR_2(\beta)\nonumber\\
&&\int_0^{\ub_1}\|\snab^{ \
2}\beb\|_{L^2(S_{\ub^\prime,u})}d\ub^\prime
\leq\delta^{1/2}\|\snab^{ \ 2}\beb\|_{L^2(C^{\ub_1}_u)}\leq\delta^2|u|^{-5}\scR_2(\beb)\nonumber\\
&&\int_0^{\ub_1}\|\snab^{ \
2}\rho\|_{L^2(S_{\ub^\prime,u})}d\ub^\prime
\leq\delta^{1/2}\|\snab^{ \ 2}\rho\|_{L^2(C^{\ub_1}_u)}\leq\delta|u|^{-4}\scR_2(\rho)\nonumber\\
&&\int_0^{\ub_1}\|\snab^{ \
2}\sigma\|_{L^2(S_{\ub^\prime,u})}d\ub^\prime
\leq\delta^{1/2}\|\snab^{ \ 2}\sigma\|_{L^2(C^{\ub_1}_u)}\leq\delta|u|^{-4}\scR_2(\sigma)\nonumber\\
&&\label{7.293}
\end{eqnarray}
by the definitions \ref{7.01}. We then deduce the following linear
integral inequality for the quantity
$\|\sd\somb\|_{L^2(S_{\ub,u})}$ on the domain $D_1^{s^*}$:
\begin{eqnarray}
\|\sd\somb\|_{L^2(S_{\ub,u})}&\leq&C\delta^{-1/2}{\cal
R}_0^\infty(\alpha)\int_0^{\ub}\|\sd\somb\|_{L^2(S_{\ub^\prime,u})}
d\ub^\prime\nonumber\\
&\s&+C\delta|u|^{-5}\left[{\cal
D}_0^\infty(\chibh)\scR_2(\beta)+(\scR_1^4(\beta)+{\cal
R}_0^\infty(\beta))^2\right]
\nonumber\\
&\s&+O(\delta^{3/2}|u|^{-6})+O(\delta^3|u|^{-8})k \label{7.294}
\end{eqnarray}
the coefficient of $k$ depending only on ${\cal D}_0^\infty$,
${\cal R}_0^\infty$. At fixed $u$ this inequality is of the form
\ref{6.21} with $C\delta^{-1/2}{\cal R}_0^\infty$ in the role of
the constant $a$. Since here condition \ref{6.158} is verified if
$\delta$ is suitably small depending on ${\cal
R}_0^\infty(\alpha)$, the result \ref{6.159} follows, which here
reads:
\begin{eqnarray}
\|\sd\somb\|_{L^2(S_{\ub,u})}&\leq& C\delta|u|^{-5}\left[{\cal
D}_0^\infty(\chibh)\scR_2(\beta)+(\scR_1^4(\beta)+{\cal
R}_0^\infty(\beta))^2\right]
\nonumber\\
&\s&+O(\delta^{3/2}|u|^{-6})+O(\delta^3|u|^{-8})k \label{7.295}
\end{eqnarray}
This holds for all $(\ub,u)\in D_1^{s^*}$. Taking the $L^2$ norm
of \ref{7.288} with respect to $\ub$ on $[0,\ub_1]$ we obtain, in
view of the last of the definitions \ref{7.01},
\begin{eqnarray}
&&\|\snab^{ \ 3}\omb\|_{L^2(C^{\ub_1}_u)}\leq C\|\sd\somb\|_{L^2(C^{\ub_1}_u)}\nonumber\\
&&\hspace{1cm}+C\delta^{3/2}|u|^{-5}[\scR_2(\beb)+\scR_1^4(\beb)+\scR_1^4(\rho)+{\cal D}_0^\infty(\chibh)\scR_1^4(\beta)\nonumber\\
&&\hspace{3cm}+(\scD_1^4(\mbox{tr}\chib)+\scD_1^4(\chibh)){\cal R}_0^\infty(\beta)]\nonumber\\
&&\hspace{3cm}+O(\delta^2|u|^{-6}) \label{7.296}
\end{eqnarray}
Noting that
$$\|\sd\somb\|_{L^2(C^{\ub_1}_u)}\leq\delta^{1/2}\sup_{\ub\in[0,\ub_1]}\|\sd\somb\|_{L^2(S_{\ub,u})}$$
and substituting the estimate \ref{7.295} then yields the lemma.

\vspace{5mm}

\noindent{\bf Lemma 7.8} \ \ \ There is a numerical constant $C$
such that with
$$k=C\Nb+O(\delta^{1/2})$$
we have $s^*=u_1$ and the estimate:
$$\|\snab^{ \ 3}\log\Omega\|_{L^2(C^{\ub_1}_u)}\leq C\delta^{3/2}|u|^{-4}\Nb+O(\delta^2|u|^{-5})$$
holds for all $u\in[u_0,u_1]$, provided that $\delta$ is suitably
small depending on ${\cal D}_0^\infty$, ${\cal R}_0^\infty$,
$\scD_1^4$, $\scR_1^4$, and $\scR_2$.

\noindent{ \em Proof:} \ Applying $\snab$ to equation \ref{6.225}
we obtain, in view of Lemma 4.1,
\begin{equation}
\Db\snab^{ \ 3}\log\Omega=\snab^{ \ 3}\omb-\Db\sGamma\cdot\snab^{
\ 2}\log\Omega-\snab\Db\sGamma\cdot\sd\log\Omega \label{7.297}
\end{equation}
Here, we denote:
\begin{eqnarray}
(\Db\sGamma\cdot\snab^{ \ 2}\log\Omega)_{ABC}&=&(\Db\sGamma)^D_{BC}(\snab^{ \ 2}\log\Omega)_{AD}\label{7.298}\\
&\s&+(\Db\sGamma)^D_{AC}(\snab^{ \
2}\log\Omega)_{BD}+(\Db\sGamma)^D_{AB}(\snab^{ \
2}\log\Omega)_{CD}\nonumber
\end{eqnarray}
and:
\begin{equation}
(\snab\Db\sGamma\cdot\sd\log\Omega)_{ABC}=\snab_A(\Db\sGamma)^D_{BC}\sd_D\log\Omega
\label{7.299}
\end{equation}
By Lemma 6.4 and the bound \ref{7.199}, the estimates of
Proposition 6.2 pertaining to $\chib$ and the bound \ref{7.210},
the second and third terms on the right in \ref{7.297} are bounded
in $L^2(S)$ by:
\begin{equation}
O(\delta^{3/2}|u|^{-6}) \label{7.300}
\end{equation}
We apply Lemma 4.7 with $p=2$ to equation \ref{7.297}. Here $r=3$,
$\nu=0$, $\gammab=0$. Taking into account the fact that $\snab^{ \
3}\log\Omega$ vanishes on $C_{u_0}$ we obtain:
\begin{eqnarray}
|u|^2\|\snab^{ \
3}\log\Omega\|_{L^2(S_{\ub,u})}&\leq&\int_{u_0}^u|u^\prime|^2\|\snab^{
\ 3}\omb\|_{L^2(S_{\ub,u^\prime})}
du^\prime\label{7.301}\\
&\s&+\int_{u_0}^u|u^\prime|^2\|\Db\sGamma\cdot\snab^{ \
2}\log\Omega+\snab\Db\sGamma\cdot\sd\omega\|_{L^2(S_{\ub,u^\prime})}
du^\prime\nonumber
\end{eqnarray}
By the bound \ref{7.300} the second integral on the right is
bounded by:
\begin{equation}
O(\delta^{3/2}|u|^{-3}) \label{7.302}
\end{equation}
We then take the $L^2$ norm of \ref{7.301} with respect to $\ub$
on $[0,\ub_1]$ to obtain:
\begin{eqnarray}
|u|^2\|\snab^{ \
3}\log\Omega\|_{L^2(C^{\ub_1}_u)}&\leq&\int_{u_0}^u|u^\prime|^2\|\snab^{
\ 3}\omb\|_{L^2(C^{\ub_1}_u)}
du^\prime\nonumber\\
&\s&+O(\delta^2|u|^{-3}) \label{7.303}
\end{eqnarray}
Substituting the estimate for $\snab^{ \ 3}\omb$ of Lemma 7.7 then
yields:
\begin{eqnarray}
\|\snab^{ \ 3}\log\Omega\|_{L^2(C^{\ub_1}_u)}&\leq&C\delta^{3/2}|u|^{-4}\Nb+O(\delta^2|u|^{-5})\nonumber\\
&\s&+O(\delta^{7/2}|u|^{-7})k \label{7.304}
\end{eqnarray}
where the coefficient of $k$ depends only on ${\cal D}_0^\infty$,
${\cal R}_0^\infty$. In reference to this estimate let:
\begin{eqnarray}
a&=&C\Nb+O(\delta^{1/2})\nonumber\\
b&=&O(1) \label{7.305}
\end{eqnarray}
so that the coefficient of $k$ in \ref{7.304} is
$\delta^{7/2}|u|^{-7}b$, with $b$ depending only on ${\cal
D}_0^\infty$, ${\cal R}_0^\infty$. The estimate \ref{7.304} then
implies:
\begin{equation}
\|\snab^{ \
3}\log\Omega\|_{L^2(C^{\ub_1}_u)}\leq\delta^{3/2}(a+\delta^2
bk)|u|^{-4} \ \ : \ \forall u\in[u_0,s^*] \label{7.306}
\end{equation}
Choosing
\begin{equation}
k=2a \label{7.307}
\end{equation}
we have
\begin{equation}
a+\delta^2 bk<2a \ \ \mbox{provided that} \ \ 2b\delta^2<1
\label{7.308}
\end{equation}
The last is a smallness condition on $\delta$ depending only on
${\cal D}_0^\infty$, ${\cal R}_0^\infty$. The estimate \ref{7.306}
then implies
\begin{equation}
\|\snab^{ \
3}\log\Omega\|_{L^2(C^{\ub_1}_u)}<k\delta^{3/2}|u|^{-4} \ \ : \
\forall u\in[u_0,s^*] \label{7.309}
\end{equation}
hence by continuity \ref{7.152} holds for some $s>s^*$
contradicting the definition of $s^*$, unless $s^*=u_1$. This
completes the proof of the lemma.

\vspace{5mm}

Since by the above lemma $D_1^{s^*}=D_1$, with $k$ as in the
statement of Lemma 7.8, the $L^2(C^{\ub_1}_u)$ estimates of Lemmas
7.5, 7.6 and 7.7 hold for all $u\in[u_0,u_1]$, in particular for $u=u_1$. Now, the right
hand sides of these inequalities are independent of $\ub_1$ or
$u_1$ and $(\ub_1,u_1)\in D^\prime$ is arbitrary.  Taking for any given
$u_1\in[u_0,c^*)$ the limit $\ub_1\rightarrow\min\{\delta,c^*-u_1\}$, we conclude that the estimates
hold with $C^{\ub_1}_{u_1}$ replaced by $C_{u_1}$, for all
$u_1\in[u_0,c^*)$. 

Moreover, by virtue of the $L^2(C^{\ub_1}_u)$
estimate for $\snab^{ \ 3}\log\Omega$ of Lemma 7.8, we can now
estimate (see \ref{7.164}):
\begin{equation}
\|\snab^{ \ 2}\rb\|_{L^2(C^{\ub_1}_u)}\leq
C\delta^{3/2}|u|^{-6}\Nb+O(\delta^2|u|^{-7}) \label{7.310}
\end{equation}
Using this bound we deduce, following the argument leading to the
estimate \ref{7.179},
\begin{equation}
\|\snab^{ \ 3}\mbox{tr}\chib^\prime\|_{L^2(C^{\ub_1}_u)}\leq
C\delta^{3/2}|u|^{-5}B^\prime+O(\delta^2|u|^{-6})
 \ \ : \ \forall u\in [u_0,u_1]
\label{7.311}
\end{equation}
where:
\begin{eqnarray}
B^\prime&=&\scD_3(\mbox{tr}\chib)+\scR_2(\beb)+\scR_1^4(\beb)+\scR_2^4(\rho)\nonumber\\
&\s&+(\scD_1^4(\mbox{tr}\chib)+\scD_1^4(\chibh)+{\cal D}_0^\infty(\chibh))(\scR_1^4(\beta)+{\cal R}_0^\infty(\beta))\nonumber\\
&\s&+{\cal
D}_0^\infty(\chibh)(\scR_2(\beta)+\scR_1^4(\beta))+(\scR_1^4(\beta)+{\cal
R}_0^\infty(\beta))^2 \label{7.312}
\end{eqnarray}

We thus arrive at the following proposition.

\vspace{5mm}

\noindent{\bf Proposition 7.3} \ \ \  The following estimates hold
for all $u\in[u_0,c^*)$:
\begin{eqnarray*}
&&\|\snab^{ \ 3}\mbox{tr}\chib^\prime\|_{L^2(C_u)}\leq C\delta^{3/2}|u|^{-5}B^\prime+O(\delta^2|u|^{-6})\\
&&\|\snab^{ \ 3}\chibh^\prime\|_{L^2(C_u)}\leq
C\delta|u|^{-4}(\scR_2(\beta)+\scR_1^4(\beta)+{\cal R}_0^\infty(\beta))+O(\delta^{3/2}|u|^{-5})\\
&&\|\snab^{ \ 3}\eta\|_{L^2(C_u)}\leq C\delta^{1/2}|u|^{-4}\tilde{M}+O(\delta|u|^{-4})\\
&&\|\snab^{ \ 3}\etb\|_{L^2(C_u)}\leq
C\delta^{1/2}|u|^{-4}\tilde{\Mb}+O(\delta|u|^{-4})\\&&\|\snab^{ \
3}\omb\|_{L^2(C_u)}\leq
C\delta^{3/2}|u|^{-5}\Nb+O(\delta^2|u|^{-6})
\end{eqnarray*}
provided that $\delta$ is suitably small depending on ${\cal
D}_0^\infty$, ${\cal R}_0^\infty$, $\scD_1^4$, $\scR_1^4$, and
$\scR_2$. Here $B^\prime$ is defined by \ref{7.312}, $\tilde{M}$
and $\tilde{\Mb}$ are defined in the statement of Lemma 7.6 and
$\Nb$ is defined in the statement of Lemma 7.7.

\section{$L^2$ estimates for $\snab^{ \ 2}\omega$ and $\snab^{ \ 3}\omega$}

As we have already remarked, the estimate of Chapter 6 for
$\snab^{ \ 2}\omega$ in $L^4(S)$ looses a factor of $\delta^{1/2}$
in behavior with respect to $\delta$ in comparison with the
$L^4(S)$ estimate for $\sd\omega$ of Chapter 4 (as well as a loss
of a factor of $|u|^{-1}$ in decay). We shall presently derive an
$L^2$ estimate for $\snab^{ \ 2}\omega$ using only the propagation
equation \ref{4.190}, in which no losses are present.

\vspace{5mm}

\noindent{\bf Proposition 7.4} \ \ \ We have:
$$\|\snab^{ \ 2}\omega\|_{L^2(C_u)}\leq C\delta^{1/2}|u|^{-3}\scR_2(\rho)+O(\delta|u|^{-4})$$
for all $u\in[u_0,c^*)$, provided that $\delta$ is suitably small
depending on ${\cal D}_0^\infty$, ${\cal R}_0^\infty$, $\scD_1^4$,
$\scR_1^4$, and $\scR_2$.

\noindent{\em Proof:} \ We consider any $(\ub_1,u_1)\in D^\prime$
and fix attention to the parameter subdomain $D_1$. We apply Lemma
4.1 to the propagation equation \ref{4.190} to obtain the
following propagation equation for $\snab^{ \ 2}\omega$:
\begin{equation}
\Db\snab^{ \ 2}\omega=\Omega^2\lb^\prime \label{7.313}
\end{equation}
where:
\begin{equation}
\lb^\prime=-\Omega^{-2}\Db\sGamma\cdot\sd\omega+2\sd\log\Omega\otimes\lb^\prime+\snab\lb
\label{7.314}
\end{equation}
To equation \ref{7.313} we apply Lemma 4.7 with $\snab^{ \
2}\omega$ in the role of $\thetab$ and $\Omega^2\lb^\prime$ in the
role of $\xib$. Then $r=2$, $\nu=0$, $\gammab=0$. In view of the
fact that $\snab^{ \ 2}\omega$ vanishes on $C_{u_0}$ we obtain,
taking $p=2$:
\begin{equation}
|u|\|\snab^{ \ 2}\omega\|_{L^2(S_{\ub,u})}\leq
C\int_{u_0}^u|u^\prime|\|\lb^\prime\|_{L^2(S_{\ub,u^\prime})}du^\prime
\label{7.315}
\end{equation}
Now, by the bound \ref{6.227} and Proposition 4.3 the first term
on the right in \ref{7.314} is bounded in $L^2(S)$ by:
\begin{equation}
O(\delta^{1/2}|u|^{-5}) \label{7.316}
\end{equation}
By the bounds \ref{7.279} and \ref{4.193} the second term on the
right in \ref{7.314} is bounded in $L^2(S)$ by:
\begin{equation}
O(\delta|u|^{-6}) \label{7.317}
\end{equation}
In regard to the last term on the right in \ref{7.314}, $\lb$ is
given by \ref{4.191}. Using the estimates of Proposition 6.2
pertaining to $\eta$, $\etb$ and the results of Chapters 3 and 4
we deduce:
\begin{equation}
\|\snab\lb\|_{L^2(S_{\ub,u})}\leq \|\snab^{ \
2}\rho\|_{L^2(S_{\ub,u})}+O(\delta^{1/2}|u|^{-5}) \label{7.318}
\end{equation}
Collecting the above results (\ref{7.316} - \ref{7.318}) we
conclude that:
\begin{equation}
\|\lb^\prime\|_{L^2(S_{\ub,u})}\leq \|\snab^{ \
2}\rho\|_{L^2(S_{\ub,u})}+O(\delta^{1/2}|u|^{-5}) \label{7.319}
\end{equation}
Substituting this in \ref{7.315} we obtain:
\begin{equation}
|u|\|\snab^{ \ 2}\omega\|_{L^2(S_{\ub,u})}\leq
C\int_{u_0}^u|u^\prime|\|\snab^{ \
2}\rho\|_{L^2(S_{\ub,u^\prime})}du^\prime +O(\delta^{1/2}|u|^{-3})
\label{7.320}
\end{equation}
Taking the $L^2$ norm with respect to $\ub$ on $[0,\ub_1]$ then
yields:
\begin{equation}
|u|\|\snab^{ \ 2}\omega\|_{L^2(C^{\ub_1}_u)}\leq C
\int_{u_0}^u|u^\prime|\|\snab^{ \
2}\rho\|_{L^2(C^{\ub_1}_{u^\prime})}du^\prime +O(\delta|u|^{-3})
\label{7.321}
\end{equation}
Since by the second of the definitions \ref{7.01}
$$\|\snab^{ \ 2}\rho\|_{L^2(C^{\ub_1}_{u^\prime})}\leq\|\snab^{ \ 2}\rho\|_{L^2(C_{u^\prime})}
\leq\delta^{1/2}|u^\prime|^{-4}\scR_2(\rho)$$ it follows that:
\begin{equation}
\|\snab^{ \ 2}\omega\|_{L^2(C^{\ub_1}_u)}\leq
C\delta^{1/2}|u|^{-3}\scR_2(\rho)+O(\delta|u|^{-4}) \label{7.322}
\end{equation}
This holds for any $u\in[u_0,u_1]$, in particular for $u=u_1$.
Since $(\ub_1,u_1)\in D^\prime$ is arbitrary, taking for any given
$u_1\in[u_0,c^*)$ the limit $\ub_1\rightarrow\min\{\delta,c^*-u_1\}$
yields the lemma.

\vspace{5mm}

We now turn to estimate $\snab^{ \ 3}\omega$ in $L^2$, using the
propagation equation \ref{6.240} for $\somega$ coupled to the
elliptic equation \ref{6.242}.

\vspace{5mm}

\noindent{\bf Proposition 7.5} \ \ \ We have:
$$\|\snab^{ \ 3}\omega\|_{L^2(C_u)}\leq C|u|^{-3}(\scR_2(\beta)+\scR_1^4(\beta))+O(\delta^{1/2}|u|^{-4})$$
for all $u\in[u_0,c^*)$, provided that $\delta$ is suitably small
depending on ${\cal D}_0^\infty$, ${\cal R}_0^\infty$, $\scD_1^4$,
$\scR_1^4$, and $\scR_2$.

\noindent{\em Proof:} \ We consider again any $(\ub_1,u_1)\in
D^\prime$ and fix attention to the parameter subdomain $D_1$.
Applying $\sd$ to the propagation equation \ref{6.240} we obtain,
in view of Lemma 1.2, the following propagation equation for the
$S$ 1-form $\sd\somega$:
\begin{equation}
\Db\sd\somega+\Omega\mbox{tr}\chib\sd\somega=-2\Omega(\chibh,\snab^{
\ 3}\omega)+m^\prime \label{7.323}
\end{equation}
where:
\begin{equation}
m^\prime=-\sd(\Omega\mbox{tr}\chib)\somega-2(\snab(\Omega\chibh),\snab^{
\ 2}\omega)+\sd m \label{7.324}
\end{equation}

We shall first estimate $m^\prime$ in $L^2(S)$. The first two
terms on the right in \ref{7.324} are estimated in $L^2(S)$ by
placing each of the two factors in $L^4(S)$ using Proposition 6.3
(see \ref{6.258}) and the results of Chapter 4. We obtain a bound
for these terms in $L^2(S)$ by:
\begin{equation}
O(|u|^{-5}) \label{7.325}
\end{equation}
To estimate $\sd m$ in $L^2(S)$, we consider the expression
\ref{6.241} for $m$. In regard to the first term on the right in
\ref{6.241} we write:
$$\sd(\sdiv(\Omega\chibh),\sd\omega)=(\sd\sdiv(\Omega\chibh),\sd\omega)+(\sdiv(\Omega\chibh),\snab^{ \ 2}\omega)$$
Placing each of the two factors of the two terms on the right in
$L^4(S)$  using Propositions 6.2 and 6.3 and the results of
Chapter 4 we obtain an $L^2(S)$ bound for the contribution of the
first term on the right in \ref{6.241} to $\sd m$ by:
\begin{equation}
O(|u|^{-5}) \label{7.326}
\end{equation}
In regard to the second term on the right in \ref{6.241} we write:
\begin{eqnarray*}
\sd\sdiv(\Omega^2\mbox{tr}\chib\beta)&=&\Omega^2\mbox{tr}\chib\sd\sdiv\beta+(\snab^{ \ 2}(\Omega\mbox{tr}\chib),\beta)\\
&\s&+\sd(\Omega^2\mbox{tr}\chib)\sdiv\beta+(\sd(\Omega^2\mbox{tr}\chib),\snab\beta)
\end{eqnarray*}
The first term on the right is bounded in $L^2(S)$ by:
\begin{equation}
C(|u|^{-1}+O(\delta|u|^{-2}))\|\snab^{ \
2}\beta\|_{L^2(S_{\ub,u})} \label{7.327}
\end{equation}
while the remaining terms are bounded in $L^2(S)$ by:
\begin{equation}
O(\delta^{1/2}|u|^{-6}) \label{7.328}
\end{equation}
using Proposition 6.2 and the results of Chapter 4. The
contributions to $\sd m$ of the third, fourth, fifth, sixth and
seventh terms on the right in \ref{6.241} are bounded in a similar
manner as the contributions to $\sd\mb$ of the corresponding terms
on the right in \ref{6.205} (see \ref{7.278}, \ref{7.280} -
\ref{7.292}, \ref{7.285} - \ref{7.286}). We obtain an $L^2(S)$
bound for the contributions of these terms to $\sd m$ by:
\begin{eqnarray}
&&O(\delta^{1/2}|u|^{-2})(\|\snab^{ \ 3}\eta\|_{L^2(S_{\ub,u})}+\|\snab^{ \ 3}\etb\|_{L^2(S_{\ub,u})})\nonumber\\
&&+O(|u|^{-3})\|\snab^{ \ 2}\etb\|_{L^2(S_{\ub,u})}\nonumber\\
&&+O(\delta^{1/2}|u|^{-2})\|\snab^{ \ 2}\beta\|_{L^2(S_{\ub,u})}\nonumber\\
&&+O(\delta^{-1/2}|u|^{-1})\|\snab^{ \ 2}\beb\|_{L^2(S_{\ub,u})}\nonumber\\
&&+O(\delta^{1/2}|u|^{-2})(\|\snab^{ \ 2}\rho\|_{L^2(S_{\ub,u})}+\|\snab^{ \ 2}\sigma\|_{L^2(S_{\ub,u})})\nonumber\\
&&+O(|u|^{-5}) \label{7.329}
\end{eqnarray}
Collecting the above results (\ref{7.325} - \ref{7.329}) we
conclude that:
\begin{eqnarray}
\|m^\prime\|_{L^2(S_{\ub,u})}&\leq&
O(\delta^{1/2}|u|^{-2})(\|\snab^{ \ 3}\eta\|_{L^2(S_{\ub,u})}+\|\snab^{ \ 3}\etb\|_{L^2(S_{\ub,u})})\nonumber\\
&\s&+O(|u|^{-3})\|\snab^{ \ 2}\etb\|_{L^2(S_{\ub,u})}\nonumber\\
&\s&+C(|u|^{-1}+O(\delta|u|^{-2}))\|\snab^{ \ 2}\beta\|_{L^2(S_{\ub,u})}\nonumber\\
&\s&+O(\delta^{-1/2}|u|^{-1})\|\snab^{ \ 2}\beb\|_{L^2(S_{\ub,u})}\nonumber\\
&\s&+O(\delta^{1/2}|u|^{-2})(\|\snab^{ \ 2}\rho\|_{L^2(S_{\ub,u})}+\|\snab^{ \ 2}\sigma\|_{L^2(S_{\ub,u})})\nonumber\\
&\s&+O(|u|^{-5}) \label{7.330}
\end{eqnarray}

Consider now equation \ref{6.242}. Setting $\theta=\sd\omega$, the
$S$ 1-form $\theta$ satisfies a Hodge system of the type
considered in Lemma 7.4 with $f=\somega-\sdiv(\Omega\beta)$,
$g=0$. By the first estimate of Lemma 7.4 we then have:
\begin{eqnarray}
\|\snab^{ \
3}\omega\|_{L^2(S_{\ub,u})}&\leq&C\left\{\|\sd\somega\|_{L^2(S_{\ub,u})}
+\|\snab^{ \ 2}\beta\|_{L^2(S_{\ub,u})}\right\}\nonumber\\
&\s&+C\delta^{-1/2}|u|^{-3}\scR_1^4(\beta)+O(|u|^{-4})
\label{7.331}
\end{eqnarray}
using the estimate \ref{6.258} (and lemma 6.4) as well as the
results of Chapter 4. It follows that in refeference to the first
term on the right in \ref{7.272} we have:
\begin{eqnarray}
&&\|\Omega(\chibh,\snab^{ \ 3}\omega)\|_{L^2(S_{\ub,u})}\leq\nonumber\\
&&\hspace{1cm}C\delta^{1/2}|u|^{-2}({\cal
D}(\chibh)+O(\delta|u|^{-3/2}))
\left\{\|\sd\somega\|_{L^2(S_{\ub,u})}+\|\snab^{ \ 2}\beta\|_{L^2(S_{\ub,u})}\right\}\nonumber\\
&&\hspace{1cm}+O(|u|^{-5}) \label{7.332}
\end{eqnarray}
We now apply Lemma 4.7 with $p=2$ to the propagation equation
\ref{7.272}. Here $r=1$, $\nu=-2$, $\gamma=0$ and we obtain:
\begin{eqnarray}
|u|^2\|\sd\somega\|_{L^2(S_{\ub,u})}&\leq&C|u_0|^2\|\somega\|_{L^2(S_{\ub,u_0})}\label{7.333}\\
&\s&+C\int_{u_0}^u|u^\prime|^2\left\|-2\Omega(\chibh,\snab^{ \
3}\omega)+m^\prime\right\|_{L^2(S_{\ub,u^\prime})}du^\prime\nonumber
\end{eqnarray}
Substituting the estimates \ref{7.332} and \ref{7.330} and noting
that in view of the fact that $\omega$ vanishes on $C_{u_0}$ we
have, from \ref{6.239},
\begin{equation}
\|\sd\somega\|_{L^2(S_{\ub,u_0})}\leq C\|\snab^{ \
2}\beta\|_{L^2(S_{\ub,u_0})}+O(\delta^{1/2}|u_0|^{-5})
\label{7.334}
\end{equation}
we obtain the following linear integral inequality for the
quantity $|u|^2\|\sd\somega\|_{L^2(S_{ub,u})}$:
\begin{equation}
|u|^2\|\sd\somega\|_{L^2(S_{\ub,u})}\leq\int_{u_0}^u\ab(u^\prime)|u^\prime|^2\|\sd\somega\|_{L^2(S_{\ub,u^\prime})}
du^\prime+\bb(\ub,u) \label{7.335}
\end{equation}
where:
\begin{eqnarray}
\ab(u)&=&C\delta^{1/2}|u|^{-2}({\cal D}_0^\infty(\chibh)+O(\delta|u|^{-3/2}))\label{7.336}\\
\bb(\ub,u)&=&\int_{u_0}^u\left\{O(\delta^{1/2})(\|\snab^{ \
3}\eta\|_{L^2(S_{\ub,u^\prime})}
+\|\snab^{ \ 3}\etb\|_{L^2(S_{\ub,u^\prime})})\right.\nonumber\\
&\s&\s\s\s+O(|u^\prime|^{-1})\|\snab^{ \ 2}\etb\|_{L^2(S_{\ub,u^\prime})}\nonumber\\
&\s&\s\s\s+C(|u^\prime|+O(\delta^{1/2}))\|\snab^{ \ 2}\beta\|_{L^2(S_{\ub,u^\prime})}\nonumber\\
&\s&\s\s\s+O(\delta^{1/2})(\|\snab^{ \
2}\rho\|_{L^2(S_{\ub,u^\prime})}
+\|\snab^{ \ 2}\sigma\|_{L^2(S_{\ub,u^\prime})})\nonumber\\
&\s&\s\s\s\left.+O(\delta^{-1/2}|u^\prime|)\|\snab^{ \ 2}\beb\|_{L^2(S_{\ub,u^\prime})}+O(|u^\prime|^{-3})\right\}du^\prime\nonumber\\
&\s&+C|u_0|^2\|\snab^{ \
2}\beta\|_{L^2(S_{\ub,u_0})}+O(\delta^{1/2}|u_0|^{-3})
\label{7.337}
\end{eqnarray}
At fixed $\ub$ the inequality \ref{7.335} is of the form
\ref{6.81} with the quantity
$|u|^2\|\sd\somega\|_{L^2(S_{\ub,u})}$ in the role of $\xb(u)$.
Note that the function $\bb(\ub,u)$ is non-decreasing in $u$.
Moreover, we have
$$\int_{u_0}^u\ab(u^\prime)du^\prime\leq\log 2$$
provided that $\delta$ is suitably small depending on ${\cal
D}_0^\infty$, ${\cal R}_0^\infty$. It follows that \ref{6.89}
holds, that is:
\begin{equation}
|u|^2\|\sd\somega\|_{L^2(S_{\ub,u})}\leq 2\bb(\ub,u) \label{7.338}
\end{equation}
taking the $L^2$ norm of this inequality with respect to $\ub$ on
$[0,\ub_1]$ we then obtain:
\begin{equation}
|u|^2\|\sd\somega\|_{L^2(C^{\ub_1}_u)}\leq
2\|\bb(\cdot,u)\|_{L^2([0,\ub_1])} \label{7.339}
\end{equation}
and from \ref{7.337} we have:
\begin{eqnarray}
\|\bb(\cdot,u)\|_{L^2([0,\ub_1])}&=&\int_{u_0}^u\left\{O(\delta^{1/2})(\|\snab^{
\ 3}\eta\|_{L^2(C^{\ub_1}_{u^\prime})}
+\|\snab^{ \ 3}\etb\|_{L^2(C^{\ub_1}_{u^\prime})})\right.\nonumber\\
&\s&\s\s\s+O(|u^\prime|^{-1})\|\snab^{ \ 2}\etb\|_{L^2(C^{\ub_1}_{u^\prime})}\nonumber\\
&\s&\s\s\s+C(|u^\prime|+O(\delta^{1/2}))\|\snab^{ \ 2}\beta\|_{L^2(C^{\ub_1}_{u^\prime})}\nonumber\\
&\s&\s\s\s+O(\delta^{1/2})(\|\snab^{ \
2}\rho\|_{L^2(C^{\ub_1}_{u^\prime})}
+\|\snab^{ \ 2}\sigma\|_{L^2(C^{\ub_1}_{u^\prime})})\nonumber\\
&\s&\s\s\s\left.+O(\delta^{-1/2}|u^\prime|)\|\snab^{ \ 2}\beb\|_{L^2(C^{\ub_1}_{u^\prime})}+O(\delta^{1/2}|u^\prime|^{-3})\right\}du^\prime\nonumber\\
&\s&+C|u_0|^2\|\snab^{ \
2}\beta\|_{L^2(C^{\ub_1}_{u_0})}+O(\delta|u_0|^{-3}) \label{7.340}
\end{eqnarray}
Substituting the second estimate of proposition 7.1, the estimates
of Proposition 7.3 pertaining to $\eta$, $\etb$, yields, in view
of the definitions \ref{7.01},
\begin{equation}
\|\bb(\cdot,u)\|_{L^2([0,\ub_1])}\leq
C|u|^{-1}\scR_2(\beta)+O(\delta^{1/2}|u|^{-2}) \label{7.341}
\end{equation}
therefore:
\begin{equation}
\|\sd\somega\|_{L^2(C^{\ub_1}_u)}\leq
C|u|^{-3}\scR_2(\beta)+O(\delta^{1/2}|u|^{-4}) \label{7.342}
\end{equation}
This holds for any $u\in[u_0,u_1]$, in particular for $u=u_1$.
Since $(\ub_1,u_1)\in D^\prime$ is arbitrary, taking for any given
$u_1\in[u_0,c^*)$ the limit $\ub_1\rightarrow\min\{\delta,c^*-u_1\}$
yields:
\begin{equation}
\|\sd\somega\|_{L^2(C_u)}\leq
C|u|^{-3}\scR_2(\beta)+O(\delta^{1/2}|u|^{-4}) \label{7.343}
\end{equation}
for all $u\in[u_0,c^*)$. Taking the $L^2$ norm of \ref{7.331} with
respect to $\ub$ on $[0,\delta)$, if $u\in[u_0,c^*-\delta]$, on
$[0,c^*-u)$, if $u\in(c^*-\delta,c^*)$, we obtain:
\begin{eqnarray}
\|\snab^{ \ 3}\omega\|_{L^2(C_u)}&\leq& C\|\sd\somega\|_{L^2(C_u)}\label{7.344}\\
&\s&+C|u|^{-3}(\scR_2(\beta)+\scR_1^4(\beta))+O(\delta^{1/2}|u|^{-4})\nonumber
\end{eqnarray}
Substituting then the estimate \ref{7.343} yields the proposition.

\section{$L^2$ estimates for $\sd D\omega$, $\sd\Db\omb$, $D^2\omega$, $\Db^2\omb$}

\noindent{\bf Proposition 7.6} \ \ \ We have:
$$\|\sd D\omega\|_{L^2(C_u)}\leq \delta^{-1/2}|u|^{-2}\scR_1(D\rho)+O(\delta^{1/2}|u|^{-3})$$
for all $u\in[u_0,c^*)$, provided that $\delta$ is suitably small
depending on ${\cal D}_0^\infty$, ${\cal R}_0^\infty$, $\scD_1^4$,
$\scR_1^4$, and $\scR_2$.

\noindent{\em Proof:} \ We consider any $(\ub_1,u_1)\in D^\prime$
and fix attention to the parameter subdomain $D_1$. Applying $\sd$
to the propagation equation \ref{4.196} we obtain, in view of
Lemma 1.2 the following propagation equation for $\sd D\omega$:
\begin{equation}
\Db(\sd D\omega)=\Omega^2\nb^\prime \label{7.345}
\end{equation}
where:
\begin{equation}
\nb^\prime=2(\sd\log\Omega)\nb+\sd\nb \label{7.346}
\end{equation}
To equation \ref{7.345} we apply Lemma 4.7 with $\sd D\omega$ in
the role of $\thetab$ and $\Omega^2\nb^\prime$ in the role of
$\xib$. Then $r=1$, $\nu=0$, $\gammab=0$. In view of the fact that
$\sd D\omega$ vanishes on $C_{u_0}$ we obtain, taking $p=2$:
\begin{equation}
\|\sd D\omega\|_{L^2(S_{\ub,u})}\leq
C\int_{u_0}^u\|\nb^\prime\|_{L^2(S_{\ub,u^\prime})}du^\prime
\label{7.347}
\end{equation}
Taking the $L^2$ norm of this inequality with respect to $\ub$ on
$[0,\ub_1]$ yields:
\begin{equation}
\|\sd D\omega\|_{L^2(C^{\ub_1}_u)}\leq
C\int_{u_0}^u\|\nb^\prime\|_{L^2(C^{\ub_1}_{u^\prime})}du^\prime
\label{7.348}
\end{equation}
Now, by the bounds \ref{7.279} and \ref{4.199} the first term on
the right in \ref{7.346} is bounded in $L^2(S)$ by:
$$O(|u|^{-4})$$
therefore in $L^2(C^{\ub_1}_u)$ by:
\begin{equation}
O(\delta^{1/2}|u|^{-4}) \label{7.349}
\end{equation}
In regard to $\sd\nb$, the second term on the right in
\ref{7.346}, $\nb$ is given by \ref{4.197}. By the first of the
definitions \ref{7.02} the contribution to $\sd\nb$ of the first
term on the right in \ref{4.197} is bounded in $L^2(C^{\ub_1}_u)$
by:
\begin{equation}
\delta^{-1/2}|u|^{-3}\scR_1(D\rho) \label{7.350}
\end{equation}
By Proposition 7.4 and the results of Chapters 3 and 4 the
contribution to $\sd\nb$ of the second term on the right in
\ref{4.197} is bounded in $L^2(C^{\ub_1}_u)$ by:
\begin{equation}
O(\delta|u|^{-5}) \label{7.351}
\end{equation}
By the results of Chapters 3 and 4 the contribution to $\sd\nb$ of
the remaining terms on the right in \ref{4.197} is bounded in
$L^2(S)$ by:
$$O(|u|^{-4})$$
therefore in $L^2(C^{\ub_1}_u)$ by:
\begin{equation}
O(\delta^{1/2}|u|^{-4}) \label{7.352}
\end{equation}
Combining \ref{7.350} - \ref{7.352} we obtain:
\begin{equation}
\|\sd\nb\|_{L^2(C^{\ub_1}_u)}\leq
\delta^{-1/2}|u|^{-3}\scR_1(D\rho)+O(\delta^{1/2}|u|^{-4})
\label{7.353}
\end{equation}
Combining the results \ref{7.349} and \ref{7.353} yields:
\begin{equation}
\|\nb^\prime\|_{L^2(C^{\ub_1}_u)}\leq
\delta^{-1/2}|u|^{-3}\scR_1(D\rho)+O(\delta^{1/2}|u|^{-4})
\label{7.354}
\end{equation}
We substitute this bound in \ref{7.348} to obtain:
\begin{equation}
\|\sd D\omega\|_{L^2(C^{\ub_1}_u)}\leq
\delta^{-1/2}|u|^{-2}\scR_1(D\rho)+O(\delta^{1/2}|u|^{-3})
\label{7.355}
\end{equation}
This holds for any $u\in[u_0,u_1]$, in particular for $u=u_1$.
Since $(\ub_1,u_1)\in D^\prime$ is arbitrary, taking for any given
$u_1\in[u_0,c^*)$ the limit $\ub_1\rightarrow\min\{\delta,c^*-u_1\}$
yields the proposition.

\vspace{5mm}

\noindent{\bf Proposition 7.7} \ \ \ We have:
$$\|\sd\Db\omb\|_{L^2(S_{\ub,u})}\leq C\delta|u|^{-4}\scR_1^4(\rho)+O(\delta^2|u|^{-5})$$
for all $(\ub,u)\in D^\prime$, provided that $\delta$ is suitably
small depending on ${\cal D}_0^\infty$, ${\cal R}_0^\infty$,
$\scD_1^4$, $\scR_1^4$, and $\scR_2$.

\noindent{\em Proof:} \ Applying $\sd$ to the propagation equation
\ref{4.200}, we obtain, in view of Lemma 1.2, the following
propagation equation for $\sd\Db\omb$:
\begin{equation}
D(\sd\Db\omb)=\Omega^2 n^\prime \label{7.356}
\end{equation}
where:
\begin{equation}
n^\prime=2(\sd\log\Omega)n+\sd n \label{7.357}
\end{equation}
To equation \ref{7.356} we apply Lemma 4.6 with $\sd\Db\omb$ in
the role of $\theta$ and $\Omega^2 n^\prime$ in the role of $\xi$.
Here $r=1$, $\nu=0$, $\gamma=0$. Taking $p=2$ we obtain:
\begin{eqnarray}
\|\sd\Db\omb\|_{L^2(S_{\ub,u})}&\leq& C\int_0^{\ub}\|n^\prime\|_{L^2(S_{\ub^\prime,u})}d\ub^\prime\nonumber\\
&\leq& C\delta^{1/2}\|n^\prime\|_{L^2(C_u)} \label{7.358}
\end{eqnarray}
Now, by the bounds \ref{7.279} and \ref{4.203} in which
\ref{4.205} is substituted, the first term on the right in
\ref{7.357} is bounded in $L^2(S)$ by:
$$O(\delta|u|^{-6})$$
therefore in $L^2(C_u)$ by:
\begin{equation}
O(\delta^{3/2}|u|^{-6}) \label{7.359}
\end{equation}
In regard to $\sd n$, the second term on the right in \ref{7.357},
$n$ is given by \ref{4.201}. The contribution to $\sd n$ of the
first term on the right in \ref{4.201} is bounded in $L^2(C_u)$
by:
\begin{equation}
\|\sd\Db\rho\|_{L^2(C_u)}\leq
C\delta^{1/2}|u|^{-4}\scR_1^4(\rho)+O(\delta^{3/2}|u|^{-5})
\label{7.360}
\end{equation}
from \ref{4.204} and the fourth of the definitions \ref{7.01}. By
the estimate for $\omb$ of Proposition 6.2 and the results of
Chapters 3 and 4 the contribution to $\sd n$ of the second term on
the right in \ref{4.201} is bounded in $L^2(S)$ by:
$$O(\delta^{3/2}|u|^{-6})$$
therefore in $L^2(C_u)$ by:
\begin{equation}
O(\delta^2|u|^{-6}) \label{7.361}
\end{equation}
By the results of Chapters 3 and 4 the contribution of the
remaining terms on the right in \ref{4.201} is bounded in $L^2(S)$
by:
$$O(\delta|u|^{-5})$$
therefore in $L^2(C_u)$ by:
\begin{equation}
O(\delta^{3/2}|u|^{-5}) \label{7.362}
\end{equation}
Combining \ref{7.360} - \ref{7.362} we obtain:
\begin{equation}
\|\sd n\|_{L^2(C_u)}\leq
C\delta^{1/2}|u|^{-4}\scR_1^4(\rho)+O(\delta^{3/2}|u|^{-5})
\label{7.363}
\end{equation}
Combining the results \ref{7.359} and \ref{7.363} yields:
\begin{equation}
\|n^\prime\|_{L^2(C_u)}\leq
C\delta^{1/2}|u|^{-4}\scR_1^4(\rho)+O(\delta^{3/2}|u|^{-5})
\label{7.364}
\end{equation}
Substituting this in \ref{7.358} yields the proposition.

\vspace{5mm}

\noindent{\bf Proposition 7.8} \ \ \ We have:
$$\|D^2\omega\|_{L^2(C_u)}\leq C\delta^{-3/2}|u|^{-1}{\cal R}_0(D^2\rho)+O(\delta^{-1/2}|u|^{-2})$$
for all $u\in[u_0,c^*)$, provided that $\delta$ is suitably small
depending on ${\cal D}_0^\infty$, ${\cal R}_0^\infty$, $\scD_1^4$,
$\scR_1^4$, and $\scR_2$.

\noindent{\em Proof:} \ We consider any $(\ub_1,u_1)\in D^\prime$
and fix attention to the parameter subdomain $D_1$. By the
commutation formula \ref{1.75} we have:
\begin{equation}
\Db D^2\omega=D\Db D\omega+2\Omega^2(\eta-\etb)^\sharp\cdot\sd
D\omega \label{7.365}
\end{equation}
Applying $D$ to equation \ref{4.196} we then obtain the following
propagation equation for $D^2\omega$:
\begin{equation}
\Db(D^2\omega)=\Omega^2\dot{\nb} \label{7.366}
\end{equation}
where:
\begin{equation}
\dot{\nb}=2(\eta-\etb)^\sharp\cdot\sd D\omega+2\omega\nb+D\nb
\label{7.367}
\end{equation}
To equation \ref{7.366} we apply Lemma 4.7 with $D^2\omega$ in the
role of $\thetab$ and $\Omega^2\dot{\nb}$ in the role of $\xib$.
Then $r=0$, $\nu=0$, $\gammab=0$. In view of the fact that
$D^2\omega$ vanishes on $C_{u_0}$ we obtain, taking $p=2$:
\begin{equation}
|u|^{-1}\|D^2\omega\|_{L^2(S_{\ub,u})}\leq
C\int_{u_0}^u|u^\prime|^{-1}\|\dot{\nb}\|_{L^2(S_{\ub,u^\prime})}du^\prime
\label{7.368}
\end{equation}
Taking the $L^2$ norm of this inequality with respect to $\ub$ on
$[0,\ub_1]$ yields:
\begin{equation}
|u|^{-1}\|D^2\omega\|_{L^2(C^{\ub_1}_u)}\leq
C\int_{u_0}^u|u^\prime|^{-1}\|\dot{\nb}\|_{L^2(C^{\ub_1}_{u^\prime})}du^\prime
\label{7.369}
\end{equation}
Now, by Proposition 7.6 and the results of Chapter 3 the first
term on the right in \ref{7.367} is bounded in $L^2(C^{\ub_1}_u)$
by:
\begin{equation}
O(|u|^{-4}) \label{7.370}
\end{equation}
Also, by the bound \ref{4.199} and the results of Chapter 3 the
second term on the right in \ref{7.367} is bounded in $L^2(S)$ by:
$$O(\delta^{-1}|u|^{-3})$$
therefore in $L^2(C^{\ub_1}_u)$ by:
\begin{equation}
O(\delta^{-1/2}|u|^{-3}) \label{7.371}
\end{equation}
By the third of the definitions \ref{7.02} the contribution to
$D\nb$ of the first term on the right in \ref{4.197} is bounded in
$L^2(C^{\ub_1}_u)$ by:
\begin{equation}
\delta^{-3/2}|u|^{-2}{\cal R}_0(D^2\rho) \label{7.372}
\end{equation}
By Proposition 7.6, equations \ref{1.66} and \ref{1.150} and the
results of Chapters 3 and 4 the contribution to $D\nb$ of the
second term on the right in \ref{4.197} is bounded in
$L^2(C^{\ub_1}_u)$ by:
\begin{equation}
O(|u|^{-4}) \label{7.373}
\end{equation}
Now, the third of the Bianchi identities of Proposition 1.2 reads:
\begin{equation}
D\beta+\frac{3}{2}\Omega\mbox{tr}\chi\beta-\Omega\chih^\sharp\cdot\beta-\omega\beta
=\Omega\left\{\sdiv\alpha+(\etb^\sharp+2\zeta^\sharp)\cdot\alpha\right\}
\label{7.374}
\end{equation}
In view of the results of Chapter 3 it follows that:
\begin{equation}
\|D\beta\|_{L^4(S_{\ub,u})}\leq
C\delta^{-3/2}|u|^{-3/2}\scR_1^4(\alpha)+O(\delta^{-1}|u|^{-5/2})
\label{7.375}
\end{equation}
Also, the seventh of the Bianchi identities of Proposition 1.2
reads:
\begin{equation}
D\rho+\frac{3}{2}\omega\mbox{tr}\chi\rho=\omega\left\{\sdiv\beta+(2\etb+\zeta,\beta)-\frac{1}{2}(\chibh,\alpha)\right\}
\label{7.376}
\end{equation}
In view of the results of Chapter 3 it follows that:
\begin{equation}
\|D\rho\|_{L^4(S_{\ub,u})}\leq C\delta^{-1}|u|^{-5/2}{\cal
D}(\chibh){\cal R}_0^\infty(\alpha)+O(\delta^{-1/2}|u|^{-5/2})
\label{7.377}
\end{equation}
Using the estimates \ref{7.375} and \ref{7.377} together with
equations \ref{1.39}, \ref{1.66} and \ref{1.150} and the results
of Chapters 3 and 4 the contribution to $D\nb$ of the remaining
terms on the right in \ref{4.197} is shown to be bounded in
$L^2(S)$ by:
$$O(\delta^{-1}|u|^{-3})$$
therefore in $L^2(C^{\ub_1}_u)$ by:
\begin{equation}
O(\delta^{-1/2}|u|^{-3}) \label{7.378}
\end{equation}
Combining \ref{7.372}, \ref{7.373} and  \ref{7.378} we obtain:
\begin{equation}
\|D\nb\|_{L^2(C^{\ub_1}_u)}\leq \delta^{-3/2}|u|^{-2}{\cal
R}_0(D^2\rho)+O(\delta^{-1/2}|u|^{-3}) \label{7.379}
\end{equation}
Combining the results \ref{7.370}, \ref{7.371} and \ref{7.379}
yields:
\begin{equation}
\|\dot{\nb}\|_{L^2(C^{\ub_1}_u)}\leq \delta^{-3/2}|u|^{-2}{\cal
R}_0(D^2\rho)+O(\delta^{-1/2}|u|^{-3}) \label{7.380}
\end{equation}
We substitute this bound in \ref{7.348} to obtain:
\begin{equation}
\|D^2\omega\|_{L^2(C^{\ub_1}_u)}\leq C\delta^{-3/2}|u|^{-1}{\cal
R}_0(D^2\rho)+O(\delta^{-1/2}|u|^{-2}) \label{7.381}
\end{equation}
This holds for any $u\in[u_0,u_1]$, in particular for $u=u_1$.
Since $(\ub_1,u_1)\in D^\prime$ is arbitrary, taking for any given
$u_1\in[u_0,c^*)$ the limit $\ub_1\rightarrow\min\{\delta,c^*-u_1\}$
yields the proposition.

\vspace{5mm}

In the following proposition we assume, in addition to the
previous assumptions on the curvature components, that also the
quantity:
\begin{equation}
{\cal R}_0^4(\Dbh\alb)=\sup_{(\ub,u)\in
D^\prime}\left(|u|^5\delta^{-3/2}\|\Dbh\alb\|_{L^4(S_{\ub,u})}\right)
\label{7.04}
\end{equation}
is finite. By the results of Chapter 2, the corresponding quantity
on $C_{u_0}$, obtained by replacing the supremum on $D^\prime$ by
the supremum on $([0,\delta]\times\{u_0\})\bigcap D^\prime$, is
bounded by a non-negative non-decreasing continuous function of
$M_7$.

\noindent{\bf Proposition 7.9} \ \ \ We have:
\begin{eqnarray*}
\|\Db^2\omb\|_{L^2(S_{\ub,u})}&\leq&C\delta|u|^{-4}{\cal
R}_0^\infty(\rho)+C\delta^2|u|^{-5}\left[\scR_1(\Db\beb)
+{\cal R}_0^\infty(\alpha){\cal R}_0^4(\Dbh\alb)\right]\\
&\s&+O(\delta^2|u|^{-5})
\end{eqnarray*}
for all $(\ub,u)\in D^\prime$, provided that $\delta$ is suitably
small depending on ${\cal D}_0^\infty$, ${\cal R}_0^\infty$,
$\scD_1^4$, $\scR_1^4$, and $\scR_2$.

\noindent{\em Proof:} \ By the commutation formula \ref{1.75} we
have:
\begin{equation}
D\Db^2\omb=\Db D\Db\omb-2\Omega^2(\eta-\etb)^\sharp\cdot\sd\Db\omb
\label{7.382}
\end{equation}
Applying $\Db$ to equation \ref{4.200} we then obtain the
following propagation equation for $\Db^2\omb$:
\begin{equation}
D(\Db^2\omb)=\Omega^2\dot{n} \label{7.383}
\end{equation}
where:
\begin{equation}
\dot{n}=-2(\eta-\etb)^\sharp\cdot\sd\Db\omb+2\omb n+\Db n
\label{7.384}
\end{equation}
To equation \ref{7.383} we apply Lemma 4.6 with $\Db^2\omb$ in the
role of $\theta$ and $\Omega^2\dot{n}$ in the role of $\xi$. Here
$r=0$, $\nu=0$, $\gamma=0$. Taking $p=2$ we obtain:
\begin{eqnarray}
\|\Db^2\omb\|_{L^2(S_{\ub,u})}&\leq& C\int_0^{\ub}\|\dot{n}\|_{L^2(S_{\ub^\prime,u})}d\ub^\prime\nonumber\\
&\leq& C\delta^{1/2}\|\dot{n}\|_{L^2(C_u)} \label{7.385}
\end{eqnarray}
Now by Proposition 7.7 and the results of Chapter 3 the first term
on the right in \ref{7.384} is bounded in $L^2(S)$ by:
$$O(\delta^{3/2}|u|^{-6})$$
therefore in $L^2(C_u)$ by:
\begin{equation}
O(\delta^2|u|^{-6}) \label{7.386}
\end{equation}
Also, by the bound \ref{4.203} in which \ref{4.205} is substituted
and the results of Chapter 3, the second term on the right in
\ref{7.384} is bounded in $L^2(S)$ by:
$$O(\delta|u|^{-6})$$
therefore in $L^2(C_u)$ by:
\begin{equation}
O(\delta^{3/2}|u|^{-6}) \label{7.387}
\end{equation}
Now, the fourth of the Bianchi identities of Proposition 1.2
reads:
\begin{equation}
\Db\beb+\frac{3}{2}\Omega\mbox{tr}\chib\beb-\Omega\chibh^\sharp\cdot\beb-\omb\beb=
-\Omega\left\{\sdiv\alb+(\eta^\sharp-2\zeta^\sharp)\cdot\alb\right\}
\label{7.388}
\end{equation}
In view of the results of Chapter 3 it follows that:
\begin{equation}
\|\Db\beb\|_{L^4(S_{\ub,u})}\leq C\delta|u|^{-9/2}{\cal
R}_0^\infty(\beb)+O(\delta^{3/2}|u|^{-5}) \label{7.389}
\end{equation}
The contribution to $\Db n$ of the first term on the right in
\ref{4.201} is $-\Db^2\rho$. This can be expressed in terms of
$\sdiv\Db\beb$, $\Db\alb$ and $\Db\rho$ by applying $\Db$ to
equation \ref{4.204} and using equations \ref{1.67}, \ref{1.147},
\ref{1.149}, \ref{3.7}, and the conjugate of the commutation
formula \ref{6.107}, which reads, for an arbitrary $S$ 1-form
$\theta$:
\begin{equation}
\Db\sdiv\theta-\sdiv\Db\theta=-2\sdiv(\Omega\chibh^\sharp\cdot\theta)-\Omega\mbox{tr}\chib\sdiv\theta
\label{7.390}
\end{equation}
Using this expression and the bound \ref{4.205} we deduce:
\begin{eqnarray}
\|\Db^2\rho\|_{L^2(C_u)}&\leq&C\delta^{1/2}|u|^{-4}{\cal
R}_0^\infty(\rho)+C\delta^{3/2}|u|^{-5}\left[\scR_1(\Db\beb)
+{\cal R}_0^\infty(\alpha){\cal R}_0^4(\Db\alb)\right]\nonumber\\
&\s&+O(\delta^{3/2}|u|^{-5})\label{7.391}
\end{eqnarray}
By Proposition 7.7, equations \ref{1.67} and \ref{1.149} and the
results of Chapters 3 and 4 the contribution to $\Db n$ of the
second term on the right in \ref{4.201} is bounded in $L^2(S)$ by:
$$O(\delta^{3/2}|u|^{-6})$$
therefore in $L^2(C_u)$ by:
\begin{equation}
O(\delta^2|u|^{-6}) \label{7.392}
\end{equation}
Using the estimates \ref{7.389} and \ref{4.205} together with
equations \ref{1.44}, \ref{1.67} and \ref{1.149} and the results
of Chapters 3 and 4 the contribution to $\Db n$ of the remaining
terms on the right in \ref{4.201} is shown to be bounded in
$L^2(S)$ by:
$$O(\delta|u|^{-5})$$
therefore in $L^2(C_u)$ by:
\begin{equation}
O(\delta^{3/2}|u|^{-5}) \label{7.393}
\end{equation}
Combining \ref{7.391} - \ref{7.393} we obtain:
\begin{eqnarray}
\|\Db n\|_{L^2(C_u)}&\leq&C\delta^{1/2}|u|^{-4}{\cal
R}_0^\infty(\rho)+C\delta^{3/2}|u|^{-5}\left[\scR_1(\Db\beb)
+{\cal R}_0^\infty(\alpha){\cal R}_0^4(\Dbh\alb)\right]\nonumber\\
&\s&+O(\delta^{3/2}|u|^{-5}) \label{7.394}
\end{eqnarray}
Combining the results \ref{7.386}, \ref{7.387} and \ref{7.394}
yields:
\begin{eqnarray}
\|\dot{n}\|_{L^2(C_u)}&\leq&C\delta^{1/2}|u|^{-4}{\cal
R}_0^\infty(\rho)+C\delta^{3/2}|u|^{-5}\left[\scR_1(\Db\beb)
+{\cal R}_0^\infty(\alpha){\cal R}_0^4(\Dbh\alb)\right]\nonumber\\
&\s&+O(\delta^{3/2}|u|^{-5}) \label{7.395}
\end{eqnarray}
Finally, substituting this in \ref{7.385} yields the proposition.

\chapter{The Multiplier Fields and the Commutation Fields}

\section{Introduction}

There are two kinds of vectorfields which play a fundamental role
in our approach. The {\em multiplier fields} and the {\em
commutation fields}. The two kinds of vectorfields play distinct
roles.

The multiplier vectorfields are used in conjunction with the {\em
Bel-Robinson tensor} associated to a {\em Weyl field} to construct
positive quantities, the {\em energies} and {\em fluxes}. This
construction shall be presented in Chapter 12. It is the energies
and fluxes which are used to control the solution. Their
positivity requires the multiplier fields to be non-spacelike and
future directed. We take as multiplier fields the vectorfields $L$
and $K$.  We are already familiar with the vectorfield $L$ and its
conjugate $\Lb$. The vectorfield $K$ is defined by:
\begin{equation}
K=u^2\Lb \label{8.1}
\end{equation}
and corresponds, in the present context, to the generator of
inverted time translations in Minkowski spacetime.

The commutation fields are used to generate the higher order Weyl
fields from the fundamental Weyl field, the curvature of the
Ricci-flat metric. These higher order Weyl fields are {\em
variations} of the fudamental Weyl field, modified Lie derivatives
of the curvature with respect to the commutation fields. The
modification is dictated by the conformal properties that a Weyl
field is to possess. This shall also be presented in Chapter 12.
The basic requirement on the set of commutation fields is that it
spans the tangent space to the spacetime manifold at each point.
We take as commutation fields the vectorfields $L$, $S$ and the
three rotation fields $O_i \ :i=1,2,3$. The vectorfield $S$ is
defined by:
\begin{equation}
S=u\Lb+\ub L \label{8.2}
\end{equation}
and corresponds to the generator of scale transformations in
Minkowski spacetime. It is spacelike in $M^\prime$, the
non-trivial part of the spacetime manifold, timelike and past
directed in the Minkowskian region $M_0$. The rotation fields,
which shall be constructed in the 3rd section of the present
chapter, generate an action of the 3-dimensional rotation group
$SO(3)$ on the spacetime manifold. They are spacelike, being
tangential to the surfaces $S_{\ub,u}$. Note on the other hand
that the vectorfields $L$, $S$ and $K$ are orthogonal to the
surfaces $S_{\ub,u}$. Note also that the vectorfield $L$ plays a
dual role, for, it plays the role of a commutation field as well
as that of a multiplier field.

The growth of the energies and fluxes shall be shown in Chapter 12
to depend on the {\em deformation tensors} of the multiplier and
commutation fields. For any vectorfield $X$ on spacetime, we
denote by $\s^{(X)}\pi$ the Lie derivative of the spacetime metric
$g$ with respect to $X$:
\begin{equation}
\s^{(X)}\pi={\cal L}_X g \label{8.3}
\end{equation}
The {\em deformation tensor} of $X$ is then the trace-free part of
$\s^{(X)}\pi$, which we denote by $\s^{(X)}\tilde{\pi}$:
\begin{equation}
\s^{(X)}\tilde{\pi}=\s^{(X)}\pi-\frac{1}{4}g\mbox{tr}\s^{(X)}\pi
\label{8.4}
\end{equation}
The deformation tensor of a vectorfield $X$ measures the variation
of the conformal geometry under the 1-parameter group of
diffeomorphisms generated by $X$, the trace part of $\s^{(X)}\pi$
giving rise only to variations of $g$ within its conformal class:
\begin{equation}
{\cal L}_X d\mu_g=\frac{1}{2}\mbox{tr}\s^{(X)}\pi d\mu_g
\label{8.5}
\end{equation}
$d\mu_g$ being the volume form of $g$. The reason why only
variations of the conformal geometry affect the growth of the
energies and fluxes lies in the conformal properties of the {\em
Bianchi equations} satisfied by a Weyl field, which are simply the
Bianchi identities for a Ricci-flat metric in the case of the
fundamental Weyl field, the spacetime curvature.

For an arbitrary vectorfield $X$ on spacetime, the associated
tensorfield $\s^{(X)}\pi$ decomposes into the symmetric
2-covariant $S$ tensorfield $\s^{(X)}\spi$, the restriction of
$\s^{(X)}\pi$ to the $S_{\ub,u}$, given on each $S_{\ub,u}$ by:
\begin{equation}
\s^{(X)}\spi(Y,Z)=\s^{(X)}\pi(Y,Z) \ \ \ : \ \forall Y,Z\in T_p
S_{\ub,u} \ \forall p\in S_{\ub,u} \label{8.6}
\end{equation}
the $S$ 1-forms $\s^{(X)}\spi_3$, $\s^{(X)}\spi_4$, given by:
\begin{equation}
\s^{(X)}\spi_3(Y)=\s^{(X)}\pi(\hat{\Lb},Y), \ \ \
\s^{(X)}\spi_4(Y)=\s^{(X)}\pi(\hat{L},Y) \ \ \ : \ \forall Y\in
T_p(S_{\ub,u}) \ \forall p\in S_{\ub,u} \label{8.7}
\end{equation}
and the functions $\s^{(X)}\pi_{33}$, $\s^{(X)}\pi_{44}$ and
$\s^{(X)}\pi_{34}$, given by:
\begin{equation}
\s^{(X)}\pi_{33}=\s^{(X)}\pi(\hat{\Lb},\hat{\Lb}), \ \ \
\s^{(X)}\pi_{44}=\s^{(X)}\pi(\hat{L},\hat{L}), \ \ \
\s^{(X)}\pi_{34}=\s^{(X)}\pi(\hat{\Lb},\hat{L}) \label{8.8}
\end{equation}
The deformation tensor of $X$ similarly decomposes into the
symmetric 2-covariant $S$ tensorfield $\s^{(X)}\tilde{\spi}$, the
$S$ 1-forms $\s^{(X)}\tilde{\spi}_3$, $\s^{(X)}\tilde{\spi}_4$,
and the functions $\s^{(X)}\tilde{\pi}_{33}$,
$\s^{(X)}\tilde{\pi}_{44}$ and $\s^{(X)}\tilde{\pi}_{34}$. We
have:
\begin{eqnarray}
&&\s^{(X)}\tilde{\spi}=\s^{(X)}\spi-\frac{1}{4}\sg\mbox{tr}\s^{(X)}\pi\nonumber\\
&&\s^{(X)}\tilde{\spi}_3=\s^{(X)}\spi_3, \ \ \ \s^{(X)}\tilde{\spi}_4=\s^{(X)}\spi_4\nonumber\\
&&\s^{(X)}\tilde{\pi}_{33}=\s^{(X)}\pi_{33}, \ \ \ \s^{(X)}\tilde{\pi}_{44}=\s^{(X)}\pi_{44}\nonumber\\
&&\s^{(X)}\tilde{\pi}_{34}=\s^{(X)}\pi_{34}+\frac{1}{2}\mbox{tr}\s^{(X)}\pi
\label{8.9}
\end{eqnarray}
Note that:
\begin{equation}
\mbox{tr}\s^{(X)}\pi=\mbox{tr}\s^{(X)}\spi-\s^{(X)}\pi_{34}
\label{8.10}
\end{equation}
In the following we denote:
\begin{eqnarray}
&&\s^{(X)}\tilde{\spi}=\s^{(X)}i\nonumber\\
&&\s^{(X)}\tilde{\spi}_3=\s^{(X)}\mb, \ \ \ \s^{(X)}\tilde{\spi}_4=\s^{(X)}m\nonumber\\
&&\s^{(X)}\tilde{\pi}_{33}=\s^{(X)}\nb, \ \ \ \s^{(X)}\tilde{\pi}_{44}=\s^{(X)}n\nonumber\\
&&\s^{(X)}\tilde{\pi}_{34}=\s^{(X)}j \label{8.11}
\end{eqnarray}
Also, we denote by $\s^{(X)}\ih$ the trace-free part of
$\s^{X)}i$. Note that since
$$\mbox{tr}\s^{(X)}\tilde{\pi}=0$$
we have:
\begin{equation}
\mbox{tr}\s^{(X)}i=\s^{(X)}j \label{8.12}
\end{equation}

\section{$L^\infty$ estimates for the deformation tensors of $L$, $K$ and $S$}

If $X,Y,Z$ are any three vectorfields on spacetime, we have:
\begin{equation}
\s^{(X)}\pi(Y,Z)=g(\nabla_Y X,Z)+g(\nabla_Z X,Y) \label{8.13}
\end{equation}

Consider the case $X=L$. Taking $Y,Z$ to be $S$-tangential
vectorfields, we obtain, in view of the definition \ref{1.24},
\begin{equation}
\s^{(L)}\spi=2\Omega\chi \label{8.14}
\end{equation}
Taking $Z=\Lbh$ and $Y$ to be an $S$ tangential vectorfield we
have:
$$\s^{(L)}\pi(\Lbh,Y)=g(\nabla_{\Lbh}L,Y)+g(\nabla_Y L,\Lbh)$$
and from the second of \ref{1.74}
$$g(\nabla_{\Lbh}L,Y)=2\Omega\eta(Y)$$
while from the second of \ref{1.68} and \ref{1.65}
$$g(\nabla_Y L,\Lbh)=g(\nabla_Y(\Omega^2 L^\prime),\Omega\Lb^\prime)=2\Omega(\eta(Y)-2Y(\log\Omega))=-2\Omega\etb(Y)$$
Since from \ref{1.65}
$$\eta-\etb=2\zeta$$
we conclude that:
\begin{equation}
\s^{(L)}\spi_3=4\Omega\zeta \label{8.15}
\end{equation}
Taking $Z=\Lh$ and $Y$ to be an $S$ tangential vectorfield we
have:
$$\s^{(L)}\pi(\Lh,Y)=g(\nabla_{\Lh}L,Y)+g(\nabla_Y L,\Lh)=0$$
hence:
\begin{equation}
\s^{(L)}\spi_4=0 \label{8.16}
\end{equation}
Taking $Y=Z=\Lbh$ we have:
$$\s^{(L)}\pi(\Lbh,\Lbh)=2g(\nabla_{\Lbh}L,\Lbh)=0$$
by the second of \ref{1.68}. Hence:
\begin{equation}
\s^{(L)}\pi_{33}=0 \label{8.17}
\end{equation}
Taking $Y=Z=\Lb$ we have:
$$\s^{(L)}\pi(\Lh,\Lh)=2g(\nabla_{\Lh}L,\Lh)=0$$
hence:
\begin{equation}
\s^{(L)}\pi_{44}=0 \label{8.18}
\end{equation}
Finally, taking $Y=\Lbh$, $Z=\Lh$ we have:
$$\s^{(L)}\pi(\Lbh,\Lh)=g(\nabla_{\Lbh}L,\Lh)+g(\nabla_{\Lh}L,\Lbh)=g(\nabla_{\Lh}L,\Lbh)=-4\omega$$
by the second of \ref{1.68} and the first of \ref{1.15}. Hence:
\begin{equation}
\s^{(L)}\pi_{34}=-4\omega \label{8.19}
\end{equation}
We collect the above results for the components of $\s^{(L)}\pi$
in the following table:
\begin{eqnarray}
&&\s^{(L)}\spi=2\Omega\chi\nonumber\\
&&\s^{(L)}\spi_3=4\Omega\zeta, \ \ \ \s^{(L)}\spi_4=0\nonumber\\
&&\s^{(L)}\pi_{33}=0, \ \ \ \s^{(L)}\pi_{44}=0\nonumber\\
&&\s^{(L)}\pi_{34}=-4\omega \label{8.20}
\end{eqnarray}
The components of $\s^{(L)}\tilde{\pi}$ are then given by the
following table:
\begin{eqnarray}
&&\s^{(L)}\ih=2\Omega\chih\nonumber\\
&&\s^{(L)}j=\Omega\mbox{tr}\chi-2\omega\nonumber\\
&&\s^{(L)}\mb=4\Omega\zeta, \ \ \ \s^{(L)}m=0\nonumber\\
&&\s^{(L)}\nb=\s^{(L)}n=0 \label{8.21}
\end{eqnarray}

Consider next the case $X=\Lb$. Proceeding along similar lines we
derive the conjugate of table \ref{8.20}, which gives the
components of $\s^{(\Lb)}\pi$:
\begin{eqnarray}
&&\s^{(\Lb)}\spi=2\Omega\chib\nonumber\\
&&\s^{(\Lb)}\spi_3=0, \ \ \ \s^{(\Lb)}\spi_4=-4\Omega\zeta\nonumber\\
&&\s^{(\Lb)}\pi_{33}=0, \ \ \ \s^{(\Lb)}\pi_{44}=0\nonumber\\
&&\s^{(\Lb)}\pi_{34}=-4\omb \label{8.22}
\end{eqnarray}

We turn to the vectorfield $K$. Now, if $X$ is an arbitrary
vectorfield and $f$ an arbitrary function on spacetime we have, in
arbitrary local coordinates,
\begin{equation}
\s^{(fX)}\pi_{\mu\nu}=f\s^{(X)}\pi_{\mu\nu}+\partial_\mu f
X_\nu+\partial_\nu f X_\mu \label{8.23}
\end{equation}
where
$$X_\mu=g_{\mu\nu}X^\nu$$
In the case of the optical functions $u$, $\ub$, we have, from
\ref{1.3},
\begin{equation}
\partial_\mu u=-\frac{1}{2}L^\prime_\mu, \ \ \ \partial_\mu\ub=-\frac{1}{2}\Lb^\prime_\mu
\label{8.24}
\end{equation}
In view of \ref{8.23}, \ref{8.24}, the definition \ref{8.1}
implies:
\begin{equation}
\s^{(K)}\pi_{\mu\nu}=u^2\s^{(\Lb)}\pi_{\mu\nu}-u(\Lbh_\mu\Lh_\nu+\Lh_\mu\Lbh_\nu)
\label{8.25}
\end{equation}
Using the table \ref{8.22} we then obtain the following table for
the components of $\s^{(K)}\pi$:
\begin{eqnarray}
&&\s^{(K)}\spi=2u^2\Omega\chib\nonumber\\
&&\s^{(K)}\spi_3=0, \ \ \ \s^{(K)}\spi_4=-4u^2\Omega\zeta\nonumber\\
&&\s^{(K)}\pi_{33}=0, \ \ \ \s^{(K)}\pi_{44}=0\nonumber\\
&&\s^{(K)}\pi_{34}=-4u^2\omb-4u \label{8.26}
\end{eqnarray}
The components of $\s^{(K)}\tilde{\pi}$ are then given by:
\begin{eqnarray}
&&\s^{(K)}\ih=2u^2\Omega\chibh\nonumber\\
&&\s^{(K)}j=u^2(\Omega\mbox{tr}\chib-2\omb)-2u\nonumber\\
&&\s^{(K)}\mb=0, \ \ \ \s^{(K)}m=-4u^2\Omega\zeta\nonumber\\
&&\s^{(K)}\nb=\s^{(K)}n=0 \label{8.27}
\end{eqnarray}

Finally, we turn to the vectorfield $S$. In view of \ref{8.23},
\ref{8.24}, the definition \ref{8.2} implies:
\begin{equation}
\s^{(S)}\pi_{\mu\nu}=u\s^{(\Lb)}\pi_{\mu\nu}+\ub\s^{(L)}\pi_{\mu\nu}-\Lbh_\mu\Lh_\nu-\Lh_\mu\Lbh_\nu
\label{8.28}
\end{equation}
Using tables \ref{8.20} and \ref{8.22} we then obtain the
following table for the components of $\s^{(S)}\pi$:
\begin{eqnarray}
&&\s^{(S)}\spi=2\Omega(u\chib+\ub\chi)\nonumber\\
&&\s^{(S)}\spi_3=4\ub\Omega\zeta, \ \ \ \s^{(S)}\spi_4=-4u\Omega\zeta\nonumber\\
&&\s^{(S)}\pi_{33}=0, \ \ \ \s^{(S)}\pi_{44}=0\nonumber\\
&&\s^{(S)}\pi_{34}=-4(u\omb+\ub\omega+1) \label{8.29}
\end{eqnarray}
the components of $\s^{(S)}\tilde{\pi}$ are then given by:
\begin{eqnarray}
&&\s^{(S)}\ih=2\Omega(u\chibh+\ub\chih)\nonumber\\
&&\s^{(S)}j=u(\Omega\mbox{tr}\chib-2\omb)+\ub(\Omega\mbox{tr}\chi-2\omega)-2\nonumber\\
&&\s^{(S)}\mb=4\ub\Omega\zeta, \ \ \ \s^{(S)}m=-4u\Omega\zeta\nonumber\\
&&\s^{(S)}\nb=\s^{(S)}n=0 \label{8.30}
\end{eqnarray}

We see from tables \ref{8.21}, \ref{8.27}, \ref{8.30} that:
\begin{equation}
\s^{(X)}\nb=\s^{(X)}n=0 \label{8.31}
\end{equation}
for each of the vectorfields $L$, $K$, $S$. The results of Chapter
3 yield, through these tables, the following $L^\infty$ estimates
for the remaining components of the deformation tensors of the
vectorfields $L$, $K$ and $S$.

\vspace{5mm}

\begin{eqnarray}
&&\|\s^{(L)}\ih\|_{L^\infty(S_{\ub,u})}\leq C\delta^{-1/2}|u|^{-1}{\cal R}_0^\infty(\alpha)\nonumber\\
&&\|\s^{(L)}j\|_{L^\infty(S_{\ub,u})}\leq O(|u|^{-1})\nonumber\\
&&\|\s^{(L)}\mb\|_{L^\infty(S_{\ub,u})}\leq O(\delta^{1/2}|u|^{-2})\nonumber\\
&&\s^{(L)}m=0 \label{8.32}
\end{eqnarray}

\vspace{2.5mm}

\begin{eqnarray}
&&\|\s^{(K)}\ih\|_{L^\infty(S_{\ub,u})}\leq C\delta^{1/2}{\cal D}_0^\infty(\chibh)+O(\delta^{3/2}|u|^{-3/2})\nonumber\\
&&\|\s^{(K)}j\|_{L^\infty(S_{\ub,u})}\leq O(\delta)\nonumber\\
&&\s^{(K)}\mb=0\nonumber\\
&&\|\s^{(K)}m\|_{L^\infty(S_{\ub,u})}\leq O(\delta^{1/2})
\label{8.33}
\end{eqnarray}

\vspace{2.5mm}

\begin{eqnarray}
&&\|\s^{(S)}\ih\|_{L^\infty(S_{\ub,u})}\leq
C\delta^{1/2}|u|^{-1}({\cal D}_0^\infty(\chibh)+{\cal
R}_0^\infty(\alpha))
+O(\delta^{3/2}|u|^{-5/2})\nonumber\\
&&\|\s^{(S)}j\|_{L^\infty(S_{\ub,u})}\leq O(\delta|u|^{-2})\nonumber\\
&&\|\s^{(S)}\mb\|_{L^\infty(S_{\ub,u})}\leq O(\delta^{3/2}|u|^{-2})\nonumber\\
&&\|\s^{(S)}m\|_{L^\infty(S_{\ub,u})}\leq O(\delta^{1/2}|u|^{-1})
\label{8.34}
\end{eqnarray}

\vspace{5mm}

Here and in the estimates to follow we specify the leading term in
behavior with respect to $\delta$ in the case of the
$\s^{(X)}\ih$ component of the deformation tensors. We do this
because this component enters the {\em borderline error integrals}
as we shall see in Chapters 13, 14 and 15. Since the above
estimates depend only on the results of Chapter 3 the symbol
$O(\delta^p|u|^r)$ may be taken here to have the more restricted
meaning of the product of $\delta^p|u|^r$ with a non-negative
non-decreasing continuous function of the quantities ${\cal
D}_0^\infty$ and ${\cal R}_0^\infty$ alone.

\section{Construction of the rotation vectorfields $O_i$}

The action of the rotation group $SO(3)$ on the spacetime manifold
$M$ is defined as follows. First we have an action of $SO(3)$ by
isometries in the Minkowskian region $M_0$, the central timelike
geodesic $\Gamma_0$ being the set of fixed points of the action
and the group orbits being the surfaces $S_{\ub,u}$. This action
in $M_0$ commutes with the flow of $L$ as well as that of $\Lb$ in
$M_0$. We extend the action to $M^\prime$, the non-trivial part of
$M$, in the following manner.

First, we extend the action to $C_{u_0}$ by conjugation with the
flow of $L$ on $C_{u_0}$:
\begin{equation}
(o\in SO(3), p\in S_{\ub,u_0})\mapsto
op=\Phi_{\ub}(o\Phi_{-\ub}p)\in S_{\ub,u_0} \label{8.35}
\end{equation}
Thus if $q=\Phi_{-\ub}p\in S_{0,u_0}$ is the point of intersection
with $S_{0,u_0}$ of the generator of $C_{u_0}$ through $p$, then
$op=\Phi_{\ub}(oq)$ is the point of intersection with
$S_{\ub,u_0}$ of the generator of $C_{u_0}$ through $oq$.

The generating rotation fields $O_i \ : i=1,2,3$ on $C_{u_0}$ are
then given by:
\begin{equation}
O_i(p)=d\Phi_{\ub}(q)\cdot O_i(q), \ q=\Phi_{-\ub}(p)\in S_{0,u_0}
\ \ :\forall p\in S_{\ub,u_0}, \ \forall \ub\in[0,\delta]
\label{8.a1}
\end{equation}
The $O_i$ are tangential to the surfaces $S_{\ub,u_0}$ and
satisfy:
\begin{equation}
[L, O_i]=0, \ \ \ [O_i, O_j]=\epsilon_{ijk}O_k \ \ \ :\mbox{on
$C_{u_0}$} \label{8.36}
\end{equation}
where $\epsilon_{ijk}$ is the fully antisymmetric 3-dimensional
symbol.

We then extend the action to $M^\prime$ by conjugation with the
flow of $\Lb$ on each $\Cb_{\ub}$ in $M^\prime$:
\begin{equation}
(o\in SO(3), p\in S_{\ub,u})\mapsto
op=\Phib_{u-u_0}(o\Phib_{u_0-u}p)\in S_{\ub,u} \label{8.37}
\end{equation}
Thus if $q=\Phib_{u_0-u}p\in S_{\ub,u_0}$ is the point of
intersection with $S_{\ub,u_0}$ of the generator of $\Cb_{\ub}$
through $p$, then $op=\Phib_{u-u_0}(oq)$ is the point of
intersection with $S_{\ub,u}$ of the generator of $\Cb_{\ub}$
through $oq$.

The generating rotation fields $O_i \ : i=1,2,3$ on $M^\prime$ are
then given by:
\begin{equation}
O_i(p)=d\Phib_{u-u_0}(q)\cdot O_i(q), \ q=\Phib_{u_0-u}(p)\in
S_{\ub,u_0} \ \ :\forall p\in S_{\ub,u}, \ \forall (\ub,u)\in
D^\prime \label{8.a2}
\end{equation}
The $O_i$ are tangential to the surfaces $S_{\ub,u}$ and satisfy:
\begin{equation}
[\Lb, O_i]=0, \ \ \ [O_i, O_j]=\epsilon_{ijk}O_k \ \ \ :\mbox{on
$M^\prime$} \label{8.38}
\end{equation}

According to the above definition, the $O_i$ are given in a
canonical coordinate system simply by:
\begin{equation}
O_i=O_i^A(\vartheta)\frac{\partial}{\partial\vartheta^A}
\label{8.a3}
\end{equation}
where the $O_i^A(\vartheta)$  are the components of $O_i$ on
$S_{0,u_0}$. In the canonical coordinates induced by the
stereographic coordinates (see Chapter 1), the $O_i$ are given in
the north polar chart on $M_{U_1}$ by:
\begin{eqnarray}
&&O_1=-\frac{1}{\left(1+\frac{1}{4}|\vartheta|^2\right)}\left\{\frac{1}{2}\vartheta^1\vartheta^2\frac{\partial}{\partial\vartheta^1}
+\left(\left(1+\frac{1}{4}(\vartheta^2)^2\right)^2-\frac{1}{16}(\vartheta^1)^4\right)\frac{\partial}{\partial\vartheta^2}\right\}\nonumber\\
&&O_2=\frac{1}{\left(1+\frac{1}{4}|\vartheta|^2\right)}\left\{\frac{1}{2}\vartheta^2\vartheta^1\frac{\partial}{\partial\vartheta^2}
+\left(\left(1+\frac{1}{4}(\vartheta^1)^2\right)^2-\frac{1}{16}(\vartheta^2)^4\right)\frac{\partial}{\partial\vartheta^1}\right\}\nonumber\\
&&O_3=\vartheta^1\frac{\partial}{\partial\vartheta^2}-\vartheta^2\frac{\partial}{\partial\vartheta^1}
\label{8.a4}
\end{eqnarray}
and in the south polar chart on $M_{U_2}$ by:
\begin{eqnarray}
&&O_1=\frac{1}{\left(1+\frac{1}{4}|\vartheta|^2\right)}\left\{\frac{1}{2}\vartheta^1\vartheta^2\frac{\partial}{\partial\vartheta^1}
+\left(\left(1+\frac{1}{4}(\vartheta^2)^2\right)^2-\frac{1}{16}(\vartheta^1)^4\right)\frac{\partial}{\partial\vartheta^2}\right\}\nonumber\\
&&O_2=-\frac{1}{\left(1+\frac{1}{4}|\vartheta|^2\right)}\left\{\frac{1}{2}\vartheta^2\vartheta^1\frac{\partial}{\partial\vartheta^2}
+\left(\left(1+\frac{1}{4}(\vartheta^1)^2\right)^2-\frac{1}{16}(\vartheta^2)^4\right)\frac{\partial}{\partial\vartheta^1}\right\}\nonumber\\
&&O_3=\vartheta^1\frac{\partial}{\partial\vartheta^2}-\vartheta^2\frac{\partial}{\partial\vartheta^1}
\label{8.a5}
\end{eqnarray}

From \ref{8.13} we have, for any pair of vectorfields $X$, $Y$ on
spacetime,
\begin{equation}
\s^{(O_i)}\pi(X,Y)=g(\nabla_X O_i,Y)+g(\nabla_Y O_i,X)
\label{8.39}
\end{equation}
Taking $Y=\Lb$ and $X$ to be an $S$ tangential vectorfield we
obtain:
\begin{eqnarray*}
&&\s^{(O_i)}\pi(\Lb,X)=g(\nabla_{\Lb}O_i,X)+g(\nabla_X O_i,\Lb)=g(\nabla_{O_i}\Lb,X)+g(\nabla_X O_i,\Lb)\\
&&\hspace{1cm}=-g(\nabla_{O_i}X,\Lb)+g(\nabla_X
O_i,\Lb)=g([X,O_i],\Lb)=0
\end{eqnarray*}
by the first of \ref{8.38} and the fact that $[X,O_i]$ is an $S$
tangential vectorfield, both $X$ and $O_i$ being $S$ tangential
vectorfields. We conclude that:
\begin{equation}
\s^{(O_i)}\spi_3=0 \label{8.40}
\end{equation}
Taking $X=Y=\Lb$ in \ref{8.39} we obtain:
$$\s^{(O_i)}\pi(\Lb,\Lb)=2g(\nabla_{\Lb}O_i,\Lb)=g(\nabla_{O_i}\Lb,\Lb)=0$$
by the first of \ref{8.38}. We conclude that:
\begin{equation}
\s^{(O_i)}\pi_{33}=0 \label{8.41}
\end{equation}
Taking $X=Y=L$ in \ref{8.39} we obtain:
$$\s^{(O_i)}\pi(L,L)=2g(\nabla_L O_i,L)=-2g(O_i,\nabla_L L)=0$$
hence:
\begin{equation}
\s^{(O_i)}\pi_{44}=0 \label{8.42}
\end{equation}
Taking $X=\Lb$, $Y=L$ in \ref{8.39} we obtain:
\begin{eqnarray*}
&&\s^{(O_i)}\pi(\Lb,L)=g(\nabla_{\Lb}O_i,L)+g(\nabla_L O_i,\Lb)=-g(O_i,\nabla_{\Lb}L)-g(O_i,\nabla_L\Lb)\\
&&\hspace{2cm}=-2\Omega^2(\eta(O_i)+\etb(O_i))=-4\Omega^2
O_i(\log\Omega)
\end{eqnarray*}
by \ref{1.68} and \ref{1.65}. We conclude that:
\begin{equation}
\s^{(O_i)}\pi_{34}=-4O_i(\log\Omega) \label{8.43}
\end{equation}

The components $\s^{(O_i)}\spi$ and $\s^{(O_i)}\spi_4$ of
$\s^{(O_i)}\pi$ remain to be considered. We shall presently show
that these components satisfy propagation equations along the
generators of the $\Cb_{\ub}$. We start from the identity:
\begin{equation}
{\cal L}_{\Lb}{\cal L}_{O_i}g-{\cal L}_{O_i}{\cal L}_{\Lb}g={\cal
L}_{[\Lb,O_i]}g \label{8.44}
\end{equation}
Here ${\cal L}_{O_i}g=\s^{(O_i)}\pi$ and the right hand side
vanishes by the first of \ref{8.38}. Thus \ref{8.44} becomes:
\begin{equation}
{\cal L}_{\Lb}\s^{(O_i)}\pi={\cal L}_{O_i}{\cal L}_{\Lb}g
\label{8.45}
\end{equation}
Let $X$, $Y$ be vectorfields defined on and tangential to
$S_{\ub,u_0}$. For each $q\in S_{\ub,u_0}$ we extend $X(q)$,
$Y(q)$ to Jacobi fields along the generator of $\Cb_{\ub}$ through
$q$ by:
\begin{equation}
X(\Phib_{u-u_0}(q))=d\Phi_{u-u_0}\cdot X(q), \ \ \
Y(\Phib_{u-u_0}(q))=d\Phi_{u-u_0}\cdot Y(q) \label{8.46}
\end{equation}
Then $X$, $Y$ are $S$ tangential vectorfields satisfying:
\begin{equation}
[\Lb,X]=[\Lb,Y]=0 \label{8.47}
\end{equation}
Let us evaluate \ref{8.45} on $X$, $Y$. We have:
\begin{eqnarray*}
&&({\cal L}_{\Lb}\s^{(O_i)}\pi)(X,Y)=\Lb(\s^{(O_i)}\pi(X,Y))=\Lb(\s^{(O_i)}\spi(X,Y))\\
&&\hspace{2cm}=({\cal
L}_{\Lb}\s^{(O_i)}\spi)(X,Y)=(\Db\s^{(O_i)}\spi)(X,Y)
\end{eqnarray*}
while:
\begin{eqnarray*}
&&({\cal L}_{O_i}{\cal L}_{\Lb}g)(X,Y)=O_i(({\cal
L}_{\Lb}g)(X,Y))-{\cal L}_{\Lb}g([O_i,X],Y)
-{\cal L}_{\Lb}g(X,[O_i,Y])\\
&&\hspace{2cm}=2\left\{O_i(\Omega\chib(X,Y))-\Omega\chib([O_i,X],Y)-\Omega\chib(X,[O_i,Y])\right\}\\
&&\hspace{2cm}=2\left(\sL_{O_i}(\Omega\chib)\right)(X,Y)
\end{eqnarray*}
We thus obtain:
$$(\Db\s^{(O_i)}\spi)(X,Y)=2\left(\sL_{O_i}(\Omega\chib)\right)(X,Y)$$
and since the restrictions of $X$, $Y$ to $S_{\ub,u_0}$ are
arbitrary vectorfields tangential to $S_{\ub,u_0}$ while
$\Phib_{u-u_0}$ is a diffeomorphism of $S_{\ub,u_0}$ onto
$S_{\ub,u}$ for each $u$, it follows that:
\begin{equation}
\Db\s^{(O_i)}\spi=2\sL_{O_i}(\Omega\chib) \label{8.48}
\end{equation}
This is the desired propagation equation for $\s^{(O_i)}\spi$.
However, this equation must be supplemented by an initial
condition on $C_{u_0}$.

To determine $\s^{(O_i)}\spi$ on $C_{u_0}$ we derive a propagation
equation for $\s^{(O_i)}\spi$ along the generators of $C_{u_0}$.
We start from the identity:
\begin{equation}
{\cal L}_L{\cal L}_{O_i}g-{\cal L}_{O_i}{\cal L}_Lg={\cal
L}_{[L,O_i]}g \label{8.49}
\end{equation}
Now by the first of \ref{8.36} the right hand side vanishes on
$C_{u_0}$ when evaluated on a pair of vectorfields tangential to
$C_{u_0}$. Let $X$, $Y$ be vectorfields defined on and tangential
to $S_{0,u_0}$. For each $q\in S_{0,u_0}$ we extend $X(q)$, $Y(q)$
to Jacobi fields along the generator of $C_{u_0}$ through $q$ by:
\begin{equation}
X(\Phi_{\ub}(q))=d\Phi_{\ub}\cdot X(q), \ \ \
Y(\Phi_{\ub}(q))=d\Phi_{\ub}\cdot Y(q) \label{8.50}
\end{equation}
Then $X$, $Y$ are $S$ tangential vectorfields along $C_{u_0}$
satisfying:
\begin{equation}
[L,X]=[L,Y]=0 \ \ \mbox{: along $C_{u_0}$} \label{8.51}
\end{equation}
Evaluating \ref{8.49} along $C_{u_0}$ on $X$, $Y$, we have:
\begin{eqnarray*}
&&({\cal L}_L\s^{(O_i)}\pi)(X,Y)=({\cal L}_L{\cal L}_{O_i}g)(X,Y)=L(\s^{(O_i)}\pi(X,Y))\\
&&\hspace{2cm}=L(\s^{(O_i)}\spi(X,Y))=({\cal
L}_L\s^{(O_i)}\spi)(X,Y)=(D\s^{(O_i)}\spi)(X,Y)
\end{eqnarray*}
and:
\begin{eqnarray*}
&&({\cal L}_{O_i}{\cal L}_L g)(X,Y)=O_i(({\cal L}_L g)(X,Y))-{\cal
L}_L g([O_i,X],Y)
-{\cal L}_L g(X,[O_i,Y])\\
&&\hspace{2cm}=2\left\{O_i(\Omega\chi(X,Y))-\Omega\chi([O_i,X],Y)-\Omega\chi(X,[O_i,Y])\right\}\\
&&\hspace{2cm}=2\left(\sL_{O_i}(\Omega\chi)\right)(X,Y)
\end{eqnarray*}
Since the right hand side of \ref{8.49} vanishes on $C_{u_0}$ when
evaluated on $X$, $Y$, we obtain:
$$(D\s^{(O_i)}\spi)(X,Y)=2\left(\sL_{O_i}(\Omega\chi)\right)(X,Y)$$
and since the restrictions of $X$, $Y$ to $S_{0,u_0}$ are
arbitrary vectorfields tangential to $S_{0,u_0}$ while
$\Phi_{\ub}$ is a diffeomorphism of $S_{0,u_0}$ onto $S_{\ub,u_0}$
for each $\ub$, it follows that:
\begin{equation}
D\s^{(O_i)}\spi=2\sL_{O_i}(\Omega\chi) \ \ \ \mbox{: on $C_{u_0}$}
\label{8.52}
\end{equation}
This is the desired propagation equation for $\s^{(O_i)}\spi$
along $C_{u_0}$. The initial condition for this equation is
simply:
\begin{equation}
\s^{(O_i)}\spi=0 \ \ \mbox{: on $S_{0,u_0}$} \label{8.53}
\end{equation}
for, the $O_i$ generate isometries in the Minkowskian region
$M_0$.

Note that if $X$, $Y$ is a pair of $S$ tangential vectorfields, we
have:
\begin{eqnarray*}
&&\s^{(O_i)}\spi(X,Y)=\s^{(O_i)}\pi(X,Y)=({\cal L}_{O_i}g)(X,Y)\\
&&\hspace{2cm}=O_i(g(X,Y))-g([O_i,X],Y)-g(X,[O_i,Y])\\
&&\hspace{2cm}=O_i(\sg(X,Y))-\sg([O_i,X],Y)-\sg(X,[O_i,Y])\\
&&\hspace{2cm}=\left(\sL_{O_i}\sg\right)(X,Y)
\end{eqnarray*}
therefore:
\begin{equation}
\s^{(O_i)}\spi=\sL_{O_i}\sg \label{8.54}
\end{equation}

Consider the propagation equation \ref{8.48}. Decomposing
$$\chib=\chibh+\frac{1}{2}\sg\mbox{tr}\chib$$
we have
$$\sL_{O_i}(\Omega\chib)=\sL_{O_i}(\Omega\chibh)+\frac{1}{2}\sg O_i(\Omega\mbox{tr}\chib)+\frac{1}{2}\Omega\mbox{tr}\chib
\sL_{O_i}\sg$$ In view of \ref{8.54} we then obtain:
\begin{equation}
\sL_{O_i}(\Omega\chib)=\sL_{O_i}(\Omega\chibh)+\frac{1}{2}\sg
O_i(\Omega\mbox{tr}\chib)
+\frac{1}{2}\Omega\mbox{tr}\chib\s^{(O_i)}\spi \label{8.55}
\end{equation}
Substituting this in \ref{8.48} brings the propagation equation
for $\s^{(O_i)}\spi$ along the $\Cb_{\ub}$ to the following form:
\begin{equation}
\Db\s^{(O_i)}\spi-\Omega\mbox{tr}\chib\s^{(O_i)}\spi=2\sL_{O_i}(\Omega\chibh)+\sg
O_i(\Omega\mbox{tr}\chib) \label{8.56}
\end{equation}
Similarly, we have:
\begin{equation}
\sL_{O_i}(\Omega\chi)=\sL_{O_i}(\Omega\chi)+\frac{1}{2}\sg
O_i(\Omega\mbox{tr}\chi)
+\frac{1}{2}\Omega\mbox{tr}\chi\s^{(O_i)}\spi \label{8.57}
\end{equation}
Substituting this in \ref{8.52} brings the propagation equation
for $\s^{(O_i)}\spi$ along $C_{u_0}$ to the following form:
\begin{equation}
D\s^{(O_i)}\spi-\Omega\mbox{tr}\chi\s^{(O_i)}\spi=2\sL_{O_i}(\Omega\chih)+\sg
O_i(\Omega\mbox{tr}\chi) \label{8.58}
\end{equation}

We proceed to derive a propagation equation for
$\s^{(O)_i}\spi_4$. We shall  do this wth the help of the
following follow-up to Lemma 1.1.

\vspace{5mm}

\noindent{\bf Lemma 8.1} \ \ \ Let $X$ be a vectorfield defined
along a given $C_u$ and tangential to its $S_{\ub,u}$ sections.
Then the vectorfield $[L,X]$, defined along $C_u$ is also
tangential to the $S_{\ub,u}$ sections and given by:
$$[L,X]=\Omega\s^{(X)}\spi_4^\sharp$$
Also, let $X$ be a vectorfield defined along a given $\Cb_{\ub}$
and tangential to its $S_{\ub,u}$ sections. Then the vectorfield
$[\Lb,X]$, defined along $\Cb_{\ub}$ is also tangential to the
$S_{\ub,u}$ sections and given by:
$$[\Lb,X]=\Omega\s^{(X)}\spi_3^\sharp$$

\noindent{\em Proof:} \ To establish the first part of the lemma,
we know from Lemma 1.1 that $[L,X]$ is $S$-tangential. Let then
$Y$ be an arbitrary $S$-tangential vectorfield. We have:
$$g([L,X],Y)=g(\nabla_L X,Y)-g(\nabla_X L,Y)
=g(\nabla_L X,Y)+g(L,\nabla_X Y)$$ and:
$$g(L,\nabla_X Y)-g(L,\nabla_Y X)=g(L,[X,Y])=0$$
$[X,Y]$ being $S$-tangential. Therefore:
$$g([L,X],Y)=g(\nabla_L X,Y)+g(\nabla_Y X,L)=\s^{(X)}\pi(L,Y)=\Omega\s^{(X)}\spi_4$$
The second part of the lemma is established in a similar manner.

\vspace{5mm}

From now on we denote:
\begin{equation}
Z_i=[L,O_i] \label{8.59}
\end{equation}
According to the above lemma we have:
\begin{equation}
\s^{(O_i)}\spi_4=\Omega^{-1}\sg\cdot Z_i \label{8.60}
\end{equation}
(see \ref{1.73}). Consider the Jacobi identity:
\begin{equation}
[\Lb,[L,O_i]]+[L,[O_i,\Lb]]+[O_i,[\Lb,L]]=0 \label{8.61}
\end{equation}
Now, by the first of \ref{8.38} the middle term vanishes while
according to \ref{1.75}:
$$[\Lb,L]=4\Omega^2\zeta^\sharp$$
In view also of the definition \ref{8.59}, the identity \ref{8.61}
reads:
\begin{equation}
[\Lb,Z_i]+4[O_i,\Omega^2\zeta^\sharp]=0 \label{8.62}
\end{equation}
or:
\begin{equation}
\Db Z_i=-4\sL_{O_i}(\Omega^2\zeta^\sharp) \label{8.63}
\end{equation}
This is the desired propagation equation for $Z_i$. According to
the first of \ref{8.36} the initial condition on $C_{u_0}$ is
simply:
\begin{equation}
Z_i=0 \ \ \mbox{: on $C_{u_0}$} \label{8.64}
\end{equation}
From the propagation equation \ref{8.63} a propagation equation
for $\s^{(O_i)}\spi_4$ can readily be derived. However we shall
work with the propagation equation \ref{8.63}, as this is simpler,
to derive estimates for $Z_i$ from which estimates for
$\s^{(O_i)}\spi_4$ shall thence be readily derived.

\section{$L^\infty$ estimates for the $O_i$ and $\snab O_i$}

\noindent{\bf Proposition 8.1} \ \ \ Given any point $p\in
S_{\ub,u}$, let $q=\Phib_{u_0-u}(p)$ be the point on $S_{\ub,u_0}$
correponding to $p$ along the same generator of $\Cb_{\ub}$ and
let $q_0=\Phi_{-\ub}(q)$ the point on $S_{0,u_0}$ corresponding to
$q$ along the same generator of $C_{u_0}$. Then, provided that
$\delta$ is suitably small depending on ${\cal D}_0^\infty$,
${\cal R}_0^\infty$, we have:
$$\frac{1}{2}\frac{|u|}{|u_0|}|O_i(q_0)|\leq |O_i(p)|\leq 2\frac{|u|}{|u_0|}|O_i(q_0)|$$
for every point $p\in M^\prime$.

\noindent{\em Proof:} \ We have, relative to an arbitary frame
field for the $S_{\ub,u}$,
\begin{equation}
|O_i|^2=\sg_{AB}O_i^A O_i^B \label{8.65}
\end{equation}
Consider first $O_i$ along $C_{u_0}$. According to the first of
\ref{8.36} we have:
\begin{equation}
DO_i=0 \ \ \mbox{: on $C_{u_0}$} \label{8.66}
\end{equation}
In view of the first of \ref{1.28}, that is
$$D\sg=2\Omega\chi,$$
and the fact that the operator $D$ satisfies the Leibniz rule, we
obtain:
\begin{eqnarray}
D(|O_i|^2)&=&(D\sg)(O_i,O_i)\nonumber\\
&=&2\Omega\chi(O_i,O_i)\nonumber\\
&=&\Omega\mbox{tr}\chi|O_i|^2+2\Omega\chih(O_i,O_i) \ \ \ \mbox{:
on $C_{u_0}$} \label{8.67}
\end{eqnarray}
Therefore:
\begin{equation}
D(|O_i|^2)\leq
2\left(\frac{1}{2}\Omega\mbox{tr}\chi+\Omega|\chih|\right)|O_i|^2
\label{8.68}
\end{equation}
and:
\begin{equation}
D(|O_i|^2)\geq
2\left(\frac{1}{2}\Omega\mbox{tr}\chi-\Omega|\chih|\right)|O_i|^2
\label{8.69}
\end{equation}
In reference to \ref{8.68}, setting:
$$v(t)=|O_i(\Phi_t(q_0))|, \ \ \ a(t)=\left(\frac{1}{2}\Omega\mbox{tr}\chi+\Omega|\chih|\right)(\Phi_t(q_0))$$
we have:
$$\frac{d}{dt}(v^2)\leq 2av^2$$
therefore Lemma 3.1 with $b=0$ applies and we obtain:
\begin{equation}
|O_i(q)|\leq\exp\left\{\int_0^{\ub}\left(\frac{1}{2}\Omega\mbox{tr}\chi+\Omega|\chih|\right)(\Phi_{\ub^\prime}(q_0))
d\ub^\prime\right\} \label{8.70}
\end{equation}
In reference to \ref{8.69} setting:
$$v(t)=|O_i(\Phi_{-t}(q))|, \ \ \ a(t)=-\left(\frac{1}{2}\Omega\mbox{tr}\chi-\Omega|\chih|\right)(\Phi_{-t}(q))$$
we again have:
$$\frac{d}{dt}(v^2)\leq 2av^2$$
therefore Lemma 3.1 applies and we obtain:
\begin{equation}
|O_i(q_0)|\leq\exp\left\{-\int_0^{\ub}\left(\frac{1}{2}\Omega\mbox{tr}\chi-\Omega|\chih|\right)(\Phi_{-\ub^\prime}(q))
d\ub^\prime\right\} \label{8.71}
\end{equation}
Now by the results of Chapter 3 the integrals in the exponentials
in \ref{8.70} and \ref{8.71} are bounded by
$O(\delta^{1/2}|u|^{-1})$. Therefore if $\delta$ is suitably small
depending on ${\cal R}_0^\infty$ these integrals do not exceed
$(1/2)\log 2$. It follows that:
\begin{equation}
\frac{1}{\sqrt{2}}|O_i(q_0)|\leq |O_i(q)|\leq \sqrt{2}|O_i(q_0)|
\label{8.72}
\end{equation}

Consider next $O_i$ along $\Cb_{\ub}$. According to the first of
\ref{8.38} we have:
\begin{equation}
\Db O_i=0 \ \ \mbox{: on $M^\prime$} \label{8.73}
\end{equation}
In view of the second of \ref{1.28}, that is
$$\Db\sg=2\Omega\chib,$$
and the fact that the operator $\Db$ satisfies the Leibniz rule,
we obtain:
\begin{eqnarray}
\Db(|O_i|^2)&=&(\Db\sg)(O_i,O_i)\nonumber\\
&=&2\Omega\chib(O_i,O_i)\nonumber\\
&=&\Omega\mbox{tr}\chib|O_i|^2+2\Omega\chibh(O_i,O_i) \ \ \
\mbox{: on $M^\prime$} \label{8.74}
\end{eqnarray}
Therefore:
\begin{equation}
\Db(|O_i|^2)\leq
2\left(\frac{1}{2}\Omega\mbox{tr}\chib+\Omega|\chibh|\right)|O_i|^2
\label{8.75}
\end{equation}
and:
\begin{equation}
\Db(|O_i|^2)\geq
2\left(\frac{1}{2}\Omega\mbox{tr}\chib-\Omega|\chibh|\right)|O_i|^2
\label{8.76}
\end{equation}
In reference to \ref{8.75}, setting:
$$v(t)=|O_i(\Phib_t(q))|, \ \ \ a(t)=\left(\frac{1}{2}\Omega\mbox{tr}\chib+\Omega|\chibh|\right)(\Phib_t(q))$$
we have:
$$\frac{d}{dt}(v^2)\leq 2av^2$$
therefore Lemma 3.1 with $b=0$ applies and we obtain:
\begin{equation}
|O_i(p)|\leq\exp\left\{\int_{u_0}^{u}\left(\frac{1}{2}\Omega\mbox{tr}\chib+\Omega|\chibh|\right)(\Phib_{u^\prime-u_0}(q))
du^\prime\right\} \label{8.77}
\end{equation}
In reference to \ref{8.76} setting:
$$v(t)=|O_i(\Phib_{-t}(p))|, \ \ \ a(t)=-\left(\frac{1}{2}\Omega\mbox{tr}\chib-\Omega|\chibh|\right)(\Phib_{-t}(p))$$
we again have:
$$\frac{d}{dt}(v^2)\leq 2av^2$$
therefore Lemma 3.1 applies and we obtain:
\begin{equation}
|O_i(q)|\leq\exp\left\{-\int_{u_0}^{u}\left(\frac{1}{2}\Omega\mbox{tr}\chib-\Omega|\chibh|\right)(\Phib_{u_0-u^\prime}(p))
du^\prime\right\} \label{8.78}
\end{equation}
Now by the results of Chapter 3 the integral in the exponential in
\ref{8.77} is bounded by
$$\log\left(\frac{|u|}{|u_0|}\right)+O(\delta^{1/2}|u|^{-1})$$
while the integral in the exponential in \ref{8.78} is bounded by
$$-\log\left(\frac{|u|}{|u_0|}\right)+O(\delta^{1/2}|u|^{-1})$$
If $\delta$ is suitably small depending on ${\cal D}_0^\infty$,
${\cal R}_0^\infty$ the terms $O(\delta^{1/2}|u|^{-1})$ do not
exceed $(1/2)\log 2$. It follows that:
\begin{equation}
\frac{1}{\sqrt{2}}\frac{|u|}{|u_0}|O_i(q)|\leq |O_i(p)|\leq
\sqrt{2}\frac{|u|}{|u_0|}|O_i(q)| \label{8.79}
\end{equation}
Combining finally \ref{8.79} with \ref{8.72} yields the
proposition.

\vspace{5mm}

We note that the three rotation fields $O_i$ satisfy on
$S_{0,u_0}$ the bound:
\begin{equation}
\|O_i\|_{L^\infty(S_{0,u_0})}\leq |u_0| \label{8.80}
\end{equation}
As $S_{0,u_0}$ lies on the boundary of the Minkowskian region
$M_0$, this follows from the expression:
\begin{equation}
O_i=\epsilon_{ijk}x^j\frac{\partial}{\partial x^k} \label{8.81}
\end{equation}
for the $O_i$ in rectangular coordinates in $M_0$, with the $x^0$
axis coinciding with the central timelike geodesic $\Gamma_0$.

\vspace{5mm}

\noindent{\bf Proposition 8.2} \ \ \ We have:
$$\|\snab O_i\|_{L^\infty(S_{\ub,u})}\leq 3\sqrt{2}+1$$
for all $(\ub,u)\in D^\prime$, provided that $\delta$ is suitably
small depending on ${\cal D}_0^\infty$, ${\cal R}_0^\infty$,
$\scD_1^4$, $\scR_1^4$ and $\scR_2$.

\noindent{\em Proof:} \ Let $X$ be an arbitrary $S$-tangential
vectorfield. In analogy with Lemma 4.1 we have the commutation
formulas:
\begin{equation}
D\snab X-\snab DX=D\sGamma\cdot X \label{8.82}
\end{equation}
and:
\begin{equation}
\Db\snab X-\snab\Db X=\Db\sGamma\cdot X \label{8.83}
\end{equation}
Here, with respect to an arbitrary local frame field for the
$S_{\ub,u}$,
$$(D\sGamma\cdot X)_A^B=(D\sGamma)^B_{AC}X^C, \ \ \ (\Db\sGamma\cdot X)_A^B=(\Db\sGamma)^B_{AC}X^C$$
The formulas \ref{8.82}, \ref{8.83} are established by the same
argument as that of Lemma 4.1.

Since $\snab O_i$ is a type $T^1_1$ type $S$ tensorfield we have,
relative to an arbitrary frame field for the $S_{\ub,u}$,
\begin{equation}
|\snab O_i|^2=\sg_{AB}(\sg^{-1})^{CD}\snab_C O_i^A\snab_D O_i^B
\label{8.84}
\end{equation}
It follows that:
\begin{eqnarray*}
D(|\snab O_i|^2)&=&2\Omega\chi_{AB}(\sg^{-1})^{CD}\snab_C O_i^A\snab_D O_i^B\\
&\s&-2\Omega\chi^{CD}\sg_{AB}\snab_C O_i^A\snab_D O_i^B\\
&\s&+2(\snab O_i,D\snab O_i)
\end{eqnarray*}
and:
\begin{eqnarray*}
\Db(|\snab O_i|^2)&=&2\Omega\chib_{AB}(\sg^{-1})^{CD}\snab_C O_i^A\snab_D O_i^B\\
&\s&-2\Omega\chib^{CD}\sg_{AB}\snab_C O_i^A\snab_D O_i^B\\
&\s&+2(\snab O_i,\Db\snab O_i)
\end{eqnarray*}
Substituting the decompositions
$$\chi=\chih+\frac{1}{2}\sg\mbox{tr}\chi, \ \ \ \chib=\chibh+\frac{1}{2}\sg\mbox{tr}\chib$$
the trace terms cancel and we obtain:
\begin{eqnarray}
D(|\snab O_i|^2)&=&2\Omega\chih_{AB}(\sg^{-1})^{CD}\snab_C O_i^A\snab_D O_i^B\nonumber\\
&\s&-2\Omega\chih^{CD}\sg_{AB}\snab_C O_i^A\snab_D O_i^B\nonumber\\
&\s&+2(\snab O_i,D\snab O_i)\label{8.85}
\end{eqnarray}
and:
\begin{eqnarray}
\Db(|\snab O_i|^2)&=&2\Omega\chibh_{AB}(\sg^{-1})^{CD}\snab_C O_i^A\snab_D O_i^B\nonumber\\
&\s&-2\Omega\chibh^{CD}\sg_{AB}\snab_C O_i^A\snab_D O_i^B\nonumber\\
&\s&+2(\snab O_i,\Db\snab O_i)\label{8.86}
\end{eqnarray}

Consider first $O_i$ along $C_{u_0}$. Since \ref{8.66} holds,
taking $X=O_i$ in \ref{8.82} we obtain:
\begin{equation}
D\snab O_i=D\sGamma\cdot O_i \ \ \mbox{: on $C_{u_0}$}
\label{8.87}
\end{equation}
Substituting in \ref{8.85} we deduce:
\begin{equation}
D(|\snab O_i|^2)\leq 2|\snab O_i|(C\Omega|\chih||\snab
O_i|+|D\sGamma||O_i|) \label{8.88}
\end{equation}
Given any point $q\in S_{\ub,u_0}$, let $q_0=\Phi_{-\ub}(q)\in
S_{0,u_0}$ be the point corresponding to $q$ along the same
generator of $C_{u_0}$. Applying Lemma 3.1 to \ref{8.88} then
yields:
\begin{eqnarray}
|(\snab O_i)(q)|&\leq&\exp\left\{\int_0^{\ub}C\Omega|\chih|(\Phi_{\ub^\prime}(q_0))d\ub^\prime\right\}|(\snab O_i)(q_0)|\label{8.89}\\
&\s&+\int_0^{\ub}\exp\left\{\int_{\ub^\prime}^{\ub}C\Omega|\chih|(\Phi_{\ub^{\prime\prime}}(q_0))d\ub^{\prime\prime}\right\}
(|D\sGamma||O_i|)(\Phi_{\ub^\prime}(q_0))d\ub^\prime\nonumber
\end{eqnarray}
Now by the results of Chapter 3 the integrals in the exponentials
in \ref{8.89} are bounded by $O(\delta^{1/2}|u|^{-1})$. Therefore
if $\delta$ is suitably small depending on ${\cal R}_0^\infty$
these integrals do not exceed $(1/2)\log 2$. Also, by the estimate
\ref{7.198},
\begin{equation}
\|D\sGamma\|_{L^\infty(S_{\ub,u})}\leq O(\delta^{-1/2}|u|^{-2})
\label{8.90}
\end{equation}
Taking also into account Proposition 8.1 and the bound \ref{8.80},
the second term on the right hand side of \ref{8.89} is bounded by
$O(\delta^{1/2}|u|^{-1})$ which does not exceed $1$ if $\delta$ is
suitably small depending on ${\cal D}_0^\infty$, ${\cal
R}_0^\infty$, $\scD_1^4$, $\scR_1^4$ and $\scR_2$. Moreover, from
the expression \ref{8.81} in the Minkowskian region $M_0$ we
deduce the following bound on $S_{0,u_0}$:
\begin{equation}
\|\snab O_i\|_{L^\infty(S_{0,u_0})}\leq\sqrt{2} \label{8.91}
\end{equation}
It follows that the first term on the right hand side of
\ref{8.89} does not exceed 2. We conclude that:
$$|(\snab O_i)(q)|\leq 3$$
and, as this holds for all $q\in S_{\ub,u_0}$,
\begin{equation}
\|\snab O_i\|_{L^\infty(S_{\ub,u_0})}\leq 3 \label{8.92}
\end{equation}

Consider next $O_i$ along $\Cb_{\ub}$. Since \ref{8.73} holds,
taking $X=O_i$ in \ref{8.82} we obtain:
\begin{equation}
\Db\snab O_i=\Db\sGamma\cdot O_i \ \ \mbox{: on $M^\prime$}
\label{8.93}
\end{equation}
Substituting in \ref{8.86} we deduce:
\begin{equation}
\Db(|\snab O_i|^2)\leq 2|\snab O_i|(C\Omega|\chibh||\snab
O_i|+|\Db\sGamma||O_i|) \label{8.94}
\end{equation}
Given any point $p\in S_{\ub,u}$, let $q=\Phib_{u_0-u}(p)\in
S_{\ub,u_0}$ be the point corresponding to $p$ along the same
generator of $\Cb_{\ub}$. Applying Lemma 3.1 to \ref{8.94} then
yields:
\begin{eqnarray}
|(\snab
O_i)(p)|&\leq&\exp\left\{\int_{u_0}^{u}C\Omega|\chibh|(\Phib_{u^\prime-u_0}(q))du^\prime\right\}|(\snab
O_i)(q)|
\label{8.95}\\
&\s&+\int_{u_0}^{u}\exp\left\{\int_{u^\prime}^{u}C\Omega|\chibh|(\Phib_{u^{\prime\prime}-u_0}(q))du^{\prime\prime}\right\}
(|\Db\sGamma||O_i|)(\Phib_{u^\prime-u_0}(q))du^\prime\nonumber
\end{eqnarray}
Now by the results of Chapter 3 the integrals in the exponentials
in \ref{8.95} are bounded by $O(\delta^{1/2}|u|^{-1})$. Therefore
if $\delta$ is suitably small depending on ${\cal D}_0^\infty$,
${\cal R}_0^\infty$ these integrals do not exceed $(1/2)\log 2$.
Also, by the estimate \ref{7.199},
\begin{equation}
\|\Db\sGamma\|_{L^\infty(S_{\ub,u})}\leq O(\delta^{1/2}|u|^{-3})
\label{8.96}
\end{equation}
Taking also into account Proposition 8.1 and the bound \ref{8.80},
the second term on the right hand side of \ref{8.95} is bounded by
$O(\delta^{1/2}|u|^{-1})$ which does not exceed $1$ if $\delta$ is
suitably small depending on ${\cal D}_0^\infty$, ${\cal
R}_0^\infty$, $\scD_1^4$, $\scR_1^4$ and $\scR_2$. Moreover, in
view of the bound \ref{8.92} the first term on the right hand side
of \ref{8.95} does not exceed $3\sqrt{2}$. We conclude that:
$$|(\snab O_i)(p)|\leq 3\sqrt{2}+1$$
and the proposition is proved.

\vspace{5mm}

\section{$L^\infty$ estimates for the deformations tensors of the $O_i$}

Consider the propagation equations \ref{8.58} and \ref{8.56}. Let
us decompose $\s^{(O_i)}\spi$ into its trace-free part
$\s^{(O_i)}\hat{\spi}$ and its trace:
\begin{equation}
\s^{(O_i)}\spi=\s^{(O_i)}\hat{\spi}+\frac{1}{2}\sg\mbox{tr}\s^{(O_i)}\spi
\label{8.97}
\end{equation}
In view of \ref{3.3} we have:
\begin{eqnarray}
\mbox{tr}(D\s^{(O_i)}\spi)&=&D\mbox{tr}\s^{(O_i)}\spi+2\Omega(\chi,\s^{(O_i)}\spi)\nonumber\\
&=&D\mbox{tr}\s^{(O_i)}\spi+\Omega\mbox{tr}\chi\mbox{tr}\s^{(O_i)}\spi+2\Omega(\chih,\s^{(O_i)}\hat{\spi})
\label{8.98}
\end{eqnarray}
and:
\begin{eqnarray}
\mbox{tr}(\Db\s^{(O_i)}\spi)&=&\Db\mbox{tr}\s^{(O_i)}\spi+2\Omega(\chib,\s^{(O_i)}\spi)\nonumber\\
&=&\Db\mbox{tr}\s^{(O_i)}\spi+\Omega\mbox{tr}\chib\mbox{tr}\s^{(O_i)}\spi+2\Omega(\chibh,\s^{(O_i)}\hat{\spi})
\label{8.99}
\end{eqnarray}
Also, since by \ref{8.54}
\begin{equation}
\sL_{O_i}\sg^{-1}=-\s^{(O_i)}\spi^{\sharp\sharp}, \label{8.100}
\end{equation}
we have:
\begin{equation}
\mbox{tr}\left(\sL_{O_i}(\Omega\chih)\right)=(\s^{(O_i)}\hat{\spi},\Omega\chih),
\ \ \
\mbox{tr}\left(\sL_{O_i}(\Omega\chibh)\right)=(\s^{(O_i)}\hat{\spi},\Omega\chibh)
\label{8.101}
\end{equation}
Taking the trace of equation \ref{8.58} and using \ref{8.98} and
the first of \ref{8.101} we obtain the following propagation
equation for $\mbox{tr}\s^{(O_i)}\spi$ along $C_{u_0}$:
\begin{equation}
D\mbox{tr}\s^{(O_i)}\spi=2O_i(\Omega\mbox{tr}\chi) \label{8.102}
\end{equation}
Also, taking the trace of equation \ref{8.56} and using \ref{8.99}
and the second of \ref{8.101} we obtain the following propagation
equation for $\mbox{tr}\s^{(O_i)}\spi$ along the $\Cb_{\ub}$:
\begin{equation}
\Db\mbox{tr}\s^{(O_i)}\spi=2O_i(\Omega\mbox{tr}\chib)
\label{8.103}
\end{equation}
Substituting in \ref{8.58} the decomposition \ref{8.97} and taking
into account \ref{8.102} we obtain the following propagation
equation for $\s^{(O_i)}\hat{\spi}$ along $C_{u_0}$:
\begin{equation}
D\s^{(O_i)}\hat{\spi}-\Omega\mbox{tr}\chi\s^{(O_i)}\hat{\spi}=-\Omega\chih\mbox{tr}\s^{(O_i)}\spi
+2\sL_{O_i}(\Omega\chih) \label{8.104}
\end{equation}
Also, substituting in \ref{8.56} the decomposition \ref{8.97} and
taking into account \ref{8.103} we obtain the following
propagation equation for $\s^{(O_i)}\hat{\spi}$ along the
$\Cb_{\ub}$:
\begin{equation}
\Db\s^{(O_i)}\hat{\spi}-\Omega\mbox{tr}\chib\s^{(O_i)}\hat{\spi}=-\Omega\chibh\mbox{tr}\s^{(O_i)}\spi
+2\sL_{O_i}(\Omega\chibh) \label{8.105}
\end{equation}

\noindent{\bf Proposition 8.3} \ \ \ The following estimates hold
for all $(\ub,u)\in D^\prime$:
$$\|\mbox{tr}\s^{(O_i)}\spi\|_{L^\infty(S_{\ub,u})}\leq O(\delta|u|^{-2})$$
$$\|\s^{(O_i)}\hat{\spi}\|_{L^\infty(S_{\ub,u})}\leq C\delta^{1/2}|u|^{-1}(\scR_1^4(\beta)+{\cal R}_0^\infty(\beta))
+O(\delta|u|^{-2})$$ provided that $\delta$ is suitably small
depending on ${\cal D}_0^\infty$, ${\cal R}_0^\infty$, $\scD_1^4$,
$\scR_1^4$ and $\scR_2$.

\noindent{ \em Proof:} \ Consider first equation \ref{8.102}.
Given any point $q\in S_{\ub,u_0}$, let $q_0=\Phi_{-\ub}(q)\in
S_{0,u_0}$ be the point corresponding to $q$ along the same
generator of $C_{u_0}$. Integrating \ref{8.102} along this
generator we obtain, in view of \ref{8.53},
\begin{equation}
(\mbox{tr}\s^{(O_i)}\spi)(q)=\int_0^{\ub}2(O_i(\Omega\mbox{tr}\chi))(\Phi_{\ub^\prime}(q_0))d\ub^\prime
\label{8.106}
\end{equation}
Now by Proposition 6.1 and the results of Chapters 3 and 4 in
conjunction with Lemma 5.2, we have
\begin{equation}
\|\sd(\Omega\mbox{tr}\chi)\|_{L^\infty(S_{\ub,u})}\leq O(|u|^{-3})
\label{8.107}
\end{equation}
which, together with Proposition 8.1 implies:
\begin{equation}
\|O_i(\Omega\mbox{tr}\chi)\|_{L^\infty(S_{\ub,u})}\leq O(|u|^{-2})
\label{8.108}
\end{equation}
Substituting this estimate for $u=u_0$ in \ref{8.106} we deduce:
\begin{equation}
\|\mbox{tr}\s^{(O_i)}\spi\|_{L^\infty(S_{\ub,u_0})}\leq
O(\delta|u_0|^{-2}) \label{8.109}
\end{equation}
Consider next equation \ref{8.103}. Given any point $p\in
S_{\ub,u}$, let $q=\Phib_{u_0-u}(p)\in S_{\ub,u_0}$ be the point
corresponding to $p$ along the same generator of $\Cb_{\ub}$.
Integrating \ref{8.103} along this generator we obtain:
\begin{equation}
(\mbox{tr}\s^{(O_i)}\spi)(p)=(\mbox{tr}\s^{(O_i)}\spi)(q)+\int_{u_0}^u
2(O_i(\Omega\mbox{tr}\chib))(\Phib_{u^\prime-u_0}(q)) du^\prime
\label{8.110}
\end{equation}
Now by Proposition 6.2 and the results of Chapters 3 and 4 in
conjunction with Lemma 5.2, we have
\begin{equation}
\|\sd(\Omega\mbox{tr}\chib)\|_{L^\infty(S_{\ub,u})}\leq
O(\delta|u|^{-4}) \label{8.111}
\end{equation}
which, together with Proposition 8.1 implies:
\begin{equation}
\|O_i(\Omega\mbox{tr}\chib)\|_{L^\infty(S_{\ub,u})}\leq
O(\delta|u|^{-3}) \label{8.112}
\end{equation}
Substituting this estimate as well as the estimate \ref{8.109} in
\ref{8.110} we then deduce:
\begin{equation}
\|\mbox{tr}\s^{(O_i)}\spi\|_{L^\infty(S_{\ub,u})}\leq
O(\delta|u|^{-2}) \label{8.113}
\end{equation}

We turn to equations \ref{8.104}, \ref{8.105}. By the first part
of Lemma 4.2 equation \ref{8.104} implies, along $C_{u_0}$,
\begin{equation}
D(|\s^{(O_i)}\hat{\spi}|^2)\leq
2|\s^{(O_i)}\hat{\spi}|\left(C\Omega|\chih||\s^{(O_i)}\hat{\spi}|
+\left|2\sL_{O_i}(\Omega\chih)-\Omega\chih\mbox{tr}\s^{(O_i)}\spi\right|\right)
\label{8.114}
\end{equation}
Given again any point $q\in S_{\ub,u_0}$, let
$q_0=\Phi_{-\ub}(q)\in S_{0,u_0}$ be the point corresponding to
$q$ along the same generator of $C_{u_0}$. Applying Lemma 3.1 to
\ref{8.114} then yields, in view of \ref{8.53},
\begin{eqnarray}
&&|(\s^{(O_i)}\hat{\spi})(q)|\leq\label{8.115}\\
&&\int_0^{\ub}\exp\left\{\int_{\ub^\prime}^{\ub}C\Omega|\chih|(\Phi_{\ub^{\prime\prime}}(q_0))d\ub^{\prime\prime}
\right\}\left|2\sL_{O_i}(\Omega\chih)-\Omega\chih\mbox{tr}\s^{(O_i)}\spi\right|(\Phi_{\ub^\prime}(q_0))d\ub^\prime\nonumber
\end{eqnarray}
By the results of Chapter 3 the integral in the exponential is
bounded by $O(\delta^{1/2}|u|^{-1})$. Therefore if $\delta$ is
suitably small depending on ${\cal R}_0^\infty$ this integral does
not exceed $\log 2$. By Proposition 6.1 in conjunction with Lemma
5.2 we have:
\begin{equation}
|u|\|\snab(\Omega\chih)\|_{L^\infty(S_{\ub,u})}+\|\Omega\chih\|_{L^\infty(S_{\ub,u})}
\leq C\delta^{-1/2}|u|^{-1}(\scR_1^4(\beta)+{\cal
R}_0^\infty(\beta)) +O(|u|^{-2}) \label{8.116}
\end{equation}
Now, for any $S$-tangential vectorfield $X$ and any 2-covariant
$S$ tensorfield $\theta$ we have, relative to an arbitrary local
frame field for the $S_{\ub,u}$,
\begin{equation}
(\sL_X\theta)_{AB}=X^C\snab_C\theta_{AB}+\theta_{CB}\snab_A
X^C+\theta_{AC}\snab_B X^C \label{8.117}
\end{equation}
Hence it holds, pointwise,
\begin{equation}
|\sL_X\theta|\leq |X||\snab\theta|+2|\theta||\snab X|
\label{8.118}
\end{equation}
Applying this taking $X=O_i$, $\theta=\Omega\chih$ we obtain, in
view of Propositions 8.1, 8.2 and the estimate \ref{8.116},
\begin{equation}
\|\sL_{O_i}(\Omega\chih)\|_{L^\infty(S_{\ub,u})}\leq
C\delta^{-1/2}|u|^{-1}(\scR_1^4(\beta)+{\cal R}_0^\infty(\beta))
+O(|u|^{-2}) \label{8.119}
\end{equation}
Substituting this estimate for $u=u_0$ as well as the estimate
\ref{8.109} in \ref{8.115} we then deduce:
\begin{equation}
\|\s^{(O_i)}\hat{\spi}\|_{L^\infty(S_{\ub,u_0})}\leq
C\delta^{1/2}|u_0|^{-1}(\scR_1^4(\beta)+{\cal R}_0^\infty(\beta))
+O(\delta|u_0|^{-2}) \label{8.120}
\end{equation}

By the second part of Lemma 4.2 equation \ref{8.105} implies,
along the $\Cb_{\ub}$,
\begin{equation}
\Db(|\s^{(O_i)}\hat{\spi}|^2)\leq
2|\s^{(O_i)}\hat{\spi}|\left(C\Omega|\chibh||\s^{(O_i)}\hat{\spi}|
+\left|2\sL_{O_i}(\Omega\chibh)-\Omega\chibh\mbox{tr}\s^{(O_i)}\spi\right|\right)
\label{8.121}
\end{equation}
Given again any point $p\in S_{\ub,u}$, let $q=\Phib_{u_0-u}(p)\in
S_{\ub,u_0}$ be the point corresponding to $p$ along the same
generator of $\Cb_{\ub}$. Applying Lemma 3.1 to \ref{8.121} then
yields:
\begin{eqnarray}
&&|(\s^{(O_i)}\hat{\spi})(p)|\leq\exp\left\{\int_{u_0}^u
C\Omega|\chibh|(\Phib_{u^\prime-u_0}(q))du^\prime\right\}
|(\s^{(O_i)}\hat{\spi})(p)|\label{8.122}\\
&&\hspace{2cm}+\int_{u_0}^u\exp\left\{\int_{u^\prime}^u
C\Omega|\chibh|(\Phib_{u^{\prime\prime}-u_0}(q))du^{\prime\prime}
\right\}\cdot\nonumber\\
&&\hspace{4cm}\cdot\left|2\sL_{O_i}(\Omega\chibh)-\Omega\chibh\mbox{tr}\s^{(O_i)}\spi\right|(\Phib_{u^\prime-u_0}(q))du^\prime\nonumber
\end{eqnarray}
By the results of Chapter 3 the integral in the exponential is
bounded by $O(\delta^{1/2}|u|^{-1})$. Therefore if $\delta$ is
suitably small depending on ${\cal D}_0^\infty$, ${\cal
R}_0^\infty$ this integral does not exceed $\log 2$. By
Proposition 6.2 in conjunction with Lemma 5.2 we have:
\begin{equation}
|u|\|\snab(\Omega\chibh)\|_{L^\infty(S_{\ub,u})}+\|\Omega\chibh\|_{L^\infty(S_{\ub,u})}
\leq C\delta^{1/2}|u|^{-2}(\scR_1^4(\beta)+{\cal
R}_0^\infty(\beta)) +O(|u|^{-3}) \label{8.123}
\end{equation}
Applying then \ref{8.118} taking $X=O_i$, $\theta=\Omega\chibh$ we
obtain, in view of Propositions 8.1, 8.2 and the estimate
\ref{8.123},
\begin{equation}
\|\sL_{O_i}(\Omega\chih)\|_{L^\infty(S_{\ub,u})}\leq
C\delta^{1/2}|u|^{-2}(\scR_1^4(\beta)+{\cal R}_0^\infty(\beta))
+O(|u|^{-3}) \label{8.124}
\end{equation}
Substituting this estimate as well as the estimates \ref{8.120}
and \ref{8.109} in \ref{8.122} we then deduce:
\begin{equation}
\|\s^{(O_i)}\hat{\spi}\|_{L^\infty(S_{\ub,u})}\leq
C\delta^{1/2}|u|^{-1}(\scR_1^4(\beta)+{\cal R}_0^\infty(\beta))
+O(\delta|u|^{-2}) \label{8.125}
\end{equation}
and the proof of the proposition is complete.

\vspace{5mm}

\noindent{\bf Proposition 8.4} \ \ \ The following estimate holds
for all $(\ub,u)\in D^\prime$:
$$\|Z_i\|_{L^\infty(S_{\ub,u})}\leq O(|u|^{-1})$$
provided that $\delta$ is suitably small depending on ${\cal
D}_0^\infty$, ${\cal R}_0^\infty$, $\scD_1^4$, $\scR_1^4$ and
$\scR_2$.

\noindent{\em Proof:} \ We consider the propagation equation
\ref{8.63} along the $\Cb_{\ub}$. We have
\begin{eqnarray}
\Db(|Z_i|^2)&=&(\Db\sg)(Z_i,Z_i)+2(Z_i,\Db Z_i)\nonumber\\
&=&2\Omega\chib(Z_i,Z_i)-8(Z_i,\sL_{O_i}(\Omega^2\zeta^\sharp))
\label{8.126}
\end{eqnarray}
hence:
\begin{equation}
\Db(|Z_i|^2)\leq
2\left(\frac{1}{2}\Omega\mbox{tr}\chib+\Omega|\chibh|\right)|Z_i|^2+8|Z_i||\sL_{O_i}(\Omega^2\zeta^\sharp)|
\label{8.127}
\end{equation}
Given any point $p\in S_{\ub,u}$, let $q=\Phib_{u_0-u}(p)\in
S_{\ub,u_0}$ be the point corresponding to $p$ along the same
generator of $\Cb_{\ub}$. In view of \ref{8.64}, applying Lemma
3.1 to \ref{8.127} yields:
\begin{eqnarray}
&&|Z_i(p)|\leq\int_{u_0}^u\exp\left\{\int_{u^\prime}^u\left(\frac{1}{2}\Omega\mbox{tr}\chib+\Omega|\chibh|\right)
(\Phib_{u^{\prime\prime}-u_0}(q))du^{\prime\prime}\right\}\cdot\nonumber\\
&&\hspace{5cm}\cdot
4\left|\sL_{O_i}(\Omega^2\zeta^\sharp)\right|(\Phib_{u^\prime-u_0}(q))du^\prime
\label{8.128}
\end{eqnarray}
By the results of Chapter 3 the integral in the exponential is
bounded by
\begin{equation}
\log\left(\frac{|u|}{|u^\prime|}\right)+O(\delta^{1/2}|u|^{-1})
\label{8.129}
\end{equation}
and if $\delta$ is suitably small depending on ${\cal
D}_0^\infty$, ${\cal R}_0^\infty$ the term
$O(\delta^{1/2}|u|^{-1})$ does not exceed $\log 2$. By Proposition
6.2 in conjunction with Lemma 5.2 we have, recalling that (see
\ref{1.65})
$$2\zeta=\eta-\etb,$$

\begin{equation}
|u|\|\snab(\Omega^2\zeta)\|_{L^\infty(S_{\ub,u})}+\|\Omega^2\zeta\|_{L^\infty(S_{\ub,u})}\leq
O(|u|^{-2}) \label{8.130}
\end{equation}
Now, for any pair of $S$-tangential vectorfields $X$, $Y$ we have:
\begin{equation}
|\sL_X Y|=|[X,Y]|=|\snab_X Y-\snab_Y X|\leq |X||\snab Y|+|Y||\snab
X| \label{8.131}
\end{equation}
Applying this taking $X=O_i$, $Y=\Omega^2\zeta^\sharp$ we obtain,
in view of Propositions 8.1, 8.2 and the estimate \ref{8.130},
\begin{equation}
\|\sL_{O_i}(\Omega^2\zeta^\sharp)\|_{L^\infty(S_{\ub,u})}\leq
O(|u|^{-2}) \label{8.132}
\end{equation}
Substituting this estimate as well as the result \ref{8.129} above
in \ref{8.128} we then deduce:
\begin{equation}
\|Z_i\|_{L^\infty(S_{\ub,u})}\leq O(|u|^{-1}) \label{8.133}
\end{equation}
and the proof of the proposition is complete.

\vspace{5mm}

According to \ref{8.41}, \ref{8.42} we have:
\begin{equation}
\s^{(O_i)}\nb=\s^{(O_i)}n=0 \label{8.134}
\end{equation}
while according to \ref{8.40} we have:
\begin{equation}
\s^{(O_i)}\mb=0 \label{8.135}
\end{equation}
The remaining components of the deformation tensor of the $O_i$
are:
\begin{equation}
\s^{(O_i)}m=\Omega^{-1}\sg\cdot Z_i \label{8.136}
\end{equation}
\begin{equation}
\s^{(O_i)}j=\frac{1}{2}\mbox{tr}\s^{(O_i)}\spi-2O_i(\log\Omega)
\label{8.137}
\end{equation}
(see \ref{8.43}) and:
\begin{equation}
\s^{(O_i)}\ih=\s^{(O_i)}\hat{\spi} \label{8.138}
\end{equation}
Propositions 8.3 and 8.4 (and the estimate \ref{7.210}) then yield
the following estimates for these components:
\begin{eqnarray}
&&\|\s^{(O_i)}\ih\|_{L^\infty(S_{\ub,u})}\leq
C\delta^{1/2}|u|^{-1}(\scR_1^4(\beta)+{\cal R}_0^\infty(\beta))
+O(\delta|u|^{-2})\nonumber\\
&&\|\s^{(O_i)}j\|_{L^\infty(S_{\ub,u})}\leq O(\delta|u|^{-2})\nonumber\\
&&\|\s^{(O_i)}m\|_{L^\infty(S_{\ub,u})}\leq
O(|u|^{-1})\label{8.139}
\end{eqnarray}

\chapter{Estimates for the Derivatives of the Deformation Tensors
of the Commutation Fields}

\section{Estimates for the 1st derivatives of the deformation tensors of $L$, $S$}

The components of the deformation tensor of the commutation field
$L$ are given by table \ref{8.21}. From this table and the results
of Chapters 3 and 4 we obtain:
\begin{eqnarray}
&&\|\snab\s^{(L)}\ih\|_{L^4(S_{\ub,u})}\leq C\delta^{-1/2}|u|^{-3/2}\scR_1^4(\alpha)+O(|u|^{-5/2})\nonumber\\
&&\|\sd\s^{(L)}j\|_{L^4(S_{\ub,u})}\leq O(|u|^{-5/2})\nonumber\\
&&\|\snab\s^{(L)}\mb\|_{L^4(S_{\ub,u})}\leq
O(\delta^{1/2}|u|^{-5/2}) \label{9.1}
\end{eqnarray}
for all $(\ub,u)\in D^\prime$.

Next, from table \ref{8.21}, equations \ref{1.61}, \ref{3.6},
\ref{3.8} and the results of Chapters 3 and 4 we deduce:
\begin{eqnarray}
&&\|\Dh\s^{(L)}\ih\|_{L^4(S_{\ub,u})}\leq
C\delta^{-3/2}|u|^{-1/2}{\cal
R}_0^\infty(\alpha)+O(\delta^{-1/2}|u|^{-5/2})
\nonumber\\
&&\|D\s^{(L)}j\|_{L^4(S_{\ub,u})}\leq O(\delta^{-1}|u|^{-3/2})\nonumber\\
&&\|D\s^{(L)}\mb\|_{L^4(S_{\ub,u})}\leq O(\delta^{-1/2}|u|^{-3/2})
\label{9.2}
\end{eqnarray}
for all $(\ub,u)\in D^\prime$.

Next, from table \ref{8.21}, equations \ref{1.63}, \ref{1.87},
\ref{4.c2}, \ref{4.c4} and the results of Chapters 3 and 4 we
deduce:
\begin{eqnarray}
&&\|\Dbh\s^{(L)}\ih\|_{L^4(S_{\ub,u})}\leq
C\delta^{-1/2}|u|^{-3/2}{\cal
R}_0^\infty(\alpha)+O(\delta^{1/2}|u|^{-5/2})
\nonumber\\
&&\|\Db\s^{(L)}j\|_{L^4(S_{\ub,u})}\leq O(|u|^{-5/2})\nonumber\\
&&\|\Db\s^{(L)}\mb\|_{L^4(S_{\ub,u})}\leq
O(\delta^{1/2}|u|^{-5/2}) \label{9.3}
\end{eqnarray}
for all $(\ub,u)\in D^\prime$.

The components of the deformation tensor of the commutation field
$S$ are given by table \ref{8.30}. Let us define the function:
\begin{equation}
\lambda=\frac{1}{2}\Omega(u\mbox{tr}\chib+\ub\mbox{tr}\chi)-1
\label{9.4}
\end{equation}
In terms of the function $\lambda$ the component $\s^{(S)}j$ is
expressed as:
\begin{equation}
\s^{(S)}j=2(\lambda-u\omb-\ub\omega) \label{9.5}
\end{equation}

By the results of Chapter 3 we have:
\begin{equation}
\|\lambda\|_{L^\infty(S_{\ub,u})}\leq O(\delta|u|^{-2})
\label{9.6}
\end{equation}
Moreover, by the results of Chapter 4 we have:
\begin{equation}
\|\sd\lambda\|_{L^4(S_{\ub,u})}\leq O(\delta|u|^{-5/2})
\label{9.7}
\end{equation}
Using equations \ref{4.c1} and \ref{3.6} we deduce:
\begin{eqnarray}
D\lambda&=&-\frac{1}{2}\Omega\mbox{tr}\chi\lambda\label{9.8}\\
&\s&+\frac{1}{2}u\Omega^2\left\{2\rho-(\chih,\chibh)+2\sdiv\etb+2|\etb|^2\right\}\nonumber\\
&\s&+\ub\Omega\omega\mbox{tr}\chi-\frac{1}{2}\ub\Omega^2|\chih|^2\nonumber
\end{eqnarray}
Also, using equations \ref{4.c2} and \ref{3.7} we deduce:
\begin{eqnarray}
\Db\lambda&=&-\frac{1}{2}\Omega\mbox{tr}\chib\lambda\label{9.9}\\
&\s&+\frac{1}{2}\ub\Omega^2\left\{2\rho-(\chih,\chibh)+2\sdiv\eta+2|\eta|^2\right\}\nonumber\\
&\s&+u\Omega\omb\mbox{tr}\chib-\frac{1}{2}u\Omega^2|\chibh|^2\nonumber
\end{eqnarray}
The results of Chapters 3 and 4 together with \ref{9.6} then
yield:
\begin{equation}
\|D\lambda\|_{L^4(S_{\ub,u})}\leq O(|u|^{-3/2}) \label{9.10}
\end{equation}
\begin{equation}
\|\Db\lambda\|_{L^4(S_{\ub,u})}\leq O(\delta|u|^{-5/2})
\label{9.11}
\end{equation}

From table \ref{8.30}, equation \ref{9.5}, the results of Chapters
3 and 4 and the estimate \ref{9.7} we obtain:
\begin{eqnarray}
&&\|\snab\s^{(S)}\ih\|_{L^4(S_{\ub,u})}\leq
C\delta^{1/2}|u|^{-3/2}(\scD_1^4(\chibh)+\scR_1^4(\alpha))+O(\delta|u|^{-5/2})
\nonumber\\
&&\|\sd\s^{(S)}j\|_{L^4(S_{\ub,u})}\leq O(\delta|u|^{-5/2})\nonumber\\
&&\|\snab\s^{(S)}\mb\|_{L^4(S_{\ub,u})}\leq O(\delta^{3/2}|u|^{-5/2})\nonumber\\
&&\|\snab\s^{(S)}m\|_{L^4(S_{\ub,u})}\leq
O(\delta^{1/2}|u|^{-3/2}) \label{9.12}
\end{eqnarray}
for all $(\ub,u)\in D^\prime$.

Next, from table \ref{8.30}, equation \ref{9.5}, together with
equations \ref{1.61}, \ref{3.6}, \ref{3.8}, \ref{4.c1},
\ref{4.c3}, the results of Chapters 3 and 4 and the estimate
\ref{9.10} we deduce:
\begin{eqnarray}
&&\|\Dh\s^{(S)}\ih\|_{L^4(S_{\ub,u})}\leq
C\delta^{-1/2}|u|^{-1/2}{\cal
R}_0^\infty(\alpha)+O(\delta^{1/2}|u|^{-3/2})
\nonumber\\
&&\|D\s^{(S)}j\|_{L^4(S_{\ub,u})}\leq O(|u|^{-3/2})\nonumber\\
&&\|D\s^{(S)}\mb\|_{L^4(S_{\ub,u})}\leq O(\delta^{1/2}|u|^{-3/2})\nonumber\\
&&\|D\s^{(S)}m\|_{L^4(S_{\ub,u})}\leq O(\delta^{-1/2}|u|^{-1/2})
\label{9.13}
\end{eqnarray}
for all $(\ub,u)\in D^\prime$.

Finally, from table \ref{8.30}, equation \ref{9.5}, together with
equations \ref{1.63}, \ref{3.7}, \ref{3.9}, \ref{4.c2},
\ref{4.c4}, the results of Chapters 3 and 4 and the estimate
\ref{9.11} we deduce:
\begin{eqnarray}
&&\|\Dbh\s^{(S)}\ih\|_{L^4(S_{\ub,u})}\leq
C\delta^{1/2}|u|^{-3/2}({\cal D}_0^\infty(\chibh)+{\cal
R}_0^\infty(\alpha))+O(\delta^{3/2}|u|^{-5/2})
\nonumber\\
&&\|\Db\s^{(S)}j\|_{L^4(S_{\ub,u})}\leq O(\delta|u|^{-5/2})\nonumber\\
&&\|\Db\s^{(S)}\mb\|_{L^4(S_{\ub,u})}\leq O(\delta^{3/2}|u|^{-5/2})\nonumber\\
&&\|\Db\s^{(S)}m\|_{L^4(S_{\ub,u})}\leq O(\delta^{1/2}|u|^{-3/2})
\label{9.14}
\end{eqnarray}
for all $(\ub,u)\in D^\prime$.

Since the above estimates  depend only on the results of Chapters
3 and 4 the symbol $O(\delta^p|u|^r)$ may be taken here to have
the more restricted meaning of the product of $\delta^p|u|^r$
with a non-negative non-decreasing continuous function of the
quantities ${\cal D}_0^\infty$, ${\cal R}_0^\infty$, $\scD_1^4$,
$\scR_1^4$.

\section{Estimates for the 1st derivatives of the deformation tensors of the $O_i$}

We begin with the following lemma.

\vspace{5mm}

\noindent{\bf Lemma 9.1} \ \ \ Let $X$ be a $S$-tangential
vectorfield and let $\theta$ be an arbitrary type $T^q_p$ $S$
tensorfield. We have, with respect to an arbitrary local frame
field for the $S_{\ub,u}$,
\begin{eqnarray*}
(\sL_X\snab\theta-\snab\sL_X\theta)^{C_1...C_q}_{AB_1...B_p}&=&-\sum_{i=1}^p\s^{(X)}\spi^D_{1,AB_i}
\theta^{C_1...C_q}_{B_1...\stackrel{D}{>B_i<}...B_p}\\
&\s&+\sum_{j=1}^q\s^{(X)}\spi^{C_j}_{1,AD}\theta^{C_1...\stackrel{D}{>C_j<}...C_q}_{B_1...B_p}
\end{eqnarray*}
where $\s^{(X)}\spi_1$ is the Lie derivative with respect to $X$
of the induced connection $\sGamma$ on the $S_{\ub,u}$, a type
$T^1_2$ $S$ tensorfield, symmetric in the lower indices, given by:
$$\s^{(X)}\spi^C_{1,AB}=\frac{1}{2}(\snab_A\s^{(X)}\spi^C_{\s B}+\snab_B\s^{(X)}\spi^C_{\s A}-\snab^C\s^{(X)}\spi_{AB})$$

\noindent{\em Proof:} \ We may restrict attention to a given
$S_{\ub,u}$. Then $X$ is simply a vectorfield and $\theta$ is a
type $T^q_p$ tensorfield on $S_{\ub,u}$. Let $X$ generate the
1-parameter group $f_t$ of diffeomorphisms of $S_{\ub,u}$.
Consider first the case $q=0$. Then
$\snab\theta=\stackrel{\sg}{\snab}\theta$ is a $p+1$ covariant
tensorfield on $S_{\ub,u}$ and we have:
\begin{equation}
f^*_t(\stackrel{\sg}{\snab}\theta)=\stackrel{f^*_t\sg}{\snab}f^*_t\theta
\label{9.c1}
\end{equation}
and:
\begin{equation}
\sL_X(\snab\theta)=\left.\frac{d}{dt}f^*_t(\snab\theta)\right|_{t=0}
\label{9.c2}
\end{equation}
Let $(\vartheta^A:A=1,2)$ be local coordinates on $S_{\ub,u}$ and
let $\stackrel{f^*_t\sg}{\sGamma^C_{AB}}$ be the connection
coefficients of $f^*_t\sg$ (Christoffel symbols) in this
coordinate system. In terms of the coordinates
$(\vartheta^A:A=1,2)$ we have:
\begin{equation}
(\stackrel{f^*_t\sg}{\snab}f^*_t\theta)_{AB_1...B_p}=\frac{\partial(f^*_t\theta)_{B_1...B_p}}{\partial\vartheta^A}
-\sum_{i=1}^p\stackrel{f^*_t\sg}{\sGamma^C_{AB_i}}(f^*_t\theta)_{B_1...\stackrel{C}{>B_i<}...B_p}
\label{9.c3}
\end{equation}
Hence:
\begin{eqnarray}
&&\left.\frac{d}{dt}(\stackrel{f^*_t\sg}{\snab}f^*_t\theta)_{AB_1...B_p}\right|_{t=0}=
\frac{\partial}{\partial\vartheta^A}\left.\frac{d}{dt}(f^*_t\theta)_{B_1...B_p}\right|_{t=0}\nonumber\\
&&\hspace{25mm}-\sum_{i=1}^p\stackrel{\sg}{\sGamma^C_{AB_i}}\left.\frac{d}{dt}(f^*_t\theta)_{B_1...\stackrel{C}{>B_i<}...B_p}\right|_{t=0}\nonumber\\
&&\hspace{25mm}-\sum_{i=1}^p\left.\frac{d}{dt}\stackrel{f^*_t\sg}{\sGamma^C_{AB_i}}\right|_{t=0}\theta_{B_1...\stackrel{C}{>B_i<}...B_p}
\label{9.c4}
\end{eqnarray}
Since
$$\left.\frac{d}{dt}(f^*_t\theta)_{B_1...B_p}\right|_{t=0}=(\sL_X\theta)_{B_1...B_p}$$
the first two terms on the right in \ref{9.c4} are simply
\begin{equation}
(\snab(\sL_X\theta))_{AB_1...B_p} \label{9.c5}
\end{equation}
Also, since
$$\stackrel{f^*_t\sg}{\sGamma^C_{AB}}=\frac{1}{2}((f^*_t\sg)^{-1})^{CD}\left(\frac{\partial(f^*_t\sg)_{BD}}{\partial\vartheta^A}
+\frac{\partial(f^*_t\sg)_{AD}}{\partial\vartheta^B}-\frac{\partial(f^*_t\sg)_{AB}}{\partial\vartheta^D}\right)$$
we obtain:
\begin{eqnarray*}
&&\left.\frac{d}{dt}\stackrel{f^*_t\sg}{\sGamma^C_{AB}}\right|_{t=0}=
\frac{1}{2}(\sg^{-1})^{CD}\left(\stackrel{\sg}{\snab}_A\left.\frac{d}{dt}(f^*_t\sg)_{BD}\right|_{t=0}
+\stackrel{\sg}{\snab}_B\left.\frac{d}{dt}(f^*_t\sg)_{AD}\right|_{t=0}\right.\\
&&\hspace{65mm}\left.-\stackrel{\sg}{\snab}_D\left.\frac{d}{dt}(f^*_t\sg)_{AB}\right|_{t=0}\right)
\end{eqnarray*}
or, since
$$\left.\frac{d}{dt}(f^*_t\sg)\right|_{t=0}=\s^{(X)}\spi$$
simply:
\begin{equation}
\left.\frac{d}{dt}\stackrel{f^*_t\sg}{\sGamma^C_{AB}}\right|_{t=0}=\frac{1}{2}(\sg^{-1})^{CD}(\snab_A\s^{(X)}\spi_{BD}
+\snab_B\s^{(X)}\spi_{AD}-\snab_D\s^{(X)}\spi_{AB})=\s^{(X)}\spi^C_{1,AB}
\label{9.c6}
\end{equation}
In view of \ref{9.c5}, \ref{9.c6}, and \ref{9.c2}, the formula
\ref{9.c4} becomes:
\begin{equation}
(\sL_X\snab\theta)_{AB_1...B_p}=(\snab\sL_X\theta)_{AB_1...B_p}-\sum_{i=1}^p\spi^C_{1,AB_i}\theta_{B_1...\stackrel{C}{>B_i<}...B_p}
\label{9.c7}
\end{equation}
which is the lemma in the case $q=0$.

To obtain the lemma in the general case of a type $T^q_p$
tensorfield on $S_{\ub,u}$, given such a tensorfield $\theta$ we
apply what we have just proved to the $p$ covariant tensorfield
$$\theta\cdot(\xi_1,...,\xi_q)$$
where $\xi_1,...,\xi_q$ are arbitrary 1-forms on $S_{\ub,u}$. The
general case then follows by virtue of the Leibniz rule for Lie
derivatives $\sL_X$ as well as covariant for derivatives $\snab$,
if we also apply what we have just proved to the 1-forms
$\xi_1,...,\xi_q$.

\vspace{5mm}

\vspace{5mm}

\noindent{\bf Proposition 9.1} \ \ \ The following estimates hold
for all $(\ub,u)\in D^\prime$:
$$\|\sd\mbox{tr}\s^{(O_i)}\spi\|_{L^4(S_{\ub,u})}\leq O(\delta|u|^{-5/2})$$
$$\|\snab\s^{(O_i)}\hat{\spi}\|_{L^4(S_{\ub,u})}\leq C\delta^{1/2}|u|^{-3/2}(\scR_1^4(\beta)+{\cal R}_0^\infty(\beta))
+O(\delta|u|^{-5/2})$$ provided that $\delta$ is suitably small
depending on ${\cal D}_0^\infty$, ${\cal R}_0^\infty$, $\scD_1^4$,
$\scR_1^4$ and $\scR_2$.

\noindent{\em Proof:} \ Applying $\sd$ to the propagation equation
\ref{8.102} we obtain, in view of Lemma 1.2, noting that for an
arbitrary $S$-tangential vectorfield $X$ and an arbitrary function
$f$ it holds:
\begin{equation}
\sd\sL_X f=\sL_X\sd f, \label{9.15}
\end{equation}
the following propagation equation for
$\sd\mbox{tr}\s^{(O_i)}\spi$ along $C_{u_0}$:
\begin{equation}
D\sd\mbox{tr}\s^{(O_i)}\spi=2\sL_{O_i}\sd(\Omega\mbox{tr}\chi)
\label{9.16}
\end{equation}
To this we apply Lemma 4.6 taking $p=4$. Here $r=1$, $\nu=0$,
$\gamma=0$ and we obtain:
\begin{eqnarray}
\|\sd\mbox{tr}\s^{(O_i)}\spi\|_{L^4(S_{\ub,u_0})}&\leq&C\int_0^{\ub}
\|\sL_{O_i}\sd(\Omega\mbox{tr}\chi)\|_{L^4(S_{\ub^\prime,u_0})}d\ub^\prime\nonumber\\
&\leq&O(\delta|u_0|^{-5/2}) \label{9.17}
\end{eqnarray}
by Proposition 6.1, Propositions 8.1 and 8.2, and the results of
Chapter 4.

Applying $\sd$ to the propagation equation \ref{8.103} we obtain,
in view of Lemma 1.2 and \ref{9.15}, the following propagation
equation for $\sd\mbox{tr}\s^{(O_i)}\spi$ along the $\Cb_{\ub}$:
\begin{equation}
\Db\sd\mbox{tr}\s^{(O_i)}\spi=2\sL_{O_i}\sd(\Omega\mbox{tr}\chib)
\label{9.18}
\end{equation}
To this we apply Lemma 4.7 taking $p=4$. Here $r=1$, $\nu=0$,
$\gammab=0$ and we obtain:
\begin{eqnarray}
|u|^{1/2}\|\sd\mbox{tr}\s^{(O_i)}\spi\|_{L^4(S_{\ub,u})}&\leq&C|u_0|^{1/2}\|\sd\mbox{tr}\s^{(O_i)}\spi\|_{L^4(S_{\ub,u_0})}
\label{9.19}\\
&\s&+C\int_{u_0}^u|u^\prime|^{1/2}\|\sL_{O_i}\sd(\Omega\mbox{tr}\chib)\|_{L^4(S_{\ub,u^\prime})}du^\prime\nonumber
\end{eqnarray}
Substituting the estimate \ref{9.17} and using Proposition 6.2
together with Propositions 8.1 and 8.2 and the results of Chapter
4 we then deduce:
\begin{equation}
\|\sd\mbox{tr}\s^{(O_i)}\spi\|_{L^4(S_{\ub,u})}\leq
O(\delta|u|^{-5/2}) \label{9.20}
\end{equation}

Next, we apply $\snab$ to the propagation equation \ref{8.104}.
Using Lemmas 4.1 and 9.1 we then obtain the following propagation
equation for $\snab\s^{(O_i)}\hat{\spi}$ along $C_{u_0}$:
\begin{eqnarray}
D\snab\s^{(O_i)}\hat{\spi}-\Omega\mbox{tr}\chi\snab^{(O_i)}\hat{\spi}&=&
-D\sGamma\cdot\s^{(O_i)}\hat{\spi}+\sd(\Omega\mbox{tr}\chi)\otimes\s^{(O_i)}\hat{\spi}\nonumber\\
&\s&+2\sL_{O_i}\snab(\Omega\chih)+2\s^{(O_i)}\spi_1\cdot\Omega\chih
\label{9.21}
\end{eqnarray}
Here, we denote:
\begin{equation}
(D\sGamma\cdot\s^{(O_i)}\hat{\spi})_{ABC}=(D\sGamma)^D_{AB}\s^{(O_i)}\hat{\spi}_{DC}
+(D\sGamma)^D_{AC}\s^{(O_i)}\hat{\spi}_{BD} \label{9.22}
\end{equation}
\begin{equation}
(\s^{(O_i)}\spi_1\cdot\Omega\chih)_{ABC}=\s^{(O_i)}\spi^D_{1,AB}\Omega\chih_{DC}+\s^{(O_i)}\spi^D_{1,AC}\Omega\chih_{BD}
\label{9.23}
\end{equation}
We decompose $\s^{(O_i)}\spi_1$ into the $T^1_2$ type $S$
tensorfields, symmetric in the lower indices,
\begin{equation}
\s^{(O_i)}\spi_1=\s^{(O_i)}\hat{\spi}_1+\mbox{tr}\s^{(O_i)}\spi_1
\label{9.24}
\end{equation}
where:
\begin{equation}
\s^{(O_i)}\hat{\spi}^C_{1,AB}=\frac{1}{2}\left\{\snab_A\s^{(O_i)}\hat{\spi}^C_{\s
B}+\snab_B\s^{(O_i)}\hat{\spi}^C_{\s A}
-\snab^C\s^{(O_i)}\hat{\spi}_{AB}\right\} \label{9.25}
\end{equation}
\begin{equation}
\mbox{tr}\s^{(O_i)}\spi^C_{1,AB}=\frac{1}{2}\left\{\delta^C_B\sd_A\mbox{tr}\s^{(O_i)}\spi
+\delta^C_A\sd_B\mbox{tr}\s^{(O_i)}\spi-\sg_{AB}\sd^C\mbox{tr}\s^{(O_i)}\spi\right\}
\label{9.26}
\end{equation}
Substituting in \ref{9.21} we bring the propagation equation for
$\snab\s^{(O_i)}\hat{\spi}$ along $C_{u_0}$ to the form:
\begin{equation}
D\snab\s^{(O_i)}\hat{\spi}-\Omega\mbox{tr}\chi\snab^{(O_i)}\hat{\spi}=\gamma\cdot\snab\s^{(O_i)}\hat{\spi}
+2\mbox{tr}\s^{(O_i)}\spi_1\cdot\Omega\chih+2\sL_{O_i}\snab(\Omega\chih)+r
\label{9.27}
\end{equation}
where:
\begin{equation}
r=-D\sGamma\cdot\s^{(O_i)}\hat{\spi}+\sd(\Omega\mbox{tr}\chi)\otimes\s^{(O_i)}\hat{\spi}
\label{9.28}
\end{equation}
In \ref{9.27},
\begin{eqnarray*}
(\gamma\cdot\snab\s^{(O_i)}\hat{\spi})_{ABC}&=&\frac{1}{2}\Omega\chih^D_C\left\{\snab_A\s^{(O_i)}\hat{\spi}_{BD}
+\snab_B\s^{(O_i)}\hat{\spi}_{AD}-\snab_D\s^{(O_i)}\hat{\spi}_{AB}\right\}\\
&\s&+\frac{1}{2}\Omega\chih^D_B\left\{\snab_A\s^{(O_i)}\hat{\spi}_{CD}+\snab_C\s^{(O_i)}\hat{\spi}_{AD}
-\snab_D\s^{(O_i)}\hat{\spi}_{AC}\right\}
\end{eqnarray*}
thus $\gamma$ is the type $T^3_3$ $S$ tensorfield given by:
\begin{eqnarray}
\gamma^{DEF}_{ABC}&=&\frac{1}{4}\Omega\left\{\chih^E_C(\delta^D_A\delta^F_B+\delta^D_B\delta^F_A)
+\chih^F_C(\delta^D_A\delta^E_B+\delta^D_B\delta^E_A)\right.\nonumber\\
&\s&\s+\chih^E_B(\delta^D_A\delta^F_C+\delta^D_C\delta^F_A)
+\chih^F_B(\delta^D_A\delta^E_C+\delta^D_C\delta^E_A)\nonumber\\
&\s&\s\left.-\chih^D_C(\delta^E_A\delta^F_B+\delta^F_A\delta^E_B)-\chih^D_B(\delta^E_A\delta^F_C+\delta^F_A\delta^E_C)
\right\}\label{9.29}
\end{eqnarray}
To \ref{9.27} we apply Lemma 4.6 taking $p=4$. Here $r=3$, $\nu=2$
and $\gamma$ is given by \ref{9.29}. We obtain:
\begin{equation}
\|\snab\s^{(O_i)}\hat{\spi}\|_{L^4(S_{\ub,u_0})}\leq C\int_0^{\ub}
\left\|2\mbox{tr}\s^{(O_i)}\spi_1\cdot\Omega\chih+2\sL_{O_i}\snab(\Omega\chih)+r\right\|_{L^4(S_{\ub^\prime,u_0})}d\ub^\prime
\label{9.30}
\end{equation}
Now, by the estimate \ref{9.17}:
\begin{equation}
\|\mbox{tr}\s^{(O_i)}\spi_1\cdot\Omega\chih\|_{L^4(S_{\ub,u_0})}\leq
O(\delta^{1/2}|u_0|^{-7/2}) \label{9.31}
\end{equation}
while by Proposition 6.1 together with Propositions 8.1 and 8.2:
\begin{equation}
\|\sL_{O_i}\snab(\Omega\chih)\|_{L^4(S_{\ub,u_0})}\leq
C\delta^{-1/2}|u_0|^{-3/2}(\scR_1^4(\beta)+{\cal
R}_0^\infty(\beta)) +O(|u_0|^{-5/2}) \label{9.32}
\end{equation}
Also, by Proposition 8.3 and the estimates \ref{8.90} and
\ref{8.107}:
\begin{equation}
\|r\|_{L^4(S_{\ub,u_0})}\leq O(|u_0|^{-5/2}) \label{9.33}
\end{equation}
Substituting then \ref{9.31} - \ref{9.33} in \ref{9.30} we
conclude that:
\begin{equation}
\|\snab\s^{(O_i)}\hat{\spi}\|_{L^4(S_{\ub,u_0})}\leq
C\delta^{1/2}|u_0|^{-3/2}(\scR_1^4(\beta)+{\cal
R}_0^\infty(\beta)) +O(\delta|u_0|^{-5/2}) \label{9.34}
\end{equation}

Applying $\snab$ to the propagation equation \ref{8.105} we
deduce, following steps analogous to those leading to \ref{9.27}
the following propagation equation for $\snab\s^{(O_i)}\hat{\spi}$
along the $\Cb_{\ub}$:
\begin{equation}
\Db\snab\s^{(O_i)}\hat{\spi}-\Omega\mbox{tr}\chib\snab^{(O_i)}\hat{\spi}=\gammab\cdot\snab\s^{(O_i)}\hat{\spi}
+2\mbox{tr}\s^{(O_i)}\spi_1\cdot\Omega\chibh+2\sL_{O_i}\snab(\Omega\chibh)+\rb
\label{9.35}
\end{equation}
where:
\begin{equation}
\rb=-\Db\sGamma\cdot\s^{(O_i)}\hat{\spi}+\sd(\Omega\mbox{tr}\chib)\otimes\s^{(O_i)}\hat{\spi}
\label{9.36}
\end{equation}
In \ref{9.35} $\gammab$ is the type $T^3_3$ $S$ tensorfield given
by:
\begin{eqnarray}
\gammab^{DEF}_{ABC}&=&\frac{1}{4}\Omega\left\{\chibh^E_C(\delta^D_A\delta^F_B+\delta^D_B\delta^F_A)
+\chibh^F_C(\delta^D_A\delta^E_B+\delta^D_B\delta^E_A)\right.\nonumber\\
&\s&\s+\chibh^E_B(\delta^D_A\delta^F_C+\delta^D_C\delta^F_A)
+\chibh^F_B(\delta^D_A\delta^E_C+\delta^D_C\delta^E_A)\nonumber\\
&\s&\s\left.-\chibh^D_C(\delta^E_A\delta^F_B+\delta^F_A\delta^E_B)-\chibh^D_B(\delta^E_A\delta^F_C+\delta^F_A\delta^E_C)
\right\}\label{9.37}
\end{eqnarray}
(this is the conjugate of $\gamma$). To \ref{9.35} we apply Lemma
4.7 taking $p=4$. Here $r=3$, $\nu=2$ and $\gammab$ is given by
\ref{9.37}. We obtain:
\begin{eqnarray}
&&|u|^{1/2}\|\snab\s^{(O_i)}\hat{\spi}\|_{L^4(S_{\ub,u})}\leq
C|u_0|^{1/2}\|\snab\s^{(O_i)}\hat{\spi}\|_{L^4(S_{\ub,u_0})}
\label{9.38}\\
&&\hspace{1cm}+C\int_{u_0}^u|u^\prime|^{1/2}\left\|2\mbox{tr}\s^{(O_i)}\spi_1\cdot\Omega\chibh
+2\sL_{O_i}\snab(\Omega\chibh)+\rb\right\|_{S_{\ub,u^\prime})}du^\prime\nonumber
\end{eqnarray}
Now, by the estimate \ref{9.17}:
\begin{equation}
\|\mbox{tr}\s^{(O_i)}\spi_1\cdot\Omega\chibh\|_{L^4(S_{\ub,u})}\leq
O(\delta^{3/2}|u|^{-9/2}) \label{9.39}
\end{equation}
while by Proposition 6.2 together with Propositions 8.1 and 8.2:
\begin{equation}
\|\sL_{O_i}\snab(\Omega\chibh)\|_{L^4(S_{\ub,u})}\leq
C\delta^{1/2}|u|^{-5/2}(\scR_1^4(\beta)+{\cal R}_0^\infty(\beta))
+O(\delta|u|^{-7/2}) \label{9.40}
\end{equation}
Also, by Proposition 8.3 and the estimates \ref{8.96} and
\ref{8.111}:
\begin{equation}
\|\rb\|_{L^4(S_{\ub,u_0})}\leq O(\delta|u|^{-7/2}) \label{9.41}
\end{equation}
Substituting then \ref{9.39} - \ref{9.41}  and the estimate
\ref{9.34} in \ref{9.38} we conclude that:
\begin{equation}
\|\snab\s^{(O_i)}\hat{\spi}\|_{L^4(S_{\ub,u})}\leq
C\delta^{1/2}|u|^{-3/2}(\scR_1^4(\beta)+{\cal R}_0^\infty(\beta))
+O(\delta|u|^{-5/2}) \label{9.42}
\end{equation}

\vspace{5mm}

To derive an $L^4(S)$ estimate for $\snab Z_i$, we must apply
$\snab$ to the propagation equation \ref{8.63}. Here we must apply
an analogue of the second part of Lemma 4.1 in the case where
$\theta$ is an $S$-tangential vectorfield. In fact, we have the
following extension of Lemma 4.1 to the general case, where
$\theta$ is an arbitrary type $T^q_p$ $S$ tensorfield.

\vspace{5mm}

\noindent{\bf Lemma 9.2} \ \ \ Let $\theta$ be an arbitrary type
$T^q_p$ $S$ tensorfield on $M^\prime$. We have, with respect to an
arbitrary frame field for the $S_{\ub,u}$,
\begin{eqnarray*}
(D\snab\theta-\snab
D\theta)^{C_1...C_q}_{AB_1...B_p}&=&-\sum_{i=1}^p(D\sGamma)^D_{AB_i}
\theta^{C_1...C_q}_{B_1...\stackrel{D}{>B_i<}...B_p}\\
&\s&+\sum_{j=1}^q(D\Gamma)^{C_j}_{AD}\theta^{C_1...\stackrel{D}{>C_j<}...C_q}_{B_1...B_p}
\end{eqnarray*}
and:
\begin{eqnarray*}
(\Db\snab\theta-\snab\Db\theta)^{C_1...C_q}_{AB_1...B_p}&=&-\sum_{i=1}^p(\Db\sGamma)^D_{AB_i}
\theta^{C_1...C_q}_{B_1...\stackrel{D}{>B_i<}...B_p}\\
&\s&+\sum_{j=1}^q(\Db\Gamma)^{C_j}_{AD}\theta^{C_1...\stackrel{D}{>C_j<}...C_q}_{B_1...B_p}
\end{eqnarray*}

\noindent{\em Proof:} In the case $q=0$ the lemma reduces to Lemma
4.1. Consider now the case $p=0$, $q=1$, namely the case that
$\theta$ is a $S$-tangential vectorfield. We work, as in the proof
of Lemma 4.1, in a Jacobi field frame. We then have (see
\ref{4.4}):
\begin{equation}
\theta=\theta^B e_B, \ \ \ D\theta=(L\theta^B)e_B \label{9.43}
\end{equation}
and (see \ref{4.7}):
\begin{equation}
(\snab
D\theta)(e_A)=\snab_{e_A}D\theta=(e_A(L\theta^B)+\sGamma^B_{AC}(L\theta^C))e_B
\label{9.44}
\end{equation}
On the other hand,
\begin{equation}
(D\snab\theta)(e_A)=[L,(\snab\theta)(e_A)] \label{9.45}
\end{equation}
while:
\begin{equation}
(\snab\theta)(e_A)=\snab_{e_A}\theta=(e_A(\theta^B)+\sGamma^B_{AC}\theta^C)e_B
\label{9.46}
\end{equation}
hence:
\begin{eqnarray}
[L,(\snab\theta)(e_A)]&=&(L(e_A(\theta^B)+\sGamma^B_{AC}\theta^C))e_B\nonumber\\
&=&(e_A(L\theta^B)+\sGamma^B_{AC}(L\theta^C)+(D\sGamma)^B_{AC}\theta^C)e_B
\label{9.47}
\end{eqnarray}
Comparing with \ref{9.44} we conclude that:
\begin{equation}
(D\snab\theta-\snab D\theta)(e_A)=(D\sGamma)^B_{AC}\theta^C e_B
\label{9.48}
\end{equation}
This is the first part lemma in the case $p=0$, $q=1$. The second
part is established in the same manner. The proof in the case
$p=0$, $q$ arbitrary, is simmilar. Finally, the general case is
obtained by combining with the argument of Lemma 4.1.

\vspace{5mm}

Applying $\snab$ to equation \ref{8.63} and using Lemmas 9.1 and
9.2 in the case $p=0$, $q=1$, we obtain the following propagation
equation for $\snab  Z_i$ along the $\Cb_{\ub}$:
\begin{equation}
\Db\snab Z_i=-4\sL_{O_i}\snab(\Omega^2\zeta^\sharp)+\sb
\label{9.49}
\end{equation}
where:
\begin{equation}
\sb=\Db\sGamma\cdot Z_i+4\s^{(O_i)}\spi_1\cdot\Omega^2\zeta^\sharp
\label{9.50}
\end{equation}
Here:
\begin{equation}
(\Db\sGamma\cdot Z_i)^B_A=(\Db\sGamma)^B_{AC}Z_i^C, \ \ \
(\s^{(O_i)}\spi_1\cdot\Omega^2\zeta^\sharp)^B_A=\s^{(O_i)}\spi^B_{1,AC}\Omega^2\zeta^C
\label{9.51}
\end{equation}
To equation \ref{9.49} we must apply an analogue of Lemma 4.7 in
the case where $\theta$ is an $T^1_1$ type $S$ tensorfield. In
fact, Lemma 4.7 can be extended to the general case, where
$\theta$ is an arbitrary type $T^s_r$ $S$ tensorfield. We first
extend  Lemma 4.2 to general case:

\vspace{5mm}

\noindent{\bf Lemma 9.3} \ \ \ Let $\theta$ be an arbitrary type
$T^s_r$ type $S$ tensorfield on $M^\prime$. We have:
\begin{eqnarray*}
D(|\theta|^2)+(p-q)\Omega\mbox{tr}\chi|\theta|^2&=&2(\theta,D\theta)-2\sum_{i=1}^p\Omega\chih^{A_i}_{B_i}
\theta^{A_1...\stackrel{B_i}{>A_i<}...A_p}_{C_1...C_q}\theta_{A_1...A_p}^{C_1...C_q}\\
&\s&\s\s\s+2\sum_{j=1}^q\Omega\chih_{C_j}^{D_j}\theta^{A_1...A_p}_{C_1...\stackrel{D_j}{>C_j<}...C_q}
\theta_{A_1...A_p}^{C_1...C_q}
\end{eqnarray*}
and:
\begin{eqnarray*}
\Db(|\theta|^2)+(p-q)\Omega\mbox{tr}\chib|\theta|^2&=&2(\theta,\Db\theta)-2\sum_{i=1}^p\Omega\chibh^{A_i}_{B_i}
\theta^{A_1...\stackrel{B_i}{>A_i<}...A_p}_{C_1...C_q}\theta_{A_1...A_p}^{C_1...C_q}\\
&\s&\s\s\s+2\sum_{j=1}^q\Omega\chibh_{C_j}^{D_j}\theta^{A_1...A_p}_{C_1...\stackrel{D_j}{>C_j<}...C_q}
\theta_{A_1...A_p}^{C_1...C_q}
\end{eqnarray*}

\noindent The proof is straightforward, along the lines of that of
Lemma 4.2, starting from the formula:
$$|\theta|^2=(\sg^{-1})^{A_1 B_1}...(\sg^{-1})^{A_p B_p}\sg_{C_1 D_1}...\sg_{C_q D_q}\theta^{C_1...C_q}_{A_1...A_p}
\theta^{D_1...D_q}_{B_1...B_p}$$

\vspace{5mm}

Using the second part of Lemma 9.3 in place of the second part of
Lemma 4.2 we establish, following the argument of Lemma 4.7, the
following more general lemma.

\vspace{5mm}

\noindent{\bf Lemma 9.4} \ \ \ Let $\thetab$ be an arbitrary type
$T^s_r$ type $S$ tensorfield on $M^\prime$ satisfying along the
generators of $\Cb_{\ub}$ the propagation equation:
$$\Db\thetab=\frac{\nu}{2}\Omega\mbox{tr}\chib\thetab+\gammab\cdot\thetab+\xib$$
where $\nu$ is a real number, $\xib$ is a type $T^s_r$ $S$
tensorfield on $M^\prime$, and $\gammab$ is a type $T^{r+s}_{r+s}$
$S$ tensorfield on $M^\prime$ satisfying, pointwise,
$$|\gammab|\leq m\Omega|\chibh|$$
where $m$ is a positive constant. Then, if $\delta$ is suitably small depending on ${\cal D}_0^\infty$, ${\cal R}_0^\infty$, for each $p\geq
1$ there is a positive constant $C$ depending only on $p, r, s,
\nu, m$ such that:
\begin{eqnarray*}
|u|^{r-s-\nu-(2/p)}\|\thetab\|_{L^p(S_{\ub,u})}&\leq&
C|u_0|^{r-s-\nu-(2/p)}\|\thetab\|_{L^p(S_{\ub,u_0})}\\
&\s& +
C\int_{u_0}^u|u^\prime|^{r-s-\nu-(2/p)}\|\xib\|_{L^p(S_{\ub,u^\prime})}du^\prime
\end{eqnarray*}

\vspace{5mm}

\noindent{\bf Proposition 9.2} \ \ \ The following estimate holds
for all $(\ub,u)\in D^\prime$:
$$\|\snab Z_i\|_{L^4(S_{\ub,u})}\leq O(|u|^{-3/2})$$
provided that $\delta$ is suitably small depending on ${\cal
D}_0^\infty$, ${\cal R}_0^\infty$, $\scD_1^4$, $\scR_1^4$ and
$\scR_2$.

\noindent{\em Proof:} \ We apply Lemma 9.4 to the propagation
equation \ref{9.49} taking $p=4$. Here $r=s=1$, $\nu=0$,
$\gammab=0$. Taking also into accound the fact that $\snab Z_i$
vanishes on $C_{u_0}$ we obtain:
\begin{equation}
|u|^{-1/2}\|\snab Z_i\|_{L^4(S_{\ub,u})}\leq
C\int_{u_0}^u|u^\prime|^{-1/2}
\left\|-4\sL_{O_i}\snab(\Omega^2\zeta^\sharp)+\sb\right\|_{L^4(S_{\ub,u^\prime})}du^\prime
\label{9.52}
\end{equation}
Now, by Proposition 6.2 together with Propositions 8.1 and 8.2:
\begin{equation}
\|\sL_{O_i}\snab(\Omega^2\zeta^\sharp)\|_{L^4(S_{\ub,u})}\leq
O(|u|^{-5/2}) \label{9.53}
\end{equation}
Also, by Proposition 8.4, the estimate \ref{8.96} and Proposition
9.1:
\begin{equation}
\|\sb\|_{L^4(S_{\ub,u})}\leq O(\delta^{1/2}|u|^{-7/2})
\label{9.54}
\end{equation}
Substituting the above in \ref{9.53} the proposition follows.

\vspace{5mm}

\noindent{\bf Proposition 9.3} \ \ \ The following estimates hold
for all $(\ub,u)\in D^\prime$:
$$\|D\mbox{tr}\s^{(O_i)}\spi\|_{L^4(S_{\ub,u})}\leq O(|u|^{-3/2})$$
$$\|D\s^{(O_i)}\hat{\spi}\|_{L^4(S_{\ub,u})}\leq C\delta^{-1/2}|u|^{-1/2}(\scR_1^4(\alpha)+{\cal R}_0^\infty(\alpha))
+O(|u|^{-3/2})$$ provided that $\delta$ is suitably small
depending on ${\cal D}_0^\infty$, ${\cal R}_0^\infty$, $\scD_1^4$,
$\scR_1^4$ and $\scR_2$.

\noindent{\em Proof:} \ By the commutation formula \ref{1.75} we
have:
\begin{equation}
\Db
D\mbox{tr}\s^{(O_i)}\spi=D\Db\mbox{tr}\s^{(O_i)}\spi+4\Omega^2\zeta^\sharp\cdot\mbox{tr}\s^{(O_i)}\spi
\label{9.55}
\end{equation}
On the other hand, applying $D$ to the propagation equation
\ref{8.103} we obtain:
\begin{eqnarray}
D\Db\mbox{tr}\s^{O_i)}\spi&=&2D(O_i(\Omega\mbox{tr}\chib))\nonumber\\
&=&2O_i(D(\Omega\mbox{tr}\chib))+2Z_i\cdot\sd(\Omega\mbox{tr}\chib)
\label{9.56}
\end{eqnarray}
by virtue of \ref{8.59}. Substituting \ref{9.56} in \ref{9.55}
yields the following propagation equation for
$D\mbox{tr}\s^{(O_i)}\spi$ along the $\Cb_{\ub}$:
\begin{equation}
\Db
D\mbox{tr}\s^{(O_i)}\spi=2O_i(D(\Omega\mbox{tr}\chib))+4\Omega^2\zeta^\sharp\cdot\sd\mbox{tr}\s^{(O_i)}\spi
+2Z_i\cdot\sd(\Omega\mbox{tr}\chib) \label{9.57}
\end{equation}
To this we apply Lemma 4.7 taking $p=4$. Here $r=0$, $\nu=0$,
$\gammab=0$ and we obtain:
\begin{eqnarray}
&&|u|^{-1/2}\|D\mbox{tr}\s^{(O_i)}\spi\|_{L^4(S_{\ub,u})}\leq
C|u_0|^{-1/2}\|D\mbox{tr}\s^{(O_i)}\spi\|_{L^4(S_{\ub,u_0})}
\label{9.58}\\
&&+C\int_{u_0}^u|u^\prime|^{-1/2}\left\|O_i(D(\Omega\mbox{tr}\chib))+2\Omega^2\zeta^\sharp\cdot\sd\mbox{tr}\s^{(O_i)}\spi
+Z_i\cdot\sd(\Omega\mbox{tr}\chib)\right\|_{L^4(S_{\ub,u^\prime})}du^\prime\nonumber
\end{eqnarray}
Now, by equation \ref{8.102}, Proposition 8.1 and the results of
Chapters 3 and 4:
\begin{equation}
\|D\mbox{tr}\s^{(O_i)}\spi\|_{L^4(S_{\ub,u_0})}\leq
O(|u_0|^{-3/2}) \label{9.59}
\end{equation}
By equation \ref{4.c1}, Proposition 6.2, Proposition 8.1 and the
results of Chapters 3 and 4:
\begin{equation}
\|O_i(D(\Omega\mbox{tr}\chib))\|_{L^4(S_{\ub,u})}\leq
O(|u|^{-5/2}) \label{9.60}
\end{equation}
Also, by the estimate \ref{8.111} and Propositions 8.4 and 9.1:
\begin{equation}
\left\|2\Omega^2\zeta^\sharp\cdot\sd\mbox{tr}\s^{(O_i)}\spi
+Z_i\cdot\sd(\Omega\mbox{tr}\chib)\right\|_{L^4(S_{\ub,u})}\leq
O(\delta|u|^{-9/2}) \label{9.61}
\end{equation}
Substituting \ref{9.59} - \ref{9.61} in \ref{9.58} we conclude
that:
\begin{equation}
\|D\mbox{tr}\s^{(O_i)}\spi\|_{L^4(S_{\ub,u})}\leq O(|u|^{-3/2})
\label{9.62}
\end{equation}

By Lemma 1.4 we have:
\begin{equation}
\Db
D\s^{(O_i)}\hat{\spi}=D\Db\s^{(O_i)}\hat{\spi}+\sL_{4\Omega^2\zeta^\sharp}\s^{(O_i)}\hat{\spi}
\label{9.63}
\end{equation}
On the other hand, applying $D$ to the propagation equation
\ref{8.105} we obtain:
\begin{eqnarray}
D\Db\s^{(O_i)}\hat{\spi}-\Omega\mbox{tr}\chib
D\s^{(O_i)}\hat{\spi}&=&D(\Omega\mbox{tr}\chib)\s^{(O_i)}\hat{\spi}
-D(\Omega\chibh)\mbox{tr}\s^{(O_i)}\spi\nonumber\\
&\s&-\Omega\chibh
D\mbox{tr}\s^{(O_i)}\spi+2D\sL_{O_i}(\Omega\chibh) \label{9.64}
\end{eqnarray}
and we have:
\begin{equation}
D\sL_{O_i}(\Omega\chibh)=\sL_{O_i}D(\Omega\chibh)+\sL_{Z_i}(\Omega\chibh)
\label{9.65}
\end{equation}
by virtue of \ref{8.59}. Substituting \ref{9.65} in \ref{9.64} and
the result in \ref{9.63} yields the following propagation equation
for $D\s^{(O_i)}\hat{\spi}$ along the $\Cb_{\ub}$:
\begin{equation}
\Db D\s^{(O_i)}\hat{\spi}-\Omega\mbox{tr}\chib
D\s^{(O_i)}\hat{\spi}=2\sL_{O_i}D(\Omega\chibh)+v \label{9.66}
\end{equation}
where:
\begin{equation}
v=2\sL_{Z_i}(\Omega\chibh)+4\sL_{\Omega^2\zeta^\sharp}\s^{(O_i)}\hat{\spi}-\Omega\chibh
D\mbox{tr}\s^{(O_i)}\spi
+D(\Omega\mbox{tr}\chib)\s^{(O_i)}\hat{\spi}-D(\Omega\chibh)\mbox{tr}\s^{(O_i)}\spi
\label{9.67}
\end{equation}
To \ref{9.66} we apply Lemma 4.7 taking $p=4$. Here $r=2$,
$\nu=2$, $\gammab=0$. We obtain:
\begin{eqnarray}
|u|^{-1/2}\|D\s^{(O_i)}\hat{\spi}\|_{L^4(S_{\ub,u})}&\leq&
C|u_0|^{-1/2}\|D\s^{(O_i)}\hat{\spi}\|_{L^4(S_{\ub,u_0})}
\label{9.68}\\
&\s&+C\int_{u_0}^u|u^\prime|^{-1/2}\left\|2\sL_{O_i}D(\Omega\chibh)+v\right\|_{L^4(S_{\ub,u^\prime})}du^\prime\nonumber
\end{eqnarray}
Now by equation \ref{8.104}, Propositions 8.1, 8.2 and 8.3 and the
results of Chapters 3 and 4:
\begin{equation}
\|D\s^{(O_i)}\hat{\spi}\|_{L^4(S_{\ub,u_0})}\leq
C\delta^{1/2}|u_0|^{-1/2}(\scR_1^4(\alpha)+{\cal
R}_0^\infty(\alpha)) +O(|u_0|^{-3/2}) \label{9.69}
\end{equation}
By equation \ref{4.c3}, Propositions 8.1 and 8.2 and the results
of Chapters 3 and 4:
\begin{equation}
\|\sL_{O_i}D(\Omega\chibh)\|_{L^4(S_{\ub,u})}\leq
C\delta^{-1/2}|u|^{-3/2}(\scR_1^4(\alpha)+{\cal
R}_0^\infty(\alpha)) +O(|u|^{-5/2}) \label{9.70}
\end{equation}
Using formula \ref{8.117} with $Z_i$ in the role of $X$ and
$\Omega\chibh$ in the role of $\theta$, we obtain, by Propositions
8.4 and 9.2,
\begin{equation}
\|\sL_{Z_i}(\Omega\chibh)\|_{L^4(S_{\ub,u})}\leq
O(\delta^{1/2}|u|^{-7/2}) \label{9.71}
\end{equation}
Also, using formula \ref{8.117} with $\Omega^2\zeta^\sharp$ in the
role of $X$ and $\s^{(O_i)}\hat{\spi}$ in the role of $\theta$, we
obtain, by Propositions 8.3 and 9.1 and the results of Chapters 3
and 4,
\begin{equation}
\|\sL_{\Omega^2\zeta^\sharp}\s^{(O_i)}\hat{\spi}\|_{L^4(S_{\ub,u})}\leq
O(\delta|u|^{-7/2}) \label{9.72}
\end{equation}
The remaining three terms on the right in \ref{9.67} are bounded
in $L^4(S)$ by $O(\delta^{1/2}|u|^{-7/2})$ using the estimate
\ref{9.62}, Proposition 8.3 and equations \ref{4.c1}, \ref{4.c3}.
Hence we obtain:
\begin{equation}
\|v\|_{L^4(S_{\ub,u})}\leq O(\delta^{1/2}|u|^{-7/2}) \label{9.73}
\end{equation}
Substituting \ref{9.69}, \ref{9.70} and \ref{9.73} in \ref{9.68}
we conclude that:
\begin{equation}
\|D\s^{(O_i)}\hat{\spi}\|_{L^4(S_{\ub,u})}\leq
C\delta^{-1/2}|u|^{-1/2}(\scR_1^4(\alpha)+{\cal
R}_0^\infty(\alpha)) +O(|u|^{-3/2}) \label{9.74}
\end{equation}

\vspace{5mm}

\noindent{\bf Proposition 9.4} \ \ \ The following estimate holds
for all $(\ub,u)\in D^\prime$:
$$\|DZ_i\|_{L^4(S_{\ub,u})}\leq O(\delta^{-1/2}|u|^{-1/2})$$
provided that $\delta$ is suitably small depending on ${\cal
D}_0^\infty$, ${\cal R}_0^\infty$, $\scD_1^4$, $\scR_1^4$ and
$\scR_2$.

\noindent{\em Proof:} \ By Lemma 1.4 we have:
\begin{equation}
\Db DZ_i=D\Db Z_i+4[\Omega^2\zeta^\sharp,Z_i] \label{9.75}
\end{equation}
On the other hand, applying $D$ to the propagation equation
\ref{8.63} we obtain:
\begin{eqnarray}
D\Db Z_i&=&-4D\sL_{O_i}(\Omega\zeta^\sharp)\nonumber\\
&=&-4\sL_{O_i}D(\Omega^2\zeta^\sharp)-4[Z_i,\Omega^2\zeta^\sharp]
\label{9.76}
\end{eqnarray}
by virtue of \ref{8.59}. Substituting \ref{9.76} in \ref{9.75}
yields the following propagation equation for $DZ_i$ along the
$\Cb_{\ub}$:
\begin{equation}
\Db
DZ_i=-4\sL_{O_i}D(\Omega^2\zeta^\sharp)+8[\Omega^2\zeta^\sharp,Z_i]
\label{9.77}
\end{equation}
To this we apply Lemma 9.4 taking $p=4$. Here $r=0$, $s=1$,
$\nu=0$, $\gammab=0$. Taking also into account the fact that
$DZ_i$ vanishes on $C_{u_0}$ we obtain:
\begin{equation}
|u|^{-3/2}\|DZ_i\|_{L^4(S_{\ub,u})}\leq
C\int_{u_0}^u|u^\prime|^{-3/2}
\left\|-\sL_{O_i}D(\Omega^2\zeta^\sharp)+2[\Omega^2\zeta^\sharp,Z_i]\right\|_{L^4(S_{\ub,u^\prime})}du^\prime
\label{9.78}
\end{equation}
Now, from equation \ref{1.61}, by Proposition 6.3 together with
Propositions 8.1, 8.2 and the results of Chapters 3 and 4 we
obtain:
\begin{equation}
\|\sL_{O_i}D(\Omega^2\zeta^\sharp)\|_{L^4(S_{\ub,u})}\leq
O(\delta^{-1/2}|u|^{-3/2}) \label{9.79}
\end{equation}
Also, by Propositions 8.4, 9.2 and the results of Chapters 3 and
4:
\begin{equation}
\|[\Omega^2\zeta^\sharp,Z_i]\|_{L^4(S_{\ub,u})}\leq
O(\delta^{1/2}|u|^{-7/2}) \label{9.80}
\end{equation}
Substituting the above in \ref{9.78} the proposition follows.

\vspace{5mm}

Estimates for $\Db\s^{(O_i)}\mbox{tr}\spi$,
$\Db\s^{(O_i)}\hat{\spi}$, and $\Db Z_i$, in $L^4(S)$, follow
directly from equations \ref{8.103}, \ref{8.105}, and \ref{8.63}
respectively. We obtain:
\begin{eqnarray}
&&\|\Db\s^{(O_i)}\mbox{tr}\spi\|_{L^4(S_{\ub,u})}\leq O(\delta|u|^{-5/2})\nonumber\\
&&\|\Db\s^{(O_i)}\hat{\spi}\|_{L^4(S_{\ub,u})}\leq
C\delta^{1/2}|u|^{-3/2}(\scR_1^4(\beta)+{\cal R}_0^\infty(\beta))
+O(\delta|u|^{-5/2})\nonumber\\
&&\|\Db Z_i\|_{L^4(S_{\ub,u})}\leq O(\delta^{1/2}|u|^{-3/2})
\label{9.81}
\end{eqnarray}

In view of the expressions \ref{8.134} - \ref{8.138} for the
components of the deformation tensor of the $O_i$, Propositions
9.1, 9.2, 9.3, 9.4 and the estimates \ref{9.81} yield the
following estimates for the 1st derivatives of the non-vanishing
components in $L^4(S)$:
\begin{eqnarray}
&&\|\snab\s^{(O_i)}\ih\|_{L^4(S_{\ub,u})}\leq
C\delta^{1/2}|u|^{-3/2}(\scR_1^4(\beta)+{\cal R}_0^\infty(\beta))
+O(\delta|u|^{-5/2})\nonumber\\
&&\|\sd\s^{(O_i)}j\|_{L^4(S_{\ub,u})}\leq O(\delta|u|^{-5/2})\nonumber\\
&&\|\snab\s^{(O_i)}m\|_{L^4(S_{\ub,u})}\leq O(|u|^{-3/2})
\label{9.82}
\end{eqnarray}
\begin{eqnarray}
&&\|\Dh\s^{(O_i)}\ih\|_{L^4(S_{\ub,u})}\leq
C\delta^{-1/2}|u|^{-1/2}(\scR_1^4(\alpha)+{\cal
R}_0^\infty(\alpha))
+O(|u|^{-3/2})\nonumber\\
&&\|D\s^{(O_i)}j\|_{L^4(S_{\ub,u})}\leq O(|u|^{-3/2})\nonumber\\
&&\|D\s^{(O_i)}m\|_{L^4(S_{\ub,u})}\leq O(\delta^{-1/2}|u|^{-1/2})
\label{9.83}
\end{eqnarray}
\begin{eqnarray}
&&\|\Dbh\s^{(O_i)}\ih\|_{L^4(S_{\ub,u})}\leq
C\delta^{1/2}|u|^{-3/2}(\scR_1^4(\beta)+{\cal R}_0^\infty(\beta))
+O(\delta|u|^{-5/2})\nonumber\\
&&\|\Db\s^{(O_i)}j\|_{L^4(S_{\ub,u})}\leq O(\delta|u|^{-5/2})\nonumber\\
&&\|\Db\s^{(O_i)}m\|_{L^4(S_{\ub,u})}\leq
O(\delta^{1/2}|u|^{-3/2}) \label{9.84}
\end{eqnarray}
In the second of \ref{9.82}, to estimate the contribution of the
2nd term on the right in \ref{8.137} we write:
\begin{equation}
\sd O_i(\log\Omega)=\sL_{O_i}\sd\log\Omega \label{9.85}
\end{equation}
and use Lemma 6.4 together with Propositions 8.1 and 8.2. In the
second of \ref{9.83} to estimate the contribution of the 2nd term
on the right in \ref{8.137} we write:
\begin{equation}
DO_i(\log\Omega)=O_i\cdot\sd\omega+Z_i\cdot\sd\log\Omega
\label{9.86}
\end{equation}
and use Propositions 4.3 and 8.4. In the second of \ref{9.84} to
estimate the contribution of the 2nd term on the right in
\ref{8.137} we write:
\begin{equation}
\Db O_i(\log\Omega)=O_i\cdot\sd\omb \label{9.87}
\end{equation}
and use the estimate for $\sd\omb$ in $L^4(S)$ of Proposition 4.2.
Finally, since
$$\Db\s^{(O_i)}m=\sg\cdot\Db Z_i+2\Omega\chib\cdot Z_i$$
the last of \ref{9.84} requires the estimate
\begin{equation}
\|Z_i\|_{L^4(S_{\ub,u})}\leq O(\delta^{1/2}|u|^{-1/2})
\label{9.a1}
\end{equation}
which follows from the propagation equation \ref{8.63} using the
results of Chapter 4.

\section{Estimates for the 2nd derivatives of the deformation tensors of $L$, $S$}

We now assume, in addition to the previous assumptions on the
curvature components, that the following quantities are finite:
\begin{eqnarray}
&&{\cal R}_0^4(\Dh\alpha)=\sup_{(\ub,u)\in D^\prime}\left(|u|^{1/2}\delta^{5/2}\|\Dh\alpha\|_{L^4(S_{\ub,u})}\right)\nonumber\\
&&{\cal R}_0^4(\Dbh\alb)=\sup_{(\ub,u)\in
D^\prime}\left(|u|^5\delta^{-3/2}\|\Dbh\alb\|_{L^4(S_{\ub,u})}\right)
\label{9.a2}
\end{eqnarray}
The second of these quantities has in fact already been introduced
(see \ref{7.04}). By the results of Chapter 2, the corresponding
quantities on $C_{u_0}$, obtained by replacing the supremum on
$D^\prime$ by the supremum on $([0,\delta]\times\{u_0\})\bigcap
D^\prime$, are bounded by a non-negative non-decreasing continuous
function of $M_7$. In the following we denote by
$O(\delta^p|u|^r)$, for real numbers $p$, $r$, the product of
$\delta^p |u|^r$ with a non-negative non-decreasing continuous
function of the quantities ${\cal D}_0^\infty$, ${\cal
R}_0^\infty$, $\scD_1^4$, $\scR_1^4$, $\scD_2^4(\mbox{tr}\chib)$,
$\scR_2$, $\scD_3(\mbox{tr}\chib)$, {\em and} ${\cal
R}_0^4(\Dh\alpha)$, ${\cal R}_0^4(\Dbh\alb)$.

The components of the deformation tensor of the commutation field
$L$ are given by table \ref{8.21}. Proposition 6.1 and
Propositions 7.1 and 7.4 imply:
\begin{eqnarray}
&&\|\snab^{ \ 2}\s^{(L)}\ih\|_{L^2(C_u)}\leq O(|u|^{-2})\nonumber\\
&&\|\snab^{ \ 2}\s^{(L)}j\|_{L^2(C_u)}\leq O(\delta^{1/2}|u|^{-3})\nonumber\\
&&\|\snab^{ \ 2}\s^{(L)}\mb\|_{L^2(C_u)}\leq O(\delta|u|^{-3})
\label{9.88}
\end{eqnarray}
for all $u\in [u_0,c^*)$.

Next, from equations \ref{1.61}, \ref{3.6}, \ref{3.8}, in view of
Propositions 7.4, 7.6 and the results of Chapters 3 and 4, we
deduce:
\begin{eqnarray}
&&\|\snab\Dh\s^{(L)}\ih\|_{L^2(C_u)}\leq O(\delta^{-1}|u|^{-1})\nonumber\\
&&\|\sd D\s^{(L)}j\|_{L^2(C_u)}\leq O(\delta^{-1/2}|u|^{-2})\nonumber\\
&&\|\snab D\s^{(L)}\mb\|_{L^2(C_u)}\leq O(|u|^{-2}) \label{9.89}
\end{eqnarray}
for all $u\in [u_0,c^*)$.

Next, from equations \ref{1.63}, \ref{1.87}, \ref{4.c2},
\ref{4.c4}, in view of Proposition 7.1, the estimate for $\snab^{
\ 2}\omb$ of Proposition 6.2 and the results of Chapters 3 and 4,
we deduce:
\begin{eqnarray}
&&\|\snab\Dbh\s^{(L)}i\|_{L^2(C_u)}\leq O(|u|^{-2})\nonumber\\
&&\|\sd\Db\s^{(L)}j\|_{L^2(C_u)}\leq O(\delta^{1/2}|u|^{-3})\nonumber\\
&&\|\snab\Db\s^{(L)}\mb\|_{L^2(C_u)}\leq O(\delta|u|^{-3})
\label{9.90}
\end{eqnarray}
for all $u\in [u_0,c^*)$.

Next, applying $D$ to equations \ref{1.61}, \ref{3.6}, \ref{3.8},
and using Propositions 7.6 and 7.8 and the results of Chapters 3
and 4, we deduce:
\begin{eqnarray}
&&\|\Dh^2\s^{(L)}\ih\|_{L^2(C_u)}\leq O(\delta^{-2})\nonumber\\
&&\|D^2\s^{(L)}j\|_{L^2(C_u)}\leq O(\delta^{-3/2}|u|^{-1})\nonumber\\
&&\|D^2\s^{(L)}\mb\|_{L^2(C_u)}\leq O(\delta^{-1}|u|^{-1})
\label{9.91}
\end{eqnarray}
for all $u\in [u_0,c^*)$.

Finally, applying $D$ to equations \ref{1.63}, \ref{1.87},
\ref{4.c2}, \ref{4.c4}, and using equations \ref{1.66},
\ref{1.86}, \ref{3.6}, \ref{3.8}, \ref{4.c1}, \ref{4.c3} and the
results of Chapters 3 and 4, we deduce:
\begin{eqnarray}
&&\|\Dh\Dbh\s^{(L)}\ih\|_{L^2(C_u)}\leq O(\delta^{-1}|u|^{-1})\nonumber\\
&&\|D\Db\s^{(L)}j\|_{L^2(C_u)}\leq O(\delta^{-1/2}|u|^{-2})\nonumber\\
&&\|D\Db\s^{(L)}\mb\|_{L^2(C_u)}\leq O(|u|^{-2}) \label{9.92}
\end{eqnarray}
for all $u\in [u_0,c^*)$. Estimates for the 2nd Lie derivative
with respect to $\Lb$ of the deformation tensor of $L$ shall not
be needed.

The components of the deformation tensor of the commutation field
$S$ are given by table \ref{8.30}. Moreover the component
$\s^{(S)}j$ is expressed in terms of the function $\lambda$ by
\ref{9.5}. From the definition \ref{9.4}, using Propositions 6.1
and 6.2 and Lemma 6.4, we deduce:
\begin{equation}
\|\snab^{ \ 2}\lambda\|_{L^2(C_u)}\leq O(\delta^{3/2}|u|^{-3})
\label{9.93}
\end{equation}
for all $u\in [u_0,c^*)$. From equations \ref{9.8} and \ref{9.9},
using Proposition 7.1, the estimate \ref{9.7} and the results of
Chapters 3 and 4, we deduce:
\begin{eqnarray}
&&\|\sd D\lambda\|_{L^2(C_u)}\leq O(\delta^{1/2}|u|^{-2})\nonumber\\
&&\|\sd\Db\lambda\|_{L^2(C_u)}\leq O(\delta^{3/2}|u|^{-3})
\label{9.94}
\end{eqnarray}
for all $u\in [u_0,c^*)$. Moreover, applying $D$ to equation
\ref{9.8} and using equations \ref{1.150}, \ref{3.6}, \ref{3.8},
\ref{4.c3}, the estimate \ref{9.10}, Proposition 7.4, as well as
the results of Chapters 3 and 4, we deduce:
\begin{equation}
\|D^2\lambda\|_{L^2(C_u)}\leq O(\delta^{-1/2}|u|^{-1})
\label{9.95}
\end{equation}
Also, applying $\Db$ to equation \ref{9.9} and using equations
\ref{1.149}, \ref{3.7}, \ref{3.9}, \ref{4.c4}, the estimate
\ref{9.11}, the estimate for $\snab^{ \ 2}\omb$ of Proposition
6.2, as well as the results of Chapters 3 and 4, we deduce:
\begin{equation}
\|\Db^2\lambda\|_{L^2(C_u)}\leq O(\delta^{3/2}|u|^{-3})
\label{9.96}
\end{equation}
for all $u\in [u_0,c^*)$. Finally, applying $D$ to equation
\ref{9.9} and using equations \ref{1.66}, \ref{1.86}, \ref{3.8},
\ref{4.c1}, \ref{4.c3}, and the results of Chapters 3 and 4, we
deduce:
\begin{equation}
\|D\Db\lambda\|_{L^2(C_u)}\leq O(\delta^{1/2}|u|^{-2})
\label{9.97}
\end{equation}
for all $u\in [u_0,c^*)$.

Propositions 6.1, 6.2, 7.4 and the estimate \ref{9.93} imply:
\begin{eqnarray}
&&\|\snab^{ \ 2}\s^{(S)}\ih\|_{L^2(C_u)}\leq O(\delta|u|^{-2})\nonumber\\
&&\|\snab^{ \ 2}\s^{(S)}j\|_{L^2(C_u)}\leq O(\delta^{3/2}|u|^{-3})\nonumber\\
&&\|\snab^{ \ 2}\s^{(S)}\mb\|_{L^2(C_u)}\leq O(\delta^2|u|^{-3})\nonumber\\
&&\|\snab^{ \ 2}\s^{(S)}m\|_{L^2(C_u)}\leq O(\delta|u|^{-2})
\label{9.98}
\end{eqnarray}
for all $u\in [u_0,c^*)$.

Next, from equations \ref{1.61}, \ref{1.86}, \ref{3.8},
\ref{4.c3}, in view of Propositions 7.1, 7.4, 7.6, the first of
the estimates \ref{9.94}, as well as the results of Chapters 3 and
4, we deduce:
\begin{eqnarray}
&&\|\snab\Dh\s^{(S)}\ih\|_{L^2(C_u)}\leq O(|u|^{-1})\nonumber\\
&&\|\sd D\s^{(S)}j\|_{L^2(C_u)}\leq O(\delta^{1/2}|u|^{-2})\nonumber\\
&&\|\snab D\s^{(S)}\mb\|_{L^2(C_u)}\leq O(\delta|u|^{-2})\nonumber\\
&&\|\snab D\s^{(S)}m\|_{L^2(C_u)}\leq O(|u|^{-1}) \label{9.99}
\end{eqnarray}
for all $u\in [u_0,c^*)$.

Next, from equations \ref{1.63}, \ref{1.87}, \ref{3.9},
\ref{4.c4}, in view of the estimate for $\snab^{ \ 2}\omb$ of
Proposition 6.2, Propositions 7.1, 7.7, the second of the
estimates \ref{9.94}, as well as the results of Chapters 3 and 4,
we deduce:
\begin{eqnarray}
&&\|\snab\Dbh\s^{(S)}\ih\|_{L^2(C_u)}\leq O(\delta|u|^{-2})\nonumber\\
&&\|\sd\Db\s^{(S)}j\|_{L^2(C_u)}\leq O(\delta^{3/2}|u|^{-3})\nonumber\\
&&\|\snab\Db\s^{(S)}\mb\|_{L^2(C_u)}\leq O(\delta^2|u|^{-3})\nonumber\\
&&\|\snab\Db\s^{(S)}m\|_{L^2(C_u)}\leq O(\delta|u|^{-2})
\label{9.100}
\end{eqnarray}
for all $u\in [u_0,c^*)$.

Next, applying $D$ to equations \ref{1.61}, \ref{1.86}, \ref{3.8},
\ref{4.c3}, and using equation \ref{1.150}, Propositions 7.4, 7.6,
7.8, the estimate \ref{9.95}, as well as the results of Chapters 3
and 4, we deduce:
\begin{eqnarray}
&&\|\Dh^2\s^{(S)}\ih\|_{L^2(C_u)}\leq O(\delta^{-1})\nonumber\\
&&\|D^2\s^{(S)}j\|_{L^2(C_u)}\leq O(\delta^{-1/2}|u|^{-1})\nonumber\\
&&\|D^2\s^{(S)}\mb\|_{L^2(C_u)}\leq O(|u|^{-1})\nonumber\\
&&\|D^2\s^{(S)}m\|_{L^2(C_u)}\leq O(\delta^{-1}) \label{9.101}
\end{eqnarray}
for all $u\in [u_0,c^*)$.

Next, applying $\Db$ to equations \ref{1.63}, \ref{1.87},
\ref{3.9}, \ref{4.c4}, and using equation \ref{1.149}, the
estimate for $\snab^{ \ 2}\omb$ of Proposition 6.2, Propositions
7.7, 7.9, the estimate \ref{9.96}, as well as the results of
Chapters 3 and 4, we deduce:
\begin{eqnarray}
&&\|\Dbh^2\s^{(S)}\ih\|_{L^2(C_u)}\leq O(\delta|u|^{-2})\nonumber\\
&&\|\Db^2\s^{(S)}j\|_{L^2(C_u)}\leq O(\delta^{3/2}|u|^{-3})\nonumber\\
&&\|\Db^2\s^{(S)}\mb\|_{L^2(C_u)}\leq O(\delta^2|u|^{-3})\nonumber\\
&&\|\Db^2\s^{(S)}m\|_{L^2(C_u)}\leq O(\delta|u|^{-2})
\label{9.102}
\end{eqnarray}
for all $u\in [u_0,c^*)$.

Finally, applying $D$ to equations \ref{1.63}, \ref{1.87},
\ref{3.9}, \ref{4.c4},and using equations \ref{1.66}, \ref{1.86},
the estimate \ref{9.97}, and the results of Chapters 3 and 4 we
deduce:
\begin{eqnarray}
&&\|\Dh\Dbh\s^{(S)}\ih\|_{L^2(C_u)}\leq O(|u|^{-1})\nonumber\\
&&\|D\Db\s^{(S)}j\|_{L^2(C_u)}\leq O(\delta^{1/2}|u|^{-2})\nonumber\\
&&\|D\Db\s^{(S)}\mb\|_{L^2(C_u)}\leq O(\delta|u|^{-2})\nonumber\\
&&\|D\Db\s^{(S)}m\|_{L^2(C_u)}\leq O(|u|^{-1}) \label{9.103}
\end{eqnarray}
for all $u\in [u_0,c^*)$.

\section{Estimates for the 2nd derivatives of the deformation tensors of the $O_i$}

We shall only require estimates for those 2nd derivatives of the
deformation tensors of the $O_i$ for which the last derivative is
tangential to the surfaces $S_{\ub,u}$.

\vspace{5mm}

\noindent{\bf Proposition 9.5} \ \ \ The following estimates hold
for all $u\in[u_0,c^*)$:
$$\|\snab^{ \ 2}\mbox{tr}\s^{(O_i)}\spi\|_{L^2(C_u)}\leq O(\delta^{3/2}|u|^{-3})$$
$$\|\snab^{ \ 2}\s^{(O_i)}\hat{\spi}\|_{L^2(C_u)}\leq O(\delta|u|^{-2})$$
provided that $\delta$ is suitably small depending on ${\cal
D}_0^\infty$, ${\cal R}_0^\infty$, $\scD_1^4$, $\scR_1^4$ and
$\scR_2$.

\noindent{\em Proof:} \ Consider any $(\ub_1,u_1)\in D^\prime$ and
fix attention to the parameter subdomain $D_1$ and the
corresponding subdomain $M_1$ of $M^\prime$ (see \ref{3.02},
\ref{3.03}).

We first apply $\snab$ to the propagation equation \ref{9.16}.
Using Lemmas 4.1 and 9.1 we then obtain the following propagation
equation for $\snab^{ \ 2}\mbox{tr}\s^{(O_i)}\spi$ along
$C_{u_0}$:
\begin{equation}
D\snab^{ \ 2}\mbox{tr}\s^{(O_i)}\spi=2\sL_{O_i}\snab^{ \
2}(\Omega\mbox{tr}\chi)
+2\s^{(O_i)}\spi_1\cdot\sd(\Omega\mbox{tr}\chi)-D\sGamma\cdot\sd\mbox{tr}\s^{(O_i)}\spi
\label{9.104}
\end{equation}
To this we apply Lemma 4.6 taking $p=2$. Here $r=2$, $\nu=0$,
$\gamma=0$ and we obtain:
\begin{eqnarray}
&&\|\snab^{ \ 2}\mbox{tr}\s^{(O_i)}\spi\|_{L^2(S_{\ub,u_0})}\leq\label{9.105}\\
&&C\int_0^{\ub}\left\|2\sL_{O_i}\snab^{ \ 2}(\Omega\mbox{tr}\chi)
+2\s^{(O_i)}\spi_1\cdot\sd(\Omega\mbox{tr}\chi)-D\sGamma\cdot\sd\mbox{tr}\s^{(O_i)}\spi\right\|_{L^2(S_{\ub^\prime,u_0})}
d\ub^\prime\nonumber
\end{eqnarray}
By Proposition 7.2, Lemma 7.8 together with Propositions 8.1 and
8.2:
\begin{equation}
\int_0^{\ub}\|\sL_{O_i}\snab^{ \
2}(\Omega\mbox{tr}\chi)\|_{L^2(S_{\ub^\prime,u_0})}d\ub^\prime
\leq\delta^{1/2}\|\sL_{O_i}\snab^{ \
2}(\Omega\mbox{tr}\chi)\|_{L^2(C_{u_0})}\leq O(\delta|u_0|^{-3})
\label{9.106}
\end{equation}
Also, by Proposition 9.1 and the estimates \ref{8.90} and
\ref{8.107},
\begin{equation}
\int_0^{\ub}\left\|2\s^{(O_i)}\spi_1\cdot\sd(\Omega\mbox{tr}\chi)-D\sGamma\cdot\sd\mbox{tr}\s^{(O_i)}\spi
\right\|_{L^2(S_{\ub^\prime,u_0})}d\ub^\prime\leq
O(\delta^{3/2}|u_0|^{-4}) \label{9.107}
\end{equation}
Substituting in \ref{9.105} we conclude that:
\begin{equation}
\|\snab^{ \ 2}\mbox{tr}\s^{(O_i)}\spi\|_{L^2(S_{\ub,u_0})}\leq
O(\delta|u_0|^{-3}) \ \ : \ \forall \ub\in[0,\delta] \label{9.108}
\end{equation}

We then apply $\snab$ to the propagation equation \ref{9.18}.
Using Lemmas 4.1 and 9.1 we then obtain the following propagation
equation for $\snab^{ \ 2}\mbox{tr}\s^{(O_i)}\spi$ along the
$\Cb_{\ub}$:
\begin{equation}
\Db\snab^{ \ 2}\mbox{tr}\s^{(O_i)}\spi=2\sL_{O_i}\snab^{ \
2}(\Omega\mbox{tr}\chib)
+2\s^{(O_i)}\spi_1\cdot\sd(\Omega\mbox{tr}\chib)-\Db\sGamma\cdot\sd\mbox{tr}\s^{(O_i)}\spi
\label{9.109}
\end{equation}
To this we apply Lemma 4.7 taking $p=2$. Here $r=2$, $\nu=0$,
$\gammab=0$ and we obtain:
\begin{eqnarray}
&&|u|\|\snab^{ \ 2}\mbox{tr}\s^{(O_i)}\spi\|_{L^2(S_{\ub,u})}\leq
C|u_0|\|\snab^{ \ 2}\mbox{tr}\spi\|_{L^2(S_{\ub,u_0})}
\label{9.110}\\
&&+C\int_{u_0}^u|u^\prime|\left\|2\sL_{O_i}\snab^{ \
2}(\Omega\mbox{tr}\chib)
+2\s^{(O_i)}\spi_1\cdot\sd(\Omega\mbox{tr}\chib)-\Db\sGamma\cdot\sd\mbox{tr}\s^{(O_i)}\spi\right\|_{L^2(S_{\ub,u^\prime})}
du^\prime\nonumber
\end{eqnarray}
Taking the $L^2$ norm with respect to $\ub$ on $[0,\ub_1]$ then
yields:
\begin{eqnarray}
&&|u|\|\snab^{ \
2}\mbox{tr}\s^{(O_i)}\spi\|_{L^2(C^{\ub_1}_u)}\leq
C|u_0|\|\snab^{ \ 2}\mbox{tr}\spi\|_{L^2(C^{\ub_1}_{u_0})}\label{9.111}\\
&&+C\int_{u_0}^u|u^\prime|\left\|2\sL_{O_i}\snab^{ \
2}(\Omega\mbox{tr}\chib)
+2\s^{(O_i)}\spi_1\cdot\sd(\Omega\mbox{tr}\chib)-\Db\sGamma\cdot
\sd\mbox{tr}\s^{(O_i)}\spi\right\|_{L^2(C^{\ub_1}_{u^\prime})}du^\prime\nonumber
\end{eqnarray}
By Proposition 7.3 and Lemma 7.8 together with Propositions 8.1
and 8.2:
\begin{equation}
\int_{u_0}^u|u^\prime|\|\sL_{O_i}\snab^{ \
2}(\Omega\mbox{tr}\chib)\|_{L^2(C^{\ub_1}_{u^\prime})}du^\prime
\leq O(\delta^{3/2}|u|^{-2}) \label{9.112}
\end{equation}
Also, by Proposition 9.1 and the estimates \ref{8.96} and
\ref{8.111},
\begin{equation}
\int_{u_0}^u|u^\prime|\left\|2\s^{(O_i)}\spi_1\cdot\sd(\Omega\mbox{tr}\chib)-\Db\sGamma\cdot
\sd\mbox{tr}\s^{(O_i)}\spi\right\|_{L^2(C^{\ub_1}_{u^\prime})}du^\prime\leq
O(\delta^2|u|^{-3}) \label{9.113}
\end{equation}
Substituting in \ref{9.111} and taking also into account the
estimate \ref{9.108} we conclude that:
\begin{equation}
\|\snab^{ \ 2}\mbox{tr}\s^{(O_i)}\spi\|_{L^2(C^{\ub_1}_u)}\leq
O(\delta^{3/2}|u|^{-3}) \ \ : \ \forall u\in[u_0,u_1]
\label{9.114}
\end{equation}

Next we apply $\snab$ to the propagation equation \ref{9.27}.
Using Lemmas 4.1 and 9.1 we then obtain the following  propagation
equation for $\snab^{ \ 2}\s^{(O_i)}\hat{\spi}$ along $C_{u_0}$:
\begin{equation}
D\snab^{ \ 2}\s^{(O_i)}\hat{\spi}-\Omega\mbox{tr}\chi\snab^{ \
2}\s^{(O_i)}\hat{\spi}= \gamma\cdot\snab^{ \
2}\s^{(O_i)}\hat{\spi}+2\snab\mbox{tr}\s^{(O_i)}\spi_1\cdot\Omega\chih
+2\sL_{O_i}\snab^{ \ 2}(\Omega\chih)+r^\prime \label{9.115}
\end{equation}
where:
\begin{eqnarray}
r^\prime&=&-D\sGamma\cdot\snab\s^{(O_i)}\hat{\spi}+\sd(\Omega\mbox{tr}\chi)\otimes\snab\s^{(O_i)}\hat{\spi}
+2\s^{(O_i)}\spi_1\cdot\snab(\Omega\chih)\nonumber\\
&\s&+\snab\gamma\cdot\snab\s^{(O_i)}\hat{\spi}
+2\mbox{tr}\s^{(O_i)}\spi_1\cdot\snab(\Omega\chih)+\snab r
\label{9.116}
\end{eqnarray}
To \ref{9.115} we apply Lemma 4.6 taking $p=2$. Here $r=4$,
$\nu=2$ and we obtain:
\begin{equation}
\|\snab^{ \ 2}\s^{(O_i)}\hat{\spi}\|_{L^2(S_{\ub,u_0})}\leq
C\int_0^{\ub}
\left\|2\snab\mbox{tr}\s^{(O_i)}\spi_1\cdot\Omega\chih+2\sL_{O_i}\snab^{
\
2}(\Omega\chih)+r^\prime\right\|_{L^2(S_{\ub^\prime,u_0})}d\ub^\prime
\label{9.117}
\end{equation}
Now from \ref{9.26} and the estimate \ref{9.108}:
\begin{equation}
\int_0^{\ub}\|\snab\mbox{tr}\s^{(O_i)}\spi_1\cdot\Omega\chih\|_{L^2(S_{\ub^\prime,u_0})}d\ub^\prime
\leq O(\delta^{3/2}|u_0|^{-4}) \label{9.118}
\end{equation}
By Proposition 7.2 and Lemma 7.8 together with Propositions 8.1
and 8.2:
\begin{eqnarray}
\int_0^{\ub}\|\sL_{O_i}\snab^{ \
2}(\Omega\chih)\|_{L^2(S_{\ub^\prime,u_0})}d\ub^\prime
&\leq&\delta^{1/2}\|\sL_{O_i}\snab^{ \ 2}(\Omega\chih)\|_{L^2(C_{u_0})}\nonumber\\
&\leq&O(\delta^{1/2}|u_0|^{-2}) \label{9.119}
\end{eqnarray}
Also, using Propositions 6.1 and 9.1 we deduce:
\begin{equation}
\int_0^{\ub}\|r^\prime\|_{L^2(S_{\ub,u_0})}d\ub^\prime\leq
O(\delta|u_0|^{-3}) \label{9.120}
\end{equation}
Substituting \ref{9.118} - \ref{9.120} in \ref{9.117} we conclude
that:
\begin{equation}
\|\snab^{ \ 2}\s^{(O_i)}\hat{\spi}\|_{L^2(S_{\ub,u_0})}\leq
O(\delta^{1/2}|u_0|^{-2}) 
\label{9.121}
\end{equation}

We then apply $\snab$ to the propagation equation \ref{9.35}.
Using Lemmas 4.1 and 9.1 we obtain the following  propagation
equation for $\snab^{ \ 2}\s^{(O_i)}\hat{\spi}$ along the
$\Cb_{\ub}$:
\begin{equation}
\Db\snab^{ \ 2}\s^{(O_i)}\hat{\spi}-\Omega\mbox{tr}\chib\snab^{ \
2}\s^{(O_i)}\hat{\spi}= \gammab\cdot\snab^{ \
2}\s^{(O_i)}\hat{\spi}+2\snab\mbox{tr}\s^{(O_i)}\spi_1\cdot\Omega\chibh
+2\sL_{O_i}\snab^{ \ 2}(\Omega\chibh)+\rb^\prime \label{9.122}
\end{equation}
where:
\begin{eqnarray}
\rb^\prime&=&-\Db\sGamma\cdot\snab\s^{(O_i)}\hat{\spi}+\sd(\Omega\mbox{tr}\chib)\otimes\snab\s^{(O_i)}\hat{\spi}
+2\s^{(O_i)}\spi_1\cdot\snab(\Omega\chibh)\nonumber\\
&\s&+\snab\gammab\cdot\snab\s^{(O_i)}\hat{\spi}
+2\mbox{tr}\s^{(O_i)}\spi_1\cdot\snab(\Omega\chibh)+\snab\rb
\label{9.123}
\end{eqnarray}
To \ref{9.122} we apply lemma 4.7 taking $p=2$. Here $r=4$,
$\nu=2$ and we obtain:
\begin{eqnarray}
&&|u|\|\snab^{ \ 2}\s^{(O_i)}\hat{\spi}\|_{L^2(S_{\ub,u})}\leq
C|u_0|\|\snab^{ \ 2}\s^{(O_i)}\hat{\spi}\|_{L^2(S_{\ub,u_0})}
\label{9.124}\\
&&\hspace{5mm}+C\int_{u_0}^u
|u^\prime|\left\|2\snab\mbox{tr}\s^{(O_i)}\spi_1\cdot\Omega\chibh
+2\sL_{O_i}\snab^{ \
2}(\Omega\chibh)+\rb^\prime\right\|_{L^2(S_{\ub,u^\prime})}du^\prime\nonumber
\end{eqnarray}
Taking the $L^2$ norm with respect to $\ub$ on $[0,\ub_1]$ then
yields:
\begin{eqnarray}
&&|u|\|\snab^{ \ 2}\s^{(O_i)}\hat{\spi}\|_{L^2(C^{\ub_1}_u)}\leq
C|u_0|\|\snab^{ \ 2}\s^{(O_i)}\hat{\spi}\|_{L^2(C^{\ub_1}_{u_0})}
\label{9.125}\\
&&\hspace{5mm}+\int_{u_0}^u|u^\prime|\left\|2\snab\mbox{tr}\s^{(O_i)}\spi_1\cdot\Omega\chibh
+2\sL_{O_i}\snab^{ \
2}(\Omega\chibh)+\rb^\prime\right\|_{L^2(C^{\ub_1}_{u^\prime})}du^\prime\nonumber
\end{eqnarray}
Now from \ref{9.26} and the estimate \ref{9.114}:
\begin{equation}
\int_{u_0}^u|u^\prime|\|\snab\mbox{tr}\s^{(O_i)}\spi_1\cdot\Omega\chibh\|_{L^2(C^{\ub_1}_{u^\prime})}du^\prime
\leq O(\delta^2|u|^{-4}) \label{9.126}
\end{equation}
By Proposition 7.3 and Lemma 7.8 together with Propositions 8.1
and 8.2:
\begin{equation}
\int_{u_0}^u|u^\prime|\|\sL_{O_i}\snab^{ \
2}(\Omega\chibh)\|_{L^2(C^{\ub_1}_{u^\prime})}du^\prime \leq
O(\delta|u|^{-1}) \label{9.127}
\end{equation}
Also, using Propositions 6.2 and 9.1 we deduce:
\begin{equation}
\int_{u_0}^u\|\rb^\prime\|_{L^2(C^{\ub_1}_{u^\prime})}du^\prime\leq
O(\delta^{3/2}|u|^{-2}) \label{9.128}
\end{equation}
Substituting \ref{9.126} - \ref{9.128} in \ref{9.125} and taking
also into account the estimate \ref{9.121} we conclude that:
\begin{equation}\
\|\snab^{ \ 2}\s^{(O_i)}\hat{\spi}\|_{L^2(C^{\ub_1}_u)}\leq
O(\delta|u|^{-2}) \ \ : \ \forall u\in[u_0,u_1] \label{9.129}
\end{equation}

Since the right hand sides of the inequalities \ref{9.114} and
\ref{9.129} are independent of $\ub_1$ or $u_1$ and
$(\ub_1,u_1)\in D^\prime$ is arbitrary the estimates hold with
$C^{\ub_1}_u$ replaced by $C_u$, for all $u\in[u_0,c^*)$ and the
proposition is established.

\vspace{5mm}

\noindent{\bf Proposition 9.6} \ \ \ The following estimate holds
for all $u\in[u_0,c^*)$:
$$\|\snab^{ \ 2}Z_i\|_{L^2(C_u)}\leq O(\delta^{1/2}|u|^{-2})$$
provided that $\delta$ is suitably small depending on ${\cal
D}_0^\infty$, ${\cal R}_0^\infty$, $\scD_1^4$, $\scR_1^4$ and
$\scR_2$.

\noindent{\em Proof:} \ Consider again any $(\ub_1,u_1)\in
D^\prime$ and fix attention to the parameter subdomain $D_1$ and
the corresponding subdomain $M_1$ of $M^\prime$. We apply $\snab$
to the propagation equation \ref{9.49}. Using Lemmas 9.1 and 9.2
in the case $p=q=1$, we obtain the following propagation equation
for $\snab^{ \ 2}Z_i$ along the $\Cb_{\ub}$:
\begin{equation}
\Db\snab^{ \ 2}Z_i=-4\sL_{O_i}\snab^{ \
2}(\Omega^2\zeta^\sharp)+\sb^\prime \label{9.130}
\end{equation}
where $\sb^\prime$ is the type $T^1_2$ $S$ tensorfield given by:
\begin{eqnarray}
\sb^{\prime A}_{BC}&=&-(\Db\sGamma)^D_{BC}\snab_D Z_i^A+(\Db\sGamma)^A_{BD}\snab_C Z_i^D\label{9.131}\\
&\s&-4\s^{(O_i)}\spi^D_{1,BC}\snab_D(\Omega^2\zeta^\sharp)^A+4\s^{(O_i)}\spi^A_{1,BD}\snab_C(\Omega^2\zeta)^D
+\snab_B\sb^A_C\nonumber
\end{eqnarray}
To \ref{9.130} we apply Lemma 9.4 taking $p=2$. Here $r=2$, $s=1$,
$\nu=0$, $\gammab=0$. Taking also into account the fact that
$\snab^{ \ 2}Z_i$ vanishes on $C_{u_0}$ we obtain:
\begin{equation}
\|\snab^{ \ 2}Z_i\|_{L^2(S_{\ub,u})}\leq
C\int_{u_0}^u\left\|-4\sL_{O_i}\snab^{ \ 2}(\Omega^2\zeta^\sharp)
+\sb^\prime\right\|_{L^2(S_{\ub,u^\prime})}du^\prime \label{9.132}
\end{equation}
Taking the $L^2$ norm with respect to $\ub$ on $[0,\ub_1]$ then
yields:
\begin{equation}
\|\snab^{ \ 2}Z_i\|_{L^2(C^{\ub_1}_u)}\leq
C\int_{u_0}^u\left\|-4\sL_{O_i}\snab^{ \ 2}(\Omega^2\zeta^\sharp)
+\sb^\prime\right\|_{L^2(C^{\ub_1}_{u^\prime})}du^\prime
\label{9.133}
\end{equation}
By Proposition 7.3 and Lemma 7.8 together with Propositions 8.1
and 8.2:
\begin{equation}
\int_{u_0}^u\|\sL_{O_i}\snab^{ \
2}(\Omega^2\zeta^\sharp)\|_{L^2(C^{\ub_1}_{u^\prime})}du^\prime\leq
O(\delta^{1/2}|u|^{-2}) \label{9.134}
\end{equation}
Also, using Propositions 6.2, 8.4, 9.1, 9.2 and 9.5 we deduce:
\begin{equation}
\int_{u_0}^u\|\sb^\prime\|_{L^2(C^{\ub_1}_{u^\prime})}du^\prime\leq
O(\delta^{3/2}|u|^{-3}) \label{9.135}
\end{equation}
Substituting \ref{9.134}, \ref{9.135} in \ref{9.133} we conclude
that:
\begin{equation}
\|\snab^{ \ 2}Z_i\|_{L^2(C^{\ub_1}_u)}\leq O(\delta^{1/2}|u|^{-2})
\ \ : \ \forall u\in[u_0,u_1] \label{9.136}
\end{equation}
Since the right hand side is independent of $\ub_1$ or $u_1$ and
$(\ub_1,u_1)\in D^\prime$ is arbitrary the estimate holds with
$C^{\ub_1}_u$ replaced by $C_u$, for all $u\in[u_0,c^*)$ and the
proposition is established.

\vspace{5mm}

\noindent{\bf Proposition 9.7} \ \ \ The following estimates hold
for all $u\in[u_0,c^*)$:
$$\|\sd D\mbox{tr}\s^{(O_i)}\spi\|_{L^2(C_u)}\leq O(\delta^{1/2}|u|^{-2})$$
$$\|\snab D\s^{(O_i)}\hat{\spi}\|_{L^2(C_u)}\leq O(|u|^{-1})$$
provided that $\delta$ is suitably small depending on ${\cal
D}_0^\infty$, ${\cal R}_0^\infty$, $\scD_1^4$, $\scR_1^4$ and
$\scR_2$.

\noindent{\em Proof:} \ Consider again any $(\ub_1,u_1)\in
D^\prime$ and fix attention to the parameter subdomain $D_1$. We
apply $\sd$ to equation \ref{9.57} to obtain the following
propagation equation for $\sd D\mbox{tr}\s^{(O_i)}\spi$ along the
$\Cb_{\ub}$:
\begin{equation}
\Db\sd D\mbox{tr}\s^{(O_i)}\spi=2\sL_{O_i}\sd
D(\Omega\mbox{tr}\chib)+\sd h \label{9.137}
\end{equation}
where $h$ is the function:
\begin{equation}
h=4\Omega^2\zeta^\sharp\cdot\sd\mbox{tr}\s^{(O_i)}\spi+2Z_i\cdot\sd(\Omega\mbox{tr}\chib)
\label{9.138}
\end{equation}
To equation \ref{9.137} we apply Lemma 4.7 taking $p=2$. here
$r=1$, $\nu=0$, $\gammab=0$ and we obtain:
\begin{eqnarray}
&&\|\sd D\mbox{tr}\s^{(O_i)}\spi\|_{L^2(S_{\ub,u})}\leq C\|\sd D\mbox{tr}\s^{(O_i)}\spi\|_{L^2(S_{\ub,u_0})}\label{9.139}\\
&&\hspace{2cm}+C\int_{u_0}^u\left\|2\sL_{O_i}\sd
D(\Omega\mbox{tr}\chib)+\sd
h\right\|_{L^2(S_{\ub,u^\prime})}du^\prime \nonumber
\end{eqnarray}
Taking the $L^2$ norm with respect to $\ub$ on $[0,\ub_1]$ then
yields:
\begin{eqnarray}
&&\|\sd D\mbox{tr}\s^{(O_i)}\spi\|_{L^2(C^{\ub_1}_u)}\leq
C\|\sd D\mbox{tr}\s^{(O_i)}\spi\|_{L^2(C^{\ub_1}_{u_0})}\label{9.140}\\
&&\hspace{2cm}+C\int_{u_0}^u\left\|2\sL_{O_i}\sd
D(\Omega\mbox{tr}\chib)+\sd
h\right\|_{L^2(C^{\ub_1}_{u^\prime})}du^\prime \nonumber
\end{eqnarray}
Now, by equation \ref{8.102} we have on $C_{u_0}$:
\begin{equation}
\sd D\mbox{tr}\s^{(O_i)}\spi=2\sL_{O_i}\sd(\Omega\mbox{tr}\chi)
\label{9.141}
\end{equation}
Proposition 6.1 together with Propositions 8.1 and 8.2 then imply:
\begin{equation}
\|\sd D\mbox{tr}\s^{(O_i)}\spi\|_{L^2(C^{\ub_1}_{u_0})}\leq
O(\delta^{1/2}|u_0|^{-2}) \label{9.142}
\end{equation}
By equation \ref{4.c1}, Proposition 7.3 together with Propositions
8.1 and 8.2 and the results of Chapter 6:
\begin{equation}
\int_{u_0}^u\|\sL_{O_i}\sd
D(\Omega\mbox{tr}\chib)\|_{L^2(C^{\ub_1}_{u^\prime})}du^\prime\leq
O(\delta^{1/2}|u|^{-2}) \label{9.143}
\end{equation}
Also, using Propositions 9.2 and 9.5 we deduce:
\begin{equation}
\int_{u_0}^u\|\sd h\|_{L^2(C^{\ub_1}_{u^\prime})}du^\prime\leq
O(\delta^{3/2}|u|^{-4}) \label{9.144}
\end{equation}
Substituting \ref{9.142} - \ref{9.143} in \ref{9.140} we conclude
that:
\begin{equation}
\|\sd D\mbox{tr}\s^{(O_i)}\spi\|_{L^2(C^{\ub_1}_u)}\leq
O(\delta^{1/2}|u|^{-2}) \ \ : \ \forall u\in[u_0,u_1]
\label{9.145}
\end{equation}

Next we apply $\snab$ to the propagation equation \ref{9.66}.
Using Lemmas 4.1 and 9.1 we then obtain the following propagation
equation for $\snab D\s^{(O_i)}\hat{\spi}$ along the $\Cb_{\ub}$:
\begin{equation}
\Db\snab D\s^{(O_i)}\hat{\spi}-\Omega\mbox{tr}\chib\snab
D\s^{(O_i)}\hat{\spi}=2\sL_{O_i}\snab D(\Omega\chibh)+v^\prime
\label{9.146}
\end{equation}
where:
\begin{equation}
v^\prime=-\Db\sGamma\cdot
D\s^{(O_i)}\hat{\spi}+\sd(\Omega\mbox{tr}\chib)\otimes
D\s^{(O_i)}\hat{\spi} +2\s^{(O_i)}\spi_1\cdot
D(\Omega\chibh)+\snab v \label{9.147}
\end{equation}
To \ref{9.147} we apply Lemma 4.7 taking $p=2$. Here $r=3$,
$\nu=2$, $\gammab=0$ and we obtain:
\begin{eqnarray}
&&\|\snab D\s^{(O_i)}\hat{\spi}\|_{L^2(S_{\ub,u})}\leq C\|\snab D\s^{(O_i)}\hat{\spi}\|_{L^2(S_{\ub,u_0})}\label{9.148}\\
&&\hspace{2cm}+C\int_{u_0}^u\left\|2\sL_{O_i}\snab
D(\Omega\chibh)+v^\prime\right\|_{L^2(S_{\ub,u^\prime})}du^\prime\nonumber
\end{eqnarray}
Taking the $L^2$ norm with respect to $\ub$ on $[0,\ub_1]$ then
yields:
\begin{eqnarray}
&&\|\snab D\s^{(O_i)}\hat{\spi}\|_{L^2(C^{\ub_1}_u)}\leq C\|\snab
D\s^{(O_i)}\hat{\spi}\|_{L^2(C^{\ub_1}_{u_0})}
\label{9.149}\\
&&\hspace{2cm}+C\int_{u_0}^u\left\|2\sL_{O_i}\snab
D(\Omega\chibh)+v^\prime\right\|_{L^2(C^{\ub_1}_{u^\prime})}du^\prime
\nonumber
\end{eqnarray}
Now, by equation \ref{8.104} we have on $C_{u_0}$:
\begin{eqnarray}
\snab
D\s^{(O_i)}\hat{\spi}&=&\Omega\mbox{tr}\chi\snab\s^{(O_i)}\hat{\spi}
+\sd(\Omega\mbox{tr}\chi)\otimes\s^{(O_i)}\hat{\spi}\label{9.150}\\
&\s&-\sd\mbox{tr}\s^{(O_i)}\spi\otimes\Omega\chih-\mbox{tr}\s^{(O_i)}\spi\snab(\Omega\chih)+2\sL_{O_i}\snab(\Omega\chih)
+2\s^{(O_i)}\spi_1\cdot\Omega\chih\nonumber
\end{eqnarray}
Propositions 6.1, 9.1 together with Propositions 8.1 and 8.2 then
imply:
\begin{equation}
\|\snab D\s^{(O_i)}\hat{\spi}\|_{L^2(C^{\ub_1}_{u_0})}\leq O(|u_0|^{-1})
\label{9.151}
\end{equation}
By equation \ref{4.c3}, Proposition 7.3 together with Propositions
8.1 and 8.2 and the results of Chapter 6:
\begin{equation}
\int_{u_0}^u\|\sL_{O_i}\snab
D(\Omega\chibh)\|_{L^2(C^{\ub_1}_{u^\prime})}du^\prime\leq
O(|u|^{-1}) \label{9.152}
\end{equation}
Also, using Propositions 9.1, 9.2, 9.3. 9.5, 9.6 and the estimate
\ref{9.145} we deduce:
\begin{equation}
\int_{u_0}^u\|v^\prime\|_{L^2(C^{\ub_1}_{u^\prime})}du^\prime\leq
O(\delta^{1/2}|u|^{-2}) \label{9.153}
\end{equation}
Substituting \ref{9.151} - \ref{9.153} in \ref{9.149} we conclude
that:
\begin{equation}
\|\snab D\s^{(O_i)}\hat{\spi}\|_{L^2(C^{\ub_1}_u)}\leq
O(|u|^{-1}) \ \ : \ \forall u\in[u_0,u_1] \label{9.154}
\end{equation}

Since the right hand sides of the inequalities \ref{9.145} and
\ref{9.154} are independent of $\ub_1$ or $u_1$ and
$(\ub_1,u_1)\in D^\prime$ is arbitrary the estimates hold with
$C^{\ub_1}_u$ replaced by $C_u$, for all $u\in[u_0,c^*)$ and the
proposition is established.

\vspace{5mm}

{\bf Proposition 9.8} \ \ \ The following estimate holds for all
$u\in[u_0,c^*)$:
$$\|\snab DZ_i\|_{L^2(C_u)}\leq O(|u|^{-1})$$
provided that $\delta$ is suitably small depending on ${\cal
D}_0^\infty$, ${\cal R}_0^\infty$, $\scD_1^4$, $\scR_1^4$ and
$\scR_2$.

\noindent{\em Proof:} \ Consider again any $(\ub_1,u_1)\in
D^\prime$ and fix attention to the parameter subdomain $D_1$. We
apply $\snab$ to equation \ref{9.77}. Using then Lemmas 9.1 and
9.2 yields the following propagation equation for $\snab DZ_i$
along the $\Cb_{\ub}$:
\begin{equation}
\Db\snab DZ_i=-4\sL_{O_i}\snab D(\Omega^2\zeta^\sharp)+w
\label{9.155}
\end{equation}
where:
\begin{equation}
w=\Db\sGamma\cdot DZ_i+4\s^{(O_i)}\spi_1\cdot
D(\Omega^2\zeta^\sharp)+8\snab[\Omega^2\zeta^\sharp,Z_i]
\label{9.156}
\end{equation}
To \ref{9.155} we apply Lemma 9.4 taking $p=2$. Here $r=s=1$,
$\nu=0$, $\gammab=0$. Taking also into account the fact that
$\snab DZ_i$ vanishes on $C_{u_0}$ we obtain:
\begin{equation}
|u|^{-1}\|\snab DZ_i\|_{L^2(S_{\ub,u})}\leq
C\int_{u_0}^u|u^\prime|^{-1} \left\|-4\sL_{O_i}\snab
D(\Omega^2\zeta^\sharp)+w\right\|_{L^2(S_{\ub,u^\prime})}du^\prime
\label{9.157}
\end{equation}
Taking the $L^2$ norm with respect to $\ub$ on $[0,\ub_1]$ then
yields:
\begin{equation}
|u|^{-1}\|\snab DZ_i\|_{L^2(C^{\ub_1}_u)}\leq
C\int_{u_0}^u|u^\prime|^{-1} \left\|-4\sL_{O_i}\snab
D(\Omega^2\zeta^\sharp)+w\right\|_{L^2(C^{\ub_1}_{u^\prime})}du^\prime
\label{9.158}
\end{equation}
Now, from equation \ref{1.61}, by Proposition 7.5 together with
Propositions 8.1, 8.2 and the results of Chapters 6 we obtain:
\begin{equation}
\int_{u_0}^u|u^\prime|^{-1} \|\sL_{O_i}\snab
D(\Omega^2\zeta^\sharp)\|_{L^2(C^{\ub_1}_{u^\prime})}du^\prime\leq
O(|u|^{-2}) \label{9.159}
\end{equation}
Also, using Propositions 9.1, 9.4 and 9.6 we deduce:
\begin{equation}
\int_{u_0}^u|u^\prime|^{-1}
\|w\|_{L^2(C^{\ub_1}_{u^\prime})}du^\prime\leq
O(\delta^{1/2}|u|^{-3}) \label{9.160}
\end{equation}
Substituting \ref{9.159}, \ref{9.160} in \ref{9.158} we conclude
that:
\begin{equation}
\|\snab DZ_i\|_{L^2(C^{\ub_1}_u)}\leq O(|u|^{-1}) \ \ : \ \forall
u\in[u_0,u_1] \label{9.161}
\end{equation}
Since the right hand side is independent of $\ub_1$ or $u_1$ and
$(\ub_1,u_1)\in D^\prime$ is arbitrary the estimate holds with
$C^{\ub_1}_u$ replaced by $C_u$, for all $u\in[u_0,c^*)$ and the
proposition is established.

\vspace{5mm}

Estimates for $\sd\Db\s^{(O_i)}\mbox{tr}\spi$,
$\snab\Db\s^{(O_i)}\hat{\spi}$, and $\snab\Db Z_i$, in $L^2(C_u)$,
follow directly from equations \ref{8.103}, \ref{8.105}, and
\ref{8.63} respectively. We obtain:
\begin{eqnarray}
&&\|\sd\Db\s^{(O_i)}\mbox{tr}\spi\|_{L^2(C_u)}\leq O(\delta^{3/2}|u|^{-3})\nonumber\\
&&\|\snab\Db\s^{(O_i)}\hat{\spi}\|_{L^2(C_u)}\leq O(\delta|u|^{-2})\nonumber\\
&&\|\snab\Db Z_i\|_{L^2(C_u)}\leq O(\delta|u|^{-2}) \label{9.162}
\end{eqnarray}
the last using Proposition 7.1.

In view of the expressions \ref{8.134} - \ref{8.138} for the
components of the deformation tensor of the $O_i$, Propositions
9.5, 9.6, 9.7, 9.8 and the estimates \ref{9.162} yield the
following estimates in $L^2(C_u)$ for the 2nd derivatives of the
non-vanishing components of the deformation tensors of the $O_i$
for which the last derivative is tangential to the surfaces
$S_{\ub,u}$:
\begin{eqnarray}
&&\|\snab^{ \ 2}\s^{(O_i)}\ih\|_{L^2(C_u)}\leq O(\delta|u|^{-2})\nonumber\\
&&\|\snab^{ \ 2}\s^{(O_i)}j\|_{L^2(C_u)}\leq O(\delta^{3/2}|u|^{-3})\nonumber\\
&&\|\snab^{ \ 2}\s^{(O_i)}m\|_{L^2(C_u)}\leq
O(\delta^{1/2}|u|^{-2}) \label{9.163}
\end{eqnarray}
\begin{eqnarray}
&&\|\snab\Dh\s^{(O_i)}\ih\|_{L^2(C_u)}\leq O(|u|^{-1})\nonumber\\
&&\|\sd D\s^{(O_i)}j\|_{L^2(C_u)}\leq O(\delta^{1/2}|u|^{-2})\nonumber\\
&&\|\snab D\s^{(O_i)}m\|_{L^2(C_u)}\leq O(|u|^{-1}) \label{9.164}
\end{eqnarray}
\begin{eqnarray}
&&\|\snab\Dbh\s^{(O_i)}\ih\|_{L^2(C_u)}\leq O(\delta|u|^{-2})\nonumber\\
&&\|\sd\Db\s^{(O_i)}j\|_{L^2(C_u)}\leq O(\delta^{3/2}|u|^{-3})\nonumber\\
&&\|\snab\Db\s^{(O_i)}m\|_{L^2(C_u)}\leq O(\delta|u|^{-2})
\label{9.165}
\end{eqnarray}
In the second of \ref{9.163}, in estimating the contribution of
the 2nd term on the right in \ref{8.137} we bound (see \ref{9.85})
$$\sL_{O_i}\snab^{ \ 2}\log\Omega$$
in $L^2(C_u)$ using Lemma 7.8. In the second of \ref{9.164} in
estimating the contribution of the 2nd term on the right in
\ref{8.137} we bound (see \ref{9.86})
$$\sL_{O_i}\sd\omega+\sd(Z_i\cdot\sd\log\Omega)$$
in $L^2(C_u)$ using Propositions 7.4 and 9.2. In the second of
\ref{9.165} to estimate the contribution of the 2nd term on the
right in \ref{8.137} we bound (see \ref{9.86})
$$\sL_{O_i}\sd\omb$$
in $L^2(C_u)$ using the estimate for $\snab^{ \ 2}\omb$ in
$L^4(S)$ of Proposition 6.2. Finally, since
$$\snab\Db\s^{(O_i)}m=\sg\cdot\snab\Db Z_i+2\snab(\Omega\chib)\cdot Z_i+2\Omega\chib\cdot\snab Z_i$$
the last of \ref{9.165} requires the estimate:
\begin{equation}
\|\snab Z_i\|_{L^2(C_u)}\leq O(\delta|u|^{-1}) \label{9.166}
\end{equation}
which follows from the propagation equation \ref{9.49} using
Proposition 7.1.

\chapter{The Sobolev Inequalities on the $C_u$ and the
$\Cb_{\ub}$}

\section{Introduction}

In this chapter we shall derive Sobolev inequalities on the $C_u$
and the $\Cb_{\ub}$ on the basis of certain bootstrap assumptions.
These Sobolev inequalities shall be applied to obtain bounds for
the spacetime curvature components in $L^\infty$ and for their 1st
derivatives in $L^4(S)$ in terms of the $L^2$ norms of up to their
2nd derivatives on the $C_u$, in the case of the components
$\alpha$, $\beta$, $\rho$, $\sigma$, $\beb$, the $L^2$ norms of up
to the 2nd derivatives on the $\Cb_{\ub}$, in the case of the
component $\alb$.

In the present chapter, we shall not make use of the estimates on
the connection coefficients derived in the previous chapters, for
these have relied on the $L^\infty$ bounds for the spacetime
curvature components and the $L^4(S)$ bounds for their 1st
derivatives which we are presently to establish. We shall rely
instead on certain bootstrap assumptions the most basic of which
control the ratio $\mbox{Area}(S_{\ub,u})/|u|^2$ and
$\mbox{I}(S_{\ub,u})$, the isoperimetric constant of the surfaces
$S_{\ub,u}$.

These most basic bootstrap assumptions are:

\vspace{5mm}

\ \ \ {\bf A1.1:} \
$\delta\Omega|\mbox{tr}\chi|\leq\frac{1}{3}\log 2$ \ : \ in
$M^\prime$

\vspace{2.5mm}

\ \ \ {\bf A1.2:} \ $\delta\Omega|\chih|\leq\frac{1}{6}\log 2$ \ :
\ in $M^\prime$

\vspace{5mm}

Let us denote by $\tilde{A}(\ub,u)$ the ratio:
\begin{equation}
\tilde{A}(\ub,u)=\frac{\mbox{Area}(S_{\ub,u})}{|u|^2} \label{10.1}
\end{equation}

\vspace{5mm}

\noindent{\bf Lemma 10.1} \ \ \ The assumption {\bf A1.1} implies:
$$2^{-1/3}\leq\frac{\tilde{A}(\ub,u)}{4\pi}\leq 2^{1/3} \ \ : \ \forall(\ub,u)\in D^\prime$$
Also, the assumptions {\bf A1.1}, {\bf A1.2} together imply:
$$\mbox{I}(S_{\ub,u})\leq\frac{1}{\pi} \ \ : \ \forall(\ub,u)\in D^\prime$$
that is, the conclusion of Lemma 5.4 holds.

\noindent{\em Proof:} \ We first revisit the proof of Lemma 5.3.
Here $u$ is fixed. From formula \ref{5.41}, assumption {\bf A1.1}
implies:
\begin{equation}
2^{-1/3}\leq\mu(\ub)\leq 2^{1/3} \label{10.2}
\end{equation}
Also, from the inequality \ref{5.54}, assumption {\bf A1.2}
implies:
\begin{equation}
\nu(\ub)\leq 2^{1/3} \label{10.3}
\end{equation}
Hence, by the first of \ref{5.56}, assumptions {\bf A1.1}, {\bf
A1.2} imply:
\begin{equation}
\lambda(\ub)\geq 2^{-2/3} \label{10.4}
\end{equation}
Now (see \ref{5.38}, \ref{5.61}),
$$\mbox{Area}(S_{\ub,u})=\int_{S_{0,u}}d\mu_{\sg(\ub)}=\int_{S_{0,u}}\mu(\ub)d\mu_{\sg(0)}$$
while
$$\mbox{Area}(S_{0,u})=\int_{S_{0,u}}d\mu_{\sg(0)}=4\pi|u|^2$$
Therefore \ref{10.2} implies:
\begin{equation}
2^{-1/3}\leq\frac{\mbox{Area}(S_{\ub,u})}{4\pi|u|^2}\leq 2^{1/3}
\label{10.5}
\end{equation}
which is the first part of the lemma.

To obtain the second part of the lemma we revisit the proof of
Lemma 5.4. Consider again a domain $U_{\ub}\subset S_{\ub,u}$ with
$C^1$ boundary $\partial U_{\ub}$, the image by $\Phi_{\ub}$ of a
domain $U_0\subset S_{0,u}$ with $C^1$ boundary $\partial U_0$.
According to \ref{5.59} we have:
\begin{equation}
\frac{\mbox{Perimeter}(\partial
U_{\ub})}{\mbox{Perimeter}(\partial
U_0)}\geq\inf_{S_{0,u}}\sqrt{\lambda(\ub)} \label{10.6}
\end{equation}
On the other hand, according to \ref{5.63} we have:
\begin{equation}
\frac{\mbox{Area}(U_{\ub})}{\mbox{Area}(U_0)}\leq\sup_{S_{0,u}}\mu(\ub)
\label{10.7}
\end{equation}
The inequalities \ref{10.6} and \ref{10.7} together imply:
\begin{equation}
\frac{\mbox{Area}(U_{\ub})}{(\mbox{Perimeter}(\partial
U_{\ub}))^2}\leq\frac{\sup_{S_{0,u}}\mu(\ub)}{\inf_{S_{0,u}}\lambda(\ub)}
\frac{\mbox{Area}(U_0)}{(\mbox{Perimeter}(\partial U_0))^2}
\label{10.8}
\end{equation}
Similarly, we obtain:
\begin{equation}
\frac{\mbox{Area}(U^c_{\ub})}{(\mbox{Perimeter}(\partial
U_{\ub}))^2}\leq\frac{\sup_{S_{0,u}}\mu(\ub)}{\inf_{S_{0,u}}\lambda(\ub)}
\frac{\mbox{Area}(U^c_0)}{(\mbox{Perimeter}(\partial U_0))^2}
\label{10.9}
\end{equation}
Therefore:
\begin{equation}
\frac{\min\{\mbox{Area}(U_{\ub}),\mbox{Area}(U^c_{\ub})\}}{(\mbox{Perimeter}(\partial
U_{\ub}))^2} \leq
\frac{\sup_{S_{0,u}}\mu(\ub)}{\inf_{S_{0,u}}\lambda(\ub)} 
\frac{\min\{\mbox{Area}(U_0),\mbox{Area}(U^c_0)\}}{(\mbox{Perimeter}(\partial
U_0))^2} \label{10.10}
\end{equation}
Taking the supremum over all domains $U_{\ub}$ with $C^1$ boundary
$\partial U_{\ub}$ in $S_{\ub,u}$ we conclude that:
\begin{equation}
\mbox{I}(S_{\ub,u})\leq\frac{\sup_{S_{0,u}}\mu(\ub)}{\inf_{S_{0,u}}\lambda(\ub)}\mbox{I}(S_{0,u})
\label{10.11}
\end{equation}
Now, inequalities \ref{10.2} and \ref{10.4} imply:
\begin{equation}
\frac{\sup_{S_{0,u}}\mu(\ub)}{\inf_{S_{0,u}}\lambda(\ub)}\leq 2
\label{10.12}
\end{equation}
In view of \ref{10.12} and \ref{5.69} the second part of the lemma
follows from \ref{10.11}.

\vspace{5mm}

\noindent{\bf Lemma 10.2} \ \ \ Let the basic bootstrap
assumptions {\bf A1.1}, {\bf A1.2} hold and let $\xi$ be an
arbitrary $p$ covariant $S$ tensorfield. Suppose that $\xi\in
W_1^2(S_{\ub,u})$ for some $(\ub,u)\in D^\prime$. Then:
$$\int_{S_{\ub,u}}|\xi|^6 d\mu_{\sg}\leq\frac{9}{\pi}\left(\int_{S_{\ub,u}}|\xi|^4 d\mu_{\sg}\right)
\left\{\frac{1}{18|u|^2}\int_{S_{\ub,u}}|\xi|^2
d\mu_{\sg}+\int_{S_{\ub,u}}|\snab\xi|^2 d\mu_{\sg}\right\}$$

\noindent{\em Proof:} \ Recall that by Lemma 5.1 $\xi\in
L^q(S_{\ub,u})$ for every $q<\infty$. The point here is to
establish the above inequality. Let $f$ be the function:
\begin{equation}
f=|\xi|^3 \label{10.13}
\end{equation}
We apply to $f$ the isoperimetric inequality \ref{5.35} on
$S_{\ub,u}$:
\begin{equation}
\int_{S_{\ub,u}}f^2 d\mu_{\sg}\leq \int_{S_{\ub,u}}\overline{f}^2
d\mu_{\sg} +\mbox{I}(S_{\ub,u})\left(\int_{S_{\ub,u}}|\sd
f|d\mu_{\sg}\right)^2 \label{10.15}
\end{equation}
Now, we have:
\begin{equation}
\sd f=3|\xi|(\xi,\snab\xi) \label{10.16}
\end{equation}
Thus,
\begin{equation}
|\sd f|\leq 3|\xi|^2|\snab\xi| \label{10.17}
\end{equation}
hence:
\begin{equation}
\left(\int_{S_{\ub,u}}|\sd f| d\mu_{\sg}\right)^2\leq
9\left(\int_{S_{\ub,u}}|\xi|^4 d\mu_{\sg}\right)
\left(\int_{S_{\ub,u}}|\snab\xi|^2 d\mu_{\sg}\right) \label{10.18}
\end{equation}
Also,
\begin{eqnarray}
&&\int_{S_{\ub,u}}\overline{f}^2
d\mu_{\sg}=\frac{1}{\mbox{Area}(S_{\ub,u})}\left(\int_{S_{\ub,u}}f
d\mu_{\sg}\right)^2
=\frac{1}{\mbox{Area}(S_{\ub,u})}\left(\int_{S_{\ub,u}}|\xi|^3 d\mu_{\sg}\right)^2\nonumber\\
&&\hspace{2cm}\leq\frac{1}{\mbox{Area}(S_{\ub,u})}\left(\int_{S_{\ub,u}}|\xi|^4
d\mu_{\sg}\right)\left(\int_{S_{\ub,u}}|\xi|^2 d\mu_{\sg}\right)
\label{10.19}
\end{eqnarray}
Substituting \ref{10.18} and \ref{10.19} in \ref{10.15}, recalling
the definition \ref{10.1}, and noting that by Lemma 10.1:
\begin{equation}
\max\left\{\mbox{I}(S_{\ub,u}),
\frac{2}{\tilde{A}(\ub,u)}\right\}\leq\frac{1}{\pi} \label{10.20}
\end{equation}
yields the inequality of the lemma.

\section{The Sobolev inequalities on the $C_u$}

\noindent{\bf Proposition 10.1} \ \ \ Let the basic bootstrap
assumptions {\bf A1.1}, {\bf A1.2} hold. Let $\xi$ be an arbitrary
$p$ covariant $S$ tensorfield vanishing on $\Cb_0$ such that
$\xi$, $D\xi$, $\snab\xi$ belong to $L^2(C_u)$ for some $u\in
[u_0,c^*)$. Then $\xi\in L^6(C_u)$ and there is a numerical
constant $C_p$, depending only on $p$, such that:
$$|u|^{2/3}\|\xi\|_{L^6(C_u)}\leq C_p\|D\xi\|^{1/3}_{L^2(C_u)}\left\{(1/18)\|\xi\|^2_{L^2(C_u)}+|u|^2\|\snab\xi\|^2_{L^2(C_u)}\right\}^{1/3}$$
Also, $\xi\in L^4(S_{\ub,u})$ for each $\ub$ and there is a
numerical constant $C^\prime_p$, depending only on $p$, such that:
$$\sup_{\ub}\left(|u|^{1/2}\|\xi\|_{L^4(S_{\ub,u})}\right)\leq  C^\prime_p\|D\xi\|^{1/2}_{L^2(C_u)}\left\{(1/18)\|\xi\|^2_{L^2(C_u)}+|u|^2\|\snab\xi\|^2_{L^2(C_u)}\right\}^{1/4}$$

\noindent{\em Proof:} By the assumptions of the proposition
$\xi\in W_1^2(S_{\ub,u})$ for almost all $\ub\in[0,\delta)$ : if
$u\in[u_0,c^*-\delta]$, for almost all $\ub\in[0,c^*-u)$ : if
$u\in(c^*-\delta,c^*)$, in the case $c^*\geq u_0+\delta$, for almost all $\ub\in[0,c^*-u)$ in the case $c^*<u_0+\delta$ 
(see \ref{5.7}, \ref{5.7a}). Thus, Lemma 10.2 applies to $\xi$. We
multiply the inequality of Lemma 10.2 by $|u|^4$ and integrate
with respect to $\ub$ to obtain:
\begin{equation}
\int_{C_u}|u|^4|\xi|^6\leq\frac{9}{\pi}\sup_{\ub}\left(\int_{S_{\ub,u}}|u|^2|\xi|^4\right)
\left\{\frac{1}{18}\int_{C_u}|\xi|^2+\int_{C_u}|u|^2|\snab\xi|^2\right\}
\label{10.22}
\end{equation}
Consider now the surface integral ($u$ is fixed):
\begin{equation}
x(\ub)=\int_{S_{\ub,u}}|u|^2|\xi|^4 d\mu_{\sg} \label{10.23}
\end{equation}
We have:
\begin{equation}
\frac{d x}{d
\ub}=\int_{S_{\ub,u}}|u|^2\left\{D(|\xi|^4)+\Omega\mbox{tr}\chi|\xi|^4\right\}d\mu_{\sg}
\label{10.24}
\end{equation}
By the first part of Lemma 4.2:
\begin{eqnarray}
&&D(|\xi|^4)+\Omega\mbox{tr}\chi|\xi|^4=4|\xi|^2(\xi,D\xi)\label{10.25}\\
&&\hspace{27mm}-(2p-1)\Omega\mbox{tr}\chi|\xi|^4-4|\xi|^2\sum_{i=1}^p\Omega\chih^{A_i}_{B_i}\xi^{A_1...\stackrel{B_i}{>A_i<}...A_p}
\xi_{A_1...A_p}\nonumber
\end{eqnarray}
It follows that:
\begin{equation}
D(|\xi|^4)+\Omega\mbox{tr}\chi|\xi|^4\leq
4|\xi|^3|D\xi|+(|2p-1|\Omega|\mbox{tr}\chi|+4p\Omega|\chih|)|\xi|^4
\label{10.26}
\end{equation}
Now, by assumptions {\bf A1.1}, {\bf A1.2},
\begin{equation}
|2p-1|\Omega|\mbox{tr}\chi|+4p\Omega|\chih|\leq
\frac{k_p}{\delta}, \ \ \mbox{where:} \ k_p=\frac{1}{3}\log
2(|2p-1|+2p) \label{10.27}
\end{equation}
Substituting \ref{10.27} in \ref{10.26} and the result in
\ref{10.24} we obtain the following ordinary differential
inequality for $x$:
\begin{equation}
\frac{d x}{d \ub}\leq \frac{k_p}{\delta} x+a \label{10.28}
\end{equation}
where:
\begin{equation}
a(\ub)=4\int_{S_{\ub,u}}|u|^2|\xi|^3|D\xi|d\mu_{\sg} \label{10.29}
\end{equation}
We have:
\begin{equation}
x(0)=0 \label{10.30}
\end{equation}
since $\xi$ vanishes on $\Cb_0$. Integrating \ref{10.28} then
yields:
\begin{eqnarray}
x(\ub)&\leq&\int_0^{\ub}e^{k_p(\ub-\ub^\prime)/\delta}a(\ub^\prime)d\ub^\prime\nonumber\\
&\leq&e^{k_p}\int_0^{\ub}a(\ub^\prime)d\ub^\prime \label{10.31}
\end{eqnarray}
Since
\begin{equation}
\int_0^{\ub}a(\ub^\prime)d\ub^\prime\leq
4\int_{C_u}|u|^2|\xi|^3|D\xi|\leq
4\left(\int_{C_u}|u|^4|\xi|^6\right)^{1/2}\left(\int_{C_u}|D\xi|^2\right)^{1/2}
\label{10.32}
\end{equation}
\ref{10.31} implies:
\begin{equation}
\sup_{\ub}\left(\int_{S_{\ub,u}}|u|^2|\xi|^4\right)\leq
4e^{k_p}\left(\int_{C_u}|u|^4|\xi|^6\right)^{1/2}\left(\int_{C_u}|D\xi|^2\right)^{1/2}
\label{10.33}
\end{equation}
Substituting in \ref{10.22} and cancelling one factor of
$$\left(\int_{C_u}|u|^4|\xi|^6\right)^{1/2}$$
from both sides we then obtain:
\begin{equation}
\left(\int_{C_u}|u|^4|\xi|^6\right)^{1/2}\leq \frac{36
e^{k_p}}{\pi}\left(\int_{C_u}|D\xi|^2\right)^{1/2}
\left\{\frac{1}{18}\int_{C_u}|\xi|^2+\int_{C_u}|u|^2|\snab\xi|^2\right\}
\label{10.34}
\end{equation}
which gives the first part of the proposition. Substituting
\ref{10.34} in turn in \ref{10.33} yields:
\begin{equation}
\sup_{\ub}\left(\int_{S_{\ub,u}}|u|^2|\xi|^4\right)\leq \frac{144
e^{2k_p}}{\pi}\left(\int_{C_u}|D\xi|^2\right)
\left\{\frac{1}{18}\int_{C_u}|\xi|^2+\int_{C_u}|u|^2|\snab\xi|^2\right\}
\label{10.35}
\end{equation}
which gives the second part of the proposition.

\vspace{5mm}

We now apply the above proposition to the spacetime curvature
componets $\alpha$, $\beta$, $\rho$, $\sigma$, $\beb$. We define:
\begin{eqnarray}
&&{\cal R}_0(\alpha)=\sup_{u\in[u_0,c^*)}\left(\delta\|\alpha\|_{L^2(C_u)}\right)\nonumber\\
&&{\cal R}_0(\Dh\alpha)=\sup_{u\in[u_0,c^*)}\left(\delta^2\|\Dh\alpha\|_{L^2(C_u)}\right)\nonumber\\
&&\scR_1(\alpha)=\sup_{u\in[u_0,c^*)}\left(\delta|u|\|\snab\alpha\|_{L^2(C_u)}\right)\label{10.36}
\end{eqnarray}
and:
\begin{equation}
{\cal R}_{[1]}(\alpha)=\max\{{\cal R}_0(\alpha),{\cal
R}_0(\Dh\alpha),\scR_1(\alpha)\} \label{10.37}
\end{equation}
We define:
\begin{eqnarray}
&&{\cal R}_0(\beta)=\sup_{u\in[u_0,c^*)}\left(|u|\|\beta\|_{L^2(C_u)}\right)\nonumber\\
&&{\cal R}_0(D\beta)=\sup_{u\in[u_0,c^*)}\left(\delta|u|\|D\beta\|_{L^2(C_u)}\right)\nonumber\\
&&\scR_1(\beta)=\sup_{u\in[u_0,c^*)}\left(|u|^2\|\snab\beta\|_{L^2(C_u)}\right)\label{10.38}
\end{eqnarray}
and:
\begin{equation}
{\cal R}_{[1]}(\beta)=\max\{{\cal R}_0(\beta),{\cal
R}_0(D\beta),\scR_1(\beta)\} \label{10.39}
\end{equation}
We define:
\begin{eqnarray}
&&{\cal R}_0(\rho)=\sup_{u\in[u_0,c^*)}\left(\delta^{-1/2}|u|^2\|\rho\|_{L^2(C_u)}\right)\nonumber\\
&&{\cal R}_0(D\rho)=\sup_{u\in[u_0,c^*)}\left(\delta^{1/2}|u|^2\|D\rho\|_{L^2(C_u)}\right)\nonumber\\
&&\scR_1(\rho)=\sup_{u\in[u_0,c^*)}\left(\delta^{-1/2}|u|^3\|\sd\rho\|_{L^2(C_u)}\right)\label{10.40}
\end{eqnarray}
and:
\begin{equation}
{\cal R}_{[1]}(\rho)=\max\{{\cal R}_0(\rho),{\cal
R}_0(D\rho),\scR_1(\rho)\} \label{10.41}
\end{equation}
Also:
\begin{eqnarray}
&&{\cal R}_0(\sigma)=\sup_{u\in[u_0,c^*)}\left(\delta^{-1/2}|u|^2\|\sigma\|_{L^2(C_u)}\right)\nonumber\\
&&{\cal R}_0(D\sigma)=\sup_{u\in[u_0,c^*)}\left(\delta^{1/2}|u|^2\|D\sigma\|_{L^2(C_u)}\right)\nonumber\\
&&\scR_1(\sigma)=\sup_{u\in[u_0,c^*)}\left(\delta^{-1/2}|u|^3\|\sd\sigma\|_{L^2(C_u)}\right)\label{10.42}
\end{eqnarray}
and:
\begin{equation}
{\cal R}_{[1]}(\sigma)=\max\{{\cal R}_0(\sigma),{\cal
R}_0(D\sigma),\scR_1(\sigma)\} \label{10.43}
\end{equation}
We define:
\begin{eqnarray}
&&{\cal R}_0(\beb)=\sup_{u\in[u_0,c^*)}\left(\delta^{-3/2}|u|^3\|\beb\|_{L^2(C_u)}\right)\nonumber\\
&&{\cal R}_0(D\beb)=\sup_{u\in[u_0,c^*)}\left(\delta^{-1/2}|u|^3\|D\beb\|_{L^2(C_u)}\right)\nonumber\\
&&\scR_1(\beb)=\sup_{u\in[u_0,c^*)}\left(\delta^{-3/2}|u|^4\|\snab\beb\|_{L^2(C_u)}\right)\label{10.44}
\end{eqnarray}
and:
\begin{equation}
{\cal R}_{[1]}(\beb)=\max\{{\cal R}_0(\beb),{\cal
R}_0(D\beb),\scR_1(\beb)\} \label{10.45}
\end{equation}

We apply Proposition 10.1 to $\xi=\alpha$. Since in regard to the
first term on the right in \ref{10.25} we have, in this case,
\begin{equation}
(\xi,D\xi)=(\xi,\Dh\xi) \label{10.46}
\end{equation}
we may replace $D\xi$ by $\Dh\xi$ on the right in \ref{10.26} and
\ref{10.29}, hence also in the conclusions of Proposition 10.1.
Defining:
\begin{equation}
{\cal R}_0^4(\alpha)=\sup_{(\ub,u)\in
D^\prime}\left(\delta^{3/2}|u|^{1/2}\|\alpha\|_{L^4(S_{\ub,u})}\right)
\label{10.47}
\end{equation}
we then conclude that there is a numerical constant $C$ such that:
\begin{equation}
{\cal R}_0^4(\alpha)\leq C{\cal R}_{[1]}(\alpha) \label{10.48}
\end{equation}
Next, we apply Proposition 10.1 to $\xi=|u|\beta$. Noting that
$D\xi=|u|D\beta$, $\snab\xi=|u|\snab\beta$ and defining:
\begin{equation}
{\cal R}_0^4(\beta)=\sup_{(\ub,u)\in
D^\prime}\left(\delta^{1/2}|u|^{3/2}\|\beta\|_{L^4(S_{\ub,u})}\right)
\label{10.49}
\end{equation}
we conclude that there is a numerical constant $C$ such that:
\begin{equation}
{\cal R}_0^4(\beta)\leq C{\cal R}_{[1]}(\beta) \label{10.50}
\end{equation}
Next, we apply Proposition 10.1 to $\xi=|u|^2\rho, |u|^2\sigma$.
Noting that, accordingly, $D\xi=|u|^2 D\rho, |u|^2D\sigma$ and
$\snab\xi=|u|^2\sd\rho, |u|^2\sd\sigma$, and defining:
\begin{eqnarray}
&&{\cal R}_0^4(\rho)=\sup_{(\ub,u)\in D^\prime}\left(|u|^{5/2}\|\rho\|_{L^4(S_{\ub,u})}\right)\nonumber\\
&&{\cal R}_0^4(\sigma)=\sup_{(\ub,u)\in
D^\prime}\left(|u|^{5/2}\|\sigma\|_{L^4(S_{\ub,u})}\right)\label{10.51}
\end{eqnarray}
we conclude that there is a numerical constant $C$ such that:
\begin{eqnarray}
&&{\cal R}_0^4(\rho)\leq C{\cal R}_{[1]}(\rho)\nonumber\\
&&{\cal R}_0^4(\sigma)\leq C{\cal R}_{[1]}(\sigma)\label{10.52}
\end{eqnarray}
Finally, we apply Proposition 10.1 to $\xi=|u|^3\beb$. Noting that
$D\xi=|u|^3D\beb$, $\snab\xi=|u|^3\snab\beb$ and defining:
\begin{equation}
{\cal R}_0^4(\beb)=\sup_{(\ub,u)\in
D^\prime}\left(\delta^{-1}|u|^{7/2}\|\beb\|_{L^4(S_{\ub,u})}\right)
\label{10.53}
\end{equation}
we conclude that there is a numerical constant $C$ such that:
\begin{equation}
{\cal R}_0^4(\beb)\leq C{\cal R}_{[1]}(\beb) \label{10.54}
\end{equation}

We proceed to the 2nd order Sobolev inequalities on the $C_u$. If
$\theta$ is an arbitrary $p$ covariant $S$ tensorfield vanishing
on $\Cb_0$, to bound
$\sup_{\ub}\left(\|\snab\theta\|_{L^4(S_{\ub,u})}\right)$ we apply
Proposition 10.1 to the $p+1$ covariant $S$ tensorfield
$\xi=\snab\theta$ which also vanishes on $\Cb_0$. In the resulting
inequalities the quantity $\|D\snab\theta\|_{L^2(C_u)}$ appears on
the right hand side. By the first part of Lemma 4.1 we have,
pointwise,
\begin{equation}
|D\snab\theta|\leq |\snab D\theta|+p|D\sGamma||\theta|
\label{10.55}
\end{equation}
Taking the $L^2$ norm of \ref{10.55} on $S_{\ub,u}$ we obtain:
\begin{equation}
\|D\snab\theta\|_{L^2(S_{\ub,u})}\leq \|\snab
D\theta\|_{L^2(S_{\ub,u})}+p\|D\sGamma\|_{L^4(S_{\ub,u})}\|\theta\|_{L^4(S_{\ub,u})}
\label{10.56}
\end{equation}
Taking in turn the $L^2$ norm of \ref{10.56} with respect to $\ub$
(on $[0,\delta)$ if $u\in[u_0,c^*-\delta]$, on $[0,c^*-u)$ if
$u\in(c^*-\delta,c^*)$, in the case $c^*\geq u_0+\delta$, on $[0,c^*-u)$, in the case $c^*<u_0+\delta$) we deduce:
\begin{equation}
\|D\snab\theta\|_{L^2(C_u)} \leq \|\snab D\theta\|_{L^2(C_u)}
+p\delta^{1/2}\sup_{\ub}\left(\|D\sGamma\|_{L^4(S_{\ub,u})}\right)\sup_{\ub}\left(\|\theta\|_{L^4(S_{\ub,u})}\right)
\label{10.57}
\end{equation}
To control the factor
$\sup_{\ub}\left(\|D\sGamma\|_{L^4(S_{\ub,u})}\right)$ we
introduce the following bootstrap assumptions:

\vspace{5mm}

\ \ \ {\bf A2.1:} \
$\delta|u|^{1/2}\|\sd(\Omega\mbox{tr}\chi)\|_{L^4(S_{\ub,u})}\leq
1$ \ : \ $\forall (\ub,u)\in D^\prime$

\vspace{2.5mm}

\ \ \ {\bf A2.2:} \
$\delta|u|^{1/2}\|\snab(\Omega\chih)\|_{L^4(S_{\ub,u})}\leq\frac{1}{2}$
\ : \ $\forall (\ub,u)\in D^\prime$

\vspace{5mm}

From the formula for $D\sGamma$ of Lemma 4.1, the assumptions {\bf
A2.1}, {\bf A2.2} imply:
\begin{equation}
\sup_{(\ub,u)\in
D^\prime}\left(\delta|u|^{1/2}\|D\sGamma\|_{L^4(S_{\ub,u})}\right)\leq
3 \label{10.58}
\end{equation}
Substituting then \ref{10.58} in \ref{10.57} yields:
\begin{equation}
\|D\snab\theta\|_{L^2(C_u)}\leq \|\snab D\theta\|_{L^2(C_u)}
+3p\delta^{-1/2}|u|^{-1/2}\sup_{\ub}\left(\|\theta\|_{L^4(S_{\ub,u})}\right)
\label{10.59}
\end{equation}
and the factor
$\sup_{\ub}\left(\|\theta\|_{L^4(S_{\ub,u})}\right)$ is bounded by
the second conclusion of Proposition 10.1 with $\theta$ in the
role of $\xi$.

Also, to bound
$\sup_{\ub}\left(\|D\theta\|_{L^4(S_{\ub,u})}\right)$, assuming
that $D\theta$ vanishes on $\Cb_0$, we simply apply Proposition
10.1 with $D\theta$ in the role of $\xi$.

We now apply the above proceedure to the spacetime curvature
components $\alpha$, $\beta$, $\rho$, $\sigma$, $\beb$. We define:
\begin{eqnarray}
&&\scR_2(\alpha)=\sup_{u\in[u_0,c^*)}\left(\delta|u|^2\|\snab^{ \ 2}\alpha\|_{L^2(C_u)}\right)\nonumber\\
&&\scR_1(\Dh\alpha)=\sup_{u\in[u_0,c^*)}\left(\delta^2|u|\|\snab\Dh\alpha\|_{L^2(C_u)}\right)\nonumber\\
&&{\cal
R}_0(\Dh^2\alpha)=\sup_{u\in[u_0,c^*)}\left(\delta^3\|\Dh^2\alpha\|_{L^2(C_u)}\right)\label{10.60}
\end{eqnarray}
and:
\begin{equation}
{\cal R}_{[2]}(\alpha)=\max\{{\cal
R}_{[1]}(\alpha),\scR_2(\alpha),\scR_1(\Dh\alpha),{\cal
R}_0(\Dh^2\alpha)\} \label{10.61}
\end{equation}
We define:
\begin{eqnarray}
&&\scR_2(\beta)=\sup_{u\in[u_0,c^*)}\left(|u|^3\|\snab^{ \ 2}\beta\|_{L^2(C_u)}\right)\nonumber\\
&&\scR_1(D\beta)=\sup_{u\in[u_0,c^*)}\left(\delta|u|^2\|\snab D\beta\|_{L^2(C_u)}\right)\nonumber\\
&&{\cal
R}_0(D^2\beta)=\sup_{u\in[u_0,c^*)}\left(\delta^2|u|\|D^2\beta\|_{L^2(C_u)}\right)\label{10.62}
\end{eqnarray}
and:
\begin{equation}
{\cal R}_{[2]}(\beta)=\max\{{\cal
R}_{[1]}(\beta),\scR_2(\beta),\scR_1(D\beta),{\cal
R}_0(D^2\beta)\} \label{10.63}
\end{equation}
We define:
\begin{eqnarray}
&&\scR_2(\rho)=\sup_{u\in[u_0,c^*)}\left(\delta^{-1/2}|u|^4\|\snab^{ \ 2}\rho\|_{L^2(C_u)}\right)\nonumber\\
&&\scR_1(D\rho)=\sup_{u\in[u_0,c^*)}\left(\delta^{1/2}|u|^3\|\snab D\rho\|_{L^2(C_u)}\right)\nonumber\\
&&{\cal
R}_0(D^2\rho)=\sup_{u\in[u_0,c^*)}\left(\delta^{3/2}|u|^2\|D^2\rho\|_{L^2(C_u)}\right)\label{10.64}
\end{eqnarray}
and:
\begin{equation}
{\cal R}_{[2]}(\rho)=\max\{{\cal
R}_{[1]}(\rho),\scR_2(\rho),\scR_1(D\rho),{\cal R}_0(D^2\rho)\}
\label{10.65}
\end{equation}
Also:
\begin{eqnarray}
&&\scR_2(\sigma)=\sup_{u\in[u_0,c^*)}\left(\delta^{-1/2}|u|^4\|\snab^{ \ 2}\sigma\|_{L^2(C_u)}\right)\nonumber\\
&&\scR_1(D\sigma)=\sup_{u\in[u_0,c^*)}\left(\delta^{1/2}|u|^3\|\snab D\sigma\|_{L^2(C_u)}\right)\nonumber\\
&&{\cal
R}_0(D^2\sigma)=\sup_{u\in[u_0,c^*)}\left(\delta^{3/2}|u|^2\|D^2\sigma\|_{L^2(C_u)}\right)\label{10.66}
\end{eqnarray}
and:
\begin{equation}
{\cal R}_{[2]}(\sigma)=\max\{{\cal
R}_{[1]}(\sigma),\scR_2(\sigma),\scR_1(D\sigma),{\cal
R}_0(D^2\sigma)\} \label{10.67}
\end{equation}
Finally, we define:
\begin{eqnarray}
&&\scR_2(\beb)=\sup_{u\in[u_0,c^*)}\left(\delta^{-3/2}|u|^5\|\snab^{ \ 2}\beb\|_{L^2(C_u)}\right)\nonumber\\
&&\scR_1(D\beb)=\sup_{u\in[u_0,c^*)}\left(\delta^{-1/2}|u|^4\|\snab D\beb\|_{L^2(C_u)}\right)\nonumber\\
&&{\cal
R}_0(D^2\beb)=\sup_{u\in[u_0,c^*)}\left(\delta^{1/2}|u|^3\|D^2\beb\|_{L^2(C_u)}\right\}\label{10.68}
\end{eqnarray}
and:
\begin{equation}
{\cal R}_{[2]}(\beb)=\max\{{\cal
R}_{[1]}(\beb),\scR_2(\beb),\scR_1(D\beb),{\cal R}_0(D^2\beb)\}
\label{10.69}
\end{equation}
The first of each of the definitions \ref{10.60}, \ref{10.62},
\ref{10.64}, \ref{10.66}, \ref{10.68} has already been introduced
in the previous chapters, the first of \ref{10.60} by \ref{5.6},
the first of \ref{10.62}, \ref{10.64}, \ref{10.66}, \ref{10.68} by
\ref{7.01}. Also, the second and third of each of the definitions
\ref{10.64}, \ref{10.66} has already been introduced by
\ref{7.02}.

We first apply the proceedure outlined above to bound
$\sup_{\ub}\left(\|\snab\theta\|_{L^4(S_{\ub,u})}\right)$ with
$\alpha$, $|u|\beta$, $|u|^2\rho$, $|u|^2\sigma$, $|u|^3\beb$ in
the role of $\theta$. In terms of the first five of the
definitions \ref{4.1}, that is:
\begin{eqnarray}
&&\scR_1^4(\alpha)=\sup_{(\ub,u)\in D^\prime}
\left(\delta^{3/2}|u|^{3/2}\|\snab\alpha\|_{L^4(S_{\ub,u})}\right)\nonumber\\
&&\scR_1^4(\beta)=\sup_{(\ub,u)\in D^\prime}
\left(\delta^{1/2}|u|^{5/2}\|\snab\beta\|_{L^4(S_{\ub,u})}\right)\nonumber\\
&&\scR_1^4(\rho)=\sup_{(\ub,u)\in D^\prime}
\left(|u|^{7/2}\|\sd\rho\|_{L^4(S_{\ub,u})}\right)\nonumber\\
&&\scR_1^4(\sigma)=\sup_{(\ub,u)\in D^\prime}
\left(|u|^{7/2}\|\sd\sigma\|_{L^4(S_{\ub,u})}\right)\nonumber\\
&&\scR_1^4(\beb)=\sup_{(\ub,u)\in D^\prime}
\left(\delta^{-1}|u|^{9/2}\|\snab\beta\|_{L^4(S_{\ub,u})}\right)\label{10.70}
\end{eqnarray}
we deduce that there are numerical constants $C$ such that:
\begin{eqnarray}
&&\scR_1^4(\alpha)\leq C{\cal R}_{[2]}(\alpha)\nonumber\\
&&\scR_1^4(\beta)\leq C{\cal R}_{[2]}(\beta)\nonumber\\
&&\scR_1^4(\rho)\leq C{\cal R}_{[2]}(\rho)\nonumber\\
&&\scR_1^4(\sigma)\leq C{\cal R}_{[2]}(\sigma)\nonumber\\
&&\scR_1^4(\beb)\leq C{\cal R}_{[2]}(\beb)\label{10.71}
\end{eqnarray}
In deducing the first of the inequalities \ref{10.71} we remark
that with $\snab\alpha$ in the role of $\xi$ in Proposition 10.1,
we may replace $D\xi$ by $\Dh\xi$, the trace-free, relative to the
last two entries, part of $D\xi$, and make use of the fact that:
\begin{eqnarray}
[\Dh,\snab]\alpha&=&[D,\snab]\alpha+\frac{1}{2}(\sd(\mbox{tr}D\alpha)-\mbox{tr}D\snab\alpha)\otimes\sg\nonumber\\
&=&[D,\snab]\alpha+(\alpha,\snab(\Omega\chih))\otimes\sg\label{10.72}
\end{eqnarray}
and the last term on the right is bounded in $L^2(C_u)$ by:
$$C\delta^{-1/2}|u|^{-1/2}\sup_{\ub}\left(\|\alpha\|_{L^2(S_{\ub,u})}\right)$$
(compare with \ref{10.59}) using assumption {\bf A2.2}.

We next apply the proceedure to bound
$\sup_{\ub}\left(\|D\theta\|_{L^4(S_{\ub,u})}\right)$ with
$|u|\beta$, $|u|^2\rho$, $|u|^2\sigma$, $|u|^3\beb$ in the role of
$\theta$. The same proceedure applies to bound
$\sup_{\ub}\left(\|\Dh\alpha\|_{L^4(S_{\ub,u})}\right)$, remarking
again that with $\Dh\alpha$ in the role of $\xi$ in Poposition
10.1, we may replace $D\xi$ by $\Dh\xi$. In terms of the
definitions:
\begin{eqnarray}
&&{\cal R}_0^4(\Dh\alpha)=\sup_{(\ub,u)\in D^\prime}\left(\delta^{5/2}|u|^{1/2}\|\Dh\alpha\|_{L^4(S_{\ub,u})}\right)\nonumber\\
&&{\cal R}_0^4(D\beta)=\sup_{(\ub,u)\in D^\prime}\left(\delta^{3/2}|u|^{3/2}\|D\beta\|_{L^4(S_{\ub,u})}\right)\nonumber\\
&&{\cal R}_0^4(D\rho)=\sup_{(\ub,u)\in D^\prime}\left(\delta|u|^{5/2}\|D\rho\|_{L^4(S_{\ub,u})}\right)\nonumber\\
&&{\cal R}_0^4(D\sigma)=\sup_{(\ub,u)\in D^\prime}\left(\delta|u|^{5/2}\|D\sigma\|_{L^4(S_{\ub,u})}\right)\nonumber\\
&&{\cal R}_0^4(D\beta)=\sup_{(\ub,u)\in
D^\prime}\left(|u|^{7/2}\|D\beb\|_{L^4(S_{\ub,u})}\right)\label{10.73}
\end{eqnarray}
(the third and fourth of which have already been introduced by
\ref{4.2}) we deduce that there are numerical constants $C$ such
that:
\begin{eqnarray}
&&{\cal R}_0^4(\Dh\alpha)\leq C{\cal R}_{[2]}(\alpha)\nonumber\\
&&{\cal R}_0^4(D\beta)\leq C{\cal R}_{[2]}(\beta)\nonumber\\
&&{\cal R}_0^4(D\rho)\leq C{\cal R}_{[2]}(\rho)\nonumber\\
&&{\cal R}_0^4(D\sigma)\leq C{\cal R}_{[2]}(\sigma)\nonumber\\
&&{\cal R}_0^4(D\beb)\leq C{\cal R}_{[2]}(\beb)\label{10.74}
\end{eqnarray}

Consider finally the first of the definitions \ref{10.70} in
conjunction with definition \ref{10.47}, the second of the
definitions \ref{10.70} in conjunction with definition
\ref{10.49}, the third and fourth of the definitions \ref{10.70}
in conjunction with ther first and second of the definitions
\ref{10.51} respectively, and the firth of the definitions
\ref{10.70} in conjunction with definition \ref{10.53}. Then in
view of Lemma 10.1, applying Lemma 5.2 taking $p=4$ and recalling
the first five of the definitions \ref{3.2}, that is:
\begin{eqnarray}
&&{\cal R}_0^\infty(\alpha)=\sup_{M^\prime}(\delta^{3/2}|u||\alpha|)\nonumber\\
&&{\cal R}_0^\infty(\beta)=\sup_{M^\prime}(\delta^{1/2}|u|^2|\beta|)\nonumber\\
&&{\cal R}_0^\infty(\rho)=\sup_{M^\prime}(|u|^3|\rho|)\nonumber\\
&&{\cal R}_0^\infty(\sigma)=\sup_{M^\prime}(|u|^3|\sigma)\nonumber\\
&&{\cal
R}_0^\infty(\beb)=\sup_{M^\prime}(\delta^{-1}|u|^{4}|\beb|)\label{10.75}
\end{eqnarray}
we deduce that there are numerical constants $C$ such that:
\begin{eqnarray}
&&{\cal R}_0^\infty(\alpha)\leq C\max\{\scR_1^4(\alpha),{\cal R}_0^4(\alpha)\}\nonumber\\
&&{\cal R}_0^\infty(\beta)\leq C\max\{\scR_1^4(\beta),{\cal R}_0^4(\beta)\}\nonumber\\
&&{\cal R}_0^\infty(\rho)\leq C\max\{\scR_1^4(\rho),{\cal R}_0^4(\rho)\}\nonumber\\
&&{\cal R}_0^\infty(\sigma)\leq C\max\{\scR_1^4(\sigma),{\cal R}_0^4(\sigma)\}\nonumber\\
&&{\cal R}_0^\infty(\beb)\leq C\max\{\scR_1^4(\beb),{\cal
R}_0^4(\beb)\}\label{10.76}
\end{eqnarray}
Therefore, combining with the inequalities \ref{10.71} and
\ref{10.48}, \ref{10.50}, \ref{10.52}, \ref{10.54}, we conclude
that there are numerical constants $C$ such that:
\begin{eqnarray}
&&{\cal R}_0^\infty(\alpha)\leq C{\cal R}_{[2]}(\alpha)\nonumber\\
&&{\cal R}_0^\infty(\beta)\leq C{\cal R}_{[2]}(\beta)\nonumber\\
&&{\cal R}_0^\infty(\rho)\leq C{\cal R}_{[2]}(\rho)\nonumber\\
&&{\cal R}_0^\infty(\sigma)\leq C{\cal R}_{[2]}(\sigma)\nonumber\\
&&{\cal R}_0^\infty(\beb)\leq C{\cal R}_{[2]}(\beb)\label{10.77}
\end{eqnarray}

\section{The Sobolev inequalities on the $\Cb_{\ub}$}

On $\Cb_{\ub}$ we have the measure:
$$d\mu_{\sg}du$$
Thus, for any $S$ tensorfield $\theta$ and any $p<\infty$ we have:
\begin{equation}
\|\theta\|^p_{L^p(\Cb_{\ub})}=\int_{\Cb_{\ub}}|\theta|^p=\int_{u_0}^{c^*-\ub}\left(\int_{S_{\ub,u}}|\theta|^p
d\mu_{\sg}\right)du \label{10.78}
\end{equation}

We introduce the following bootstrap assumptions:

\vspace{5mm}

\ \ \ {\bf A3.1:} \
$\left|\Omega\mbox{tr}\chib+\frac{2}{|u|}\right|\leq\frac{\log
2}{3}\frac{1}{|u|^{3/2}}$ \ : \ in $M^\prime$

\vspace{2.5mm}

\ \ \ {\bf A3.2:} \ $\Omega|\chibh|\leq\frac{\log
2}{6}\frac{1}{|u|^{3/2}}$ \ : \ in $M^\prime$

\vspace{5mm}

\noindent{\bf Proposition 10.2} \ \ \ Let the basic bootstrap
assumptions {\bf A1.1}, {\bf A1.2}, as well as the bootstrap
assumptions {\bf A3.1}, {\bf A3.2} hold. Let $\xi$ be an arbitrary
$p$ covariant $S$ tensorfield such that for some
$\ub\in[0,\delta)$ $|u|^{q-1} \xi$, $|u|^q\Db\xi$, $|u|^q\snab\xi$
belong to $L^2(\Cb_{\ub})$ for some constant
$$q\geq p-\frac{1}{2}$$
Moreover, let  $\xi\in L^4(S_{\ub,u_0})$. Then  $|u|^q\xi\in
L^6(\Cb_u)$ and there is a numerical constant $C_p$, depending
only on $p$ such that:
\begin{eqnarray*}
&&\||u|^q\xi\|_{L^6(\Cb_{\ub})}\leq\\ &&\hspace{5mm}C_p\left\{(1/18)\||u|^{q-1}\xi\|^2_{L^2(\Cb_{\ub})}+\||u|^q\snab\xi\|^2_{L^2(\Cb_{\ub})}\right\}^{1/6}\cdot\\
&&\hspace{7mm}\cdot\left\{\left(|u_0|^q\|\xi\|_{L^4(S_{\ub,u_0})}\right)^{2/3}+\right.\\
&&\hspace{10mm}\left.\left(\||u|\Db\xi\|_{L^2(\Cb_{\ub})}\right)^{1/3}\left\{(1/18)\||u|^{q-1}\xi\|^2_{L^2(\Cb_{\ub})}+\||u|^q\snab\xi\|^2_{L^2(\Cb_{\ub})}\right\}^{1/6}\right\}
\end{eqnarray*}
Also, $\xi\in L^4(S_{\ub,u})$ for each $u$ and there is a
numerical constant $C^\prime_p$ depending only on $p$ such that:
\begin{eqnarray*}
&&\sup_{u}\left(|u|^q\|\xi\|_{L^4(S_{\ub,u})}\right)\\
&&\hspace{1cm}\leq
C^\prime_p\left\{|u_0|^q\|\xi\|_{L^4(S_{\ub,u_0})}
+\||u|^q\Db\xi\|^{1/2}_{L^2(\Cb_{\ub})}\cdot\right.\\
&&\hspace{3cm}\left.\cdot\left\{(1/18)\||u|^{q-1}\xi\|^2_{L^2(\Cb_{\ub})}+\||u|^q\snab\xi\|^2_{L^2(\Cb_{\ub})}\right\}^{1/4}\right\}
\end{eqnarray*}

\noindent{\em Proof:} \ By the assumptions of the proposition
$\xi\in W_1^2(S_{\ub,u})$ for almost all $u\in[u_0,c^*-\ub)$. Thus
Lemma 10.2 applies to $\xi$. We multiply the inequality of Lemma
10.2 by $|u|^{6q}$ and integrate with respect to $u$ to obtain:
\begin{equation}
\int_{\Cb_{\ub}}|u|^{6q}|\xi|^6\leq\frac{9}{\pi}\sup_u\left(\int_{S_{\ub,u}}|u|^{4q}|\xi|^4\right)
\left\{\frac{1}{18}\int_{\Cb_{\ub}}|u|^{2q-2}|\xi|^2+\int_{\Cb_{\ub}}|u|^{2q}|\snab\xi|^2\right\}
\label{10.79}
\end{equation}
Consider now the surface integral ($\ub$ is fixed):
\begin{equation}
\xb(u)=\int_{S_{\ub,u}}|u|^{4q}|\xi|^4 d\mu_{\sg} \label{10.80}
\end{equation}
We have:
\begin{equation}
\frac{d\xb}{du}=\int_{S_{\ub,u}}|u|^{4q}\left\{\Db(|\xi|^4)+\left(\Omega\mbox{tr}\chib-\frac{4q}{|u|}\right)|\xi|^4\right\}d\mu_{\sg}
\label{10.81}
\end{equation}
By the second part of Lemma 4.2:
\begin{eqnarray}
&&\Db(|\xi|^4)+\Omega\mbox{tr}\chib|\xi|^4=4|\xi|^2(\xi,\Db\xi)\label{10.82}\\
&&\hspace{27mm}-(2p-1)\Omega\mbox{tr}\chib|\xi|^4
-4|\xi|^2\sum_{i=1}^p\Omega\chibh^{A_i}_{B_i}\xi^{A_1...\stackrel{B_i}{>A_i<}...A_p}\xi_{A_1...A_p}\nonumber
\end{eqnarray}
It follows that:
\begin{equation}
\Db(|\xi|^4)+\left(\Omega\mbox{tr}\chib-\frac{4q}{|u|}\right)|\xi|^4
\leq 4|\xi|^3|\Db\xi|+\nu|\xi|^4 \label{10.83}
\end{equation}
where:
\begin{equation}
\nu=-(2p-1)\Omega\mbox{tr}\chib-\frac{4q}{|u|}+4p\Omega|\chibh|
\label{10.84}
\end{equation}
Now, by virtue of assumptions {\bf A3.1}, {\bf A3.2} we have:
\begin{eqnarray}
\nu&\leq&\frac{4(p-q)-2}{|u|}+|2p-1|\left|\Omega\mbox{tr}\chib+\frac{2}{|u|}\right|+4p\Omega|\chibh|\nonumber\\
&\leq&\frac{k_p}{|u|^{3/2}} \label{10.85}
\end{eqnarray}
Substituting \ref{10.85} in \ref{10.83} and the result in
\ref{10.81} we obtain the following ordinary differential
inequality for $\xb$:
\begin{equation}
\frac{d\xb}{du}\leq \frac{k_p}{|u|^{3/2}}\xb+\ab \label{10.86}
\end{equation}
where:
\begin{equation}
\ab=4\int_{S_{\ub,u}}|u|^{4q}|\xi|^3|\Db\xi|d\mu_{\sg}
\label{10.87}
\end{equation}
Integrating \ref{10.86} from $u_0$ we obtain:
\begin{eqnarray}
\xb(u)&\leq&\exp\left(\int_{u_0}^u\frac{k_p}{|u^\prime|^{3/2}}du^\prime\right)\xb(u_0)\label{10.88}\\
&\s&+\int_{u_0}^u\exp\left(\int_{u^\prime}^u\frac{k_p}{|u^{\prime\prime}|^{3/2}}du^{\prime\prime}\right)\ab(u^\prime)du^\prime\nonumber
\end{eqnarray}
which, since the integrals in the exponentials are bounded by
$2k_p$, implies:
\begin{equation}
\xb(u)\leq
e^{2k_p}\left(\xb(u_0)+\int_{u_0}^u\ab(u^\prime)du^\prime\right)
\label{10.89}
\end{equation}
Since
\begin{equation}
\int_{u_0}^u\ab(u^\prime)du^\prime\leq
4\int_{\Cb_{\ub}}|u|^{4q}|\xi|^3|\Db\xi| \leq
4\left(\int_{\Cb_{\ub}}|u|^{6q}|\xi|^6\right)^{1/2}\left(\int_{\Cb_{\ub}}|u|^2|\Db\xi|^2\right)^{1/2}
\label{10.90}
\end{equation}
\ref{10.89} implies:
\begin{equation}
\sup_u\xb(u)\leq e^{2k_p}\left(\xb(u_0)+A^{1/2}\yb^{1/2}\right)
\label{10.91}
\end{equation}
where:
\begin{equation}
\yb=\int_{\Cb_{\ub}}|u|^{6q}|\xi|^6 \label{10.92}
\end{equation}
and:
\begin{equation}
A=16\int_{\Cb_{\ub}}|u|^{2q}|\Db\xi|^2 \label{10.93}
\end{equation}
Substituting \ref{10.91} in \ref{10.79} we obtain, with:
\begin{equation}
B=\frac{1}{18}\int_{\Cb_{\ub}}|u|^{2(q-1)}|\xi|^2+\int_{\Cb_{\ub}}|u|^{2q}|\snab\xi|^2
\label{10.94}
\end{equation}
the inequality:
\begin{equation}
\yb\leq m_p\left(\xb(u_0)+A^{1/2}\yb^{1/2}\right)B \ \ \
\mbox{where:} \ m_p=\frac{9e^{2k_p}}{\pi} \label{10.95}
\end{equation}
This implies:
\begin{equation}
\yb\leq m_p B\left(2\xb(u_0)+m_p AB\right) \label{10.96}
\end{equation}
which yields the first conclusion of the proposition. Moreover,
\ref{10.96} implies:
\begin{eqnarray}
A^{1/2}\yb^{1/2}&\leq& (2m_p AB\xb(u_0))^{1/2}+m_p AB\nonumber\\
&\leq&\frac{1}{2}\xb(u_0)+2m_p AB\label{10.97}
\end{eqnarray}
Substituting then \ref{10.97} in \ref{10.91} we obtain:
\begin{equation}
\sup_u\xb(u)\leq e^{2k_p}\left(\frac{3}{2}\xb(u_0)+2m_p AB\right)
\label{10.98}
\end{equation}
which yields the second conclusion of the proposition.

\vspace{5mm}

We now apply the above proposition to the spacetime curvature
component $\alb$. We define:
\begin{eqnarray}
&&\cRb_0(\alb)=\sup_{\ub\in[0,\delta)}\left(\delta^{-3/2}\||u|^3\alb\|_{L^2(\Cb_{\ub})}\right)\nonumber\\
&&\cRb_0(\Dbh\alb)=\sup_{\ub\in[0,\delta)}\left(\delta^{-3/2}\||u|^4\Dbh\alb\|_{L^2(\Cb_{\ub})}\right)\nonumber\\
&&\scRb_1(\alb)=\sup_{\ub\in[0,\delta)}\left(\delta^{-3/2}\||u|^4\snab\alb\|_{L^2(\Cb_{\ub})}\right)\label{10.99}
\end{eqnarray}
and:
\begin{equation}
\cRb_{[1]}(\alb)=\max\{\cRb_0(\alb),\cRb_0(\Dbh\alb),\scRb_1(\alb)\}
\label{10.100}
\end{equation}
We also define:
\begin{equation}
{\cal D}^{\prime
4}_0(\alb)=\sup_{\ub\in[0,\delta)}\left(\delta^{-3/2}|u_0|^4\|\alb\|_{L^4(S_{\ub,u_0})}\right)
\label{10.101}
\end{equation}
Note that here, as in the case of the initial data quantities ${\cal D}_0^\infty$, $\scD_1^4$, $\scD_2^4(\mbox{tr}\chib)$, 
and $\scD_3(\mbox{tr}\chib)$ we are considering the whole of $C_{u_0}$, not only the part lying in 
$M^\prime$. For $c^*\geq u_0+\delta$ this is the same, but not for $c^*<u_0+\delta$. In the latter case the part of 
$C_{u_0}$ lying in $M^\prime$ corresponds to $\ub < c^*-u_0$. By the results of Chapter 2 ${\cal D}^{\prime 4}_0(\alb)$ 
is bounded by a non-negative non-decreasing continuous function of $M_5$. 

We apply Proposition 10.2 to $\xi=\alb$, taking $q=4$. Since in
regard to the first term on the right in \ref{10.80} we have, in
this case,
$$(\xi,\Db\xi)=(\xi,\Dbh\xi)$$
we may replace $\Db\xi$ by $\Dbh\xi$ on the right in \ref{10.83}
and \ref{10.87}, hence also in the conclusions of Proposition
10.2. Defining:
\begin{equation}
{\cal R}_0^4(\alb)=\sup_{(\ub,u)\in
D^\prime}\left(\delta^{-3/2}|u|^4\|\alb\|_{L^4(S_{\ub,u})}\right)
\label{10.102}
\end{equation}
we then conclude that there is a numerical constant $C$ such that:
\begin{equation}
{\cal R}_0^4(\alb)\leq C\max\{{\cal D}^{\prime
4}_0(\alb),\cRb_{[1]}(\alb)\} \label{10.103}
\end{equation}

We proceed to the 2nd order Sobolev inequalities on the
$\Cb_{\ub}$. If $\theta$ is an arbitrary $p$ covariant $S$
tensorfield defined on some $\Cb_{\ub}$, to bound
$\sup_{u}\left(|u|^{q+1}\|\snab\theta\|_{L^4(S_{\ub,u})}\right)$
we apply Proposition 10.2 to the $p+1$ covariant tensorfield
$\xi=\snab\theta$, replacing $q$ by $q+1$. In the resulting
inequalities the quantity
$\||u|^{q+1}\Db\snab\theta\|_{L^2(\Cb_{\ub})}$ appears on the
right hand side. By the second part of Lemma 4.1 we have,
pointwise,
\begin{equation}
|\Db\snab\theta|\leq|\snab\Db\theta|+p|\Db\sGamma||\theta|
\label{10.104}
\end{equation}
Taking the $L^2$ norm of \ref{10.104} on $S_{\ub,u}$ we obtain:
\begin{equation}
\|\Db\snab\theta\|_{L^2(S_{\ub,u})}\leq\|\snab\Db\theta\|_{L^2(S_{\ub,u})}
+p\|\Db\sGamma\|_{L^4(S_{\ub,u})}\|\theta\|_{L^4(S_{\ub,u})}
\label{10.105}
\end{equation}
Multiplying by $|u|^{q+1}$ and taking the $L^2$ norm with respect
to $|u|$ on $[u_0,c^*-\ub)$ we then deduce:
\begin{eqnarray}
&&\||u|^{q+1}\Db\snab\theta\|_{L^2(\Cb_{\ub})}\leq\||u|^{q+1}\snab\Db\theta\|_{L^2(\Cb_{\ub})}\label{10.106}\\
&&+p\left(\int_{u_0}^{c^*-\ub}|u|^2\|\Db\sGamma\|^2_{L^4(S_{\ub,u})}du\right)^{1/2}\sup_u\left(|u|^q\|\theta\|_{L^4(S_{\ub,u})}\right)\nonumber
\end{eqnarray}
To control the first factor in the second term on the right we
introduce the following bootstrap assumptions:

\vspace{5mm}

\ \ \ {\bf A4.1:} \
$|u|^2\|\sd(\Omega\mbox{tr}\chib)\|_{L^4(S_{\ub,u})}\leq 1$ \ : \
$\forall (\ub,u)\in D^\prime$

\vspace{2.5mm}

\ \ \ {\bf A4.2:} \
$|u|^2\|\snab(\Omega\chibh)\|_{L^4(S_{\ub,u})}\leq\frac{1}{2}$ \ :
\ $\forall (\ub,u)\in D^\prime$

\vspace{5mm}

From the formula for $\Db\sGamma$ of Lemma 4.1, the assumptions
{\bf A4.2}, {\bf A4.2} imply:
\begin{equation}
\left(\int_{u_0}^{c^*-\ub}|u|^2\|\Db\sGamma\|^2_{L^4(S_{\ub,u})}du\right)^{1/2}\leq
3 \label{10.107}
\end{equation}
Substituting then \ref{10.107} in \ref{10.106} yields:
\begin{equation}
\||u|^{q+1}\Db\snab\theta\|_{L^2(\Cb_{\ub})}\leq\||u|^{q+1}\snab\Db\theta\|_{L^2(\Cb_{\ub})}
+3p\sup_u\left(|u|^q\|\theta\|_{L^4(S_{\ub,u})}\right)
\label{10.108}
\end{equation}
and the factor
$\sup_u\left(|u|^q\|\theta\|_{L^4(S_{\ub,u})}\right)$ is bounded
by the second conclusion of Proposition 10.2 with $\theta$ in the
role of $\xi$.

Also, to bound
$\sup_u\left(|u|^{q+1}\|\Db\theta\|_{L^4(S_{\ub,u})}\right)$ we
simply apply Proposition 10.2 with $D\theta$ in the role of $\xi$,
replacing $q$ by $q+1$.

We now apply the above procedure to the spacetime curvature
component $\alb$. We define:
\begin{eqnarray}
&&\scRb_2(\alb)=\sup_{\ub\in[0,\delta)}\left(\delta^{-3/2}\||u|^5\snab^{ \ 2}\alb\|_{L^2(\Cb_{\ub})}\right)\nonumber\\
&&\scRb_1(\Dbh\alb)=\sup_{\ub\in[0,\delta)}\left(\delta^{-3/2}\||u|^5\snab\Dbh\alb\|_{L^2(\Cb_{\ub})}\right)\nonumber\\
&&\cRb_0(\Dbh^2\alb)=\sup_{\ub\in[0,\delta)}\left(\delta^{-3/2}\||u|^5\Dbh^2\alb\|_{L^2(\Cb_{\ub})}\right)\label{10.109}
\end{eqnarray}
and:
\begin{equation}
\cRb_{[2]}(\alb)=\max\{\cRb_{[1]}(\alb),\scRb_2(\alb),\scRb_1(\Dbh\alb),\cRb_0(\Dbh^2\alb)\}
\label{10.110}
\end{equation}
We also define:
\begin{eqnarray}
&&\scD^{ \ \prime
4}_1(\alb)=\sup_{\ub\in[0,\delta)}\left(\delta^{-3/2}|u_0|^5\|\snab\alb\|_{L^4(S_{\ub,u_0})}\right)
\nonumber\\
&&{\cal D}^{\prime
4}_0(\Dbh\alb)=\sup_{\ub\in[0,\delta)}\left(\delta^{-3/2}|u_0|^5\|\Dbh\alb\|_{L^4(S_{\ub,u_0})}\right)
\label{10.111}
\end{eqnarray}
and:
\begin{equation}
{\cal D}^{\prime 4}_{[1]}(\alb)=\max\{{\cal D}^{\prime
4}_0(\alb),\scD^{ \ \prime 4}_1(\alb),{\cal D}^{\prime
4}_0(\Dbh\alb)\} \label{10.112}
\end{equation}
Note that here we are again considering the whole of $C_{u_0}$, not only the part lying in 
$M^\prime$. For $c^*\geq u_0+\delta$ this is the same, but not for $c^*<u_0+\delta$. In the latter case the part of 
$C_{u_0}$ lying in $M^\prime$ corresponds to $\ub < c^*-u_0$. By the results of Chapter 2 ${\cal D}^{\prime 4}_{[1]}(\alb)$ 
is bounded by a non-negative non-decreasing continuous function of $M_7$.

We first apply the proceedure outlined above to bound
$\sup_u\left(|u|^{q+1}\|\snab\theta\|_{L^4(S_{\ub,u})}\right)$
with $\alb$ in the role of $\theta$, taking $q=4$. In terms of the
fifth of the definitions \ref{4.1}, that is:
\begin{equation}
\scR_1^4(\alb)=\sup_{(\ub,u)\in
D^\prime}\left(\delta^{-3/2}|u|^{-5}\|\snab\alb\|_{L^4(S_{\ub,u})}\right)
\label{10.113}
\end{equation}
we deduce that there is a numerical constant $C$ such that:
\begin{equation}
\scR_1^4(\alb)\leq C\max\{{\cal D}^{\prime
4}_{[1]}(\alb),\cRb_{[2]}(\alb)\} \label{10.114}
\end{equation}
where In deducing \ref{10.114} we remark that with $\snab\alb$ in
the role of $\xi$ in Proposition 10.2, we may replace $\Db\xi$ by
$\Dbh\xi$, the trace-free, relative to the last two entries, part
of $\Db\xi$, and make use of the fact that:
\begin{eqnarray}
[\Dbh,\snab]\alb&=&[\Db,\snab]\alb+\frac{1}{2}(\sd(\mbox{tr}\Db\alb)-\mbox{tr}\Db\snab\alb)\otimes\sg\nonumber\\
&=&[\Db,\snab]\alb+(\alb,\snab(\Omega\chibh))\otimes\sg
\label{10.115}
\end{eqnarray}
and the last term on the right, multiplied by $|u|^5$, is bounded
in $L^2(\Cb_{\ub})$ by:
$$C\sup_u\left(|u|^4\|\alb\|_{L^4(S_{\ub,u})}\right)$$
(compare with \ref{10.108} with $q=4$) using assumption {\bf
A4.2}.

We next apply the proceedure outlined above to obtain an estimate
for $\sup_u\left(|u|^{5}\|\Dbh\alb\|\right)$ remarking again that
with $\Dbh\alb$ in the role of $\xi$ in Proposition 10.2, we may
replace $\Db\xi$ by $\Dbh\xi$. In terms of the definition
\ref{7.04}, that is:
\begin{equation}
{\cal R}_0^4(\Dbh\alb)=\sup_{(\ub,u)\in
D^\prime}\left(\delta^{-3/2}|u|^5\|\Dbh\alb\|\right)
\label{10.116}
\end{equation}
we deduce that there is a numerical constant $C$ such that:
\begin{equation}
{\cal R}_0^4(\Dbh\alb)\leq C\max\{{\cal D}^{\prime
4}_{[1]}(\alb),\cRb_{[2]}(\alb)\} \label{10.117}
\end{equation}

Consider finally definition \ref{10.113} in conjunction with
definition \ref{10.102}. Then in view of Lemma 10.1, applying
Lemma 5.2 taking $p=4$ and recalling the fifth of the definitions
\ref{3.2}, that is:
\begin{equation}
{\cal
R}_0^\infty(\alb)=\sup_{M^\prime}(\delta^{-3/2}|u|^{9/2}|\alb|)
\label{10.118}
\end{equation}
we deduce that there is a numerical constant $C$ such that:
\begin{equation}
{\cal R}_0^\infty(\alb)\leq C\max\{\scR_1^4(\alb),{\cal
R}_0^4(\alb)\} \label{10.119}
\end{equation}
Therefore, combining with the inequalities \ref{10.103} and
\ref{10.114}, we conclude that there is a numerical constant $C$
such that:
\begin{equation}
{\cal R}_0^\infty(\alb)\leq C\max\{{\cal D}^{\prime
4}_{[1]}(\alb),\cRb_{[2]}(\alb)\} \label{10.120}
\end{equation}

\chapter{The $S$-tangential Derivatives and the Rotational Lie
Derivatives}

\section{Introduction and preliminaries}

In the present chapter we shall establish certain coercivity
inequalities for the Lie derivatives of $S$ tensorfields with
respect to the rotation fields $O_i \ :i=1,2,3$. These
inequalities show that for any covariant $S$ tensorfield $\theta$
the sum $\sum_i|\sL_{O_i}\theta|^2$ bounds pointwise
$|\snab\theta|^2$.

We define the rescaled induced metric $\tilde{\sg}$ on the
$S_{\ub,u}$ by:
\begin{equation}
\tilde{\sg}=|u|^{-2}\sg \label{11.1}
\end{equation}
Since $S_{0,u}$ is a round sphere of radius $|u|$ in Euclidean
3-dimensional space,
$(S_{0,u},\left.\tilde{\sg}\right|_{S_{\ub,u}})$ is isometric to
the unit sphere in Euclidean 3-dimensional space. Now, for each
$\ub\in[0,\delta]$, $\Phi_{\ub}$ defines a diffeomorphism of
$S_{0,u_0}$ onto $S_{\ub,u_0}$ and for each $(\ub,u)\in D^\prime$,
$\Phib_{u-u_0}$ defines a diffeomorphism of $S_{\ub,u_0}$ onto
$S_{\ub,u}$. Thus, for each $(\ub,u)\in D^\prime$,
$\Phib_{u-u_0}\circ\Phi_{\ub}$ defines a diffeomorphism of
$S_{0,u_0}$ onto $S_{\ub,u}$. We consider the pullback:
\begin{equation}
\tilde{\sg}(\ub,u)=(\Phib_{u-u_0}\circ\Phi_{\ub})^*\left.\tilde{\sg}\right|_{S_{\ub,u}}
=\Phi_{\ub}^*\Phib_{u-u_0}^*\left.\tilde{\sg}\right|_{S_{\ub,u}}
\label{11.2}
\end{equation}
a metric on $S_{0,u_0}$. We have:
$$\tilde{\sg}(0,u_0)=\left.\tilde{\sg}\right|_{S_{0,u_0}}$$
and $(S_{\ub,u_0},\left.\tilde{\sg}\right|_{S_{0,u_0}})$ may be
identified with $(S^2,\up{\sg})$, the unit sphere in Euclidean
$\Re^3$. As a preliminary step we establish bounds on the
eigenvalues of $\tilde{\sg}(\ub,u)$ with respect to $\up{\sg}$
under the bootstrap assumptions {\bf A1.1}, {\bf A1.2}, {\bf
A3.1}, {\bf A3.2} of Chapter 10.

Let $\theta$ be an arbitrary $p$ covariant $S$ tensorfield on
$M^\prime$ and let $\theta(\ub,u)$ be the pullback:
\begin{equation}
\theta(\ub,u)=(\Phib_{u-u_0}\circ\Phi_{\ub})^*\left.\theta\right|_{S_{\ub,u}}
=\Phi_{\ub}^*\Phib_{u-u_0}^*\left.\theta\right|_{S_{\ub,u}}
\label{11.3}
\end{equation}
a $p$ covariant tensorfield on $S_{0,u_0}$. Then according to the
discussion in the paragraph which preceeds Lemma 4.3 we have:
\begin{equation}
\frac{\partial\theta}{\partial\ub}(\ub,u_0)=\Phi_{\ub}^*\left.D\theta\right|_{S_{\ub,u_0}}
\label{11.4}
\end{equation}
and:
\begin{equation}
\frac{\partial\theta}{\partial
u}(\ub,u)=\Phi_{\ub}^*\Phib_{u-u_0}^*\left.\Db\theta\right|_{S_{\ub,u}}
\label{11.5}
\end{equation}
From equations \ref{1.28} we then obtain, in the case of the
rescaled induced metric $\tilde{\sg}$:
\begin{equation}
\frac{\partial\tilde{\sg}}{\partial\ub}(\ub,u_0)=(2|u|^{-2}\Omega\chi)(\ub,u_0)
\label{11.6}
\end{equation}
and:
\begin{equation}
\frac{\partial\tilde{\sg}}{\partial
u}(\ub,u)=(2|u|^{-2}\Omega\chib+2|u|^{-1}\tilde{\sg})(\ub,u)
\label{11.7}
\end{equation}

\vspace{5mm}

\noindent{\bf Lemma 11.1} \ \ \ Let $\lambda(\ub,u)$ and
$\Lambda(\ub,u)$ be respectively the smallest and largest
eigenvalues of $\tilde{\sg}(\ub,u)$ relative to
$\tilde{g}(0,u_0)=\up{\sg}$. Then, under the bootstrap assumptions
{\bf A1.1}, {\bf A1.2}, {\bf A3.1}, {\bf A3.2} of Chapter 10 we
have:
$$\frac{1}{4}\leq\lambda(\ub,u)\leq\Lambda(\ub,u)\leq 4$$
for all $(\ub,u)\in D^\prime$.

\noindent{\em Proof:} \ The proof is along the lines of that of
Lemma 5.3. We now define:
\begin{equation}
\mu(\ub,u)=\frac{d\mu_{\tilde{\sg}(\ub,u)}}{d\mu_{\tilde{\sg}(0,u_0)}}=\sqrt{\lambda(\ub,u)\Lambda(\ub,u)}
\label{11.8}
\end{equation}
and:
\begin{equation}
\nu(\ub,u)=\frac{1}{\mu(\ub,u)}\sup_{|X|_{\tilde{\sg}(0,u_0)}=1}\tilde{\sg}(\ub,u)(X,X)=\frac{\Lambda(\ub,u)}{\mu(\ub,u)}
=\sqrt{\frac{\Lambda(\ub,u)}{\lambda(\ub,u)}} \label{11.9}
\end{equation}
By \ref{11.6} and \ref{11.7} we have:
\begin{equation}
\frac{\partial\mu}{\partial\ub}(\ub,u_0)=\frac{1}{2}\mbox{tr}_{\tilde{\sg}(\ub,u_0)}\left(\frac{\partial\tilde{\sg}}{\partial\ub}(\ub,u_0)\right)\mu(\ub,u_0)=(\Omega\mbox{tr}\chi)(\ub,u_0)\mu(\ub,u_0)
\label{11.10}
\end{equation}
and:
\begin{equation}
\frac{\partial\mu}{\partial
u}(\ub,u)=\frac{1}{2}\mbox{tr}{\tilde{\sg}(\ub,u)}\left(\frac{\partial\tilde{\sg}}{\partial
u}(\ub,u)\right)
\mu(\ub,u)=(\Omega\mbox{tr}\chib+2|u|^{-1})(\ub,u)\mu(\ub,u)
\label{11.11}
\end{equation}
Integrating \ref{11.10} with respect to $\ub$ and noting that
$\mu(0,u_0)=1$ we obtain:
\begin{equation}
\mu(\ub,u_0)=\exp\left(\int_0^{\ub}(\Omega\mbox{tr}\chi)(\ub^\prime,u_0)d\ub^\prime\right)
\label{11.12}
\end{equation}
The bootstrap assumption {\bf A1.1} then implies:
\begin{equation}
2^{-1/3}\leq\mu(\ub,u_0)\leq 2^{1/3} \label{11.13}
\end{equation}
Integrating \ref{11.11} with respect to $u$ we obtain:
\begin{equation}
\mu(\ub,u)=\exp\left(\int_{u_0}^u(\Omega\mbox{tr}\chib+2|u|^{-1})(\ub,u^\prime)du^\prime\right)\mu(\ub,u_0)
\label{11.14}
\end{equation}
The bootstrap assumption {\bf A3.1} then implies:
\begin{equation}
2^{-2/3}\leq\frac{\mu(\ub,u)}{\mu(\ub,u_0)}\leq 2^{2/3}
\label{11.15}
\end{equation}
Combining \ref{11.13} and \ref{11.15} we conclude that:
\begin{equation}
\frac{1}{2}\leq\mu(\ub,u)\leq 2 \label{11.16}
\end{equation}
To estimate $\nu(\ub,u)$ we set:
\begin{equation}
\hat{\sg}(\ub,u)=\frac{\tilde{\sg}(\ub,u)}{\mu(\ub,u)}
\label{11.17}
\end{equation}
Then according to the definition \ref{11.9}:
\begin{equation}
\sup_{|X|_{\tilde{\sg}(0,u_0)}=1}\hat{\sg}(X,X)=\nu(\ub,u)
\label{11.18}
\end{equation}
From \ref{11.6} and \ref{11.10} we deduce:
\begin{equation}
\frac{\partial\hat{\sg}}{\partial\ub}(\ub,u_0)=(2|u|^{-2}\mu^{-1}\Omega\chih)(\ub,u_0)
\label{11.19}
\end{equation}
while from \ref{11.7} and \ref{11.11} we deduce:
\begin{equation}
\frac{\partial\hat{\sg}}{\partial
u}(\ub,u)=(2|u|^{-2}\mu^{-1}\Omega\chibh)(\ub,u) \label{11.20}
\end{equation}
Consider any tangent vector $X$ to $S_{0,u_0}$ such that
$|X|_{\tilde{\sg}(0,u_0)}=1$. Then:
\begin{equation}
\frac{\partial\hat{\sg}}{\partial\ub}(\ub,u_0)(X,X)=\frac{2}{|u_0|^2\mu(\ub,u_0)}(\Omega\chih)(\ub,u_0)(X,X)
\label{11.21}
\end{equation}
and:
\begin{equation}
\frac{\partial\hat{\sg}}{\partial
u}(\ub,u)(X,X)=\frac{2}{|u|^2\mu(\ub,u)}(\Omega\chibh)(\ub,u)(X,X)
\label{11.22}
\end{equation}
Integrating \ref{11.21} with respect to $\ub$ and noting that
$\hat{\sg}(0,u_0)(X,X)=1$ yields:
\begin{equation}
\hat{\sg}(\ub,u_0)(X,X)\leq
1+2\int_0^{\ub}\frac{|(\Omega\chih)(\ub^\prime,u_0)(X,X)|}{|u_0|^2\mu(\ub^\prime,u_0)}d\ub^\prime
\label{11.23}
\end{equation}
Also, integrating \ref{11.22} with respect to $u$ yields:
\begin{equation}
\hat{\sg}(\ub,u)(X,X)\leq\hat{\sg}(\ub,u_0)(X,X)+2\int_{u_0}^u\frac{|(\Omega\chibh)(\ub,u^\prime)(X,X)|}{|u^\prime|^2\mu(\ub,u^\prime)}du^\prime
\label{11.24}
\end{equation}
Now if $\xi$ is an arbitrary type $T^q_p$ tensorfield on
$S_{0,u_0}$ then, in components with respect to an arbitrary local
frame field on $S_{0,u_0}$,
\begin{eqnarray}
&&|\xi|^2_{\tilde{\sg}(\ub,u)}=(\tilde{\sg}(\ub,u))_{A_1
B_1}...(\tilde{\sg}(\ub,u))_{A_q B_q}
(\tilde{\sg}^{-1}(\ub,u))^{C_1 D_1}...(\tilde{\sg}^{-1}(\ub,u))^{C_p D_p}\nonumber\\
&&\hspace{5cm}\xi^{A_1...A_q}_{C_1..C_p}\xi^{B_1...B_q}_{D_1...D_p}
\label{11.25}
\end{eqnarray}
It follows that:
\begin{equation}
(\lambda(\ub,u))^q(\Lambda(\ub,u))^{-p}|\xi|^2_{\tilde{\sg}(0,u_0)}\leq
|\xi|^2_{\tilde{\sg}(\ub,u)} \leq
(\Lambda(\ub,u))^q(\lambda(\ub,u))^{-p}|\xi|^2_{\tilde{\sg}(0,u_0)}
\label{11.26}
\end{equation}
In particular, taking $\xi=(\Omega\chih)(\ub,u_0)$ ($q=0$, $p=2$),
we have:
\begin{equation}
|(\Omega\chih)(\ub,u_0)|_{\tilde{\sg}(\ub,u_0)}\geq(\Lambda(\ub,u_0))^{-1}|(\Omega\chih)(\ub,u_0)|_{\tilde{\sg}(0,u_0)}
\label{11.27}
\end{equation}
hence:
\begin{eqnarray}
&&|(\Omega\chih)(\ub,u_0)(X,X)|\leq|(\Omega\chih)(\ub,u_0)|_{\tilde{\sg}(0,u_0)}|X|^2_{\tilde{\sg}(0,u_0)}\label{11.28}\\
&&\hspace{3cm}=|(\Omega\chih)(\ub,u_0)|_{\tilde{\sg}(0,u_0)}\leq\Lambda(\ub,u_0)|(\Omega\chih)(\ub,u_0)|_{\tilde{\sg}(\ub,u_0)}\nonumber
\end{eqnarray}
Also, taking $\xi=(\Omega\chibh)(\ub,u)$ ($q=0$, $p=2$), we have:
\begin{equation}
|(\Omega\chibh)(\ub,u)|_{\tilde{\sg}(\ub,u)}\geq(\Lambda(\ub,u))^{-1}|(\Omega\chibh)(\ub,u)|_{\tilde{\sg}(0,u_0)}
\label{11.29}
\end{equation}
hence:
\begin{eqnarray}
&&|(\Omega\chibh)(\ub,u)(X,X)|\leq|(\Omega\chibh)(\ub,u)\|_{\tilde{\sg}(0,u_0)}|X|^2_{\tilde{\sg}(0,u_0)}\label{11.30}\\
&&\hspace{3cm}=|(\Omega\chibh)(\ub,u)|_{\tilde{\sg}(0,u_0)}\leq\Lambda(\ub,u)|(\Omega\chibh)(\ub,u)|_{\tilde{\sg}(\ub,u)}\nonumber
\end{eqnarray}
Moreover, for an arbitrary type $T^q_p$ tensorfield $\xi$ on
$S_{0,u_0}$ we have:
\begin{equation}
|\xi|_{\tilde{\sg}(\ub,u)}=|u|^{p-q}|\xi|_{\sg(\ub,u)}
\label{11.31}
\end{equation}
Thus \ref{11.28} and \ref{11.30} imply
\begin{equation}
|(\Omega\chih)(\ub,u_0)(X,X)|\leq\Lambda(\ub,u_0)|u_0|^2|(\Omega\chih)(\ub,u_0)|_{\sg(\ub,u_0)}
\label{11.32}
\end{equation}
and
\begin{equation}
|(\Omega\chibh)(\ub,u)(X,X)|\leq\Lambda(\ub,u)|u|^2|(\Omega\chibh)(\ub,u)|_{\sg(\ub,u)}
\label{11.33}
\end{equation}
Substituting \ref{11.32} in \ref{11.23}, taking the supremum over
$X\in T_q S_{0,u_0}$ such that $|X|_{\tilde{\sg}(0,u_0)}=1$ at
each $q\in S_{0,u_0}$, and recalling \ref{11.9} and \ref{11.18} we
obtain the following linear integral inequality for
$\nu(\ub,u_0)$:
\begin{equation}
\nu(\ub,u_0)\leq
1+2\int_0^{\ub}|(\Omega\chih)(\ub^\prime,u_0)|_{\sg(\ub^\prime,u_0)}\nu(\ub^\prime,u_0)d\ub^\prime
\label{11.34}
\end{equation}
This implies:
\begin{equation}
\nu(\ub,u_0)\leq
\exp\left(2\int_0^{\ub}|(\Omega\chih)(\ub^\prime,u_0)|_{\sg(\ub^\prime,u_0)}d\ub^\prime\right)
\label{11.35}
\end{equation}
Also, substituting \ref{11.33} in \ref{11.24}, taking the supremum
over $X\in T_q S_{0,u_0}$ such that $|X|_{\tilde{\sg}(0,u_0)}=1$
at each $q\in S_{0,u_0}$, and recalling \ref{11.9} and \ref{11.18}
we obtain the following linear integral inequality for
$\nu(\ub,u)$:
\begin{equation}
\nu(\ub,u)\leq\nu(\ub,u_0)+2\int_{u_0}^u|(\Omega\chibh)(\ub,u^\prime)|_{\sg(\ub,u^\prime)}du^\prime
\label{11.36}
\end{equation}
This implies:
\begin{equation}
\nu(\ub,u)\leq\nu(\ub,u_0)\exp\left(2\int_{u_0}^u|(\Omega\chibh)(\ub,u^\prime)|_{\sg(\ub,u^\prime)}du^\prime\right)
\label{11.37}
\end{equation}
Now, we have:
\begin{equation}
|(\Omega\chih)(\ub,u)|_{\sg(\ub,u)}=(|\Omega\chih|_{\sg})(\ub,u),
\ \ \
|(\Omega\chibh)(\ub,u)|_{\sg(\ub,u)}=(|\Omega\chibh|_{\sg})(\ub,u)
\label{11.38}
\end{equation}
Thus assumption {\bf A1.2} yields through \ref{11.35}:
\begin{equation}
\nu(\ub,u_0)\leq 2^{1/3} \label{11.39}
\end{equation}
while assumption {\bf A3.2} yields through \ref{11.37}:
\begin{equation}
\nu(\ub,u)\leq 2^{2/3}\nu(\ub,u_0) \label{11.40}
\end{equation}
Combining, we conclude that:
\begin{equation}
\nu(\ub,u)\leq 2 \label{11.41}
\end{equation}
Since by the definitions \ref{11.8} and \ref{11.9}:
\begin{equation}
\lambda(\ub,u)=\frac{\mu(\ub,u)}{\nu(\ub,u)}, \ \ \
\Lambda(\ub,u)=\mu(\ub,u)\nu(\ub,u) \label{11.42}
\end{equation}
\ref{11.16} together with \ref{11.41} imply the lemma.

\vspace{5mm}

Let $\sGamma(\ub,u)$ be the connection of $\sg(\ub,u)$ and
$\sGamma(0,u_0)$ the connection of $\sg(0,u_0)$ (both are
connections on $TS_{0,u_0}$). Then $\sGamma(\ub,u)-\sGamma(0,u_0)$
is a $T^1_2$-type $S$ tensorfield symmetric in the lower indices.
By \ref{11.4} and \ref{11.5} we have:
\begin{equation}
\frac{\partial\sGamma}{\partial\ub}(\ub,u_0)=\Phi^*_{\ub}\left.D\sGamma\right|_{S_{\ub,u_0}}=(D\sGamma)(\ub,u_0)
\label{11.43}
\end{equation}
and:
\begin{equation}
\frac{\partial\sGamma}{\partial
u}(\ub,u_0)=\Phi^*_{\ub}\Phi^*_{u-u_0}\left.\Db\sGamma\right|_{S_{\ub,u}}
=(\Db\sGamma)(\ub,u) \label{11.44}
\end{equation}
Integrating \ref{11.43} with respect to $\ub$ we obtain:
\begin{equation}
\sGamma(\ub,u_0)-\sGamma(0,u_0)=\int_0^{\ub}(D\sGamma)(\ub^\prime,u_0)d\ub^\prime
\label{11.45}
\end{equation}
Also, integrating \ref{11.44} with respect to $u$ we obtain:
\begin{equation}
\sGamma(\ub,u)-\sGamma(\ub,u_0)=\int_{u_0}^u(\Db\sGamma)(\ub,u^\prime)du^\prime
\label{11.46}
\end{equation}
Let $\tilde{\sGamma}(\ub,u)$ be the connection of
$\tilde{\sg}(\ub,u)$. Since the metrics $\tilde{\sg}(\ub,u)$ and
$\sg(\ub,u)$ differ by a constant scale factor their connections
coincide:
\begin{equation}
\tilde{\sGamma}(\ub,u)=\sGamma(\ub,u) \label{11.47}
\end{equation}
Thus, the left hand sides of \ref{11.45} and \ref{11.46} may be
replaced by $\tilde{\sGamma}(\ub,u_0)-\tilde{\sGamma}(0,u_0)$ and
$\tilde{\sGamma}(\ub,u)-\tilde{\sGamma}(\ub,u_0)$ respectively.
These equations then imply
\begin{equation}
|\tilde{\sGamma}(\ub,u_0)-\tilde{\sGamma}(0,u_0)|_{\tilde{\sg}(0,u_0)}\leq
\int_0^{\ub}|(D\sGamma)(\ub^\prime,u_0)|_{\tilde{\sg}(0,u_0)}d\ub^\prime
\label{11.48}
\end{equation}
and
\begin{equation}
|\tilde{\sGamma}(\ub,u)-\tilde{\sGamma}(\ub,u_0)|_{\tilde{\sg}(0,u_0)}
\leq\int_{u_0}^u|(\Db\sGamma)(\ub,u^\prime)|_{\tilde{\sg}(0,u_0)}du^\prime
\label{11.49}
\end{equation}
respectively. Now with $(D\sGamma)(\ub,u)$ and
$(\Db\sGamma)(\ub,u)$ in the role of $\xi$ ($q=1$, $p=2$),
\ref{11.26} and \ref{11.31} give:
\begin{eqnarray}
&&|(D\sGamma)(\ub,u)|_{\tilde{\sg}(0,u_0}\leq\frac{\Lambda(\ub,u)}{\sqrt{\lambda(\ub,u)}}
|(D\sGamma)(\ub,u)|_{\tilde{\sg}(\ub,u)}\label{11.50}\\
&&\hspace{2cm}=|u|\frac{\Lambda(\ub,u)}{\sqrt{\lambda(\ub,u)}}|(D\Gamma)(\ub,u)|_{\sg(\ub,u)}
=|u|\frac{\Lambda(\ub,u)}{\sqrt{\lambda(\ub,u)}}(|D\sGamma|_{\sg})(\ub,u)\nonumber
\end{eqnarray}
and:
\begin{eqnarray}
&&|(\Db\sGamma)(\ub,u)|_{\tilde{\sg}(0,u_0}\leq\frac{\Lambda(\ub,u)}{\sqrt{\lambda(\ub,u)}}
|(\Db\sGamma)(\ub,u)|_{\tilde{\sg}(\ub,u)}\label{11.51}\\
&&\hspace{2cm}=|u|\frac{\Lambda(\ub,u)}{\sqrt{\lambda(\ub,u)}}|(\Db\Gamma)(\ub,u)|_{\sg(\ub,u)}
=|u|\frac{\Lambda(\ub,u)}{\sqrt{\lambda(\ub,u)}}(|\Db\sGamma|_{\sg})(\ub,u)\nonumber
\end{eqnarray}
Hence, by Lemma 11.1:
\begin{equation}
|(D\sGamma)(\ub,u)|_{\tilde{\sg}(0,u_0)}\leq
8|u|(|D\sGamma|_{\sg})(\ub,u) \label{11.52}
\end{equation}
and:
\begin{equation}
|(\Db\sGamma)(\ub,u)|_{\tilde{\sg}(0,u_0)}\leq
8|u|(|\Db\sGamma|_{\sg})(\ub,u) \label{11.53}
\end{equation}
In view of \ref{11.52}, \ref{11.53}, the inequalities \ref{11.48}
and \ref{11.49} imply:
\begin{equation}
|\tilde{\sGamma}(\ub,u_0)-\tilde{\sGamma}(0,u_0)|_{\tilde{\sg}(0,u_0)}\leq
8|u_0|\int_0^{\ub}(|D\sGamma|_{\sg})(\ub^\prime,u)d\ub^\prime
\label{11.54}
\end{equation}
and:
\begin{equation}
|\tilde{\sGamma}(\ub,u)-\tilde{\sGamma}(\ub,u_0)|_{\tilde{\sg}(0,u_0)}\leq
8\int_{u_0}^u|u^\prime|(|\Db\sGamma|_{\sg})(\ub,u^\prime)du^\prime
\label{11.55}
\end{equation}
respectively.

Now, the estimate \ref{8.90} yields, through \ref{11.54}:
\begin{equation}
|\tilde{\sGamma}(\ub,u_0)-\tilde{\sGamma}(0,u_0)|_{\tilde{\sg}(0,u_0)}\leq
O(\delta^{1/2}|u_0|^{-1}) \label{11.56}
\end{equation}
while the estimate \ref{8.96} yields, through \ref{11.55}:
\begin{equation}
|\tilde{\sGamma}(\ub,u)-\tilde{\sGamma}(\ub,u_0)|_{\tilde{\sg}(0,u_0)}\leq
O(\delta^{1/2}|u|^{-1}) \label{11.57}
\end{equation}
Combining then yields the conclusion:
\begin{equation}
|\tilde{\sGamma}(\ub,u)-\tilde{\sGamma}(0,u_0)|_{\tilde{\sg}(0,u_0)}\leq
O(\delta^{1/2}|u|^{-1}) \label{11.58}
\end{equation}
Thus if $\delta$ is suitably small depending on ${\cal
D}_0^\infty$, ${\cal R}_0^\infty$, $\scD_1^4$, $\scR_1^4$ we have:
\begin{equation}
|\tilde{\sGamma}(\ub,u)-\tilde{\sGamma}(0,u_0)|_{\tilde{\sg}(0,u_0)}\leq
1 \label{11.59}
\end{equation}

However, the estimates \ref{8.90}, \ref{8.96} depend on the
results of Chapter 6, and these in turn depend on the results of
Chapters  3, 4 and 5, which rely on the boundedness of the
quantities ${\cal R}_0^\infty$, $\scR_1^4$ and $\scR_2$. The
present chapter on the other hand is to be antecedent to the
establishment of bounds for these quantities in the logic of the
proof of the existence theorem. We thus introduce in present
chapter as a bootstrap assumption:

\vspace{5mm}

\ \ \ {\bf B1:} \
$\sup_{S_{0,u_0}}|\tilde{\sGamma}(\ub,u)-\tilde{\sGamma}(0,u_0)|_{\tilde{\sg}(0,u_0)}\leq
1$ \ : \  for all $(\ub,u)\in D^\prime$

\vspace{5mm}

\section{The coercivity inequalities on the standard sphere}

The coercivity inequalities on
$(S_{\ub,u},\left.\sg\right|_{S_{\ub,u}})$ shall be established on
the basis of the coercivity inequalities on
$(S_{0,u_0},\left.\tilde{\sg}\right|_{S_{0,u_0}})$, which, as we
have seen, can be identified with $(S^2,\up{\sg})$, the unit
sphere in Euclidean 3-dimensional space. Consider the rotation
fields $O_i : i=1,2,3$ in Euclidean 3-dimensional space. These are
given in rectangular coordinates by:
\begin{equation}
O_i=\epsilon_{ijk}x^j\frac{\partial}{\partial x^k} \label{11.60}
\end{equation}
and are tangential to $S^2$. In the following lemma we denote by
$( \ , \ )$ and $| \ |$ the Euclidean inner product and norm. Then
$\up{\sg}$ and $\up{\snab}$ are the induced metric and induced
covariant derivative on
$$S^2=\{x\in \Re^3 \ : \ |x|=1\}$$
Also, we consider $p$ covariant tensorfields on $S^2$ as $p$
covariant tesorfields on Euclidean 3-dimensional space defined
along $S^2$ and vanishing if one of the entries is normal to
$S^2$. Thus, if $\xi$ is such a tensorfield its components
$\xi_{a_1...a_p}$ in rectangular coordinates satisfy:
\begin{equation}
\xi_{a_1...\stackrel{b}{>a_i<}...a_p}x^b=0, \ \ : \ i=1,...,p
\label{11.61}
\end{equation}
Moreover, if $\xi$ is any $p$ covariant tensorfield on $S^2$ we
have:
\begin{equation}
|\xi|_{\up{\sg}}=|\xi| \label{11.a1}
\end{equation}
and if $\xi$, $\zeta$ are any two $p$ covariant tensorfields on
$S^2$ we have:
\begin{equation}
(\xi,\zeta)_{\up{\sg}}=(\xi,\zeta) \label{11.a2}
\end{equation}
Repeated indices are as usual summed, nevertheless we shall
explicitly denote the summation over the three rotation fields
$O_i : i=1,2,3$.

\vspace{5mm}

\noindent{\bf Lemma 11.2} \ \ \ Let $f$ be a function on $S^2$.
Then we have:
$$\sum_i(O_i f)^2=|\sd f|_{\up{\sg}}^2$$
Let $\xi$ be a 1-form on $S^2$. Then we have:
$$\sum_i|\sL_{O_i}\xi|_{\up{\sg}}^2=|\up\snab\xi|_{\up{\sg}}^2+|\xi|_{\up{\sg}}^2$$
Let $\xi$ be a $p$ covariant tensorfield on $S^2$ with $p\geq 2$.
Then we have:
$$\sum_i|\sL_{O_i}\xi|_{\up{\sg}}^2=|\up\snab\xi|_{\up{\sg}}^2+p|\xi|^2
+\sum_{k\neq l}(\xi,\stackrel{k\sim l}{\xi})_{\up{\sg}}
-\sum_{k\neq l}|\stackrel{i,j}{\mbox{tr}}\xi|_{\up{\sg}}^2$$ Here
for $k\neq l$ we denote by $\stackrel{k\sim l}{\xi}$ the transpose
of $\xi$ with respect to the $k$th and $l$th index:
$$\stackrel{k\sim l}{\xi}_{a_1...a_k...a_l...a_p}=\xi_{a_1...a_l...a_k...a_p}$$
if $k<l$ and similarly with the roles of $k$ and $l$ reversed if
$k>l$. Also, for $k\neq l$ we denote by
$\stackrel{k,l}{\mbox{tr}}\xi$ the $p-2$ covariant tensorfield on
$S^2$ obtained by tracing the $k$th and $l$th index:
$$\stackrel{k,l}{\mbox{tr}}{\xi}_{a_1...>a_k<...>a_l<...a_p}=\xi_{a_1...b...b...a_p}$$
if $k<l$ and similary with the roles of $k$ and $l$ reversed if
$k>l$. Obviously,
$$\stackrel{k\sim l}\xi=\stackrel{l\sim k}{\xi} \ \ \mbox{and} \ \
\stackrel{k,l}{\mbox{tr}}\xi=\stackrel{j,k}{\mbox{tr}}\xi$$

\noindent{\em Proof :} \ We have:
\begin{eqnarray}
&&\sum_i(O_i)^m(O_i)^n=\sum_i\epsilon_{ijm}x^j\epsilon_{ikn}x^k
=(\delta_{jk}\delta_{mn}-\delta_{jn}\delta_{km})x^j x^k\nonumber\\
&&\hspace{2cm}=\delta_{mn}-x^m x^n=\Pi^m_n \label{11.62}
\end{eqnarray}
where $\Pi^m_n=\Pi^n_m$ are the components of the orthogonal
projection to $S^2$. The first part of the lemma follows directly
from this formula:
$$\sum_i(O_i f)^2=\sum_i(O_i)^m\frac{\partial f}{\partial x^m}(O_i)^n\frac{\partial f}{\partial x^n}
=\Pi^m_n\frac{\partial f}{\partial x^m}\frac{\partial f}{\partial
x^n}=|\sd f|^2=|\sd f|_{\up{\sg}}^2$$

To establish the rest of the lemma  we note that
$\up{\snab}O_i=\Pi\cdot((\Pi\cdot\nabla) O_i)$, where $\nabla$ is the
covariant derivative in Euclidean space. By \ref{11.60}:
\begin{equation}
(\nabla O_i)^n_j=\frac{\partial (O_i)^n}{\partial
x^j}=\epsilon_{ijn} \label{11.63}
\end{equation}
hence:
\begin{equation}
(\up{\snab} O_i)^m_a=\Pi^j_a\Pi^m_n\epsilon_{ijn} \label{11.64}
\end{equation}
We then have:
\begin{eqnarray}
&&\sum_i(O_i)^n(\up{\snab}
O_i)^m_a=\sum_i\epsilon_{ikn}x^k\Pi^j_a\Pi^m_l\epsilon_{ijl}
=(\delta_{kj}\delta_{nl}-\delta_{kl}\delta_{nj})x^k\Pi^j_a\Pi^m_l\nonumber\\
&&\hspace{27mm}=x^j\Pi^j_a\Pi^m_n-x^l\Pi^n_a\Pi^m_l=0
\label{11.65}
\end{eqnarray}
and:
\begin{eqnarray}
&&\sum_i(\up{\snab}O_i)^m_a(\up{\snab}O_i)^n_b=\sum_i\Pi^c_a\Pi^m_k\epsilon_{ick}\Pi^d_b\Pi^n_l\epsilon_{idl}
\nonumber\\
&&\hspace{17mm}=(\delta_{cd}\delta_{kl}-\delta_{cl}\delta_{dk})\Pi^c_a\Pi^m_k\Pi^d_b\Pi^n_l\nonumber\\
&&\hspace{17mm}=\Pi^c_a\Pi^c_b\Pi^m_k\Pi^n_k-\Pi^l_a\Pi^k_b\Pi^m_k\Pi^n_l=\Pi^a_b\Pi^m_n-\Pi^n_a\Pi^m_b
\label{11.66}
\end{eqnarray}

Let then $\xi$ be a 1-form on $S^2$. We have:
\begin{equation}
(\sL_{O_i}\xi)_a=(\up{\snab}_{O_i}\xi)_a+\xi_m(\up{\snab} O_i)^m_a
\label{11.67}
\end{equation}
and:
\begin{equation}
(\up{\snab}_{O_i}\xi)_a=(O_i)^m(\up{\snab}\xi)_{ma} \label{11.68}
\end{equation}
Hence:
\begin{eqnarray}
&&\sum_i|\sL_{O_i}\xi|_{\up{\sg}}^2=\sum_i (O_i)^m(\up{\snab}\xi)_{ma}(O_i)^n(\up{\snab}\xi)_{na}\label{11.69}\\
&&\hspace{18mm}+2\sum_i(O_i)^n(\up{\snab}\xi)_{na}\xi_m(\up{\snab}O_i)^m_a
+\sum_i\xi_m(\up{\snab}O_i)^m_a\xi_n(\up{\snab}O_i)^n_a\nonumber
\end{eqnarray}
By \ref{11.62} the first term on the right is:
\begin{equation}
\Pi^m_n(\up{\snab}\xi)_{ma}(\up{\snab}\xi)_{na}=|\up{\snab}\xi|^2
\label{11.70}
\end{equation}
By \ref{11.65} the second term on the right in \ref{11.69}
vanishes, while by \ref{11.66}, which implies:
\begin{equation}
\sum_i(\up{\snab}O_i)^m_a(\up{\snab}O_i)^n_a=\Pi^m_n \label{11.71}
\end{equation}
the third term on the right in \ref{11.69} is equal to:
\begin{equation}
\Pi^m_n\xi_m\xi_n=|\xi|^2 \label{11.72}
\end{equation}
We thus obtain:
$$\sum_i|\sL_{O_i}\xi|^2=|\up{\snab}\xi|^2+|\xi|^2$$
which, in view of \ref{11.a1}, is the second part of the lemma.

Let finally $\xi$ be a $p$ covariant tensorfield on $S^2$, $p\geq
2$. We have:
\begin{equation}
(\sL_{O_i}\xi)_{a_1...a_p}=(\up{\snab}_{O_i}\xi)_{a_1...a_p}+\sum_{k=1}^p\xi_{a_1...\stackrel{m}{>a_k<}...a_p}
(\up{\snab}O_i)^m_{a_k} \label{11.73}
\end{equation}
and:
\begin{equation}
(\up{\snab}_{O_i}\xi)_{a_1...a_p}=(O_i)^m(\up{\snab}\xi)_{ma_1...a_p}
\label{11.74}
\end{equation}
Hence:
\begin{eqnarray}
&&\sum_i|\sL_{O_i}\xi|^2=\sum_i(O_i)^m(\up{\snab}\xi)_{ma_1...a_p}(O_i)^n(\up{\snab}\xi)_{na_1...a_p}\label{11.75}\\
&&\hspace{18mm}+2\sum_i\sum_{k=1}^p
(O_i)^n(\up{\snab}\xi)_{na_1...a_p}\xi_{a_1...\stackrel{m}{>a_k<}...a_p}(\up{\snab}O_i)^m_{a_k}
\nonumber\\
&&\hspace{18mm}+\sum_i\sum_{k,l=1}^p\xi_{a_1...\stackrel{m}{>a_k<}...a_p}(\up{\snab}O_i)^m_{a_k}
\xi_{a_1...\stackrel{n}{>a_l<}...a_p}(\up{\snab}O_i)^n_{a_l}\nonumber
\end{eqnarray}
By \ref{11.62} the first term on the right is:
\begin{equation}
\Pi^m_n(\up{\snab}\xi)_{ma_1...a_p}(\up{\snab}\xi)_{na_1...a_p}=|\up{\snab}\xi|^2
\label{11.76}
\end{equation}
By \ref{11.65} the second term on the right in \ref{11.75}
vanishes. The third term on the right in \ref{11.75} splits into:
\begin{eqnarray}
&&\sum_i\sum_{k=1}^p\xi_{a_1...\stackrel{m}{>a_k<}...a_p}(\up{\snab}O_i)^m_{a_k}
\xi_{a_1...\stackrel{n}{>a_k<}...a_p}(\up{\snab}O_i)^n_{a_k}\nonumber\\
&&+\sum_i\sum_{k\neq
l=1}^p\xi_{a_1...\stackrel{m}{>a_k<}...a_p}(\up{\snab}O_i)^m_{a_k}
\xi_{a_1...\stackrel{n}{>a_l<}...a_p}(\up{\snab}O_i)^n_{a_l}
\label{11.77}
\end{eqnarray}
By \ref{11.71} the first sum is equal to:
\begin{equation}
\sum_{k=1}^p\Pi^m_n\xi_{a_1...\stackrel{m}{>a_k<}...a_p}\xi_{a_1...\stackrel{n}{>a_k<}...a_p}=p|\xi|^2
\label{11.78}
\end{equation}
while by \ref{11.66} the second sum is equal to:
\begin{eqnarray}
&&\sum_{k\neq
l=1}^p(\Pi^{a_k}_{a_l}\Pi^m_n-\Pi^m_{a_l}\Pi^n_{a_k})
\xi_{a_1...\stackrel{m}{>a_k}...a_p}\xi_{a_1...\stackrel{n}{>a_l}...a_p}\nonumber\\
&&=\sum_{k\neq l=1}^p(\delta_{a_k
a_l}\delta_{mn}-\delta_{ma_l}\delta_{na_k})
\xi_{a_1...\stackrel{m}{>a_k}...a_p}\xi_{a_1...\stackrel{n}{>a_l}...a_p}\nonumber\\
&&=\sum_{k\neq l=1}^p\left\{(\xi,\stackrel{k\sim
l}{\xi})-(\stackrel{k,l}{\mbox{tr}}\xi)^2\right\} \label{11.79}
\end{eqnarray}
We thus obtain:
$$\sum_i|\sL_{O_i}\xi|^2=|\up{\snab}\xi|^2+p|\xi|^2+\sum_{k\neq l}(\xi,\stackrel{k\sim l}\xi)
-\sum_{k\neq l}(\stackrel{k,l}{\mbox{tr}}\xi)^2$$ which, in view
of \ref{11.a1}, \ref{11.a2}, is the third part of the lemma.

\vspace{5mm}

Given a $p$ covariant tensorfield $\xi$ on $S^2$ with $p\geq 2$,
for each $k\neq l=1,...,p$,  we may decompose $\xi$ into a part
$\stackrel{k,l}{\theta}$ which is symmetric in the $k$th and $l$th
index and a part $\stackrel{k,l}{\omega}$ which is antisymmetric
in the $k$th and $l$th index:
\begin{equation}
\xi=\stackrel{k,l}{\theta}+\stackrel{k,l}{\omega}, \ \ \
\stackrel{k,l}{\theta}=\frac{1}{2}(\xi+\stackrel{k\sim l}{\xi}), \
\ \ \stackrel{k,l}{\omega}=\frac{1}{2}(\xi-\stackrel{k\sim
l}{\xi}) \label{11.80}
\end{equation}
The symmetric part $\stackrel{k,l}{\theta}$ may be further
decomposed into its trace-free part $\stackrel{k,l}{\sigma}$ and
its trace part:
\begin{equation}
\stackrel{k,l}{\theta}=\stackrel{k,l}{\sigma}+\frac{1}{2}\up{\sg}\stackrel{k,l}{\mbox{tr}}\xi
\label{11.81}
\end{equation}
For each $k\neq l=1,...,p$ we have:
\begin{equation}
|\xi|^2=|\stackrel{k,l}{\theta}|^2+|\stackrel{k,l}{\omega}|^2=|\stackrel{k,l}{\sigma}|^2+|\stackrel{k,l}{\omega}|^2
+\frac{1}{2}|\stackrel{k,l}{\mbox{tr}}\xi|^2 \label{11.82}
\end{equation}
Hence:
\begin{equation}
p(p-1)|\xi|^2=\sum_{k\neq
l}\left\{|\stackrel{k,l}{\sigma}|^2+|\stackrel{k,l}{\omega}|^2
+\frac{1}{2}|\stackrel{k,l}{\mbox{tr}}\xi|^2\right\} \label{11.83}
\end{equation}
On the other hand we have, for each $k\neq l=1,...,p$,
\begin{equation}
(\xi,\stackrel{k\sim
l}{\xi})=|\stackrel{k,l}{\theta}|^2-|\stackrel{k,l}{\omega}|^2
=|\stackrel{k,l}{\sigma}|^2-|\stackrel{k,l}{\omega}|^2+\frac{1}{2}|\stackrel{k,l}{\mbox{tr}}\xi|^2
\label{11.84}
\end{equation}
Hence:
\begin{equation}
\sum_{k\neq l}\left\{(\xi,\stackrel{k\sim
l}{\xi})-|\stackrel{k,l}{\mbox{tr}}\xi|^2\right\} =\sum_{k\neq
l}\left\{|\stackrel{k,l}{\sigma}|^2-|\stackrel{k,l}{\omega}|^2
-\frac{1}{2}|\stackrel{k,l}{\mbox{tr}}\xi|^2\right\} \label{11.85}
\end{equation}
Comparing \ref{11.85} with \ref{11.83} we conclude that:
\begin{equation}
\sum_{k\neq l}\left\{(\xi,\stackrel{k\sim
l}{\xi})-|\stackrel{k,l}{\mbox{tr}}\xi|^2\right\} \geq
-p(p-1)|\xi|^2 \label{11.86}
\end{equation}
Therefore the third conclusion of Lemma 11.2 implies that for any
$p$ covariant tensorfield on $S^2$ with $p\geq 2$ we have:
\begin{equation}
\sum_i|\sL_{O_i}\xi|_{\up{\sg}}^2\geq
|\up{\snab}\xi|_{\up{\sg}}^2-p(p-2)|\xi|_{\up{\sg}}^2
\label{11.87}
\end{equation}
In view of the first two parts of the lemma this also holds in the
cases $p=0,1$.

\section{The coercivity inequalities on $S_{\ub,u}$}

We now derive on the basis of Lemmas 11.1 and 11.2 and the
bootstrap assumption {\bf B1}, the coercivity inequalities on
$S_{\ub,u}$.

\vspace{5mm}

\noindent{\bf Proposition 11.1} \ \ \ Let the bootstrap
assumptions {\bf A1.1}, {\bf A1.2}, {\bf A3.1}, {\bf A3.2} and
{\bf B1} hold. Let $f$ be an arbitrary function defined on
$M^\prime$. Then for every $(\ub,u)\in D^\prime$ we have,
pointwise on $S_{\ub,u}$:
$$\sum_i(O_i f)^2\geq 4^{-1}|u|^2|\sd f|_{\sg}^2$$
Let $\xi$ be a $p$ covariant $S$ tensorfield defined on
$M^\prime$, $p\geq 1$. Then for every $(\ub,u)\in D^\prime$ we
have, pointwise on $S_{\ub,u}$:
$$\sum_i|\sL_{O_i}\xi|_{\sg}^2\geq 2^{-1}4^{-1-2p}|u|^2|\stackrel{\sg}{\snab}\xi|_{\sg}^2-2p(p-1)|\xi|_{\sg}^2$$

\noindent{\em Proof :} \ Given a function $f$ on $M^\prime$, we
consider the function
\begin{equation}
f(\ub,u)=(\Phib_{u-u_0}\circ\Phib_{\ub})^*\left.f\right|_{S_{\ub,u}}
=\Phi_{\ub}^*\Phi_{u-u_0}^*\left.f\right|_{S_{\ub,u}}
=\left.f\right|_{S_{\ub,u}}\circ\Phib_{u-u_0}\circ\Phi_{\ub}
\label{11.88}
\end{equation}
on $S_{0,u_0}$. To this function we apply the first conclusion of
Lemma 11.2 to obtain, in view of the identification of
$(S_{0,u_0},\tilde{\sg}(0,u_0))$ with $(S^2,\up{\sg})$,
\begin{equation}
\sum_i(O_i(f(\ub,u)))^2=|\sd f(\ub,u)|_{\tilde{\sg}(0,u_0)}^2
\label{11.89}
\end{equation}
Now, by \ref{8.a2} we have, for every $p\in S_{\ub,u}$:
\begin{equation}
O_i(p)f=(d\Phib_{u-u_0}(q)\cdot
O_i(q))f=O_i(q)(f\circ\Phib_{u-u_0}), \ q=\Phib_{u_0-u}(p)\in
S_{\ub,u_0} \label{11.90}
\end{equation}
while by \ref{8.a1} we have, for every $q\in S_{\ub,u_0}$:
\begin{eqnarray}
&&O_i(q)(f\circ\Phib_{u-u_0})=(d\Phi_{\ub}(q_0)\cdot O_i(q_0))(f\circ\Phib_{u-u_0})\nonumber\\
&&\hspace{1cm}=O_i(q_0)(f\circ\Phib_{u-u_0}\circ\Phi_{\ub}), \
q_0=\Phi_{-\ub}(q)\in S_{0,u_0} \label{11.91}
\end{eqnarray}
hence, combining:
\begin{equation}
O_i(p)f=O_i(q_0)f(\ub,u), \ q_0=\Phi_{-\ub}(\Phib_{u_0-u}(p)) \ \
: \ \forall p\in S_{\ub,u} \label{11.92}
\end{equation}
In terms of the function $(O_i
f)(\ub,u)=(\Phib_{u-u_0}\circ\Phi_{\ub})^*\left.(O_i
f)\right|_{S_{\ub,u}}$, this reads:
\begin{equation}
(O_i f)(\ub,u)=O_i (f(\ub,u)) \label{11.93}
\end{equation}
On the other hand, by Lemma 11.1 and inequality \ref{11.26} with
$\sd f(\ub,u)$ in the role of $\xi$ ($p=1$, $q=0$) we have:
\begin{equation}
|\sd f(\ub,u)|^2_{\tilde{\sg}(0,u_0)}\geq \lambda(\ub,u)|\sd
f(\ub,u)|^2_{\tilde{\sg}(\ub,u)} \geq 4^{-1}|\sd
f(\ub,u)|^2_{\tilde{\sg}(\ub,u)} \label{11.94}
\end{equation}
while by \ref{11.31}:
\begin{equation}
|\sd f(\ub,u)|^2_{\tilde{\sg}(\ub,u)}=|u|^2|\sd
f(\ub,u)|^2_{\sg(\ub,u)}=|u|^2(|\sd f|^2_{\sg})(\ub,u)
\label{11.95}
\end{equation}
hence, combining:
\begin{equation}
|\sd f(\ub,u)|^2_{\tilde{\sg}(0,u_0)}\geq 4^{-1}|u|^2(|\sd
f|^2_{\sg})(\ub,u) \label{11.96}
\end{equation}
In view of \ref{11.93} and \ref{11.96} we deduce from \ref{11.89}:
\begin{equation}
\sum_i((O_i f)(\ub,u))^2\geq 4^{-1}|u|^2(|\sd f|^2_{\sg})(\ub,u)
\label{11.97}
\end{equation} which yields the first part of the proposition.

Next, given a $p$ covariant $S$ tensorfield $\xi$ on $M^\prime$,
$p\geq 1$, we consider the $p$ covariant tensorfield
\begin{equation}
\xi(\ub,u)=(\Phib_{u-u_0}\circ\Phi_{\ub})^*\left.\xi\right|_{S_{\ub,u}}
=\Phi_{\ub}^*\Phib_{u-u_0}^*\left.\xi\right|_{S_{\ub,u}}
\label{11.98}
\end{equation}
To this tensorfield we apply \ref{11.87}, a corollary of Lemma
11.2, to obtain, in view of the identification of
$(S_{0,u_0},\tilde{\sg}(0,u_0))$ with $(S^2,\up{\sg})$,
\begin{equation}
\sum_i|\sL_{O_i}(\xi(\ub,u))|^2_{\tilde{\sg}(0,u_0)}\geq
|\stackrel{\tilde{\sg}(0,u_0)}{\snab}\xi(\ub,u)|^2_{\tilde{\sg}(0,u_0)}
-p(p-2)|\xi(\ub,u)|^2_{\tilde{\sg}(0,u_0)} \label{11.99}
\end{equation}

Now, let $\Psi_{i,s}$ be the 1-parameter rotation group generated
by $O_i$. According to the definition of the action of the
rotation group of Chapter 8 we have, for every $(\ub,u)\in
D^\prime$:
\begin{equation}
\Psi_{i,s}\circ\Phib_{u-u_0}=\Phib_{u-u_0}\circ\Psi_{i,s} \ \ : \
\mbox{on $S_{\ub,u_0}$} \label{11.100}
\end{equation}
and:
\begin{equation}
\Psi_{i,s}\circ\Phi_{\ub}=\Phi_{\ub}\circ\Psi_{i,s} \ \ : \mbox{on
$S_{0,u_0}$} \label{11.101}
\end{equation}
Composing \ref{11.100} on the right by $\Phi_{\ub}$ acting on
$S_{0,u_0}$ we obtain, by virtue of \ref{11.101},
\begin{equation}
\Psi_{i,s}\circ(\Phib_{u-u_0}\circ\Phi_{\ub})=(\Phib_{u-u_0}\circ\Phi_{\ub})\circ\Psi_{i,s}
\ \ : \mbox{on $S_{0,u_0}$} \label{11.102}
\end{equation}
This expresses the equality of two diffeomorphisms of $S_{0,u_0}$
onto $S_{\ub,u}$. Therefore the corresponding pullbacks, which map
$p$ covariant tensorfields on $S_{\ub,u}$ to $p$ covariant
tensorfields on $S_{0,u_0}$, coincide:
$$(\Psi_{i,s}\circ(\Phib_{u-u_0}\circ\Phi_{\ub}))^*=((\Phib_{u-u_0}\circ\Phi_{\ub})\circ\Psi_{i,s})^*$$
or:
\begin{equation}
(\Phib_{u-u_0}\circ\Phi_{\ub})^*\Psi_{i,s}^*=\Psi_{i,s}^*(\Phib_{u-u_0}\circ\Phi_{\ub})^*
\label{11.103}
\end{equation}
Applying this to $\left.\xi\right|_{S_{\ub,u}}$ we have:
\begin{equation}
(\Phib_{u-u_0}\circ\Phi_{\ub})^*\Psi_{i,s}^*\left.\xi\right|_{S_{\ub,u}}
=\Psi_{i,s}^*(\Phib_{u-u_0}\circ\Phi_{\ub})^*\left.\xi\right|_{S_{\ub,u}}
=\Psi_{i,s}^*(\xi(\ub,u)) \label{11.104}
\end{equation}
Taking the derivative of this with respect to $s$ at $s=0$, we
have on the left hand side the pullback by
$\Phib_{u-u_0}\circ\Phi_{\ub}$ of:
\begin{equation}
\left(\frac{d}{ds}\Psi_{i,s}^*\left.\xi\right|_{S_{\ub,u}}\right)_{s=0}=\left.(\sL_{O_i}\xi)\right|_{S_{\ub,u}}
\label{11.105}
\end{equation}
and on the right hand side:
\begin{equation}
\left(\frac{d}{ds}\Psi_{i,s}^*(\xi(\ub,u))\right)_{s=0}=\sL_{O_i}(\xi(\ub,u))
\label{11.106}
\end{equation}
We thus obtain:
$$(\Phib_{u-u_0}\circ\Phi_{\ub})^*\left.(\sL_{O_i}\xi)\right|_{S_{\ub,u}}=\sL_{O_i}(\xi(\ub,u))$$
or:
\begin{equation}
(\sL_{O_i}\xi)(\ub,u)=\sL_{O_i}(\xi(\ub,u)) \label{11.107}
\end{equation}

In view of the equality \ref{11.107}, \ref{11.99} is equivalent
to:
\begin{equation}
\sum_i|(\sL_{O_i}\xi)(\ub,u))|^2_{\tilde{\sg}(0,u_0)}\geq
|\stackrel{\tilde{\sg}(0,u_0)}{\snab}\xi(\ub,u)|^2_{\tilde{\sg}(0,u_0)}
-p(p-2)|\xi(\ub,u)|^2_{\tilde{\sg}(0,u_0)} \label{11.108}
\end{equation}
Consider next $\stackrel{\tilde{\sg}(\ub,u)}{\snab}\xi$. We have,
in an arbitrary local frame field on $S_{0,u_0}$<
\begin{eqnarray*}
&&(\stackrel{\tilde{\sg}(\ub,u)}{\snab}\xi(\ub,u))_{AB_1...B_p}=
(\stackrel{\tilde{\sg}(0,u_0)}{\snab}\xi(\ub,u))_{AB_1...B_p}\\
&&\hspace{3cm}-\sum_{k=1}^p(\tilde{\sGamma}(\ub,u)-\tilde{\sGamma}(0,u_0))^C_{AB_k}\xi(\ub,u)_{B_1...\stackrel{C}{>B_k<}...B_p}
\end{eqnarray*}
Hence:
\begin{eqnarray}
&&|\stackrel{\tilde{\sg}(\ub,u)}{\snab}\xi(\ub,u)|_{\tilde{\sg}(0,u_0)}\leq
|\stackrel{\tilde{\sg}(0,u_0)}{\snab}\xi(\ub,u)|_{\tilde{\sg}(0,u_0)}\nonumber\\
&&\hspace{3cm}+p|\tilde{\sGamma}(\ub,u)-\tilde{\sGamma}(0,u_0)|_{\tilde{\sg}(0,u_0)}|\xi(\ub,u)|_{\tilde{\sg}(0,u_0)}\nonumber\\
&&\hspace{3cm}\leq|\stackrel{\tilde{\sg}(0,u_0)}{\snab}\xi(\ub,u)|_{\tilde{\sg}(0,u_0)}
+p|\xi(\ub,u)|_{\tilde{\sg}(0,u_0)} \label{11.109}
\end{eqnarray}
by virtue of assumption {\bf B1}. Thus, we have:
$$|\stackrel{\tilde{\sg}(\ub,u)}{\snab}\xi(\ub,u)|_{\tilde{\sg}(0,u_0)}\leq
2|\stackrel{\tilde{\sg}(0,u_0)}{\snab}\xi(\ub,u)|^2_{\tilde{\sg}(0,u_0)}+2p^2|\xi(\ub,u)|^2_{\tilde{\sg}(0,u_0)}$$
and combining with \ref{11.108} we obtain:
\begin{equation}
\sum_i|(\sL_{O_i}\xi)(\ub,u))|^2_{\tilde{\sg}(0,u_0)}\geq
\frac{1}{2}|\stackrel{\tilde{\sg}(\ub,u)}{\snab}\xi(\ub,u)|^2_{\tilde{\sg}(0,u_0)}
-2p(p-1)|\xi(\ub,u)|^2_{\tilde{\sg}(0,u_0)} \label{11.110}
\end{equation}
On the other hand, by Lemma 11.1 and inequality \ref{11.26}
applied to $\xi(\ub,u)$ and to $(\sL_{O_i}\xi)(\ub,u)$:
\begin{equation}
|\xi(\ub,u)|^2_{\tilde{\sg}(0,u_0)}\leq
(\Lambda(\ub,u))^p|\xi(\ub,u)|^2_{\tilde{\sg}(\ub,u)} \leq
4^p|\xi(\ub,u)|^2_{\tilde{\sg}(\ub,u)} \label{11.111}
\end{equation}
and:
\begin{equation}
|(\sL_{O_i}\xi)(\ub,u)|^2_{\tilde{\sg}(0,u_0)}\leq
(\Lambda(\ub,u))^p|(\sL_{O_i}\xi)(\ub,u)|^2_{\tilde{\sg}(\ub,u)}
\leq 4^p|(\sL_{O_i}\xi)(\ub,u)|^2_{\tilde{\sg}(\ub,u)}
\label{11.112}
\end{equation}
Also, by Lemma 11.1 and inequality \ref{11.26} with
$\stackrel{\tilde{\sg}(\ub,u)}{\snab}\xi(\ub,u)$ in the role of
$\xi$ (and $(p+1,0)$ in the role of $(p,q)$):
\begin{eqnarray}
&&|\stackrel{\tilde{\sg}(\ub,u)}{\snab}\xi(\ub,u)|^2_{\tilde{\sg}(0,u_0)}\geq
(\lambda(\ub,u))^{p+1}|\stackrel{\tilde{\sg}(\ub,u)}{\snab}\xi(\ub,u)|_{\tilde{\sg}(\ub,u)}\nonumber\\
&&\hspace{32mm}\geq
4^{-p-1}|\stackrel{\tilde{\sg}(\ub,u)}{\snab}\xi(\ub,u)|_{\tilde{\sg}(\ub,u)}
\label{11.113}
\end{eqnarray}
Moreover, by \ref{11.31} we have:
\begin{equation}
|\xi(\ub,u)|^2_{\tilde{\sg}(\ub,u)}=|u|^{2p}|\xi(\ub,u)|_{\sg(\ub,u)},
\ \ \
|(\sL_{O_i}\xi)(\ub,u)|^2_{\tilde{\sg}(\ub,u)}=|u|^{2p}|(\sL_{O_i}\xi)(\ub,u)|^2_{\sg(\ub,u)}
\label{11.114}
\end{equation}
and, in view of \ref{11.47}:
\begin{equation}
|\stackrel{\tilde{\sg}(\ub,u)}{\snab}\xi(\ub,u)|_{\tilde{\sg}(\ub,u)}
=|u|^{2p+2}|\stackrel{\sg(\ub,u)}{\snab}\xi(\ub,u)|_{\sg(\ub,u)}
\label{11.115}
\end{equation}
Finally, for any diffeomorphism $\Omega$ of $S_{0,u_0}$ onto
$S_{\ub,u}$ and any $p$ covariant tensorfield $\xi$ on $S_{\ub,u}$
we have:
\begin{equation}
\stackrel{\Omega^*\sg}{\snab}(\Omega^*\xi)=\Omega^*(\stackrel{\sg}{\snab}\xi)
\label{11.116}
\end{equation}
In particular this holds in the case
$\Omega=\Phib_{u-u_0}\circ\Phi_{\ub}$. Hence:
\begin{equation}
\stackrel{\sg(\ub,u)}{\snab}\xi(\ub,u)=(\stackrel{\sg}{\snab}\xi)(\ub,u)
\label{11.117}
\end{equation}
In view of \ref{11.111} - \ref{11.115}, \ref{11.117}, the
inequality \ref{11.110} implies:
\begin{equation}
\sum_i|(\sL_{O_i}\xi)(\ub,u)|^2_{\sg(\ub,u)}\geq
\frac{1}{2}4^{-2p-1}|u|^2|(\stackrel{\sg}{\snab}\xi)(\ub,u)|^2_{\sg(\ub,u)}-2p(p-1)|\xi(\ub,u)|^2_{\sg(\ub,u)}
\label{11.118}
\end{equation}
Since
$$|(\sL_{O_i}\xi)(\ub,u)|_{\sg(\ub,u)}=(|\sL_{O_i}\xi|_{\sg})(\ub,u), \ \ \
|\xi(\ub,u)|_{\sg(\ub,u)}=(|\xi|_{\sg})(\ub,u)$$ and
$$|(\stackrel{\sg}{\snab}\xi)(\ub,u)|_{\sg(\ub,u)}=(|\stackrel{\sg}{\snab}\xi|_{\sg})(\ub,u)$$
the inequality \ref{11.118} yields the second part of the
proposition.

\vspace{5mm}

Let $\xi$ be a $p$-covariant $S$ tensorfield defined on $M^\prime$
with $p\geq 2$. We denote by $\mbox{tr}_{\sg}\xi$ the $p-2$
covariant $S$ tensorfield obtained by tracing the last two
indices. In terms of components in an arbitrary local frame field
for $S_{\ub,u}$:
\begin{equation}
(\mbox{tr}_{\sg}\xi)_{A_1...A_{p-2}}=(\sg^{-1})^{BC}\xi_{A_1...A_{p-2}BC}
\label{11.119}
\end{equation}
Suppose now that $\xi$ is a $p$ covariant $S$ tensorfield on
$M^\prime$, $p\geq 2$, which is symmetric and trace-free in the
last two indices. Then we have:
\begin{eqnarray*}
&&(\mbox{tr}_{\sg}\sL_{O_i}\xi)_{A_1...A_{p-2}}=(\sL_{O_i}\mbox{tr}_{\sg}\xi)_{A_1...A_{p-2}}
-(\sL_{O_i}\sg^{-1})^{BC}\xi_{A_1...A_{p-2}BC}\\
&&\hspace{27mm}=\s^{(O_i)}\spi^{BC}\xi_{A_1...A_{p-2}BC}
\end{eqnarray*}
or:
\begin{equation}
\mbox{tr}_{\sg}\sL_{O_i}\xi=(\s^{(O_i)}\spi,\xi)_{\sg}
\label{11.120}
\end{equation}
Decomposing then $\sL_{O_i}\xi$ into $\sLh_{O_i}\xi$, its
trace-free part with respect to the last two indices, and
$(1/2)\mbox{tr}_{\sg}\xi\otimes\sg$, its trace part with respect
to the last two indices,
\begin{equation}
\sL_{O_i}\xi=\sLh_{O_i}\xi+\frac{1}{2}\mbox{tr}_{\sg}\xi\otimes\sg
\label{11.121}
\end{equation}
we have:
\begin{eqnarray}
&&\sum_i|\sL_{O_i}\xi|^2_{\sg}=
\sum_i\left\{|\sLh_{O_i}\xi|^2_{\sg}+\frac{1}{2}|\mbox{tr}_{\sg}\sL_{O_i}\xi|^2_{\sg}\right\}\nonumber\\
&&\hspace{17mm}=\sum_i\left\{|\sLh_{O_i}\xi|^2_{\sg}+\frac{1}{2}|\s^{(O_i)}\spi|^2_{\sg}|\xi|^2_{\sg}\right\}
\label{11.122}
\end{eqnarray}
therefore the bootstrap assumption:

\hspace{5mm}

\ \ \ {\bf B2:} \ $|\s^{(O_i)}\spi|\leq 1$ \  i=1,2,3 \ in $M^\prime$ 

\hspace{5mm}

implies:
\begin{equation}
\sum_i|\sLh_{O_i}\xi|^2_{\sg}\geq\sum_i|\sL_{O_i}\xi|^2_{\sg}-\frac{3}{2}|\xi|^2_{\sg}
\label{11.123}
\end{equation}
It follows that in the case of a $p$ covariant $S$ tensorfield
which is symmetric and trace-free in the last two indices
Proposition 11.1 takes the form:
\begin{eqnarray}
&&\sum_i|\sLh_{O_i}\xi|^2_{\sg}\geq 2^{-1}4^{-1-2p}|u|^2|\stackrel{\sg}{\snab}\xi|^2_{\sg}-C_p|\xi|^2_{\sg}\nonumber\\
&&\hspace{1cm}\mbox{: pointwise on $S_{\ub,u}$, for every
$(\ub,u)\in D^\prime$} \label{11.124}
\end{eqnarray}
where:
\begin{equation}
C_p=2p(p-1)+\frac{3}{2} \label{11.125}
\end{equation}

We proceed to derive from Proposition 11.1 2nd order coercivity
inequalities on $S_{\ub,u}$. Consider first the case of a function
$f$, defined on $M^\prime$. By the first part of Proposition 11.1
applied to the function $O_i f$ we have:
\begin{equation}
|u|^2|\sd O_i f|^2_{\sg}\leq 4\sum_j(O_j O_i f)^2 \label{11.126}
\end{equation}
hence, summing over $i=1,2,3$,
\begin{equation}
|u|^2\sum_i|\sd O_i f|^2_{\sg}\leq 4\sum_{i,j}(O_j O_i f)^2
\label{11.127}
\end{equation}
Now, we have:
\begin{equation}
\sd O_i f=\sL_{O_i}\sd f \label{11.128}
\end{equation}
Thus, applying the second part of Proposition 11.1 in the case
$p=1$ to the $S$ 1-form $\sd f$ we obtain:
\begin{equation}
|u|^2|\snab^{ \ 2} f|^2_{\sg}\leq 2.4^3\sum_i|\sd O_i f|^2_{\sg}
\label{11.129}
\end{equation}
Combining \ref{11.129} and \ref{11.127} we conclude that:
\begin{equation}
|u|^4|\snab^{ \ 2} f|^2_{\sg}\leq 2.4^4\sum_{i,j}(O_j O_i f)^2
\label{11.130}
\end{equation}

Consider next the case of a $p$ covariant $S$ tensorfield $\xi$
defined on $M^\prime$. In the following, we shall denote by $C_p$
various numerical constants depending only on $p$. By the second
part of  Proposition 11.1 applied to the $p$ covariant $S$
tensorfield $\sL_{O_i}\xi$ we have:
\begin{equation}
|u|^2|\snab\sL_{O_i}\xi|^2_{\sg}\leq
C_p\left\{\sum_j|\sL_{O_j}\sL_{O_i}\xi|^2_{\sg}+|\sL_{O_i}\xi|^2_{\sg}\right\}
\label{11.131}
\end{equation}
hence, summing over $i=1,2,3$,
\begin{equation}
|u|^2\sum_i|\snab\sL_{O_i}\xi|^2_{\sg}\leq
C_p\left\{\sum_{i,j}|\sL_{O_j}\sL_{O_i}\xi|^2_{\sg}
+\sum_i|\sL_{O_i}\xi|^2_{\sg}\right\} \label{11.133}
\end{equation}
We now express $\sL_{O_i}\snab\xi$ in terms of
$\snab\sL_{O_i}\xi$. By Lemma 9.1 we have, with respect to an
arbitrary local frame field for the $S_{\ub,u}$,
\begin{equation}
(\sL_{O_i}\snab\xi-\snab\sL_{O_i}\xi)_{AB_1...B_p}=-\sum_{k=1}^p\s^{(O_i)}\spi_{1,AB_k}^C
\xi_{B_1...\stackrel{C}{>B_k<}...B_k} \label{11.134}
\end{equation}
It follows that:
\begin{equation}
\sum_i|\sL_{O_i}\snab\xi|^2\leq
2\sum_i|\snab\sL_{O_i}\xi|^2+2p^2\left(\sum_i|\s^{(O_i)}\spi_1|^2_{\sg}\right)|\xi|^2_{\sg}
\label{11.135}
\end{equation}
Applying the second part of Proposition 11.1 to the $p+1$
covariant $S$ tensorfield $\snab\xi$ we obtain:
\begin{equation}
|u|^2|\snab^{ \ 2}\xi|^2_{\sg}\leq
C_p\left\{\sum_i|\sL_{O_i}\snab\xi|^2_{\sg}+|\snab\xi|^2_{\sg}\right\}
\label{11.136}
\end{equation}
Moreover, by Proposition 11.1 applied to $\xi$ itself we have:
\begin{equation}
|u|^2|\snab\xi|^2\leq
C_p\left\{\sum_i|\sL_{O_i}\xi|^2_{\sg}+|\xi|^2_{\sg}\right\}
\label{11.137}
\end{equation}
hence, substituting in \ref{11.136},
\begin{equation}
|u|^4|\snab^{ \ 2}\xi|^2_{\sg}\leq
C_p\left\{|u|^2\sum_i|\sL_{O_i}\snab\xi|^2_{\sg}
+\sum_i|\sL_{O_i}\xi|^2_{\sg}+|\xi|^2_{\sg}\right\} \label{11.138}
\end{equation}
Combining \ref{11.138}, \ref{11.135} and \ref{11.133} we conclude
that:
\begin{eqnarray}
&&|u|^4|\snab^{ \ 2}\xi|^2_{\sg}\leq
C_p\left\{\sum_{i,j}|\sL_{O_j}\sL_{O_i}\xi|^2_{\sg}
+\sum_i|\sL_{O_i}\xi|^2_{\sg}+|\xi|^2_{\sg}\right.\nonumber\\
&&\hspace{25mm}\left.+|u|^2\left(\sum_i|\s^{(O_i)}\spi_1|^2_{\sg}\right)|\xi|^2_{\sg}\right\}
\label{11.140}
\end{eqnarray}
We shall integrate this inequality on $S_{\ub,u}$. The integral on
$S_{\ub,u}$ of the last term in parenthesis on the right is
bounded by:
\begin{eqnarray}
&&|u|^2\left(\int_{S_{\ub,u}}\left(\sum_i|\s^{(O_i)}\spi_1|^2_{\sg}\right)^2\right)^{1/2}
\left(\int_{S_{\ub,u}}|\xi|^4_{\sg}\right)^{1/2}\nonumber\\
&&\hspace{1cm}\leq
|u|^2\sum_i\|\s^{(O_i)}\spi_1\|^2_{L^4(S_{\ub,u})}\|\xi\|^2_{L^4(S_{\ub,u})}
\label{11.141}
\end{eqnarray}
Now by Lemma 5.1 with $p=4$ in that lemma and Lemma 10.1 there is
a numerical constant $C$ such that:
\begin{eqnarray}
&&|u|\|\xi\|^2_{L^4(S_{\ub,u})}\leq
C\left\{|u|^2\|\snab\xi\|^2_{L^2(S_{\ub,u})}+\|\xi\|^2_{L^2(S_{\ub,u})}\right\}
\nonumber\\
&&\hspace{22mm}\leq
C_p\int_{S_{\ub,u}}\left\{\sum_i|\sL_{O_i}\xi|^2_{\sg}+|\xi|^2_{\sg}\right\}d\mu_{\sg}
\label{11.142}
\end{eqnarray}
the last step by \ref{11.137} integrated on $S_{\ub,u}$. At this
point we introduce the bootstrap assumption:

\hspace{5mm}

\ \ \ {\bf B3:} \
$|u|^{1/2}\|\snab\s^{(O_i)}\spi\|_{L^4(S_{\ub,u})}\leq 1$ \ \ : \
for all $(\ub,u)\in D^\prime$

\hspace{5mm}

The assumption {\bf B3} implies that:
\begin{equation}
|u|^{1/2}\|\s^{(O_i)}\spi_1\|_{L^4(S_{\ub,u})}\leq\frac{3}{2} \ \
: \ \forall(\ub,u)\in D^\prime \label{11.139}
\end{equation}
Then in view of \ref{11.142}, the right hand side of \ref{11.141}
is bounded by:
\begin{equation}
C_p\int_{S_{\ub,u}}\left\{\sum_i|\sL_{O_i}\xi|^2_{\sg}+|\xi|^2_{\sg}\right\}d\mu_{\sg}
\label{11.143}
\end{equation}
Integrating \ref{11.140} on $S_{\ub,u}$ we then conclude that:
\begin{equation}
\int_{S_{\ub,u}}|u|^4|\snab^{ \ 2}\xi|^2_{\sg}d\mu_{\sg}\leq
C_p\int_{S_{\ub,u}}
\left\{\sum_{i,j}|\sL_{O_j}\sL_{O_i}\xi|^2_{\sg}+\sum_i|\sL_{O_i}\xi|^2_{\sg}+|\xi|^2_{\sg}\right\}d\mu_{\sg}
\label{11.144}
\end{equation}

Consider finally the case of a $p$ covariant $S$ tensorfield $\xi$
defined on $M^\prime$, $\geq 2$, which is symmetric and trace-free
in the last two indices. Then by \ref{11.124} applied to the $p$
covariant $S$ tensorfield $\sLh_{O_i}\xi$ which is also symmetric
and trace-free in the last two indices we have:
\begin{equation}
|u|^2|\snab\sLh_{O_i}\xi|^2_{\sg}\leq
C_p\left\{\sum_j|\sLh_{O_j}\sLh_{O_i}\xi|^2_{\sg}+|\sLh_{O_i}\xi|^2_{\sg}\right\}
\label{11.145}
\end{equation}
hence, summing over $i=1,2,3$,
\begin{equation}
|u|^2\sum_i|\snab\sLh_{O_i}\xi|^2_{\sg}\leq
C_p\left\{\sum_{i,j}|\sLh_{O_j}\sLh_{O_i}|\xi|^2_{\sg}
+\sum_i|\sLh_{O_i}\xi|^2_{\sg}\right\} \label{11.146}
\end{equation}
We now express $\sLh_{O_i}\snab\xi$ in terms of
$\snab\sLh_{O_i}\xi$. From \ref{11.120}, \ref{11.121} applied to
$\xi$ and to $\snab\xi$ we deduce:
\begin{equation}
\sLh_{O_i}\snab\xi-\snab\sLh_{O_i}\xi=\sL_{O_i}\snab\xi-\snab\sL_{O_i}\xi+\frac{1}{2}(\snab\s^{(O_i)}\spi,\xi)_{\sg}\otimes\sg
\label{11.147}
\end{equation}
Together with the formula \ref{11.134} this implies that:
\begin{equation}
\sum_i|\sLh_{O_i}\snab\xi|^2\leq
2\sum_i|\snab\sLh_{O_i}\xi|^2+2\left(\sum_i(2p^2|\s^{(O_i)}\spi_1|^2
+|\snab\s^{(O_i)}\spi|^2)\right)|\xi|^2_{\sg} \label{11.148}
\end{equation}
Applying \ref{11.124} to the $p+1$ covariant $S$ tensorfield
$\snab\xi$ which is likewise symmetric and trace-free in the last
two indices we obtain:
\begin{equation}
|u|^2|\snab^{\ 2}\xi|^2_{\sg}\leq
C_p\left\{\sum_i|\sLh_{O_i}\snab\xi|^2_{\sg}+|\snab\xi|^2_{\sg}\right\}
\label{11.149}
\end{equation}
Moreover, by \ref{11.124} applied to $\xi$ itself we have:
\begin{equation}
|u|^2|\snab\xi|^2_{\sg}\leq
C_p\left\{\sum_i|\sLh_{O_i}\xi|^2_{\sg}+|\xi|^2_{\sg}\right\}
\label{11.150}
\end{equation}
hence, substituting in \ref{11.149},
\begin{equation}
|u|^4|\snab^{ \ 2}\xi|^2_{\sg}\leq
C_p\left\{|u|^2\sum_i|\sLh_{O_i}\snab\xi|^2_{\sg}
+\sum_i|\sLh_{O_i}\xi|^2_{\sg}+|\xi|^2_{\sg}\right\}
\label{11.151}
\end{equation}
Combining \ref{11.151}, \ref{11.148} and \ref{11.146} we conclude
that:
\begin{eqnarray}
&&|u|^4|\snab^{ \ 2}\xi|^2_{\sg}\leq
C_p\left\{\sum_{i,j}|\sLh_{O_j}\sLh_{O_i}\xi|^2_{\sg}
+\sum_i|\sLh_{O_i}\xi|^2_{\sg}+|\xi|^2_{\sg}\right.\nonumber\\
&&\hspace{25mm}\left.+|u|^2\left(\sum_i(|\s^{(O_i)}\spi_1|^2_{\sg}+|\snab\s^{(O_i)}\spi|^2_{\sg})\right)|\xi|^2_{\sg}\right\}
\label{11.152}
\end{eqnarray}
Now the integral on $S_{\ub,u}$ of the last term in parenthesis on
the right is bounded by:
\begin{equation}
|u|^2\sum_i(\|\s^{(O_i)}\spi_1\|^2_{L^4(S_{\ub,u})}+\|\snab\s^{(O_i)}\spi|^2_{L^4(S_{\ub,u})})
\|\xi\|^2_{L^4(S_{\ub,u})} \label{11.153}
\end{equation}
Again, by Lemma 5.1 with $p=4$ in that lemma and Lemma 10.1 there
is a numerical constant $C$ such that:
\begin{eqnarray}
&&|u|\|\xi\|^2_{L^4(S_{\ub,u})}\leq
C\left\{|u|^2\|\snab\xi\|^2_{L^2(S_{\ub,u})}+\|\xi\|^2_{L^2(S_{\ub,u})}\right\}
\nonumber\\
&&\hspace{22mm}\leq
C_p\int_{S_{\ub,u}}\left\{\sum_i|\sLh_{O_i}\xi|^2_{\sg}+|\xi|^2_{\sg}\right\}d\mu_{\sg}
\label{11.154}
\end{eqnarray}
the last step by \ref{11.150} integrated on $S_{\ub,u}$. Then by
virtue of assumption {\bf B3} \ref{11.153} is bounded by:
\begin{equation}
C_p\int_{S_{\ub,u}}\left\{\sum_i|\sLh_{O_i}\xi|^2_{\sg}+|\xi|^2_{\sg}\right\}d\mu_{\sg}
\label{11.155}
\end{equation}
Integrating \ref{11.152} on $S_{\ub,u}$ we then conclude that in
the case that $\xi$ is a $p$ covariant $S$ tensorfield on
$M^\prime$ which is symmetric and trace-free in the last two
indices we have:
\begin{equation}
\int_{S_{\ub,u}}|u|^4|\snab^{ \ 2}\xi|^2_{\sg}d\mu_{\sg}\leq
C_p\int_{S_{\ub,u}}
\left\{\sum_{i,j}|\sLh_{O_j}\sLh_{O_i}\xi|^2_{\sg}+\sum_i|\sLh_{O_i}\xi|^2_{\sg}+|\xi|^2_{\sg}\right\}d\mu_{\sg}
\label{11.156}
\end{equation}

\chapter{Weyl Fields and Currents. The Existence Theorem}

\section{Weyl fields and Bianchi equations. Weyl currents}

We begin by recalling certain fundamental concepts from [C-K]. A
{\em Weyl field} on a general 4-dimensional spacetime manifold
$(M,g)$ is defined to be a tensorfield on $(M,g)$ with the same
algebraic properties as the Weyl - or conformal - curvature
tensor, that is, in components relative to an arbitrary local
frame field, the antisymmetry in the first two as well as in the
last two indices:
\begin{equation}
W_{\beta\alpha\gamma\delta}=W_{\alpha\beta\delta\gamma}=-W_{\alpha\beta\gamma\delta}
\label{12.1}
\end{equation}
the cyclic condition:
\begin{equation}
W_{\alpha[\beta\gamma\delta]}:=W_{\alpha\beta\gamma\delta}+W_{\alpha\gamma\delta\beta}+W_{\alpha\delta\beta\gamma}=0
\label{12.2}
\end{equation}
and the trace condition:
\begin{equation}
(g^{-1})^{\mu\nu}W_{\mu\alpha\nu\beta}=0 \label{12.3}
\end{equation}
As is well known, the first two properties imply the symmetry
under exchange of the first and second pair of indices:
\begin{equation}
W_{\gamma\delta\alpha\beta}=W_{\alpha\beta\gamma\delta}
\label{12.4}
\end{equation}

Given a Weyl field $W$ we can define a {\em right dual} $W^*$ as
well as a {\em left dual} $\s^* W$. The right dual is defined as
\begin{equation}
W^*_{\alpha\beta\gamma\delta}=\frac{1}{2}W_{\alpha\beta}^{\s\s\mu\nu}\epsilon_{\mu\nu\gamma\delta}
\label{12.5}
\end{equation}
by freezing the first pair of indices and considering $W$ as a
2-form in the second pair. The left dual is defined as
\begin{equation}
\s^*W_{\alpha\beta\gamma\delta}=\frac{1}{2}W^{\mu\nu}_{\s\s\gamma\delta}\epsilon_{\mu\nu\alpha\beta}
\label{12.6}
\end{equation}
by freezing the second pair of indices and considering $W$ as a
2-form in the first pair. In \ref{12.5}, \ref{12.6}
$\epsilon_{\alpha\beta\gamma\delta}$ are the components of the
volume 4-form of $(M,g)$ and the indices are raised  with respect
to $g$. Now, by virtue of the algebraic properties of $W$ the two
duals coincide:
\begin{equation}
\s^*W=W^* \label{12.7}
\end{equation}
We shall thus only write $\s^*W$ in the following. Moreover,
$\s^*W$ is also a Weyl field. In fact, the cyclic condition for
$\s^*W$ is equivalent - modulo the other conditions - to the trace
condition for $W$, and vise-versa.

Given a Weyl field $W$ and a vectorfield $X$, the Lie derivative
with respect to $X$ of $W$, ${\cal L}_X W$, is not in general a
Weyl field, because it does not in general satisfy the trace
condition \ref{12.3}. We can however define a modified Lie
derivative, $\tilde{{\cal L}}_X W$, which is a Weyl field:
\begin{eqnarray}
&&(\tilde{{\cal L}}_X W)_{\alpha\beta\gamma\delta}=({\cal L}_X W)_{\alpha\beta\gamma\delta}\nonumber\\
&&\hspace{6mm}-\frac{1}{2}(\s^{(X)}\tilde{\pi}_{\alpha}^{\s\mu}W_{\mu\beta\gamma\delta}
+\s^{(X)}\tilde{\pi}_{\beta}^{\s\mu}W_{\alpha\mu\gamma\delta}
+\s^{(X)}\tilde{\pi}_{\gamma}^{\s\mu}W_{\alpha\beta\mu\delta}
+\s^{(X)}\tilde{\pi}_{\delta}^{\s\mu}W_{\alpha\beta\gamma\mu})\nonumber\\
&&\hspace{23mm}-\frac{1}{8}\mbox{tr}\s^{(X)}\pi
W_{\alpha\beta\gamma\delta} \label{12.8}
\end{eqnarray}
Here $\s^{(X)}\tilde{\pi}$ is the deformation tensor of $X$ (see
Chapter 8). The modified Lie derivative commutes with duality:
\begin{equation}
\tilde{{\cal L}}_X(\s^*W)=\s^*(\tilde{{\cal L}}_X W) \label{12.9}
\end{equation}
The fundamental Weyl field $W$ shall be the curvature tensor $R$
of our solution $(M,g)$ of the vacuum Einstein equations. By
virtue of the vanishing of the Ricci curvature the curvature
tensor reduces to Weyl tensor. The derived Weyl fields shall be
the iterated modified Lie derivatives of $R$ with respect to the
commutation fields $L$, $S$ and $O_i:i=1,2,3$. In fact we shall
consider the following derived Weyl fields:
\begin{eqnarray}
\mbox{1st order:}&&\hspace{5mm} \tcL_L R, \ \ \tcL_S R, \ \ \tcL_{O_i}R:i=1,2,3\nonumber\\
\mbox{2nd order:}&&\hspace{5mm} \tcL_L\tcL_L R, \ \ \tcL_S\tcL_S R, \ \ \tcL_{O_j}\tcL_{O_i} R:i,j=1,2,3,\nonumber\\
&&\hspace{5mm} \tcL_{O_i}\tcL_L R:i=1,2,3, \ \ \tcL_{O_i}\tcL_S
R:i=1,2,3 \ \ \label{12.10}
\end{eqnarray}

The {\em  homogeneous Bianchi equations} for a Weyl field $W$ are
the equations:
\begin{equation}
\nabla_{[\alpha}W_{\beta\gamma]\delta\epsilon}=0 \label{12.11}
\end{equation}
where, as usual, we denote by $[\alpha\beta\gamma]$ the cyclic
sum. We can write these equations as:
\begin{equation}
DW=0 \label{12.12}
\end{equation}
to emphasize the analogy with the exterior derivative. Thus the
homogeneous Bianchi equations are seen to be analogues of the
homogeneous Maxwell equations for the electromagnetic field $F$, a
2-form on $(M,g)$:
$$d F=0$$
However, $D$ is not an exterior differential operator, so $D^2\neq
0$. The equation
$$D^2 W=0,$$
a differential consequence of the homogenenous Bianchi equations,
reduces in fact to the following algebraic condition:
\begin{equation}
R_{\mu}^{\s\alpha\beta\gamma}\s^*W_{\nu\alpha\beta\gamma}-R_{\nu}^{\s\alpha\beta\gamma}\s^*W_{\mu\alpha\beta\gamma}
=0 \label{12.13}
\end{equation}
Here $R$ is the curvature tensor of the underlying metric $g$. One
may inquire about the analogues of the other Maxwell equations,
which are in general inhomogeneous, but which in the absence of
sources read:
$$d\s^*F=0$$
The remarkable fact here, which follows from the equality
\ref{12.7}, is that the equations:
\begin{equation}
D\s^*W=0 \label{12.14}
\end{equation}
are {\em equivalent} to the equations $DW=0$. In components, the
equations
$$D\s^*W=0$$
read:
\begin{equation}
\nabla^{\alpha}W_{\alpha\beta\gamma\delta}=0 \label{12.15}
\end{equation}

The fundamental Weyl field, the curvature tensor $R$ of our
solution of the vacuum Einstein equations, satisfies the
homogeneous Bianchi equations, for, these become in this case
simply the Bianchi identities for the curvature. However the
derived Weyl fields do not satisfy the homogeneous Bianchi
equations but rather equations of the form:
\begin{equation}
\nabla^{\alpha}W_{\alpha\beta\gamma\delta}=J_{\beta\gamma\delta}
\label{12.16}
\end{equation}
We call these equations {\em inhomogeneous Bianchi equations} and
the right hand side $J$ we call {\em Weyl current}. In fact, given
a Weyl field $W$ equation \ref{12.16} may be thought of as simply
the definition of the corresponding Weyl current $J$. The
algebraic properties of a Weyl current follow from those of a Weyl
field and are the following. The antisymmetry in the last two
indices:
\begin{equation}
J_{\beta\delta\gamma}=-J_{\beta\gamma\delta} \label{12.17}
\end{equation}
the cyclic condition:
\begin{equation}
J_{[\beta\gamma\delta]}:=J_{\beta\gamma\delta}+J_{\gamma\delta\beta}+J_{\delta\beta\gamma}=0
\label{12.18}
\end{equation}
and the trace condition:
\begin{equation}
(\mbox{tr}
J)_{\gamma}:=(g^{-1})^{\beta\delta}J_{\beta\gamma\delta}=0
\label{12.19}
\end{equation}
The inhomogeneous Bianchi equations \ref{12.16} are equivalent to
the equations:
\begin{equation}
\nabla^{\alpha}\s^*W_{\alpha\beta\gamma\delta}=J^*_{\beta\gamma\delta}
\label{12.20}
\end{equation}
where the dual Weyl current $J^*$ is defined by:
\begin{equation}
J^*_{\beta\gamma\delta}=\frac{1}{2}J_{\beta}^{\s\mu\nu}\epsilon_{\mu\nu\gamma\delta}
\label{12.21}
\end{equation}
Equations \ref{12.20} are in turn equivalent to the equations:
\begin{equation}
\nabla_{[\alpha}W_{\beta\gamma]\delta\epsilon}=\epsilon_{\mu\alpha\beta\gamma}J^{*\mu}_{\s\s\delta\epsilon}
\label{12.22}
\end{equation}
while equations \ref{12.16} are equivalent to the equations:
\begin{equation}
\nabla_{[\alpha}\s^*W_{\beta\gamma]\delta\epsilon}=-\epsilon_{\mu\alpha\beta\gamma}J^{\mu}_{\s\delta\epsilon}
\label{12.23}
\end{equation}
Thus, all four sets of equations are equivalent.

The dual $J^*$ of a Weyl current $J$ is also a Weyl current. In
fact, the cyclic condition for $J^*$ is equivalent - modulo the
remaining condition - equivalent to the trace condition for $J$
and vice-versa.

Given a Weyl current $J$ and a vectorfield $X$, the Lie derivative
with respect to $X$ of $J$, ${\cal L}_X J$, is not in general a
Weyl current, because it does not in general satisfy the trace
condition \ref{12.19}. We can however define a modifed Lie
derivative, $\tcL_X J$, which is a Weyl current:
\begin{eqnarray}
&&(\tcL_X J)_{\beta\gamma\delta}=({\cal L}_X J)_{\beta\gamma\delta}\nonumber\\
&&\hspace{6mm}-\frac{1}{2}(\s^{(X)}\tilde{\pi}_{\beta}^{\s\mu}J_{\mu\gamma\delta}
+\s^{(X)}\tilde{\pi}_{\gamma}^{\s\gamma}J_{\beta\mu\delta}
+\s^{(X)}\tilde{\pi}_{\delta}^{\s\mu}J_{\beta\gamma\mu})\nonumber\\
&&\hspace{23mm}+\frac{1}{8}\mbox{tr}\s^{(X)}\pi
J_{\beta\gamma\delta} \label{12.24}
\end{eqnarray}
Again, the modified Lie derivative commutes with duality:
\begin{equation}
\tcL_X (J^*)=(\tcL_X J)^* \label{12.25}
\end{equation}

We consider next the conformal properties of the Bianchi
equations. If the Weyl field $W$ is a solution of the Bianchi
equations \ref{12.16} on $(M,g)$ with Weyl current $J$, then, for
any conformal factor $\Omega$, setting:
\begin{equation}
g^\prime=\Omega^{-2}g, \ \ W^\prime=\Omega^{-1}W, \ \
J^\prime=\Omega J \label{12.26}
\end{equation}
the Weyl field $W^\prime$ satisfies the Bianchi equations on
$(M,g^\prime)$ with Weyl current $J^\prime$:
\begin{equation}
\nabla^{\prime\alpha}W^\prime_{\alpha\beta\gamma\delta}=J^\prime_{\beta\gamma\delta}
\label{12.27}
\end{equation}
It follows that if $f$ is a conformal isometry of $(M,g)$, that
is, we have:
\begin{equation}
f^* g=\Omega^2 g \label{12.28}
\end{equation}
for some conformal factor $\Omega$, and if the Weyl field $W$ is a
solution of the Bianchi equations on $(M,g)$ with Weyl current
$J$, then the Weyl field $W^\prime=\Omega^{-1}f^* W$  is a
solution of the same equations on the same manifold $(M,g)$ with
Weyl current $J^\prime=\Omega f^*J$. Suppose now that $X$ is a
vectorfield generating a 1-parameter group $f_t$ of conformal
isometries of $(M,g)$, that is, a conformal Killing field. Then if
$W$ is a solution of the Bianchi equations on $(M,g)$ with Weyl
current $J$, then the Weyl field $\Omega_t^{-1}f_t^*W$ is a
solution of the same equations on the same manifold $(M,g)$ with
Weyl current $\Omega_t f^*J$. It follows, in view of the linearity
of the Bianchi equations that the Weyl field
\begin{equation}
\left.\frac{d}{dt}\Omega_t^{-1}f_t^*W\right|_{t=0}=\tcL_X W
\label{12.29}
\end{equation}
is likewise a solution of the same equations with Weyl current:
\begin{equation}
\left.\frac{d}{dt}\Omega_tf_t^*J\right|_{t=0}=\tcL_X J
\label{12.30}
\end{equation}
We see that the terms $-(1/8)\mbox{tr}\s^{(X)}\pi W$ in $\tcL_X W$
and $(1/8)\mbox{tr}\s^{(X)}\pi J$ in $\tcL_X J$ come from the
conformal weights $\Omega^{-1}$ and $\Omega$ respectively.

Off course $(M,g)$ does not in general admit a non-trivial
conformal Killing field. The following fundamental proposition
from [C-K] shows the commutation properties of the modified Lie
derivative with respect to an arbitrary vectorfield with the
Bianchi equations.

\vspace{5mm}

\noindent{\bf Proposition 12.1} \ \ \ Let $W$ be a Weyl field and
$J$ the corresponding Weyl current, so the Bianchi equations
$$\nabla^{\alpha}W_{\alpha\beta\gamma\delta}=J_{\beta\gamma\delta}$$
hold. Let also $X$ be an arbitrary vectorfield with deformation
tensor $\s^{(X)}\tilde{\pi}$. Then the derived Weyl field $\tcL_X
W$ satisfies the Bianchi equations
$$\nabla^{\alpha}(\tcL_X W)_{\alpha\beta\gamma\delta}=\s^{(X)}J(W)_{\beta\gamma\delta}$$
where the derived Weyl current $\s^{(X)}J(W)$ is given by:
\begin{eqnarray*}
&&\s^{(X)}J(W)_{\beta\gamma\delta}=(\tcL_X J)_{\beta\gamma\delta}
+\frac{1}{2}\s^{(X)}\tilde{\pi}^{\mu\nu}\nabla_{\nu}W_{\mu\beta\gamma\delta}\\
&&\hspace{6mm}+\frac{1}{2}\s^{(X)}p_{\mu}W^{\mu}_{\s\beta\gamma\delta}
+\frac{1}{2}(\s^{(X)}q_{\mu\beta\nu}W^{\mu\nu}_{\s\s\gamma\delta}
+\s^{(X)}q_{\mu\gamma\nu}W^{\mu\s\nu}_{\s\beta\s\delta}+\s^{(X)}q_{\mu\delta\nu}W^{\mu\s\s\nu}_{\s\beta\gamma})
\end{eqnarray*}
Here:
$$\s^{(X)}p_{\beta}=\nabla^{\alpha}\s^{(X)}\tilde{\pi}_{\alpha\beta}$$
and:
$$\s^{(X)}q_{\alpha\beta\gamma}=\nabla_{\beta}\s^{(X)}\tilde{\pi}_{\gamma\alpha}
-\nabla_{\gamma}\s^{(X)}\tilde{\pi}_{\beta\alpha}
+\frac{1}{3}(\s^{(X)}p_{\beta}g_{\alpha\gamma}-\s^{(X)}p_{\gamma}g_{\alpha\beta})$$
The 3-covariant tensorfield $\s^{(X)}q$ has the algebraic
properties of a Weyl current.

\vspace{5mm}

\section{Null decompositions of Weyl fields and currents}

Le $(e_A \ : \ A=1,2)$ be an arbitrary local frame field for the
$S_{\ub,u}$. We complement $(e_A \ : \ A=1,2)$ with the
vectorfields $e_3=\Lbh$ , $e_4=\Lh$ to obtain a frame field
$(e_{\mu} \ : \ \mu=1,2,3,4)$ for $M$, as in Chapter 1. The
algebraically independent components of an arbitrary Weyl field
$W$ on $M$ are, as in the case of the fundamental Weyl field $R$,
the curvature tensor of our solution $(M,g)$ of the vacuum
Einstein equations, the trace-free symmetric 2-covariant $S$
tensorfields $\alpha(W)$ and $\alb(W)$ given by:
\begin{equation}
W_{A3B3}=\alb_{AB}(W), \ \ \ W_{A4B4}=\alpha_{AB}(W) \label{12.31}
\end{equation}
the $S$ 1-forms $\beta(W)$ and $\beb(W)$ given by:
\begin{equation}
W_{A334}=2\beb_A(W), \ \ \ W_{A434}=2\beta_A(W) \label{12.32}
\end{equation}
and the functions $\rho(W)$ and $\sigma(W)$ given by:
\begin{equation}
W_{3434}=4\rho(W), \ \ \ W_{AB34}=2\sigma(W)\seps_{AB}
\label{12.33}
\end{equation}
The remaining components of $W$ are expressed in terms of these
by:
\begin{eqnarray}
&W_{A3BC}=\s^*\beb_A(W)\seps_{BC}=\sg_{AB}\beb_C(W)-\sg_{AC}\beb_B(W) \nonumber\\
&W_{A4BC}=-\s^*\beta_A(W)\seps_{BC}=-\sg_{AB}\beta_C(W)+\sg_{AC}\beta_B(W)\nonumber\\
&W_{A3B4}=-\rho(W)\sg_{AB}+\sigma(W)\seps_{AB}\nonumber\\
&W_{ABCD}=-\rho(W)\seps_{AB}\seps_{CD}=-\rho(W)(\sg_{AC}\sg_{BD}-\sg_{AD}\sg_{BC})
\label{12.34}
\end{eqnarray}
(compare with \ref{1.154}). In the case of the fundamental Weyl
field $R$, we shall, as hitherto, omit the reference to $W$.

The components of $\s^*W$, the dual of $W$, are expressed in terms
of the components of $W$ by:
\begin{eqnarray}
&\alb(\s^*W)=\s^*\alb(W), \ \ \ \alpha(\s^*W)=-\s^*\alpha(W)\nonumber\\
&\beb(\s^*W)=\s^*\beb(W), \ \ \ \beta(\s^*W)=-\s^*\beta(W)\nonumber\\
&\rho(\s^*W)=\sigma(W), \ \ \ \sigma(\s^*W)=-\rho(W) \label{12.35}
\end{eqnarray}
Here $\s^*\alb$, $\s^*\alpha$, $\s^*\beb$, $\s^*\beta$ denote the
left duals relative to the $S_{\ub,u}$ of $\alb$, $\alpha$,
$\beb$, $\beta$ respectively. The left dual $\s^*\theta$ of a
trace-free symmetric 2-covariant $S$ tensorfield $\theta$ is given
by:
\begin{equation}
\s^*\theta_{AB}=\seps_A^{\s C}\theta_{CB} \label{12.36}
\end{equation}
We note that $\s^*\theta$ is also a trace-free symmetric
2-covariant $S$ tensorfield. In fact, the symmetry condition for
$\s^*\theta$ is equivalent to the trace condition for $\theta$ and
vice-versa.

We shall now analyze the relationship, for a commutation field $Y$
and Weyl field $W$, between the null components of $\tcL_Y W$ and
$\sL_Y$ applied to the null components of $W$. We shall need the
following lemma. Let us define:
\begin{equation}
\s^{(Y)}\nu=\left\{
\begin{array}{lll}
Y\log\Omega&:&\mbox{for $Y=L,O_i:i=1,2,3$}\\
1+Y\log\Omega&:&\mbox{for $Y=S$}
\end{array}
\right. \label{12.37}
\end{equation}

\noindent{\bf Lemma 12.1} \ \ \ For all five commutation fields
$L$, $O_i:i=1,2,3$, $S$ we have:
\begin{eqnarray*}
&&[Y,e_A]=\Pi[Y,e_A]\\
&&[Y,e_3]=-\s^{(Y)}\mb^A e_A-\s^{(Y)}\nu e_3\\
&&[Y,e_4]=-\s^{(Y)}m^A e_A-\s^{(Y)}\nu e_4
\end{eqnarray*}

\noindent{\em Proof:} \ According to Lemma 1.1 and the commutation
formula \ref{1.75} we have:
\begin{eqnarray}
&&[L,e_A]=\Pi[L,e_A]\nonumber\\
&&[L,e_3]=-4\Omega\zeta^A e_A-(L\log\Omega)e_3\nonumber\\
&&[L,e_4]=-(L\log\Omega)e_4 \label{12.38}
\end{eqnarray}
and:
\begin{eqnarray}
&&[\Lb,e_A]=\Pi[\Lb,e_A]\nonumber\\
&&[\Lb,e_3]=-(\Lb\Omega)e_3\nonumber\\
&&[\Lb,e_4]=4\Omega\zeta^A e_A-(\Lb\log\Omega)e_4 \label{12.39}
\end{eqnarray}
In the case $Y=L$ the lemma follows immediately from \ref{12.38}
in conjunction with the table \ref{8.21}. In the case $Y=S$ the
lemma follows from \ref{12.38}, \ref{12.39} and the facts that
$$\Lb u=1, \ \ Lu=e_A u=0; \ \ \ L\ub=1, \Lb\ub=e_A\ub=0$$
in conjunction with table \ref{8.30}.  Finally, in the case $Y=O_i$
we have, since both $O_i$ and $e_A$ are $S$-tangential
vectorfields:
\begin{equation}
[O_i,e_A]=\Pi[O_i,e_A] \label{12.40}
\end{equation}
Also, according to the first of \ref{8.38}:
\begin{equation}
[O_i,e_3]=-(O_i\log\Omega)e_3 \label{12.41}
\end{equation}
and according to \ref{8.59}:
\begin{equation}
[O_i,e_4]=-\Omega^{-1}Z_i-(O_i\log\Omega)e_4 \label{12.42}
\end{equation}
The lemma in the case $Y=O_i$ then follows from \ref{12.40} -
\ref{12.42} in conjunction with \ref{8.135}, \ref{8.136}.

\vspace{5mm}

\noindent{\bf Proposition 12.2} \ \ \ Let $Y$ be any of the five
commutation fields $L$, $O_i:i=1,2,3$, $S$ and let $W$ be an
arbitrary Weyl field. Then the null components of $\tcL_Y W$ are
given by:
\begin{eqnarray*}
&&\alpha_{AB}(\tcL_Y W)=(\sLh_Y\alpha(W))_{AB}+\s^{(Y)}\nu\alpha_{AB}(W)-\frac{1}{4}\s^{(Y)}j\alpha_{AB}(W)\\
&&\hspace{22mm}+(\s^{(Y)}m\oth\beta(W))_{AB}\\
&&\alb_{AB}(\tcL_Y W)=(\sLh_Y\alb(W))_{AB}+\s^{(Y)}\nu\alb_{AB}(W)-\frac{1}{4}\s^{(Y)}j\alb_{AB}(W)\\
&&\hspace{22mm}-(\s^{(Y)}\mb\oth\beb(W))_{AB}\\
&&\beta_A(\tcL_Y
W)=(\sL_Y\beta(W))_A+2\s^{(Y)}\nu\beta_A(W)+\frac{1}{4}\s^{(Y)}j\beta_A(W)
-\frac{1}{2}\s^{(Y)}\ih_A^{\s B}\beta_B(W)\\
&&\hspace{20mm}+\frac{1}{4}\s^{(Y)}\mb^B\alpha_{AB}(W)+\frac{3}{4}\s^{(Y)}m_A\rho(W)
+\frac{3}{4}\s^{*(Y)}m_A\sigma(W)\\
&&\beb_A(\tcL_Y
W)=(\sL_Y\beb(W))_A+2\s^{(Y)}\nu\beb_A(W)+\frac{1}{4}\s^{(Y)}j\beb_A(W)
-\frac{1}{2}\s^{(Y)}\ih_A^{\s B}\beb_B(W)\\
&&\hspace{20mm}-\frac{1}{4}\s^{(Y)}m^B\alb_{AB}(W)-\frac{3}{4}\s^{(Y)}\mb_A\rho(W)
+\frac{3}{4}\s^{*(Y)}\mb_A\sigma(W)\\
&&\rho(\tcL_Y W)=Y\rho(W)+3\s^{(Y)}\nu\rho(W)+\frac{3}{4}\s^{(Y)}j\rho(W)\\
&&\hspace{18mm}-\frac{1}{2}\s^{(Y)}m^A\beb_A(W)+\frac{1}{2}\s^{(Y)}\mb^A\beta_A(W)\\
&&\sigma(\tcL_Y W)=Y\sigma(W)+3\s^{(Y)}\nu\sigma(W)+\frac{3}{4}\s^{(Y)}j\sigma(W)\\
&&\hspace{18mm}+\frac{1}{2}\s^{*(Y)}m^A\beb_A(W)+\frac{1}{2}\s^{*(Y)}\mb^A\beta_A(W)
\end{eqnarray*}

\noindent{\em Proof:} \ From the last of \ref{8.20}, \ref{8.29}
and from \ref{8.43} we have:
\begin{equation}
\s^{(Y)}\pi_{34}=-4\s^{(Y)}\nu \label{12.46}
\end{equation}
for all five commutation fields $Y$. It follows that:
\begin{equation}
\mbox{tr}\s^{(Y)}\pi=2\s^{(Y)}j+8\s^{(Y)}\nu \label{12.47}
\end{equation}
and:
\begin{equation}
\mbox{tr}\s^{(Y)}\spi=2\s^{(Y)}j+4\s^{(Y)}\nu \label{12.49}
\end{equation}
for all five commutation fields $Y$.

Using Lemma 12.1, the fact that
\begin{equation}
\sL_Y\seps=\frac{1}{2}\mbox{tr}\s^{(Y)}\spi\seps \label{12.50}
\end{equation}
and \ref{12.49}, we derive the following formulas:
\begin{eqnarray}
&&({\cal L}_Y W)_{A4B4}=(\sL_Y\alpha(W))_{AB}+2\s^{(Y)}\nu\alpha_{AB}(W)\nonumber\\
&&\hspace{22mm}+\s^{(Y)}m_A\beta_B(W)+\s^{(Y)}m_B\beta_A(W)-2\s^{(Y)}m^C\beta_C(W)\sg_{AB}\nonumber\\
&&({\cal L}_Y W)_{A3B3}=(\sL_Y\alb(W))_{AB}+2\s^{(Y)}\nu\alb_{AB}(W)\nonumber\\
&&\hspace{22mm}-\s^{(Y)}\mb_A\beb_B(W)-\s^{(Y)}\mb_B\beb_A(W)+2\s^{(Y)}\mb^C\beb_C(W)\sg_{AB}\nonumber\\
&&\frac{1}{2}({\cal L}_Y W)_{A434}=(\sL_Y\beta(W))_A+3\s^{(Y)}\nu\beta_A(W)\nonumber\\
&&\hspace{24mm}+\frac{1}{2}\s^{(Y)}\mb^B\alpha_{AB}(W)+\frac{1}{2}\s^{(Y)}m_A\rho(W)+\frac{3}{2}\s^{*(Y)}m_A\sigma(W)
\nonumber\\
&&\frac{1}{2}({\cal L}_Y W)_{A334}=(\sL_Y\beb(W))_A+3\s^{(Y)}\nu\beb_A(W)\nonumber\\
&&\hspace{24mm}-\frac{1}{2}\s^{(Y)}m^B\alb_{AB}(W)-\frac{1}{2}\s^{(Y)}\mb_A\rho(W)+\frac{3}{2}\s^{*(Y)}\mb_A\sigma(W)
\nonumber\\
&&\frac{1}{4}({\cal L}_Y W)_{3434}=Y\rho(W)+4\s^{(Y)}\nu\rho(W)\nonumber\\
&&\hspace{24mm}+\s^{(Y)}\mb^A\beta_A(W)-\s^{(Y)}m^A\beb_A(W)\nonumber\\
&&\frac{1}{4}\seps^{AB}({\cal L}_Y W)_{AB34}=Y\sigma(W)+4\s^{(Y)}\nu\sigma(W)+\s^{(Y)}j\sigma(W)\nonumber\\
&&\hspace{28mm}+\frac{1}{2}\s^{*(Y)}\mb^A\beta_A(W)+\frac{1}{2}\s^{*(Y)}m^A\beb_A(W)
\label{12.43}
\end{eqnarray}

Consider next the expression
\begin{equation}
\s^{(Y)}[W]_{\alpha\beta\gamma\delta}=\s^{(Y)}\tilde{\pi}_{\alpha}^{\s\mu}W_{\mu\beta\gamma\delta}
+\s^{(Y)}\tilde{\pi}_{\beta}^{\s\mu}W_{\alpha\mu\gamma\delta}
+\s^{(Y)}\tilde{\pi}_{\gamma}^{\s\mu}W_{\alpha\beta\mu\delta}
+\s^{(Y)}\tilde{\pi}_{\delta}^{\s\mu}W_{\alpha\beta\gamma\mu}
\label{12.44}
\end{equation}
in formula \ref{12.8}. By a straightforward computation we find:
\begin{eqnarray}
&&\s^{(Y)}[W]_{A4B4}=\s^{(Y)}\ih_A^{\s
C}\alpha_{CB}(W)+\s^{(Y)}\ih_B^{\s C}\alpha_{CA}(W)
-2\s^{(Y)}m^C\beta_C(W)\sg_{AB}\nonumber\\
&&\s^{(Y)}[W]_{A3B3}=\s^{(Y)}\ih_A^{\s
C}\alb_{CB}(W)+\s^{(Y)}\ih_B^{\s C}\alb_{CA}(W)
+2\s^{(Y)}\mb^C\beb_C(W)\sg_{AB}\nonumber\\
&&\frac{1}{2}\s^{(Y)}[W]_{A434}=-\s^{(Y)}j\beta_A(W)+\s^{(Y)}\ih_A^{\s B}\beta_B(W)\nonumber\\
&&\hspace{22mm}+\frac{1}{2}\s^{(Y)}\mb^B\alpha_{AB}(W)-\frac{1}{2}\s^{(Y)}m_A\rho(W)
+\frac{3}{2}\s^{*(Y)}m^B\sigma(W)\nonumber\\
&&\frac{1}{2}\s^{(Y)}[W]_{A334}=-\s^{(Y)}j\beb_A(W)+\s^{(Y)}\ih_A^{\s B}\beb_B(W)\nonumber\\
&&\hspace{22mm}-\frac{1}{2}\s^{(Y)}m^B\alb_{AB}(W)+\frac{1}{2}\s^{(Y)}\mb_A\rho(W)
+\frac{3}{2}\s^{*(Y)}\mb_A\sigma(W)\nonumber\\
&&\frac{1}{4}\s^{(Y)}[W]_{3434}=-2\s^{(Y)}j\rho(W)+\s^{(Y)}\mb^A\beta_A(W)-\s^{(Y)}m^A\beb_A(W)\nonumber\\
&&\frac{1}{4}\seps^{AB}\s^{(Y)}[W]_{AB34}=0 \label{12.45}
\end{eqnarray}

In view of formulas \ref{12.43}, \ref{12.45}, \ref{12.46}, the
proposition follows through formula \ref{12.8}, if we also take
into account the fact that by virtue of the identity \ref{1.163}
for any trace-free symmatric 2-covariant $S$ tensorfield $\theta$
we have:
\begin{equation}
(\sL_Y\theta)_{AB}-\frac{1}{2}\s^{(Y)}\ih_A^{\s
C}\theta_{CB}-\frac{1}{2}\s^{(Y)}\ih_B^{\s C}\theta_{CA}
=(\sL_Y\theta)_{AB}-\frac{1}{2}(\s^{(Y)}\ih,\theta)\sg_{AB}=(\sLh_Y\theta)_{AB}
\label{12.48}
\end{equation}

\vspace{5mm}

The algebraically independent components of an arbitrary Weyl
current $J$ on $M$, or more generally of a 3-covariant tensorfield
on $M$  with the algebraic properties \ref{12.17} - \ref{12.19} of
a Weyl current, are the $S$ 1-forms $\Xi(J)$ and $\Xib(J)$ given
by:
\begin{equation}
\frac{1}{2}J_{33A}=\Xib_A(J), \ \ \ \frac{1}{2}J_{44A}=\Xi_A(J)
\label{12.51}
\end{equation}
the trace-free symmetric 2-covariant $S$ tensorfields $\Theta(J)$
and $\Thetab(J)$ to be defined below, the functions $\Lambda(J)$,
$K(J)$ and $\Lambdab(J)$, $\Kb(J)$ given by:
\begin{equation}
\frac{1}{4}J_{343}=\Lambdab(J), \ \ \
\frac{1}{4}J_{434}=\Lambda(J) \label{12.52}
\end{equation}
and:
\begin{equation}
\frac{1}{4}\seps^{AB}J_{3AB}=\Kb(J), \ \ \
\frac{1}{4}\seps^{AB}J_{4AB}=K(J) \label{12.53}
\end{equation}
and the $S$ 1-forms $I(J)$ and $\Ib(J)$ given by:
\begin{equation}
\frac{1}{2}J_{43A}=\Ib_A(J), \ \ \ \frac{1}{2}J_{34A}=I_A(J)
\label{12.54}
\end{equation}
To define the components $\Thetab(J)$ and $\Theta(J)$ we consider
the 2-covariant $S$ tensorfields $\sJ_3$ and $\sJ_4$ given by:
\begin{equation}
J_{A3B}=(\sJ_3)_{AB}, \ \ \ J_{A4B}=(\sJ_4)_{AB} \label{12.55}
\end{equation}
Then $\Thetab(J)$ and $\Theta(J)$ are the trace-free symmetric
parts of $\sJ_3$ and $\sJ_4$ respectively:
\begin{eqnarray}
&\Thetab(J)_{AB}=\frac{1}{2}((\sJ_3)_{AB}+(\sJ_3)_{BA}-\sg_{AB}\mbox{tr}\sJ_3)\nonumber\\
&\Theta(J)_{AB}=\frac{1}{2}((\sJ_4)_{AB}+(\sJ_4)_{BA}-\sg_{AB}\mbox{tr}\sJ_4)
\label{12.56}
\end{eqnarray}
Note that by the trace condition \ref{12.19} we have:
\begin{equation}
\mbox{tr}\sJ_3=-2\Lambdab(J), \ \ \ \mbox{tr}\sJ_4=-2\Lambda(J)
\label{12.57}
\end{equation}
while by the cyclic condition \ref{12.18} the antisymmetric parts
of $\sJ_3$ and $\sJ_4$ are given by:
\begin{equation}
\frac{1}{2}((\sJ_3)_{AB}-(\sJ_3)_{BA})=\Kb(J)\seps_{AB}, \ \ \
\frac{1}{2}((\sJ_4)_{AB}-(\sJ_4)_{BA})=K(J)\seps_{AB}
\label{12.58}
\end{equation}
Therefore the 2-covariant $S$ tensorfields $\sJ_3$ and $\sJ_4$ are
expressed as:
\begin{eqnarray}
&\sJ_3=\Thetab(J)-\Lambdab(J)\sg+\Kb(J)\seps\nonumber\\
&\sJ_4=\Theta(J)-\Lambda(J)\sg+K(J)\seps \label{12.59}
\end{eqnarray}
The remaining components of $J$ are axpressed in terms of the
above components by:
\begin{eqnarray}
&J_{A34}=2\Ib_A(J)-2I_A(J), \ \ \ J_{A43}=2I_A(J)-2\Ib_A(J)\nonumber\\
&J_{ABC}=(\s^*I_A(J)+\s^*\Ib_A(J))\seps_{BC} \label{12.60}
\end{eqnarray}

The components of $J^*$, the dual of $J$, are expressed in terms
of the components of $J$ by:
\begin{eqnarray}
&\Xib(J^*)=\s^*\Xib(J), \ \ \ \Xi(J^*)=-\s^*\Xi(J)\nonumber\\
&\Thetab(J^*)=\s^*\Thetab(J), \ \ \ \Theta(J^*)=-\s^*\Theta(J)\nonumber\\
&\Lambdab(J^*)=-\Kb(J), \ \ \ \Lambda(J^*)=K(J)\nonumber\\
&\Kb(J^*)=\Lambdab(J), \ \ \ K(J^*)=-\Lambda(J)\nonumber\\
&\Ib(J^*)=\s^*I(J), \ \ \ I(J^*)=-\s^*I(J) \label{12.61}
\end{eqnarray}

The following proposition specifies the relationship, for a
commutation field $Y$ and Weyl curent $J$, between the null
components of $\tcL_Y J$ and $\sL_Y$ applied to the null
components of $J$.

\vspace{5mm}

\noindent{\bf Proposition 12.3} \ \ \ Let $Y$ be any of the five
commutation fields $L$, $O_i:i=1,2,3$, $S$ and let $J$ be an
arbitrary Weyl current, or more generally an arbitrary 3-covariant
tensorfield with the algebraic properties of a Weyl current. Then
the null components of $\tcL_Y J$ are given by:
\begin{eqnarray*}
&&\Xi_A(\tcL_Y
J)=(\sL_Y\Xi(J))_A+3\s^{(Y)}\nu\Xi_A(J)+\frac{1}{2}\s^{(Y)}j\Xi_A(J)
-\frac{1}{2}\s^{(Y)}\ih_A^{\s B}\Xi_B(J)\\
&&\hspace{20mm}+\frac{1}{4}\s^{(Y)}m^B\Theta_{AB}(J)-\frac{3}{4}\s^{(Y)}m_A\Lambda(J)-\frac{3}{4}\s^{*(Y)}m_A K(J)\\
&&\Xib_A(\tcL_Y
J)=(\sL_Y\Xib(J))_A+3\s^{(Y)}\nu\Xib_A(J)+\frac{1}{2}\s^{(Y)}j\Xib_A(J)
-\frac{1}{2}\s^{(Y)}\ih_A^{\s B}\Xib_B(J)\\
&&\hspace{20mm}+\frac{1}{4}\s^{(Y)}\mb^B\Thetab_{AB}(J)-\frac{3}{4}\s^{(Y)}\mb_A\Lambdab(J)-\frac{3}{4}\s^{*(Y)}\mb_A\Kb(J)\\
&&\Theta_{AB}(\tcL_Y J)=(\sLh_Y\Theta(J))_{AB}+2\s^{(Y)}\nu\Theta_{AB}(J)\\
&&\hspace{22mm}+\frac{1}{4}(\s^{(Y)}\mb\oth\Xi(J))_{AB}+\frac{3}{4}(\s^{(Y)}m\oth I(J))_{AB}\\
&&\Thetab_{AB}(\tcL_Y J)=(\sLh_Y\Thetab(J))_{AB}+2\s^{(Y)}\nu\Thetab_{AB}(J)\\
&&\hspace{22mm}+\frac{1}{4}(\s^{(Y)}m\oth\Xib(J))_{AB}+\frac{3}{4}(\s^{(Y)}\mb\oth\Ib(J))_{AB}\\
&&\Lambda(\tcL_Y J)=Y\Lambda(J)+4\s^{(Y)}\nu\Lambda(J)+\s^{(Y)}j\Lambda(J)\\
&&\hspace{18mm}-\frac{1}{4}(\s^{(Y)}\mb,\Xi(J))+\frac{1}{2}(\s^{(Y)}m,\Ib(J))-\frac{1}{4}(\s^{(Y)}m,I(J))\\
&&\Lambdab(\tcL_Y J)=Y\Lambdab(J)+4\s^{(Y)}\nu\Lambdab(J)+\s^{(Y)}j\Lambdab(J)\\
&&\hspace{18mm}-\frac{1}{4}(\s^{(Y)}m,\Xib(J))+\frac{1}{2}(\s^{(Y)}\mb,I(J))-\frac{1}{4}(\s^{(Y)}\mb,\Ib(J))\\
&&K(\tcL_Y J)=YK(J)+4\s^{(Y)}\nu K(J)+\s^{(Y)}jK(J)\\
&&\hspace{18mm}-\frac{1}{4}(\s^{*(Y)}\mb,\Xi(J))-\frac{1}{2}(\s^{*(Y)}m,\Ib(J))-\frac{1}{4}(\s^{*(Y)}m,I(J))\\
&&\Kb(\tcL_Y J)=Y\Kb(J)+4\s^{(Y)}\nu\Kb(J)+\s^{(Y)}j\Kb(J)\\
&&\hspace{18mm}-\frac{1}{4}(\s^{*(Y)}m,\Xib(J))-\frac{1}{2}(\s^{*(Y)}\mb,I(J))
-\frac{1}{4}(\s^{*(Y)}\mb,\Ib(J))\\
&&I_A(\tcL_Y J)=(\sL_Y I(J))_A+3\s^{(Y)}\nu
I_A(J)+\frac{1}{2}\s^{(Y)}j I_A(J)
-\frac{1}{2}\s^{(Y)}\ih_A^{\s B}I_B(J)\\
&&\hspace{20mm}+\frac{1}{4}\s^{(Y)}\mb^B\Theta_{AB}(J)-\frac{1}{4}\s^{(Y)}\mb_A\Lambda(J)-\frac{1}{4}\s^{*(Y)}\mb_A K(J)\\
&&\hspace{20mm}+\frac{1}{2}\s^{(Y)}m_A\Lambdab(J)-\frac{1}{2}\s^{*(Y)}m_A\Kb(J)\\
&&\Ib_A(\tcL_Y J)=(\sL_Y \Ib(J))_A+3\s^{(Y)}\nu
\Ib_A(J)+\frac{1}{2}\s^{(Y)}j\Ib_A(J)
-\frac{1}{2}\s^{(Y)}\ih_A^{\s B}\Ib_B(J)\\
&&\hspace{20mm}+\frac{1}{4}\s^{(Y)}m^B\Thetab_{AB}(J)-\frac{1}{4}\s^{(Y)}m_A\Lambdab(J)-\frac{1}{4}\s^{*(Y)}m_A\Kb(J)\\
&&\hspace{20mm}+\frac{1}{2}\s^{(Y)}\mb_A\Lambda(J)-\frac{1}{2}\s^{*(Y)}\mb_A
K(J)
\end{eqnarray*}

\vspace{5mm}

The proof is  similar to that of Proposition 12.2.

\vspace{5mm}

\noindent{\bf Proposition 12.4} \ \ \ The following system of
equations constitute the inhomogeneous Bianchi equations
\ref{12.16}:
\begin{eqnarray*}
&&\Omega^{-1}(\Dbh\alpha(W)+2\omb\alpha(W))-\frac{1}{2}\mbox{tr}\chib\alpha(W)-\snab\oth\beta(W)
-(4\eta+\zeta)\oth\beta(W)\\
&&\hspace{55mm}+3\chih\rho(W)+3\s^*\chih\sigma(W)
=-2\Theta(J)\\
&&\Omega^{-1}(\Dh\alb(W)+2\omega\alb(W))-\frac{1}{2}\mbox{tr}\chi\alb(W)+\snab\oth\beb(W)+(4\etb-\zeta)\oth\beb(W)\\
&&\hspace{55mm}+3\chibh\rho(W)-3\s^*\chibh\sigma(W)
=-2\Thetab(J)\\
&&\Omega^{-1}(D\beta(W)-\omega\beta(W))-\chi^\sharp\cdot\beta(W)+2\mbox{tr}\chi\beta(W)-\sdiv\alpha(W)\\
&&\hspace{55mm}-(\etb^\sharp+2\zeta^\sharp)\cdot\alpha(W)=2\Xi(J)\\
&&\Omega^{-1}(\Db\beb(W)-\omb\beb(W))-\chib^\sharp\cdot\beb(W)+2\mbox{tr}\chib\beb(W)+\sdiv\alb(W)\\
&&\hspace{55mm}+(\eta^\sharp-2\zeta^\sharp)\cdot\alb(W)=-2\Xib(J)\\
&&\Omega^{-1}(\Db\beta(W)+\omb\beta(W))-\chib^\sharp\cdot\beta(W)+\mbox{tr}\chib\beta(W)-\sd\rho(W)-\s^*\sd\sigma(W)\\
&&\hspace{35mm}-3\eta\rho(W)-3\s^*\eta\sigma(W)-2\chih^\sharp\cdot\beb(W)=-2I(J)\\
&&\Omega^{-1}(D\beb(W)+\omega\beb(W))-\chi^\sharp\cdot\beb+\mbox{tr}\chi\beb(W)+\sd\rho(W)-\s^*\sd\sigma(W)\\
&&\hspace{35mm}+3\etb\rho(W)-3\s^*\etb\sigma(W)-2\chibh^\sharp\cdot\beta(W)=2\Ib_A(J)\\
&&\Omega^{-1}D\rho(W)+\frac{3}{2}\mbox{tr}\chi\rho(W)-\sdiv\beta(W)-(2\etb+\zeta,\beta(W))+\frac{1}{2}(\chibh,\alpha(W))\\
&&\hspace{80mm}=-2\Lambda(J)\\
&&\Omega^{-1}\Db\rho(W)+\frac{3}{2}\mbox{tr}\chib\rho(W)+\sdiv\beb(W)+(2\eta-\zeta,\beb(W))+\frac{1}{2}(\chih,\alb(W))\\
&&\hspace{80mm}=-2\Lambdab(J)\\
&&\Omega^{-1}D\sigma(W)+\frac{3}{2}\mbox{tr}\chi\sigma(W)+\scurl\beta(W)+(2\etb+\zeta)\wedge\beta(W)-\frac{1}{2}\chibh\wedge\alpha(W)\\
&&\hspace{80mm}=-2K(W)\\
&&\Omega^{-1}\Db\sigma(W)+\frac{3}{2}\mbox{tr}\chib\sigma(W)+\scurl\beb(W)+(2\eta-\zeta)\wedge\beb(W)+\frac{1}{2}\chih\wedge\alb(W)\\
&&\hspace{80mm}=2\Kb(W)
\end{eqnarray*}

\vspace{5mm}

The proof is similar to that of Proposition 1.2.

\vspace{5mm}

\section{The Bel-Robinson tensor. The energy-momentum density vectorfields}

Given a Weyl field $W$, the {\em Bel-Robinson tensor} associated
to $W$ is the 4-covariant tensorfield:
\begin{equation}
Q(W)_{\alpha\beta\gamma\delta}=W_{\alpha\rho\gamma\sigma}W_{\beta\s\delta}^{\s\rho\s\sigma}
+\s^*W_{\alpha\rho\gamma\sigma}\s^*W_{\beta\s\delta}^{\s\rho\s\sigma}
\label{12.62}
\end{equation}
This is the analogue of the Maxwell energy-momentum stress tensor
$T(F)$ associated to an electromagnetic field $F$:
$$T(F)_{\alpha\beta}=F_{\alpha\rho}F_{\beta}^{\s\rho}+\s^*F_{\alpha\rho}\s^*F_{\beta}^{\s\rho}$$
We recall the following basic facts from [C-K].

\vspace{5mm}

\noindent{\bf Proposition 12.5} \ \ \ Given an arbitrary Weyl
field $W$, the associated Bel-Robinson tensor $Q(W)$ has the
following algebraic properties:

 \ 1. \ $Q(W)$ is symmetric and trace-free in all pairs of indices.

 \ 2. \ $Q(W)(X_1,X_2,X_3,X_4)$ is non-negative for any tetrad $X_1, X_2, X_3, X_4$ of future-directed
 non-spacelike vectors at a point.

 \vspace{5mm}

 \noindent{\bf Proposition 12.6} \ \ \ Given a Weyl field $W$ satisfying the Bianchi equations \ref{12.16},
 $$(\mbox{div}Q(W))_{\beta\gamma\delta}=\nabla^{\alpha}Q_{\alpha\beta\gamma\delta}$$
 the divergence of the associated Bel-Robinson tensor $Q(W)$, a 3-covariant tensorfield which is symmetric and trace-free
 in all pairs of indices, is given by:
 $$(\mbox{div}Q(W))_{\beta\gamma\delta}=W_{\beta\s\delta}^{\s\mu\s\nu}J_{\mu\gamma\nu}
 +W_{\beta\s\gamma}^{\s\mu\s\nu}J_{\mu\delta\nu}
 +\s^*W_{\beta\s\delta}^{\s\mu\s\nu}J^*_{\mu\gamma\nu}+\s^*W_{\beta\s\gamma}^{\s\mu\s\nu}J^*_{\mu\delta\nu}$$

 \vspace{5mm}

 We shall need expressions for those components of $Q(W)$ relative to the frame field $(e_{\mu}:\mu=1,2,3,4)$
 for which at least two of the indices are from the set $\{3,4\}$. These are given by the following lemma
 from [C-K].

 \vspace{5mm}

 \noindent{\bf Lemma 12.2} \ \ \ We have:
 \begin{eqnarray*}
 &Q(W)_{3333}=2|\alb(W)|^2, \ \ \ Q(W)_{4444}=2|\alpha(W)|^2\\
 &Q(W)_{4333}=4|\beb(W)|^2, \ \ \ Q(W)_{3444}=4|\beta(W)|^2\\
 &Q(W)_{3344}=4(\rho(W)^2+\sigma(W)^2)
 \end{eqnarray*}
 Also:
 \begin{eqnarray*}
 &&Q(W)_{A333}=-4\alb_{AB}(W)\beb^B(W)\\
 &&Q(W)_{A444}=4\alpha_{AB}(W)\beta^B(W)\\
 &&Q(W)_{A433}=-4\rho(W)\beb_A(W)-4\sigma(W)\s^*\beb_A(W)\\
 &&Q(W)_{A344}=4\rho(W)\beta_A(W)-4\sigma(W)\s^*\beta_A(W)\\
 &&Q(W)_{AB33}=2\sg_{AB}|\beb(W)|^2+2\rho(W)\alb_{AB}(W)+2\sigma(W)\s^*\alb_{AB}(W)\\
 &&Q(W)_{AB44}=2\sg_{AB}|\beta(W)|^2+2\rho(W)\alpha_{AB}(W)-2\sigma(W)\s^*\alpha_{AB}(W)\\
 &&Q(W)_{AB34}=-2(\beb(W)\oth\beta(W))_{AB}+2\sg_{AB}(\rho(W)^2+\sigma(W)^2)
 \end{eqnarray*}

 \vspace{5mm}

 We shall also need expressions for those components of $\mbox{div}Q(W)$ for which all the indices are from the set $\{3,4\}$.
 These are given by the following lemma from [C-K], which follows directly from Proposition 12.6.

 \vspace{5mm}

 \noindent{\bf Lemma 12.3} \ \ \ We have:
 \begin{eqnarray*}
 &&(\mbox{div}Q(W))_{333}=4(\alb(W),\Thetab(J))+8(\beb(W),\Xib(J))\\
 &&(\mbox{div}Q(W))_{444}=4(\alpha(W),\Theta(J))-8(\beta(W),\Xi(J))\\
 &&(\mbox{div}(Q(W))_{433}=8\rho(W)\Lambdab(J)-8\sigma(W)\Kb(J)-8(\beb(W),\Ib(J))\\
 &&(\mbox{div}(Q(W))_{344}=8\rho(W)\Lambda(J)+8\sigma(W)K(J)+8(\beta(W),I(J))
 \end{eqnarray*}

 \vspace{5mm}

 Given now a Weyl field $W$ and three future directed non-spacelike vectorfields $X,Y,Z$, we define the {\em energy-momentum density}
 vectorfield $P(W;X,Y,Z)$ associated to $W$ and to the triplet $X,Y,Z$ by:
 \begin{equation}
 P(W;X,Y,Z)^{\alpha}=-Q(W)^{\alpha}_{\s\beta\gamma\delta}X^{\beta}Y^{\gamma}Z^{\delta}
 \label{12.63}
 \end{equation}
 Then by the first statement of Proposition 12.5,
 $$\mbox{div}P(W;X,Y,Z)=\nabla_{\alpha}P(W;X,Y,Z)^{\alpha},$$
 the divergence of the energy-momentum density vectorfield $P(W;X,Y,Z)$, is given by:
 \begin{eqnarray}
 &&\mbox{div}P(W;X,Y,Z)=-(\mbox{div}Q(W))(X,Y,Z)\label{12.64}\\
 &&\hspace{20mm}-\frac{1}{2}Q(W)_{\alpha\beta\gamma\delta}(\s^{(X)}\tilde{\pi}^{\alpha\beta}Y^{\gamma}Z^{\delta}
 +\s^{(Y)}\tilde{\pi}^{\alpha\beta}X^{\gamma}Z^{\delta}+\s^{(Z)}\tilde{\pi}^{\alpha\beta}X^{\gamma}Y^{\delta})\nonumber
 \end{eqnarray}

\section{The divergence theorem in spacetime}

Consider the equation
\begin{equation}
\mbox{div}P=\tau \label{12.65}
\end{equation}
for an arbitrary vectorfield $P$ and function $\tau$ on a general
spacetime manifold $(M,g)$. In arbitrary local coordinates this
equation reads:
\begin{equation}
\frac{1}{\sqrt{-\mbox{det}g}}\frac{\partial}{\partial
x^\mu}(\sqrt{-\mbox{det}g}P^\mu)=\tau \label{12.66}
\end{equation}
Consider then equation \ref{12.65} on our spacetime manifold
$(M^\prime,g)$. We express this equation in canonical coordinates
$(\ub,u;\vartheta^A:A=1,2)$ (see Chapter 1, Section 4). Expanding
$P$ in the associated coordinate frame field,
$$\left(\frac{\partial}{\partial u}, \frac{\partial}{\partial\ub};\frac{\partial}{\partial\vartheta^A}:A=1,2\right),$$
\begin{equation}
P=P^u\frac{\partial}{\partial
u}+P^{\ub}\frac{\partial}{\partial\ub}+(P^{\vartheta})^A\frac{\partial}{\partial\vartheta^A}
\label{12.67}
\end{equation}
the equation takes the form (see \ref{1.180}, \ref{1.181}):
\begin{equation}
\frac{1}{\sqrt{\mbox{det}\sg}}\left\{\frac{\partial}{\partial
u}(2\Omega^2\sqrt{\mbox{det}\sg}P^u)+
\frac{\partial}{\partial\ub}(2\Omega^2\sqrt{\mbox{det}\sg}P^{\ub})\right\}+\sdiv(2\Omega^2
M)=2\Omega^2\tau \label{12.68}
\end{equation}
where $M$ is the $S$-tangential vectorfield:
\begin{equation}
M=(P^{\vartheta})^A\frac{\partial}{\partial\vartheta^A}
\label{12.69}
\end{equation}
We integrate this equation on the $S_{\ub,u}$ with respect to the
measure $d\mu_{\sg}$ to obtain the equation:
\begin{equation}
\frac{\partial}{\partial u}\left(\int_{S_{\ub,u}}2\Omega^2 P^u
d\mu_{\sg}\right)
+\frac{\partial}{\partial\ub}\left(\int_{S_{\ub,u}}2\Omega^2
P^{\ub} d\mu_{\sg}\right)= \int_{S_{\ub,u}}2\Omega^2\tau
d\mu_{\sg} \label{12.70}
\end{equation}
Consider any $(\ub_1,u_1)\in D^\prime$. We integrate equation
\ref{12.70} with respect to $(\ub,u)$ on the parameter domain
$$D_1=[0,\ub_1]\times[u_0,u_1]\subset D^\prime$$
(see \ref{3.02}), to obtain, under the assumption that $P^{\ub}$
vanishes on $\Cb_0$,
\begin{equation}
E^{\ub_1}(u_1)-E^{\ub_1}(u_0)+F^{u_1}(\ub_1)=\int_{M_1}\tau d\mu_g
\label{12.71}
\end{equation}
Here $E^{\ub_1}(u_1)$ is the ``energy":
\begin{equation}
E^{\ub_1}(u_1)=\int_{C^{\ub_1}_{u_1}}2\Omega^2
P^u:=\int_0^{\ub_1}\left(\int_{S_{\ub,u}}2\Omega^2 P^u
d\mu_{\sg}\right)d\ub \label{12.72}
\end{equation}
and $F^{u_1}(\ub_1)$ is the ``flux":
\begin{equation}
F^{u_1}(\ub_1)=\int_{\Cb^{u_1}_{\ub_1}}2\Omega^2
P^{\ub}:=\int_{u_0}^{u_1}\left(\int_{S_{\ub,u}}2\Omega P^{\ub}
d\mu_{\sg}\right)du \label{12.73}
\end{equation}
Also, $M_1$ is the subdomain of $M^\prime$ corresponding to $D_1$:
$$M_1=\bigcup_{(\ub,u)\in D_1}S_{\ub,u}$$
(see \ref{3.03}), and we have (see \ref{1.180}):
\begin{equation}
\int_{M_1}\tau
d\mu_{g}=\int\int_{D_1}\left(\int_{S_{\ub,u}}2\Omega^2\tau
d\mu_{\sg}\right)d\ub du \label{12.74}
\end{equation}

Now, the vectorfield $P$ can also be expanded in the frame field
$(e_{\mu}:\mu=1,2,3,4)$ where
\begin{equation}
e_A=\frac{\partial}{\partial\vartheta^A} \ : \ A=1,2 \label{12.75}
\end{equation}
We thus have:
\begin{equation}
P= P^{\mu} e_{\mu} \label{12.76}
\end{equation}
Now by \ref{1.171} and \ref{1.6}:
\begin{equation}
e_3=\Omega^{-1}\frac{\partial}{\partial u}, \ \ \
e_4=\Omega^{-1}\left(\frac{\partial}{\partial\ub}
+b^A\frac{\partial}{\partial\vartheta^A}\right) \label{12.77}
\end{equation}
Substituting in \ref{12.76} and comparing coefficients with the
expansion \ref{12.67} yields:
\begin{equation}
P^u=\Omega^{-1}P^3, \ \ \ P^{\ub}=\Omega^{-1}P^4 \label{12.78}
\end{equation}
and:
\begin{equation}
(P^{\vartheta})^A=P^A+\Omega^{-1}b^A P^4 \label{12.79}
\end{equation}
In conclusion, we have arrived at the following lemma.

\vspace{5mm}

\noindent{\bf Lemma 12.4} \ \ \ Let $P$ be a vectorfield and
$\tau$ a function defined on $(M^\prime,g)$ and satisfying the
equation:
$$\mbox{div}P=\tau$$
Let also $P^4$ vanish on $\Cb_0$. Then, for every $(\ub_1,u_1)\in
D^\prime$ we have:
$$E^{\ub_1}(u_1)-E^{\ub_1}(u_0)+F^{u_1}(\ub_1)=\int_{M_1}\tau d\mu_{g}$$
where $E^{\ub_1}(u_1)$ is the ``energy":
$$E^{\ub_1}(u_1)=\int_{C^{\ub_1}_{u_1}}2\Omega P^3
:=\int_0^{\ub_1}\left(\int_{S_{\ub,u}}2\Omega P^3
d\mu_{\sg}\right)d\ub$$ and $F^{u_1}(\ub_1)$ is the ``flux":
$$F^{u_1}(\ub_1)=\int_{\Cb^{u_1}_{\ub_1}}2\Omega P^4:=\int_{u_0}^{u_1}\left(\int_{S_{\ub,u}}2\Omega P^4
d\mu_{\sg}\right)du$$

\vspace{5mm}

\section{The energies and fluxes. The quantity ${\cal P}_2$}

We now define the energies and fluxes on which our entire approach
is based.  We consider the energy-momentum density vectorfield
$P(W;X,Y,Z)$, defined by \ref{12.63} with the multiplier fields
$L,K$ in the role of the future directed non-spacelike
vectorfields $X,Y,Z$. Taking in \ref{12.63} the fundamental Weyl
field $W=R$ we consider the energy-momentum density vectorfields:
\begin{equation}
P(R;L,L,L), \ \ P(R;K,L,L,), \ \ P(R;K,K,L), \ \ P(R; K,K,K)
\label{12.80}
\end{equation}
Considering, for each $u\in[u_0,c^*)$ the entire $C_u$ in
$M^\prime$ we define the corresponding energies
$\stackrel{(n)}{E}_0(u) \ : \ n=0,1,2,3$:
\begin{eqnarray}
&&\stackrel{(0)}{E}_0(u)=\int_{C_u}2\Omega P^3(R;L,L,L)\nonumber\\
&&\stackrel{(1)}{E}_0(u)=\int_{C_u}2\Omega P^3(R;K,L,L)\nonumber\\
&&\stackrel{(2)}{E}_0(u)=\int_{C_u}2\Omega P^3(R;K,K,L)\nonumber\\
&&\stackrel{(3)}{E}_0(u)=\int_{C_u}2\Omega P^3(R;K,K,K)
\label{12.81}
\end{eqnarray}
Also, considering for each $\ub\in[0,\delta)$ the entire
$\Cb_{\ub}$ in $M^\prime$ we define the corresponding fluxes
$\stackrel{(n)}{F}_0(\ub) \ : \ n=0,1,2,3$:
\begin{eqnarray}
&&\stackrel{(0)}{F}_0(\ub)=\int_{\Cb_{\ub}}2\Omega P^4(R;L,L,L)\nonumber\\
&&\stackrel{(1)}{F}_0(\ub)=\int_{\Cb_{\ub}}2\Omega P^4(R;K,L,L)\nonumber\\
&&\stackrel{(2)}{F}_0(\ub)=\int_{\Cb_{\ub}}2\Omega P^4(R;K,K,L)\nonumber\\
&&\stackrel{(3)}{F}_0(\ub)=\int_{\Cb_{\ub}}2\Omega P^4(R;K,K,K)
\label{12.82}
\end{eqnarray}
According to the definitions \ref{12.63} and \ref{8.1}, and Lemma
12.2, we have:
\begin{eqnarray}
&&\stackrel{(0)}{E}_0(u)=\int_{C_u}2\Omega^4|\alpha|^2\nonumber\\
&&\stackrel{(1)}{E}_0(u)=\int_{C_u}4\Omega^4|u|^2|\beta|^2\nonumber\\
&&\stackrel{(2)}{E}_0(u)=\int_{C_u}4\Omega^4|u|^4(|\rho|^2+|\sigma|^2)\nonumber\\
&&\stackrel{(3)}{E}_0(u)=\int_{C_u}4\Omega^4|u|^6|\beb|^2
\label{12.83}
\end{eqnarray}
Also, we have:
\begin{equation}
\stackrel{(3)}{F}_0(\ub)=\int_{\Cb_{\ub}}2\Omega^4|u|^6|\alb|^2
\label{12.84}
\end{equation}
The above are the 0th order energies and fluxes.

Next, we consider the 1st order energies and fluxes. Taking in
\ref{12.63} the Weyl field $W=\tcL_L R$ we consider the
energy-momentum density vectorfields:
\begin{equation}
P(\tcL_L R;L,L,L), \ \ P(\tcL_L R;K,L,L), \ \ P(\tcL_L R;K,K,L), \
\ P(\tcL_L R;K,K,K) \label{12.85}
\end{equation}
We then define the corresponding energies
$\s^{(L)}\stackrel{(n)}{E}(u) \ : \ n=0,1,2,3$:
\begin{eqnarray}
&&\s^{(L)}\stackrel{(0)}{E}(u)=\int_{C_u}2\Omega P^3(\tcL_L R;L,L,L)\nonumber\\
&&\s^{(L)}\stackrel{(1)}{E}(u)=\int_{C_u}2\Omega P^3(\tcL_L R;K,L,L)\nonumber\\
&&\s^{(L)}\stackrel{(2)}{E}(u)=\int_{C_u}2\Omega P^3(\tcL_L R;K,K,L)\nonumber\\
&&\s^{(L)}\stackrel{(3)}{E}(u)=\int_{C_u}2\Omega P^3(\tcL_L
R;K,K,K) \label{12.86}
\end{eqnarray}
and fluxes $\s^{(L)}\stackrel{(n)}{F}(\ub) \ : \ n=0,1,2,3$:
\begin{eqnarray}
&&\s^{(L)}\stackrel{(0)}{F}(\ub)=\int_{\Cb_{\ub}}2\Omega P^4(\tcL_L R;L,L,L)\nonumber\\
&&\s^{(L)}\stackrel{(1)}{F}(\ub)=\int_{\Cb_{\ub}}2\Omega P^4(\tcL_L R;K,L,L)\nonumber\\
&&\s^{(L)}\stackrel{(2)}{F}(\ub)=\int_{\Cb_{\ub}}2\Omega P^4(\tcL_L R;K,K,L)\nonumber\\
&&\s^{(L)}\stackrel{(3)}{F}(\ub)=\int_{\Cb_{\ub}}2\Omega
P^4(\tcL_L R;K,K,K) \label{12.87}
\end{eqnarray}
According to the definitions \ref{12.63} and \ref{8.1}, and Lemma
12.2, we have:
\begin{eqnarray}
&&\s^{(L)}\stackrel{(0)}{E}(u)=\int_{C_u}2\Omega^4|\alpha(\tcL_L R)|^2\nonumber\\
&&\s^{(L)}\stackrel{(1)}{E}(u)=\int_{C_u}4\Omega^4|u|^2|\beta(\tcL_L R)|^2\nonumber\\
&&\s^{(L)}\stackrel{(2)}{E}(u)=\int_{C_u}4\Omega^4|u|^4(|\rho(\tcL_L R)|^2+|\sigma(\tcL_L R)|^2)\nonumber\\
&&\s^{(L)}\stackrel{(3)}{E}(u)=\int_{C_u}4\Omega^4|u|^6|\beb(\tcL_L
R)|^2 \label{12.88}
\end{eqnarray}
Also, we have:
\begin{equation}
\s^{(L)}\stackrel{(3)}{F}(u)=\int_{\Cb_{\ub}}2\Omega^4|u|^6|\alb(\tcL_L
R)|^2 \label{12.89}
\end{equation}

Taking in \ref{12.63} the Weyl fields $W=\tcL_{O_i} R \ :i=1,2,3$
we consider the energy-momentum density vectorfields:
\begin{eqnarray}
&P(\tcL_{O_i} R;L,L,L), \  P(\tcL_{O_i} R;K,L,L), \  P(\tcL_{O_i} R; K,K,L), \  P(\tcL_{O_i} R; K,K,K) \nonumber\\
&: \ i=1,2,3 \label{12.90}
\end{eqnarray}
We then define the corresponding energies
$\s^{(O)}\stackrel{(n)}{E}(u) \ : \ n=0,1,2,3$:
\begin{eqnarray}
&&\s^{(O)}\stackrel{(0)}{E}(u)=\int_{C_u}2\Omega\sum_i P^3(\tcL_{O_i} R;L,L,L)\nonumber\\
&&\s^{(O)}\stackrel{(1)}{E}(u)=\int_{C_u}2\Omega\sum_i P^3(\tcL_{O_i} R;K,L,L)\nonumber\\
&&\s^{(O)}\stackrel{(2)}{E}(u)=\int_{C_u}2\Omega\sum_i P^3(\tcL_{O_i} R;K,K,L)\nonumber\\
&&\s^{(O)}\stackrel{(3)}{E}(u)=\int_{C_u}2\Omega\sum_i
P^3(\tcL_{O_i} R;K,K,K) \label{12.91}
\end{eqnarray}
and fluxes $\s^{(O)}\stackrel{(n)}{F}(\ub) \ : \ n=0,1,2,3$:
\begin{eqnarray}
&&\s^{(O)}\stackrel{(0)}{F}(\ub)=\int_{\Cb_{\ub}}2\Omega\sum_i P^4(\tcL_{O_i} R;L,L,L)\nonumber\\
&&\s^{(O)}\stackrel{(1)}{F}(\ub)=\int_{\Cb_{\ub}}2\Omega\sum_i P^4(\tcL_{O_i} R;K,L,L)\nonumber\\
&&\s^{(O)}\stackrel{(2)}{F}(\ub)=\int_{\Cb_{\ub}}2\Omega\sum_i P^4(\tcL_{O_i} R;K,K,L)\nonumber\\
&&\s^{(O)}\stackrel{(3)}{F}(\ub)=\int_{\Cb_{\ub}}2\Omega\sum_i
P^4(\tcL_{O_i} R;K,K,K) \label{12.92}
\end{eqnarray}
According to the definitions \ref{12.63} and \ref{8.1}, and Lemma
12.2, we have:
\begin{eqnarray}
&&\s^{(O)}\stackrel{(0)}{E}(u)=\int_{C_u}2\Omega^4\sum_i|\alpha(\tcL_{O_i} R)|^2\nonumber\\
&&\s^{(O)}\stackrel{(1)}{E}(u)=\int_{C_u}4\Omega^4|u|^2\sum_i|\beta(\tcL_{O_i} R)|^2\nonumber\\
&&\s^{(O)}\stackrel{(2)}{E}(u)=\int_{C_u}4\Omega^4|u|^4\sum_i(|\rho(\tcL_{O_i} R)|^2+|\sigma(\tcL_{O_i} R)|^2)\nonumber\\
&&\s^{(O)}\stackrel{(3)}{E}(u)=\int_{C_u}4\Omega^4|u|^6\sum_i|\beb(\tcL_{O_i}
R)|^2 \label{12.93}
\end{eqnarray}
Also, we have:
\begin{equation}
\s^{(O)}\stackrel{(3)}{F}(\ub)=\int_{\Cb_{\ub}}2\Omega^4|u|^6\sum_i|\alb(\tcL_{O_i}
R)|^2 \label{12.94}
\end{equation}

Taking in \ref{12.63} the Weyl field $W=\tcL_S R$ we consider the
energy-momentum density vectorfield:
\begin{equation}
P(\tcL_S R;K,K,K) \label{12.95}
\end{equation}
We then define the corresponding energy:
\begin{equation}
\s^{(S)}\stackrel{(3)}{E}(u)=\int_{C_u}2\Omega P^3(\tcL_S R;K,K,K)
\label{12.96}
\end{equation}
and flux:
\begin{equation}
\s^{(S)}\stackrel{(3)}{F}(\ub)=\int_{\Cb_{\ub}}2\Omega P^4(\tcL_S
R;K,K,K) \label{12.97}
\end{equation}
According to the definitions \ref{12.63} and \ref{8.1}, and Lemma
12.2, we have:
\begin{equation}
\s^{(S)}\stackrel{(3)}{E}(u)=\int_{C_u}4\Omega^4|u|^6|\beb(\tcL_S
R)|^2 \label{12.98}
\end{equation}
and:
\begin{equation}
\s^{(S)}\stackrel{(3)}{F}(\ub)=\int_{\Cb_{\ub}}2\Omega^4|u|^6|\alb(\tcL_S
R)|^2 \label{12.99}
\end{equation}
The above are the 1st order energies and fluxes.

Finally, we consider the 2nd order energies and fluxes. Taking in
\ref{12.63} the Weyl field $W=\tcL_L\tcL_L R$ we consider the
energy-momentum density vectorfields:
\begin{eqnarray}
&P(\tcL_L\tcL_L R;L,L,L), \ \ P(\tcL_L\tcL_L R;K,L,L), \nonumber\\
&P(\tcL_L\tcL_L R;K,K,L), \ \ P(\tcL_L\tcL_L R;K,K,K)
\label{12.100}
\end{eqnarray}
We then define the corresponding energies
$\s^{(LL)}\stackrel{(n)}{E} \ : \ n=0,1,2,3$:
\begin{eqnarray}
&&\s^{(LL)}\stackrel{(0)}{E}(u)=\int_{C_u}2\Omega P^3(\tcL_L\tcL_L R;L,L,L)\nonumber\\
&&\s^{(LL)}\stackrel{(1)}{E}(u)=\int_{C_u}2\Omega P^3(\tcL_L\tcL_L R;K,L,L)\nonumber\\
&&\s^{(LL)}\stackrel{(2)}{E}(u)=\int_{C_u}2\Omega P^3(\tcL_L\tcL_L R;K,K,L)\nonumber\\
&&\s^{(LL)}\stackrel{(3)}{E}(u)=\int_{C_u}2\Omega P^3(\tcL_L\tcL_L
R;K,K,K) \label{12.101}
\end{eqnarray}
and fluxes $\s^{(LL)}\stackrel{(n)}{F} \ : \ n=0,1,2,3$:
\begin{eqnarray}
&&\s^{(LL)}\stackrel{(0)}{F}(\ub)=\int_{\Cb_{\ub}}2\Omega P^4(\tcL_L\tcL_L R;L,L,L)\nonumber\\
&&\s^{(LL)}\stackrel{(1)}{F}(\ub)=\int_{\Cb_{\ub}}2\Omega P^4(\tcL_L\tcL_L R;K,L,L)\nonumber\\
&&\s^{(LL)}\stackrel{(2)}{F}(\ub)=\int_{\Cb_{\ub}}2\Omega P^4(\tcL_L\tcL_L R;K,K,L)\nonumber\\
&&\s^{(LL)}\stackrel{(3)}{F}(\ub)=\int_{\Cb_{\ub}}2\Omega
P^4(\tcL_L\tcL_L R;K,K,K) \label{12.102}
\end{eqnarray}
According to the definitions \ref{12.63} and \ref{8.1}, and Lemma
12.2, we have:
\begin{eqnarray}
&&\s^{(LL)}\stackrel{(0)}{E}(u)=\int_{C_u}2\Omega^4|\alpha(\tcL_L\tcL_L R)|^2\nonumber\\
&&\s^{(LL)}\stackrel{(1)}{E}(u)=\int_{C_u}4\Omega^4|u|^2|\beta(\tcL_L\tcL_L R)|^2\nonumber\\
&&\s^{(LL)}\stackrel{(2)}{E}(u)=\int_{C_u}4\Omega^4|u|^4(|\rho(\tcL_L\tcL_L R)|^2+|\sigma(\tcL_L\tcL_L R)|^2)\nonumber\\
&&\s^{(LL)}\stackrel{(3)}{E}(u)=\int_{C_u}4\Omega^4|u|^6|\beb(\tcL_L\tcL_L
R)|^2 \label{12.103}
\end{eqnarray}
Also, we have:
\begin{equation}
\s^{(LL)}\stackrel{(3)}{F}(\ub)=\int_{\Cb_{\ub}}2\Omega^4|u|^6|\alb(\tcL_L\tcL_L
R)|^2 \label{12.104}
\end{equation}

Taking in \ref{12.63} the Weyl fields $W=\tcL_{O_i}\tcL_L R \ : \
i=1,2,3$ we consider the energy-momentum density vectorfields:
\begin{eqnarray}
&P(\tcL_{O_i}\tcL_L R;L,L,L), \ P(\tcL_{O_i}\tcL_L R;K,L,L)\nonumber\\
&P(\tcL_{O_i}\tcL_L R;K,K,L), \ P(\tcL_{O_i}\tcL_L R;K,K,K)\nonumber\\
&: \ i=1,2,3 \label{12.105}
\end{eqnarray}
We then define the corresponding energies
$\s^{(OL)}\stackrel{(n)}{E}(u) \ : \ n=0,1,2,3$:
\begin{eqnarray}
&&\s^{(OL)}\stackrel{(0)}{E}(u)=\int_{C_u}2\Omega\sum_i P^3(\tcL_{O_i}\tcL_L R;L,L,L)\nonumber\\
&&\s^{(OL)}\stackrel{(1)}{E}(u)=\int_{C_u}2\Omega\sum_i P^3(\tcL_{O_i}\tcL_L R;K,L,L)\nonumber\\
&&\s^{(OL)}\stackrel{(2)}{E}(u)=\int_{C_u}2\Omega\sum_i P^3(\tcL_{O_i}\tcL_L R;K,K,L)\nonumber\\
&&\s^{(OL)}\stackrel{(3)}{E}(u)=\int_{C_u}2\Omega\sum_i
P^3(\tcL_{O_i}\tcL_L R;K,K,K) \label{12.106}
\end{eqnarray}
and fluxes $\s^{(OL)}\stackrel{(n)}{F}(\ub) \ : \ n=0,1,2,3$:
\begin{eqnarray}
&&\s^{(OL)}\stackrel{(0)}{F}(\ub)=\int_{\Cb_{\ub}}2\Omega\sum_i P^4(\tcL_{O_i}\tcL_L R;L,L,L)\nonumber\\
&&\s^{(OL)}\stackrel{(1)}{F}(\ub)=\int_{\Cb_{\ub}}2\Omega\sum_i P^4(\tcL_{O_i}\tcL_L R;K,L,L)\nonumber\\
&&\s^{(OL)}\stackrel{(2)}{F}(\ub)=\int_{\Cb_{\ub}}2\Omega\sum_i P^4(\tcL_{O_i}\tcL_L R;K,K,L)\nonumber\\
&&\s^{(OL)}\stackrel{(3)}{F}(\ub)=\int_{\Cb_{\ub}}2\Omega\sum_i
P^4(\tcL_{O_i}\tcL_L R;K,K,K) \label{12.107}
\end{eqnarray}
According to the definitions \ref{12.63} and \ref{8.1}, and Lemma
12.2, we have:
\begin{eqnarray}
&&\s^{(OL)}\stackrel{(0)}{E}(u)=\int_{C_u}2\Omega^4\sum_i|\alpha(\tcL_{O_i}\tcL_L R)|^2\nonumber\\
&&\s^{(OL)}\stackrel{(1)}{E}(u)=\int_{C_u}4\Omega^4|u|^2\sum_i|\beta(\tcL_{O_i}\tcL_L R)|^2\nonumber\\
&&\s^{(OL)}\stackrel{(2)}{E}(u)=\int_{C_u}4\Omega^4|u|^4\sum_i(|\rho(\tcL_{O_i}\tcL_L
R)|^2
+|\sigma(\tcL_{O_i}\tcL_L R)|^2)\nonumber\\
&&\s^{(OL)}\stackrel{(3)}{E}(u)=\int_{C_u}4\Omega^4|u|^6\sum_i|\beb(\tcL_{O_i}\tcL_L
R)|^2 \label{12.108}
\end{eqnarray}
Also, we have:
\begin{equation}
\s^{(OL)}\stackrel{(3)}{F}(\ub)=\int_{\Cb_{\ub}}2\Omega^4|u|^6\sum_i|\alb(\tcL_{O_i}\tcL_L
R)|^2 \label{12.109}
\end{equation}

Taking in \ref{12.63} the Weyl fields $W=\tcL_{O_j}\tcL_{O_i}R \ :
i,j=1,2,3$ we consider the energy-momentum density vectorfields:
\begin{eqnarray}
&P(\tcL_{O_j}\tcL_{O_i}R;L,L,L), \ P(\tcL_{O_j}\tcL_{O_i}R;K,L,L)\nonumber\\
&P(\tcL_{O_j}\tcL_{O_i}R;K,K,L), \ P(\tcL_{O_j}\tcL_{O_i}R;K,K,K)\nonumber\\
&: \ i,j=1,2,3 \label{12.110}
\end{eqnarray}
We then define the corresponding energies
$\s^{(OO)}\stackrel{(n)}{E}(u) \ : \ i,j=1,2,3$:
\begin{eqnarray}
&&\s^{(OO)}\stackrel{(0)}{E}(u)=\int_{C_u}2\Omega\sum_{i,j}P^3(\tcL_{O_j}\tcL_{O_i}R;L,L,L)\nonumber\\
&&\s^{(OO)}\stackrel{(1)}{E}(u)=\int_{C_u}2\Omega\sum_{i,j}P^3(\tcL_{O_j}\tcL_{O_i}R;K,L,L)\nonumber\\
&&\s^{(OO)}\stackrel{(2)}{E}(u)=\int_{C_u}2\Omega\sum_{i,j}P^3(\tcL_{O_j}\tcL_{O_i}R;K,K,L)\nonumber\\
&&\s^{(OO)}\stackrel{(3)}{E}(u)=\int_{C_u}2\Omega\sum_{i,j}P^3(\tcL_{O_j}\tcL_{O_i}R;K,K,K)
\label{12.111}
\end{eqnarray}
and fluxes $\s^{(OO)}\stackrel{(n)}{F}(\ub) \ : \ n=0,1,2,3$:
\begin{eqnarray}
&&\s^{(OO)}\stackrel{(0)}{F}(\ub)=\int_{\Cb_{\ub}}\sum_{i,j}P^4(\tcL_{O_j}\tcL_{O_i}R;L,L,L)\nonumber\\
&&\s^{(OO)}\stackrel{(1)}{F}(\ub)=\int_{\Cb_{\ub}}\sum_{i,j}P^4(\tcL_{O_j}\tcL_{O_i}R;K,L,L)\nonumber\\
&&\s^{(OO)}\stackrel{(2)}{F}(\ub)=\int_{\Cb_{\ub}}\sum_{i,j}P^4(\tcL_{O_j}\tcL_{O_i}R;K,K,L)\nonumber\\
&&\s^{(OO)}\stackrel{(3)}{F}(\ub)=\int_{\Cb_{\ub}}\sum_{i,j}P^4(\tcL_{O_j}\tcL_{O_i}R;K,K,K)
\label{12.112}
\end{eqnarray}
According to the definitions \ref{12.63} and \ref{8.1}, and Lemma
12.2, we have:
\begin{eqnarray}
&&\s^{(OO)}\stackrel{(0)}{E}(u)=\int_{C_u}2\Omega^4\sum_{i,j}|\alpha(\tcL_{O_j}\tcL_{O_i}R)|^2\nonumber\\
&&\s^{(OO)}\stackrel{(1)}{E}(u)=\int_{C_u}4\Omega^4|u|^2\sum_{i,j}|\beta(\tcL_{O_j}\tcL_{O_j}R)|^2\nonumber\\
&&\s^{(OO)}\stackrel{(2)}{E}(u)=\int_{C_u}4\Omega^4|u|^4\sum_{i,j}(|\rho(\tcL_{O_j}\tcL_{O_i}R)|^2
+|\sigma(\tcL_{O_i}\tcL_L R)|^2)\nonumber\\
&&\s^{(OO)}\stackrel{(3)}{E}(u)=\int_{C_u}4\Omega^4|u|^6\sum_{i,j}|\beb(\tcL_{O_j}\tcL_{O_i}R)|^2
\label{12.113}
\end{eqnarray}
Also, we have:
\begin{equation}
\s^{(OO)}\stackrel{(3)}{F}(\ub)=\int_{\Cb_{\ub}}2\Omega^4|u|^6\sum_{i,j}|\alb(\tcL_{O_j}\tcL_{O_i}R)|^2
\label{12.114}
\end{equation}

Taking in \ref{12.63} the Weyl fields $W=\tcL_{O_i}\tcL_S R$ we
consider the energy-momentum density vectorfields:
\begin{equation}
P(\tcL_{O_i}\tcL_S R;K,K,K) \ : \ i=1,2,3 \label{12.115}
\end{equation}
We then define the corresponding energy:
\begin{equation}
\s^{(OS)}\stackrel{(3)}{E}(u)=\int_{C_u}2\Omega\sum_i
P^3(\tcL_{O_i}\tcL_S R;K,K,K) \label{12.116}
\end{equation}
and flux:
\begin{equation}
\s^{(OS)}\stackrel{(3)}{F}(\ub)=\int_{\Cb_{\ub}}2\Omega\sum_i
P^4(\tcL_{O_i}\tcL_S R;K,K,K) \label{12.117}
\end{equation}
According to the definitions \ref{12.63} and \ref{8.1}, and Lemma
12.2, we have:
\begin{equation}
\s^{(OS)}\stackrel{(3)}{E}(u)=\int_{C_u}4\Omega^4|u|^6\sum_i|\beb(\tcL_{O_i}\tcL_S
R)|^2 \label{12.118}
\end{equation}
and:
\begin{equation}
\s^{(OS)}\stackrel{(3)}{F}(u)=\int_{\Cb_{\ub}}2\Omega^4|u|^6\sum_i|\alb(\tcL_{O_i}\tcL_S
R)|^2 \label{12.119}
\end{equation}

Finally, taking in \ref{12.63} the Weyl field $W=\tcL_S\tcL_S R$
we consider the energy-momentum density vectorfield:
\begin{equation}
P(\tcL_S\tcL_S R;K,K,K) \label{12.120}
\end{equation}
We then define the corresponding energy:
\begin{equation}
\s^{(SS)}\stackrel{(3)}{E}(u)=\int_{C_u}2\Omega P^3(\tcL_S\tcL_S
R;K,K,K) \label{12.121}
\end{equation}
and flux:
\begin{equation}
\s^{(SS)}\stackrel{(3)}{F}(u)=\int_{\Cb_{\ub}}2\Omega
P^4(\tcL_S\tcL_S R;K,K,K) \label{12.122}
\end{equation}
According to the definitions \ref{12.63} and \ref{8.1}, and Lemma
12.2, we have:
\begin{equation}
\s^{(SS)}\stackrel{(3)}{E}(u)=\int_{C_u}4\Omega^4|u|^6|\beb(\tcL_S\tcL_S
R)|^2 \label{12.123}
\end{equation}
and:
\begin{equation}
\s^{(SS)}\stackrel{(3)}{F}(\ub)=\int_{\Cb_{\ub}}2\Omega^4|u|^6|\alb(\tcL_S\tcL_S
R)|^2 \label{12.124}
\end{equation}
The above are the 2nd order energies and fluxes.

We note that by the second statement of Pproposition 12.5 the
integrants of all the energies and fluxes are non-negative.

We now define the total 1st order energies $\stackrel{(n)}{E}_1(u)
\ : n=0,1,2,3$ by:
\begin{eqnarray}
&&\stackrel{(0)}{E}_1(u)=\stackrel{(0)}{E}_0(u)+\delta^2\s^{(L)}\stackrel{(0)}{E}(u)+\s^{(O)}\stackrel{(0)}{E}(u)\nonumber\\
&&\stackrel{(1)}{E}_1(u)=\stackrel{(1)}{E}_0(u)+\delta^2\s^{(L)}\stackrel{(1)}{E}(u)+\s^{(O)}\stackrel{(1)}{E}(u)
\nonumber\\
&&\stackrel{(2)}{E}_1(u)=\stackrel{(2)}{E}_0(u)+\delta^2\s^{(L)}\stackrel{(2)}{E}(u)+\s^{(O)}\stackrel{(2)}{E}(u)
\label{12.125}
\end{eqnarray}
and:
\begin{equation}
\stackrel{(3)}{E}_1(u)=\stackrel{(3)}{E}_0(u)+\delta^2\s^{(L)}\stackrel{(3)}{E}(u)+\s^{(O)}\stackrel{(3)}{E}(u)
+\s^{(S)}\stackrel{(3)}{E}(u) \label{12.126}
\end{equation}
We also define the total 1st order fluxes $\stackrel{(n)}{F}_1(u)
\ : n=0,1,2,3$ by:
\begin{eqnarray}
&&\stackrel{(0)}{F}_1(\ub)=\stackrel{(0)}{F}_0(\ub)+\delta^2\s^{(L)}\stackrel{(0)}{F}(\ub)+\s^{(O)}\stackrel{(0)}{F}(\ub)\nonumber\\
&&\stackrel{(1)}{F}_1(\ub)=\stackrel{(1)}{F}_0(\ub)+\delta^2\s^{(L)}\stackrel{(1)}{F}(\ub)+\s^{(O)}\stackrel{(1)}{F}(\ub)\nonumber\\
&&\stackrel{(2)}{F}_1(\ub)=\stackrel{(2)}{F}_0(\ub)+\delta^2\s^{(L)}\stackrel{(2)}{F}(\ub)+\s^{(O)}\stackrel{(2)}{F}
(\ub) \label{12.127}
\end{eqnarray}
and:
\begin{equation}
\stackrel{(3)}{F}_1(\ub)=\stackrel{(3)}{F}_0(\ub)+\delta^2\s^{(L)}\stackrel{(3)}{F}(\ub)+\s^{(O)}\stackrel{(3)}{F}(\ub)
+\s^{(S)}\stackrel{(3)}{F}(\ub) \label{12.128}
\end{equation}

We then define the total 2nd order energies
$\stackrel{(n)}{E}_2(u) \ : \ n=0,1,2,3$ by:
\begin{eqnarray}
&&\stackrel{(0)}{E}_2(u)=\stackrel{(0)}{E}_1(u)+\delta^4\s^{(LL)}\stackrel{(0)}{E}(u)
+\delta^2\s^{(OL)}\stackrel{(0)}{E}(u)+\s^{(OO)}\stackrel{(0)}{E}(u)\nonumber\\
&&\stackrel{(1)}{E}_2(u)=\stackrel{(1)}{E}_1(u)+\delta^4\s^{(LL)}\stackrel{(1)}{E}(u)
+\delta^2\s^{(OL)}\stackrel{(1)}{E}(u)+\s^{(OO)}\stackrel{(1)}{E}(u)\nonumber\\
&&\stackrel{(2)}{E}_2(u)=\stackrel{(2)}{E}_1(u)+\delta^4\s^{(LL)}\stackrel{(2)}{E}(u)
+\delta^2\s^{(OL)}\stackrel{(2)}{E}(u)+\s^{(OO)}\stackrel{(2)}{E}(u)\nonumber\\
&&\label{12.129}
\end{eqnarray}
and:
\begin{eqnarray}
&&\stackrel{(3)}{E}_2(u)=\stackrel{(3)}{E}_1(u)+\delta^4\s^{(LL)}\stackrel{(3)}{E}(u)
+\delta^2\s^{(OL)}\stackrel{(3)}{E}(u)+\s^{(OO)}\stackrel{(3)}{E}(u)\nonumber\\
&&\hspace{20mm}+\s^{(OS)}\stackrel{(3)}{E}(u)+\s^{(SS)}\stackrel{(3)}{E}(u)
\label{12.130}
\end{eqnarray}
We also define the total 2nd order fluxes
$\stackrel{(n)}{F}_2(\ub) \ : \ n=0,1,2,3$ by:
\begin{eqnarray}
&&\stackrel{(0)}{F}_2(\ub)=\stackrel{(0)}{F}_1(\ub)+\delta^4\s^{(LL)}\stackrel{(0)}{F}(\ub)
+\delta^2\s^{(OL)}\stackrel{(0)}{F}(\ub)+\s^{(OO)}\stackrel{(0)}{F}(\ub)\nonumber\\
&&\stackrel{(1)}{F}_2(\ub)=\stackrel{(1)}{F}_1(\ub)+\delta^4\s^{(LL)}\stackrel{(1)}{F}(\ub)
+\delta^2\s^{(OL)}\stackrel{(1)}{F}(\ub)+\s^{(OO)}\stackrel{(1)}{F}(\ub)\nonumber\\
&&\stackrel{(2)}{F}_2(\ub)=\stackrel{(2)}{F}_1(\ub)+\delta^4\s^{(LL)}\stackrel{(2)}{F}(\ub)
+\delta^2\s^{(OL)}\stackrel{(2)}{F}(\ub)+\s^{(OO)}\stackrel{(2)}{F}(\ub)\nonumber\\
&&\label{12.131}
\end{eqnarray}
and:
\begin{eqnarray}
&&\stackrel{(3)}{F}_2(\ub)=\stackrel{(3)}{F}_1(\ub)+\delta^4\s^{(LL)}\stackrel{(3)}{F}(\ub)
+\delta^2\s^{(OL)}\stackrel{(3)}{F}(\ub)+\s^{(OO)}\stackrel{(3)}{F}(\ub)\nonumber\\
&&\hspace{20mm}+\s^{(OS)}\stackrel{(3)}{F}(\ub)+\s^{(SS)}\stackrel{(3)}{F}(\ub)
\label{12.132}
\end{eqnarray}

We now define the 0th order quantities:
\begin{eqnarray}
&&\stackrel{(0)}{{\cal E}}_0=\sup_{u\in[u_0,c^*)}\left(\delta^2\stackrel{(0)}{E}_0(u)\right)\nonumber\\
&&\stackrel{(1)}{{\cal E}}_0=\sup_{u\in[u_0,c^*)}\left(\stackrel{(1)}{E}_0(u)\right)\nonumber\\
&&\stackrel{(2)}{{\cal E}}_0=\sup_{u\in[u_0,c^*)}\left(\delta^{-1}\stackrel{(2)}{E}_0(u)\right)\nonumber\\
&&\stackrel{(3)}{{\cal
E}}_0=\sup_{u\in[u_0,c^*)}\left(\delta^{-3}\stackrel{(3)}{E}_0(u)\right)
\label{12.133}
\end{eqnarray}
and:
\begin{equation}
\stackrel{(3)}{{\cal
F}}_0=\sup_{\ub\in[0,\min\{\delta,c^*-u_0\})}\left(\delta^{-3}\stackrel{(3)}{F}_0(\ub)\right)
\label{12.134}
\end{equation}
and the quantity ${\cal P}_0$ by:
\begin{equation}
{\cal P}_0=\max\{\stackrel{(0)}{{\cal E}}_0,\stackrel{(1)}{{\cal
E}}_0,\stackrel{(2)}{{\cal E}}_0, \stackrel{(3)}{{\cal E}}_0;
\stackrel{(3)}{{\cal F}}_0\} \label{12.135}
\end{equation}
Next, we define the 1st order quantities:
\begin{eqnarray}
&&\stackrel{(0)}{{\cal E}}_1=\sup_{u\in[u_0,c^*)}\left(\delta^2\stackrel{(0)}{E}_1(u)\right)\nonumber\\
&&\stackrel{(1)}{{\cal E}}_1=\sup_{u\in[u_0,c^*)}\left(\stackrel{(1)}{E}_1(u)\right)\nonumber\\
&&\stackrel{(2)}{{\cal E}}_1=\sup_{u\in[u_0,c^*)}\left(\delta^{-1}\stackrel{(2)}{E}_1(u)\right)\nonumber\\
&&\stackrel{(3)}{{\cal
E}}_1=\sup_{u\in[u_0,c^*)}\left(\delta^{-3}\stackrel{(3)}{E}_1(u)\right)
\label{12.136}
\end{eqnarray}
and:
\begin{equation}
\stackrel{(3)}{{\cal
F}}_1=\sup_{\ub\in[0,\min\{\delta,c^*-u_0\})}\left(\delta^{-3}\stackrel{(3)}{F}_1(\ub)\right)
\label{12.137}
\end{equation}
and the quantity ${\cal P}_1$ by:
\begin{equation}
{\cal P}_1=\max\{\stackrel{(0)}{{\cal E}}_1,\stackrel{(1)}{{\cal
E}}_1,\stackrel{(2)}{{\cal E}}_1, \stackrel{(3)}{{\cal E}}_1;
\stackrel{(3)}{{\cal F}}_1\} \label{12.138}
\end{equation}

Finally, we define the 2nd order quantities:
\begin{eqnarray}
&&\stackrel{(0)}{{\cal E}}_2=\sup_{u\in[u_0,c^*)}\left(\delta^2\stackrel{(0)}{E}_2(u)\right)\nonumber\\
&&\stackrel{(1)}{{\cal E}}_2=\sup_{u\in[u_0,c^*)}\left(\stackrel{(1)}{E}_2(u)\right)\nonumber\\
&&\stackrel{(2)}{{\cal E}}_2=\sup_{u\in[u_0,c^*)}\left(\delta^{-1}\stackrel{(2)}{E}_2(u)\right)\nonumber\\
&&\stackrel{(3)}{{\cal
E}}_2=\sup_{u\in[u_0,c^*)}\left(\delta^{-3}\stackrel{(3)}{E}_2(u)\right)
\label{12.139}
\end{eqnarray}
and:
\begin{equation}
\stackrel{(3)}{{\cal
F}}_2=\sup_{\ub\in[0,\min\{\delta,c^*-u_0\})}\left(\delta^{-3}\stackrel{(3)}{F}_2(\ub)\right)
\label{12.140}
\end{equation}

and the quantity ${\cal P}_1$ by:
\begin{equation}
{\cal P}_2=\max\{\stackrel{(0)}{{\cal E}}_2,\stackrel{(1)}{{\cal
E}}_2,\stackrel{(2)}{{\cal E}}_2, \stackrel{(3)}{{\cal E}}_2;
\stackrel{(3)}{{\cal F}}_2\} \label{12.141}
\end{equation}

\section{The controlling quantity ${\cal Q}_2$. Bootstrap assumptions and the comparison lemma}

In reference to the quantities defined in Chapter 10, we set:
\begin{equation}
{\cal Q}_1=\max\{{\cal R}_{[1]}(\alpha),{\cal R}_{[1]}(\beta),
{\cal R}_{[1]}(\rho),{\cal R}_{[1]}(\sigma), {\cal
R}_{[1]}(\beb);\cRb_{[1]}(\alb),{\cal D}_0^{\prime 4}(\alb)\}
\label{12.142}
\end{equation}
and:
\begin{equation}
{\cal Q}_2=\max\{{\cal R}_{[2]}(\alpha),{\cal R}_{[2]}(\beta),
{\cal R}_{[2]}(\rho), {\cal R}_{[2]}(\sigma), {\cal
R}_{[2]}(\beb);\cRb_{[2]}(\alb), {\cal D}^{\prime 4}_{[1]}(\alb)\}
\label{12.143}
\end{equation}
Also, in reference to the quantity $\scR_1(\Db\beb)$ defined by
\ref{7.03}, we set:
\begin{equation}
{\cal Q}_2^\prime=\max\{{\cal Q}_2,\scR_1(\Db\beb)\}
\label{12.144}
\end{equation}

Let us define:
\begin{equation}
{\cal R}_0^4=\max\{{\cal R}_0^4(\alpha),{\cal R}_0^4(\beta),{\cal
R}_0^4(\rho),{\cal R}_0^4(\sigma), {\cal R}_0^4(\beb), {\cal
R}_0^4(\alb)\} \label{12.145}
\end{equation}
By the results of Chapter 10, under the bootstrap assumptions {\bf
A1.1}, {\bf A1.2}, {\bf A3.1}, {\bf A3.2} there is a numerical
constant $C$ such that:
\begin{equation}
{\cal R}_0^4\leq C{\cal Q}_1 \label{12.146}
\end{equation}
Also, under the additional bootstrap assumptions {\bf A2.1}, {\bf
A2.2}, {\bf A4.1} {\bf A4.2}, there are numerical constants $C$
such that:
\begin{equation}
{\cal R}_0^\infty\leq C{\cal Q}_2 \label{12.147}
\end{equation}
and:
\begin{equation}
\scR_1^4\leq C{\cal Q}_2 \label{12.148}
\end{equation}
Moreover, there are numerical constants $C$ such that:
\begin{equation}
\max\{{\cal R}_0^4(\Dh\alpha),{\cal R}_0^4(D\beta),{\cal
R}_0^4(D\rho),{\cal R}_0^4(D\sigma),{\cal R}_0^4(D\beb)\} \leq
C{\cal Q}_2 \label{12.149}
\end{equation}
and:
\begin{equation}
{\cal R}_0^4(\Dbh\alb)\leq C{\cal Q}_2 \label{12.150}
\end{equation}
Reviewing Chapters 3 - 7, as well as Chapters 8 - 9, we see that
the quantity ${\cal Q}^\prime_2$ bounds all the curvature norms
which enter the estimates for the connection coefficients and,
through these, the estimates for the deformation tensors of the
multiplier fields and the commutation fields.

Our present purpose is to compare the quantity $({\cal
Q}_2^\prime)^2$ to the quantity ${\cal P}_2$. We begin by
comparing the quantity $({\cal Q}_1)^2$ to the quantity ${\cal
P}_1$. To effect this comparison we introduce the following
bootstrap assumptions for the deformation tensors of the
commutation fields.

\vspace{5mm}

\ \ \ {\bf C1.1:} \
$\delta^{3/4}\|\s^{(L)}\ih\|_{L^\infty(S_{\ub,u})}\leq 1 \ : \
\forall (\ub,u)\in D^\prime$

\vspace{2.5mm}

\ \ \ {\bf C1.2:} \
$\delta^{3/4}\|\s^{(L)}j\|_{L^\infty(S_{\ub,u})}\leq 1 \ : \
\forall (\ub,u)\in D^\prime$

\vspace{2.5mm}

\ \ \ {\bf C1.3:} \
$\delta^{3/4}\|\s^{(L)}\nu\|_{L^\infty(S_{\ub,u})}\leq 1 : \
\forall (\ub,u)\in D^\prime$

\vspace{2.5mm}

\ \ \ {\bf C1.4:} \
$\delta^{-1/4}|u|\|\s^{(L)}\mb\|_{L^\infty(S_{\ub,u})}\leq 1 : \
\forall (\ub,u)\in D^\prime$

\vspace{2.5mm}

\ \ \ {\bf C2.1:} \
$\delta^{-1/4}\|\s^{(O_i)}\ih\|_{L^\infty(S_{\ub,u})}\leq 1 \ : \
\forall (\ub,u)\in D^\prime$

\vspace{2.5mm}

\ \ \ {\bf C2.2:} \
$\delta^{-1/4}\|\s^{(O_i)}j\|_{L^\infty(S_{\ub,u})}\leq 1 \ : \
\forall (\ub,u)\in D^\prime$

\vspace{2.5mm}

\ \ \ {\bf C2.3:} \
$\delta^{-1/4}\|\s^{(O_i)}\nu\|_{L^\infty(S_{\ub,u})}\leq 1 \ : \
\forall (\ub,u)\in D^\prime$

\vspace{2.5mm}

\ \ \ {\bf C2.4:} \
$\delta^{1/4}|u|^{-1/2}\|\s^{(O_i)}m\|_{L^\infty(S_{\ub,u})}\leq 1
\ : \ \forall (\ub,u)\in D^\prime$

\vspace{2.5mm}

\ \ \ {\bf C3.1:} \
$\delta^{-1/4}\|\s^{(S)}\ih\|_{L^\infty(S_{\ub,u})}\leq 1 \ : \
\forall (\ub,u)\in D^\prime$

\vspace{2.5mm}

\ \ \ {\bf C3.2:} \
$\delta^{-1/4}\|\s^{(S)}j\|_{L^\infty(S_{\ub,u})}\leq 1 \ : \
\forall (\ub,u)\in D^\prime$

\vspace{2.5mm}

\ \ \ {\bf C3.3:} \
$\delta^{-1/4}\|\s^{(S)}\nu-1\|_{L^\infty(S_{\ub,u})}\leq 1 \ : \
\forall (\ub,u)\in D^\prime$

\vspace{2.5mm}

\ \ \ {\bf C3.4:} \
$\delta^{1/4}|u|^{-1/2}\|\s^{(S)}m\|_{L^\infty(S_{\ub,u})}\leq 1 \
: \ \forall (\ub,u)\in D^\prime$

\vspace{2.5mm}

\ \ \ {\bf C3.5:} \
$\delta^{-5/4}|u|\|\s^{(S)}\mb\|_{L^\infty(S_{\ub,u})}\leq 1 \ : \
\forall (\ub,u)\in D^\prime$

\vspace{5mm}

(Recall from Chapter 8 that $\s^{(L)}m=\s^{(O_i)}\mb=0$.)
Comparing with the results of Chapter 8 we see that there is room
of a factor of $\delta^{1/4}$ at least in the above bootstrap
assumptions, a fact which will allow us to show that the
inequalities in these bootstrap assumptions are not saturated.

\vspace{5mm}

\noindent{\bf Lemma 12.5} \ \ \ Under the bootstrap assumptions
{\bf A0}, {\bf A1.1}, {\bf A1.2}, {\bf A3.1}, {\bf A3.2}, {\bf
B1}, {\bf B2}, and {\bf C1.1} - {\bf C1.4}, {\bf C2.1} - {\bf
C2.4}, {\bf C3.1} - {\bf C3.5}, there is a numerical constant $C$
such that:
$$({\cal Q}_1)^2\leq C\max\{{\cal P}_1,({\cal D}^{\prime 4}_0(\alb))^2\}$$
provided that $\delta$ does not exceed a certain positive numerical constant. Moreover, we have:
\begin{eqnarray*}
&&({\cal R}_{[1]}(\alpha))^2\leq C\stackrel{(0)}{{\cal E}}_1+C\delta^{3/2}\max\{{\cal P}_1,({\cal D}^{\prime 4}_0(\alb))^2\}\\
&&({\cal R}_{[1]}(\beta))^2\leq C\stackrel{(1)}{{\cal E}}_1+C\delta^{1/2}\max\{{\cal P}_1,({\cal D}^{\prime 4}_0(\alb))^2\}\\
&&({\cal R}_{[1]}(\rho))^2,({\cal R}_{[1]}(\sigma))^2\leq
C\stackrel{(2)}{{\cal E}}_1
+C\delta^{3/2}\max\{{\cal P}_1,({\cal D}^{\prime 4}_0(\alb))^2\}\\
&&({\cal R}_{[1]}(\beb))^2\leq C\stackrel{(3)}{{\cal
E}}_1+C\delta^{1/2}\max\{{\cal P}_1,({\cal D}^{\prime
4}_0(\alb))^2\}
\end{eqnarray*}
and:
$$(\cRb_{[1]}(\alb))^2\leq C\stackrel{(3)}{{\cal F}}_1+C\delta^{3/2}\max\{{\cal P}_1,({\cal D}^{\prime 4}_0(\alb))^2\}$$

\noindent{\em Proof:} Taking $Y=L, O_i:i=1,2,3$ and $W=R$ in
Proposition 12.2 and using the bootstrap assumptions {\bf C1.1} -
{\bf C1.4}, {\bf C2.1} -{\bf C2.4}, as well as the basic bootstrap
assumption {\bf A0}, we deduce, in view of \ref{12.83},
\ref{12.86}, \ref{12.93}, and \ref{12.125}, \ref{12.126},
\begin{eqnarray}
&&\int_{C_u}\left\{|\alpha|^2+\delta^2|\Dh\alpha|^2+\sum_i|\sLh_{O_i}\alpha|^2\right\}
\leq C\stackrel{(0)}{E}_1(u)+C\delta^{-1/2}\int_{C_u}|u|^2|\beta|^2\nonumber\\
&&\int_{C_u}|u|^2\left\{|\beta|^2+\delta^2|D\beta|^2+\sum_i|\sL_{O_i}\beta|^2\right\}\nonumber\\
&&\hspace{20mm}\leq C\stackrel{(1)}{E}_1(u)+C\delta^{5/2}\int_{C_u}|\alpha|^2+C\delta^{-1/2}\int_{C_u}|u|^4(|\rho|^2+|\sigma|^2)\nonumber\\
&&\int_{C_u}|u|^4\left\{|\rho|^2+|\sigma|^2+\delta^2(|D\rho|^2+|D\sigma|^2)+\sum_i(|O_i\rho|^2+|O_i\sigma|^2\right\}
\nonumber\\
&&\hspace{20mm}\leq
C\stackrel{(2)}{E}_1(u)+C\delta^{5/2}\int_{C_u}|u|^2|\beta|^2
+C\delta^{-1/2}\int_{C_u}|u|^6|\beb|^2\nonumber\\
&&\int_{C_u}|u|^6\left\{|\beb|^2+\delta^2|D\beb|^2+\sum_i|\sL_{O_i}\beb|^2\right\}\nonumber\\
&&\hspace{20mm}\leq
C\stackrel{(3)}{E}_1(u)+C\delta^{5/2}\int_{C_u}|u|^4(|\rho|^2+|\sigma|^2)
+C\delta^{-1/2}\int_{C_u}|u|^7|\alb|^2\nonumber\\
&&\label{12.151}
\end{eqnarray}
Now,
\begin{eqnarray}
&&\int_{C_u}|\alpha|^2\leq \delta^{-2}({\cal R}_0(\alpha))^2\leq \delta^{-2}({\cal Q}_1)^2\nonumber\\
&&\int_{C_u}|u|^2|\beta|^2\leq ({\cal R}_0(\beta))^2\leq ({\cal Q}_1)^2\nonumber\\
&&\int_{C_u}|u|^4(|\rho|^2+|\sigma|^2)\leq \delta(({\cal
R}_0(\rho))^2+({\cal R}_0(\sigma))^2)
\leq 2\delta({\cal Q}_1)^2\nonumber\\
&&\int_{C_u}|u|^6|\beb|^2\leq \delta^3({\cal R}_0(\beb))^2\leq
\delta^3({\cal Q}_1)^2 \label{12.152}
\end{eqnarray}
and in view of the definition \ref{10.102} and \ref{12.146} we
have:
\begin{equation}
\int_{C_u}|u|^7|\alb|^2\leq C\delta^4({\cal R}_0^4(\alb))^2\leq
C^\prime\delta^4({\cal Q}_1)^2 \label{12.153}
\end{equation}
Under the bootstrap assumptions {\bf B1}, {\bf B2}, we have, from
the coercivity inequality \ref{11.124}, a corollary of Proposition
11.1, applied to $\alpha$ and $\alb$, and from Proposition 11.1
itself applied to $\beta$, $\beb$, $\rho$, $\sigma$:
\begin{eqnarray}
&&|u|^2|\snab\alpha|^2\leq C\left(\sum_i|\sLh_{O_i}\alpha|^2+|\alpha|^2\right)\nonumber\\
&&|u|^2|\snab\alb|^2\leq C\left(\sum_i|\sLh_{O_i}\alb|^2+|\alb|^2\right)\nonumber\\
&&|u|^2|\snab\beta|^2\leq C\sum_i|\sL_{O_i}\beta|^2\nonumber\\
&&|u|^2|\snab\beb|^2\leq C\sum_i|\sL_{O_i}\beb|^2\nonumber\\
&&|u|^2|\sd\rho|^2\leq C\sum_i|O_i\rho|^2\nonumber\\
&&|u|^2|\sd\sigma|^2\leq C\sum_i|O_i\sigma|^2 \label{12.154}
\end{eqnarray}
Substituting \ref{12.154} (except the second), \ref{12.152} and
\ref{12.153}, in \ref{12.151} we then obtain:
\begin{eqnarray}
&&\int_{C_u}\left\{|\alpha|^2+\delta^2|\Dh\alpha|^2+|u|^2|\snab\alpha|^2\right\}
\leq C\stackrel{(0)}{E}_1(u)+C\delta^{-1/2}({\cal Q}_1)^2\nonumber\\
&&\int_{C_u}|u|^2\left\{|\beta|^2+\delta^2|D\beta|^2+|u|^2|\snab\beta|^2\right\}\nonumber\\
&&\hspace{20mm}\leq C\stackrel{(1)}{E}_1(u)+C\delta^{1/2}({\cal Q}_1)^2\nonumber\\
&&\int_{C_u}|u|^4\left\{|\rho|^2+|\sigma|^2+\delta^2(|D\rho|^2+|D\sigma|^2)+|u|^2(|\sd\rho|^2+|\sd\sigma|^2\right\}
\nonumber\\
&&\hspace{20mm}\leq C\stackrel{(2)}{E}_1(u)+C\delta^{5/2}({\cal Q}_1)^2\nonumber\\
&&\int_{C_u}|u|^6\left\{|\beb|^2+\delta^2|D\beb|^2+|u|^2|\snab\beb|^2\right\}\nonumber\\
&&\hspace{20mm}\leq C\stackrel{(3)}{E}_1(u)+C\delta^{7/2}({\cal
Q}_1)^2\label{12.155}
\end{eqnarray}
Multiplying the first of \ref{12.155} by $\delta^2$, the second by
1, the third by $\delta^{-1}$, and the fourth by $\delta^{-3}$,
and taking the supremum over $u\in[u_0,c^*)$, we obtain, in view
of the definitions \ref{10.36} - \ref{10.45} and \ref{12.136},
\begin{eqnarray}
&&({\cal R}_{[1]}(\alpha))^2\leq C\stackrel{(0)}{{\cal E}}_1+C\delta^{3/2}({\cal Q}_1)^2\nonumber\\
&&({\cal R}_{[1]}(\beta))^2\leq C\stackrel{(1)}{{\cal E}}_1+C\delta^{1/2}({\cal Q}_1)^2\nonumber\\
&&({\cal R}_{[1]}(\rho))^2,({\cal R}_{[1]}(\sigma))^2\leq
C\stackrel{(2)}{{\cal E}}_1+C\delta^{3/2}({\cal Q}_1)^2
\nonumber\\
&&({\cal R}_{[1]}(\beb))^2\leq C\stackrel{(3)}{{\cal
E}}_1+C\delta^{1/2}({\cal Q}_1)^2 \label{12.156}
\end{eqnarray}

Next, taking $Y=L, O_i:i=1,2,3, S$ and $W=R$ in Proposition 12.2
and using the bootstrap assumptions {\bf C1.1} - {\bf C1.4}, {\bf
C2.1} -{\bf C2.4}, {\bf C3.1} - {\bf C3.5}, as well as the basic
bootstrap assumption {\bf A0}, noting that for any trace-free
symmetric 2-covariant $S$ tensorfield $\theta$ we have:
\begin{equation}
\sLh_S\theta=u\Dbh\theta+\ub\Dh\theta \label{12.157}
\end{equation}
we deduce, in view of \ref{12.84}, the last of \ref{12.87},
\ref{12.94}, and \ref{12.99}, and \ref{12.128},
\begin{eqnarray}
&&\int_{\Cb_{\ub}}|u|^6\left\{|\alb|^2+\delta^2|\Dh\alb|^2+|u|^2|\Dbh\alb|^2+\sum_i|\sLh_{O_i}\alb|^2\right\}\nonumber\\
&&\hspace{20mm}\leq
C\stackrel{(3)}{F}_1(\ub)+C\delta^{5/2}\int_{\Cb_{\ub}}|u|^4|\beb|^2
\label{12.158}
\end{eqnarray}
Now, in view of the definition \ref{10.53} and \ref{12.146} we
have:
\begin{equation}
\int_{\Cb_{\ub}}|u|^4|\beb|^2\leq C\delta^2({\cal
R}_0^4(\beb))^2\leq C^\prime\delta^2({\cal Q}_1)^2 \label{12.159}
\end{equation}
Substituting the second of \ref{12.154} and \ref{12.159} in
\ref{12.158} we obtain:

\begin{eqnarray}
&&\int_{\Cb_{\ub}}|u|^6\left\{|\alb|^2+\delta^2|\Dh\alb|^2+|u|^2|\Dbh\alb|^2+|u|^2|\snab\alb|^2
\right\}\nonumber\\
&&\hspace{20mm}\leq C\stackrel{(3)}{F}_1(\ub)+C\delta^{9/2}({\cal
Q}_1)^2 \label{12.160}
\end{eqnarray}
Multiplying by $\delta^{-3}$ and taking the supremum over
$\ub$ yields, in view of the definitions \ref{10.99},
\ref{10.100} and \ref{12.137},
\begin{equation}
(\cRb_{[1]}(\alb))^2\leq C\stackrel{(3)}{{\cal
F}}_1+\delta^{3/2}({\cal Q}_1)^2 \label{12.161}
\end{equation}
The inequalities \ref{12.156} and \ref{12.161} together imply,
recalling the definitions \ref{12.138} and \ref{12.142},
\begin{equation}
({\cal Q}_1)^2\leq C\max\{{\cal P}_1,({\cal D}^{\prime
4}_0(\alb))^2\}+C\delta^{1/2}({\cal Q}_1)^2 \label{12.162}
\end{equation}
which, if $\delta$ is suitably small implies:
\begin{equation}
({\cal Q}_1)^2\leq C^\prime\max\{{\cal P}_1,({\cal D}^{\prime
4}_0(\alb))^2\} \label{12.163}
\end{equation}
This is the first conclusion of the lemma. Substituting this bound
in \ref{12.156} and \ref{12.161} then yields the remaining
conlusions of the lemma.

\vspace{5mm}

We proceed to compare the quantity $({\cal Q}_2^\prime)^2$ to the
quantity ${\cal P}_2$. To effect this comparison we introduce the
following additional bootstrap assumptions for the deformation
tensors of the commutation fields.

\vspace{5mm}

\ \ \ {\bf C4.1:} \
$\delta^{7/4}|u|^{-1/2}\|\sLh_L\s^{(L)}\ih\|_{L^4(S_{\ub,u})}\leq
1 \ : \ \forall (\ub,u)\in D^\prime$

\vspace{2.5mm}

\ \ \ {\bf C4.2:} \
$\delta^{7/4}|u|^{-1/2}\|L\s^{(L)}j\|_{L^4(S_{\ub,u})}\leq 1 \ : \
\forall (\ub,u)\in D^\prime$

\vspace{2.5mm}

\ \ \ {\bf C4.3:} \
$\delta^{7/4}|u|^{-1/2}\|L\s^{(L)}\nu\|_{L^4(S_{\ub,u})}\leq 1 : \
\forall (\ub,u)\in D^\prime$

\vspace{2.5mm}

\ \ \ {\bf C4.4:} \
$\delta^{3/4}|u|^{1/2}\|\sL_L\s^{(L)}\mb\|_{L^4(S_{\ub,u})}\leq 1
: \ \forall (\ub,u)\in D^\prime$

\vspace{2.5mm}

\ \ \ {\bf C4.5:} \
$\delta^{3/4}|u|^{-1/2}\|\sLh_{O_i}\s^{(L)}\ih\|_{L^4(S_{\ub,u})}\leq
1 \ : \ \forall (\ub,u)\in D^\prime$

\vspace{2.5mm}

\ \ \ {\bf C4.6:} \
$\delta^{3/4}|u|^{-1/2}\|O_i\s^{(L)}j\|_{L^4(S_{\ub,u})}\leq 1 \ :
\ \forall (\ub,u)\in D^\prime$

\vspace{2.5mm}

\ \ \ {\bf C4.7:} \
$\delta^{3/4}|u|^{-1/2}\|O_i\s^{(L)}\nu\|_{L^4(S_{\ub,u})}\leq 1 \
: \ \forall (\ub,u)\in D^\prime$

\vspace{2.5mm}

\ \ \ {\bf C4.8:} \
$\delta^{-1/4}|u|^{1/2}\|\sL_{O_i}\s^{(L)}\mb\|_{L^4(S_{\ub,u})}\leq
1 \ : \ \forall (\ub,u)\in D^\prime$

\vspace{2.5mm}

\ \ \ {\bf C5.1:} \
$\delta^{-1/4}|u|^{-1/2}\|\sLh_{O_j}\s^{(O_i)}\ih\|_{L^4(S_{\ub,u})}\leq
1 \ : \ \forall (\ub,u)\in D^\prime$

\vspace{2.5mm}

\ \ \ {\bf C5.2:} \
$\delta^{-1/4}|u|^{-1/2}\|O_j\s^{(O_i)}j\|_{L^4(S_{\ub,u})}\leq 1
\ : \ \forall (\ub,u)\in D^\prime$

\vspace{2.5mm}

\ \ \ {\bf C5.3:} \
$\delta^{-1/4}|u|^{-1/2}\|O_j\s^{(O_i)}\nu\|_{L^4(S_{\ub,u})}\leq
1 \ : \ \forall (\ub,u)\in D^\prime$

\vspace{2.5mm}

\ \ \ {\bf C5.4:} \
$\delta^{1/4}|u|^{-1}\|\sL_{O_j}\s^{(O_i)}m\|_{L^4(S_{\ub,u})}\leq
1 \ : \ \forall (\ub,u)\in D^\prime$

\vspace{2.5mm}

\ \ \ {\bf C6.1:} \
$\delta^{-1/4}|u|^{-1/2}\|\sLh_{O_i}\s^{(S)}\ih\|_{L^4(S_{\ub,u})}\leq
1 \ : \ \forall (\ub,u)\in D^\prime$

\vspace{2.5mm}

\ \ \ {\bf C6.2:} \
$\delta^{-1/4}|u|^{-1/2}\|O_i\s^{(S)}j\|_{L^4(S_{\ub,u})}\leq 1 \
: \ \forall (\ub,u)\in D^\prime$

\vspace{2.5mm}

\ \ \ {\bf C6.3:} \
$\delta^{-1/4}|u|^{-1/2}\|O_i\s^{(S)}\nu\|_{L^4(S_{\ub,u})}\leq 1
\ : \ \forall (\ub,u)\in D^\prime$

\vspace{2.5mm}

\ \ \ {\bf C6.4:} \
$\delta^{1/4}|u|^{-1}\|\sL_{O_i}\s^{(S)}m\|_{L^4(S_{\ub,u})}\leq 1
\ : \ \forall (\ub,u)\in D^\prime$

\vspace{2.5mm}

\ \ \ {\bf C6.5:} \
$\delta^{-5/4}|u|^{1/2}\|\sL_{O_i}\s^{(S)}\mb\|_{L^4(S_{\ub,u})}\leq
1 \ : \ \forall (\ub,u)\in D^\prime$

\vspace{2.5mm}

\ \ \ {\bf C6.6:} \
$\delta^{-1/4}|u|^{-1/2}\|\sLh_S\s^{(S)}\ih\|_{L^4(S_{\ub,u})}\leq
1 \ : \ \forall (\ub,u)\in D^\prime$

\vspace{2.5mm}

\ \ \ {\bf C6.7:} \
$\delta^{-1/4}|u|^{-1/2}\|S\s^{(S)}j\|_{L^4(S_{\ub,u})}\leq 1 \ :
\ \forall (\ub,u)\in D^\prime$

\vspace{2.5mm}

\ \ \ {\bf C6.8:} \
$\delta^{-1/4}|u|^{-1/2}\|S\s^{(S)}\nu\|_{L^4(S_{\ub,u})}\leq 1 \
: \ \forall (\ub,u)\in D^\prime$

\vspace{2.5mm}

\ \ \ {\bf C6.9:} \
$\delta^{1/4}|u|^{-1}\|\sL_S\s^{(S)}m\|_{L^4(S_{\ub,u})}\leq 1 \ :
\ \forall (\ub,u)\in D^\prime$

\vspace{2.5mm}

\ \ \ {\bf C6.10:} \
$\delta^{-5/4}|u|^{1/2}\|\sL_S\s^{(S)}\mb\|_{L^4(S_{\ub,u})}\leq 1
\ : \ \forall (\ub,u)\in D^\prime$

\vspace{5mm}

Comparing with the results of Chapter 9 we see that there is room
of a factor of $\delta^{1/4}$ at least in the above bootstrap
assumptions, a fact which will allow us to show that the
inequalities in these bootstrap assumptions are not saturated.

We shall also make use of the following supplementary bootstrap
assumptions on the rotation fields.

\vspace{5mm}

\ \ \ {\bf D0:} \ $|u|^{-1}\|O_i\|_{L^\infty(S_{\ub,u})}\leq 4 \ :
\ \forall (\ub,u)\in D^\prime$

\vspace{2.5mm}

\ \ \ {\bf D1:} \ $\|\snab O_i\|_{L^\infty(S_{\ub,u})}\leq
2(3\sqrt{2}+1) \ : \ \forall (\ub,u)\in D^\prime$

\vspace{5mm}

Comparing with Proposition 8.1 together with the inequality
\ref{8.83}, and Proposition 8.2 we see that there is an extra
factor of 2 in the above assumptions a fact which will allow us to
show that the inequalities of these bootstrap assumptions are not
saturated. We shall also need the following supplementary
bootstrap assumptions on the connection coefficients.

\vspace{5mm}

\ \ \ {\bf D2.1:}
$\delta^{3/4}|u|^{1/2}\|\mbox{tr}\chi\|_{L^\infty(S_{\ub,u})}\leq
1 \ : \ \forall (\ub,u)\in D^\prime$

\vspace{2.5mm}

\ \ \ {\bf D2.2:} $\delta^{3/4}\|\chih\|_{L^\infty(S_{\ub,u})}\leq
1 \ : \ \forall (\ub,u)\in D^\prime$

\vspace{2.5mm}

\ \ \ {\bf D3.1:} $|u|\|\mbox{tr}\chib\|_{L^\infty(S_{\ub,u})}\leq
4 \ : \forall (\ub,u)\in D^\prime$

\vspace{2.5mm}

\ \ \ {\bf D3.2:}
$\delta^{-1/4}|u|^2\|\chibh\|_{L^\infty(S_{\ub,u})}\leq 1 \ :
\forall (\ub,u)\in D^\prime$

\vspace{2.5mm}

\ \ \ {\bf D4.1:}
$\delta^{1/4}|u|\|\eta\|_{L^\infty(S_{\ub,u})}\leq 1 \ : \ \forall
(\ub,u)\in D^\prime$

\vspace{2.5mm}

\ \ \ {\bf D4.2:}
$\delta^{1/4}|u|\|\etb\|_{L^\infty(S_{\ub,u})}\leq 1 \ : \ \forall
(\ub,u)\in D^\prime$

\vspace{2.5mm}

\ \ \ {\bf D5:}
$\delta^{1/4}|u|\|\omega\|_{L^\infty(S_{\ub,u})}\leq 1  \ : \
\forall (\ub,u)\in D^\prime$

\vspace{2.5mm}

\ \ \ {\bf D6:}
$\delta^{-1/2}|u|^2\|\omb\|_{L^\infty(S_{\ub,u})}\leq 1 \ : \
\forall (\ub,u)\in D^\prime$

\vspace{2.5mm}

\ \ \ {\bf D7:}
$\delta^{-1/2}|u|^{3/2}\|\sd\Omega\|_{L^4(S_{\ub,u})}\leq 1 \ : \
\forall (\ub,u)\in D^\prime$

\vspace{2.5mm}

\ \ \ {\bf D8:}
$\delta^{3/4}|u|\|\Db\mbox{tr}\chi\|_{L^4(S_{\ub,u})}\leq 1 \ : \
\forall (\ub,u)\in D^\prime$

\vspace{2.5mm}

\ \ \ {\bf D9.1:}
$|u|^{3/2}\|\sd\mbox{tr}\chib\|_{L^4(S_{\ub,u})}\leq 1 \ : \
\forall (\ub,u)\in D^\prime$

\vspace{2.5mm}

\ \ \ {\bf D9.2:}
$\delta^{-1/4}|u|^{5/2}\|\snab\chibh\|_{L^4(S_{\ub,u})}\leq 1 \ :
\ \forall (\ub,u)\in D^\prime$

\vspace{2.5mm}

\ \ \ {\bf D9.3:}
$\delta^{-3/4}|u|^{5/2}\|\Dbh\chibh\|_{L^4(S_{\ub,u})}\leq 1 \ : \
\forall (\ub,u)\in D^\prime$

\vspace{2.5mm}

\ \ \ {\bf D10.1:}
$\delta^{1/4}|u|^{3/2}\|\snab\eta\|_{L^4(S_{\ub,u})}\leq 1 \ : \
\forall (\ub,u)\in D^\prime$

\vspace{2.5mm}

\ \ \ {\bf D10.2:}
$\delta^{1/4}|u|^{3/2}\|\snab\etb\|_{L^4(S_{\ub,u})}\leq 1 \ :
\forall (\ub,u)\in D^\prime$

\vspace{2.5mm}

\ \ \ {\bf D10.3:}
$\delta^{1/4}|u|^{3/2}\|\Db\eta\|_{L^4(S_{\ub,u})}\leq 1 \ : \
\forall (\ub,u)\in D^\prime$

\vspace{2.5mm}

\ \ \ {\bf D10.4:}
$\delta^{1/4}|u|^{3/2}\|\Db\etb\|_{L^4(S_{\ub,u})}\leq 1 \ : \
\forall (\ub,u)\in D^\prime$

\vspace{2.5mm}

\ \ \ {\bf D11:}
$\delta^{1/4}|u|^{3/2}\|\Db\omega\|_{L^4(S_{\ub,u})}\leq 1 \ :
\forall (\ub,u)\in D^\prime$

\vspace{2.5mm}

\ \ \ {\bf D12:}
$\delta^{-1/2}|u|^{5/2}\|\sd\omb\|_{L^4(S_{\ub,u})}\leq 1 \ :
\forall (\ub,u)\in D^\prime$

\vspace{5mm}

Comparing with the results of Chapters 3 and 4 we see that there
is room of a factor of $\delta^{1/4}$ at least in the above
bootstrap assumptions, a fact which will allow us to show that the
inequalities in these bootstrap assumptions are not saturated.

\vspace{5mm}

\noindent{\bf Lemma 12.6} \ \ \ Under the bootstrap assumptions
{\bf A0}, {\bf A1.1}, {\bf A1.2}, {\bf A2.1}, {\bf A2.2}, {\bf
A3.1}, {\bf A3.2}, {\bf A4.1}, {\bf A4.2}, {\bf B1}, {\bf B2},
{\bf B3}, and {\bf C1.1} - {\bf C1.4}, {\bf C2.1} - {\bf C2.4},
{\bf C3.1} - {\bf C3.5}, {\bf C4.1} - {\bf C4.8}, {\bf C5.1} -
{\bf C5.4}, {\bf C6.1} - {\bf C6.10}, as well as {\bf D0}, {\bf
D1}, {\bf D2.1}, {\bf D2.2}, {\bf D3.1}, {\bf D3.2}, {\bf D4.1},
{\bf D4.2}, {\bf D5}, {\bf D6}, {\bf D7}, {\bf D8}, {\bf D9.1} -
{\bf D9.3}, {\bf D10.1} - {\bf D10.4}, {\bf D11}, {\bf D 12},
there is a numerical constant $C$ such that:
$$({\cal Q}^\prime_2)^2\leq C\max\{{\cal P}_2,({\cal D}^{\prime 4}_{[1]}(\alb))^2\}$$
provided that $\delta$ does not exceed a certain positive numerical constant. Moreover, we have:
\begin{eqnarray*}
&&({\cal R}_{[2]}(\alpha))^2\leq C\stackrel{(0)}{{\cal E}}_2+C\delta^{1/2}\max\{{\cal P}_1,({\cal D}^{\prime 4}_0(\alb))^2\}\\
&&({\cal R}_{[2]}(\beta))^2\leq C\stackrel{(1)}{{\cal E}}_2+C\delta^{1/2}\max\{{\cal P}_1,({\cal D}^{\prime 4}_0(\alb))^2\}\\
&&({\cal R}_{[2]}(\rho))^2,({\cal R}_{[1]}(\sigma))^2\leq
C\stackrel{(2)}{{\cal E}}_2
+C\delta^{1/2}\max\{{\cal P}_1,({\cal D}^{\prime 4}_0(\alb))^2\}\\
&&({\cal R}_{[2]}(\beb))^2\leq C\stackrel{(3)}{{\cal E}}_2
+C\delta^{1/2}\max\{{\cal P}_2,({\cal D}^{\prime
4}_{[1]}(\alb))^2\} +C\max\{{\cal P}_1,({\cal D}^{\prime
4}_0(\alb))^2\}
\end{eqnarray*}
and:
$$(\cRb_{[2]}(\alb))^2\leq C\stackrel{(3)}{{\cal F}}_2
+C\delta\max\{{\cal P}_2,({\cal D}^{\prime 4}_{[1]}(\alb))^2\}
+C\delta^{1/2}\max\{{\cal P}_1,({\cal D}^{\prime 4}_0(\alb))^2\}$$

\noindent{\em Proof:} Taking $Y=L, O_i:i=1,2,3$ and $W=\tcL_L R$
and also $Y=O_j:j=1,2,3$ and $W=\tcL_{O_i}R:i=1,2,3$ in
Proposition 12.2 and using the bootstrap assumptions {\bf C1.1} -
{\bf C1.4}, {\bf C2.1} -{\bf C2.4}, {\bf C4.1} - {\bf C4.8}, {\bf
C5.1} - {\bf C5.4}, as well as the basic bootstrap assumption {\bf
A0}, we deduce, in view of \ref{12.103}, \ref{12.108},
\ref{12.113}, and \ref{12.129}, \ref{12.130}, and  \ref{12.146},
using also \ref{12.155},
\begin{eqnarray}
&&\int_{C_u}\left\{|\alpha|^2+\delta^2|\Dh\alpha|^2+\sum_i|\sLh_{O_i}\alpha|^2\right.\nonumber\\
&&\hspace{17mm}\left.+\delta^4|\Dh\Dh\alpha|^2+\delta^2\sum_i|\sLh_{O_i}\Dh\alpha|^2
+\sum_{i,j}|\sLh_{O_j}\sLh_{O_i}\alpha|^2\right\}\nonumber\\
&&\hspace{7mm}\leq
C\stackrel{(0)}{E}_2(u)+C\delta^{-1/2}\int_{C_u}|u|^2\left\{\delta^2|D\beta|^2+\sum_i|\sL_{O_i}\beta|^2\right\}
+\delta^{-3/2}({\cal Q}_1)^2\nonumber\\
&&\int_{C_u}|u|^2\left\{|\beta|^2+\delta^2|D\beta|^2+\sum_i|\sL_{O_i}\beta|^2\right.\nonumber\\
&&\hspace{17mm}\left.+\delta^4|DD\beta|^2+\delta^2\sum_i|\sL_{O_i}D\beta|^2
+\sum_{i,j}|\sL_{O_j}\sL_{O_i}\beta|^2\right\}\nonumber\\
&&\hspace{7mm}\leq C\stackrel{(1)}{E}_2(u)+C\delta^{5/2}\int_{C_u}\left\{\delta^2|\Dh\alpha|^2+\sum_i|\sLh_{O_i}\alpha|^2\right\}\nonumber\\
&&\hspace{7mm}+C\delta^{-1/2}\int_{C_u}|u|^4\left\{\delta^2(|D\rho|^2+|D\sigma|^2)+\sum_i(|O_i\rho|^2+|O_i\sigma|^2)\right\}
+C\delta^{1/2}({\cal Q}_1)^2\nonumber\\
&&\int_{C_u}|u|^4\left\{|\rho|^2+|\sigma|^2+\delta^2(|D\rho|^2+|D\sigma|^2)
+\sum_i(|O_i\rho|^2+|O_i\sigma|^2)\right.\nonumber\\
&&\hspace{17mm}+\delta^4(|DD\rho|^2+|DD\sigma|^2)+\delta^2\sum_i(|O_i D\rho|^2+|O_i D\sigma|^2)\nonumber\\
&&\hspace{17mm}\left.+\sum_{i,j}(|O_j O_i\rho|^2+|O_j O_i\sigma|^2)\right\}\nonumber\\
&&\hspace{7mm}\leq\stackrel{(2)}{E}_2(u)+C\delta^{5/2}\int_{C_u}|u|^2\left\{\delta^2|D\beta|^2
+\sum_i|\sL_{O_i}\beta|^2\right\}\nonumber\\
&&\hspace{7mm}+C\delta^{-1/2}\int_{C_u}|u|^6\left\{\delta^2|D\beb|^2+\sum_i|\sL_{O_i}\beb|^2\right\}
+C\delta^{3/2}({\cal Q}_1)^2\nonumber\\
&&\int_{C_u}|u|^6\left\{|\beb|^2+\delta^2|D\beb|^2+\sum_i|\sL_{O_i}\beb|^2\right.\nonumber\\
&&\hspace{17mm}\left.+\delta^4|DD\beb|^2+\delta^2\sum_i|\sL_{O_i}D\beb|^2+\sum_{i,j}|\sL_{O_j}\sL_{O_i}\beb|^2\right\}
\nonumber\\
&&\hspace{7mm}\leq
C\stackrel{(3)}{E}_2(u)+C\delta^{5/2}\int_{C_u}|u|^4\left\{\delta^2(|D\rho|^2+|D\sigma|^2)
+\sum_i(|O_i\rho|^2+|O_i\sigma|^2)\right\}\nonumber\\
&&\hspace{7mm}+C\delta^{-1/2}\int_{C_u}|u|^7\left\{\delta^2|\Dh\alb|^2+\sum_i|\sLh_{O_i}\alb|^2\right\}
+C\delta^{7/2}({\cal Q}_1)^2 \label{12.164}
\end{eqnarray}
We note here that by assumptions {\bf D0}, {\bf D1}:
\begin{eqnarray}
&&\sum_i|\sLh_{O_i}\alpha|^2\leq C(|u|^2|\snab\alpha|^2+|\alpha|^2)\nonumber\\
&&\sum_i|\sLh_{O_i}\alb|^2\leq C(|u|^2|\snab\alb|^2+|\alb|^2)\nonumber\\
&&\sum_i|\sL_{O_i}\beta|^2\leq C(|u|^2|\snab\beta|^2+|\beta|^2)\nonumber\\
&&\sum_i|\sL_{O_i}\beb|^2\leq C(|u|^2|\snab\beb|^2+|\beb|^2)\nonumber\\
&&\sum_i(|O_i\rho|^2+|O_i\sigma|^2)\leq
C|u|^2(|\sd\rho|^2+|\sd\sigma|^2) \label{12.165}
\end{eqnarray}
Thus, in reference to the integrals on the right hand sides of the
inequalities \ref{12.164} we have:
\begin{eqnarray}
&&\int_{C_u}\left\{\delta^2|\Dh\alpha|^2+\sum_i|\sLh_{O_i}\alpha|^2\right\}\leq
C\delta^{-2}({\cal R}_{[1]}(\alpha))^2
\leq C\delta^{-2}({\cal Q}_1)^2\nonumber\\
&&\int_{C_u}|u|^2\left\{\delta^2|D\beta|^2+\sum_i|\sL_{O_i}\beta|^2\right\}\leq
C({\cal R}_{[1]}(\beta))^2
\leq C({\cal Q}_1)^2\nonumber\\
&&\int_{C_u}|u|^4\left\{\delta^2(|D\rho|^2+|D\sigma|^2)+\sum_i(|O_i\rho|^2+|O_i\sigma|^2)\right\}\nonumber\\
&&\hspace{30mm}
\leq C\delta(({\cal R}_{[1]}(\rho))^2+({\cal R}_{[1]}(\sigma))^2)\leq 2C\delta({\cal Q}_1)^2\nonumber\\
&&\int_{C_u}|u|^6\left\{\delta^2|D\beb|^2+\sum_i|\sL_{O_i}\beb|^2\right\}
\leq C\delta^3({\cal R}_{[1]}(\beb))^2\leq C\delta^3({\cal Q}_1)^2\nonumber\\
&&\label{12.166}
\end{eqnarray}
Also, by \ref{12.148} and \ref{12.146}:
\begin{eqnarray}
&&\int_{C_u}|u|^7\sum_i|\sLh_{O_i}\alb|^2\leq C\int_{C_u}|u|^7\left\{|u|^2|\snab\alb|^2+|\alb|^2\right\}\nonumber\\
&&\hspace{20mm}\leq C\delta^4((\scR_1^4(\alb))^2+({\cal
R}_0^4(\alb))^2)\leq C^\prime\delta^4({\cal Q}_2)^2 \label{12.167}
\end{eqnarray}
The integral
$$\int_{C_u}|u|^7|\Dh\alb|^2$$
remains to be appropriately bounded. This shall be done through a
suitable bound for $\|\Dh\alb\|_{L^4(S_{\ub,u})}$. To derive this
bound we consider the second of the Bianchi identities, given by
Proposition 1.2:
\begin{equation}
\Dh\alb-\frac{1}{2}\Omega\mbox{tr}\chi\alb+2\omega\alb=-\Omega\left\{\snab\oth\beb+(4\etb-\zeta)\oth\beb
+3\chibh\rho-3\s^*\chibh\sigma\right\} \label{12.168}
\end{equation}
By \ref{12.148},
\begin{equation}
\|\snab\oth\beb\|_{L^4(S_{\ub,u})}\leq
C\delta|u|^{-9/2}\scR_1^4(\beb)\leq C^\prime\delta|u|^{-9/2}{\cal
Q}_2 \label{12.169}
\end{equation}
By assumption {\bf D3.2} and \ref{12.146}:
\begin{eqnarray}
&&\|\chibh\rho-\s^*\chibh\sigma\|_{L^4(S_{\ub,u})}\leq
C\delta^{1/4}|u|^{-2}\left(\|\rho\|_{L^4(S_{\ub,u})}
+\|\sigma\|_{L^4(S_{\ub,u})}\right)\nonumber\\
&&\hspace{10mm}\leq C\delta^{1/4}|u|^{-9/2}({\cal
R}_0^4(\rho)+{\cal R}_0^4(\sigma)) \leq
C^\prime\delta^{1/4}|u|^{-9/2}{\cal Q}_1 \label{12.170}
\end{eqnarray}
Also, by assumptions {\bf D4.1}, {\bf D4.2} and \ref{12.146},
\begin{eqnarray}
&&\|(4\etb-\zeta)\oth\beb\|_{L^4(S_{\ub,u})}\leq C\delta^{-1/4}|u|^{-1}\|\beb\|_{L^4(S_{\ub,u})}\nonumber\\
&&\hspace{10mm}\leq C\delta^{3/4}|u|^{-9/2}{\cal R}_0^4(\beb)\leq
C^\prime\delta^{3/4}|u|^{-9/2}{\cal Q}_1 \label{12.171}
\end{eqnarray}
Moreover, by assumptions {\bf D2.1}, {\bf D5} and \ref{12.146},
\begin{eqnarray}
&&\|\mbox{tr}\chi\alb\|_{L^4(S_{\ub,u})}\leq \delta^{-3/4}|u|^{-1/2}\|\alb\|_{L^4(S_{\ub,u})}\nonumber\\
&&\hspace{10mm}\leq \delta^{3/4}|u|^{-9/2}{\cal R}_0^4(\alb)\leq
C\delta^{3/4}|u|^{-9/2}{\cal Q}_1 \label{12.172}
\end{eqnarray}
\begin{eqnarray}
&&\|\omega\alb\|_{L^4(S_{\ub,u})}\leq \delta^{-1/4}|u|^{-1}\|\alb\|_{L^4(S_{\ub,u})}\nonumber\\
&&\hspace{10mm}\leq \delta^{5/4}|u|^{-5}{\cal R}_0^4(\alb)\leq
C\delta^{5/4}|u|^{-5}{\cal Q}_1 \label{12.173}
\end{eqnarray}
In view of the bounds \ref{12.169} - \ref{12.173}, we conclude
from \ref{12.168} that:
\begin{equation}
\|\Dh\alb\|_{L^4(S_{\ub,u})}\leq C\delta|u|^{-9/2}{\cal
Q}_2+C\delta^{1/4}|u|^{-9/2}{\cal Q}_1 \label{12.174}
\end{equation}
It then follows that:
\begin{equation}
\int_{C_u}|u|^7|\Dh\alb|^2\leq C\delta^3({\cal
Q}_2)^2+C\delta^{3/2}({\cal Q}_1)^2 \label{12.175}
\end{equation}

Under the bootstrap assumptions {\bf B1}, {\bf B2}, {\bf B3}, we
have, from the coercivity inequality \ref{11.156} applied to
$\alpha$, $\alb$, the coercivity inequality \ref{11.144} applied
to $\beta$, $\beb$ and the coercivity inequality \ref{11.130}
applied to $\rho$, $\sigma$:
\begin{eqnarray}
&&\int_{S_{\ub,u}}|u|^4|\snab^{ \ 2}\alpha|^2\leq
C\int_{S_{\ub,u}}\left\{\sum_{i,j}|\sLh_{O_j}\sLh_{O_i}\alpha|^2
+\sum_i|\sLh_{O_i}\alpha|^2+|\alpha|^2\right\}\nonumber\\
&&\int_{S_{\ub,u}}|u|^4|\snab^{ \ 2}\alb|^2\leq
C\int_{S_{\ub,u}}\left\{\sum_{i,j}|\sLh_{O_j}\sLh_{O_i}\alb|^2
+\sum_i|\sLh_{O_i}\alb|^2+|\alb|^2\right\}\nonumber\\
&&\int_{S_{\ub,u}}|u|^4|\snab^{ \ 2}\beta|^2\leq
C\int_{S_{\ub,u}}\left\{\sum_{i,j}|\sL_{O_j}\sL_{O_i}\beta|^2
+\sum_i|\sL_{O_i}\beta|^2+|\beta|^2\right\}\nonumber\\
&&\int_{S_{\ub,u}}|u|^4|\snab^{ \ 2}\beb|^2\leq
C\int_{S_{\ub,u}}\left\{\sum_{i,j}|\sL_{O_j}\sL_{O_i}\beb|^2
+\sum_i|\sL_{O_i}\beb|^2+|\beb|^2\right\}\nonumber\\
&&\hspace{10mm}|u|^4|\snab^{ \ 2}\rho|^2\leq C\sum_{i,j}|O_j O_i\rho|^2\nonumber\\
&&\hspace{10mm}|u|^4|\snab^{ \ 2}\sigma|^2\leq C\sum_{i,j}|O_j
O_i\sigma|^2 \label{12.176}
\end{eqnarray}
Also, under the bootstrap assumptions {\bf B1}, {\bf B2}, we have,
from the coercivity inequality \ref{11.124}, applied to
$\Dh\alpha$, $\Dh\alb$ and $\Dbh\alb$ and from Proposition 11.1
itself applied to $D\beta$, $D\beb$ and $\Db\beb$, $\rho$,
$\sigma$:
\begin{eqnarray}
&&|u|^2|\snab\Dh\alpha|^2\leq C\left(\sum_i|\sLh_{O_i}\Dh\alpha|^2+|\Dh\alpha|^2\right)\nonumber\\
&&|u|^2|\snab\Dh\alb|^2\leq C\left(\sum_i|\sLh_{O_i}\Dh\alb|^2+|\Dh\alb|^2\right)\nonumber\\
&&|u|^2|\snab\Dbh\alb|^2\leq C\left(\sum_i|\sLh_{O_i}\Dbh\alb|^2+|\Dbh\alb|^2\right)\nonumber\\
&&|u|^2|\snab D\beta|^2\leq C\sum_i|\sL_{O_i}D\beta|^2\nonumber\\
&&|u|^2|\snab D\beb|^2\leq C\sum_i|\sL_{O_i}D\beb|^2\nonumber\\
&&|u|^2|\snab\Db\beb|^2\leq C\sum_i|\sL_{O_i}\Db\beb|^2\nonumber\\
&&|u|^2|\sd D\rho|^2\leq C\sum_i|O_i D\rho|^2\nonumber\\
&&|u|^2|\sd D\sigma|^2\leq C\sum_i|O_i D\sigma|^2 \label{12.177}
\end{eqnarray}

Substituting \ref{12.166}, \ref{12.167} and \ref{12.175}, as well
as \ref{12.154} (except the second) and \ref{12.176} (except the
second), \ref{12.177} (except the second, third and sixth) in
\ref{12.164} we obtain:
\begin{eqnarray}
&&\int_{C_u}\left\{|\alpha|^2+\delta^2|\Dh\alpha|^2+|u|^2|\snab\alpha|^2\right.\nonumber\\
&&\hspace{17mm}\left.+\delta^4|\Dh\Dh\alpha|^2+\delta^2|u|^2|\snab\Dh\alpha|^2
+|u|^4|\snab^{ \ 2}\alpha|^2\right\}\nonumber\\
&&\hspace{27mm}\leq C\stackrel{(0)}{E}_2(u)
+\delta^{-3/2}({\cal Q}_1)^2\nonumber\\
&&\int_{C_u}|u|^2\left\{|\beta|^2+\delta^2|D\beta|^2+|u|^2|\snab\beta|^2\right.\nonumber\\
&&\hspace{17mm}\left.+\delta^4|DD\beta|^2+\delta^2|u|^2|\snab
D\beta|^2
+|u|^4|\snab^{ \ 2}\beta|^2\right\}\nonumber\\
&&\hspace{27mm}\leq C\stackrel{(1)}{E}_2(u)
+C\delta^{1/2}({\cal Q}_1)^2\nonumber\\
&&\int_{C_u}|u|^4\left\{|\rho|^2+|\sigma|^2+\delta^2(|D\rho|^2+|D\sigma|^2)
+|u|^2(|\sd\rho|^2+|\sd\sigma|^2)\right.\nonumber\\
&&\hspace{17mm}+\delta^4(|DD\rho|^2+|DD\sigma|^2)+\delta^2|u|^2(|\sd D\rho|^2+|\sd D\sigma|^2)\nonumber\\
&&\hspace{17mm}\left.+|u|^4(|\snab^{ \ 2}\rho|^2+|\snab^{ \ 2}\sigma|^2)\right\}\nonumber\\
&&\hspace{27mm}\leq\stackrel{(2)}{E}_2(u)
+C\delta^{3/2}({\cal Q}_1)^2\nonumber\\
&&\int_{C_u}|u|^6\left\{|\beb|^2+\delta^2|D\beb|^2+|u|^2|\snab\beb|^2\right.\nonumber\\
&&\hspace{17mm}\left.+\delta^4||DD\beb|^2+\delta^2|u|^2|\snab
D\beb|^2+|u|^4|\snab^{ \ 2}\beb|^2\right\}
\nonumber\\
&&\hspace{17mm}\leq C\stackrel{(3)}{E}_2(u)+C\delta^{7/2}({\cal
Q}_2)^2 +C\delta^3({\cal Q}_1)^2 \label{12.178}
\end{eqnarray}
Multiplying the first of \ref{12.178} by $\delta^2$, the second by
1, the third by $\delta^{-1}$, and the fourth by $\delta^{-3}$,
and taking the supremum over $u\in[u_0,c^*)$, we obtain, in view
of the definitions \ref{10.60} - \ref{10.67} and \ref{12.139},
\begin{eqnarray}
&&({\cal R}_{[2]}(\alpha))^2\leq C\stackrel{(0)}{{\cal E}}_2+C\delta^{1/2}({\cal Q}_1)^2\nonumber\\
&&({\cal R}_{[2]}(\beta))^2\leq C\stackrel{(1)}{{\cal E}}_2+C\delta^{1/2}({\cal Q}_1)^2\nonumber\\
&&({\cal R}_{[2]}(\rho))^2,({\cal R}_{[2]}(\sigma))^2\leq
C\stackrel{(2)}{{\cal E}}_2+C\delta^{1/2}({\cal Q}_1)^2
\nonumber\\
&&({\cal R}_{[2]}(\beb))^2\leq C\stackrel{(3)}{{\cal
E}}_2+C\delta^{1/2}({\cal Q}_2)^2+C({\cal Q}_1)^2 \label{12.179}
\end{eqnarray}

Next, taking in Proposition 12.2 $Y=L,O_i:i=1,2,3$ and $W=\tcL_L
R$, and $Y=O_j:j=1,2,3$ and $W=\tcL_{O_i}R:i=1,2,3$, and also
$W=\tcL_S R$ and $Y=O_i:i=1,2,3, S$, and using the bootstrap
assumptions {\bf C1.1} - {\bf C1.4}, {\bf C2.1} - {\bf C2.4}, {\bf
C3.1} - {\bf C3.5}, {\bf C4.1} - {\bf C4.8}, {\bf C5.1} - {\bf
C5.4}, {\bf C6.1} - {\bf C6.10}, in view of \ref{12.104},
\ref{12.109}, \ref{12.114}, \ref{12.132}, and \ref{12.146}, using
also \ref{12.160}, and noting that by \ref{12.157} for any
trace-free symmetric 2-covariant  tensorfield $\theta$ we have
\begin{equation}
\sLh_{O_i}\sLh_S\theta=u\sLh_{O_i}\Dbh\theta+\ub\sLh_{O_i}\Dh\theta,
\label{12.180}
\end{equation}
we deduce:
\begin{eqnarray}
&&\int_{\Cb_{\ub}}|u|^6\left\{|\alb|^2+\delta^2|\Dh\alb|^2+|u|^2|\Dbh\alb|^2+\sum_i|\sLh_{O_i}\alb|^2\right.\nonumber\\
&&\hspace{15mm}+\delta^4|\Dh\Dh\alb|^2+\delta^2\sum_i|\sLh_{O_i}\Dh\alb|^2+|u|^2\sum_i|\sLh_{O_i}\Dbh\alb|^2\nonumber\\
&&\hspace{15mm}\left.+\sum_{i,j}|\sLh_{O_j}\sLh_{O_i}\alb|^2+|\sLh_S\sLh_S\alb|^2\right\}\nonumber\\
&&\hspace{10mm}\leq
C\stackrel{(3)}{F}_2(\ub)+C\delta^{5/2}\int_{\Cb_{\ub}}|u|^4\left\{\delta^2|D\beb|^2
+|\sL_S\beb|^2+\sum_i|\sL_{O_i}\beb|^2\right\}\nonumber\\
&&\hspace{10mm}+C\delta^{7/2}({\cal Q}_1)^2 \label{12.181}
\end{eqnarray}

Now, by the fourth of \ref{12.165} and \ref{12.148}, \ref{12.146}:
\begin{eqnarray}
&&\int_{\Cb_{\ub}}|u|^4\sum_i|\sL_{O_i}\beb|^2\leq C\int_{\Cb_{\ub}}|u|^4(|u|^2|\snab\beb|^2+|\beb|^2)\nonumber\\
&&\hspace{20mm}\leq C\delta^2((\scR_1^4(\beb))^2+({\cal
R}_0^4(\beb))^2)\leq C^\prime\delta^2({\cal Q}_2)^2 \label{12.182}
\end{eqnarray}
Also, by \ref{12.149}:
\begin{equation}
\int_{\Cb_{\ub}}|u|^4|D\beb|^2\leq C({\cal R}_0^4(D\beb))^2\leq
C^\prime({\cal Q}_2)^2 \label{12.183}
\end{equation}
Since for any $S$ 1-form $\xi$ we have;
\begin{equation}
\sL_S\xi=u\Db\xi+\ub D\xi \label{12.184}
\end{equation}
to bound the integral
$$\int_{\Cb_{\ub}}|u|^4|\sL_S\beb|^2$$
what remains to be bounded is the integral
$$\int_{\Cb_{\ub}}|u|^6|\Db\beb|^2$$
For this purpose we consider the fourth of the Bianchi identities,
given by Proposition 1.2:
\begin{equation}
\Db\beb+\frac{3}{2}\Omega\mbox{tr}\chib\beb-\Omega\chibh^\sharp\cdot\beb-\omb\beb=-\Omega\left\{\sdiv\alb
+(\eta^\sharp-2\zeta^\sharp)\cdot\alb\right\} \label{12.185}
\end{equation}
By \ref{12.148},
\begin{equation}
\|\sdiv\alb\|_{L^4(S_{\ub,u})}\leq
C\delta^{3/2}|u|^{-5}\scR_1^4(\alb)\leq
C^\prime\delta^{3/2}|u|^{-5}{\cal Q}_2 \label{12.186}
\end{equation}
By assumptions {\bf D3.1}, {\bf D3.2} and \ref{12.146},
\begin{eqnarray}
&&\|\mbox{tr}\chib\beb\|_{L^4(S_{\ub,u})}\leq
4|u|^{-1}\|\beb\|_{L^4(S_{\ub,u})}\leq 4\delta|u|^{-9/2}{\cal
R}_0^4(\beb)
\leq C\delta|u|^{-9/2}{\cal Q}_1\nonumber\\
&&\|\chibh^\sharp\cdot\beb\|_{L^4(S_{\ub,u})}\leq
\delta^{1/4}|u|^{-3/2}\|\beb\|_{L^4(S_{\ub,u})}
\leq \delta^{5/4}|u|^{-5}{\cal R}_0^4(\beb)\leq C\delta^{5/4}|u|^{-5}{\cal Q}_1\nonumber\\
&&\label{12.187}
\end{eqnarray}
Also, by assumption {\bf D6} and \ref{12.146},
\begin{equation}
\|\omb\beb\|_{L^4(S_{\ub,u})}\leq
\delta^{1/2}|u|^{-2}\|\beb\|_{L^4(S_{\ub,u})} \leq
\delta^{3/2}|u|^{-11/2}{\cal
R}_0^4(\beb)\leq\delta^{3/2}|u|^{-11/2}{\cal Q}_1 \label{12.188}
\end{equation}
Moreover, by assumptions {\bf D4.1}, {\bf D4.2} and \ref{12.146},
\begin{eqnarray}
&&\|(\eta^\sharp-2\zeta^\sharp)\cdot\alb\|_{L^4(S_{\ub,u})}\leq
C\delta^{-1/4}|u|^{-1}\|\alb\|_{L^4(S_{\ub,u})}
\nonumber\\
&&\hspace{10mm}\leq C\delta^{5/4}|u|^{-5}{\cal R}_0^4(\alb)\leq
C^\prime\delta^{5/4}|u|^{-5}{\cal Q}_1 \label{12.189}
\end{eqnarray}
In view of the bound \ref{12.186} - \ref{12.189}, we conclude from
\ref{12.185} that:
\begin{equation}
\|\Db\beb\|_{L^4(S_{\ub,u})}\leq C\delta^{3/2}|u|^{-5}{\cal
Q}_2+C\delta|u|^{-9/2}{\cal Q}_1 \label{12.190}
\end{equation}
It then follows that:
\begin{equation}
\int_{\Cb_{\ub}}|u|^6|\Db\beb|^2\leq C\delta^3({\cal
Q}_2)^2+C\delta^2({\cal Q}_1)^2 \label{12.191}
\end{equation}
which together with \ref{12.183} implies:
\begin{equation}
\int_{\Cb_{\ub}}|u|^4|\sL_S\beb|^2\leq C\delta^2({\cal Q}_2)^2
\label{12.192}
\end{equation}

Substituting \ref{12.182}, \ref{12.183}, \ref{12.192}, as well as
the second of \ref{12.154} and \ref{12.176} and the second and
third of \ref{12.177} in \ref{12.181} we obtain:
\begin{eqnarray}
&&\int_{\Cb_{\ub}}|u|^6\left\{|\alb|^2+\delta^2|\Dh\alb|^2+|u|^2|\Dbh\alb|^2+|u|^2|\snab\alb|^2\right.\nonumber\\
&&\hspace{15mm}+\delta^4|\Dh\Dh\alb|^2+\delta^2|u|^2|\snab\Dh\alb|^2+|u|^4|\snab\Dbh\alb|^2\nonumber\\
&&\hspace{15mm}\left.+|u|^4|\snab^{ \ 2}\alb|^2+|\sLh_S\sLh_S\alb|^2\right\}\nonumber\\
&&\hspace{10mm}\leq C\stackrel{(3)}{F}_2(\ub)+C\delta^{9/2}({\cal
Q}_2)^2+C\delta^{7/2}({\cal Q}_1)^2 \label{12.193}
\end{eqnarray}

We shall now derive, using the bound \ref{12.193}, an appropriate
bound for the integral
\begin{equation}
\int_{\Cb_{\ub}}|u|^{10}|\Dbh\Dbh\alb|^2 \label{12.a1}
\end{equation}
Using \ref{12.157} and \ref{1.9} we obtain:
\begin{eqnarray}
&&\sLh_S\sLh_S\alb=u^2\Dbh\Dbh\alb+\ub^2\Dh\Dh\alb+u\Dbh\alb+\ub\Dh\alb\nonumber\\
&&\hspace{15mm}+\ub u(\Dbh\Dh\alb+\Dh\Dbh\alb) \label{12.194}
\end{eqnarray}
Now, according to Lemma 1.4 for any $S$ tensorfield $\theta$ we
have:
\begin{equation}
\Db D\theta-D\Db\theta=4\sL_{\Omega^2\zeta^\sharp}\theta
\label{12.195}
\end{equation}
For any trace-free symmetric 2-covariant $S$ tensorfield $\theta$
we have:
\begin{eqnarray}
&&D\theta=\Dh\theta+\frac{1}{2}\sg(\Omega\chih,\theta)\nonumber\\
&&\Db\theta=\Dbh\theta+\frac{1}{2}\sg(\Omega\chibh,\theta)
\label{12.196}
\end{eqnarray}
It follows that in the case of a trace-free 2-covariant $S$
tensorfield $\theta$ the following commutation formula holds:
\begin{eqnarray}
&&\Dbh\Dh\theta-\Dh\Dbh\theta=4\sLh_{\Omega^2\zeta^\sharp}\theta\label{12.197}\\
&&\hspace{25mm}-\frac{1}{2}\Omega\chibh(\Omega\chih,\theta)+\frac{1}{2}\Omega\chih(\Omega\chibh,\theta)\nonumber
\end{eqnarray}
We consider this formula with $\alb$ in the role of $\theta$.
Using assumptions {\bf D7} and {\bf D10.1}, {\bf D10.2} we deduce:
\begin{equation}
\|\sLh_{\Omega^2\zeta^\sharp}\alb\|_{L^2(S_{\ub,u})}\leq
C\delta^{-1/4}|u|^{-1}\|\snab\alb\|_{L^2(S_{\ub,u})}
+C\delta^{-1/4}|u|^{-3/2}\|\alb\|_{L^4(S_{\ub,u})} \label{12.198}
\end{equation}
Taking into account the fact that by \ref{12.146}
\begin{equation}
\|\alb\|_{L^4(S_{\ub,u})}\leq \delta^{3/2}|u|^{-4}{\cal
R}_0^4(\alb)\leq C\delta^{3/2}|u|^{-4}{\cal Q}_1 \label{12.199}
\end{equation}
it follows that:
\begin{equation}
\int_{\Cb_{\ub}}|u|^8|\sLh_{\Omega^2\zeta^\sharp}\alb|^2\leq
C\delta^{5/2}({\cal Q}_1)^2 \label{12.200}
\end{equation}
Moreover, by assumptions {\bf D2.2}, {\bf D3.2} we have:
\begin{eqnarray}
&&\int_{\Cb_{\ub}}|u|^8|\chibh(\chih,\alb)|^2,\int_{\Cb_{\ub}}|u|^8|\chih(\chibh,\alb)|^2\nonumber\\
&&\hspace{10mm}\leq\delta^{-1}\int_{\Cb_{\ub}}|u|^5|\alb|^2\leq\delta^2({\cal
Q}_1)^2 \label{12.201}
\end{eqnarray}
We then conclude from \ref{12.197} with $\alb$ in the role of
$\theta$ that:
\begin{equation}
\int_{\Cb_{\ub}}|u|^8|\Dbh\Dh\alb-\Dh\Dbh\alb|^2\leq
C\delta^2({\cal Q}_1)^2 \label{12.202}
\end{equation}

In view of the formula \ref{12.194} and the bounds \ref{12.193}
and \ref{12.201} the required bound for the integral \ref{12.a1}
will follow if we can appropriately estimate the integral
\begin{equation}
\int_{\Cb_{\ub}}|u|^8|\Dbh\Dh\alb|^2 \label{12.a2}
\end{equation}
For this purpose we apply $\Dbh$ to the Bianchi identity
\ref{12.168} to obtain an expression for $\Dbh\Dh\alb$. In
deriving this expression we make use of the following three
elementary facts. First, that for an arbitrary $S$ 1-form $\xi$ we
have:
\begin{equation}
\Dbh\snab\oth\xi=\snab\oth\Db\xi-2\hat{\Db\sGamma}\cdot\xi-2\Omega\chibh\sdiv\xi
\label{12.203}
\end{equation}
This formula follows in a straightforward manner from the second
part of Lemma 4.1 together with the conjugate of formula
\ref{6.107}. The second fact, which also follows in a
straightforward manner, is that for any pair  of $S$ 1-forms
$\xi$, $\theta$ we have:
\begin{equation}
\Dbh(\xi\oth\theta)=\Db\xi\oth\theta+\xi\oth\Db\theta-2\Omega\chibh(\xi,\theta)
\label{12.204}
\end{equation}
Finally, the third fact is that for any trace-free symmetric
2-covariant $S$ tensorfield $\theta$ we have:
\begin{equation}
\Dbh\s^*\theta=\s^*\Dbh\theta \label{12.205}
\end{equation}
(Note that an analogous formula does not hold for $S$ 1-forms). To
derive \ref{12.205} we begin from the fact that:
\begin{equation}
\Db\seps=\Omega\mbox{tr}\chib\seps \label{12.206}
\end{equation}
This implies that, with the notations of Chapter 1 (see
\ref{1.118}),
\begin{equation}
\Db\seps^\sharp=-2\Omega\s^*\chibh^\sharp \label{12.207}
\end{equation}
Let then $\theta$ be an arbitrary trace-free symmetric 2-covariant
$S$ tensorfield. With respect to an arbitrary local frame field
for the $S_{\ub,u}$ we then have:
\begin{equation}
(\Db\s^*\theta)_{AB}=\seps^{\sharp\s
C}_A(\Db\theta)_{CB}-2\Omega\s^*\chibh^{\sharp\s C}_A\theta_{CB}
\label{12.208}
\end{equation}
This is symmetric, for,
$$\seps^{\sharp\sharp AB}(\seps^{\sharp\s C}_A(\Db\theta)_{CB}-2\Omega\s^*\chibh^{\sharp\s C}_A\theta_{CB})
=(\sg^{-1})^{BC}(\Db\theta)_{CB}-2\Omega\chibh^{\sharp\sharp
BC}\theta_{CB}=0$$ Thus we can write:
\begin{eqnarray}
&&(\Db\s^*\theta)_{AB}=\frac{1}{2}(\seps^{\sharp\s C}_A(\Db\theta)_{CB}+\seps^{\sharp\s C}_B(\Db\theta)_{CA})\nonumber\\
&&\hspace{18mm}-\Omega(\s^*\chibh^{\sharp\s C}_A\theta_{CB}+\s^*\chibh^{\sharp\s C}_B\theta_{CA})\nonumber\\
&&\hspace{15mm}=\frac{1}{2}(\seps^{\sharp\s
C}_A(\Dbh\theta)_{CB}+\seps^{\sharp\s
C}_B(\Dbh\theta)_{CA})-\Omega\sg_{AB}(\s^*\chibh,\theta)
\label{12.209}
\end{eqnarray}
by the identity \ref{1.163}. Taking the trace free parts on both
sides, and recalling that the dual of a trace-free symmetric
2-covariant $S$ tensorfield is also symmetric and trace-free, then
yields \ref{12.205}.

Applying $\Dbh$ to the Bianchi identity \ref{12.168} and adding
$\frac{3}{2}\Omega\mbox{tr}\chib$ times the Bianchi identity
\ref{12.168}, and using the above three facts we deduce the
following expression for
$\Dbh\Dh\alb+\frac{3}{2}\Omega\mbox{tr}\chib\Dh\alb$:
\begin{eqnarray}
&&\Dbh\Dh\alb+\frac{3}{2}\Omega\mbox{tr}\chib\Dh\alb=\frac{1}{2}\Omega\mbox{tr}\chi\Dbh\alb-2\omega\Dbh\alb
+\frac{1}{2}\alb\Db(\Omega\mbox{tr}\chi)-2\alb\Db\omega\nonumber\\
&&\hspace{29mm}+\Omega\left(\frac{1}{2}\omb\mbox{tr}\chi-3\omega\mbox{tr}\chib
+\frac{3}{4}\Omega\mbox{tr}\chi\mbox{tr}\chib\right)\alb\nonumber\\
&&\hspace{29mm}-\Omega\left\{\snab\oth\Db\beb-2\hat{\Db\sGamma}\cdot\beb-2\omega\chibh\sdiv\beb\right.\nonumber\\
&&\hspace{29mm}+(4\Db\etb-\Db\zeta)\oth\beb+(4\etb-\zeta)\oth\Db\beb-2\Omega\chibh(4\etb-\zeta,\beb)\nonumber\\
&&\hspace{29mm}+3\rho\Dbh\chibh-3\sigma\s^*\Dbh\chibh\nonumber\\
&&\hspace{29mm}\left.+3\chibh\left(\Db\rho+\frac{3}{2}\Omega\mbox{tr}\chib\rho\right)
-3\s^*\chibh\left(\Db\sigma+\frac{3}{2}\Omega\mbox{tr}\chib\sigma\right)\right\}\nonumber\\
&&\hspace{29mm}-\Omega\left(\frac{3}{2}\Omega\mbox{tr}\chib+\omb\right)\left\{\snab\oth\beb+(4\etb-\zeta)\oth\beb\right\}\nonumber\\
&&\hspace{29mm}-3\Omega\omb(\chibh\rho-\s^*\chibh\sigma)\label{12.210}
\end{eqnarray}

Consider first the principal term on the right. This is the term
$\snab\oth\Db\beb$ in the second parenthesis. Now $\Db\beb$ is
given by the Bianchi identity \ref{12.185}. The principal term in
$\Db\beb$ is the term $-\Omega\sdiv\alb$. The contibution of this
term to the integral
\begin{equation}
\int_{\Cb_{\ub}}|u|^8|\snab\oth\Db\beb|^2 \label{12.211}
\end{equation}
is bounded through \ref{12.193} by:
\begin{equation}
C\stackrel{(3)}{F}_2(\ub)+C\delta^{9/2}({\cal
Q}_2)^2+C\delta^{7/2}({\cal Q}_1)^2 \label{12.212}
\end{equation}
Using assumptions {\bf D3.1}, {\bf D3.2}, {\bf D4.1}, {\bf D4.2},
{\bf D6}, {\bf D7}, {\bf D9.1}, {\bf D9.2}, {\bf D10.1}, {\bf
D10.2}, and {\bf D12}, we deduce that the contributions of the
remaining terms in $\Db\beb$ to $\snab\oth\Db\beb$ are bounded in
$L^2(S_{\ub,u})$ by:
\begin{equation}
C\delta|u|^{-5}(\scR_1^4+{\cal R}_0^4)\leq
C^\prime\delta|u|^{-5}{\cal Q}_2 \label{12.213}
\end{equation}
by \ref{12.148} and \ref{12.146}. We then conclude that:
\begin{equation}
\int_{\Cb_{\ub}}|u|^8|\snab\oth\Db\beb|^2\leq
C\stackrel{(3)}{F}_2(\ub)+C\delta^2({\cal Q}_2)^2 \label{12.214}
\end{equation}

Consider next the terms
$$3\chibh\left(\Db\rho+\frac{3}{2}\Omega\mbox{tr}\chib\rho\right)-3\s^*\chibh\left(\Db\sigma+\frac{3}{2}\mbox{tr}\chib\sigma\right)$$
in the second parenthesis on the right in \ref{12.210}. The eighth
and tenth Bianchi identities of Proposition 1.2 read:
\begin{eqnarray}
&&\Db\rho+\frac{3}{2}\Omega\mbox{tr}\chib\rho=-\Omega\left\{\sdiv\beb+(2\eta-\zeta,\beb)+\frac{1}{2}(\chih,\alb)\right\}\nonumber\\
&&\Db\sigma+\frac{3}{2}\Omega\mbox{tr}\chib\sigma=-\Omega\left\{\scurl\beb+(2\eta-\zeta,\s^*\beb)+\frac{1}{2}\chih\wedge\alb\right\}
\label{12.215}
\end{eqnarray}
By assumptions {\bf D4.1}, {\bf D4.2}, {\bf D2.2} and
\ref{12.148}, \ref{12.146} the right hand sides of equations
\ref{12.215} are bounded in $L^4(S_{\ub,u})$ by:
\begin{equation}
C\delta^{3/4}|u|^{-4}(\scR_1^4+{\cal R}_0^4)\leq
C^\prime\delta^{3/4}|u|^{-4}{\cal Q}_2 \label{12.217}
\end{equation}
Taking also into account assumption {\bf D3.2} we then conclude
that:
\begin{equation}
\left\|\chibh\left(\Db\rho+\frac{3}{2}\Omega\mbox{tr}\chib\rho\right)-\s^*\chibh\left(\Db\sigma+\frac{3}{2}\Omega\mbox{tr}\chib\sigma\right)\right\|_{L^2(S_{\ub,u})}\leq
C\delta|u|^{-11/2}{\cal Q}_2 \label{12.218}
\end{equation}

From the expression for $\Db\sGamma$ of the second part of Lemma
4.1 and assumptions {\bf D9.1}, {\bf D9.2} and {\bf D3.1}, {\bf
D3.2} and {\bf D7} we deduce:
\begin{equation}
\|\hat{\Db\sGamma}\|_{L^4(S_{\ub,u})}\leq C|u|^{-3/2}
\label{12.219}
\end{equation}
Using assumptions {\bf D3.2}, {\bf D5}, {\bf D9.3}, {\bf D10.3},
{\bf D10.4} as well as the bounds \ref{12.190} and \ref{12.219} we
deduce that the remaining terms in the second parenthesis on the
right in \ref{12.210} are bounded in $L^2(S_{\ub,u})$ by:
\begin{equation}
C\delta|u|^{-11/2}{\cal Q}_2+C\delta^{3/4}|u|^{-5}{\cal Q}_1
\label{12.220}
\end{equation}
Also, using assumptions {\bf D3.2}, {\bf D4.1}, {\bf D4.2}, and
{\bf D6} we deduce that the last two terms in \ref{12.210} are
bounded in $L^2(S_{\ub,u})$ by:
\begin{equation}
C\delta|u|^{-5}{\cal Q}_2+C\delta^{3/4}|u|^{-5}{\cal Q}_1
\label{12.221}
\end{equation}

The first five terms on the right in \ref{12.210} remain to be
considered. By assumptions {\bf D2.1}, {\bf D5}, {\bf D6} and the
bound \ref{12.193} the contributions to the integral \ref{12.a2}
of the first and second terms in \ref{12.210} are bounded by:
\begin{equation}
C\delta^{-3/2}\stackrel{(3)}{F}_2(\ub)+C\delta^3({\cal
Q}_2)^2+C\delta^2({\cal Q}_1)^2 \label{12.222}
\end{equation}
Also, by assumptions {\bf D8}, {\bf D11} and \ref{12.146} the
third and fourth terms in \ref{12.210} are bounded in
$L^2(S_{\ub,u})$ by:
\begin{equation}
C\delta^{3/4}|u|^{-5}{\cal Q}_1 \label{12.223}
\end{equation}
Finally, by assumptions {\bf D2.1}, {\bf D3.1}, {\bf D5}, {\bf D6}
the fifth term in \ref{12.210} is bounded in $L^2(S_{\ub,u})$ by:
\begin{equation}
C\delta^{3/4}|u|^{-5}{\cal Q}_1 \label{12.224}
\end{equation}

Collecting the above results we conclude that:
\begin{equation}
\int_{\Cb_{\ub}}|u|^8\left|\Dbh\Dh\alb+\frac{3}{2}\Omega\mbox{tr}\chib\Dh\alb\right|^2\leq
C\delta^{-3/2}\stackrel{(3)}{F}_2(\ub)+C\delta^2({\cal
Q}_2)^2+C\delta^{3/2}({\cal Q}_1)^2 \label{12.225}
\end{equation}
Combining with the bound
\begin{equation}
\int_{\Cb_{\ub}}|u|^8\left|\frac{3}{2}\Omega\mbox{tr}\chib\Dh\alb\right|^2\leq
C\delta^{-2}\stackrel{(3)}{F}_2(\ub)+C\delta^{5/2}({\cal
Q}_2)^2+C\delta^{3/2}({\cal Q}_1)^2 \label{12.226}
\end{equation}
which follows from the estimate \ref{12.193} and assumption {\bf
D3.1}, we then obtain:
\begin{equation}
\int_{\Cb_{\ub}}|u|^8|\Dbh\Dh\alb|^2\leq
C\delta^{-2}\stackrel{(3)}{F}_2(\ub)+C\delta^2({\cal
Q}_2)^2+C\delta^{3/2}({\cal Q}_1)^2 \label{12.227}
\end{equation}
This estimate together with the estimate \ref{12.202} implies:
\begin{equation}
\int_{\Cb_{\ub}}|\ub|^2|u|^8|\Dbh\Dh\alb+\Dh\Dbh\alb|^2\leq
C\stackrel{(3)}{F}_2(\ub)+C\delta^4({\cal Q}_2)^2
+C\delta^{7/2}({\cal Q}_1)^2 \label{12.228}
\end{equation}
Using equation \ref{12.194} to expresses $u^2\Dbh\Dbh\alb$ in
terms of $\sLh_S\sLh_S\alb$, $\ub u(\Dbh\Dh\alb+\Dh\Dbh\alb)$,
$\ub^2\Dh\Dh\alb$, and $u\Dbh\alb$, $\ub\Dh\alb$, we conclude from
the estimates \ref{12.193} and \ref{12.228} that:
\begin{eqnarray}
&&\int_{\Cb_{\ub}}|u|^6\left\{|\alb|^2+\delta^2|\Dh\alb|^2+|u|^2|\Dbh\alb|^2+|u|^2|\snab\alb|^2\right.\nonumber\\
&&\hspace{15mm}+\delta^4|\Dh\Dh\alb|^2+\delta^2|u|^2|\snab\Dh\alb|^2+|u|^4|\snab\Dbh\alb|^2\nonumber\\
&&\hspace{15mm}\left.+|u|^4|\snab^{ \ 2}\alb|^2+|u|^4|\Dbh\Dbh\alb|^2\right\}\nonumber\\
&&\hspace{10mm}\leq C\stackrel{(3)}{F}_2(\ub)+C\delta^4({\cal
Q}_2)^2+C\delta^{7/2}({\cal Q}_1)^2 \label{12.229}
\end{eqnarray}
Multiplying by $\delta^{-3}$ and taking the supremum over
$\ub$ yields, in view of the definitions
\ref{10.107}, \ref{10.108} and \ref{12.140},
\begin{equation}
(\cRb_{[2]}(\alb))^2\leq C\stackrel{(3)}{{\cal F}}_2+C\delta({\cal
Q}_2)^2+C\delta^{1/2}({\cal Q}_1)^2 \label{12.230}
\end{equation}
The inequalities \ref{12.178} and \ref{12.230} together imply,
recalling the definitions \ref{12.141} and \ref{12.142},
\begin{equation}
({\cal Q}_2)^2\leq C\max\{{\cal P}_2,({\cal D}^{\prime
4}_{[1]}(\alb))^2\}+C\delta^{1/2}({\cal Q}_2)^2+C({\cal Q}_1)^2
\label{12.231}
\end{equation}
which, if $\delta$ is suitably small implies, taking also into
account Lemma 12.5,
\begin{equation}
({\cal Q}_2)^2\leq C^\prime\max\{{\cal P}_2,({\cal D}^{\prime
4}_{[1]}(\alb))^2\} \label{12.232}
\end{equation}
Substituting this bound in \ref{12.178} and \ref{12.230} then
yields the conclusions of the lemma with the exception of the
conclusion that there is a numerical constant $C$ such that:
\begin{equation}
\scR_1(\Db\beb)\leq C\max\{{\cal P}_2,({\cal D}^{\prime
4}_{[1]}(\alb))^2\} \label{12.233}
\end{equation}
To obtain the remaining conclusion we take again $Y=O_i:i=1,2,3$
and $W=\tcL_S R$ in Proposition 12.2 and use the bootstrap
assumptions {\bf C2.1} - {\bf C2.4}, {\bf C3.1} - {\bf C3.5}, {\bf
C6.1} - {\bf C6.5}, to deduce, in view of \ref{12.118},
\ref{12.130}, and the last of \ref{12.139}:
\begin{eqnarray}
&&\int_{C_u}|u|^6\sum_i|\sL_{O_i}\sL_S\beb|^2\leq
C\stackrel{(3)}{E}_2(u)
+C\delta^{1/2}\int_{C_u}|u|^6\left\{|\sL_S\beb|^2+\sum_i|\sL_{O_i}\beb|^2\right\}\nonumber\\
&&\hspace{28mm}+C\delta^{-1/2}\int_{C_u}|u|^7\left\{|\sLh_S\alb|^2+|\sLh_{O_i}\alb|^2\right\}\nonumber\\
&&\hspace{28mm}+C\delta^{5/2}\int_{C_u}|u|^4\sum_i(|O_i\rho|^2+|O_i\sigma|^2)\nonumber\\
\label{12.234}
\end{eqnarray}
Now, from \ref{12.184} together with \ref{12.149} and \ref{12.190}
we deduce:
\begin{equation}
\int_{C_u}|u|^6|\sL_S\beb|^2=\int_{C_u}|u|^6|u\Db\beb+\ub
D\beb|^2\leq C\delta^3({\cal Q}_2)^2 \label{12.235}
\end{equation}
From \ref{12.157} together with \ref{12.150} and \ref{12.174} we
deduce:
\begin{equation}
\int_{C_u}|u|^7|\sLh_S\alb|^2=\int_{C_u}|u|^7|u\Dbh\alb+\ub\Dh\alb|^2\leq
C\delta^{7/2}({\cal Q}_2)^2 \label{12.236}
\end{equation}
Also, by the fourth of \ref{12.165} together with \ref{12.148} and
\ref{12.146}:
\begin{equation}
\int_{C_u}|u|^6\sum_i|\sL_{O_i}\beb|^2\leq C\delta^3({\cal Q}_2)^2
\label{12.237}
\end{equation}
By the second of \ref{12.165} together with \ref{12.148} and
\ref{12.146}:
\begin{equation}
\int_{C_u}|u|^7\sum_i|\sLh_{O_i}\alb|^2\leq C\delta^4({\cal
Q}_2)^2 \label{12.238}
\end{equation}
and by the last of \ref{12.165}:
\begin{equation}
\int_{C_u}|u|^4\sum_i(|O_i\rho|^2+|O_i\sigma|^2)\leq C\delta({\cal
Q}_1)^2 \label{12.239}
\end{equation}
Substituting the above in \ref{12.234} we obtain:
\begin{equation}
\int_{C_u}|u|^6\sum_i|\sL_{O_i}\sL_S\beb|^2\leq
C\stackrel{(3)}{E}_2(u)+C\delta^3({\cal Q}_2)^2 \label{12.240}
\end{equation}
Now, by Proposition 11.1 applied to $\sL_S\beb$:
\begin{equation}
|u|^2|\snab\sL_S\beb|^2\leq C\sum_i|\sL_{O_i}\sL_S\beb|^2
\label{12.241}
\end{equation}
It follows that:
\begin{equation}
\int_{C_u}|u|^8|\snab\sL_S\beb|^2\leq
C\stackrel{(3)}{E}_2(u)+C\delta^3({\cal Q}_2)^2 \label{12.242}
\end{equation}
From \ref{12.184},
$$\snab\sL_S\beb=u\snab\Db\beb+\ub\snab D\beb$$
and we have:
\begin{equation}
\int_{C_u}|u|^8|\ub|^2|\snab D\beb|^2\leq
\delta^3(\scR_1(D\beb))^2\leq\delta^3({\cal Q}_2)^2 \label{12.243}
\end{equation}
Therefore \ref{12.242} implies:
\begin{equation}
\int_{C_u}|u|^{10}|\snab\Db\beb|^2\leq
C\stackrel{(3)}{E}_2(u)+C\delta^{3}({\cal Q}_2)^2 \label{12.244}
\end{equation}
Multiplying this by $\delta^{-3}$, taking the supremum over
$u\in[u_0,c^*)$, recalling the definition \ref{7.03}, and taking
into account the bound \ref{12.232}, then yields the desired bound
\ref{12.233}. This completes the proof of Lemma 12.6.

\vspace{5mm}

\section{Statement of the existence theorem. Outline of the continuity argument}

Let us denote by $\stackrel{(n)}{D} \ : \ n=0,1,2,3$ the following
quantities, which depend only on the initial data on $C_{u_0}$:
\begin{eqnarray}
&&\stackrel{(0)}{D}=\delta^2\stackrel{(0)}{E}_2(u_0)\nonumber\\
&&\stackrel{(1)}{D}=\stackrel{(1)}{E}_2(u_0)\nonumber\\
&&\stackrel{(2)}{D}=\delta^{-1}\stackrel{(2)}{E}_2(u_0)\nonumber\\
&&\stackrel{(3)}{D}=\delta^{-3}\stackrel{(3)}{E}_2(u_0)
\label{12.245}
\end{eqnarray}

The spacetime manifold $(M,g)$, solution of the vacuum Einstein
equations, which we have hitherto been considering, corresponds to
the real number $c^*\in(u_0,-1]$. Thus for the sake of clarity we
denote it presently by $(M_{c^*},g)$. We may place any real number
$c\in(u_0,-1]$ in the role of $c^*$ and denote the corresponding
spacetime manifold by $(M_c,g)$. We also denote by $M^\prime_c$
the non-trivial part of $M_c$ and by $D_c$ and $D^\prime_c$ the
parameter domains to which $M_c$ and $M^\prime_c$ respectively
correspond. The symbol $c^*$ shall from now on be reserved for the
particular real number in $(u_0,-1]$ to be defined below.

We must first clarify the following point. The initial data
quantities $\stackrel{(n)}{D} \ : \ n=0,1,2,3$, like the initial
data quantities ${\cal D}_0^\infty$, $\scD_1^4$,
$\scD_2(\mbox{tr}\chib)$, $\scD_3(\mbox{tr}\chib)$, and ${\cal
D}^{\prime 4}_{[1]}(\alb)$ are defined relative to the whole of
$C_{u_0}$, even though we may be considering a spacetime manifold
$M_c$ with $c\in(u_0,u_0+\delta)$, which contains only the part
$C_{u_0}^{c-u_0}$ of $C_{u_0}$. That is, given the pair of smooth
mappings $\psi_0$, $\psi_0^\prime$ of
$[0,1]\times\oD_{2\rho}\rightarrow\hat{S}$ (see \ref{2.37}) and
the numbers $\delta\in(0,1]$, $u_0<-1$, the procedure of Chapter 2
determines the full set of initial data along $C_{u_0}$, that is
the metric $\sg$, the 2nd fundamental forms $\chi$ and $\chib$,
the torsion $\zeta$, the function $\omb$, its first transversal
derivative $\Db\omb$, all curvature components $\alpha$, $\beta$,
$\rho$, $\sigma$, $\beb$, $\alb$, and the first transversal
derivative $\Db\alb$ of $\alb$. These suffice to determine the
first transversal derivatives of all connection coefficients along
$C_{u_0}$ directly through the optical structure equations and the
first transversal derivatives of all curvature componets directly
through the Bianchi equations. Moreover, the second transversal
derivative $\Db^2\beb$ is also directly determined, in terms of
$\snab\Db\alb$, by applying $\Db$ to the fourth Bianchi identity
of Proposition 1.2. To be sure, the Einstein equations are used to
make this determination, and if we are given a solution on $M_c$
for some $c\in(u_0,u_0+\delta)$ then these equations are satisfied
by our solution only on the part $C_{u_0}^{c-u_0}$ of $C_{u_0}$.
Nevertheless, our procedure determines entities on the whole of
$C_{u_0}$, which must coincide on $C_{u_0}^{c-u_0}$ with the
corresponding entities induced by our solution. Now the components
of the deformation tensors of the commutation fields $L$ and $S$
depend only on the connection coefficients (see \ref{8.21},
\ref{8.30}) so these deformation tensors as well as their
transversal derivatives are determined in the above sense on all
of $C_{u_0}$. By \ref{8.134} - \ref{8.138} and \ref{8.64} the only
components of the deformation tensors of the commutation fields
$O_i$ which do not vanish identically on $C_{u_0}$ are:
\begin{equation}
\s^{(O_i)}j=\frac{1}{2}\s^{(O_i)}\spi \ \ \mbox{and} \ \
\s^{(O_i)}\ih=\s^{(O_i)}\hat{\spi} \label{12.b1}
\end{equation}
Since $\s^{(O_i)}\spi=\sL_{O_i}\sg$ and $\sL_{O_i}\up{\sg}=0$, the
$O_i$ being Killing fields of the standard sphere, it follows from
\ref{2.110} that:
\begin{equation}
\s^{(O_i)}\spi={\bf M}_1(\delta^{1/2}|u_0|^{-1}) \ \ \mbox{: on
$C_{u_0}$} \label{12.b2}
\end{equation}
Moreovever, in canonical coordinates we have, along $C_{u_0}$:
\begin{equation}
\mbox{tr}\s^{(O_i)}\spi=\frac{2}{\sqrt{\mbox{det}{\sg}}}\frac{\partial(O_i^A\sqrt{\mbox{det}\sg})}{\partial\vartheta^A}
\label{12.b3}
\end{equation}
and by \ref{2.7}, \ref{2.5} (see also \ref{2.112}):
\begin{equation}
\sqrt{\mbox{det}\sg}=\phi^2|u_0|^2\sqrt{\mbox{det}\up{\sg}}
\label{12.b4}
\end{equation}
hence:
$$\mbox{tr}\s^{(O_i)}\spi=\frac{2}{\phi^2\sqrt{\mbox{det}\up{\sg}}}\frac{\partial}{\partial\vartheta^A}
(O_i^A\phi^2\sqrt{\mbox{det}\up{\sg}})=\frac{2}{\phi^2}O_i^A\frac{\partial\phi^2}{\partial\vartheta^A}$$
that is:
\begin{equation}
\mbox{tr}\s^{O_i)}\spi=\frac{4}{\phi}O_i\phi \ \ \mbox{: on
$C_{u_0}$} \label{12.b5}
\end{equation}
The first of \ref{12.109} then gives:
\begin{equation}
\mbox{tr}\s^{(O_i)}\spi={\bf M}_1(\delta|u_0|^{-2}) \label{12.b6}
\end{equation}
Thus, according to the construction of Chapter 2 we have on
$C_{u_0}$:
\begin{equation}
\s^{(O_i)}j={\bf M}_1(\delta|u_0|^{-2}), \ \ \ \s^{(O_i)}\ih={\bf
M_1}(\delta^{1/2}|u_0|^{-1}) \label{12.b7}
\end{equation}
In view of Proposition 12.2, it follows that the energy densities
are defined everywhere on $C_{u_0}$, so the energy integrals
\ref{12.245} on the whole of $C_{u_0}$ are well defined. Moreover,
the results of Chapter 2 imply that there is a non-negative
non-decreasing continuous function of $M_8$ such that:
\begin{equation}
\max\{\stackrel{(0)}{D},\stackrel{(1)}{D},\stackrel{(2)}{D},\stackrel{(3)}{D}\}\leq
F(M_8) \label{12.b8}
\end{equation}
the entity which requires $M_k$ with the highest $k$ being
$\Db^2\beb$, which occurs in $\s^{(SS)}\stackrel{(3)}{E}(u_0)$ and
is expressed in terms of
\begin{equation}
\snab\Dbh\alb={\bf M}_8(\delta^{3/2}|u_0|^{-7}) \label{12.b9}
\end{equation}
by \ref{2.215}. Also, by \ref{2.188} and \ref{2.215} the quantity
${\cal D}^{\prime 4}_{[1]}(\alb)$ is bounded by a non-negative
non-decreasing continuous function of $M_7$ (in fact by
$|u_0|^{-1}$ times such a function), while by \ref{2.146},
\ref{2.153} and \ref{2.164}, the quantities ${\cal D}_0^\infty$,
$\scD_1^4$, $\scD_2^4(\mbox{tr}\chib)$, $\scD_3(\mbox{tr}\chib)$
are bounded by non-negative non-decreasing continous functions of
$M_3$, $M_4$, $M_5$, $M_6$, respectively. Thus all of the
quantities $\stackrel{(n)}{D} \ : \ n=0,1,2,3$, ${\cal D}^{\prime
4}_{[1]}(\alb)$, ${\cal D}_0^\infty$, $\scD_1^4$,
$\scD_2^4(\mbox{tr}\chib)$, $\scD_3(\mbox{tr}\chib)$ are bounded
by a non-negative non-decreasing continuous function of $M_8$.

The aim of the next four chapters is to establish the following
theorem.

\vspace{5mm}

\noindent{\bf Theorem 12.1} \ \ \ Let us be given smooth initial
data on $C_{u_0}$ as described in Chapter 2. Let us define $c^*$
to be the supremum of the set ${\cal A}$ of real numbers $c$ in
the interval $(u_0,-1]$ such that to $c$ there corresponds a
spacetime manifold $(M_c,g)$, as defined in Chapter 1, smooth
solution of the vacuum Einstein equations, taking the given
initial data along $C_{u_0}$, and such that:

\begin{enumerate}

\item The generators of the null hypersurfaces $C_u$ and $\Cb_{\ub}$, as defined in Chapter 1, have no end points in $M_c\setminus\Gamma_0$.
Thus, the $C_u$ contain no conjugate or cut points in $M_c$ and
the $\Cb_{\ub}$ contain no focal or cut points in
$M_c\setminus\Gamma_0$.

\item The bootstrap assumptions {\bf A0}, {\bf A1.1}, {\bf A1.2}, {\bf A2.1}, {\bf A2.2}, {\bf A3.1},
{\bf A3.2}, {\bf A4.1}, {\bf A4.2}, {\bf B1}, {\bf B2}, {\bf B3},
and {\bf C1.1} - {\bf C1.4}, {\bf C2.1} - {\bf C2.4}, {\bf C3.1} -
{\bf C3.5}, {\bf C4.1} - {\bf C4.8}, {\bf C5.1} - {\bf C5.4}, {\bf
C6.1} - {\bf C6.10}, as well as {\bf D0}, {\bf D1}, {\bf D2.1},
{\bf D2.2}, {\bf D3.1}, {\bf D3.2}, {\bf D4.1}, {\bf D4.2}, {\bf
D5}, {\bf D6}, {\bf D7}, {\bf D8}, {\bf D9.1} - {\bf D9.3}, {\bf
D10.1} - {\bf D10.4}, {\bf D11}, {\bf D 12}, hold on $M^\prime_c$
and $D^\prime_c$.

\item The quantity ${\cal P}_2$ corresponding to $M^\prime_c$ satisfies the bound:
$${\cal P}_2\leq G(\stackrel{(0)}{D},\stackrel{(1)}{D}, \stackrel{(2)}{D}, \stackrel{(3)}{D})$$
where $G$ is a positive continuous function, non-decreasing in
each of its arguments, which shall be specified in the sequel.

\end{enumerate}

Then, if $\delta$ is suitably small depending on the initial data
quantities:
$$\stackrel{(0)}{D},\stackrel{(1)}{D},\stackrel{(2)}{D},\stackrel{(3)}{D}
\ \mbox{and} \ {\cal D}^{\prime 4}_{[1]}(\alb)$$ as well as:
$${\cal D}_0^\infty,\scD_1^4 \ \mbox{and} \ \scD_2^4(\mbox{tr}\chib),\scD_3(\mbox{tr}\chib)$$
we have:
$$c^*=-1\in{\cal A}$$

\vspace{5mm}

We remark that condition 1 of the theorem may be restated as
follows. With $M_c\setminus\Gamma_0=(D_c\setminus A_0)\times S^2$
we have a smooth solution in {\em canonical coordinates} on
$M_c\setminus\Gamma_0$, the subdomain of $M_c\setminus\Gamma$
where $\ub\leq 0$ being isometric to the corresponding domain in
Minkowski spacetime.

We also remark that by the local existence theorem of Chapter 2
and the argument to be presented in the 3rd section of Chapter 16,
the set ${\cal A}$ is not empty.

The proof of the theorem begins with the observation that $c^*\in
{\cal A}$ and ${\cal A}=(u_0,c^*]$. This is because $c\in {\cal
A}$ implies $(u_0,c]\subset{\cal A}$, hence ${\cal A}\supset
(u_0,c^*)$ while $M_c$ was defined in Chapter 1 to include its
past boundary but not its future boundary, hence:
\begin{equation}
M_{c^*}=\bigcup_{c\in(u_0,c^*)}M_c \label{12.246}
\end{equation}
It follows that the 1st and 2nd of the above conditions for
membership in the set ${\cal A}$ hold for $c^*$. Moreover, since
$(u_0,c^*)\subset{\cal A}$, for every $c\in(u_0,c^*)$ the energies
$\stackrel{(n)}{E}_2 \ : \ n=0,1,2,3$ and flux
$\stackrel{(3)}{F}_2$ corresponding to $M_c$ satisfy:
\begin{eqnarray*}
&&\sup_{u\in[u_0,c)}\left(\delta^2\stackrel{(0)}{E}_2(u)\right)\leq
G(\stackrel{(0)}{D},\stackrel{(1)}{D}, \stackrel{(2)}{D}, \stackrel{(3)}{D})\\
&&\sup_{u\in[u_0,c)}\left(\stackrel{(1)}{E}_2(u)\right)\leq
G(\stackrel{(0)}{D},\stackrel{(1)}{D}, \stackrel{(2)}{D}, \stackrel{(3)}{D})\\
&&\sup_{u\in[u_0,c)}\left(\delta^{-1}\stackrel{(2)}{E}_2(u)\right)\leq
G(\stackrel{(0)}{D},\stackrel{(1)}{D}, \stackrel{(2)}{D}, \stackrel{(3)}{D})\\
&&\sup_{u\in[u_0,c)}\left(\delta^{-3}\stackrel{(3)}{E}_2(u)\right)\leq
G(\stackrel{(0)}{D},\stackrel{(1)}{D}, \stackrel{(2)}{D},
\stackrel{(3)}{D})
\end{eqnarray*}
and:
$$\sup_{\ub\in[0,\min\{\delta,c-u_0\})}\left(\delta^{-3}\stackrel{(3)}{F}_2(\ub)\right)\leq
G(\stackrel{(0)}{D},\stackrel{(1)}{D}, \stackrel{(2)}{D},
\stackrel{(3)}{D})$$ it follows that the energies
$\stackrel{(n)}{E}_2 \ : \ n=0,1,2,3$ and flux
$\stackrel{(3)}{F}_2$ corresponding to $M_{c^*}$ likewise satisfy:
\begin{eqnarray*}
&&\sup_{u\in[u_0,c^*)}\left(\delta^2\stackrel{(0)}{E}_2(u)\right)\leq
G(\stackrel{(0)}{D},\stackrel{(1)}{D}, \stackrel{(2)}{D}, \stackrel{(3)}{D})\\
&&\sup_{u\in[u_0,c^*)}\left(\stackrel{(1)}{E}_2(u)\right)\leq
G(\stackrel{(0)}{D},\stackrel{(1)}{D}, \stackrel{(2)}{D}, \stackrel{(3)}{D})\\
&&\sup_{u\in[u_0,c^*)}\left(\delta^{-1}\stackrel{(2)}{E}_2(u)\right)\leq
G(\stackrel{(0)}{D},\stackrel{(1)}{D}, \stackrel{(2)}{D}, \stackrel{(3)}{D})\\
&&\sup_{u\in[u_0,c^*)}\left(\delta^{-3}\stackrel{(3)}{E}_2(u)\right)\leq
G(\stackrel{(0)}{D},\stackrel{(1)}{D}, \stackrel{(2)}{D},
\stackrel{(3)}{D})
\end{eqnarray*}
and:
$$\sup_{\ub\in[0,\min\{\delta,c^*-u_0\})}\left(\delta^{-3}\stackrel{(3)}{F}_2(\ub)\right)\leq
G(\stackrel{(0)}{D},\stackrel{(1)}{D}, \stackrel{(2)}{D},
\stackrel{(3)}{D})$$ hence the quantity ${\cal P}_2$ corresponding
to $M_{c^*}$ satisfies:
\begin{equation}
{\cal P}_2\leq G(\stackrel{(0)}{D},\stackrel{(1)}{D},
\stackrel{(2)}{D}, \stackrel{(3)}{D}) \label{12.247}
\end{equation}
Thus also the 3rd condition for membership in the set ${\cal A}$
holds for $c^*$, consequently $c^*\in{\cal A}$. Form this point
and for the next four chapters our spacetime manifold shall be
$(M_{c^*},g)$ with $c^*$ defined as in the statement of Theorem
12.1.

Since 2nd condition above holds with $c^*$ in role of $c$, the
assumptions of Lemmas 12.5 and 12.6 hold for $M^\prime_{c^*}$.
Since also the 3rd condition holds with $c^*$ in the role of $c$,
the conclusion of Lemma 12.6 yields:
\begin{equation}
({\cal Q}^\prime_2)^2\leq
C\max\{G(\stackrel{(0)}{D},\stackrel{(1)}{D}, \stackrel{(2)}{D},
\stackrel{(3)}{D}), ({\cal D}^{\prime 4}_{[1]}(\alb))^2\}
\label{12.248}
\end{equation}
More precisely, to obtain the results of Lemmas 12.5 and 12.6 for
$M^\prime_{c^*}$, we first apply Lemmas 12.5 and 12.6 with the
subdomain $M^\prime_{c^*,\varepsilon}$ of $M^\prime_{c^*}$,
defined by the restrictions $\ub\leq \delta-\varepsilon$ and
$\ub+u\leq c^*-\varepsilon$, in the role of $M^\prime_{c^*}$, the
$L^2$ norms on $C_u$ and $\Cb_{\ub}$ in the definitions of the
quantities ${\cal Q}_1$ and ${\cal Q}^\prime_2$ being replaced by
the corresponding $L^2$ norms on $C_u\bigcap
M^\prime_{c^*,\varepsilon}$ and  $\Cb_{\ub}\bigcap
M^\prime_{c^*,\varepsilon}$. The quantities ${\cal Q}_1$ and
${\cal Q}^\prime_2$ referring to $M^\prime_{c^*,\varepsilon}$ so
defined, are finite, since $M^\prime_{c^*,\varepsilon}$ is a
compact subdomain of $M^\prime_{c^*}$ and we have a smooth
solution in canonical coordinates in $M^\prime_{c^*}$. Lemmas 12.5
and 12.6 then give us bounds for these quantities in terms of the
quantities ${\cal P}_1$ and ${\cal P}_2$ referring to
$M^\prime_{c^*}$, and these bounds are independent of
$\varepsilon$. Taking then the limit $\varepsilon\rightarrow 0$ we
obtain the results of Lemmas 12.5 and 12.6 for $M^\prime_{c^*}$
itself.

As we have already remarked, the quantity ${\cal Q}^\prime_2$
bounds all the curvature norms which enter the estimates for the
connection coefficients. These estimates depend additionally on
the initial data quantities ${\cal D}_0^\infty$, $\scD_1^4$ and
$\scD_2^4(\mbox{tr}\chib)$, $\scD_3(\mbox{tr}\chib)$. Thus if in a
non-negative non-decreasing continuous function of the quantities
${\cal R}_0^\infty$, $\scR_1^4$, $\scR_2$, and the quantities
${\cal D}_0^\infty$, $\scD_1^4$, $\scD_2^4(\mbox{tr}\chib)$,
$\scD_3(\mbox{tr}\chib)$, the first four quantities are first
replaced by their bounds in terms of ${\cal Q}^\prime_2$, (see
\ref{12.147}, \ref{12.148}) and then ${\cal Q}^\prime_2$ is
replaced by its bound \ref{12.248}, the value of the function will
not be decreased and a non-negative non-decreasing continuous
function of the quantities $\stackrel{(n)}{D} \ : \ n=0,1,2,3$;
${\cal D}^{\prime 4}_{[1]}(\alb)$ and the quantities ${\cal
D}_0^\infty$, $\scD_1^4$, $\scD_2^4(\mbox{tr}\chib)$,
$\scD_3(\mbox{tr}\chib)$ will result (see last section of Chapter
3). Moreover, the same holds if the original function depends
additionally on $\scRb_2(\alb)$, the quantities:
$${\cal R}_0^4(\Dh\alpha),{\cal R}_0^4(D\beta),{\cal R}_0^4(D\rho),{\cal R}_0^4(D\sigma),{\cal R}_0^4(D\beb);{\cal R}_0^4(\Dbh\alb),$$
the quantities:
$$\scR_1(\Dh\alpha),\scR_1(D\beta),\scR_1(D\rho),\scR_1(D\sigma),\scR_1(D\beb);\scRb_1(\Dbh\alb),$$
as well as the quantities:
$${\cal R}_0(\Dh^2\alpha),{\cal R}_0(D^2\beta),{\cal R}_0(D^2\rho),{\cal R}_0(D^2\sigma),{\cal R}(D^2\beb);\cRb_0(\Dbh^2\alb)$$
and $\scR_1(\Db\beb)$. {\em We thus denote from now on by
$O(\delta ^p|u|^r)$, for real numbers $p$, $r$, the product of
$\delta^p|u|^r$ with a non-negative non-decreasing continuous
function of the quantities $\stackrel{(n)}{D} \ : \ n=0,1,2,3$;
${\cal D}^{\prime 4}_{[1]}(\alb)$ and ${\cal D}_0^\infty$,
$\scD_1^4$, $\scD_2^4(\mbox{tr}\chib)$, $\scD_3(\mbox{tr}\chib)$.
Also, we denote simply by $O(\delta^p)$, for a real number $p$,
the product of $\delta^p$ with a non-negative non-decreasing
continuous function of the quantities $\stackrel{(n)}{D} \ : \
n=0,1,2,3$; ${\cal D}^{\prime 4}_{[1]}(\alb)$ and ${\cal
D}_0^\infty$, $\scD_1^4$, $\scD_2^4(\mbox{tr}\chib)$,
$\scD_3(\mbox{tr}\chib)$}. Then, in all the estimates for the
connection coefficients (Chapters 3 - 7) and all the estimates for
the deformation tensors of the multiplier fields and the
commutation fields (Chapters 8 - 9), the symbol $O(\delta^p|u|^r)$
may be taken in the new sense. Moreover, the smallness condition
on $\delta$ required for these estimates to hold, a smallness
condition depending on the quantities ${\cal R}_0^\infty$,
$\scR_1^4$, $\scR_2$, and the quantities ${\cal D}_0^\infty$,
$\scD_1^4$, $\scD_2^4(\mbox{tr}\chib)$, $\scD_3(\mbox{tr}\chib)$,
may be replaced by a smallness condition depending on the
quantities $\stackrel{(n)}{D} \ : \ n=0,1,2,3$; ${\cal D}^{\prime
4}_{[1]}(\alb)$  and the quantities ${\cal D}_0^\infty$,
$\scD_1^4$, $\scD_2^4(\mbox{tr}\chib)$, $\scD_3(\mbox{tr}\chib)$.
For, the former smallness condition may be reduced to the form:
$$\delta F({\cal R}_0^\infty,\scR_1^4,\scR_2,{\cal D}_0^\infty,\scD_1^4,\scD_2^4(\mbox{tr}\chib),\scD_3(\mbox{tr}(\chib))\leq 1$$
where $F$ is a non-negative non-decreasing continuous function of
its arguments (see last section of Chapter 3).

Thus, under the assumptions of the theorem all the aforementioned
estimates hold on $M^\prime_{c^*}$. It follows that with
$M^\prime_{c^*}$ and $D^\prime_{c^*}$ in the role of $M^\prime$
and $D^\prime$ respectively, none of the inequalities in the
bootstrap assumptions is saturated, provided that $\delta$ is
suitably small depending on $\stackrel{(n)}{D} \ : \ n=0,1,2,3$,
${\cal D}_0^\infty$, $\scD_1^4$, $\scD_2^4(\mbox{tr}\chib)$,
$\scD_3(\mbox{tr}\chib)$. In particular, in place of {\bf A0} we
have:
\begin{equation}
\frac{2}{3}\leq\Omega\leq\frac{3}{2} \ : \ \mbox{in} \
M^\prime_{c^*} \label{12.249}
\end{equation}
in place of {\bf D3.1} we have:
\begin{equation}
|u|\|\mbox{tr}\chib\|_{L^\infty(S_{\ub,u})}\leq 3 \ : \ \forall
(\ub,u)\in D^\prime_{c^*} \label{12.250}
\end{equation}
and the inequalities of all the other bootstrap assumptions hold
on $M^\prime_{c^*}$ and $D^\prime_{c^*}$ with their right hand
sides halved. The aim of the next three chapters and the first
section of the fourth is to show that the quantity ${\cal P}_2$
corresponding to $M^\prime_{c^*}$ actually satisfies:
\begin{equation}
{\cal P}_2\leq \frac{1}{2}G(\stackrel{(0)}{D},\stackrel{(1)}{D},
\stackrel{(2)}{D}, \stackrel{(3)}{D}) \label{12.251}
\end{equation}
so the inequality \ref{12.247} is not saturated either. Then in
the third section of Chapter 16 we shall show that the solution
extends to a larger domain $M_{c}$ for some $c>c^*$ and the three
conditions in the statement of the theorem hold for the extended
solution as well, thereby contradicting the definition of $c^*$,
unless $c^*=-1$.

Going back to the proof of Lemma 12.5 and replacing assumptions
{\bf C1.1} - {\bf C1.4}, {\bf C2.1} - {\bf C2.4}, {\bf C3.1} -
{\bf C3.5} by the corresponding estimates from Chapter 8, which
improve the former by a factor of at least $\delta^{1/4}$, the
secondary conclusions of Lemma 12.5 may be replaced by:
\begin{eqnarray}
&&({\cal R}_{[1]}(\alpha))^2\leq C\stackrel{(0)}{{\cal E}}_1+O(\delta^2)\nonumber\\
&&({\cal R}_{[1]}(\beta))^2\leq C\stackrel{(1)}{{\cal E}}_1+O(\delta)\nonumber\\
&&({\cal R}_{[1]}(\rho))^2,({\cal R}_{[1]}(\sigma))^2\leq
C\stackrel{(2)}{{\cal E}}_1
+O(\delta^2)\nonumber\\
&&({\cal R}_{[1]}(\beb))^2\leq C\stackrel{(3)}{{\cal E}}_1+O(\delta)\nonumber\\
&&(\cRb_{[1]}(\alb))^2\leq C\stackrel{(3)}{{\cal F}}_1+O(\delta^2)
\label{12.a3}
\end{eqnarray}
Also, going back to the proof of Lemma 12.6 and replacing
assumptions {\bf C4.1} - {\bf C4.8}, {\bf C5.1} - {\bf C5.4}, {\bf
C6.1} - {\bf C6.10} by the corresponding estimates from Chapter 9,
which improve the former by a factor of at least $\delta^{1/4}$,
as well as assumptions {\bf D2.1}, {\bf D2.2}, {\bf D3.1}, {\bf
D3.2}, {\bf D4.1}, {\bf D4.2}, {\bf D5}, {\bf D6}, {\bf D7}, {\bf
D8}, {\bf D9.1} - {\bf D9.3}, {\bf D10.1} - {\bf D10.4}, {\bf
D11}, {\bf D12} by the corresponding estimates from Chapters 3 and
4, which improve the former by a factor of at least
$\delta^{1/4}$, the secondary conclusions of Lemma 12.6 may be
replaced by:
\begin{eqnarray}
&&({\cal R}_{[2]}(\alpha))^2\leq C\stackrel{(0)}{{\cal E}}_2+O(\delta)\nonumber\\
&&({\cal R}_{[2]}(\beta))^2\leq C\stackrel{(1)}{{\cal E}}_2+O(\delta)\nonumber\\
&&({\cal R}_{[2}(\rho))^2,({\cal R}_{[2]}(\sigma))^2\leq C\stackrel{(2)}{{\cal E}}_2+O(\delta)\nonumber\\
&&({\cal R}_{[2]}(\beb))^2\leq C\stackrel{(3)}{{\cal E}}_2+O(\delta)\nonumber\\
&&(\cRb_{[2]}(\alb))^2\leq C\stackrel{(3)}{{\cal F}}_2+O(\delta)
\label{12.a4}
\end{eqnarray}

Let us now apply Lemma 12.4 to the energy-momentum density
vectorfields \ref{12.80}, \ref{12.85}, \ref{12.90}, \ref{12.95},
\ref{12.100}, \ref{12.105}, \ref{12.110}, \ref{12.115} and
\ref{12.120}. Let us denote the corresponding divergences by:
\begin{eqnarray}
&&\stackrel{(0)}{\tau}=\mbox{div}P(R;L,L,L) \ \ \ \ \stackrel{(1)}{\tau}=\mbox{div}P(R;K,L,L)\nonumber\\
&&\stackrel{(2)}{\tau}=\mbox{div}P(R;K,K,L) \ \ \ \
\stackrel{(3)}{\tau}=\mbox{div}P(R;K,K,K) \label{12.252}
\end{eqnarray}
\begin{eqnarray}
&&\s^{(L)}\stackrel{(0)}{\tau}=\mbox{div}P(\tcL_L R;L,L,L) \ \ \ \ \s^{(L)}\stackrel{(1)}{\tau}=\mbox{div}P(\tcL_L R;K,L,L)\nonumber\\
&&\s^{(L)}\stackrel{(2)}{\tau}=\mbox{div}P(\tcL_L R;K,K,L) \ \ \ \ \s^{(L)}\stackrel{(3)}{\tau}=\mbox{div}P(\tcL_L R;K,K,K)\nonumber\\
&&\label{12.253}
\end{eqnarray}
\begin{eqnarray}
&&\s^{(O)}\stackrel{(0)}{\tau}=\sum_i\mbox{div}P(\tcL_{O_i}R;L,L,L)
\ \ \ \
\s^{(O)}\stackrel{(1)}{\tau}=\sum_i\mbox{div}P(\tcL_{O_i}R;K,L,L)\nonumber\\
&&\s^{(O)}\stackrel{(2)}{\tau}=\sum_i\mbox{div}P(\tcL_{O_i}R;K,K,L)
\ \ \ \
\s^{(O)}\stackrel{(3)}{\tau}=\sum_i\mbox{div}P(\tcL_{O_i}R;K,K,K)\nonumber\\
&&\label{12.254}
\end{eqnarray}
\begin{equation}
\s^{(S)}\stackrel{(3)}{\tau}=\mbox{div}P(\tcL_S R;K,K,K)
\label{12.255}
\end{equation}
\begin{eqnarray}
&&\s^{(LL)}\stackrel{(0)}{\tau}=\mbox{div}P(\tcL_L\tcL_L R;L,L,L)
\ \ \ \
\s^{(LL)}\stackrel{(1)}{\tau}=\mbox{div}P(\tcL_L\tcL_L R;K,L,L)\nonumber\\
&&\s^{(LL)}\stackrel{(2)}{\tau}=\mbox{div}P(\tcL_L\tcL_L R;K,K,L)
\ \ \ \
\s^{(LL)}\stackrel{(3)}{\tau}=\mbox{div}P(\tcL_L\tcL_L R;K,K,K)\nonumber\\
&&\label{12.256}
\end{eqnarray}
\begin{eqnarray}
&&\s^{(OL)}\stackrel{(0)}{\tau}=\sum_i\mbox{div}P(\tcL_{O_i}\tcL_L R;L,L,L)\nonumber\\
&&\s^{(OL)}\stackrel{(1)}{\tau}=\sum_i\mbox{div}P(\tcL_{O_i}\tcL_L R;K,L,L)\nonumber\\
&&\s^{(OL)}\stackrel{(2)}{\tau}=\sum_i\mbox{div}P(\tcL_{O_i}\tcL_L R;K,K,L)\nonumber\\
&&\s^{(OL)}\stackrel{(3)}{\tau}=\sum_i\mbox{div}P(\tcL_{O_i}\tcL_L
R;K,K,K) \label{12.257}
\end{eqnarray}
\begin{eqnarray}
&&\s^{(OO)}\stackrel{(0)}{\tau}=\sum_{i,j}\mbox{div}P(\tcL_{O_j}\tcL_{O_i}R;L,L,L)\nonumber\\
&&\s^{(OO)}\stackrel{(1)}{\tau}=\sum_{i,j}\mbox{div}P(\tcL_{O_j}\tcL_{O_i}R;K,L,L)\nonumber\\
&&\s^{(OO)}\stackrel{(2)}{\tau}=\sum_{i,j}\mbox{div}P(\tcL_{O_j}\tcL_{O_i}R;K,K,L)\nonumber\\
&&\s^{(OO)}\stackrel{(3)}{\tau}=\sum_{i,j}\mbox{div}P(\tcL_{O_j}\tcL_{O_i}R;K,K,K)
\label{12.258}
\end{eqnarray}
\begin{equation}
\s^{(OS)}\stackrel{(3)}{\tau}=\sum_i\mbox{div}P(\tcL_{O_i}\tcL_S
R;K,K,K) \label{12.259}
\end{equation}
and:
\begin{equation}
\s^{(SS)}\stackrel{(3)}{\tau}=\mbox{div}P(\tcL_S\tcL_S R;K,K,K)
\label{12.260}
\end{equation}
We consider the total 1st order energy-momentum density
vecctorfields $\stackrel{(n)}{P}_1 \ : \ n=0,1,2,3$:
\begin{eqnarray}
&&\stackrel{(0)}{P}_1=P(R;L,L,L)+\delta^2 P(\tcL_L R;L,L,L)+\sum_i P(\tcL_{O_i}R;L,L,L)\nonumber\\
&&\stackrel{(1)}{P}_1=P(R;K,L,L)+\delta^2 P(\tcL_L R;K,L,L)+\sum_i P(\tcL_{O_i}R;K,L,L)\nonumber\\
&&\stackrel{(2)}{P}_1=P(R;K,K,L)+\delta^2 P(\tcL_L R;K,K,L)+\sum_i P(\tcL_{O_i}R;K,K,L)\nonumber\\
&&\stackrel{(3)}{P}_1=P(R;K,K,K)+\delta^2 P(\tcL_L R;K,K,K)+\sum_i P(\tcL_{O_i}R;K,K,K)\nonumber\\
&&\hspace{10mm}+P(\tcL_S R;K,K,K) \label{12.261}
\end{eqnarray}
We then consider the total 2nd order energy-momentum vectorfields
$\stackrel{(n)}{P}_2 \ : \ n=0,1,2,3$:
\begin{eqnarray}
&&\stackrel{(0)}{P}_2=\stackrel{(0)}{P}_1+\delta^4 P(\tcL_L\tcL_L R;L,L,L)+\delta^2\sum_i P(\tcL_{O_i}\tcL_L R;L,L,L)\nonumber\\
&&\hspace{10mm}+\sum_{i,j}P(\tcL_{O_j}\tcL_{O_i}R;L,L,L)\nonumber\\
&&\stackrel{(1)}{P}_2=\stackrel{(1)}{P}_1+\delta^4 P(\tcL_L\tcL_L R;K,L,L)+\delta^2\sum_i P(\tcL_{O_i}\tcL_L R;K,L,L)\nonumber\\
&&\hspace{10mm}+\sum_{i,j}P(\tcL_{O_j}\tcL_{O_i}R;K,L,L)\nonumber\\
&&\stackrel{(2)}{P}_2=\stackrel{(2)}{P}_1+\delta^4 P(\tcL_L\tcL_L R;K,K,L)+\delta^2\sum_i P(\tcL_{O_i}\tcL_L R;K,K,L)\nonumber\\
&&\hspace{10mm}+\sum_{i,j}P(\tcL_{O_j}\tcL_{O_i}R;K,K,K)\nonumber\\
&&\stackrel{(3)}{P}_2=\stackrel{(3)}{P}_1+\delta^4 P(\tcL_L\tcL_L R;K,K,K)+\delta^2\sum_i P(\tcL_{O_i}\tcL_L R;K,K,K)\nonumber\\
&&\hspace{10mm}+\sum_{i,j}P(\tcL_{O_j}\tcL_{O_i}R;K,K,K)\nonumber\\
&&\hspace{10mm}+\sum_i P(\tcL_{O_i}\tcL_S R;K,K,K)+P(\tcL_S\tcL_S
R;K,K,K) \label{12.262}
\end{eqnarray}
The total 1st order energy-momentum density vectorfields satisfy:
\begin{equation}
\mbox{div}\stackrel{(n)}{P}_1=\stackrel{(n)}{\tau}_1 \ : \
n=0,1,2,3 \label{12.263}
\end{equation}
where the $\stackrel{(n)}{\tau}_1 \ : \ n=0,1,2,3$ are the total
1st order divergences:
\begin{eqnarray}
&&\stackrel{(0)}{\tau}_1=\stackrel{(0)}{\tau}+\delta^2\s^{(L)}\stackrel{(0)}{\tau}+\s^{(O)}\stackrel{(0)}{\tau}\nonumber\\
&&\stackrel{(1)}{\tau}_1=\stackrel{(1)}{\tau}+\delta^2\s^{(L)}\stackrel{(1)}{\tau}+\s^{(O)}\stackrel{(1)}{\tau}\nonumber\\
&&\stackrel{(2)}{\tau}_1=\stackrel{(2)}{\tau}+\delta^2\s^{(L)}\stackrel{(2)}{\tau}+\s^{(O)}\stackrel{(2)}{\tau}\nonumber\\
&&\stackrel{(3)}{\tau}_1=\stackrel{(3)}{\tau}+\delta^2\s^{(L)}\stackrel{(3)}{\tau}+\s^{(O)}\stackrel{(3)}{\tau}
+\s^{(S)}\stackrel{(3)}{\tau} \label{12.264}
\end{eqnarray}
The total 2nd order energy-momentum density vectorfields satisfy:
\begin{equation}
\mbox{div}\stackrel{(n)}{P}_2=\stackrel{(n)}{\tau}_2 \ : \
n=0,1,2,3 \label{12.265}
\end{equation}
where the $\stackrel{(n)}{\tau}_2 \ : \ n=0,1,2,3$ are the total
2nd order divergences:
\begin{eqnarray}
&&\stackrel{(0)}{\tau}_2=\stackrel{(0)}{\tau}_1+\delta^4\s^{(LL)}\stackrel{(0)}{\tau}+\delta^2\s^{(OL)}\stackrel{(0)}{\tau}
+\s^{(OO)}\stackrel{(0)}{\tau}\nonumber\\
&&\stackrel{(1)}{\tau}_2=\stackrel{(1)}{\tau}_1+\delta^4\s^{(LL)}\stackrel{(1)}{\tau}+\delta^2\s^{(OL)}\stackrel{(1)}{\tau}
+\s^{(OO)}\stackrel{(1)}{\tau}\nonumber\\
&&\stackrel{(2)}{\tau}_2=\stackrel{(2)}{\tau}_1+\delta^4\s^{(LL)}\stackrel{(2)}{\tau}+\delta^2\s^{(OL)}\stackrel{(2)}{\tau}
+\s^{(OO)}\stackrel{(2)}{\tau}\nonumber\\
&&\stackrel{(3)}{\tau}_2=\stackrel{(3)}{\tau}_1+\delta^4\s^{(LL)}\stackrel{(3)}{\tau}+\delta^2\s^{(OL)}\stackrel{(3)}{\tau}
+\s^{(OO)}\stackrel{(3)}{\tau}\nonumber\\
&&\hspace{10mm}+\s^{(OS)}\stackrel{(3)}{\tau}+\s^{(SS)}\stackrel{(3)}{\tau}
\label{12.266}
\end{eqnarray}

We apply Lemma 12.4 to \ref{12.265} to obtain, in view of the
definitions of the energies and fluxes \ref{12.81}, \ref{12.82},
\ref{12.86}, \ref{12.87}, \ref{12.91}, \ref{12.92}, \ref{12.96},
\ref{12.97}, \ref{12.101}, \ref{12.102}, \ref{12.106},
\ref{12.107}, \ref{12.111}, \ref{12.112}, \ref{12.116},
\ref{12.117}, \ref{12.121}, \ref{12.122}, and \ref{12.125} -
\ref{12.132}, for any $(\ub_1,u_1)\in D^\prime_{c^*}$:
\begin{equation}
\stackrel{(n)}{E^{\ub_1}}_2(u_1)-\stackrel{(n)}{E^{\ub_1}}_2(u_0)+\stackrel{(n)}{F^{u_1}}_2(\ub_1)=\int_{M_1}\stackrel{(n)}{\tau}_2
d\mu_g \ \ \ : \ n=0,1,2,3 \label{12.267}
\end{equation}
Since the energies and fluxes are integrals, on the $C_u$ and
$\Cb_{\ub}$ respectively, of non-negative functions this implies:
\begin{equation}
\stackrel{(n)}{E^{\ub_1}}_2(u_1)\leq\stackrel{(n)}{E}_2(u_0)+\int_{M^\prime_{c^*}}|\stackrel{(n)}{\tau}_2|d\mu_g
\ \ \ : \ n=0,1,2,3 \label{12.268}
\end{equation}
and:
\begin{equation}
\stackrel{(n)}{F^{u_1}}_2(\ub_1)\leq\stackrel{(n)}{E}_2(u_0)+\int_{M^\prime_{c^*}}|\stackrel{(n)}{\tau}_2|d\mu_g
\ \ \ : \ n=0,1,2,3 \label{12.269}
\end{equation}
for all $(\ub_1,u_1)\in D^\prime_{c^*}$.

Consider \ref{12.268} with $u_1$ fixed. Since the energy
$\stackrel{(n)}{E}(u_1)$ is the integral over $C_{u_1}$ of a
non-negative function and the inequality \ref{12.268} holds for
any $\ub_1\in[0,\delta)$ if $u_1\in[u_0,c^*-\delta)$ and any
$\ub_1\in[0,c^*-u_1)$ if $u_1\in(c^*-\delta,c^*)$, in the case $c^*\geq u_0+\delta$, any $\ub_1\in[0,c^*-u_1)$, in the 
case $c^*<u_0+\delta$, it follows
that:
\begin{equation}
\stackrel{(n)}{E}_2(u_1)\leq\stackrel{(n)}{E}_2(u_0)+\int_{M^\prime_{c^*}}|\stackrel{(n)}{\tau}_2|d\mu_g
\ \ \ : \ n=0,1,2,3 \label{12.270}
\end{equation}
Consider \ref{12.269} with $\ub_1$ fixed. Since the flux
$\stackrel{(n)}{F}(\ub_1)$ is the integral over $\Cb_{\ub_1}$ of a
non-negative function and the inequality \ref{12.269} holds for
any $u_1\in[u_0,c^*-\ub_1)$, it follows that:
\begin{equation}
\stackrel{(n)}{F}_2(\ub_1)\leq\stackrel{(n)}{E}_2(u_0)+\int_{M^\prime_{c^*}}|\stackrel{(n)}{\tau}_2|d\mu_g
\ \ \ : \ n=0,1,2,3 \label{12.271}
\end{equation}
Multiplying \ref{12.270} by $\delta^2$ in the case $n=0$, by 1 in
the case $n=1$, by $\delta^{-1}$ in the case $n=2$, and by
$\delta^{-3}$ in the case $n=3$, and taking the supremum over
$u_1\in[u_0,c^*)$ yields, recalling the definitions \ref{12.139}
and \ref{12.245},
\begin{eqnarray}
&&\stackrel{(0)}{{\cal E}}_2\leq\stackrel{(0)}{D}+\delta^2\int_{M^\prime_{c^*}}|\stackrel{(0)}{\tau}_2|d\mu_g\nonumber\\
&&\stackrel{(1)}{{\cal E}}_2\leq\stackrel{(1)}{D}+\int_{M^\prime_{c^*}}|\stackrel{(1)}{\tau}_2|d\mu_g\nonumber\\
&&\stackrel{(2)}{{\cal E}}_2\leq\stackrel{(2)}{D}+\delta^{-1}\int_{M^\prime_{c^*}}|\stackrel{(2)}{\tau}_2|d\mu_g\nonumber\\
&&\stackrel{(3)}{{\cal
E}}_2\leq\stackrel{(3)}{D}+\delta^{-3}\int_{M^\prime_{c^*}}|\stackrel{(3)}{\tau}_2|d\mu_g
\label{12.272}
\end{eqnarray}
Also, multiplying \ref{12.271} in the case $n=3$ by $\delta^{-3}$
and taking the supremum over $\ub_1\in[0,\min\{\delta,c^*-u_0\})$ yields,
recalling the definition \ref{12.140} and \ref{12.245},
\begin{equation}
\stackrel{(3)}{{\cal
F}}_2\leq\stackrel{(3)}{D}+\delta^{-3}\int_{M^\prime_{c^*}}|\stackrel{(3)}{\tau}_2|d\mu_g
\label{12.273}
\end{equation}

Let us define the exponents $q_n \ : \ n=0,1,2,3$ by:
\begin{equation}
q_0=1, \ \ \ q_1=0, \ \ \ q_2=-\frac{1}{2}, \ \ \ q_3=-\frac{3}{2}
\label{12.274}
\end{equation}
Then inequalities \ref{12.272}, \ref{12.273} take the form:
\begin{eqnarray}
&&\stackrel{(n)}{{\cal
E}}_2\leq\stackrel{(n)}{D}+\delta^{2q_n}\int_{M^\prime_{c^*}}|\stackrel{(n)}{\tau}_2|d\mu_g
\ \ : \ n=0,1,2,3
\nonumber\\
&&\stackrel{(3)}{{\cal
F}}_2\leq\stackrel{(3)}{D}+\delta^{2q_3}\int_{M^\prime_{c^*}}|\stackrel{(3)}{\tau}_2|d\mu_g
\label{12.275}
\end{eqnarray}
In view of the above, we shall obtain a closed system of
inequalities for the quantities $\stackrel{(0)}{{\cal E}}_2$,
$\stackrel{(1)}{{\cal E}}_2$, $\stackrel{(2)}{{\cal E}}_2$,
$\stackrel{(3)}{{\cal E}}_2$ and $\stackrel{(3)}{{\cal F}}_2$, if
we succeed in estimating appropriately the {\em error integrals}:
\begin{equation}
\delta^{2q_n}\int_{M^\prime_{c^*}}|\stackrel{(n)}{\tau}_2|d\mu_{\sg}
\ \ : \ n=0,1,2,3 \label{12.276}
\end{equation}
in terms of the quantities $\stackrel{(0)}{{\cal E}}_2$,
$\stackrel{(1)}{{\cal E}}_2$, $\stackrel{(2)}{{\cal E}}_2$,
$\stackrel{(3)}{{\cal E}}_2$, $\stackrel{(3)}{{\cal F}}_2$
themselves. This shall be our aim in the next three chapters.

\chapter{The Multiplier Error Estimates}

\section{Preliminaries}

Let us introduce the following definitions.

\vspace{5mm}

\noindent{\bf Definition 13.1} \ \ \ Let $\xi$ be a $S$
tensorfield defined on $M^\prime_{c^*}$. We write:
$$\xi={\cal O}^\infty(\delta^r|u|^p)$$
for real numbers $r$, $p$, if:
$$\|\xi\|_{L^\infty(S_{\ub,u})}\leq O(\delta^r|u|^p) \ \ : \ \forall (\ub,u)\in D^\prime_{c^*}$$

\vspace{2.5mm}

\noindent{\bf Definition 13.2} \ \ \ Let $\xi$ be a $S$
tensorfield defined on $M^\prime_{c^*}$. We write:
$$\xi={\cal O}^4(\delta^r|u|^p)$$
for real numbers $r$, $p$, if:
$$\|\xi\|_{L^4(S_{\ub,u})}\leq O(\delta^r|u|^{p+\frac{1}{2}}) \ \ : \ \forall (\ub,u)\in D^\prime_{c^*} $$

\vspace{2.5mm}

\noindent{\bf Definition 13.3} \ \ \ Let $\xi$ be a $S$
tensorfield defined on $M^\prime_{c^*}$. We write:
$$\xi={\bf O}(\delta^{r}|u|^p)$$
for real numbers $r$, $p$, if:
$$\|\xi\|_{L^2(C_u)}\leq O(\delta^{r+\frac{1}{2}}|u|^{p+1}) \ \  : \ \forall u\in[u_0,c^*)$$
\vspace{2.5mm}

\noindent{\bf Definition 13.4} \ \ \ Let $\xi$ be a $S$
tensorfield defined on $M^\prime_{c^*}$. We write:
$$\xi=\bfob(\delta^r|u|^p)$$
for real numbers $r$, $p$, if:
$$\||u|^{-p-\frac{3}{2}}\xi\|_{L^2(\Cb_{\ub})}\leq O(\delta^r) \ \ : \ \forall \ub\in[0,\delta)$$

\vspace{5mm}

In estimating the error integrals \ref{12.276} we shall make
repeated use of the following lemma.

\vspace{5mm}

\noindent{\bf Lemma 13.1} \ \ \ Let $\{\xi_1,\xi_2,\xi_3\}$ be a
scalar trilinear expression in the $S$ tensorfields
$\xi_1,\xi_2,\xi_3$, defined on $M^\prime_{c^*}$, with
coefficients depending only on $\sg$ and $\seps$. Consider the
following five cases.

Case 1: \ \ \ $\xi_1={\cal O}^\infty(\delta^{r_1}|u|^{p_1}), \ \
\xi_2={\bf O}(\delta^{r_2}|u|^{p_2}), \ \ \xi_3={\bf
O}(\delta^{r_3}|u|^{p_3})$

Case 2: \ \ \ $\xi_1={\cal O}^\infty(\delta^{r_1}|u|^{p_1}), \ \
\xi_2={\bf O}(\delta^{r_2}|u|^{p_2}), \ \
\xi_3=\bfob(\delta^{r_3}|u|^{p_3})$

Case 3: \ \ \ $\xi_1={\cal O}^\infty(\delta^{r_1}|u|^{p_1}), \ \
\xi_2=\bfob(\delta^{r_2}|u|^{p_2}), \ \
\xi_3=\bfob(\delta^{r_3}|u|^{p_3})$

Case 4: \ \ \ $\xi_1={\cal O}^4(\delta^{r_1}|u|^{p_1}), \ \
\xi_2={\cal O}^4(\delta^{r_2}|u|^{p_2}), \ \ \xi_3={\bf
O}(\delta^{r_3}|u|^{p_3})$

Case 5: \ \ \ $\xi_1={\cal O}^4(\delta^{r_1}|u|^{p_1}), \ \
\xi_2={\cal O}^4(\delta^{r_2}|u|^{p_2}), \ \
\xi_3=\bfob(\delta^{r_3}|u|^{p_3})$

\noindent Then in each of the five cases
$$p_1+p_2+p_3+3<0$$
implies:
$$\int_{M^\prime_c}|\{\xi_1,\xi_2,\xi_3\}|d\mu_g\leq O(\delta^{r_1+r_2+r_3+1})$$

\noindent{\em Proof:} We begin by remarking that there is a
numerical constant $C$ such that we have, pointwise,
\begin{equation}
|\{\xi_1,\xi_2,\xi_3\}|\leq C|\xi_1||\xi_2||\xi_3| \label{13.1}
\end{equation}
We shall show that in each of the five cases
$$p_1+p_2+p_3+3<0$$
implies:
\begin{equation}
\int_{M^\prime_{c^*}}|\xi_1||\xi_2||\xi_3|d\mu_g\leq
O(\delta^{r_1+r_2+r_3+1}) \label{13.2}
\end{equation}

Consider first Case 1. In this case we estimate:
\begin{eqnarray}
&&\int_{M^\prime_{c^*}}|\xi_1||\xi_2||\xi_3|d\mu_g\leq C\int_{u_0}^{c^*}\left\{\int_{C_u}|\xi_1||\xi_2||\xi_3|\right\}du\nonumber\\
&&\hspace{20mm}\leq O(\delta^{r_1})\int_{u_0}^{c^*}|u|^{p_1}\left\{\int_{C_u}|\xi_2||\xi_3|\right\}du\nonumber\\
&&\hspace{20mm}\leq O(\delta^{r_1})\int_{u_0}^{c^*}|u|^{p_1}\|\xi_2\|_{L^2(C_u)}\|\xi_3\|_{L^2(C_u)}du\nonumber\\
&&\hspace{20mm}\leq O(\delta^{r_1})\int_{u_0}^{c^*}|u|^{p_1}O(\delta^{r_2+\frac{1}{2}}|u|^{p_2+1})O(\delta^{r_3+\frac{1}{2}}|u|^{p_3+1})du\nonumber\\
&&\hspace{20mm}=O(\delta^{r_1+r_2+r_3+1})\int_{u_0}^{c^*}|u|^{p_1+p_2+p_3+2}du\nonumber\\
&&\hspace{20mm}\leq O(\delta^{r_1+r_2+r_3+1}) \ \ : \ \mbox{if
$p_1+p_2+p_3+3<0$} \label{13.3}
\end{eqnarray}

Consider next Case 2. In this case we write:
\begin{eqnarray}
&&\int_{M^\prime_{c^*}}|\xi_1||\xi_2||\xi_3|d\mu_g\label{13.4}\\
&&\hspace{10mm}\leq\left(\int_{M^\prime_{c^*}}|u|^{2p_3+3}|\xi_1|^2|\xi_2|^2
d\mu_g\right)^{1/2}
\left(\int_{M^\prime_{c^*}}|u|^{-2p_3-3}|\xi_3|^2
d\mu_g\right)^{1/2}\nonumber
\end{eqnarray}
and we have:
\begin{eqnarray}
&&\int_{M^\prime_{c^*}}|u|^{-2p_3-3}|\xi_3|^2 d\mu_g\leq C\int_0^{\min\{\delta,c^*-u_0\}}\||u|^{-p_3-\frac{3}{2}}\xi_3\|^2_{L^2(\Cb_{\ub})}d\ub\nonumber\\
&&\hspace{35mm}\leq O(\delta^{2r_3+1}) \label{13.5}
\end{eqnarray}
while:
\begin{eqnarray}
&&\int_{M^\prime_{c^*}}|u|^{2p_3+3}|\xi_1|^2|\xi_2|^2 d\mu_g
\leq C\int_{u_0}^{c^*}|u|^{2p_3+3}\left\{\int_{C_u}|\xi_1|^2|\xi_2|^2\right\}du\nonumber\\
&&\hspace{20mm}\leq O(\delta^{2r_1})\int_{u_0}^{c^*}|u|^{2p_1+2p_3+3}\|\xi_2\|^2_{L^2(C_u)}du\nonumber\\
&&\hspace{20mm}\leq O(\delta^{2r_1+2r_2+1})\int_{u_0}^{c^*}|u|^{2(p_1+p_2+p_3)+5}du\nonumber\\
&&\hspace{20mm}\leq O(\delta^{2r_1+2r_2+1}) \ \ : \ \mbox{if
$p_1+p_2+p_3+3<0$} \label{13.6}
\end{eqnarray}
Substituting \ref{13.5} and \ref{13.6} in \ref{13.4} yields again
\ref{13.2}.

Consider next Case 3. In this case we estimate:
\begin{eqnarray}
&&\int_{M^\prime_{c^*}}|\xi_1||\xi_2||\xi_3|d\mu_{\sg}\leq\sup_{M^\prime_{c^*}}\left\{|u|^{p_2+p_3+3}|\xi_1|\right\}\label{13.7}\\
&&\hspace{30mm}\cdot\left(\int_{M^\prime_{c^*}}|u|^{-2p_2-3}|\xi_2|^2
d\mu_g\right)^{1/2} \left(\int_{M^\prime_{c^*}}|u|^{-2p_3-3}|\xi_3|^2
d\mu_g\right)^{1/2}\nonumber
\end{eqnarray}
and we have:
\begin{eqnarray}
&&\sup_{M^\prime_{c^*}}\left\{|u|^{p_2+p_3+3}|\xi_1|\right\}
=\sup_{(\ub,u)\in D^\prime_{c^*}}\left\{|u|^{p_2+p_3+3}\|\xi_1\|_{L^\infty(S_{\ub,u})}\right\}\nonumber\\
&&\hspace{33mm}\leq O(\delta^{r_1})\sup_{u\in[u_0,c^*)}\left\{|u|^{p_1+p_2+p_3+3}\right\}\nonumber\\
&&\hspace{33mm}\leq O(\delta^{r_1}) \ \ : \ \mbox{if
$p_1+p_2+p_3+3\leq 0$} \label{13.8}
\end{eqnarray}
while the two integrals on the right in \ref{13.7} can be
estimated in a similar manner to \ref{13.5} above. Substituting
then in \ref{13.7} yields again \ref{13.2}.

Consider next Case 4. In this case we write:
\begin{eqnarray}
&&\int_{M^\prime_{c^*}}|\xi_1||\xi_2||\xi_3|\leq C\int_{u_0}^{c^*}\left\{\int_{C_u}|\xi_1||\xi_2||\xi_3|\right\}du\nonumber\\
&&\hspace{20mm}\leq
C\int_{u_0}^{c^*}\||\xi_1||\xi_2|\|_{L^2(C_u)}\|\xi_3\|_{L^2(C_u)}du
\label{13.9}
\end{eqnarray}
Now,
\begin{eqnarray}
&&\||\xi_1||\xi_2|\|^2_{L^2(C_u)}=\int\left\{\int_{S_{\ub,u}}|\xi_1|^2|\xi_2|^2\right\}d\ub\nonumber\\
&&\hspace{20mm}\leq\int\|\xi_1\|^2_{L^4(S_{\ub,u})}\|\xi_2\|^2_{L^4(S_{\ub,u})}d\ub\nonumber\\
&&\hspace{20mm}\leq \delta O(\delta^{2r_1}|u|^{2p_1+1})O(\delta^{2r_2}|u|^{2p_2+1})\nonumber\\
&&\hspace{20mm}=O(\delta^{2(r_1+r_2)+1}|u|^{2(p_1+p_2+1)})
\label{13.10}
\end{eqnarray}
and:
\begin{equation}
\|\xi_3\|_{L^2(C_u)}\leq O(\delta^{r_3+\frac{1}{2}}|u|^{p_3+1})
\label{13.11}
\end{equation}
Hence the right hand side of \ref{13.9} is bounded by:
\begin{eqnarray}
&&O(\delta^{r_1+r_2+r_3+1})\int_{u_0}^{c^*}|u|^{p_1+p_2+p_3+2}du\nonumber\\
&&\leq O(\delta^{r_1+r_2+r_3+1}) \ \ : \ \mbox{if
$p_1+p_2+p_3+3<0$} \label{13.12}
\end{eqnarray}
thus \ref{13.2} results once again.

Consider finally Case 5. In this case we write:
\begin{eqnarray}
&&\int_{M^\prime_{c^*}}|\xi_1||\xi_2||\xi_3|d\mu_g\label{13.13}\\
&&\hspace{10mm}\leq\left(\int_{M^\prime_{c^*}}|u|^{2p_3+3}|\xi_1|^2|\xi_2|^2d\mu_g\right)^{1/2}
\left(\int_{M^\prime_{c^*}}|u|^{-2p_3-3}|\xi_3|^2
d\mu_g\right)^{1/2}\nonumber
\end{eqnarray}
We have:
\begin{eqnarray}
&&\int_{M^\prime_{c^*}}|u|^{2p_3+3}|\xi_1|^2|\xi_2|^2 d\mu_g\leq
C\int_{u_0}^{c^*}|u|^{2p_3+3}\left\{\int_{C_u}|\xi_1|^2|\xi_2|^2\right\}du
\nonumber\\
&&\hspace{30mm}\leq O(\delta^{2(r_1+r_2)+1})\int_{u_0}^{c^*}|u|^{2(p_1+p_2+p_3)+5}du\nonumber\\
&&\hspace{30mm}\leq O(\delta^{2(r_1+r_2)+1}) \ \ : \ \mbox{if
$p_1+p_2+p_3+3<0$}\nonumber\\
&&\label{13.14}
\end{eqnarray}
where we have substituted for
$$\int_{C_u}|\xi_1|^2|\xi_2|^2=\||\xi_1||\xi_2|\|^2_{L^2(C_u)}$$
the bound \ref{13.10}. Substituting \ref{13.14} and \ref{13.5} in
\ref{13.13} yields again \ref{13.2}.

\vspace{5mm}

Let us recall from Chapter 12 that the Weyl fields we are
considering are:
\begin{eqnarray}
\mbox{0th order:}&&\hspace{5mm} W=R\nonumber\\
\mbox{1st order:}&&\hspace{5mm} W=\tcL_L R,  \ \ W=\tcL_{O_i}R:i=1,2,3, \ \ W=\tcL_S R,\nonumber\\
\mbox{2nd order:}&&\hspace{5mm} W=\tcL_L\tcL_L R, \ \ W=\tcL_{O_i}\tcL_L R:i=1,2,3,\nonumber\\
&&\hspace{17mm}W=\tcL_{O_j}\tcL_{O_i} R:i,j=1,2,3,\nonumber\\
&&\hspace{5mm} W=\tcL_{O_i}\tcL_S R:i=1,2,3, \ \ W=\tcL_S\tcL_S R\
\ \label{13.15}
\end{eqnarray}
We assign to such a Weyl field $W$ the index $l$ which is the
number of $\tcL_L$ operators in the definition of $W$ in terms of
$R$. Thus:
\begin{eqnarray}
&&l=0 \ \mbox{for} \ W=R, \ W=\tcL_{O_i}R: i=1,2,3, \ W=\tcL_S R \nonumber\\
&&\hspace{2mm}\mbox{and for} \ W=\tcL_{O_j}\tcL_{O_i}R:i,j=1,2,3, \nonumber\\
&&\hspace{14mm}W=\tcL_{O_i}\tcL_S R:i=1,2,3, \ W=\tcL_S\tcL_S R\nonumber\\
&&l=1 \ \mbox{for} \ W=\tcL_L R, \ W=\tcL_{O_i}\tcL_L R:i=1,2,3\nonumber\\
&&l=2 \ \mbox{for} \ W=\tcL_L\tcL_L R \label{13.16}
\end{eqnarray}

Let us also recall from \ref{12.264}, \ref{12.266} that in
$\stackrel{(n)}{\tau}_1$, $\stackrel{(n)}{\tau}_2$,
$\s^{(X)}\stackrel{(n)}{\tau}$ and $\s^{(YX)}\stackrel{(n)}{\tau}$
enter multiplied by the factor $\delta^{2l}$ where $l$ is the
number of $L$'s in $(X)$ and in $(YX)$ respectively. Thus in the
case of $\s^{(X)}\stackrel{(n)}{\tau}$ we have $l=1$ for
$(X)=(L)$, $l=0$ for $(X)=(O)$ and $(X)=(S)$, and in the case of
$\s^{(YX)}\stackrel{(n)}{\tau}$ we have $l=2$ for $(YX)=(LL)$,
$l=1$ for $(YX)=(OL)$, $l=0$ for $(YX)=(OO)$, $(YX)=(OS)$ and
$(YX)=(SS)$. In the case of $\stackrel{(n)}{\tau}$ itself we have
$l=0$. Bounds for the error integrals:
\begin{eqnarray}
&&\mbox{0th order:} \ \delta^{2q_n+2l}\int_{M^\prime_{c^*}}|\stackrel{(n)}{\tau}|d\mu_g\label{13.17}\\
&&\mbox{1st order:} \ \delta^{2q_n+2l}\int_{M^\prime_{c^*}}|\s^{(X)}\stackrel{(n)}{\tau}|d\mu_g\nonumber\\
&&\hspace{35mm} : \  \mbox{for} \ (X)=(L),(O),(S)\nonumber\\
&&\mbox{2nd order:} \ \delta^{2q_n+2l}\int_{M^\prime_{c^*}}|\s^{(YX)}\stackrel{(n)}{\tau}|d\mu_g\nonumber\\
&&\hspace{35mm} : \ \mbox{for} \
(YX)=(LL),(OL),(OO),(OS),(SS)\nonumber
\end{eqnarray}
then imply corresponding bounds for the error integrals
\ref{12.276}.

From \ref{12.83}, \ref{12.84}, \ref{12.88}, \ref{12.89},
\ref{12.93}, \ref{12.94}, \ref{12.98}, \ref{12.99}, \ref{12.103},
\ref{12.104}, \ref{12.108}, \ref{12.109}, \ref{12.113},
\ref{12.114}, \ref{12.118}, \ref{12.119}, \ref{12.123},
\ref{12.124}, \ref{12.125}, \ref{12.126}, \ref{12.128},
\ref{12.129}, \ref{12.130}, \ref{12.132}, \ref{12.139},
\ref{12.140}, \ref{12.141}, in conjunction with the inequality
\ref{12.247}, we have:
\begin{eqnarray}
&&\alpha(W)={\bf O}(\delta^{-\frac{1}{2}-q_0-l}|u|^{-1})\nonumber\\
&&\beta(W)={\bf O}(\delta^{-\frac{1}{2}-q_1-l}|u|^{-2})\nonumber\\
&&(\rho,\sigma)(W)={\bf O}(\delta^{-\frac{1}{2}-q_2-l}|u|^{-3})\nonumber\\
&&\beb(W)={\bf O}(\delta^{-\frac{1}{2}-q_3-l}|u|^{-4})
\label{13.18}
\end{eqnarray}
and:
\begin{equation}
\alb(W)=\bfob(\delta^{-q_3-l}|u|^{-9/2}) \label{13.19}
\end{equation}
for all Weyl fields $W$ in \ref{13.15} except for the Weyl fields
$$W=\tcL_S R, \ W=\tcL_{O_i}\tcL_S R:i=1,2,3, \ W=\tcL_S\tcL_S R$$
and \ref{13.19} and the last of \ref{13.18} hold for these three
Weyl fields as well.

\section{The multiplier error estimates}

Let us recall from \ref{12.64} that $\tau(W;X,Y,Z)$, the
divergence of the energy-momentum density vectorfield $P(W;X,Y,Z)$
associated to the Weyl field $W$ and the multiplier fields $X,Y,Z$
is expressed as the sum:
\begin{equation}
\tau(W;X,Y,Z)=\tau_c(W;X,Y,Z)+\tau_m(W;X,Y,Z) \label{13.20}
\end{equation}
where:
\begin{equation}
\tau_m(W;X,Y,Z)=-\frac{1}{2}Q(W)_{\alpha\beta\gamma\delta}(\s^{(X)}\tilde{\pi}^{\alpha\beta}Y^\gamma
Z^\delta +\s^{(Y)}\tilde{\pi}^{\alpha\beta}X^\gamma
Z^\delta+\s^{(Z)}\tilde{\pi}^{\alpha\beta}X^\gamma Y^\delta)
\label{13.21}
\end{equation}
is generated by the deformation tensors of the multiplier fields,
while:
\begin{equation}
\tau_c(W;X,Y,Z)=-(\mbox{div}Q(W))(X,Y,Z) \label{13.22}
\end{equation}
is generated (see Proposition 12.6) by the Weyl current $J$
corresponding to $W$. Thus the divergences
\begin{eqnarray*}
&&\mbox{0th order:} \ \stackrel{(n)}{\tau} \ : \ n=0,1,2,3,\\
&&\mbox{1st order:} \ \s^{(X)}\stackrel{(n)}{\tau} \ : \ (X)=(L),(O),(S); \ n=0,1,2,3\\
&&\mbox{2nd order:} \ \s^{(YX)}\stackrel{(n)}{\tau} \ : \
(YX)=(LL),(OL),(OO),(OS),(SS); \ n=0,1,2,3
\end{eqnarray*}
correspondingly split into the sums:
\begin{eqnarray}
&&\stackrel{(n)}{\tau}=\stackrel{(n)}{\tau}_c+\stackrel{(n)}{\tau}_m\nonumber\\
&&\s^{(X)}\stackrel{(n)}{\tau}=\s^{(X)}\stackrel{(n)}{\tau}_c+\s^{(X)}\stackrel{(n)}{\tau}_m\nonumber\\
&&\s^{(YX)}\stackrel{(n)}{\tau}=\s^{(YX)}\stackrel{(n)}{\tau}_c+\s^{(YX)}\stackrel{(n)}{\tau}_m
\label{13.23}
\end{eqnarray}
In the present chapter we shall estimate the contributions of the
{\em multiplier error terms} $\stackrel{(n)}{\tau}_m$,
$\s^{(X)}\stackrel{(n)}{\tau}_m$,
$\s^{(YX)}\stackrel{(n)}{\tau}_m$, to the error integrals
\ref{13.17}. These error terms are all of the form:
\begin{eqnarray}
&&\stackrel{(0)}{\tau}_m(W)=-\frac{3}{2}\s^{(L)}\tilde{\pi}^{\mu\nu}Q_{\mu\nu\kappa\lambda}(W)L^\kappa L^\lambda\nonumber\\
&&\stackrel{(1)}{\tau}_m(W)=-\s^{(L)}\tilde{\pi}^{\mu\nu}Q_{\mu\nu\kappa\lambda}(W)K^\kappa
L^\lambda
-\frac{1}{2}\s^{(K)}\tilde{\pi}^{\mu\nu}Q_{\mu\nu\kappa\lambda}(W)L^\kappa L^\lambda\nonumber\\
&&\stackrel{(2)}{\tau}_m(W)=-\frac{1}{2}\s^{(L)}\tilde{\pi}^{\mu\nu}Q_{\mu\nu\kappa\lambda}(W)K^\kappa
K^\lambda
-\s^{(K)}\tilde{\pi}^{\mu\nu}Q_{\mu\nu\kappa\lambda}(W)K^\kappa L^\lambda\nonumber\\
&&\stackrel{(3)}{\tau}_m(W)=-\frac{3}{2}\s^{(K)}\tilde{\pi}^{\mu\nu}Q_{\mu\nu\kappa\lambda}(W)K^\kappa
K^\lambda \label{13.24}
\end{eqnarray}
where $W$ is each one of the Weyl fields \ref{13.15}. The
integrals to be estimated, the {\em multiplier error integrals},
are then:
\begin{equation}
\delta^{2q_n+2l}\int_{M^\prime_{c^*}}|\stackrel{(n)}{\tau}_m(W)|d\mu_g
\label{13.25}
\end{equation}
where $l$ is the index associated to $W$.

Expanding the above expressions in the frame $(e_\mu \ : \
\mu=1,2,3,4)$ using \ref{8.11}, \ref{8.12} and Lemma 12.2 we find:
\begin{eqnarray}
&&\stackrel{(0)}{\tau}_m(W)=-3\Omega^2\left\{2\s^{(L)}j|\beta(W)|^2-2\s^{(L)}\mb^\sharp\cdot\alpha(W)\cdot\beta^\sharp(W)\right.\nonumber\\
&&\hspace{27mm}\left.+\rho(W)(\s^{(L)}\ih,\alpha(W))-\sigma(W)\s^{(L)}\ih\wedge\alpha(W)\right\}
\label{13.26}
\end{eqnarray}
\begin{eqnarray}
&&\stackrel{(1)}{\tau}_m(W)=-2\Omega^2\left\{2|u|^2\s^{(L)}j((\rho(W))^2+(\sigma(W))^2)\right.\nonumber\\
&&\hspace{27mm}-2|u|^2\s^{(L)}\mb^\sharp\cdot(\rho(W)\beta(W)-\sigma(W)\s^*\beta(W))\nonumber\\
&&\hspace{27mm}-|u|^2(\s^{(L)}\ih,\beta(W)\oth\beb(W))\nonumber\\
&&\hspace{27mm}+\s^{(K)}j|\beta(W)|^2-\s^{(K)}m^\sharp\cdot(\rho(W)\beta(W)-\sigma(W)\s^*\beta(W))\nonumber\\
&&\hspace{27mm}\left.+\frac{1}{2}\rho(W)(\s^{(K)}\ih,\alpha(W))-\frac{1}{2}\sigma(W)\s^{(K)}\ih\wedge\alpha(W)\right\}
\label{13.27}
\end{eqnarray}
\begin{eqnarray}
&&\stackrel{(2)}{\tau}_m(W)=-2\Omega^2\left\{|u|^4\s^{(L)}j|\beb(W)|^2\right.\nonumber\\
&&\hspace{27mm}+|u|^4\s^{(L)}\mb^\sharp\cdot(\rho(W)\beb(W)+\sigma(W)\s^*\beb(W))\nonumber\\
&&\hspace{27mm}+\frac{1}{2}|u|^4\rho(W)(\s^{(L)}\ih,\alb(W))+\frac{1}{2}|u|^4\sigma(W)\s^{(L)}\ih\wedge\alb(W)\nonumber\\
&&\hspace{27mm}+2|u|^2\s^{(K)}j((\rho(W))^2+(\sigma(W))^2)\nonumber\\
&&\hspace{27mm}+2|u|^2\s^{(K)}m^\sharp\cdot(\rho(W)\beb(W)+\sigma(W)\s^*\beb(W))\nonumber\\
&&\hspace{27mm}\left.-|u|^2(\s^{(K)}\ih,\beta(W)\oth\beb(W))\right\}\nonumber\\
&&\label{13.28}
\end{eqnarray}
\begin{eqnarray}
&&\stackrel{(3)}{\tau}_m(W)=-3\Omega^2\left\{2|u|^4\s^{(K)}j|\beb(W)|^2+2|u|^4\s^{(K)}m^\sharp\cdot\alb(W)\cdot\beb^\sharp(W)\right.\nonumber\\
&&\hspace{27mm}\left.+|u|^4\rho(W)(\s^{(K)}\ih,\alb(W))+|u|^4\sigma(W)\s^{(K)}\ih\wedge\alb(W)\right\}\nonumber\\
\label{13.29}
\end{eqnarray}
In the above expressions each term is a trilinear expression
$\{\xi_1,\xi_2,\xi_3\}$ with coefficients depending only on $\sg$
and $\seps$, where $\xi$ is a component of the deformation tensors
of $L$ or $K$, multiplied by $\Omega^2$ and the appropriate power
of $|u|^2$, while $\xi_1$ and $\xi_2$ are components of $W$.
Recalling from Chapter 8 that the components of the deformation
tensors of $L$ and $K$ satisfy the $L^\infty$ bounds \ref{8.32}
and \ref{8.33}, we have:
\begin{eqnarray}
&&\s^{(L)}\ih={\cal O}^\infty(\delta^{-1/2}|u|^{-1})\nonumber\\
&&\s^{(L)}j={\cal O}^\infty(|u|^{-1})\nonumber\\
&&\s^{(L)}\mb={\cal O}^\infty(\delta^{1/2}|u|^{-2}) \label{13.30}
\end{eqnarray}
and:
\begin{eqnarray}
&&\s^{(K)}\ih={\cal O}^\infty(\delta^{1/2})\nonumber\\
&&\s^{(K)}j={\cal O}^\infty(\delta)\nonumber\\
&&\s^{(K)}m={\cal O}^\infty(\delta^{1/2}) \label{13.31}
\end{eqnarray}
while the components of $W$ satisfy \ref{13.18}, \ref{13.19}. Thus
to each term on the right in \ref{13.26} - \ref{13.29} we can
apply one of the first two cases of Lemma 13.1. The third case
does not occur because there are no terms involving two $\alb(W)$
factors. Comparing with what is required to obtain a bound for
\ref{13.25} by $O(1)$, we define the {\em excess index} $e$ of the
contribution of a given term to \ref{13.25} by:
\begin{equation}
e=2q_n+2l+r_1+r_2+r_3+1 \label{13.32}
\end{equation}
Then the contribution of the given term to \ref{13.25} is bounded
by $O(\delta^e)$, provided that the {\em integrability index} $s$
of that term, defined by:
\begin{equation}
s=p_1+p_2+p_3+3 \label{13.33}
\end{equation}
is negative so that Lemma 13.1 applies. We obtain in this way the
following tables. The ordinals in these tables refer to the terms
on the right hand side of each of \ref{13.26} - \ref{13.29}. We
consider a pair of similar terms one involving $\rho(W)$ and the
other $\sigma(W)$ as a single term. Thus, we consider the 2nd line
of \ref{13.26}, the 5th line of \ref{13.27}, the 3rd line of
\ref{13.28} and the 2nd line of \ref{13.29}, each as a single
term.

\vspace{5mm}

\hspace{45mm}{\bf Case $n=0$:}
\begin{equation}
\begin{array}{l|ll}
\mbox{term}&e&s\\ \hline
\mbox{1st}&2&-2\\
\mbox{2nd}&3/2&-2\\
\mbox{3rd}&1&-2
\end{array}
\label{13.34}
\end{equation}

\vspace{2.5mm}

\hspace{45mm}{\bf Case $n=1$:}
\begin{equation}
\begin{array}{l|ll}
\mbox{term}&e&s\\ \hline
\mbox{1st}&1&-2\\
\mbox{2nd}&1&-2\\
\mbox{3rd}&1&-2\\
\mbox{4th}&1&-1\\
\mbox{5th}&1&-2\\
\mbox{6th}&0&-1
\end{array}
\label{13.35}
\end{equation}

\vspace{2.5mm}

\hspace{45mm}{\bf Case $n=2$:}
\begin{equation}
\begin{array}{l|ll}
\mbox{term}&e&s\\ \hline
\mbox{1st}&2&-2\\
\mbox{2nd}&3/2&-2\\
\mbox{3rd}&1&-3/2\\
\mbox{4th}&1&-1\\
\mbox{5th}&3/2&-2\\
\mbox{6th}&1&-1
\end{array}
\label{13.36}
\end{equation}

\vspace{2.5mm}

\hspace{45mm}{\bf Case $n=3$:}
\begin{equation}
\begin{array}{l|ll}
\mbox{term}&e&s\\ \hline
\mbox{1st}&1&-1\\
\mbox{2nd}&1&-3/2\\
\mbox{3rd}&0&-1/2
\end{array}
\label{13.37}
\end{equation}

\vspace{5mm}

We see that all terms have negative integrability index so Lemma
13.1 indeed applies. All terms have non-negative excess index. The
terms with vanishing excess index play a crucial role because they
give rise to {\em borderline error integrals}. These must be
analyzed in more detail. They occur only in the cases $n=1$ and
$n=3$. The borderline terms are the 6th term in \ref{13.27} and
the 3rd term in \ref{13.29}.

Consider the 6th term in \ref{13.27}. Replacing the first of
\ref{13.31} by the more precise statement:
\begin{equation}
\|\s^{(K)}\ih\|_{L^\infty(S_{\ub,u})}\leq C\delta^{1/2}({\cal
D}_0^\infty(\chibh)+O(\delta)) \ \ : \ \forall (\ub,u)\in
D^\prime_{c^*} \label{13.38}
\end{equation}
implied by the first of \ref{8.33}, replacing also the first and
third of \ref{13.18} by the more precise statements:
\begin{eqnarray}
&&\|\alpha(W)\|_{L^2(C_u)}\leq C(\stackrel{(0)}{{\cal E}}_2)^{1/2}\delta^{-1-l} \ \ : \ \forall u\in[u_0,c^*)\nonumber\\
&&\|(\rho,\sigma)(W)\|_{L^2(C_u)}\leq C(\stackrel{(2)}{{\cal
E}}_2)^{1/2}\delta^{1/2-l}|u|^{-2} \ \ : \forall u\in[u_0,c^*)
\label{13.39}
\end{eqnarray}
implied by the first and third of each of \ref{12.83},
\ref{12.88}, \ref{12.93}, \ref{12.103}, \ref{12.108},
\ref{12.113}, \ref{12.125},  \ref{12.129}, and \ref{12.139}, and
following the proof of Case 1 of Lemma 13.1, we deduce that the
contribution of the term in question to the error integral
\ref{13.25} with $n=1$ is bounded by:
\begin{equation}
C{\cal D}_0^\infty(\chibh)(\stackrel{(0)}{{\cal
E}}_2)^{1/2}(\stackrel{(2)}{{\cal E}}_2)^{1/2}+O(\delta)
\label{13.40}
\end{equation}

Consider the 3rd term in \ref{13.29}. Replacing again the first of
\ref{13.31} by \ref{13.38}, the third of \ref{13.18} by the second
of \ref{13.39}, and \ref{13.19} by the more precise statement:
\begin{equation}
\||u|^3\alb(W)\|_{L^2(\Cb_{\ub})}\leq C(\stackrel{(3)}{{\cal
F}}_2)^{1/2}\delta^{3/2} \ \ : \ \forall \ub\in[0,\delta)
\label{13.41}
\end{equation}
implied by \ref{12.84}, \ref{12.89}, \ref{12.94}, \ref{12.99},
\ref{12.104}, \ref{12.109}, \ref{12.114}, \ref{12.119},
\ref{12.124}, \ref{12.128}, \ref{12.132}, \ref{12.140}, and
following the proof of Case 2 of Lemma 13.1, we deduce that the
contribution of the term in question to the error integral
\ref{13.25} with $n=3$ is bounded by:
\begin{equation}
C{\cal D}_0^\infty(\chibh)(\stackrel{(2)}{{\cal
E}}_2)^{1/2}(\stackrel{(3)}{{\cal F}}_2)^{1/2}+O(\delta)
\label{13.42}
\end{equation}

However, the above have not been shown to hold in the case of the
three Weyl fields which involve $S$:
$$W=\tcL_S R, \ W=\tcL_{O_i}\tcL_S R:i=1,2,3, \ W=\tcL_S\tcL_S R$$
For these only the case $n=3$ occurs so only \ref{13.29} is to be
considered. The point here is that the second of \ref{13.39} does
not quite hold for these three Weyl fields. The aim of the
remainder of the present chapter is to show that for these three
Weyl fields it nevertheless holds:
\begin{equation}
\|(\rho,\sigma)(W)\|_{L^2(C_u)}\leq C\delta^{1/2}|u|^{-2}((\stackrel{(2)}{{\cal
E}}_2)^{1/2}+O(\delta^{1/2}))\label{13.43}
\end{equation}
(here $l=0$) so that \ref{13.42} is simply replaced by:
\begin{equation}
C{\cal D}_0^\infty(\chibh)(\stackrel{(2)}{{\cal
E}}_2)^{1/2}(\stackrel{(3)}{{\cal F}}_2)^{1/2}+O(\delta^{1/2})
\label{13.44}
\end{equation}

Consider first the case $W=\tcL_S R$. From Proposition 12.2 we
have:
\begin{eqnarray}
&&\rho(\tcL_S
R)=S\rho+3\s^{(S)}\nu\rho+\frac{3}{4}\s^{(S)}j\rho-\frac{1}{2}\s^{(S)}m^\sharp\cdot\beb
+\frac{1}{2}\s^{(S)}\mb^\sharp\cdot\beta\nonumber\\
&&\sigma(\tcL_S
R)=S\sigma+3\s^{(S)}\nu\sigma+\frac{3}{4}\s^{(S)}j\sigma+\frac{1}{2}\s^{*(S)}m^\sharp\cdot\beb
+\frac{1}{2}\s^{*(S)}\mb^\sharp\cdot\beta \nonumber\\
&&\label{13.45}
\end{eqnarray}
In view of the estimates \ref{8.34} together with the fact that by
\ref{12.37}
\begin{equation}
\s^{(S)}\nu-1=S\log\Omega=u\omb+\ub\omega \label{13.46}
\end{equation}
and the estimates of Chapter 3 we have:
\begin{equation}
\|\s^{(S)}\nu-1\|_{L^\infty(S_{\ub,u})}\leq O(\delta|u|^{-2})
\label{13.47}
\end{equation}
we deduce:
\begin{eqnarray}
&&\|\rho(\tcL_S R)\|_{L^2(C_u)}\leq \|S\rho\|_{L^2(C_u)}+3\|\rho\|_{L^2(C_u)}+O(\delta^{3/2}|u|^{-3})\nonumber\\
&&\|\sigma(\tcL_S R)\|_{L^2(C_u)}\leq
\|S\sigma\|_{L^2(C_u)}+3\|\sigma\|_{L^2(C_u)}+O(\delta^{3/2}|u|^{-3})
\label{13.48}
\end{eqnarray}

We have:
\begin{equation}
(S\rho,S\sigma)=(u\Db\rho+\ub D\rho,u\Db\sigma+\ub D\sigma)
\label{13.49}
\end{equation}
and:
\begin{eqnarray}
&&\|\ub D\rho\|_{L^2(C_u)}\leq \delta\|D\rho\|_{L^2(C_u)}\leq \delta^{1/2}|u|^{-2}{\cal R}_{[1]}(\rho)\nonumber\\
&&\hspace{20mm}\leq \delta^{1/2}|u|^{-2}\left\{C(\stackrel{(2)}{{\cal E}}_1)^{1/2}+O(\delta)\right\}\nonumber\\
&&\|\ub D\sigma\|_{L^2(C_u)}\leq \delta\|D\sigma\|_{L^2(C_u)}\leq \delta^{1/2}|u|^{-2}{\cal R}_{[1]}(\sigma)\nonumber\\
&&\hspace{20mm}\leq\delta^{1/2}|u|^{-2}\left\{C(\stackrel{(2)}{{\cal
E}}_1)^{1/2}+O(\delta)\right\} \label{13.50}
\end{eqnarray}
by \ref{12.a3}. To obtain appropriate estimates for
$\|u\Db\rho\|_{L^2(C_u)}$, $\|u\Db\sigma\|_{L^2(C_u)}$, we
consider the eighth and tenth of the Bianchi identities, given by
Proposition 1.2:
\begin{eqnarray}
&&\Db\rho+\frac{3}{2}\Omega\mbox{tr}\chib\rho=-\Omega\{\sdiv\beb+(2\eta-\zeta,\beb)+\frac{1}{2}(\chih,\alb)\}\nonumber\\
&&\Db\sigma+\frac{3}{2}\Omega\mbox{tr}\chib\sigma=-\Omega\{\scurl\beb+(2\eta-\zeta,\s^*\beb)+\frac{1}{2}\chih\wedge\alb\}
\label{13.51}
\end{eqnarray}
By the results of Chapter 3 the right hand sides are bounded in
$L^2(C_u)$ by $O(\delta^{3/2}|u|^{-4})$. Hence we obtain:
\begin{eqnarray}
&&\|u\Db\rho\|_{L^2(C_u)}\leq C\|\rho\|_{L^2(C_u)}+O(\delta^{3/2}|u|^{-3})\nonumber\\
&&\|u\Db\sigma\|_{L^2(C_u)}\leq
C\|\sigma\|_{L^2(C_u)}+O(\delta^{3/2}|u|^{-3}) \label{13.52}
\end{eqnarray}
Combining \ref{13.50} and \ref{13.52} and taking into account the
fact that:
\begin{equation}
\|\rho\|_{L^2(C_u)},\|\sigma\|_{L^2(C_u)}\leq
C\delta^{1/2}|u|^{-2}(\stackrel{(2)}{{\cal E}}_0)^{1/2}
\label{13.53}
\end{equation}
(see third of \ref{12.83}, \ref{12.133}) we conclude that:
\begin{equation}
\|S\rho\|_{L^2(C_u)},\|S\sigma\|_{L^2(C_u)}\leq
C\delta^{1/2}|u|^{-2}(\stackrel{(2)}{{\cal
E}}_1)^{1/2}+O(\delta^{3/2}|u|^{-2}) \label{13.54}
\end{equation}
Substituting \ref{13.54} and \ref{13.53} in \ref{13.48} then
yields:
\begin{equation}
\|(\rho,\sigma)(\tcL_S R)\|_{L^2(C_u)}\leq
C\delta^{1/2}|u|^{-2}(\stackrel{(2)}{{\cal
E}}_1)^{1/2}+O(\delta^{3/2}|u|^{-2}) \label{13.55}
\end{equation}

Consider next the case $W=\tcL_{O_i}\tcL_S R:i=1,2,3$. From
Proposition 12.2 we have:
\begin{eqnarray}
&&\rho(\tcL_{O_i}\tcL_S R)=O_i\rho(\tcL_S R)+3\s^{(O_i)}\nu\rho(\tcL_S R)+\frac{3}{4}\s^{(O_i)}j\rho(\tcL_S R)\nonumber\\
&&\hspace{23mm}-\frac{1}{2}\s^{(O_i)}m^\sharp\cdot\beb(\tcL_S R)\nonumber\\
&&\sigma(\tcL_{O_i}\tcL_S R)=O_i\sigma(\tcL_S R)+3\s^{(O_i)}\nu\sigma(\tcL_S R)+\frac{3}{4}\s^{(O_i)}j\sigma(\tcL_S R)\nonumber\\
&&\hspace{23mm}+\frac{1}{2}\s^{*(O_i)}m^\sharp\cdot\beb(\tcL_S R)
\label{13.56}
\end{eqnarray}
In view of the estimates \ref{8.139} together with the fact that
by \ref{12.37}
\begin{equation}
\s^{(O_i)}\nu=O_i(\log\Omega) \label{13.57}
\end{equation}
and the estimate \ref{7.210} we have:
\begin{equation}
\|\s^{(O_i)}\nu\|_{L^\infty(S_{\ub,u})}\leq O(\delta|u|^{-2})
\label{13.58}
\end{equation}
we deduce, taking also into account \ref{13.55},
\begin{eqnarray}
&&\|\rho(\tcL_{O_i}\tcL_S R)\|_{L^2(C_u)}\leq \|O_i\rho(\tcL_S R)\|_{L^2(C_u)}+O(\delta^{3/2}|u|^{-4})\nonumber\\
&&\|\sigma(\tcL_{O_i}\tcL_S R)\|_{L^2(C_u)}\leq \|O_i\sigma(\tcL_S
R)\|_{L^2(C_u)}+O(\delta^{3/2}|u|^{-4}) \label{13.59}
\end{eqnarray}
Moreover, applying $O_i$ to the expressions \ref{13.45} and using
the estimates \ref{9.12} we deduce:
\begin{eqnarray}
\|O_i\rho(\tcL_S R)\|_{L^2(C_u)}\leq \|O_i S\rho\|_{L^2(C_u)}+3\|O_i\rho\|_{L^2(C_u)}+O(\delta^{3/2}|u|^{-3})\nonumber\\
\|O_i\sigma(\tcL_S R)\|_{L^2(C_u)}\leq \|O_i
S\sigma\|_{L^2(C_u)}+3\|O_i\sigma\|_{L^2(C_u)}+O(\delta^{3/2}|u|^{-3})
\label{13.60}
\end{eqnarray}

We have:
\begin{equation}
(O_i S\rho,O_i S\sigma)=(uO_i\Db\rho+\ub
O_iD\rho,uO_i\Db\sigma+\ub O_iD\sigma) \label{13.61}
\end{equation}
and:
\begin{eqnarray}
&&\|\ub O_i D\rho\|_{L^2(C_u)}\leq \delta\|O_i D\rho\|_{L^2(C_u)}\leq \delta^{1/2}|u|^{-2}{\cal R}_{[2]}(\rho)\nonumber\\
&&\hspace{20mm}\leq \delta^{1/2}|u|^{-2}\left\{C(\stackrel{(2)}{{\cal E}}_2)^{1/2}+O(\delta^{1/2})\right\}\nonumber\\
&&\|\ub O_i D\sigma\|_{L^2(C_u)}\leq \delta\|D\sigma\|_{L^2(C_u)}\leq \delta^{1/2}|u|^{-2}{\cal R}_{[2]}(\sigma)\nonumber\\
&&\hspace{20mm}\leq\delta^{1/2}|u|^{-2}\left\{C(\stackrel{(2)}{{\cal
E}}_2)^{1/2}+O(\delta^{1/2})\right\} \label{13.62}
\end{eqnarray}
by \ref{12.a4}. To obtain appropriate estimates for
$\|uO_i\Db\rho\|_{L^2(C_u)}$, $\|uO_i\Db\sigma\|_{L^2(C_u)}$, we
apply $O_i$ to the Bianchi identities \ref{13.51}. Using the
results of Chapters 3 and 4 we then deduce:
\begin{eqnarray}
&&\|uO_i\Db\rho\|_{L^2(C_u)}\leq C\|O_i\rho\|_{L^2(C_u)}+O(\delta^{3/2}|u|^{-3})\nonumber\\
&&\|uO_i\Db\sigma\|_{L^2(C_u)}\leq
C\|O_i\sigma\|_{L^2(C_u)}+O(\delta^{3/2}|u|^{-3}) \label{13.63}
\end{eqnarray}
Combining \ref{13.62} and \ref{13.63} and taking into account the
fact that by \ref{12.a3}:
\begin{eqnarray}
\|O_i\rho\|_{L^2(C_u)}\leq C\delta^{1/2}|u|^{-2}{\cal
R}_{[1]}(\rho)\leq C\delta^{1/2}|u|^{-2}
\left\{(\stackrel{(2)}{{\cal E}}_1)^{1/2}+O(\delta)\right\}\nonumber\\
\|O_i\sigma\|_{L^2(C_u)}\leq C\delta^{1/2}|u|^{-2}{\cal
R}_{[1]}(\sigma)\leq C\delta^{1/2}|u|^{-2}
\left\{(\stackrel{(2)}{{\cal E}}_1)^{1/2}+O(\delta)\right\}
\label{13.64}
\end{eqnarray}
we conclude that:
\begin{equation}
\|O_i S\rho\|_{L^2(C_u)},\|O_i S\sigma\|_{L^2(C_u)}\leq
C\delta^{1/2}|u|^{-2}(\stackrel{(2)}{{\cal
E}}_2)^{1/2}+O(\delta|u|^{-2}) \label{13.65}
\end{equation}
Substituting \ref{13.65} and \ref{13.64} in \ref{13.60} yields:
\begin{eqnarray}
&&\|O_i\rho(\tcL_S R)\|_{L^2(C_u)}\leq C\delta^{1/2}|u|^{-2}(\stackrel{(2)}{{\cal E}}_2)^{1/2}+O(\delta|u|^{-2})\nonumber\\
&&\|O_i\sigma(\tcL_S R\|_{L^2(C_u)}\leq
C\delta^{1/2}|u|^{-2}(\stackrel{(2)}{{\cal
E}}_2)^{1/2}+O(\delta|u|^{-2}) \label{13.66}
\end{eqnarray}
Substituting these in turn in \ref{13.59} we conclude that:
\begin{equation}
\|(\rho,\sigma)(\tcL_{O_i}\tcL_S R)\|_{L^2(C_u)}\leq
C\delta^{1/2}|u|^{-2}(\stackrel{(2)}{{\cal
E}}_2)^{1/2}+O(\delta|u|^{-2}) \label{13.67}
\end{equation}

Consider finally the case $W=\tcL_S\tcL_S R$. From Proposition
12.2 we have:
\begin{eqnarray}
&&\rho(\tcL_S\tcL_S R)=S\rho(\tcL_S R)+3\s^{(S)}\nu\rho(\tcL_S R)+\frac{3}{4}\s^{(S)}j\rho(\tcL_S R)\nonumber\\
&&\hspace{20mm}-\frac{1}{2}\s^{(S)}m^\sharp\cdot\beb(\tcL_S R)
+\frac{1}{2}\s^{(S)}\mb^\sharp\cdot\beta(\tcL_S R)\nonumber\\
&&\sigma(\tcL_S\tcL_S R)=S\sigma(\tcL_S
R)+3\s^{(S)}\nu\sigma(\tcL_S R)
+\frac{3}{4}\s^{(S)}j\sigma(\tcL_S R)\nonumber\\
&&\hspace{20mm}+\frac{1}{2}\s^{*(S)}m^\sharp\cdot\beb(\tcL_S R)
+\frac{1}{2}\s^{*(S)}\mb^\sharp\cdot\beta(\tcL_S R) \label{13.68}
\end{eqnarray}
and:
\begin{eqnarray}
&&\beta(\tcL_S R)=\sL_S\beta+2\s^{(S)}\nu\beta+\frac{1}{4}\s^{(S)}j\beta-\frac{1}{2}\s^{(S)}\ih^\sharp\cdot\beta\nonumber\\
&&\hspace{20mm}+\frac{1}{4}\s^{(S)}\mb^\sharp\cdot\alpha+\frac{3}{4}\s^{(S)}m\rho+\frac{3}{4}\s^{*(S)}m\sigma
\label{13.69}
\end{eqnarray}
Taking into account the estimates \ref{8.34} and \ref{13.47} we
obtain:
\begin{equation}
\|\beta(\tcL_S R)\|_{L^2(C_u)}\leq
\|\sL_S\beta\|_{L^2(C_u)}+2\|\beta\|_{L^2(C_u)}+O(\delta^{1/2}|u|^{-2})
\label{13.70}
\end{equation}
We have:
\begin{equation}
\sL_S\beta=u\Db\beta+\ub D\beta \label{13.71}
\end{equation}
and:
\begin{equation}
\|\ub D\beta\|_{L^2(C_u)}\leq\delta\|D\beta\|_{L^2(C_u)}\leq
|u|^{-1}{\cal R}_{[1]}(\beta)\leq O(|u|^{-1}) \label{13.72}
\end{equation}
To obtain an appropriate estimate for $\|u\Db\beta\|_{L^2(C_u)}$
we consider the fifth of the Bianchi identities of Proposition
1.2:
\begin{equation}
\Db\beta+\frac{1}{2}\Omega\mbox{tr}\chib\beta+\omb\beta=\Omega\{\chibh^\sharp\cdot\beta+\sd\rho+\s*\sd\sigma+3\eta\rho+3\s^*\eta\sigma
+2\chih^\sharp\cdot\beb\} \label{13.73}
\end{equation}
Using the results of Chapter 3 we then deduce:
\begin{equation}
\|u\Db\beta\|_{L^2(C_u)}\leq O(|u|^{-1}) \label{13.74}
\end{equation}
Combining with \ref{13.72} yields:
\begin{equation}
\|\sL_S\beta\|_{L^2(C_u)}\leq O(|u|^{-1}) \label{13.75}
\end{equation}
Substituting in \ref{13.70} we then obtain:
\begin{equation}
\|\beta(\tcL_S R)\|_{L^2(C_u)}\leq O(|u|^{-1}) \label{13.76}
\end{equation}
Going back to \ref{13.68} and using the estimates \ref{8.34},
\ref{13.47} and \ref{13.76} we deduce:
\begin{eqnarray}
&&\|\rho(\tcL_S\tcL_S R)\|_{L^2(C_u)}\leq \|S\rho(\tcL_S R)\|_{L^2(C_u)}+3\|\rho(\tcL_S R)\|_{L^2(C_u)}+O(\delta^{3/2}|u|^{-3})\nonumber\\
&&\|\sigma(\tcL_S\tcL_S R)\|_{L^2(C_u)}\leq \|S\sigma(\tcL_S
R)\|_{L^2(C_u)}+3\|\sigma(\tcL_S
R)\|_{L^2(C_u)}+O(\delta^{3/2}|u|^{-3})
\nonumber\\
&&\label{13.77}
\end{eqnarray}

Applying $S$ to the expressions \ref{13.45} and using the
estimates \ref{12.235}, \ref{13.75} and the estimates \ref{8.34}
as well as the estimates:
\begin{eqnarray}
&&\|\sLh_S\s^{(S)}\ih\|_{L^4(S_{\ub,u})}\leq O(\delta^{1/2}|u|^{-1/2})\nonumber\\
&&\|S\s^{(S)}j\|_{L^4(S_{\ub,u})}\leq O(\delta|u|^{-3/2})\nonumber\\
&&\|\sL_S\s^{(S)}\mb\|_{L^4(S_{\ub,u})}\leq O(\delta^{3/2}|u|^{-3/2})\nonumber\\
&&\|\sL_S\s^{(S)}m\|_{L^4(S_{\ub,u})}\leq
O(\delta^{1/2}|u|^{-1/2}) \label{13.78}
\end{eqnarray}
which follow by combining the estimates \ref{9.13} and \ref{9.14},
and  in addition the estimate:
\begin{equation}
\|S\s^{(S)}\nu\|_{L^4(S_{\ub,u})}\leq O(\delta|u|^{-3/2})
\label{13.79}
\end{equation}
which follows from the estimates of Chapter 4 (see \ref{13.46}),
we deduce:
\begin{eqnarray}
&&\|S\rho(\tcL_S R)\|_{L^2(C_u)}\leq \|S^2\rho\|_{L^2(C_u)}+3\|S\rho\|_{L^2(C_u)}+O(\delta^{3/2}|u|^{-3})\nonumber\\
&&\|S\sigma(\tcL_S R)\|_{L^2(C_u)}\leq
\|S^2\sigma\|_{L^2(C_u)}+3\|S\sigma\|_{L^2(C_u)}+O(\delta^{3/2}|u|^{-3})\nonumber\\
&&\label{13.80}
\end{eqnarray}

Now, for any function $f$ defined on $M^\prime_{c^*}$ we have:
\begin{equation}
S^2 f=u^2\Db^2 f+u\ub(\Db Df+D\Db f)+\ub^2 D^2 f+Sf \label{13.81}
\end{equation}
In particular, this holds for the functions $\rho, \sigma$. We
have:
\begin{eqnarray}
&&\|\ub^2 D^2\rho\|_{L^2(C_u)}\leq \delta^2\|D^2\rho\|_{L^2(C_u)}\leq \delta^{1/2}|u|^{-2}{\cal R}_{[2]}(\rho)\nonumber\\
&&\hspace{20mm}\leq \delta^{1/2}|u|^{-2}\left\{C(\stackrel{(2)}{{\cal E}}_2)^{1/2}+O(\delta^{1/2})\right\}\nonumber\\
&&\|\ub^2 D^2\sigma\|_{L^2(C_u)}\leq \delta^2\|D^2\sigma\|_{L^2(C_u)}\leq \delta^{1/2}|u|^{-2}{\cal R}_{[2]}(\sigma)\nonumber\\
&&\hspace{20mm}\leq
\delta^{1/2}|u|^{-2}\left\{C(\stackrel{(2)}{{\cal
E}}_2)^{1/2}+O(\delta^{1/2})\right\} \label{13.82}
\end{eqnarray}
To obtain appropriate estimates for
$$\|u\ub D\Db\rho\|_{L^2(C_u)},\|u\ub D\Db\sigma\|_{L^2(C_u)},\|u^2\Db^2\rho\|_{L^2(C_u)},\|u^2\Db^2\sigma\|_{L^2(C_u)}$$
we consider again the Bianchi identities \ref{13.51}. Applying $D$
to these identities we obtain:
\begin{eqnarray}
&&D\Db\rho+\frac{3}{2}\Omega\mbox{tr}\chib D\rho+\frac{3}{2}D(\Omega\mbox{tr}\chib)\rho\nonumber\\
&&\hspace{20mm}=-\Omega\left\{\sdiv D\beb+(2D\eta-D\zeta,\beb)+(2\eta-\zeta,D\beb)\right.\nonumber\\
&&\hspace{45mm}\left.+\frac{1}{2}(\Dh\chih,\alb)+\frac{1}{2}(\chih,\Dh\alb)\right\}\nonumber\\
&&\hspace{35mm}+\Omega\left\{2\sdiv(\Omega\chih^\sharp\cdot\beb)+\Omega\mbox{tr}\chi\sdiv\beb\right.\nonumber\\
&&\hspace{20mm}\left.+\Omega\mbox{tr}\chi(2\eta-\zeta,\beb)+2\Omega(2\eta-\zeta)^\sharp\cdot\chih\cdot\beb^\sharp
+\Omega\mbox{tr}\chi(\chih,\alb)\right\}\nonumber\\
&&\hspace{35mm}-\Omega\omega\left\{\sdiv\beb+(2\eta-\zeta,\beb)+\frac{1}{2}(\chih,\alb)\right\}\nonumber\\
&&D\Db\sigma+\frac{3}{2}\Omega\mbox{tr}\chib D\sigma+\frac{3}{2}D(\Omega\mbox{tr}\chib)\sigma\nonumber\\
&&\hspace{20mm}=-\Omega\left\{\scurl D\beb+(2D\eta-D\zeta,\s^*\beb)+(2\eta-\zeta,\s^*D\beb)\right.\nonumber\\
&&\hspace{45mm}\left.+\frac{1}{2}\Dh\chih\wedge\alb+\frac{1}{2}\chih\wedge\Dh\alb\right\}\nonumber\\
&&\hspace{35mm}+\Omega(\Omega\mbox{tr}\chi)\left\{\scurl\beb+(2\eta-\zeta,\s^*\beb)+\chih\wedge\alb\right\}\nonumber\\
&&\hspace{35mm}-\Omega\omega\left\{\scurl\beb+(2\eta-\zeta,\s^*\beb)+\frac{1}{2}\chih\wedge\alb\right\}
\label{13.83}
\end{eqnarray}
To derive the above formulas we have made use of the following
eight facts. First, the commutation formula \ref{6.107}. Second,
the fact that for any pair of trace-free symmetric 2-covariant $S$
tensorfields $\theta, \theta^\prime$ we have:
\begin{equation}
D(\theta,\theta^\prime)=(\Dh\theta,\theta^\prime)+(\theta,\Dh\theta^\prime)-2\Omega\mbox{tr}\chi(\theta,\theta^\prime)
\label{13.84}
\end{equation}
This is obtained using the identity \ref{1.163} which implies that
for any triplet of trace-free symmetric 2-covariant $S$
tensorfields $\theta, \theta^\prime, \theta^{\prime\prime}$ we
have:
\begin{equation}
(\theta,\theta^\prime\times\theta^{\prime\prime})=0 \label{13.85}
\end{equation}
Third, the fact, which follows from \ref{12.208}, that for any $S$
1-form $\xi$ we have:
\begin{equation}
D^*\xi=\s^*D\xi-2\Omega\s^*\chih^\sharp\cdot\xi \label{13.86}
\end{equation}
Fourth, the fact that if $\theta$ is a trace-free symmetric
2-covariant $S$ tensorfield and $\xi$ an $S$ 1-form, we have:
\begin{equation}
\s^*\theta^\sharp\cdot\xi+\theta^\sharp\cdot\s^*\xi=0
\label{13.87}
\end{equation}
This follows from the symmetry of $\s^*\theta$. Remarking that if
$\xi$ is an arbitrary $S$ 1-form we have:
\begin{equation}
\scurl\xi=\sdiv\s^*\xi \label{13.88}
\end{equation}
and using the formula \ref{6.107} as well as the third and fourth
facts just mentioned, we deduce the commutation formula:
\begin{equation}
D\scurl\xi-\scurl D\xi=-\Omega\mbox{tr}\chi\scurl\xi \label{13.89}
\end{equation}
for an arbitrary $S$ 1-form $\xi$. This is the fifth fact. Using
the third and fourth facts we deduce that for any pair of $S$
1-forms $\xi, \xi^\prime$ we have:
\begin{equation}
D(\xi,\xi^\prime)=(D\xi,\s^*\xi^\prime)+(\xi,\s^*D\xi^\prime)-\Omega\mbox{tr}\chi(\xi,\xi^\prime)
\label{13.90}
\end{equation}
This is the sixth fact. Finally, remarking that for any pair of
trace-free symmetric 2-covariant $S$ tensorfields $\theta,
\theta^\prime$ we have:
\begin{equation}
\theta\wedge\theta^\prime=(\theta,\s^*\theta^\prime) \label{13.91}
\end{equation}
and using the second of the above facts as well as the commutation
formula \ref{12.207}, which we think of as the seventh fact, we
deduce:
\begin{equation}
D(\theta\wedge\theta^\prime)=\Dh\theta\wedge\theta^\prime+\theta\wedge\Dh\theta^\prime-2\Omega\mbox{tr}\chi\theta\wedge\theta^\prime
\label{13.92}
\end{equation}
This is the eighth fact.

We express in equations \ref{13.83}, $D(\Omega\mbox{tr}\chib)$ by
\ref{4.c1}, $D\eta$ by \ref{1.66}, $D\zeta$ by \ref{1.61}, that
is:
\begin{equation}
D\zeta=-\sd\omega+\Omega(\chi^\sharp\cdot\etb-\beta) \label{13.93}
\end{equation}
 and $\Dh\chih$ by \ref{3.8}. We also remark that if we use, in place of the bootstrap assumptions, the results of Chapters 3 and 4
(which improve the former by a factor of at least $\delta^{1/4}$)
to derive from the Bianchi identity \ref{12.168} an estimate for
$\|\Dh\alb\|_{L^4(S_{\ub,u})}$, we obtain, in place of
\ref{12.174}, the estimate:
\begin{equation}
\|\Dh\alb\|_{L^4(S_{\ub,u})}\leq O(\delta^{1/2}|u|^{-4})
\label{13.94}
\end{equation}
Then the right hand sides of \ref{13.83} are seen to be bounded in
$L^2(C_u)$ by $O(\delta^{1/2}|u|^{-4}$ and we obtain:
\begin{eqnarray}
&&\|u\ub D\Db\rho\|_{L^2(C_u)}\leq 3\|\ub D\rho\|_{L^2(C_u)}+O(\delta^{3/2}|u|^{-3})\nonumber\\
&&\hspace{25mm}\leq 3\delta^{1/2}|u|^{-2}{\cal R}_{[1]}(\rho)
+O(\delta^{3/2}|u|^{-3})\nonumber\\
&&\hspace{25mm}\leq \delta^{1/2}|u|^{-2}\left\{C(\stackrel{(2)}{{\cal E}}_1)^{1/2}+O(\delta)\right\}\nonumber\\
&&\|u\ub D\Db\sigma\|_{L^2(C_u)}\leq 3\|\ub D\sigma\|_{L^2(C_u)}+O(\delta^{3/2}|u|^{-3})\nonumber\\
&&\hspace{25mm}\leq 3\delta^{1/2}|u|^{-2}{\cal R}_{[1]}(\sigma)
+O(\delta^{3/2}|u|^{-3})\nonumber\\
&&\hspace{25mm}\leq
\delta^{1/2}|u|^{-2}\left\{C(\stackrel{(2)}{{\cal
E}}_1)^{1/2}+O(\delta)\right\} \label{13.95}
\end{eqnarray}
by \ref{12.a3}.

Applying $\Db$ to the Bianchi identities \ref{13.51} and using the
conjugates of the eight facts mentioned above we obtain:
\begin{eqnarray}
&&\Db^2\rho+\frac{3}{2}\Omega\mbox{tr}\chib\Db\rho+\frac{3}{2}\Db(\Omega\mbox{tr}\chib)\rho\nonumber\\
&&\hspace{20mm}=-\Omega\left\{\sdiv\Db\beb+(2\Db\eta-\Db\zeta,\beb)+(2\eta-\zeta,\Db\beb)\right.\nonumber\\
&&\hspace{45mm}\left.+\frac{1}{2}(\Dbh\chih,\alb)+\frac{1}{2}(\chih,\Dbh\alb)\right\}\nonumber\\
&&\hspace{35mm}+\Omega\left\{2\sdiv(\Omega\chibh^\sharp\cdot\beb)+\Omega\mbox{tr}\chib\sdiv\beb\right.\nonumber\\
&&\hspace{20mm}\left.+\Omega\mbox{tr}\chib(2\eta-\zeta,\beb)+2\Omega(2\eta-\zeta)^\sharp\cdot\chibh\cdot\beb^\sharp
+\Omega\mbox{tr}\chib(\chih,\alb)\right\}\nonumber\\
&&\hspace{35mm}-\Omega\omb\left\{\sdiv\beb+(2\eta-\zeta,\beb)+\frac{1}{2}(\chih,\alb)\right\}\nonumber\\
&&\Db^2\sigma+\frac{3}{2}\Omega\mbox{tr}\chib\Db\sigma+\frac{3}{2}\Db(\Omega\mbox{tr}\chib)\sigma\nonumber\\
&&\hspace{20mm}=-\Omega\left\{\scurl\Db\beb+(2\Db\eta-\Db\zeta,\s^*\beb)+(2\eta-\zeta,\s^*\Db\beb)\right.\nonumber\\
&&\hspace{45mm}\left.+\frac{1}{2}\Dbh\chih\wedge\alb+\frac{1}{2}\chih\wedge\Dbh\alb\right\}\nonumber\\
&&\hspace{35mm}+\Omega(\Omega\mbox{tr}\chib)\left\{\scurl\beb+(2\eta-\zeta,\s^*\beb)+\chih\wedge\alb\right\}\nonumber\\
&&\hspace{35mm}-\Omega\omb\left\{\scurl\beb+(2\eta-\zeta,\s^*\beb)+\frac{1}{2}\chih\wedge\alb\right\}
\label{13.96}
\end{eqnarray}
We express in these equations $\Db(\Omega\mbox{tr}\chib)$ by
\ref{3.7}, $\Db\eta$ by \ref{1.149}, $\Db\zeta$ by \ref{1.63},
that is:
\begin{equation}
-\Db\zeta=-\sd\omb+\Omega(\chib^\sharp\cdot\eta+\beb)
\label{13.97}
\end{equation}
and $\Dbh\chih$ by \ref{4.c4}. Then, by the results of Chapters 3
and 4 and the estimate \ref{12.190}, the right hand sides of
\ref{13.96} are seen to be bounded in $L^2(C_u)$ by
$O(\delta^{3/2}|u|^{-5})$ and we obtain:
\begin{eqnarray}
&&\|u^2\Db^2\rho\|_{L^2(C_u)}\leq 3\|u\Db\rho\|_{L^2(C_u)}+3\|\rho\|_{L^2(C_u)}+O(\delta^{3/2}|u|^{-3})\nonumber\\
&&\hspace{25mm}\leq C\delta^{1/2}|u|^{-2}{\cal R}_{[1]}(\rho)+O(\delta^{3/2}|u|^{-3})\nonumber\\
&&\hspace{25mm}\leq \delta^{1/2}|u|^{-2}\left\{C(\stackrel{(2)}{{\cal E}}_1)^{1/2}+O(\delta)\right\}\nonumber\\
&&\|u^2\Db^2\sigma\|_{L^2(C_u)}\leq 3\|u\Db\sigma\|_{L^2(C_u)}+3\|\sigma\|_{L^2(C_u)}+O(\delta^{3/2}|u|^{-3})\nonumber\\
&&\hspace{25mm}\leq C\delta^{1/2}|u|^{-2}{\cal R}_{[1]}(\sigma)+O(\delta^{3/2}|u|^{-3})\nonumber\\
&&\hspace{25mm}\leq
\delta^{1/2}|u|^{-2}\left\{C(\stackrel{(2)}{{\cal
E}}_1)^{1/2}+O(\delta)\right\} \label{13.98}
\end{eqnarray}
by \ref{13.52}.

Combining \ref{13.82}, \ref{13.95}, \ref{13.98} and \ref{13.54},
and noting that by the commutation formula \ref{1.75}:
\begin{equation}
\|u\ub(\Db
D-D\Db)(\rho,\sigma)\|_{L^2(C_u)}=\|4u\ub\Omega^2\zeta^\sharp\cdot\sd(\rho,\sigma)\|_{L^2(C_u)}\leq
O(\delta^2|u|^{-4}) \label{13.99}
\end{equation}
we conclude through \ref{13.81} that:
\begin{eqnarray}
&&\|S^2\rho\|_{L^2(C_u)}\leq\delta^{1/2}|u|^{-2}\left\{C(\stackrel{(2)}{{\cal E}}_2)^{1/2}+O(\delta^{1/2})\right\}\nonumber\\
&&\|S^2\sigma\|_{L^2(C_u)}\leq\delta^{1/2}|u|^{-2}\left\{C\stackrel{(2)}{{\cal
E}}_2)^{1/2}+O(\delta^{1/2})\right\} \label{13.100}
\end{eqnarray}
Combining this in turn with \ref{13.80}, \ref{13.77}, \ref{13.55}
and \ref{13.55}, we conclude that:
\begin{equation}
\|(\rho,\sigma)(\tcL_S\tcL_S R)\|_{L^2(C_u)}\leq
C\delta^{1/2}|u|^{-2}(\stackrel{(2)}{{\cal
E}}_2)^{1/2}+O(\delta|u|^{-2}) \label{13.101}
\end{equation}
We have thus established \ref{13.43} for all three Weyl fields
involving the commutation field $S$.

We summarize the results of this chapter in the following
proposition.

\vspace{5mm}

\noindent{\bf Proposition 13.1} \ \ \ The multiplier error
integrals \ref{13.25} satisfy the estimates:
\begin{eqnarray*}
&&\delta^{2q_0+2l}\int_{M^\prime_{c^*}}|\stackrel{(0)}{\tau}_m|d\mu_g\leq O(\delta)\\
&&\delta^{2q_1+2l}\int_{M^\prime_{c^*}}|\stackrel{(1)}{\tau}_m|d\mu_g\leq
C{\cal D}_0^\infty(\chibh)(\stackrel{(0)}{{\cal E}}_2)^{1/2}(\stackrel{(2)}{{\cal E}}_2)^{1/2}+O(\delta)\\
&&\delta^{2q_2+2l}\int_{M^\prime_{c^*}}|\stackrel{(2)}{\tau}_m|d\mu_g\leq O(\delta)\\
&&\delta^{2q_3+2l}\int_{M^\prime_{c^*}}|\stackrel{(3)}{\tau}_m|d\mu_q\leq
C{\cal D}_0^\infty(\chibh)(\stackrel{(2)}{{\cal
E}}_2)^{1/2}(\stackrel{(3)}{{\cal F}}_2)^{1/2}+O(\delta^{1/2})
\end{eqnarray*}

\chapter{The 1st Order Weyl Current Error Estimates}

\section{Introduction}

In the present chapter and the next we shall estimate the
contributions of the {\em Weyl current error terms}:
\begin{eqnarray}
&\stackrel{(n)}{\tau}_c \ : \ n=0,1,2,3\label{14.1}\\
&\s^{(X)}\stackrel{(n)}{\tau}_c \ : (X)=(L),(O),(S); \ n=0,1,2,3\nonumber\\
&\s^{(YX)}\stackrel{(n)}{\tau}_c \ :
(YX)=(LL),(OL),(OO),(OS),(SS); \ n=0,1,2,3\nonumber
\end{eqnarray}
the Weyl current error term $\tau_c(W;X,Y,Z)$ associated to a Weyl
field $W$ and the multiplier fields $X,Y,Z$ being given by
\ref{13.22}:
\begin{equation}
\tau_c(W;X,Y,Z)=-(\mbox{div}Q(W))(X,Y,Z) \label{14.2}
\end{equation}
This is generated by the Weyl current $J$ corresponding to $W$,
for, according to Proposition 12.6 we have:
\begin{equation}
(\mbox{div}Q(W))_{\beta\gamma\delta}=W^{\s\mu\s\nu}_{\beta\s\delta}J_{\mu\gamma\nu}+W^{\s\mu\s\nu}_{\beta\s\gamma}J_{\mu\delta\nu}
+\s^*W^{\s\mu\s\nu}_{\beta\s\delta}J^*_{\mu\gamma\nu}+\s^*W^{\s\mu\s\nu}_{\beta\s\gamma}J^*_{\mu\delta\nu}
\label{14.3}
\end{equation}
Now the fundamental Weyl field $W=R$ satisfies the homogeneous
Bianchi equations \ref{12.15}. Thus, the corresponding Weyl
current vanishes, hence:
\begin{equation}
\stackrel{(n)}{\tau}_c=0 \ : \ n=0,1,2,3 \label{14.4}
\end{equation}
Only to the derived Weyl fields there correspond non-trivial Weyl
currents. These are given by Proposition 12.1. Given a Weyl field
$W$ and a commutation field $X$ let us define the Weyl currents
$\s^{(X)}J^1(W)$, $\s^{(X)}J^2(W)$, $\s^{(X)}J^3(W)$ by:
\begin{eqnarray}
&&\s^{(X)}J^1(W)_{\beta\gamma\delta}=\frac{1}{2}\s^{(X)}\tilde{\pi}^{\mu\nu}\nabla_\nu W_{\mu\beta\gamma\delta}\label{14.5}\\
&&\s^{(X)}J^2(W)_{\beta\gamma\delta}=\frac{1}{2}\s^{(X)}p_\mu W^\mu_{\s\beta\gamma\delta}\nonumber\\
&&\s^{(X)}J^3(W)_{\beta\gamma\delta}=\frac{1}{2}(\s^{(X)}q_{\mu\beta\nu}W^{\mu\nu}_{\s\s\gamma\delta}+\s^{(X)}q_{\mu\gamma\nu}W^{\mu\s\nu}_{\s\beta\delta}
+\s^{(X)}q_{\mu\delta\nu}W^{\mu\s\s\nu}_{\s\beta\gamma})\nonumber
\end{eqnarray}
Here:
\begin{equation}
\s^{(X)}p_\beta=\nabla^\alpha\s^{(X)}\tilde{\pi}_{\alpha\beta}
\label{14.6}
\end{equation}
and:
\begin{equation}
\s^{(X)}q_{\alpha\beta\gamma}=\nabla_\beta\s^{(X)}\tilde{\pi}_{\gamma\alpha}-\nabla_\gamma\s^{(X)}\tilde{\pi}_{\beta\alpha}
+\frac{1}{3}(\s^{(X)}p_\beta g_{\alpha\gamma}-\s^{(X)}p_\gamma
g_{\alpha\beta}) \label{14.7}
\end{equation}
the 3-covariant tensorfield $\s^{(X)}q$ having the algebraic
properties of a Weyl current.

Let us denote by $\s^{(X)}J$ the Weyl current corresponding to the
1st order derived Weyl field $\tcL_X R$. Then according to
Proposition 12.1 $\s^{(X)}J$ is given by:
\begin{equation}
\s^{(X)}J=\s^{(X)}J^1(R)+\s^{(X)}J^2(R)+\s^{(X)}J^3(R)
\label{14.8}
\end{equation}
The 1st order Weyl current error terms are:
\begin{eqnarray}
&&\s^{(L)}\stackrel{(0)}{\tau}_c=-(\mbox{div}Q(\tcL_L R))(L,L,L)\nonumber\\
&&\s^{(L)}\stackrel{(1)}{\tau}_c=-(\mbox{div}Q(\tcL_L R))(K,L,L)\nonumber\\
&&\s^{(L)}\stackrel{(2)}{\tau}_c=-(\mbox{div}Q(\tcL_L R))(K,K,L)\nonumber\\
&&\s^{(L)}\stackrel{(3)}{\tau}_c=-(\mbox{div}Q(\tcL_L R))(K,K,K)
\label{14.9}
\end{eqnarray}
\begin{eqnarray}
&&\s^{(O)}\stackrel{(0)}{\tau}_c=-\sum_i(\mbox{div}Q(\tcL_{O_i}R))(L,L,L)\nonumber\\
&&\s^{(O)}\stackrel{(1)}{\tau}_c=-\sum_i(\mbox{div}Q(\tcL_{O_i}R))(K,L,L)\nonumber\\
&&\s^{(O)}\stackrel{(2)}{\tau}_c=-\sum_i(\mbox{div}Q(\tcL_{O_i}R))(K,K,L)\nonumber\\
&&\s^{(O)}\stackrel{(3)}{\tau}_c=-\sum_i(\mbox{div}Q(\tcL_{O_i}R))(K,K,K)
\label{14.10}
\end{eqnarray}
and:
\begin{equation}
\s^{(S)}\stackrel{(3)}{\tau}_c=-(\mbox{div}Q(\tcL_S R))(K,K,K)
\label{14.11}
\end{equation}
By Lemma 12.3 the above are given by:
\begin{eqnarray}
&&\s^{(L)}\stackrel{(0)}{\tau}_c=-4\Omega^3\left\{(\Theta(\s^{(L)}J),\alpha(\tcL_L R))-2(\Xi(\s^{(L)}J),\beta(\tcL_L R))\right\}\nonumber\\
&&\s^{(L)}\stackrel{(1)}{\tau}_c=-8\Omega^3|u|^2\left\{\Lambda(\s^{(L)}J)\rho(\tcL_L R)-K(\s^{(L)}J)\sigma(\tcL_L R)\right.\nonumber\\
&&\hspace{30mm}\left.+(I(\s^{(L)}J),\beta(\tcL_L R))\right\}\nonumber\\
&&\s^{(L)}\stackrel{(2)}{\tau}_c=-8\Omega^3|u|^4\left\{\Lambdab(\s^{(L)}J)\rho(\tcL_L R)-\Kb(\s^{(L)}J)\sigma(\tcL_L R)\right.\nonumber\\
&&\hspace{30mm}\left.-(\Ib(\s^{(L)}J),\beb(\tcL_L R))\right\}\nonumber\\
&&\s^{(L)}\stackrel{(3)}{\tau}_c=-4\Omega^3|u|^6\left\{(\Thetab(\s^{(L)}J),\alb(\tcL_L
R))+2(\Xib(\s^{(L)}J),\beb(\tcL_L R))\right\} \nonumber\\
&&\label{14.12}
\end{eqnarray}
\begin{eqnarray}
&&\s^{(O)}\stackrel{(0)}{\tau}_c=-4\sum_i\Omega^3\left\{(\Theta(\s^{(O_i)}J),\alpha(\tcL_{O_i}R))-2(\Xi(\s^{(O_i)}J),\beta(\tcL_{O_i}R))\right\}\nonumber\\
&&\s^{(O)}\stackrel{(1)}{\tau}_c=-8\sum_i\Omega^3|u|^2\left\{\Lambda(\s^{(O_i)}J)\rho(\tcL_{O_i}R)-K(\s^{(O_i)}J)\sigma(\tcL_{O_i}R)\right.\nonumber\\
&&\hspace{30mm}\left.+(I(\s^{(O_i)}J,\beta(\tcL_{O_i}R))\right\}\nonumber\\
&&\s^{(O)}\stackrel{(2)}{\tau}_c=-8\sum_i\Omega^3|u|^4\left\{\Lambdab(\s^{(O_i)}J)\rho(\tcL_{O_i}R)-\Kb(\s^{(O_i)}J)\sigma(\tcL_{O_i}R)\right.\nonumber\\
&&\hspace{30mm}\left.-(\Ib(\s^{(O_i)}J),\beb(\tcL_{O_i}R))\right\}\nonumber\\
&&\s^{(O)}\stackrel{(3)}{\tau}_c=-4\sum_i\Omega^3|u|^6\left\{(\Thetab(\s^{(O_i)}J),\alb(\tcL_{O_i}R))+2(\Xib(\s^{(O_i)}J),\beb(\tcL_{O_i}R))\right\}\nonumber\\
&&\label{14.13}
\end{eqnarray}
and:
\begin{equation}
\s^{(S)}\stackrel{(3)}{\tau}_c=-4\Omega^3|u|^6\left\{(\Thetab(\s^{(S)}J),\alb(\tcL_S
R))+2(\Xib(\s^{(S)}J),\beb(\tcL_S R))\right\} \label{14.14}
\end{equation}

Let us denote by $\s^{(YX)}J$ the Weyl current corresponding to
the 2nd order derived Weyl field $\tcL_Y\tcL_X R$. Then according
to Proposition 12.1 $\s^{(YX)}J$ is given by:
\begin{equation}
\s^{(YX)}J=\tcL_Y \s^{(X)}J+\s^{(Y)}J^1(\tcL_X
R)+\s^{(Y)}J^2(\tcL_X R)+\s^{(Y)}J^3(\tcL_X R) \label{14.15}
\end{equation}
The 2nd order Weyl current error terms are:
\begin{eqnarray}
&&\s^{(LL)}\stackrel{(0)}{\tau}_c=-(\mbox{div}Q(\tcL_L\tcL_L R))(L,L,L)\nonumber\\
&&\s^{(LL)}\stackrel{(1)}{\tau}_c=-(\mbox{div}Q(\tcL_L\tcL_L R))(K,L,L)\nonumber\\
&&\s^{(LL)}\stackrel{(2)}{\tau}_c=-(\mbox{div}Q(\tcL_L\tcL_L R))(K,K,L)\nonumber\\
&&\s^{(LL)}\stackrel{(3)}{\tau}_c=-(\mbox{div}Q(\tcL_L\tcL_L
R))(K,K,K) \label{14.16}
\end{eqnarray}
\begin{eqnarray}
&&\s^{(OL)}\stackrel{(0)}{\tau}_c=-\sum_i(\mbox{div}Q(\tcL_{O_i}\tcL_L R))(L,L,L)\nonumber\\
&&\s^{(OL)}\stackrel{(1)}{\tau}_c=-\sum_i(\mbox{div}Q(\tcL_{O_i}\tcL_L R))(K,L,L)\nonumber\\
&&\s^{(OL)}\stackrel{(2)}{\tau}_c=-\sum_i(\mbox{div}Q(\tcL_{O_i}\tcL_L R))(K,K,L)\nonumber\\
&&\s^{(OL)}\stackrel{(3)}{\tau}_c=-\sum_i(\mbox{div}Q(\tcL_{O_i}\tcL_L
R))(K,K,K) \label{14.17}
\end{eqnarray}
\begin{eqnarray}
&&\s^{(OO)}\stackrel{(0)}{\tau}_c=-\sum_{i,j}(\mbox{div}Q(\tcL_{O_j}\tcL_{O_i}R))(L,L,L)\nonumber\\
&&\s^{(OO)}\stackrel{(1)}{\tau}_c=-\sum_{i,j}(\mbox{div}Q(\tcL_{O_j}\tcL_{O_i}R))(K,L,L)\nonumber\\
&&\s^{(OO)}\stackrel{(2)}{\tau}_c=-\sum_{i,j}(\mbox{div}Q(\tcL_{O_j}\tcL_{O_i}R))(K,K,L)\nonumber\\
&&\s^{(OO)}\stackrel{(3)}{\tau}_c=-\sum_{i,j}(\mbox{div}Q(\tcL_{O_j}\tcL_{O_i}R))(K,K,K)
\label{14.18}
\end{eqnarray}
and:
\begin{equation}
\s^{(OS)}\stackrel{(3)}{\tau}_c=-\sum_i(\mbox{div}Q(\tcL_{O_i}\tcL_S
R))(K,K,K) \label{14.19}
\end{equation}
\begin{equation}
\s^{(SS)}\stackrel{(3)}{\tau}_c=-(\mbox{div}Q(\tcL_S\tcL_S
R))(K,K,K) \label{14.20}
\end{equation}
By Lemma 12.3 the above are given by:
\begin{eqnarray}
&&\s^{(LL)}\stackrel{(0)}{\tau}_c=-4\Omega^3\left\{(\Theta(\s^{(LL)}J),\alpha(\tcL_L\tcL_L R))-2(\Xi(\s^{(LL)}J),\beta(\tcL_L\tcL_L R))\right\}\nonumber\\
&&\s^{(LL)}\stackrel{(1)}{\tau}_c=-8\Omega^3|u|^2\left\{\Lambda(\s^{(LL)}J)\rho(\tcL_L\tcL_L R)-K(\s^{(LL)}J)\sigma(\tcL_L\tcL_L R)\right.\nonumber\\
&&\hspace{35mm}\left.+(I(\s^{(LL)}J),\beta(\tcL_L\tcL_L R))\right\}\nonumber\\
&&\s^{(LL)}\stackrel{(2)}{\tau}_c=-8\Omega^3|u|^4\left\{\Lambdab(\s^{(LL)}J)\rho(\tcL_L\tcL_L R)-\Kb(\s^{(LL)}J)\sigma(\tcL_L\tcL_L R)\right.\nonumber\\
&&\hspace{35mm}\left.-(\Ib(\s^{(LL)}J),\beb(\tcL_L\tcL_L R))\right\}\nonumber\\
&&\s^{(LL)}\stackrel{(3)}{\tau}_c=-4\Omega^3|u|^6\left\{(\Thetab(\s^{(LL)}J),\alb(\tcL_L\tcL_L R))+2(\Xib(\s^{(LL)}J),\beb(\tcL_L\tcL_L R))\right\}\nonumber\\
&&\label{14.21}
\end{eqnarray}
\begin{eqnarray}
&&\s^{(OL)}\stackrel{(0)}{\tau}_c=-4\sum_i\Omega^3\left\{(\Theta(\s^{(O_i L)}J),\alpha(\tcL_{O_i}\tcL_L R))\right.\nonumber\\
&&\hspace{40mm}\left.-2(\Xi(\s^{(O_i L)}J),\beta(\tcL_{O_i}\tcL_L R))\right\}\nonumber\\
&&\s^{(OL)}\stackrel{(1)}{\tau}_c=-8\sum_i\Omega^3|u|^2\left\{\Lambda(\s^{(O_i L)}J)\rho(\tcL_{O_i}\tcL_L R)-K(\s^{(O_i L)}J)\sigma(\tcL_{O_i}\tcL_L R)\right.\nonumber\\
&&\hspace{40mm}\left.+(I(\s^{(O_i L)}J,\beta(\tcL_{O_i}\tcL_L R))\right\}\nonumber\\
&&\s^{(OL)}\stackrel{(2)}{\tau}_c=-8\sum_i\Omega^3|u|^4\left\{\Lambdab(\s^{(O_i L)}J)\rho(\tcL_{O_i}\tcL_L R)-\Kb(\s^{(O_i L)}J)\sigma(\tcL_{O_i}\tcL_L R)\right.\nonumber\\
&&\hspace{40mm}\left.-(\Ib(\s^{(O_i L)}J),\beb(\tcL_{O_i}\tcL_L R))\right\}\nonumber\\
&&\s^{(OL)}\stackrel{(3)}{\tau}_c=-4\sum_i\Omega^3|u|^6\left\{(\Thetab(\s^{(O_i L)}J),\alb(\tcL_{O_i}\tcL_L R))\right.\nonumber\\
&&\hspace{40mm}\left.+2(\Xib(\s^{(O_i L)}J),\beb(\tcL_{O_i}\tcL_L
R))\right\} \label{14.22}
\end{eqnarray}
\begin{eqnarray}
&&\s^{(OO)}\stackrel{(0)}{\tau}_c=-4\sum_{i,j}\Omega^3\left\{(\Theta(\s^{(O_j O_i)}J),\alpha(\tcL_{O_j}\tcL_{O_i}R))\right.\nonumber\\
&&\hspace{40mm}\left.-2(\Xi(\s^{(O_j O_i)}J),\beta(\tcL_{O_j}\tcL_{O_i}R))\right\}\nonumber\\
&&\s^{(OO)}\stackrel{(1)}{\tau}_c=-8\sum_{i,j}\Omega^3|u|^2\left\{\Lambda(\s^{(O_j O_i)}J)\rho(\tcL_{O_j}\tcL_{O_i}R)-K(\s^{(O_j O_i)}J)\sigma(\tcL_{O_j}\tcL_{O_i}R)\right.\nonumber\\
&&\hspace{40mm}\left.+(I(\s^{(O_j O_i)}J,\beta(\tcL_{O_j}\tcL_{O_i}R))\right\}\nonumber\\
&&\s^{(OO)}\stackrel{(2)}{\tau}_c=-8\sum_{i,j}\Omega^3|u|^4\left\{\Lambdab(\s^{(O_j O_i)}J)\rho(\tcL_{O_j}\tcL_{O_i}R)-\Kb(\s^{(O_j O_i)}J)\sigma(\tcL_{O_j}\tcL_{O_i}R)\right.\nonumber\\
&&\hspace{40mm}\left.-(\Ib(\s^{(O_j O_i)}J),\beb(\tcL_{O_j}\tcL_{O_i}R))\right\}\nonumber\\
&&\s^{(OO)}\stackrel{(3)}{\tau}_c=-4\sum_{i,j}\Omega^3|u|^6\left\{(\Thetab(\s^{(O_j O_i)}J),\alb(\tcL_{O_j}\tcL_{O_i} R))\right.\nonumber\\
&&\hspace{40mm}\left.+2(\Xib(\s^{(O_j
O_i)}J),\beb(\tcL_{O_j}\tcL_{O_i}R))\right\} \label{14.23}
\end{eqnarray}
and:
\begin{eqnarray}
&&\s^{(OS)}\stackrel{(3)}{\tau}_c=-4\sum_i\Omega^3|u|^6\left\{(\Thetab(\s^{(O_i S)}J),\alb(\tcL_{O_i}\tcL_S R))\right.\nonumber\\
&&\hspace{40mm}\left.+2(\Xib(\s^{(O_i S)}J),\beb(\tcL_{O_i}\tcL_S
R))\right\} \label{14.24}
\end{eqnarray}
\begin{equation}
\s^{(SS)}\stackrel{(3)}{\tau}_c=-4\Omega^3|u|^6\left\{(\Thetab(\s^{(SS)}J),\alb(\tcL_S\tcL_S
R)) +2(\Xib(s^{(SS)}J),\beb(\tcL_S\tcL_S R))\right\} \label{14.25}
\end{equation}

The aim of the next two chapters is to obtain bounds for the {\em
Weyl current error integrals}:
\begin{eqnarray}
&&\mbox{1st order:} \ \delta^{2q_n+2l}\int_{M^\prime_{c^*}}|\s^{(X)}\stackrel{(n)}{\tau}_c|d\mu_g\label{14.26}\\
&&\hspace{35mm} : \  \mbox{for} \ (X)=(L),(O),(S)\nonumber\\
&&\mbox{2nd order:} \ \delta^{2q_n+2l}\int_{M^\prime_{c^*}}|\s^{(YX)}\stackrel{(n)}{\tau}_c|d\mu_g\nonumber\\
&&\hspace{35mm} : \ \mbox{for} \
(YX)=(LL),(OL),(OO),(OS),(SS)\nonumber
\end{eqnarray}
In the present chapter we shall estimate the 1st order Weyl
current error integrals. The 2nd order Weyl current error
integrals shall be estimated in the next chapter.

\section{The error estimates arising from $J^1$}

We begin with the following proposition which is deduced in a
straightforward manner using the table \ref{1.151} of connection
coefficients of the frame $(e_\mu \ : \ \mu=1,2,3,4)$ and noting
the basic identities \ref{1.163} and \ref{13.85}.

\noindent{\bf Lemma 14.1} \ \ \ The components of the Weyl current
$\s^{(X)}J^1(W)$ associated to the commutation field $X$ and to
the Weyl field $W$ are given by:
\begin{eqnarray*}
&&4\Xi_A(\s^{(X)}J^1(W))=-\frac{1}{2}\s^{(X)}j\left\{\Omega^{-1}(D\beta(W))_A+(\sdiv\alpha(W))_A
-\chi_A^{\s B}\beta_B(W)\right.\\
&&\hspace{25mm}\left.-\mbox{tr}\chi\beta_A(W)-\Omega^{-1}\omega\beta_A(W)+(2\zeta^B-\etb^B)\alpha_{AB}(W)\right\}\\
&&\hspace{25mm}+\frac{1}{2}\s^{(X)}m^B\left\{2\snab_B\beta_A(W)+\Omega^{-1}(\Dbh\alpha(W))_{AB}-\chib_B^{\s C}\alpha_{AC}(W)\right.\\
&&\hspace{25mm}-\mbox{tr}\chib\alpha_{AB}(W)+2\Omega^{-1}\omb\alpha_{AB}(W)+2((\zeta-2\eta)\oth\beta(W))_{AB}\\
&&\hspace{40mm}\left.-3(\chi_{AB}\rho(W)+\s^*\chi_{AB}\sigma(W))\right\}\\
&&\hspace{25mm}+\frac{1}{2}\s^{(X)}\mb^B\left\{\Omega^{-1}(\Dh\alpha(W))_{AB}-\mbox{tr}\chi\alpha_{AB}(W)\right.\\
&&\hspace{40mm}\left.-2\Omega^{-1}\omega\alpha_{AB}(W)\right\}\\
&&\hspace{25mm}-\s^{(X)}\ih^{BC}\left\{\snab_C\alpha_{AB}(W)-(\chi\oth\beta(W))_{CAB}+2\zeta_C\alpha_{AB}(W)\right\}\\
&&4\Xib_A(\s^{(X)}J^1(W))=-\frac{1}{2}\s^{(X)}j\left\{-\Omega^{-1}(\Db\beb(W))_A+(\sdiv\alb(W))_A
+\chib_A^{\s B}\beb_B(W)\right.\\
&&\hspace{25mm}\left.+\mbox{tr}\chib\beb_A(W)+\Omega^{-1}\omb\beb_A(W)-(2\zeta^B+\eta^B)\alb_{AB}(W)\right\}\\
&&\hspace{25mm}+\frac{1}{2}\s^{(X)}\mb^B\left\{-2\snab_B\beb_A(W)+\Omega^{-1}(\Dh\alb(W))_{AB}-\chi_B^{\s C}\alb_{AC}(W)\right.\\
&&\hspace{25mm}-\mbox{tr}\chi\alb_{AB}(W)+2\Omega^{-1}\omega\alb_{AB}(W)+2((\zeta+2\etb)\oth\beb(W))_{AB}\\
&&\hspace{40mm}\left.-3(\chib_{AB}\rho(W)-\s^*\chib_{AB}\sigma(W))\right\}\\
&&\hspace{25mm}+\frac{1}{2}\s^{(X)}m^B\left\{\Omega^{-1}(\Dbh\alb(W))_{AB}-\mbox{tr}\chib\alb_{AB}(W)\right.\\
&&\hspace{40mm}\left.-2\Omega^{-1}\omb\alb_{AB}(W)\right\}\\
&&\hspace{25mm}-\s^{(X)}\ih^{BC}\left\{\snab_C\alb_{AB}(W)+(\chib\oth\beb(W))_{CAB}-2\zeta_C\alb_{AB}(W)\right\}\\
&&4\Theta_{AB}(\s^{(X)}J^1(W))=\frac{1}{2}\s^{(X)}j\left\{\Omega^{-1}(\Dbh\alpha(W))_{AB}+(\snab\oth\beta(W))_{AB}
-\frac{3}{2}\mbox{tr}\chib\alpha(W)\right.\\
&&\hspace{40mm}\left.+2\Omega^{-1}\omb\alpha_{AB}(W)+((\zeta-4\eta)\oth\beta(W))_{AB}\right.\\
&&\hspace{55mm}\left.-3(\chih_{AB}\rho(W)+\s^*\chih_{AB}\sigma(W))\right\}\\
&&\hspace{25mm}-\frac{1}{2}\left\{\s^{(X)}m\oth\left(\Omega^{-1}\Db\beta(W)-\chib^\sharp\cdot\beta(W)+\Omega^{-1}\omb\beta(W)\right)\right\}_{AB}\\
&&\hspace{25mm}+\frac{3}{2}\left(\s^{(X)}m\oth\eta\right)_{AB}\rho(W)+\frac{3}{2}\left(\s^{*(X)}m\oth\eta\right)_{AB}\sigma(W)\\
&&\hspace{25mm}-\frac{1}{2}\left\{\s^{(X)}\mb\oth\left(\Omega^{-1}D\beta(W)-\chih^\sharp\cdot\beta(W)-\Omega^{-1}\omega\beta(W)\right)\right\}_{AB}\\
&&\hspace{25mm}-\frac{1}{2}\left\{\eta\oth\left(\s^{(X)}\mb^\sharp\cdot\alpha(W)\right)\right\}_{AB}
+\frac{1}{2}\left(\s^{(X)}\mb,\eta\right)\alpha_{AB}(W)\\
&&\hspace{25mm}-\frac{1}{2}\s^{(X)}\mb^C\left\{\snab_C\alpha_{AB}(W)-(\chi\oth\beta(W))_{CAB}+2\zeta_C\alpha_{AB}(W)\right\}\\
&&\hspace{25mm}+\s^{(X)}\ih_A^{\s D}\snab_D\beta_B(W)+\s^{(X)}\ih_B^{\s D}\snab_D\beta_A(W)-\sg_{AB}\s^{(X)}\ih^{CD}\snab_D\beta_C(W)\\
&&\hspace{25mm}+\frac{1}{2}\s^{(X)}\ih^{CD}\left(\chibh_{CA}\alpha_{DB}(W)+\chibh_{CB}\alpha_{DA}(W)-\sg_{AB}\chibh_C^{\s
E}\alpha_{DE}(W)
\right)\\
&&\hspace{25mm}-(\s^{(X)}\ih,\chibh)\alpha_{AB}(W)+\frac{1}{2}\left\{(\s^{(X)}\ih^\sharp\cdot\zeta)\oth\beta(W)\right\}_{AB}\\
&&\hspace{25mm}-\frac{3}{2}\mbox{tr}\chi\left\{\s^{(X)}\ih_{AB}\rho(W)+\s^{*(X)}\ih_{AB}\sigma(W)\right\}\\
&&4\Thetab_{AB}(\s^{(X)}J^1(W))=\frac{1}{2}\s^{(X)}j\left\{\Omega^{-1}(\Dh\alb(W))_{AB}-(\snab\oth\beb(W))_{AB}
-\frac{3}{2}\mbox{tr}\chi\alb(W)\right.\\
&&\hspace{40mm}\left.+2\Omega^{-1}\omega\alb_{AB}(W)+((\zeta+4\etb)\oth\beb(W))_{AB}\right.\\
&&\hspace{55mm}\left.-3(\chibh_{AB}\rho(W)-\s^*\chibh_{AB}\sigma(W))\right\}\\
&&\hspace{25mm}-\frac{1}{2}\left\{\s^{(X)}\mb\oth\left(-\Omega^{-1}D\beb(W)+\chi^\sharp\cdot\beb(W)-\Omega^{-1}\omega\beb(W)\right)\right\}_{AB}\\
&&\hspace{25mm}+\frac{3}{2}\left(\s^{(X)}\mb\oth\etb\right)_{AB}\rho(W)-\frac{3}{2}\left(\s^{*(X)}\mb\oth\etb\right)_{AB}\sigma(W)\\
&&\hspace{25mm}-\frac{1}{2}\left\{\s^{(X)}m\oth\left(-\Omega^{-1}\Db\beb(W)+\chibh^\sharp\cdot\beb(W)+\Omega^{-1}\omb\beb(W)\right)\right\}_{AB}\\
&&\hspace{25mm}-\frac{1}{2}\left\{\etb\oth\left(\s^{(X)}m^\sharp\cdot\alb(W)\right)\right\}_{AB}
+\frac{1}{2}\left(\s^{(X)}m,\etb\right)\alb_{AB}(W)\\
&&\hspace{25mm}-\frac{1}{2}\s^{(X)}m^C\left\{\snab_C\alb_{AB}(W)+(\chib\oth\beb(W))_{CAB}-2\zeta_C\alb_{AB}(W)\right\}\\
&&\hspace{25mm}-\s^{(X)}\ih_A^{\s D}\snab_D\beb_B(W)-\s^{(X)}\ih_B^{\s D}\snab_D\beb_A(W)+\sg_{AB}\s^{(X)}\ih^{CD}\snab_D\beb_C(W)\\
&&\hspace{25mm}+\frac{1}{2}\s^{(X)}\ih^{CD}\left(\chih_{CA}\alb_{DB}(W)+\chih_{CB}\alb_{DA}(W)-\sg_{AB}\chih_C^{\s
E}\alb_{DE}(W)
\right)\\
&&\hspace{25mm}-(\s^{(X)}\ih,\chih)\alb_{AB}(W)+\frac{1}{2}\left\{(\s^{(X)}\ih^\sharp\cdot\zeta)\oth\beb(W)\right\}_{AB}\\
&&\hspace{25mm}-\frac{3}{2}\mbox{tr}\chib\left\{\s^{(X)}\ih_{AB}\rho(W)-\s^{*(X)}\ih_{AB}\sigma(W)\right\}\\
&&4\Lambda(\s^{(X)}J^1(W))=\frac{1}{2}\s^{(X)}j\left\{\Omega^{-1}D\rho(W)+\sdiv\beta(W)-\frac{3}{2}\mbox{tr}\chi\rho(W)\right.\\
&&\hspace{40mm}\left.+(\zeta-2\etb,\beta(W))-\frac{1}{2}(\chibh,\alpha)\right\}\\
&&\hspace{25mm}-\frac{1}{2}\left(\s^{(X)}m,\Omega^{-1}\Db\beta(W)-\chib^\sharp\cdot\beta(W)+\Omega^{-1}\omb\beta(W)\right)\\
&&\hspace{40mm}+\frac{3}{2}\left(\s^{(X)}m,\eta\rho(W)+\s^*\eta\sigma(W)\right)\\
&&\hspace{40mm}-\left(\s^{(X)}m,\sd\rho(W)+\chi^\sharp\cdot\beb(W)-\chib^\sharp\cdot\beta(W)\right)\\
&&\hspace{25mm}-\left(\s^{(X)}\mb,\Omega^{-1}D\beta(W)-\chi^\sharp\cdot\beta(W)-\Omega^{-1}\omega\beta(W)\right)\\
&&\hspace{40mm}+\left(\s^{(X)}\mb,\etb^\sharp\cdot\alpha(W)\right)\\
&&\hspace{25mm}+\frac{1}{2}\left(\s^{(X)}\ih,\snab\oth\beta(W)+\zeta\oth\beta(W)-3(\chih\rho(W)+\s^*\chih\sigma(W))\right)\\
&&\hspace{40mm}-\frac{1}{4}\mbox{tr}\chib(\s^{(X)}\ih,\alpha(W))\\
&&4\Lambdab(\s^{(X)}J^1(W))=\frac{1}{2}\s^{(X)}j\left\{\Omega^{-1}\Db\rho(W)-\sdiv\beb(W)-\frac{3}{2}\mbox{tr}\chib\rho(W)\right.\\
&&\hspace{40mm}\left.+(\zeta+2\eta,\beb(W))-\frac{1}{2}(\chih,\alb)\right\}\\
&&\hspace{25mm}-\frac{1}{2}\left(\s^{(X)}\mb,-\Omega^{-1}D\beb(W)+\chi^\sharp\cdot\beb(W)-\Omega^{-1}\omega\beb(W)\right)\\
&&\hspace{40mm}+\frac{3}{2}\left(\s^{(X)}\mb,\etb\rho(W)-\s^*\etb\sigma(W)\right)\\
&&\hspace{40mm}-\left(\s^{(X)}\mb,\sd\rho(W)-\chib^\sharp\cdot\beta(W)+\chi^\sharp\cdot\beb(W)\right)\\
&&\hspace{25mm}-\left(\s^{(X)}m,-\Omega^{-1}\Db\beb(W)+\chib^\sharp\cdot\beb(W)+\Omega^{-1}\omb\beb(W)\right)\\
&&\hspace{40mm}+\left(\s^{(X)}m,\eta^\sharp\cdot\alb(W)\right)\\
&&\hspace{25mm}+\frac{1}{2}\left(\s^{(X)}\ih,-\snab\oth\beb(W)+\zeta\oth\beb(W)-3(\chibh\rho(W)-\s^*\chibh\sigma(W))\right)\\
&&\hspace{40mm}-\frac{1}{4}\mbox{tr}\chi(\s^{(X)}\ih,\alb(W))\\
&&4K(\s^{(X)}J^1(W))=\frac{1}{2}\s^{(X)}j\left\{\Omega^{-1}D\sigma(W)-\scurl\beta(W)-\frac{3}{2}\mbox{tr}\chi\sigma(W)\right.\\
&&\hspace{40mm}\left.-(\zeta-2\etb)\wedge\beta(W)+\frac{1}{2}\chibh\wedge\alpha(W)\right\}\\
&&\hspace{25mm}+\frac{1}{2}\s^{(X)}m\wedge\left\{\Omega^{-1}\Db\beta(W)-\chib^\sharp\cdot\beta(W)+\Omega^{-1}\omb\beta(W)\right\}\\
&&\hspace{40mm}-\frac{3}{2}\s^{(X)}m\wedge\left\{\eta\rho(W)+\s^*\eta\sigma(W)\right\}\\
&&\hspace{40mm}-\left(\s^{(X)}m,\sd\sigma(W)+\chi^\sharp\cdot\s^*\beb(W)+\chib^\sharp\cdot\s^*\beta(W)\right)\\
&&\hspace{25mm}+\s^{(X)}\mb\wedge\left\{\Omega^{-1}D\beta(W)-\chi^\sharp\cdot\beta(W)-\Omega^{-1}\omega\beta(W)\right\}\\
&&\hspace{40mm}-\left(\s^{(X)}\mb,\etb^\sharp\cdot\s^*\alpha(W)\right)\\
&&\hspace{25mm}-\frac{1}{2}\left(\s^{(X)}\ih,\snab\oth\s^*\beta(W)+\zeta\oth\s^*\beta(W)-3(\s^*\chih\rho(W)-\chih\sigma(W))\right)\\
&&\hspace{40mm}+\frac{1}{4}\mbox{tr}\chib\s^{(X)}\ih\wedge\alpha(W)\\
&&4\Kb(\s^{(X)}J^1(W))=\frac{1}{2}\s^{(X)}j\left\{-\Omega^{-1}\Db\sigma(W)+\scurl\beb(W)+\frac{3}{2}\mbox{tr}\chib\sigma(W)\right.\\
&&\hspace{40mm}\left.-(\zeta+2\eta)\wedge\beb(W)+\frac{1}{2}\chih\wedge\alb(W)\right\}\\
&&\hspace{25mm}+\frac{1}{2}\s^{(X)}\mb\wedge\left\{-\Omega^{-1}D\beb(W)+\chi^\sharp\cdot\beb(W)-\Omega^{-1}\omega\beb(W)\right\}\\
&&\hspace{40mm}-\frac{3}{2}\s^{(X)}\mb\wedge\left\{\etb\rho(W)-\s^*\etb\sigma(W)\right\}\\
&&\hspace{40mm}+\left(\s^{(X)}\mb,\sd\sigma(W)+\chib^\sharp\cdot\s^*\beta(W)+\chi^\sharp\cdot\s^*\beb(W)\right)\\
&&\hspace{25mm}+\s^{(X)}m\wedge\left\{-\Omega^{-1}\Db\beb(W)+\chib^\sharp\cdot\beb(W)+\Omega^{-1}\omb\beb(W)\right\}\\
&&\hspace{40mm}-\left(\s^{(X)}m,\eta^\sharp\cdot\s^*\alb(W)\right)\\
&&\hspace{25mm}+\frac{1}{2}\left(\s^{(X)}\ih,\snab\oth\s^*\beb(W)-\zeta\oth\s^*\beb(W)+3(\s^*\chibh\rho(W)+\chibh\sigma(W))\right)\\
&&\hspace{40mm}+\frac{1}{4}\mbox{tr}\chi\s^{(X)}\ih\wedge\alb(W)\\
&&4I_A(\s^{(X)}J^1(W))=\frac{1}{2}\s^{(X)}j\left\{\Omega^{-1}(\Db\beta(W))_A+\sd_A\rho(W)+\s^*\sd_A\sigma((W)\right.\\
&&\hspace{25mm}-\mbox{tr}\chib\beta_A(W)+\Omega^{-1}\omb\beta_A(W)-3(\eta_A\rho(W)+\s^*\eta_A\sigma(W))\\
&&\hspace{55mm}\left.-\chib_A^{\s B}\beta_B(W)+2\chih_A^{\s B}\beb_B(W)\right\}\\
&&\hspace{25mm}-\frac{1}{2}\s^{(X)}m^B\left\{\sg_{AB}\Omega^{-1}\Db\rho(W)+\seps_{AB}\Omega^{-1}\Db\sigma(W)\right.\\
&&\hspace{40mm}\left.+2(\eta_A\beb_B(W)-\eta_B\beb_A(W)+\sg_{AB}\eta^C\beb_C(W))\right\}\\
&&\hspace{25mm}-\frac{1}{2}\s^{(X)}\mb^B\left\{\sg_{AB}\Omega^{-1}D\rho(W)+\seps_{AB}\Omega^{-1}D\sigma(W)\right.\\
&&\hspace{40mm}\left.-2(\etb_B\beta_A(W)-\etb_A\beta_B(W)+\sg_{AB}\etb^C\beta_C(W)\right\}\\
&&\hspace{25mm}-\frac{1}{2}\s^{X)}\mb^B\left\{2\snab_B\beta_A(W)+2\zeta_B\beta_A(W)-\chib_B^{\s C}\alpha_{AC}(W)\right.\\
&&\hspace{40mm}\left.-3(\chi_{AB}\rho(W)+\s^*\chi_{AB}\sigma(W))\right\}\\
&&\hspace{25mm}+\s^{(X)}\ih_A^{\s B}\sd_B\rho(W)+\s^{*(X)}\ih_A^{\s B}\sd_B\sigma(W)\\
&&\hspace{25mm}+\s^{(X)}\ih^{BC}\left\{\chibh_{CA}\beta_B(W)-\chibh_{CB}\beta_A(W)-\sg_{AB}\chibh_C^{\s D}\beta_D(W)\right.\\
&&\hspace{40mm}\left.+\chih_{CA}\beb_B(W)-\chih_{CB}\beb_A(W)+\sg_{AB}\chih_C^{\s D}\beb_D(W)\right\}\\
&&\hspace{55mm}+2\mbox{tr}\chi\s^{(X)}\ih_A^{\s B}\beb_B(W)\\
&&4\Ib_A(\s^{(X)}J^1(W))=\frac{1}{2}\s^{(X)}j\left\{\Omega^{-1}(-D\beb(W))_A+\sd_A\rho(W)-\s^*\sd_A\sigma((W)\right.\\
&&\hspace{25mm}+\mbox{tr}\chi\beb_A(W)-\Omega^{-1}\omega\beb_A(W)-3(\etb_A\rho(W)-\s^*\etb_A\sigma(W))\\
&&\hspace{55mm}\left.+\chi_A^{\s B}\beb_B(W)-2\chibh_A^{\s B}\beta_B(W)\right\}\\
&&\hspace{25mm}-\frac{1}{2}\s^{(X)}\mb^B\left\{\sg_{AB}\Omega^{-1}D\rho(W)-\seps_{AB}\Omega^{-1}D\sigma(W)\right.\\
&&\hspace{40mm}\left.-2(\etb_A\beta_B(W)-\etb_B\beta_A(W)+\sg_{AB}\etb^C\beta_C(W))\right\}\\
&&\hspace{25mm}-\frac{1}{2}\s^{(X)}m^B\left\{\sg_{AB}\Omega^{-1}\Db\rho(W)+\seps_{AB}\Omega^{-1}\Db\sigma(W)\right.\\
&&\hspace{40mm}\left.+2(\eta_B\beb_A(W)-\eta_A\beb_B(W)+\sg_{AB}\eta^C\beb_C(W)\right\}\\
&&\hspace{25mm}-\frac{1}{2}\s^{(X)}m^B\left\{-2\snab_B\beb_A(W)+2\zeta_B\beb_A(W)-\chi_B^{\s C}\alb_{AC}(W)\right.\\
&&\hspace{40mm}\left.-3(\chib_{AB}\rho(W)-\s^*\chib_{AB}\sigma(W))\right\}\\
&&\hspace{25mm}+\s^{(X)}\ih_A^{\s B}\sd_B\rho(W)-\s^{*(X)}\ih_A^{\s B}\sd_B\sigma(W)\\
&&\hspace{25mm}+\s^{(X)}\ih^{BC}\left\{-\chih_{CA}\beb_B(W)+\chih_{CB}\beb_A(W)+\sg_{AB}\chih_C^{\s D}\beb_D(W)\right.\\
&&\hspace{40mm}\left.-\chibh_{CA}\beta_B(W)+\chibh_{CB}\beta_A(W)-\sg_{AB}\chibh_C^{\s D}\beta_D(W)\right\}\\
&&\hspace{55mm}-2\mbox{tr}\chib\s^{(X)}\ih_A^{\s B}\beta_B(W)
\end{eqnarray*}

\vspace{5mm}

In the expressions for the components of $\s^{(X)}J^1(W)$ given by
the above proposition we now substitute for:
$$\Dbh\alpha(W),\Db\beta(W),\Db\rho(W),\Db\sigma(W),\Db\beb(W) \ \mbox{and} \ \Dh\alb(W)$$
from the inhomogeneous Bianchi equations of Proposition 12.4. When
considering the 1st order Weyl current error estimates, $W$ is the
fundamental Weyl field $R$ and these equations reduce to the
corresponding homogeneous equations (the Bianchi identities of
Proposition 1.2). When considering, on the other hand, the  2nd
order Weyl current error estimates, $W$ is one of the derived Weyl
fields $\tcL_L R$, $\tcL_{O_i}R:i=1,2,3$, and $\tcL_SR$, and on
the right hand sides of the inhomegeneous Bianchi equations we
have the components of the corresponding Weyl currents
$\s^{(L)}J$, $\s^{(O_i)}J:i=1,2,3$, and $\s^{(S)}J$, respectively.
Therefore the consideration of the 2nd order Weyl current error
estimates arising from $J^1$, requires that we first assess the
1st order Weyl currents. We thus confine ourselves in this chapter
to the 1st order Weyl current error estimates. The 2nd order Weyl
current error estimates shall be addressed in the next chapter.

The expression for each component of $\s^{(X)}J^1(W)$ given by
Lemma 14.1 is a sum of terms one factor of which is a component of
$\s^{(X)}\tilde{\pi}$. We consider the terms with the same such
factor as a single term. We thus consider the expression for each
component of $\s^{(X)}J^1(W)$ as consisting of four terms,
proportional to $\s^{(X)}j$, $\s^{(X)}m$, $\s^{(X)}\mb$, and
$\s^{(X)}\ih$, in the order in which these appear in the
expression given by Lemma 14.1.

Now, each component of $\s^{(X)}\tilde{\pi}$ is ${\cal
O}^\infty(\delta^{r_1}|u|^{p_1})$ for some $r_1, p_1$ according to
the estimates of Chapter 8. In the case $W=R$, to which our
attention is at present confined, after the substitution for
$$\Dbh\alpha,\Db\beta,\Db\rho,\Db\sigma,\Db\beb \ \mbox{and} \ \Dh\alb$$
from the homogeneous Bianchi equations, the other factor of each
term in the expression of a given component of $\s^{(X)}J^1(R)$
becomes the sum of a principal part, which is a sum of 1st
derivatives of components of $R$, and a non-principal part, which
is a sum of terms consisting of two factors, one of which is a
connection coefficient and the other a component of $R$. Viewing
these terms in a way which would be valid also in the case of the
2nd order Weyl current error estimates, we see each term as being
either ${\bf O}(\delta^r|u|^p)$ in the case of the terms involving
the components $\alpha,\beta,\rho,\sigma,\beb$ and their first
derivatives (which are either $\snab$ or $D$ derivatives after the
substitution), or $\bfob(\delta^r|u|^p)$ in the case of the terms
involving the component $\alb$ and its first derivatives (which
are either $\snab$ or $\Db$ derivatives after the substitution).
The values of $r$ and $p$ assigned are those implied by the bound
on the quantity ${\cal Q}_1$, taking also into account the
$L^\infty$ estimates of Chapter 3 for the connection coefficients
in the case of the non-principal terms. We then define $r_2$ to be
the minimal $r$ and $p_2$ to be the maximal $p$ occuring in the
terms of the other factor of a term involving a given component of
$\s^{(X)}\tilde{\pi}$ in the expression of a given component of
$\s^{(X)}J^1(R)$. We then set, for each term involving a given
component of $\s^{(X)}\tilde{\pi}$ in the expression of a given
component of $\s^{(X)}J^1(R)$,
\begin{equation}
r^\prime=r_1+r_2, \ \ \ p^\prime=p_1+p_2 \label{14.27}
\end{equation}
We then assign, to each component of $\s^{(X)}J^1(R)$, the pair
$r^*,p^*$, where $r^*$ is the minimal $r^\prime$ and $p^*$ is the
maximal $p^\prime$ occuring in the four terms constituting the
expression of that component. We obtain in this way the following
tables.

\vspace{10mm}

\hspace{40mm}{\large{1. {\bf Case} $X=L$}}

\vspace{5mm}

\hspace{30mm}$\Xi(\s^{(L)}J^1(R)):r^*=-2, p^*=-3$

\begin{equation}
\begin{array}{l|ll}
\mbox{term}&r^\prime&p^\prime\\ \hline
\mbox{1st}&-3/2&-3\\
\mbox{3rd}&-2&-3\\
\mbox{4th}&-2&-3
\end{array}
\label{14.28}
\end{equation}
\hspace{40mm}(the 2nd term vanishes)

\hspace{2.5mm}

\hspace{30mm}$\Xib(\s^{(L)}J^1(R)):r^*=1/2, p^*=-6$

\begin{equation}
\begin{array}{l|ll}
\mbox{term}&r^\prime&p^\prime\\ \hline
\mbox{1st}&1&-6\\
\mbox{2nd}&1/2&-6\\
\mbox{4th}&1/2&-6
\end{array}
\label{14.29}
\end{equation}
\hspace{40mm}(the 3rd term vanishes)

\hspace{2.5mm}

\hspace{30mm}$\Theta(\s^{(L)}J^1(R)):r^*=-3/2, p^*=-3$

\begin{equation}
\begin{array}{l|ll}
\mbox{term}&r^\prime&p^\prime\\ \hline
\mbox{1st}&-3/2&-3\\
\mbox{3rd}&-1&-4\\
\mbox{4th}&-3/2&-3
\end{array}
\label{14.30}
\end{equation}
\hspace{40mm}(the 2nd term vanishes)

\hspace{2.5mm}

\hspace{30mm}$\Thetab(\s^{(L)}J^1(R)):r^*=-1/2, p^*=-5$

\begin{equation}
\begin{array}{l|ll}
\mbox{term}&r^\prime&p^\prime\\ \hline
\mbox{1st}&1/2&-6\\
\mbox{2nd}&1/2&-6\\
\mbox{4th}&-1/2&-5
\end{array}
\label{14.31}
\end{equation}
\hspace{40mm}(the 3rd term vanishes)

\noindent This gives rise, as we shall see, to a {\em borderline
error integral}. The borderline contribution comes from the last
part of the 4th term namely from the terms:
\begin{equation}
-\frac{3}{2}\mbox{tr}\chib\left\{\s^{(L)}\ih\rho-\s^{*(L)}\ih\sigma\right\}
\label{14.32}
\end{equation}
The contribution of these shall be estimated more precisely in the
sequel. If these terms are removed from the 4th term the values
$r^\prime=1/2$, $p^\prime=-6$ would be assigned to the remainder,
hence if we denote by $\Thetab^\prime(\s^{(L)}J^1(R))$ what
results if we remove from $\Thetab(\s^{(L)}J^1(R))$ the terms
\ref{14.32} then to $\Thetab^\prime(\s^{(L)}J^1(R))$ is to be
assigned $r^*=1/2$, $p^*=-6$.

\hspace{2.5mm}

\hspace{30mm}$\Lambda(\s^{(L)}J^1(R)):r^*=-2, p^*=-3$

\begin{equation}
\begin{array}{l|ll}
\mbox{term}&r^\prime&p^\prime\\ \hline
\mbox{1st}&-1&-4\\
\mbox{3rd}&-1&-4\\
\mbox{4th}&-2&-3
\end{array}
\label{14.33}
\end{equation}
\hspace{40mm}(the 2nd term vanishes)

\noindent This gives rise, as we shall see, to a {\em borderline
error integral}. The borderline contribution comes from the last
part of the 4th term namely from the term:
\begin{equation}
-\frac{1}{4}\mbox{tr}\chib(\s^{(L)}\ih,\alpha) \label{14.34}
\end{equation}
The contribution of this shall be estimated more precisely in the
sequel. If this term is removed from the 4th term the values
$r^\prime=-1$, $p^\prime=-4$ would be assigned to the remainder,
hence if we denote by $\Lambda^\prime(\s^{(L)}J^1(R))$ what
results if we remove from $\Lambda(\s^{(L)}J^1(R))$ the term
\ref{14.34} then to $\Lambda^\prime(\s^{(L)}J^1(R))$ is to be
assigned $r^*=-1$, $p^*=-4$.

\hspace{2.5mm}

\hspace{30mm}$\Lambdab(\s^{(L)}J^1(R)):r^*=0, p^*=-5$

\begin{equation}
\begin{array}{l|ll}
\mbox{term}&r^\prime&p^\prime\\ \hline
\mbox{1st}&0&-5\\
\mbox{2nd}&0&-5\\
\mbox{4th}&0&-6
\end{array}
\label{14.35}
\end{equation}
\hspace{40mm}(the 3rd term vanishes)

\hspace{2.5mm}

\hspace{30mm}$K(\s^{(L)}J^1(R)):r^*=-2, p^*=-3$

\begin{equation}
\begin{array}{l|ll}
\mbox{term}&r^\prime&p^\prime\\ \hline
\mbox{1st}&-1&-4\\
\mbox{3rd}&-1&-4\\
\mbox{4th}&-2&-3
\end{array}
\label{14.36}
\end{equation}
\hspace{40mm}(the 2nd term vanishes)

\noindent This gives rise, as we shall see, to a {\em borderline
error integral}. The borderline contribution comes from the last
part of the 4th term namely from the term:
\begin{equation}
\frac{1}{4}\mbox{tr}\chib\s^{(L)}\ih\wedge\alpha \label{14.37}
\end{equation}
The contribution of this shall be estimated more precisely in the
sequel. If this term is removed from the 4th term the values
$r^\prime=-1$, $p^\prime=-4$ would be assigned to the remainder,
hence if we denote by $K^\prime(\s^{(L)}J^1(R))$ what results if
we remove from $K(\s^{(L)}J^1(R))$ the term \ref{14.37} then to
$K^\prime(\s^{(L)}J^1(R))$ is to be assigned $r^*=-1$, $p^*=-4$.

\hspace{2.5mm}

\hspace{30mm}$\Kb(\s^{(L)}J^1(R)):r^*=0, p^*=-5$

\begin{equation}
\begin{array}{l|ll}
\mbox{term}&r^\prime&p^\prime\\ \hline
\mbox{1st}&0&-5\\
\mbox{2nd}&0&-5\\
\mbox{4th}&0&-6
\end{array}
\label{14.38}
\end{equation}
\hspace{40mm}(the 3rd term vanishes)

\hspace{2.5mm}

\hspace{30mm}$I(\s^{(L)}J^1(R)):r^*=-1, p^*=-4$

\begin{equation}
\begin{array}{l|ll}
\mbox{term}&r^\prime&p^\prime\\ \hline
\mbox{1st}&-1/2&-4\\
\mbox{3rd}&-1&-4\\
\mbox{4th}&-1/2&-5
\end{array}
\label{14.39}
\end{equation}
\hspace{40mm}(the 2nd term vanishes)

\hspace{2.5mm}

\hspace{30mm}$\Ib(\s^{(L)}J^1(R)):r^*=-1, p^*=-4$

\begin{equation}
\begin{array}{l|ll}
\mbox{term}&r^\prime&p^\prime\\ \hline
\mbox{1st}&0&-5\\
\mbox{2nd}&-1/2&-5\\
\mbox{4th}&-1&-4
\end{array}
\label{14.40}
\end{equation}
\hspace{40mm}(the 3rd term vanishes)

\vspace{10mm}

\hspace{30mm}{\large{2. {\bf Case} $X=O_i:i=1,2,3$}}

\vspace{5mm}

\hspace{30mm}$\Xi(\s^{(O_i)}J^1(R)):r^*=-3/2, p^*=-3$

\begin{equation}
\begin{array}{l|ll}
\mbox{term}&r^\prime&p^\prime\\ \hline
\mbox{1st}&-1/2&-4\\
\mbox{2nd}&-3/2&-3\\
\mbox{4th}&-1&-3
\end{array}
\label{14.41}
\end{equation}
\hspace{40mm}(the 3rd term vanishes)

\hspace{2.5mm}

\hspace{30mm}$\Xib(\s^{(O_i)}J^1(R)):r^*=3/2, p^*=-6$

\begin{equation}
\begin{array}{l|ll}
\mbox{term}&r^\prime&p^\prime\\ \hline
\mbox{1st}&2&-7\\
\mbox{3rd}&3/2&-13/2\\
\mbox{4th}&3/2&-6
\end{array}
\label{14.42}
\end{equation}
\hspace{40mm}(the 2nd term vanishes)

\hspace{2.5mm}

\hspace{30mm}$\Theta(\s^{(O_i)}J^1(R)):r^*=-1/2, p^*=-4$

\begin{equation}
\begin{array}{l|ll}
\mbox{term}&r^\prime&p^\prime\\ \hline
\mbox{1st}&-1/2&-4\\
\mbox{2nd}&-1/2&-4\\
\mbox{4th}&-1/2&-4
\end{array}
\label{14.43}
\end{equation}
\hspace{40mm}(the 3rd term vanishes)

\hspace{2.5mm}

\hspace{30mm}$\Thetab(\s^{(O_i)}J^1(R)):r^*=1/2, p^*=-5$

\begin{equation}
\begin{array}{l|ll}
\mbox{term}&r^\prime&p^\prime\\ \hline
\mbox{1st}&3/2&-7\\
\mbox{3rd}&1&-6\\
\mbox{4th}&1/2&-5
\end{array}
\label{14.44}
\end{equation}
\hspace{40mm}(the 2nd term vanishes)

\noindent This gives rise, as we shall see, to a {\em borderline
error integral}. The borderline contribution comes from the last
part of the 4th term namely from the terms:
\begin{equation}
-\frac{3}{2}\mbox{tr}\chib\left\{\s^{(O_i)}\ih\rho-\s^{*(O_i)}\ih\sigma\right\}
\label{14.45}
\end{equation}
The contribution of these shall be estimated more precisely in the
sequel. If these terms are removed from the 4th term the values
$r^\prime=3/2$, $p^\prime=-6$ would be assigned to the remainder,
hence if we denote by $\Thetab^\prime(\s^{(O_i)}J^1(R))$ what
results if we remove from $\Thetab(\s^{(O_i)}J^1(R))$ the terms
\ref{14.45} then to $\Thetab^\prime(\s^{(O_i)}J^1(R))$ is to be
assigned $r^*=1$, $p^*=-6$.

\hspace{2.5mm}

\hspace{30mm}$\Lambda(\s^{(O_i)}J^1(R)):r^*=-1, p^*=-3$

\begin{equation}
\begin{array}{l|ll}
\mbox{term}&r^\prime&p^\prime\\ \hline
\mbox{1st}&0&-5\\
\mbox{2nd}&-1/2&-4\\
\mbox{4th}&-1&-3
\end{array}
\label{14.46}
\end{equation}
\hspace{40mm}(the 3rd term vanishes)

\noindent This gives rise, as we shall see, to a {\em borderline
error integral}. The borderline contribution comes from the last
part of the 4th term namely from the term:
\begin{equation}
-\frac{1}{4}\mbox{tr}\chib(\s^{(O_i)}\ih,\alpha) \label{14.47}
\end{equation}
The contribution of this shall be estimated more precisely in the
sequel. If this term is removed from the 4th term the values
$r^\prime=0$, $p^\prime=-4$ would be assigned to the remainder,
hence if we denote by $\Lambda^\prime(\s^{(O_i)}J^1(R))$ what
results if we remove from $\Lambda(\s^{(O_i)}J^1(R))$ the term
\ref{14.47} then to $\Lambda^\prime(\s^{O_i)}J^1(R))$ is to be
assigned $r^*=-1/2$, $p^*=-4$.

\hspace{2.5mm}

\hspace{30mm}$\Lambdab(\s^{(O_i)}J^1(R)):r^*=1, p^*=-6$

\begin{equation}
\begin{array}{l|ll}
\mbox{term}&r^\prime&p^\prime\\ \hline
\mbox{1st}&1&-6\\
\mbox{3rd}&1&-6\\
\mbox{4th}&3/2&-6
\end{array}
\label{14.48}
\end{equation}
\hspace{40mm}(the 2nd term vanishes)

\hspace{2.5mm}

\hspace{30mm}$K(\s^{(O_i)}J^1(R)):r^*=-1, p^*=-3$

\begin{equation}
\begin{array}{l|ll}
\mbox{term}&r^\prime&p^\prime\\ \hline
\mbox{1st}&0&-5\\
\mbox{2nd}&-1/2&-4\\
\mbox{4th}&-1&-3
\end{array}
\label{14.49}
\end{equation}
\hspace{40mm}(the 3rd term vanishes)

\noindent This gives rise, as we shall see, to a {\em borderline
error integral}. The borderline contribution comes from the last
part of the 4th term namely from the term:
\begin{equation}
\frac{1}{4}\mbox{tr}\chib\s^{(O_i)}\ih\wedge\alpha \label{14.50}
\end{equation}
The contribution of this shall be estimated more precisely in the
sequel. If this term is removed from the 4th term the values
$r^\prime=0$, $p^\prime=-4$ would be assigned to the remainder,
hence if we denote by $K^\prime(\s^{(O_i)}J^1(R))$ what results if
we remove from $K(\s^{(O_i)}J^1(R))$ the term \ref{14.50} then to
$K^\prime(\s^{(O_i)}J^1(R))$ is to be assigned $r^*=-1/2$,
$p^*=-4$.

\hspace{2.5mm}

\hspace{30mm}$\Kb(\s^{(O_i)}J^1(R)):r^*=1, p^*=-6$

\begin{equation}
\begin{array}{l|ll}
\mbox{term}&r^\prime&p^\prime\\ \hline
\mbox{1st}&1&-6\\
\mbox{3rd}&1&-6\\
\mbox{4th}&1&-6
\end{array}
\label{14.51}
\end{equation}
\hspace{40mm}(the 2nd term vanishes)

\hspace{2.5mm}

\hspace{30mm}$I(\s^{(O_i)}J^1(R)):r^*=0, p^*=-5$

\begin{equation}
\begin{array}{l|ll}
\mbox{term}&r^\prime&p^\prime\\ \hline
\mbox{1st}&1/2&-5\\
\mbox{2nd}&0&-5\\
\mbox{4th}&1/2&-5
\end{array}
\label{14.52}
\end{equation}
\hspace{40mm}(the 3rd term vanishes)

\hspace{2.5mm}

\hspace{30mm}$\Ib(\s^{(O_i)}J^1(R)):r^*=0, p^*=-4$

\begin{equation}
\begin{array}{l|ll}
\mbox{term}&r^\prime&p^\prime\\ \hline
\mbox{1st}&1&-6\\
\mbox{3rd}&0&-5\\
\mbox{4th}&0&-4
\end{array}
\label{14.53}
\end{equation}
\hspace{40mm}(the 2nd term vanishes)

\vspace{10mm}

\hspace{40mm}{\large{3. {\bf Case} $X=S$}}

\vspace{2.5mm}

In this case only the case $n=3$ occurs, therefore we only have to
consider $\Thetab(\s^{(S)}J^1(R))$ and $\Xib(\s^{(S)}J^1(R))$ (see
\ref{14.14}).

\hspace{2.5mm}

\hspace{30mm}$\Xib(\s^{(S)}J^1(R)):r^*=3/2, p^*=-6$

\begin{equation}
\begin{array}{l|ll}
\mbox{term}&r^\prime&p^\prime\\ \hline
\mbox{1st}&2&-7\\
\mbox{2nd}&3/2&-6\\
\mbox{3rd}&2&-13/2\\
\mbox{4th}&3/2&-6
\end{array}
\label{14.54}
\end{equation}
\hspace{40mm}

\hspace{2.5mm}

\hspace{30mm}$\Thetab(\s^{(S)}J^1(R)):r^*=1/2, p^*=-5$

\begin{equation}
\begin{array}{l|ll}
\mbox{term}&r^\prime&p^\prime\\ \hline
\mbox{1st}&3/2&-7\\
\mbox{2nd}&3/2&-6\\
\mbox{3rd}&3/2&-6\\
\mbox{4th}&1/2&-5
\end{array}
\label{14.55}
\end{equation}

\noindent This gives rise, as we shall see, to a {\em borderline
error integral}. The borderline contribution comes from the last
part of the 4th term namely from the terms:
\begin{equation}
-\frac{3}{2}\mbox{tr}\chib\left\{\s^{(S)}\ih\rho-\s^{*(S)}\ih\sigma\right\}
\label{14.56}
\end{equation}
The contribution of these shall be estimated more precisely in the
sequel. If these terms are removed from the 4th term the values
$r^\prime=3/2$, $p^\prime=-6$ would be assigned to the remainder,
hence if we denote by $\Thetab^\prime(\s^{(S)}J^1(R))$ what
results if we remove from $\Thetab(\s^{(S)}J^1(R))$ the terms
\ref{14.56} then to $\Thetab^\prime(\s^{(S)}J^1(R))$ is to be
assigned $r^*=3/2$, $p^*=-6$.

We have completed the investigation of the components of the Weyl
currents $\s^{(L)}J^1(R)$, $\s^{(O_i)}J^1(R):i=1,2,3$,  and
$\s^{(S)}J^1(R)$. These currents contribute to the error terms
$\s^{(L)}\stackrel{(n)}{\tau}_c:n=0,1,2,3$,
$\s^{(O)}\stackrel{(n)}{\tau}_c:n=0,1,2,3$, and
$\s^{(S)}\stackrel{(3)}{\tau}_c$, respectively, according to
\ref{14.12}, \ref{14.13}, and \ref{14.14}. We call these
contributions $\s^{(L)}\stackrel{(n)}{\tau}_{c,1}:n=0,1,2,3$,
$\s^{(O)}\stackrel{(n)}{\tau}_{c,1}:n=0,1,2,3$, and
$\s^{(S)}\stackrel{(3)}{\tau}_{c,1}$, respectively.

Each component of the Weyl currents $\s^{(L)}J^1(R)$,
$\s^{(O_i)}J^1(R):i=1,2,3$, and $\s^{(S)}J^1(R)$ being written as
a sum of terms in the manner discussed above, and these
expressions being substituted into \ref{14.12}, \ref{14.13}, and
\ref{14.14} respectively, sums of trilinear terms result, two of
the factors in each term being contributed by the expression for a
component of the Weyl current $\s^{(L)}J^1(R)$ in the case of
\ref{14.12}, $\s^{(O_i)}J^1(R):i=1,2,3$ in the case of
\ref{14.13}, and $\s^{(S)}J^1(R)$ in the case of \ref{14.14}, and
the other factor being a component of $\tcL_L R$,
$\tcL_{O_i}R:i=1,2,3$, and $\tcL_S R$, respectively, multiplied by
$\Omega^3$ and the appropriate power of $|u|^2$. Viewing these
third factors in a way which would be valid also in the case of
the 2nd order Weyl current error estimates, we see each third
factor as being either ${\bf O}(\delta^{r_3}|u|^{p_3})$ in the
case of factors involving $\alpha(\tcL_X R),\beta(\tcL_X
R),\rho(\tcL_X R),\sigma(\tcL_X R),\beb(\tcL_X R)$, or
$\bfob(\delta^{r_3}|u|^{p_3})$ in the case of factors involving
$\alb(\tcL_X R)$. The values of $r_3$ and $p_3$ assigned are those
implied by the bound on the quantity ${\cal P}_1$. To each
trilinear term we can then apply accordingly one of the first
three cases of Lemma 13.1. Since the pair $r^\prime,p^\prime$
corresponding to the first two factors of each trilinear term may
be replaced by the corresponding $r^*,p^*$, we then obtain a bound
for the contribution of all terms resulting from the product of a
given component of $\s^{(X)}J^1$ with a given component of $\tcL_X
R$ to the corresponding error integral
\begin{equation}
\delta^{2q_n+2l}\int_{M^\prime_{c^*}}|\s^{(X)}\stackrel{(n)}{\tau}_{c,1}|d\mu_g
\label{14.57}
\end{equation}
by $O(\delta^e)$, where $e$ is the excess index (see \ref{13.32}):
\begin{equation}
e=2q_n+2l+r^*+r_3+1 \label{14.58}
\end{equation}
provided that the integrability index $s$, defined by (see
\ref{13.33}):
\begin{equation}
s=p^*+p_3+3 \label{14.59}
\end{equation}
is negative so that Lemma 13.1 applies. We obtain in this way the
following tables. The ordinals in these tables refer to the terms
on the right hand side of each of \ref{14.12} - \ref{14.14}.

\vspace{10mm}

\hspace{30mm}{\large{1. {\bf Case} $X=L \ : \ l=1$}}

\vspace{2.5mm}

\hspace{45mm}{\large{$\s^{(L)}\stackrel{(0)}{\tau}_{c,1}$}}

\begin{equation}
\begin{array}{l|ll}
\mbox{term}&e&s\\ \hline
\mbox{1st}&1&-1\\
\mbox{2nd}&3/2&-2\\
\end{array}
\label{14.60}
\end{equation}

\vspace{2.5mm}

\hspace{45mm}{\large{$\s^{(L)}\stackrel{(1)}{\tau}_{c,1}$}}

\begin{equation}
\begin{array}{l|ll}
\mbox{term}&e&s\\ \hline
\mbox{1st}&0&-1\\
\mbox{2nd}&0&-1\\
\mbox{3rd}&1/2&-1
\end{array}
\label{14.61}
\end{equation}

\vspace{2.5mm}

\hspace{45mm}{\large{$\s^{(L)}\stackrel{(2)}{\tau}_{c,1}$}}

\begin{equation}
\begin{array}{l|ll}
\mbox{term}&e&s\\ \hline
\mbox{1st}&1&-1\\
\mbox{2nd}&1&-1\\
\mbox{3rd}&1&-1
\end{array}
\label{14.62}
\end{equation}

\vspace{2.5mm}

\hspace{45mm}{\large{$\s^{(L)}\stackrel{(3)}{\tau}_{c,1}$}}

\begin{equation}
\begin{array}{l|ll}
\mbox{term}&e&s\\ \hline
\mbox{1st}&0&-1/2\\
\mbox{2nd}&1/2&-1\\
\end{array}
\label{14.63}
\end{equation}

\vspace{10mm}

\hspace{30mm}{\large{2. {\bf Case} $X=O \ : \ l=0$}}

\vspace{2.5mm}

\hspace{45mm}{\large{$\s^{(O)}\stackrel{(0)}{\tau}_{c,1}$}}

\begin{equation}
\begin{array}{l|ll}
\mbox{term}&e&s\\ \hline
\mbox{1st}&1&-2\\
\mbox{2nd}&1&-2\\
\end{array}
\label{14.64}
\end{equation}

\vspace{2.5mm}

\hspace{45mm}{\large{$\s^{(O)}\stackrel{(1)}{\tau}_{c,1}$}}

\begin{equation}
\begin{array}{l|ll}
\mbox{term}&e&s\\ \hline
\mbox{1st}&0&-1\\
\mbox{2nd}&0&-1\\
\mbox{3rd}&1/2&-2
\end{array}
\label{14.65}
\end{equation}

\vspace{2.5mm}

\hspace{45mm}{\large{$\s^{(O)}\stackrel{(2)}{\tau}_{c,1}$}}

\begin{equation}
\begin{array}{l|ll}
\mbox{term}&e&s\\ \hline
\mbox{1st}&1&-2\\
\mbox{2nd}&1&-2\\
\mbox{3rd}&1&-1
\end{array}
\label{14.66}
\end{equation}

\vspace{2.5mm}

\hspace{45mm}{\large{$\s^{(O)}\stackrel{(3)}{\tau}_{c,1}$}}

\begin{equation}
\begin{array}{l|ll}
\mbox{term}&e&s\\ \hline
\mbox{1st}&0&-1/2\\
\mbox{2nd}&1/2&-1\\
\end{array}
\label{14.67}
\end{equation}

\vspace{10mm}

\hspace{30mm}{\large{3. {\bf Case} $X=S \ : \ l=0$}}

\vspace{2.5mm}

\hspace{45mm}{\large{$\s^{(S)}\stackrel{(3)}{\tau}_{c,1}$}}

\begin{equation}
\begin{array}{l|ll}
\mbox{term}&e&s\\ \hline
\mbox{1st}&0&-1/2\\
\mbox{2nd}&1/2&-1\\
\end{array}
\label{14.68}
\end{equation}

\vspace{10mm}

We see that all terms have negative integrability index so Lemma
13.1 indeed applies. All terms have non-negative excess index. The
terms with vanishing excess index play a crucial role because they
give rise to {\em borderline error integrals}. These must be
analyzed in more detail. They occur only in the cases $n=1$ and
$n=3$. The borderline terms are the 1st term in each of
\ref{14.63}, \ref{14.67}, and \ref{14.68}, and the 1st and 2nd
terms in each of \ref{14.61}, \ref{14.65}.

Replacing $\Thetab(\s^{(L)}J^1(R)$, $\Thetab(\s^{(O_i)}J^1(R))$,
and $\Thetab(\s^{(S)}J^1(R)$ by $\Thetab^\prime(\s^{(L)}J^1(R))$,
$\Thetab^\prime(\s^{(O_i)}J^1(R))$ and
$\Thetab^\prime(\s^{(S)}J^1(R))$ in the 1st term of each of
\ref{14.63}, \ref{14.67}, and \ref{13.68}, respectively, terms
with excess indices of 1, 1/2, and 1, respectively, would result.
Thus the actual borderline error terms in
$\s^{(L)}\stackrel{(3)}{\tau}_{c,1}$,
$\s^{(O)}\stackrel{(3)}{\tau}_{c,1}$, and
$\s^{(S)}\stackrel{(3)}{\tau}_{c,1}$, are the terms:
\begin{equation}
-6\Omega^3|u|^6\mbox{tr}\chib((\s^{(L)}\ih\rho-\s^{*(L)}\ih\sigma),\alb(\tcL_L
R)) \label{14.69}
\end{equation}
\begin{equation}
-6\Omega^3|u|^6\mbox{tr}\chib\sum_i((\s^{(O_i)}\ih\rho-\s^{*(O_i)}\ih\sigma),\alb(\tcL_{O_i}R))
\label{14.70}
\end{equation}
and:
\begin{equation}
-6\Omega^3|u|^6\mbox{tr}\chib((\s^{(S)}\ih\rho-\s^{*(S)}\ih\sigma),\alb(\tcL_S
R)) \label{14.71}
\end{equation}
contributed by \ref{14.32}, \ref{14.45}, and \ref{14.56},
respectively.

Using the precise bound:
\begin{equation}
\|\s^{(L)}\ih\|_{L^\infty(S_{\ub,u})}\leq
C\delta^{-1/2}|u|^{-1}{\cal R}_0^\infty(\alpha) \ \ : \ \forall
(\ub,u)\in D^\prime_{c^*} \label{14.72}
\end{equation}
(see first of \ref{8.32}), using also the facts that:
\begin{eqnarray}
\|\mbox{tr}\chib\rho\|_{L^2(C_u)}\leq C\delta^{1/2}|u|^{-3}(\stackrel{(2)}{{\cal E}}_0)^{1/2} \ \ : \ \forall u\in[u_0,c^*)\nonumber\\
\|\mbox{tr}\chib\sigma\|_{L^2(C_u)}\leq
C\delta^{1/2}|u|^{-3}(\stackrel{(2)}{{\cal E}}_0)^{1/2} \ \ : \
\forall u\in[u_0,c^*) \label{14.73}
\end{eqnarray}
as well as the fact that:
\begin{equation}
\|\Omega^3|u|^3\alb(\tcL_L R)\|_{L^2(\Cb_{\ub})}\leq
C\delta^{1/2}(\stackrel{(3)}{{\cal F}}_1)^{1/2} \ \ : \ \forall
\ub\in[0,\delta) \label{14.74}
\end{equation}
(see \ref{12.89}, \ref{12.128} and \ref{12.137}), and following
the proof of Case 2 of Lemma 13.1, we deduce that the contribution
of the term \ref{14.69} to the error integral \ref{14.57} with
$X=L$ ($l=1$) and $n=3$ is bounded by:
\begin{equation}
C{\cal R}_0^\infty(\alpha)(\stackrel{(2)}{{\cal
E}}_0)^{1/2}(\stackrel{(3)}{{\cal F}}_1)^{1/2} \label{14.75}
\end{equation}

Using the more precise bound:
\begin{equation}
\|\s^{(O_i)}\ih\|_{L^\infty(S_{\ub,u})}\leq
C\delta^{1/2}|u|^{-1}(\scR_1^4(\beta)+{\cal
R}_0^\infty(\beta)+O(\delta^{1/2}))
 \ \ : \ \forall (\ub,u)\in D^\prime_{c^*}
\label{14.76}
\end{equation}
implied by the first of \ref{8.139}, using also \ref{14.73} as
well as the fact that:
\begin{equation}
\|\Omega^3|u|^3\alb((\tcL_{O_i}R)\|_{L^2(\Cb_{\ub})}\leq
C\delta^{3/2}(\stackrel{(3)}{{\cal F}}_1)^{1/2} \ \ : \ \forall
\ub\in [0,\delta) \label{14.77}
\end{equation}
(see \ref{12.94}, \ref{12.128} and \ref{12.137}), and following
the proof of Case 2 of Lemma 13.1, we deduce that the contribution
of the term \ref{14.70} to the error integral \ref{14.57} with
$X=O$ ($l=0$) and $n=3$ is bounded by:
\begin{equation}
C(\scR_1^4(\beta)+{\cal R}_0^\infty(\beta))(\stackrel{(2)}{{\cal
E}}_0)^{1/2}(\stackrel{(3)}{{\cal F}}_1)^{1/2}+O(\delta^{1/2})
\label{14.78}
\end{equation}

Using the more precise bound:
\begin{equation}
\|\s^{(S)}\ih\|_{L^\infty(S_{\ub,u})}\leq
C\delta^{1/2}|u|^{-1}({\cal D}_0^\infty(\chibh)+{\cal
R}_0^\infty(\alpha)+O(\delta)) \ \ : \ \forall (\ub,u)\in
D^\prime_{c^*} \label{14.79}
\end{equation}
implied by the first of \ref{8.34}, using also \ref{14.73} as well
as the fact that:
\begin{equation}
\|\Omega^3|u|^3\alb(\tcL_S R)\|_{L^2(\Cb_{\ub})}\leq
C\delta^{3/2}(\stackrel{(3)}{{\cal F}}_1)^{1/2} \ \ : \ \forall
\ub\in [0,\delta) \label{14.80}
\end{equation}
(see \ref{12.99}, \ref{12.128} and \ref{12.137}), and following
the proof of Case 2 of Lemma 13.1, we deduce that the contribution
of the term \ref{14.71} to the error integral \ref{14.57} with
$X=S$ ($l=0$) and $n=3$ is bounded by:
\begin{equation}
C({\cal D}_0^\infty(\chibh)+{\cal
R}_0^\infty(\alpha))(\stackrel{(2)}{{\cal
E}}_0)^{1/2}(\stackrel{(3)}{{\cal F}}_1)^{1/2}+O(\delta)
\label{14.81}
\end{equation}

Now, by the first of \ref{10.75} and the first of \ref{12.a4}:
\begin{equation}
{\cal R}_0^\infty(\alpha)\leq C{\cal R}_{[2]}(\alpha)\leq
C^\prime(\stackrel{(0)}{{\cal E}}_2)^{1/2}+O(\delta^{1/2})
\label{14.82}
\end{equation}
It follows that \ref{14.75} is in turn bounded by:
\begin{equation}
C(\stackrel{(0)}{{\cal E}}_2)^{1/2}(\stackrel{(2)}{{\cal
E}}_0)^{1/2}(\stackrel{(3)}{{\cal F}}_1)^{1/2}+O(\delta^{1/2})
\label{14.84}
\end{equation}
By the second of \ref{10.68}, the second of \ref{10.75} and the
second of \ref{12.a4}:
\begin{eqnarray}
&&\scR_1^4(\beta)\leq C{\cal R}_{[2]}(\beta)\leq C^\prime(\stackrel{(1)}{{\cal E}}_2)^{1/2}+O(\delta^{1/2})\nonumber\\
&&{\cal R}_0^\infty(\beta)\leq C{\cal R}_{[2]}(\beta)\leq
C^\prime(\stackrel{(1)}{{\cal E}}_2)^{1/2}+O(\delta^{1/2})
\label{14.85}
\end{eqnarray}
It follows that \ref{14.78} is in turn bounded by:
\begin{equation}
C(\stackrel{(1)}{{\cal E}}_2)^{1/2}(\stackrel{(2)}{{\cal
E}}_0)^{1/2}(\stackrel{(3)}{{\cal F}}_1)^{1/2}+O(\delta^{1/2})
\label{14.86}
\end{equation}
Moreover, \ref{14.81} is in turn bounded by, in view of
\ref{14.82},
\begin{equation}
C\left\{{\cal D}_0^\infty(\chibh)+(\stackrel{(0)}{{\cal
E}}_2)^{1/2}\right\}(\stackrel{(2)}{{\cal
E}}_0)^{1/2}(\stackrel{(3)}{{\cal F}}_1)^{1/2}+O(\delta^{1/2})
\label{14.87}
\end{equation}

Replacing $\Lambda(\s^{(L)}J^1(R)$, $K(\s^{(L)}J^1(R))$, and
$\Lambda(\s^{(O_i)}J^1(R))$, $K(\s^{(O_i)}J^1(R))$ by
$\Lambda^\prime(\s^{(L)}J^1(R))$, $K^\prime(\s^{(L)}J^1(R))$ and
$\Lambda^\prime(\s^{(O_i)}J^1(R))$, $K^\prime(\s^{(O_i)}J^1(R))$
in the 1st and 2nd terms of each of \ref{14.61} and \ref{14.65},
respectively, terms with excess indices of 1 and 1/2,
respectively, would result. Thus the actual borderline error terms
in $\s^{(L)}\stackrel{(1)}{\tau}_{c,1}$ and
$\s^{(O)}\stackrel{(1)}{\tau}_{c,1}$, are the terms:
\begin{equation}
2\Omega^3|u|^2\mbox{tr}\chib\left\{(\s^{(L)}\ih,\alpha)\rho(\tcL_L
R)+\s^{(L)}\ih\wedge\alpha\sigma(\tcL_L R)\right\} \label{14.88}
\end{equation}
and:
\begin{equation}
2\Omega^3|u|^2\mbox{tr}\chib\sum_i\left\{(\s^{(O_i)}\ih,\alpha)\rho(\tcL_{O_i}R)+\s^{(O_i)}\ih\wedge\alpha\sigma(\tcL_{O_i}R)\right\}
\label{14.89}
\end{equation}
contributed by \ref{14.34}, \ref{14.37} and \ref{14.47},
\ref{14.50} respectively.

Using \ref{14.72}, the fact that:
\begin{equation}
\|\mbox{tr}\chib\alpha\|_{L^2(C_u)}\leq
C\delta^{-1}|u|^{-1}(\stackrel{(0)}{{\cal E}}_0)^{1/2} \ \ : \
\forall u\in[u_0,c^*) \label{14.90}
\end{equation}
as well as the facts that:
\begin{eqnarray}
&&\|\Omega^3|u|^2\rho(\tcL_L R)\|_{L^2(C_u)}\leq C\delta^{-1/2}(\stackrel{(2)}{{\cal E}}_1)^{1/2} \ \ : \ \forall u\in[u_0,c^*)\nonumber\\
&&\|\Omega^3|u|^2\sigma(\tcL_L R)\|_{L^2(C_u)}\leq
C\delta^{-1/2}(\stackrel{(2)}{{\cal E}}_1)^{1/2} \ \ : \ \forall
u\in[u_0,c^*) \label{14.91}
\end{eqnarray}
(see \ref{12.88}, \ref{12.125} and \ref{12.136}), and following
the proof of Case 1 of Lemma 13.1 we deduce that the contribution
of the term \ref{14.88} to the error integral \ref{14.57} with
$X=L$ ($l=1$) and $n=1$ is bounded by:
\begin{equation}
C{\cal R}_0^\infty(\alpha)(\stackrel{(0)}{{\cal
E}}_0)^{1/2}(\stackrel{(2)}{{\cal E}}_1)^{1/2} \label{14.92}
\end{equation}

Using \ref{14.76}, \ref{14.90}, as well as the facts that:
\begin{eqnarray}
&&\|\Omega^3|u|^2\rho(\tcL_{O_i}R)\|_{L^2(C_u)}\leq C\delta^{1/2}(\stackrel{(2)}{{\cal E}}_1)^{1/2} \ \ : \ \forall u\in[u_0,c^*)\nonumber\\
&&\|\Omega^3|u|^2\sigma(\tcL_{O_i}R)\|_{L^2(C_u)}\leq
C\delta^{1/2}(\stackrel{(2)}{{\cal E}}_1)^{1/2} \ \ : \ \forall
u\in[u_0,c^*) \label{14.93}
\end{eqnarray}
(see \ref{12.93}, \ref{12.125} and \ref{12.136}), and following
the proof of Case 1 of Lemma 13.1 we deduce that the contribution
of the term \ref{14.89} to the error integral \ref{14.57} with
$X=O$ ($l=0$) and $n=1$ is bounded by:
\begin{equation}
C(\scR_1^4(\beta)+{\cal R}_0^\infty(\beta))(\stackrel{(0)}{{\cal
E}}_0)^{1/2}(\stackrel{(2)}{{\cal E}}_1)^{1/2} \label{14.94}
\end{equation}
In view of \ref{14.82} we conclude that \ref{14.92} is in turn
bounded by:
\begin{equation}
C(\stackrel{(0)}{{\cal E}}_2)^{1/2}(\stackrel{(0)}{{\cal
E}}_0)^{1/2}(\stackrel{(2)}{{\cal E}}_1)^{1/2}+O(\delta^{1/2})
\label{14.96}
\end{equation}
Moreover, in view of \ref{14.85} we conclude that \ref{14.94} is
in turn bounded by:
\begin{equation}
C(\stackrel{(1)}{{\cal E}}_2)^{1/2}(\stackrel{(0)}{{\cal
E}}_0)^{1/2}(\stackrel{(2)}{{\cal E}}_1)^{1/2}+O(\delta^{1/2})
\label{14.97}
\end{equation}

We summarize the results of this section in the following
proposition.

\vspace{5mm}

\noindent{\bf Proposition 14.1} The 1st order Weyl current error
integrals arising from $J^1$, \ref{14.57}, satisfy the estimates:
\begin{eqnarray*}
&&\delta^{2q_0+2}\int_{M^\prime_{c^*}}|\s^{(L)}\stackrel{(0)}{\tau}_{c,1}|d\mu_g\leq O(\delta)\\
&&\delta^{2q_1+2}\int_{M^\prime_{c^*}}|\s^{(L)}\stackrel{(1)}{\tau}_{c,1}|d\mu_g\leq
C(\stackrel{(0)}{{\cal E}}_2)^{1/2}(\stackrel{(0)}{{\cal E}}_0)^{1/2}(\stackrel{(2)}{{\cal E}}_1)^{1/2}+O(\delta^{1/2})\\
&&\delta^{2q_2+2}\int_{M^\prime_{c^*}}|\s^{(L)}\stackrel{(2)}{\tau}_{c,1}|d\mu_g\leq O(\delta)\\
&&\delta^{2q_3+2}\int_{M^\prime_{c^*}}|\s^{(L)}\stackrel{(3)}{\tau}_{c,1}|d\mu_g\leq
C(\stackrel{(0)}{{\cal E}}_2)^{1/2}(\stackrel{(2)}{{\cal
E}}_0)^{1/2}(\stackrel{(3)}{{\cal F}}_1)^{1/2}+O(\delta^{1/2})
\end{eqnarray*}
\begin{eqnarray*}
&&\delta^{2q_0}\int_{M^\prime_{c^*}}|\s^{(O)}\stackrel{(0)}{\tau}_{c,1}|d\mu_g\leq O(\delta)\\
&&\delta^{2q_1}\int_{M^\prime_{c^*}}|\s^{(O)}\stackrel{(1)}{\tau}_{c,1}|d\mu_g\leq
C(\stackrel{(1)}{{\cal E}}_2)^{1/2}(\stackrel{(0)}{{\cal E}}_0)^{1/2}(\stackrel{(2)}{{\cal E}}_1)^{1/2}+O(\delta^{1/2})\\
&&\delta^{2q_2}\int_{M^\prime_{c^*}}|\s^{(O)}\stackrel{(2)}{\tau}_{c,1}|d\mu_g\leq O(\delta)\\
&&\delta^{2q_3}\int_{M^\prime_{c^*}}|\s^{(O)}\stackrel{(3)}{\tau}_{c,1}|d\mu_g\leq
C(\stackrel{(1)}{{\cal E}}_2)^{1/2}(\stackrel{(2)}{{\cal
E}}_0)^{1/2}(\stackrel{(3)}{{\cal F}}_1)^{1/2}+O(\delta^{1/2})
\end{eqnarray*}
and:
$$\delta^{2q_3}\int_{M^\prime_{c^*}}|\s^{(S)}\stackrel{(3)}{\tau}_{c,1}|d\mu_g\leq
C\left\{{\cal D}_0^\infty(\chibh)+(\stackrel{(0)}{{\cal
E}}_2)^{1/2}\right\}(\stackrel{(2)}{{\cal
E}}_0)^{1/2}(\stackrel{(3)}{{\cal F}}_1)^{1/2}+O(\delta^{1/2})$$

\vspace{5mm}

\section{The error estimates arising from $J^2$}

For any commutation field $X$, the associated 1-form $\s^{(X)}p$
defined by \ref{14.6} decomposes into the $S$ 1-form
$\s^{(X)}\sp$, the restriction of $\s^{(X)}p$ to the $S_{\ub,u}$,
given on each $S_{\ub,u}$ by:
\begin{equation}
\s^{(X)}\sp(Y)=\s^{(X)}p(Y) \ \ : \ \forall Y\in T_q S_{\ub,u} \
\forall q\in S_{\ub,u} \label{14.98}
\end{equation}
and the functions $\s^{(X)}p_3$, $\s^{(X)}p_4$, given by:
\begin{equation}
\s^{(X)}p_3=\s^{(X)}p(\Lbh), \ \ \ \s^{(X)}p_4=\s^{(X)}p(\Lh)
\label{14.99}
\end{equation}

The components of the Weyl current $\s^{(X)}J^2(W)$ associated to
the commutation field $X$ and to the Weyl field $W$ are then given
by:
\begin{eqnarray}
&&4\Xi(\s^{(X)}J^2(W))=-\s^{(X)}p_4\beta(W)-\s^{(X)}\sp^\sharp\cdot\alpha(W)\nonumber\\
&&4\Xib(\s^{(X)}J^2(W))=\s^{(X)}p_3\beb(W)-\s^{(X)}\sp^\sharp\cdot\alb(W)\nonumber\\
&&4\Theta(\s^{(X)}J^2(W))=-\s^{(X)}p_3\alpha(W)+\s^{(X)}\sp\oth\beta(W)\nonumber\\
&&4\Thetab(\s^{(X)}J^2(W))=-\s^{(X)}p_4\alb(W)-\s^{(X)}\sp\oth\beb(W)\nonumber\\
&&4\Lambda(\s^{(X)}J^2(W))=-\s^{(X)}p_4\rho(W)+(\s^{(X)}\sp,\beta(W))\nonumber\\
&&4\Lambdab(\s^{(X)}J^2(W))=-\s^{(X)}p_3\rho(W)-(\s^{(X)}\sp,\beb(W))\nonumber\\
&&4K(\s^{(X)}J^2(W))=-\s^{(X)}p_4\sigma(W)-\s^{(X)}\sp\wedge\beta(W)\nonumber\\
&&4\Kb(\s^{(X)}J^2(W))=\s^{(X)}p_3\sigma(W)+\s^{(X)}\sp\wedge\beb(W)\nonumber\\
&&4I(\s^{(X)}J^2(W))=-\s^{(X)}p_3\beta(W)+\s^{(X)}\sp\rho(W)+\s^{*(X)}\sp\sigma(W)\nonumber\\
&&4\Ib(\s^{(X)}J^2(W))=\s^{(X)}p_4\beb(W)+\s^{(X)}\sp\rho(W)-\s^{*(X)}\sp\sigma(W)
\label{14.100}
\end{eqnarray}

Using the table \ref{1.151} of connection coefficients of the
frame $(e_\mu \ : \ \mu=1,2,3,4)$ we obtain:
\begin{eqnarray}
&&\s^{(X)}p_4=-\frac{1}{2}\Omega^{-1}D\s^{(X)}j+\sdiv\s^{(X)}m\nonumber\\
&&\hspace{15mm}-\mbox{tr}\chi\s^{(X)}j+\frac{1}{2}((\etb+5\eta),\s^{(X)}m)-(\chih,\s^{(X)}\ih)
\label{14.101}
\end{eqnarray}
\begin{eqnarray}
&&\s^{(X)}p_3=-\frac{1}{2}\Omega^{-1}\Db\s^{(X)}j+\sdiv\s^{(X)}\mb\nonumber\\
&&\hspace{15mm}-\mbox{tr}\chib\s^{(X)}j+\frac{1}{2}((\eta+5\etb),\s^{(X)}\mb)-(\chibh,\s^{(X)}\ih)
\label{14.102}
\end{eqnarray}
\begin{eqnarray}
&&\s^{(X)}\sp=-\frac{1}{2}(\Db\s^{(X)}m+D\s^{(X)}\mb)+\frac{1}{2}\sd\s^{(X)}j+\sdiv\s^{(X)}\ih\nonumber\\
&&\hspace{15mm}-\frac{1}{2}\Omega^{-1}\omb\s^{(X)}m-\frac{1}{2}\Omega^{-1}\omega\s^{(X)}\mb
-\frac{1}{2}\mbox{tr}\chib\s^{(X)}m-\frac{1}{2}\mbox{tr}\chi\s^{(X)}\mb\nonumber\\
&&\hspace{15mm}+(\eta+\etb)\s^{(X)}j+(\eta+\etb)^\sharp\cdot\s^{(X)}\ih
\label{14.103}
\end{eqnarray}

Using the estimates \ref{9.1} - \ref{9.3} as well as \ref{8.32}
and the estimates of Chapter 3 we deduce, for $X=L$:
\begin{eqnarray}
&&\s^{(L)}p_4={\cal O}^4(\delta^{-1}|u|^{-2})\nonumber\\
&&\s^{(L)}p_3={\cal O}^4(|u|^{-2})\nonumber\\
&&\s^{(L)}\sp={\cal O}^4(\delta^{-1/2}|u|^{-2}) \label{14.104}
\end{eqnarray}
Using the estimates \ref{9.82} - \ref{9.84} as well as \ref{8.139}
and the estimates of Chapter 3 we deduce, for $X=O_i:i=1,2,3$:
\begin{eqnarray}
&&\s^{(O_i)}p_4={\cal O}^4(|u|^{-2})\nonumber\\
&&\s^{(O_i)}p_3={\cal O}^4(\delta|u|^{-3})\nonumber\\
&&\s^{(O_i)}\sp={\cal O}^4(\delta^{1/2}|u|^{-2}) \label{14.105}
\end{eqnarray}
Also, using the estimates \ref{9.12} - \ref{9.14} as well as
\ref{8.34} we deduce, for $X=S$:
\begin{eqnarray}
&&\s^{(S)}p_4={\cal O}^4(|u|^{-2})\nonumber\\
&&\s^{(S)}p_3={\cal O}^4(\delta|u|^{-3})\nonumber\\
&&\s^{(S)}\sp={\cal O}^4(\delta^{1/2}|u|^{-2}) \label{14.106}
\end{eqnarray}

The expression for each component of $\s^{(X)}J^2(W)$ given by
\ref{14.100} is a sum of terms one factor of which is a component
of $\s^{(X)}p$ and the other factor a component of $W$. According
to the above, each component of $\s^{(X)}p$ is ${\cal
O}^4(\delta^{r_1}|u|^{p_1})$ for some $r_1, p_1$. Our attention is
at present confined to the case $W=R$. Viewing the components of
$W=R$ in a way which would be valid also in the case of the 2nd
order Weyl current error estimates we see each component as being
${\cal O}^4(\delta^{r_2}|u|^{p_2})$ according to:
\begin{eqnarray}
&&\alpha={\cal O}^4(\delta^{-3/2}|u|^{-1})\nonumber\\
&&\beta={\cal O}^4(\delta^{-1/2}|u|^{-2})\nonumber\\
&&\rho,\sigma={\cal O}^4(|u|^{-3})\nonumber\\
&&\beb={\cal O}^4(\delta|u|^{-4})\nonumber\\
&&\alb={\cal O}^4(\delta^{3/2}|u|^{-9/2}) \label{14.107}
\end{eqnarray}
implied by \ref{12.146} and the bound on the quantity ${\cal
Q}_1$. We set, for each term in the expression of a given
component of $\s^{(X)}J^2(R)$,
\begin{equation}
r^\prime=r_1+r_2, \ \ \ p^\prime=p_1+p_2 \label{14.108}
\end{equation}
We then assign, to each component of $\s^{(X)}J^2(R)$, the pair
$r^*, p^*$, where $r^*$ is the minimal $r^\prime$ and $p^*$ is the
maximal $p^\prime$ occuring in the terms constituting the
expression of that component. We obtain in this way the following
tables. The ordinals in these tables refer to the terms in the
expressions \ref{14.100}.

\vspace{10mm}

\hspace{40mm}{\large{1. {\bf Case} $X=L$}}

\vspace{5mm}

\hspace{30mm}$\Xi(\s^{(L)}J^2(R)):r^*=-2, p^*=-3$

\begin{equation}
\begin{array}{l|ll}
\mbox{term}&r^\prime&p^\prime\\ \hline
\mbox{1st}&-3/2&-4\\
\mbox{2nd}&-2&-3
\end{array}
\label{14.109}
\end{equation}

\hspace{2.5mm}

\hspace{30mm}$\Xib(\s^{(L)}J^2(R)):r^*=1, p^*=-6$

\begin{equation}
\begin{array}{l|ll}
\mbox{term}&r^\prime&p^\prime\\ \hline
\mbox{1st}&1&-6\\
\mbox{2nd}&1&-13/2
\end{array}
\label{14.110}
\end{equation}

\hspace{2.5mm}

\hspace{30mm}$\Theta(\s^{(L)}J^2(R)):r^*=-3/2, p^*=-3$

\begin{equation}
\begin{array}{l|ll}
\mbox{term}&r^\prime&p^\prime\\ \hline
\mbox{1st}&-3/2&-3\\
\mbox{2nd}&-1&-4
\end{array}
\label{14.111}
\end{equation}

\hspace{2.5mm}

\hspace{30mm}$\Thetab(\s^{(L)}J^2(R)):r^*=1/2, p^*=-6$

\begin{equation}
\begin{array}{l|ll}
\mbox{term}&r^\prime&p^\prime\\ \hline
\mbox{1st}&1/2&-13/2\\
\mbox{2nd}&1/2&-6
\end{array}
\label{14.112}
\end{equation}

\hspace{2.5mm}

\hspace{30mm}$\Lambda(\s^{(L)}J^2(R)):r^*=-1, p^*=-4$

\begin{equation}
\begin{array}{l|ll}
\mbox{term}&r^\prime&p^\prime\\ \hline
\mbox{1st}&-1&-5\\
\mbox{2nd}&-1&-4
\end{array}
\label{14.113}
\end{equation}

\hspace{2.5mm}

\hspace{30mm}$\Lambdab(\s^{(L)}J^2(R)):r^*=0, p^*=-5$

\begin{equation}
\begin{array}{l|ll}
\mbox{term}&r^\prime&p^\prime\\ \hline
\mbox{1st}&0&-5\\
\mbox{2nd}&1/2&-6
\end{array}
\label{14.114}
\end{equation}

\hspace{2.5mm}

\hspace{30mm}$K(\s^{(L)}J^2(R)):r^*=-1, p^*=-4$

\begin{equation}
\begin{array}{l|ll}
\mbox{term}&r^\prime&p^\prime\\ \hline
\mbox{1st}&-1&-5\\
\mbox{2nd}&-1&-4
\end{array}
\label{14.115}
\end{equation}

\hspace{2.5mm}

\hspace{30mm}$\Kb(\s^{(L)}J^2(R)):r^*=0, p^*=-5$

\begin{equation}
\begin{array}{l|ll}
\mbox{term}&r^\prime&p^\prime\\ \hline
\mbox{1st}&0&-5\\
\mbox{2nd}&1/2&-6
\end{array}
\label{14.116}
\end{equation}

\hspace{2.5mm}

\hspace{30mm}$I(\s^{(L)}J^2(R)):r^*=-1/2, p^*=-4$

\begin{equation}
\begin{array}{l|ll}
\mbox{term}&r^\prime&p^\prime\\ \hline
\mbox{1st}&-1/2&-4\\
\mbox{2nd}&-1/2&-5
\end{array}
\label{14.117}
\end{equation}

\hspace{2.5mm}

\hspace{30mm}$\Ib(\s^{(L)}J^2(R)):r^*=-1/2, p^*=-5$

\begin{equation}
\begin{array}{l|ll}
\mbox{term}&r^\prime&p^\prime\\ \hline
\mbox{1st}&0&-6\\
\mbox{2nd}&-1/2&-5
\end{array}
\label{14.118}
\end{equation}

\vspace{10mm}

\hspace{30mm}{\large{2. {\bf Case} $X=O_i:i=1,2,3$}}

\vspace{5mm}

\hspace{30mm}$\Xi(\s^{(O_i)}J^2(R)):r^*=-1, p^*=-3$

\begin{equation}
\begin{array}{l|ll}
\mbox{term}&r^\prime&p^\prime\\ \hline
\mbox{1st}&-1/2&-4\\
\mbox{2nd}&-1&-3
\end{array}
\label{14.119}
\end{equation}

\hspace{2.5mm}

\hspace{30mm}$\Xib(\s^{(O_i)}J^2(R)):r^*=2, p^*=-13/2$

\begin{equation}
\begin{array}{l|ll}
\mbox{term}&r^\prime&p^\prime\\ \hline
\mbox{1st}&2&-7\\
\mbox{2nd}&2&-13/2
\end{array}
\label{14.120}
\end{equation}

\hspace{2.5mm}

\hspace{30mm}$\Theta(\s^{(O_i)}J^2(R)):r^*=-1/2, p^*=-4$

\begin{equation}
\begin{array}{l|ll}
\mbox{term}&r^\prime&p^\prime\\ \hline
\mbox{1st}&-1/2&-4\\
\mbox{2nd}&0&-4
\end{array}
\label{14.121}
\end{equation}

\hspace{2.5mm}

\hspace{30mm}$\Thetab(\s^{(O_i)}J^2(R)):r^*=3/2, p^*=-6$

\begin{equation}
\begin{array}{l|ll}
\mbox{term}&r^\prime&p^\prime\\ \hline
\mbox{1st}&3/2&-13/2\\
\mbox{2nd}&3/2&-6
\end{array}
\label{14.122}
\end{equation}

\hspace{2.5mm}

\hspace{30mm}$\Lambda(\s^{(O_i)}J^2(R)):r^*=0, p^*=-4$

\begin{equation}
\begin{array}{l|ll}
\mbox{term}&r^\prime&p^\prime\\ \hline
\mbox{1st}&0&-5\\
\mbox{2nd}&0&-4
\end{array}
\label{14.123}
\end{equation}

\hspace{2.5mm}

\hspace{30mm}$\Lambdab(\s^{(O_i)}J^2(R)):r^*=1, p^*=-6$

\begin{equation}
\begin{array}{l|ll}
\mbox{term}&r^\prime&p^\prime\\ \hline
\mbox{1st}&1&-6\\
\mbox{2nd}&3/2&-6
\end{array}
\label{14.124}
\end{equation}

\hspace{2.5mm}

\hspace{30mm}$K(\s^{(O_i)}J^2(R)):r^*=0, p^*=-4$

\begin{equation}
\begin{array}{l|ll}
\mbox{term}&r^\prime&p^\prime\\ \hline
\mbox{1st}&0&-5\\
\mbox{2nd}&0&-4
\end{array}
\label{14.125}
\end{equation}

\hspace{2.5mm}

\hspace{30mm}$\Kb(\s^{(O_i)}J^2(R)):r^*=1, p^*=-6$

\begin{equation}
\begin{array}{l|ll}
\mbox{term}&r^\prime&p^\prime\\ \hline
\mbox{1st}&1&-6\\
\mbox{2nd}&3/2&-6
\end{array}
\label{14.126}
\end{equation}

\hspace{2.5mm}

\hspace{30mm}$I(\s^{(O_i)}J^2(R)):r^*=1/2, p^*=-5$

\begin{equation}
\begin{array}{l|ll}
\mbox{term}&r^\prime&p^\prime\\ \hline
\mbox{1st}&1/2&-5\\
\mbox{2nd}&1/2&-5
\end{array}
\label{14.127}
\end{equation}

\hspace{2.5mm}

\hspace{30mm}$\Ib(\s^{(O_i)}J^2(R)):r^*=1/2, p^*=-5$

\begin{equation}
\begin{array}{l|ll}
\mbox{term}&r^\prime&p^\prime\\ \hline
\mbox{1st}&1&-6\\
\mbox{2nd}&1/2&-5
\end{array}
\label{14.128}
\end{equation}

\vspace{10mm}

\hspace{40mm}{\large{3. {\bf Case} $X=S$}}

\vspace{2.5mm}

In this case only the case $n=3$ occurs, therefore we only have to
consider $\Thetab(\s^{(S)}J^2(R))$ and $\Xib(\s^{(S)}J^2(R))$ (see
\ref{14.14}).

\hspace{2.5mm}

\hspace{30mm}$\Xib(\s^{(S)}J^2(R)):r^*=2, p^*=-13/2$

\begin{equation}
\begin{array}{l|ll}
\mbox{term}&r^\prime&p^\prime\\ \hline
\mbox{1st}&2&-7\\
\mbox{2nd}&2&-13/2
\end{array}
\label{14.129}
\end{equation}
\hspace{40mm}

\hspace{2.5mm}

\hspace{30mm}$\Thetab(\s^{(S)}J^2(R)):r^*=3/2, p^*=-6$

\begin{equation}
\begin{array}{l|ll}
\mbox{term}&r^\prime&p^\prime\\ \hline
\mbox{1st}&3/2&-13/2\\
\mbox{2nd}&3/2&-6
\end{array}
\label{14.130}
\end{equation}

We have completed the investigation of the components of the Weyl
currents $\s^{(L)}J^2(R)$, $\s^{(O_i)}J^2(R):i=1,2,3$,  and
$\s^{(S)}J^2(R)$. These currents contribute to the error terms
$\s^{(L)}\stackrel{(n)}{\tau}_c:n=0,1,2,3$,
$\s^{(O)}\stackrel{(n)}{\tau}_c:n=0,1,2,3$, and
$\s^{(S)}\stackrel{(3)}{\tau}_c$, respectively, according to
\ref{14.12}, \ref{14.13}, and \ref{14.14}. We call these
contributions $\s^{(L)}\stackrel{(n)}{\tau}_{c,2}:n=0,1,2,3$,
$\s^{(O)}\stackrel{(n)}{\tau}_{c,2}:n=0,1,2,3$, and
$\s^{(S)}\stackrel{(3)}{\tau}_{c,2}$, respectively.

Each component of the Weyl currents $\s^{(L)}J^2(R)$,
$\s^{(O_i)}J^2(R):i=1,2,3$, and $\s^{(S)}J^2(R)$ being expressed
as a sum of terms by \ref{14.100}, and these expressions being
substituted into \ref{14.12}, \ref{14.13}, and \ref{14.14}, sums
of trilinear terms result, two of the factors in each term being
contributed by the expression for a component of the Weyl current
$\s^{(L)}J^2(R)$ in the case of \ref{14.12},
$\s^{(O_i)}J^2(R):i=1,2,3$ in the case of \ref{14.13}, and
$\s^{(S)}J^2(R)$ in the case of \ref{14.14}, and the other factor
being a component of $\tcL_L R$, $\tcL_{O_i}R:i=1,2,3$, and
$\tcL_S R$, respectively, multiplied by $\Omega^3$ and the
appropriate power of $|u|^2$. Viewing these third factors in a way
which would be valid also in the case of the 2nd order Weyl
current error estimates, we see each third factor as being either
${\bf O}(\delta^{r_3}|u|^{p_3})$ in the case of factors involving
$\alpha(\tcL_X R),\beta(\tcL_X R),\rho(\tcL_X R),\sigma(\tcL_X
R),\beb(\tcL_X R)$, or $\bfob(\delta^{r_3}|u|^{p_3})$ in the case of factors
involving $\alb(\tcL_X R)$. The values of $r_3$ and $p_3$ assigned
are those implied by the bound on the quantity ${\cal P}_1$. To
each trilinear term we can then apply accordingly one of the last
two cases of Lemma 13.1. Since the pair $r^\prime,p^\prime$
corresponding to the first two factors of each trilinear term may
be replaced by the corresponding $r^*,p^*$, we then obtain a bound
for the contribution of all terms resulting from the product of a
given component of $\s^{(X)}J^2$ with a given component of $\tcL_X
R$ to the corresponding error integral
\begin{equation}
\delta^{2q_n+2l}\int_{M^\prime_{c^*}}|\s^{(X)}\stackrel{(n)}{\tau}_{c,2}|d\mu_g
\label{14.131}
\end{equation}
by $O(\delta^e)$, where $e$ is the excess index (see \ref{13.32}):
\begin{equation}
e=2q_n+2l+r^*+r_3+1 \label{14.132}
\end{equation}
provided that the integrability index $s$, defined by (see
\ref{13.33}):
\begin{equation}
s=p^*+p_3+3 \label{14.133}
\end{equation}
is negative so that Lemma 13.1 applies. We obtain in this way the
following tables. The ordinals in these tables refer to the terms
on the right hand side of each of \ref{14.12} - \ref{14.14}.

\vspace{10mm}

\hspace{30mm}{\large{1. {\bf Case} $X=L \ : \ l=1$}}

\vspace{2.5mm}

\hspace{45mm}{\large{$\s^{(L)}\stackrel{(0)}{\tau}_{c,2}$}}

\begin{equation}
\begin{array}{l|ll}
\mbox{term}&e&s\\ \hline
\mbox{1st}&1&-1\\
\mbox{2nd}&3/2&-2\\
\end{array}
\label{14.134}
\end{equation}

\vspace{2.5mm}

\hspace{45mm}{\large{$\s^{(L)}\stackrel{(1)}{\tau}_{c,2}$}}

\begin{equation}
\begin{array}{l|ll}
\mbox{term}&e&s\\ \hline
\mbox{1st}&1&-2\\
\mbox{2nd}&1&-2\\
\mbox{3rd}&1&-1
\end{array}
\label{14.135}
\end{equation}

\vspace{2.5mm}

\hspace{45mm}{\large{$\s^{(L)}\stackrel{(2)}{\tau}_{c,2}$}}

\begin{equation}
\begin{array}{l|ll}
\mbox{term}&e&s\\ \hline
\mbox{1st}&1&-1\\
\mbox{2nd}&1&-1\\
\mbox{3rd}&3/2&-2
\end{array}
\label{14.136}
\end{equation}

\vspace{2.5mm}

\hspace{45mm}{\large{$\s^{(L)}\stackrel{(3)}{\tau}_{c,2}$}}

\begin{equation}
\begin{array}{l|ll}
\mbox{term}&e&s\\ \hline
\mbox{1st}&1&-3/2\\
\mbox{2nd}&1&-1\\
\end{array}
\label{14.137}
\end{equation}

\vspace{10mm}

\hspace{30mm}{\large{2. {\bf Case} $X=O \ : \ l=0$}}

\vspace{2.5mm}

\hspace{45mm}{\large{$\s^{(O)}\stackrel{(0)}{\tau}_{c,2}$}}

\begin{equation}
\begin{array}{l|ll}
\mbox{term}&e&s\\ \hline
\mbox{1st}&1&-2\\
\mbox{2nd}&3/2&-2\\
\end{array}
\label{14.138}
\end{equation}

\vspace{2.5mm}

\hspace{45mm}{\large{$\s^{(O)}\stackrel{(1)}{\tau}_{c,2}$}}

\begin{equation}
\begin{array}{l|ll}
\mbox{term}&e&s\\ \hline
\mbox{1st}&1&-2\\
\mbox{2nd}&1&-2\\
\mbox{3rd}&1&-2
\end{array}
\label{14.139}
\end{equation}

\vspace{2.5mm}

\hspace{45mm}{\large{$\s^{(O)}\stackrel{(2)}{\tau}_{c,2}$}}

\begin{equation}
\begin{array}{l|ll}
\mbox{term}&e&s\\ \hline
\mbox{1st}&1&-2\\
\mbox{2nd}&1&-2\\
\mbox{3rd}&3/2&-2
\end{array}
\label{14.140}
\end{equation}

\vspace{2.5mm}

\hspace{45mm}{\large{$\s^{(O)}\stackrel{(3)}{\tau}_{c,2}$}}

\begin{equation}
\begin{array}{l|ll}
\mbox{term}&e&s\\ \hline
\mbox{1st}&1&-3/2\\
\mbox{2nd}&1&-3/2\\
\end{array}
\label{14.141}
\end{equation}

\vspace{10mm}

\hspace{30mm}{\large{3. {\bf Case} $X=S \ : \ l=0$}}

\vspace{2.5mm}

\hspace{45mm}{\large{$\s^{(S)}\stackrel{(3)}{\tau}_{c,2}$}}

\begin{equation}
\begin{array}{l|ll}
\mbox{term}&e&s\\ \hline
\mbox{1st}&1&-3/2\\
\mbox{2nd}&1&-3/2\\
\end{array}
\label{14.142}
\end{equation}

\vspace{10mm}

We see that all terms have negative integrability index so that
Lemma 13.1 indeed applies. All terms have positive excess index,
so there are no borderline terms.

We summarize the results of this section in the following
proposition.

\vspace{5mm}

\noindent{\bf Proposition 14.2} The 1st order Weyl current error
integrals arising from $J^2$, \ref{14.131}, satisfy the estimates:
\begin{eqnarray*}
&&\delta^{2q_0+2}\int_{M^\prime_{c^*}}|\s^{(L)}\stackrel{(0)}{\tau}_{c,2}|d\mu_g\leq O(\delta)\\
&&\delta^{2q_1+2}\int_{M^\prime_{c^*}}|\s^{(L)}\stackrel{(1)}{\tau}_{c,2}|d\mu_g\leq O(\delta)\\
&&\delta^{2q_2+2}\int_{M^\prime_{c^*}}|\s^{(L)}\stackrel{(2)}{\tau}_{c,2}|d\mu_g\leq O(\delta)\\
&&\delta^{2q_3+2}\int_{M^\prime_{c^*}}|\s^{(L)}\stackrel{(3)}{\tau}_{c,2}|d\mu_g\leq
O(\delta)
\end{eqnarray*}
\begin{eqnarray*}
&&\delta^{2q_0}\int_{M^\prime_{c^*}}|\s^{(O)}\stackrel{(0)}{\tau}_{c,2}|d\mu_g\leq O(\delta)\\
&&\delta^{2q_1}\int_{M^\prime_{c^*}}|\s^{(O)}\stackrel{(1)}{\tau}_{c,2}|d\mu_g\leq O(\delta)\\
&&\delta^{2q_2}\int_{M^\prime_{c^*}}|\s^{(O)}\stackrel{(2)}{\tau}_{c,2}|d\mu_g\leq O(\delta)\\
&&\delta^{2q_3}\int_{M^\prime_{c^*}}|\s^{(O)}\stackrel{(3)}{\tau}_{c,2}|d\mu_g\leq
O(\delta)
\end{eqnarray*}
and:
$$\delta^{2q_3}\int_{M^\prime_{c^*}}|\s^{(S)}\stackrel{(3)}{\tau}_{c,2}|d\mu_g\leq O(\delta)$$

\vspace{5mm}

\section{The error estimates arising from $J^3$}

As we remarked in Chapter 12, for any commutation field $X$ the
associated 3-covariant tensorfield $\s^{(X)}q$ defined by
\ref{14.7} has the algebraic properties of a Weyl current. It
therefore decomposes into the components $\Xi(\s^{(X)}q)$,
$\Xib(\s^{(X)}q)$, $\Theta(\s^{(X)}q)$, $\Thetab(\s^{(X)}q)$,
$\Lambda(\s^{(X)}q)$, $\Lambdab(\s^{(X)}q)$, $K(\s^{(X)}q)$,
$\Kb(\s^{(X)}q)$, $I(\s^{(X)}q)$, $\Ib(\s^{(X)}q)$. The following
lemma is established in a straightforward manner.

\vspace{5mm}

\noindent{\bf Lemma 14.2} \ \ \ The components of the Weyl current
$\s^{(X)}J^3$ associated to the commutation field $X$ and to the
Weyl field $W$ are given by:
\begin{eqnarray*}
&&4\Xi(\s^{(X)}J^3(W))=-3\Ib^\sharp(\s^{(X)}q)\cdot\alpha(W)+3\Lambda(\s^{(X)}q)\beta(W)+6K(\s^{(X)}q)\s^*\beta(W)\\
&&\hspace{28mm}+3\Xi(\s^{(X)}q)\rho(W)+2\s^*\Xi(\s^{(X)}q)\sigma(W)\\
&&4\Xib(\s^{(X)}J^3(W))=-3I^\sharp(\s^{(X)}q)\cdot\alb(W)-3\Lambdab(\s^{(X)}q)\beb(W)-6\Kb(\s^{(X)}q)\s^*\beb(W)\\
&&\hspace{28mm}+3\Xib(\s^{(X)}q)\rho(W)-2\s^*\Xib\sigma(W)\\
&&4\Theta(\s^{(X)}J^3(W))=3\Lambdab(\s^{(X)}q)\alpha(W)-3\Kb(\s^{(X)}q)\s^*\alpha(W)-6I(\s^{(X)}q)\oth\beta(W)\\
&&\hspace{28mm}+3\Theta(\s^{(X)}q)\rho(W)+3\s^*\Theta(\s^{(X)}q)\sigma(W)\\
&&4\Thetab(\s^{(X)}J^3(W))=3\Lambda(\s^{(X)}q)\alb(W)-3K(\s^{(X)}q)\s^*\alb(W)+6\Ib(\s^{(X)}q)\oth\beb(W)\\
&&\hspace{28mm}+3\Thetab(\s^{(X)}q)\rho(W)-3\s^*\Thetab(\s^{(X)}q)\sigma(W)\\
&&4\Lambda(\s^{(X)}J^3(W))=\frac{1}{2}(\Thetab(\s^{(X)}q),\alpha(W))-3\Lambda(\s^{(X)}q)\rho(W)+3K(\s^{(X)}q)\sigma(W)\\
&&\hspace{28mm}+(\Xi(\s^{(X)}q),\beb(W))\\
&&4\Lambdab(\s^{(X)}J^3(W))=\frac{1}{2}(\Theta(\s^{(X)}q),\alb(W))-3\Lambdab(\s^{(X)}q)\rho(W)-3\Kb(\s^{(X)}q)\sigma(W)\\
&&\hspace{28mm}-(\Xib(\s^{(X)}q),\beta(W))\\
&&4K(\s^{(X)}J^3(W))=-\frac{1}{2}\Thetab(\s^{(X)}q)\wedge\alpha(W)+2\Ib(\s^{(X)}q)\wedge\beta(W)\\
&&\hspace{28mm}-3\Lambda(\s^{(X)}q)\sigma(W)-3K(\s^{(X)}q)\rho(W)+\Xi(\s^{(X)}q)\wedge\beb(W)\\
&&4\Kb(\s^{(X)}J^3(W))=-\frac{1}{2}\Theta(\s^{(X)}q)\wedge\alb(W)-2I(\s^{(X)}q)\wedge\beb(W)\\
&&\hspace{28mm}+3\Lambdab(\s^{(X)}q)\sigma(W)-3\Kb(\s^{(X)}q)\rho(W)-\Xib(\s^{(X)}q)\wedge\beta(W)\\
&&4I(\s^{(X)}J^3(W))=-\Xib^\sharp(\s^{(X)}q)\cdot\alpha(W)-4\Lambdab(\s^{(X)}q)\beta(W)\\
&&\hspace{28mm}-3I(\s^{(X)}q)\rho(W)-3\s^*I(\s^{(X)}q)\sigma(W)-2\Theta(\s^{(X)}q)\cdot\beb^\sharp(W)\\
&&4\Ib(\s^{(X)}J^3(W))=-\Xi^\sharp(\s^{(X)}q)\cdot\alb(W)+4\Lambda(\s^{(X)}q)\beb(W)\\
&&\hspace{28mm}-3\Ib(\s^{(X)}q)\rho(W)+3\s^*\Ib(\s^{(X)}q)\sigma(W)+2\Thetab(\s^{(X)}q)\cdot\beta^\sharp(W)
\end{eqnarray*}

\vspace{5mm}

Using the table \ref{1.151} of connection coefficients of the
frame $(e_\mu \ : \ \mu=1,2,3,4)$ and taking into account the
basic identity \ref{1.163} we obtain:
\begin{equation}
2\Xi(\s^{(X)}q)=\Omega^{-1}D\s^{(X)}m+\chi^\sharp\cdot
m-\Omega^{-1}\omega m \label{14.143}
\end{equation}
\begin{equation}
2\Xib(\s^{(X)}q)=\Omega^{-1}\Db\s^{(X)}\mb+\chib^\sharp\cdot\mb-\Omega^{-1}\omb\mb
\label{14.144}
\end{equation}
\begin{eqnarray}
&&2\Theta(\s^{(X)}q)=2\Omega^{-1}\left(\Dh\s^{(X)}\ih-\frac{1}{2}\Omega\mbox{tr}\chi\s^{(X)}\ih\right)-\snab\oth\s^{(X)}m\nonumber\\
&&\hspace{20mm}+2\chih\s^{(X)}j-(2\etb+\zeta)\oth\s^{(X)}m
\label{14.145}
\end{eqnarray}
\begin{eqnarray}
&&2\Thetab(\s^{(X)}q)=2\Omega^{-1}\left(\Dbh\s^{(X)}\ih-\frac{1}{2}\Omega\mbox{tr}\chib\s^{(X)}\ih\right)-\snab\oth\s^{(X)}\mb\nonumber\\
&&\hspace{20mm}+2\chibh\s^{(X)}j-(2\eta-\zeta)\oth\s^{(X)}\mb
\label{14.146}
\end{eqnarray}
\begin{eqnarray}
&&2\Lambda(\s^{(X)}q)=-\Omega^{-1}D\s^{(X)}j-\mbox{tr}\chi\s^{(X)}j+\sdiv\s^{(X)}m\nonumber\\
&&\hspace{20mm}-(\chih,\s^{(X)}\ih)+(2\etb+\zeta,\s^{(X)}m)-\frac{2}{3}\s^{(X)}p_4
\label{14.147}
\end{eqnarray}
\begin{eqnarray}
&&2\Lambdab(\s^{(X)}q)=-\Omega^{-1}\Db\s^{(X)}j-\mbox{tr}\chib\s^{(X)}j+\sdiv\s^{(X)}\mb\nonumber\\
&&\hspace{20mm}-(\chibh,\s^{(X)}\ih)+(2\eta-\zeta,\s^{(X)}\mb)-\frac{2}{3}\s^{(X)}p_3
\label{14.148}
\end{eqnarray}
\begin{equation}
2K(\s^{(X)}q)=\scurl\s^{(X)}m+\chih\wedge\s^{(X)}\ih+\zeta\wedge\s^{(X)}m
\label{14.149}
\end{equation}
\begin{equation}
2\Kb(\s^{(X)}q)=\scurl\s^{X)}\mb+\chibh\wedge\s^{(X)}\ih-\zeta\wedge\s^{(X)}\mb
\label{14.150}
\end{equation}
\begin{eqnarray}
&&2I(\s^{(X)}q)=\Omega^{-1}D\s^{(X)}\mb-\Omega^{-1}\omega\s^{(X)}\mb-\sd\s^{(X)}j-2\etb\s^{(X)}j\nonumber\\
&&\hspace{20mm}-2\etb^\sharp\cdot\s^{(X)}\ih+\chib^\sharp\cdot\s^{(X)}m+\frac{2}{3}\s^{(X)}\sp
\label{14.151}
\end{eqnarray}
\begin{eqnarray}
&&2\Ib(\s^{(X)}q)=\Omega^{-1}\Db\s^{X)}m-\Omega^{-1}\omb\s^{(X)}m-\sd\s^{(X)}j-2\eta\s^{(X)}j\nonumber\\
&&\hspace{20mm}-2\eta^\sharp\cdot\s^{(X)}\ih+\chi^\sharp\cdot\s^{(X)}\mb+\frac{2}{3}\s^{(X)}\sp
\label{14.152}
\end{eqnarray}

We consider as a single term the first term on the right hand side
of \ref{14.146} because there is a {\em crucial cancellation}
involved here as we shall presently see. For the sake of symmetry
we also consider as a single term the first term on the right hand
side of \ref{14.145} although there is no cancellation involved.

Consider first the case $X=L$. Then by the first of \ref{9.3}:
\begin{equation}
\Dbh\s^{(L)}\ih={\cal O}^4(\delta^{-1/2}|u|^{-2}) \label{14.153}
\end{equation}
However, since according to the first of \ref{8.22}:
\begin{equation}
\s^{(L)}\ih=2\Omega\chih, \label{14.154}
\end{equation}
according to \ref{4.c4} we have:
\begin{equation}
\Dbh\s^{(L)}\ih-\frac{1}{2}\Omega\mbox{tr}\chib\s^{(L)}\ih=\Omega^2\{2\snab\oth\eta+2\eta\oth\eta-\mbox{tr}\chi\chibh\}
\label{14.155}
\end{equation}
The estimates of Chapter 4 for the right hand side then give:
\begin{equation}
\Dbh\s^{(L)}\ih-\frac{1}{2}\Omega\mbox{tr}\chib\s^{(L)}\ih={\cal
O}^4(\delta^{1/2}|u|^{-3}) \label{14.156}
\end{equation}

Consider next the case $X=S$. Then by the first of \ref{9.14}:
\begin{equation}
\Dbh\s^{(S)}\ih={\cal O}^4(\delta^{1/2}|u|^{-2}) \label{14.157}
\end{equation}
However, according to the first of \ref{8.30}:
\begin{equation}
\s^{(S)}\ih=2\Omega(u\chibh+\ub\chih), \label{14.158}
\end{equation}
hence:
\begin{eqnarray}
&&\Dbh\s^{(S)}\ih-\frac{1}{2}\Omega\mbox{tr}\chib\s^{(S)}\ih=2\Omega^2 u\Dbh(\Omega^{-1}\chibh)\nonumber\\
&&\hspace{35mm}+2\ub\left(\Dbh(\Omega\chih)-\frac{1}{2}\Omega\mbox{tr}\chib(\Omega\chih)\right)\nonumber\\
&&\hspace{35mm}+4\Omega
u\omb\chibh+2\Omega\chibh\left(1-\frac{1}{2}u\Omega\mbox{tr}\chib\right)
\label{14.159}
\end{eqnarray}
Now, according to \ref{3.9}:
\begin{equation}
\Dbh(\Omega^{-1}\chibh)=-\alb \label{14.160}
\end{equation}
It follows that the first term on the right in \ref{14.159} is
${\cal O}^4(\delta^{3/2}|u|^{-7/2})$. By \ref{14.156} the second
term  is ${\cal O}^4(\delta^{3/2}|u|^{-3})$. Also, by the
estimates of Chapter 3 the third and fourth terms are ${\cal
O}^4(\delta^{3/2}|u|^{-4})$ and ${\cal O}^4(\delta^{3/2}|u|^{-3})$
respectively. We conclude that:
\begin{equation}
\Dbh\s^{(S)}\ih-\frac{1}{2}\Omega\mbox{tr}\chib\s^{(S)}\ih={\cal
O}^4(\delta^{3/2}|u|^{-3}) \label{14.161}
\end{equation}

Consider finally the case $X=O_i:i=1,2,3$. Then by the first of
\ref{9.84}:
\begin{equation}
\Dbh\s^{(O_i)}\ih={\cal O}^4(\delta^{1/2}|u|^{-2}) \label{14.162}
\end{equation}
However $\s^{(O_i)}\ih=\s^{(O_i)}\hat{\spi}$ satisfies the
propagation equation \ref{8.105}:
\begin{equation}
\Dbh\s^{(O_i)}\hat{\spi}-\Omega\mbox{tr}\chib\s^{(O_i)}\hat{\spi}=-\Omega\chibh\mbox{tr}\s^{(O_i)}\spi+2\sLh_{O_i}(\Omega\chibh)
\label{14.163}
\end{equation}
hence we have:
\begin{equation}
\Dbh\s^{(O_i)}\ih-\frac{1}{2}\Omega\mbox{tr}\chib\s^{(O_i)}\ih=\frac{1}{2}\Omega\mbox{tr}\chib\s^{(O_i)}\hat{\spi}
+2\sLh_{O_i}(\Omega\chibh)-\Omega\chibh\mbox{tr}\s^{(O_i)}\spi
\label{14.164}
\end{equation}
Let us define the symmetric 2-covariant $S$ tensorfields:
\begin{equation}
\vartheta_i=\s^{(O_i)}\spi+2u\sL_{O_i}(\Omega\chibh)
\label{14.165}
\end{equation}
Let us also set:
\begin{equation}
\varepsilon=\frac{1}{2}u\Omega\mbox{tr}\chib-1 \label{14.166}
\end{equation}
Then, taking into account the fact that:
\begin{equation}
\mbox{tr}\sL_{O_i}(\Omega\chibh)=\sg^{-1}\cdot\sL_{O_i}(\Omega\chibh)=-\Omega\chibh\cdot\sL_{O_i}(\sg^{-1})
=(\Omega\chibh,\s^{(O_i)}\hat{\spi}) \label{14.167}
\end{equation}
\ref{14.164} takes the form:
\begin{eqnarray}
&&u\left(\Dbh\s^{(O_i)}\ih-\frac{1}{2}\Omega\mbox{tr}\chib\s^{(O_i)}\ih\right)=
\vartheta_i+\varepsilon\s^{(O_i)}\hat{\spi}-u\sg(\Omega\chibh,\s^{(O_i)}\hat{\spi})\nonumber\\
&&\hspace{45mm}-\frac{1}{2}\sg\mbox{tr}\s^{(O_i)}\spi-u\Omega\chibh\mbox{tr}\s^{(O_i)}\spi
\label{14.168}
\end{eqnarray}
Now, by the estimates of Chapter 3:
\begin{equation}
\varepsilon={\cal O}^\infty(\delta|u|^{-1}) \label{14.169}
\end{equation}
hence by Proposition 8.3 the second term on the right in
\ref{14.168} is ${\cal O}^4(\delta^{3/2}|u|^{-2})$. Moreover, by
the same proposition and the results of Chapter 3 the third,
fourth and fifth terms on the right in \ref{14.168} are ${\cal
O}^4(\delta|u|^{-2})$, ${\cal O}^4(\delta|u|^{-2})$ and ${\cal
O}^4(\delta^{3/2}|u|^{-3})$ respectively. It follows that we will
have shown that:
\begin{equation}
\Dbh\s^{(O_i)}\ih-\frac{1}{2}\Omega\mbox{tr}\chib\s^{(O_i)}\ih={\cal
O}^4(\delta|u|^{-3}) \label{14.170}
\end{equation}
once we establish the following lemma.

\vspace{5mm}

\noindent{\bf Lemma 14.3} \ \ \ We have:
$$\vartheta_i={\cal O}^4(\delta|u|^{-2})$$

\noindent{\em Proof:} \ Consider first $D\vartheta_i$ along
$C_{u_0}$. Since $[L,O_i]=0$ along $C_{u_0}$, we have:
\begin{equation}
D\sL_{O_i}(\Omega\chibh)=\sL_{O_i}D(\Omega\chibh) \ \ : \
\mbox{along $C_{u_0}$} \label{14.171}
\end{equation}
Now, taking into account the fact that
\begin{equation}
\mbox{tr}D(\Omega\chibh)=\sg^{-1}\cdot
D(\Omega\chibh)=-\Omega\chibh\cdot
D(\sg^{-1})=2(\Omega\chibh,\Omega\chih) \label{14.172}
\end{equation}
\ref{4.c3} gives:
\begin{equation}
D(\Omega\chibh)=\Omega^2\left\{\snab\oth\etb+\etb\oth\etb+\frac{1}{2}\mbox{tr}\chi\chibh-\frac{1}{2}\mbox{tr}\chib\chih
+\sg(\chih,\chibh)\right\} \label{14.173}
\end{equation}
Hence:
\begin{equation}
\sL_{O_i}D(\Omega\chibh)=\Omega^2\left\{-\frac{1}{2}\mbox{tr}\chib\sL_{O_i}\chih+\kappa_i\right\}
\label{14.174}
\end{equation}
where the $\kappa_i$ are the symmetric 2-covariant $S$
tensorfields:
\begin{eqnarray}
&&\kappa_i=\sL_{O_i}(\snab\oth\etb+\etb\oth\etb)+\sL_{O_i}\left(\frac{1}{2}\mbox{tr}\chi\chibh+\sg(\chih,\chibh)\right)-\frac{1}{2}(O_i\mbox{tr}\chib)\chih\nonumber\\
&&\hspace{8mm}
+2(O_i\log\Omega)\left\{\snab\oth\etb+\etb\oth\etb+\frac{1}{2}\mbox{tr}\chi\chibh-\frac{1}{2}\mbox{tr}\chib\chih
+\sg(\chih,\chibh)\right\} \label{14.175}
\end{eqnarray}
By Proposition 6.2 and the estimates of Chapters 3 and 4 together
with Propositions 8.1 and 8.2 the first term on the right in
\ref{14.175} is seen to be ${\cal O}^4(|u|^{-3})$. Also, by the
estimates of Chapters 3 and 4 and the estimate \ref{7.210}
together with Propositions 8.1 and 8.2 the remaining terms on the
right hand side of \ref{14.175} are seen to be ${\cal
O}^4(|u|^{-3})$ as well. We conclude that:
\begin{equation}
\kappa_i={\cal O}^4(|u|^{-3}) \label{14.176}
\end{equation}

Considering $D\vartheta_i$ along $C_{u_0}$ and substituting for
$D\sL_{O_i}(\Omega\chibh)$ along $C_{u_0}$ from \ref{14.171},
\ref{14.174} and for $D\s^{(O_i)}\spi$ along $C_{u_0}$ from
\ref{8.58} (and recalling the definition \ref{14.166} and the fact
that $Du=0$), we obtain the following propagation equation for
$\vartheta_i$ along $C_{u_0}$:
\begin{equation}
D\vartheta_i-\Omega\mbox{tr}\chi\vartheta_i=\varphi_i \ \ : \
\mbox{along $C_{u_0}$} \label{14.177}
\end{equation}
where the $\varphi_i$ are the symmetric 2-covariant $S$
tensorfields:
\begin{eqnarray}
&&\varphi_i=-2\varepsilon\sL_{O_i}(\Omega\chih)+2(1+\varepsilon)(O_i\Omega)\chih\nonumber\\
&&\hspace{8mm}-2u\Omega\mbox{tr}\chi\sL_{O_i}(\Omega\chibh)+\sg
O_i(\Omega\mbox{tr}\chi)+2u\Omega^2\kappa_i \label{14.178}
\end{eqnarray}
By the results of Chapters 3 and 4 together with Propositions 8.1
and 8.2 the first four terms on the right in \ref{14.178} are seen
to be ${\cal O}^4(\delta^{1/2}|u|^{-2})$, ${\cal
O}^4(\delta^{1/2}|u|^{-3})$, ${\cal O}^4(\delta^{1/2}|u|^{-3})$,
${\cal O}^4(|u|^{-2})$, respectively. The last term is ${\cal
O}^4(|u|^{-2})$ by \ref{14.176}. We conclude that:
\begin{equation}
\varphi_i={\cal O}^4(|u|^{-2}) \label{14.179}
\end{equation}
Noting that the $\vartheta_i$ vanish on $\Cb_0$, we apply Lemma
4.6 to the propagation equation \ref{14.177} along $C_{u_0}$,
taking $p=4$. Here $r=2$, $\nu=2$ and$\gamma=0$. We obtain:
\begin{equation}
\|\vartheta_i\|_{L^4(S_{\ub,u_0})}\leq
C\int_0^{\ub}\|\varphi_i\|_{L^4(S_{\ub^\prime,u_0})}d\ub^\prime
\label{14.180}
\end{equation}
Substituting \ref{14.179} then yields:
\begin{equation}
\|\vartheta_i\|_{L^4(S_{\ub,u_0})}\leq O(\delta|u_0|^{-3/2})
\label{14.181}
\end{equation}

Consider next $\Db\vartheta_i$ along each $\Cb_{\ub}$. Since
$[\Lb,O_i]=0$ on $M^\prime$ we have:
\begin{equation}
\Db\sL_{O_i}(\Omega\chibh)=\sL_{O_i}\Db(\Omega\chibh) \ \ : \
\mbox{on $M^\prime$} \label{14.182}
\end{equation}
Now, taking into account the fact that:
\begin{equation}
\mbox{tr}\Db(\Omega^{-1}\chibh)=\sg^{-1}\cdot\Db(\Omega^{-1}\chibh)=-\Omega^{-1}\chibh\cdot\Db(\sg^{-1})
=2|\chibh|^2 \label{14.183}
\end{equation}
\ref{14.160} gives:
\begin{equation}
\Db(\Omega\chibh)=\Omega^2\left(-\alb+\sg|\chibh|^2\right)+2\Omega\omb\chibh
\label{14.184}
\end{equation}
Considering $\Db\vartheta_i$ along each $\Cb_{\ub}$ and
substituting for $\Db\sL_{O_i}(\Omega\chibh)$ from \ref{14.182}
and $\sL_{O_i}$ applied to \ref{14.184} and for
$\Db\s^{(O_i)}\spi$ along $C_{u_0}$ from \ref{8.56} (and recalling
the definition \ref{14.166} and the fact that $\Db u=1$), we
obtain the following propagation equation for $\vartheta_i$ along
the $\Cb_{\ub}$:
\begin{equation}
\Db\vartheta_i-\Omega\mbox{tr}\chib\vartheta_i=\vphb_i
\label{14.185}
\end{equation}
where the $\vphb_i$ are the symmetric 2-covariant $S$
tensorfields:
\begin{eqnarray}
&&\vphb_i=-4\varepsilon\sL_{O_i}(\Omega\chibh)+\sg O_i(\Omega\mbox{tr}\chib)\nonumber\\
&&\hspace{8mm}+2u\sL_{O_i}\left\{\Omega^2\left(-\alb+\sg|\chibh|^2\right)+2\Omega\omb\chibh\right\}
\label{14.186}
\end{eqnarray}
The results of Chapters 3 and 4 together with Propositions 8.1 and
8.2 imply:
\begin{equation}
\vphb_i={\cal O}^4(\delta|u|^{-3}) \label{14.187}
\end{equation}
To the propagation equation \ref{14.185} we apply Lemma 4.7 taking
$p=4$. Here $r=2$, $\nu=2$, $\gammab=0$ and we obtain:
\begin{equation}
|u|^{-1/2}\|\vartheta_i\|_{L^4(S_{\ub,u})}\leq
C|u_0|^{-1/2}\|\vartheta_i\|_{L^4(S_{\ub,u_0})}
+C\int_{u_0}^u|u^\prime|^{-1/2}\|\vphb_i\|_{L^4(S_{\ub,u^\prime})}du^\prime
\label{14.188}
\end{equation}
Substituting \ref{14.181} and \ref{14.187} then yields:
\begin{equation}
\|\vartheta_i\|_{L^4(S_{\ub,u})}\leq O(\delta|u|^{-3/2}) \ \ : \
\forall (\ub,u)\in D^\prime_{c^*} \label{14.189}
\end{equation}
and the lemma is established.

\vspace{5mm}

Let us now go back to expressions \ref{14.143} - \ref{14.152}.
Consider first the case $X=L$. Using the estimates \ref{9.1} -
\ref{9.3} and \ref{14.156} as well as \ref{14.104} and the
estimates \ref{8.32} together with the estimates of Chapter 3, we
deduce:
\begin{eqnarray}
&&\Xi(\s^{(L)}q)=0\nonumber\\
&&\Xib(\s^{(L)}q)={\cal O}^4(\delta^{1/2}|u|^{-3})\nonumber\\
&&\Theta(\s^{(L)}q)={\cal O}^4(\delta^{-3/2}|u|^{-1})\nonumber\\
&&\Thetab(\s^{(L)}q)={\cal O}^4(\delta^{1/2}|u|^{-3})\nonumber\\
&&\Lambda(\s^{(L)}q)={\cal O}^4(\delta^{-1}|u|^{-2})\nonumber\\
&&\Lambdab(\s^{(L)}q)={\cal O}^4(|u|^{-2})\nonumber\\
&&K(\s^{(L)}q)=0\nonumber\\
&&\Kb(\s^{(L)}q)={\cal O}^4(|u|^{-3})\nonumber\\
&&I(\s^{(L)}q)={\cal O}^4(\delta^{-1/2}|u|^{-2})\nonumber\\
&&\Ib(\s^{(L)}q)={\cal O}^4(\delta^{-1/2}|u|^{-2}) \label{14.190}
\end{eqnarray}
Consider next the case $X=O_i:i=1,2,3$. Using the estimates
\ref{9.82} - \ref{9.84} and \ref{14.170} as well as \ref{14.105}
and the estimates \ref{8.139} together with the estimates of
Chapter 3, we deduce:
\begin{eqnarray}
&&\Xi(\s^{(O_i)}q)={\cal O}^4(\delta^{-1/2}|u|^{-1})\nonumber\\
&&\Xib(\s^{(O_i)}q)=0\nonumber\\
&&\Theta(\s^{(O_i)}q)={\cal O}^4(\delta^{-1/2}|u|^{-1})\nonumber\\
&&\Thetab(\s^{(O_i)}q)={\cal O}^4(\delta|u|^{-3})\nonumber\\
&&\Lambda(\s^{(O_i)}q)={\cal O}^4(|u|^{-2})\nonumber\\
&&\Lambdab(\s^{(O_i)}q)={\cal O}^4(\delta|u|^{-3})\nonumber\\
&&K(\s^{(O_i)}q)={\cal O}^4(|u|^{-2})\nonumber\\
&&\Kb(\s^{(O_i)}q)={\cal O}^4(\delta|u|^{-3})\nonumber\\
&&I(\s^{(O_i)}q)={\cal O}^4(\delta^{1/2}|u|^{-2})\nonumber\\
&&\Ib(\s^{(O_i)}q)={\cal O}^4(\delta^{1/2}|u|^{-2}) \label{14.191}
\end{eqnarray}
Consider finally the case $X=S$. Using the estimates \ref{9.12} -
\ref{9.14} and \ref{14.161} as well as \ref{14.106}  and the
estimates \ref{8.34} together with the estimates of Chapter 3, we
deduce:
\begin{eqnarray}
&&\Xi(\s^{(S)}q)={\cal O}^4(\delta^{-1/2}|u|^{-1})\nonumber\\
&&\Xib(\s^{(S)}q)={\cal O}^4(\delta^{3/2}|u|^{-3})\nonumber\\
&&\Theta(\s^{(S)}q)={\cal O}^4(\delta^{-1/2}|u|^{-1})\nonumber\\
&&\Thetab(\s^{(S)}q)={\cal O}^4(\delta^{3/2}|u|^{-3})\nonumber\\
&&\Lambda(\s^{(S)}q)={\cal O}^4(|u|^{-2})\nonumber\\
&&\Lambdab(\s^{(S)}q)={\cal O}^4(\delta|u|^{-3})\nonumber\\
&&K(\s^{(S)}q)={\cal O}^4(|u|^{-2})\nonumber\\
&&\Kb(\s^{(S)}q)={\cal O}^4(\delta|u|^{-3})\nonumber\\
&&I(\s^{(S)}q)={\cal O}^4(\delta^{1/2}|u|^{-2})\nonumber\\
&&\Ib(\s^{(S)}q)={\cal O}^4(\delta^{1/2}|u|^{-2}) \label{14.192}
\end{eqnarray}

The expression for each component of $\s^{(X)}J^3(W)$ given by
Lemma 14.2 is a sum of terms one factor of which is a component of
$\s^{(X)}q$ and the other factor a component of $W$. According to
the above, each component of $\s^{(X)}q$ is ${\cal
O}^4(\delta^{r_1}|u|^{p_1})$ for some $r_1, p_1$. Our attention is
at present confined to the case $W=R$. Viewing the components of
$W=R$ in a way which would be valid also in the case of the 2nd
order Weyl current error estimates we see each component as being
${\cal O}^4(\delta^{r_2}|u|^{p_2})$ according to \ref{14.107}. We
set, for each term in the expression of a given component of
$\s^{(X)}J^3(R)$,
\begin{equation}
r^\prime=r_1+r_2, \ \ \ p^\prime=p_1+p_2 \label{14.193}
\end{equation}
We then assign, to each component of $\s^{(X)}J^3(R)$, the pair
$r^*, p^*$, where $r^*$ is the minimal $r^\prime$ and $p^*$ is the
maximal $p^\prime$ occuring in the terms constituting the
expression of that component. We obtain in this way the following
tables. The ordinals in these tables refer to the terms in the
expressions of Lemma 14.2.

\vspace{10mm}

\hspace{40mm}{\large{1. {\bf Case} $X=L$}}

\vspace{5mm}

\hspace{30mm}$\Xi(\s^{(L)}J^3(R)):r^*=-2, p^*=-3$

\begin{equation}
\begin{array}{l|ll}
\mbox{term}&r^\prime&p^\prime\\ \hline
\mbox{1st}&-2&-3\\
\mbox{2nd}&-3/2&-4
\end{array}
\label{14.194}
\end{equation}
\hspace{40mm}(the remaining terms vanish)

\hspace{2.5mm}

\hspace{30mm}$\Xib(\s^{(L)}J^3(R)):r^*=1/2, p^*=-6$

\begin{equation}
\begin{array}{l|ll}
\mbox{term}&r^\prime&p^\prime\\ \hline
\mbox{1st}&1&-13/2\\
\mbox{2nd}&1&-6\\
\mbox{3rd}&1&-7\\
\mbox{4th}&1/2&-6\\
\mbox{5th}&1/2&-6
\end{array}
\label{14.195}
\end{equation}

\hspace{2.5mm}

\hspace{30mm}$\Theta(\s^{(L)}J^3(R)):r^*=-3/2, p^*=-3$

\begin{equation}
\begin{array}{l|ll}
\mbox{term}&r^\prime&p^\prime\\ \hline
\mbox{1st}&-3/2&-3\\
\mbox{2nd}&-3/2&-4\\
\mbox{3rd}&-1&-4\\
\mbox{4th}&-3/2&-4\\
\mbox{5th}&-3/2&-4
\end{array}
\label{14.196}
\end{equation}

\hspace{2.5mm}

\hspace{30mm}$\Thetab(\s^{(L)}J^3(R)):r^*=1/2, p^*=-6$

\begin{equation}
\begin{array}{l|ll}
\mbox{term}&r^\prime&p^\prime\\ \hline
\mbox{1st}&1/2&-13/2\\
\mbox{3rd}&1/2&-6\\
\mbox{4th}&1/2&-6\\
\mbox{5th}&1/2&-6
\end{array}
\label{14.197}
\end{equation}
\hspace{40mm}(the 2nd term vanishes)

\hspace{2.5mm}

\hspace{30mm}$\Lambda(\s^{(L)}J^3(R)):r^*=-1, p^*=-4$

\begin{equation}
\begin{array}{l|ll}
\mbox{term}&r^\prime&p^\prime\\ \hline
\mbox{1st}&-1&-4\\
\mbox{2nd}&-1&-5
\end{array}
\label{14.198}
\end{equation}
\hspace{40mm}(the remaining terms vanish)

\hspace{2.5mm}

\hspace{30mm}$\Lambdab(\s^{(L)}J^3(R)):r^*=0, p^*=-5$

\begin{equation}
\begin{array}{l|ll}
\mbox{term}&r^\prime&p^\prime\\ \hline
\mbox{1st}&0&-11/2\\
\mbox{2nd}&0&-5\\
\mbox{3rd}&0&-6\\
\mbox{4th}&0&-5
\end{array}
\label{14.199}
\end{equation}

\hspace{2.5mm}

\hspace{30mm}$K(\s^{(L)}J^3(R)):r^*=-1, p^*=-4$

\begin{equation}
\begin{array}{l|ll}
\mbox{term}&r^\prime&p^\prime\\ \hline
\mbox{1st}&-1&-4\\
\mbox{2nd}&-1&-4\\
\mbox{3rd}&-1&-5
\end{array}
\label{14.200}
\end{equation}
\hspace{40mm}(the remaining terms vanish)

\hspace{2.5mm}

\hspace{30mm}$\Kb(\s^{(L)}J^3(R)):r^*=0, p^*=-5$

\begin{equation}
\begin{array}{l|ll}
\mbox{term}&r^\prime&p^\prime\\ \hline
\mbox{1st}&0&-11/2\\
\mbox{2nd}&1/2&-6\\
\mbox{3rd}&0&-5\\
\mbox{4th}&0&-6\\
\mbox{5th}&0&-5
\end{array}
\label{14.201}
\end{equation}

\hspace{2.5mm}

\hspace{30mm}$I(\s^{(L)}J^3(R)):r^*=-1, p^*=-4$

\begin{equation}
\begin{array}{l|ll}
\mbox{term}&r^\prime&p^\prime\\ \hline
\mbox{1st}&-1&-4\\
\mbox{2nd}&-1/2&-4\\
\mbox{3rd}&-1/2&-5\\
\mbox{4th}&-1/2&-5\\
\mbox{5th}&-1/2&-5
\end{array}
\label{14.202}
\end{equation}

\hspace{2.5mm}

\hspace{30mm}$\Ib(\s^{(L)}J^3(R)):r^*=-1/2, p^*=-5$

\begin{equation}
\begin{array}{l|ll}
\mbox{term}&r^\prime&p^\prime\\ \hline
\mbox{2nd}&0&-6\\
\mbox{3rd}&-1/2&-5\\
\mbox{4th}&-1/2&-5\\
\mbox{5th}&0&-5
\end{array}
\label{14.203}
\end{equation}
\hspace{40mm}(the 1st term vanishes)

\vspace{10mm}

\hspace{30mm}{\large{2. {\bf Case} $X=O_i:i=1,2,3$}}

\vspace{5mm}

\hspace{30mm}$\Xi(\s^{(O_i)}J^3(R)):r^*=-1, p^*=-3$

\begin{equation}
\begin{array}{l|ll}
\mbox{term}&r^\prime&p^\prime\\ \hline
\mbox{1st}&-1&-3\\
\mbox{2nd}&-1/2&-4\\
\mbox{3rd}&-1/2&-4\\
\mbox{4th}&-1/2&-4\\
\mbox{5th}&-1/2&-4
\end{array}
\label{14.204}
\end{equation}

\hspace{2.5mm}

\hspace{30mm}$\Xib(\s^{(O_i)}J^3(R)):r^*=2, p^*=-13/2$

\begin{equation}
\begin{array}{l|ll}
\mbox{term}&r^\prime&p^\prime\\ \hline
\mbox{1st}&2&-13/2\\
\mbox{2nd}&2&-7\\
\mbox{3rd}&2&-7
\end{array}
\label{14.205}
\end{equation}
\hspace{40mm}(the remaining terms vanish)

\hspace{2.5mm}

\hspace{30mm}$\Theta(\s^{(O_i)}J^3(R)):r^*=-1/2, p^*=-4$

\begin{equation}
\begin{array}{l|ll}
\mbox{term}&r^\prime&p^\prime\\ \hline
\mbox{1st}&-1/2&-4\\
\mbox{2nd}&-1/2&-4\\
\mbox{3rd}&0&-4\\
\mbox{4th}&-1/2&-4\\
\mbox{5th}&-1/2&-4
\end{array}
\label{14.206}
\end{equation}

\hspace{2.5mm}

\hspace{30mm}$\Thetab(\s^{(O_i)}J^3(R)):r^*=1, p^*=-6$

\begin{equation}
\begin{array}{l|ll}
\mbox{term}&r^\prime&p^\prime\\ \hline
\mbox{1st}&3/2&-13/2\\
\mbox{2nd}&3/2&-13/2\\
\mbox{3rd}&3/2&-6\\
\mbox{4th}&1&-6\\
\mbox{5th}&1&-6
\end{array}
\label{14.207}
\end{equation}

\hspace{2.5mm}

\hspace{30mm}$\Lambda(\s^{(O_i)}J^3(R)):r^*=-1/2, p^*=-4$

\begin{equation}
\begin{array}{l|ll}
\mbox{term}&r^\prime&p^\prime\\ \hline
\mbox{1st}&-1/2&-4\\
\mbox{2nd}&0&-5\\
\mbox{3rd}&0&-5\\
\mbox{4th}&1/2&-5
\end{array}
\label{14.208}
\end{equation}

\hspace{2.5mm}

\hspace{30mm}$\Lambdab(\s^{(O_i)}J^3(R)):r^*=1, p^*=-11/2$

\begin{equation}
\begin{array}{l|ll}
\mbox{term}&r^\prime&p^\prime\\ \hline
\mbox{1st}&1&-11/2\\
\mbox{2nd}&1&-6\\
\mbox{3rd}&1&-6
\end{array}
\label{14.209}
\end{equation}
\hspace{40mm}(the 4th term vanishes)

\hspace{2.5mm}

\hspace{30mm}$K(\s^{(O_i)}J^3(R)):r^*=-1/2, p^*=-4$

\begin{equation}
\begin{array}{l|ll}
\mbox{term}&r^\prime&p^\prime\\ \hline
\mbox{1st}&-1/2&-4\\
\mbox{2nd}&0&-4\\
\mbox{3rd}&0&-5\\
\mbox{4th}&0&-5\\
\mbox{5th}&1/2&-5
\end{array}
\label{14.210}
\end{equation}

\hspace{2.5mm}

\hspace{30mm}$\Kb(\s^{(O_i)}J^3(R)):r^*=1, p^*=-11/2$

\begin{equation}
\begin{array}{l|ll}
\mbox{term}&r^\prime&p^\prime\\ \hline
\mbox{1st}&1&-11/2\\
\mbox{2nd}&3/2&-6\\
\mbox{3rd}&1&-6\\
\mbox{4th}&1&-6
\end{array}
\label{14.211}
\end{equation}
\hspace{40mm}(the 5th term vanishes)

\hspace{2.5mm}

\hspace{30mm}$I(\s^{(O_i)}J^3(R)):r^*=1/2, p^*=-5$

\begin{equation}
\begin{array}{l|ll}
\mbox{term}&r^\prime&p^\prime\\ \hline
\mbox{2nd}&1/2&-5\\
\mbox{3rd}&1/2&-5\\
\mbox{4th}&1/2&-5\\
\mbox{5th}&1/2&-5
\end{array}
\label{14.212}
\end{equation}
\hspace{40mm}(the 1st term vanishes)

\hspace{2.5mm}

\hspace{30mm}$\Ib(\s^{(O_i)}J^3(R)):r^*=1/2, p^*=-5$

\begin{equation}
\begin{array}{l|ll}
\mbox{term}&r^\prime&p^\prime\\ \hline
\mbox{1st}&1&-11/2\\
\mbox{2nd}&1&-6\\
\mbox{3rd}&1/2&-5\\
\mbox{4th}&1/2&-5\\
\mbox{5th}&1/2&-5
\end{array}
\label{14.213}
\end{equation}

\vspace{10mm}

\hspace{40mm}{\large{3. {\bf Case} $X=S$}}

\vspace{2.5mm}

In this case only the case $n=3$ occurs, therefore we only have to
consider $\Thetab(\s^{(S)}J^3(R))$ and $\Xib(\s^{(S)}J^3(R))$ (see
\ref{14.14}).

\hspace{2.5mm}

\hspace{30mm}$\Xib(\s^{(S)}J^3(R)):r^*=3/2, p^*=-6$

\begin{equation}
\begin{array}{l|ll}
\mbox{term}&r^\prime&p^\prime\\ \hline
\mbox{1st}&2&-13/2\\
\mbox{2nd}&2&-7\\
\mbox{3rd}&2&-7\\
\mbox{4th}&3/2&-6\\
\mbox{5th}&3/2&-6
\end{array}
\label{14.214}
\end{equation}
\hspace{40mm}

\hspace{2.5mm}

\hspace{30mm}$\Thetab(\s^{(S)}J^3(R)):r^*=3/2, p^*=-6$

\begin{equation}
\begin{array}{l|ll}
\mbox{term}&r^\prime&p^\prime\\ \hline
\mbox{1st}&3/2&-13/2\\
\mbox{2nd}&3/2&-13/2\\
\mbox{3rd}&3/2&-6\\
\mbox{4th}&3/2&-6\\
\mbox{5th}&3/2&-6
\end{array}
\label{14.215}
\end{equation}

We have completed the investigation of the components of the Weyl
currents $\s^{(L)}J^3(R)$, $\s^{(O_i)}J^3(R):i=1,2,3$,  and
$\s^{(S)}J^3(R)$. These currents contribute to the error terms
$\s^{(L)}\stackrel{(n)}{\tau}_c:n=0,1,2,3$,
$\s^{(O)}\stackrel{(n)}{\tau}_c:n=0,1,2,3$, and
$\s^{(S)}\stackrel{(3)}{\tau}_c$, respectively, according to
\ref{14.12}, \ref{14.13}, and \ref{14.14}. We call these
contributions $\s^{(L)}\stackrel{(n)}{\tau}_{c,3}:n=0,1,2,3$,
$\s^{(O)}\stackrel{(n)}{\tau}_{c,3}:n=0,1,2,3$, and
$\s^{(S)}\stackrel{(3)}{\tau}_{c,3}$, respectively.

Each component of the Weyl currents $\s^{(L)}J^3(R)$,
$\s^{(O_i)}J^3(R):i=1,2,3$, and $\s^{(S)}J^3(R)$ being expressed
as a sum of terms by Lemma 14.2, and these expressions being
substituted into \ref{14.12}, \ref{14.13}, and \ref{14.14}, sums
of trilinear terms result, two of the factors in each term being
contributed by the expression for a component of the Weyl current
$\s^{(L)}J^3(R)$ in the case of \ref{14.12},
$\s^{(O_i)}J^3(R):i=1,2,3$ in the case of \ref{14.13}, and
$\s^{(S)}J^3(R)$ in the case of \ref{14.14}, and the other factor
being a component of $\tcL_L R$, $\tcL_{O_i}R:i=1,2,3$, and
$\tcL_S R$, respectively, multiplied by $\Omega^3$ and the
appropriate power of $|u|^2$. Viewing these third factors in a way
which would be valid also in the case of the 2nd order Weyl
current error estimates, we see each third factor as being either
${\bf O}(\delta^{r_3}|u|^{p_3})$ in the case of factors involving
$\alpha(\tcL_X R),\beta(\tcL_X R),\rho(\tcL_X R),\sigma(\tcL_X
R),\beb(\tcL_X R)$, or $\bfob(\delta^{r_3}|u|^{p_3})$ in the case of factors
involving $\alb(\tcL_X R)$. The values of $r_3$ and $p_3$ assigned
are those implied by the bound on the quantity ${\cal P}_1$. To
each trilinear term we can then apply accordingly one of the last
two cases of Lemma 13.1. Since the pair $r^\prime,p^\prime$
corresponding to the first two factors of each trilinear term may
be replaced by the corresponding $r^*,p^*$, we then obtain a bound
for the contribution of all terms resulting from the product of a
given component of $\s^{(X)}J^3$ with a given component of $\tcL_X
R$ to the corresponding error integral
\begin{equation}
\delta^{2q_n+2l}\int_{M^\prime_{c^*}}|\s^{(X)}\stackrel{(n)}{\tau}_{c,3}|d\mu_g
\label{14.216}
\end{equation}
by $O(\delta^e)$, where $e$ is the excess index, defined as in
\ref{14.132}, provided that the integrability index $s$, defined
as in \ref{14.133}, is negative so that Lemma 13.1 applies. We
obtain in this way the following tables. The ordinals in these
tables refer to the terms on the right hand side of each of
\ref{14.12} - \ref{14.14}.

\vspace{10mm}

\hspace{30mm}{\large{1. {\bf Case} $X=L \ : \ l=1$}}

\vspace{2.5mm}

\hspace{45mm}{\large{$\s^{(L)}\stackrel{(0)}{\tau}_{c,3}$}}

\begin{equation}
\begin{array}{l|ll}
\mbox{term}&e&s\\ \hline
\mbox{1st}&1&-1\\
\mbox{2nd}&3/2&-2\\
\end{array}
\label{14.217}
\end{equation}

\vspace{2.5mm}

\hspace{45mm}{\large{$\s^{(L)}\stackrel{(1)}{\tau}_{c,3}$}}

\begin{equation}
\begin{array}{l|ll}
\mbox{term}&e&s\\ \hline
\mbox{1st}&1&-2\\
\mbox{2nd}&1&-2\\
\mbox{3rd}&1/2&-1
\end{array}
\label{14.218}
\end{equation}

\vspace{2.5mm}

\hspace{45mm}{\large{$\s^{(L)}\stackrel{(2)}{\tau}_{c,3}$}}

\begin{equation}
\begin{array}{l|ll}
\mbox{term}&e&s\\ \hline
\mbox{1st}&1&-1\\
\mbox{2nd}&1&-1\\
\mbox{3rd}&3/2&-2
\end{array}
\label{14.219}
\end{equation}

\vspace{2.5mm}

\hspace{45mm}{\large{$\s^{(L)}\stackrel{(3)}{\tau}_{c,3}$}}

\begin{equation}
\begin{array}{l|ll}
\mbox{term}&e&s\\ \hline
\mbox{1st}&1&-3/2\\
\mbox{2nd}&1/2&-1\\
\end{array}
\label{14.220}
\end{equation}

\vspace{10mm}

\hspace{30mm}{\large{2. {\bf Case} $X=O \ : \ l=0$}}

\vspace{2.5mm}

\hspace{45mm}{\large{$\s^{(O)}\stackrel{(0)}{\tau}_{c,3}$}}

\begin{equation}
\begin{array}{l|ll}
\mbox{term}&e&s\\ \hline
\mbox{1st}&1&-2\\
\mbox{2nd}&3/2&-2\\
\end{array}
\label{14.221}
\end{equation}

\vspace{2.5mm}

\hspace{45mm}{\large{$\s^{(O)}\stackrel{(1)}{\tau}_{c,3}$}}

\begin{equation}
\begin{array}{l|ll}
\mbox{term}&e&s\\ \hline
\mbox{1st}&1/2&-2\\
\mbox{2nd}&1/2&-2\\
\mbox{3rd}&1&-2
\end{array}
\label{14.222}
\end{equation}

\vspace{2.5mm}

\hspace{45mm}{\large{$\s^{(O)}\stackrel{(2)}{\tau}_{c,3}$}}

\begin{equation}
\begin{array}{l|ll}
\mbox{term}&e&s\\ \hline
\mbox{1st}&1&-3/2\\
\mbox{2nd}&1&-3/2\\
\mbox{3rd}&3/2&-2
\end{array}
\label{14.223}
\end{equation}

\vspace{2.5mm}

\hspace{45mm}{\large{$\s^{(O)}\stackrel{(3)}{\tau}_{c,3}$}}

\begin{equation}
\begin{array}{l|ll}
\mbox{term}&e&s\\ \hline
\mbox{1st}&1/2&-3/2\\
\mbox{2nd}&1&-3/2\\
\end{array}
\label{14.224}
\end{equation}

\vspace{10mm}

\hspace{30mm}{\large{3. {\bf Case} $X=S \ : \ l=0$}}

\vspace{2.5mm}

\hspace{45mm}{\large{$\s^{(S)}\stackrel{(3)}{\tau}_{c,3}$}}

\begin{equation}
\begin{array}{l|ll}
\mbox{term}&e&s\\ \hline
\mbox{1st}&1&-3/2\\
\mbox{2nd}&1/2&-1\\
\end{array}
\label{14.225}
\end{equation}

\vspace{10mm}

We see that all terms have negative integrability index so that
Lemma 13.1 indeed applies. All terms have positive excess index,
so there are no borderline terms.

We summarize the results of this section in the following
proposition.

\vspace{5mm}

\noindent{\bf Proposition 14.3} The 1st order Weyl current error
integrals arising from $J^3$, \ref{14.216}, satisfy the estimates:
\begin{eqnarray*}
&&\delta^{2q_0+2}\int_{M^\prime_{c^*}}|\s^{(L)}\stackrel{(0)}{\tau}_{c,3}|d\mu_g\leq O(\delta)\\
&&\delta^{2q_1+2}\int_{M^\prime_{c^*}}|\s^{(L)}\stackrel{(1)}{\tau}_{c,3}|d\mu_g\leq O(\delta^{1/2})\\
&&\delta^{2q_2+2}\int_{M^\prime_{c^*}}|\s^{(L)}\stackrel{(2)}{\tau}_{c,3}|d\mu_g\leq O(\delta)\\
&&\delta^{2q_3+2}\int_{M^\prime_{c^*}}|\s^{(L)}\stackrel{(3)}{\tau}_{c,3}|d\mu_g\leq
O(\delta^{1/2})
\end{eqnarray*}
\begin{eqnarray*}
&&\delta^{2q_0}\int_{M^\prime_{c^*}}|\s^{(O)}\stackrel{(0)}{\tau}_{c,3}|d\mu_g\leq O(\delta)\\
&&\delta^{2q_1}\int_{M^\prime_{c^*}}|\s^{(O)}\stackrel{(1)}{\tau}_{c,3}|d\mu_g\leq O(\delta^{1/2})\\
&&\delta^{2q_2}\int_{M^\prime_{c^*}}|\s^{(O)}\stackrel{(2)}{\tau}_{c,3}|d\mu_g\leq O(\delta)\\
&&\delta^{2q_3}\int_{M^\prime_{c^*}}|\s^{(O)}\stackrel{(3)}{\tau}_{c,3}|d\mu_g\leq
O(\delta^{1/2})
\end{eqnarray*}
and:
$$\delta^{2q_3}\int_{M^\prime_{c^*}}|\s^{(S)}\stackrel{(3)}{\tau}_{c,3}|d\mu_g\leq O(\delta^{1/2})$$

\vspace{5mm}

We have now completed the 1st order Weyl current error estimates.

\chapter{The 2nd Order Weyl Current Error Estimates}

\section{The 2nd order estimates which are of the same form as the 1st order estimates}

In the present section we shall estimate the contributions to the
2nd order Weyl current error integrals of the terms
$$\s^{(Y)}J^1(\tcL_X R)+\s^{(Y)}J^2(\tcL_X R)+\s^{(Y)}J^3(\tcL_X R)$$
in $\s^{(YX)}J$,
$(Y,X)=(L,L),(O_i,L):i=1,2,3,(O_j,O_i):i,j=1,2,3,(O_i,S):i=1,2,3,(S,S)$
(see \ref{14.15}). These contributions are identical in form to
the 1st order Weyl current error integrals. Thus the contribution
to $\s^{(LL)}\stackrel{(n)}{\tau}_c$ is identical in form to
$\s^{(L)}\stackrel{(n)}{\tau}_c$ but with the derived Weyl field
$\tcL_L R$ in the role of the fundamental Weyl field $R$. The
contribution to $\s^{(OL)}\stackrel{(n)}{\tau}_c$ is identical in
form to $\s^{(O)}\stackrel{(n)}{\tau}_c$ but with $\tcL_L R$ in
the role of $R$. The contribution to
$\s^{(OO)}\stackrel{(n)}{\tau}_c$ is identical in form to
$\s^{(O)}\stackrel{(n)}{\tau}_c$ but with the derived Weyl fields
$\tcL_{O_i}R:i=1,2,3$ in the role of the fundamental Weyl field
$R$. Finally, the contributions to
$\s^{(OS)}\stackrel{(3)}{\tau}_c$ and to
$\s^{(SS)}\stackrel{(3)}{\tau}_c$ are identical in form to
$\s^{(O)}\stackrel{(3)}{\tau}_c$ and
$\s^{(S)}\stackrel{(3)}{\tau}_c$ but with the derived Weyl field
$\tcL_S R$ in the role of the fundamental Weyl field $R$. The {\em
genuine} 2nd order Weyl current error integrals, namely the
contributions to the 2nd order Weyl current error integrals of the
term
$$\tcL_Y\s^{(X)}J$$
in $\s^{(YX)}J$ shall be estimated in the next section. We pay
special attention to the terms giving rise to borderline error
integrals.

We first consider the contributions to the 2nd order Weyl current
error integrals of the term
$$\s^{(Y)}J^1(\tcL_X R)$$
in $\s^{(YX)}J$,
$(Y,X)=(L,L),(O_i,L):i=1,2,3,(O_j,O_i):i,j=1,2,3,(O_i,S):i=1,2,3,(S,S)$.
As we have seen in the previous chapter, it is the Weyl current
$J^1$ which contributes borderline error integrals. We consider
the expressions for the components of $\s^{(Y)}J^1(\tcL_X R)$
given by Proposition 14.1 with the vectorfield $Y$ in the role of
the vectorfield $X$ and the derived Weyl field $\tcL_X R$ in the
role of the Weyl field $W$. The expression for each component of
$\s^{(Y)}J^1(\tcL_X R)$ is a sum of terms one factor of which is a
component of $\s^{(Y)}\tilde{\pi}$. We consider, as before, the
terms with the same such factor as a single term.

Consider first the cases $X=L, O_i:i=1,2,3$. As we discussed in
the previous chapter, in these expressions we are to substitute
for:
$$\Dbh\alpha(\tcL_X R),\Db\beta(\tcL_X R),\Db\rho(\tcL_X R),\Db\sigma(\tcL_X R), \Db\beb(\tcL_X R)$$
from the inhomogeneous Bianchi equations of Proposition 12.4 with
$\tcL_X R$ in the role of the Weyl field $W$ and $\s^{(X)}J$ in
the role of the corresponding Weyl current $J$. Then the above
will involve:
$$\Theta(\s^{(X)}J),I(\s^{(X)}J),\Lambdab(\s^{(X)}J),\Kb(\s^{(X)}J),\Xib(\s^{(X)}J)$$
respectively. After this substitution, the other factor of each
term in the expression of a given component of $\s^{(Y)}J^1(\tcL_X
R)$ becomes the sum of a principal part, which is a sum of $\snab$
and $D$ derivatives of components of $\tcL_X R$, and
$\Dbh\alb(\tcL_X R)$,
 a non-principal part, which is a
sum of terms involving two factors, one of which is a connection
coefficient and the other a component of $\tcL_X R$, and terms
involving the above components of $\s^{(X)}J$.

We estimate
$$\snab\alpha(\tcL_X R),\snab\beta(\tcL_X R),\sd\rho(\tcL_X R),\sd\sigma(\tcL_X R),\snab\beb(\tcL_X R),\snab\alb(\tcL_X R)$$
through Proposition 11.1 (see also \ref{11.124}) in terms of
$$\sLh_{O_j}\alpha(\tcL_X R),\sL_{O_j}\beta(\tcL_X R),O_j\rho(\tcL_X R),O_j\sigma(\tcL_X R),\sL_{O_j}\beb(\tcL_X R),
\sLh_{O_j}\alb(\tcL_X R)$$ ($j=1,2,3$) respectively, and the
latter through Proposition 12.2, using also the $L^\infty$ bounds
for the components of $\s^{(O_j)}\tilde{\pi}$ to estimate the
remainders, in terms of
$$\alpha(\tcL_{O_j}\tcL_X R),\beta(\tcL_{O_j}\tcL_X R),\rho(\tcL_{O_j}\tcL_X R),\sigma(\tcL_{O_j}\tcL_X R),\beb(\tcL_{O_j}\tcL_X R),
\alb(\tcL_{O_j}\tcL_X R)$$ ($j=1,2,3$) respectively. The last are
bounded according to the bound on the quantity ${\cal P}_2$.

In the case $X=L$ we estimate
$$\Dh\alpha(\tcL_L R),D\beta(\tcL_L R),D\rho(\tcL_L R),D\sigma(\tcL_L R),D\beb(\tcL_L R),D\alb(\tcL_L R)$$
through Proposition 12.2, using also the $L^\infty$ bounds for the
components of $\s^{(L)}\tilde{\pi}$ to estimate the remainders, in
terms of
$$\alpha(\tcL_L\tcL_L R),\beta(\tcL_L\tcL_L R),\rho(\tcL_L\tcL_L R),\sigma(\tcL_L\tcL_L R),\beb(\tcL_L\tcL_L R),\alb(\tcL_L\tcL_L R)$$
respectively. The latter are bounded according to the bound on the
quantity ${\cal P}_2$.

In the case $X=O_i:i=1,2,3$, to estimate
$$\Dh\alpha(\tcL_{O_i}R),D\beta(\tcL_{O_i}R),D\rho(\tcL_{O_i}R),D\sigma(\tcL_{O_i}R),D\beb(\tcL_{O_i}R),\Dh\alb(\tcL_{O_i}R)$$
we first express these though Proposition 12.2 in terms of
$$\Dh\sLh_{O_i}\alpha,D\sL_{O_i}\beta,DO_i\rho,DO_i\sigma,D\sL_{O_i}\beb,\Dh\sLh_{O_i}\alb$$
respectively. The remainders are bounded using the $L^\infty$
bounds for the components of $\s^{(O_i)}\tilde{\pi}$ and the
$L^4(S)$ bounds for their $D$ derivatives. We then express, using
Lemma 1.3 and the commutator \ref{8.59},
\begin{eqnarray}
&&\Dh\sLh_{O_i}\alpha=\sLh_{O_i}\Dh\alpha+\sLh_{Z_i}\alpha+\s^{(O_i)}\ih(\Omega\chih,\alpha)-\Omega\chih(\s^{(O_i)}\ih,\alpha)\nonumber\\
&&D\sL_{O_i}\beta=\sL_{O_i}D\beta+\sL_{Z_i}\beta\nonumber\\
&&DO_i\rho=O_i D\rho+Z_i\rho\nonumber\\
&&DO_i\sigma=O_i D\sigma+Z_i\sigma\nonumber\\
&&D\sL_{O_i}\beb=\sL_{O_i}D\beb+\sL_{Z_i}\beb\nonumber\\
&&\Dh\sLh_{O_i}\alb=\sLh_{O_i}\Dh\alb+\sLh_{Z_i}\alb+\s^{(O_i)}\ih(\Omega\chih,\alb)-\Omega\chih(\s^{(O_i)}\ih,\alb)
\label{15.1}
\end{eqnarray}
The second term on the right in each of the above is estimated
using Propositions 8.4 and 9.2 and the bound on the quantity
${\cal Q}_1$. The third term present on the right in the first and
last of the above is estimated using the $L^\infty$ bounds for
$\s^{(O_i)}\ih$ and $\Omega\chih$. The first terms on the right in
the first five of \ref{15.1} are bounded through the inequalities
\ref{12.164} after substituting on the right the inequalities
\ref{12.166}, \ref{12.167} and \ref{12.175}. In view of Lemma 12.6
we then obtain bounds in terms of the quantity ${\cal P}_2$. The
first term in the last of \ref{15.1} is bounded through the
inequality \ref{12.181} after substituting on the right the
inequalities \ref{12.182}, \ref{12.83} and \ref{12.192}. In view
of Lemma 12.6 we then obtain bounds in terms of the quantity
${\cal P}_2$.

Considering still the cases $X=L, O_i:i=1,2,3$, we must also
appropriately estimate $\Dbh\alb(\tcL_X R)$. In the case $X=L$ we
express $\Dbh\alb(\tcL_L R)$ through Proposition 12.2 in terms of
$\Dbh\Dh\alb$. The remainder is bounded using the $L^\infty$
bounds for the components of $\s^{(L)}\tilde{\pi}$ and the
$L^4(S)$ estimates for their $\Db$ derivatives. Then $\Dbh\Dh\alb$
is bounded through the inequality \ref{12.227}. In view of Lemma
12.6 we then obtain a bound in terms of the quantity ${\cal P}_2$.
In the case $X=O_i:i=1,2,3$ we express $\Dbh\alb(\tcL_{O_i}R)$ in
terms of $\Dh\alb(\tcL_{O_i}R)$ and $\sLh_S\alb(\tcL_{O_i}R)$. We
already discussed how the former is estimated. To estimate
$\sLh_S\alb(\tcL_{O_i}R)$ we express it through Proposition 12.2
in terms of $\sLh_S\sLh_{O_i}\alb$. The remainder is bounded using
the $L^\infty$ bounds for the components of
$\s^{(O_i)}\tilde{\pi}$ and the $L^4(S)$ bounds for their $\sL_S$
derivatives (which are expressed in terms of $D$ and $\Db$
derivatives). Using the commutation relation
\begin{equation}
[S,O_i]=\ub Z_i \label{15.2}
\end{equation}
which follows from \ref{8.59} and the first of \ref{8.38}, we then
express, using Lemma 1.3,
\begin{equation}
\sLh_S\sLh_{O_i}\alb=\sLh_{O_i}\sLh_S\alb+\ub\sLh_{Z_i}\alb+\s^{(O_i)}\ih(\s^{(S)}\ih,\alb)-\s^{(S)}\ih(\s^{(O_i)}\ih,\alb)
\label{15.3}
\end{equation}
The second term on the right is estimated using Propositions 8.4
and 9.2 and the bound on the quantity ${\cal Q}_1$. The third term
on the right is estimated using the $L^\infty$ bounds for
$\s^{(O_i)}\ih$ and $\s^{(S)}\ih$. Finally, in view of
\ref{12.180}, the first term on the right is estimated through the
inequality \ref{12.181} after substituting on the right the
inequalities \ref{12.182}, \ref{12.83} and \ref{12.192}. In view
of Lemma 12.6 we then obtain a  bound in terms of the quantity
${\cal P}_2$.

Consider finally the case $X=S$. In this case only the case $n=3$
occurs, so we only have to consider the expressions for
$$\Xib(\s^{(Y)}J^1(\tcL_S R)) \ \ \mbox{and} \ \ \Thetab(\s^{(Y)}J^1(\tcL_S R))$$
given by Proposition 14.1 with $Y=S, O_i:i=1,2,3$ in the role of
$X$ and $\tcL_S R$ in the role of $W$. These expressions contain
\begin{eqnarray*}
&\snab\beb(\tcL_S R),\snab\alb(\tcL_S R),\\
&D\beb(\tcL_S R),\Dh\alb(\tcL_S R),\Db\beb(\tcL_S
R),\Dbh\alb(\tcL_S R)
\end{eqnarray*}
We estimate
$$\snab\beb(\tcL_S R),\snab\alb(\tcL_S R)$$
through Proposition 11.1 (see also \ref{11.124}) in terms of
$$\sL_{O_i}\beb(\tcL_S R),\sLh_{O_i}\alb(\tcL_S R) \ ; i=1,2,3$$
respectively, and the latter through Proposition 12.2, using also
the $L^\infty$ bounds for the components of
$\s^{(O_i)}\tilde{\pi}$ to estimate the remainders, in terms of
$$\beb(\tcL_{O_i}\tcL_S R),\alb(\tcL_{O_i}\tcL_S R) \ ; i=1,2,3$$
respectively. The last are bounded according to the bound on the
quantity ${\cal P}_2$.

To estimate $D\beb(\tcL_S R)$ we express it through the
inhomogeneous Bianchi equations (Proposition 12.4), with $\tcL_S R$
in the role of $W$ and $\s^{(S)}J$ in the role of $J$, in terms of
$\sd\rho(\tcL_S R),\sd\sigma(\tcL_S R)$, which are estimated
through Proposition 11.1 in terms of $O_i\rho(\tcL_S
R),O_i\sigma(\tcL_S R) ;i=1,2,3$ respectively. The last are
estimated by \ref{13.66}. There is also a source term
$\Ib(\s^{(S)}J)$ in the inhomogeneous Bianchi equation in
question, which is estimated following the approach of Chapter 14.

To estimate $\Dh\alb(\tcL_S R)$ we express it through the
inhomogeneous Bianchi equations, with $\tcL_S R$ in the role of $W$
and $\s^{(S)}J$ in the role of $J$, in terms of $\snab\beb(\tcL_S
R)$, which is estimated through Proposition 11.1 in terms of
$\sL_{O_i}\beb(\tcL_S R)$. The latter is in turn estimated through
Proposition 12.2 in terms of $\beb(\tcL_{O_i}\tcL_S R)$. The last
is bounded according to the bound on the quantity ${\cal P}_2$.
There is also a source term $\Thetab(\s^{(S)}J)$ in the
inhomogeneous Bianchi equation in question, which has already been
estimated in Chapter 14.

To estimate $\Dbh\beb(\tcL_S R)$ we express it through the
inhomogeneous Bianchi equations (Proposition 12.4) with $\tcL_S R$
in the role of $W$ and $\s^{(S)}J$ in the role of $J$ in terms of
$\snab\alb(\tcL_S R)$. We have already discussed how the latter is
estimated. There is also a source term $\Xib(\s^{(S)}J)$ in the
inhomogeneous Bianchi equation in question, which has already been
estimated in Chapter 14.

Finally, to estimate $\Dbh\alb(\tcL_S R)$ we express it in terms
of $\Dh\alb(\tcL_S R)$ and $\sLh_S\alb(\tcL_S R)$. We have already
discussed how the former is estimated. We estimate the latter
through Proposition 12.2 in terms of $\alb(\tcL_S\tcL_S R)$. The
last is bounded according to the bound on the quantity ${\cal
P}_2$.

According to the above discussion, the other factor in the
expression of a given component of $\s^{(Y)}J^1(\tcL_X R)$ is a
sum of terms each of which may be viewed as being either ${\bf
O}(\delta^r|u|^p)$ or $\bfob(\delta^r|u|^p)$. We then define $r_2$
to be the minimal $r$ and $p_2$ to be the maximal $p$ occuring in
the terms of the other factor of a term involving a given
component of $\s^{(Y)}\tilde{\pi}$ in the expression of a given
component of $\s^{(Y)}J^1(\tcL_X R)$. We then set, for each term
involving a given component of $\s^{(Y)}\tilde{\pi}$, which is
${\cal O}^\infty(\delta^{r_1}|u|^{p_1})$, in the expression of a
given component of $\s^{(Y)}J^1(\tcL_X R)$,
\begin{equation}
r^\prime=r_1+r_2, \ \ \ p^\prime=p_1+p_2 \label{15.4}
\end{equation}
We then assign, to each component of $\s^{(Y)}J^1(\tcL_X R)$, the pair
$r^*,p^*$, where $r^*$ is the minimal $r^\prime$ and $p^*$ is the
maximal $p^\prime$ occuring in the four terms constituting the
expression of that component. We obtain in this way tables which
are similar to the tables \ref{14.28} - \ref{14.31}, \ref{14.33},
\ref{14.35}, \ref{14.36}, \ref{14.38} - \ref{14.40}, and
\ref{14.41} - \ref{14.44}, \ref{14.46}, \ref{14.48}, \ref{14.49},
\ref{14.51} - \ref{14.53}, and \ref{14.54}, \ref{14.55}. In fact,
the tables for the case $Y=X=L$ ($l=2$) are identical to the
tables for the case $X=L$ ($l=1$) of Chapter 14, but with the
values of $r^\prime$ and $r^*$ decreased by 1. The tables for the
case $Y=O_i:i=1,2,3$, $X=L$ ($l=1$) are identical to the tables
for the case $X=O_i:i=1,2,3$ ($l=0$) of Chapter 14 but with the
values of $r^\prime$ and $r^*$ decreased by 1. The tables for the
case $Y=O_j:j=1,2,3$, $X=O_i:i=1,2,3$ ($l=0$) are identical to the
tables for the case $X=O_j:j=1,2,3 $ ($l=0$) of Chapter 14. The
tables for the case $Y=O_i:i=1,2,3$, $X=S$ ($l=0$) are identical
to the tables for the case $X=O_i:i=1,2,3$ ($l=0$) of Chapter 14.
Finally, the tables for the case $Y=X=S$ ($l=0$) are identical to
the tables for the case $X=S$ ($l=0$) of Chapter 14.

Each component of the Weyl currents $\s^{(L)}J^1(\tcL_L R)$,
$\s^{(O_i)}J^1(\tcL_L R):i=1,2,3$,
$\s^{(O_j)}J^1(\tcL_{O_i}R);i,j=1,2,3$ and $\s^{(O_i)}J^1(\tcL_S
R):i=1,2,3$, $\s^{(S)}J^1(\tcL_S R)$ being written as a sum of
terms in the manner discussed above, and these expressions being
substituted into \ref{14.21}, \ref{14.22}, \ref{14.23} and
\ref{14.24}, \ref{14.25} respectively, sums of trilinear terms
result, two of the factors in each term being contributed by the
expression for a component of the Weyl current $\s^{(L)}J^1(\tcL_L
R)$ in the case of \ref{14.21}, $\s^{(O_i)}J^1(\tcL_L R):i=1,2,3$
in the case of \ref{14.22}, $\s^{(O_j)}J^1(\tcL_{O_i}R):i,j=1,2,3$
in the case of \ref{14.23} and $\s^{(O_i)}J^1(\tcL_S R)$ in the
case of \ref{14.24}, $\s^{(S)}J^1(\tcL_S R)$ in the case of
\ref{14.25}, and the other factor being a component of
$\tcL_L\tcL_L R$, $\tcL_{O_i}\tcL_L R:i=1,2,3$,
$\tcL_{O_j}\tcL_{O_i}R:i,j=1,2,3$ and $\tcL_{O_i}\tcL_S
R:i=1,2,3$, $\tcL_S\tcL_S R$ respectively, multiplied by
$\Omega^3$ and the appropriate power of $|u|^2$. Each third factor
is either ${\bf O}(\delta^{r_3}|u|^{p_3})$ in the case of factors
involving $\alpha(\tcL_Y\tcL_X R)$, $\beta(\tcL_Y\tcL_X
R)$, $\rho(\tcL_Y\tcL_X R)$, $\sigma(\tcL_Y\tcL_X R)$, $\beb(\tcL_Y\tcL_X R)$, or
$\bfob(\delta^{r_3}|u|^{p_3})$ in the case of factors involving
$\alb(\tcL_Y\tcL_X R)$. The values of $r_3$ and $p_3$ assigned are
those implied by the bound on the quantity ${\cal P}_2$. To each
trilinear term we can then apply accordingly one of the first
three cases of Lemma 13.1. Since the pair $r^\prime,p^\prime$
corresponding to the first two factors of each trilinear term may
be replaced by the corresponding $r^*,p^*$, we then obtain a bound
for the contribution of all terms resulting from the product of a
given component of $\s^{(Y)}J^1(\tcL_X R)$ with a given component
of $\tcL_Y\tcL_X R$ to the corresponding error integral
\begin{equation}
\delta^{2q_n+2l}\int_{M^\prime_{c^*}}|\s^{(YX)}\stackrel{(n)}{\tau}_c|d\mu_g
\label{15.5}
\end{equation}
by $O(\delta^e)$, where $e$ is the excess index, defined as in
\ref{14.58}, provided that the integrability index $s$, defined as
in \ref{14.59}, is negative so that Lemma 13.1 applies. We obtain
in this way tables which are similar to the tables \ref{14.60} -
\ref{14.68}. In fact, the tables for the case $(YX)=(LL)$ ($l=2$)
are identical to the tables for the case $X=L$ ($l=1$) of Chapter
14. The tables for the case $(YX)=(OL)$ ($l=1$) are identical to
the tables for the case $X=O$ ($l=0$) of Chapter 14. The tables
for the case $(YX)=(OO)$ ($l=0$) are identical to the tables for
the case $X=O$ ($l=0$) of Chapter 14. Finally, the table for the
case $(YX)=(OS)$ ($l=0$) is identical to the table for the case
$X=O$ ($l=0$) of Chapter 14 and the table for the case $(YX)=(SS)$
($l=0$) is identical to the table for the case $X=S$ ($l=0$) of
Chapter 14.

We must now analyze in more detail the terms with vanishing excess
index as they give rise to {\em borderline error integrals}. They
occur only in the cases $n=1$ and $n=3$. The terms which give
borderline contributions to $\s^{(LL)}\stackrel{(3)}{\tau}_c,
\s^{(OL)}\stackrel{(3)}{\tau}_c,\s^{(OO)}\stackrel{(3)}{\tau}_c$
and
$\s^{(OS)}\stackrel{(3)}{\tau}_c,\s^{(SS)}\stackrel{(3)}{\tau}_c$
are the terms:
\begin{equation}
-6\Omega^3|u|^6\mbox{tr}\chib((\s^{(L)}\ih\rho(\tcL_L
R)-\s^{*(L)}\ih\sigma(\tcL_L R)),\alb(\tcL_L\tcL_L R))
\label{15.6}
\end{equation}
\begin{equation}
-6\Omega^3|u|^6\mbox{tr}\chib\sum_i((\s^{(O_i)}\ih\rho(\tcL_L
R)-\s^{*(O_i)}\ih\sigma(\tcL_L R)),\alb(\tcL_{O_i}\tcL_L R)
\label{15.7}
\end{equation}
\begin{equation}
-6\Omega^3|u|^6\mbox{tr}\chib\sum_{i,j}((\s^{(O_j)}\ih\rho(\tcL_{O_i}R)-\s^{*(O_j)}\ih\sigma(\tcL_{O_i}R)),\alb(\tcL_{O_j}\tcL_{O_i}R))
\label{15.8}
\end{equation}
and:
\begin{equation}
-6\Omega^3|u|^6\mbox{tr}\chib\sum_i((\s^{(O_i)}\ih\rho(\tcL_S
R)-\s^{*(O_i)}\ih\sigma(\tcL_S R)),\alb(\tcL_{O_i}\tcL_S R))
\label{15.9}
\end{equation}
\begin{equation}
-6\Omega^3|u|^6\mbox{tr}\chib((\s^{(S)}\ih\rho(\tcL_S
R)-\s^{*(S)}\ih\sigma(\tcL_S R)),\alb(\tcL_S\tcL_S R))
\label{15.10}
\end{equation}
respectively.

Using the precise bound \ref{14.72}, the facts that:
\begin{eqnarray}
&&\|\mbox{tr}\chib\rho(\tcL_L R)\|_{L^2(C_u)}\leq
C\delta^{-1/2}|u|^{-3}(\stackrel{(2)}{{\cal E}}_1)^{1/2} \ \ : \
\forall u\in[u_0,c^*)
\nonumber\\
&&\|\mbox{tr}\chib\sigma(\tcL_L R)\|_{L^2(C_u)}\leq
C\delta^{-1/2}|u|^{-3}(\stackrel{(2)}{{\cal E}}_1)^{1/2} \ \ : \
\forall u\in[u_0,c^*) \label{15.11}
\end{eqnarray}
as well as the fact that:
\begin{equation}
\|\Omega^3|u|^3\alb(\tcL_L\tcL_L R)\|_{L^2(\Cb_{\ub})}\leq
C\delta^{-1/2}(\stackrel{(3)}{{\cal F}}_2)^{1/2} \ \ : \
\forall\ub\in[0,\delta) \label{15.12}
\end{equation}
(see \ref{12.104}, \ref{12.132} and \ref{12.140}), and following
the proof of Case 2 of Lemma 13.1, we deduce that the contribution
of the term \ref{15.6} to the error integral \ref{15.5} with
$(YX)=(LL)$ ($l=2$) and $n=3$ is bounded by:
\begin{equation}
C{\cal R}_0^\infty(\alpha)(\stackrel{(2)}{{\cal
E}}_1)^{1/2}(\stackrel{(3)}{{\cal F}}_2)^{1/2} \label{15.13}
\end{equation}

Using the precise bound \ref{14.76}, as well as \ref{15.11} and
\begin{eqnarray}
&&\|\mbox{tr}\chib\rho(\tcL_{O_i}R)\|_{L^2(C_u)}\leq
C\delta^{1/2}|u|^{-3}(\stackrel{(2)}{{\cal E}}_1)^{1/2} \ \ : \
\forall u\in[u_0,c^*)
\nonumber\\
&&\|\mbox{tr}\chib\sigma(\tcL_{O_i}R)\|_{L^2(C_u)}\leq
C\delta^{1/2}|u|^{-3}(\stackrel{(2)}{{\cal E}}_1)^{1/2} \ \ : \
\forall u\in[u_0,c^*) \label{15.14}
\end{eqnarray}
and:
\begin{eqnarray}
&&\|\mbox{tr}\chib\rho(\tcL_ S R)\|_{L^2(C_u)}\leq
C\delta^{1/2}|u|^{-3}((\stackrel{(2)}{{\cal
E}}_1)^{1/2}+O(\delta)) \ \ : \ \forall u\in[u_0,c^*)
\nonumber\\
&&\|\mbox{tr}\chib\sigma(\tcL_S R)\|_{L^2(C_u)}\leq
C\delta^{1/2}|u|^{-3}((\stackrel{(2)}{{\cal
E}}_1)^{1/2}+O(\delta)) \ \ : \ \forall u\in[u_0,c^*)\nonumber\\
&&\label{15.15}
\end{eqnarray}
which follow from \ref{13.55}, and the facts that:
\begin{eqnarray}
&&\|\Omega^3|u|^3\alb(\tcL_{O_i}\tcL_L R)\|_{L^2(\Cb_{\ub})}\leq C\delta^{1/2}(\stackrel{(3)}{{\cal F}}_2)^{1/2} \ \ : \ \forall\ub\in[0,\delta)\label{15.16}\\
&&\|\Omega^3|u|^3\alb(\tcL_{O_j}\tcL_{O_i}R\|_{L^2(\Cb_{\ub})}\leq
C\delta^{3/2}(\stackrel{(3)}{{\cal F}}_2)^{1/2} \ \ : \
\forall\ub\in[0,\delta)\label{15.17}\\
&&\|\Omega^3|u|^3\alb(\tcL_{O_i}\tcL_S R)\|_{L^2(\Cb_{\ub})}\leq
C\delta^{3/2}(\stackrel{(3)}{{\cal F}}_2)^{1/2} \ \ : \
\forall\ub\in[0,\delta)\label{15.18}
\end{eqnarray}
(see \ref{12.109}, \ref{12.114}, \ref{12.119}, \ref{12.132} and
\ref{12.140}), and following the proof of Case 2 of Lemma 13.1, we
deduce that the contribution of the term \ref{15.7} to the error
integral \ref{15.5} with $(YX)=(OL)$ ($l=1$) and $n=3$, the
contribution of the term \ref{15.8} to the error integral
\ref{15.5} with $(YX)=(OO)$ ($l=0$) and $n=3$, and the
contribution of the term \ref{15.9} to the error integral
\ref{15.5} with $(YX)=(OS)$ ($l=0$) and $n=3$, are all three
bounded by:
\begin{equation}
C(\scR_1^4(\beta)+{\cal R}_0^\infty(\beta))(\stackrel{(2)}{{\cal
E}}_1)^{1/2}(\stackrel{(3)}{{\cal F}}_2)^{1/2}+O(\delta^{1/2})
\label{15.19}
\end{equation}

Using the precise bound \ref{14.79}, as well as \ref{15.15} and
the fact that:
\begin{equation}
\|\Omega^3|u|^3\alb(\tcL_S\tcL_S R)\|_{L^2(\Cb_{\ub})}\leq
C\delta^{3/2}(\stackrel{(3)}{{\cal F}}_2)^{1/2} \ \ : \
\forall\ub\in[0,\delta) \label{15.20}
\end{equation}
(see \ref{12.124}, \ref{12.132} and \ref{12.140}), and following
the proof of Case 2 of Lemma 13.1, we deduce that the contribution
of the term \ref{15.10} to the error integral \ref{15.5} with
$(YX)=(SS)$ ($l=0$) and $n=3$ is bounded by:
\begin{equation}
C({\cal D}_0^\infty(\chibh)+{\cal
R}_0^\infty(\alpha))(\stackrel{(2)}{{\cal
E}}_1)^{1/2}(\stackrel{(3)}{{\cal F}}_2)^{1/2}+O(\delta)
\label{15.21}
\end{equation}

In view of \ref{14.82}, \ref{14.85}, the expressions \ref{15.13},
\ref{15.19} and \ref{15.21} are in turn bounded by
\begin{equation}
C(\stackrel{(0)}{{\cal E}}_2)^{1/2}(\stackrel{(2)}{{\cal
E}}_1)^{1/2}(\stackrel{(3)}{{\cal F}}_2)^{1/2}+O(\delta^{1/2})
\label{15.22}
\end{equation}
\begin{equation}
C(\stackrel{(1)}{{\cal E}}_2)^{1/2}(\stackrel{(2)}{{\cal
E}}_1)^{1/2}(\stackrel{(3)}{{\cal F}}_2)^{1/2}+O(\delta^{1/2})
\label{15.23}
\end{equation}
and
\begin{equation}
C\left\{{\cal D}_0^\infty(\chibh)+(\stackrel{(0)}{{\cal
E}}_2)^{1/2}\right\}(\stackrel{(2)}{{\cal
E}}_1)^{1/2}(\stackrel{(3)}{{\cal F}}_2)^{1/2}+O(\delta^{1/2})
\label{15.24}
\end{equation}
respectively.

The terms which give borderline contributions to
$\s^{(LL)}\stackrel{(1)}{\tau}_c,
\s^{(OL)}\stackrel{(1)}{\tau}_c,\s^{(OO)}\stackrel{(1)}{\tau}_c$
are the terms:
\begin{equation}
2\Omega^3|u|^2\mbox{tr}\chib\left\{(\s^{(L)}\ih,\alpha(\tcL_L
R))\rho(\tcL_L\tcL_L R) +\s^{(L)}\ih\wedge\alpha(\tcL_L
R)\sigma(\tcL_L\tcL_L R)\right\} \label{15.25}
\end{equation}
\begin{equation}
2\Omega^3|u|^2\mbox{tr}\chib\sum_i\left\{(\s^{(O_i)}\ih,\alpha(\tcL_L
R))\rho(\tcL_{O_i}\tcL_L R) +\s^{(O_i)}\ih\wedge\alpha(\tcL_L
R)\sigma(\tcL_{O_i}\tcL_L R)\right\} \label{15.26}
\end{equation}
\begin{equation}
2\Omega^3|u|^2\mbox{tr}\chib\sum_{i,j}\left\{(\s^{(O_j)}\ih,\alpha(\tcL_{O_i}R))\rho(\tcL_{O_j}\tcL_{O_i}R)
+\s^{(O_j)}\ih\wedge\alpha(\tcL_{O_i}R)\sigma(\tcL_{O_j}\tcL_{O_i}R)\right\}
\label{15.27}
\end{equation}
respectively.

Using \ref{14.72}, the fact that:
\begin{equation}
\|\mbox{tr}\chib\alpha(\tcL_L R)\|_{L^2(C_u)}\leq
C\delta^{-2}|u|^{-1}(\stackrel{(0)}{{\cal E}}_1)^{1/2} \ \ : \
\forall u\in[u_0,c^*) \label{15.28}
\end{equation}
as well as the facts that:
\begin{eqnarray}
&&\|\Omega^3|u|^2\rho(\tcL_L R\tcL_L R)\|_{L^2(C_u)}\leq
C\delta^{-3/2}(\stackrel{(2)}{{\cal E}}_2)^{1/2} \ \ : \ \forall
u\in[u_0,c^*)
\nonumber\\
&&\|\Omega^3|u|^2\sigma(\tcL_L R\tcL_L R)\|_{L^2(C_u)}\leq
C\delta^{-3/2}(\stackrel{(2)}{{\cal E}}_2)^{1/2} \ \ : \ \forall
u\in[u_0,c^*) \label{15.29}
\end{eqnarray}
(see \ref{12.103}, \ref{12.129} and \ref{12.139}), and following
the proof of Case 1 of Lemma 13.1 we deduce that the contribution
of the term \ref{15.25} to the error integral \ref{15.5} with
$(YX)=(LL)$ ($l=2$) and $n=1$ is bounded by:
\begin{equation}
C{\cal R}_0^\infty(\alpha)(\stackrel{(0)}{{\cal
E}}_1)^{1/2}(\stackrel{(2)}{{\cal E}}_2)^{1/2} \label{15.30}
\end{equation}

Using \ref{14.76}, as well as \ref{15.28} and
\begin{equation}
\|\mbox{tr}\chib\alpha(\tcL_{O_i}R)\|_{L^2(C_u)}\leq
C\delta^{-1}|u|^{-1}(\stackrel{(0)}{{\cal E}}_1)^{1/2} \ \ : \
\forall u\in[u_0,c^*) \label{15.31}
\end{equation}
and the facts that:
\begin{eqnarray}
&&\|\Omega^3|u|^2\rho(\tcL_{O_i}\tcL_L R)\|_{L^2(C_u)}\leq
C\delta^{-1/2}(\stackrel{(2)}{{\cal E}}_2)^{1/2} \ \ : \ \forall
u\in[u_0,c^*)
\nonumber\\
&&\|\Omega^3|u|^2\sigma(\tcL_{O_i}\tcL_L R)\|_{L^2(C_u)}\leq
C\delta^{-1/2}(\stackrel{(2)}{{\cal E}}_2)^{1/2} \ \ : \ \forall
u\in[u_0,c^*)
\label{15.32}\\
&&\|\Omega^3|u|^2\rho(\tcL_{O_j}\tcL_{O_i}R)\|_{L^2(C_u)}\leq
C\delta^{1/2}(\stackrel{(2)}{{\cal E}}_2)^{1/2} \ \ : \ \forall
u\in[u_0,c^*)
\nonumber\\
&&\|\Omega^3|u|^2\sigma(\tcL_{O_j}\tcL_{O_i}R)\|_{L^2(C_u)}\leq
C\delta^{1/2}(\stackrel{(2)}{{\cal E}}_2)^{1/2} \ \ : \ \forall
u\in[u_0,c^*) \label{15.33}
\end{eqnarray}
(see \ref{12.108}, \ref{12.113}, \ref{12.129} and \ref{12.139}),
and following the proof of Case 1 of Lemma 13.1, we deduce that
the contribution of the term \ref{15.26} to the error integral
\ref{15.5} with $(YX)=(OL)$ ($l=1$) and $n=1$ and the contribution
of the term \ref{15.27} to the error integral \ref{15.5} with
$(YX)=(OO)$ ($l=0$) and $n=1$, are both bounded by:
\begin{equation}
C(\scR_1^4(\beta)+{\cal R}_0^\infty(\beta))(\stackrel{(0)}{{\cal
E}}_1)^{1/2}(\stackrel{(2)}{{\cal E}}_2)^{1/2}+O(\delta^{1/2})
\label{15.34}
\end{equation}

In view of \ref{14.82}, \ref{14.85} the expressions \ref{15.30},
\ref{15.34} are in turn bounded by
\begin{equation}
C(\stackrel{(0)}{{\cal E}}_2)^{1/2}(\stackrel{(0)}{{\cal
E}}_1)^{1/2}(\stackrel{(2)}{{\cal E}}_2)^{1/2}+O(\delta^{1/2})
\label{15.35}
\end{equation}
and
\begin{equation}
C(\stackrel{(1)}{{\cal E}}_2)^{1/2}(\stackrel{(0)}{{\cal
E}}_1)^{1/2}(\stackrel{(2)}{{\cal E}}_2)^{1/2}+O(\delta^{1/2})
\label{15.36}
\end{equation}
respectively.

We turn to consider the contributions to the 2nd order Weyl
current error integrals of the terms
$$\s^{(Y)}J^2(\tcL_X R) \ \ \mbox{and} \ \ \s^{(Y)}J^3(\tcL_X R)$$
in $\s^{(YX)}J$,
$(Y,X)=(L,L),(O_i,L):i=1,2,3,(O_j,O_i):i,j=1,2,3,(O_i,S):i=1,2,3,(S,S)$.
Now, the contributions of $\s^{(L)}J^2(\tcL_L R)$ and
$\s^{(L)}J^3(\tcL_L R)$ to $\s^{(LL)}\stackrel{(n)}{\tau}_c$ are
identical in form to $\s^{(L)}\stackrel{(n)}{\tau}_{c,2}$ and
$\s^{(L)}\stackrel{(n)}{\tau}_{c,3}$ respectively, but with
$\tcL_L R$ in the role of $R$. The contributions of
$\s^{(O_i)}J^2(\tcL_L R)$ and $\s^{(O_i)}J^3(\tcL_L R)$ to
$\s^{(OL)}\stackrel{(n)}{\tau}_c$ are indentical in form to
$\s^{(O)}\stackrel{(n)}{\tau}_{c,2}$ and
$\s^{(O)}\stackrel{(n)}{\tau}_{c,3}$ respectively, but with
$\tcL_L R$ in the role of $R$. The contributions of
$\s^{(O_j)}J^2(\tcL_{O_i}R)$ and $\s^{(O_j)}J^3(\tcL_{O_i}R)$ to
$\s^{(OO)}\stackrel{(n)}{\tau}_c$ are identical in form to
$\s^{(O)}\stackrel{(n)}{\tau}_{c,2}$ and
$\s^{(O)}\stackrel{(n)}{\tau}_{c,3}$ respectively, but with
$\tcL_{O_i}R$ in the role of $R$. The contributions of
$\s^{(O_i)}J^2(\tcL_S R)$ and $\s^{(O_i)}J^3(\tcL_S R)$ to
$\s^{(OS)}\stackrel{(3)}{\tau}_c$ are identical in form to
$\s^{(O)}\stackrel{(3)}{\tau}_{c,2}$ and
$\s^{(O)}\stackrel{(3)}{\tau}_{c,3}$ respectively, but with
$\tcL_S R$ in the role of $R$. Finally, the contributions of
$\s^{(S)}J^2(\tcL_S R)$ and $\s^{(S)}J^3(\tcL_S R)$ to
$\s^{(SS)}\stackrel{(3)}{\tau}_c$ are identical in form to
$\s^{(S)}\stackrel{(3)}{\tau}_{c,2}$ and
$\s^{(S)}\stackrel{(3)}{\tau}_{c,3}$ respectively, but with
$\tcL_S R$ in the role of $R$.

Now, by \ref{12.149}, \ref{12.174} and the bounds on the quantity
${\cal Q}_2$, we have:
\begin{eqnarray}
&&\Dh\alpha={\cal O}^4(\delta^{-5/2}|u|^{-1})\nonumber\\
&&D\beta={\cal O}^4(\delta^{-3/2}|u|^{-2})\nonumber\\
&&D\rho={\cal O}^4(\delta^{-1}|u|^{-3})\nonumber\\
&&D\sigma={\cal O}^4(\delta^{-1}|u|^{-3})\nonumber\\
&&D\beb={\cal O}^4(|u|^{-4})\nonumber\\
&&\Dh\alb={\cal O}^4(\delta^{1/2}|u|^{-9/2}) \label{15.37}
\end{eqnarray}
Taking also into account \ref{14.107} and the $L^\infty$ bounds on
the components of $\s^{(L)}\tilde{\pi}$, we then deduce from
Proposition 12.2 that:
\begin{eqnarray}
&&\alpha(\tcL_L R)={\cal O}^4(\delta^{-5/2}|u|^{-1})\nonumber\\
&&\beta(\tcL_L R)={\cal O}^4(\delta^{-3/2}|u|^{-2})\nonumber\\
&&\rho(\tcL_L R)={\cal O}^4(\delta^{-1}|u|^{-3})\nonumber\\
&&\sigma(\tcL_L R)={\cal O}^4(\delta^{-1}|u|^{-3})\nonumber\\
&&\beb(\tcL_L R)={\cal O}^4(|u|^{-4})\nonumber\\
&&\alb(\tcL_L R)={\cal O}^4(\delta^{1/2}|u|^{-9/2}) \label{15.38}
\end{eqnarray}
These are similar to \ref{14.107} but with the exponents of
$\delta$ reduced by 1. It follows that the tables for the
components of $\s^{(L)}J^2(\tcL_L R)$ and $\s^{(L)}J^3(\tcL_L R)$
($l=2$) are similar to the tables \ref{14.109} - \ref{14.118} and
\ref{14.194} - \ref{14.203} for the components of $\s^{(L)}J^2(R)$
and $\s^{(L)}J^3(R)$ ($l=1$), respectively, but with the values of
$r^\prime$ and $r^*$ reduced by 1. It then follows that the tables
for the contributions of $\s^{(L)}J^2(\tcL_L R)$ and
$\s^{(L)}J^3(\tcL_L R)$ to $\s^{(LL)}\stackrel{(n)}{\tau}_c$ are
identical to the tables \ref{14.134} - \ref{14.137} and
\ref{14.217} - \ref{14.220} for
$\s^{(L)}\stackrel{(n)}{\tau}_{c,2}$ and
$\s^{(L)}\stackrel{(n)}{\tau}_{c,3}$ respectively. Also, the
tables for the components of $\s^{(O_i)}J^2(\tcL_L R)$ and
$\s^{(O_i)}J^3(\tcL_L R)$ ($l=1$) are similar to the tables
\ref{14.119} - \ref{14.128} and \ref{14.204} - \ref{14.213} for
the components of $\s^{(O_i)}J^2(R)$ and $\s^{(O_i)}J^3(R)$
($l=0$), respectively, but with the values of $r^\prime$ and $r^*$
reduced by 1. Then the tables for the contributions of
$\s^{(O_i)}J^2(\tcL_L R)$ and $\s^{(O_i)}J^3(\tcL_L R)$ to
$\s^{(OL)}\stackrel{(n)}{\tau}_c$ are indentical to the tables
\ref{14.138} - \ref{14.141} and \ref{14.221} - \ref{14.224} for
$\s^{(O)}\stackrel{(n)}{\tau}_{c,2}$ and
$\s^{(O)}\stackrel{(n)}{\tau}_{c,3}$ respectively.

By \ref{12.148} and the bound on the quantity ${\cal Q}_2$,
together with \ref{14.107} and Propositions 8.1 and 8.2, we have:
\begin{eqnarray}
&&\sLh_{O_i}\alpha={\cal O}^4(\delta^{-3/2}|u|^{-1})\nonumber\\
&&\sL_{O_i}\beta={\cal O}^4(\delta^{-1/2}|u|^{-2})\nonumber\\
&&O_i\rho={\cal O}^4(|u|^{-3})\nonumber\\
&&O_i\sigma={\cal O}^4(|u|^{-3})\nonumber\\
&&\sL_{O_i}\beb={\cal O}^4(\delta|u|^{-4})\nonumber\\
&&\sLh_{O_i}\alb={\cal O}^4(\delta^{3/2}|u|^{-9/2}) \label{15.39}
\end{eqnarray}
Taking also into account \ref{14.107} and the $L^\infty$ bounds on
the components of $\s^{(O_i)}\tilde{\pi}$, we then deduce from
Proposition 12.2 that:
\begin{eqnarray}
&&\alpha(\tcL_{O_i}R)={\cal O}^4(\delta^{-3/2}|u|^{-1})\nonumber\\
&&\beta(\tcL_{O_i}R)={\cal O}^4(\delta^{-1/2}|u|^{-2})\nonumber\\
&&\rho(\tcL_{O_i}R)={\cal O}^4(|u|^{-3})\nonumber\\
&&\sigma(\tcL_{O_i}R)={\cal O}^4(|u|^{-3})\nonumber\\
&&\beb(\tcL_{O_i}R)={\cal O}^4(\delta|u|^{-4})\nonumber\\
&&\alb(\tcL_{O_i}R)={\cal O}^4(\delta^{3/2}|u|^{-9/2})
\label{15.40}
\end{eqnarray}
These are similar to \ref{14.107}. It follows that the tables for
the components of $\s^{(O_j)}J^2(\tcL_{O_i}R)$ and
$\s^{(O_j)}J^3(\tcL_{O_i}R$ ($l=0$) are similar to the tables
\ref{14.119} - \ref{14.128} and \ref{14.204} - \ref{14.213} for
the components of $\s^{(O_j)}J^2(R)$ and $\s^{(O_j)}J^3(R)$
($l=0$), respectively. It then follows that the tables for the
contributions of $\s^{(O_j)}J^2(\tcL_{O_i}R)$ and
$\s^{(O_j)}J^3(\tcL_{O_i}R)$ to $\s^{(OO)}\stackrel{(n)}{\tau}_c$
are identical to the tables \ref{14.138} - \ref{14.141} and
\ref{14.221} - \ref{14.224} for
$\s^{(O)}\stackrel{(n)}{\tau}_{c,2}$ and
$\s^{(O)}\stackrel{(n)}{\tau}_{c,3}$ respectively.

Expressing
$$\Dbh\alpha,\Db\beta,\Db\rho,\Db\sigma,\Db\beb$$
by means of the homogeneous Bianchi equations and using
\ref{12.148} and the bound on the quantity ${\cal Q}_2$ as well as
\ref{14.107} and the $L^\infty$ bounds of Chapter 3 for the
connection coefficients we deduce:
\begin{eqnarray}
&&\Dbh\alpha={\cal O}^4(\delta^{-3/2}|u|^{-2})\nonumber\\
&&\Db\beta={\cal O}^4(\delta^{-1/2}|u|^{-3})\nonumber\\
&&\Db\rho={\cal O}^4(|u|^{-4})\nonumber\\
&&\Db\sigma={\cal O}^4(|u|^{-4})\nonumber\\
&&\Db\beb={\cal O}^4(\delta|u|^{-5}) \label{15.41}
\end{eqnarray}
Also, according to \ref{12.150} and the bound on the quantity
${\cal Q}_2$:
\begin{equation}
\Dbh\alb={\cal O}^4(\delta^{3/2}|u|^{-11/2}) \label{15.42}
\end{equation}
Combining \ref{15.41} and \ref{15.42} with \ref{15.37} we conclude
that:
\begin{eqnarray}
&&\sLh_S\alpha={\cal O}^4(\delta^{-3/2}|u|^{-1})\nonumber\\
&&\sL_S\beta={\cal O}^4(\delta^{-1/2}|u|^{-2})\nonumber\\
&&S\rho={\cal O}^4(|u|^{-3})\nonumber\\
&&S\sigma={\cal O}^4(|u|^{-3})\nonumber\\
&&\sL_S\beb={\cal O}^4(\delta|u|^{-4})\nonumber\\
&&\sLh_S\alb={\cal O}^4(\delta^{3/2}|u|^{-9/2}) \label{15.43}
\end{eqnarray}
Taking also into account \ref{14.107} and the $L^\infty$ bounds on
the components of $\s^{(S)}\tilde{\pi}$, we then deduce from
Proposition 12.2 that:
\begin{eqnarray}
&&\alpha(\tcL_S R)={\cal O}^4(\delta^{-3/2}|u|^{-1})\nonumber\\
&&\beta(\tcL_S R)={\cal O}^4(\delta^{-1/2}|u|^{-2})\nonumber\\
&&\rho(\tcL_S R)={\cal O}^4(|u|^{-3})\nonumber\\
&&\sigma(\tcL_S R)={\cal O}^4(|u|^{-3})\nonumber\\
&&\beb(\tcL_S R)={\cal O}^4(\delta|u|^{-4})\nonumber\\
&&\alb(\tcL_S R)={\cal O}^4(\delta^{3/2}|u|^{-9/2}) \label{15.44}
\end{eqnarray}
These are similar to \ref{14.107}. It follows that the tables for
the components of $\s^{(O_i)}J^2(\tcL_S R)$ and
$\s^{(O_i)}J^3(\tcL_S R)$ ($l=0$) are similar to the tables
\ref{14.119} - \ref{14.128} and \ref{14.204} - \ref{14.213} for
the components of $\s^{(O_i)}J^2(R)$ and $\s^{(O_i)}J^3(R)$
($l=0$), respectively. It then follows that the tables for the
contributions of $\s^{(O_i)}J^2(\tcL_S R)$ and
$\s^{(O_i)}J^3(\tcL_S R)$ to $\s^{(OS)}\stackrel{(3)}{\tau}_c$ are
identical to the tables \ref{14.141} and \ref{14.224} for
$\s^{(O)}\stackrel{(S)}{\tau}_{c,2}$ and
$\s^{(O)}\stackrel{(S)}{\tau}_{c,3}$ respectively. Also, the
tables for the components of $\s^{(S)}J^2(\tcL_S R)$ and
$\s^{(S)}J^3(\tcL_S R)$ ($l=0$) are similar to the tables
\ref{14.129} - \ref{14.130} and \ref{14.214} - \ref{14.215} for
the components of $\s^{(S)}J^2(R)$ and $\s^{(S)}J^3(R)$ ($l=0$),
respectively. Then the tables for the contributions of
$\s^{(S)}J^2(\tcL_S R)$ and $\s^{(S)}J^3(\tcL_S R)$ to
$\s^{(SS)}\stackrel{(3)}{\tau}_c$ are identical to the tables
\ref{14.142} and \ref{14.225} for
$\s^{(S)}\stackrel{(3)}{\tau}_{c,2}$ and
$\s^{(S)}\stackrel{(3)}{\tau}_{c,3}$ respectively.

We summarize the results of this section in the following
Proposition.

\vspace{5mm}

\noindent{\bf Proposition 15.1} \ \ \ The contributions to the
error integral
$$\delta^{2q_n+2l}\int_{M^\prime_{c^*}}|\s^{(YX))}\stackrel{(n)}{\tau}_c|d\mu_g$$
of the terms
$$\s^{(Y)}J^1(\tcL_X R)+\s^{(Y)}J^2(\tcL_X R)+J^3\s^{(Y)}(\tcL_X R)$$
are bounded as follows.

\vspace{3mm}

1. For $(YX)=(LL)$.

\vspace{2mm}

\noindent In the case $n=0$ by:
$$O(\delta)$$

\noindent In the case $n=1$ by:
$$C(\stackrel{(0)}{{\cal E}}_2)^{1/2}(\stackrel{(0)}{{\cal E}}_1)^{1/2}(\stackrel{(2)}{{\cal E}}_2)^{1/2}+O(\delta^{1/2})$$

\noindent In the case $n=2$ by:
$$O(\delta)$$

\noindent In the case $n=3$ by:
$$C(\stackrel{(0)}{{\cal E}}_2)^{1/2}(\stackrel{(2)}{{\cal E}}_1)^{1/2}(\stackrel{(3)}{{\cal F}}_2)^{1/2}+O(\delta^{1/2})$$

\vspace{3mm}

2. For $(YX)=(OL)$.

\vspace{2mm}

\noindent In the case $n=0$ by:

$$O(\delta)$$

\noindent In the case $n=1$ by:
$$C(\stackrel{(1)}{{\cal E}}_2)^{1/2}(\stackrel{(0)}{{\cal E}}_1)^{1/2}(\stackrel{(2)}{{\cal E}}_2)^{1/2}+O(\delta^{1/2})$$

\noindent In the case $n=2$ by:
$$O(\delta)$$

\noindent In the case $n=3$ by:
$$C(\stackrel{(1)}{{\cal E}}_2)^{1/2}(\stackrel{(2)}{{\cal E}}_1)^{1/2}(\stackrel{(3)}{{\cal F}}_2)^{1/2}+O(\delta^{1/2})$$

\vspace{3mm}

3. For $(YX)=(OO)$.

\vspace{2mm}

\noindent In the case $n=0$ by:
$$O(\delta)$$

\noindent In the case $n=1$ by:
$$C(\stackrel{(1)}{{\cal E}}_2)^{1/2}(\stackrel{(0)}{{\cal E}}_1)^{1/2}(\stackrel{(2)}{{\cal E}}_2)^{1/2}+O(\delta^{1/2})$$

\noindent In the case $n=2$ by:

$$O(\delta)$$

\noindent In the case $n=3$ by:
$$C(\stackrel{(1)}{{\cal E}}_2)^{1/2}(\stackrel{(2)}{{\cal E}}_1)^{1/2}(\stackrel{(3)}{{\cal F}}_2)^{1/2}+O(\delta^{1/2})$$

\vspace{3mm}

4. For $(YX)=(OS)$.

\noindent Here we only have the case $n=3$, where we have a bound
by:
$$C(\stackrel{(1)}{{\cal E}}_2)^{1/2}(\stackrel{(2)}{{\cal E}}_1)^{1/2}(\stackrel{(3)}{{\cal F}}_2)^{1/2}+O(\delta^{1/2})$$

5. For $(YX)=(SS)$.

\noindent Here we only have the case $n=3$, where we have a bound
by:
$$C\left\{{\cal D}_0^\infty(\chibh)+(\stackrel{(0)}{{\cal E}}_2)^{1/2}\right\}(\stackrel{(2)}{{\cal E}}_1)^{1/2}(\stackrel{(3)}{{\cal F}}_2)^{1/2}+O(\delta^{1/2})$$

\vspace{5mm}

\section{The genuine 2nd order estimates}

In the present section we shall estimate the contribution to the
2nd order Weyl current error integrals of the term
$$\tcL_Y\s^{(X)}J$$
in $\s^{(YX)}J, (Y,X)=(L,L), (O_i,L):i=1,2,3, (O_j,O_i):i,j=1,2,3,
(O_i,S):i=1,2,3, (S,S)$ (see \ref{14.15}). Again, we pay special
attention to the terms giving rise to borderline error integrals.

We first consider the contribution to the 2nd order Weyl current
error integrals of the term
$$\tcL_Y\s^{(X)}J^1(R)$$
Special care is needed here, because, as we have seen in the
previous chapter, it is the Weyl current $J^1$ which contributes
borderline error integrals. We consider the expressions for the
components of $\s^{(X)}J^1(R)$ given by Proposition 14.1 with the
fundamental Weyl field $R$ in the role of the Weyl field $W$. The
expression for each component of $\s^{(X)}J^1(R)$ is a sum of
terms one factor of which is a component of $\s^{(X)}\tilde{\pi}$,
and we consider the terms with the same such factor as a single
term. After the substitution for
$$\Dbh\alpha,\Db\beta,\Db\rho,\Db\sigma,\Db\beb \ \ \mbox{and} \ \ \Dh\alb$$
from the homogeneous Bianchi equations, as discussed in the
previous chapter, the other factor of each term in the expression
of a given component of $\s^{(X)}J^1(R)$ becomes a sum of a
principal part, which is a sum of 1st derivatives of components of
$R$, and a non-principal part, which is a sum of terms consisting
of two factors, one of which is a connection coefficient and the
other a component of $R$. Moreover, only $\snab$ or $D$
derivatives of the components $\alpha,\beta,\rho,\sigma,\beb$ and
only $\snab$ or $\Db$ derivatives of the component $\alb$ occur.
Now, the components of $\tcL_Y\s^{(X)}J^1(R)$ are given by
Proposition 12.3 in terms of the $\sL_Y$ derivatives of the
components of $\s^{(X)}J^1(R)$. The remainders are estimated using
the $L^\infty$ bounds for the components of $\s^{(Y)}\tilde{\pi}$
and the estimates for the components of $\s^{(X)}J^1(R)$ of
Chapter 14. Now if $\sL_Y$ is applied to a term in the expression
of a given component of $\s^{(X)}J^1(R)$ a sum of two terms will
result, one of which will have as a first factor the $\sL_Y$
derivative of the corresponding component of $\s^{(X)}\tilde{\pi}$
and the same second factor as the original term, and the other
term will have the component of $\s^{(X)}\tilde{\pi}$ as a first
factor and $\sL_Y$ applied to the second factor of the original
term, plus possibly a bilinear expression in $\s^{(Y)}\spi$ and
the second factor of the original term, as the other factor. The
bilinear expression results from applying $\sL_Y$ to the
coefficients of the expression constituting the given original
term, which may involve $\sg$ and $\seps$. The first of the two
resulting terms is to be estimated by placing the first factor in
$L^4(S)$ using $L^4(S)$ estimates for the 1st derivatives of the
components of $\s^{(X)}\tilde{\pi}$, and the second factor in
$L^4(S)$  using the $L^4(S)$ estimates for the 1st derivatives of
the componets of $R$. The second of the two resulting terms is to
be estimated by placing the first factor in $L^\infty$ using the
$L^\infty$ estimates for the components of $\s^{(X)}\tilde{\pi}$,
the principal part of the second factor in $L^2(C_u)$ or
$L^2(\Cb_{\ub})$ as shall be described in more detail below, and
the remainder of the second factor in $L^2(C_u)$. In fact, the
estimates of Chapters 3 and 4 together with  the $L^\infty$
estimates for the components of $R$ and the $L^4(S)$ estimates for
their 1st derivatives  (and $L^\infty$ bounds for $\s^{(Y)}\spi$)
give an $L^4(S)$ estimate for the remainder, which implies a
corresponding $L^2(C_u)$ estimate.

The principal part of the second factor of the second of the two
terms resulting from a given term in the expression of a given
component of $\s^{(X)}J^1(R)$ by applying $\sL_Y$, is a term or a
sum of terms of the form:
\begin{equation}
\sLh_Y\snab\alpha,\sL_Y\snab\beta,\sL_Y\sd\rho,\sL_Y\sd\sigma,\sL_Y\snab\beb
\ \ \mbox{and} \ \ \sLh_Y\snab\alb \label{15.45}
\end{equation}
(the first and last of these being the trace-free parts relative
to the last two indices of $\sL_Y\snab\alpha$ and $\sL_Y\snab\alb$
respectively) and:
\begin{equation}
\sLh_Y\Dh\alpha,\sL_Y D\beta,YD\rho,YD\sigma,\sL_Y D\beb \ \
\mbox{and} \ \ \sLh_Y\Dbh\alb \label{15.46}
\end{equation}
The first five of each of \ref{15.45}, \ref{15.46} is to be
estimated in $L^2(C_u)$, while the last is to be estimated in
$L^2(\Cb_{\ub})$, in the manner to be presently discussed.

Consider first the case $Y=L$. Then by Lemma 4.1 (see also
\ref{10.70} and a similar identity with $\alb$ in the role of
$\alpha$) each of \ref{15.45} is equal to
\begin{equation}
\snab\Dh\alb,\snab D\beta,\sd D\rho,\sd D\sigma,\snab D\beb \ \
\mbox{and} \ \ \snab\Dh\alb \label{15.47}
\end{equation}
respectively, plus a remainder which can be estimated in $L^4(S)$,
therefore also in $L^2(C_u)$ using the estimates of Chapter 4
together with the $L^\infty$ estimates for the components of $R$.
Also, each of \ref{15.46} is equal to
\begin{equation}
\Dh^2\alpha,D^2\beta,D^2\rho,D^2\sigma,D^2\beb \ \ \mbox{and} \ \
\Dh\Dbh\alb \label{15.48}
\end{equation}
respectively. The first five of each of \ref{15.47}, \ref{15.48}
are estimated in $L^2(C_u)$ according to the bound on the quantity
${\cal Q}_2$. The last of \ref{15.47} is estimated in
$L^2(\Cb_{\ub})$ through inequality \ref{12.193} and the bounds on
the quantities ${\cal P}_2$ and ${\cal Q}_2$. The last of
\ref{15.48} is estimated in $L^2(\Cb_{\ub})$ through inequalities
\ref{12.202} and \ref{12.227} and the bounds on the quantities
${\cal P}_2$ and ${\cal Q}_2$. In this way we obtain:
\begin{eqnarray}
&&\sLh_L\snab\alpha={\bf O}(\delta^{-5/2}|u|^{-2})\nonumber\\
&&\sL_L\snab\beta={\bf O}(\delta^{-3/2}|u|^{-3})\nonumber\\
&&\sL_L\sd\rho={\bf O}(\delta^{-1}|u|^{-4})\nonumber\\
&&\sL_L\sd\sigma={\bf O}(\delta^{-1}|u|^{-4})\nonumber\\
&&\sL_L\snab\beb={\bf O}(|u|^{-5})\nonumber\\
&&\sLh_L\snab\alb=\bfob(\delta^{1/2}|u|^{-11/2}) \label{15.49}
\end{eqnarray}
and:
\begin{eqnarray}
&&\sLh_L\Dh\alpha={\bf O}(\delta^{-7/2}|u|^{-1})\nonumber\\
&&\sL_L D\beta={\bf O}(\delta^{-5/2}|u|^{-2})\nonumber\\
&&LD\rho={\bf O}(\delta^{-2}|u|^{-3})\nonumber\\
&&LD\sigma={\bf O}(\delta^{-2}|u|^{-3})\nonumber\\
&&\sL_L D\beb={\bf O}(\delta^{-1}|u|^{-4})\nonumber\\
&&\sLh_L\Dbh\alb=\bfob(\delta^{1/2}|u|^{-11/2}) \label{15.50}
\end{eqnarray}

Consider next the case $Y=O_j:j=1,2,3$. Then each of \ref{15.45}
is equal to
\begin{equation}
\sLh_{O_j}\snab\alpha,\sL_{O_j}\snab\beta,\sL_{O_j}\sd\rho,\sL_{O_j}\sd\sigma,\sL_{O_j}\snab\beb
\ \ \mbox{and} \ \ \sLh_{O_j}\snab\alb \label{15.51}
\end{equation}
respectively, and each of \ref{15.46} is equal to
\begin{equation}
\sLh_{O_j}\Dh\alpha,\sL_{O_j}D\beta,O_j D\rho,O_j
D\sigma,\sL_{O_j}D\beb \ \ \mbox{and} \ \ \sLh_{O_j}\Dbh\alb
\label{15.52}
\end{equation}
respectively. In view of Propositions 8.1, 8.2, the first five of
each of \ref{15.50}, \ref{15.51} are estimated in $L^2(C_u)$ and
the last of each of \ref{15.51}, \ref{15.52} are estimated in
$L^2(\Cb_{\ub})$, according to the bound on the quantity ${\cal
Q}_2$. In this way we obtain:
\begin{eqnarray}
&&\sLh_{O_j}\snab\alpha={\bf O}(\delta^{-3/2}|u|^{-2})\nonumber\\
&&\sL_{O_j}\snab\beta={\bf O}(\delta^{-1/2}|u|^{-3})\nonumber\\
&&\sL_{O_j}\sd\rho={\bf O}(|u|^{-4})\nonumber\\
&&\sL_{O_j}\sd\sigma={\bf O}(|u|^{-4})\nonumber\\
&&\sL_{O_j}\snab\beb={\bf O}(\delta|u|^{-5})\nonumber\\
&&\sLh_{O_j}\snab\alb=\bfob(\delta^{3/2}|u|^{-11/2}) \label{15.53}
\end{eqnarray}
and:
\begin{eqnarray}
&&\sLh_{O_j}\Dh\alpha={\bf O}(\delta^{-5/2}|u|^{-1})\nonumber\\
&&\sL_{O_j}D\beta={\bf O}(\delta^{-3/2}|u|^{-2})\nonumber\\
&&O_jD\rho={\bf O}(\delta^{-1}|u|^{-3})\nonumber\\
&&O_jD\sigma={\bf O}(\delta^{-1}|u|^{-3})\nonumber\\
&&\sL_{O_j}D\beb={\bf O}(|u|^{-4})\nonumber\\
&&\sLh_{O_j}\Dbh\alb=\bfob(\delta^{3/2}|u|^{-11/2}) \label{15.54}
\end{eqnarray}

Consider finally the case $Y=S$. Then $X=S$, and since in this
case only the case $n=3$ occurs, we only have to consider the
terms resulting from $\sL_S\Xib(\s^{(S)}J^1(R))$ and
$\sLh_S\Thetab(\s^{(S)}J^1(R))$. From Proposition 14.1 we see that
these involve only
\begin{equation}
\sL_S\snab\beb \ \ \mbox{and} \ \ \sLh_S\snab\alb \label{15.55}
\end{equation}
(the second being the trace-free part relative to the last two
indices of $\sL_S\snab\alb$) and
\begin{equation}
\sL_S D\beb \ \ \mbox{and} \ \ \sLh_S\Dbh\alb \label{15.56}
\end{equation}
By Lemma 4.1 the first and second of \ref{15.55} are equal to
\begin{equation}
\snab\sL_S\beb \ \ \mbox{and} \ \ \snab\sLh_S\alb \label{15.57}
\end{equation}
plus remainders which can be estimated in $L^4(S)$, therefore also
in $L^2(C_u)$ using the estimates of Chapter 4 together with the
$L^\infty$ estimates for the components of $R$. The first of
\ref{15.57} is estimated through inequality \ref{12.242} and the
bounds on the quantities ${\cal P}_2$ and ${\cal Q}_2$. The second
of \ref{15.57} is estimated through inequality \ref{12.193} and
the bounds on the quantities ${\cal P}_2$ and ${\cal Q}_2$. In
this way we obtain:
\begin{eqnarray}
&&\sL_S\snab\beb={\bf O}(\delta|u|^{-5})\nonumber\\
&&\sLh_S\snab\alb=\bfob(\delta^{3/2}|u|^{-11/2}) \label{15.58}
\end{eqnarray}

The first of \ref{15.56} is equal to:
\begin{equation}
\sL_S D\beb=u\Db D\beb+\ub D^2\beb \label{15.59}
\end{equation}
The second term is estimated in $L^2(C_u)$ according to the bound
on the quantity ${\cal Q}_2$. To estimate the first term we appeal
to the sixth of the Bianchi identities, given by Proposition 1.2:
\begin{equation}
D\beb+\frac{1}{2}\Omega\mbox{tr}\chi\beb-\Omega\chih^\sharp\cdot\beb+\omega\beb=\Omega\left\{-\sd\rho+\s^*\sd\sigma
-3\etb\rho+3\s^*\etb\sigma+2\chibh^\sharp\cdot\beta\right\}
\label{15.60}
\end{equation}
Applying $\Db$to this identity  yields the following expression
for $\Db D\beb$:
\begin{eqnarray}
&&\Db D\beb=-\frac{1}{2}\Omega\mbox{tr}\chi\Db\beb+\Omega\chih^\sharp\cdot\Db\beb-\omega\Db\beb\nonumber\\
&&\hspace{13mm}-\frac{1}{2}\Db(\Omega\mbox{tr}\chi)\beb+\Dbh(\Omega\chih)\cdot\beb^\sharp-(\Db\omega)\beb\nonumber\\
&&\hspace{13mm}+\Omega^2\left\{(\chih,\chibh)\beb-2(\chih\times\chibh)\cdot\beb^\sharp\right\}\nonumber\\
&&\hspace{13mm}+\Omega\left\{-\sd\Db\rho+\s^*\sd\Db\sigma-3\etb\Db\rho+3\s^*\etb\Db\sigma\right.\nonumber\\
&&\hspace{18mm}-3(\Db\etb)\rho+3\s^*(\Db\etb)\sigma\nonumber\\
&&\hspace{18mm}+\left.2\chibh^\sharp\cdot\Db\beta+2(\Dbh\chibh)\cdot\beta^\sharp\right\}\nonumber\\
&&\hspace{13mm}-2\Omega^2\s^*\chibh^\sharp\cdot(\sd\sigma+3\etb\sigma)\nonumber\\
&&\hspace{13mm}+\Omega\omb\left\{-\sd\rho+\s^*\sd\sigma-3\etb\rho+3\s^*\etb\sigma+2\chibh^\sharp\cdot\beta\right\}
\label{15.61}
\end{eqnarray}
Consider first the principal terms on the right. These are the
terms $-\sd\Db\rho+\s^*\sd\Db\sigma$ in the second parenthesis.
Now $\Db\rho$, $\Db\sigma$ are given by the Bianchi identities
\ref{13.51}. The principal terms in $\Db\rho$ and $\Db\sigma$ are
the terms $-\Omega\sdiv\beb$ and $-\Omega\scurl\beb$ respectively.
Thus the principal part of the right hand side of \ref{15.61} is:
\begin{equation}
\Omega^2(\sd\sdiv\beb-\s^*\sd\scurl\beb)={\bf O}(\delta|u|^{-6})
\label{15.62}
\end{equation}
according to the bound on the quantity ${\cal Q}_2$. Using
\ref{15.41}, the $L^\infty$ bounds for the components of $R$ and
the results of Chapters 3 and 4 we find that the remaining terms
on the right in \ref{15.61} are ${\cal O}^4(|u|^{-5})$, therefore
also ${\bf O}(|u|^{-5})$. We then conclude that:
\begin{equation}
\Db D\beb={\bf O}(|u|^{-5}) \label{15.63}
\end{equation}
It then follows that:
\begin{equation}
\sL_S D\beb={\bf O}(|u|^{-4}) \label{15.64}
\end{equation}

The second of \ref{15.56} is equal to:
\begin{equation}
\sLh_S\Dbh\alb=u\Dbh^2\alb+\ub\Dh\Dbh\alb \label{15.65}
\end{equation}
The first term is estimated in $L^2(\Cb_{\ub})$ according to the
bound on the quantity ${\cal Q}_2$, while the second term is
estimated in $L^2(\Cb_{\ub})$ through inequalities \ref{12.202}
and \ref{12.227} and the bounds on the quantities ${\cal P}_2$ and
${\cal Q}_2$. We then obtain:
\begin{equation}
\sLh_S\Dbh\alb=\bfob(\delta^{3/2}|u|^{-11/2}) \label{15.66}
\end{equation}

According to the preceding discussion the first term resulting by
applying $\sL_Y$ to a given term in the expression of a given
component of $\s^{(X)}J^1(R)$ consists of two factors the first of
which is ${\cal O}^4(\delta^{r_{1,1}}|u|^{p_{1,1}})$ and the other
is ${\cal O}^4(\delta^{r_{2,1}}|u|^{p_{2,1}})$. We set:
\begin{equation}
r^\prime_1=r_{1,1}+r_{2,1}, \ \ \ p^\prime_1=p_{1,1}+p_{2,1}
\label{15.67}
\end{equation}
The second term resulting by applying $\sL_Y$ to a term in the
expression of a given component of $\s^{(X)}J^1(R)$ consists of
two factors the first of which is ${\cal
O}^\infty(\delta^{r_{1,2}}|u|^{p_{1,2}})$ and the second is a sum
of terms each of which is either ${\bf O}(\delta^r |u|^p)$ or
$\bfob(\delta^r |u|^p)$. We define $r_{2,2}$ to be the minimal $r$
and $p_{2,2}$ to be the maximal $p$ occuring in the terms of the
other factor. We set:
\begin{equation}
r^\prime_2=r_{1,2}+r_{2,2}, \ \ \ p^\prime_2=p_{1,2}+p_{2,2}
\label{15.68}
\end{equation}
We then set:
\begin{equation}
r^\prime=\min\{r^\prime_1,r^\prime_2\}, \ \ \
p^\prime=\max\{p^\prime_1,p^\prime_2\} \label{15.69}
\end{equation}
In this way a pair $r^\prime$, $p^\prime$ is assigned to the pair
of terms resulting by applying $\sL_Y$ to a given term in the
expression of a given component of $\s^{(X)}J^1(R)$. Let $r^*_1$
be the minimal $r^\prime$ and $p^*_1$ be the maximal $p^\prime$
occuring for a given component of $\s^{(X)}J^1(R)$. Now, each
component of $\tcL_Y\s^{(X)}J^1(R)$ is expressed by Proposition
12.3 as $\sL_Y$ applied to the corresponding component of
$\s^{(X)}J^1(R)$ plus a remainder which is a sum of bilinear
terms, the first factor of each term being a  component of
$\s^{(Y)}\tilde{\pi}$ and the second factor a component of
$\s^{(X)}J^1(R)$. For each term the first factor is ${\cal
O}^\infty(\delta^{r_{1,3}}|u|^{p_{1,3}})$ and the second factor is
a sum of terms which may be viewed as being either ${\bf
O}(\delta^r |u|^p)$ or $\bfob(\delta^r |u|^p)$ according to the
analysis of Chapter 14. We define $r_{2,3}$ to be the minimal $r$
and $p_{2,3}$ to be the maximal $p$ occuring in the terms of the
second factor. The pair $r_{2,3},p_{2,3}$ is the pair $r^*,p^*$
assigned to that component of $\s^{(X)}J^1(R)$ in Chapter 14. We set:
\begin{equation}
r^\prime_3=r_{1,3}+r_{2,3}, \ \ \ p^\prime_3=p_{1,3}+p_{2,3}
\label{15.70}
\end{equation}
We then define $r^*_2$ to be the minimal $r^\prime_3$ and $p^*_2$
to be the maximal $p^\prime_3$ occuring in the terms of the
remainder. Finally, we assign to the given component of
$\tcL_Y\s^{(X)}J^1(R)$ the pair $r^*,p^*$ where:
\begin{equation}
r^*=\min\{r^*_1,r^*_2\}, \ \ \ p^*=\max\{p^*_1,p^*_2\}
\label{15.71}
\end{equation}
It turns out that for all components:
\begin{equation}
r^*=r^*_1, \ \ \ p^*=p^*_1
\label{15.b1}
\end{equation}
We obtain in this way tables which are similar to the tables
\ref{14.28} - \ref{14.31}, \ref{14.33}, \ref{14.35}, \ref{14.36},
\ref{14.38} - \ref{14.40}, and \ref{14.41} - \ref{14.44},
\ref{14.46}, \ref{14.48}, \ref{14.49}, \ref{14.51} - \ref{14.53},
and \ref{14.54}, \ref{14.55}. In fact, the tables for the case
$Y=X=L$ ($l=2$) are identical to the tables for the case $X=L$
($l=1$) of Chapter 14, but with the values of $r^\prime$ and $r^*$
decreased by 1. The tables for the case $Y=O_i:i=1,2,3$, $X=L$
($l=1$) are identical to the tables for the case $X=L$ ($l=1$) of
Chapter 14. The tables for the case $Y=O_j:j=1,2,3$,
$X=O_i:i=1,2,3$ ($l=0$) are identical to the tables for the case
$X=O_i:i=1,2,3 $ ($l=0$) of Chapter 14. The tables for the case
$Y=O_i:i=1,2,3$, $X=S$ ($l=0$) are identical to the tables for the
case $X=S$ ($l=0$) of Chapter 14. Finally, the tables for the case
$Y=X=S$ ($l=0$) are identical to the tables for the case $X=S$
($l=0$) of Chapter 14.

Each component of the Weyl currents $\tcL_L\s^{(L)}J^1(R)$,
$\tcL_{O_i}\s^{(L)}J^1(R):i=1,2,3$,
$\tcL_{O_j}\s^{(O_i)}J^1(R);i,j=1,2,3$ and
$\tcL_{O_i}\s^{(S)}J^1(R):i=1,2,3$, $\tcL_S\s^{(S)}J^1(R)$ being
written as a sum of terms in the manner discussed above, and these
expressions being substituted into \ref{14.21}, \ref{14.22},
\ref{14.23} and \ref{14.24}, \ref{14.25} respectively, sums of
trilinear terms result, two of the factors in each term being
contributed by the expression for a component of the Weyl current
$\tcL_L\s^{(L)}J^1(R)$ in the case of \ref{14.21},
$\tcL_{O_i}\s^{(L)}J^1(R):i=1,2,3$ in the case of \ref{14.22},
$\tcL_{O_j}\s^{(O_i)}J^1(R):i,j=1,2,3$ in the case of \ref{14.23}
and $\tcL_{O_i}\s^{(S)}J^1(R)$ in the case of \ref{14.24},
$\tcL_S\s^{(S)}J^1(R)$ in the case of \ref{14.25}, and the other
factor being a component of $\tcL_L\tcL_L R$, $\tcL_{O_i}\tcL_L
R:i=1,2,3$, $\tcL_{O_j}\tcL_{O_i}R:i,j=1,2,3$ and
$\tcL_{O_i}\tcL_S R:i=1,2,3$, $\tcL_S\tcL_S R$ respectively,
multiplied by $\Omega^3$ and the appropriate power of $|u|^2$.
Each third factor is either ${\bf O}(\delta^{r_3}|u|^{p_3})$ in
the case of factors involving $\alpha(\tcL_Y\tcL_X
R)$, $\beta(\tcL_Y\tcL_X R)$, $\rho(\tcL_Y\tcL_X R)$, $\sigma(\tcL_Y\tcL_X
R)$, $\beb(\tcL_Y\tcL_X R)$, or $\bfob(\delta^{r_3}|u|^{p_3})$ in the case of factors
involving $\alb(\tcL_Y\tcL_X R)$. We then obtain a bound for the
contribution of all terms resulting from the product of a given
component of $\tcL_Y\s^{(X)}J^1(R)$ with a given component of
$\tcL_Y\tcL_X R$ to the corresponding error integral \ref{15.5} by
$O(\delta^e)$, where $e$ is the excess index, defined as in
\ref{14.58}, provided that the integrability index $s$, defined as
in \ref{14.59}, is negative so that Lemma 13.1 applies. We obtain
in this way tables which are similar to the tables \ref{14.60} -
\ref{14.68}. In fact, the tables for the case $(YX)=(LL)$ ($l=2$)
are identical to the tables for the case $X=L$ ($l=1$) of Chapter
14. The tables for the case $(YX)=(OL)$ ($l=1$) are identical to
the tables for the case $X=L$ ($l=1$) of Chapter 14. The tables
for the case $(YX)=(OO)$ ($l=0$) are identical to the tables for
the case $X=O$ ($l=0$) of Chapter 14. Finally, the tables for the
cases $(YX)=(OS)$ ($l=0$) and $(YX)=(SS)$ ($l=0$) are identical to
the table for the case $X=S$ ($l=0$) of Chapter 14.

We must now analyze in more detail the terms with vanishing excess
index as they give rise to {\em borderline error integrals}. They
occur only in the cases $n=1$ and $n=3$. The terms which give
borderline contributions to $\s^{(LL)}\stackrel{(3)}{\tau}_c,
\s^{(OL)}\stackrel{(3)}{\tau}_c,\s^{(OO)}\stackrel{(3)}{\tau}_c$
and
$\s^{(OS)}\stackrel{(3)}{\tau}_c,\s^{(SS)}\stackrel{(3)}{\tau}_c$
are the terms:
\begin{eqnarray}
&&-6\Omega^3|u|^6\mbox{tr}\chib(((\Dh\s^{(L)}\ih)\rho-\s^*(\Dh\s^{(L)}\ih)\sigma),\alb(\tcL_L\tcL_L R))\nonumber\\
&&-6\Omega^3|u|^6\mbox{tr}\chib((\s^{(L)}\ih D\rho-\s^{*(L)}\ih
D\sigma),\alb(\tcL_L\tcL_L R)) \label{15.72}
\end{eqnarray}
\begin{eqnarray}
&&-6\Omega^3|u|^6\mbox{tr}\chib\sum_i(((\sLh_{O_i}\s^{(L)}\ih)\rho-\s^*(\sLh_{O_i}\s^{(L)}\ih)\sigma),\alb(\tcL_{O_i}\tcL_L R))\nonumber\\
&&-6\Omega^3|u|^6\mbox{tr}\chib\sum_i((\s^{(L)}\ih
O_i\rho-\s^{*(L)}\ih O_i\sigma),\alb(\tcL_{O_i}\tcL_L R))
\label{15.73}
\end{eqnarray}
\begin{eqnarray}
&&-6\Omega^3|u|^6\mbox{tr}\chib\sum_{i,j}(((\sLh_{O_j}\s^{(O_i)}\ih)\rho-\s^*(\sLh_{O_j}\s^{(O_i)}\ih)\sigma),\alb(\tcL_{O_j}\tcL_{O_i}R))
\nonumber\\
&&-6\Omega^3|u|^6\mbox{tr}\chib\sum_{i,j}((\s^{(O_i)}\ih
O_j\rho-\s^{*(O_i)}\ih O_j\sigma),\alb(\tcL_{O_j}\tcL_{O_i}R))
\label{15.74}
\end{eqnarray}
and:
\begin{eqnarray}
&&-6\Omega^3|u|^6\mbox{tr}\chib\sum_i(((\sLh_{O_i}\s^{(S)}\ih)\rho-\s^*(\sLh_{O_i}\s^{(S)}\ih)\sigma),\alb(\tcL_{O_i}\tcL_S R))\nonumber\\
&&-6\Omega^3|u|^6\mbox{tr}\chib\sum_i((\s^{(S)}\ih
O_i\rho-\s^{*(S)}\ih O_i\sigma),\alb(\tcL_{O_i}\tcL_S R))
\label{15.75}
\end{eqnarray}
\begin{eqnarray}
&&-6\Omega^3|u|^6\mbox{tr}\chib(((\sLh_S\s^{(S)}\ih)\rho-\s^*(\sLh_S\s^{(S)}\ih)\sigma),\alb(\tcL_S\tcL_S R))\nonumber\\
&&-6\Omega^3|u|^6\mbox{tr}\chib((\s^{(S)}\ih S\rho-\s^{*(S)}\ih S\sigma),\alb(\tcL_S\tcL_S R))\nonumber\\
&&-6\Omega^3|u|^6(S\mbox{tr}\chib+2\mbox{tr}\chib)((\s^{(S)}\ih\rho-\s^{*(S)}\ih\sigma),\alb(\tcL_S\tcL_S
R)) \label{15.76}
\end{eqnarray}
respectively. The last results from the term
$2\s^{(S)}\nu\Thetab(J)$ in the expression for $\Thetab(\tcL_S J)$
given by Proposition 12.3 with $S$ in the role of $Y$ (see
\ref{12.37}). It is the only borderline term contributed by the remainders in Proposition 12.3. 

Using the more precise bound:
\begin{equation}
\|\Dh\s^{(L)}\ih\|_{L^4(S_{\ub,u})}\leq
C\delta^{-3/2}|u|^{-1/2}({\cal R}_0^\infty(\alpha)+O(\delta)) \ \
: \ \forall (\ub,u)\in D^\prime_{c^*} \label{15.77}
\end{equation}
implied by the first of \ref{9.2}, the facts that:
\begin{eqnarray}
&&\|\mbox{tr}\chib\rho\|_{L^4(S_{\ub,u})}\leq C|u|^{-7/2}{\cal R}_0^4(\rho) \ \ : \ \forall (\ub,u)\in D^\prime_{c^*}\nonumber\\
&&\|\mbox{tr}\chib\sigma\|_{L^4(S_{\ub,u})}\leq C|u|^{-7/2}{\cal
R}_0^4(\sigma) \ \ : \ \forall (\ub,u)\in D^\prime_{c^*}
\label{15.78}
\end{eqnarray}
as well as \ref{15.12}, and following the proof of Case 5 of Lemma
13.1, we deduce that the contribution of the first of the terms
\ref{15.72} to the error integral \ref{15.5} with $(YX)=(LL)$
($l=2$) and $n=3$ is bounded by:
\begin{equation}
C{\cal R}_0^\infty(\alpha)({\cal R}_0^4(\rho)+{\cal
R}_0^4(\sigma))(\stackrel{(3)}{{\cal F}}_2)^{1/2}+O(\delta)
\label{15.79}
\end{equation}
Also, using the precise bound \ref{14.72}, the facts that:
\begin{eqnarray}
&&\|\mbox{tr}\chib D\rho\|_{L^2(C_u)}\leq C\delta^{-1/2}|u|^{-3}{\cal R}_0(D\rho) \ \ : \ \forall u\in[u_0,c^*)\nonumber\\
&&\|\mbox{tr}\chib D\sigma\|_{L^2(C_u)}\leq
C\delta^{-1/2}|u|^{-3}{\cal R}_0(D\sigma) \ \ : \ \forall
u\in[u_0,c^*) \label{15.80}
\end{eqnarray}
as well as \ref{15.12}, and following the proof of Case 2 of lemma
13.1, we deduce that the contribution of the second of the terms
\ref{15.72} to the same error integral is bounded by:
\begin{equation}
C{\cal R}_0^\infty(\alpha)({\cal R}_0(D\rho)+{\cal
R}_0(D\sigma))(\stackrel{(3)}{{\cal F}}_2)^{1/2} \label{15.81}
\end{equation}
By \ref{14.82}, and by \ref{10.50} and the third of \ref{12.a3}
which together imply:
\begin{equation}
{\cal R}_0^4(\rho),{\cal R}_0^4(\sigma)\leq C(\stackrel{(2)}{{\cal
E}}_1)^{1/2}+O(\delta) \label{15.82}
\end{equation}
\ref{15.78} is in turn bounded by:
\begin{equation}
C(\stackrel{(0)}{{\cal E}}_2)^{1/2}(\stackrel{(2)}{{\cal
E}}_1)^{1/2}(\stackrel{(3)}{{\cal F}}_2)^{1/2}+O(\delta^{1/2})
\label{15.83}
\end{equation}
Also, by \ref{14.82} and the third of \ref{12.a3}, \ref{15.80} is
in turn bounded by \ref{15.83} as well.

The first of the bounds \ref{9.1} together with the first of the
bounds \ref{8.32} and Propositions 8.1 and 8.2 implies:
\begin{equation}
\|\sLh_{O_i}\s^{(L)}\ih\|_{L^4(S_{\ub,u})}\leq
\delta^{-1/2}|u|^{-1/2}(\scR_1^4(\alpha)+{\cal
R}_0^\infty(\alpha)+O(\delta^{1/2})) \ \ : \ \forall (\ub,u)\in
D^\prime_{c^*} \label{15.85}
\end{equation}
Using this bound, \ref{15.78}, as well as \ref{15.16}, and
following the proof of Case 5 of Lemma 13.1, we deduce that the
contribution of the first of the terms \ref{15.73} to the error
integral \ref{15.5} with $(YX)=(OL)$ ($l=1$) and $n=3$ is bounded
by:
\begin{equation}
C(\scR_1^4(\alpha)+{\cal R}_0^\infty(\alpha))({\cal
R}_0^4(\rho)+{\cal R}_0^4(\sigma))(\stackrel{(3)}{{\cal
F}}_2)^{1/2}+O(\delta^{1/2}) \label{15.86}
\end{equation}
Also, using the precise bound \ref{14.72}, the facts that by
Proposition 8.1:
\begin{eqnarray}
&&\|\mbox{tr}\chib O_i\rho\|_{L^2(C_u)}\leq C\delta^{1/2}|u|^{-3}\scR_1(\rho) \ \ : \ \forall u\in[u_0,c^*)\nonumber\\
&&\|\mbox{tr}\chib O_i\sigma\|_{L^2(C_u)}\leq
C\delta^{1/2}|u|^{-3}\scR_1(\sigma) \ \ : \ \forall u\in[u_0,c^*)
\label{15.87}
\end{eqnarray}
as well as \ref{15.16}, and following the proof of Case 2 of lemma
13.1, we deduce that the contribution of the second of the terms
\ref{15.73} to the same error integral is bounded by:
\begin{equation}
C{\cal
R}_0^\infty(\alpha)(\scR_1(\rho)+\scR_1(\sigma))(\stackrel{(3)}{{\cal
F}}_2)^{1/2} \label{15.88}
\end{equation}
By the first of \ref{10.69} and the first of \ref{10.75} together
with the first of \ref{12.a4}:
\begin{equation}
\scR_1^4(\alpha)+{\cal R}_0^\infty(\alpha)\leq
C(\stackrel{(0)}{{\cal E}}_2)^{1/2}+O(\delta^{1/2}) \label{15.89}
\end{equation}
Taking also into account \ref{15.82}, we conclude that \ref{15.86}
is in turn bounded by:
\begin{equation}
C(\stackrel{(0)}{{\cal E}}_2)^{1/2}(\stackrel{(2)}{{\cal
E}}_1)^{1/2}(\stackrel{(3)}{{\cal F}}_2)^{1/2}+O(\delta^{1/2})
\label{15.90}
\end{equation}
Also, by \ref{14.82} and the third of \ref{12.a3}, \ref{15.88} is
in turn bounded by \ref{15.90} as well.

The first of the bounds \ref{9.82} together with the first of the
bounds \ref{8.139} and Propositions 8.1 and 8.2 implies:
\begin{equation}
\|\sLh_{O_j}\s^{(O_i)}\ih\|_{L^4(S_{\ub,u})}\leq
C\delta^{1/2}|u|^{-1/2}(\scR_1^4(\beta)+{\cal
R}_0^\infty(\beta)+O(\delta^{1/2})) \ \ : \ \forall (\ub,u)\in
D^\prime_{c^*} \label{15.92}
\end{equation}
Using this bound, \ref{15.78}, as well as \ref{15.17}, and
following the proof of Case 5 of Lemma 13.1, we deduce that the
contribution of the first of the terms \ref{15.74} to the error
integral \ref{15.5} with $(YX)=(OO)$ ($l=0$) and $n=3$ is bounded
by:
\begin{equation}
C(\scR_1^4(\beta)+{\cal R}_0^\infty(\beta))({\cal
R}_0^4(\rho)+{\cal R}_0^4(\sigma))(\stackrel{(3)}{{\cal
F}}_2)^{1/2}+O(\delta^{1/2}) \label{15.93}
\end{equation}
Also, using the more precice bound \ref{14.76}, \ref{15.87} as
well as \ref{15.17}, and following the proof of Case 2 of lemma
13.1, we deduce that the contribution of the second of the terms
\ref{15.74} to the same error integral is bounded by:
\begin{equation}
C(\scR_1^4(\beta)+{\cal
R}_0^\infty(\beta))(\scR_1(\rho)+\scR_1(\sigma))(\stackrel{(3)}{{\cal
F}}_2)^{1/2}+O(\delta^{1/2}) \label{15.94}
\end{equation}
By \ref{14.85}, \ref{15.82} and the third of \ref{12.a3},
\ref{15.93} and \ref{15.94} are both in turn bounded by:
\begin{equation}
C(\stackrel{(1)}{{\cal E}}_2)^{1/2}(\stackrel{(2)}{{\cal
E}}_1)^{1/2}(\stackrel{(3)}{{\cal F}}_2)^{1/2}+O(\delta^{1/2})
\label{15.95}
\end{equation}

The first of the bounds \ref{9.12} together with the first of the
bounds \ref{8.34} and Propositions 8.1 and 8.2 implies:
\begin{eqnarray}
&&\|\sLh_{O_i}\s^{(S)}\ih\|_{L^4(S_{\ub,u})}\leq
C\delta^{1/2}(\scD_1^4(\chibh)+{\cal D}_0^\infty(\chibh)
+\scR_1^4(\alpha)+{\cal R}_0^\infty(\alpha)+O(\delta^{1/2}))\nonumber\\
&&\hspace{30mm}: \ \forall (\ub,u)\in D^\prime_{c^*} \label{15.96}
\end{eqnarray}
Using this bound, \ref{15.78}, as well as \ref{15.18}, and
following the proof of Case 5 of Lemma 13.1, we deduce that the
contribution of the first of the terms \ref{15.75} to the error
integral \ref{15.5} with $(YX)=(OS)$ ($l=0$) and $n=3$ is bounded
by:
\begin{equation}
C(\scD_1^4(\chibh)+{\cal D}_0^\infty(\chibh)
+\scR_1^4(\alpha)+{\cal R}_0^\infty(\alpha))({\cal
R}_0^4(\rho)+{\cal R}_0^4(\sigma))(\stackrel{(3)}{{\cal
F}}_2)^{1/2}+O(\delta^{1/2}) \label{15.97}
\end{equation}
Also, using the more precice bound \ref{14.79}, \ref{15.87} as
well as \ref{15.18}, and following the proof of Case 2 of lemma
13.1, we deduce that the contribution of the second of the terms
\ref{15.75} to the same error integral is bounded by:
\begin{equation}
C({\cal D}_0^\infty(\chibh)+{\cal
R}_0^\infty(\alpha))(\scR_1(\rho)+\scR_1(\sigma))(\stackrel{(3)}{{\cal
F}}_2)^{1/2}+O(\delta) \label{15.98}
\end{equation}
By \ref{15.89}, \ref{15.82} and the third of \ref{12.a3},
\ref{15.97} and \ref{15.98} are in turn bounded by:
\begin{equation}
C(\scD_1^4(\chibh)+{\cal D}_0^\infty(\chibh)+(\stackrel{(0)}{{\cal
E}}_2)^{1/2})(\stackrel{(2)}{{\cal E}}_1)^{1/2}
(\stackrel{(3)}{{\cal F}}_2)^{1/2}+O(\delta^{1/2}) \label{15.99}
\end{equation}
and:
\begin{equation}
C({\cal D}_0^\infty(\chibh)+(\stackrel{(0)}{{\cal
E}}_2)^{1/2})(\stackrel{(2)}{{\cal
E}}_1)^{1/2}(\stackrel{(3)}{{\cal F}}_2)^{1/2} +O(\delta^{1/2})
\label{15.100}
\end{equation}
respectively.

The first of the bounds \ref{9.13}, \ref{9.14} together imply:
\begin{equation}
\|\sLh_S\s^{(S)}\ih\|_{L^4(S_{\ub,u})}\leq
C\delta^{1/2}|u|^{-1/2}({\cal D}_0^\infty(\chibh)+{\cal R}_0^\infty(\alpha)+O(\delta))
\label{15.101}
\end{equation}
Using this bound, \ref{15.78}, as well as \ref{15.20}, and
following the proof of Case 5 of Lemma 13.1, we deduce that the
contribution of the first of the terms \ref{15.76} to the error
integral \ref{15.5} with $(YX)=(SS)$ ($l=0$) and $n=3$ is bounded
by:
\begin{equation}
C({\cal D}_0^\infty(\chibh)+{\cal R}_0^\infty(\alpha))({\cal
R}_0^4(\rho)+{\cal R}_0^4(\sigma)) (\stackrel{(3)}{{\cal
F}}_2)^{1/2}+O(\delta^{1/2}) \label{15.102}
\end{equation}
Also, using the more precise bound \ref{14.79}, the bounds
\ref{13.54} which imply:
\begin{eqnarray}
&&\|\mbox{tr}\chib S\rho\|_{L^2(C_u)}\leq
C\delta^{1/2}|u|^{-3}((\stackrel{(2)}{{\cal
E}}_1)^{1/2}+O(\delta))
\ \ : \ \forall u\in[u_0,c^*)\nonumber\\
&&\|\mbox{tr}\chib S\rho\|_{L^2(C_u)}\leq
C\delta^{1/2}|u|^{-3}((\stackrel{(2)}{{\cal
E}}_1)^{1/2}+O(\delta)) \ \ : \ \forall u\in[u_0,c^*)\nonumber\\
&&\label{15.103}
\end{eqnarray}
as well as \ref{15.20}, and following the proof of Case 2 of lemma
13.1, we deduce that the contribution of the second of the terms
\ref{15.76} to the same error integral is bounded by:
\begin{equation}
C({\cal D}_0^\infty(\chibh)+{\cal
R}_0^\infty(\alpha))(\stackrel{(2)}{{\cal
E}}_1)^{1/2}(\stackrel{(3)}{{\cal F}}_2)^{1/2}+O(\delta)
\label{15.104}
\end{equation}
By \ref{14.82} and \ref{15.82}, \ref{15.102} and \ref{15.104} are
both in turn bounded by:
\begin{equation}
C({\cal D}_0^\infty(\chibh)+(\stackrel{(0)}{{\cal
E}}_2)^{1/2})(\stackrel{(2)}{{\cal
E}}_1)^{1/2}(\stackrel{(3)}{{\cal F}}_2)^{1/2}+O(\delta^{1/2})
\label{15.105}
\end{equation}
Finally, we consider the contribution of the third of the terms
\ref{15.76}. Using equations \ref{3.7} and \ref{4.c1} we deduce
the following expression for $S\mbox{tr}\chib+\mbox{tr}\chib$:
\begin{eqnarray}
&&S\mbox{tr}\chib+\mbox{tr}\chib=(u\omb-\ub\omega-\varepsilon)\mbox{tr}\chib\nonumber\\
&&\hspace{20mm}-u\Omega|\chibh|^2+\ub\Omega\left\{2\sdiv\etb+2|\etb|^2-(\chih,\chibh)+2\rho\right\}
\label{15.106}
\end{eqnarray}
where $\varepsilon$ is the function defined by \ref{14.166}. Using
the results of Chapters 3 and 4 we obtain:
\begin{equation}
S\mbox{tr}\chib+\mbox{tr}\chib={\cal O}^4(\delta|u|^{-2})
\label{15.107}
\end{equation}
It follows that the contribution comming from the factor
$S\mbox{tr}\chib+\mbox{tr}\chi$ is not borderline, being bounded
by $O(\delta)$. We are left with the contribution of
\begin{equation}
-6\Omega^3|u|^6\mbox{tr}\chib((\s^{(S)}\ih\rho-\s^{*(S)}\ih\sigma),\alb(\tcL_S\tcL_S
R)) \label{15.108}
\end{equation}
which is similar to that of \ref{14.71}, but with
$\alb(\tcL_S\tcL_S R)$ in the role of $\alb(\tcL_S R)$. Using
\ref{5.20} in place of \ref{14.80} we conclude that the
contribution in question is bounded by:
\begin{equation}
C\left\{{\cal D}_0^\infty(\chibh)+(\stackrel{(0)}{{\cal
E}}_2)^{1/2}\right\}(\stackrel{(2)}{{\cal E}}_0)^{1/2}
(\stackrel{(3)}{{\cal F}}_2)^{1/2}+O(\delta^{1/2}) \label{15.109}
\end{equation}
(compare with \ref{14.87}).

The terms which give borderline contributions to
$\s^{(LL)}\stackrel{(1)}{\tau}_c,
\s^{(OL)}\stackrel{(1)}{\tau}_c,\s^{(OO)}\stackrel{(1)}{\tau}_c$
are the terms:
\begin{eqnarray}
&&2\Omega^3|u|^2\mbox{tr}\chib\left\{(\Dh\s^{(L)}\ih,\alpha)\rho(\tcL_L\tcL_L
R)
+\Dh\s^{(L)}\ih\wedge\alpha\sigma(\tcL_L\tcL_L R)\right\}\nonumber\\
&&+2\Omega^3|u|^2\mbox{tr}\chib\left\{(\s^{(L)}\ih,\Dh\alpha)\rho(\tcL_L\tcL_L
R) +\s^{(L)}\ih\wedge\Dh\alpha\sigma(\tcL_L\tcL_L R)\right\}\nonumber\\
&&\label{15.110}
\end{eqnarray}
\begin{eqnarray}
&&2\Omega^3|u|^2\mbox{tr}\chib\sum_i\left\{(\sLh_{O_i}\s^{(L)}\ih,\alpha)\rho(\tcL_{O_i}\tcL_L
R)
+\sLh_{O_i}\s^{(L)}\ih\wedge\alpha\sigma(\tcL_{O_i}\tcL_L R)\right\}\nonumber\\
&&+2\Omega^3|u|^2\mbox{tr}\chib\sum_i\left\{\s^{(L)}\ih,\sLh_{O_i}\alpha)\rho(\tcL_{O_i}\tcL_L
R)
+\s^{(L)}\ih\wedge\sLh_{O_i}\alpha\sigma(\tcL_{O_i}\tcL_L R)\right\}\nonumber\\
&&\label{15.111}
\end{eqnarray}
\begin{eqnarray}
&&2\Omega^3|u|^2\mbox{tr}\chib\sum_{i,j}\left\{(\sLh_{O_j}\s^{(O_i)}\ih,\alpha)\rho(\tcL_{O_j}\tcL_{O_i}R)
+\sLh_{O_j}\s^{(O_i)}\ih\wedge\alpha\sigma(\tcL_{O_j}\tcL_{O_i}R)\right\}\nonumber\\
&&+2\Omega^3|u|^2\mbox{tr}\chib\sum_{i,j}\left\{(\s^{O_i)}\ih,\sLh_{O_j}\alpha)\rho(\tcL_{O_j}\tcL_{O_i}R)
+\s^{(O_i)}\ih\wedge\sLh_{O_j}\alpha\sigma(\tcL_{O_j}\tcL_{O_i}R)\right\}\nonumber\\
&&\label{15.112}
\end{eqnarray}
respectively.

Using \ref{15.77}, the fact that:
\begin{equation}
\|\mbox{tr}\chib\alpha\|_{L^4(S_{\ub,u})}\leq
C\delta^{-3/2}|u|^{-3/2}{\cal R}_0^4(\alpha) \ \ : \ \forall
(\ub,u)\in D^\prime_{c^*} \label{15.113}
\end{equation}
as well as \ref{15.29}, and following the proof of Case 4 of Lemma
13.1, we deduce that the contribution of the first of the terms
\ref{15.110} to the error integral \ref{15.5} with $(YX)=(LL)$
($l=2$) and $n=1$ is bounded by:
\begin{equation}
C{\cal R}_0^\infty(\alpha){\cal
R}_0^4(\alpha)(\stackrel{(2)}{{\cal E}}_2)^{1/2}+O(\delta)
\label{15.115}
\end{equation}
Also, using the precise bound \ref{14.72}, the fact that:
\begin{equation}
\|\mbox{tr}\chib D\alpha\|_{L^2(C_u)}\leq
C\delta^{-2}|u|^{-1}{\cal R}_0(D\alpha) \ \ : \ \forall
u\in[u_0,c^*) \label{15.116}
\end{equation}
as well as \ref{15.29}, and following the proof of Case 1 of lemma
13.1, we deduce that the contribution of the second of the terms
\ref{15.110} to the same error integral is bounded by:
\begin{equation}
C{\cal R}_0^\infty(\alpha){\cal R}_0(D\alpha)(\stackrel{(2)}{{\cal
E}}_2)^{1/2} \label{15.117}
\end{equation}
By \ref{14.82}, and by \ref{10.46} and the first of \ref{12.a3}
which together imply:
\begin{equation}
{\cal R}_0^4(\alpha)\leq C(\stackrel{(0)}{{\cal
E}}_1)^{1/2}+O(\delta) \label{15.118}
\end{equation}
\ref{15.115} is in turn bounded by:
\begin{equation}
C(\stackrel{(0)}{{\cal E}}_2)^{1/2}(\stackrel{(0)}{{\cal
E}}_1)^{1/2}(\stackrel{(2)}{{\cal E}}_2)^{1/2}+O(\delta^{1/2})
\label{15.119}
\end{equation}
Also, by \ref{14.82} and the first of \ref{12.a3}, \ref{15.117} is
in turn bounded by \ref{15.119} as well.

Using the bound \ref{15.85}, \ref{15.113}, as well as \ref{15.32},
and following the proof of Case 4 of Lemma 13.1, we deduce that
the contribution of the first of the terms \ref{15.111} to the
error integral \ref{15.5} with $(YX)=(OL)$ ($l=1$) and $n=1$ is
bounded by:
\begin{equation}
C(\scR_1^4(\alpha)+{\cal R}_0^\infty(\alpha)){\cal
R}_0^4(\alpha)(\stackrel{(2)}{{\cal E}}_2)^{1/2}+O(\delta^{1/2})
\label{15.121}
\end{equation}
Also, using the bound \ref{14.72}, the facts that by Propositions
8.1 and 8.2:
\begin{equation}
\|\mbox{tr}\chib\sLh_{O_i}\alpha\|_{L^2(C_u)}\leq
C\delta^{-1}|u|^{-1}\max\{\scR_1(\alpha),{\cal R}_0(\alpha)\} \ \
: \ \forall u\in[u_0,c^*) \label{15.122}
\end{equation}
as well as \ref{15.32}, and following the proof of Case 1 of lemma
13.1, we deduce that the contribution of the second of the terms
\ref{15.111} to the same error integral is bounded by:
\begin{equation}
C{\cal R}_0^\infty(\alpha)\max\{\scR_1(\alpha),{\cal
R}_0(\alpha)\}(\stackrel{(2)}{{\cal E}}_2)^{1/2} \label{15.123}
\end{equation}
By \ref{15.89} and \ref{15.118}, \ref{15.121} is in turn bounded
by:
\begin{equation}
C(\stackrel{(0)}{{\cal E}}_2)^{1/2}(\stackrel{(0)}{{\cal
E}}_1)^{1/2}(\stackrel{(2)}{{\cal E}}_2)^{1/2}+O(\delta^{1/2})
\label{15.124}
\end{equation}
Also, by \ref{14.82} and the first of \ref{12.a3}, \ref{15.123} is
in turn bounded by \ref{15.124} as well.

Using \ref{15.92}, \ref{15.113}, as well as \ref{15.33}, and
following the proof of Case 4 of Lemma 13.1, we deduce that the
contribution of the first of the terms \ref{15.112} to the error
integral \ref{15.5} with $(YX)=(OO)$ ($l=0$) and $n=1$ is bounded
by:
\begin{equation}
C(\scR_1^4(\beta)+{\cal R}_0^\infty(\beta)){\cal
R}_0^4(\alpha)(\stackrel{(2)}{{\cal E}}_2)^{1/2}+O(\delta^{1/2})
\label{15.125}
\end{equation}
Also, using the bound \ref{14.76}, \ref{15.122} as well as
\ref{15.33}, and following the proof of Case 1 of lemma 13.1, we
deduce that the contribution of the second of the terms
\ref{15.112} to the same error integral is bounded by:
\begin{equation}
C(\scR_1^4(\beta)+{\cal
R}_0^\infty(\beta))\max\{\scR_1(\alpha),{\cal
R}_0(\alpha)\}(\stackrel{(2)}{{\cal E}}_2)^{1/2}+O(\delta^{1/2})
\label{15.126}
\end{equation}
By \ref{14.85}, \ref{15.118} and the first of \ref{12.a3},
\ref{15.125} and \ref{15.126} are both in turn bounded by:
\begin{equation}
C(\stackrel{(1)}{{\cal E}}_2)^{1/2}(\stackrel{(0)}{{\cal
E}}_1)^{1/2}(\stackrel{(2)}{{\cal E}}_2)^{1/2}+O(\delta^{1/2})
\label{15.127}
\end{equation}

We now turn to consider the contribution to the 2nd order Weyl
current error integrals of the term
$$\tcL_Y\s^{(X)}J^2(R)$$
We consider the expressions for the components of $\s^{(X)}J^2(R)$
given by \ref{14.100} with the fundamental Weyl field $R$ in the
role of the Weyl field $W$. The expression for each component of
$\s^{(X)}J^2(R)$ is a sum of terms with two factors, one factor
being a component of $\s^{(X)}p$ and the other a component of $R$.
Now, the components of $\tcL_Y\s^{(X)}J^2(R)$ are given by
Proposition 12.3 in terms of the $\sL_Y$ derivatives of the
components of $\s^{(X)}J^2(R)$. The remainders are estimated using
the $L^\infty$ bounds for the components of $\s^{(Y)}\tilde{\pi}$
and the estimates for the components of $\s^{(X)}J^2(R)$ of
Chapter 14. Now if $\sL_Y$ is applied to a term in the expression
of a given component of $\s^{(X)}J^2(R)$ a sum of two terms will
result, one of which will have as a first factor the $\sL_Y$
derivative of the corresponding component of $\s^{(X)}p$ and the
same second factor as the original term, and the other term will
have the component of $\s^{(X)}p$ as a first factor and $\sL_Y$
applied to the second factor of the original term, plus possibly a
bilinear expression in $\s^{(Y)}\spi$ and the second factor of the
original term, as the other factor. The bilinear expression
results from applying $\sL_Y$ to the coefficients of the
expression constituting the given original term, which may involve
$\sg$ and $\seps$. The first of the two resulting terms is to be
estimated by placing the first factor in $L^2(C_u)$ using the
estimates to be given below, and the second factor in $L^\infty$
using the $L^\infty$ bounds for the components of $R$. The second
of the two resulting terms is to be estimated by placing the first
factor in $L^4(S)$ using the estimates \ref{14.104} -
\ref{14.106}, and the second factor in $L^4(S)$ using the
estimates \ref{15.37}, \ref{15.39}, and \ref{15.43}.

The estimates \ref{9.88} - \ref{9.92}, \ref{9.98} - \ref{9.103},
\ref{9.163} - \ref{9.165}, together with the estimates \ref{9.1} -
\ref{9.3}, \ref{9.12} - \ref{9.14}, \ref{9.82} - \ref{9.84} and
\ref{8.32}, \ref{8.34}, \ref{8.139}, the results of Chapters 3 and
4 and Propositions 8.1 and 8.2, imply through the formulas
\ref{14.101} - \ref{14.103} the following $L^2(C_u)$ estimates for
$\sL_Y$ applied to the components of $\s^{(X)}p$:
\begin{eqnarray}
&&L\s^{(L)}p_4={\bf O}(\delta^{-2}|u|^{-2})\nonumber\\
&&L\s^{(L)}p_3={\bf O}(\delta^{-1}|u|^{-2})\nonumber\\
&&\sL_L\s^{(L)}\sp={\bf O}(\delta^{-3/2}|u|^{-2}) \label{15.128}
\end{eqnarray}
\begin{eqnarray}
&&O_i\s^{(L)}p_4={\bf O}(\delta^{-1}|u|^{-2})\nonumber\\
&&O_i\s^{(L)}p_3={\bf O}(|u|^{-2})\nonumber\\
&&\sL_{O_i}\s^{(L)}\sp={\bf O}(\delta^{-1/2}|u|^{-2})
\label{15.129}
\end{eqnarray}
\begin{eqnarray}
&&O_j\s^{(O_i)}p_4={\bf O}(|u|^{-2})\nonumber\\
&&O_j\s^{(O_i)}p_3={\bf O}(\delta|u|^{-3})\nonumber\\
&&\sL_{O_j}\s^{(O_i)}\sp={\bf O}(\delta^{1/2}|u|^{-2})
\label{15.130}
\end{eqnarray}
\begin{eqnarray}
&&O_i\s^{(S)}p_4={\bf O}(|u|^{-2})\nonumber\\
&&O_i\s^{(S)}p_3={\bf O}(\delta|u|^{-3})\nonumber\\
&&\sL_{O_i}\s^{(S)}\sp={\bf O}(\delta^{1/2}|u|^{-2})
\label{15.131}
\end{eqnarray}
\begin{eqnarray}
&&S\s^{(S)}p_4={\bf O}(|u|^{-2})\nonumber\\
&&S\s^{(S)}p_3={\bf O}(\delta|u|^{-3})\nonumber\\
&&\sL_S\s^{(S)}\sp={\bf O}(\delta^{1/2}|u|^{-2}) \label{15.132}
\end{eqnarray}

According to the preceding discussion the first term resulting by
applying $\sL_Y$ to a given term in the expression of a given
component of $\s^{(X)}J^2(R)$ consists of two factors the first of
which is ${\bf O}(\delta^{r_{1,1}}|u|^{p_{1,1}})$ and the other is
${\cal O}^\infty(\delta^{r_{2,1}}|u|^{p_{2,1}})$. We set:
\begin{equation}
r^\prime_1=r_{1,1}+r_{2,1}, \ \ \ p^\prime_1=p_{1,1}+p_{2,1}
\label{15.133}
\end{equation}
The second term resulting by applying $\sL_Y$ to a term in the
expression of a given component of $\s^{(X)}J^2(R)$ consists of
two factors the first of which is ${\cal
O}^4(\delta^{r_{1,2}}|u|^{p_{1,2}})$ and the second is ${\cal
O}^4(\delta^{r_{2,2}}|u|^{p_{2,2}})$. We set:
\begin{equation}
r^\prime_2=r_{1,2}+r_{2,2}, \ \ \ p^\prime_2=p_{1,2}+p_{2,2}
\label{15.134}
\end{equation}
We then set:
\begin{equation}
r^\prime=\min\{r^\prime_1,r^\prime_2\}, \ \ \
p^\prime=\max\{p^\prime_1,p^\prime_2\} \label{15.135}
\end{equation}
In this way a pair $r^\prime$, $p^\prime$ is assigned to the pair
of terms resulting by applying $\sL_Y$ to a given term in the
expression of a given component of $\s^{(X)}J^2(R)$. Let $r^*_1$
be the minimal $r^\prime$ and $p^*_1$ be the maximal $p^\prime$
occuring for a given component of $\s^{(X)}J^2(R)$. Now, each
component of $\tcL_Y\s^{(X)}J^2(R)$ is expressed by Proposition
12.3 as $\sL_Y$ applied to the corresponding component of
$\s^{(X)}J^2(R)$ plus a remainder which is a sum of bilinear
terms, the first factor of each term being a  component of
$\s^{(Y)}\tilde{\pi}$ and the second factor a component of
$\s^{(X)}J^2(R)$. For each term the first factor is ${\cal
O}^\infty(\delta^{r_{1,3}}|u|^{p_{1,3}})$ and the second factor is
${\bf O}(\delta^{r_{2,3}}|u|^{p_{2,3}})$ according to the analysis
of Chapter 14. The pair $r_{2,3},p_{2,3}$ is the pair $r^*,p^*$
assigned to that component of $\s^{(X)}J^2(R)$ in Chapter 14. For, if
$\{\xi_1,\xi_2\}$ is an $S$ tensorfield which is a bilinear
expression in the $S$ tensorfields $\xi_1$ and $\xi_2$ with
coefficients depending only on $\sg$ and $\seps$ and $\xi_1={\cal
O}^4(\delta^{r_1}|u|^{p_1})$ while $\xi_2={\cal
O}^4(\delta^{r_2}|u|^{p_2})$, then $\{\xi_1,\xi_2\}={\bf
O}(\delta^{r_1+r_2}|u|^{p_1+p_2})$.  We set:
\begin{equation}
r^\prime_3=r_{1,3}+r_{2,3}, \ \ \ p^\prime_3=p_{1,3}+p_{2,3}
\label{15.136}
\end{equation}
We then define $r^*_2$ to be the minimal $r^\prime_3$ and $p^*_2$
to be the maximal $p^\prime_3$ occuring in the terms of the
remainder. Finally, we assign to the given component of
$\tcL_Y\s^{(X)}J^2(R)$ the pair $r^*,p^*$ where:
\begin{equation}
r^*=\min\{r^*_1,r^*_2\}, \ \ \ p^*=\max\{p^*_1,p^*_2\}
\label{15.137}
\end{equation}
It turns out that for all components:
\begin{equation}
r^*=r^*_1, \ \ \ p^*=p^*_1
\label{15.b2}
\end{equation}
We obtain in this way tables which are similar to the tables
\ref{14.109} - \ref{14.118}, \ref{14.119} - \ref{14.128}, and
\ref{14.129} - \ref{14.130}. In fact, the tables for the case
$Y=X=L$ ($l=2$) are identical to the tables for the case $X=L$
($l=1$) of Chapter 14, but with the values of $r^\prime$ and $r^*$
decreased by 1. The tables for the case $Y=O_i:i=1,2,3$, $X=L$
($l=1$) are identical to the tables for the case $X=L$ ($l=1$) of
Chapter 14. The tables for the case $Y=O_j:j=1,2,3$,
$X=O_i:i=1,2,3$ ($l=0$) are identical to the tables for the case
$X=O_i:i=1,2,3 $ ($l=0$) of Chapter 14. The tables for the case
$Y=O_i:i=1,2,3$, $X=S$ ($l=0$) are identical to the tables for the
case $X=S$ ($l=0$) of Chapter 14. Finally, the tables for the case
$Y=X=S$ ($l=0$) are identical to the tables for the case $X=S$
($l=0$) of Chapter 14.

Each component of the Weyl currents $\tcL_L\s^{(L)}J^2(R)$,
$\tcL_{O_i}\s^{(L)}J^2(R):i=1,2,3$,
$\tcL_{O_j}\s^{(O_i)}J^2(R);i,j=1,2,3$ and
$\tcL_{O_i}\s^{(S)}J^2(R):i=1,2,3$, $\tcL_S\s^{(S)}J^2(R)$ being
written as a sum of terms in the manner discussed above, and these
expressions being substituted into \ref{14.21}, \ref{14.22},
\ref{14.23} and \ref{14.24}, \ref{14.25} respectively, sums of
trilinear terms result, two of the factors in each term being
contributed by the expression for a component of the Weyl current
$\tcL_L\s^{(L)}J^2(R)$ in the case of \ref{14.21},
$\tcL_{O_i}\s^{(L)}J^2(R):i=1,2,3$ in the case of \ref{14.22},
$\tcL_{O_j}\s^{(O_i)}J^2(R):i,j=1,2,3$ in the case of \ref{14.23}
and $\tcL_{O_i}\s^{(S)}J^2(R)$ in the case of \ref{14.24},
$\tcL_S\s^{(S)}J^2(R)$ in the case of \ref{14.25}, and the other
factor being a component of $\tcL_L\tcL_L R$, $\tcL_{O_i}\tcL_L
R:i=1,2,3$, $\tcL_{O_j}\tcL_{O_i}R:i,j=1,2,3$ and
$\tcL_{O_i}\tcL_S R:i=1,2,3$, $\tcL_S\tcL_S R$ respectively,
multiplied by $\Omega^3$ and the appropriate power of $|u|^2$.
Each third factor is either ${\bf O}(\delta^{r_3}|u|^{p_3})$ in
the case of factors involving $\alpha(\tcL_Y\tcL_X
R)$, $\beta(\tcL_Y\tcL_X R)$, $\rho(\tcL_Y\tcL_X R)$, $\sigma(\tcL_Y\tcL_X
R)$, $\beb(\tcL_Y\tcL_X R)$, or $\bfob(\delta^{r_3}|u|^{p_3})$ in the case of factors
involving $\alb(\tcL_Y\tcL_X R)$. We then obtain a bound for the
contribution of all terms resulting from the product of a given
component of $\tcL_Y\s^{(X)}J^2(R)$ with a given component of
$\tcL_Y\tcL_X R$ to the corresponding error integral \ref{15.5} by
$O(\delta^e)$, where $e$ is the excess index, defined as in
\ref{14.58}, provided that the integrability index $s$, defined as
in \ref{14.59}, is negative so that Lemma 13.1 applies. We obtain
in this way tables which are similar to the tables \ref{14.134} -
\ref{14.142}. In fact, the tables for the case $(YX)=(LL)$ ($l=2$)
are identical to the tables for the case $X=L$ ($l=1$) of Chapter
14. The tables for the case $(YX)=(OL)$ ($l=1$) are identical to
the tables for the case $X=L$ ($l=1$) of Chapter 14. The tables
for the case $(YX)=(OO)$ ($l=0$) are identical to the tables for
the case $X=O$ ($l=0$) of Chapter 14. Finally, the tables for the
cases $(YX)=(OS)$ ($l=0$) and $(YX)=(SS)$ ($l=0$) are identical to
the table for the case $X=S$ ($l=0$) of Chapter 14. All terms have
positive excess index so there are no borderline terms.

We finally consider the contribution to the 2nd order Weyl current
error integrals of the term
$$\tcL_Y\s^{(X)}J^3(R)$$
We consider the expressions for the components of $\s^{(X)}J^3(R)$
given by Lemma 14.2 with the fundamental Weyl field $R$ in the
role of the Weyl field $W$. The expression for each component of
$\s^{(X)}J^3(R)$ is a sum of terms with two factors, one factor
being a component of $\s^{(X)}q$, which has the algebraic
properties of a Weyl current, and the other a component of $R$.
Now, the components of $\tcL_Y\s^{(X)}J^3(R)$ are given by
Proposition 12.3 in terms of the $\sL_Y$ derivatives of the
components of $\s^{(X)}J^3(R)$. The remainders are estimated using
the $L^\infty$ bounds for the components of $\s^{(Y)}\tilde{\pi}$
and the estimates for the components of $\s^{(X)}J^3(R)$ of
Chapter 14. Now if $\sL_Y$ is applied to a term in the expression
of a given component of $\s^{(X)}J^3(R)$ a sum of two terms will
result, one of which will have as a first factor the $\sL_Y$
derivative of the corresponding component of $\s^{(X)}q$ and the
same second factor as the original term, and the other term will
have the component of $\s^{(X)}q$ as a first factor and $\sL_Y$
applied to the second factor of the original term, plus possibly a
bilinear expression in $\s^{(Y)}\spi$ and the second factor of the
original term, as the other factor. The bilinear expression
results from applying $\sL_Y$ to the coefficients of the
expression constituting the given original term, which may involve
$\sg$ and $\seps$. The first of the two resulting terms is to be
estimated by placing the first factor in $L^2(C_u)$ using the
estimates to be given below, and the second factor in $L^\infty$
using the $L^\infty$ bounds for the components of $R$. The second
of the two resulting terms is to be estimated by placing the first
factor in $L^4(S)$ using the estimates \ref{14.104} -
\ref{14.106}, and the second factor in $L^4(S)$ using the
estimates \ref{15.37}, \ref{15.39}, and \ref{15.43}.

In deducing the $L^2(C_u)$ estimates for $\sL_Y$ applied to the
componets of $\s^{(X)}q$ to be given below, we must take into
account the {\em crucial cancellation} involved in
\begin{equation}
\Dbh\s^{(X)}\ih-\frac{1}{2}\Omega\mbox{tr}\chib\s^{(X)}\ih
\label{15.138}
\end{equation}
the first term on the right in \ref{14.146}, as discussed in
Chapter 14.

In the case $X=L$, \ref{15.138} is given by equation \ref{14.155}.
In view of equation \ref{1.66} the results of Chapters 3 and 4
imply:
\begin{equation}
\sLh_L\left(\Dbh\s^{(L)}\ih-\frac{1}{2}\Omega\mbox{tr}\chib\s^{(L)}\ih\right)={\bf
O}(\delta^{-1/2}|u|^{-3}) \label{15.139}
\end{equation}
Also, Proposition 7.1 together with Propositions 8.1 and 8.2 and
the results of Chapters 3 and 4 implies:
\begin{equation}
\sLh_{O_i}\left(\Dbh\s^{(L)}\ih-\frac{1}{2}\Omega\mbox{tr}\chib\s^{(L)}\ih\right)={\bf
O}(\delta^{1/2}|u|^{-3}) \label{15.140}
\end{equation}

In the case $X=S$, \ref{15.138} is given by equation \ref{14.156}.
Propositions 8.1 and 8.2 together with the results of Chapters 3
and 4 and \ref{15.140} imply:
\begin{equation}
\sLh_{O_i}\left(\Dbh\s^{(S)}\ih-\frac{1}{2}\Omega\mbox{tr}\chib\s^{(S)}\ih\right)={\bf
O}(\delta^{3/2}|u|^{-3}) \label{15.141}
\end{equation}
Taking into account the fact that by equation \ref{1.149} and the
$L^4(S)$ estimate for $\snab^{ \ 2}\omb$ of Proposition 6.2 we have:
\begin{equation}
\snab\Db\eta={\cal O}^4(\delta^{1/2}|u|^{-4}) \label{15.142}
\end{equation}
Noting moreover that the function
$$\varepsilon=\frac{1}{2}\Omega u\mbox{tr}\chib-1$$
(see \ref{14.166}) satifies:
\begin{equation}
S\varepsilon=(1+\varepsilon)S\log\Omega+\frac{1}{2}\Omega
u(S\mbox{tr}\chib+\mbox{tr}\chib)={\cal O}^4(\delta|u|^{-2})
\label{15.143}
\end{equation}
by \ref{15.105}, and taking also into account the last of
\ref{15.43}, we deduce:
\begin{equation}
\sLh_S\left(\Dbh\s^{(S)}\ih-\frac{1}{2}\Omega\mbox{tr}\chib\s^{(S)}\ih\right)={\bf
O}(\delta^{3/2}|u|^{-3}) \label{15.144}
\end{equation}

In the case $X=O_i$, \ref{15.138} is given by equation
\ref{14.168}. In view of Proposition 9.1 and the fact that
\begin{equation}
O_j\varepsilon={\cal O}^\infty(\delta|u|^{-2}) \label{15.145}
\end{equation}
we will have shown that:
\begin{equation}
\sLh_{O_j}\left(\Dbh\s^{(O_i)}\ih-\frac{1}{2}\Omega\mbox{tr}\chib\s^{(O_i)}\ih\right)={\bf
O}(\delta|u|^{-2}) \label{15.146}
\end{equation}
once we establish the following Lemma.

\vspace{5mm}

\noindent{\bf Lemma 15.1} We have:
$$\sL_{O_j}\vartheta_i={\bf O}(\delta|u|^{-2})$$

\noindent{\em Proof:} We first consider the propagation equation
\ref{14.177} along $C_{u_0}$. Since $[L,O_j]=0$ along $C_{u_0}$,
applying $\sL_{O_j}$ to this equation we obtain, in view of Lemma
1.3,
\begin{equation}
D\sL_{O_j}\vartheta_i-\Omega\mbox{tr}\chi\sL_{O_j}\vartheta_i=\sL_{O_j}\varphi_i+O_j(\Omega\mbox{tr}\chi)\vartheta_i
\ \ : \ \mbox{along $C_{u_0}$} \label{15.147}
\end{equation}
Now, $\varphi$ is given by \ref{14.178}, with $\kappa_i$ given by
\ref{14.75}. In applying $\sL_{O_j}$ to \ref{14.178} and
\ref{14.175} terms of the form $\sL_{O_j}\sL_{O_i}\xi$ result,
where $\xi$ is a symmetric trace-free 2-covariant $S$ tensorfield,
one of
$$\Omega\chih,\Omega\chibh,\frac{1}{2}\mbox{tr}\chi\chibh+\sg(\chih,\chibh) \ \ \mbox{and} \ \ \snab\oth\etb+\etb\oth\etb$$
We shall presently show how $\sL_{O_j}\sL_{O_i}\xi$ is to be
estimated in $L^2(C_u)$. We have:
\begin{equation}
\sL_{O_i}\xi=O_i\cdot\snab\xi+\snab O_i\cdot\xi+\xi\cdot\snab O_i
\label{15.148}
\end{equation}
where, in terms of components in an arbitrary local frame field
for $S_{\ub,u_0}$,
$$(\snab O_i\cdot\xi)_{AB}=\snab_A O_i^C\xi_{CB}, \ \ \ (\xi\cdot\snab O_i)=\xi_{AC}\snab_B O_i^C$$
It follows that:
\begin{eqnarray}
&&\sL_{O_j}\sL_{O_i}\xi=\sL_{O_j}O_i\cdot\snab\xi+O_i\cdot\sL_{O_j}\snab\xi
+(\sL_{O_j}\snab O_i)\cdot\xi +\xi\cdot(\sL_{O_j}\snab O_i)\nonumber\\
&&\hspace{17mm}+\snab O_i\cdot\sL_{O_j}\xi +\sL_{O_j}\xi\cdot\snab
O_i \label{15.149}
\end{eqnarray}
In view of the fact that
$\sL_{O_j}O_i=[O_j,O_i]=\epsilon_{jik}O_k$ the first term on the
right is $\epsilon_{jik}O_k\cdot\snab\xi$. Also, by Lemma 9.1 in
the case $p=0$, $q=1$:
\begin{equation}
\sL_{O_j}\snab O_i=\snab\sL_{O_j}O_i+\s^{(O_j)}\spi_1\cdot
O_i=\epsilon_{jik}\snab O_k+\s^{(O_j)}\spi_1\cdot O_i
\label{15.150}
\end{equation}
By Propositions 8.1 and 9.1 we have:
\begin{equation}
\|\s^{(O_j)}\spi_1\cdot O_i\|_{L^4(S_{\ub,u})}\leq
O(\delta^{1/2}|u|^{-1/2}) \label{15.151}
\end{equation}
Taking also into account Proposition 8.2, it follows that:
\begin{eqnarray}
&&\|\sL_{O_j}\sL_{O_i}\xi\|_{L^2(S_{\ub,u})}\leq C(|u|^2\|\snab^{
\ 2}\xi\|_{L^2(S_{\ub,u})}+|u|\|\snab\xi\|_{L^2(S_{\ub,u})}
+\|\xi\|_{L^2(S_{\ub,u})})\nonumber\\
&&\hspace{32mm}+O(\delta^{1/2}|u|^{-1/2})\|\xi\|_{L^4(S_{\ub,u})}
\label{15.152}
\end{eqnarray}
hence:
\begin{eqnarray}
&&\|\sL_{O_j}\sL_{O_i}\xi\|_{L^2(C_u)}\leq C(|u|^2\|\snab^{ \ 2}\xi\|_{L^2(C_u)}+|u|\|\snab\xi\|_{L^2(C_u)}+\|\xi\|_{L^2(C_u)})\nonumber\\
&&\hspace{32mm}+O(\delta|u|^{-1/2})\sup_{\ub}\|\xi\|_{L^4(S_{\ub,u})}
\label{15.153}
\end{eqnarray}
Using Propositions 7.3 and 6.1, 6.2 we then obtain:
\begin{equation}
\sL_{O_j}\kappa_i={\bf O}(|u|^{-3}) \label{15.154}
\end{equation}
and:
\begin{equation}
\sL_{O_j}\varphi={\bf O}(|u|^{-2}) \label{15.155}
\end{equation}
We now apply Lemma 4.6, taking $p=2$, to the propagation equation
\ref{15.147} along $C_{u_0}$, noting that the
$\sL_{O_j}\vartheta_i$ vanish along $\Cb_0$. Here $r=2$, $\nu=2$
and $\gamma=0$. We obtain:
\begin{eqnarray}
&&\|\sL_{O_j}\vartheta_i\|_{L^2(S_{\ub,u_0})}\leq
C\int_0^{\ub}\|\sL_{O_j}\varphi
+O_j(\Omega\mbox{tr}\chi)\vartheta_i\|_{L^2(S_{\ub^\prime,u_0})}d\ub^\prime\nonumber\\
&&\hspace{27mm}\leq C\delta^{1/2}\|\sL_{O_j}\varphi
+O_j(\Omega\mbox{tr}\chi)\vartheta_i\|_{L^2(C_{u_0})}
\label{15.156}
\end{eqnarray}
Substituting \ref{15.155} and the estimate \ref{14.181} yields:
\begin{equation}
\|\sL_{O_j}\vartheta_i\|_{L^2(S_{\ub,u_0})}\leq
O(\delta|u_0|^{-1}) \label{15.157}
\end{equation}

We then consider the propagation equation \ref{14.185} along the
$\Cb_{\ub}$. Since $[\Lb,O_j]=0$ on $M^\prime$ applying
$\sL_{O_j}$ to this equation we obtain, in view of Lemma 1.3,
\begin{equation}
\Db\sL_{O_j}\vartheta_i-\Omega\mbox{tr}\chib\sL_{O_j}\vartheta_i=\sL_{O_j}\varphi_i+O_j(\Omega\mbox{tr}\chib)\vartheta_i
\label{15.158}
\end{equation}
We apply Lemma 4.7 to this equation taking $p=2$. Here $r=2$,
$\nu=2$, $\gammab=0$ and we obtain:
\begin{eqnarray}
&&|u|^{-1}\|\sL_{O_j}\vartheta_i\|_{L^2(S_{\ub,u})}\leq C|u_0|^{-1}\|\sL_{O_j}\vartheta_i\|_{L^2(S_{\ub,u_0})}\label{15.159}\\
&&\hspace{33mm}+C\int_{u_0}^u|u^\prime|^{-1}\|\sL_{O_j}\varphi_i+O_j(\Omega\mbox{tr}\chib)\vartheta_i\|_{L^2(S_{\ub,u^\prime})}du^\prime
\nonumber
\end{eqnarray}
Now the symmetric 2covariant $S$ tensorfields $\varphi_i$ are
given by \ref{14.186}. We estimate the contribution of the term
$-2u\Omega^2\sL_{O_j}\sL_{O_i}\alb$ in $\sL_{O_j}\varphi$ to the
integral on the right in \ref{15.159} by:
\begin{eqnarray}
&&C\int_{u_0}^u\|\sL_{O_j}\sL_{O_i}\alb\|_{L^2(S_{\ub,u^\prime})}du^\prime
\leq C\left(\int_{u_0}^u|u^\prime|^{-6}du^\prime\right)^{1/2}\||u|^3\sL_{O_j}\sL_{O_i}\alb\|_{L^2(\Cb_{\ub})}\nonumber\\
&&\hspace{45mm}\leq O(\delta^{3/2}|u|^{-5/2}) \label{15.160}
\end{eqnarray}
The contributions of the remaining terms are estimated using
\ref{15.152} and Proposition 6.2. Substituting also the estimate
\ref{14.189} we obtain a bound for the integral on the right in
\ref{15.159} by $O(\delta|u|^{-2})$. In view also of the estimate
\ref{15.157}, we then conclude that:
\begin{equation}
\|\sL_{O_j}\vartheta_i\|_{L^2(S_{\ub,u})}\leq O(\delta|u|^{-1})
\label{15.161}
\end{equation}
from which the lemma follows.

\vspace{5mm}

Let us now go back to expressions \ref{14.143} - \ref{14.152}.
Consider first the case $X=L$. Using the estimates \ref{9.88} -
\ref{9.92} and \ref{15.139}, \ref{15.140} and \ref{15.128},
\ref{15.129}, as well as the estimates \ref{9.1} - \ref{9.3},
\ref{14.156}, \ref{14.104}, \ref{8.32}, Propositions 8.1, 8.2 and
the results of Chapters 3 and 4, we deduce:
\begin{eqnarray}
&&\sL_L\Xib(\s^{(L)}q={\bf O}(\delta^{-1/2}|u|^{-3})\nonumber\\
&&\sLh_L\Theta(\s^{(L)}q)={\bf O}(\delta^{-5/2}|u|^{-1})\nonumber\\
&&\sLh_L\Thetab(\s^{(L)}q)={\bf O}(\delta^{-1/2}|u|^{-3})\nonumber\\
&&L\Lambda(\s^{(L)}q)={\bf O}(\delta^{-2}|u|^{-2})\nonumber\\
&&L\Lambdab(\s^{(L)}q)={\bf O}(\delta^{-1}|u|^{-2})\nonumber\\
&&L\Kb(\s^{(L)}q)={\bf O}(\delta^{-1}|u|^{-3})\nonumber\\
&&\sL_L I(\s^{(L)}q)={\bf O}(\delta^{-3/2}|u|^{-2})\nonumber\\
&&\sL_L\Ib(\s^{(L)}q)={\bf O}(\delta^{-3/2}|u|^{-2})
\label{15.162}
\end{eqnarray}
and:
\begin{eqnarray}
&&\sL_{O_i}\Xib(\s^{(L)}q)={\bf O}(\delta^{1/2}|u|^{-3})\nonumber\\
&&\sLh_{O_i}\Theta(\s^{(L)}q)={\bf O}(\delta^{-3/2}|u|^{-1})\nonumber\\
&&\sLh_{O_i}\Thetab(\s^{(L)}q)={\bf O}(\delta^{1/2}|u|^{-3})\nonumber\\
&&O_i\Lambda(\s^{(L)}q)={\bf O}(\delta^{-1}|u|^{-2})\nonumber\\
&&O_i\Lambdab(\s^{(L)}q)={\bf O}(|u|^{-2})\nonumber\\
&&O_i\Kb(\s^{(L)}q)={\bf O}(|u|^{-3})\nonumber\\
&&\sL_{O_i}I(\s^{(L)}q)={\bf O}(\delta^{-1/2}|u|^{-2})\nonumber\\
&&\sL_{O_i}\Ib(\s^{(L)}q)={\bf O}(\delta^{-1/2}|u|^{-2})
\label{15.163}
\end{eqnarray}
(recall from \ref{14.190} that $\Xi(\s^{(L)}q)$, $K(\s^{(L)}q)$
vanish).

\noindent Consider next the case $X=O_i:i=1,2,3$. Using the
estimates \ref{9.163} - \ref{9.165} and \ref{15.146} and
\ref{15.130} as well as the estimates \ref{9.82} - \ref{9.84},
\ref{14.170}, \ref{14.105}, \ref{8.139}, Propositions 8.1, 8.2 and
the results of Chapters 3 and 4, we deduce:
\begin{eqnarray}
&&\sL_{O_j}\Xi(\s^{(O_i)}q)={\bf O}(\delta^{-1/2}|u|^{-1})\nonumber\\
&&\sLh_{O_j}\Theta(\s^{(O_i)}q)={\bf O}(\delta^{-1/2}|u|^{-1})\nonumber\\
&&\sLh_{O_j}\Thetab(\s^{(O_i)}q)={\bf O}(\delta|u|^{-3})\nonumber\\
&&O_j\Lambda(\s^{(O_i)}q)={\bf O}(|u|^{-2})\nonumber\\
&&O_j\Lambdab(\s^{(O_i)}q)={\bf O}(\delta|u|^{-3})\nonumber\\
&&O_j K(\s^{(O_i)}q)={\bf O}(|u|^{-2})\nonumber\\
&&O_j\Kb(\s^{(O_i)}q)={\bf O}(\delta|u|^{-3})\nonumber\\
&&\sL_{O_j}I(\s^{(O_i)}q)={\bf O}(\delta^{1/2}|u|^{-2})\nonumber\\
&&\sL_{O_j}\Ib(\s^{(O_i)}q)={\bf O}(\delta^{1/2}|u|^{-2})
\label{15.164}
\end{eqnarray}
(recall from \ref{14.191} that $\Xib(\s^{(O_i)}q)$ vanishes).

\noindent Consider finally the case $X=S$. Using the estimates
\ref{9.98} - \ref{9.103} and \ref{15.141}, \ref{15.144} and
\ref{15.131}, \ref{15.132} as well as the estimates \ref{9.12} -
\ref{9.14}, \ref{14.161}, \ref{14.106}, \ref{8.34}, Propositions
8.1, 8.2 and the results of  Chapters 3 and 4, we deduce:
\begin{eqnarray}
&&\sL_{O_i}\Xi(\s^{(S)}q)={\bf O}(\delta^{-1/2}|u|^{-1})\nonumber\\
&&\sL_{O_i}\Xib(\s^{(S)}q)={\bf O}(\delta^{3/2}|u|^{-3})\nonumber\\
&&\sLh_{O_i}\Theta(\s^{(S)}q)={\bf O}(\delta^{-1/2}|u|^{-1})\nonumber\\
&&\sLh_{O_i}\Thetab(\s^{(S)}q)={\bf O}(\delta^{3/2}|u|^{-3})\nonumber\\
&&O_i\Lambda(\s^{(S)}q)={\bf O}(|u|^{-2})\nonumber\\
&&O_i\Lambdab(\s^{(S)}q)={\bf O}(\delta|u|^{-3})\nonumber\\
&&O_i K(\s^{(S)}q)={\bf O}(|u|^{-2})\nonumber\\
&&O_i\Kb(\s^{(S)}q)={\bf O}(\delta|u|^{-3})\nonumber\\
&&\sL_{O_i}I(\s^{(S)}q)={\bf O}(\delta^{1/2}|u|^{-2})\nonumber\\
&&\sL_{O_i}\Ib(\s^{(S)}q)={\bf O}(\delta^{1/2}|u|^{-2})
\label{15.165}
\end{eqnarray}
and:
\begin{eqnarray}
&&\sL_S\Xi(\s^{(S)}q)={\bf O}(\delta^{-1/2}|u|^{-1})\nonumber\\
&&\sL_S\Xib(\s^{(S)}q)={\bf O}(\delta^{3/2}|u|^{-3})\nonumber\\
&&\sLh_S\Theta(\s^{(S)}q)={\bf O}(\delta^{-1/2}|u|^{-1})\nonumber\\
&&\sLh_S\Thetab(\s^{(S)}q)={\bf O}(\delta^{3/2}|u|^{-3})\nonumber\\
&&S\Lambda(\s^{(S)}q)={\bf O}(|u|^{-2})\nonumber\\
&&S\Lambdab(\s^{(S)}q)={\bf O}(\delta|u|^{-3})\nonumber\\
&&SK(\s^{(S)}q)={\bf O}(|u|^{-2})\nonumber\\
&&S\Kb(\s^{(S)}q)={\bf O}(\delta|u|^{-3})\nonumber\\
&&\sL_S I(\s^{(S)}q)={\bf O}(\delta^{1/2}|u|^{-2})\nonumber\\
&&\sL_S\Ib(\s^{(S)}q)={\bf O}(\delta^{1/2}|u|^{-2}) \label{15.166}
\end{eqnarray}

According to the discussion of the paragraph preceding
\ref{15.138} the first term resulting by applying $\sL_Y$ to a
given term in the expression of a given component of
$\s^{(X)}J^3(R)$ consists of two factors the first of which is
${\bf O}(\delta^{r_{1,1}}|u|^{p_{1,1}})$ and the other is ${\cal
O}^\infty(\delta^{r_{2,1}}|u|^{p_{2,1}})$. We set:
\begin{equation}
r^\prime_1=r_{1,1}+r_{2,1}, \ \ \ p^\prime_1=p_{1,1}+p_{2,1}
\label{15.167}
\end{equation}
The second term resulting by applying $\sL_Y$ to a term in the
expression of a given component of $\s^{(X)}J^3(R)$ consists of
two factors the first of which is ${\cal
O}^4(\delta^{r_{1,2}}|u|^{p_{1,2}})$ and the second is ${\cal
O}^4(\delta^{r_{2,2}}|u|^{p_{2,2}})$. We set:
\begin{equation}
r^\prime_2=r_{1,2}+r_{2,2}, \ \ \ p^\prime_2=p_{1,2}+p_{2,2}
\label{15.168}
\end{equation}
We then set:
\begin{equation}
r^\prime=\min\{r^\prime_1,r^\prime_2\}, \ \ \
p^\prime=\max\{p^\prime_1,p^\prime_2\} \label{15.169}
\end{equation}
In this way a pair $r^\prime$, $p^\prime$ is assigned to the pair
of terms resulting by applying $\sL_Y$ to a given term in the
expression of a given component of $\s^{(X)}J^3(R)$. Let $r^*_1$
be the minimal $r^\prime$ and $p^*_1$ be the maximal $p^\prime$
occuring for a given component of $\s^{(X)}J^3(R)$. Now, each
component of $\tcL_Y\s^{(X)}J^3(R)$ is expressed by Proposition
12.3 as $\sL_Y$ applied to the corresponding component of
$\s^{(X)}J^3(R)$ plus a remainder which is a sum of bilinear
terms, the first factor of each term being a  component of
$\s^{(Y)}\tilde{\pi}$ and the second factor a component of
$\s^{(X)}J^3(R)$. For each term the first factor is ${\cal
O}^\infty(\delta^{r_{1,3}}|u|^{p_{1,3}})$ and the second factor is
${\bf O}(\delta^{r_{2,3}}|u|^{p_{2,3}})$ according to the analysis
of Chapter 14. The pair $r_{2,3},p_{2,3}$ is the pair $r^*,p^*$
assigned to that component of $\s^{(X)}J^3(R)$ in Chapter 14. We set:
\begin{equation}
r^\prime_3=r_{1,3}+r_{2,3}, \ \ \ p^\prime_3=p_{1,3}+p_{2,3}
\label{15.170}
\end{equation}
We then define $r^*_2$ to be the minimal $r^\prime_3$ and $p^*_2$
to be the maximal $p^\prime_3$ occuring in the terms of the
remainder. Finally, we assign to the given component of
$\tcL_Y\s^{(X)}J^3(R)$ the pair $r^*,p^*$ where:
\begin{equation}
r^*=\min\{r^*_1,r^*_2\}, \ \ \ p^*=\max\{p^*_1,p^*_2\}
\label{15.171}
\end{equation}
It turns out that for all components:
\begin{equation}
r^*=r^*_1, \ \ \ p^*=p^*_1
\label{15.b3}
\end{equation}
We obtain in this way tables which are similar to the tables
\ref{14.194} - \ref{14.203}, \ref{14.204} - \ref{14.213}, and
\ref{14.214} - \ref{14.215}. In fact, the tables for the case
$Y=X=L$ ($l=2$) are identical to the tables for the case $X=L$
($l=1$) of Chapter 14, but with the values of $r^\prime$ and $r^*$
decreased by 1. The tables for the case $Y=O_i:i=1,2,3$, $X=L$
($l=1$) are identical to the tables for the case $X=L$ ($l=1$) of
Chapter 14. The tables for the case $Y=O_j:j=1,2,3$,
$X=O_i:i=1,2,3$ ($l=0$) are identical to the tables for the case
$X=O_i:i=1,2,3 $ ($l=0$) of Chapter 14. The tables for the case
$Y=O_i:i=1,2,3$, $X=S$ ($l=0$) are identical to the tables for the
case $X=S$ ($l=0$) of Chapter 14. Finally, the tables for the case
$Y=X=S$ ($l=0$) are identical to the tables for the case $X=S$
($l=0$) of Chapter 14.

Each component of the Weyl currents $\tcL_L\s^{(L)}J^3(R)$,
$\tcL_{O_i}\s^{(L)}J^3(R):i=1,2,3$,
$\tcL_{O_j}\s^{(O_i)}J^3(R);i,j=1,2,3$ and
$\tcL_{O_i}\s^{(S)}J^3(R):i=1,2,3$, $\tcL_S\s^{(S)}J^3(R)$ being
written as a sum of terms in the manner discussed above, and these
expressions being substituted into \ref{14.21}, \ref{14.22},
\ref{14.23} and \ref{14.24}, \ref{14.25} respectively, sums of
trilinear terms result, two of the factors in each term being
contributed by the expression for a component of the Weyl current
$\tcL_L\s^{(L)}J^3(R)$ in the case of \ref{14.21},
$\tcL_{O_i}\s^{(L)}J^3(R):i=1,2,3$ in the case of \ref{14.22},
$\tcL_{O_j}\s^{(O_i)}J^3(R):i,j=1,2,3$ in the case of \ref{14.23}
and $\tcL_{O_i}\s^{(S)}J^3(R)$ in the case of \ref{14.24},
$\tcL_S\s^{(S)}J^3(R)$ in the case of \ref{14.25}, and the other
factor being a component of $\tcL_L\tcL_L R$, $\tcL_{O_i}\tcL_L
R:i=1,2,3$, $\tcL_{O_j}\tcL_{O_i}R:i,j=1,2,3$ and
$\tcL_{O_i}\tcL_S R:i=1,2,3$, $\tcL_S\tcL_S R$ respectively,
multiplied by $\Omega^3$ and the appropriate power of $|u|^2$.
Each third factor is either ${\bf O}(\delta^{r_3}|u|^{p_3})$ in
the case of factors involving $\alpha(\tcL_Y\tcL_X
R)$, $\beta(\tcL_Y\tcL_X R)$, $\rho(\tcL_Y\tcL_X R)$, $\sigma(\tcL_Y\tcL_X
R)$, $\beb(\tcL_Y\tcL_X R)$, or $\bfob(\delta^{r_3}|u|^{p_3})$ in the case of factors
involving $\alb(\tcL_Y\tcL_X R)$. We then obtain a bound for the
contribution of all terms resulting from the product of a given
component of $\tcL_Y\s^{(X)}J^3(R)$ with a given component of
$\tcL_Y\tcL_X R$ to the corresponding error integral \ref{15.5} by
$O(\delta^e)$, where $e$ is the excess index, defined as in
\ref{14.58}, provided that the integrability index $s$, defined as
in \ref{14.59}, is negative so that Lemma 13.1 applies. We obtain
in this way tables which are similar to the tables \ref{14.217} -
\ref{14.225}. In fact, the tables for the case $(YX)=(LL)$ ($l=2$)
are identical to the tables for the case $X=L$ ($l=1$) of Chapter
14. The tables for the case $(YX)=(OL)$ ($l=1$) are identical to
the tables for the case $X=L$ ($l=1$) of Chapter 14. The tables
for the case $(YX)=(OO)$ ($l=0$) are identical to the tables for
the case $X=O$ ($l=0$) of Chapter 14. Finally, the tables for the
cases $(YX)=(OS)$ ($l=0$) and $(YX)=(SS)$ ($l=0$) are identical to
the table for the case $X=S$ ($l=0$) of Chapter 14. All terms have
positive excess index so there are no borderline terms.

We summarize the results of this section in the following
proposition.

\vspace{5mm}

\noindent{\bf Proposition 15.2} \ \ \ The contribution to the
error integral
$$\delta^{2q_n+2l}\int_{M^\prime_{c^*}}|\s^{(YX)}\stackrel{(n)}{\tau}_c|d\mu_g$$
of the term
$$\tcL_Y\s^{(X)}J$$
is bounded as follows.

\vspace{3mm}

1. For $(YX)=(LL)$.

\vspace{2mm}

\noindent In the case $n=0$ by:
$$O(\delta)$$

\noindent In the case $n=1$ by:
$$C(\stackrel{(0)}{{\cal E}}_2)^{1/2}(\stackrel{(0)}{{\cal E}}_1)^{1/2}(\stackrel{(2)}{{\cal E}}_2)^{1/2}+O(\delta^{1/2})$$

\noindent In the case $n=2$ by:
$$O(\delta)$$

\noindent In the case $n=3$ by:
$$C(\stackrel{(0)}{{\cal E}}_2)^{1/2}(\stackrel{(2)}{{\cal E}}_1)^{1/2}(\stackrel{(3)}{{\cal F}}_2)^{1/2}+O(\delta^{1/2})$$

\vspace{3mm}

2. For $(YX)=(OL)$.

\vspace{2mm}

\noindent In the case $n=0$ by:

$$O(\delta)$$

\noindent In the case $n=1$ by:
$$C(\stackrel{(0)}{{\cal E}}_2)^{1/2}(\stackrel{(0)}{{\cal E}}_1)^{1/2}(\stackrel{(2)}{{\cal E}}_2)^{1/2}+O(\delta^{1/2})$$

\noindent In the case $n=2$ by:
$$O(\delta)$$

\noindent In the case $n=3$ by:
$$C(\stackrel{(0)}{{\cal E}}_2)^{1/2}(\stackrel{(2)}{{\cal E}}_1)^{1/2}(\stackrel{(3)}{{\cal F}}_2)^{1/2}+O(\delta^{1/2})$$

\vspace{3mm}

3. For $(YX)=(OO)$.

\vspace{2mm}

\noindent In the case $n=0$ by:
$$O(\delta)$$

\noindent In the case $n=1$ by:
$$C(\stackrel{(1)}{{\cal E}}_2)^{1/2}(\stackrel{(0)}{{\cal E}}_1)^{1/2}(\stackrel{(2)}{{\cal E}}_2)^{1/2}+O(\delta^{1/2})$$

\noindent In the case $n=2$ by:

$$O(\delta)$$

\noindent In the case $n=3$ by:
$$C(\stackrel{(1)}{{\cal E}}_2)^{1/2}(\stackrel{(2)}{{\cal E}}_1)^{1/2}(\stackrel{(3)}{{\cal F}}_2)^{1/2}+O(\delta^{1/2})$$

\vspace{3mm}

4. For $(YX)=(OS)$.

\noindent Here we only have the case $n=3$, where we have a bound
by:
$$C\left\{\scD_1^4(\chibh)+{\cal D}_0^\infty(\chibh)+(\stackrel{(0)}{{\cal E}}_2)^{1/2}\right\}
(\stackrel{(2)}{{\cal E}}_1)^{1/2}(\stackrel{(3)}{{\cal
F}}_2)^{1/2}+O(\delta^{1/2})$$

5. For $(YX)=(SS)$.

\noindent Here we only have the case $n=3$, where we have a bound
by:
$$C\left\{{\cal D}_0^\infty(\chibh)+(\stackrel{(0)}{{\cal E}}_2)^{1/2}\right\}
(\stackrel{(2)}{{\cal E}}_1)^{1/2}(\stackrel{(3)}{{\cal
F}}_2)^{1/2}+O(\delta^{1/2})$$

\chapter{The Energy-Flux Estimates. Completion of the Continuity
Argument}

\section{The energy-flux estimates}

We now consider the error integrals
\begin{equation}
\delta^{2q_n}\int_{M^\prime_{c^*}}|\stackrel{(n)}{\tau}_2|d\mu _g
\ \ : \ n=0,1,2,3 \label{16.1}
\end{equation}
which appear on the right hand sides of the energy-flux
inequalities \ref{12.272} - \ref{12.273} (see definitions
\ref{12.274}). Let us recall the definitions \ref{12.264},
\ref{12.266}. Combining the results of Propositions 13.1, 14.1,
14.2, 14.3, 15.1 and 15.2 we arrive at the following bounds for
the error integrals \ref{16.1}:
\begin{eqnarray}
&&\delta^{2q_0}\int_{M^\prime_{c^*}}|\stackrel{(0)}{\tau}_2|d\mu_g\leq O(\delta)\nonumber\\
&&\delta^{2q_1}\int_{M^\prime_{c^*}}|\stackrel{(1)}{\tau}_2|d\mu_g\leq
C\left\{{\cal D}_0^\infty(\chibh)+(\stackrel{(0)}{{\cal
E}}_2)^{1/2}+(\stackrel{(1)}{{\cal E}}_2)^{1/2}\right\}
(\stackrel{(0)}{{\cal E}}_2)^{1/2}(\stackrel{(2)}{{\cal E}}_2)^{1/2}\nonumber\\
&&\hspace{35mm}+O(\delta^{1/2})\nonumber\\
&&\delta^{2q_2}\int_{M^\prime_{c^*}}|\stackrel{(2)}{\tau}_2|d\mu_g\leq O(\delta)\nonumber\\
&&\delta^{2q_3}\int_{M^\prime_{c^*}}|\stackrel{(3)}{\tau}_2|d\mu_g\leq\nonumber\\
&&\hspace{21mm}C\left\{\scD_1^4(\chibh)+{\cal
D}_0^\infty(\chibh)+(\stackrel{(0)}{{\cal
E}}_2)^{1/2}+(\stackrel{(1)}{{\cal E}}_2)^{1/2}\right\}
(\stackrel{(2)}{{\cal E}}_2)^{1/2}(\stackrel{(3)}{{\cal F}}_2)^{1/2}\nonumber\\
&&\hspace{35mm}+O(\delta^{1/2}) \label{16.2}
\end{eqnarray}

Now, the initial data quantities ${\cal D}_0^\infty(\chibh)$, $\scD_1^4(\chibh)$ come from the bound \ref{13.38} for 
$\|\s^{(K)}\ih\|_{L^\infty(S_{\ub,u})}$, the bound \ref{14.79} for $\|\s^{(S)}\ih\|_{L^\infty(S_{\ub,u})}$, the bound 
\ref{15.96} for $\|\sLh_{O_i}\s^{(S)}\ih\|_{L^4(S_{\ub,u})}$, and the bound \ref{15.101} for 
$\|\sLh_S\s^{(S)}\ih\|_{L^4(S_{\ub,u})}$. The contributions of the initial data quantities in question to these bounds 
is through the bounds of Chapters 3 and 4 for $\|\chibh\|_{L^\infty(S_{\ub,u})}$ and $\|\snab\chibh\|_{L^4(S_{\ub,u}}$ 
(see first of \ref{8.27}, \ref{8.30}). Evidently, the bounds of Chapters 3 and 4 hold equally well if in the definitions 
\ref{3.37}, \ref{3.108} $C_{u_0}$ is replaced by its part lying in $M^\prime_{c^*}$ and in the definitions \ref{4.80}, \ref{4.81} the supremum over $[0,\delta]$ is replaced by the supremum over $[0,\min\{\delta,c^*-u_0\}]$. Thus, in the inequalities \ref{16.2} we may replace ${\cal D}_0^\infty(\chibh)$ and $\scD_1^4(\chibh)$ by:
\begin{equation}
\sup_{\ub\in[0,\min\{\delta,c^*-u_0\}]}\left(|u_0|^2\delta^{-1/2}\|\chibh\|_{L^\infty(S_{\ub,u_0})}\right)
\label{16.b1}
\end{equation}
and:
\begin{equation}
\sup_{\ub\in[0,\min\{\delta,c^*-u_0\}]}\left(|u_0|^{5/2}\delta^{-1/2}\|\chibh\|_{L^4(S_{\ub,u_0})}\right)
\label{16.b2}
\end{equation}
respectively.

Consider the estimate of Proposition 6.2 for $\chibh$:
\begin{eqnarray}
&&|u|\|\snab^{ \ 2}\chibh\|_{L^4(S_{\ub,u})}+\|\snab\chibh\|_{L^4(S_{\ub,u})}+|u|^{-1}\|\chibh\|_{L^4(S_{\ub,u})}\nonumber\\
&&\hspace{20mm}\leq C\delta^{1/2}|u|^{-5/2}(\scR_1^4(\beta)+{\cal
R}_0^\infty(\beta))+O(\delta|u|^{-7/2}) \label{16.3}
\end{eqnarray}
in $D^\prime_{c^*}$. This estimate holds in particular at $u=u_0$. Through Lemma 5.2
with $p=4$, it implies (see definitions \ref{3.37}, \ref{4.81}):
\begin{eqnarray}
&&\sup_{\ub\in[0,\min\{\delta,c^*-u_0\}]}\left(|u_0|^2\delta^{-1/2}\|\chibh\|_{L^\infty(S_{\ub,u_0})}\right)\leq
C(\scR_1^4(\beta)+{\cal R}_0^\infty(\beta))+O(\delta^{1/2})\nonumber\\
&&\sup_{\ub\in[0,\min\{\delta,c^*-u_0\}]}\left(|u_0|^{5/2}\delta^{-1/2}\|\chibh\|_{L^4(S_{\ub,u_0})}\right)\leq
C(\scR_1^4(\beta)+{\cal R}_0^\infty(\beta))+O(\delta^{1/2})\nonumber\\
&&\label{16.4}
\end{eqnarray}
hence, by \ref{14.85}:
\begin{eqnarray}
&&\sup_{\ub\in[0,\min\{\delta,c^*-u_0\}]}\left(|u_0|^2\delta^{-1/2}\|\chibh\|_{L^\infty(S_{\ub,u_0})}\right)\leq
C(\stackrel{(1)}{{\cal E}}_2)^{1/2}+O(\delta^{1/2})\nonumber\\
&&\sup_{\ub\in[0,\min\{\delta,c^*-u_0\}]}\left(|u_0|^{5/2}\delta^{-1/2}\|\chibh\|_{L^4(S_{\ub,u_0})}\right)\leq
C(\stackrel{(1)}{{\cal E}}_2)^{1/2}+O(\delta^{1/2})\nonumber\\
&&\label{16.5}
\end{eqnarray}
In view of \ref{16.5},  the inequalities  \ref{16.2} simplify to:
\begin{eqnarray}
&&\delta^{2q_0}\int_{M^\prime_{c^*}}|\stackrel{(0)}{\tau}_2|d\mu_g\leq O(\delta)\nonumber\\
&&\delta^{2q_1}\int_{M^\prime_{c^*}}|\stackrel{(1)}{\tau}_2|d\mu_g\leq
C\left\{(\stackrel{(0)}{{\cal E}}_2)^{1/2}+(\stackrel{(1)}{{\cal
E}}_2)^{1/2}\right\}
(\stackrel{(0)}{{\cal E}}_2)^{1/2}(\stackrel{(2)}{{\cal E}}_2)^{1/2}+O(\delta^{1/2})\nonumber\\
&&\delta^{2q_2}\int_{M^\prime_{c^*}}|\stackrel{(2)}{\tau}_2|d\mu_g\leq O(\delta)\nonumber\\
&&\delta^{2q_3}\int_{M^\prime_{c^*}}|\stackrel{(3)}{\tau}_2|d\mu_g\leq
C\left\{(\stackrel{(0)}{{\cal E}}_2)^{1/2}+(\stackrel{(1)}{{\cal
E}}_2)^{1/2}\right\}
(\stackrel{(2)}{{\cal E}}_2)^{1/2}(\stackrel{(3)}{{\cal F}}_2)^{1/2}+O(\delta^{1/2})\nonumber\\
&&\label{16.6}
\end{eqnarray}
The constant $C$ in the second and fourth bound may be taken to be
the same constant.

Substituting the bounds \ref{16.6} in the energy-flux inequalities
\ref{12.272}, \ref{12.273}, we obtain the inequalities:
\begin{eqnarray}
&&\stackrel{(0)}{{\cal E}}_2\leq\stackrel{(0)}{D}+O(\delta)\nonumber\\
&&\stackrel{(1)}{{\cal E}}_2\leq
\stackrel{(1)}{D}+C\left\{(\stackrel{(0)}{{\cal
E}}_2)^{1/2}+(\stackrel{(1)}{{\cal E}}_2)^{1/2}\right\}
(\stackrel{(0)}{{\cal E}}_2)^{1/2}(\stackrel{(2)}{{\cal E}}_2)^{1/2}+O(\delta^{1/2})\nonumber\\
&&\stackrel{(2)}{{\cal E}}_2\leq \stackrel{(2)}{D}+O(\delta)\nonumber\\
&&\stackrel{(3)}{{\cal E}}_2\leq
\stackrel{(3)}{D}+C\left\{(\stackrel{(0)}{{\cal
E}}_2)^{1/2}+(\stackrel{(1)}{{\cal E}}_2)^{1/2}\right\}
(\stackrel{(2)}{{\cal E}}_2)^{1/2}(\stackrel{(3)}{{\cal F}}_2)^{1/2}+O(\delta^{1/2})\nonumber\\
&&\label{16.7}
\end{eqnarray}
and:
\begin{equation}
\stackrel{(3)}{{\cal
F}}_2\leq\stackrel{(3)}{D}+C\left\{(\stackrel{(0)}{{\cal
E}}_2)^{1/2}+(\stackrel{(1)}{{\cal E}}_2)^{1/2}\right\}
(\stackrel{(2)}{{\cal E}}_2)^{1/2}(\stackrel{(3)}{{\cal
F}}_2)^{1/2}+O(\delta^{1/2}) \label{16.8}
\end{equation}
Choosing $\delta$ suitably small depending on the quantities
$\stackrel{(n)}{D} \ : \ n=0,1,2,3$; ${\cal D}^{\prime
4}_{[1]}(\alb)$  and the quantities ${\cal D}_0^\infty$,
$\scD_1^4$, $\scD_2^4(\mbox{tr}\chib)$, $\scD_3(\mbox{tr}\chib)$,
we can make each of the $O(\delta^{1/2})$ and $O(\delta)$ terms on
the right hand sides of the inequalities \ref{16.7} and \ref{16.8}
less than or equal to 1. We then obtain the following closed
system of inequalities for the quantities $\stackrel{(0)}{{\cal
E}}_2,\stackrel{(1)}{{\cal E}}_2,\stackrel{(2)}{{\cal
E}}_2,\stackrel{(3)}{{\cal E}}_2$ and $\stackrel{(3)}{{\cal
F}}_2$:
\begin{eqnarray}
&&\stackrel{(0)}{{\cal E}}_2\leq\stackrel{(0)}{D}+1\nonumber\\
&&\stackrel{(1)}{{\cal E}}_2\leq
\stackrel{(1)}{D}+C\left\{(\stackrel{(0)}{{\cal
E}}_2)^{1/2}+(\stackrel{(1)}{{\cal E}}_2)^{1/2}\right\}
(\stackrel{(0)}{{\cal E}}_2)^{1/2}(\stackrel{(2)}{{\cal E}}_2)^{1/2}+1\nonumber\\
&&\stackrel{(2)}{{\cal E}}_2\leq \stackrel{(2)}{D}+1\nonumber\\
&&\stackrel{(3)}{{\cal E}}_2\leq
\stackrel{(3)}{D}+C\left\{(\stackrel{(0)}{{\cal
E}}_2)^{1/2}+(\stackrel{(1)}{{\cal E}}_2)^{1/2}\right\}
(\stackrel{(2)}{{\cal E}}_2)^{1/2}(\stackrel{(3)}{{\cal F}}_2)^{1/2}+1\nonumber\\
&&\label{16.9}
\end{eqnarray}
and:
\begin{equation}
\stackrel{(3)}{{\cal
F}}_2\leq\stackrel{(3)}{D}+C\left\{(\stackrel{(0)}{{\cal
E}}_2)^{1/2}+(\stackrel{(1)}{{\cal E}}_2)^{1/2}\right\}
(\stackrel{(2)}{{\cal E}}_2)^{1/2}(\stackrel{(3)}{{\cal
F}}_2)^{1/2}+1 \label{16.10}
\end{equation}

Substituting the first and third of inequalities \ref{16.9} in the
second, we obtain:
\begin{eqnarray}
&&\stackrel{(1)}{{\cal E}}_2\leq \stackrel{(1)}{D}+C(\stackrel{(0)}{D}+1)(\stackrel{(2)}{D}+1)^{1/2}+1\nonumber\\
&&\hspace{8mm}+C(\stackrel{(0)}{D}+1)^{1/2}(\stackrel{(2)}{D}+1)^{1/2}(\stackrel{(1)}{{\cal
E}}_2)^{1/2} \label{16.11}
\end{eqnarray}
Writing
$$C(\stackrel{(0)}{D}+1)^{1/2}(\stackrel{(2)}{D}+1)^{1/2}(\stackrel{(1)}{{\cal E}}_2)^{1/2}\leq
\frac{1}{2}C^2(\stackrel{(0)}{D}+1)(\stackrel{(2)}{D}+1)+\frac{1}{2}\stackrel{(1)}{{\cal
E}}_2$$ \ref{16.11} implies:
\begin{equation}
\stackrel{(1)}{{\cal E}}_2\leq \stackrel{(1)}{D}+A \label{16.12}
\end{equation}
where:
\begin{equation}
A=1+(\stackrel{(1)}{D}+1)+2C(\stackrel{(0)}{D}+1)(\stackrel{(2)}{D}+1)^{1/2}
+C^2(\stackrel{(0)}{D}+1)(\stackrel{(2)}{D}+1) \label{16.13}
\end{equation}
is a positive continuous non-decreasing function of
$\stackrel{(0)}{D},\stackrel{(1)}{D},\stackrel{(2)}{D}$. It
follows that:
\begin{eqnarray}
C(\stackrel{(0)}{D}+1)(\stackrel{(2)}{D}+1)^{1/2}+1+C(\stackrel{(0)}{D}+1)^{1/2}(\stackrel{(2)}{D}+1)^{1/2}(\stackrel{(1)}{{\cal
E}}_2)^{1/2} \leq A
\nonumber\\
&&\label{16.14}
\end{eqnarray}
which in turn through \ref{16.11} implies \ref{16.12}.

Substituting the first and third of inequalities \ref{16.9} and
\ref{16.12} in inequality \ref{16.10} we obtain:
\begin{equation}
\stackrel{(3)}{{\cal F}}_2\leq
\stackrel{(3)}{D}+1+B(\stackrel{(3)}{{\cal F}}_2)^{1/2}
\label{16.15}
\end{equation}
where:
\begin{equation}
B=C\left\{(\stackrel{(0)}{D}+1)^{1/2}+(\stackrel{(1)}{D}+A)^{1/2}\right\}(\stackrel{(2)}{D}+1)^{1/2}
\label{16.16}
\end{equation}
is a positive continuous non-decreasing function of
$\stackrel{(0)}{D},\stackrel{(1)}{D},\stackrel{(2)}{D}$. Writing
$$B(\stackrel{(3)}{{\cal F}}_2)^{1/2}\leq \frac{1}{2}B^2+\frac{1}{2}\stackrel{(3)}{{\cal F}}_2$$
\ref{16.14} implies:
\begin{equation}
\stackrel{(3)}{{\cal F}}_2\leq 2(\stackrel{(3)}{D}+1)+B^2
\label{16.17}
\end{equation}
It follows that:
\begin{equation}
1+B(\stackrel{(3)}{{\cal F}}_2)^{1/2}\leq
1+(\stackrel{(3)}{D}+1)+B^2 \label{16.18}
\end{equation}
which in turn through \ref{16.15} implies \ref{16.17}.
Substituting the first and third of the inequalities  \ref{16.9}
and \ref{16.12} in the fourth of the inequalities \ref{16.9} we
obtain:
\begin{equation}
\stackrel{(3)}{{\cal E}}_2\leq
\stackrel{(3)}{D}+1+B(\stackrel{(3)}{{\cal F}}_2)^{1/2}
\label{16.19}
\end{equation}
hence by \ref{16.18}, also:
\begin{equation}
\stackrel{(3)}{{\cal E}}_2\leq 2(\stackrel{(3)}D+1)+B^2
\label{16.20}
\end{equation}

We conclude from the above that:
\begin{equation}
{\cal P}_2\leq \max\{\stackrel{(0)}{D}+1,
\stackrel{(1)}{D}+A(\stackrel{(0)}{D},\stackrel{(1)}{D},\stackrel{(2)}{D}),
\stackrel{(2)}{D}+1,2(\stackrel{(3)}{D}+1)+B^2(\stackrel{(0)}{D},\stackrel{(1)}{D},\stackrel{(2)}{D})\}
\label{16.21}
\end{equation}
Moreover, in view of \ref{16.14} and \ref{16.18} the error
integrals \ref{16.1} are bounded according to:
\begin{eqnarray}
&&\delta^{2q_0}\int_{M^\prime_{c^*}}|\stackrel{(0)}{\tau}_2|d\mu_g\leq 1\nonumber\\
&&\delta^{2q_1}\int_{M^\prime_{c^*}}|\stackrel{(1)}{\tau}_2|d\mu_g\leq A\nonumber\\
&&\delta^{2q_2}\int_{M^\prime_{c^*}}|\stackrel{(2)}{\tau}_2|d\mu_g\leq 1\nonumber\\
&&\delta^{2q_3}\int_{M^\prime_{c^*}}|\stackrel{(3)}{\tau}_2|d\mu_g\leq
1+(\stackrel{(3)}{D}+1)+B^2 \label{16.22}
\end{eqnarray}
and the error integral bounds \ref{16.21} imply through
inequalities \ref{12.272} - \ref{12.273} the energy-flux bounds,
namely the first and third of \ref{16.9}, \ref{16.12}, \ref{16.17}
and \ref{16.20}. Let us define
$G(\stackrel{(0)}{D},\stackrel{(1)}{D},\stackrel{(2)}{D},
\stackrel{(3)}{D})$ by:
\begin{eqnarray}
&&G(\stackrel{(0)}{D},\stackrel{(1)}{D},\stackrel{(2)}{D},\stackrel{(3)}{D})\label{16.23}\\
&&\hspace{10mm}=2\max\{\stackrel{(0)}{D}+1,
\stackrel{(1)}{D}+A(\stackrel{(0)}{D},\stackrel{(1)}{D},\stackrel{(2)}{D}),
\stackrel{(2)}{D}+1,2(\stackrel{(3)}{D}+1)+B^2(\stackrel{(0)}{D},\stackrel{(1)}{D},\stackrel{(2)}{D})\}\nonumber
\end{eqnarray}
Then
$G(\stackrel{(0)}{D},\stackrel{(1)}{D},\stackrel{(2)}{D},\stackrel{(3)}{D})$
is a positive non-decreasing continuous function of
$\stackrel{(0)}{D},\stackrel{(1)}{D},\stackrel{(2)}{D},\stackrel{(3)}{D}$
and we have:
\begin{equation}
{\cal
P}_2\leq\frac{1}{2}G(\stackrel{(0)}{D},\stackrel{(1)}{D},\stackrel{(2)}{D},\stackrel{(3)}{D})
\label{16.a1}
\end{equation}
Thus, with the choice \ref{16.23} for the function
$G(\stackrel{(0)}{D},\stackrel{(1)}{D},\stackrel{(2)}{D},\stackrel{(3)}{D})$
we have established inequality \ref{12.251} of Chapter 12 which
has been our aim.

\section{Higher order bounds}

In this section we shall derive uniform higher order bounds for
the solution in the domain $M^\prime_{c^*}$.

Consider the spacelike hypersurfaces $H_t \ : \ t\in (u_0,c^*)$
corresponding to $\ub+u=t$ (see \ref{1.2}, \ref{1.3}). We denote
by $H^\prime_t$ the part of $H_t$ which lies in the non-trivial
part of the spacetime manifold:
\begin{equation}
H^\prime_t=\{p\in H_t \ : \ \ub(p)\geq 0\} \label{16.24}
\end{equation}
Setting
\begin{equation}
\lambda=\ub-u \label{16.25}
\end{equation}
the induced metric $\og$ on the $H_t$ is given in terms of
canonical coordinates (see \ref{1.177}) $(\lambda,
\vartheta^1,\vartheta^2)$ by:
\begin{equation}
\og=\Omega^2d\lambda\otimes d\lambda
+\sg_{AB}(d\vartheta^A-\frac{1}{2}b^A
d\lambda)\otimes(d\vartheta^B-\frac{1}{2}b^B d\lambda)
\label{16.26}
\end{equation}
The volume form $d\mu_{\og}$ of $\og$ is given by:
\begin{equation}
d\mu_{\og}=\Omega\sqrt{\mbox{det}\sg}d\lambda\wedge
d\vartheta^1\wedge d\vartheta^2=\Omega d\lambda\wedge d\mu_{\sg}
\label{16.27}
\end{equation}
The future directed timelike vectorfield
\begin{equation}
T=\frac{1}{2}(\Lb+L) \label{16.28}
\end{equation}
generates a flow $F_\tau$, which maps $H_t$ onto $H_{t+\tau}$.
For,
\begin{equation}
Tt=1 \label{16.29}
\end{equation}
We have:
\begin{equation}
T=\Omega\Th \label{16.30}
\end{equation}
where $\Th$ is the future directed timelike unit normal to the
$H_t$.

We shall first derive $L^2(H^\prime_t)$ bounds, uniform in
$t\in(u_0,c^*)$, for the derivatives of the curvature componets of
all orders. This will allow us to show that extending the manifold
$M_{c^*}$ by attaching the future boundary $H_{c^*}$, the metric
extends smoothly to $M_{c^*}\bigcup H_{c^*}$. Only rough bounds
are needed here, the dependence of the bounds on $\delta$ or $u_0$
being of no consequence for our argument. Since the derivation of
these bounds is standard, we shall only outline the arguments
involved.

For $c\in(u_0,c^*)$ the spacetime domain $M^\prime_c$ is foliated
by the $H^\prime_t \ : \ t\in (u_0,c)$:
\begin{equation}
M^\prime_c\setminus S_{0,u_0}=\bigcup_{t\in(u_0,c)}H^\prime_t
\label{16.31}
\end{equation}
Given any $\ub_1\in(0,\delta)$, let us denote by
$D^{\prime\ub_1}_c$ the subdomain of the parameter domain
$D^\prime_c$ corresponding to $\ub\leq \ub_1$ and let $M^{\prime
\ub_1}_c$ be the corresponding spacetime domain:
\begin{equation}
M^{\prime\ub_1}_c=\bigcup_{(\ub,u)\in D^{\prime\ub_1}_c}S_{\ub,u}
\label{16.32}
\end{equation}
a subdomain of the spacetime domain $M^\prime_c$. We also denote
by $H^{\prime\ub_1}_c$, the part of $H^\prime_c$ where
$\ub\leq\ub_1$:
\begin{equation}
H^{\prime\ub_1}_c=\{p\in H^\prime_c \ : \ \ub(p)\leq\ub_1\}
\label{16.33}
\end{equation}
The past boundary of $M^{\prime\ub_1}_c$ is $\Cb_0^c\bigcup
C_{u_0}^{\ub_1}$ and its future boundary is
$H^{\prime\ub_1}_c\bigcup\Cb_{\ub_1}^{c-\ub_1}$. We have the
following form of the divergence theorem in spacetime (see Lemma
12.4).

\vspace{5mm}

\noindent{\bf Lemma 16.1} \ \ \ Let $P$ be a vectorfield and
$\tau$ a function defined on $M^\prime_{c^*}$ and satisfying the
equation:
$$\mbox{div}P=\tau$$
Let also $P^4$ vanish on $\Cb_0$. Then for $c\in(u_0,u_0+\ub_1]$
we have:
$$E^{\prime\ub_1}(c)-E^{c-u_0}(u_0)=\int_{M^{\prime\ub_1}_c}\tau d\mu_g$$
and for $c\in(u_0+\ub_1,c^*)$ we have:
$$E^{\prime\ub_1}(c)-E^{\ub_1}(u_0)+F^{c-\ub_1}(\ub_1)=\int_{M^{\prime\ub_1}_c}\tau d\mu_g$$
Here $E^{\prime\ub_1}$ is the ``energy"' associated to
$H^{\prime\ub_1}_c$:
$$E^{\prime\ub_1}(c)=\int_{H^{\prime\ub_1}_c}P^{\Th}d\mu_{\og}=\int_{H^{\prime\ub_1}_c}(P^3+P^4)d\mu_{\og}$$

\noindent{\em Proof:} \ We integrate equation \ref{12.70} with
respect to $(\ub,u)$ on the parameter domain $D^{\prime\ub_1}_c$.
In view of the relations \ref{12.78}, the condition of vanishing
of $P^4$ on $\Cb_0$, definition \ref{12.72} with $u_0$ in the role
of $u_1$, definition \ref{12.73}, the above definition of
$E^{\prime\ub_1}(c)$, and \ref{16.27}, the lemma follows.

\vspace{5mm}

Let us now consider the case where the components $P^3$ and $P^4$
of $P$ are non-negative functions, which is the case for the
energy-momentum density vectorfields. Then Lemma 16.1 implies, for
all $c\in(u_0,c^*)$:
\begin{equation}
E^{\prime\ub_1}(c)\leq E(u_0)+\int_{M^\prime_c}|\tau|d\mu_g
\label{16.34}
\end{equation}
and since this holds for every $\ub_1\in(0,\delta)$, taking the
limit $\ub_1\rightarrow\delta$ we obtain, for all $c\in(u_0,c^*)$:
\begin{equation}
E^\prime(c)\leq E(u_0)+\int_{M^\prime_c}|\tau|d\mu_g \label{16.35}
\end{equation}
where $E^\prime(c)$ is the energy associated to $H^\prime_c$:
\begin{equation}
E^\prime(c)=\int_{H^\prime_c}P^{\Th}d\mu_{\og}=\int_{H^\prime_c}(P^3+P^4)d\mu_{\og}
\label{16.36}
\end{equation}

We denote by $\stackrel{(n)}{E^\prime}_0(t) \ : \ n=0,1,2,3$ the
energies associated to the energy-momentum density vectorfields
\ref{12.80} and to $H^\prime_t$. They are given by:
\begin{eqnarray}
&&\stackrel{(0)}{E^\prime}_0(t)=\int_{H_t}\Omega^3(|\alpha|^2+2|\beta|^2)d\mu_{\og}\nonumber\\
&&\stackrel{(1)}{E^\prime}_0(t)=\int_{H_t}\Omega^3|u|^2(2|\beta|^2+2|\rho|^2+2|\sigma|^2)d\mu_{\og}\nonumber\\
&&\stackrel{(2)}{E^\prime}_0(t)=\int_{H_t}\Omega^3|u|^4(2|\rho|^2+2|\sigma|^2+2|\beb|^2)d\mu_{\og}\nonumber\\
&&\stackrel{(3)}{E^\prime}_0(t)=\int_{H_t}\Omega^3|u|^6(2|\beb|^2+|\alb|^2)d\mu_{\og}
\label{16.37}
\end{eqnarray}
We denote by $\s^{(L)}\stackrel{(n)}{E^\prime}(t) \ : \ n=0,1,2,3$
the energies associated to the energy-momentum density
vectorfields \ref{12.85} and to $H^\prime_t$. They are given by:
\begin{eqnarray}
&&\s^{(L)}\stackrel{(0)}{E^\prime}(t)=\int_{H^\prime_t}\Omega^3(|\alpha(\tcL_L R)|^2+2|\beta(\tcL_L R)|^2)d\mu_{\og}\nonumber\\
&&\s^{(L)}\stackrel{(1)}{E^\prime}(t)=\int_{H^\prime_t}\Omega^3|u|^2(2|\beta(\tcL_L R)|^2+2|\rho(\tcL_L R)|^2+2|\sigma(\tcL_L R)|^2)d\mu_{\og}\nonumber\\
&&\s^{(L)}\stackrel{(2)}{E^\prime}(t)=\int_{H^\prime_t}\Omega^3|u|^4(2|\rho(\tcL_L R)|^2+2|\sigma(\tcL_L R)|^2+2|\beb(\tcL_L R)|^2)d\mu_{\og}\nonumber\\
&&\s^{(L)}\stackrel{(3)}{E^\prime}(t)=\int_{H^\prime_t}\Omega^3|u|^6(2|\beb(\tcL_L
R)|^2+|\alb(\tcL_L R)|^2)d\mu_{\og} \label{16.38}
\end{eqnarray}
We denote by $\s^{(O)}\stackrel{(n)}{E^\prime}(t) \ : \ n=0,1,2,3$
the energies associated to the energy-momentum density
vectorfields \ref{12.90} and to $H^\prime_t$. They are given by:
\begin{eqnarray}
&&\s^{(O)}\stackrel{(0)}{E^\prime}(t)=\int_{H^\prime_t}\Omega^3\sum_i(|\alpha(\tcL_{O_i}R)|^2+2|\beta(\tcL_{O_i}R)|^2)d\mu_{\og}\nonumber\\
&&\s^{(O)}\stackrel{(1)}{E^\prime}(t)=\int_{H^\prime_t}\Omega^3|u|^2\sum_i(2|\beta(\tcL_{O_i}R)|^2+2|\rho(\tcL_{O_i}R)|^2
+2|\sigma(\tcL_{O_i}R)|^2)d\mu_{\og}\nonumber\\
&&\s^{(O)}\stackrel{(2)}{E^\prime}(t)=\int_{H^\prime_t}\Omega^3|u|^4\sum_i(2|\rho(\tcL_{O_i}R)|^2+2|\sigma(\tcL_{O_i}R)|^2
+2|\beb(\tcL_{O_i}R)|^2)d\mu_{\og}\nonumber\\
&&\s^{(O)}\stackrel{(3)}{E^\prime}(t)=\int_{H^\prime_t}\Omega^3|u|^6\sum_i(2|\beb(\tcL_{O_i}R)|^2+|\alb(\tcL_{O_i}R|^2)d\mu_{\og}
\label{16.39}
\end{eqnarray}
We denote by $\s^{(S)}\stackrel{(3)}{E^\prime}(t)$ the energy
associated to the energy-momentum density vectorfield \ref{12.95}
and to $H^\prime_t$. It is given by:
\begin{equation}
\s^{(S)}\stackrel{(3)}{E^\prime}(t)=\int_{H^\prime_t}\Omega^3|u|^6(2|\beb(\tcL_S
R)|^2+|\alb(\tcL_S R)|^2)d\mu_{\og} \label{16.40}
\end{equation}
We denote by $\s^{(LL)}\stackrel{(n)}{E^\prime}(t) \ : \
n=0,1,2,3$ the energies associated to the energy-momentum density
vectorfields \ref{12.100} and to $H^\prime_t$. They are given by:
\begin{eqnarray}
&&\s^{(LL)}\stackrel{(0)}{E^\prime}(t)=\int_{H^\prime_t}\Omega^3(|\alpha(\tcL_L\tcL_L R)|^2+2|\beta(\tcL_L\tcL_L R)|^2)d\mu_{\og}\nonumber\\
&&\s^{(LL)}\stackrel{(1)}{E^\prime}(t)=\int_{H^\prime_t}\Omega^3|u|^2(2|\beta(\tcL_L\tcL_L R)|^2\nonumber\\
&&\hspace{40mm}+2|\rho(\tcL_L\tcL_L R)|^2+2|\sigma(\tcL_L\tcL_L R)|^2)d\mu_{\og}\nonumber\\
&&\s^{(LL)}\stackrel{(2)}{E^\prime}(t)=\int_{H^\prime_t}\Omega^3|u|^4(2|\rho(\tcL_L\tcL_L R)|^2+2|\sigma(\tcL_L\tcL_L  R)|^2\nonumber\\
&&\hspace{62mm}+2|\beb(\tcL_L\tcL_L R)|^2)d\mu_{\og}\nonumber\\
&&\s^{(LL)}\stackrel{(3)}{E^\prime}(t)=\int_{H^\prime_t}\Omega^3|u|^6(2|\beb(\tcL_L\tcL_L
R)|^2+|\alb(\tcL_L\tcL_L R)|^2)d\mu_{\og} \label{16.41}
\end{eqnarray}
We denote by $\s^{(OL)}\stackrel{(n)}{E^\prime}(t) \ : \
n=0,1,2,3$ the energies associated to the energy-momentum density
vectorfields \ref{12.105} and to $H^\prime_t$. They are given by:
\begin{eqnarray}
&&\s^{(OL)}\stackrel{(0)}{E^\prime}(t)=\int_{H^\prime_t}\Omega^3\sum_i(|\alpha(\tcL_{O_i}\tcL_L R)|^2+2|\beta(\tcL_{O_i}\tcL_L R)|^2)d\mu_{\og}\nonumber\\
&&\s^{(OL)}\stackrel{(1)}{E^\prime}(t)=\int_{H^\prime_t}\Omega^3|u|^2\sum_i(2|\beta(\tcL_{O_i}\tcL_L R)|^2\nonumber\\
&&\hspace{45mm}+2|\rho(\tcL_{O_i}\tcL_L R)|^2+2|\sigma(\tcL_{O_i}\tcL_L R)|^2)d\mu_{\og}\nonumber\\
&&\s^{(OL)}\stackrel{(2)}{E^\prime}(t)=\int_{H^\prime_t}\Omega^3|u|^4\sum_i(2|\rho(\tcL_{O_i}\tcL_L R)|^2+2|\sigma(\tcL_{O_i}\tcL_L R)|^2\nonumber\\
&&\hspace{69mm}+2|\beb(\tcL_{O_i}\tcL_L R)|^2)d\mu_{\og}\nonumber\\
&&\s^{(OL)}\stackrel{(3)}{E^\prime}(t)=\int_{H^\prime_t}\Omega^3|u|^6\sum_i(2|\beb(\tcL_{O_i}\tcL_L R)|^2+|\alb(\tcL_{O_i}\tcL_L R)|^2)d\mu_{\og}\nonumber\\
&&\label{16.42}
\end{eqnarray}
We denote by $\s^{(OO)}\stackrel{(n)}{E^\prime}(t) \ : \
n=0,1,2,3$ the energies associated to the energy-momentum density
vectorfields \ref{12.110} and to $H^\prime_t$. They are given by:
\begin{eqnarray}
&&\s^{(OO)}\stackrel{(0)}{E^\prime}(t)=\int_{H^\prime_t}\Omega^3\sum_{i,j}(|\alpha(\tcL_{O_j}\tcL_{O_i}R)|^2+2|\beta(\tcL_{O_j}\tcL_{O_i}R)|^2)d\mu_{\og}\nonumber\\
&&\s^{(OO)}\stackrel{(1)}{E^\prime}(t)=\int_{H^\prime_t}\Omega^3|u|^2\sum_{i,j}(2|\beta(\tcL_{O_j}\tcL_{O_i}R)|^2\nonumber\\
&&\hspace{45mm}+2|\rho(\tcL_{O_j}\tcL_{O_i}R)|^2+2|\sigma(\tcL_{O_j}\tcL_{O_i}R)|^2)d\mu_{\og}\nonumber\\
&&\s^{(OO)}\stackrel{(2)}{E^\prime}(t)=\int_{H^\prime_t}\Omega^3|u|^4\sum_{i,j}(2|\rho(\tcL_{O_j}\tcL_{O_i}R)|^2+2|\sigma(\tcL_{O_j}\tcL_{O_i}R)|^2\nonumber\\
&&\hspace{69mm}+2|\beb(\tcL_{O_J}\tcL_{O_i}R)|^2)d\mu_{\og}\nonumber\\
&&\s^{(OO)}\stackrel{(3)}{E^\prime}(t)=\int_{H^\prime_t}\Omega^3|u|^6\sum_{i,j}(2|\beb(\tcL_{O_j}\tcL_{O_i}R)|^2+|\alb(\tcL_{O_j}\tcL_{O_i}R)|^2)d\mu_{\og}\nonumber\\
&&\label{16.43}
\end{eqnarray}
We denote by $\s^{(OS)}\stackrel{(3)}{E^\prime}(t)$ the energy
associated to the energy-momentum density vectorfields
\ref{12.115} and to $H^\prime_t$. It is given by:
\begin{equation}
\s^{(OS)}\stackrel{(3)}{E^\prime}(t)=\int_{H^\prime_t}\Omega^3|u|^6\sum_i(2|\beb(\tcL_{O_i}\tcL_S
R)|^2+|\alb(\tcL_{O_i}\tcL_S R)|^2)d\mu_{\og} \label{16.44}
\end{equation}
Finally, we denote by $\s^{(SS)}\stackrel{(3)}{E^\prime}(t)$ the
energy associated to the energy-momentum density vectorfield
\ref{12.120} and to $H^\prime_t$. It is given by:
\begin{equation}
\s^{(SS)}\stackrel{(3)}{E^\prime}(t)=\int_{H^\prime_t}\Omega^3|u|^6(2|\beb(\tcL_S\tcL_S
R)|^2+|\alb(\tcL_S\tcL_S R)|^2)d\mu_{\og} \label{16.45}
\end{equation}

Next, we denote by $\stackrel{(n)}{E^\prime}_1(t) \ : \ n=0,1,2,3$
the total 1st order energies associated to $H^\prime_t$. They are
given by:
\begin{eqnarray}
&&\stackrel{(0)}{E^\prime}_1(t)=\stackrel{(0)}{E^\prime}_0(t)+\delta^2\s^{(L)}\stackrel{(0)}{E^\prime}(t)+\s^{(O)}\stackrel{(0)}{E^\prime}(t)
\nonumber\\
&&\stackrel{(1)}{E^\prime}_1(t)=\stackrel{(1)}{E^\prime}_0(t)+\delta^2\s^{(L)}\stackrel{(1)}{E^\prime}(t)+\s^{(O)}\stackrel{(1)}{E^\prime}(t)
\nonumber\\
&&\stackrel{(2)}{E^\prime}_1(t)=\stackrel{(2)}{E^\prime}_0(t)+\delta^2\s^{(L)}\stackrel{(2)}{E^\prime}(t)+\s^{(O)}\stackrel{(2)}{E^\prime}(t)
\nonumber\\
&&\stackrel{(3)}{E^\prime}_1(t)=\stackrel{(3)}{E^\prime}_0(t)+\delta^2\s^{(L)}\stackrel{(3)}{E^\prime}(t)+\s^{(O)}\stackrel{(3)}{E^\prime}(t)
+\s^{(S)}\stackrel{(3)}{E^\prime}(t) \label{16.46}
\end{eqnarray}
We denote by $\stackrel{(n)}{E^\prime}_2(t) \ : \ n=0,1,2,3$ the
total 2nd order energies associated to $H^\prime_t$. They are
given by:
\begin{eqnarray}
&&\stackrel{(0)}{E^\prime}_2(t)=\stackrel{(0)}{E^\prime}_1(t)+\delta^4\s^{(LL)}\stackrel{(0)}{E^\prime}(t)
+\delta^2\s^{(OL)}\stackrel{(0)}{E^\prime}(t)+\s^{(OO)}\stackrel{(0)}{E^\prime}(t)\nonumber\\
&&\stackrel{(1)}{E^\prime}_2(t)=\stackrel{(1)}{E^\prime}_1(t)+\delta^4\s^{(LL)}\stackrel{(1)}{E^\prime}(t)
+\delta^2\s^{(OL)}\stackrel{(1)}{E^\prime}(t)+\s^{(OO)}\stackrel{(1)}{E^\prime}(t)\nonumber\\
&&\stackrel{(2)}{E^\prime}_2(t)=\stackrel{(2)}{E^\prime}_1(t)+\delta^4\s^{(LL)}\stackrel{(2)}{E^\prime}(t)
+\delta^2\s^{(OL)}\stackrel{(2)}{E^\prime}(t)+\s^{(OO)}\stackrel{(2)}{E^\prime}(t)\nonumber\\
&&\stackrel{(3)}{E^\prime}_2(t)=\stackrel{(3)}{E^\prime}_1(t)+\delta^4\s^{(LL)}\stackrel{(3)}{E^\prime}(t)
+\delta^2\s^{(OL)}\stackrel{(3)}{E^\prime}(t)+\s^{(OO)}\stackrel{(3)}{E^\prime}(t)\nonumber\\
&&\hspace{26mm}+\s^{(OS)}\stackrel{(3)}{E^\prime}(t)+\s^{(SS)}\stackrel{(3)}{E^\prime}(t)
\label{16.47}
\end{eqnarray}
The 2nd order energies associated to $H^\prime_t$ control the
$L^2$ norms on $H^\prime_t$ of all 2nd derivatives of all
curvature components.

Finally, we define the quantities:
\begin{equation}
\stackrel{(n)}{{\cal
E}^\prime}_2=\sup_{t\in(u_0,c^*)}\left(\delta^{2q_n}\stackrel{(n)}{E^\prime}_2(t)\right)
\ \ : \ n=0,1,2,3 \label{16.48}
\end{equation}
(the exponents $q_n$ being defined by \ref{12.274}). Then in view
of the inequality \ref{16.35} we obtain:
\begin{equation}
\stackrel{(n)}{{\cal
E}^\prime}_2\leq\stackrel{(n)}{D}+\delta^{2q_n}\int_{M^\prime_{c^*}}|\stackrel{(n)}{\tau}_2|d\mu_g
\ \ : \ n=0,1,2,3 \label{16.49}
\end{equation}
The bounds \ref{16.22} for the error integrals then imply the
following sharp bounds for the quantities $\stackrel{(n)}{{\cal
E}^\prime}_2$:
\begin{eqnarray}
&&\stackrel{(0)}{{\cal E}^\prime}_2\leq\stackrel{(0)}{D}+1\nonumber\\
&&\stackrel{(1)}{{\cal E}^\prime}_2\leq\stackrel{(1)}{D}+A\nonumber\\
&&\stackrel{(2)}{{\cal E}^\prime}_2\leq\stackrel{(2)}{D}+1\nonumber\\
&&\stackrel{(3)}{{\cal E}^\prime}_2\leq 2(\stackrel{(3)}{D}+1)+B^2
\label{16.50}
\end{eqnarray}

We now begin the outline of the derivation of the rough
$L^2(H^\prime_t)$ bounds for the higher derivatives of the
curvature components. Here we use only the future directed
timelike vectorfield $T$ (see \ref{16.27}) as a multiplier field,
and the vectorfields $L$ and $\Lb$ as commutation fields. We thus
consider the derived Weyl fields:
\begin{equation}
W_{m,n}=(\tcL_{\Lb})^m(\tcL_L)^n R \label{16.51}
\end{equation}
where $m$ and $n$ are non-negative integers. The {\em order} of
$W_{m,n}$ is $k=m+n$. We associate to $W_{m,n}$ the
energy-momentum density vectorfields:
\begin{equation}
P_{m,n}=P(W_{m,n};T,T,T) \label{16.52}
\end{equation}
(see \ref{12.63}). We then consider the energies
$E^\prime_{m,n}(t)$ associated to $P_{m,n}$ and to $H^\prime_t$:
\begin{equation}
E^\prime_{m,n}=\int_{H^\prime_t}(P^3_{m,n}+P^4_{m,n})d\mu_{\og}
\label{16.53}
\end{equation}
(see \ref{16.36}). Now, by Lemma 12.2, the energy $E^\prime[W](t)$
associated to $P(W;T,T,T)$ and to $H^\prime_t$, that is:
\begin{equation}
E^\prime[W](t)=\int_{H^\prime_t}(P^3(W;T,T,T)+P^4(W;T,T,T))d\mu_{\og}
\label{16.54}
\end{equation}
is given by:
\begin{eqnarray}
&&E^\prime[W](t)=\int_{H^\prime_t}\left\{\frac{1}{8}|\alpha(W)|^2+|\beta(W)|^2+\frac{3}{2}(|\rho(W)|^2+|\sigma(W)|^2)\right.\nonumber\\
&&\hspace{50mm}\left.+|\beb(W)|^2+\frac{1}{8}|\alb(W)|^2\right\}d\mu_{\og}
\label{16.55}
\end{eqnarray}
Thus $E^\prime[W](t)$ is equivalent to the sum of the squares of
the $L^2$ norms on $H^\prime_t$ of all components of $W$. The
total $k$th order energy $E^\prime_k(t)$ associated to
$H^\prime_t$ is then defined by:
\begin{equation}
E^\prime_k(t)=\sum_{m+n=k}E^\prime_{m,n}(t) \label{16.56}
\end{equation}

Now, from \ref{12.64} $\tau_{m,n}$, the divergence of $P_{m,n}$,
is given by:
\begin{equation}
\tau_{m,n}=-(\mbox{div}Q(W_{m,n})(T,T,T)-\frac{3}{2}Q(W_{m,n})_{\alpha\beta\gamma\delta}\s^{(T)}\tilde{\pi}_{\alpha\beta}T^\gamma
T^\delta \label{16.57}
\end{equation}
the second term on the right being the ``multiplier"' part and the
first term the ``Weyl current"' part, given according to
Proposition 12.6 by:
\begin{equation}
-2((W_{m,n})^{\s\mu\s\nu}_{\beta\s\delta}(J_{m,n})_{\mu\gamma\nu}+(\s^*W_{m,n})^{\s\mu\s\nu}_{\beta\s\delta}(J^*_{m,n})_{\mu\gamma\nu})T^\beta
T^\gamma T^\delta \label{16.58}
\end{equation}
Here $J_{m,n}$ is the Weyl current corresponding to the Weyl field
$W_{m,n}$. By Proposition 12.1 the $J_{m,n}$ satisfy the following
recursion relations:
\begin{eqnarray}
&&(J_{0,n+1})_{\beta\gamma\delta}=(\tcL_L
J_{0,n})_{\beta\gamma\delta}
+\frac{1}{2}\s^{(L)}\tilde{\pi}^{\mu\nu}\nabla_\nu(W_{0,n})_{\mu\beta\gamma\delta}
+\frac{1}{2}\s^{(L)}p_\mu(W_{0,n})^\mu_{\s\beta\gamma\delta}\nonumber\\
&&\hspace{10mm}+\frac{1}{2}(\s^{(L)}q_{\mu\beta\nu}(W_{0,n})^{\mu\nu}_{\s\s\gamma\delta}
+\s^{(L)}q_{\mu\gamma\nu}(W_{0,n})^{\mu\s\nu}_{\s\beta\s\delta}+\s^{(L)}q_{\mu\delta\nu}(W_{0,n})^{\mu\s\s\nu}_{\s\beta\gamma})\nonumber\\
&&\label{16.59}
\end{eqnarray}
\begin{eqnarray}
&&(J_{m+1,n})_{\beta\gamma\delta}=(\tcL_{\Lb}J_{m,n})_{\beta\gamma\delta}
+\frac{1}{2}\s^{(\Lb)}\tilde{\pi}^{\mu\nu}\nabla_\nu(W_{m,n})_{\mu\beta\gamma\delta}
+\frac{1}{2}\s^{(\Lb)}p_{\mu}(W_{m,n})^\mu_{\s\beta\gamma\delta}\nonumber\\
&&\hspace{10mm}+\frac{1}{2}(\s^{(\Lb)}q_{\mu\beta\nu}(W_{m,n})^{\mu\nu}_{\s\s\gamma\delta}
+\s^{(\Lb)}q_{\mu\gamma\nu}(W_{m,n})^{\mu\s\nu}_{\s\beta\s\delta}+\s^{(\Lb)}q_{\mu\delta\nu}(W_{m,n})^{\mu\s\s\nu}_{\s\beta\gamma})\nonumber\\
&&\label{16.60}
\end{eqnarray}
To determine the $k$th order currents, we first obtain $J_{0,n}$
for $n=1,...,k$ from the recursion relation \ref{16.59} and the
fact that $J_{0,0}=0$. We then obtain $J_{m,n}$ for $m=1,...,k-n$,
$n=0,...,k$ from the recursion relation \ref{16.60}. The leading
terms in the expression for any component of $J_{m,n}$ are of the
following four types. First, terms which are bilinear expressions,
with coefficients depending only on $\sg$ and $\seps$, the first
factor of which is $\Db^{m_1}D^{i_1}$ of a component of
$\s^{(L)}\tilde{\pi}$, and the second factor is a ($D,\Db$ or
$\snab$) derivative of a component of $W_{m_2,n-1+i_2-i}$, where
$m_1+m_2=m$, $i_1+i_2=i$ and $i=0,...,n-1$. Second, terms which
are bilinear expressions, with coefficients depending only $\sg$
and $\seps$, the first factor of which is a ($D,\Db$ or $\snab$)
derivative of $\Db^{m_1}D^{i_1}$ of a component of
$\s^{(L)}\tilde{\pi}$, and the second factor is a component of
$W_{m_2,n-1+i_2-i}$, where $m_1+m_2=m$, $i_1+i_2=i$ and
$i=0,...,n-1$. Third, terms which are bilinear expressions with
coefficients depending only on $\sg$ and $\seps$, the first factor
of which is $\Db^{j_1}$ of a component of $\s^{(\Lb)}\tilde{\pi}$,
and the second factor is a ($D,\Db$ or $\snab$) derivative of a
component of $W_{m-1+j_2-j,n}$, where $j_1+j_2=j$ and
$j=0,...,m-1$. Fourth, terms which are bilinear expressions, with
coefficients depending only on $\sg$ and $\seps$, the first factor
of which is a ($D,\Db$ or $\snab$) derivative of $\Db^{j_1}$ of a
component of $\s^{(\Lb)}\tilde{\pi}$, and the second factor is a
component of $W_{m-1+j_2-j,n}$, where $j_1+j_2=j$ and
$j=0,...,m-1$. Since the components of $\s^{(L)}\pi$,
$\s^{(\Lb)}\pi$ are expressed by \ref{8.21}, \ref{8.22} in terms
of the connection coefficients, we can simply say that the leading
terms in the expression of a component of $J_{m,n}$ are terms
which are bilinear expressions, with coefficients depending only
on $\sg$ and $\seps$, the first factor of which is a $i$th order
derivative of a connection coefficient, and the second factor a
$j$th order derivative of a component of $R$, where $i+j=m+n=k$.
The lower order terms in the expression of a component of
$J_{m,n}$ are terms which are are $l$-linear expressions, $l\geq
3$, with coefficients depending only on $\sg$ and $\seps$, the
first $l-1$ factors of which are $i_{p}$th order derivatives of
connection coefficients, $p=1,...,l-1$ respectively, and the last
factor is $j$th order derivative of a component of $R$, where
$i_1+...+i_{l-1}=i-l+2$ and $i+j=m+n=k$. Note that $l\leq i+2$.

As noted previously, we already have sharp $L^2$ bounds on the
$H^\prime_t$ for the 2nd derivatives of the curvature components.
By the results of Chapter 6 we already have sharp $L^4(S)$ bounds
for the 2nd derivatives of the connection coefficients, except for
$\sd D\omega$, $\sd\Db\omb$, $D^2\omega$, $\Db^2\omb$. For
$\sd\Db\omb$, $\Db^2\omb$ we have sharp $L^2(S)$ bounds from
Chapter 7. The $L^2(S)$ and $L^4(S)$ bounds imply $L^2$ bounds on
the $H^\prime_t$. For $\sd D\omega$, $D^2\omega$ we have sharp
$L^2$ bounds on the $C_u$ from Chapter 7. We shall presently show
how sharp $L^2$ bounds on the $H^\prime_t$ may also be obtained
for these. Here and in the following we shall make use of the
following elementary lemma. We denote by $\lambda^*(t)$ the
supremum of $\lambda$ on $H^\prime_t$:
\begin{equation}
\lambda^*(t)=\sup_{H^\prime_t}\lambda=\left\{
\begin{array}{lll}
2\delta-t&:&\mbox{for $t\in(u_0+\delta,c^*)$}\\
t-2u_0&:&\mbox{for $t\in(u_0,u_0+\delta]$}
\end{array}\right.
\label{16.61}
\end{equation}
The infimum of $\lambda$ on $H^\prime_t$ is $-t$.

\vspace{5mm}

\noindent{\bf Lemma 16.2} \ \ \ Let $t\in(u_0,c^*)$ and let
$f(\ub,u)$ be a non-negative function which is square integrable
on $D^\prime_t$. Then we have:
\begin{eqnarray*}
&&\left\{\int_{-t}^{\lambda^*(t)}\left(\int_{u_0}^{\frac{1}{2}(t-\lambda)}f\left(\frac{1}{2}(t+\lambda),u^\prime\right)du^\prime\right)^2d\lambda
\right\}^{1/2}\\
&&\hspace{30mm}\leq
\int_{u_0}^t\left(\int_{-t^\prime}^{\lambda^*(t^\prime)}f^2\left(\frac{1}{2}(t^\prime+\lambda^\prime),\frac{1}{2}(t^\prime-\lambda^\prime)\right)
d\lambda^\prime\right)^{1/2}dt^\prime
\end{eqnarray*}
and:
\begin{eqnarray*}
&&\left\{\int_{-t}^{\lambda^*(t)}\left(\int_0^{\frac{1}{2}(t+\lambda)}f\left(\ub^\prime,\frac{1}{2}(t-\lambda)\right)d\ub^\prime\right)^2d\lambda
\right\}^{1/2}\\
&&\hspace{25mm}\leq
\int_{\max\{t-\delta,u_0\}}^t\left(\int_{-t^\prime}^{\lambda^*(t^\prime)}f^2\left(\frac{1}{2}(t^\prime+\lambda^\prime),\frac{1}{2}(t^\prime-\lambda^\prime)\right)
d\lambda^\prime\right)^{1/2}dt^\prime
\end{eqnarray*}

\noindent{\em Proof:} To establish the first inequality we define
the function $\tilde{f}(\ub,u)$ to be the exension of $f$ by zero
to $u<u_0$ (see Figure 16.1)

\begin{figure}[htbp]
\begin{center}
\input{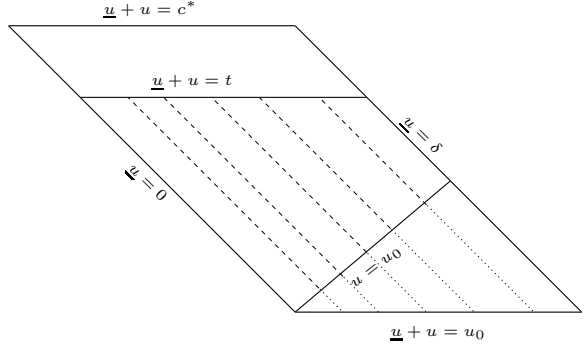}
\caption{The domain considered in regard to the first inequality}
\label{fig16.1}
\end{center}
\end{figure}

\noindent and write:

\begin{equation}
\int_{u_0}^{\frac{1}{2}(t-\lambda)}f\left(\frac{1}{2}(t+\lambda),u^\prime\right)du^\prime
=\int_{u_0}^t\tilde{f}\left(\frac{1}{2}(t+\lambda),t^\prime-\frac{1}{2}(t+\lambda)\right)dt^\prime
\label{16.62}
\end{equation}
Setting:
\begin{equation}
g(t^\prime,\lambda)=\tilde{f}\left(\frac{1}{2}(t+\lambda),t^\prime-\frac{1}{2}(t+\lambda)\right)
\label{16.63}
\end{equation}
the left hand side of the inequality is:
\begin{equation}
\left\{\int_{-t}^{\lambda^*(t)}\left(\int_{u_0}^t
g(t^\prime,\lambda)dt^\prime\right)^2d\lambda\right\}^{1/2}
=\left\|\int_{u_0}^t
g(t^\prime,\cdot)dt^\prime\right\|_{L^2(-t,\lambda^*(t))}
\label{16.64}
\end{equation}
and we have:
\begin{equation}
\left\|\int_{u_0}^t
g(t^\prime,\cdot)dt^\prime\right\|_{L^2(-t,\lambda^*(t))}\leq
\int_{u_0}^t\|g(t^\prime,\cdot)\|_{L^2(-t,\lambda^*(t))}dt^\prime
\label{16.65}
\end{equation}
Now,
\begin{equation}
\|g(t^\prime,\cdot)\|^2_{L^2(-t,\lambda^*(t))}=\int_{-t}^{\lambda^*(t)}\tilde{f}^2\left(\frac{1}{2}(t+\lambda),t^\prime-\frac{1}{2}(t+\lambda)\right)d\lambda
\label{16.66}
\end{equation}
Setting $\lambda^\prime=t-t^\prime+\lambda$ the last integral is:
\begin{equation}
\int_{-t^\prime}^{t-t^\prime+\lambda^*(t)}\tilde{f}^2\left(\frac{1}{2}(t^\prime+\lambda^\prime),\frac{1}{2}(t^\prime-\lambda^\prime)\right)d\lambda^\prime
\label{16.67}
\end{equation}
Now, for $t^\prime>u_0+\delta$ (hence also $t>u_0+\delta$) we
have, by \ref{16.61}:
$$t-t^\prime+\lambda^*(t)=t-t^\prime+2\delta-t=2\delta-t^\prime=\lambda^*(t^\prime)$$
hence, since $\tilde{f}$ coincides with $f$ in $D^\prime_t$,
\ref{16.67} is:
\begin{equation}
\int_{-t^\prime}^{\lambda^*(t^\prime)}f^2\left(\frac{1}{2}(t^\prime+\lambda^\prime),\frac{1}{2}(t^\prime-\lambda^\prime)\right)d\lambda^\prime
\label{16.68}
\end{equation}
For $t^\prime\leq u_0+\delta$ and $t>u_0+\delta$ we have, by
\ref{16.61}:
$$t-t^\prime+\lambda^*(t)=2\delta-t^\prime\geq t^\prime-2u_0=\lambda^*(t^\prime)$$
hence, since $\tilde{f}$ coincides with $f$ in $D^\prime_t$ and
vanishes for $u<u_0$, \ref{16.67} is equal to \ref{16.68}.
Finally, for $t\leq u_0+\delta$ (hence also $t^\prime\leq
u_0+\delta$) we have, by \ref{16.61}:
$$t-t^\prime+\lambda^*(t)=2t-t^\prime-2u_0\geq t^\prime-2u_0=\lambda^*(t^\prime)$$
hence again, since $\tilde{f}$ coincides with $f$ in $D^\prime_t$
and vanishes for $u<u_0$, \ref{16.67} is equal to \ref{16.68}.
Thus, we have, in general,
\begin{equation}
\|g(t^\prime,\cdot)\|_{L^2(-t,\lambda^*(t))}=\left(\int_{-t^\prime}^{\lambda(t^\prime)}f^2\left(\frac{1}{2}(t^\prime+\lambda^\prime),\frac{1}{2}(t^\prime-\lambda^\prime)\right)d\lambda^\prime\right)^{1/2}
\label{16.69}
\end{equation}
and the first inequality of the lemma is established.

To establish the second inequality we define the function
$\tilde{f}$ to be the extension of $f$ by zero to $\ub<0$ (see Figures 16.2, 16.3)

\begin{figure}[htbp]
\begin{center}
\input{bh.fig3.pstex_t}
\caption{The domain considered in regard to the second inequality, case  $t< u_0+\delta$}
\label{fig16.2}
\end{center}
\end{figure}

\begin{figure}[htbp]
\begin{center}
\input{bh.fig4.pstex_t}
\caption{The domain considered in regard to the second inequality, case  $t\geq u_0+\delta$}
\label{fig16.3}
\end{center}
\end{figure}

\noindent and,
noting that by \ref{16.61}:
$$\frac{1}{2}(t-\lambda)\geq\frac{1}{2}(t-\lambda^*(t))=\max\{t-\delta,u_0\}$$
we write:
\begin{eqnarray}
&&\int_0^{\frac{1}{2}(t+\lambda)}f\left(\ub^\prime,\frac{1}{2}(t-\lambda)\right)d\ub^\prime=
\int_{\frac{1}{2}(t-\lambda)}^tf\left(t^\prime-\frac{1}{2}(t-\lambda),\frac{1}{2}(t-\lambda)\right)dt^\prime\nonumber\\
&&\hspace{45mm}=\int_{\max\{t-\delta,u_0\}}^t\tilde{f}\left(t^\prime-\frac{1}{2}(t-\lambda),\frac{1}{2}(t-\lambda)\right)dt^\prime
\nonumber\\
&&\label{16.70}
\end{eqnarray}
Setting:
\begin{equation}
g(t^\prime,\lambda)=\tilde{f}\left(t^\prime-\frac{1}{2}(t-\lambda),\frac{1}{2}(t-\lambda)\right)
\label{16.71}
\end{equation}
the left hand side of the inequality is:
\begin{equation}
\left\{\int_{-t}^{\lambda^*(t)}\left(\int_{\max\{t-\delta,u_0\}}^t
g(t^\prime,\lambda)dt^\prime\right)^2d\lambda\right\}^{1/2}
=\left\|\int_{\max\{t-\delta,u_0\}}^tg(t^\prime,\cdot)dt^\prime\right\|_{L^2(-t,\lambda^*(t))}
\label{16.72}
\end{equation}
and we have:
\begin{equation}
\left\|\int_{\max\{t-\delta,u_0\}}^t
g(t^\prime,\cdot)dt^\prime\right\|_{L^2(-t,\lambda^*(t))}\leq
\int_{\max\{t-\delta,u_0\}}^t\|g(t^\prime,\cdot)\|_{L^2(-t,\lambda^*(t))}dt^\prime
\label{16.73}
\end{equation}
Now,
\begin{equation}
\|g(t^\prime,\cdot)\|^2_{L^2(-t,\lambda^*(t))}=\int_{-t}^{\lambda^*(t)}\tilde{f}^2\left(t^\prime-\frac{1}{2}(t-\lambda),\frac{1}{2}(t-\lambda)\right)d\lambda
\label{16.74}
\end{equation}
Setting $\lambda^\prime=t^\prime-t+\lambda$ the last integral is:
\begin{equation}
\int_{t^\prime-2t}^{t^\prime-t+\lambda^*(t)}\tilde{f}^2\left(\frac{1}{2}(t^\prime+\lambda^\prime),\frac{1}{2}(t^\prime-\lambda^\prime)\right)d\lambda^\prime
\label{16.75}
\end{equation}
Since $t^\prime-2t\leq-t^\prime$ and $\tilde{f}$ coincides with
$f$ on $D^\prime_t$ and vanishes for $\ub<0$, \ref{16.75} is equal
to:
\begin{equation}
\int_{-t^\prime}^{t^\prime-t+\lambda^*(t)}f^2\left(\frac{1}{2}(t^\prime+\lambda^\prime),\frac{1}{2}(t^\prime-\lambda^\prime)\right)d\lambda^\prime
\label{16.76}
\end{equation}
Now, for $t>u_0+\delta$ we have, by \ref{16.61}:
$$t^\prime-t+\lambda^*(t)=t^\prime-2t+2\delta\leq 2\delta-t^\prime=\lambda^*(t^\prime)$$
while for $t\leq u_0+\delta$ (hence also $t^\prime\leq
u_0+\delta$) we have, by \ref{16.61}:
$$t^\prime-t+\lambda^*(t)=t^\prime-2u_0=\lambda^*(t^\prime)$$
In both cases \ref{16.76} is less than or equal to:
\begin{equation}
\int_{-t^\prime}^{\lambda^*(t^\prime)}f^2\left(\frac{1}{2}(t^\prime+\lambda^\prime),\frac{1}{2}(t^\prime-\lambda^\prime)\right)d\lambda^\prime
\label{16.77}
\end{equation}
Thus:
\begin{equation}
\|g(t^\prime,\cdot)\|_{L^2(-t,\lambda^*(t))}\leq
\left(\int_{-t^\prime}^{\lambda^*(t^\prime)}f^2\left(\frac{1}{2}(t^\prime+\lambda^\prime),\frac{1}{2}(t^\prime-\lambda^\prime)\right)d\lambda^\prime\right)^{1/2}
\label{16.78}
\end{equation}
and the second inequality of the lemma is established as well.

\vspace{5mm}

Let now $\theta$ be an arbitrary $S$ tensorfield. By \ref{16.27}
we have:
\begin{equation}
\|\theta\|^2_{L^2(H^\prime_t)}:=\int_{H^\prime_t}|\theta|^2d\mu_{\og}=\int_{-t}^{\lambda^*(t)}\left(
\int_{S_{\frac{1}{2}(t+\lambda),\frac{1}{2}(t-\lambda)}}|\theta|^2\Omega
d\mu_{\sg}\right)d\lambda \label{16.79}
\end{equation}
hence:
\begin{equation}
C^{-1}\int_{-t}^{\lambda^*(t)}\|\theta\|^2_{L^2(S_{\frac{1}{2}(t+\lambda),\frac{1}{2}(t-\lambda)})}d\lambda
\leq\|\theta\|^2_{L^2(H^\prime_t)}\leq
C\int_{-t}^{\lambda^*(t)}\|\theta\|^2_{L^2(S_{\frac{1}{2}(t+\lambda),\frac{1}{2}(t-\lambda)})}d\lambda
\label{16.80}
\end{equation}
Aplying Lemma 16.2 to the function
$$f(\ub,u)=\|\theta\|_{L^2(S_{\ub,u})}$$
we then obtain:
\begin{equation}
\left\{\int_{-t}^{\lambda^*(t)}\left(\int_{u_0}^{\frac{1}{2}(t-\lambda)}\|\theta\|_{L^2(S_{\frac{1}{2}(t+\lambda),u^\prime})}du^\prime\right)^2d\lambda
\right\}^{1/2}\leq
C\int_{u_0}^t\|\theta\|_{L^2(H^\prime_{t^\prime})}dt^\prime
\label{16.81}
\end{equation}
and:
\begin{equation}
\left\{\int_{-t}^{\lambda^*(t)}\left(\int_0^{\frac{1}{2}(t+\lambda)}\|\theta\|_{L^2(S_{\ub^\prime,\frac{1}{2}(t-\lambda)})}d\ub^\prime\right)^2d\lambda
\right\}^{1/2} \leq
\int_{\max\{t-\delta,u_0\}}^t\|\theta\|_{L^2(H^\prime_{t^\prime})}dt^\prime
\label{16.82}
\end{equation}

We now revisit the proofs of Propositions 7.6 and 7.8. Setting
$\ub=\frac{1}{2}(t+\lambda)$, $u=\frac{1}{2}(t-\lambda)$ in
\ref{7.347}, squaring, integrating with respect to $\lambda$ on
$(-t,\lambda^*(t))$, and taking the square root, we obtain, by
\ref{16.81} and \ref{16.80}:
\begin{equation}
\|\sd D\omega\|_{L^2(H^\prime_t)}\leq
C\int_{u_0}^t\|\nb^\prime\|_{L^2(H^\prime_{t^\prime})}dt^\prime
\label{16.83}
\end{equation}
Substituting for $\nb^\prime$ we now obtain an estimate for $\sd
D\omega$ in $L^2(H^\prime_t)$ in terms of
\begin{equation}
\int_{u_0}^t\|\sd D\rho\|_{L^2(H^\prime_{t^\prime})}dt^\prime
\label{16.84}
\end{equation}
plus lower order terms. Now, $\sd D\rho$ is bounded in
$L^2(H^\prime_t)$ through the bounds \ref{16.50}. We obtain in
this way the following sharp bound for $\sd D\omega$ in
$L^2(H^\prime_t)$:
\begin{equation}
\|\sd D\omega\|_{L^2(H^\prime_t)}\leq O(\delta^{-1/2}|t|^{-2}) \ \
: \ \forall t\in(u_0,c^*) \label{16.85}
\end{equation}
Setting $\ub=\frac{1}{2}(t+\lambda)$, $u=\frac{1}{2}(t-\lambda)$
in \ref{7.368}, squaring, integrating with respect to $\lambda$ on
$(-t,\lambda^*(t))$, and taking the square root, we obtain, by
\ref{16.81} and \ref{16.80}:
\begin{equation}
|t|^{-1}\|D^2\omega\|_{L^2(H^\prime_t)}\leq
C\int_{u_0}^t|t^\prime|^{-1}\|\dot{\nb}\|_{L^2(H^\prime_{t^\prime})}dt^\prime
\label{16.86}
\end{equation}
Substituting for $\dot{\nb}$ we now obtain an estimate for
$D^2\omega$ in $L^2(H^\prime_t)$ in terms of
\begin{equation}
\int_{u_0}^t\|D^2\rho\|_{L^2(H^\prime_{t^\prime})}dt^\prime
\label{16.87}
\end{equation}
plus lower order terms. Now, $D^2\rho$ is bounded in
$L^2(H^\prime_t)$ through the bounds \ref{16.50}. We obtain in
this way the following sharp bound for $D^2\omega$ in
$L^2(H^\prime_t)$:
\begin{equation}
\|D^2\omega\|_{L^2(H^\prime_t)}\leq O(\delta^{-3/2}|t|^{-1}) \ \ :
\ \forall t\in(u_0,c^*) \label{16.88}
\end{equation}

Let us now consider the $k$th order energy $E^\prime_k(t)$
associated to $H^\prime_t$ (see \ref{16.51} - \ref{16.56}). It
seems at first sight that $E^\prime_k(t)$ controls the $L^2$ norm
on $H^\prime_t$ of only the $D,\Db$ derivatives of the curvature
components of of order $k$ and no $\snab$ derivatives.
Nevertheless, we shall presently show that the $E^\prime_l(t)$ for
$l=3,...,k$ together with the sharp bounds which we have already
obtained, actually control the $L^2$ norms on $H^\prime_t$ of all
$D,\Db$ and $\snab$ derivatives of the curvature components of
order up to $k$. This is by virtue of the Bianchi identities.

Consider in fact the Bianchi identities, given by Proposition 1.2.
Denoting by $\simeq$ equality up to lower order terms (here
products of connection coefficients with curvature components),
these equations read:
\begin{eqnarray}
&&\Omega^{-1}\Dbh\alpha\simeq\snab\oth\beta\nonumber\\
&&\Omega^{-1}\Dh\alb\simeq-\snab\oth\beb\nonumber\\
&&\Omega^{-1}D\beta\simeq\sdiv\alpha\nonumber\\
&&\Omega^{-1}\Db\beb\simeq-\sdiv\alb\nonumber\\
&&\Omega^{-1}\Db\beta\simeq\sd\rho+\s^*\sd\sigma\nonumber\\
&&\Omega^{-1}D\beb\simeq-\sd\rho+\s^*\sd\sigma\nonumber\\
&&\Omega^{-1}D\rho\simeq\sdiv\beta\nonumber\\
&&\Omega^{-1}D\sigma\simeq-\scurl\beta\nonumber\\
&&\Omega^{-1}\Db\rho\simeq-\sdiv\beb\nonumber\\
&&\Omega^{-1}\Db\sigma\simeq-\scurl\beb \label{16.89}
\end{eqnarray}
In the third and fourth of the above equations we have the
elliptic operator $\sdiv$ from trace-free symmetric 2-covariant
tensorfields to 1-forms on $S_{\ub,u}$, analyzed in Chapter 5 (and
also in Chapter 7). The $L^2$-adjoint of this operator is the
operator
\begin{equation}
\xi \mapsto-\frac{1}{2}\snab\oth\xi \label{16.90}
\end{equation}
from 1-forms $\xi$ to trace-free symmetric 2-covariant
tensorfields on $S_{\ub,u}$, and we have:
\begin{equation}
\snab\oth\sdiv\theta=\slap\theta-2K\theta \label{16.91}
\end{equation}
for any trace-free symmetric 2-covariant tensorfield $\theta$ on
$S_{\ub,u}$. Therefore, applying $\Dh$ to the first of equations
\ref{16.89}  and substituting from the third we obtain:
\begin{equation}
\Omega^{-2}\Dbh\Dh\alpha\simeq\slap\alpha \label{16.92}
\end{equation}
Also, applying $\Dbh$ to the second of equations \ref{16.89} and
substituting from the fourth we obtain:
\begin{equation}
\Omega^{-2}\Dbh\Dh\alb\simeq\slap\alb \label{16.93}
\end{equation}
In the last two pairs of equations \ref{16.89}  we have the
elliptic operator $(\sdiv,\scurl)$ from 1-forms to pairs of
functions on $S_{\ub,u}$, analyzed in Chapter 5 (and also in
Chapter 7). The $L^2$-adjoint of this operator is the operator
\begin{equation}
(f,g)\mapsto-\sd f+\s^*\sd g \label{16.94}
\end{equation}
from pairs of functions $(f,g)$ to 1-forms on $S_{\ub,u}$, and we
have:
\begin{equation}
\sd\sdiv\xi-\s^*\sd\scurl\xi=\slap\xi-K\xi \label{16.95}
\end{equation}
for any 1-form $\xi$ on $S_{\ub,u}$, and:
\begin{equation}
(\sdiv(\sd f-\s^*\sd g),\scurl(\sd f-\s^*\sd g))=(\slap f,\slap g)
\label{16.96}
\end{equation}
for any pair of functions $(f,g)$ on $S_{\ub,u}$. Therefore
applying $D$ to the fifth of equations \ref{16.89} and
substituting from the seventh and eighth we obtain:
\begin{equation}
\Omega^{-1}\Db D\beta\simeq\slap\beta \label{16.97}
\end{equation}
and applying $\Db$ to the sixth of equations \ref{16.89} and
substituting from the ninth and tenth we obtain:
\begin{equation}
\Omega^{-2}\Db D\beb\simeq\slap\beb \label{16.98}
\end{equation}
Also, applying $\Db$ to the senth and eighth of equations
\ref{16.89} and substituting from the fifth we obtain:
\begin{eqnarray}
&&\Omega^{-2}\Db D\rho\simeq\slap\rho\nonumber\\
&&\Omega^{-2}\Db D\sigma\simeq\slap\sigma \label{16.99}
\end{eqnarray}
Applying $\Db^{m}D^n$ to equations \ref{16.89} yields:
\begin{eqnarray}
&&\Omega^{-1}\Dbh^{m+1}D^n\alpha\simeq\snab\oth(\Db^m D^n\beta)\nonumber\\
&&\Omega^{-1}\Dbh^m\Dh^{n+1}\alb\simeq-\snab\oth(\Db^m D^n\beb)\nonumber\\
&&\Omega^{-1}\Db^m D^{n+1}\beta\simeq\sdiv(\Dbh^m\Dh^n\alpha)\nonumber\\
&&\Omega^{-1}\Db^{m+1} D^n\beb\simeq-\sdiv(\Dbh^m\Dh^n\alb)\nonumber\\
&&\Omega^{-1}\Db^{m+1}\Dh^n\beta\simeq\sd(\Db^m D^n\rho)+\s^*\sd(\Db^m D^n\sigma)\nonumber\\
&&\Omega^{-1}\Db^m D^{n+1}\beb\simeq-\sd(\Db^m D^n\rho)+\s^*\sd(\Db^m D^n\sigma)\nonumber\\
&&\Omega^{-1}\Db^m D^{n+1}\rho\simeq\sdiv(\Db^m D^n\beta)\nonumber\\
&&\Omega^{-1}\Db^m D^{n+1}\sigma\simeq-\scurl(\Db^m D^n\beta)\nonumber\\
&&\Omega^{-1}\Db^{m+1}D^n\rho\simeq-\sdiv(\Db^m D^n\beb)\nonumber\\
&&\Omega^{-1}\Db^{m+1}D^n\sigma\simeq-\scurl(\Db^m D^n\beb)
\label{16.100}
\end{eqnarray}
Here we take $m+n=k-1$. In view of the identity \ref{5.149}, the
third and fourth of the above give us bounds for
$\|\snab\Dbh^m\Dh^n\alpha\|^2_{L^2(S_{\ub,u})}$ and
$\|\snab\Dbh^m\Dh^n\alb\|^2_{L^2(S_{\ub,u})}$ in terms of $\|\Db^m
D^{n+1}\beta\|^2_{L^2(S_{\ub,u})}$ and
$\|\Db^{m+1}D^n\beb\|^2_{L^2(S_{\ub,u})}$, respectively, plus
lower terms. In view of the identity \ref{5.179}, the seventh and
eighth of \ref{16.100} give us a bound for $\|\snab\Db^m
D^n\beta\|^2_{L^2(S_{\ub,u})}$ in terms of $\|\Db^m
D^{n+1}\rho\|^2_{L^2(S_{\ub,u})}$, $\|\Db^m
D^{n+1}\sigma\|^2_{L^2(S_{\ub,u})}$ plus lower order terms, while
the ninth and tenth of \ref{16.100} give us a bound for
$\|\snab\Db^m D^n\beb\|^2_{L^2(S_{\ub,u})}$ in terms of
$\|\Db^{m+1}D^n\rho\|^2_{L^2(S_{\ub,u})}$ and
$\|\Db^{m+1}D^n\sigma\|^2_{L^2(S_{\ub,u})}$ plus lower order
terms. Finally, in view of the identity:
\begin{equation}
\int_{S_{\ub,u}}\left\{|\sd f|^2+|\s^*\sd
g|^2\right\}d\mu_{\sg}=\int_{S_{\ub,u}}|\sd f-\s^*\sd
g|^2d\mu_{\sg} \label{16.101}
\end{equation}
for any pair of functions $f,g$ on $S_{\ub,u}$, the fifth and sixth
of \ref{16.100} gives us bounds for $\|\sd\Db^m
D^n\rho\|^2_{L^2(S_{\ub,u})}$, $\|\sd\Db^m
D^n\sigma\|^2_{L^2(S_{\ub,u})}$ in terms of
$\|\Db^{m+1}D^n\beta\|^2_{L^2(S_{\ub,u})}$ and $\|\Db^m
D^{n+1}\beb\|^2_{L^2(S_{\ub,u})}$ respectively, plus lower order
terms.

Applying $\Db^m\Db^n$ to equations \ref{16.92}, \ref{16.93},
\ref{16.97}, \ref{16.98}, \ref{6.99} yields:
\begin{eqnarray}
&&\Omega^{-2}\Dbh^{m+1}\Dh^{n+1}\alpha\simeq\slap(\Dbh^m\Dh^n\alpha)\nonumber\\
&&\Omega^{-2}\Dbh^{m+1}\Dh^{m+1}\alb\simeq\slap(\Dbh^m\Dh^n\alb)\nonumber\\
&&\Omega^{-2}\Db^{m+1}D^{n+1}\beta\simeq\slap(\Db^m D^n\beta)\nonumber\\
&&\Omega^{-2}\Db^{m+1}D^{n+1}\beb\simeq\slap(\Db^m D^n\beb)\nonumber\\
&&\Omega^{-2}\Db^{m+1}D^{n+1}\rho\simeq\slap(\Db^m D^n\rho)\nonumber\\
&&\Omega^{-2}\Db^{m+1}D^{n+1}\sigma\simeq\slap(\Db^m D^n\sigma)
\label{16.102}
\end{eqnarray}
Here we take $m+n=k-l$, $l=2,...,k$. Standard elliptic estimates
for the Laplacian $\slap$ on $S_{\ub,u}$ then give us bounds for
\begin{eqnarray}
&&\|\snab^{ \ l}\Dbh^m\Dh^n\alpha\|^2_{L^2(S_{\ub,u})}\nonumber\\
&&\|\snab^{ \ l}\Dbh^m\Dh^n\alb\|^2_{L^2(S_{\ub,u})}\nonumber\\
&&\|\snab^{ \ l}\Db^m D^n\beta\|^2_{L^2(S_{\ub,u})}\nonumber\\
&&\|\snab^{ \ l}\Db^m D^n\beb\|^2_{L^2(S_{\ub,u})}\nonumber\\
&&\|\snab^{ \ l}\Db^m D^n\rho\|^2_{L^2(S_{\ub,u})}\nonumber\\
&&\|\snab^{ \ l}\Db^m D^n\sigma\|^2_{L^2(S_{\ub,u})} \label{16.103}
\end{eqnarray}
in terms of
\begin{eqnarray}
&&\|\snab^{ \ l-2}\Dbh^{m+1}\Dh^{n+1}\alpha\|^2_{L^2(S_{\ub,u})}\nonumber\\
&&\|\snab^{ \ l-2}\Dbh^{m+1}\Dh^{n+1}\alb\|^2_{L^2(S_{\ub,u})}\nonumber\\
&&\|\snab^{ \ l-2}\Db^{m+1}D^{n+1}\beta\|^2_{L^2(S_{\ub,u})}\nonumber\\
&&\|\snab^{ \ l-2}\Db^{m+1}D^{n+1}\beb\|^2_{L^2(S_{\ub,u})}\nonumber\\
&&\|\snab^{ \ l-2}\Db^{m+1}D^{n+1}\rho\|^2_{L^2(S_{\ub,u})}\nonumber\\
&&\|\snab^{ \ l-2}\Db^{m+1}D^{n+1}\sigma\|^2_{L^2(S_{\ub,u})}
\label{16.104}
\end{eqnarray}
respectively, plus lower order terms. We thus have a recursion
which for $l$ even gives us bounds for \ref{16.103} in terms of
\begin{eqnarray}
&&\|\Dbh^{m+(l/2)}\Dh^{n+(l/2)}\alpha\|^2_{L^2(S_{\ub,u})}\nonumber\\
&&\|\Dbh^{m+(l/2)}\Dh^{n+(l/2)}\alb\|^2_{L^2(S_{\ub,u})}\nonumber\\
&&\|\Db^{m+(l/2)}D^{n+(l/2)}\beta\|^2_{L^2(S_{\ub,u})}\nonumber\\
&&\|\Db^{m+(l/2)}D^{n+(l/2)}\beb\|^2_{L^2(S_{\ub,u})}\nonumber\\
&&\|\Db^{m+(l/2)}D^{n+(l/2)}\rho\|^2_{L^2(S_{\ub,u})}\nonumber\\
&&\|\Db^{m+(l/2)}D^{n+(l/2)}\sigma\|^2_{L^2(S_{\ub,u})}
\label{16.105}
\end{eqnarray}
respectively, plus lower order terms, while for $l$ odd gives us
bounds for \ref{16.103} in terms of
\begin{eqnarray}
&&\|\snab\Dbh^{m+((l-1)/2)}\Dh^{n+((l-1)/2)}\alpha\|^2_{L^2(S_{\ub,u})}\nonumber\\
&&\|\snab\Dbh^{m+((l-1)/2)}\Dh^{n+((l-1)/2)}\alb\|^2_{L^2(S_{\ub,u})}\nonumber\\
&&\|\snab\Db^{m+((l-1)/2)}D^{n+((l-1)/2)}\beta\|^2_{L^2(S_{\ub,u})}\nonumber\\
&&\|\snab\Db^{m+((l-1)/2)}D^{n+((l-1)/2)}\beb\|^2_{L^2(S_{\ub,u})}\nonumber\\
&&\|\snab\Db^{m+((l-1)/2)}D^{n+((l-1)/2)}\rho\|^2_{L^2(S_{\ub,u})}\nonumber\\
&&\|\snab\Db^{m+((l-1)/2)}D^{n+((l-1)/2)}\sigma\|^2_{L^2(S_{\ub,u})}
\label{16.106}
\end{eqnarray}
respectively, plus lower order terms. According to the previous
discussion the latter are bounded in terms of
\begin{eqnarray}
&&\|\Db^{m+((l-1)/2)}D^{n+((l-1)/2)+1}\beta\|^2_{L^2(S_{\ub,u})}\nonumber\\
&&\|\Db^{m+((l-1)/2)+1}D^{n+((l-1)/2)}\beb\|^2_{L^2(S_{\ub,u})}\nonumber\\
&&\|\Db^{m+((l-1)/2)}D^{n+((l-1)/2)+1}\rho\|^2_{L^2(S_{\ub,u})}\nonumber\\
&&\hspace{10mm}+\|\Db^{m+((l-1)/2)}D^{n+((l-1)/2)+1}\sigma\|^2_{L^2(S_{\ub,u})}\nonumber\\
&&\|\Db^{m+((l-1)/2)+1}D^{n+((l-1)/2)}\rho\|^2_{L^2(S_{\ub,u})}\nonumber\\
&&\hspace{10mm}+\|\Db^{m+((l-1)/2)+1}D^{n+((l-1)/2)}\sigma\|^2_{L^2(S_{\ub,u})}\nonumber\\
&&\|\Db^{m+((l-1)/2)+1}D^{n+((l-1)/2)}\beta\|^2_{L^2(S_{\ub,u})}\nonumber\\
&&\hspace{10mm}\mbox{or} \ \
\|\Db^{m+((l-1))/2}D^{n+((l-1)/2)+1}\beb\|^2_{L^2(S_{\ub,u})}
\label{16.107}
\end{eqnarray}
respectively, plus lower order terms. Setting
$\ub=\frac{1}{2}(t+\lambda)$, $u=\frac{1}{2}(t-\lambda)$ and
integrating the inequalities obtained in this way with respect to
$\lambda$ on $(-t,\lambda^*(t))$, we then obtain a bound for the
sum of the squares of the $L^2$ norms on $H^\prime_t$ of all
derivatives of order $k$ of all curvature componets in terms of
the $k$th order energy $E^\prime_k(t)$ associated to $H^\prime_t$
plus lower order terms. The lower order terms are bounded in terms
of the sum of the squares of the $L^2$ norms on $H^\prime_t$ of
the derivatives of the curvature components of order up to $k-1$
and the sum of the squares of the $L^2$ norms on $H^\prime_t$ of
the derivatives of the connection coefficients of order up to
$k-1$. As we shall presently show, any derivative of order
$k^\prime$ of any connection coefficient is bounded in
$L^2(H^\prime_t)$ by the supremum for $t^\prime\in(u_0,t]$ of the
sum of the $L^2$ norms on $H^\prime_{t^\prime}$ of the derivatives
of order $k^\prime$ of the curvature components, plus lower order
terms. We are thus able to establish inductively that
$E^\prime_k(t)$ bounds the sum of the squares of the $L^2$ norms
on $H^\prime_t$ of all derivatives of order $k$ of all curvature
components, up to terms which are bounded in terms of
$\sup_{t^\prime\in(u_0,t]}E^\prime_{k^\prime}(t^\prime)$: for
$k^\prime=3,...,k-1$ and the sharp bounds already obtained for up
to the second derivatives of the curvature components and the
connection coefficients. Here the Sobolev inequality on the
$H^\prime_t$ is used, which is similar to the Sobolev inequality
on the $C_u$, Proposition 10.1, but with the $D$ derivative
replaced by the $\sL_N$ derivative, where $N$ is the vectorfield
\begin{equation}
N=\frac{1}{2}(L-\Lb) \hspace{20mm} N\lambda=1 \label{16.108}
\end{equation}
which is tangential to $H^\prime_t$ and generates a flow
$G_\sigma$ on $H^\prime_t$ which maps
$S_{\frac{1}{2}(t+\lambda),\frac{1}{2}(t-\lambda)}$ onto
$S_{\frac{1}{2}(t+\lambda+\sigma),\frac{1}{2}(t-\lambda-\sigma)}$.
Note that on $H^\prime_t$ we have: $1-\delta<|t|/|u|\leq 1$.

We shall now show that any derivative of order $k$ of any
connection coefficient is bounded in $L^2(H^\prime_t)$ by a
constant times $\sqrt{\oE^\prime_k(t)}$, where
\begin{equation}
\oE^\prime_k(t)=\sup_{t^\prime\in(u_0,t]}E^\prime_k(t^\prime)
\label{16.109}
\end{equation}
plus  a lower order term which is bounded in terms of the supremum
for $t^\prime\in(u_0,t]$ of the sum of the $L^2$ norms on
$H^\prime_{t^\prime}$ of the derivatives of the curvature
components of order up to $k-1$ and the supremum for
$t^\prime\in(u_0,t]$ of the sum of the $L^2$ norms on
$H^\prime_{t^\prime}$ of the derivatives of the connection
coefficients of order up to $k-1$, and the sharp bounds for the
derivatives of the curvature components and the connection
coefficients of up to second order which have already been
obtained. This shall be accomplished by appealing only to the
propagation equations.

It will then follow by induction that the sum of the $L^2$ norms
on $H^\prime_t$ of the derivatives of the connection coefficients
of order $k$ is bounded by a constant times
$\sqrt{\oE^\prime_k(t)}$ up to terms which are bounded in terms of
$\sqrt{\oE^\prime_{k^\prime}(t)}$: for $k^\prime=3,...,k-1$ and
the sharp bounds already obtained for up to the second derivatives
of the connection coefficients and the curvature components.

We first consider the propagation equations for
$\snab^{ \ k}\mbox{tr}\chi^\prime$, $\snab^{ \ k}\chih^\prime$. These are
deduced by applying $\snab^{ \ k-2}$ to the equations \ref{5.8} and
using Lemma 4.1. We obtain:
\begin{eqnarray}
&&D\snab^{ \ k}\mbox{tr}\chi^\prime=f\cdot\snab^{ \ k}\mbox{tr}\chi^\prime+g\cdot\snab^{ \ k}\chih^\prime+r_k\nonumber\\
&&D\snab^{ \ k}\chih^\prime=h\cdot\snab^{ \ k}\mbox{tr}\chi^\prime+i\cdot\snab^{ \ k}\chih^\prime+s_k
\label{16.110}
\end{eqnarray}
The coefficients $f$, $g$, $h$, $i$ are given by \ref{4.54}, and
we have:
\begin{eqnarray}
&&r_k\simeq 0\nonumber\\
&&s_k\simeq-\snab^{ \ k}\alpha \label{16.111}
\end{eqnarray}
Here, we denote by $\simeq$ equality up to lower order terms
involving the $\snab$ derivatives of $\mbox{tr}\chi^\prime$,
$\chih^\prime$ and $\sd\log\Omega=(1/2)(\eta+\etb)$ of order up to
$k-1$.

We apply Lemma 4.6 to the above propagation equations taking
$p=2$. In the case of the first of \ref{16.110} we have
$\snab^{ \ k}\mbox{tr}\chi^\prime$ in the role of $\theta$,
$g\cdot\snab^{ \ k}\chih^\prime+r_k$ in the role of $\xi$, $r=k$,
$\nu=-2$ and $\gamma=0$. In the case of the second of \ref{16.110}
we have $\snab^{ \ k}\chih^\prime$ in the role of $\theta$,
$h\cdot\snab^{ \ k}\mbox{tr}\chi^\prime+s_k$ in the role of $\xi$,
$r=k+2$, $\nu=0$ and $\gamma=i$. In view of the estimates of
Chapter 3 we obtain:
\begin{eqnarray}
&&\|\snab^{ \ k}\mbox{tr}\chi^\prime\|_{L^2(S_{\ub,u})}\leq
C\int_0^{\ub}\|g\cdot\snab^{ \ k}\chih^\prime+r_k\|_{L^2(S_{\ub^\prime,u})}d\ub^\prime
\nonumber\\
&&\hspace{26mm}\leq
C^\prime\int_0^{\ub}\|\snab^{ \ k}\chih^\prime\|_{L^2(S_{\ub^\prime,u})}d\ub^\prime
+C\int_0^{\ub}\|r_k\|_{L^2(S_{\ub^\prime,u})}d\ub^\prime\nonumber\\
&&\|\snab^{ \ k}\chih^\prime\|_{L^2(S_{\ub,u})}\leq C\int_0^{\ub}\|h\cdot\snab^{ \ k}\mbox{tr}\chi^\prime+s_k\|_{L^2(S_{\ub^\prime,u})}\nonumber\\
&&\hspace{23mm}\leq
C^\prime\int_0^{\ub}\|\snab^{ \ k}\mbox{tr}\chi^\prime\|_{L^2(S_{\ub^\prime,u})}d\ub^\prime
+C\int_0^{\ub}\|s_k\|_{L^2(S_{\ub^\prime,u})}d\ub^\prime\nonumber\\
&&\label{16.112}
\end{eqnarray}
Here and in the following {\em we allow the constants to depend on
$\delta$, $u_0$, and the quantities $\stackrel{(n)}{D} \ : \
n=0,1,2,3$, ${\cal D}^{\prime 4}_{[1]}(\alb)$, ${\cal
D}_0^\infty$, $\scD_1^4$, $\scD_2^4(\mbox{tr}\chib)$, and
$\scD_3(\mbox{tr}\chib)$}, as only rough bounds are required.

Setting $\ub=\frac{1}{2}(t+\lambda)$, $u=\frac{1}{2}(t-\lambda)$
in each of \ref{16.112}, squaring, integrating with respect to
$\lambda$ on $(-t,\lambda^*(t))$, and taking the square root, we
obtain, by \ref{16.82} and \ref{16.80},
\begin{eqnarray}
&&\|\snab^{ \ k}\mbox{tr}\chi^\prime\|_{L^2(H^\prime_t)}\leq
C^\prime\int_{u_0}^t\|\snab^{ \ k}\chih^\prime\|_{L^2(H^\prime_{t^\prime})}dt^\prime
+C\int_{u_0}^t\|r_k\|_{L^2(H^\prime_{t^\prime})}dt^\prime\nonumber\\
&&\|\snab^{ \ k}\chih^\prime\|_{L^2(H^\prime_t)}\leq
C^\prime\int_{u_0}^t\|\snab^{ \ k}\mbox{tr}\chi^\prime\|_{L^2(H^\prime_{t^\prime})}dt^\prime
+C\int_{u_0}^t\|s_k\|_{L^2(H^\prime_{t^\prime})}dt^\prime\nonumber\\
&&\label{16.113}
\end{eqnarray}
Summing these two inequalities we obtain a linear integral
inequality for the quantity
$$\|\snab^{ \ k}\mbox{tr}\chi^\prime\|_{L^2(H^\prime_t)}+\|\snab^{ \ k}\chih^\prime\|_{L^2(H^\prime_t)}$$
which implies:
\begin{equation}
\|\snab^{ \ k}\mbox{tr}\chi^\prime\|_{L^2(H^\prime_t)}+\|\snab^{ \ k}\chih^\prime\|_{L^2(H^\prime_t)}\leq
C^{\prime\prime}\int_{u_0}^t\left\{\|r_k\|_{L^2(H^\prime_{t^\prime})}
+\|s_k\|_{L^2(H^\prime_{t^\prime})}\right\}dt^\prime
\label{16.114}
\end{equation}
Now by \ref{16.111} the integral
\begin{equation}
\int_{u_0}^t\|r_k\|_{L^2(H^\prime_{t^\prime})}dt^\prime
\label{16.115}
\end{equation}
is bounded in terms of the supremum for $t^\prime\in(u_0,t]$ of
the sum of the $L^2$ norms of the derivatives of the connection
coefficients of order up to $k-1$, and the sharp bounds for the
derivatives of up to the 2nd order of the connection coefficients
already obtained. Also, the integral
\begin{equation}
\int_{u_0}\|s_k\|_{L^2(H^\prime_{t^\prime})}dt^\prime
\label{16.116}
\end{equation}
is bounded by the integral
\begin{equation}
\int_{u_0}\|\snab^{ \ k}\alpha\|_{L^2(H^\prime_{t^\prime})}dt^\prime
\label{16.117}
\end{equation}
plus a lower order term which is bounded in terms of the supremum
for $t^\prime\in(u_0,t]$ of the sum of the $L^2$ norms of the
derivatives of the connection coefficients of order up to $k-1$,
and the sharp bounds for the derivatives of up to the 2nd order of
the connection coefficients already obtained. Moreover, according
to what has previously been demonstrated, the integral
\ref{16.117} is bounded by a constant times
$\sqrt{\oE^\prime_k}(t)$ plus a lower order term which is bounded
in terms of $\sqrt{\oE^\prime_{k^\prime}(t)}$: for
$k^\prime=3,...,k-1$, the supremum for $t^\prime\in(u_0,t]$ of the
sum of the $L^2$ norms of the derivatives of the connection
coefficients of order up to $k-1$, and the sharp bounds for the
derivatives of up to the 2nd order of the connection coefficients
already obtained. We have thus shown what was required in the case
of $\snab^{ \ k}\mbox{tr}\chi$, $\snab^{ \ k}\chih$.

We turn to $\snab^{ \ k}\mbox{tr}\chi^\prime$, $\snab^{ \ k}\chibh^\prime$.
These statisfy the conjugates of the propagation equations
\ref{16.110}:

\begin{eqnarray}
&&\Db\snab^{ \ k}\mbox{tr}\chib^\prime=\fb\cdot\snab^{ \ k}\mbox{tr}\chib^\prime+\gb\cdot\snab^{ \ k}\chibh^\prime+\rb_k\nonumber\\
&&\Db\snab^{ \ k}\chibh^\prime=\hb\cdot\snab^{ \ k}\mbox{tr}\chib^\prime+\ib\cdot\snab^{ \ k}\chibh^\prime+\sb_k
\label{16.118}
\end{eqnarray}
The coefficients $\fb$, $\gb$, $\hb$, $\ib$ are given by
\ref{4.85}, and we have:
\begin{eqnarray}
&&\rb_k\simeq 0\nonumber\\
&&\sb_k\simeq-\snab^{ \ k}\alb \label{16.119}
\end{eqnarray}
Here, we denote by $\simeq$ equality up to lower order terms
involving the $\snab$ derivatives of $\mbox{tr}\chib^\prime$,
$\chibh^\prime$ and $\sd\log\Omega=(1/2)(\eta+\etb)$ of order up
to $k-1$.

We apply Lemma 4.7 to the above propagation equations taking
$p=2$. In the case of the first of \ref{16.118} we have
$\snab^{ \ k}\mbox{tr}\chib^\prime$ in the role of $\thetab$,
$\gb\cdot\snab^{ \ k}\chibh^\prime+\rb_k$ in the role of $\xib$, $r=k$,
$\nu=-2$ and $\gammab=0$. In the case of the second of
\ref{16.118} we have $\snab^{ \ k}\chibh^\prime$ in the role of
$\thetab$, $\hb\cdot\snab^{ \ k}\mbox{tr}\chib^\prime+\sb_k$ in the
role of $\xib$, $r=k+2$, $\nu=0$ and $\gammab=\ib$. We obtain:
\begin{eqnarray}
&&|u|^{k+1}\|\snab^{ \ k}\mbox{tr}\chib^\prime\|_{L^2(S_{\ub,u})}\leq C|u_0|^{k+1}\|\snab^{ \ k}\mbox{tr}\chib^\prime\|_{L^2(S_{\ub,u_0})}\nonumber\\
&&\hspace{30mm}+ C\int_{u_0}^u |u^\prime|^{k+1}\|\gb\cdot\snab^{ \ k}\chibh^\prime+\rb_k\|_{L^2(S_{\ub,u^\prime})}du^\prime\nonumber\\
&&|u|^{k+1}\|\snab^{ \ k}\chih^\prime\|_{L^2(S_{\ub,u})}\leq
C|u_0|^{k+1}\|\snab^{ \ k}\chibh^\prime\|_{L^2(S_{\ub,u_0})}\nonumber\\
&&\hspace{30mm}+C\int_{u_0}^u|u^\prime|^{k+1}\|\hb\cdot\snab^{ \ k}\mbox{tr}\chib^\prime+\sb_k\|_{L^2(S_{\ub,u^\prime})}
\label{16.120}
\end{eqnarray}
In view of the estimates of Chapter 3 these imply:
\begin{eqnarray}
&&\|\snab^{ \ k}\mbox{tr}\chib^\prime\|_{L^2(S_{\ub,u})}\leq C^\prime\|\snab^{ \ k}\mbox{tr}\chib^\prime\|_{L^2(S_{\ub,u_0})}\nonumber\\
&&\hspace{25mm}+C^\prime\int_{u_0}^u\|\snab^{ \ k}\chibh^\prime\|_{L^2(S_{\ub,u^\prime})}du^\prime
+C^\prime\int_{u_0}^u\|\rb_k\|_{L^2(S_{\ub,u^\prime})}du^\prime\nonumber\\
&&\|\snab^{ \ k}\chibh^\prime\|_{L^2(S_{\ub,u})}\leq C^\prime\|\snab^{ \ k}\chibh^\prime\|_{L^2(S_{\ub,u_0})}\nonumber\\
&&\hspace{20mm}+C^\prime\int_{u_0}^u\|\snab^{ \ k}\mbox{tr}\chib^\prime\|_{L^2(S_{\ub,u^\prime})}du^\prime
+C^\prime\int_{u_0}^u\|\sb_k\|_{L^2(S_{\ub,u^\prime})}du^\prime\nonumber\\
&&\label{16.121}
\end{eqnarray}

Setting $\ub=\frac{1}{2}(t+\lambda)$, $u=\frac{1}{2}(t-\lambda)$
in each of \ref{16.121}, squaring, integrating with respect to
$\lambda$ on $(-t,\lambda^*(t))$, and taking the square root, we
obtain, by \ref{16.81} and \ref{16.80},
\begin{eqnarray}
&&\|\snab^{ \ k}\mbox{tr}\chib^\prime\|_{L^2(H^\prime_t)}\leq C^\prime\|\snab^{ \ k}\mbox{tr}\chib^\prime\|_{L^2(C_{u_0})}\nonumber\\
&&\hspace{20mm}+C^\prime\int_{u_0}^t\|\snab^{ \ k}\chibh^\prime\|_{L^2(H^\prime_{t^\prime})}dt^\prime
+C^\prime\int_{u_0}^t\|\rb_k\|_{L^2(H^\prime_{t^\prime})}dt^\prime\nonumber\\
&&\|\snab^{ \ k}\chibh^\prime\|_{L^2(H^\prime_t)}\leq C^\prime\|\snab^{ \ k}\chibh^\prime\|_{L^2(C_{u_0})}\nonumber\\
&&\hspace{20mm}+C^\prime\int_{u_0}^t\|\snab^{ \ k}\mbox{tr}\chib^\prime\|_{L^2(H^\prime_{t^\prime})}dt^\prime
+C^\prime\int_{u_0}^t\|\sb_k\|_{L^2(H^\prime_{t^\prime})}dt^\prime\nonumber\\
&&\label{16.122}
\end{eqnarray}
Summing these two inequalities we obtain a linear integral
inequality for the quantity
$$\|\snab^{ \ k}\mbox{tr}\chib^\prime\|_{L^2(H^\prime_t)}+\|\snab^{ \ k}\chibh^\prime\|_{L^2(H^\prime_t)}$$
which implies:
\begin{eqnarray}
\|\snab^{ \ k}\mbox{tr}\chib^\prime\|_{L^2(H^\prime_t)}+\|\snab^{ \ k}\chibh^\prime\|_{L^2(H^\prime_t)}\leq
C^{\prime\prime}\left\{\|\snab^{ \ k}\mbox{tr}\chib^\prime\|_{L^2(C_{u_0})}+\|\snab^{ \ k}\chibh^\prime\|_{L^2(C_{u_0})}\right\}\nonumber\\
\hspace{40mm}+C^{\prime\prime}\int_{u_0}^t\left\{\|\rb_k\|_{L^2(H^\prime_{t^\prime})}
+\|\sb_k\|_{L^2(H^\prime_{t^\prime})}\right\}dt^\prime\nonumber\\
&&\label{16.123}
\end{eqnarray}
Now by \ref{16.119} the integral
\begin{equation}
\int_{u_0}^t\|\rb_k\|_{L^2(H^\prime_{t^\prime})}dt^\prime
\label{16.124}
\end{equation}
is bounded in terms of the supremum for $t^\prime\in(u_0,t]$ of
the sum of the $L^2$ norms of the derivatives of the connection
coefficients of order up to $k-1$, and the sharp bounds for the
derivatives of up to the 2nd order of the connection coefficients
already obtained. Also, the integral
\begin{equation}
\int_{u_0}\|\sb_k\|_{L^2(H^\prime_{t^\prime})}dt^\prime
\label{16.125}
\end{equation}
is bounded by the integral
\begin{equation}
\int_{u_0}\|\snab^{ \ k}\alb\|_{L^2(H^\prime_{t^\prime})}dt^\prime
\label{16.126}
\end{equation}
plus a lower order term which is bounded in terms of the supremum
for $t^\prime\in(u_0,t]$ of the sum of the $L^2$ norms of the
derivatives of the connection coefficients of order up to $k-1$,
and the sharp bounds for the derivatives of up to the 2nd order of
the connection coefficients already obtained. Moreover, according
to what has previously been demonstrated, the integral
\ref{16.126} is bounded by a constant times
$\sqrt{\oE^\prime_k}(t)$ plus a lower order term which is bounded
in terms of $\sqrt{\oE^\prime_{k^\prime}(t)}$: for
$k^\prime=3,...,k-1$, the supremum for $t^\prime\in(u_0,t]$ of the
sum of the $L^2$ norms of the derivatives of the connection
coefficients of order up to $k-1$, and the sharp bounds for the
derivatives of up to the 2nd order of the connection coefficients
already obtained. We have thus shown what was required also in the
case of $\snab^{ \ k}\mbox{tr}\chib$, $\snab^{ \ k}\chibh$.

We turn to $\snab^{ \ k}\eta$,$\snab^{ \ k}\etb$. These satisfy propagation
equations deduced by applying $\snab^{ \ k-1}$ to the propagation
equations \ref{4.110} and using Lemma 4.1. We obtain:
\begin{eqnarray}
&&D\snab^{ \ k}\eta=a\cdot\snab^{ \ k}\etb+b_k\nonumber\\
&&\Db\snab^{ \ k}\etb=\ab\cdot\snab^{ \ k}\eta+\bb_k \label{16.127}
\end{eqnarray}
The coefficients $a$, $\ab$ are given by \ref{4.111} and we have:
\begin{eqnarray}
&&b_k\simeq-\Omega\snab^{ \ k}\beta+\Omega(\snab^{ \ k}\chi)\cdot\etb-(\snab^{ \ k-1}D\sGamma)\cdot\eta\nonumber\\
&&\bb_k\simeq\Omega\snab^{ \ k}\beb+\Omega(\snab^{ \ k}\chib)\cdot\eta-(\snab^{ \ k-1}\Db\Gamma)\cdot\etb
\label{16.128}
\end{eqnarray}
Here we denote by $\simeq$ equality up to lower order terms
involving the $\snab$ derivatives of $\chi$, $\chib$, $\eta$,
$\etb$ of order up to $k-1$ as well as the $\snab$ derivatives of
$\beta$, $\beb$ of order up to $k-1$.

To the first of the propagation equations \ref{16.127} we apply
Lemma 4.6 taking $p=2$, with $\snab^{ \ k}\eta$ in the role of $\theta$ and
$a\cdot\snab^{ \ k}\etb+b_k$ in the role of $\xi$. Here $r=k+1$,
$\nu=0$, $\gamma=0$ and we obtain:
\begin{equation}
\|\snab^{ \ k}\eta\|_{L^2(S_{\ub,u})}\leq
C\int_0^{\ub}\|a\cdot\snab^{ \ k}\etb+b_k\|_{L^2(S_{\ub^\prime,u})}d\ub^\prime
\label{16.129}
\end{equation}
To the second of the propagation equations \ref{16.127} we apply
Lemma 4.7 taking $p=2$, with $\snab^{ \ k}\etb$ in the role of $\thetab$ and
$\ab\cdot\snab^{ \ k}\eta+\bb_k$ in the role of $\xib$. Here $r=k+1$,
$\nu=0$, $\gammab=0$ and we obtain:
\begin{eqnarray}
&&|u|^k\|\snab^{ \ k}\etb\|_{L^2(S_{\ub,u})}\leq C|u_0|^k\|\snab^{ \ k}\etb\|_{L^2(S_{\ub,u_0})}\nonumber\\
&&\hspace{25mm}+C\int_{u_0}^u|u^\prime|^k\|\ab\cdot\snab^{ \ k}\eta+\bb_k\|_{L^2(S_{\ub,u^\prime})}du^\prime
\label{16.130}
\end{eqnarray}
In view of the estimates of Chapter 3, \ref{16.129} and
\ref{16.130} imply:
\begin{eqnarray}
&&\|\snab^{ \ k}\eta\|_{L^2(S_{\ub,u})}\leq
C^\prime\int_0^{\ub}\|\snab^{ \ k}\etb\|_{L^2(S_{\ub^\prime,u})}d\ub^\prime
+C^\prime\int_0^{\ub}\|b_k\|_{L^2(S_{\ub^\prime,u})}d\ub^\prime\nonumber\\
&&\|\snab^{ \ k}\etb\|_{L^2(S_{\ub,u})}\leq C^\prime\|\snab^{ \ k}\etb\|_{L^2(S_{\ub,u_0})}\nonumber\\
&&\hspace{22mm}+C^\prime\int_{u_0}^u\|\snab^{ \ k}\eta\|_{L^2(S_{\ub,u^\prime})}du^\prime
+C^\prime\int_{u_0}^u\|\bb_k\|_{L^2(S_{\ub,u^\prime})}du^\prime\nonumber\\
&&\label{16.131}
\end{eqnarray}
Setting $\ub=\frac{1}{2}(t+\lambda)$, $u=\frac{1}{2}(t-\lambda)$
in each of \ref{16.121}, squaring, integrating with respect to
$\lambda$ on $(-t,\lambda^*(t))$, and taking the square root, we
obtain, by \ref{16.82}, \ref{16.81} and \ref{16.80},
\begin{eqnarray}
&&\|\snab^{ \ k}\eta\|_{L^2(H^\prime_t)}\leq
C^\prime\int_{u_0}^t\|\snab^{ \ k}\etb\|_{L^2(H^\prime_{t^\prime})}dt^\prime
+C^\prime\int_{u_0}^t\|b_k\|_{L^2(H^\prime_{t^\prime})}dt^\prime\nonumber\\
&&\|\snab^{ \ k}\etb\|_{L^2(H^\prime_t)}\leq C^\prime\|\snab^{ \ k}\etb\|_{L^2(C_{u_0})}\nonumber\\
&&\hspace{22mm}+C^\prime\int_{u_0}^t\|\snab^{ \ k}\eta\|_{L^2(H^\prime_{t^\prime})}dt^\prime
+C^\prime\int_{u_0}^t\|\bb_k\|_{L^2(H^\prime_{t^\prime})}dt^\prime\nonumber\\
&&\label{16.132}
\end{eqnarray}
Summing these two inequalities we obtain a linear integral
inequality for the quantity
$$\|\snab^{ \ k}\eta\|_{L^2(H^\prime_t)}+\|\snab^{ \ k}\etb\|_{L^2(H^\prime_t)}$$
which implies:
\begin{eqnarray}
&&\|\snab^{ \ k}\eta\|_{L^2(H^\prime_t)}+\|\snab^{ \ k}\etb\|_{L^2(H^\prime_t)}\leq  C^{\prime\prime}\|\snab^{ \ k}\etb\|_{L^2(C_{u_0})}\nonumber\\
&&\hspace{35mm}+C^{\prime\prime}\int_{u_0}^t\left\{\|b_k\|_{L^2(H^\prime_{t^\prime})}+\|\bb_k\|_{L^2(H^\prime_{t^\prime})}\right\}dt^\prime\nonumber\\
&&\label{16.133}
\end{eqnarray}
Now, in view of \ref{16.128}, the integral
\begin{equation}
\int_{u_0}^t\left\{\|b_k\|_{L^2(H^\prime_{t^\prime})}+\|\bb_k\|_{L^2(H^\prime_{t^\prime})}\right\}dt^\prime
\label{16.134}
\end{equation}
is bounded by
\begin{eqnarray}
&&C\int_{u_0}^t\left\{\|\snab^{ \ k} \beta\|_{L^2(H^\prime_{t^\prime})}+\|\snab^{ \ k}\beb\|_{L^2(H^\prime_{t^\prime})}\right\}dt^\prime\nonumber\\
&&+C^\prime\int_{u_0}^t\left\{\|\snab^{ \ k}\chi\|_{L^2(H^\prime_{t^\prime})}+\|\snab^{ \ k}\chib\|_{L^2(H^\prime_{t^\prime})}\right\}dt^\prime
\label{16.135}
\end{eqnarray}
plus a lower order term which is bounded in terms of the supremum
for $t^\prime\in(u_0,t]$ of the sum of the $L^2$ norms of the
derivatives of the connection coefficients and of the curvature
components of order up to $k-1$, and the sharp bounds for the
derivatives of up to the 2nd order of the connection coefficients
and of  the curvature components already obtained. Moreover,
according to what has previously been demonstrated, the first of
the integrals \ref{16.135} is bounded by a constant times
$\sqrt{\oE^\prime_k}(t)$ plus a lower order term which is bounded
in terms of $\sqrt{\oE^\prime_{k^\prime}(t)}$: for
$k^\prime=3,...,k-1$, the supremum for $t^\prime\in(u_0,t]$ of the
sum of the $L^2$ norms of the derivatives of the connection
coefficients of order up to $k-1$, and the sharp bounds for the
derivatives of up to the 2nd order of the connection coefficients
already obtained. Also, the second of the integrals \ref{16.135}
has already been appropriately estimated above. We have thus shown
what was required also in the case of $\snab^{ \ k}\eta$,
$\snab^{ \ k}\etb$.

We turn to $\snab^{ \ l} D^m\omega \ : \ l+m=k$. Applying $\snab^{ \ k}$ to
the propagation equation \ref{1.87} and using Lemmas 1.2 and 4.1
we obtain the following propagation equation for $\snab^{ \ k}\omega$:
\begin{equation}
\Db\snab^{ \ k}\omega=\nb_{k,0} \label{16.136}
\end{equation}
where
\begin{equation}
\nb_{k,0}\simeq
\Omega^2\left\{2(\etb,\snab^{ \ k}\eta)+2(\eta-\etb,\snab^{ \ k}\etb)-\snab^{ \ k}\rho\right\}
\label{16.137}
\end{equation}
and we denote by $\simeq$ equality up to lower order terms
involving the $\snab$ derivarives of $\chib$, $\eta$, $\etb$ and
$\rho$ of order up to $k-1$. For $n\geq 1$, we apply $D^{m-1}$ to
the propagation equation \ref{4.197}. In view of the commutation
formula \ref{1.75} we obtain:
\begin{equation}
\Db D^m\omega=\nb_{0,m} \label{16.138}
\end{equation}
where the functions $\nb_{0,m}$ satisfy the recursion relation:
\begin{equation}
\nb_{0,m}=D\nb_{0,m-1}+2\Omega^2(\eta-\etb)^\sharp\cdot\sd
D^{m-1}\omega \label{16.139}
\end{equation}
and we have:
\begin{equation}
\nb_{0,0}=\Omega^2\{2(\eta,\etb)-|\etb|^2-\rho\} \label{16.140}
\end{equation}
Therefore the functions $\nb_{0,m}$ are given by:
\begin{equation}
\nb_{0,m}=D^m\nb_{0,0}+\sum_{m^\prime=0}^{m-1}D^{m^\prime}\left\{2\Omega^2(\eta-\etb)^\sharp\cdot\sd
D^{m-1-m^\prime}\omega\right\} \label{16.141}
\end{equation}
It follows that:
\begin{equation}
\nb_{0,m}\simeq\Omega^2\left\{2(\etb,D^m\eta)+2(\eta-\etb,D^m\etb)-D^m\rho\right\}+2m\Omega^2(\eta-\etb)^\sharp\cdot\sd
D^{m-1}\omega \label{16.142}
\end{equation}
where we denote by $\simeq$ equality up to lower order terms
involving the $D$ and $\snab$ derivatives of the connection
coefficients and of the curvature components of order up to $m-1$.
Moreover, applying $D^{m-1}$ to equations \ref{1.66}, \ref{1.150}
we obtain:
\begin{eqnarray}
&&D^m\eta\simeq 0\nonumber\\
&&D^m\etb\simeq 2\sd D^{m-1}\omega \label{16.143}
\end{eqnarray}
We conclude that:
\begin{equation}
\nb_{0,m}\simeq 2(m+2)\Omega^2(\eta-\etb)^\sharp\cdot\sd
D^{m-1}\omega-\Omega^2D^m\rho \label{16.144}
\end{equation}
We now apply $\snab^{ \ l}$, for $l+m=k$, to the propagation equation
\ref{16.138}. Using Lemmas 1.2 and 4.1 we then deduce the
following propagation equation for $\snab^{ \ l} D^m\omega$:
\begin{equation}
\Db\snab^{ \ l} D^m\omega=\nb_{l,m} \label{16.145}
\end{equation}
where
\begin{equation}
\nb_{l,m}\simeq\snab^{ \ l}\nb_{0,m}\simeq
2(m+2)\Omega^2(\eta-\etb)^\sharp\cdot\snab^{ \ l+1}D^{m-1}\omega-\Omega^2\snab^{ \ l}D^m\rho 
\label{16.146}
\end{equation}
and we denote by $\simeq$ equality up to lower order terms
involving the $D$ and $\snab$ derivatives of the connection
coefficients and of the curvature components of order up to $k-1$.

To the propagation equation \ref{16.136} we apply Lemma 4.7 taking
$p=2$. Here $r=k$, $\nu=0$, $\gammab=0$ and we obtain, in view of
the fact that $\snab^{ \ k}\omega$ vanishes on $C_{u_0}$,
\begin{equation}
|u|^{k-1}\|\snab^{ \ k}\omega\|_{L^2(S_{\ub,u})}\leq
C\int_{u_0}^{u}|u^\prime|^{k-1}\|\nb_{k,0}\|_{L^2(S_{\ub,u^\prime})}du^\prime
\label{16.147}
\end{equation}
This implies:
\begin{equation}
\|\snab^{ \ k}\omega\|_{L^2(S_{\ub,u})}\leq
C^\prime\int_{u_0}^u\|\nb_{k,0}\|_{L^2(S_{\ub,u^\prime})}du^\prime
\label{16.148}
\end{equation}
Setting $\ub=\frac{1}{2}(t+\lambda)$, $u=\frac{1}{2}(t-\lambda)$,
squaring, integrating with respect to $\lambda$ on
$(-t,\lambda^*(t))$, and taking the square root, we obtain, by
\ref{16.81} and \ref{16.80},
\begin{equation}
\|\snab^{ \ k}\omega\|_{L^2(H^\prime_t)}\leq
C^\prime\int_{u_0}^t\|\nb_{k,0}\|_{L^2(H^\prime_{t^\prime})}dt^\prime
\label{16.149}
\end{equation}
Now, in view of \ref{16.137} the integral on the right in
\ref{16.149} is bounded by
\begin{equation}
C\int_{u_0}^t\|\snab^{ \ k}\rho\|_{L^2(H^\prime_{t^\prime})}dt^\prime+
C^\prime\int_{u_0}^t\left\{\|\snab^{ \ k}\eta\|_{L^2(H^\prime_{t^\prime})}+\|\snab^{ \ k}\etb\|_{L^2(H^\prime_{t^\prime})}\right\}dt^\prime
\label{16.150}
\end{equation}
plus a lower order term which is bounded in terms of the supremum
for $t^\prime\in(u_0,t]$ of the sum of the $L^2$ norms of the
$\snab$  derivatives of the connection coefficients and of the
curvature components of order up to $k-1$, and the sharp bounds
for the derivatives of up to the 2nd order of the connection
coefficients and of  the curvature components already obtained.
Moreover, according to what has previously been demonstrated, the
first of the integrals \ref{16.149} is bounded by a constant times
$\sqrt{\oE^\prime_k}(t)$ plus a lower order term which is bounded
in terms of $\sqrt{\oE^\prime_{k^\prime}(t)}$: for
$k^\prime=3,...,k-1$, the supremum for $t^\prime\in(u_0,t]$ of the
sum of the $L^2$ norms of the derivatives of the connection
coefficients of order up to $k-1$, and the sharp bounds for the
derivatives of up to the 2nd order of the connection coefficients
already obtained. Also, the second of the integrals \ref{16.149}
has already been appropriately estimated above. We have thus shown
what was required in the case of $\snab^{ \ k}\omega$.

We proceed to derive the appropriate estimates for $\snab^{ \ l}
D^m\omega \ : \ l+m=k$ for $m=1,...,k$ by induction on $m$.
Suppose then that $\snab^{ \ l+1} D^{m-1}\omega$ has been
appropriately estimated. We apply Lemma 4.7 to the propagation
equation \ref{16.145}, taking $p=2$. Here $r=l$, $\nu=0$,
$\gammab=0$ and we obtain, in view of the fact that $\snab^{ \ l}
D^m\omega$ vanishes on $C_{u_0}$,
\begin{equation}
|u|^{l-1}\|\snab^{ \ l} D^m\omega\|_{L^2(S_{\ub,u})}\leq
C\int_{u_0}^{u}|u^\prime|^{l-1}\|\nb_{l,m}\|_{L^2(S_{\ub,u^\prime})}du^\prime
\label{16.151}
\end{equation}
This implies:
\begin{equation}
\|\snab^{ \ l} D^m\omega\|_{L^2(S_{\ub,u})}\leq
C^\prime\int_{u_0}^u\|\nb_{l,m}\|_{L^2(S_{\ub,u^\prime})}du^\prime
\label{16.152}
\end{equation}
Setting $\ub=\frac{1}{2}(t+\lambda)$, $u=\frac{1}{2}(t-\lambda)$,
squaring, integrating with respect to $\lambda$ on
$(-t,\lambda^*(t))$, and taking the square root, we obtain, by
\ref{16.81} and \ref{16.80},
\begin{equation}
\|\snab^{ \ l} D^m\omega\|_{L^2(H^\prime_t)}\leq
C^\prime\int_{u_0}^t\|\nb_{l,m}\|_{L^2(H^\prime_{t^\prime})}dt^\prime
\label{16.153}
\end{equation}
Now, in view of \ref{16.146} the integral on the right in
\ref{16.153} is bounded by
\begin{equation}
C\int_{u_0}^t\|\snab^{ \ l}
D^m\rho\|_{L^2(H^\prime_{t^\prime})}dt^\prime+
C^\prime\int_{u_0}^t\|\snab^{ \ l+1}D^{m-1}\omega\|_{L^2(H^\prime_{t^\prime})}dt^\prime
\label{16.154}
\end{equation}
plus a lower order term which is bounded in terms of the supremum
for $t^\prime\in(u_0,t]$ of the sum of the $L^2$ norms of the
$\snab$ and $D$ derivatives of the connection coefficients and of
the curvature components of order up to $k-1$, and the sharp
bounds for the derivatives of up to the 2nd order of the
connection coefficients and of  the curvature components already
obtained. By the inductive hypothesis the second of the integrals
\ref{16.154} has been appropriately estimated. Moreover, according
to what has previously been demonstrated, the first of the
integrals \ref{16.154} is bounded by a constant times
$\sqrt{\oE^\prime_k}(t)$ plus a lower order term which is bounded
in terms of $\sqrt{\oE^\prime_{k^\prime}(t)}$: for
$k^\prime=3,...,k-1$, the supremum for $t^\prime\in(u_0,t]$ of the
sum of the $L^2$ norms of the derivatives of the connection
coefficients of order up to $k-1$, and the sharp bounds for the
derivatives of up to the 2nd order of the connection coefficients
already obtained. The inductive step is thus complete and we have
established the required estimates for $\snab^{ \ l} D^m\omega$, for
all $l+m=k$.

The required estimates for $\snab^{ \ l}\Db^m\omb$ are deduced in a
similar manner. Indeed $\snab^{ \ k}\omb$ satisfies the conjugate of
the propagation equation \ref{16.136}:
\begin{equation}
D\snab^{ \ k}\omb=n_{k,0} \label{16.155}
\end{equation}
where
\begin{equation}
n_{k,0}\simeq
\Omega^2\left\{2(\eta,\snab^{ \ k}\etb)+2(\etb-\eta,\snab^{ \ k}\eta)-\snab^{ \ k}\rho\right\}
\label{16.156}
\end{equation}
and we denote by $\simeq$ equality up to lower order terms
involving the $\snab$ derivarives of $\chi$, $\eta$, $\etb$ and
$\rho$ of order up to $k-1$. Also $\snab^{ \ l}\Db^m\omb \ : \ l+m=k$,
for $m\geq 1$, satisfies the conjugate of the propagation equation
\ref{16.145}:
\begin{equation}
D\snab^{ \ l}\Db^m\omb=n_{l,m} \label{16.157}
\end{equation}
where
\begin{equation}
n_{l,m}\simeq
2(m+2)\Omega^2(\etb-\eta)^\sharp\cdot\snab^{ \ l+1}\Db^{m-1}\omb-\Omega^2\snab^{ \ l}\Db^m\rho
\label{16.158}
\end{equation}
and we denote by $\simeq$ equality up to lower order terms
involving the $\Db$ and $\snab$ derivatives of the connection
coefficients and of the curvature components of order up to $k-1$.

Noting that $\snab^{ \ k}\omb$ vanishes on $\Cb_0$, we apply Lemma 4.6
to the propagation equation \ref{16.155}, taking $p=2$ . Here
$r=k$, $\nu=0$, $\gamma=0$ and we obtain:
\begin{equation}
\|\snab^{ \ k}\omb\|_{L^2(S_{\ub,u})}\leq
C\int_0^{\ub}\|n_{k,0}\|_{L^2(S_{\ub^\prime,u})}d\ub^\prime
\label{16.159}
\end{equation}
Setting $\ub=\frac{1}{2}(t+\lambda)$, $u=\frac{1}{2}(t-\lambda)$,
squaring, integrating with respect to $\lambda$ on
$(-t,\lambda^*(t))$, and taking the square root, we obtain, by
\ref{16.82} and \ref{16.80},
\begin{equation}
\|\snab^{ \ k}\omb\|_{L^2(H^\prime_t)}\leq
C^\prime\int_{u_0}^t\|n_{k,0}\|_{L^2(H^\prime_{t^\prime})}dt^\prime
\label{16.160}
\end{equation}
Now, in view of \ref{16.156} the integral on the right in
\ref{16.160} is bounded by
\begin{equation}
C\int_{u_0}^t\|\snab^{ \ k}\rho\|_{L^2(H^\prime_{t^\prime})}dt^\prime+
C^\prime\int_{u_0}^t\left\{\|\snab^{ \ k}\eta\|_{L^2(H^\prime_{t^\prime})}+\|\snab^{ \ k}\etb\|_{L^2(H^\prime_{t^\prime})}\right\}dt^\prime
\label{16.161}
\end{equation}
which has already been appropriately estimated above. We have thus
shown what was required in the case of $\snab^{ \ k}\omb$.

Finally, we derive the appropriate estimates for $\snab^{ \ l}\Db^m\omb
\ : \ l+m=k$ for $m=1,...,k$ by induction on $m$. Suppose then
that $\snab^{ \ l+1}\Db^{m-1}\omb$ has been appropriately estimated.
Noting that $\snab^{ \ l}\Db^m\omb$ vanishes on $\Cb_0$, we apply Lemma
4.6 to the propagation equation \ref{16.145}, taking $p=2$. Here
$r=l$, $\nu=0$, $\gamma=0$ and we obtain:
\begin{equation}
\|\snab^{ \ l}\Db^m\omb\|_{L^2(S_{\ub,u})}\leq
C\int_0^{\ub}\|n_{l,m}\|_{L^2(S_{\ub^\prime,u})}d\ub^\prime
\label{16.162}
\end{equation}
Setting $\ub=\frac{1}{2}(t+\lambda)$, $u=\frac{1}{2}(t-\lambda)$,
squaring, integrating with respect to $\lambda$ on
$(-t,\lambda^*(t))$, and taking the square root, we obtain, by
\ref{16.82} and \ref{16.80},
\begin{equation}
\|\snab^{ \ l}\Db^m\omb\|_{L^2(H^\prime_t)}\leq
C^\prime\int_{u_0}^t\|n_{l,m}\|_{L^2(H^\prime_{t^\prime})}dt^\prime
\label{16.163}
\end{equation}
Now, in view of \ref{16.158} the integral on the right in
\ref{16.153} is bounded by
\begin{equation}
C\int_{u_0}^t\|\snab^{ \ l}\Db^m\rho\|_{L^2(H^\prime_{t^\prime})}dt^\prime+
C^\prime\int_{u_0}^t\|\snab^{ \ l+1}\Db^{m-1}\omb\|_{L^2(H^\prime_{t^\prime})}dt^\prime
\label{16.164}
\end{equation}
plus a lower order term which is bounded in terms of the supremum
for $t^\prime\in(u_0,t]$ of the sum of the $L^2$ norms of the
$\snab$ and $\Db$ derivatives of the connection coefficients and
of the curvature components of order up to $k-1$, and the sharp
bounds for the derivatives of up to the 2nd order of the
connection coefficients and of  the curvature components already
obtained. By the inductive hypothesis the second of the integrals
\ref{16.164} has been appropriately estimated. Moreover, according
to what has previously been demonstrated, the first of the
integrals \ref{16.164} is bounded by a constant times
$\sqrt{\oE^\prime_k}(t)$ plus a lower order term which is bounded
in terms of $\sqrt{\oE^\prime_{k^\prime}(t)}$: for
$k^\prime=3,...,k-1$, the supremum for $t^\prime\in(u_0,t]$ of the
sum of the $L^2$ norms of the derivatives of the connection
coefficients of order up to $k-1$, and the sharp bounds for the
derivatives of up to the 2nd order of the connection coefficients
already obtained. The inductive step is thus complete and we have
established the required estimates for $\snab^{ \ l}\Db^m\omb$, for all
$l+m=k$, as well.

Having established appropriate estimates for $\snab^{ \ k}\chi$,
$\snab^{ \ k}\chib$, $\snab^{ \ k}\eta$, $\snab^{ \ k}\etb$, and $\snab^{ \ l}
D^m\omega \ : \ l+m=k$, $\snab^{ \ l}\Db^m\omb \ : \ l+m=k$,
appropriate estimates for all $D,\Db$ and $\snab$ derivatives of
order $k$ of all connection coefficients follow in a
straightforward manner from equations \ref{1.39}, \ref{1.44},
\ref{1.66}, \ref{1.67}, \ref{1.86}, \ref{1.87}, \ref{1.146},
\ref{1.147}, \ref{1.149}, \ref{1.150}.

We are now ready to derive bounds for the energies $E^\prime_k(t)$
which are uniform in $t\in(u_0,c^*)$, for any $k\geq 3$. The
energy inequality \ref{16.35} reads, in the case of the
energy-momentum vectorfield
\begin{equation}
P_k=\sum_{m+n=k}P_{m,n} \label{16.165}
\end{equation}
(see \ref{16.52}),
\begin{equation}
E^\prime_k(t)\leq E_k(u_0)+\int_{M^\prime_t}|\tau_k|d\mu_g
\label{16.166}
\end{equation}
where
\begin{equation}
\tau_k=\sum_{m+n=k}\tau_{m,n} \label{16.167}
\end{equation}
and $\tau_{m,n}$ is given by \ref{16.57}. In view of the fact that
$d\mu_g$, the volume form of $(M^\prime_{c^*},g)$ is given in
terms of $d\mu_{\og}$, the volume form of $(H^\prime_t,\og)$ by:
\begin{equation}
d\mu_g=\Omega dt\wedge d\mu_{\og} \label{16.168}
\end{equation}
we have:
\begin{equation}
\int_{M^\prime_t}|\tau_k|d\mu_g\leq
C\int_{u_0}^t\|\tau_k\|_{L^1(H^\prime_{t^\prime})}dt^\prime
\label{16.169}
\end{equation}
Starting from the sharp bounds for up to the 2nd derivatives of
the connection coefficients and of the curvature components on
$M^\prime_{c^*}$ which we have previously obtained, we shall show
by induction that the $E^\prime_k(t)$ are uniformly bounded in
$t\in(u_0,c^*)$ for any $k\geq 3$. The inductive hypothesis for
$k\geq 4$ is that the $E^\prime_{k^\prime}(t)$ are uniformly
bounded in $t\in(u_0,c^*)$ for $k^\prime=3,...,k-1$, and there is
no hypothesis for $k=3$. {\em In establishing the inductive step
we allow the constants to depend not only on $\delta$, $u_0$ and
the ininial data quantities, but also on the uniform bounds in
$t\in(u_0,c^*)$ for the $E^\prime_{k^\prime}(t)$ for
$k^\prime=3,...,k-1$ provided by the inductive hypothesis.}

Now by the $L^\infty$ bounds on the components of
$\s^{(T)}\tilde{\pi}$ which follow from the $L^\infty$ bounds for
the connection cefficients, the multiplier part of $\tau_{m,n}$ is
bounded in $L^1(H^\prime_t)$ by:
\begin{equation}
CE^\prime_{m,n}(t) \label{16.170}
\end{equation}
while the Weyl current part of $\tau_{m,n}$, given by \ref{16.58}
is bounded in $L^1(H^\prime_t)$ by:
\begin{equation}
C\sqrt{E^\prime_{m,n}(t)}\|J_{m,n}\|_{L^2(H^\prime_t)}
\label{16.171}
\end{equation}
where we denote by $\|J_{m,n}\|_{L^2(H^\prime_t)}$ the square root
of the sum of the squares of the $L^2$ norms on $H^\prime_t$ of
the components of $J_{m,n}$ (in the null decomposition). According
to the discussion following the recursion relations \ref{16.59},
\ref{16.60}, the leading terms in the expression for any component
of $J_{m,n}$ are of four types. All four types are bilinear
expressions, with coefficients depending only on $\sg$ and
$\seps$, the first factor of which is a derivative of a connection
coefficient of order $k_1$, and the second factor a derivative of
a curvature component of order $k_2$, where $k_1+k_2=k$. In the
case $k_1=0$, $k_2=k$, we place the first factor in
$L^\infty(H^\prime_t)$ using the results of Chapter 3, and the
second factor in $L^2(H^\prime_t)$ the resulting quantity being
bounded in terms of $\sqrt{E^\prime_k(t)}$. In the case $k_1=k$,
$k_2=0$ we place the first factor in $L^2(H^\prime_t)$ using the
estimates, just obtained,  for the $k$th order derivatives of the
connection coefficients in $L^2(H^\prime_t)$ in terms of
$\sqrt{\oE^\prime_k(t)}$, and the second factor in
$L^\infty(H^\prime_t)$. In the case $k_1=1$, $k_2=k-1$, we place
the first factor in
$L^4(S_{\frac{1}{2}(t+\lambda),\frac{1}{2}(t-\lambda)})$ using the
results of Chapter 4, and the second factor also in
$L^4(S_{\frac{1}{2}(t+\lambda),\frac{1}{2}(t-\lambda)})$ the
resulting quantity being bounded in terms of
$\sqrt{E^\prime_k(t)}$ by virtue of the Sobolev inequality on
$H^\prime_t$ (the analogue of the second statement of Proposition
10.1). In the case $k_1=k-1$, $k_2=1$, we place the first factor
in $L^4(S_{\frac{1}{2}(t+\lambda),\frac{1}{2}(t-\lambda)})$ the
resulting quantity being bounded in terms of the $L^2$ norm on
$H^\prime_t$ of the $k$th order derivatives of the connection
coefficients (by virtue of the Sobolev inequality just mentioned),
which is in turn bounded in terms of $\sqrt{\oE^\prime_k(t)}$, and
the second factor also in
$L^4(S_{\frac{1}{2}(t+\lambda),\frac{1}{2}(t-\lambda)})$. Finally,
in the remaining cases $2\leq k_1,k_2\leq k-2$ (which exist only
for $k\geq 4$) we may place both factors in
$L^4(S_{\frac{1}{2}(t+\lambda),\frac{1}{2}(t-\lambda)})$, the
resulting quantities being bounded by constants in the above
sense. The contributions of lower order terms in $J_{m,n}$ to
$\|J_{m,n}\|_{L^2(H^\prime_t)}$ are similarly bounded by constants
in the above sense. We thus arrive at the following conclusion:
\begin{equation}
\|J_{m,n}\|_{L^2(H^\prime_t)}\leq C\sqrt{\oE^\prime_k(t)}+C^\prime
\label{16.172}
\end{equation}
Substituting in \ref{16.171} and combining with \ref{16.170} we
conclude that:
\begin{equation}
\|\tau_k\|_{L^1(H^\prime_t)}\leq C\oE^\prime_k(t) +C^\prime
\label{16.173}
\end{equation}
(for different constants $C$ and $C^\prime$). Substituting in
\ref{16.169} and \ref{16.166} then yields:
\begin{equation}
E^\prime_k(t)\leq
E_k(u_0)+C\int_{u_0}^t\oE^\prime_k(t^\prime)dt^\prime+C^\prime \ \
: \ \forall t\in(u_0,c^*) \label{16.174}
\end{equation}
Since the right hand side is a nondecreasing function of $t$, this
implies that also:
\begin{equation}
E^\prime_k(t^\prime)\leq
E_k(u_0)+C\int_{u_0}^t\oE^\prime(t^\prime)dt^\prime+C^\prime \ \ :
\ \forall t^\prime\in(u_0,t],  \ \forall t\in(u_0,c^*)
\label{16.175}
\end{equation}
Taking the supremum with respect to $t^\prime\in(u_0,t]$ then
yields the following {\em linear} integral inequality for
$\oE^\prime_k$:
\begin{equation}
\oE^\prime_k(t)\leq
E_k(u_0)+C\int_{u_0}^t\oE^\prime_k(t^\prime)dt^\prime+C^\prime
\label{16.176}
\end{equation}
This implies:
\begin{equation}
\oE^\prime_k(t)\leq CE_k(u_0)+C^\prime \label{16.177}
\end{equation}
(for different constants $C$ and $C^\prime$), which shows that
$E^\prime_k$ is uniformly bounded on $(u_0,c^*)$ completing the
inductive step.

Having establised that the $E^\prime_k$ are uniformly bounded in
$(u_0,c^*)$ for all $k$, it follows that the $L^2$ norms on
$H^\prime_t$ of all the ($D,\Db$ and $\snab$) derivatives of any
order of all the curvature components and of all the connection
coefficients are uniformly bounded in $t\in(u_0,c^*)$. By the
Sobolev inequality on the $H^\prime_t$ (the analogue of the second
statement of Proposition 10.1), it then follows that the $L^4$
norms on $S_{\ub,u}$ of all the derivatives of any order of all
the curvature components and of all the connection coefficients
are uniformly bounded in $(\ub,u)\in D^\prime_{c^*}$. Then by
Lemma 5.2 with $p=4$ all the derivatives of any order of all the
curvature components and of all the connection coefficients  are
uniformly bounded in $M^\prime_{c^*}$.

\section{Completion of the continuity argument}

In this section we shall show that the solution extends as a
smooth solution to a larger domain $M_c$ for some $c>c^*$ and the
three conditions in the statement of Theorem 12.1 hold for the
extended solution as well, thereby contradicting the definition of
$c^*$ unless $c^*=-1$. The proof of Theorem 12.1 will then be
complete.

It follows from what has just been established that the metric
components in a canonical coordinate system extend smoothly to the
future boundary $H_{c^*}$ of $M_{c^*}$ as do the connection
coefficients and curvature components. Thus $H_c^*$ is endowed
with a smooth induced metric $\og$ (see \ref{16.26}) and a smooth
2nd fundamental form $k$. We denote
\begin{equation}
\oM_{c^*}=M_{c^*}\bigcup H_{c^*}, \ \ \ \ \
\oM^\prime_{c^*}=M^\prime_{c^*}\bigcup H^\prime_{c^*}
\label{16.178}
\end{equation}
and:
\begin{equation}
\lambda^*(c^*)=\sup_{H^\prime_{c^*}}\lambda=\left\{
\begin{array}{lll}
2\delta-c^*&:&\mbox{for $c^*\in(u_0+\delta, -1)$}\nonumber\\
c^*-2u_0&:&\mbox{for $c^*\in(u_0,u_0+\delta]$}
\end{array}
\right. \label{16.a2}
\end{equation}
(see \ref{16.24}, \ref{16.61}). In the Minkowskian region $M_0$,
which corresponds to $\ub\leq 0$, $\og$ coincides with the
Euclidean metric $e$ in the same $(\lambda,\vartheta^A:A=1,2)$
coordinates and $k$ vanishes. The 2nd fundamental form of the
$H_t$, in particular of $H_{c^*}$, is given by:
\begin{equation}
k(X,Y)=g(\nabla_X\Th,Y) \label{16.179}
\end{equation}
where $X,Y$ is any pair of vectors tangent to $H_t$ at a point.
Let us supplement the frame field
$(e_A=\partial/\partial\vartheta^A:A=1,2)$ for the
$S_{\frac{1}{2}(t+\lambda),\frac{1}{2}(t-\lambda)}$ with the
vectorfield
\begin{equation}
\Nh=\Omega^{-1}N \label{16.180}
\end{equation}
the outward unit normal to the
$S_{\frac{1}{2}(t+\lambda),\frac{1}{2}(t-\lambda)}$, to a frame
field for $H_t$. From \ref{16.108} and \ref{1.171} we have:
\begin{equation}
\Nh=\frac{1}{2}(e_4-e_3)=\Omega^{-1}\left(\frac{\partial}{\partial\lambda}+\frac{1}{2}b^A\frac{\partial}{\partial\vartheta^A}\right)
\label{16.181}
\end{equation}
From \ref{16.28}, \ref{16.30} and the table \ref{1.151} we then
find, for the components of $k$ in the frame $(\Nh,e_A:A=1,2)$:
\begin{eqnarray}
&&k(\Nh,\Nh)=\frac{1}{2}\Omega^{-1}(\omega+\omb)=\Th\log\Omega\nonumber\\
&&k(e_A,\Nh)=-\zeta_A\nonumber\\
&&k(e_A,e_B)=\frac{1}{2}(\chi_{AB}+\chib_{AB}) \label{16.182}
\end{eqnarray}
Therefore $k$ is given in canonical coordinates
$(\lambda,\vartheta^A:A=1,2)$ on $H_t$ by:
\begin{eqnarray}
&&k=\frac{1}{2}\Omega(\omega+\omb)d\lambda\otimes d\lambda\nonumber\\
&&\hspace{5mm}-\Omega\zeta_A(d\vartheta^A-\frac{1}{2}b^A
d\lambda)\otimes d\lambda
-\Omega\zeta_A d\lambda\otimes(d\vartheta^A-\frac{1}{2}b^A d\lambda)\nonumber\\
&&\hspace{5mm}+\frac{1}{2}(\chi_{AB}+\chib_{AB})(d\vartheta^A-\frac{1}{2}b^A
d\lambda)\otimes(d\vartheta^B-\frac{1}{2}b^B d\lambda)
\label{16.183}
\end{eqnarray}

Now by virtue of the definition of a canonical coordinate system,
the metric $\sg_{AB}(\ub,u)d\vartheta^A\otimes d\vartheta^B$  is
precisely the pullback metric
$(\Phib_{u-u_0}\circ\Phi_{\ub})^*\left.\sg\right|_{S_{\ub,u}}$ on
$S_{0,u_0}$, while the metric
$\left.\sg\right|_{S_{0,u_0}}=|u_0|^2\up{\sg}$ where $\up{\sg}$ is
the standard metric on $S^2$. Then by Lemma 11.1, $\lambda(\ub,u)$
and $\Lambda(\ub,u)$, respectively the smallest and largest
eigenvalues of $|u|^{-2}\sg_{AB}(\ub,u)$ relative to
$\up{\sg}_{AB}$, satisfy:
\begin{equation}
\frac{1}{4}\leq\lambda(\ub,u)\leq\Lambda(\ub,u)\leq 4
\label{16.184}
\end{equation}
for all $(\ub,u)\in D^\prime_{c^*}$, hence by continuity also for
all $(\ub,u)\in \oD^\prime_{c^*}$, where
\begin{equation}
\oD^\prime_{c^*}=D^\prime_{c^*}\bigcup\{(\ub,u) \ : \ \ub+u=c^*, \
\ub\in[0,\ub^*(c^*))\} \label{16.185}
\end{equation}
where for any $t\in(u_0,c^*]$ we denote:
\begin{equation}
\ub^*(t)=\sup_{H_t}\ub=\left\{
\begin{array}{lll}
\delta&:&\mbox{for $t\in(u_0+\delta,c^*)$}\\
t-u_0&:&\mbox{for $t\in(u_0,u_0+\delta]$}
\end{array}\right.
\label{16.186}
\end{equation}
The left inequality in \ref{16.183} implies that the null
hypersurfaces $\Cb_{\ub}$ do not contain focal points on
$H^\prime_{c^*}$, the future boundary of $M^\prime_{c^*}$, either.
Let now $\lambda^\prime(\ub,u)$ and $\Lambda^\prime(\ub,u)$ be
respectively the smallest and largest eigevalues of the pullback
metric $\Phi^*_{\ub}\left.\sg\right|_{S_{\ub,u}}$ on $S_{0,u}$
with respect to the metric $\left.\sg\right|_{S_{0,u}}
=|u|^2\up{\sg}$. Then by Lemma 5.3, taking $\delta$ suitably
small, we have:
\begin{equation}
\lambda^\prime(\ub,u)\geq\frac{1}{2}, \ \ \
\Lambda^\prime(\ub,u)\leq 2 \label{16.187}
\end{equation}
for all $(\ub,u)\in D^\prime_{c^*}$, hence by continuity also for
all $(\ub,u)\in \oD^\prime_{c^*}$. The left inequality implies
that the null hypersurfaces $C_u$ do not contain conjugate points
on $H^\prime_{c^*}$ either.

Now the induced metric $\og$ on $H^\prime_{c^*}$, given by
\ref{16.26}, is equivalent to the Euclidean metric $e$ on
$H^\prime_{c^*}$:
\begin{equation}
e=d\lambda\otimes
d\lambda+\lambda^2\up{\sg}_{AB}d\vartheta^A\otimes d\vartheta^B
\label{16.188}
\end{equation}
and, as mentioned above, $\og=e$ on $H_{c^*}\bigcap M_0$. It
follows that the distance relative to $e$ between any pair of
points on $H_{c^*}$ is bounded from above by a constant multiple
of the distance between the same points relative to $\og$, and
conversely. Hence, if $p^\prime$ and $p^{\prime\prime}$ are points
on $S_{\ub,c^*-\ub}$, $\ub\in[0,\ub^*(c^*))$, which correspond to
distinct points $\vartheta^\prime$ and $\vartheta^{\prime\prime}$
on $S^2$, then their distance relative to $\og$ is positive. As a
consequence, the null hypersurfaces $\Cb_{\ub}$ do not contain cut
points on $H^\prime_{c^*}$, the future boundary of
$M^\prime_{c^*}$ either. The generators of the $C_u$ are the
integral curves of $L$, which is given in canonical coordinates by
\ref{1.171}. Thus, the generator of $C_u$ initiating at a point
$q\in S_{0,u}$ corresponding to the point $\vartheta_0\in S^2$
represented by the coordinates $(\vartheta^1_0,\vartheta^2_0)$, is
represented in canonical coordinates by:
$$\ub\mapsto (\ub,u,\vartheta(\ub;\vartheta_0))$$
where $\ub\mapsto\vartheta(\ub;\vartheta_0)$ is the solution of:
\begin{equation}
\frac{d\vartheta^A}{d\ub}=b^A(\ub,u,\vartheta), \ \ \
\vartheta^A(0;\vartheta)=\vartheta^A_0 \label{16.189}
\end{equation}
a non-autonomous system of ordinary differential equations on
$S^2$. The vectorfield
\begin{equation}
b=b^A\frac{\partial}{\partial\vartheta^A} \label{16.190}
\end{equation}
being smooth on $\oM_{c^*}$, it follows that
$\vartheta_0^\prime\neq\vartheta_0^{\prime\prime}$ implies
$\vartheta(\ub;\vartheta_0^\prime)\neq\vartheta(\ub;\vartheta_0^{\prime\prime})$
for all $\ub\in [0,c^*-u]$, if $u>c^*-\delta$, for all $\ub\in
[0,\delta)$, if $u\leq c^*-\delta$, in the case $c^*\geq u_0+\delta$, 
for all $\ub\in[0,c^*-u]$, in the case $c^*<u_0+\delta$. Consequently, the null
hypersurfaces $C_u$ do not contain cut points on $H^\prime_{c^*}$
either. Moreover, $H^\prime_{c^*}$ is smoothly foliated by the
$\{S_{\frac{1}{2}(c^*+\lambda),\frac{1}{2}(c^*-\lambda)} \ : \
\lambda\in (-c^*,\lambda^*(c^*))\}$, the lapse function $\Omega$
(see \ref{16.180}) being bounded above and below by positive
constants.

We now define ``Cartesian" coordinates $x^1,x^2,x^3$ on $H_{c^*}$
as follows. In the Minkowskian region $M_0$ these coincide with
the Cartesian coordinates $x^1,x^2,x^3$ previously defined. To
extend these coordinates to $H^\prime_{c^*}$ we simply stipulate
that they are given in terms of the ``polar" coordinates
$\lambda,\vartheta^1,\vartheta^2$ by the same formulas as in the
Minkowskian region $M_0$. That is, the $x^i:i=1,2,3$ are given in
the north polar chart by:
\begin{equation}
x^1=\frac{\lambda\vartheta^1}{1+\frac{1}{4}|\vartheta|^2}, \ \ \
x^2=\frac{\lambda\vartheta^2}{1+\frac{1}{4}|\vartheta|^2}, \ \ \
x^3=\lambda\left(\frac{1-\frac{1}{4}|\vartheta|^2}{1+\frac{1}{4}|\vartheta|^2}\right)
\label{16.191}
\end{equation}
and in the south polar chart by:
\begin{equation}
x^1=\frac{\lambda\vartheta^1}{1+\frac{1}{4}|\vartheta|^2}, \ \ \
x^2=\frac{\lambda\vartheta^2}{1+\frac{1}{4}|\vartheta|^2}, \ \ \
x^3=-\lambda\left(\frac{1-\frac{1}{4}|\vartheta|^2}{1+\frac{1}{4}|\vartheta|^2}\right)
\label{16.192}
\end{equation}
the inverse of the transformation \ref{16.191} being:
\begin{equation}
\lambda=|x|, \ \ \ \vartheta^1=\frac{2x^1}{|x|+x^3}, \ \ \
\vartheta^2=\frac{2x^2}{|x|+x^3} \label{16.193}
\end{equation}
and the inverse of the transformation \ref{16.192} being:
\begin{equation}
\lambda=|x|, \ \ \ \vartheta^1=\frac{2x^1}{|x|-x^3}, \ \ \
\vartheta^2=\frac{2x^2}{|x|-x^3} \label{16.194}
\end{equation}
(see \ref{1.183}, \ref{1.184}) where
$$|x|=\sqrt{(x^1)^2+(x^2)^2+(x^3)^2}$$
In the north polar chart we have:
\begin{eqnarray}
&&\frac{\partial\lambda}{\partial x^A}=\frac{\vartheta^A}{1+\frac{1}{4}|\vartheta|^2} \ : \ A=1,2 \ \nonumber\\
&&\frac{\partial\lambda}{\partial x^3}=\frac{1-\frac{1}{4}|\vartheta|^2}{1+\frac{1}{4}|\vartheta|^2}\nonumber\\
&&\lambda\frac{\partial\vartheta^A}{\partial
x^B}=\left(1+\frac{1}{4}|\vartheta|^2\right)\delta_{AB}-\frac{1}{2}\vartheta^A\vartheta^B
\ : \ A,B=1,2\nonumber\\
&&\lambda\frac{\partial\vartheta^A}{\partial x^3}=-\vartheta^A \ :
\ A=1,2 \label{16.195}
\end{eqnarray}
and in the south polar chart we have:
\begin{eqnarray}
&&\frac{\partial\lambda}{\partial x^A}=\frac{\vartheta^A}{1+\frac{1}{4}|\vartheta|^2} \ : \ A=1,2 \ \nonumber\\
&&\frac{\partial\lambda}{\partial x^3}=-\frac{1-\frac{1}{4}|\vartheta|^2}{1+\frac{1}{4}|\vartheta|^2}\nonumber\\
&&\lambda\frac{\partial\vartheta^A}{\partial
x^B}=\left(1+\frac{1}{4}|\vartheta|^2\right)\delta_{AB}-\frac{1}{2}\vartheta^A\vartheta^B
\ : \ A,B=1,2\nonumber\\
&&\lambda\frac{\partial\vartheta^A}{\partial x^3}=\vartheta^A \ :
\ A=1,2 \label{16.196}
\end{eqnarray}
Note that the right hand sides of \ref{16.195}, \ref{16.196} are
analytic functions of $\vartheta=(\vartheta^1,\vartheta^2)$ alone.

Let us write:
\begin{equation}
\og=e+h \label{16.197}
\end{equation}
The components $e_{ij}$ of the Euclidean metric in Cartesian
coordinates being simply $\delta_{ij}$, we have:
\begin{equation}
e_{ij}=\frac{\partial\lambda}{\partial
x^i}\frac{\partial\lambda}{\partial x^j}+\lambda^2\up{\sg}_{AB}
\frac{\partial\vartheta^A}{\partial
x^i}\frac{\partial\vartheta^B}{\partial x^j}=\delta_{ij}
\label{16.198}
\end{equation}
Therefore, the components $\og_{ij}$ of the metric $\og$ in
Cartesian coordinates are given by:
\begin{equation}
\og_{ij}=\delta_{ij}+h_{ij} \label{16.199}
\end{equation}
Here $h_{ij}$ are the components of the tensorfield $h$ in
Cartesian coordinates:
\begin{eqnarray}
&&h_{ij}=\left(\Omega^2-1+\frac{1}{4}|b|^2\right)\frac{\partial\lambda}{\partial x^i}\frac{\partial\lambda}{\partial x^j}\nonumber\\
&&\hspace{8mm}-\frac{1}{2}b_A\left(\frac{\partial\lambda}{\partial
x^i}\frac{\partial\vartheta^A}{\partial x^j}
+\frac{\partial\vartheta^A}{\partial x^i}\frac{\partial\lambda}{\partial x^j}\right)\nonumber\\
&&\hspace{8mm}+(\sg_{AB}-\lambda^2\up{\sg}_{AB})\frac{\partial\vartheta^A}{\partial
x^i}\frac{\partial\vartheta^B}{\partial x^j} \label{16.200}
\end{eqnarray}
where:
\begin{equation}
b_A=\sg_{AB}b^B \label{16.201}
\end{equation}

Letting again $\lambda(\ub,u)$ and $\Lambda(\ub,u)$ be
respectively the smallest and largest eigenvalues of
$|u|^{-2}\sg_{AB}(\ub,u)$ relative to $\up{\sg}_{AB}$, following
the proof of Lemma 11.1, and using the estimates of Chapter 3 in
place of the bootstrap assumptions {\bf A1.1}, {\bf A1.2}, {\bf
A3.1}, {\bf A3.2}, we now obtain:
\begin{equation}
\lambda(\ub,u)\geq 1-O(\delta^{1/2}|u|^{-1}), \ \ \ \ \
\Lambda(\ub,u)\leq 1+O(\delta^{1/2}|u|^{-1}) \label{16.202}
\end{equation}
It follows that:
\begin{eqnarray}
&&||u|^{-2}\sg(\ub,u)-\up{\sg}|_{\up{\sg}}=\sqrt{(\lambda(\ub,u)-1)^2+(\Lambda(\ub,u)-1)^2}\nonumber\\
&&\hspace{30mm}\leq O(\delta^{1/2}|u|^{-1}) \ \ : \
\forall(\ub,u)\in\oD^\prime_{c^*} \label{16.203}
\end{eqnarray}
Also, the $L^\infty$ bound on $\zeta$ from Chapter 3 yields
through equation \ref{1.173} the following $L^\infty$ bound on
$b$:
\begin{equation}
\|b\|_{L^\infty(S_{\ub,u})}\leq O(\delta^{1/2})|u|^{-1}) \ \ : \
\forall(\ub,u)\in\oD^\prime_{c^*} \label{16.204}
\end{equation}
The bounds \ref{16.203}, \ref{16.204}, together with the bound on
$\log\Omega$ for Chapter 3, yield through \ref{16.200}, in view of
the formulas \ref{16.195}, \ref{16.196}, the following bound for
the functions $h_{ij}$ on $H^\prime_{c^*}$:
\begin{equation}
\sup_{H^\prime_{c^*}}|h_{ij}|\leq O(\delta^{1/2}|c^*|^{-1}) \ : \
i,j=1,2,3 \label{16.205}
\end{equation}

The components of the tensorfield $k$ in Cartesian coordinates are
given by:
\begin{eqnarray}
&&k_{ij}=\frac{1}{2}\Omega(\omega+\omb)\frac{\partial\lambda}{\partial x^i}\frac{\partial\lambda}{\partial x^j}
\label{16.206}\\
&&\hspace{8mm}-\Omega\zeta_A\left(\frac{\partial\vartheta^A}{\partial
x^i}-\frac{1}{2}b^A\frac{\partial\lambda}{\partial x^i}\right)
\frac{\partial\lambda}{\partial x^j}\nonumber\\
&&\hspace{8mm}-\Omega\zeta_A\frac{\partial\lambda}{\partial x^i}\left(\frac{\partial\vartheta^A}{\partial x^j}-\frac{1}{2}b^A\frac{\partial\lambda}{\partial x^j}\right)\nonumber\\
&&\hspace{8mm}+\frac{1}{2}(\chi_{AB}+\chib_{AB})
\left(\frac{\partial\vartheta^A}{\partial
x^i}-\frac{1}{2}b^A\frac{\partial\lambda}{\partial x^i}\right)
\left(\frac{\partial\vartheta^A}{\partial
x^j}-\frac{1}{2}b^A\frac{\partial\lambda}{\partial x^j}\right)\nonumber
\end{eqnarray}
The estimates of Chapter 3 together with the bound \ref{16.204},
imply, in view of the formulas \ref{16.195}, \ref{16.196}, the
following bound for the functions $k_{ij}$ on $H^\prime_{c^*}$:
\begin{equation}
\sup_{H^\prime_{c^*}}|k_{ij}|\leq O(\delta^{-1/2}|c^*|^{-1}) \ : \
i,j=1,2,3 \label{16.207}
\end{equation}

The functions $h_{ij}$ and $k_{ij}$, considered as functions of
the Cartesian coordinates $x^1,x^2,x^3$, are smooth functions on
the open ball $B_{\lambda^*(c^*)}$ of radius $\lambda^*(c^*)$ in
$\Re^3$ extending to smooth functions on the closed ball
$\oB_{\lambda^*(c^*)}$, their derivatives of all orders being
bounded. We apply to these functions a radial {\em total extension
operator} $E$ for $B_{\lambda^*(c^*)}$. This yields smooth
functions $Eh_{ij}$ and $Ek_{ij}$ on the closed ball
$\oB_{\lambda^*(c^*)+1}$, and there is a numerical constant $C$
such that:
\begin{equation}
\sup_{\oB_{\lambda^*(c^*)+1}}|Eh_{ij}|\leq
C\sup_{B_{\lambda^*(c^*)}}|h_{ij}|, \ \ \ \ \
\sup_{\oB_{\lambda^*(c^*)+1}}|Ek_{ij}|\leq
C\sup_{B_{\lambda^*(c^*)}}|k_{ij}| \label{16.208}
\end{equation}
In particular, the bound \ref{16.205} implies:
\begin{equation}
\sup_{\oB_{\lambda^*(c^*)+1}}|Eh_{ij}|\leq O(\delta^{1/2})
\label{16.209}
\end{equation}
hence, if $\delta$ is suitably small (depending on the quantities
$\stackrel{(n)}{D} \ : \ n=0,1,2,3$; ${\cal D}^{\prime
4}_{[1]}(\alb)$  and the quantities ${\cal D}_0^\infty$,
$\scD_1^4$, $\scD_2^4(\mbox{tr}\chib)$, $\scD_3(\mbox{tr}\chib)$)
the tensorfield with components:
\begin{equation}
\delta_{ij}+Eh_{ij} \label{16.210}
\end{equation}
defines a positive-definite metric on $\oB_{\lambda^*(c^*)+1}$.
From now on we denote this metric simply by $\og_{ij}$ and we
denote $Ek_{ij}$ simply by $k_{ij}$.

We shall now construct a local future development
$V^\prime_{\varepsilon_0}$ of the initial data $(\og,k)$ on
$H_{c^*}$. This shall be done by the classical method of
Choquet-Bruhat (see [Cho1], [Cho2], [Cho3]), which uses ``wave" coordinates (also called
``harmonic" coordinates) adapted to the initial spacelike
hypersurface. We shall presently briefly review this method. The
wave coordinates, which we denote by $x^\mu:\mu=0,1,2,3$ are
functions $\phi$ on $V^\prime_{\varepsilon_0}$ which are solutions
of the wave equation
\begin{equation}
\triangle\phi=0 \label{16.211}
\end{equation}
on $V^\prime_{\varepsilon_0}$. Such a solution is determined by
its initial data on $H_{c^*}$ which is $\phi$ together with
$\Th\phi$ on $H_{c^*}$. Here, we set:
\begin{equation}
x^0=c^* \ : \ \mbox{on} \ H_{c^*}, \ \ \ x^i=x^i:i=1,2,3 \ \mbox{:
the given Cartesian coordinates on} \ H_{c^*} \label{16.212}
\end{equation}
and:
\begin{equation}
\Th x^0=1 \ \mbox{: on $H_{c^*}$}, \ \ \ \Th x^i=0:i=1,2,3 \
\mbox{: on $H_{c^*}$} \label{16.213}
\end{equation}
These conditions state that the coordinate system is Gaussian
normal along $H_{c^*}$. For, they state that:
\begin{equation}
\Th=\frac{\partial}{\partial x^0} \ \mbox{: along $H_{c^*}$}
\label{16.214}
\end{equation}
Since
$$g(\Th,\Th)=-1, \ \ \ g\left(\Th,\frac{\partial}{\partial x^i}\right)=0$$
the vectorfields $\partial/\partial x^i$ being by \ref{16.212}
tangential to $H_{c^*}$, we have:
\begin{equation}
g_{00}=-1, \ \ \ g_{0i}=0:i=1,2,3 \ \ \mbox{: along $H_{c^*}$}
\label{16.215}
\end{equation}
Also:
\begin{equation}
g_{ij}=\og_{ij}:i,j=1,2,3 \ \ \mbox{: along $H_{c^*}$}
\label{16.216}
\end{equation}
By the domain of dependence theorem, the domain of dependence of
$H_{c^*}\bigcap M_0$ in $V^\prime_{\varepsilon_0}$ shall be the
Minkowskian region $V^\prime_{\varepsilon_0}\bigcap M_0$, the
coordinates $x^\mu:\mu=0,1,2,3$ coinciding with the original
Cartesian coordinates in $M_0$.

The wave equation \ref{16.211} expressed in an arbitrary
coordinate system reads:
\begin{equation}
\triangle\phi=(g^{-1})^{\alpha\beta}\partial_\alpha\partial_\beta\phi-\Gamma^\mu\partial_\mu\phi=0
\ \ \mbox{where} \ \
\Gamma^\mu=(g^{-1})^{\alpha\beta}\Gamma^\mu_{\alpha\beta}
\label{16.217}
\end{equation}
Thus the condition \ref{16.211} imposed on the coordinate
functions $x^\mu$ is equivalent to:
\begin{equation}
\Gamma^\mu=0 \label{16.218}
\end{equation}

In view of \ref{16.215}, along $H_{c^*}$ we have:
\begin{equation}
\partial_0 g_{ij}=2k_{ij}:i,j=1,2,3 \ \ \mbox{: along $H_{c^*}$}
\label{16.219}
\end{equation}
The $\partial_0 g_{0i}:i=1,2,3$ and $\partial_0 g_{00}$ along
$H_{c^*}$ are chosen so as to satisfy the conditions:
\begin{equation}
\Gamma^\mu=0 \ \ \mbox{: along $H_{c^*}$} \label{16.220}
\end{equation}
A short calculation shows that:
\begin{equation}
\partial_0 g_{0i}=\overline{\Gamma}_i:i=1,2,3 \ \ \ \partial_0 g_{00}=2\mbox{tr}k \ \ \mbox{: along $H_{c^*}$}
\label{16.221}
\end{equation}
where $\overline{\Gamma}_i=\og_{ij}\overline{\Gamma}^j$ and
$\overline{\Gamma}^j=(\og^{-1})^{mn}\overline{\Gamma}^j_{mn}$ are
the corresponding 3-dimensional functions for the induced metric
$\og_{ij}$. Also, $\mbox{tr}k=(\og^{-1})^{mn}k_{mn}$. Let us
denote by $H_{c^*,1}$ the closed ball $\oB_{\lambda^*(c^*)+1}$ on
the coordinate hyperplane $x^0=c^*$ where the extended functions
$\og_{ij}$ and $k_{ij}$ are defined. We then define $g_{\mu\nu}$
and $\partial_0 g_{\mu\nu}$, $\mu,\nu=0,1,2,3$, on $H_{c^*,1}$, so
as to satisfy \ref{16.215}, \ref{16.216}, \ref{16.219}, and
\ref{16.221}, relative to the extended functions $\og_{ij}$ and
$k_{ij}$.

The Ricci curvature components in an arbitrary coordinate system
are given by:
\begin{equation}
Ric_{\mu\nu}=H_{\mu\nu}+\frac{1}{2}S_{\mu\nu} \label{16.222}
\end{equation}
Here:
\begin{equation}
H_{\mu\nu}=-\frac{1}{2}(g^{-1})^{\alpha\beta}\partial_\alpha\partial_\beta
g_{\mu\nu}+B^{\alpha\beta\kappa\lambda\rho\sigma}_{\mu\nu}
\partial_\alpha g_{\kappa\lambda}\partial_\beta g_{\rho\sigma}
\label{16.223}
\end{equation}
where $B$ is a homogeneous rational function of the metric
components. Also:
\begin{equation}
S_{\mu\nu}=\partial_\mu\Gamma_\nu+\partial_\nu\Gamma_\mu, \ \ \
\Gamma_\mu=g_{\mu\nu}\Gamma^\nu \label{16.224}
\end{equation}
Thus in wave coordinates the vaccuum Einstein equations reduce to
the quasilinear system of wave equations:
\begin{equation}
H_{\mu\nu}=0 \label{16.225}
\end{equation}
for the metric components, which are called {\em reduced
equations}. Let us denote:
\begin{equation}
\tilde{Ric}_{\mu\nu}=Ric_{\mu\nu}-\frac{1}{2}g_{\mu\nu}\mbox{tr}Ric
\label{16.226}
\end{equation}
$\mbox{tr}Ric$ being the scalar curvature:
\begin{equation}
\mbox{tr}Ric=(g^{-1})^{\alpha\beta}Ric_{\alpha\beta}
\label{16.227}
\end{equation}
Similarly, we denote:
\begin{equation}
\tilde{S}_{\mu\nu}=S_{\mu\nu}-\frac{1}{2}g_{\mu\nu}\mbox{tr}S, \ \
\ \mbox{tr}S=(g^{-1})^{\alpha\beta}S_{\alpha\beta} \label{16.228}
\end{equation}
The twice contracted Bianchi identities
\begin{equation}
\nabla^\nu\tilde{Ric}_{\mu\nu}=0 \label{16.229}
\end{equation}
imply that every solution of the reduced equations satisfies:
\begin{equation}
\nabla^\nu\tilde{S}_{\mu\nu}=0 \label{16.230}
\end{equation}
The equations \ref{16.230} constitute a homogeneous linear system
of wave equations for the $\Gamma_\mu$:
\begin{equation}
(g^{-1})^{\alpha\beta}\partial_\alpha\partial_\beta\Gamma_\mu+A^{\alpha\beta}_\mu\partial_\alpha\Gamma_\beta=0
\label{16.231}
\end{equation}
where $A$ is a linear form in the partial derivatives of the
metric components, with coefficients which are homogeneous
rational functions of the metric components. Moreover, the
constraint equations:
\begin{equation}
\tilde{Ric}_{0i}=0:i=1,2,3 \ \ \ \tilde{Ric}_{00}=0 \ \ \mbox{:
along $H_{c^*}$} \label{16.232}
\end{equation}
imply that every solution of the reduced equations which satisfies
\ref{16.220} along $H_{c^*}$ also satisfies:
\begin{equation}
\partial_0\Gamma_\mu=0 \ \ \mbox{: along $H_{c^*}$}
\label{16.233}
\end{equation}
Now, the constraint equations \ref{16.232} are the contracted
Codazzi and twice contracted Gauss equations of the embedding of
$H_{c^*}$ in spacetime:
\begin{eqnarray}
&&\overline{\nabla}^j k_{ij}-\partial_i\mbox{tr}k=0\nonumber\\
&&\mbox{tr}\overline{Ric}-|k|^2+(\mbox{tr}k)^2=0 \label{16.234}
\end{eqnarray}
Here $\overline{\nabla}$ is the covariant derivative operator on
$H_{c^*}$ associated to $\og$, $\overline{Ric}$ is the Ricci
curvature of $\og$, and
$\mbox{tr}\overline{Ric}=(\og^{-1})^{mn}\overline{Ric}_{mn}$ the
scalar curvature of $\og$. Also, $|k|$ is the magnitude of $k$
with respect to $\og$:
$|k|=\sqrt{(\og^{-1})^{mi}(\og^{-1})^{nj}k_{mn}k_{ij}}$. The
constraint equations \ref{16.234} are satisfied along $H_{c^*}$ by
virtue of the fact that we have a smooth solution of the vacuum
Einstein equations on $M^\prime_{c^*}$ which smoothly extends to
its future boundary $H_{c^*}$.

Let then $(V_{\varepsilon_0},g)$ be a {\em development} of the
extended initial data $(g_{\mu\nu},\partial_0 g_{\mu\nu})$ on
$H_{c^*,1}$, that is, a domain $V_{\varepsilon_0}$, with past
boundary $H_{c^*,1}$ and a solution $g_{\mu\nu}$ of the {\em
reduced equations} \ref{16.225} on $V_{\varepsilon_0}$, taking the
given initial data along $H_{c^*,1}$, such that for each point
$x\in V_{\varepsilon_0}$ each past directed causal curve (with
respect to $g_{\mu\nu}$) initiating at $x$ terminates at a point
of $H_{c^*,1}$. Then equations \ref{16.231} are satisfied on
$V_{\varepsilon_0}$ and conditions \ref{16.220} and \ref{16.233}
are satisfied along $H_{c^*}$. It follows that the $\Gamma_\mu$
vanish on $V^\prime_{\varepsilon_0}$, the {\em domain of
dependence} of $H_{c^*}$ in $(V_{\varepsilon_0},g)$. Consequently,
the restriction of $g_{\mu\nu}$ to $V^\prime_{\varepsilon_0}$ is a
solution of the vacuum Einstein equations on
$V^\prime_{\varepsilon_0}$.

Let $f$ be a non-negative non-increasing $C^\infty$ function on
the real line such that:
\begin{equation}
f(r)=\left\{\begin{array}{lll}
0&:&r\geq 1\\
1&:&r\leq\frac{1}{2}
\end{array}\right.
\label{16.235}
\end{equation}
For any $\varepsilon>0$ we define the domain $V_{\varepsilon}$ by:
\begin{equation}
V_{\varepsilon}=\{(x^0,x^1,x^2,x^3)\in \Re^4 \ : \ 0\leq
x^0-c^*\leq\varepsilon f(|x|-\lambda^*(c^*)),
|x|\leq\lambda^*(c^*)+1\} \label{16.236}
\end{equation}
and the local existence theorem for the reduced equations
\ref{16.225} states that given any smooth initial data
$(g_{\mu\nu},\partial_0 g_{\mu\nu})$ on $H_{c^*,1}$ where
$g_{\mu\nu}$ is a Lorentzian metric defined along $H_{c^*,1}$, in
particular for any smooth initial data of the form \ref{16.215},
\ref{16.216}, \ref{16.219}, \ref{16.221}, there is an
$\varepsilon_0>0$ and a development $(V_{\varepsilon_0},g)$,
taking the given initial data along $H_{c^*,1}$. Moreover, for any
$\varepsilon_1\in(0,\varepsilon_0]$ the future boundary of
$V_{\varepsilon_1}$ in $V_{\varepsilon_0}$ is a spacelike
hypersurface.

The vectorfields $L^{\prime}$, $\Lb^{\prime}$ satisfy everywhere on
$M_{c^*}\bigcup H_{c^*}$, in particular along $H_{c^*}$ the
formulas \ref{1.3}:
$$L^{\prime\mu}=-2(g^{-1})^{\mu\nu}\partial_\nu u, \ \ \ \Lb^{\prime\mu}=-2(g^{-1})^{\mu\nu}\partial_\nu\ub$$
Thus, in view of \ref{16.214}, \ref{16.215}, \ref{16.216} and the
facts that $2u=c^*-\lambda$, $2\ub=c^*+\lambda$ along $H_{c^*}$,
the components of $L^\prime$ and $\Lb^{\prime}$ along $H_{c^*}$ in
the wave coordinate system are given by:
\begin{eqnarray}
&&L^{\prime 0}=2\partial_0 u=2\Th u=\Omega^{-1}\nonumber\\
&&L^{\prime i}=-2(\og^{-1})^{ij}\partial_j u=(\og^{-1})^{ij}\frac{x^j}{|x|}\nonumber\\
&&\Lb^{\prime 0}=2\partial_0\ub=2\Th\ub=\Omega^{-1}\nonumber\\
&&\Lb^{\prime
i}=-2(\og^{-1})^{ij}\partial_j\ub=-(\og^{-1})^{ij}\frac{x^j}{|x|}\label{16.237}
\end{eqnarray}
and the function $\Omega$ is determined along $H_{c^*}$ by the
conditions:
$$g_{\mu\nu}L^{\prime\mu}L^{\prime\nu}=g_{\mu\nu}\Lb^{\prime\mu}\Lb^{\prime\nu}=0$$
which yield:
\begin{equation}
\Omega^{-2}=(\og^{-1})^{ij}\frac{x^i}{|x|}\frac{x^j}{|x|}
\label{16.238}
\end{equation}
We extend $L^\prime$ and $\Lb^\prime$ to $H_{c^*,1}$ by
stipulating that \ref{16.237} and \ref{16.238} hold everywhere
along $H_{c^*,1}$. Then $L^\prime$ and $\Lb^\prime$ are
respectively outgoing and incoming null normal fields to the
surfaces of constant $\lambda$ on $H_{c^*,1}$.

Let us denote, for any $\eta\in(0,1]$, by $H_{c^*,\eta}$ the
closed ball $\oB_{\lambda^*(c^*)+\eta}$ on the coordinate
hyperplane $x^0=c^*$. Let us also denote by $\oH^\prime_{c^*}$ the
closed annular region $\oB_{\lambda^*(c^*)}\setminus B_{-c^*}$ on
the coordinate hyperplane $x^0=c^*$. This is $H^\prime_{c^*}$
together with its outer boundary in $H_{c^*,1}$.

We now consider, for each point $p\in H_{c^*,1/4}\setminus
B_{-c^*-1/4}$ the null geodesic initiating at $p$ with initial
tangent vector $L^\prime(p)$. This is the first member of the
solution
\begin{equation}
x^\mu=x^\mu(s;p), \ \ \ L^{\prime\mu}=L^{\prime\mu}(s;p)
\label{16.239}
\end{equation}
of the first order system of ordinary differential equations:
\begin{eqnarray}
&&\frac{dx^\mu}{ds}=L^{\prime\mu}\nonumber\\
&&\frac{dL^{\prime\mu}}{ds}=-\Gamma^\mu_{\alpha\beta}(x(s;p))L^{\prime\alpha}L^{\prime\beta}
\label{16.240}
\end{eqnarray}
corresponding to the initial conditions:
\begin{equation}
x^0(0;p)=c^*, \ x^i(0;p)=x_p^i: i=1,2,3 \ \ \
L^{\prime\mu}(0;p)=L^{\prime\mu}(p) \label{16.241}
\end{equation}
the Cartesian coordinates of $p$ being $(c^*,x_p^1,x_p^2,x_p^3)$.
We call this family of null geodesics the ``outgoing" family. We
shall show in the following that there is a
$\varepsilon_1\in(0,\varepsilon_0]$ such that each null geodesic
is defined as long as it remains in $V_{\varepsilon_1}$. Thus the
outgoing null geodesic initiating at $p$ is defined for all
$s\in[0,s^*(p)]$ where
\begin{equation}
x^0(s^*(p);p)=c^*+\varepsilon_1 f(|x(s^*(p);p)|-\lambda^*(c^*))
\label{16.242}
\end{equation}
that is, $x(s^*(p);p)$ lies on the future boundary of
$V_{\varepsilon_1}$. The coefficients $\Gamma^\mu_{\alpha\beta}$
being smooth functions on $V_{\varepsilon_0}$, the mapping
\begin{equation}
(s,p)\mapsto x(s;p) \label{16.243}
\end{equation}
is a smooth mapping of the domain:
\begin{equation}
K_{\varepsilon_1}=\{(s,p)\in \Re\times(H_{c^*,1/4}\setminus
B_{-c^*-1/4})  \ : \ s\in[0,s^*(p)], \ p\in H_{c^*,1/4}\setminus
B_{-c^*-1/4}\} \label{16.244}
\end{equation}
into $V_{\varepsilon_1}$. Assigning polar coordinates
$(\lambda,\vartheta)$ to $p$, and setting $\lambda=c^*-2u$, the
mapping \ref{16.243} may be described as a smooth mapping
\begin{equation}
(s,u,\vartheta)\mapsto x(s;u,\vartheta) \label{16.245}
\end{equation}
of the domain:
\begin{eqnarray}
&&K_{\varepsilon_1}=\{(s,u,\vartheta)\in \Re^2\times S^2 \ : \ \label{16.246}\\
&&\hspace{15mm}s\in[0,s^*(c^*-2u,\vartheta)], \
2u\in[c^*-\lambda^*(c^*)-1/4),2c^*+1/4], \ \vartheta\in
S^2\}\nonumber
\end{eqnarray}
into $V_{\varepsilon_1}$. The vectorfield $L^\prime$ is thus
extended to a null geodesic vectorfield and $s$ is the corresponding affine
parameter function which vanishes on the coordinate hyperplane
$x^0=c^*$. An outgoing null geodesic initiating at a point $p\in
H_{c^*}\bigcap M_0$ is described by:
\begin{equation}
x^0(s;p)=c^*+s, \ \ \ x^i(s,p)=x_p^i+s\frac{x_p^i}{|x_p|} \
i=1,2,3 \ \ \ \ \ L^{\prime\mu}(s;p)=L^{\prime\mu}(p)
\label{16.247}
\end{equation}
as long as it remains in $M_0$.

Similarly, we consider, for each point $p\in\oH^\prime_{c^*}$ the
null geodesic initiating at $p$ with initial tangent vector
$\Lb^{\prime}(p)$. This is the first member of the solution
\begin{equation}
x^\mu=x^\mu(\sb;p), \ \ \ \Lb^{\prime\mu}=\Lb^{\prime\mu}(\sb;p)
\label{16.248}
\end{equation}
of the first order system of ordinary differential equations:
\begin{eqnarray}
&&\frac{dx^\mu}{d\sb}=\Lb^{\prime\mu}\nonumber\\
&&\frac{d\Lb^{\prime\mu}}{d\sb}=-\Gamma^\mu_{\alpha\beta}(x(\sb;p))\Lb^{\prime\alpha}\Lb^{\prime\beta}
\label{16.249}
\end{eqnarray}
corresponding to the initial conditions:
\begin{equation}
x^0(0;p)=c^*, \ x^i(0;p)=x_p^i: i=1,2,3 \ \ \
\Lb^{\prime\mu}(0;p)=\Lb^{\prime\mu}(p) \label{16.250}
\end{equation}
the Cartesian coordinates of $p$ being $(c^*,x_p^1,x_p^2,x_p^3)$.
We call this family of null geodesics the ``incoming"' family.
Again,  we shall show in the following that each null geodesic is
defined as long as it remains in $V_{\varepsilon_1}$. Thus the
incoming null geodesic initiating at $p$ is defined for all
$\sb\in[0,\sb^*(p)]$ where
\begin{equation}
x^0(\sb^*(p);p)=c^*+\varepsilon_1
f(|x(\sb^*(p);p)|-\lambda^*(c^*)) \label{16.251}
\end{equation}
that is, $x(\sb^*(p);p)$ lies on the future boundary of
$V_{\varepsilon_1}$. The coefficients $\Gamma^\mu_{\alpha\beta}$
being smooth functions on $V_{\varepsilon_0}$, the mapping
\begin{equation}
(\sb,p)\mapsto x(\sb;p) \label{16.252}
\end{equation}
is a smooth mapping of the domain:
\begin{equation}
\Kb_{\varepsilon_1}=\{(s,p)\in \Re\times\oH_{c^*} \ : \
\sb\in[0,\sb^*(p)], \ p\in\oH^\prime_{c^*}\} \label{16.253}
\end{equation}
into $V_{\varepsilon_1}$. Assigning polar coordinates
$(\lambda,\vartheta)$ to $p$, and setting $\lambda=2\ub-c^*$, the
mapping \ref{16.243} may be described as a smooth mapping
\begin{equation}
(\sb,\ub,\vartheta)\mapsto x(\sb;\ub,\vartheta) \label{16.254}
\end{equation}
of the domain:
\begin{equation}
\Kb_{\varepsilon_1}=\{(\sb,\ub,\vartheta)\in \Re^2\times S^2 \ : \
\sb\in[0,\sb^*(2\ub-c^*,\vartheta)], \
2\ub\in[0,c^*+\lambda^*(c^*)], \ \vartheta\in S^2\} \label{16.255}
\end{equation}
into $V_{\varepsilon_1}$. The vectorfield $\Lb^\prime$ is thus
extended to a null geodesic vectorfield and $\sb$ is the corresponding affine
parameter function which vanishes on the coordinate hyperplane
$x^0=c^*$.

We shall now derive certain bounds in $V_{\varepsilon_1}$. In the
following, for any multiplet of real numbers of the form
$A^{\mu_1...\mu_q}_{\nu_1...\nu_p}:\mu_1,...,\mu_q;\nu_1,...,\nu_p=0,1,2,3$
we denote by $|A|$ the ``magnitude":
\begin{equation}
|A|=\sqrt{\sum_{\mu_1,...,\mu_q;\nu_1,...,\nu_p}(A^{\mu_1...\mu_q}_{\nu_1..\nu_p})^2}
\label{16.256}
\end{equation}
Since the $\Gamma^\mu_{\alpha\beta}$ are smooth functions on
$V_{\varepsilon_0}$, there is a positive constant $M$ such that:
\begin{equation}
\sup_{x\in V_{\varepsilon_0}}|\Gamma(x)|=M \label{16.257}
\end{equation}
where, in accordance with the above notation,
$$|\Gamma(x)|=\sqrt{\sum_{\mu,\alpha,\beta}(\Gamma^\mu_{\alpha\beta}(x))^2}$$
We shall presently show that if $\varepsilon_1$ is chosen suitably
small depending on $M$, then for the outgoing family of null
geodesics we have:
\begin{equation}
|L^\prime(s;p)|\leq 2, \ \ \ L^{\prime 0}(s;p)\geq\frac{1}{2} \ \
\ \mbox{: for all $(s;p)\in K_{\varepsilon_1}$} \label{16.258}
\end{equation}
and for the incoming family of null geodesics we similarly have:
\begin{equation}
|\Lb^\prime(\sb;p)|\leq 2, \ \ \ \Lb^{\prime
0}(\sb;p)\geq\frac{1}{2} \ \ \ \mbox{: for all
$(\sb;p)\in\Kb_{\varepsilon_1}$} \label{16.259}
\end{equation}
Here, in accordance with the notation \ref{16.256},
$$|L^\prime(s;p)|=\sqrt{\sum_\mu(L^{\prime\mu})^2}, \ \ \ |\Lb^\prime(s;p)|=\sqrt{\sum_\mu(\Lb^{\prime\mu})^2}$$

Consider first \ref{16.258}. By \ref{16.237}, \ref{16.238} and
\ref{16.205} along $H_{c^*,1}$ we have:
\begin{equation}
|L^\prime|\leq\sqrt{2}+O(\delta^{1/2}|c^*|^{-1}), \ \ \ L^0\geq
1-O(\delta^{1/2}|c^*|^{-1}) \label{16.260}
\end{equation}
therefore \ref{16.258} hold at $s=0$ provided that $\delta$ is
suitably small. Consider now a given $p\in H_{c^*,1/4}\setminus
B_{-c^*-1/4}$. By continuity \ref{16.258} hold at the given $p$
and for $s$ suitably small. Let $\os$ be the largest value of $s$
in the interval $(0,s^*(p)]$ for which \ref{16.258} holds at the given
$p$ and for all smaller values of $s$. Then either $\os=s^*(p)$,
in which case what is required has been demonstrated, or one of
the inequalities must be saturated at $\os$. By \ref{16.257} and
the second of \ref{16.240} we have:
\begin{equation}
\left|\frac{dL^\prime}{ds}\right|\leq 4M \ \ : \ \mbox{on
$[0,\os]$} \label{16.261}
\end{equation}
hence:
\begin{equation}
|L^\prime(\os;p)|-|L^\prime(0;p)|\leq 4M\os \label{16.262}
\end{equation}
Moreover, again by \ref{16.257} and the second of \ref{16.240}:
\begin{equation}
|L^{\prime 0}(\os;p)-L^{\prime 0}(0;p)|\leq 4M\os \label{16.263}
\end{equation}
Now by the first of \ref{16.240} and the second of the
inequalities \ref{16.258} at $p$, which holds for all
$s\in[0,\os]$ we have:
\begin{equation}
\frac{dx^0}{ds}\geq\frac{1}{2} \label{16.264}
\end{equation}
hence:
\begin{equation}
x^0(s;p)-c^*\geq \frac{s}{2} \ \ : \ \forall s\in[0,\os]
\label{16.265}
\end{equation}
Since in $V^\prime_{\varepsilon_1}$ we have
$x^0-c^*\leq\varepsilon_1$, it follows that:
\begin{equation}
\os\leq 2\varepsilon_1 \label{16.266}
\end{equation}
Substituting this bound in \ref{16.262} and \ref{16.263} we
obtain:
\begin{equation}
|L^\prime(\os;p)|-|L^\prime(0;p)|\leq 8M\varepsilon_1
\label{16.267}
\end{equation}
and:
\begin{equation}
|L^{\prime 0}(\os;p)-L^{\prime 0}(0;p)|\leq 8M\varepsilon_1
\label{16.268}
\end{equation}
Comparing with \ref{16.260} we conclude that neither of the two
inequalities \ref{16.258} is saturated at $\os$ provided that:
\begin{equation}
\varepsilon_1\leq\varepsilon_{1,0}:=\frac{1}{32M} \label{16.269}
\end{equation}
and $\delta$ is suitably small (depending on the quantities
$\stackrel{(n)}{D} \ : \ n=0,1,2,3$; ${\cal D}^{\prime
4}_{[1]}(\alb)$  and the quantities ${\cal D}_0^\infty$,
$\scD_1^4$, $\scD_2^4(\mbox{tr}\chib)$, $\scD_3(\mbox{tr}\chib)$).
Therefore $\os=s^*(p)$ and we have established \ref{16.258}. The
inequalities \ref{16.259} are established in a similar manner. It
follows that the mapping \ref{16.243} is indeed a smooth mapping
of $K_{\varepsilon_1}$ into $V_{\varepsilon_1}$ and the mapping
\ref{16.252} is indeed a smooth mapping of $\Kb_{\varepsilon_1}$
into $V_{\varepsilon_1}$, and these statements hold for any
$\varepsilon_1\in(0,\varepsilon_{1,0}]$.

The first of the inequalities \ref{16.258} and \ref{16.259}
together with the bounds (see \ref{16.266}):
\begin{equation}
s^*(p)\leq 2\varepsilon_1 \ : \ \forall p\in H_{c^*,1/4}\setminus
B_{-c^*-1/4}, \ \ \ \ \ \sb^*(p)\leq 2\varepsilon_1 \ : \ \forall
p\in \oH^\prime_{c^*} \label{16.270}
\end{equation}
imply, integrating the first of equations \ref{16.240} and
\ref{16.249},
\begin{equation}
|x(s;p)-x_p|\leq 4\varepsilon_1 \ : \ \forall s\in[0,s^*(p)], \
\forall p\in H_{c^*,1/4}\setminus B_{-c^*-1/4} \label{16.271}
\end{equation}
and:
\begin{equation}
|x(\sb;p)-x_p|\leq 4\varepsilon_1 \ : \ \forall\sb\in[0,\sb^*(p)],
\ \forall p\in \oH^\prime_{c^*} \label{16.272}
\end{equation}
Here, as previously, and in contrast to the notation \ref{16.256},
we denote by $|x|$ the magnitude of the spatial part of $x$:
$$|x|=\sqrt{\sum_{i=1}^3 (x^i)^2}$$
Let then $(s_1,p_1)$ and $(s_2,p_2)$ be two points of
$K_{\varepsilon_1}$ which are mapped by the mapping \ref{16.243}
to the same point of  $V_{\varepsilon_1}$. Then by \ref{16.271}
the Euclidean distance of the points $p_1$ and $p_2$ is at most:
$8\varepsilon_1$ Similarly, let $(\sb_1,p_1)$ and $(\sb_2,p_2)$ be
two points of $\Kb_{\varepsilon_1}$ which are mapped by the
mapping \ref{16.252} to the same point of $V_{\varepsilon_1}$.
Then by \ref{16.272} the Euclidean distance of the points $p_1$
and $p_2$ is at most $8\varepsilon_1$ as well.

The inequality \ref{16.271} also implies that:
\begin{equation}
|x(s^*(p);p)|\leq \lambda^*(c^*)+\frac{1}{4}+4\varepsilon_1
\label{16.274}
\end{equation}
Therefore by \ref{16.242} and the definition \ref{16.235} of the
function $f$:
\begin{equation}
x^0(s^*(p);p)-c^*=\varepsilon_1 \ \ : \ \forall p\in
H_{c^*,1/4}\setminus B_{-c^*-1/4} \label{16.275}
\end{equation}
provided that:
\begin{equation}
\varepsilon_1\leq \frac{1}{16} \label{16.276}
\end{equation}
Similarly, the inequality \ref{16.272} also implies that:
\begin{equation}
|x(\sb^*(p);p)|\leq \lambda^*(c^*)+4\varepsilon_1 \label{16.277}
\end{equation}
therefore \ref{16.276} a fortiori implies:
\begin{equation}
x^0(\sb^*(p);p)-c^*=\varepsilon_1 \ \ : \ \forall p\in
\oH^\prime_{c^*} \label{16.278}
\end{equation}
Let us set:
\begin{equation}
\varepsilon_{2,0}=\min\{\varepsilon_{1,0},1/16\} \label{16.284}
\end{equation}
Then all the above hold in reference to $V_{\varepsilon_1}$ for
any $\varepsilon_1\in(0,\varepsilon_{2,0}]$. By the first of
\ref{16.240} and the first of the inequalities \ref{16.258}:
\begin{equation}
\frac{dx^0}{ds}\leq 2 \label{16.279}
\end{equation}
hence integrating on $[0,s^*(p)]$ we obtain, by \ref{16.275}:
\begin{equation}
s^*(p)\geq\frac{\varepsilon_1}{2} \ \ : \ \forall p\in
H_{c^*,1/4}\setminus B_{-c^*-1/4} \label{16.280}
\end{equation}
Using \ref{16.278} we obtain, in a similar manner:
\begin{equation}
\sb^*(p)\geq\frac{\varepsilon_1}{2} \ \ : \ \forall p\in \oH_{c^*}
\label{16.281}
\end{equation}

Consider now the Jacobian of the mapping \ref{16.243} at $s=0$.
Let us assign Cartesian coordinates $(y^1,y^2,y^3)$ to $p$. Since
$s=0$ corresponds to $H_{c^*,1/4}\setminus B_{-c^*-1/4}\subset
H_{c^*,1}$, we have, by \ref{16.241}:
\begin{equation}
\left.\frac{\partial x^\mu}{\partial
s}\right|_{s=0}=\left.L^{\prime\mu}\right|_{H_{c^*,1}}
\label{16.282}
\end{equation}
and, for $j=1,2,3$:
\begin{equation}
\left.\frac{\partial x^\mu}{\partial y^j
}\right|_{s=0}=\left.\frac{\partial x^\mu}{\partial
y^j}\right|_{H_{c^*,1}}=\left\{
\begin{array}{lll}
0&:&\mbox{for $\mu=0$}\\
\delta_{ij}&:&\mbox{for $\mu=i=1,2,3$}
\end{array}\right.
\label{16.283}
\end{equation}
Taking into account \ref{16.237}, we then conclude that the
Jacobian of the mapping \ref{16.243} at $s=0$ is simply:
\begin{equation}
\left.\frac{\partial(x^0,x^1,x^2,x^3)}{\partial(s,y^1,y^2,y^3)}\right|_{s=0}=\left.\Omega^{-1}\right|_{H_{c^*,1}}
\label{16.285}
\end{equation}
The Jacobian of the mapping \ref{16.245} is then the product of
the above by $\left.\partial\lambda/\partial
u\right|_{H_{c^*,1}}=-2$ times the Jacobian of the transformation
from polar to Cartesian coordinates on $H_{c^*,1}$.

Consider next the Jacobian of the mapping \ref{16.252} at $\sb=0$.
Let us again assign Cartesian coordinates $(y^1,y^2,y^3)$ to $p$.
Since $\sb=0$ corresponds to $\oH_{c^*}\subset H_{c^*,1}$, we
have, by \ref{16.250}:
\begin{equation}
\left.\frac{\partial
x^\mu}{\partial\sb}\right|_{\sb=0}=\left.\Lb^{\prime\mu}\right|_{H_{c^*,1}}
\label{16.286}
\end{equation}
and, for $j=1,2,3$:
\begin{equation}
\left.\frac{\partial x^\mu}{\partial y^j
}\right|_{\sb=0}=\left.\frac{\partial x^\mu}{\partial
y^j}\right|_{H_{c^*,1}}=\left\{
\begin{array}{lll}
0&:&\mbox{for $\mu=0$}\\
\delta_{ij}&:&\mbox{for $\mu=i=1,2,3$}
\end{array}\right.
\label{16.287}
\end{equation}
Taking into account \ref{16.237}, we then conclude that the
Jacobian of the mapping \ref{16.252} at $\sb=0$ is simply:
\begin{equation}
\left.\frac{\partial(x^0,x^1,x^2,x^3)}{\partial(\sb,y^1,y^2,y^3)}\right|_{\sb=0}=\left.\Omega^{-1}\right|_{H_{c^*,1}}
\label{16.288}
\end{equation}
The Jacobian of the mapping \ref{16.254} is then the product of
the above by
$\left.\partial\lambda/\partial\ub\right|_{H_{c^*,1}}=2$ times the
Jacobian of the transformation from polar to Cartesian coordinates
on $H_{c^*,1}$.

By \ref{16.238} $\Omega^{-1}$ is bounded from below by a positive
constant on $H_{c^*,1}$. In fact, by \ref{16.209}:
\begin{equation}
\Omega^{-1}\geq 1-O(\delta^{1/2})\geq \frac{1}{2} \ \ \mbox{on
$H_{c^*,1}$} \label{16.289}
\end{equation}
if $\delta$ is suitably small depending on the quantities
$\stackrel{(n)}{D} \ : \ n=0,1,2,3$; ${\cal D}^{\prime
4}_{[1]}(\alb)$  and the quantities ${\cal D}_0^\infty$,
$\scD_1^4$, $\scD_2^4(\mbox{tr}\chib)$, $\scD_3(\mbox{tr}\chib)$.
We are then in a position to apply the {\em implicit function
theorem} to conclude that there is a
$\varepsilon_{3,0}\in(0,\varepsilon_{2,0}/2]$ and a $\eta_0>0$
such that for every $p\in H_{c^*,1/4}\setminus B_{-c^*-1/4}$ the
mapping \ref{16.243}, with $\varepsilon_{2,0}$ in the role of
$\varepsilon_1$, restricted to $[0,\varepsilon_{3,0}]\times
\oB_{\eta_0}(p)$, is a diffeomorphism onto its image in
$V_{\varepsilon_{2,0}}$, and for every $p\in\oH^\prime_{c^*}$ the
mapping \ref{16.252}, with $\varepsilon_{2,0}$ in the role of
$\varepsilon_1$, restricted to $[0,\varepsilon_{3,0}]\times
\oB_{\eta_0}(p)$, is a diffeomorphism onto its image in
$V_{\varepsilon_{2,0}}$. Here we denote by $\oB_{\eta_0}(p)$ the
closed Euclidean ball on $H_{c^*,1}$ with center at $p$,
intersection $H_{c^*,1/4}\setminus B_{-c^*-1/4}$ and
$\oH^\prime_{c^*}$ respectively. We choose the upper bound
$\varepsilon_{2,0}/2$ for $\varepsilon_{3,0}$, so that by inequalities
\ref{16.280} and \ref{16.281} the domain of the restricted
mappings is indeed contained in the domains
$K_{\varepsilon_{2,0}}$ and $\Kb_{\varepsilon_{2,0}}$
respectively. Consider now the mappings \ref{16.243} and
\ref{16.252} with $\varepsilon_1$ satisfying:
\begin{equation}
\varepsilon_1\leq\frac{\varepsilon_{3,0}}{2} \label{16.290}
\end{equation}
Then by virtue of the inequalities \ref{16.270} we have:
\begin{equation}
\bigcup_{p\in H_{c^*,1/4}\setminus
B_{-c^*-1/4}}[0,\varepsilon_{3,0}]\times\oB_{\eta_0}(p)\supset
K_{\varepsilon_1} \ \ \ \mbox{and} \ \ \
\bigcup_{p\in\oH^\prime_{c^*}}[0,\varepsilon_{3,0}]\times\oB_{\eta_0}(p)\supset\Kb_{\varepsilon_1}
\label{16.291}
\end{equation}
respectively. Let then $(s_1,p_1)$ and $(s_2,p_2)$ be two points
of $K_{\varepsilon_1}$ which are mapped by the mapping
\ref{16.243} two the same point of $V_{\varepsilon_1}$. Then as we
have shown earlier, the Euclidean distance of the points $p_1$ and
$p_2$ is at most $8\varepsilon_1$. Therefore if $\varepsilon_1$
satisfies:
\begin{equation}
\varepsilon_1\leq\frac{\eta_0}{16} \label{16.292}
\end{equation}
the points $p_1$ and $p_2$ are both contained in one of the balls
$\oB_{\eta_0}$. Moreover, by the first of the inequalities
\ref{16.270}, $s_1,s_2\leq 2\varepsilon_1$, therefore if
$\varepsilon_1$ satisfies also:
\begin{equation}
\varepsilon_1\leq\frac{\varepsilon_{3,0}}{2} \label{16.293}
\end{equation}
the points $(s_1,p_1)$ and $(s_2,p_2)$ are both contained in one
of the domains $[0,\varepsilon_{3,0}]\times\oB_{\eta_0}(p)$. It
then follows that $(s_1,p_1)=(s_2,p_2)$. We have thus shown that
setting:
\begin{equation}
\varepsilon_1=\min\{\varepsilon_{3,0}/2,\eta_0/16\} \label{16.294}
\end{equation}
the mapping \ref{16.243} of $K_{\varepsilon_1}$ into
$V_{\varepsilon_1}$ is one to one. Being locally a diffeomorphism
this mapping is then a diffeomorphism onto its image in
$V_{\varepsilon_1}$. A similar argument shows that the mapping
\ref{16.252} of $\Kb_{\varepsilon_1}$ into $V_{\varepsilon_1}$ is
a diffeomorphism onto its image in $V_{\varepsilon_1}$. Moreover,
it readily follows from the inequalities \ref{16.271} and
\ref{16.272} that the range of the mapping \ref{16.243} includes
the range of the mapping \ref{16.252}.

Now the mapping \ref{16.245}, which for the sake of clarity we
denote from now on by $\phi$:
\begin{equation}
x=\phi(s,u,\vartheta) \label{16.295}
\end{equation}
is the mapping \ref{16.243} composed on the right with the
mapping:
\begin{equation}
(s,u,\vartheta)\mapsto (s,p(c^*-2u,\vartheta)) \label{16.296}
\end{equation}
the mapping $(\lambda,\vartheta)\mapsto p(\lambda,\vartheta)$
being simply the transformation from polar to Cartesian
coordinates on $H_{c^*,1}$. Since the last is a diffeomorphism of
$[-c^*-1/4,\lambda^*(c^*)+1/4]\times S^2$ onto
$H_{c^*,1/4}\setminus B_{-c^*-1/4}$, it follows that the mapping
\ref{16.295} is a diffeomorphism of $K_{\varepsilon_1}$, as
defined by \ref{16.246}, onto its image in $V_{\varepsilon_1}$.
Similarly, the mapping \ref{16.254}, which for the sake of clarity
we denote from now on by $\phib$:
\begin{equation}
x=\phib(\sb,\ub,\vartheta) \label{16.297}
\end{equation}
is the mapping \ref{16.252} composed on the right with the
mapping:
\begin{equation}
(\sb,\ub,\vartheta)\mapsto (\sb,p(2\ub-c^*,\vartheta))
\label{16.298}
\end{equation}
The transformation from polar to Cartesian coordinates on
$H_{c^*,1}$ being a diffeomorphism of $[-c^*,\lambda^*(c^*)]\times
S^2$ onto $\oH^\prime_{c^*}$, it follows that the mapping
\ref{16.297} is a diffeomorphism of $\Kb_{\varepsilon_1}$, as
defined by \ref{16.255}, onto its image in $V_{\varepsilon_1}$.
Moreover, the range $W_{\varepsilon_1}$ of the mapping
\ref{16.295} includes the range $\Wb_{\varepsilon_1}$ of the
mapping \ref{16.297}.

It follows from the above that $u$ is a smooth function of $x$ on
$W_{\varepsilon_1}$ . Now, by construction each level set $C_u$ of
$u$ in $W_{\varepsilon_1}$ is generated by the congruence of
outgoing null geodesic normals to the surface $S_{c^*-u,u}$ in
$\oH^\prime_{c^*}$. The hypersurface $C_u$ is smooth, being the
image by the diffeomorphism $\phi$ of the ``cylinder"
$$\{(s,u,\vartheta) \ : \ s\in[0,s^*(c^*-2u,\vartheta)], \ \vartheta\in S^2\}\subset K_{\varepsilon_1}$$
and the generators all terminate on the part of the future
boundary of $V_{\varepsilon_1}$ where $x^0=c^*+\varepsilon_1$. It
follows that the vectorfield $-2(g^{-1})^{\mu\nu}\partial_\nu u$
is collinear to $L^{\prime\mu}$, therefore it is null and $u$ is a
solution of the eikonal equation
\begin{equation}
(g^{-1})^{\mu\nu}\partial_\mu u\partial_\nu u=0 \label{16.299}
\end{equation}
in $W_{\varepsilon_1}$. It then follows that the vectorfield
$-2(g^{-1})^{\mu\nu}\partial_\nu u$ is a null geodesic
vectorfield, whose integral curves must coincide with those of
$L^{\prime\mu}$. Since the two vectorfields coincide along
$H_{c^*,1/4}\setminus B_{-c^*-1/4}$ it follows that:
\begin{equation}
-2(g^{-1})^{\mu\nu}\partial_\nu u=L^{\prime\mu} \label{16.300}
\end{equation}
in $W_{\varepsilon_1}$. Similarly, by construction each level set
$\Cb_{\ub}$ of $\ub$ in $\Wb_{\varepsilon_1}$ is generated by the
congruence of incoming null geodesic normals to the surface
$S_{\ub,c^*-\ub}$ in $\oH^\prime_{c^*}$. The hypersurface
$\Cb_{\ub}$ is smooth, being the image by the diffeomorphism
$\phi$ of the ``cylinder"
$$\{(s,\ub,\vartheta) \ : \ s\in[0,s^*(2\ub-c^*,\vartheta)], \ \vartheta\in S^2\}\subset\Kb_{\varepsilon_1}$$
and the future end points of the generators all lie on the part of
the future boundary of $V_{\varepsilon_1}$ where
$x^0=c^*+\varepsilon_1$. It follows that the vectorfield
$-2(g^{-1})^{\mu\nu}\partial_\nu\ub$ is collinear to
$\Lb^{\prime\mu}$, therefore it is null and $\ub$ is a solution of
the eikonal equation
\begin{equation}
(g^{-1})^{\mu\nu}\partial_\mu\ub\partial_\nu\ub=0 \label{16.301}
\end{equation}
in $\Wb_{\varepsilon_1}$. It then follows that the vectorfield
$-2(g^{-1})^{\mu\nu}\partial_\nu\ub$ is a null geodesic
vectorfield, whose integral curves must coincide with those of
$\Lb^{\prime\mu}$. Since the two vectorfields coincide along
$\oH^\prime_{c^*}$ it follows that:
\begin{equation}
-2(g^{-1})^{\mu\nu}\partial_\nu\ub=\Lb^{\prime\mu} \label{16.302}
\end{equation}
in $\Wb_{\varepsilon_1}$.

Now, since in particular $\Cb_{(c^*+\lambda^*(c^*))/2}$, the outer
boundary of $\Wb_{\varepsilon_1}$, is a smooth null hypersurface
generated by the congruence of incoming null geodesic normals to
$S_{(c^*+\lambda^*(c^*))/2,(c^*-\lambda^*(c^*))/2}$, the outer
boundary of $\oH^\prime_{c^*}$, and the generators all terminate
on the part of the future boundary of $V_{\varepsilon_1}$ where
$x^0=c^*+\varepsilon_1$, it follows that the closure of
$V^\prime_{\varepsilon_1}$, the domain of dependence of $H_{c^*}$
in $(V_{\varepsilon_1},g)$, is:
\begin{equation}
(M_0\bigcap V_{\varepsilon_1})\bigcup\Wb_{\varepsilon_1}
\label{16.303}
\end{equation}
and its outer boundary is the outer boundary of
$\Wb_{\varepsilon_1}$. We shall restrict attention from now on to
this domain, where we have a smooth solution $g_{\mu\nu}$ of the
vacuum Einstein equations. Moreover, since in $M_0\bigcap
V_{\varepsilon_1}$ $g_{\mu\nu}$ is simply the Minkowski metric in
Cartesian coordinates, we shall focus on the domain
$\Wb_{\varepsilon_1}$ where we have a non-trivial solution.

Consider on $\Wb_{\varepsilon_1}$ the function $\Omega$ defined
by:
\begin{equation}
\Omega^{-2}=-\frac{1}{2}g_{\mu\nu}\Lb^{\prime\mu}L^{\prime\nu}
\label{16.304}
\end{equation}
This is a smooth positive function, the vectorfields $\Lb^\prime$
and $L^\prime$ being future directed null vectorfields which are
nowhere collinear. By \ref{16.237}, the function $\Omega$
coincides along $\oH^\prime_{c^*}$ to the function previously
defined there by \ref{16.238}. Now $u$ is a smooth function of $x$
in $\Wb_{\varepsilon_1}$. For the sake of clarity we write:
\begin{equation}
u=f(x) \label{16.305}
\end{equation}
Substituting $x=\phib(\sb,\ub,\vartheta)$, that is, considering
the composition:
\begin{equation}
h=f\circ\phib \label{16.306}
\end{equation}
a smooth function on $\Kb_{\varepsilon_1}$, we write:
\begin{equation}
u=h(\sb,\ub,\vartheta)
\end{equation}
We then have:
\begin{equation}
\frac{\partial h}{\partial\sb}=(\partial_\mu
f)\circ\phib\frac{\partial\phib^\mu}{\partial\sb} \label{16.307}
\end{equation}
which in view of the fact that:
$$\frac{\partial\phib^\mu}{\partial\sb}=\Lb^{\prime\mu}\circ\phib$$
becomes:
\begin{equation}
\frac{\partial h}{\partial\sb}=(\Lb^{\prime\mu}\partial_\mu
u)\circ\phi=\Omega^{-2}\circ\phi \label{16.308}
\end{equation}
by \ref{16.300} and \ref{16.304}.Therefore $\partial
u/\partial\sb$ is a smooth positive function on
$\Kb_{\varepsilon_1}$. It then follows that the mapping $\pi$ of
$\Kb_{\varepsilon_1}$ onto the domain:
\begin{eqnarray}
&&\tilde{N}_{\varepsilon_1}=\{(u,\ub,\vartheta)\in \Re^2\times S^2 \ : \ \label{16.309}\\
&&\hspace{15mm}u\in
[c^*-\ub,h(\sb^*(2\ub-c^*,\vartheta),\ub,\vartheta)], \
2\ub\in[0,c^*+\lambda^*(c^*)], \ \vartheta\in S^2\}\nonumber
\end{eqnarray}
by:
\begin{equation}
(\sb,\ub,\vartheta)\mapsto(h(\sb,\ub,\vartheta),\ub,\vartheta)
\label{16.310}
\end{equation}
is a diffeomorphism. Then:
\begin{equation}
\psi=\phib\circ\pi^{-1} \label{16.311}
\end{equation}
is a diffeomorphism of $\tilde{N}_{\varepsilon_1}$ onto
$\Wb_{\varepsilon_1}$. The coordinates $(u,\ub,\vartheta)$ are
{\em canonical coordinates} on $\Wb_{\varepsilon_1}$, which
smoothly extend the canonical coordinates on $M^\prime_{c^*}$. The
mapping $\psi$ being a diffeomorphism, we have smooth metric in
canonical coordinates on $\tilde{N}_{\varepsilon_1}$, a smooth
extension of the metric in the same coordinates on
$M^\prime_{c^*}$. Let finally
\begin{equation}
\Omega_m^{-2}=\inf_{\Wb_{\varepsilon_1}}\Omega^{-2}>0
\label{16.312}
\end{equation}
Then along the generators of each $\Cb_{\ub}$,
$\ub\in[0,(c^*+\lambda^*(c^*))/2]$, we have:
\begin{equation}
\frac{\partial h}{\partial\sb}\geq\Omega_m^{-2} \label{16.313}
\end{equation}
Integrating we obtain:
\begin{equation}
h(\sb^*(2\ub-c^*,\vartheta),\ub,\vartheta)-(c^*-\ub)\geq
\Omega_m^{-2}\sb^*(2\ub-c^*,\vartheta)\geq\frac{1}{2}\varepsilon_1\Omega_m^{-2}
\label{16.314}
\end{equation}
by \ref{16.281}. Hence the domain $\tilde{N}_{\varepsilon_1}$
includes the domain:
\begin{equation}
N_{\varepsilon_2}=\{(u,\ub,\vartheta) \ : \
u+\ub\in[c^*,c^*+\varepsilon_2], \ 2\ub\in[0,c^*+\lambda^*(c^*)],
\ \vartheta\in S^2\} \label{16.315}
\end{equation}
where:
\begin{equation}
\varepsilon_2=\frac{1}{2}\varepsilon_1\Omega_m^{-2} \label{16.316}
\end{equation}

Now, if $c^*\in[u_0+\delta,-1)$, then by \ref{16.a2}
$\lambda^*(c^*)=2\delta-c^*$ and we have:
\begin{equation}
M^\prime_{c^*}\bigcup N_{\varepsilon_2}\supset
M^\prime_{c^*+\varepsilon_2} \label{16.317}
\end{equation}
If on the other hand $c^*\in(u_0,u_0+\delta)$, then by \ref{16.a2}
$\lambda^*(c^*)=c^*-2u_0$ and we consider the characteristic
initial value problem with initial hypersurfaces one of which, the
inner one, is the part of $\Cb_{c^*-u_0}$ which constitutes the
outer boundary of $N_{\varepsilon_2}$, and the other, the outer
one, is the part of $C_{u_0}$ which lies to the future of
$S_{c^*-u_0,u_0}$. The intersection of these two null
hypersurfaces is the surface $S_{c^*-u_0,u_0}$. On the inner null
hypersurface we have as characteristic initial data the data
induced from the solution on $N_{\varepsilon_2}$, and on the outer
null hypersurace we have as characteristic initial data the
original initial data on $C_{u_0}$. We then apply the theorem of
Rendall [R] which constructs a smooth solution of this characteristic
initial value problem for the vacuum Einstein equations in a
neighborhood of the surface $S_{c^*-u_0,u_0}$ bounded in the past
by the two null hypersurfaces. The solution is again obtained in a
wave coordinate system, adapted in this case to the two
intersecting null hypersurfaces. An entirely analogous argument to
the one presented above then demonstrates that there is a
\begin{equation}
\varepsilon^\prime_2\in(0,\min\{\varepsilon_2, u_0+\delta-c^*\})
\label{16.318}
\end{equation}
and a smooth solution in canonical coordinates on the domain:
\begin{equation}
N^\prime_{\varepsilon^\prime_2}=\{(u,\ub,\vartheta) \ : \
u+\ub\in[c^*,c^*+\varepsilon^\prime_2], \ u\geq u_0, \ \ub\geq
c^*-u_0, \ \vartheta\in S^2\} \label{16.319}
\end{equation}
Then
\begin{equation}
M^\prime_{c^*}\bigcup N_{\varepsilon_2}\bigcup
N^\prime_{\varepsilon^\prime_2}\supset
M^\prime_{c^*+\varepsilon^\prime_2} \label{16.320}
\end{equation}

The same argument shows that there is a subdomain of the initial
domain $U$ (see last section of Chapter 2) of the form
$M^\prime_{u_0+\eta_0}$, for some $\eta_0>0$, where we have a
smooth metric in canonical coordinates.

Since we have a smooth metric in canonical coordinates on
$N_{\varepsilon_2}$, in the case $c^*\in[u_0+\delta,-1]$, and on
$N_{\varepsilon_2}\bigcup N^\prime_{\varepsilon^\prime_2}$, in the
case $c^*\in(u_0,u_0+\delta)$, all the quantities which appear in
the bootstrap assumptions {\bf A0}, {\bf A1.1}, {\bf A1.2}, {\bf
A2.1}, {\bf A2.2}, {\bf A3.1}, {\bf A3.2}, {\bf A4.1}, {\bf A4.2},
{\bf B1}, {\bf B2}, {\bf B3}, and {\bf C1.1} - {\bf C1.4}, {\bf
C2.1} - {\bf C2.4}, {\bf C3.1} - {\bf C3.5}, {\bf C4.1} - {\bf
C4.8}, {\bf C5.1} - {\bf C5.4}, {\bf C6.1} - {\bf C6.10}, as well
as {\bf D0}, {\bf D1}, {\bf D2.1}, {\bf D2.2}, {\bf D3.1}, {\bf
D3.2}, {\bf D4.1}, {\bf D4.2}, {\bf D5}, {\bf D6}, {\bf D7}, {\bf
D8}, {\bf D9.1} - {\bf D9.3}, {\bf D10.1} - {\bf D10.4}, {\bf
D11}, {\bf D 12}, are continuous on $N_{\varepsilon_2}$ and
$N_{\varepsilon_2}\bigcup N^\prime_{\varepsilon^\prime_2}$,
respectively. Since the inequalities involved are not saturated on
$H^\prime_{c^*}$, it follows that there is a $\varepsilon_3$,
$\varepsilon_3\in(0,\varepsilon_2]$ in the case
$c^*\in[u_0+\delta,-1]$,
$\varepsilon_3\in(0,\varepsilon^\prime_2]$ in the case
$c^*\in(u_0,u_0+\delta)$, such that all the above bootstrap
assumptions hold on $M_{c^*+\varepsilon_3}$ as well.

The same argument shows that there is a $\eta_1\in(0,\eta_0]$ such
that the bootstrap assumptions hold on
$M^\prime_{u_0+\eta_1}\subset M^\prime_{u_0+\eta_0}$.

Finally, the error integrants $\delta^{2q_n}\stackrel{(n)}{\tau}_2
\ : \ n=0,1,2,3$ are integrable functions on $N_{\varepsilon_2}$,
in the case $c^*\in[u_0+\delta,-1]$, and on
$N_{\varepsilon_2}\bigcup N^\prime_{\varepsilon^\prime_2}$, in the
case $c^*\in(u_0,u_0+\delta)$. It follows that there is a
$\varepsilon_4\in(0,\varepsilon_3]$ such that:
\begin{equation}
\delta^{2q_n}\int_{M^\prime_{c^*+\varepsilon_4}\setminus
M^\prime_{c^*}}|\stackrel{(n)}{\tau}_2|d\mu_g\leq 1 \ \ : \
n=0,1,2,3 \label{16.321}
\end{equation}
hence by the energy-flux inequalities \ref{12.275} with
$M^\prime_{c^*+\varepsilon_4}$ in the role of $M^\prime_{c^*}$ we
have:
\begin{eqnarray}
&&\stackrel{(n)}{{\cal
E}}_2\leq\stackrel{(n)}{D}+\delta^{2q_n}\int_{M^\prime_{c^*}}|\stackrel{(n)}{\tau}_2|d\mu_g+1
\ \ : \ n=0,1,2,3
\nonumber\\
&&\stackrel{(3)}{{\cal
F}}_2\leq\stackrel{(3)}{D}+\delta^{2q_3}\int_{M^\prime_{c^*}}|\stackrel{(3)}{\tau}_2|d\mu_g+1
\label{16.322}
\end{eqnarray}
{\em where now the quantities $\stackrel{(n)}{{\cal E}}_2 \
n=0,1,2,3$ and $\stackrel{(3)}{{\cal F}}_2$ refer to
$M^\prime_{c^*+\varepsilon_4}$}. Combining with the bounds
\ref{16.22} we then conclude that:
\begin{equation}
{\cal P}_2\leq
G(\stackrel{(0)}{D},\stackrel{(1)}{D},\stackrel{(2)}{D},\stackrel{(3)}{D})
\label{16.323}
\end{equation}
{\em where now the quantity ${\cal P}_2$ refers to
$M^\prime_{c^*+\varepsilon_4}$}.

Similarly, in reference to the domain $M^\prime_{u_0+\eta_1}$,
there is a $\eta_2\in(0,\eta_1]$ such that:
\begin{equation}
\delta^{2q_n}\int_{M^\prime_{u_0+\eta_2}}|\stackrel{(n)}{\tau}_2|d\mu_g\leq
1 \ \ : \ n=0,1,2,3 \label{16.a3}
\end{equation}
This implies that the quantity ${\cal P}_2$ corresponding to
$M^\prime_{u_0+\eta_2}$ satisfies the bound required in the 3rd
condition of the statement of Theorem 12.1. We have thus shown
that $u_0+\eta_2\in {\cal A}$, so the set ${\cal A}$ is indeed
non-empty.

Returning now to the inequality \ref{16.323}, which refers to
$M^\prime_{c^*+\varepsilon_4}$, we conclude that all three
conditions of Theorem 12.1 are satisfied with $c^*+\varepsilon_4$
in the role of $c$. This contradictcs the definition of $c^*$
unless $c^*=-1$. This completes the proof of Theorem 12.1.

\section{Restatement of the existence theorem}

According to the discussion preceding the statement of Theorem
12.1 the quantities $\stackrel{(n)}{D} \ : \ n=0,1,2,3$, ${\cal
D}^{\prime 4}_{[1]}(\alb)$, ${\cal D}_0^\infty$, $\scD_1^4$,
$\scD_2^4(\mbox{tr}\chib)$, $\scD_3(\mbox{tr}\chib)$ are all
bounded by a non-negative non-decreasing continuous function of
$M_8$. The theorem can now be restated as follows.

\vspace{5mm}

\noindent{\bf Theorem 16.1} \ \ \ Let us be given smooth initial
data on $C_{u_0}$ as described in Chapter 2. There is a non-negative non-decreasing
continuous function $F$ on the non-negative real line such that
if:
$$\delta F(M_8)\leq 1$$
the following hold:

\begin{enumerate}

\item With $M_{-1}\setminus\Gamma_0=(D_{-1}\setminus A_0)\times S^2$ there is a smooth solution $g$ of the vacuum Einstein equations in canonical coordinates on $M_{-1}\setminus\Gamma_0$, taking the given initial data along $C_{u_0}$, the subdomain of $M_{-1}\setminus\Gamma_0$ where
$\ub\leq 0$ being isometric to the corresponding domain in
Minkowski spacetime. In particular, the null hypersurfaces $C_u$
and $\Cb_{\ub}$ contain no focal or cut points in $M_{-1}\setminus
\Gamma_0$.

\item The assumptions {\bf A0}, {\bf A1.1}, {\bf A1.2}, {\bf A2.1}, {\bf A2.2}, {\bf A3.1}, {\bf A3.2}, {\bf A4.1}, {\bf A4.2}
hold on $M^\prime_{-1}$.

\item There is a non-negative non-decreasing continuous function $G$ on the non-negative real line such that:
$${\cal Q}_2^\prime\leq G(M_8)$$
the quantity $Q^\prime_2$ referring to $M^\prime_{-1}$. It
follows, by the results of Chapter 10, that:
$${\cal R}_0^\infty\leq CG(M_8), \ \ \scR_1^4\leq CG(M_8)$$
$$\max\{{\cal R}_0^4(\Dh\alpha),{\cal R}_0^4(D\beta),{\cal R}_0^4(D\sigma),{\cal R}_0^4(D\beb)\}\leq CG(M_8)$$
$${\cal R}_0^4(\Dbh\alb)\leq CG(M_8)$$
all quantities referring to $M^\prime_{-1}$.

\item The results of Chapters 3 - 7, all hold on $M^\prime_{-1}$ with the symbol $O(\delta^p|u|^r)$ re-interpreted to mean
the product of $\delta^p|u|^r$ with a non-negative non-decreasing
continuous function of $M_8$.

\end{enumerate}

\chapter{Trapped Surface Formation}

We are now ready to reach the aim of this work, namely the
analysis of the formation of trapped surfaces. 

By Theorem 16.1 and the results of Chapter 3 we have:
\begin{equation}
\left|\mbox{tr}\chib+\frac{2}{|u|}\right|\leq O(\delta|u|^{-2}) \ \ \mbox{: on $M^\prime_{-1}$}
\label{17.a1}
\end{equation}
Therefore if $\delta$ is suitably small depending on $M_8$, $\mbox{tr}\chib$ is everywhere negative on $M^\prime_{-1}$. Consequently, a surface $S_{\ub,u}$ 
contained in $M^\prime_{-1}$ is a {\em trapped sphere} if  and only if everywhere on this surface $\mbox{tr}\chi<0$, or equivalently $\mbox{tr}\chi^\prime<0$.

Consider then equation \ref{4.c4}:
\begin{equation}
\Dbh(\Omega\chih)=\Omega^2\left\{\snab\oth\eta+\eta\oth\eta+\frac{1}{2}\mbox{tr}\chib\chih-\frac{1}{2}\mbox{tr}\chi\chibh
\right\} \label{17.1}
\end{equation}
In view of the fact that $\Db\log\Omega=\omb$, this equation takes
the form:
\begin{equation}
\Dbh\chih-\frac{1}{2}\Omega\mbox{tr}\chib\chih=\theta \label{17.2}
\end{equation}
where $\theta$ is the trace-free symmetric 2-covariant $S$
tensorfield:
\begin{equation}
\theta=\Omega\left\{\snab\oth\eta+\eta\oth\eta-\frac{1}{2}\mbox{tr}\chi\chibh\right\}-\omb\chih
\label{17.3}
\end{equation}
We apply the second part of Lemma 4.2 to obtain:
$$\Db(|\chih|^2)+2\Omega\mbox{tr}\chib|\chih|^2=2(\chih,\Dbh\chih)-4\Omega(\chibh,\chih\times\chih)$$
In view of the identity \ref{1.a5} the last term vanishes and we
obtain simply:
\begin{equation}
\Db(|\chih|^2)+2\Omega\mbox{tr}\chib|\chih|^2=2(\chih,\Dbh\chih)
\label{17.4}
\end{equation}
Substituting for $\Db\chih$ from \ref{17.2} yields:
\begin{equation}
\Db(|\chi|^2)+\Omega\mbox{tr}\chib|\chih|^2=2(\chih,\theta)
\label{17.5}
\end{equation}
We then consider the function:
\begin{equation}
f=|u|^2|\chih|^2 \label{17.6}
\end{equation}
We have (recall that $u<0$):
\begin{eqnarray*}
&&\Db f=|u|^2\Db(|\chih|^2)-2|u||\chih|^2\\
&&\hspace{8mm}=|u|^2\left\{\Db(|\chih|^2)+\Omega\mbox{tr}\chib|\chih|^2\right\}-|u|^2\left(\Omega\mbox{tr}\chib+\frac{2}{|u|}\right)|\chih|^2
\end{eqnarray*}
Substituting from \ref{17.5} then yields the equation:
\begin{equation}
\Db f=g \label{17.7}
\end{equation}
where $g$ is the function:
\begin{equation}
g=|u|^2\left\{-\left(\Omega\mbox{tr}\chib+\frac{2}{|u|}\right)|\chih|^2+2(\chih,\theta)\right\}
\label{17.8}
\end{equation}
Now, by Theorem 16.1 and the results of Chapter 3:
\begin{equation}
\left|\Omega\mbox{tr}\chib+\frac{2}{|u|}\right|\leq
O(\delta|u|^{-2}) \ \ \mbox{: on $M^\prime_{-1}$} \label{17.9}
\end{equation}
and:
\begin{equation}
|\chih|^2\leq O(\delta^{-1}|u|^{-2}) \ \ \mbox{: on
$M^\prime_{-1}$} \label{17.10}
\end{equation}
with the symbol $O(\delta^p|u|^r)$ interpreted from this point on
as in the 4th statement of Theorem 16.1.

Consider next $\theta$. From Theorem 16.1 and Proposition 6.2:
\begin{equation}
|u|\|\snab^{ \
2}\eta\|_{L^4(S_{\ub,u})}+\|\snab\eta\|_{L^4(S_{\ub,u})}\leq
O(|u|^{-5/2}) \label{17.11}
\end{equation}
Applying Lemma 5.2 with $p=4$ we then obtain, in view of Lemma
10.1,
\begin{equation}
|\snab\eta|\leq O(|u|^{-3}) \ \ \mbox{: on $M^\prime_{-1}$}
\label{17.12}
\end{equation}
Also, by Theorem 16.1 and the results of Chapter 3:
\begin{eqnarray}
&&|\eta|^2\leq O(\delta|u|^{-4}) \ \ \mbox{: on $M^\prime_{-1}$}\nonumber\\
&&|\mbox{tr}\chi||\chibh|\leq O(\delta^{1/2}|u|^{-3}) \ \ \mbox{: on $M^\prime_{-1}$}\nonumber\\
&&|\omb||\chih|\leq O(\delta^{1/2}|u|^{-4}) \ \ \mbox{: on
$M^\prime_{-1}$} \label{17.13}
\end{eqnarray}
It then follows that:
\begin{equation}
|\theta|\leq O(|u|^{-3}) \ \ \mbox{: on $M^\prime_{-1}$}
\label{17.14}
\end{equation}

Since $|(\chih,\theta)|\leq|\chih||\theta|$, the bounds
\ref{17.9}, \ref{17.10}, \ref{17.14}, imply:
\begin{equation}
|g|\leq O(\delta^{-1/2}|u|^{-2}) \ \ \mbox{: on $M^\prime_{-1}$}
\label{17.15}
\end{equation}

Let us work in a canonical coordinate system. Then equation
\ref{17.7} reads simply:
\begin{equation}
\frac{\partial f}{\partial u}=g \label{17.16}
\end{equation}
Integrating with respect to $u$ on $[u_0,-1-\delta]$ we then
obtain:
\begin{equation}
f(\ub,-1-\delta,\vartheta)=f(\ub,u_0,\vartheta)+\int_{u_0}^{-1-\delta}g(\ub,u,\vartheta)du
\label{17.17}
\end{equation}
By \ref{17.15} the integral on the right is bounded from below by:
\begin{equation}
-\int_{u_0}^{-1-\delta}|g(\ub,u,\vartheta|du\geq -O(\delta^{-1/2})
\label{17.18}
\end{equation}
We conclude that:
\begin{equation}
f(\ub,-1-\delta,\vartheta)\geq
f(\ub,u_0,\vartheta)-O(\delta^{-1/2}) \label{17.19}
\end{equation}

Now, according to equation \ref{3.6} we have on each $C_u$, in
particular on $C_{-1-\delta}$:
\begin{equation}
D\mbox{tr}\chi^\prime=-\frac{1}{2}(\mbox{tr}\chi)^2-|\chih|^2
\label{17.20}
\end{equation}
From the definition \ref{17.6}, on $C_{-1-\delta}$ this implies:
\begin{equation}
D\mbox{tr}\chi^\prime\leq -(1+\delta)^{-2}f \label{17.21}
\end{equation}
Consider now each generator of $C_{-1-\delta}$, integral curve of
$L$. If $(0,-1-\delta,\vartheta_0)$ is the point where a generator
intersects $S_{0,-1-\delta}$, the generator is given in canonical
coordinates by:
\begin{equation}
\ub\mapsto (\ub,-1-\delta,\vartheta(\ub;\vartheta_0))
\label{17.22}
\end{equation}
where $\vartheta(\ub;\vartheta_0)$ is the solution of the ordinary
differential equation (see \ref{1.171}):
\begin{equation}
\frac{d\vartheta^A(\ub;\vartheta_0)}{d\ub}=b^A(\ub,-1-\delta,\vartheta(\ub;\vartheta_0))
\label{17.23}
\end{equation}
corresponding to the initial condition:
\begin{equation}
\vartheta(0,\vartheta_0)=\vartheta_0 \label{17.24}
\end{equation}
Equation \ref{17.23} represents a non-autonomous flow on $S^2$,
$b$ at $u=-1-\delta$ being a vectorfield on $S^2$ depending on
$\ub$. This is simply the flow $\Phi_{\ub}$ on $C_{-1-\delta}$.

Since
$$\left.\mbox{tr}\chi^\prime\right|_{S_{0,-1-\delta}}=\frac{2}{1+\delta}$$
integrating equation \ref{17.21} along the generator of
$C_{-1-\delta}$ originating at $(0,-1-\delta,\vartheta_0)\in
S_{0,u_0}$ we obtain:
\begin{equation}
\mbox{tr}\chi^\prime(\ub,-1-\delta,\vartheta(\ub;\vartheta_0))\leq
\frac{2}{1+\delta}-\frac{1}{(1+\delta)^2}\int_0^{\ub}
f(\ub^\prime,-1-\delta,\vartheta(\ub^\prime;\vartheta_0))d\ub^\prime
\label{17.25}
\end{equation}
Suppose now that:
\begin{equation}
\int_0^\delta f(\ub,-1-\delta,\vartheta(\ub;\vartheta_0))d\ub>
2(1+\delta) \ \ \mbox{: for all $\vartheta_0\in S^2$}
\label{17.26}
\end{equation}
Then there is a $\ub^*\in(0,\delta)$ such that for all $\ub\in
(\ub^*,\delta)$ we have:
\begin{equation}
\mbox{tr}\chi^\prime(\ub,-1-\delta,\vartheta(\ub;\vartheta_0))<0 \
\ \mbox{: for all $\vartheta_0\in S^2$} \label{17.27}
\end{equation}
therefore, since $\Phi_{\ub}$ is a diffeomorphism of
$S_{0,-1-\delta}$ onto $S_{\ub,-1-\delta}$,
$$\mbox{tr}\chi^\prime<0 \  \ : \ \mbox{everywhere on $S_{\ub,-1-\delta}$}$$
that is, $S_{\ub,-1-\delta}$ is a {\em trapped sphere}.

By \ref{17.19} the condition \ref{17.26} is satisfied if:
\begin{equation}
\int_0^\delta
f(\ub,u_0,\vartheta(\ub;\vartheta_0))d\ub>2+O(\delta^{1/2}) \ \
\mbox{: for all $\vartheta_0\in S^2$} \label{17.28}
\end{equation}
(where we have absorbed $\delta$ into $O(\delta^{1/2})$.)
Recalling the definition \ref{2.64} of the function $e$ on
$C_{u_0}$ we see that:
\begin{equation}
f(\ub,u_0,\vartheta)=2|u_0|^2 e(\ub,u_0,\vartheta) \label{17.29}
\end{equation}
therefore condition \ref{17.28} reads, in terms of the function
$e$:
\begin{equation}
|u_0|^2\int_0^\delta
e(\ub,u_0,\vartheta(\ub;\vartheta_0))d\ub>1+O(\delta^{1/2}) \ \
\mbox{: for all $\vartheta_0\in S^2$} \label{17.30}
\end{equation}

Now condition \ref{17.30} is not natural as it stands, because the
integral is not an integral along the generators of $C_{u_0}$, but
rather along the images on $C_{u_0}$ by $\Phib_{u_0+1+\delta}$ of
the generators of $C_{-1-\delta}$. To turn it into a natural
condition, we must first estimate the deviation of the image on
$C_{u_0}$ by $\Phi_{u_0+1+\delta}$ of a generator of
$C_{-1-\delta}$ from the corresponding generator of $C_{u_0}$,
namely the generator originating at the same point on $S_{0,u_0}$.
Analytically, the problem is to estimate:
$$\up{d}(\vartheta(\ub;\vartheta_0),\vartheta_0)$$
the distance $\up{d}$, relative to the standard metric $\up{\sg}$
on $S^2$, of the points $\vartheta(\ub;\vartheta_0)$ and
$\vartheta_0$. Now, the $S$ tangential vectorfield $b$ is the
solution of \ref{1.174}:
\begin{equation}
\Db b=4\Omega^2\zeta^\sharp \label{17.31}
\end{equation}
with the initial condition \ref{1.173}:
\begin{equation}
b=0 \ \ \mbox{: on $C_{u_0}$} \label{17.32}
\end{equation}
We have:
\begin{equation}
\Db(|b|^2)=\Db(\sg(b,b))=2\Omega\chib(b,b)+2\sg(b,\Db
b)=\Omega\mbox{tr}\chib|b|^2+\Omega\chibh(b,b)+8\Omega^2(b,\zeta)
\label{17.33}
\end{equation}
hence:
\begin{equation}
\Db(|b|^2)\leq
2|b|\left\{\left(\frac{1}{2}\Omega\mbox{tr}\chib+\Omega|\chibh|\right)|b|+4\Omega^2|\zeta|\right\}
\label{17.34}
\end{equation}
Applying Lemma 3.1 we then obtain:
\begin{eqnarray}
&&|b(\ub,u,\vartheta)|\leq
4\int_{u_0}^u\exp\left(\int_{u^\prime}^u((1/2)\Omega\mbox{tr}\chib+\Omega|\chibh|)
(\ub,u^{\prime\prime},\vartheta)du^{\prime\prime}\right)\nonumber\\
&&\hspace{65mm}\cdot(\Omega^2|\zeta|)(\ub,u^\prime,\vartheta)du^\prime
\label{17.35}
\end{eqnarray}
By Theorem 16.1 and the results of Chapter 3 the integral in the
exponential does not exceed:
\begin{equation}
\log\left(\frac{|u|}{|u^\prime|}\right)+O(\delta^{1/2}|u|^{-1})
\label{17.36}
\end{equation}
Moreover, if $\delta$ is suitably small depending on $M_8$ the
$O(\delta^{1/2}|u|^{-1})$ term does not exceed $\log 2$. It
follows that \ref{17.35} implies:
\begin{equation}
|u|^{-1}||b(\ub,u,\vartheta)|\leq
8\int_{u_0}^u|u^\prime|^{-1}(\Omega^2|\zeta|)(\ub,u^\prime,\vartheta)du^\prime
\label{17.37}
\end{equation}
Since by Theorem 16.1 and the results of Chapter 3 we have:
\begin{equation}
\Omega^2|\zeta|\leq O(\delta^{1/2}|u|^{-2}) \ \ \mbox{: on
$M^\prime_{-1}$} \label{17.38}
\end{equation}
it follows that:
\begin{equation}
|b|\leq O(\delta^{1/2}|u|^{-1}) \ \ \mbox{: on $M^\prime_{-1}$}
\label{17.39}
\end{equation}
In particular, this holds on $C_{-1-\delta}$ and we obtain:
\begin{equation}
|\left.
b\right|_{S_{\ub,-1-\delta}}|_{\left.\sg\right|_{S_{\ub,-1-\delta}}}\leq
O(\delta^{1/2}) \label{17.40}
\end{equation}
Now, by Lemma 11.1 and the inequality \ref{11.26} applied to
$\left. b\right|_{S_{\ub,-1-\delta}}$ there is a numerical
constant $C$ such that:
\begin{equation}
|b(\ub,-1-\delta,\vartheta)|_{\up{\sg}}\leq C|\left.
b\right|_{S_{\ub,-1-\delta}}|_{\left.\sg\right|_{S_{\ub,-1-\delta}}}
\label{17.41}
\end{equation}
Hence also:
\begin{equation}
\sup_{\vartheta\in S^2}|b(\ub,-1-\delta,\vartheta)|_{\up{\sg}}\leq
O(\delta^{1/2}) \label{17.42}
\end{equation}
This is a bound for the speed, relative to $\up{\sg}$, of the flow
defined by \ref{17.23} on $S^2$. It follows that the length,
relative to $\up{\sg}$ of the arc \ref{17.22} ($\ub\in[0,\delta]$)
does not exceed $O(\delta^{3/2})$, therefore:
\begin{equation}
\up{d}(\vartheta(\ub;\vartheta_0,\vartheta_0))\leq O(\delta^{3/2})
\label{17.43}
\end{equation}

Now, by \ref{2.67}:
\begin{equation}
|\sd e|_{\up{\sg}}\leq O(\delta^{-1}|u_0|^{-2}) \label{17.44}
\end{equation}
The bounds \ref{17.43} and \ref{17.44} together imply:
\begin{equation}
|u_0|^2|e(\ub,u_0,\vartheta(\ub;\vartheta_0))-e(\ub,u_0,\vartheta_0)|\leq
O(\delta^{1/2}) \label{17.45}
\end{equation}
Consequently, condition \ref{17.30} is satisfied if:
\begin{equation}
|u_0|^2\int_0^\delta e(\ub,u_0,\vartheta_0)d\ub> 1+O(\delta^{1/2})
\ \ \mbox{: for all $\vartheta_0\in S^2$} \label{17.46}
\end{equation}
Since $O(\delta^{1/2})$ is of the form $\delta^{1/2}F(M_8)$ where
$F$ is a non-negative non-decreasing continuous function on the
non-negative real line, given any constant $k>1$,
\begin{equation}
|u_0|^2\int_0^\delta e(\ub,u_0,\vartheta_0)d\ub\geq k \ \ \mbox{:
for all $\vartheta_0\in S^2$} \label{17.47}
\end{equation}
and:
\begin{equation}
\delta^{1/2}F(M_8)<k-1 \label{17.48}
\end{equation}
imply \ref{17.46}. The condition \ref{17.47} is natural: it simply
says that the integral of $|u_0|^2 e$ along any generator of
$C_{u_0}$, with respect to the affine parameter, is not less than
$k$.

Finally, from \ref{2.35}, \ref{2.36}, and \ref{2.65}, we have:
\begin{equation}
e=\frac{1}{8}(m^{-1})^{AC}(m^{-1})^{BD}\frac{\partial
m_{AB}}{\partial\ub}\frac{\partial m_{CD}}{\partial\ub}
\label{17.49}
\end{equation}
in the north polar chart, and:
\begin{equation}
e=\frac{1}{8}(m^{\prime-1})^{AC}(m^{\prime-1})^{BD}\frac{\partial
m^\prime_{AB}}{\partial\ub}\frac{\partial
m^\prime_{CD}}{\partial\ub} \label{17.50}
\end{equation}
in the south polar chart. Then by \ref{2.33}, \ref{2.34},
\ref{2.41}, and the fact that $\exp$ is an analytic map with
derivative at the origin equal to $I$, we have:
\begin{equation}
\left|\delta |u_0|^2
e(\ub,u_0,\vartheta)-\frac{1}{8}\left|\frac{\partial\psi_0}{\partial
s}\left(\frac{\ub}{\delta},\vartheta\right)\right|^2\right| \leq
O(\delta^{1/2}|u_0|^{-1}) \label{17.51}
\end{equation}
in the north polar chart, and:
\begin{equation}
\left|\delta |u_0|^2
e(\ub,u_0,\vartheta)-\frac{1}{8}\left|\frac{\partial\psi^\prime_0}{\partial
s}\left(\frac{\ub}{\delta},\vartheta\right)\right|^2\right| \leq
O(\delta^{1/2}|u_0|^{-1}) \label{17.52}
\end{equation}
in the south polar chart. It follows that, given any constant
$k>1$,
\begin{equation}
\frac{1}{8}\int_0^1\left|\frac{\partial\psi_0}{\partial
s}(s,\vartheta_0)\right|^2ds\geq k \ \ \mbox{for all
$\vartheta_0\in \oD_{2\rho}$} \label{17.53}
\end{equation}
and:
\begin{equation}
\frac{1}{8}\int_0^1\left|\frac{\partial\psi^\prime_0}{\partial
s}(s,\vartheta_0)\right|^2ds\geq k \ \ \mbox{for all
$\vartheta_0\in \oD_{2\rho}$} \label{17.54}
\end{equation}
(here $|\s|$ denotes the Euclidean norm on $\Re^2$, as in
\ref{2.42}) then if also $\delta^{1/2}|u_0|^{-1}$ is suitably
small depending on $M_1$, the condition \ref{17.47} is satisfied
with $1+(1/2)(k-1)>1$ in the role of $k$.

We have thus established the following theorem.

\vspace{5mm}

\noindent{\bf Theorem 17.1} \ \ \ Let $2\geq k>1$. With initial
data on $C_{u_0}$ as in Theorem 16.1 there is a non-negative
non-decreasing continous function $F$ on the non-negative real
line such that if:
$$\delta^{1/2} F(M_8)< k-1$$
and one of the following two conditions holds:
$$1.\hspace{10mm}|u_0|^2\int_0^\delta e(\ub,u_0,\vartheta)d\ub\geq k \ \ \mbox{: for all $\vartheta_0\in S^2$}$$

\noindent or:

$$2.\hspace{10mm}\frac{1}{8}\int_0^1\left|\frac{\partial\psi_0}{\partial s}(s,\vartheta)\right|^2ds\geq k \ \ \mbox{for all $\vartheta\in \oD_{2\rho}$}$$
and:
$$\hspace{12mm}\frac{1}{8}\int_0^1\left|\frac{\partial\psi^\prime_0}{\partial s}(s,\vartheta)\right|^2ds\geq k \ \ \mbox{for all $\vartheta\in \oD_{2\rho}$}$$

\noindent then there is a $\ub^*\in(0,\delta)$ such that for all
$\ub\in(\ub^*,\delta)$ the surfaces $S_{\ub,-1-\delta}\subset
C_{-1-\delta}\subset M^\prime_{-1}$ are {\em trapped spheres}.
Moreover, their areas satisfy:
$$|\mbox{Area}(S_{\ub,-1-\delta})-4\pi|\leq O(\delta)$$

\vspace{5mm}

Now the initial data on $C_{u_0}$ can be extended to regular
initial data on a {\em complete} cone $C_o$. This can be done, for
example, in the following way. We recall that the  mappings
$\psi_0$ and $\psi^\prime_0$ extend smoothly by zero to $s<0$. We
may then apply a total extension operator $E$ for $(-\infty,1]$ to
$\psi_0(\cdot,\vartheta), \psi^\prime_0(\cdot,\vartheta) \ : \
\vartheta\in \oD_{2\rho}$, obtaining smooth mappings $E\psi_0$,
$E\psi^\prime_0$ of $\Re\times\oD_{2\rho}$ into $\hat{S}$, with
support in  $[0,2]\times\oD_{2\rho}$, satisfying the
transformation rule \ref{2.38}, \ref{2.39}. This gives
asymptotically flat data along $C_o$. Let $M^*$ be the unique
maximal future development of the data on $C_o$. Then since the
data on the part of $C_o$ which corresponds to $\ub\leq \delta$
coincides with the data we have been considering, the development
$M_{-1}$ is a subdevelopment of $M^*$. Therefore $M^*$ contains
trapped spheres. We may then apply the theorem of Penrose, with
$C_o$ in the role of the non-compact Cauchy hypersurface in that
theorem (see Prologue), to conclude the following.

\vspace{5mm}

\noindent{Theorem 17.2} \ \ \ Consider smooth regular initial data
on a complete cone $C_o$ which extends initial data satisfying the
conditions of Theorem 17.1. Then the maximal future development of
the initial data on $C_o$ is future null geodesically incomplete.

\vspace{5mm}

We conclude the present monograph with the following remarks. With
a fixed pair $(\psi_0,\psi^\prime_0)$ of smooth mappings of
$[0,1]\times\oD_{2\rho}$ into $\hat{S}$ as in Chapter 2, and fixed
$\delta$ satisfying the smallness conditions of Theorems 16.1 and
17.1 relative to $M_8$, we may consider the sequence of
characteristic initial value problems corresponding to a
decreasing sequence $u_{0,n}\rightarrow-\infty$. The fact that we
have obtained bounds independent of $u_0$ allows us to extract a
subsequence such that the corresponding sequence of solutions
converges in the future of $C_{c}$ for any $c<-1$. We may
construct in this way a solution such that the generators of the
$\Cb_{\ub}$ for $\ub\in[0,\delta)$ are complete toward the past,
and the interior of $\Cb_0$ is  the Minkowskian region which is
the complete past of the point $e$ (see first section of Chapter
1). For this solution we have:
\begin{equation}
\phi(\ub,u,\vartheta)\rightarrow 1 \ \ \mbox{: as
$u\rightarrow-\infty$} \label{17.55}
\end{equation}
and:
\begin{equation}
|u|^{-1}\chih_{AB}(\ub,u,\vartheta)\rightarrow
\frac{w^2(\vartheta)}{2\delta^{1/2}}\left(\frac{\partial(\psi_0)_{AB}}{\partial
s}\right) \left(\frac{\ub}{\delta},\vartheta)\right) \ \ \mbox{:
as $u\rightarrow-\infty$} \label{17.56}
\end{equation}
in the north polar chart, and similarly in the south polar chart,
uniformly in $[0,\delta]\times\oD_{2\rho}$. Moreover, it can be
shown, using the method of Friedrich [Fr], that this solution is the unique solution of this type
satisfying \ref{17.56} and its analogue, for a given pair
$(\psi_0,\psi^\prime_0)$. We thus have a unique solution of the
asymptotic characteristic initial value problem with initial data
at {\em past null infinity} determined by the pair
$(\psi_0,\psi^\prime_0)$. The limit \ref{17.56} may be thought of
as the {\em incoming radiative amplitude}. It is a 2-covariant
symmetric tensorfield on $S^2$ (see \ref{2.27}), depending on
$\ub$, which is trace-free relative to the standard metric
$\up{\sg}$.

Furthermore, if the data on $C_{u_{0,n}}$ are extended to regular
asymptotically flat data on a complete cone in the manner
described in the paragraph preceding Theorem 17.2, we can show
that if $|u_{0,n}|$ is sufficiently large the maximal development
contains complete cones $C_u$ for all $u\in[u_{0,n},c]$, for some $c<0$ with $|c|$ 
suitably large. It can then be shown that the corresponding
limiting solution with initial data at past null infinity contains
complete cones $C_u$ for all $u\leq c$. The solution then contains a piece of future null infinity 
which is a neighborhood of spacelike infinity.

As a consequence of \ref{17.56} we have:
\begin{equation}
|u|^2 e(\ub,u,\vartheta)\rightarrow e_\infty(\ub,\vartheta) \ \
\mbox{as $u\rightarrow-\infty$} \label{17.57}
\end{equation}
where:
\begin{equation}
e_\infty(\ub,\vartheta)=\frac{1}{8\delta}\left|\frac{\partial\psi_0}{\partial
s}\left(\frac{\ub}{\delta},\vartheta\right)\right|^2 \label{17.58}
\end{equation}
in the north polar chart, and similarly in the south polar chart,
uniformly in $[0,\delta]\times\oD_{2\rho}$. Note here that since
 $O$ is an orthogonal transformation (see
\ref{2.b3}), the transformation rule \ref{2.38} implies that for
all $(s,\vartheta)\in [0,1]\times\oA_{\rho}$:
\begin{equation}
\left|\frac{\partial\psi_0}{\partial
s}\left(s,\vartheta\right)\right|^2=
\left|\frac{\partial\psi^\prime_0}{\partial
s}\left(s,\vartheta^\prime\right)\right|^2, \ \
\vartheta^\prime=f(\vartheta) \label{17.59}
\end{equation}
hence if $e_\infty(\ub,\vartheta)$ and
$e_\infty^\prime(\ub,\vartheta^\prime)$ are the functions
corresponding to the north polar and south polar charts
respectively, then we have
\begin{equation}
e_\infty(\ub,\vartheta)=e^\prime_\infty(\ub,\vartheta^\prime), \ \
\vartheta^\prime=f(\vartheta) \ \ : \ \ \forall (s,\vartheta)\in
[0,\delta]\times\oA_{\rho} \label{17.60}
\end{equation}
as we should, $e_\infty$ and $e^\prime_\infty$ being
representations of the same function on $[0,1]\times S^2$.

The function $e_\infty$, as a function on $[0,\delta]\times S^2$,
has a simple physical meaning: it is $8\pi$ times the {\em
incoming radiative power per unit solid angle}, a function of the
{\em advanced time} $\ub$ and the direction, the last being
represented by a point on $S^2$. (This statement holds in units
where Newton's graviational constant is set equal to 1). Thus the
limiting form of both conditions of Theorem 17.1, namely:
\begin{equation}
\int_0^\delta e_\infty(\ub,\vartheta)d\ub\geq k \ \ \mbox{: for
all $\vartheta\in S^2$} \label{17.61}
\end{equation}
means simply that {\em the incoming energy per unit solid angle in each direction in
the advanced time interval $[0,\delta]$ is not less than
$k/8\pi$}. This corresponds to a total incoming energy ${\cal E}$
in the advanced time interval $[0,\delta]$ of at least $k/2$. The
``area radius" of $S_{\ub,u}$ being defined by
$r(S_{\ub,u})=\sqrt{\mbox{Area}(S_{\ub,u})/4\pi}$, according to
Theorem 17.1 the area radius of the trapped spheres which form
satisfies $|r-1|\leq O(\delta)$. Therefore, in the limit
$k\rightarrow 1$, $\delta\rightarrow 0$, we recover the inequality
$2{\cal E}/r\geq 1$, familiar from the spherically symmetric case.
However in the general case the requirement for the formation of a
trapped sphere is not on the total incoming energy but rather on
the incoming energy in each direction.

If the above condition is satisfied for some, but not all,
$\vartheta\in S^2$, we can still conclude that some generator of
$C_{-1-\delta}$ has a segment on which $\mbox{tr}\chi^\prime<0$.
It then follows that if that generator is extended in the maximal
development then either there is a value $s^*$ of the affine
parameter and a (non-zero) Jacobi field $X$ along the generator in
question such that $|X(s)|\rightarrow 0$ as $s\rightarrow s^*$, or
the generator does not extend to the interval $[0,s^*)$ of the
affine parameter, in which case the maximal development is future
null geodesically incomplete anyway. If the first alternative
holds, then either the generator extends to $s^*$ itself and the
end point is a point conjugate to $o$ along the generator in
question, or the generator does not extend to $s^*$, the end point
being a singular point. The latter would necessarily be the case
if, for example, we can show that $\alpha(X(s),X(s))$ does not
tend to zero as $s\rightarrow s^*$. But this cannot be shown
without analyzing the solution in the maximal development beyond
$M_{-1}$. And off course the nature of the future ``boundary" of
the maximal development, when incompletess holds, remains an open
question.

\pagebreak

\section*{\huge{Bibliography}}

\vspace{10mm}

\noindent [Be] \ \ Bel, L. ``Introduction d'un tenseur du quatri\`{e}me ordre", {\em C. R. Acad. Sci. Paris} {\bf 248}, 1297 - 1300 (1959). 

\noindent [Bie] \ \ Bieri, L. {\em An Extension of the Stability Theorem of the Minkowski Space in General Relativity}, Ph. D. thesis, Mathematics Department, ETH Zurich, 2007. 

\noindent [Bir] \ \ Birkoff, G. D. {\em Relativity and Modern Physics}, Harvard University Press, Cambridge Mass., 1923. 

\noindent [B-L] \ \ Boyer, R. H. and Lindquist, R. W. ``Maximal analytic extension of the Kerr metric", J. Math. Phys. {\bf 8}, 265 - 281 (1967). 

\noindent [Cho1] \ \ Choquet-Bruhat, Y. ``Theorem d'existence pour certain systems d'equations aux derive\'{e}s partielles nonlineaires", {\em Acta Mathematica} {\bf 88},
141 - 225 (1952).

\noindent [Cho2] \ \ Choquet-Bruhat, Y. ``The Cauchy problem", in {\em Gravitation: An Introduction to Current Research}, L. Witten (editor), Wiley, New York, 1962.

\noindent [Cho3] \ \ Choquet-Bruhat, Y. ``Solutions $C^\infty$ d'equations hyperboliques non lin\'{e}aires", {\em C. R. Acad. Sci. Paris} {\bf 272}, 386 - 388 (1968).

\noindent [Cho4] \ \ Choquet-Bruhat, Y. ``Probleme des condition initiales sur un conoide characteristique", {\em C. R. Acad. Sci. Paris}, 3971 - 3973 (1963). 

\noindent [C-G] \ \ Choquet-Bruhat, Y. and Geroch, R. P. ``Global aspects of the Cauchy problem in in general relativity", {\em Commun. Math. Phys.} {\bf 14}, 329 - 335 (1969). 

\noindent [Chr1] \ \ Christodoulou, D. ``Violation of cosmic censorship in the gravitational collapse of a dust cloud", {\em Commun. Math. Phys.} {\bf 93}, 171 -195 (1984). 

\noindent [Chr2] \ \ Christodoulou, D. ``The problem of a self-gravitating scalar field", {\em Commun. Math. Phys.} {\bf 105}, 337 - 361 (1986). 

\noindent [Chr3] \ \ Christodoulou, D. {\em Investigations in Gravitational Collapse and the Physics of Black Holes}, Ph. D. thesis, Physics Department, Princeton University, 1971. 

\noindent [Chr4] \ \ Christodoulou, D. ``Global existence of generalized solutions of the spherically symmetric Einstein-scalar equations in the large", 
{\em Commun. Math. Phys.} {\bf 106}, 587 - 621 (1986). 

\noindent [Chr5] \ \ Christodoulou, D. ``A mathematical theory of gravitational collapse", {\em Commun. Math. Phys.} {\bf 109}, 613 - 647 (1987). 

\noindent [Chr6] \ \ Christodoulou, D. ``The formation of black holes and singularities in spherically symmetric gravitational collapse", {\em Comm. Pure Appl. Math.} 
{\bf 44}, 339 -373 (1991).

\noindent [Chr7] \ \ Christodoulou, D. ``Bounded variation solutions of the spherically symmetric Einstein-scalar field equations", {\em Comm. Pure Appl. Math.} 
{\bf 46}, 1131 - 1220 (1993). 

\noindent [Chr8] \ \ Christodoulou, D. ``Examples of naked singularity formation in the gravitational collapse of a scalar field", {\em Ann. of Math.} {\bf 140}, 
607 - 653 (1994).

\noindent [Chr9] \ \ Christodoulou, D. ``The instability of naked singularities in the gravitational collapse of a scalar field", {\em Ann. of Math.} {\bf 149}, 
183 - 217 (1999). 

\noindent [Chr10] \ \ Christodoulou, D. ``On the global initial value problem and the issue of singularities", {\em Class. Quantum Grav.} {\bf 16} A, 23 - 35 (1999). 

\noindent [Chr11] \ \ Christodoulou, D. ``Self-gravitating relativistic fluids: a two phase model", {\em Arch. Rational Mech. Anal.} {\bf 130}, 343 - 400 (1995).

\noindent [Chr12] \ \ Christodoulou, D. ``Self-gravitating relativistic fluids: the continuation and termination of a free phase boundary", {\em Arch. Rational Mech. Anal.} {\bf 133}, 333 - 398 (1996).

\noindent [Chr13] \ \ Christodoulou, D. ``Self-gravitating relativistic fluids: the formation of a free phase boundary in the phase transition from soft to hard", 
{\em Arch. Rational Mech. Anal.} {\bf 134}, 97 - 154 (1996).  

\noindent [Chr14] \ \ Christodoulou, D. {\em The Formation of Shocks in 3-Dimensional Fluids}, EMS Monographs in Mathematics, EMS Publishing House, 2007. 

\noindent [Chr15] \ \ Christodoulou, D. ``Nonlinear nature of gravitation and gravitational wave experiments", {\em Phys. Rev. Lett.} {\bf 67}, 1486 - 1489 (1991). 

\noindent [Chr16] \ \ Christodoulou, D. ``Notes on the geometry of null hypersurfaces", U.S. Copyright Office, Library of Congress, Registration Number TXu 832-728, 
\copyright 1991, Demetrios Christodoulou, All Rights Reserved.

\noindent [Chr17] \ \ Christodoulou, D. {\em The Action Principle and Partial Differential Equations}, Ann. Math. Stud. {\bf 146}, Princeton University Press, 2000.

\noindent [C-K] \ \ Christodoulou, D. and Klainerman, S. {\em The Global Nonlinear Stability of the Minkowski Space}, Princeton Mathematical Series {\bf 41}, 
Princeton University Press, 1993. 

\noindent [Co] \ \ Constantin, P. ``On the Euler equations of incompressible fluids", {Bull. Amer. Math. Soc.} {\bf 44}, 603 - 621 (2007). 

\noindent [D1] \ \ Dafermos, M. ``Stability and instability of the Cauchy horizon for the spherically symmetric Einstein-Maxwell-scalar field equations", 
{\em Ann. Math.} {\bf 158}, 875 - 928 (2003).

\noindent [D2] \ \ Dafermos, M. ``The interior of charged black holes and the problem of uniqueness in general relativity", {\em Comm. Pure Appl. Math.} {\bf 58}, 
445 - 504 (2005). 

\noindent [Ed] \ \ Eddington, A. S. ``A comparison of Whitehead's and Einstein's formulas", {\em Nature} {\bf 113}, 192 (1924).

\noindent [Ei1] \ \ Einstein, A. ``Z\"{u}r algemeinen Relativit\"{a}tstheorie", {\em Preuss. Akad. Wiss. Berlin, Sitzber.} {\bf 44}, 778 - 786; 799-801 (1915).

\noindent [Ei2] \ \ Einstein, A. ``Die Feldgleichungen der Gravitation",{\em Preuss. Akad. Wiss. Berlin, Sitzber.} {\bf 50}, 844 - 847 (1915).

\noindent [Fi] \ \ Finkelstein, D. ``Past-future asymmetry of the gravitational field of a point particle", {\em Phys. Rev.} {\bf 110}, 965 - 967 (1958). 

\noindent [Fr] \ \ Friedrich, H. ``The asymptotic characteristic initial value problem for Einstein's vacuum field equations as an initial value probelm for 
a first order quasilinear symmetric hyperbolic system", {\em Proc. Roy. Soc. Lond. A} {\bf 378}, 401 - 421 (1981). 

\noindent [Ge] \ \ Geroch, R. P. ``The domain of dependence", {\em J. Math. Phys.} {\bf 11}, 437 - 439 (1970). 

\noindent [G-K-P] \ \ R. P. Geroch, E. H. Kronheimer, and R. Penrose ``Ideal points in spacetime", {\em Proc. Roy. Soc. Lond. A} {\bf 327}, 545 - 567 (1972). 

\noindent [G-T] \ \ Gilbarg, D. and Trudinger, N. S. {\em Elliptic Partial Differential Equations of Second Order}, Springer-Verlag, 1977. 

\noindent [H-E] \ \ Hawking, S. W. and Ellis, G. F. R. {\em The Large Scale Structure of Spacetime}, Cambridge Monographs on Mathematical Physics, 
Cambridge University Press, Cambridge, 1973. 

\noindent [Ke] \ \ Kerr, R. P. ``Gravitational field of a spinning mass as an example of algebraically special metrics", {\em Phys. Rev. Lett.} {\bf 11}, 237 - 238 (1963). 

\noindent [Kr] \ \ Kruskal, M. D. ``Maximal extension of the Schwarzschild metric", {\em Phys. Rev.} {\bf 119}, 1743-1745 (1960). 

\noindent [K-N] \ \ Klainerman, S. and Nicol\`{o}, F. {\em The Evolution Problem in General Relativity}, Progress in Math. Phys. {\bf 25}, Birkh\"{a}user, Boston, 2003.

\noindent [L] \ \ Lema\^{i}tre, G.  ``L'univers en expansion", {\em Ann. Soc. Sci. Bruxelles I} {\bf A53}, 51 - 85 (1933). 

\noindent [M] \ \ M\"{u}ller zum Hagen, H. ``Characteristic initial value problem for hyperbolic systems of second order differential equations", 
{\em Ann. Inst. Henri Poincar\'{e}, Phys. Theor.} {\bf 53}, 159 - 216 (1990). 

\noindent [M-S] \ \ M\"{u}ller zum Hagen, H. and Seifert, H. J. ``On Characteristic initial value and mixed problems", {\em Gen, Rel. Grav.} {\bf 8}, 259 - 301 (1977)

\noindent [M-T-W] \ \ Misner, C. W., Thorne, K.S., and Wheeler, J. A. {\em Gravitation}, W. H. Freeman \& Company, San Franscisco, 1973. 

\noindent [O] \ \ Osserman, R. ``The isoperimetric inequality", {\em Bull. Amer. Math. Soc.} {\bf 84}, 1182 - 1238 (1978). 

\noindent [O-S] \ \ Oppenheimer, J. R. and Snyder, H. ``On continued gravitational contraction", {\em Phys. Rev.} {\bf 56}, 455 - 459 (1939). 

\noindent [P1] \ \ Penrose, R. ``Conformal treatment of infinity", in {\em Relativity, Groups and Topology}, C. M. de Witt and B. de Witt (editors), 
Les Houches Summer School 1963, Gordon and Breach, New York, 1964. 

\noindent [P2] \ \ Penrose, R. ``Gravitational collapse and space-time singularities", {\em Phys. Rev. Lett.} {\bf 14}, 57 - 59 (1965). 

\noindent [P3] \ \ Penrose, R. ``Structure of spacetime", in {\em Battelle Rencontres}, C. M. de Witt and J. A. Wheeler (editors), Benjamin, New York, 1968.

\noindent [P4] \ \ Penrose, R.  {\em Techniques of Differential Topology in Relativity}, Regional Conference Series in Applied mathematics, SIAM, 1972.

\noindent [P5] \ \ Penrose, R. ``Gravitational collapse: the role of general relativity" {\em Noovo Cimento} {\bf 1}, 252 - 276 (1969). 

\noindent [P6] \ \ Penrose, R. ``Singularities and time-asymmetry", in {\em General Relativity, an Einstein Centenary Survey}, S. W. Hawking and W. Israel (editors), 
Cambridge University Press, Cambridge, 1979. 

\noindent [R] \ \ Rendall, A. D. ``Reduction of the characteristic initial value problem to the Cauchy problem and its applications to the Einstein equations", 
{\em Proc. Roy. Soc. Lond. A} {\bf 427}, 221 - 239 (1990). 

\noindent [Sc] \ \ Schwarzschild, K. ``\"{U}ber das Gravitationsfeld eines massenpunktes nack der Einsteinschen Theorie", {\em Sitz. Deut. Akad. Wiss. Berlin, 
Kl. Math. - Phys. Tech.}, 189 - 196 (1916). 

\noindent [Sy] \ \ Synge, J. L. ``The gravitational field of a particle", {\em Proc. R. Irish Acad. A} {\bf 53}, 83 - 114 (1950). 

\noindent [Sz] \ \ Szekeres, G. ``On the singularities of a Riemannian manifold", {\em Publ. Mat. Debrecen} {\bf 7}, 285 - 301 (1960).

\noindent [S-Y1] \ \ Schoen, R. and Yau, S. T. ``The existence of a black hole due to condensation of matter", {\em Commun. Math. Phys.} {\bf 90}, 575 - 579 (1983). 

\noindent [S-Y2] \ \ Schoen, R. and Yau, S. T. ``Proof of the positive mass theorem II", {\em Commun. Math. Phys.} {\bf 79}, 231 - 260 (1981). 

\noindent [S-Y3] \ \ Schoen, R. and Yau, S. T. ``On the proof of the positive mass conjecture in general relativity", {\em Commun. Math. Phys.} {\bf 65}, 45 - 76 
(1979). 

\noindent [T] \ \ Tolman, R. C. ``Effect of inhomogeneity on cosmologicak models" {\em Proc. Nat. Acad. Sci. U. S. } {\bf 20}, 169 - 176 (1934). 

\noindent [Wh] \ \ Wheeler, J. A. {\em Geons, Black Holes \& Quantum Foam}, W. W. Norton \& Company, New York, London, 1998. 

\noindent [Wi] \ \ Witten, E. ``A new proof of the positive energy theorem" {\em Commun. Math. Phys.} {\bf 80}, 381 - 402 (1981). 

\noindent [Z] \ \ Zipser, N. {\em The Global Nonlinear Stability of the Trivial Solution of the Einstein-Maxwell Equations}, Ph. D. thesis, Harvard University, 2000.

\end{document}